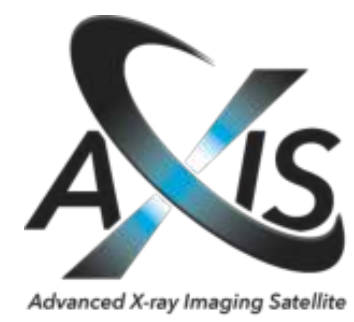

# AXIS Community Science Book

*A Vision for High-Resolution X-ray Astronomy in the 2030s*

Edited by Michael Koss, Nico Cappelluti, Brad Cenko, Lia Corrales, Adi Foord, Daryl Haggard, Helen Russell, Samar Safi-Harb, and the AXIS Science Team

## a. Introduction

We are excited to present the *AXIS Community Science Book*, a compendium of science cases for the **Advanced X-ray Imaging Satellite (AXIS)** mission concept. This volume represents the culmination of a community-driven effort to envision transformative science enabled by AXIS. Over the past year, we have worked together through regular online meetings and both online and in-person workshops to develop these science cases. The process was truly collaborative, from virtual brainstorming sessions to the **AXIS Community Science Conference** held on May 14–16, 2025, in Annapolis, Maryland, ensuring that a broad range of ideas and voices shaped the mission's scientific vision.

### Community Collaboration and Science Case Development

The science case development for AXIS has been a grassroots effort that has engaged scientists worldwide. **A total of 592 scientists from 27 countries and 31 U.S. states contributed to this Science Book**, reflecting a broad geographical reach and diverse expertise. Notably, **35% of the contributors were first-time participants in a mission study**, and approximately **10% were students** (advanced undergraduate or master's level), bringing fresh perspectives.

Crucially, AXIS is conceived as a **community-driven observatory** in its operations. More than 70% of the observing time on AXIS will be allocated to the community via a competitive Guest Observer (GO) program [1]. In practice, this amounts to **over $10^8$ seconds (100 Ms) of open observing time** over the prime mission, available for astronomers worldwide to pursue their ideas. By supporting such a large GO program, AXIS will empower a wide range of investigators, from seasoned X-ray astronomers to those new to the field, to advance their science and make new discoveries. This philosophy of broad community access, together with the collaborative development of the science case, makes AXIS **a mission built by the community and for the community**.

**AXIS Science Book Overview**

The result of this collective effort is a comprehensive science document detailing how AXIS will address many of the most compelling questions in astrophysics. The Science Book contains **138 detailed science use cases**, each describing a unique investigation enabled by AXIS. These use cases were distilled from over 250 initial science ideas submitted by the community, which is a clear testament to the excitement surrounding AXIS.

The Science Book spans 595 pages and includes 225 figures and over **2,300 bibliographic references**, reflecting the depth of analysis and scholarship behind each scientific case study. The content is organized into **five major chapters**, corresponding to the broad astrophysical themes of our science working groups:

1. **Galaxies and Galaxy Clusters**: Star formation, feedback, and the cosmic baryon cycle.
2. **Active Galactic Nuclei (AGN)**: Supermassive black hole growth, accretion, and feedback across cosmic time.
3. **Time-Domain and Multimessenger Astronomy**: Transient X-ray phenomena and gravitational wave or neutrino source follow-up.
4. **Compact Objects and Supernova Remnants**: Neutron stars, black holes, White Dwarfs, their environments, and extreme physics.
5. **Stars and Exoplanets**: Stellar coronae, flares, and planetary X-ray environments.

| Science Working Group | Leads | Submitted Ideas | Final Cases | Figures | Pages |
|---|---|---|---|---|---|
| Galaxies and Clusters | Helen Russell & Laura Lopez | 98 | 35 | 57 | 150 |
| AGN and SMBH Evolution | Nico Cappelluti & Adi Foord | 64 | 35 | 45 | 154 |
| TDAMM | Brad Cenko & Daryl Haggard | 70 | 17 | 40 | 68 |
| Compact Objects and SNRs | Samar Safi-Harb & Kevin Burdge | 43 | 37 | 64 | 160 |
| Stars and Exoplanets | Lia Corrales & Keivan Stassun | 33 | 14 | 20 | 55 |
| **Total** | | **308** | **138** | **226** | **587** |

**Table 1.** Science Working Groups, their leads, and contributions to the *AXIS Community Science Book*. All groups met weekly or biweekly from November 2024 to May 2025 to coordinate writing, figures, and cross-group integration. The columns list the number of submitted ideas, finalized science cases, total figures, and total pages contributed by each group. Totals reflect the combined effort across all groups.

These studies demonstrate strong synergy with multiwavelength and multimessenger facilities (optical, NIR, MIR, submillimeter, radio, gravitational waves, and neutrinos), citing more than 70 ground- and space-based observatories. They include hundreds of mentions of critical science with US facilities (JWST, Roman, Rubin, ngVLA, ELTs, CMB-S4, IceCube-Gen2), leveraging both high-resolution imaging and time domain surveys. Collectively, the studies outlined in the GO Science Book amount to nearly 100 Ms of observing time, which is comparable to the GO time in a full five-year mission with 95% observing efficiency.

**Building on the legacy of past X-ray missions**

AXIS builds directly on the legacy of **Chandra, XMM-Newton, and Swift** while moving far beyond their current capabilities. Chandra established the power of sub-arcsecond X-ray imaging; however, its small collecting area and the gradual loss of sensitivity below 1 keV due to contamination now limit its efficiency. XMM-Newton has provided excellent throughput and spectroscopy, but cannot match Chandra's sharp imaging. Swift has been the workhorse for rapid response to targets of opportunity, but it has a modest effective area. AXIS is designed to enhance and go beyond the best aspects of these observatories into a single mission, delivering a transformative leap for X-ray astronomy and an X-ray mission needed for multiwavelength, multimessenger studies in the 2030s and beyond.

The observatory will combine a sharp point-spread function of $< 2''$ across a wide $24'$ field of view with a collecting area that is 5–12 times greater than *Chandra*'s over the 0.3–10 keV band at launch [1]. Through careful contamination control, *AXIS* will maintain its full soft X-ray sensitivity, enabling efficient studies of faint sources and diffuse emission. This combination of high angular resolution and throughput translates into survey speeds roughly 50–100 times faster than *Chandra*'s for equivalent point-source sensitivities. These gains arise from both the larger collecting area and the ability to cover large sky regions with far fewer pointings, thereby reducing total observing time. For example, the entire *Chandra* Cycle 27 program could, in principle, be achieved in less than a week of *AXIS* observations, given the increased survey speed and sensitivity. More importantly, *AXIS* will not merely reproduce *Chandra* programs more efficiently but will also expand the discovery space to new populations of faint and diffuse X-ray sources.

AXIS is also designed to be highly responsive and agile, with the ability to slew to a new target in minutes and begin scientific observations within about one hour. This responsiveness makes it at least 50 times faster than Chandra for targets of opportunity, opening a new window for time-domain and multimessenger astrophysics. Together, these advances in throughput, sensitivity, and responsiveness ensure that AXIS will provide the sharpest, deepest, and most efficient X-ray view of the Universe in the 2030s.

**Synergies with Athena**

The European Space Agency's Athena (Advanced Telescope for High-ENergy Astrophysics) mission is currently studying an X-ray observatory planned for the late 2030s, optimized to address the "Hot and Energetic Universe." Equipped with two powerful instruments, the Wide Field Imager (WFI) and the X-ray Integral Field Unit (X-IFU), Athena will combine a large collecting area, high spectral resolution, and survey efficiency over wide sky regions. The mission is currently in its study phase, with final adoption expected in mid-2027, a launch planned for the late 2030s ($\approx$2037), followed by at least 5 years of operations.

If both are launched on similar timescales, Athena and AXIS will be deeply complementary in both capability and scientific scope. The WFI will provide imaging spectroscopy over a $40'$ FOV with $9''$ HPD while the X-IFU will have a $4'$ FOV. Athena's X-IFU will deliver non-dispersive spectroscopy with a spectral resolution of 4.0eV, enabling precise measurements of plasma temperatures, turbulence, and metal enrichment in bright sources. These data will provide spectral context for AXIS's high-resolution imaging, which will resolve these systems on smaller scales. For example, Athena will measure the velocity structure and ionization states of AGN outflows, cluster cores, and supernova remnants, while AXIS will map the morphology and spatial coupling of those outflows with their environments. Similarly, Athena's wide-field WFI surveys will benefit from AXIS's high spatial resolution for extended or confused sources, enabling efficient identification of dual AGN, groups, and clusters emission separate from AGN across large sky areas.

Together, AXIS and Athena could provide the most complete picture of the hot and dynamic universe. Athena's spectral precision will reveal the energetics of cosmic feedback, while AXIS's imaging sharpness will show its spatial structure and evolution. Their combined datasets will transform our understanding of how black holes, stars, and galaxies exchange energy and matter, creating a unified, multi-scale view of high-energy astrophysics in the late 2030s and beyond.

**Outlook**

In summary, the AXIS Science Book showcases ambitious scientific goals enabled by AXIS's unprecedented capabilities. Developed by a vibrant community of over 500 scientists, it demonstrates how AXIS will carry forward the legacy of past missions and open new discovery space. AXIS builds on the legacy of **Chandra, XMM-Newton, and Swift** while aligning with the astrophysics landscape of the 2030s.

It will operate alongside **JWST**, **Rubin**, **Roman**, **ngVLA**, **SKA**, **ELTs**, and **LISA**, and the next generation of gravitational wave and neutrino observatories, maximizing synergies across the full spectrum and messenger domains. Looking ahead, **we envision AXIS as a cornerstone of 2030s astrophysics**, one that will **carry forward the legacy of Chandra, XMM-Newton, and Swift** while **opening new frontiers** with coordinated multiwavelength campaigns alongside the newest observatories of the coming decade.

— *The AXIS Science Team (2025)*

1. Reynolds, C. S., Kara, E. A., Mushotzky, R. F., et al. 2023, in Society of Photo-Optical Instrumentation Engineers (SPIE) Conference Series, Vol. 12678, UV, X-Ray, and Gamma-Ray Space Instrumentation for Astronomy XXIII, ed. O. H. Siegmund & K. Hoadley, 126781E

**US AXIS Scientists**

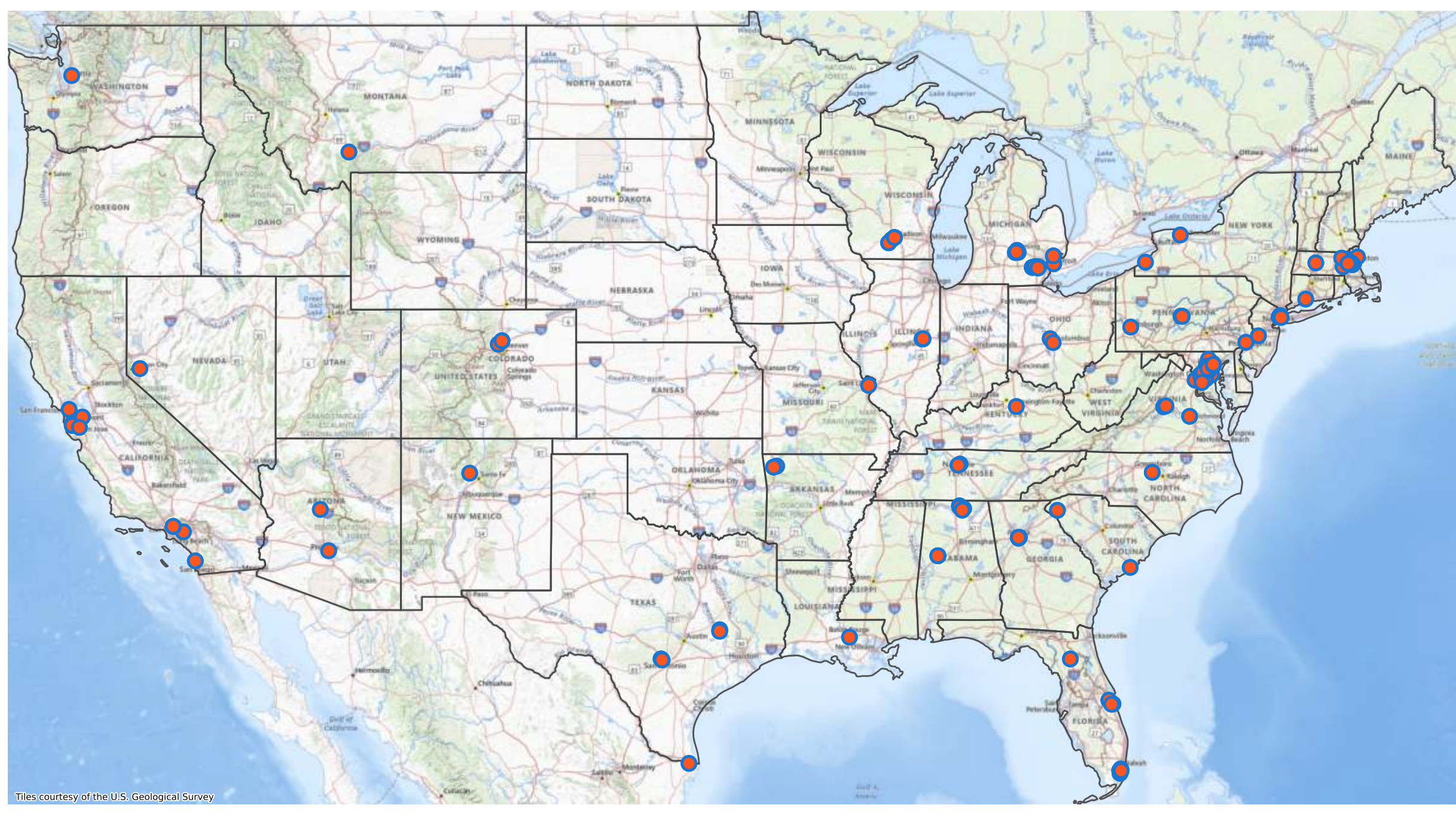

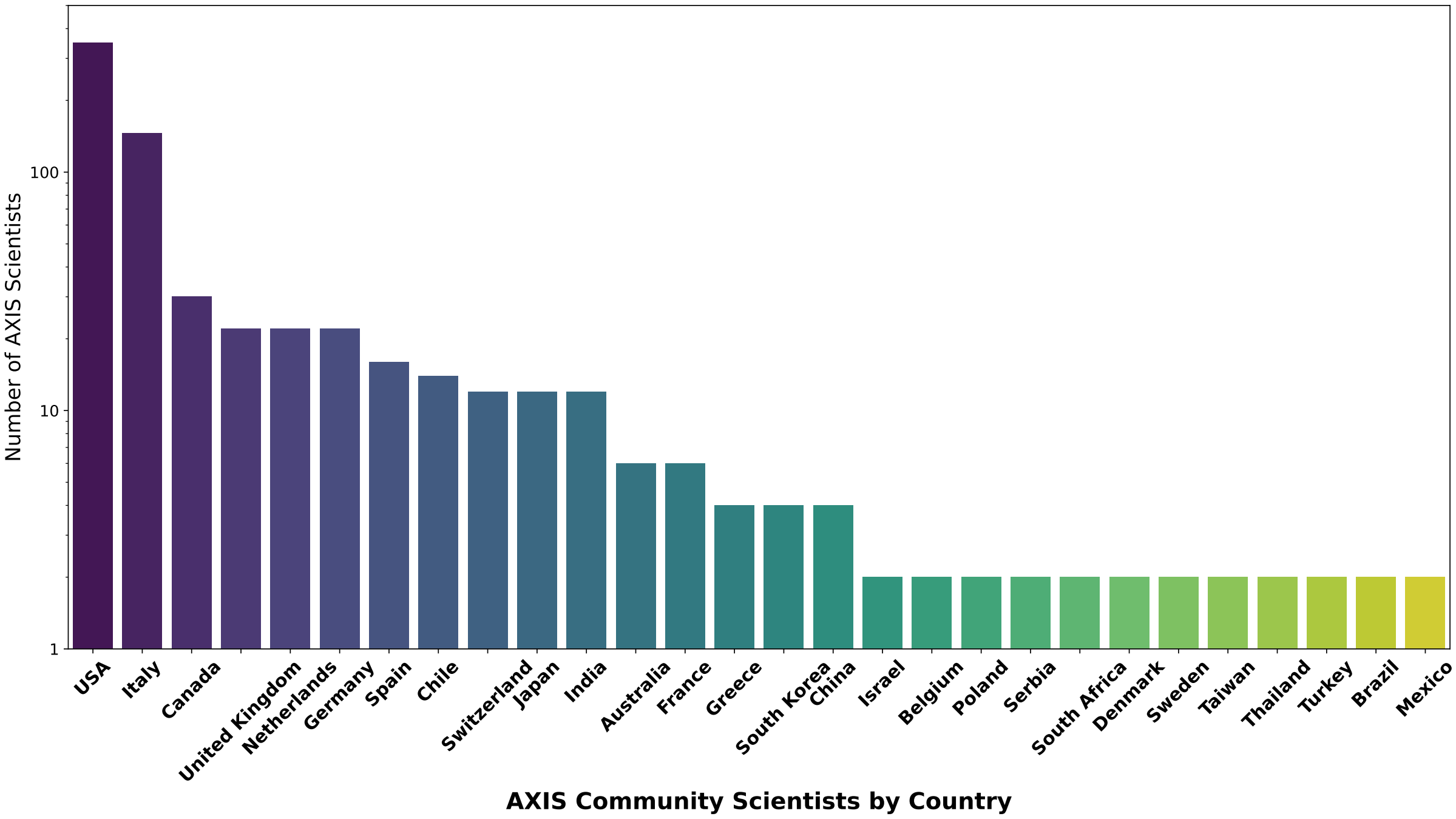

**Figure 1. AXIS enjoys broad community support across the United States and internationally. Top:** Geographic distribution of US-based AXIS Science Book contributors, representing institutions across 28 states and demonstrating widespread engagement from coast to coast. **Bottom:** International participation in AXIS, with over 280 scientists from the US and substantial contributions from Italy, Canada, the United Kingdom, Germany, Spain, and 13 additional countries. This broad, geographically distributed community reflects a strong enthusiasm for AXIS science, positioning the mission to leverage expertise from leading institutions worldwide.

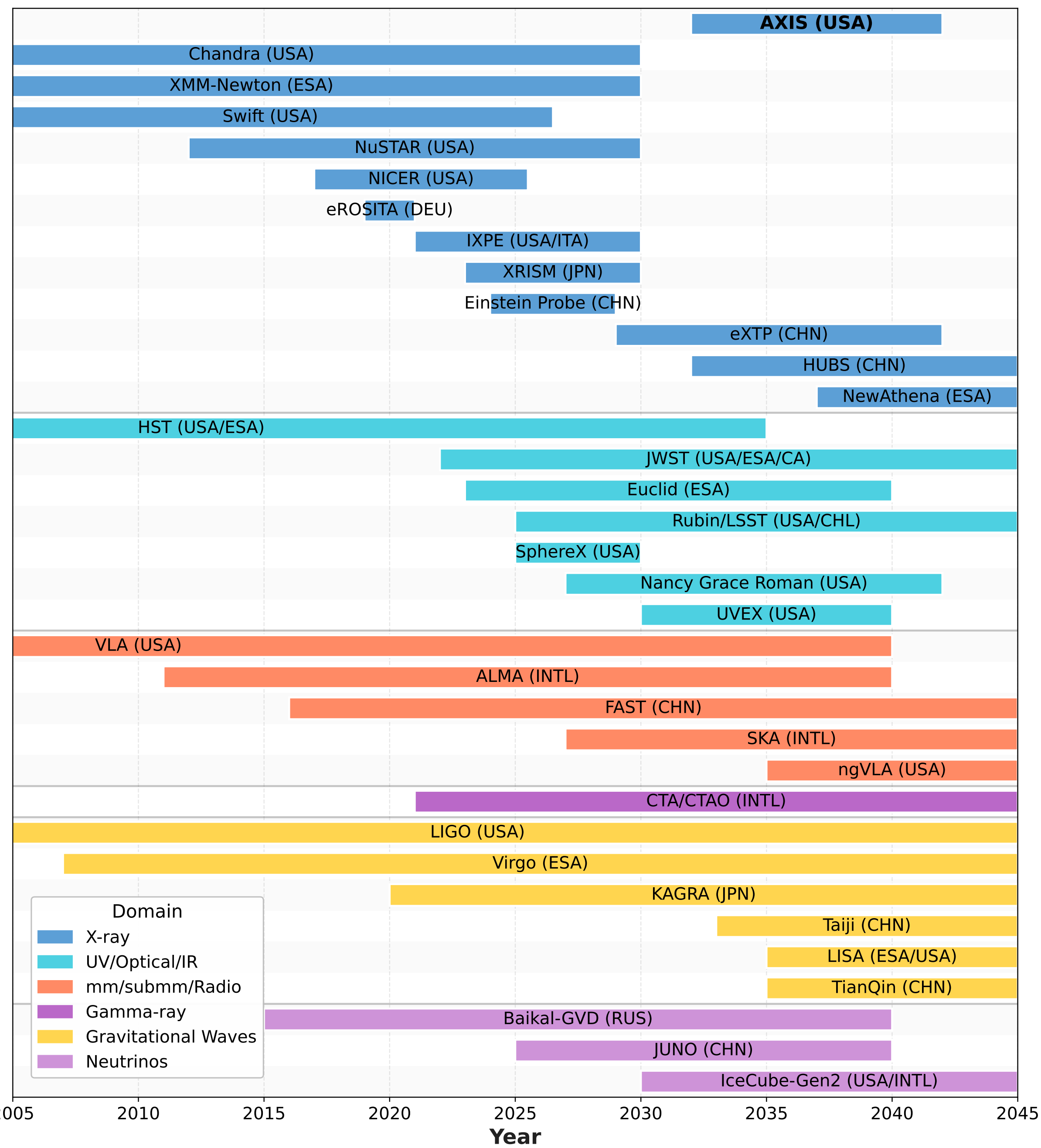

**Figure 2. Planned timeline of AXIS relative to international multiwavelength capabilities.** The 2030s represent a pivotal decade: as US X-ray missions reach end-of-life, China is rapidly expanding its X-ray astronomy program (Einstein Probe, eXTP, HUBS), while ESA maintains continuous capabilities (XMM-Newton, NewAthena). Without AXIS, the US will cede X-ray astronomy leadership during the most scientifically productive era in history, with simultaneous access to JWST, Roman, Rubin, next-generation ground telescopes (ELT/GMT/TMT), radio arrays (SKA, ngVLA), and gravitational wave detectors (LIGO, Virgo, LISA), enabling groundbreaking multi-wavelength and multi-messenger science impossible at any other time. Without AXIS, the US risks a decade-long gap in flagship X-ray capabilities during the most productive era of astrophysics.

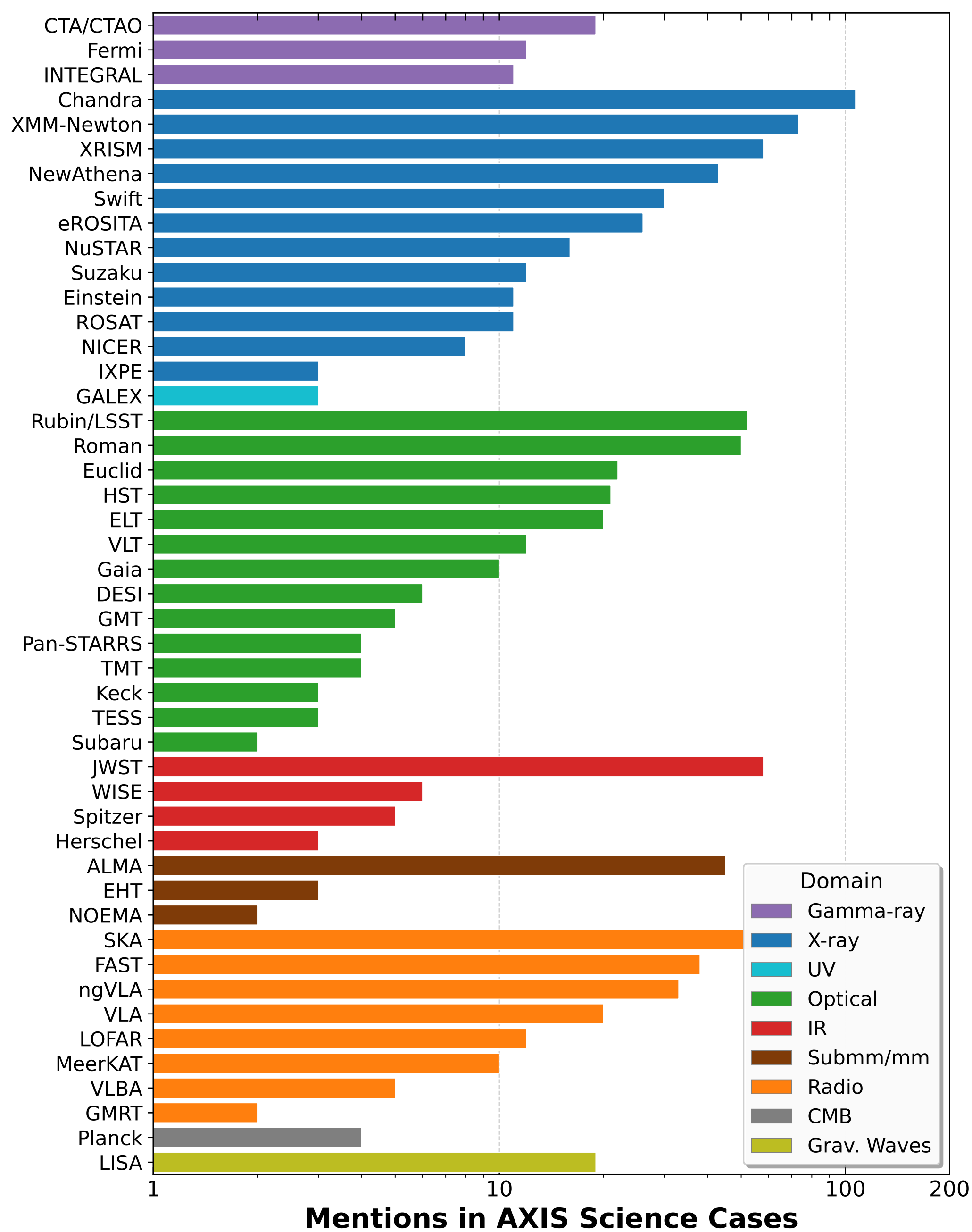

**Figure 3. AXIS demonstrates exceptional cross-facility synergy, with major observatories across the electromagnetic spectrum referenced extensively in the 138 AXIS science cases.** The breadth of facilities cited, including JWST, ALMA, Rubin/LSST, Nancy Grace Roman, SKA, CTA/CTAO, and LISA, underscores AXIS's competitive advantage as a powerful complement to the multi-wavelength and multi-messenger astronomy landscape of the 2030s. AXIS builds upon the transformative legacies of current flagship X-ray observatories (e.g., Chandra, XMM-Newton) while positioning itself as an essential facility for coordinated observations spanning gamma rays to gravitational waves, enabling breakthrough science that is impossible with any single facility alone.

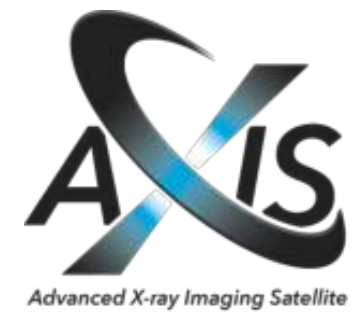

# AXIS Galaxies GO Science

**B. Alcalde Pampliega[1,2], S. W. Allen[3,4,5], M. Balboni[6,7], I. Bartalucci[8], A. Basu-Zych[9,10], R. M. Batalha[11,12], E. Bertola[13], V. S. Bessa[12], N. Biava[14,7], E. L. Blanton[15], A. Bonafede[6,7], A. Botteon[7], J. Bregman[16], F. Brighenti[6,17], E. Bulbul[18], M. Calzadilla[19], P. Chakraborty[19,20], G. Chartas[21], P. P. Choudhury[22], W. Cui[23], S. De Grandi[24], A. Del Popolo[25], R. Dupke[16,12], D. Eckert[26], E. Fedorova[27], A. M. Flores[3,4,28], N. Foo[29], B. Frye[30], C. Garcia Diaz[31], M. Gaspari[32], F. Gastaldello[8], S. Ghizzardi[8], S. Giacintucci[33], A. Gill[34], R. Gilli[35], M. Gitti[6,7], L. Gu[36], K. C. Harrington[37,38], J. Hlavacek-Larrondo[39], E. Hodges-Kluck[10], R. Huang[16], A. Ignesti[40], J. A. Irwin[41], E. F. Jimenez-Andrade[42], Y. Jiménez-Teja[12,43], S. Johnson[16], P. S. Kamieneski[29], W. Lee[44], B. Lehmer[20], E. Lentini[24], M. Lepore[13], J. Li[45], D. Liu[45], N. Locatelli[24], L. Lopez[46], S. Lopez[46], L. Lovisari[8], J. D. Lowenthal[47], A. B. Mantz[3], S. Marchesi[6], M. Markevitch[10], M. McDonald[48], E. D. Miller[48], M. S. Mirakhor[49], S. Molendi[8], E. B. Monson[50], R. G. Morris[3,5], E. O'Sullivan[19], A. Y. Pan[3,4], M. Pascale[51], E. Perlman[52], E. Piconcelli[27], A. Pillepich[53], G. Ponti[24,18,54], M. Prunier[39,53], K. Rajpurohit[19], S. W. Randall[19], G. Riva[8], M. Rossetti[8], H. R. Russell[55], F. Salvestrini[56], A. Sarkar[48,20], M. Sasaki[57], G. Schellenberger[19], E. M. Schlegel[58], T. Shimwell[59,60], T. Somboonpanyakul[61], H. Stueber[3,4], Y. Su[62], F. Tombesi[63], P. Tozzi[13], A. Tumer[9,10], F. Ubertosi[6,7], B. Vigneron[39], A. Vishwas[64], F. Vito[35], S. A. Walker[49], Q. D. Wang[31], F. Wang[16], R. van Weeren[60], A. Wolter[65], K.-W. Wong[66], J. Yang[16], M. Yeung[18], M. Yukita[67], M. S. Yun[31], C. Zhang[68,69], X. Zheng[18], D. Zhou[70], J. ZuHone[19]**

[1] ESO Vitacura, Alonso de Córdova 3107, Vitacura, Casilla, 19001 Santiago de Chile, Chile
[2] SKA Observatory, Jodrell Bank, SK11 9FT, UK
[3] Kavli Institute for Particle Astrophysics and Cosmology, Stanford University, 452 Lomita Mall, Stanford, CA 94305, USA
[4] Department of Physics, Stanford University, 382 Via Pueblo Mall, Stanford, CA 94305, USA
[5] SLAC National Accelerator Laboratory, 2575 Sand Hill Road, Menlo Park, CA 94025, USA
[6] Dipartimento di Fisica e Astronomia, Università di Bologna, via Gobetti 93/2, I-40129 Bologna, Italy
[7] INAF - Istituto di Radioastronomia, via P. Gobetti 101, 40129 Bologna, Italy
[8] INAF/IASF - Milano, Via A. Corti 12, 20133 Milano, Italy
[9] Center for Space Sciences and Technology, University of Maryland, Baltimore County (UMBC), Baltimore, MD 21250, USA
[10] NASA/GSFC, Greenbelt, MD 20771, USA
[11] Université Paris-Saclay, Université Paris Cité, CEA, CNRS, AIM, 91191 Gif-sur-Yvette, France
[12] Observatório Nacional, Rua General José Cristino, 77 – Bairro Imperial de São Cristóvão, Rio de Janeiro 20921-400, Brazil
[13] INAF - Osservatorio Astrofisico di Arcetri, largo E. Fermi 5, 50125, Firenze, Italy
[14] Thüringer Landessternwarte, Sternwarte 5, 07778 Tautenburg, Germany
[15] Department of Astronomy, Boston University, Boston, MA 02215, USA
[16] Department of Astronomy, University of Michigan, 311 West Hall, 1085 S. University Avenue, Ann Arbor, MI 48109-1107, USA
[17] University of California Observatories/Lick Observatory, Department of Astronomy and Astrophysics, Santa Cruz, CA 95064, USA
[18] Max-Planck-Institut für Extraterrestrische Physik (MPE), Giessenbachstr. 1, D-85748 Garching bei München, Germany
[19] Center for Astrophysics | Harvard & Smithsonian, 60 Garden St., Cambridge, MA 02138, USA
[20] Department of Physics, University of Arkansas, Fayetteville, AR 72701, USA

[21] Department of Physics and Astronomy, College of Charleston, Charleston, SC, 29424, USA
[22] Department of Physics, University of Oxford, Parks Road, OX1 3PU, UK
[23] Centro de Investigación Avanzada en Física Fundamental (CIAFF), Universidad Autónoma de Madrid, Cantoblanco, E-28049 Madrid, Spain
[24] INAF - Osservatorio Astronomico di Brera, Via E. Bianchi 46, 23807 Merate (LC), Italy
[25] Dipartimento di Fisica e Astronomia, University of Catania, Viale Andrea Doria 6, 95125 Catania, Italy
[26] Department of Astronomy, University of Geneva, Ch. d'Ecogia 16, CH-1290 Versoix, Switzerland
[27] INAF - Osservatorio Astronomico di Roma, Via Frascati 33, I-00040 Monte Porzio Catone, Italy
[28] Department of Physics and Astronomy, Rutgers University, 136 Frelinghuysen Road, Piscataway, NJ 08854, USA
[29] School of Earth and Space Exploration, Arizona State University, PO Box 876004, Tempe, AZ 85287-6004, USA
[30] Department of Astronomy/Steward Observatory, University of Arizona, 933 N Cherry Ave., Tucson, AZ 85721-0009, USA
[31] Department of Astronomy, University of Massachusetts, Amherst, MA 01003, USA
[32] Department of Physics, Informatics and Mathematics, University of Modena and Reggio Emilia, 41125 Modena, Italy
[33] Naval Research Laboratory, 4555 Overlook Avenue SW, Code 7213, Washington, DC 20375, USA
[34] Department of Aeronautics and Astronautics, Massachusetts Institute of Technology, 77 Massachusetts Avenue, Cambridge, MA 02139, USA
[35] INAF – Osservatorio di Astrofisica e Scienza dello Spazio di Bologna, Via P. Gobetti 93/3, 40129 Bologna, Italy
[36] SRON Space Research Organisation Netherlands, Niels Bohrweg 4, 2333 CA Leiden, The Netherlands
[37] Joint ALMA Observatory, Alonso de Córdova 3107, Vitacura, Casilla 19001, Santiago de Chile, Chile
[38] National Astronomical Observatory of Japan, Los Abedules 3085 Oficina 701, Vitacura 763 0414, Santiago, Chile
[39] Département de Physique, Université de Montréal, Succ. Centre-Ville, Montréal, Québec H3C 3J7, Canada
[40] INAF - Padova Astronomical Observatory, Vicolo dell'Osservatorio 5, I-35122 Padova, Italy
[41] Department of Physics and Astronomy, University of Alabama, Box 870324, Tuscaloosa, AL 35487, USA
[42] Instituto de Radioastronomía y Astrofísica, Universidad Nacional Autónoma de México, Antigua Carretera a Pátzcuaro 8701, Ex-Hda. San José de la Huerta, Morelia, Michoacán, C.P. 58089, México
[43] Instituto de Astrofísica de Andalucía-CSIC, Glorieta de la Astronomía s/n, E-18008 Granada, Spain
[44] Yonsei University, Department of Astronomy, Seoul, Republic of Korea
[45] Purple Mountain Observatory, Chinese Academy of Sciences, 10 Yuanhua Road, Nanjing 210023, People's Republic of China
[46] Department of Astronomy, The Ohio State University, 140 W. 18th Ave., Columbus, OH 43210, USA
[47] Smith College, Northampton, MA 01063, USA
[48] Kavli Institute for Astrophysics and Space Research, Massachusetts Institute of Technology, Cambridge, MA 02139, USA
[49] Department of Physics and Astronomy, The University of Alabama in Huntsville, 301 Sparkman Drive, Huntsville, AL 35899, USA
[50] Department of Physics and Astronomy, Middle Tennessee State University, 1301 E. Main Street Box 71, Murfreesboro, TN 37312, USA
[51] Department of Astronomy, University of California, 501 Campbell Hall 3411, Berkeley, CA 94720, USA
[52] Aerospace, Physics and Space Sciences Department, Florida Institute of Technology, 150 W. University Boulevard, Melbourne, FL 32901, USA
[53] Max-Planck-Institut für Astronomie, Königstuhl 17, D-69117 Heidelberg, Germany
[54] Como Lake Center for Astrophysics (CLAP), DiSAT, Universitá degli Studi dell'Insubria, via Valleggio 11, 22100 Como, Italy
[55] School of Physics & Astronomy, University of Nottingham, University Park, Nottingham NG7 2RD, UK
[56] INAF - Osservatorio Astronomico di Trieste, Via G. Tiepolo 11, 34143 Trieste, Italy
[57] Dr. Karl Remeis Observatory, Erlangen Centre for Astroparticle Physics, Friedrich-Alexander-Universität Erlangen-Nürnberg, Sternwartstr. 7, 96049 Bamberg, Germany
[58] Department of Physics and Astronomy, University of Texas at San Antonio, San Antonio, TX 78249, USA
[59] ASTRON, The Netherlands Institute for Radio Astronomy, Postbus 2, NL-7990 AA Dwingeloo, The Netherlands
[60] Leiden Observatory, Leiden University, PO Box 9513, 2300 RA Leiden, The Netherlands

[61] Department of Physics, Faculty of Science, Chulalongkorn University, 254 Phyathai Road, Patumwan, Bangkok 10330, Thailand
[62] Department of Physics and Astronomy, University of Kentucky, 505 Rose Street, Lexington, KY 40506, USA
[63] Dipartimento di Fisica, Università degli Studi di Roma "Tor Vergata", Via della Ricerca Scientifica 1, 00133 Roma, Italy
[64] Cornell Center for Astrophysics and Planetary Science, Cornell University, Space Sciences Building, Ithaca, NY 14853, USA
[65] INAF - Osservatorio Astronomico di Brera, Via Brera 28. 20121 Milano (MI), Italy
[66] Department of Physics, SUNY Brockport, Brockport, NY, 14420, USA
[67] Johns Hopkins University, Baltimore, MD 21218, USA
[68] Department of Theoretical Physics and Astrophysics, Faculty of Science, Masaryk University, Kotlářská, Brno, 61137, Czech Republic
[69] Department of Astronomy and Astrophysics, The University of Chicago, Chicago, IL 60637, USA
[70] Department of Physics and Astronomy, University of British Columbia, 6225 Agricultural Road, Vancouver V6T 1Z1, Canada

---

**a. Stellar Feedback**

*1. The Origin of Extraplanar Multi-phase Filaments*

**Science Area:** Stellar feedback, multi-phase galactic outflow

**First Author:** Edmund Hodges-Kluck (NASA/GSFC)

**Co-authors:** Mihoko Yukita (Johns Hopkins University), Priyanka Chakraborty (Center for Astrophysics | Harvard & Smithsonian), Laura A. Lopez (Ohio State University), Sebastian Lopez (Ohio State University), Manami Sasaki (Friedrich-Alexander-University Erlangen Nürnberg)

**Abstract:** Galaxy outflows are complex webs of interlinked, multiphase filaments whose origins are immediately relevant to understanding the impact of these outflows. Deep Chandra images of nearby winds and fountains reveal a strong *morphological* relationship between X-ray filaments and those seen in H$\alpha$, dust continuum, and CO, but to fully understand the physical connection between the hot and cool gas – and thus the survival of cool clouds, radiative cooling rate, and mass loading of winds – we need thermodynamic measurements. X-ray and H$\alpha$ surface brightness maps already suggest diverse origins for filaments, so we must make these measurements for individual filaments with characteristic widths of 50 pc. AXIS will provide the sensitivity required to measure temperature, density, metallicity, and charge exchange intensity from individual filaments in multiple outflows of different intensity from a sample of nearby, edge-on galaxies.

**Science:** Most galactic outflows are driven by fast stellar winds and clustered supernovae (SNe) within massive star clusters. The spatial and temporal coincidence of many SNe prevents the formation of radiatively cooling remnants. Instead, they convert their kinetic energy to thermal energy via weak shocks in a hot ($10^{6-7}$ K) bubble evacuated by winds and the first generation of SNe [255,479,571]. The bubble expands until it breaks through the gas disk, whereupon it breaks up and its pressure drives gas out of the disk and into the galaxy halo [294]. While outflows from individual star clusters are fountains in which the matter will eventually return to the disk [64], starbursts can power strong winds that blow a large amount of metals into the intergalactic medium [539].

These outflows circulate gas and regulate star formation, perhaps even explaining the Kennicutt-Schmidt law [372] or the low-mass end of the galaxy luminosity function [345]. Yet we do not understand how they transport the multiphase gas that we see, how much of that gas escapes the galaxy, and why, in very low-mass halos, they do not quench star formation. Models that reproduce the gross aspects of winds disagree on crucial details. For example, many hydrodynamic models (e.g., [163,254,255]) predict that most outflowing gas is neutral or warm ionized gas, and this appears to be the case in the few winds where we can measure the mass budget (e.g., M82 [282,465,540,563]). On the other hand, models differ on the *origin* of this cool gas: is it condensation out of radiatively cooling gas [203,499] or acceleration and compression of cool interstellar medium [83]? Do cool clouds evaporate faster than they form? Does the wind primarily shock itself or the ambient gas?

To answer these questions, we cannot rely on direct measurements of the gas that drives outflows but instead must look at interaction zones. The luminosity of the very hot ($T > 10^7$ K) gas in star clusters or starbursts [95] is low and scales with the star-formation rate surface density [480], and rapidly decreases away from the massive stars due to adiabatic expansion. As such, this gas has only been detected in a few nearby starbursts [339,479] and there are no prospects for observing it in extragalactic H II regions. Cosmic-ray protons may also drive outflows [142,602] but cannot be directly observed.

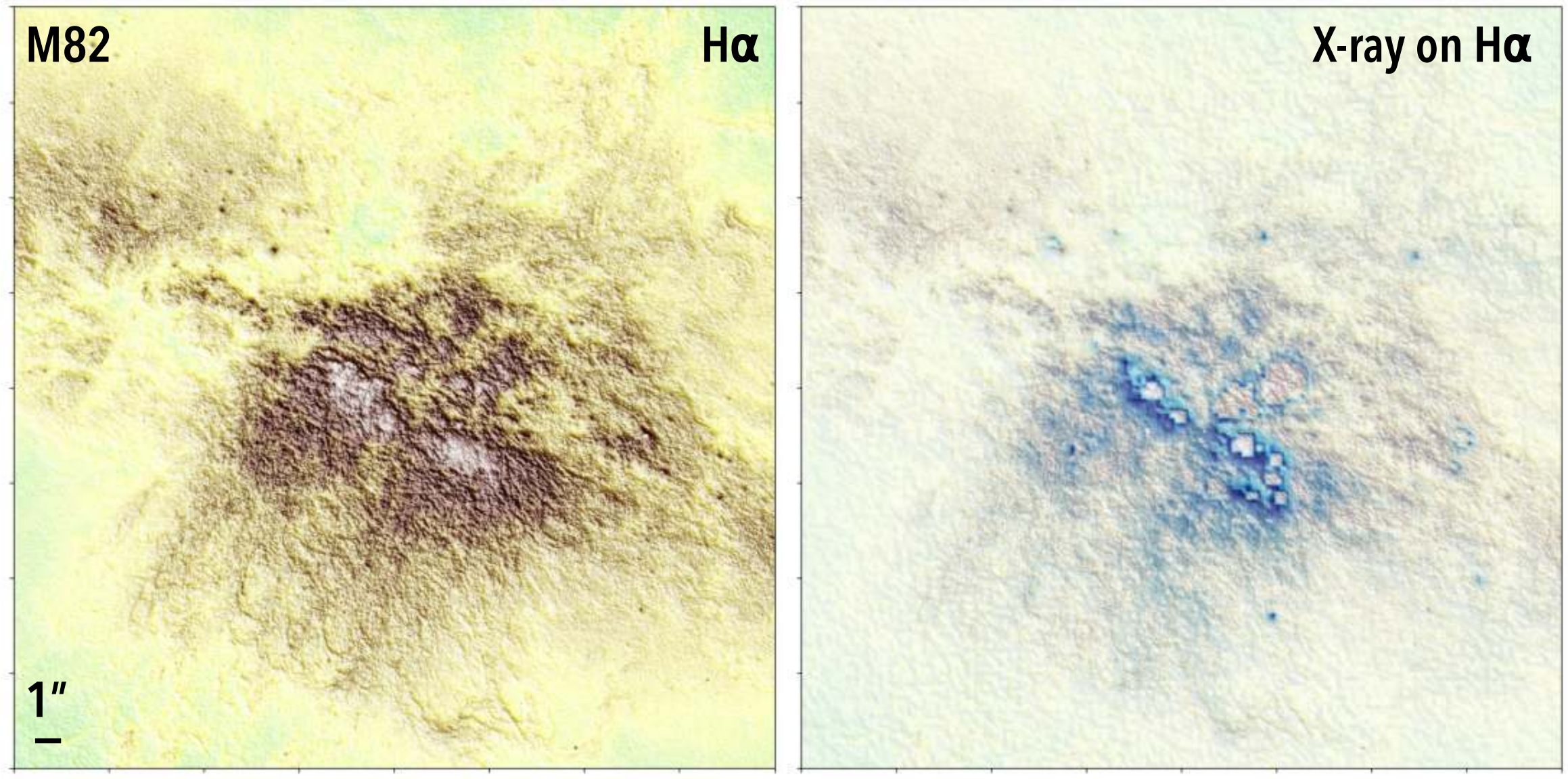

**Figure 1. AXIS will probe the link between X-ray and H$\alpha$ filaments.** *Left*: The HST H$\alpha$ map of M82 reveals a complex network of filaments and arcs that extend over several kpc and with characteristic widths of <50 pc. *Right*: An overlay of the Chandra image (blue) on the H$\alpha$ map shows that there is a broad correspondence between the X-ray and H$\alpha$ nebula. But some H$\alpha$ filaments coincide with X-ray filaments, while others lack X-rays or even appear to be anti-correlated. We must measure the temperature, density, ionization state, and cooling time of individual filaments to understand the different behavior.

The most prominent interaction zones are filaments that may trace shocks or radiative cooling [108]. Indeed, most emission from winds originates from filaments. In the few cases where we have deep, high-resolution X-ray images, this is also true in the soft X-rays. Figure 1 shows the H$\alpha$ nebula around M82 and the broad, and sometimes detailed, correspondence between the H$\alpha$ and X-ray emission. X-ray measurements of these filaments are critical because they can show how rapidly gas is cooling (or how much it has been shock-heated). The combination of X-ray, H$\alpha$, and molecular line measurements can diagnose the nature of the interaction, but we have made meager progress with Chandra and reached its limits.

Beyond M82, we can barely identify X-ray filaments (Figure 2), much less measure their luminosities, temperatures, and masses. Even in M82, with nearly 1 Ms of useful Chandra data [282], we lack the signal to extract spectra from individual filaments beyond the brightest part of the nebula. Complicating matters is the prospect that charge exchange may contribute significantly; high signal imaging spectroscopy of individual filaments is necessary to determine the emission mechanism and the physical conditions. As the filaments have widths of order an arcsecond and tend to have soft, emission-line dominated spectra ($E < 2$ keV), further progress with Chandra is impossible, and other facilities, like XMM-Newton, XRISM, or eROSITA, cannot resolve individual filaments.

But AXIS can resolve filaments and obtain enough signal to measure their **temperature, density, and cooling time**, as well as their **ionization state and charge exchange contribution**. These are enabled by AXIS' high resolution, large effective area, and Suzaku-like spectral resolution, which is sufficient to reveal non-equilibrium plasma and charge exchange through line (complex) ratios. By surveying a population of X-ray filaments in outflows of different intensities (galactic fountains, galactic superbubbles, and galactic superwinds) and, within those outflows, different spatial relationships to cooler gas, we can finally determine how winds deposit mass and energy into their environments.

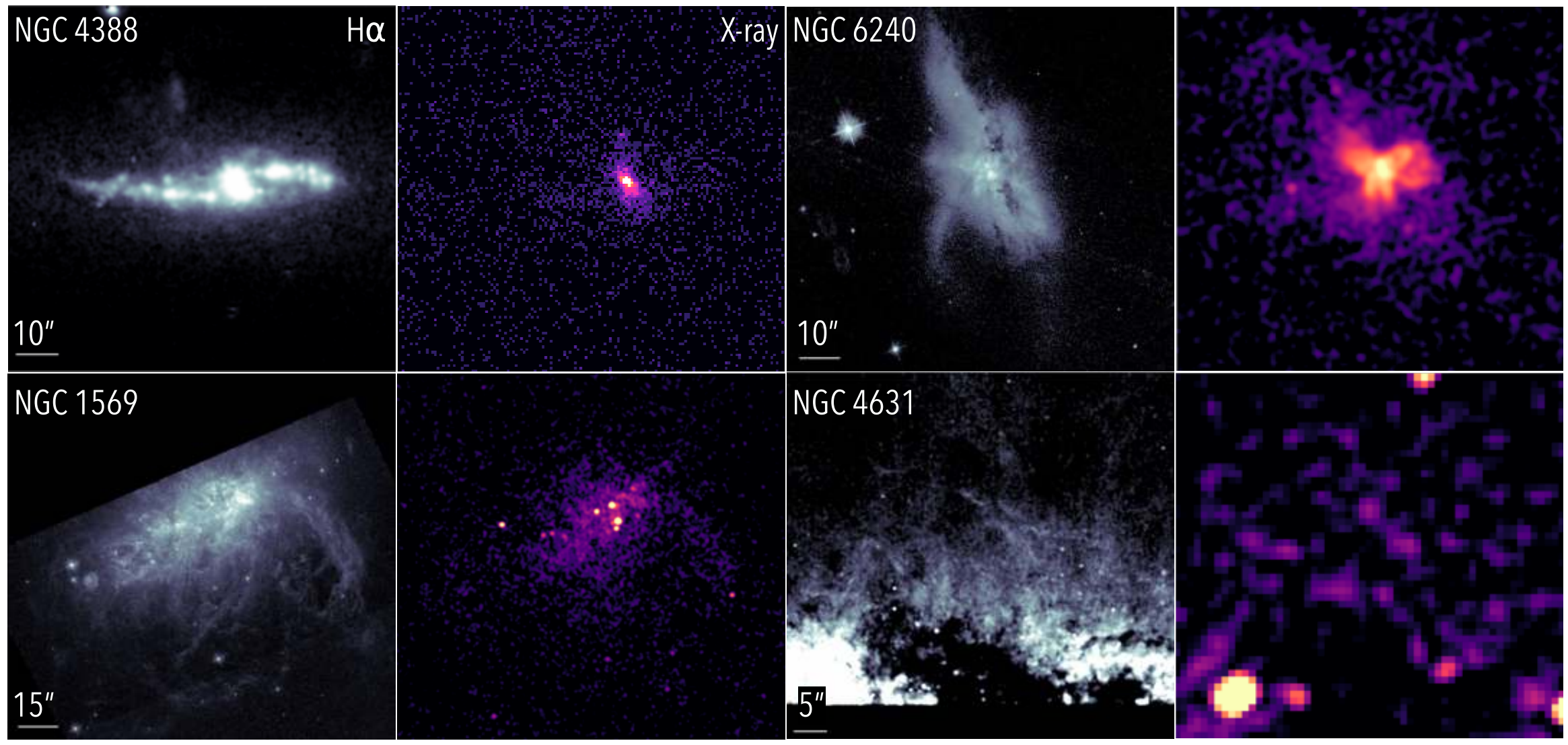

**Figure 2. AXIS will provide M82-like quality for filaments in other systems.** Here we show four nearby, edge-on galaxies in different regimes, each with striking H$\alpha$ filaments and poorly defined X-ray filaments seen by Chandra. *Top Left*: NGC 4388, a Seyfert 2 AGN with a potential starburst. *Top Right*: NGC 6240, a major merger with a double AGN and nuclear starburst. *Bottom Left*: NGC 1569, a dwarf starburst galaxy. *Bottom Right*: A close-up of the nuclear region of NGC 4631, a sub-$L_*$ "distributed starburst" galaxy with a giant superbubble. None of the X-ray filaments are well defined with Chandra, but modest AXIS exposures can bring them into sharp relief.

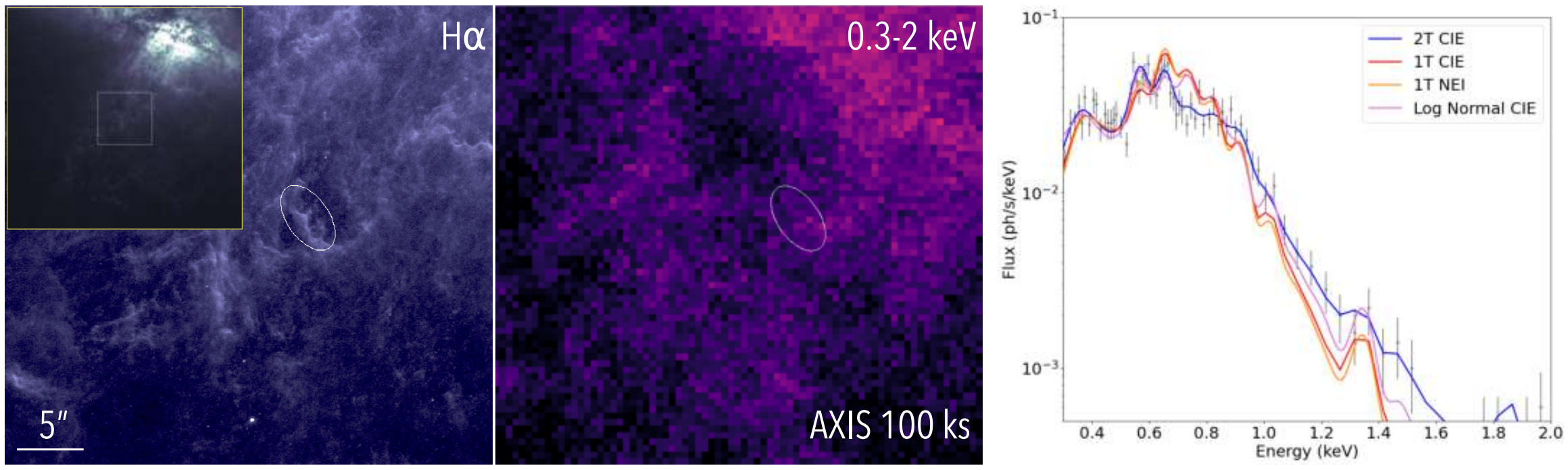

**Figure 3. AXIS spectra will enable unique measurements from individual filaments.** An H$\alpha$ filament a few kpc from the nucleus as seen by HST (*left*) can also be detected by AXIS (center), with a spectrum (*right*) with enough signal and definition in the key line complexes to tell the difference between one- and two-temperature thermal plasma models, or non-equilibrium ionization or charge exchange models.

Figure 3 shows that AXIS would measure spatially resolved spectra on the scale of H$\alpha$ filaments with sufficient signal to distinguish between basic plasma emission models. This is a challenge even in M82, given the existing Chandra data, due to the degrading soft response throughout the mission.

**Exposure time (ks):** 600 ks

**Observing description:** Here we propose a small survey of nearby, edge-on galaxies (plus 30 Doradus in the Large Magellanic Cloud) with filamentary H$\alpha$ nebulae to measure the temperature, density, and cooling time of filaments in different environments (individual star-forming regions, starbursts, and AGN, in galaxies of different masses). This legacy survey would leverage the enormous multiwavelength investment in nearby galaxies to determine how the hot gas relates to cooler gas in galactic outflows.

Each galaxy would fit within the AXIS field of view and requires a single exposure. Some point sources in each galaxy may be piled up, but the windowed mode is not a suitable option. The exposure times are based on simulations of filaments with a characteristic width of 2 arcsec and a characteristic length of 6 arcsec, with "typical" brightness for each galaxy based on the existing Chandra maps. We require sufficient counts to distinguish between a one- and two-temperature model in each filament at 99% confidence, and/or to identify the presence of charge exchange at 95% confidence, provided that charge exchange contributes at least 50% of the signal. With this signal, we will measure the temperature to 25% precision and the density to 50% precision (for a single-temperature model), with the uncertainty limited by the unknown true filament volume. The most significant uncertainty lies in the emission model, which determines the exposure requirement.

It is difficult to estimate the number of useful X-ray filaments these observations would detect, based on the limited signal in existing Chandra data and the range in "filament" size, but at a minimum, we would detect about 300 distinct filaments with sufficient signal to measure the temperature and density as described above. We would also stack data at the locations of very faint H$\alpha$ filaments to determine the average X-ray signal from these regions.

Chandra observations of M82, NGC 6240, and other galaxies show that there is no truly diffuse "background," only unresolved filaments. Thus, for filaments bright enough to measure properties with AXIS, and sufficiently outside of the galaxy disk, we do not need to subtract a background (the amount of Galactic foreground in front of any one filament is negligible).

- 30 Doradus; $t_{exp} = 80$ ks
- NGC 1569; $t_{exp} = 80$ ks
- NGC 4631; $t_{exp} = 120$ ks
- NGC 891; $t_{exp} = 80$ ks
- NGC 4388; $t_{exp} = 90$ ks
- NGC 6240; $t_{exp} = 150$ ks

**Joint Observations and synergies with other observatories in the 2030s:** Joint observations are not required. There is good synergy with wide-field optical IFUs.

**Special Requirements:** None.

*2. Morphology and Metal Content of Star Formation Driven Outflows*

**Science Area:** stellar feedback, galactic outflows

**First Author:** Sebastian Lopez (Ohio State University)

**Co-authors:** Edmund Hodges-Kluck (NASA/GSFC), Laura Lopez (Ohio State University), Manami Sasaki (Friedrich-Alexander-University Erlangen Nürnberg), Priyanka Chakraborty (Center for Astrophysics | Harvard & Smithsonian)

**Abstract:** Starbursts are galaxies undergoing an intense period of star formation and are characterized by their prominent kiloparsec-scale winds powered by stellar feedback (e.g., supernovae). These winds are responsible for driving gas and metals out of the disk, affecting the metallicity and future star formation of their host galaxies. The hot wind ($> 10^6$ K) phase emits predominantly in the soft (<2 keV) X-ray band at distances larger than the starburst radius. The superb soft X-ray sensitivity of AXIS will enable more in-depth mapping of these outflows, not only in popular targets like M82 but also in the less-studied and fainter dwarf starburst galaxies. Sampling galaxies across an extensive mass range will constrain how the metal loading factor changes with stellar mass, allowing for comparisons with similar optical studies of warm gas outflows and how hot metal loading affects or explains the stellar mass-metallicity relationship.

**Science:** We will study the morphology and metal content of galactic winds spanning several orders of stellar mass ($10^7 - 10^{10}$ $M_\odot$). The goal is to observe whether these various wind properties depend on the characteristics of their host galaxy.

*Comparisons with Galactic Wind Models*: One of the main observables of galactic winds in X-ray studies is their temperature and density profiles with distance. These gradients can then be compared to galactic wind models, whether analytical (as shown in Figure 4b; Nguyen & Thompson 355) or multidimensional hydrodynamic simulations [454], to ascertain the physics and evolution of the outflow. Recent work by [282, 283,401] has shown that analytical wind models must incorporate contributions from cool mass-loading and non-spherical wind geometry to accurately match observed temperature and density profiles. The classical [95] model (CC85) has both temperature and density falling off with distance from the disk much faster than the observed profiles. Mass loading allows the wind to maintain its temperature and density by converting the loss of kinetic energy (from the wind slowing with the added mass) to thermal energy, maintaining the $> 10^6$ K wind temperatures. The added mass into the wind keeps the density elevated. The geometry has effects on how quickly the wind cools: a large expansion area (e.g., spherical) cools faster than a more confined geometry (e.g., conical, as observed). Due to the dynamical nature of starburst-driven outflows, mass loading of the ISM is expected. Less clear is the non-spherical nature of galactic winds where this analysis has been conducted, such as NGC 253, NGC 4945, and M82. An explanation may be the presence of super star clusters that are driving the wind. For example, in NGC 253, these SSCs are arranged in a ring-like manner [272], which can cause a collimation effect on the wind, creating a conical geometry (see [356] for models testing this). Both M82 and NGC 4945 also contain SSCs, although their arrangement is less clear.

While the collimation effect is observed in larger galaxies, it is not clear in dwarf-like galaxies. Dwarf galaxies typically have less well-defined disks, and their lack of molecular gas precludes the formation of future SSCs. Both of these observations suggest that outflows from dwarf galaxies are unlikely to be collimated, either through a dense gas disk or ring-like SSC arrangements. This raises the question: Are these winds more like the theoretical CC85-predicted ones? If so, that implies different wind dynamics in these systems than in larger systems. A limitation of X-ray observations of dwarf galaxies is that they are generally dimmer. In Figure 4c we show an image of I Zw 15 [214], a nearby low-mass starburst dwarf galaxy that contains expanding shells of cool gas. Absent from this system is an X-ray counterpart [378] that must exist for the outflow to drive out the cooler gas phases. With the added sensitivity of AXIS and its higher spatial resolution than XMM-Newton, systems like I Zw 18 could be observed in the soft X-ray band and conclusively decide whether they contain outflows or not.

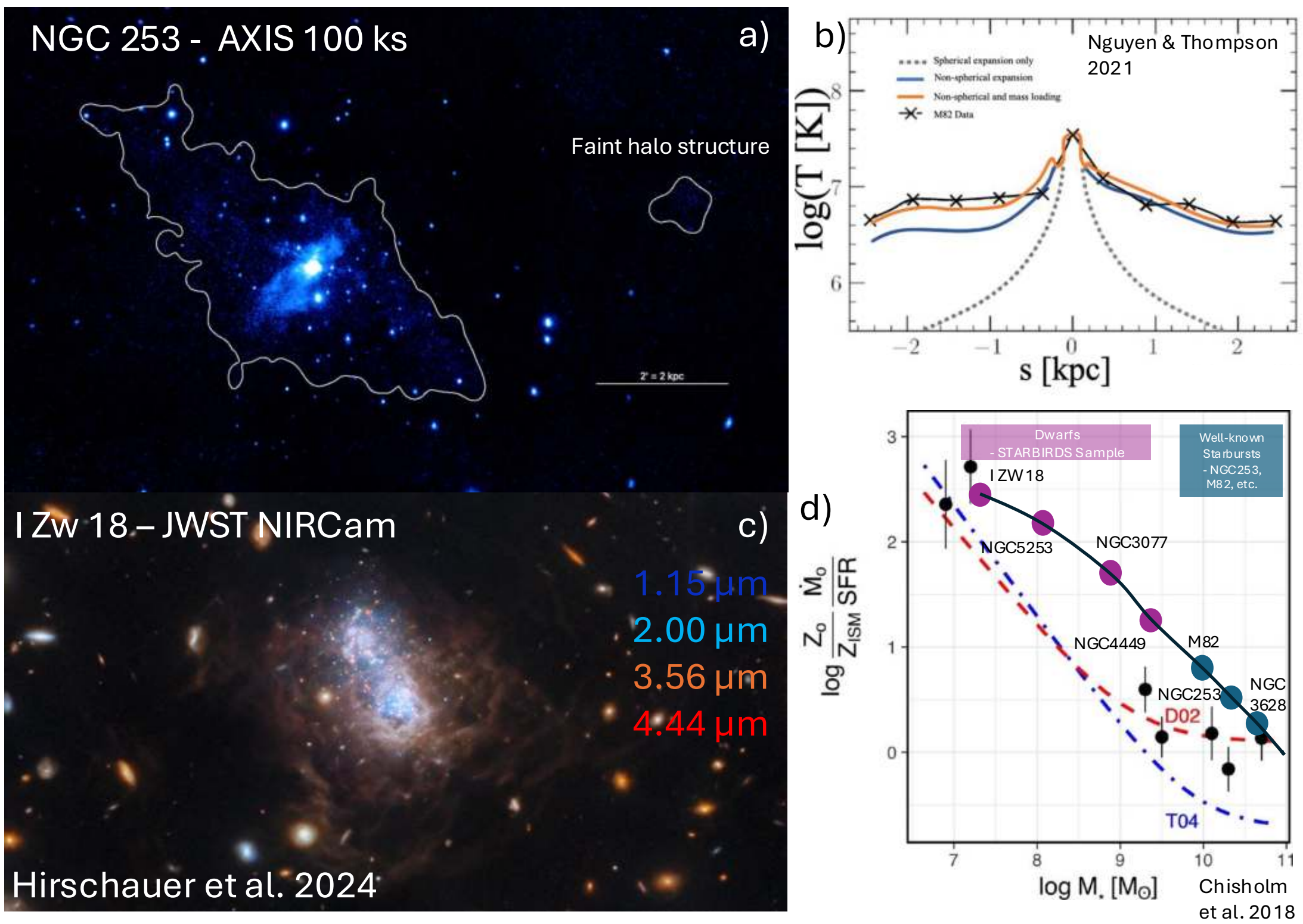

**Figure 4.** a) Simulated 100 ks AXIS observation of NGC 253. The contours highlight the extent of the disk visible in the X-ray and a notable faint halo feature. b) Various semi-analytical models from [355] compared to the observed X-ray derived temperature profile of M82. The gray line represents the spherical CC85 model, the blue line uses a non-spherical expansion geometry, and the orange line incorporates non-spherical expansion geometry with mass loading, providing the best fit to the observations. c) Four-color image of I Zw 15 from [214] showing the star-forming clusters at the center and expanding shells of dust that trace cool gas. d) Plot of metal loading factor from [98] as a function of galaxy stellar mass. The red dashed and blue dot-dashed lines represent different models for the stellar mass-metallicity relationship [512]. Black dots are the galaxies studied in [98]. Highlighted are regions where other notable starbursts occupy and whose metal loading factor in the hot phase is worth comparing to the optical. The purple and blue points indicate the targets for this science case, which span several magnitudes in stellar mass.

*The Hot Phase Metal Loading Factor*: Galactic winds in starburst galaxies are driven primarily through supernova explosions. As a result, this hot X-ray emitting ejecta is metal-rich and, with sufficient signal (>5000 counts), the abundance of these metals can be constrained. This process, in which metals are stripped from galactic disks, can have major effects on the overall chemical abundance of the wind host, thereby influencing future stellar populations and star formation, making galactic winds a pivotal part of galactic evolution.

The stellar mass-metallicity relationship [512] has shown that lower-mass galaxies have lower metallicity than large, Milky Way-type galaxies. A common explanation is that smaller galaxies are more efficient at removing metals from their disks than hosts with larger gravitational potentials. This metric is called the metal loading factor. Work in the optical regime studying this metal loading factor [98] found that there is an inverse correlation between galaxy stellar mass and its metal loading factor, as shown in Figure 4d. Whether this holds in the hot phase of the wind, where most of the metals are contained, remains a mystery. The theoretical work of [349]and [383], using different versions of the FIRE simulations, has both found that most of the metals are in the hot phase of the wind for galaxies with $M_* \gtrsim 10^9$ $M_\odot$. This suggests that, at least for the higher-mass galaxies in the [98] sample, the most significant metal loading factor is associated with the hot phase, as constrained by X-rays. Recent observational studies of metal gradients [282,283] in starburst outflows have only sampled galaxies with stellar masses of order $10^{10}$ $M_\odot$, unable to quantify how the metal loading factor changes with mass. With the superb sensitivity of AXIS, we can study smaller, fainter, dwarf starbursts like those identified in the STARBIRDS [327] survey, and assess whether these galaxies are ejecting more metals than their well-studied, larger counterparts, and if the simulations of [349,383] correctly predict that the metals are mostly in the warm optical emitting phase rather than the hot phase.

*Hot Galactic Haloes*: Simulations and models of galactic feedback often have the winds extending to tens of kiloparsecs. Such behavior is notably shown by the "Cap" in M82. However, other notable wind hosts, such as NGC 4945, do not have prominent winds that reach such scales; rather, the bulk of the emission is confined to $\sim 2$ kpc. [481] showed the halo emission for several nearby wind host systems. While large-scale structures are evident, they depend on smoothing areas of low signal that could introduce artifacts rather than concrete detections. In Figure 4a, we show NGC 253 and highlight the disk along with a halo structure 5 kpc away from the disk, not explicitly studied in the literature. [482] studied the halo of NGC 253, noting a X-like morphology in the shallow 13 ks data available at the time. They speculate that the soft X-ray emission could be a result of shocks into an ambient HI halo medium. Since then, [287] observed the HI emission around NGC 253 with the KAT-7 telescope and found neutral gas emission out to $\sim 9-10$ kpc from the disk. They also noted the overlap between this gas and ROSAT observations of the halo. Thus, it may be that the halo structure we highlight in Figure 4 may be a result of the shocks [482] predicted of the hot gas ramming into the cooler halo. With the added sensitivity of AXIS, these large-scale structures can be mapped more accurately, providing insights into how far galactic feedback occurs and can affect the CGM. We will also be able to measure its properties, like temperature and density, to see how they compare with the inner wind and if the temperature and density are consistent with hot gas that has cooled or has a different origin.

**Exposure time (ks):** Total of 500 ks

**Observing description:** Our targets are star-forming systems with galactic winds detected in the X-ray or cooler phases. These targets span a large range of stellar masses, $10^7 - 10^{10}$ $M_\odot$, in order to study how various galactic wind properties, such as morphology and metal loading, vary with galaxy host characteristics, including mass. The targets are the following, from least to most massive, and with their estimated exposure times:

- I Zw 18; $t_{exp} = 80$ ks
- NGC 5253; $t_{exp} = 60$ ks
- NGC 3077; $t_{exp} = 60$ ks
- NGC 4449; $t_{exp} = 60$ ks
- M82; $t_{exp} = 100$ ks
- NGC 253; $t_{exp} = 80$ ks
- NGC 3628; $t_{exp} = 60$ ks

These estimated exposure times will produce at least 20,000 counts, allowing at least three regions (one nucleus and two outflows) along the minor axis to constrain metallicity, where the required counts are of order 5000. Due to their well-studied nature, potential for complementary science in other wavelengths, and synergy with other science cases, we allocate more exposure time to M82 and NGC 253. Conversely, due to the understudied nature of I Zw 18 and outflow evidence in other phases, we devote more time to this system.

**Joint Observations and synergies with other observatories in the 2030s:** This AXIS science case will have synergies with JWST, ngVLA, and HWO. JWST acts as a high-resolution tracer of cold material due to the dust that it is mixed with. As a result, we can study how the cool outflow phase relates to the hot X-ray emitting one AXIS traces, and map well where it interacts and what effects it has on the hot wind plasma emission (mass loading, charge-exchange emission, etc.). However, it is unclear which phase of the cold gas regime the dust traces, which is why, with the construction of the ngVLA, we will be able to probe molecular and atomic gas in extragalactic sources with MW GMC resolution. This will enable us to map both small filamentary and large-scale diffuse structures within and surrounding galactic winds, allowing us to observe how the hot wind transports colder phases into the CGM and IGM. HWO will provide us with higher resolution and sensitivity than what is possible with HST. This will allow us to probe the drivers of the hot winds we observe with AXIS, such as super star clusters, their arrangement, and how they can affect the morphology of galactic winds.

**Special Requirements:** None.

*3. Hot Gas, Supernovae, and Novae in Nearby Spiral Galaxies*

**Science Area:** galaxies, hot ISM

**First Author:** Eric M Schlegel, Univ of Texas at San Antonio

**Co-authors:** Manami Sasaki (Friedrich-Alexander-University Erlangen-Nürnberg), Laura Lopez (Ohio State University)

**Abstract:** Deep observations of nearby spiral galaxies have led to significant scientific understanding about the point source population, particularly ultraluminous sources. The same advances are possible for the hot diffuse gas present in the galaxy's interstellar medium (ISM) and its SNR population. Intervening dust and gas make such advances difficult for studies of the Milky Way. The hot, X-ray-emitting gas in nearby spiral galaxies is difficult to analyze because there are typically too few photons for high-resolution spatial analysis and comparison with images obtained at other wavelengths. AXIS will be significant for studies of hot gas, given the ∼10x larger collecting area compared to *Chandra*. Supernovae are expected, and have been argued, to provide the energy input to the hot gas. They are almost certainly necessary, but are they sufficient? Novae occur more often and are spread around a spiral, so they may also be necessary contributors to heating the gas. Our proposed science goals include a study of the hot diffuse emission component through correlations with emission features in other bandpasses and a detailed examination of SNRs at luminosities fainter than those available in the archival data. The multi-wavelength correlations are expected to advance our understanding of the impact of SNRs on the hot diffuse component as well as investigate the physics of the hot diffuse gas.

**Science:** The study of the X-ray emission from normal galaxies commenced with the Einstein observations of galaxies (e.g., [143]). The total X-ray emission from a spiral originates primarily from the point source emission of X-ray binaries (low- and high-mass), supernova remnants, and OB star clusters, plus the hot, diffuse component of the ISM.

This last component, first proposed by Spitzer [474], is expected to be present based upon the more detailed theoretical work of [324] and [360] among many others. It forms a real discovery space for X-ray observations. Indeed, a lasting legacy of AXIS could be the integration of the diffuse ISM component into studies of nearby galaxies.

The spatial resolution of *Einstein*, ∼1', prevented the separation of the X-ray-emitting constituents for all but the Local Group galaxies. ROSAT improved the resolution by roughly a factor of three, providing views of galaxies a short distance beyond the Local Group (e.g., [450] for NGC 6946). *Chandra*, with its sharp resolution (∼1") showed what is possible for studies of the hot component, but lacked sufficiently large effective area to make observations 'proposal-able' for a reasonable time request: several galaxies were awarded ∼1 Ms exposure times, but sufficiently late in the mission that the effective area below ∼0.8 keV had been diminished.

*Benefits to Astronomy:* Both the SNR and diffuse emission goals represent significant contributions to astrophysics. Long observations of nearby spirals will provide sufficient counts in the diffuse component (e.g., surface brightness) to enable variations in the diffuse emission to be matched against data from other bands, particularly radio observations, at spatial scales of order a few arc seconds. Detections of X-ray-emitting SNRs lead to a comparison of the late phases of stellar evolution.

Spatially-resolved spectroscopy on fine spatial scales was demonstrated with the Antennae data (e.g., [144]). At these scales, correlations between wavebands show us variations in the spectral energy distributions with spatial position. For more nearby galaxies, such as face-on spirals, data exist for most of the other bandpasses and require deep observations in the X-ray band. This is a crucial and significant step in integrating X-rays into studies of the hot, diffuse gas. For example, such a study could resolve the not-yet-understood correlation between far-IR and radio luminosities by providing a third dimension (e.g., if X-ray emission from shocks correlates with both far-IR and radio emission), or demonstrate that the correlation only holds because variations are integrated out.

For SNRs, a long observation provides real data for the luminosity function, rather than relying on an extrapolation. Ideally, observations of spirals out to 10 Mpc that reach $10^{34}$ erg $s^{-1}$ would address previous limitations. Real data leads to a comparison with the Milky Way distribution, testing the completeness of the MW SNR population. It also directly tests whether previous galaxy observations have been insufficiently sensitive, failing to reach the bulk of a galaxy's SNR population. This is a critical test given the importance of SNRs to ISM studies and the expectation that the SNR data sets in the X-ray and radio should be approximately identical.

High-resolution radio and optical observations of nearby spirals yield candidate SNRs via the radio spectral index and the optical [S II]/H$\alpha$ ratio (e.g., [385] for NGC 300; [200] for M33). This approach led to the realization that both bands are possibly identifying different populations [384]. Pannuti, Schlegel, & Lacey (2007) showed, in a study of five nearby spirals, that the X-ray band may add a third population, as only two SNRs are matched in all three bands from a total of 185 SNRs [386]. Did that occur because of population differences or sufficiently insensitive observations?

From the AXIS simulations and sensitivity estimates, a 100-ksec observation for each galaxy would reach a 3$\sigma$ completeness value of $\sim 10^{-16}$ erg/sec/cm$^2$ in the 0.5 – 2 keV band. That value is an order of magnitude lower than obtained with *Chandra*. At a distance of 10 Mpc, that flux level would translate to a luminosity of $\sim 10^{35}$ erg/sec – admittedly, above the ideal value but significantly closer to it.

**Exposure time (ks):** If AXIS has 10x the effective area of *Chandra*, then 100 ks pointings would be relatively 'typical.' Such an exposure time would require adjustments to compensate for lower metallicity, mass, or the known distance differences. Regardless, nearby galaxies could be observed more 'routinely' with AXIS than currently done with *Chandra*, leading to studies of time-variable objects as well as stacking for more sensitivity.

**Observing description:** nearby face-on spirals: NGC 5194/5195, NGC 6946, NGC 6744, NGC 3938, M74, M83, M101, NGC 7793, NGC 2403, NGC 300, IC 342, NGC 1313, IC 5332 for a combined exposure time of $\approx$1.5 Msec.

**Joint Observations and synergies with other observatories in the 2030s:** JWST, HWO, ALMA, ngVLA, Euclid, Rubin, Roman (about the only band missing is the UV): each of these instruments has sufficient spatial resolution to provide a good match to AXIS for pursuing physics, in much the same manner as the optical image of the Cat's Eye Nebula was a great match to the Chandra image and immediately demonstrated the physics of the nebula. Such a demonstration should be possible for star formation (IR vs X-ray) and supernova remnant impact on the ISM (IR, radio vs X-ray), to name two examples.

**Special Requirements:** Potentially subdivide the observations to pursue time-variable X-ray point sources. Pile-up would be expected *only* for extremely bright point sources, perhaps one per galaxy for the nearer galaxies.

*4. Time Capsules from Cosmic Dawn: Probing Reionization with Green Pea Galaxies*

**Science Area:** star-forming galaxies

**First Author:** Yuanyuan Su (University of Kentucky)

**Co-authors:** Jimmy Irwin (University of Alabama)

**Abstract:** Green Pea galaxies—low-redshift, compact, low-metallicity, low-dust, starburst systems exhibiting Lyman continuum leakage—serve as local analogs to high-redshift Ly$\alpha$ emitters believed to contribute to the reionization of the early Universe. Their elevated specific star formation rates (sSFRs) and extreme [OIII]$\lambda$5007 emission lines suggest prolific high-mass X-ray binary (HMXB) formation, a potential source of the high-energy photons required for ionizing the intergalactic medium (IGM). Pilot studies with XMM-Newton have detected X-ray emission in two out of three Green Pea galaxies, with conclusive AGN identification in one of them. We propose a deep AXIS survey of 20 Green Pea galaxies at $z = 0.2 \sim 0.4$ with published Lyman continuum measurements to obtain a representative census of their X-ray luminosities. This program will provide key insights into the role of X-ray sources in cosmic reionization, as well as the physics of stellar feedback and early galaxy evolution.

**Science:** The heating and re-ionization of neutral hydrogen in the intergalactic medium (IGM) of the early ($z = 6 - 20$) Universe is intimately tied to our understanding of the subsequent galaxy evolution and star formation rate of the Universe (e.g., [27,100]) with obvious cosmological implications. However, the source of the high-energy photons (blueward of 912 Å) needed to re-ionize the neutral IGM remains unclear. They are thought to be created in high-star formation rate, high-redshift dwarf galaxies.Such high-redshift Ly$\alpha$ emitters (LAEs) (e.g., [127,513]) are characterized by low metallicity ($0.1 \sim 0.4\,Z_\odot$), compact sizes ($1 \sim 2$ kpc), modest stellar masses ($\sim 10^8\,M_\odot$), and high specific star formation rates (sSFR) ($10^{-8} \sim 10^{-7}\,\mathrm{yr}^{-1}$), and can represent 60% of all Lyman break galaxies at z$\sim$6 [475].

A growing number of studies have highlighted the potential role of X-ray sources—such as high-mass X-ray binaries (HMXBs) and low-luminosity AGNs (LLAGNs)—in cosmic reionization (e.g., [263]). The difficulty of observing such high-redshift galaxies—and, by extension, their HMXB and LLAGN populations—has prompted efforts to identify low-redshift analogs. The most similar local counterparts of high-z LAEs appear to be so-called 'Green Pea' galaxies that share high-z LAEs compact sizes, low stellar masses, high sSFRs, and low metallicities (Figure 5). Recent JWST observations have revealed that Green Pea galaxies are remarkably similar to cosmic dawn galaxies at z=8, in terms of physical sizes, morphologies, and chemical fingerprints [417]. Furthermore, Lyman continuum leaking is found to be common among Green Pea galaxies [13,235], making them stand out as local analogs of cosmic dawn galaxies that are responsible for reionization. Green Pea galaxies were first identified in Sloan Digital Sky Survey (SDSS) images of very compact, mostly low-mass $z \sim 0.2$–$0.4$ galaxies with very strong emission in the $r$-band, giving the galaxies a green appearance in *gri* color images [81]. The $r$-band emission results from strong [O III]$\lambda$5007 line emission redshifted into the $r$-band of SDSS. Subsequent studies have revealed a variety of food-related galaxies, including Blueberry galaxies [586], and Purple Grape galaxies [71]. They can all be regarded as 'Green Pea' galaxies at different redshifts given their commonality of very strong [O III] emission lines indicative of high rates of specific star formation. More generally, they can all be referred to as 'extreme emission line galaxies' (EELGs). While the exact definition of extreme emission line galaxies varies, the median value from various studies defines an EELG as having an [OIII]$\lambda$5007 line equivalent width of at least 300 Å [14,241,586].

A variety of efforts have been made to relate the HMXB population to the star formation rate (SFR) of local galaxies, although many of them are not necessarily EELG galaxies. Earlier studies found a power-law relation with an exponent of roughly one for galaxies with SFR above a few $M_\odot$ yr$^{-1}$ between the (typically H$\alpha$–derived) star formation rates and summed HMXB X-ray luminosity ($L_{X,HMXB}$) from galaxies with a range of SFR [202,269,336]. [30] demonstrated that the $L_X$-SFR relation evolves with redshift, which likely results from the HMXB/metallicity dependence. Later efforts specifically incorporated the effects that low galactic metallicity has on the SFR/$L_{X,HMXB}$ relation, in the sense that lower galactic metallicity leads to a higher $L_{X,HMXB}$ at a fixed SFR. This is because a lower stellar metallicity decreases mass-loss due to line-driven winds of stars throughout their lifetimes, which increases the mass of black holes formed from massive stars [168]. The increased number of formed black holes increases the fraction of binary systems that can become HMXBs [168]. This metallicity dependence is quite important, given the low metallicity of high-redshift Ly$\alpha$ emitters, which are believed to be responsible for reionizing the Universe. The age of the starburst also plays a role in the importance of HMXBs, as [191] find that $L_X/M_{star}$ declines by nearly an order of magnitude as stellar populations age from 10 Myr to 100 Myr. Although less studied, LLAGNs have also been investigated in nearby star-forming galaxies [26,333,413]. Their detection rate is low, but this may be biased by observational limitations and confusion with stellar processes. Time-domain surveys and stacking analyses suggest that LLAGNs may be more common than previously thought [25,334].

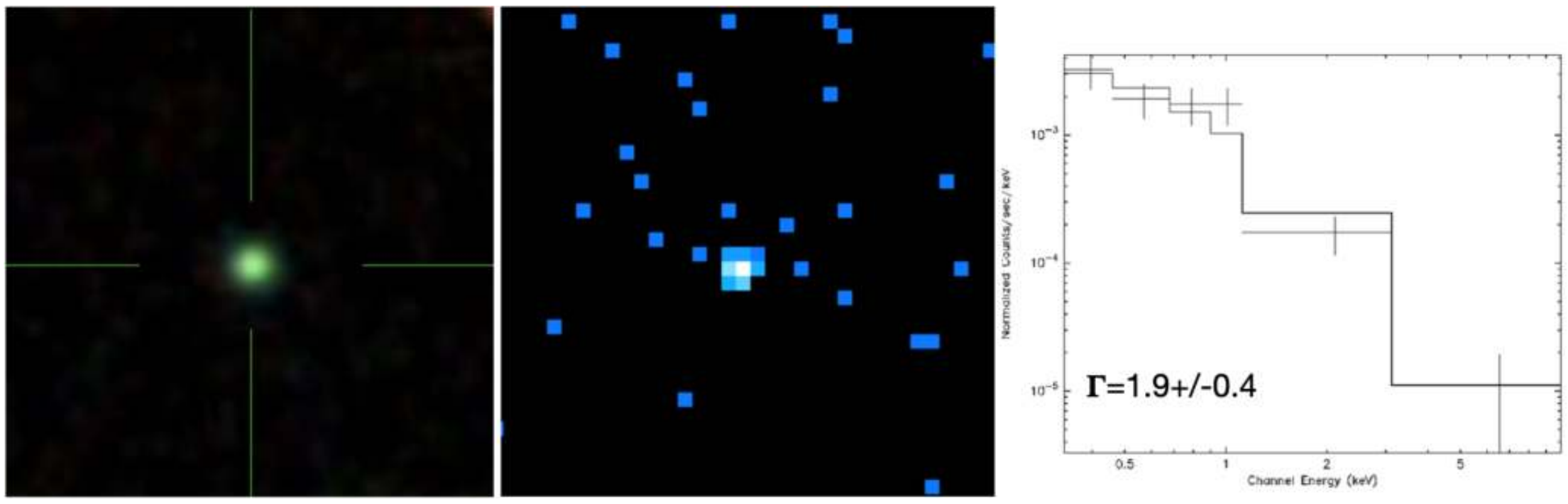

**Figure 5. left**: SDSS image of GP3, a Green Pea galaxy at z=0.2. **middle**: X-ray image of this galaxy from 25 ksec synthetic AXIS observation. **right**: Synthetic spectrum of the AXIS observation fitted with a Galactic absorbed powerlaw model.

[492] has presented a pilot XMM-Newton study of three Green Pea galaxies. X-ray was detected in two of them, GP1 and GP2, both exhibiting higher X-ray luminosities than expected from high-mass X-ray binaries (HMXBs), suggesting the presence of additional sources. Notably, [53] has confirmed that GP2 harbors a LLAGN with a 2.0-10.0 keV luminosity of $10^{42}$ erg/s. In the third galaxy, GP3, X-ray emission was not detected, but the upper limit is consistent with expectations from HMXBs based on its star formation rate and metallicity. [4] has presented XMM-Newton observations of 7 Blueberry galaxies, which are nearby analogs of Green Pea galaxies and typically have even lower masses and metallicities. X-ray emission was detected in two of them, while the remaining five have upper limits below the expected level from HMXBs. These Blueberry galaxies may be too young to have developed significant HMXB populations.

AXIS will enable effective X-ray observations of a large sample of Green Pea galaxies with a reasonable total exposure time. Our overarching goal is to quantify the role of X-ray sources in cosmic reionization, either through their direct contribution to ionizing photons or by creating interstellar medium (ISM) conditions that facilitate the escape of energetic UV photons. More specifically, we aim to achieve the following objectives:

1. Understanding what shapes the HMXB population in Green Pea galaxies, via examining the relationship between $L_X$ and their SFR, metallicity, and age.

**Table 1.** Target properties and requested exposure time for AXIS.

| Name | $f_{esc}$ (1) | $z$ (2) | log(SFR) (3) | 12+log(O/H) (4) | log$L_X$ (5) erg/s | Expo (6) ksec |
|---|---|---|---|---|---|---|
| J115449+244333 | 0.625 | 0.3690 | $1.134 \pm 0.027$ | $7.710 \pm 0.043$ | 41.2 | 89 |
| J115205+340050 | 0.177 | 0.3400 | $1.450 \pm 0.021$ | $8.025 \pm 0.034$ | 41.3 | 88 |
| J091704+315221 | 0.161 | 0.3000 | $1.354 \pm 0.02$ | $8.459 \pm 0.035$ | 41.0 | 55 |
| J144231-020952 | 0.120 | 0.2936 | $1.387 \pm 0.023$ | $7.939 \pm 0.035$ | 41.3 | 52 |
| J092532+140313 | 0.092 | 0.3012 | $1.390 \pm 0.023$ | $8.022 \pm 0.037$ | 41.3 | 55 |
| J101138+194721 | 0.090 | 0.3322 | $1.456 \pm 0.020$ | $7.918 \pm 0.034$ | 41.4 | 69 |
| J124835+123403 | 0.047 | 0.2634 | $1.253 \pm 0.023$ | $8.163 \pm 0.036$ | 41.1 | 41 |
| J012217+052044 | 0.038 | 0.3656 | $0.971 \pm 0.041$ | $7.799 \pm 0.064$ | 41.0 | 87 |
| J090146+211928 | 0.026 | 0.2993 | $1.131 \pm 0.027$ | $8.102 \pm 0.045$ | 41.0 | 55 |
| J091113+183108 | 0.023 | 0.2622 | $1.440 \pm 0.022$ | $8.060 \pm 0.037$ | 41.3 | 40 |
| J113304+651341 | 0.022 | 0.2414 | $0.910 \pm 0.024$ | $7.983 \pm 0.039$ | 40.8 | 34 |
| J011309+000223 | 0.022 | 0.3062 | $0.699 \pm 0.076$ | $8.329 \pm 0.115$ | 40.4 | 59 |
| J095838+202508 | 0.019 | 0.3017 | $1.194 \pm 0.021$ | $7.801 \pm 0.037$ | 41.2 | 56 |
| J131037+214817 | 0.016 | 0.2831 | $1.116 \pm 0.024$ | $8.354 \pm 0.044$ | 40.8 | 48 |
| J004743+015440 | 0.013 | 0.3537 | $1.336 \pm 0.024$ | $8.029 \pm 0.037$ | 41.2 | 80 |
| J105331+523753 | 0.012 | 0.2526 | $1.450 \pm 0.020$ | $8.251 \pm 0.032$ | 41.2 | 37 |
| J144010+461937 | 0.005 | 0.3008 | $1.558 \pm 0.020$ | $8.206 \pm 0.033$ | 41.3 | 55 |
| J003601+003307 | $< 0.029$ | 0.3475 | $1.184 \pm 0.024$ | $7.781 \pm 0.037$ | 41.2 | 77 |
| J081409+211459 | $< 0.007$ | 0.2272 | $1.218 \pm 0.020$ | $8.060 \pm 0.032$ | 41.1 | 29 |
| J154050+572442 | $< 0.001$ | 0.2944 | $1.399 \pm 0.025$ | $8.396 \pm 0.043$ | 41.1 | 53 |

(1) Lyman continuum escape fraction taken from [13]. (2) SDSS spectroscopic redshift [165]. (3) Star formation rate traced by H$\beta$ [165]. (4) Oxygen abundance [165]. (5) Expected HMXB luminosity in 0.5-8.0 keV based on the scaling relation between $L_X$ and SFR and metallicity [68]. (6) Requested exposure time for AXIS to obtain $> 40$ net counts from each source.

2. Characterizing whether AGN are common among Green Pea galaxies and what factors may be linked to their AGN occupation.

3. Determining whether there is a correlation between their X-ray content and their LyC escape fraction.

**Exposure time (ks):** 1159 ksec

**Observing description:** We aim to characterize the X-ray properties of Green Pea galaxies as a population. Specifically, we seek to detect X-ray emission from Green Pea galaxies that lack AGN, where the X-ray luminosity is expected to be predominantly produced by HMXBs. If any of these galaxies harbor AGN, their $L_X$ will exceed the expected level, allowing for tighter constraints. We use GP3 as an example to demonstrate the advantage of AXIS over XMM-Newton for detecting such sources. GP3 is a Green Pea galaxy at $z = 0.2$, which was not detected in a previous 25 ksec XMM-Newton observation. We simulated a 25 ksec AXIS observation for this galaxy, assuming that it has an X-ray luminosity of $L_{0.5-8.0\mathrm{keV}} = 10^{41}$ ergs/s – a value below the upper limit placed by XMM-Newton and consistent with that expected for its SFR and metallicity. We adopted a photon index of 2, typical for HMXB. As shown in Figure 5, the AXIS synthetic observation yields more than 40 net counts and provides a meaningful constraint on the photon index.

To achieve our scientific goals, we select 20 Green Pea galaxies at $z = 0.2 - 0.4$ with published Lyman continuum escape (LCE) fraction measurements, as listed in Table 1. Observing LyC at lower redshifts is impossible due to severe absorption by neutral hydrogen in the Milky Way's interstellar medium (ISM). Following the classification scheme of [165], our sample includes 6 galaxies classified as strong LCEs, 11 as weak LCEs, and 3 as non-LCEs. These galaxies also span a wide range of star formation rates and metallicities. We estimate their expected $L_X$ from the $L_X$—SFR—metallicity plane [68]. We will obtain the same level of constraints for these 20 Green Pea galaxies as demonstrated for GP3 (see Figure 5) with a total exposure time of 1159 ksec. This observational campaign can be carried out over two observational cycles, allowing us to assess the variation of their X-ray emission and verify a possible AGN origin.

**Joint Observations and synergies with other observatories in the 2030s:**

- UVEX (UV) - Direct LyC detection, UV continuum slope of Green Pea galaxies
- ELT (optical) - Spatially resolved emission line maps and dynamics of Green Pea galaxies
- JWST (IR) - Rest-optical properties of 'early peas' during the epoch of reionization: ionization parameter, metallicity, electron density, dust attenuation, stellar masses, ages, star formation histories, morphology.
- SKA/ngVLA (radio) - Atomic and molecular hydrogen gas content of Green Pea galaxies

In combination with AXIS, next-generation observatories in the 2030s will provide comprehensive electromagnetic coverage, enabling multi-wavelength studies of Green Pea galaxies and offering new insights into photon escape mechanisms in high-redshift 'early peas' and their role in cosmic reionization.

**Special Requirements:** None

## b. Black Hole Feedback: Quasar Mode

*5. Exploring the impact of powerful SMBH winds on galaxy ecosystems*

**Science Area:** AGN winds, ultra-fast outflows, AGN feedback

**First Author:** F. Tombesi (Tor Vergata University of Rome)

**Co-authors:** M. Gaspari (U. of Modena & Reggio Emilia), Weiguang Cui (University Autonoma de Madrid), Francesco Salvestrini (INAF - Astronomical Observatory of Trieste), E. Piconcelli (INAF - Astronomical Observatory of Rome)

**Abstract:** Powerful winds driven by supermassive black holes (SMBHs) significantly impact galaxy evolution by redistributing energy and material across different scales. These winds, which manifest as ultra-fast outflows (UFOs), provide a critical link between accretion processes near the SMBH and large-scale feedback on the host galaxy and surrounding intergalactic medium. The proposed observations will leverage the unprecedented spatial and spectral resolution of NASA's AXIS mission to investigate the dynamical properties of these winds and their direct impact on multi-phase gas structures and extended regions within galaxy ecosystems. By combining AXIS capabilities with multiwavelength data, we aim to unravel the interplay between UFOs, star formation, and gas regulation in active galaxies across a range of redshifts and luminosities.

**Science:** Mergers between gas-rich galaxies are widely recognized as a key mechanism in driving significant evolutionary processes in galaxies and their central supermassive black holes (SMBHs). Such mergers are believed to trigger major starbursts, the formation and growth of SMBHs, and, eventually, the transition from gas-rich galaxies to gas-poor elliptical galaxies. The role of mergers in the formation of these objects has been explored in numerous studies (e.g., [222]). In this scenario, dust and gas are gradually dispersed, transforming a completely obscured Ultra-Luminous Infrared Galaxy (ULIRG) into a dusty quasar, and eventually, into a fully exposed quasar. One crucial aspect of this evolutionary path is the impact of galactic winds, driven by the central active galactic nucleus (AGN) and/or the surrounding starburst. These winds are thought to play a fundamental role in regulating both the growth of the SMBH and the stellar spheroid component, which ultimately explains the tight relationship observed between the mass of the SMBH and the stellar spheroid [467].

Understanding the precise mechanisms behind these feedback processes is of crucial importance for comprehending galaxy evolution and the growth of SMBHs. One of the most important feedback processes in this context is the effect of outflows on star formation. For outflows to effectively quench star formation, they must be able to affect the molecular gas from which stars form. Observations of ULIRGs with *Herschel* in the far-infrared have made a breakthrough in identifying and analyzing these massive molecular outflows. Recent OH-absorption observations have revealed molecular outflows on kiloparsec scales, with velocities reaching up to $\sim$1,000 km s$^{-1}$, suggesting mass outflow rates as high as $\sim$1,000 M$_\odot$ yr$^{-1}$ (e.g., [483]; [541]; [473]). These observations indicate that the outflows are driven by a fast and energetic mechanism, such as an AGN accretion disk wind, which interacts with the host galaxy's interstellar medium (ISM) to drive large-scale outflows.

Theoretical models have proposed that the origin of these massive outflows is a fast, high-velocity ($v_{\rm out} \sim 0.1c$) wind originating from the AGN accretion disk. This wind drives a shock into the host galaxy's ISM, pushing the molecular material outward in an energy-conserving flow (e.g., [611]; [158]). The shock interaction between the fast AGN wind and the ISM yields two distinct types of outflows: momentum-conserving flows and energy-conserving flows. In momentum-conserving flows, most of the wind's kinetic energy is radiated away, and the gas is pushed outward by the ram pressure of the wind. In contrast, energy-conserving flows occur when the shocked wind gas is not efficiently cooled, allowing the gas to expand adiabatically as a hot bubble. In this case, the momentum flux of the outflow is expected to be significantly higher than that of the radiation, with momentum flux values reaching up to $\dot{P}_w \sim 20L_{AGN}/c$, which have been observed in several ULIRGs (e.g., [483]).

Despite the theoretical developments, conclusive observational evidence for the connection between the fast AGN wind and the large-scale molecular outflow remains elusive. Blueshifted absorption lines in the X-ray spectra of several AGNs have been detected, indicating the presence of ultrafast outflows (UFOs) with velocities reaching up to ~0.1c ([503]; [502]; [501]). These winds are observed at sub-pc scales from the central black hole and are consistent with an accretion disk wind model. The strength of these winds suggests they are powerful enough to have a significant impact on the host galaxy environment, potentially quenching star formation, cooling flows, and regulating the growth of the SMBH (e.g., [175,556]). The detection of a disk wind with a velocity of $v_{out} \sim 0.25c$ in the ULIRG IRAS F11119+3257, along with the observation of similar winds in other ULIRGs and quasars (e.g., [504]; [505]; [162]; [472]; [268]), has strengthened the case for AGN-driven outflows playing a crucial role in shaping galaxy evolution.

However, conclusive evidence linking these powerful accretion disk winds with the large-scale molecular outflows observed in ULIRGs and quasars is still lacking. Therefore, the detection and characterization of these accretion disk winds are essential to understanding the mechanisms behind AGN feedback and its role in galaxy evolution [177].

The main objectives of this proposal are as follows:

- *Study the energetics and dynamics of ultra-fast outflows in nearby ULIRGs and quasars:* We aim to quantify the efficiency of feedback from AGN-driven outflows and assess their role in regulating the growth of both the SMBH and the host galaxy. By analyzing the energetics and dynamics of these outflows, we will gain insights into the mechanisms responsible for quenching star formation and suppressing further accretion onto the SMBH. While many of these targets have existing CCD and grating observations, the superior sensitivity and resolution of AXIS will enable more precise measurements of outflow properties, surpassing the limitations of current datasets.
- *Investigate the spatial distribution and ionization state of the wind components:* To constrain wind-launching mechanisms, we will study the spatial distribution and ionization state of different outflow components (e.g., molecular, neutral, and ionized gas). AXIS's high angular resolution and spectral capabilities will allow us to resolve these components more effectively than previous instruments, providing key information about acceleration regions and the suppression of star formation and SMBH growth.
- *Analyze the interaction of SMBH winds with multi-phase gas:* We will examine the interaction between SMBH-driven winds and the multi-phase gas in the host galaxy, including molecular, neutral, and ionized components. AXIS's combination of very high sensitivity, high angular resolution, and spectroscopy will facilitate detailed studies of these interactions, offering insights into the impact of AGN feedback on the galaxy's ISM and its role in driving large-scale outflows that affect star formation and gas content.
- *Assess the role of SMBH winds in regulating star formation and gas content:* By studying the relationship between SMBH winds and molecular gas reservoirs in ULIRGs and quasars, we aim to assess the role of AGN feedback in regulating star formation rates and the overall gas content of the host galaxy. AXIS's capabilities will enable us to probe these relationships with greater accuracy, contributing to our understanding of how AGN feedback shapes galaxy evolution, particularly in the context of mergers and the transition from gas-rich to gas-poor elliptical galaxies.

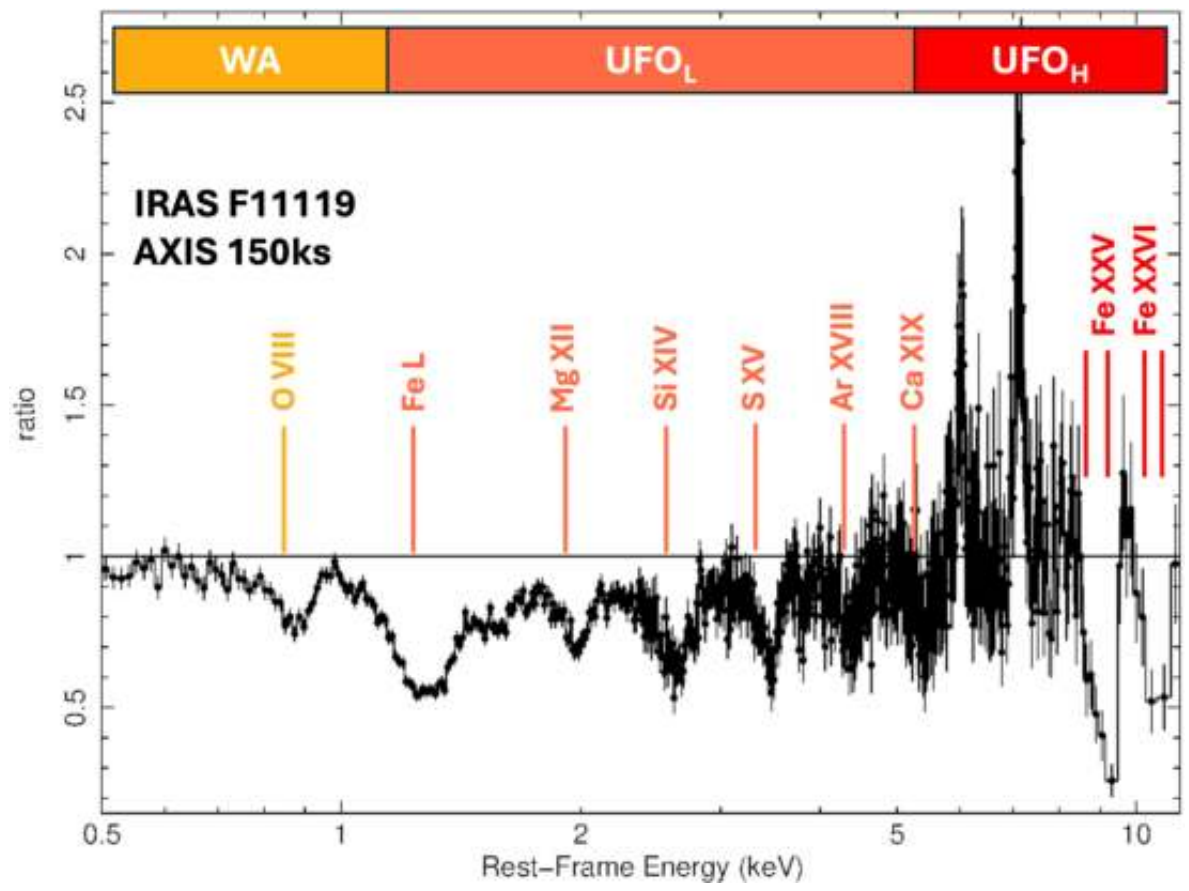

**Figure 6.** Simulated 150 ks AXIS spectrum of IRAS F11119+3257 showing the multi-phase structure of the three-component X-ray winds. This is the prototypical ULIRG hosting multi-phase and multi-scale AGN-driven outflows, which is one of the prime targets for multi-wavelength observations.

Through this research, we aim to provide conclusive evidence linking the fast AGN winds observed at small scales with the large-scale molecular outflows in ULIRGs and quasars. This will enable us to understand better the role of AGN feedback in galaxy evolution and the growth of SMBHs, as well as to probe AGN evolutionary models.

**Exposure time (ks):** 1.5 Ms for 10 targets.

**Observing Description:** Our feasibility analysis focuses on IRAS F11119+3257, a prototypical local ultra-luminous infrared galaxy (ULIRG) in a late-stage merger, located in the field (i.e., not in a galaxy cluster). It exhibits powerful, multi-phase nuclear X-ray winds and galaxy-scale outflows, making it an ideal laboratory for studying AGN feedback. We adopt the most recent results from *XMM-Newton* and *NuSTAR* [268] to characterize the complex nuclear wind in this source, which includes multiple absorption components with distinct ionization and kinematic properties.

The warm absorber (WA) has an ionization parameter of $\log \xi \approx 0.05$ and column density $\log(N_H) \approx$ 22.5, tracing relatively low-ionization gas. Two faster components correspond to ultra-fast outflows (UFOs): a soft UFO (UFOL) with $\log(N_H) \approx 21.5$, $\log \xi \approx 2.57$, and $v_out \approx 0.30c$; and a hard UFO (UFOH) with $\log(N_H) \simeq 24$, $\log \xi \approx 4.95$, and $v_out \approx 0.27c$. These components illustrate a highly stratified wind structure with the potential to drive galaxy-scale feedback.

Although previous missions, such as *Chandra*, *XMM-Newton*, *NuSTAR*, and *XRISM*, have revealed the existence of these outflows, *AXIS* brings a critical advantage for feedback studies. Its unprecedented effective area in the soft X-ray band (below 5 keV) allows for the sensitive detection of low- and mid-ionization absorption and emission features, which are essential for characterizing the energetics and geometry of the wind. The modest energy resolution ($E/\Delta E \sim$ 50–150) is ideally suited to isolate multiple absorption components and resolve P-Cygni-like profiles associated with outflowing, shocked material. This capability is crucial for interpreting wind-driven feedback and directly addresses the referee's point regarding the need to resolve feedback-related shocks in AGN.

A 150 ks exposure with *AXIS* will enable high signal-to-noise detection of soft X-ray features in IRAS F11119+3257, including lines from O, Ne, Mg, Si, S, and Fe at different ionization stages (M-, L-, and K-shell). These diagnostics are sensitive to gas ionized by both direct AGN radiation and by shocks. Therefore, AXIS observations will constrain the column density, ionization parameter, and velocity structure of each wind component, allowing us to estimate the mass outflow rate, momentum flux, and kinetic power—key quantities for evaluating feedback strength.

Importantly, the *AXIS* angular resolution of 1.5 arcsec corresponds to $\simeq$ kpc scales at the redshift of IRAS F11119+3257 ($z \approx 0.189$). This is sufficient to distinguish compact nuclear emission from galaxy-scale structures in nearby ULIRGs and quasars. In particular, it enables the separation of circumnuclear regions influenced by AGN-driven shocks from more extended soft X-ray emission associated with star formation or large-scale outflows. In systems like IRAS F11119+3257 and Mrk 231, where kpc-scale outflows are observed in optical, IR, and millimeter bands, AXIS will uniquely trace the X-ray counterparts of these structures. This multi-wavelength synergy is essential to map the propagation of AGN-driven winds and determine where and how they interact with the interstellar medium, potentially quenching star formation—one of the central issues in AGN feedback theory.

Furthermore, AXIS will benefit from a highly stable, orders-of-magnitude lower and less variable particle background compared to *XMM-Newton*. This low-background environment is critical for detecting faint, soft X-ray emission from extended gas and for robustly characterizing low-contrast absorption features. The combination of high sensitivity and low background makes AXIS uniquely capable of revealing the signatures of shock-heated and photoionized gas over galactic scales.

To maximize the scientific return, we include 10 ULIRGs and quasars with known X-ray UFOs and multi-phase, galaxy-wide winds. These targets, selected from Tozzi et al. [509], are: IRAS 17020+4544, PG 1126-041, MCG-03-58-007, I Zw 1, MR 2251-178, IRAS 05189-2524, Mrk 231, IRAS F11119+3257, PDS 456, and APM 08279+5255. These sources have substantial archival data—for example, PDS 456 has 1.2 Ms with *XMM-Newton*, 300 ks with HETG, and 260 ks with *XRISM*—yet *AXIS*'s sensitivity below 5 keV will reveal features unresolved in past data, especially from warm absorbers and shocked gas.

Our observing strategy is as follows:

- *Target AGN with known ultra-fast outflows*: We will focus on objects where the outflows' properties—velocity, ionization, and energetics—are already partially constrained, enabling robust modeling and follow-up.
- *Sample a range of black hole masses and AGN luminosities*: This allows investigation of how AGN wind properties and feedback scale with galaxy mass and accretion power.
- *Tailor exposure times based on X-ray brightness and wind strength*: We will optimize S/N across the sample, accommodating for long-term variability in AGN flux.
- *Integrate with multi-wavelength observations*: Coordination with ALMA, JWST, ELT, and GMT will help connect small-scale AGN winds to large-scale galaxy evolution by tracing the impact of outflows on gas reservoirs and star formation.

By leveraging *AXIS*'s strengths, we will directly address how ultra-fast outflows influence their host galaxies, how shocks propagate into the ISM, and whether AGN winds can regulate star formation on galactic scales. These results will provide key constraints on models of black hole–galaxy co-evolution.

**Joint Observations and synergies with other observatories in the 2030s:** Future observatories, such as the Extremely Large Telescope (ELT), the Square Kilometre Array (SKA), ALMA, and NewAthena, will provide complementary data on different physical scales and wavelengths. The combination of AXIS and these facilities will allow multi-phase and multi-scale studies of AGN winds and their feedback on extended structures.

**Special Requirements:** None.

*6. X-ray Einstein rings of lensed quasars: Probing relativistic quasar winds and ISM interactions at cosmic noon*

**Science Area:** AGN, Quasar Winds, AGN Feedback

**First Author:** George Chartas (College of Charleston)

**Co-authors:** Elena Bertola (INAF - Osservatorio Astrofisico di Arcetri), Massimo Gaspari (U. Modena & Reggio Emilia), Francesco Salvestrini (INAF - Astronomical Observatory of Trieste)

**Abstract:** Highly ionized ultrafast outflows (UFOs) of AGN are now considered one of the primary mechanisms regulating the evolution of galaxies through a feedback process. To obtain direct evidence of the effectiveness of AGN winds in providing feedback, it is essential to search for the interaction of AGN winds with the interstellar medium (ISM) of high-redshift galaxies closer to the peak of star formation and AGN activity. We propose to employ gravitational lensing to detect UFOs in high-z quasars and resolve the extended shocked X-ray emission from the interaction of the UFOs with the ISM. Studying UFOs in gravitationally lensed quasars offers a significant advantage due to the lensing magnification, which typically ranges between 5–100. Extended shocked emission around the lensed quasar forms a ring (i.e., Einstein ring) that connects the individual images. AXIS observations of a sample of high-$z$ lensed quasars will allow us to study both the central drivers of galaxy growth at the smallest scales (1–100 $r_g$) and the interaction of the winds with the ISM at the mesoscale ($\sim$ pc $- \sim$ kpc; [177]).

**Study the interaction of UFOs with the ISM as manifested in X-ray Einstein rings**

The observed evolution of the stellar mass of galaxies [e.g., 295] indicates that approximately 75% of the stellar mass had already formed before $z \sim 1.3$. It is therefore important to determine the energetics and morphology of multi-phase outflows/winds at high redshift, near the peak of AGN and star formation activity. One prediction of numerical models of ultrafast outflows (UFOs) is that they interact with the ISM, producing shocks and extended X-ray emission at the shock front [e.g., 158,611]. The extent of the shocked ISM gas is predicted to be of the order of $\sim 0.1-1$ kpc. At $z = 2$, this corresponds to a range of angular sizes of 0.01 $\sim$ 0.1 arcsec that any current X-ray telescope cannot resolve. A direct detection of the interaction of a small-scale UFO originating at $10 - 100 r_g$, where $r_g = GM_{\rm BH}/c^2$, with kpc-scale gas would confirm that UFOs transfer a significant amount of their energy to the host galaxy and are thus an important component of galaxy feedback.

When gravitationally lensed, a compact quasar source forms multiple images, whilst any extended emission around the quasar forms a ring (aka Einstein ring) that connects the individual images. A technique that has been successfully employed to resolve extended optical and IR emission around quasars involves lens modeling the Einstein rings of gravitationally lensed quasars. Finding lensed extended emission is easier than finding unlensed extended emission because the lens magnification enormously reduces the contrast between the extended emission and the central quasar [259].

The detections of X-ray Einstein rings associated with UFOs in distant lensed quasars are rare due to their X-ray weakness and the limited number of known lensed quasars with relativistic outflows [92]. The detection of both the UFO from the spectra of the lensed images and the extended emission from the shock can only be accomplished with an X-ray telescope with a PSF of the order of 1 arcsec (required to resolve the lensed images and Einstein ring) and an effective area of about $>$ 10 times Chandra (required to obtain high signal-to-noise UFO and shocked ISM spectra). The AXIS probe mission meets these requirements.

A hint of the detection of both a UFO and an Einstein ring is provided in Chandra observations of the lensed quasar PG 1115+080 [91]. PG 1115+080 is a quadruply lensed quasar containing a powerful UFO ($v \sim 0.4c$). Our preliminary spatial analysis of the available Chandra observations of PG1115+080 shows hints of a partial X-ray Einstein ring. In Fig. 7, we show the deconvolved Chandra image obtained by stacking all available Chandra observations (total exposure time of 173 ks). The extended emission forming a partial Einstein ring is visible in both the stacked observation and in a single 30-ks observation of PG115+080. The Chandra spectrum of the partial Einstein ring is also shown in Fig.7. An acceptable spectral model includes a plane-parallel shock plasma model and an absorbed power-law model. However, due to the low signal-to-noise ratio, a model consisting only of an absorbed power-law also provides a satisfactory fit to the data. We simulated an AXIS spectrum with a matching exposure time (173 ks) using the best-fit model to the Chandra spectrum of the partial Einstein ring (Fig. 7). The simulated AXIS observation of PG1115+080 shows that we will unambiguously detect the presence of shocked emission from the interaction of the UFO with the ISM.

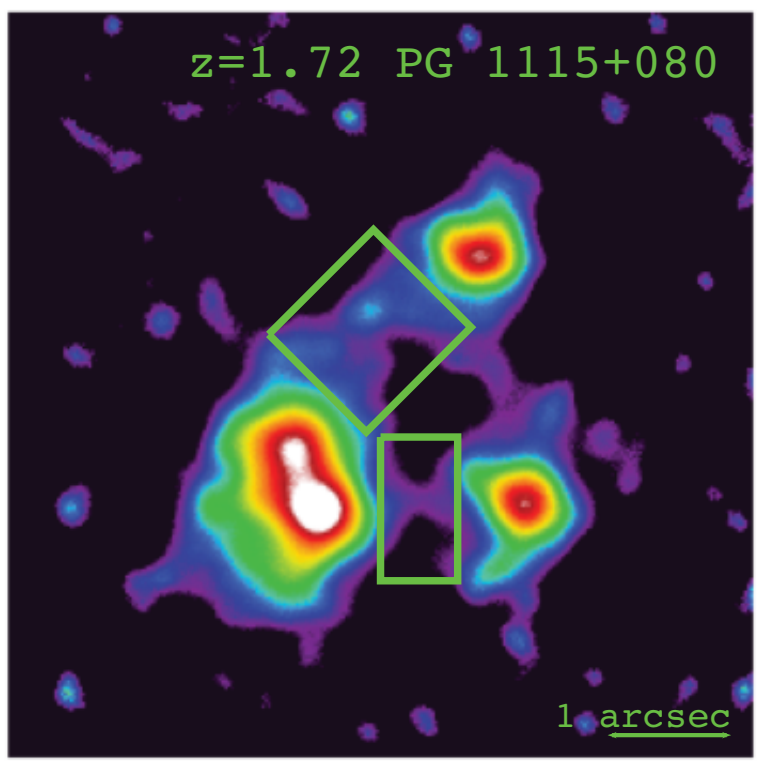

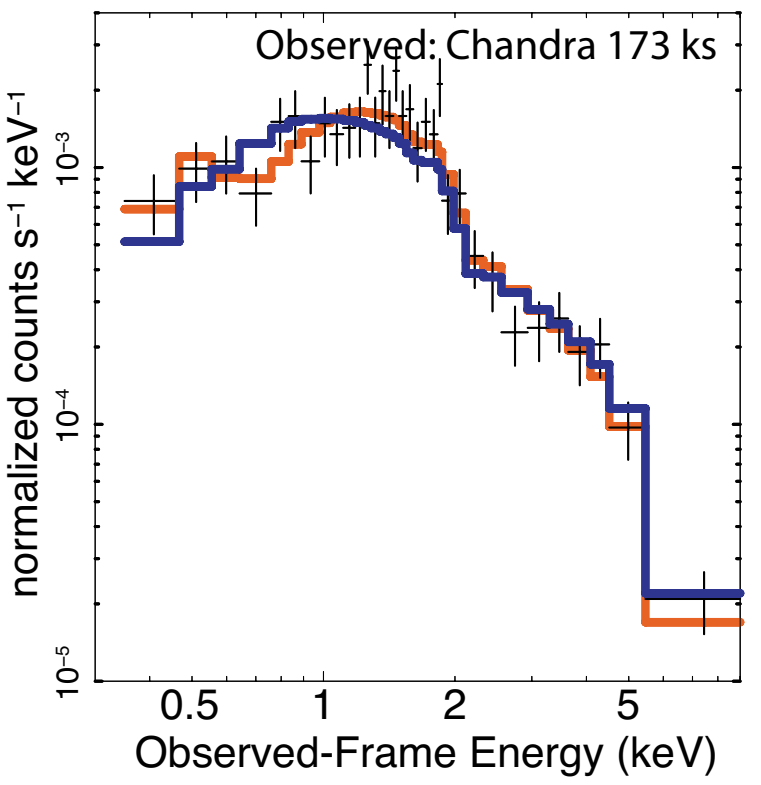

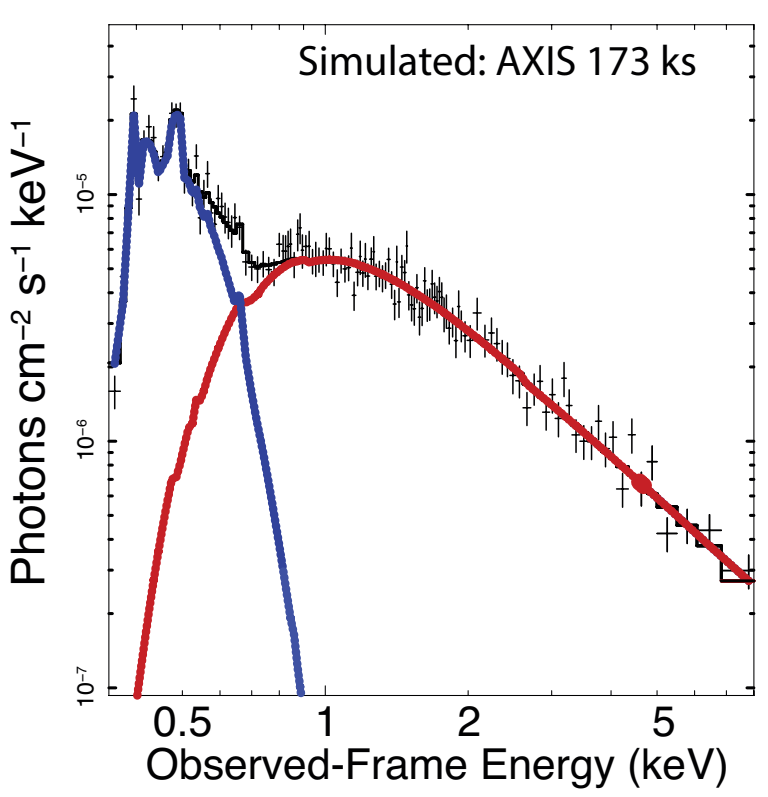

**Figure 7.** Left: A deconvolved Chandra image obtained by stacking all available observations of PG1115+080 (total exposure time of 173 ks). A partial X-ray Einstein ring is resolved between the lensed images. Center: The stacked 173 ks Chandra spectrum extracted from the regions (left) covering the partial X-ray Einstein ring. A model that includes a plane-parallel shock plasma component, along with a power-law component, is depicted in red. The spectrum can be equivalently fit by only a power-law model (blue). Right: A simulated 173 ks unfolded AXIS spectrum of the partial X-ray Einstein ring clearly resolves the shocked plasma component (blue) from the power-law component (red).

There is an important synergy between AXIS and other multi-wavelength observatories that will improve our understanding of galaxy feedback through multi-phase winds. Spectra and images from mm observatories (e.g., ALMA) will be used to determine the properties of the macro-scale winds. Numerical models of UFOs [611**?** ] predict that they drive powerful winds that heat and expel cold gas from the galaxy on larger scales. At present, there are few examples where the momentum flux of the AGN wind can be linked to a large-scale molecular outflow, but a combination of AXIS and ALMA can increase the sample and allow us to test models of this process. The Vera C. Rubin Observatory is predicted to discover $>$ 10,000 gravitationally lensed quasars, many of which will be resolved with AXIS [[366], see Fig.8]. The planned observing window of AXIS ($\sim 2032 - 2037$) will overlap with the multi-band photometric optical surveys of Rubin ($\sim 2025{-}2035$) and EUCLID. The eROSITA all-sky survey will provide the X-ray fluxes of the lensed quasars discovered by Rubin and other surveys (J-PAS, Pan-STARRS1, DESI legacy, and Dark Energy Survey), but cannot resolve the lensed images or extended emission. AXIS will target lensed systems that are X-ray bright enough to detect UFOs and Einstein rings. AXIS observations of a sample of high-$z$ lensed quasars will allow us to study both the central drivers of galaxy growth at the smallest scales ($1{-}100\ r_g$) and the interaction of the winds with the ISM at mesoscale ($\sim$ pc $-$ kpc).

The primary goals of our AXIS proposal are to:

(a) Detect absorption features related to UFOs in individual images. The images will be resolved with AXIS. This will allow us to study the variability of the outflow by taking advantage of the fact that the spectra of the images provide the outflow properties at different epochs separated by the time delays. The high signal-to-noise (S/N) spectra of the resolved images will allow us to determine the energetics and geometry of a UFO at a moderate redshift. The lensing magnification enables the study of outflows from quasars that would otherwise be too faint to detect. The relatively high redshifts of the quasars shift the absorption lines associated with the UFOs to energies where the effective area of AXIS peaks. We will search for a dependence of the outflow velocity with quasar luminosity and the ionization parameter of the outflowing wind.

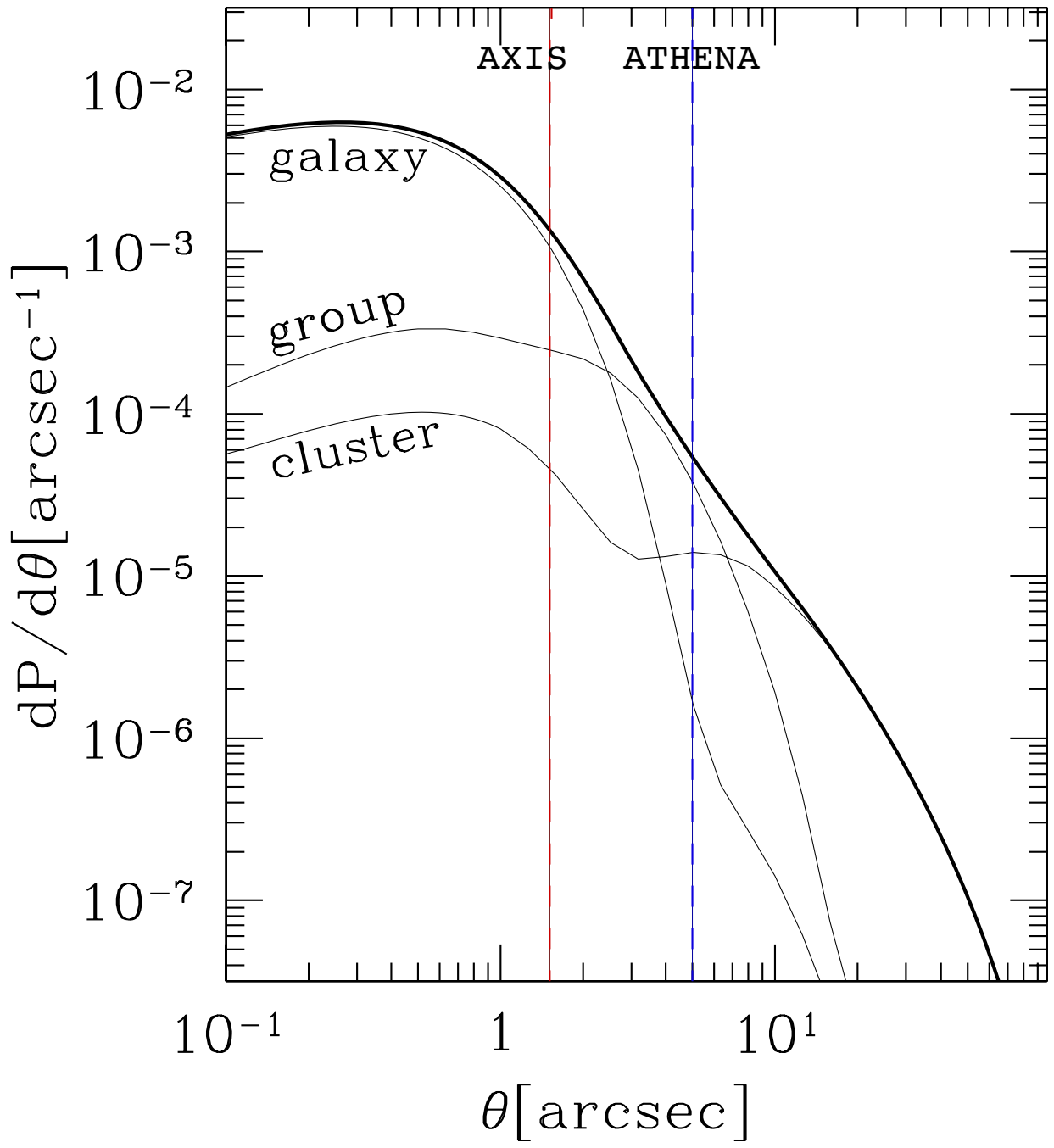

**Figure 8.** The distribution of lens image separations for three different lens types: galaxy, group, and cluster-scales, predicted by a halo model. The total distribution is shown by the thick line. The vertical lines represent the current best estimates of the half-power diameters of AXIS and Athena.

(b) Constrain the properties of a possible X-ray Einstein ring produced by the interaction of a small-scale quasar outflow interacting with kpc-scale gas. A direct detection of the interaction of a small-scale UFO originating at 100 $r_g$ with kpc-scale gas would confirm that UFOs are an important component of galaxy feedback. Finding lensed extended emission is easier than finding unlensed extended emission because the lens magnification enormously improves the contrast between the extended emission and the central quasar. Lens inversion of the partial X-ray Einstein ring will be used to infer the actual shape of the extended emission source.

**Exposure time (ks):** The total proposed exposure time is 300 ks.

**Observing description:** To select the best targets to study AGN feedback, we started from a list of 306 known lensed quasars, including ones detected in recent Gaia surveys. We found that 156/306 of the lensed quasars in our list fall in the area of the sky (with German data rights) covered in the 4 passes of the eROSITA survey (eRASS:4). By using a positional match within 10 arcsec ($\sim 2\sigma$ of the eROSITA positional uncertainty), we found 120/156 detections in eRASS:4. From this target, we filtered for lensed quasars with image separations greater than 1.5 arcsec and with combined 2 −10 keV fluxes exceeding 1 $\times$ $10^{-13}$ erg s$^{-1}$ cm$^{-2}$. This filtering resulted in 18 lensed quasars. Of these 18 lensed quasars, we selected to observe the four X-ray bright gravitationally lensed quasars PG1115+080 ($z$ = 1.72), WGD2038−4008 ($z$ = 0.777), WISEJ0259−1635 ($z = 2.16$), and J0921+2854 ($z = 1.41$). The proposed lensed quasars are known to contain powerful relativistic outflows and have shown hints of X-ray Einstein rings in Chandra observations.

**Joint Observations and synergies with other observatories in the 2030s:** ALMA, Rubin, Euclid (see above discussion)

**Special Requirements:** None

*7. Short time delays in gravitationally lensed quasar images*

**Science Area:** galaxies, AGN, quasars, gravitational lensing

**First Author:** Elena Fedorova (INAF-OAR, Italy)

**Co-authors:** Antonio del Popolo (Dipartimento di Fisica e Astronomia, University of Catania, Italy)

**Abstract:** Time delays in the images of gravitationally lensed quasars play a crucial role in understanding the geometry and physical content of the gravitational lens systems (GLS). In case of the short time delays on hourly/daily timescales, correlating the X-ray/gamma-ray data is the best way to determine them, as the variability of quasars at these energies is usually faster than at lower energies. Here we demonstrate the use of our cross-correlation/autocorrelation web tool to determine time delays in the famous GLS Q2237+0305 ("Huchra lens"/"Einstein Cross"), applying it to the Chandra and XMM-Newton lightcurves. We also propose a list of other GLS with delays presumably on the same timescales, to which this method can also be applied at the same energy range. The spatial resolution of AXIS, being able to resolve the quasar images in GLS with (sub)arcsecond angular dimensions, makes this task more promising even for such GLS.

**Science:** Gravitational Lensing (GL) of distant extragalactic sources such as quasars causes the appearance of multiple (two or more) images of the same object [412]. These images have similar spectral properties, but there are delays in the time of the signal arrival from different images caused by the differences in light paths deflected by the gravitational field of a lens. Measuring these time delays could provide important information about the structure of a gravitational lens system (GLS).

Time delays in real extragalactic GLS cover a wide span of values from hours or days to years. The longest time delay of more than six years was found in SDSS J1004+4112 [347]. The shortest time delays on hourly timescales are known in the famous Q2237+0305 "The Einstein Cross" [532]. Recently, GAIA GRAL disclosed several GLS with predicted daily time delays according to their lens density model[577].

Correlation analysis is one of the most effective tools for determining time delays in GLS from the light curves of the images of gravitationally lensed systems. For yearly to weekly timescales of delays, optical monitoring, such as OGLE https://ogle.astrouw.edu.pl/cont/4_main/len/huchra/ or COSMOGRAIL [496], can provide the best clue to estimate them with a high level of accuracy. However, for shorter time delays on hourly to daily timescales, X-ray photometry is necessary because the variability of quasars at these energies is typically faster. In this work, we demonstrate our web tool for correlation and autocorrelation analysis to estimate the time delays in Q2237+0305, "The Einstein Cross," using Chandra and XMM-Newton photometry data.

**Q2237+0305 the Einstein Cross:** The gravitational lens system (GLS) Q2237+0305 (Einstein Cross, [224]) consists of a quadruply-imaged quasar (with the redshift $z_Q = 1.695$) and the nearest known lensing galaxy (with the redshift $z_G = 0.0395$). The time delays between the images in the Einstein Cross are on hourly (less than daily) timescales. The timescales of the inner optical variations in the quasar are on the order of days; therefore, it is unlikely to improve this level of accuracy based on optical lightcurves. The detection of a variable X-ray flux from this system will lead to better-quality estimates of the relative time delays $\Delta\tau$ between the images, both due to faster temporal X-ray variability and shorter time bins in X-ray observations.

ROSAT made the first X-ray detection of Q2237+0305 in 1997 [564]; these observations revealed no significant flux variability. After ROSAT, the Chandra X-ray Observatory observed this GLS several times in 2000 and 2001 with the Advanced CCD Imaging Spectrometer. The longest of these two observations lasting 30 ks, carried out in September 2000, had detected the hourly-timescale inner flux variability which had given a clue to determine the time delay of $\Delta\tau_{BA} = 2.7^{+0.5}_{-0.9}$ hours between two of the four

GLS images [115]. The first XMM-Newton observation of Q2237+0305 had not revealed any signs of variability [159]. Later observations of the "Einstein Cross" by Chandra were synchronized in time with the strong microlensing events observed by OGLE. These observations revealed bright microlensing-induced peaks on weekly timescales in the quasar image light curves [93], but did not demonstrate inner quasar variability.

**Auto/cross-correlation functions and the web tool for correlation analysis:** Cross-correlation functions (CCF) are used quite extensively in astronomy to trace the links between variable parameters. For two discrete sets of random observed values **X** and **Y**, the cross-correlation function is: $CF_n(\mathbf{X},\mathbf{Y}) = \frac{\sum_{\mathbf{k=0}}^{\mathbf{K}}(\mathbf{X_k}-\overline{\mathbf{X}}_\mathbf{0})(\mathbf{Y_{k+n}}-\overline{\mathbf{Y}}_\mathbf{n})}{\mathbf{var_0}(\mathbf{X})\mathbf{var_n}(\mathbf{Y})}$, where $\overline{X}_0 = \frac{1}{K}\sum_{k=0}^{K} X_k$, $\overline{Y}_n = \frac{1}{K}\sum_{k=0}^{K} Y_{k+n}$ are the mean values of X and Y over the subset of K values starting from 0 and the n*th* ones, and $Var_n(Y) = \sqrt{\frac{1}{K}\sum_{k=0}^{K} Y_{k+n}^2 - \overline{Y}_n^2}$, $Var_0(X) = \sqrt{\frac{1}{K}\sum_{k=0}^{K} X_k^2 - \overline{X}_0^2}$ are the variances of Y and X over the subset of K values starting from the n*th* and 0 ones.

CCF represents the degree of similarity between the two series, one of which is lagged. It can be very useful for determining the time delay between two signals, for instance, time delays between images of the same emission source in a GLS. The maximum value of the cross-correlation function indicates the value of the time delay when the signals are best aligned; thus, the time delay between the two signals is determined by the argument of the maximum of the cross-correlation, namely $\tau = arg_max_{t\in\mathbb{R}}((\mathbf{X} * \mathbf{Y})(t))$.

The autocorrelation function (ACF) is similar to CCF, but uses the same series twice instead of two different ones (one beginning from its origin and the other lagged with a varying value of the lag). The autocorrelation function is a helpful tool for revealing periodicity (in this case, the ACF would also be periodic) or the superposition of several signals lagged one with respect to another (the case of GL images). The tool is under development at https://www.ict.inaf.it/gitlab/olena.fedorova/autocorrelationfunctions.

**Time delays in Q2237+0305 images:** Angular resolution of Chandra ensures the separation of the images of the quasar in this GLS. This ensures that the delay we can determine from the CCF is indeed the one between the two images whose lightcurves we analyze. However, the longest exposures provided by Chandra are on the order of 30 ksec, and thus we cannot observe the longer time delays from Chandra light curves and corresponding CCFs. This renders them useless for the C image in Q2237+0305, and thus, we excluded the light curves of this image from our consideration of Chandra light curves. XMM-Newton, on the other hand, has significantly longer exposures (up to 150 ksec), and longer time delays can also be observed. However, on the other hand, its angular resolution is significantly worse than that of Chandra, and the images cannot be resolved with it, resulting in a single total light curve for all images. That is why we cannot be sure when attributing the time delays found in XMM-Newton all-images light curve. Based on our consideration, we excluded Chandra observations with exposures shorter than 15 ks and those with variances in all images in 1000 sec-bin light curves that were lower than 1.0. The CCF example is shown in Fig.9.

The averaged values of the time delay between images A and B are compatible within the error bars with those found in [115] and [33]. The time delays between images D and A (with D leading) and between D and B are also compatible with those found in [33].

All three observations of XMM-Newton demonstrate enough variability in 1000sec bin lightcurves, so we performed the ACF analysis for all of them, despite the fact that the exposure of the second one is quite short (25 ksec). CCF of the longest observation is shown in Fig.10.

Time delays of 3.1$\pm$0.3 h (1st observation) and 2.9$\pm$0.3 h (2nd observation) found in the first two observations can be interpreted rather as the delay between B and A images, as well as the delay of $1.1^{+0.8}_{-0.3}$ hours found in the 2nd observation's ACF only can be attributed to the delay between D and B images. The time delay of 3.6±0.4 h (3rd observation) can be attributed to the delays between B and or A and D images (or to some combination of them); a 4.6±0.3 hours delay found in the CCF of the second observation can also be interpreted as the delay between D and A. Furthermore, in the first light curve, two additional delays were detected: 6.0±0.3 hours and 8.1±0.3 hours. Those, as well as the second delay value found in the ACF of the third XMM-Newton observation, 6.9±0.6 h, can be compared with the values predicted for D and B images by the SIE and Bar Accounted lens model for this GLS [33]. The longer time delays found in the third and longest XMM-Newton light curves are $17.8^{+4.4}_{-3.9}$ hours, 25.0±1.7 hours, and 29.7±0.8 hours. The first of them agrees with the value found for images C (leading) and A in [33]; the interpretation of the other two is unclear.

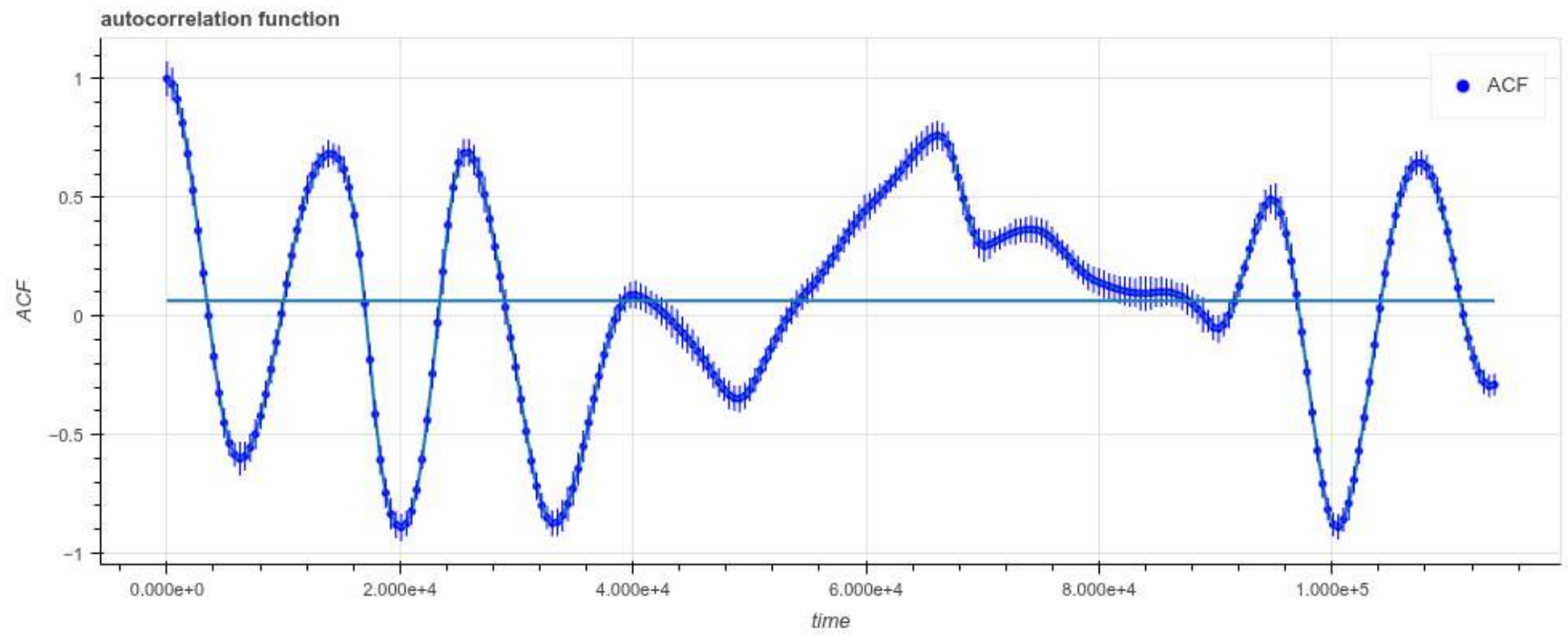

**Figure 9.** Chandra CCF for images A and B during the first observation (ID 431) in 2000.

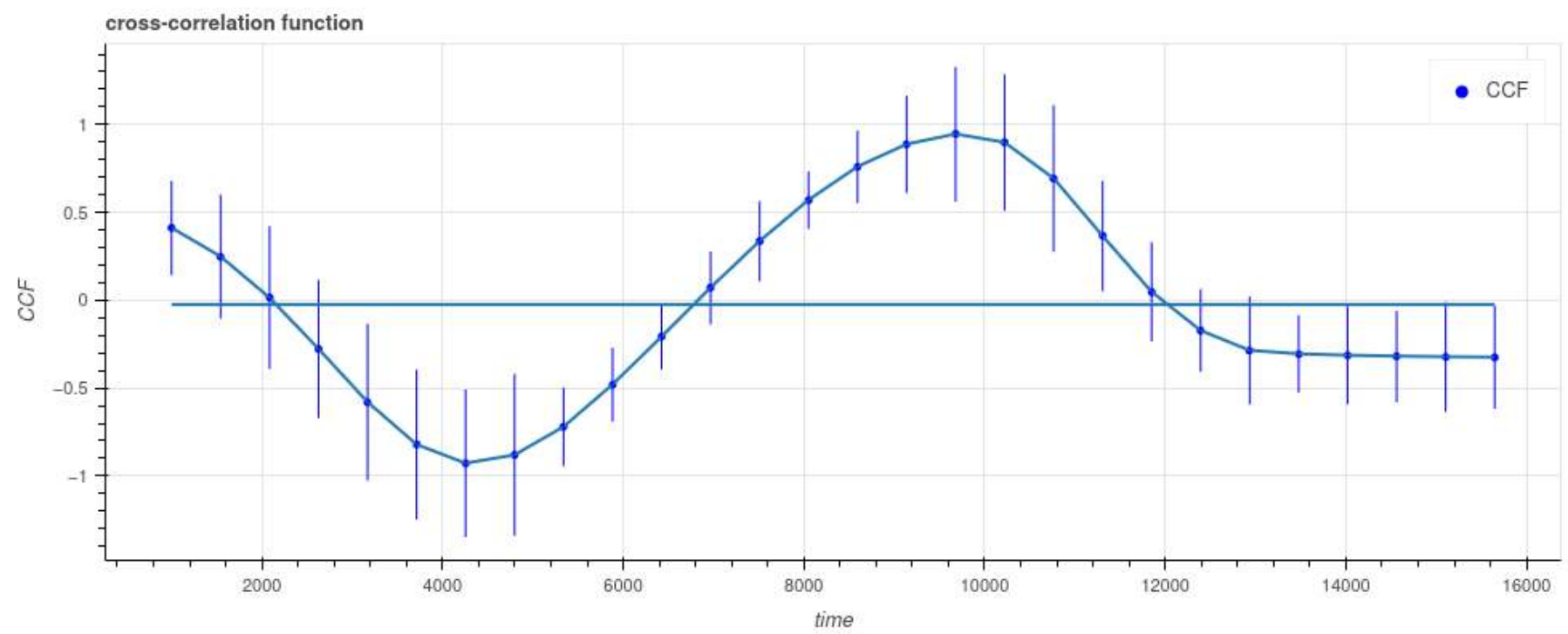

**Figure 10.** ACF to the EPIC lightcurve during the third XMM-Newton observation of Q2237+0305.

**Exposure time (ks):** Requested exposure times are $40 - 400$ ks. See Table 2. Total exposure time requested is 2.65 Ms.

**Observing description:** The XMM-Newton and Chandra datasets discussed here demonstrate the values of some of the time delays in the GLS Q227+0305 "Einstein Cross". Thanks to the outstanding angular resolution of Chandra, the images of quasars in this GLS can be separated, but the exposures of Chandra observations are too short to determine the rest of them. Longer XMM-Newton exposures enable us to determine longer time delays, but not to attribute them to specific images, as the angular resolution of XMM-Newton is not sufficient to resolve them. Longer exposures with AXIS would be very desirable to determine time delays not only in Q2237+0305, but in other GLS with hourly to daily timescales of delays. In Table 2, we demonstrate a sample of such objects; the images separated by the angular distances shown in the fourth column could be distinguished by AXIS with its 1.5"-2.0" PSF (for comparison, the PSF of Chandra/ACIS is near 0.5" at 1 keV, but grows up to 8" with the energy). The 5" PSF provided by ATHENA is not enough to distinguish the majority of images in Table 2, but it can be used in combination with AXIS, especially for the observations scheduled soon after each other.

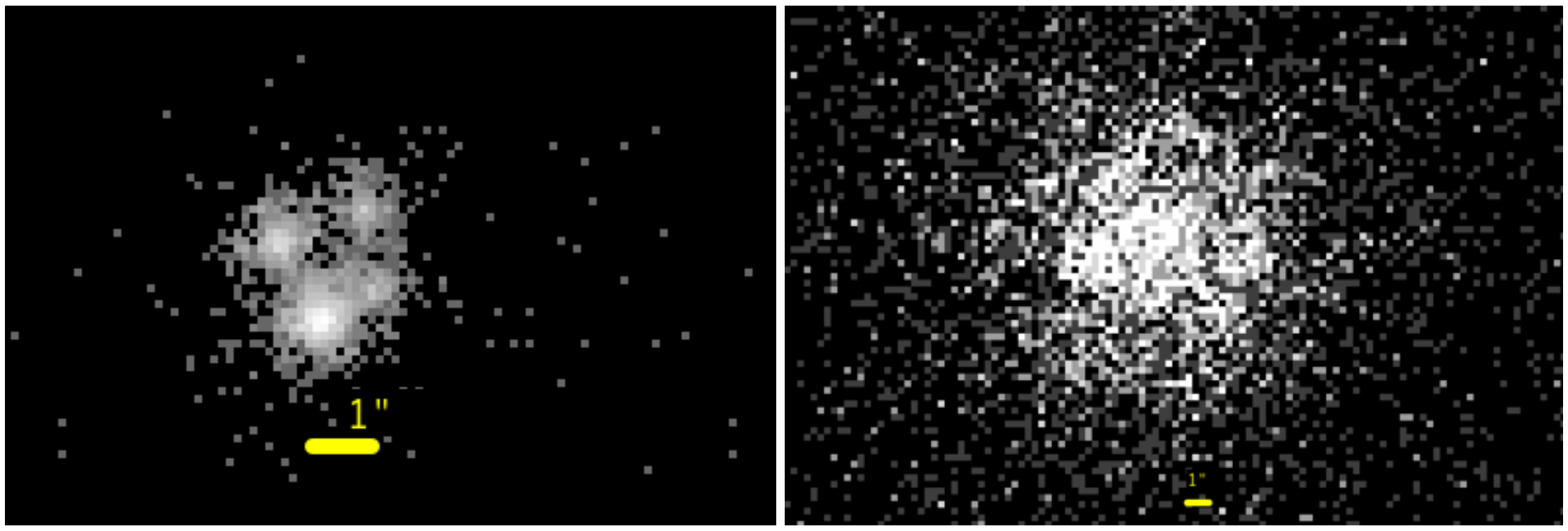

**Figure 11.** Chandra/ACIS (left) and XMM-Newton/EPIC PN (right) images of the Einstein Cross obtained during the first Chandra observation (ID 431, August 2000) and the last XMM-Newton observation (ID 0823730101, May 2018).

**Table 2.** GLS with (lens model predicted) short time delays.

| Name | $N_{img}$ | images | angular distance | (predicted) time delay | obs. exposure time |
|---|---|---|---|---|---|
| Q2237+0305 | 4 | A-B | 1.83" | 2-6 hours | 15 hours |
| | | C-A | 1.36" | 14-35 hours | 40 hours |
| | | D-A | 1.01" | 2-6 hours | 15 hours |
| | | C-B | 1.4" | 19-45 hours | 50 hours |
| | | C-D | 1.66" | 11-35 hours | 40 hours |
| | | D-B | 1.17" | 3-8 hours | 15 hours |
| RX J1131-1231 | 4 | A-B | 1.1" | 0.7±1.4 days | 60 hours |
| | | B-C | 2.34" | 0.4±2.0 days | 50 hours |
| B1422+231 | 2 | A-B | 0.5" | 1.5±1.4 days | 75 hours |
| HE 0435-1223 | 4 | C-A | 2.54" | 2.1±0.7 days | 75 hours |
| PS1 J0147+4630 | 2 | B-A | 0.5" | 2.2±2.1 days | 100 hours |
| WFI 2026-4536 | 4 | A1-A2 | 0.33 " | 2.4 hours | 10 hours |
| WFI 2033-4723 | 4 | A1-A2 | 0.72" | 1 day | 30 hours |
| PG1115+080 | 4 | A1-A2 | 0.43" | 3.6±0.5 hours | 10 hours |
| B1608+656 | 4 | A-C | 0.87" | -9..0 days | 50 hours |
| RX 0911+0551 | 4 | A-C | 0.96" | -24..36 days | 50 hours |
| | | B-C | 0.57" | -49..59 days | 50 hours |

**Joint Observations and synergies with other observatories in the 2030s:** Joint monitoring of AXIS and ATHENA in X-rays not only for time delays estimation, but also to obtain long-duration lightcurves of images and trace out microlensing events. Additionally, optical monitoring of these GLS would be desirable to estimate the sizes of the emitting regions (such as COSMOGRAIL or OGLE).

**Special Requirements:** None

## c. Black Hole Feedback: Radio Mode

### *8. AGN Feedback and Heating Mechanisms in Galaxy Groups*

**Science Area: galaxy groups, intracluster medium, AGN feedback, shock heating**

**First Author:** Scott W. Randall (Center for Astrophysics | Harvard & Smithsonian)

**Co-authors:** Yuanyuan Su (Univ of Kentucky), Arnab Sarkar (MIT)

**Abstract:** One of the most important revelations in the field of galaxy evolution over the past several decades has been an understanding of the importance of feedback in galaxies, galaxy groups, and galaxy clusters. For example, cosmological simulations are unable to reproduce fundamental features of the observable universe without including feedback, although implementation methods and details tend to vary from simulation to simulation. Galaxy groups represent a particularly interesting population for studying AGN feedback, as they contain the bulk of the baryons in the local Universe, and their relatively shallow gravitational potentials lead to feedback having a greater impact on the intra-group medium (IGrM) compared to rich galaxy clusters. Despite the importance of AGN feedback, the details regarding how the kinetic energy output of the AGN couples to the IGrM are relatively poorly understood. Detailed studies of the imprints of AGN feedback on the surrounding diffuse gas, in the form of cavities evacuated by jets and shocks driven by the expanding cavities, have and will continue to play an important role in solving this issue. Although exploring AGN feedback in groups with existing facilities has been possible, detailed studies are limited to relatively nearby systems, with only one well-studied case with clear detections of outburst shocks [407]. With its high effective area, especially below 1-2 keV, and high angular resolution, AXIS will dramatically open up the discovery space for detailed feedback studies in groups. Here, we consider AXIS's ability to detect cavities and shocks resulting from AGN feedback at relatively large distances and suggest potential paths forward to identify targets of interest.

**Science:** Centrally located low temperature gas and associated star formation in galaxy clusters and groups is not observed at the level expected based on the observed radiative cooling rates in the intracluster medium (ICM)/IGrM (the so-called "cooling flow" problem) [326]. The generally accepted solution to this puzzle is AGN feedback, where a negative feedback loop is established between the central AGN, which outbursts and heats the diffuse gas, and the ICM/IGrM, which cools and flows onto the AGN, driving outbursts. In general, the central AGN has shown that there is enough kinetic energy output to compensate for the cooling of ICM / IGrM [146], although the details of how the energy is transferred from the AGN to the ICM/IGrM are poorly understood.

Galaxy groups are of particular interest for understanding AGN feedback and its impact on cosmic evolution. First, the bulk of baryons in the universe is contained within groups, as opposed to more massive (and better-studied) galaxy clusters. Second, groups are expected to be impacted more strongly by feedback, since the energy available from the central AGN is comparable to that of the IGrM. This means that studies of AGN feedback in rich clusters cannot be simply extrapolated to the group scale, for example, when considering methods for incorporating feedback into cosmological simulations.

Despite the incomplete picture of the details of IGrM heating in AGN feedback, it is clear that the formation of cavities and shocks in the IGrM (driven by AGN jets) plays a significant role. Although there are many examples of the former, there are relatively few confirmed cases of the latter due to the difficulties in detecting weak outburst shocks. In one very well-studied case, NGC 5813 [407], repeated outburst shocks alone are sufficient to compensate for radiative cooling of the IGrM within the cooling radius. This raises the question of where the energy associated with the observed cavities ultimately ends up and how that impacts cosmic evolution. This energy may extend to heat the IGrM beyond the cooling radius, potentially even driving some of this gas outside the virial radius, thereby providing the surrounding large-scale environment with a source of metal-enriched gas that has been processed within galaxy groups. It is desirable to identify and characterize more systems like NGC 5813 to establish a scaling relation between the energy in shocks and that required for offset cooling.

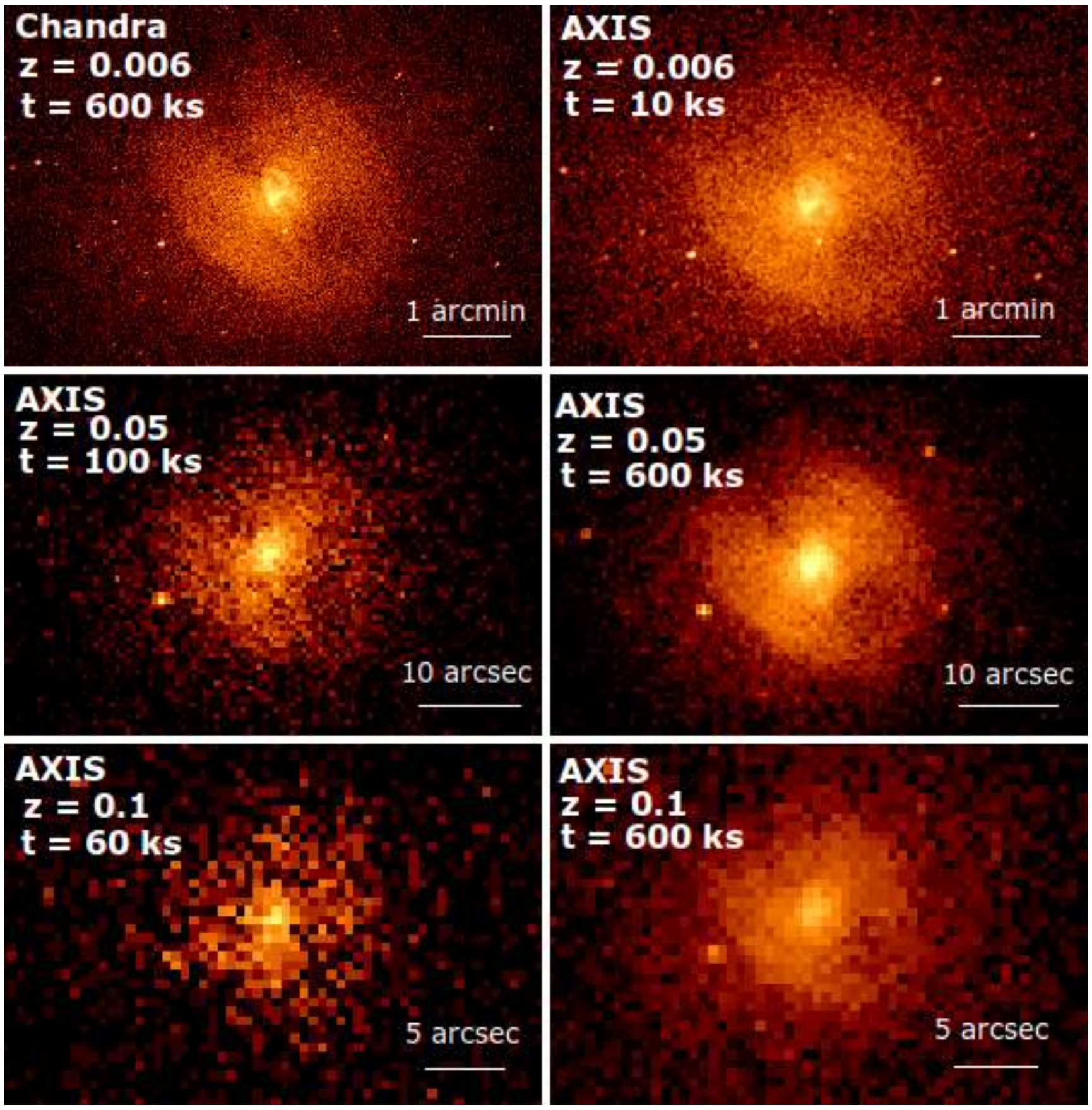

**Figure 12.** The existing 600 ks 0.3–3 keV band Chandra image of NGC 5813 is shown in the upper left. Other panels display the AXIS simulated view of the same system at various redshifts and with different exposure times (inset). The 0.3-3 keV source count rates are 6.7 cts/s at $z = 0.006$, 0.11 cts/s at $z = 0.05$, and 0.022 cts/s at $z = 0.1$.

With its high angular resolution across the FOV and high effective area at low energies, AXIS is extremely well suited to detecting cavities and shocks, as well as mapping the metal distribution, in galaxy groups. At the same time, with proper modeling of the instrumental and sky backgrounds, there is the potential to measure thermodynamic profiles of the IGrM out to large radii and to search for signs of global entropy injection driven by AGN feedback. In particular, AXIS will be able to extend this work to higher redshifts than is possible with currently available facilities, probing the scale and evolution of feedback in groups across cosmic time.

As a case study to demonstrate what AXIS will be capable of, we consider the well-known low-mass galaxy group NGC 5813. This system is a nearby ($z \approx 0.006$), low-mass group, with $kT \approx 0.7$ keV. It shows three pairs of cavities, with each pair associated with an elliptical outburst shock front. As such, it is extraordinarily well-suited to study the cumulative effects of AGN feedback in groups. [407] shows that, in this system, the cumulative heating of outburst shocks alone is sufficient to offset radiative cooling within the cooling radius. This is the only known galaxy group where this analysis is possible. AXIS will significantly expand the redshift over which systems like NGC 5813 can be identified and studied with practical exposure times.

To demonstrate this point, we used SIXTE [118] to perform simulated AXIS observations of NGC 5813. The results are shown in Fig. 12. The existing 600 ks Chandra image is shown in the upper left. The panel in the upper right shows that with just a 10 ks AXIS snapshot observation, the salient features of AGN feedback are detected, with the inner two pairs of cavities and their associated shocks clearly visible. The third, outer pair of cavities and the associated shock edge are also detected (not shown here due to space restrictions). As shown in the second row of Fig 12, at the $\approx$ 8 times higher redshift of $z = 0.05$, AXIS can detect cavities and shock fronts in a 100 ks observation (although the innermost cavities and shock edge are not well resolved). In 600 ks at $z = 0.05$, all outer features are clearly detected, and the structure of the innermost outburst features is somewhat visible. Finally, as shown in the bottom row, at $z = 0.1$, the intermediate shock front is only marginally suggested in a 60 ks observation (although the group is strongly detected, with $\approx$ 1500 net counts). In a 600 ks observation, the inner outburst features are not resolved, but the intermediate and outer features are clearly detected. This difference in redshift, from $z \approx 0.006$ to $z = 0.1$, corresponds to more than a factor of 300 increase in the comoving volume explored. Despite this, we will focus on groups in the nearby universe to maximize observing efficiency.

**Exposure time (ks):** 2 Msec

**Observing description:** The proposed observations will include a TBD sample of groups, e.g., chosen from eROSITA catalogues. The exact composition of this sample will depend on trade-offs between the ranges probed for radius, redshift, mass, and large-scale environment versus the overall exposure time. For example, the simulations presented above show that for as little as 200 ks we could probe detailed signatures of AGN feedback, detecting any apparent cavities **and** shocks, in at least 20 low-mass groups in the local Universe. Our results also show that we can identify targets of interest serendipitously in moderate exposure ($\sim$ 50 ks) observations out to at least $z = 0.1$, and clearly detect salient feedback features at this redshift with deep (targeted or serendipitous) $\sim$ 600 ks observations, even in a low-mass $kT \approx 0.7$ keV group like NGC 5813. Here, we propose to observe a sample of 20 groups out to $z = 0.05$, with an exposure time of up to 100 ks per group for a total of 2 Msec.

**Joint Observations and synergies with other observatories in the 2030s:** Next generation radio observatories (e.g., SKA, ngVLA) will reveal the jet emission. Multi-wavelength observations of cool gas nebulae will allow us to understand the triggers of thermal instability (ALMA, JWST, ELTs). NewAthena will track the motions of hot gas driven by bubbles.

**Special Requirements:** None

*9. Detecting ghost bubbles in nearby galaxy clusters and groups*

**Science Area: Galaxy clusters, intracluster medium**

**First Author:** Congyao Zhang (Masaryk University, The University of Chicago)

**Abstract:** Radio-mode AGN feedback is a key mechanism for regulating the thermal state of the ICM. X-ray cavities inflated by radio jets play a central role in distributing energy within cluster cool cores. However, the processes by which this energy is transferred and thermalized remain unclear. *Chandra* has revealed X-ray cavities at the centers of numerous cool core clusters. In the brightest systems, multiple pairs of cavities that rise buoyantly to distances of tens of kpc from the central AGN have often been observed. Although their radio emission spectrally ages, X-ray observations reveal that these 'ghost' bubbles remain intact. AXIS, with its dramatic expansion in sensitivity and sharp spatial resolution across the field of view, will uniquely enable the ability to reveal the fate of these ghost bubbles: how far they travel and distribute energy, whether they break up, merge or pile up at a particular radius, and the history of AGN activity heating cluster atmospheres. AXIS will thereby provide critical insights into how AGN feedback operates over cosmic timescales and maintains the heating–cooling balance in galaxy clusters.

**Science:** Radio-mode AGN feedback has been widely accepted as one of the most plausible heating mechanisms in galaxy clusters [146]. Fed by accretion, the central SMBHs power radio jets and inflate bubbles of relativistic plasma. Current X-ray observations have shown that most of the energy from the SMBHs is deposited into the bubbles [166,494], implying an essential role of these bubbles in preventing the ICM from catastrophic overcooling [103]. However, through which physical process(es) such energy is distributed and thermalized in the atmosphere is still an open question. The answer heavily depends on the dynamical behaviors of the bubble-ICM interaction. For example, long-lived bubbles generate efficient turbulence and g-waves in their wakes, and uplift a large amount of low-entropy gas from the cluster center [598,599,609]. On the contrary, if bubbles break shortly after their formation, they would quickly release cosmic rays, mixing with the surrounding atmosphere near the cluster center [437].

Searching for bubbles and imaging their morphology at large cluster radii, ideally within the entire cool-core region ($\simeq 100 - 200$ kpc in massive clusters), would be crucial for understanding how AGN feedback works in the ICM. An essential observational feature of ancient AGN bubbles is a deficit in X-ray surface brightness, known as ghost X-ray cavities. The bubbles' radio counterparts unfortunately fade away rapidly at high frequencies due to their short cooling timescale, and often mix with the complex diffusive radio halo emissions at low frequencies (e.g., van Weeren et al. [535]).

Detecting ghost bubbles is only possible with deep, high-resolution X-ray imaging observations, which began with Chandra. One of the ideal targets is the Perseus Cluster, well-known for its bright X-ray emission and ongoing black hole activity [e.g., 101,167,531,535]. Bubbles $\simeq 30 - 40$ kpc away from the Perseus center have been identified unambiguously in the $\simeq 1$ Ms exposure Chandra map [150]. Some faint features tentatively suggest the presence of even older bubbles at larger radii ($\sim 100 - 200$ kpc) [153]. AXIS will offer unprecedented capabilities to advance our detection of the ghost bubbles, pushing the detection limit to larger radii and resolving the bubble's sharp edges, given its large effective area ($\sim 6\times$ larger than ACIS/Chandra at 1 keV), high angular resolution, and uniform PSF across the entire FOV. Fig. 13 shows a comparison of mock Chandra and AXIS images of a Perseus-like cluster with 500 ks exposure each. As the scientific goal is to uncover new, faint features, estimating the required exposure time is inherently non-trivial. Scaling with the existing Chandra data of Perseus, $\sim 5 - 10$ Ms Chandra-equivalent exposure would be generally needed to significantly advance our current detection, which approximately corresponds to $\sim 1$ Ms AXIS exposure.

Probing ghost bubbles provides a unique opportunity to trace the evolution of the AGN activity over hundreds of millions of years in individual clusters. Deep AXIS observations will enable us to quantify the number of bubbles, their radial distributions, and bubble morphologies in our nearby clusters and groups. Combining the gas density and pressure radial profiles of the ICM, the age and total enthalpy of the bubbles will be estimated [e.g., 40,500], characterizing the history of the cluster's AGN power, which is key to understanding the heating-cooling balance of the ICM. Note that bubble age is inversely proportional to its terminal velocity [598], strongly depending on the bubble size and morphology. Detailed imaging of bubble boundaries is vital for this analysis.

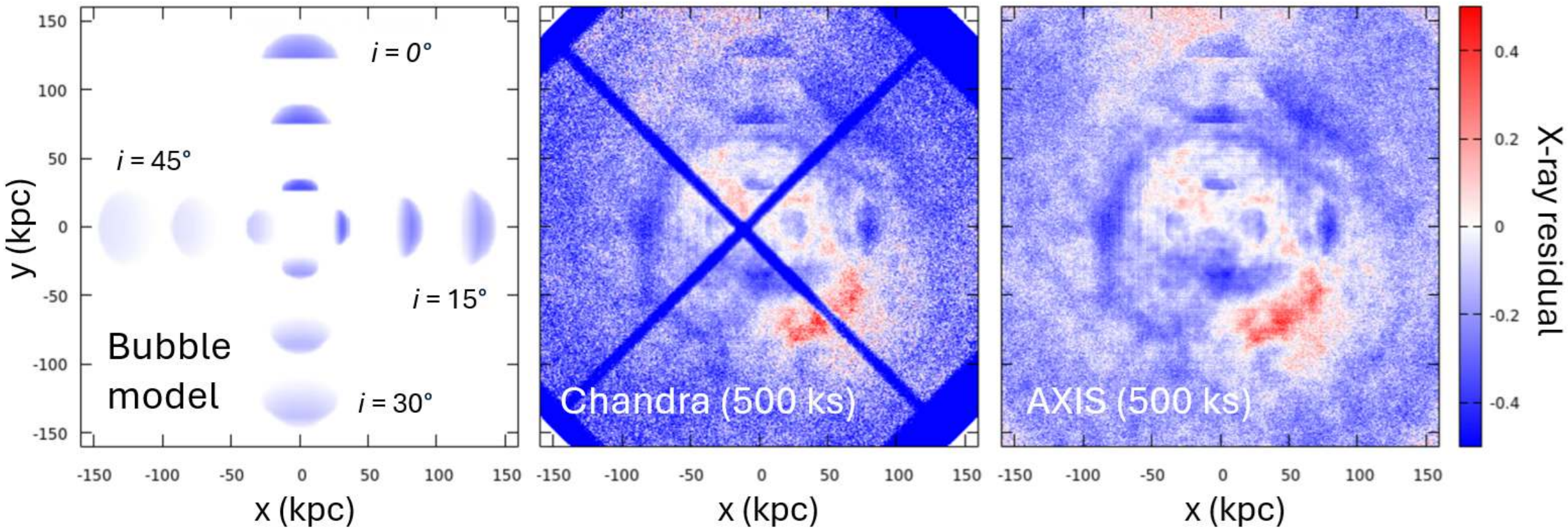

**Figure 13.** A comparison of Chandra (middle) and AXIS (right) mock residual images of a Perseus-like cluster with 500 ks exposure. The left panel indicates the input bubble models. All bubbles are intrinsically in the shape of a spherical cap with an aspect ratio of 3. The mock image displays multiple bubbles at various cluster radii, differing in size and inclination angle. X-ray surface brightness fluctuations caused by stratified turbulence are included in the mock to mimic the complex X-ray residuals of galaxy clusters, motivated by Perseus.

**Exposure time (ks):** Total 1.8 Ms

**Observing description:** Priority targets would be those with multiple pairs of cavities revealed in existing *Chandra* observations, including Perseus, Hydra A, and NGC 5813. To significantly advance current bubble measurements, we estimate that exposures approximately 5–10 times deeper than existing *Chandra* data are required. Based on this, we request AXIS exposures of 1000 ks for Perseus, 200 ks for Hydra A, and 600 ks for NGC 5813.

**Joint Observations and synergies with other observatories in the 2030s:** Next generation radio observatories (e.g. SKA, ngVLA), particularly at low frequencies for ghost bubbles, to reveal the aging radio-emitting plasma. Multi-wavelength observations of cool gas nebulae may reveal filamentary structures, ongoing gas cooling, or star formation at large radii triggered by the bubbles' passage (ALMA, JWST, ELTs). HUBS and NewAthena will track the motions of hot and warm gas driven by bubbles.

**Special Requirements:** None

*10. Heating the intracluster medium with AGN-driven shock fronts*

**Science Area: galaxy clusters, intracluster medium, AGN feedback**

**First Author:** Francesco Ubertosi (University of Bologna)

**Co-authors:** Myriam Gitti (University of Bologna), Fabrizio Brighenti (University of Bologna), Massimo Gaspari (University of Modena & Reggio Emilia)

**Abstract:** The brightest central galaxies (BCGs) of cool core galaxy clusters and groups are largely affected by the activity of the supermassive black hole (SMBH) they host. The footprints of active galactic nucleus (AGN) feedback, predominantly observed in X-ray emissions as X-ray cavities and shock fronts, trace the thermodynamic regulation of the surrounding intracluster or intragroup medium (ICM/IGrM). However, there is a significant imbalance in the study of feedback signatures – with shock fronts being largely overlooked, also due to the observational challenges associated with their detection. Existing studies and numerical results suggest that the AGN energy budget may be underestimated when relying solely on cavities; therefore, it is also important to study the effects of shock and sound waves, which influence the isotropic heating of the ICM. AXIS, with its high angular resolution and sensitivity, will transform our understanding of these phenomena. With deep exposures of exemplary feedback cases, we can finally open a window on the connection between shocks, radio jets, and X-ray cavities, and detect low-Mach-number waves. Furthermore, an AXIS survey of cool core clusters and groups could provide a comprehensive census of the occurrence, properties, and impact on the feedback cycle of shock fronts, as well as address existing uncertainties regarding the relationship between the mechanical power and radio luminosity of relativistic jets.

**Science:** It is well established that the central active galactic nucleus (AGN) of cool core galaxy clusters and groups deposits energy in the intracluster or intragroup medium (ICM/IGrM) on tens and hundreds of kpc. This *AGN feedback* mechanism is mostly imprinted in the X-ray-emitting ICM/IGrM. Specifically, the AGN drives relativistic, radio-emitting jets that act as supersonic pistons on the hot gas, excavating bubbles (X-ray cavities) and driving shock waves in the plasma (e.g., [122,146,193,325,326]). Among the footprints of feedback, AGN-inflated cavities are the most evident outcome of radio lobe expansion in the ICM. Cavities appear to be present in ∼46% of mass-selected cool cores [368] and in 60%-90% of X-ray-selected cool cores [41,146,382]. With more than a hundred known cases of galaxy clusters or groups with X-ray cavities (e.g., [463]), their global role in heating the environment has been extensively investigated. For instance, since the bubble enthalpy is a proxy for the energy deposited by the jet, the study of X-ray cavities has provided constraints on, for example, the balance between the mechanical heating power of the AGN and radiative cooling losses of the ICM; the relation between the synchrotron radio power and the kinetic power of the jet; and the duty cycle of AGN activity in the cool cores of groups and clusters [40,122,146,193,325,326,374].

*Unlike cavities, AGN-driven shock fronts have been extremely overlooked*. Due to the observational difficulty of detecting them – requiring both high spatial resolution (≈ arcsec) to identify the edges in surface brightness, and high sensitivity to collect enough counts to significantly measure the shallow associated deprojected temperature jumps ($T_{in}/T_{out} \sim 1.5$, corresponding to a Mach number $\mathcal{M} \sim 1.5$) – only a dozen systems with shock fronts are known (see tab. 3 in [523]). As such, aside from individual case studies, only a few observational works have focused on the properties of jet-driven shocks [281,362,523]. The known shocks have Mach numbers in the range $\mathcal{M} \sim 1.2 - 2.0$, and are usually spatially correlated with X-ray cavities and radio lobes (e.g., [195,526]). However, while several clusters and groups with multiple X-ray cavities at different distances from the center are known, tracing successive AGN outbursts, there are only three clusters with detections of more than one shock front (see Figure 14). *Overall, the role of*

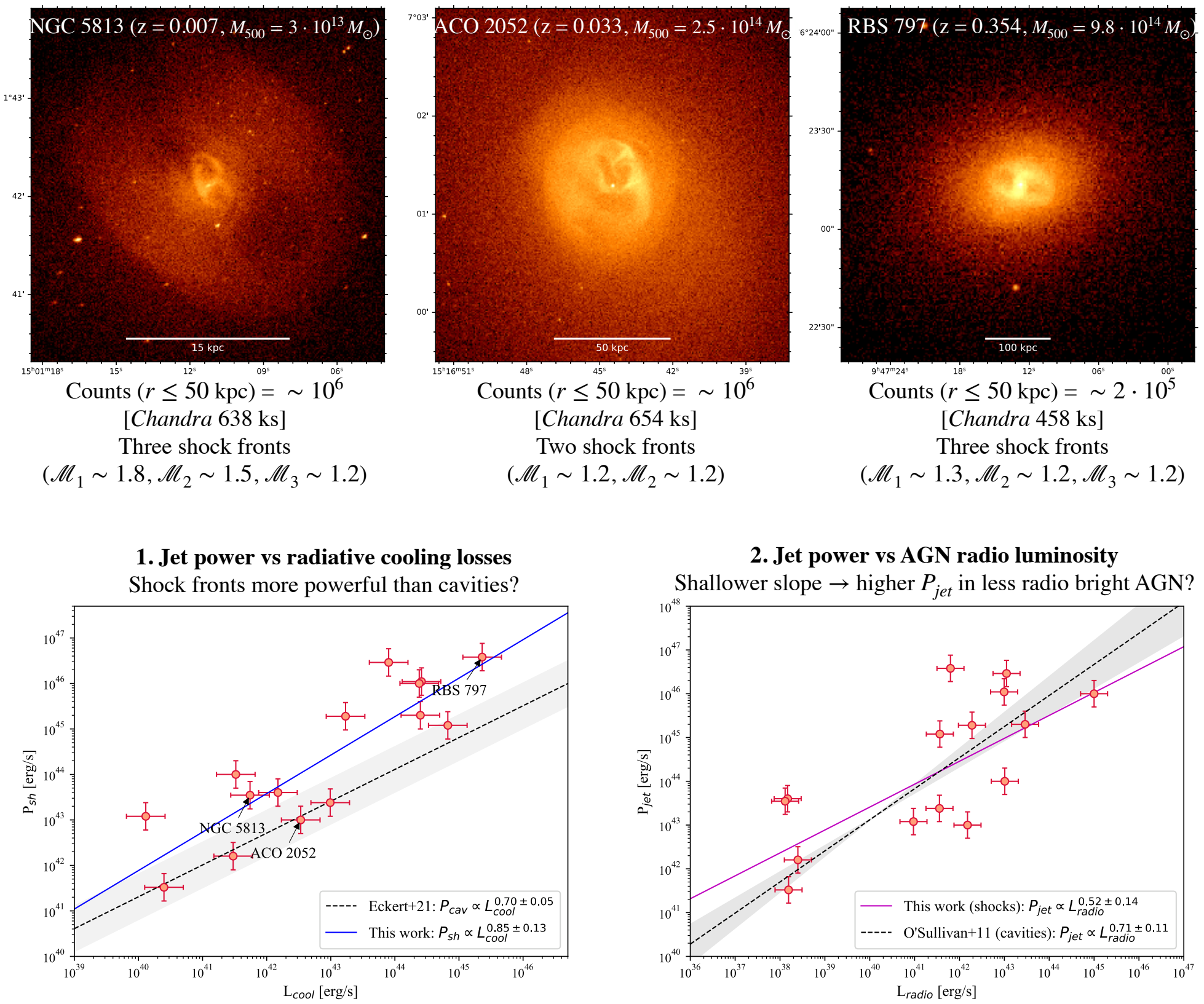

**Figure 14.** *Top panels: Chandra* 0.5 - 7 keV images of the known cases of multiple AGN-driven shock fronts [43,407,523]. *Bottom panels: Left:* Jet power measured using shock fronts versus radiative losses of the hot gas (bolometric X-ray luminosity within a radius where the cooling time is lower than 7.7 Gyr). The best-fit blue solid line is compared with the relation found by [133] using the jet power measured using X-ray cavities (black dashed line). *Right:* Jet power from shock fronts versus the 10 MHz - 100 GHz radio luminosity. The best-fit pink solid line is compared with the relation found by [374] using the jet power measured from X-ray cavities (black dashed line). To account for the different methods employed in the literature to measure these quantities, we assumed a 50% relative uncertainty. These plots suggest that $P_{cav}$ underestimates the AGN energy budget; the relation between jet power and cooling luminosity may be steeper than previously thought; and the jet power may be a shallower function of the synchrotron luminosity, compared to results obtained from cavities. However, more systems are needed to verify these points robustly.

*shock fronts in the AGN feedback cycle remains, at least observationally, poorly constrained*. Despite this, several predictions exist from numerical simulations: various works showed that both high- and low-power jets can drive shock waves in the surrounding medium that become mildly supersonic at kpc scales, and typically assume the shape of cocoons surrounding X-ray cavities (e.g., [66,69,104,210,314,582]). Moreover, turbulence can broaden shock fronts [362], and vice versa shocks can generate and amplify gas turbulence (e.g., [104,582]). As such, spatially resolved studies of shock fronts are necessary to inform future spectrally resolved studies of AGN-driven turbulence in the ICM/IGrM. While initially expanding at supersonic speed, shock waves progressively slow down and broaden into sound waves (e.g., [62]). Shock and sound waves are not only expected to contribute to the total energy released in the ICM by the AGN outburst, but can also provide a relatively simple solution as to how the ICM may be isotropically heated, considering that jets and X-ray cavities are instead intrinsically directional (e.g., [587]).

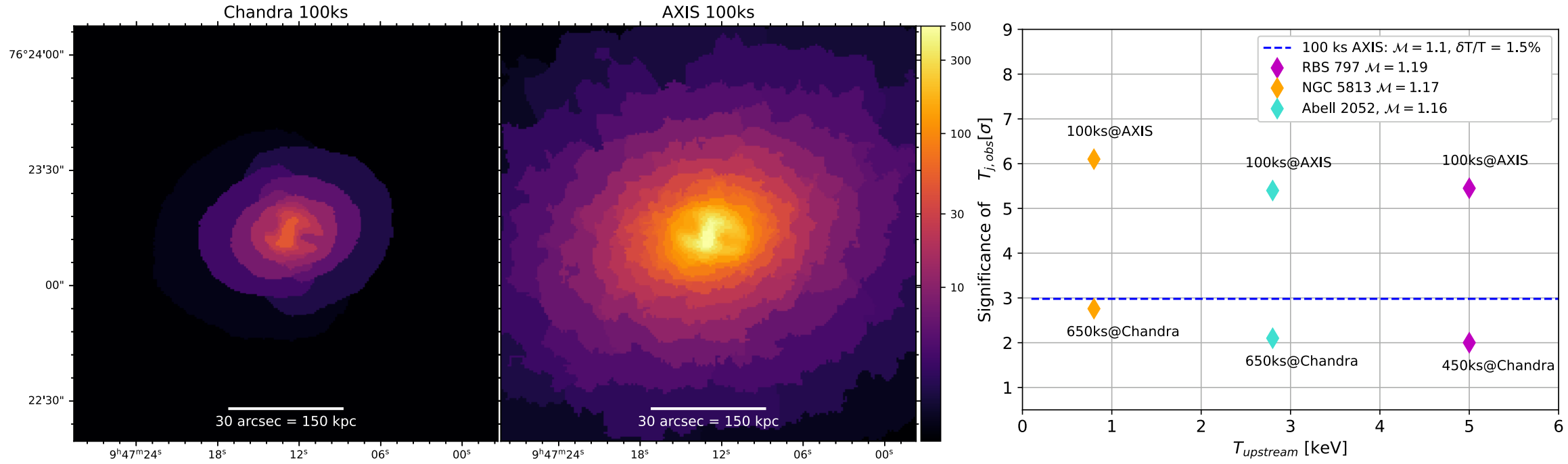

**Figure 15.** *Chandra* image (*left panel*) and AXIS (*middle panel*) simulated image of RBS 797 (100 ks) with matching colorbar and field of view, in counts/pixel, binned to 5000 counts per region. AXIS will provide $\geq$100 independent regions where to measure temperature (with relative uncertainties $\Delta T/T \sim 3\%$), density, and pressure (critical to identify shock fronts), compared to the $\sim$10 *Chandra* regions (see [436] for another example). *Right:* significance of temperature jumps as a function of upstream gas temperature, with the current *Chandra* measurements and expected AXIS performances for the cases shown in Figure 14. With individual 100 ks exposures, AXIS will provide $\geq 5\sigma$ temperature jump detections for $\mathcal{M} \geq 1.2$, and $3\sigma$ detection of temperature jumps from weaker, $\mathcal{M} \sim 1.1$ shock fronts (Goal 1). We expect $\sim$50 ks exposures to collect more than 200 kcounts for $r \leq 50$ kpc, providing $3\sigma$ detections of $\mathcal{M} \geq 1.2$ shock fronts (Goal 2).

The dramatic lack of demographics on AGN-driven shock fronts raises the following questions:

- *What are the fraction of cool core galaxy clusters with shocks and the average properties of jet-driven shock fronts?* The shock fronts known so far likely represent the tip of the iceberg of AGN-driven waves in the ICM, with low Mach number shocks ($\mathcal{M} \sim 1.1$) and sound waves being largely missed in existing observations [149,441]. As such, it is unclear how efficient jets are at driving shocks of variable strength.

- *What is the relation between cavities and shock fronts?* A fraction of clusters and groups hosts multiple X-ray cavities at different radii, tracing successive episodes of radio activity in the BCG, each separated by a few tens of Myr (e.g., [122,146,193,325,326]). Whether this relatively high duty cycle is also reflected by shock fronts, with only 3 known cases of multiple shocks (NGC 5813, ACO 2052, RBS 797; see Figure 14), is currently uncertain. For instance, numerical works suggest that a single cavity-inflating jet episode may drive multiple shock fronts (see e.g., [69]).

- *Are the jet mechanical energy and power comparable when computed using shocks or cavities?* Shock fronts could offer an independent confirmation that the enthalpy of X-ray cavities is a good proxy for the AGN mechanical power. By analyzing a collection of 13 objects with detected shocks and X-ray cavities, [281] found that the shock energy roughly scales with the cavity energy. [523] further found that, in 15 objects, the jet power measured using shock fronts, $P_{sh}$, is proportional to the radiative losses of the ICM, $L_{cool}$ (see Figure 14). This supports the analogous conclusion obtained from X-ray cavities (first by [40]). However, comparison with the $P_{cav} - L_{cool}$ relation might suggest that, on average, $P_{jet,sh} > P_{jet,cav}$, raising questions on whether a large fraction of the energy budget is being missed.

- *Is the mechanical power - radio luminosity relation well calibrated?* The relationship between the AGN mechanical power and BCG radio emission can probe both the physical nature of the jet (e.g., [201,580]) and the energy budget of AGN based on more easily acquired radio data (e.g., [34]; see also [374]). All existing studies are based on cavity powers; a critical next step is to perform an independent calibration of such a relation using shock fronts as gauges of $P_{jet}$. Using available data for the systems listed in [281,523], we would find that $P_{jet,sh} \overset{\propto}{\sim} L_{radio}^{0.5}$ (Figure 14), compared to $P_{jet,cav} \propto L_{radio}^{0.7}$ from [374], but again with large uncertainties.

AXIS can significantly improve our understanding of these phenomena by providing us with:

1. **A census of AGN feedback footprints from the perspective of shock fronts**: deep AXIS observations of three exemplary halos with ongoing and recurrent AGN feedback (NGC 5813, ACO 2052, RBS 797) will allow us to obtain $\geq 3\sigma$ detection of temperature jumps from $\mathcal{M} \sim 1.1$ shock fronts, completely missed by current observations (see Figure 15). This will reveal how far from the center shock fronts travel (beyond the cool core; see [362]) before broadening into sound waves, and a comparison with X-ray cavities in these systems will clarify whether individual outbursts produce single or multiple shock fronts. Overall, deep AXIS exposures of these three objects will reveal the demographics of shock waves over a large halo mass range ($10^{13} - 10^{15}$ $M_{\odot}$).
2. **Robust measurements of shock-driving efficiency, AGN energy balance and cooling regulation with shock fronts**: a transformative ~2.5 Ms AXIS survey of about 50 systems from a complete cluster sample would collect enough counts to detect density and temperature jumps from $\mathcal{M} \geq 1.2$ shock fronts (as well as cavities) in the ICM/IGrM. Such observations will unveil the fraction of cool cores with shock fronts, clarify the energy balance between cavity and shocks, and their relation with radiative losses of the ICM/IGrM, and reveal if scaling relations between radio luminosity and mechanical power change when using shocks as gauges of the jet power.

**Exposure time (ks):** 300 ks + 2.5 Ms

**Observing description:** Detecting shocks in the X-rays requires (i) identifying an edge in surface brightness, (ii) confirming the presence of a density jump from the surface brightness profile, and then (iii) spectroscopically measuring temperature and pressure jumps. Measuring a negative or positive temperature gradient across a front is essential to discriminate between a shock or a cool cavity rim/cold front, respectively (see e.g., [85]). We consider systems with detected shock fronts (see [281,523]), and especially those in Figure 14, as the basis of the feasibility.

• **Goal 1:** As shown in Figure 15 (*left* and *middle* panels), 100 ks of AXIS observations of e.g., RBS 797 will provide 10× more spatially-resolved regions where to measure the gas temperature at high significance ($\Delta T/T \sim 5\%$) compared to an equally long *Chandra* exposure. For comparison, NGC 5813 and ACO 2052 have higher X-ray flux and will provide a similar, if not higher, number of counts. To determine the exposure required to meet our goals above, we have leveraged the existing observations of NGC 5813, ACO 2052, and RBS 797, to study the dependence of temperature jump significance as a function of shock Mach number. Figure 15 shows that, if the shocks in these three clusters were not of $1.2 \leq \mathcal{M} \leq 1.7$ but rather $\mathcal{M} \sim 1.1$ (approaching the regime of sound waves), 100 ks of AXIS exposure would still provide $\geq 3\sigma$ detections of the associated temperature jump ($T_{in}/T_{out} \sim 1.1$, $\Delta T/T \sim 1.5\%$).

• **Goal 2:** To minimize selection effects in determining the fraction of systems with shocks and the relation between $P_{jet}$ and other quantities, we must use a sample that is as complete as possible. We consider the work of [41], which compiled a sample of 50 cool core groups and clusters from complete cluster samples (HIFLUGCS and B55). Existing *Chandra* data of systems with shocks (see [281,523]) point to a minimum of ~200 kcounts (0.5 - 7 keV) in the inner 50 kpc to detect a shock front with $\mathcal{M} \geq 1.2$. For comparison, at least 20 kcounts in the inner 20 kpc are needed to detect a cavity [382], reflecting the difference between the number of systems with shocks and of those with cavities. Using the 0.5 – 7 keV counts and X-ray flux of the three clusters in Figure 14, scaled by the X-ray flux of systems in [41] and extrapolated to AXIS counts (via XSPEC and simx), we estimate that ~50 ks per system is required to reach at least 200 kcounts. Thus, to observe the full sample of 50 objects, we require a total AXIS exposure time of $\sim$2.5 Ms.

**Joint Observations and synergies with other observatories in the 2030s:** Radio observations of the AGN jets and lobes are required to understand the connection between the X-ray features in the gas and the activity of the AGN. Sensitive data from SKA and its pathfinders (e.g., MeerKAT), the next generation VLA (ngVLA), and LOFAR2.0 will provide an excellent synergy with the AXIS data. Additionally, AGN-driven shock fronts identified with AXIS will be followed up using high-resolution X-ray spectroscopy (with XRISM and NewAthena), which will return the velocity at which shock fronts travel and cavities expand, as well as the ICM/IGrM turbulent velocity dispersion at the location of these feedback footprints.

**Special Requirements:** None

*11. The Milky Way: Probing the effects of feedback in a quiescent galaxy*

**Science Area: Milky Way, feedback, circumgalactic medium**

**First Author:** Nicola Locatelli (INAF OAB) & Gabriele Ponti (INAF OAB, MPE)

**Co-authors:** Michael Yeung (MPE), Xueying Zheng (MPE), Elisa Lentini (INAF OAB)

**Abstract:** All-sky soft X-ray observations have revealed large excesses of emission north and south of the Galactic center, known as the eROSITA bubbles. This emission is attributed to a hot Galactic wind, with plasma heated by an adiabatic shock to $kT \sim 0.3$ keV. However, Suzaku spectra challenge this model, suggesting an isothermal transition with a density increase. A deep AXIS scan across the bubble edges would improve spectral analysis, resolving whether the plasma primarily heats or compresses. In addition, preliminary eROSITA maps indicate unexpected temperature variations, hinting at off-equilibrium ionization. AXIS observations would clarify these shocks and refine our understanding of Galactic feedback.

**Science:** All sky observations in the soft X-ray band have revealed excesses of X-ray emission north and south of the Galactic center, covering a large portion of the sky, which have been dubbed "eROSITA bubbles". The emission from these bubbles is thought to be produced by the hot plasma associated with the hot Galactic wind. In particular, the X-ray emission is attributed to an adiabatic shock that heats the plasma to a temperature of $kT \sim 0.3$ keV, tracing a shock with a Mach number of approximately 1.5 [252]. This shock is thought to heat the plasma inside the bubbles to temperatures exceeding those typically found in the Galactic halo ($kT \sim 0.2$ keV). However, this model has recently been challenged by the analysis of Suzaku spectra [206], which shows that an isothermal transition with an equal density increase also fits the data.

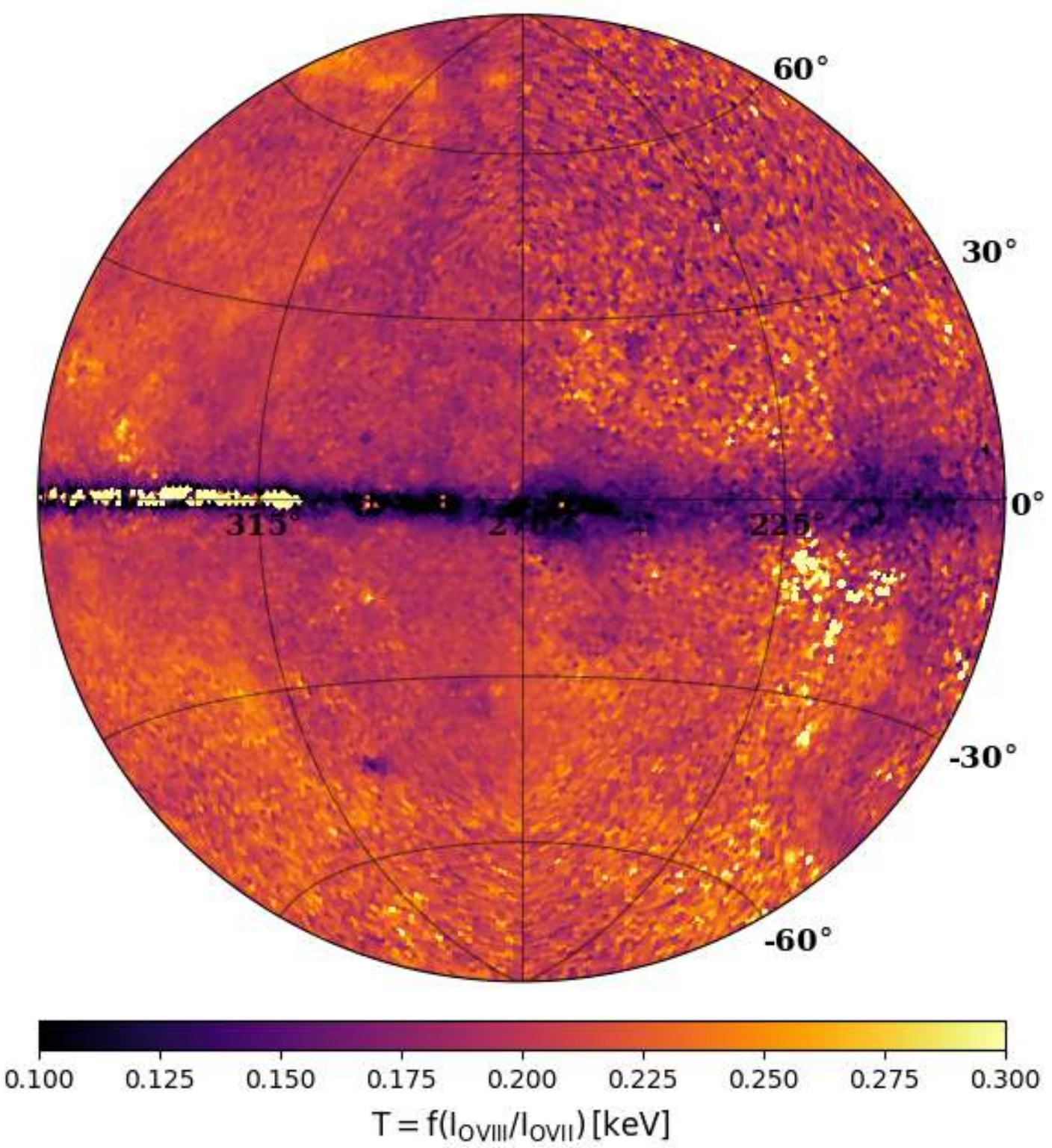

**Figure 16.** Temperature map of the hot gas component in the Western Galactic hemisphere, as derived by the OVIII/OVII line ratio measured by eROSITA.

A deep ($\sim 1$ ks) scan across the edge of the eROSITA bubbles would enable a spatially resolved spectral analysis with statistical precision roughly 10 times greater than what eROSITA achieved during its first all-sky survey [588, eRASS1]. With such observations, the AXIS spectra would provide crucial insights into the variations in surface brightness, temperature, density, and other properties of the hot plasma as it transitions from the bubble interiors to the unshocked circumgalactic medium. This would allow us to determine whether the plasma primarily undergoes an increase in temperature or density as it moves through the shock, thereby shedding light on the interaction between the Galactic outflow and the outer circumgalactic medium.

Preliminary results from the shallow eROSITA half-sky OVII/OVIII maps (Fig. 16) reveal a high degree of complexity within $\sim 10^{\circ}$ of the eROSITA bubble edges [607]. These observations suggest a puzzling trend, where the temperature initially decreases (an unexpected finding) before increasing again, possibly indicating that the ionization state of the plasma might be off-equilibrium [607]. A targeted AXIS scan across the bubble edges would provide critical information on the nature of the shocks at these boundaries.

By resolving these questions, such observations would enhance our understanding of feedback mechanisms in the Milky Way, shedding light on how energy and matter are cycled through the Galactic ecosystem.

**Exposure time (ks):** $\sim$600ks

**Observing description:** A scan of the edges of the eROSITA bubbles, covering approximately 600 square degrees at a depth of 1 ks, would allow us to obtain spectra with a statistics $\sim$10 times better than eRASS:1 in the study of the diffuse emission and significantly better in the removal of the contribution from point sources.

**Joint Observations and synergies with other observatories in the 2030s:** Radio observations are crucial to constrain the non-thermal particle component of the Galactic emission. In particular, they are very useful for tracking the location of particle acceleration at shocks, where the thermal X-ray-emitting gas provides the source for the CR population to be accelerated. A large field of view combined with high sensitivity to diffuse emission and moderate angular resolution will be crucial. While the Meerkat and ASKAP have just started to provide such requirements, the SKA-low and -mid will prove excellent to be combined with AXIS X-ray observations of the eROSITA bubble edges, providing a comprehensive view of the emission mechanism and source particles.

**Special Requirements:** None

*12. Detecting eRosita-like bubbles in the circumgalactic medium of external Milky Way-mass galaxies*

**Science Area: eRosita bubbles, feedback, circumgalactic medium**

**First Author:** Marine Prunier (UdeM, MPIA)

**Co-authors:** Julie Hlavacek-Larrondo (UdeM), Annalisa Pillepich (MPIA)

**Abstract:** The *eROSITA*-bubbles are remarkable, soft X-ray-emitting features extending ∼14 kpc above and below the Milky Way (MW) Galactic plane. While their origin - stellar feedback vs. supermassive black hole (SMBH) outbursts - remains debated, they offer direct evidence of feedback shaping the circumgalactic medium (CGM). What are the physical mechanisms that inflated such structures, and are they unique to our Milky Way?
Cosmological simulations of galaxies like TNG50 of the IllustrisTNG project predict similar *eRosita*-like X-ray bubbles to be widespread around MW–**like** galaxies, driven by low-accreting SMBH mechanical feedback. However, detecting them in external systems remains challenging: their faint emission requires >1 Ms exposures with current observatories such as *Chandra*, and current hints of their possible existence rely on stacking large samples (>90 galaxies). With its high angular resolution and large effective area, *AXIS* is uniquely capable of resolving such structures in individual galaxies. Mock *AXIS* observations of TNG50 analogs show that 100–200 ks exposures of nearby disky galaxies ($z < 0.01$) are sufficient to detect CGM emission within ∼30 kpc.
We propose a targeted survey of ∼20 nearby, edge-on spiral MW-like systems ($M_\star \sim 10^{10.5}$–$10^{11.2}\ M_\odot$, $i > 75°$, $D < 100$ Mpc), selected from the HECATE catalog for which SMBH mass estimates and star formation are available. This sample will enable a systematic search for soft X-ray bubbles and allow us to investigate their origin - AGN-driven vs. stellar feedback. Assessing the prevalence and properties of these structures provides direct insight into how feedback shapes the CGM in disk star-forming galaxies. Ultimately, the detection (or absence) of MW-like bubbles beyond our Galaxy will serve as a powerful test of current galaxy formation models and of the uniqueness (or not) of our Galaxy.

**Science**: The *eROSITA*-bubbles are a prominent pair of shell-like X-ray structures [404], emerging from the Galactic center at energies of 0.6–1.0 keV. This feature extends approximately 14 kpc above and below the Galactic plane and spatially coincides with the previously discovered Fermi bubbles [484], a pair of 8–10 kpc lobes emitting in $\gamma$-rays. The physical origin of the *eRosita* and Fermi bubbles remains debated, with proposed scenarios including starburst-driven outflows [e.g., 358,592] or SMBH feedback [e.g., 184]. Their discovery raises fundamental questions. **What mechanisms inflated such X-ray structures? Are similar feedback-driven *eRosita* bubbles a common feature in other disky star-forming galaxies?**

The IllustrisTNG cosmological simulations of galaxy formation [e.g., 354] suggest that feedback-driven bubbles are common in Milky Way–like galaxies [395]. In TNG50, a population of MW-mass systems is found to host spectacular pairs of X-ray bubbles carved out by the central SMBH. These features share many similarities with the *eROSITA* bubbles observed in the Milky Way: they extend perpendicular to the galactic disk, both above and below, with sizes ranging from a few to several tens of kpc and with surface brightness in the range of $10^{35-36}$ erg s$^{-1}$ kpc$^{-2}$. In the TNG model, the largest majority of these bubbles seem to be caused by the SMBH mechanical feedback operating at low-accretion rates, and the simulation predicts typical Mach 2-4 shocks at the bubble boundaries and expansion velocities up to 1000-2000 km $s^{-1}$. However, while TNG and many idealised numerical simulations [e.g., 346,447], predict feedback-driven X-ray bubbles in the CGM of MW galaxies, no definitive detection of these structures has yet been made in external galaxies.

Attempts to directly detect *eROSITA*-like bubbles in the inner CGM (<30 kpc) of external galaxies using current X-ray facilities like *Chandra*, *XMM-Newton*, and *eROSITA* are limited by sensitivity and resolution, requiring exposures $\geq$1 Ms for individual galaxies [371,395]. While diffuse CGM emission has been detected around a few massive disks [e.g., 274], these studies found no clear evidence for asymmetries or bubble-like structures. Nearby galaxies like M31 and M104 show hints of anisotropic feedback-driven features [45,279], but limited angular resolution prevents morphological analysis. Stacking studies (e.g., eFEDs [88]; *eRosita* [604], and *Chandra* [438]) have statistically detected large-scale CGM emission and/or anisotropies in the stacked soft X-ray emission. However, the inherent averaging process of stacking washes out spatial details, making it impossible to resolve these kpc-scale structures and shock fronts.

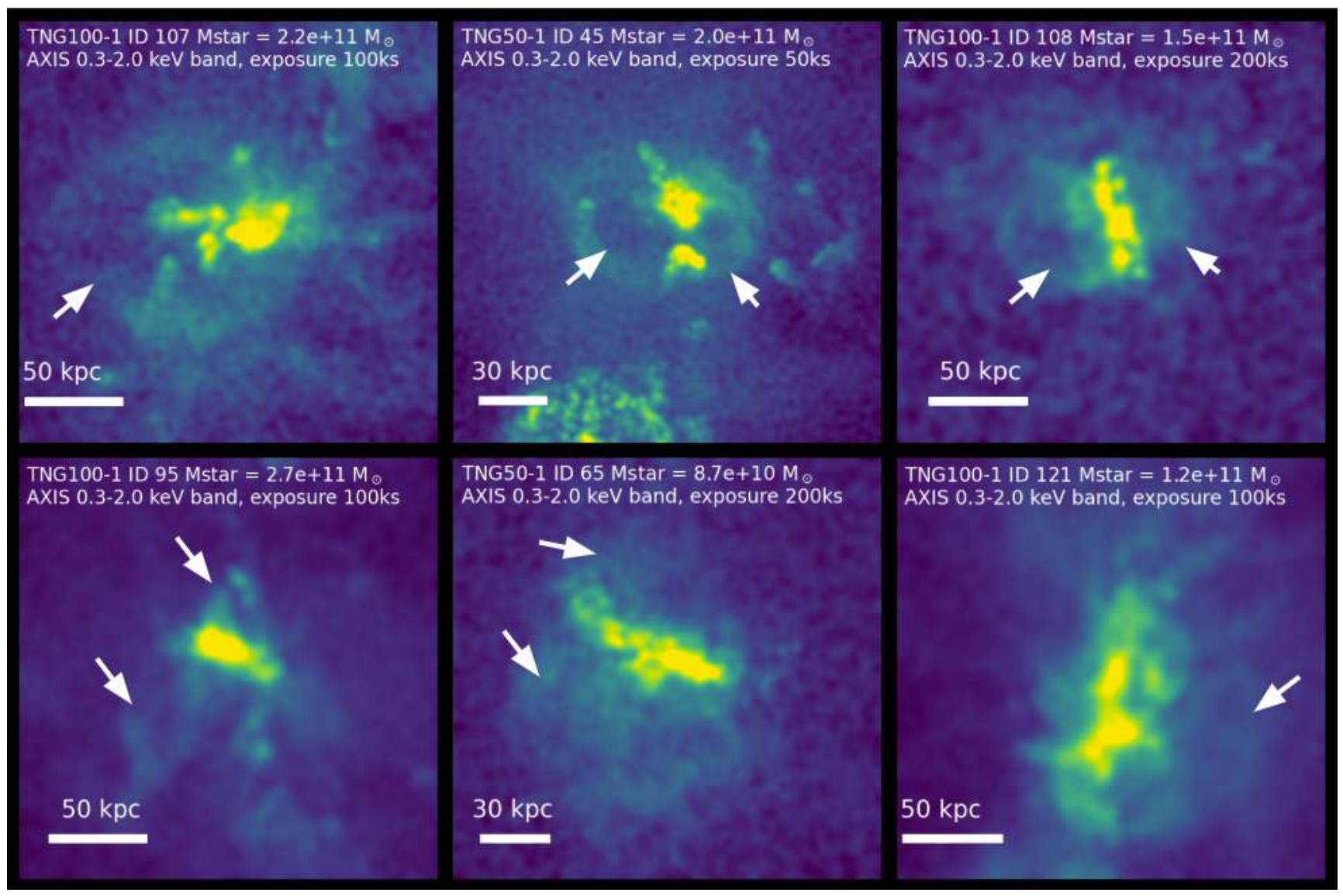

**Figure 17.** Mock AXIS observations 200 $\times$ 200kpc a side of six Milky Way–like and more massive disky galaxies in TNG (generated with pyXSIM and SOXS, including instrumental noise and Galactic foreground). These simulated galaxies show prominent 10–50 kpc *eROSITA*-like bubbles extending perpendicular to the disk's major axis. Despite the faint soft X-ray emission from the bubbles, their morphology and the surrounding CGM structure are clearly detectable with a 100 ks exposure time, and even within just 50 ks for some more massive galaxies (top middle).

***AXIS* has the unique capability to transform this field by enabling, for the first time, the direct detection and mapping of *eRosita*-like X-ray bubbles in individual external galaxies.** With its combination of arcsecond imaging constant over a wide 24' field of view, and more than 10$\times$ the sensitivity of *Chandra*, *AXIS* can distinguish faint thermal emission from the foreground contribution from the MW. Its capability will allow us to resolve the morphology, extent, and thermodynamic structure of feedback-driven bubbles out to tens of kiloparsecs. The *AXIS* angular resolution will also allow efficient spatial masking of the background point sources and resolving the predicted shocks at the edge of *eROSITA*-like bubbles. At redshift $\sim$ 0.01, its field-of-view ($\sim$ 300 kpc with 300 pc pixel-resolution) enables complete coverage of many nearby Milky Way-mass galaxies in a single deep pointing. We test the detectability of *eROSITA*-like bubbles with AXIS by generating mock *AXIS* observations of TNG50 and TNG100 galaxies that exhibit such features with various sizes. Using `pyXSIM` [613], we simulate the intrinsic X-ray emission, and with `SOXS` [617] we simulate an *AXIS* observation including instrumental effects (PSF, detector response, instrumental noise) and galactic foreground. For exposure times ranging from 50 to 200 ks, we find that X-ray bubbles can be clearly detected in many systems. Six representative examples are shown in Figure 17.

To detect and study *eROSITA*-like bubbles with *AXIS*, we propose to target a carefully selected sample of nearby, edge-on, star-forming disk galaxies with Milky Way-like stellar masses. These galaxies are chosen to maximize the likelihood of detection, based on recent stacking analyses. In particular, we focus on a subsample of galaxies from the *Chandra* stacking study by [438], which revealed extended soft X-ray emission up to ∼14 kpc, with a marked enhancement along the galactic minor axis. Observing these systems individually with *AXIS* will enable the detection of potential X-ray bubbles and target the following key science objectives:

- Provide a direct detection of *eRosita*-like bubbles in nearby MW-like galaxies.
- Map the morphology, spatial extent, and (thermo)dynamical properties of the X-ray bubbles.
- Provide powerful constraints on the nature (SMBH outflows vs. starburst), strength, and timescale of the feedback processes inflating them.
- Enable a critical test of galaxy formation models, as their detection and characteristics will support or invalidate predictions from these models.

**In summary, detecting *eRosita*-like X-ray bubbles beyond the Milky Way would be a groundbreaking result, a major step in understanding how galaxies interact with their circumgalactic environments.**

**Exposure time (ks):** 100-200 ks per target

**Observing description:** We propose a survey of ∼20 nearby, edge-on spiral galaxies selected to match the Milky Way and M31 in stellar mass ($10^{10.5}$–$10^{11.2}$ $M_{\odot}$) and star-forming properties. The sample is drawn from [438], which selected galaxies in the HECATE catalog ([264] all-sky catalog of 200,000 galaxies within D $<$ 200 Mpc) with inclination $i > 75°$, known star formation rates, and no association with bright X-ray environments such as groups or clusters that could outshine the CGM. SMBH masses for these systems are estimated via the $M_{\rm BH}$–$\sigma$ relation [319], with velocity dispersions from HyperLeda [296]. Among this subsample of 93 galaxies we further select the closest 20 galaxies with distances <100 Mpc (z=0.025, 15 kpc ≈ 30”) having the lowest galactic foreground absorption (using the NASA HEARSC tool that calculates $n_{\rm H}$ based on the sky position of the system [250,497]). This setup allows us to 1) systematically search for soft X-ray bubbles in the CGM of the sample of nearby MW-like galaxies and 2) use stellar mass, SFR, and central SMBH of these objects to disentangle their debated origin.

**Joint Observations and synergies with other observatories in the 2030s:** AXIS observations in synergy with NewAthena X-IFU will enable the high-resolution spectroscopic measurements of CGM velocities, temperatures, and metal abundances. Together, the facilities will provide crucial insights into the role of stellar and AGN feedback in driving bubbles by mapping inflow and outflow velocities in the CGM and comparing them with predictions from idealized numerical and cosmological simulations. Complementary SKA observations will detect radio emission from active SMBHs in galaxies hosting X-ray bubbles, helping to connect bubble formation with ongoing AGN activity.

**Special Requirements:** None.

*13. Jet reorientation in central galaxies of clusters and groups*

**Science Area: galaxy clusters, galaxy groups, intracluster medium, AGN feedback, radio jets**

**First Author:** Francesco Ubertosi (University of Bologna)

**Co-authors:** Gerrit Schellenberger (SAO/Center for Astrophysics), Ewan O'Sullivan (SAO/Center for Astrophysics), Myriam Gitti (University of Bologna), Eric Perlman (Florida Institute of Technology), Massimo Gaspari (University of Modena & Reggio Emilia).

**Abstract:** A key aspect of AGN outbursts is the directionality of the radio jets. In galaxy clusters and groups, this is evident both in the radio band, where jets extend into the hot gas for tens of kpc, and in the X-ray band, where the corresponding X-ray cavities map the direction in which the jet was pointing in the past. Older and younger pairs of X-ray cavities are usually aligned along a common axis, suggesting that the jet has been pointing in a similar direction over time. In other instances, jets pointing away from the cavities, or multiple misaligned X-ray cavities, suggest that jets can change their orientation. A combination of X-ray and radio observations has enabled initial studies on the occurrence of these misalignments, but only for a few exemplary cases. This is primarily due to the difficulty of systematically detecting cavities, which requires long X-ray exposures and makes large-sample studies challenging. We discuss how the high spatial resolution and sensitivity of AXIS would be critical to measure the position angle of X-ray cavities in a large number of systems. In particular, we propose to observe 46 new clusters with AXIS, complementing 40 archival cases, to build a statistically robust sample to study both the causes of reorientation and the effects that reorienting jets have on their cooling halos.

**Science:** *AGN feedback in galaxy clusters and groups: directionality of successive outbursts -* In the past 25 years, high angular resolution X-ray and radio observations have routinely showed that powerful jets launched by supermassive black holes (SMBHs) in central galaxies of clusters and groups can heat the intracluster/intragroup medium (ICM/IGrM; e.g., [122,146,193,325,326]). The active galactic nuclei (AGN), powered by the supermassive black hole (SMBH), and the hot atmosphere are linked through a self-regulating feeding-feedback mechanism. Gas cools out of the hot intracluster medium (ICM), sinks towards the SMBH, and fuels relativistic jets and outflows (via e.g., chaotic cold accretion, see [177,178]). These jets heat the surrounding environment and ultimately quench cooling flows [325,326]. The clearest footprints of jet activity in the ICM are bubbles – named X-ray cavities – carved by the central radio jets in the hot gas (e.g., [40,41]). Multiple pairs of cavities were also found in some cases, tracing successive jet outbursts pushing aside the surrounding hot gas (e.g., [407]). A key aspect of AGN activity is the directionality of the jet outbursts. In several cases, older and younger pairs of X-ray cavities are aligned along a common axis, suggesting that the jet has been pointing in a similar direction over time (e.g., [521,581]). In other instances, jets pointing away from the X-ray cavities, or multiple X-ray cavities spread over the entire azimuth, suggest that either the jets can change their orientation over time or external (environmental) effects can move cavities away from where the jet that inflated them was originally pointing. Well-known cases of misalignment include the cluster RBS 797 and the group NGC 5044. In RBS 797, two pairs of orthogonal radio lobes are present in the central 50 kpc [194,524], and deep X-ray *Chandra* data revealed four X-ray cavities corresponding to the perpendicular lobes [522]. In NGC 5044, X-ray cavities provide evidence of two successive outbursts, both aligned northeast-southwest. However, VLBA data reveal pc-scale jets aligned orthogonal to the cavity axis ([448] and references therein; see also Figure 18, *middle panel*). The different hypotheses for explaining these pc-kpc mismatches between jets and cavities include (see also [525]): jet precession; backflows from the primary jets of the radio galaxy; quasar-mode accretion at near-Eddington rates; projection effects creating the illusion of a large shift in direction; binary black holes powering jets in different directions; environmental effects such as sloshing moving cavities around the hot gas. These have been carefully investigated for a few objects, such as RBS 797 and NGC 5044, which have deep, high-quality multiwavelength data. However, to understand the cause and effect of these geometries, it is key to undertake a systematic investigation of jet (mis)alignment for a compilation of cool cores where AGN feedback activity is ongoing and vigorous.

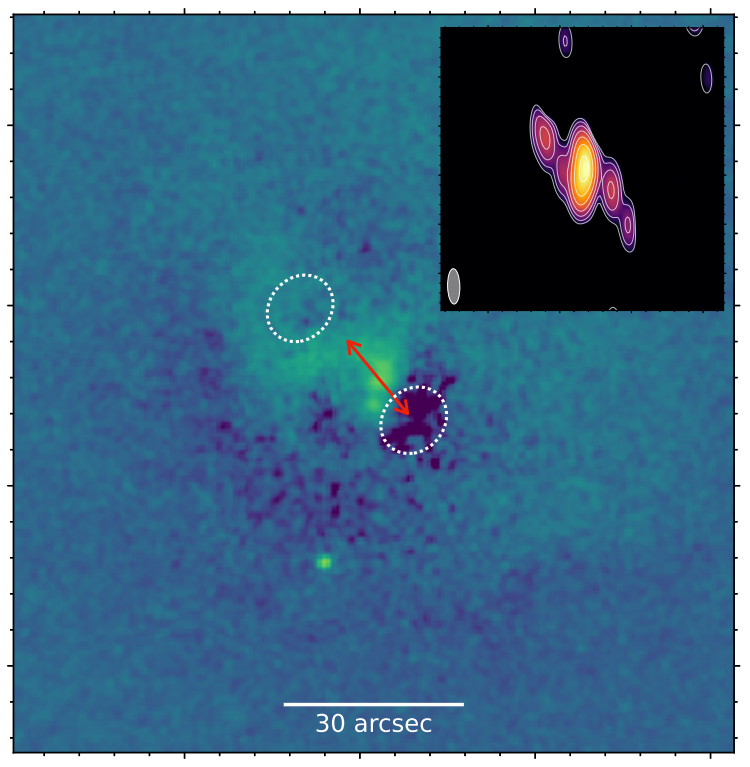

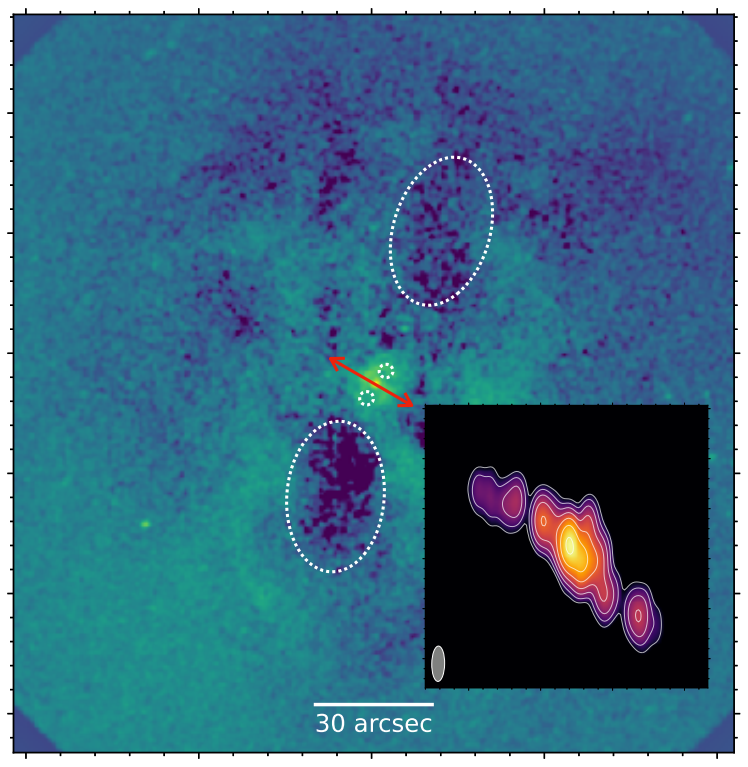

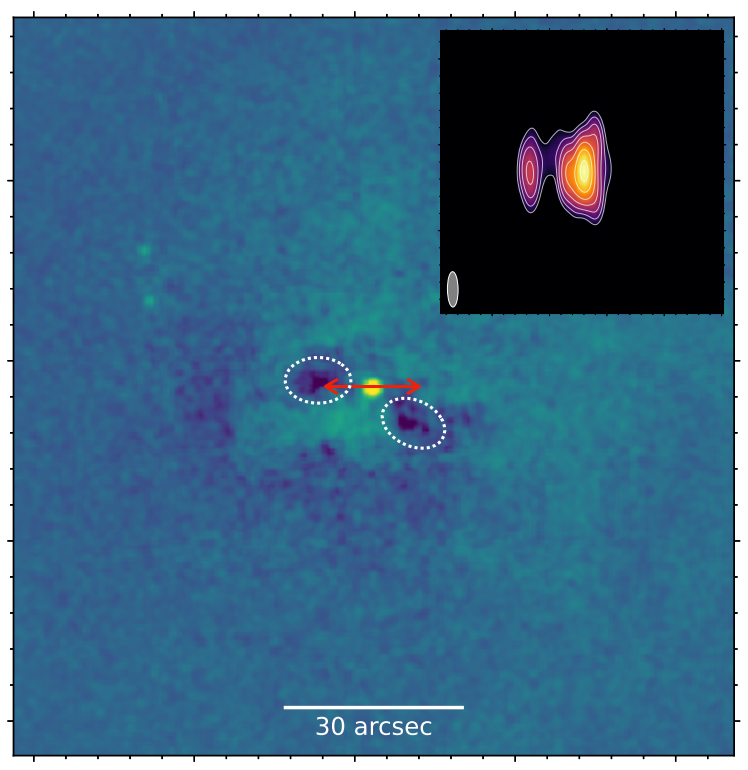

**Figure 18.** Examples of jet-cavity (mis)alignment in clusters and groups (see [525] for the original work). The main panels show the *Chandra* images of Abell 496 (left), NGC 5044 (middle), and Abell 3581 (right), with white ellipses tracing the cavities on kpc scales. The red arrows trace the direction of jets on pc scales. This was measured from the VLBA images shown in the insets, where jets are seen extending from the bright radio core. The resolution of the radio images is indicated by an ellipse in the bottom left corner of the figure. In Abell 496 and Abell 3581 the cavities are roughly aligned with the pc-scale jets, while NGC 5044 displays a prominent $\sim 90^\circ$ misalignment.

*The current picture based on Chandra and VLBA observations -* Very recently, a combination of high-resolution X-ray (Chandra) and radio (VLBA) observations has allowed a pilot study of the occurrence of these misalignments. [525] measured and compared the parsec-scale position angle of the jets and the position angle of the X-ray cavities in a sample of 16 galaxy clusters and groups (see three examples in Figure 18). In these systems, around 30% of the jets show misalignments by more than $45^\circ$ relative to the X-ray cavities. Additionally, projection and environmental effects cannot explain the majority of misaligned cases. Ultimately, the distribution of misalignment angle from [525] shows a peak in the number of sources with misalignments $\sim 90^\circ$ which cannot be explained by environmental factors or experimental uncertainties, and may provide an important constraint to reorientation models (e.g., [226]; see also [79,390]). Clearly, a larger investigation is required to explore these mechanisms further. Additionally, beyond the causes of misalignment, their effects are also of interest: in cooling atmospheres, continuously reorienting jets may achieve a more efficient, isotropic feedback. Simulations suggest that without frequent jet reorientation (e.g., in the simulations of [104] the reorientation timescale is 2 Myr, compatible with spin-flip events or precession; see [522,524]) or some mechanism to distribute the energy (e.g., cocoon shock fronts), AGN jets fail to heat their hot atmospheres efficiently. To understand how AGN jets distribute their energy, a comparison of the gas cooling efficiency between systems with and without reoriented jets is needed.

*An AXIS survey of jet reorientation in central galaxies of clusters and groups -* Achieving the above goals requires having both high-resolution (milli-arcsec) parsec-scale imaging of jets and high-resolution (arcsec) kpc-scale imaging of X-ray cavities. The VLBA archive already contains sufficient observations of suitable cluster-dominant galaxies to expand the sample of [525] by a factor of 5. In the near future, new sensitive radio observatories (ngVLA, SKA, and their precursors) will enable the imaging of pc-scale jets from a much larger number of cluster-central radio galaxies. As such, it is desirable to have similarly optimal

X-ray coverage. However, it is unlikely that these numbers will ever be matched by *Chandra*, since it requires long exposures to detect X-ray cavities (from a few tens of ks to 100s of ks). *The high spatial resolution and sensitivity of AXIS is ideally suited to mapping the X-ray cavities of a large sample of clusters with measurements of the parsec-scale position angle of jets from VLBA data*. From a parent catalog of 59 BCGs (see [220,221]), [525] selected the 16 cool core systems with clearly defined, parsec-scale jets in VLBA data and good quality, archival Chandra observations. To expand this selection, we searched the VLBA archive and identified 86 galaxy clusters and groups in the local Universe ($z \leq 0.5$) with uniform 5 GHz VLBA observations that detect jets on parsec scales (see examples in Figure 19). These VLBA observations at 5 GHz match the observational setup of the programs published in [220,221], enabling the definition of a uniformly covered radio sample. Among these 86 objects, 16 have been published in [525], and 24 have sufficiently deep Chandra data for the proposed study. However, the majority (46/86) lack high-resolution X-ray observations. Notably, the 24/86 objects with high-quality *Chandra* data include many systems that are well-known test cases for feedback models (e.g., A1835, NGC 813, ZwCl 0335+096), supporting our selection criteria.

Overall, *by observing 46 galaxy clusters and groups with AXIS, we will analyze the pc and kpc scales of feedback in a uniform sample of 86 galaxy clusters and groups, and determine:*

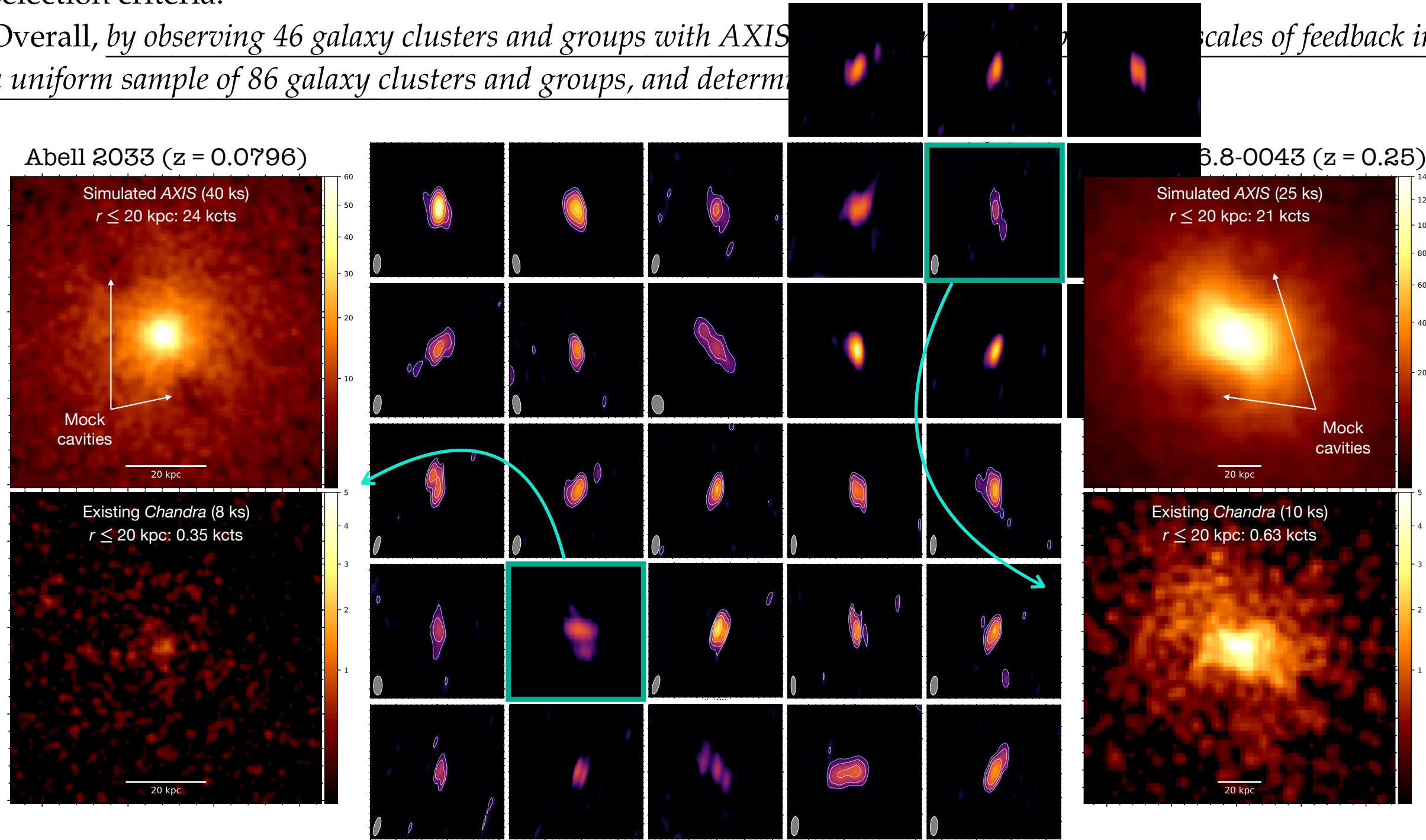

**Figure 19.** Proposed investigation of jet-cavity misalignment with AXIS. *Middle panels:* VLBA radio images on parsec-scale showing resolved jet features of 25 cluster-central AGN as an example of the 46 systems to be observed with AXIS (see Observing description). For Abell 2033, *left*, and MCS J0326.8-0043, *right* we show the available *Chandra* exposures (*bottom panel*; $\sim$10 ks each) and the simulated AXIS exposures (*top panel*), smoothed with a Gaussian of 3 pixels (see the Observing description). In the *Chandra* and AXIS exposures, we have manually placed a pair of 30% X-ray depressions at $\sim$20 kpc from the center and with a $\sim$ 10 kpc diameter. The simulated exposures demonstrate that AXIS will collect $\geq$ 20 kcounts within $r = 20$ kpc in a few 10s of ks, thus enabling the detection of X-ray cavities at a signal-to-noise ratio SNR $\geq 4$. For comparison, *Chandra* could collect as many counts only with exposures of 100s of ks.

1. **The drivers of jet-cavity misalignment.** The detection of X-ray cavities for a number of clusters and groups with pc-scale measurement of the jet direction would enable the measurement of the cavities position angle with respect to the AGN (e.g., [525]). Then, from the distribution of the angle of misalignments, we will verify, as suggested by the pilot study [525], if orthogonal misalignments are favored, providing a strong observational constraint to reorientation models. We will also measure the age of small, central cavities ($r \leq 10$ kpc) that already appear misaligned by large angles ($\geq 45°$) from the pc-scale jets; this will return the timescale over which the reorientation mechanism occurs.

2. **The outcomes of jet-cavity misalignment.** By using the X-ray imaging and spectroscopic data provided via new AXIS (and archival *Chandra*), we will search for global and spatially-resolved differences between aligned and misaligned systems in terms of hot gas properties, such as entropy, cooling time, and X-ray luminosity of the cooling gas. We will also measure azimuthal fluctuations in thermodynamic quantities related to feeding and feedback (metals, temperature, density), including diagnostics of cooling efficiency (e.g., ratios between cooling and dynamical times, [175,178,550]). These tests will allow us to verify whether cooling is more quenched in misaligned systems or not, and test the predictions of high-resolution hydrodynamical simulations (e.g., [104,582]).

Ultimately, a larger sample will sensibly increase the number of systems with multiple pairs of detected cavities (9/16 of the original sample from [525] have more than 1 pair of cavities). This is important to study the environmental effects at play (e.g., bulk motions) on the oldest pairs of X-ray cavities. *Are outer cavities more often misaligned from the jets than inner ones, especially in dynamically disturbed systems?* In this sense, spatial correlation with sloshing signatures (cold fronts) will reveal the influence of gas bulk motions on buoyancy. Simulations predict that bulk motions can strongly influence the evolution of radio bubbles (e.g., [121,612]), but observational evidence is so far limited to hints in a handful of cases (e.g., [60,151]).

**Exposure time (ks):** 1.5 Ms.

**Observing description:** We aim to observe with AXIS 46 galaxy clusters and groups selected based on a jet detection in VLBA data (see examples in Figure 19). Our primary aim, which drives the sensitivity and the estimated exposure time, is to detect X-ray cavities on kpc scales and measure their position angle. As demonstrated by [382], at least 20 kcounts in the inner 20 kpc of a galaxy cluster or group are needed to provide a secure cavity detection (signal-to-noise ratio SNR $\geq 3$). We thus consider this requirement to estimate the necessary exposure time. Specifically, we demonstrate in Figure 19 the observational strategy. We have simulated AXIS exposures (via XSPEC and simx) for two clusters (Abell 2033 and MCS J0326.8-0043) with $\sim$10 ks of *Chandra* data, which only collect a few hundred counts in the innermost 20 kpc. In both the existing *Chandra* observations and in the simulated *AXIS* exposure, we have manually placed 30% X-ray depressions, of 10 kpc diameter, at a distance of about 20 kpc from the center. Figure 19 shows that only 40 ks and 25 ks ks exposures for Abell 2033 and MCS J0326.8-0043, respectively, are enough to collect more than 20 kcounts in the innermost 20 kpc and clearly detect X-ray cavities (at SNR $\sim 4$). For comparison, it would require more than 300 ks with *Chandra* to reach the same sensitivity. Then, we have considered the 0.5 – 7 keV flux of the remaining 44 targets, scaled by that of the two examples shown in Figure 19, to estimate the average exposure required to collect enough counts. The simulated AXIS exposures show that 30 ks provide 20 kcounts in the innermost 20 kpc for sources with X-ray flux of $3 - 6 \times 10^{-12}$ erg/s/cm$^2$. Scaling to the new 46 targets, we foresee that 1.5 Ms of AXIS observations will allow us to complete the X-ray observations of the 86 systems and reveal the drivers and outcomes of jet reorientation.

**Joint Observations and synergies with other observatories in the 2030s:** Multi-scale radio observations of several cluster-central AGN are desirable to both map the direction of jets on pc-scales, and to trace the position of lobes on kpc scales, thus aiding cavity identification. Parsec-scale imaging of jets is accessible with Very Long Baseline Interferometry (VLBI), using e.g., the VLBA, EVN, eMerlin (currently), or SKA VLBI (in the future). Mapping of kpc-scale lobes, instead, requires sensitive arrays like MeerKAT,

JVLA, uGMRT, and LOFAR at present, with the next-generation VLA (ngVLA) and LOFAR2.0 enhancing capabilities in the future. In this context, AXIS is the only telescope capable of spatially resolving the X-ray footprints of these jets and lobes in the surrounding hot atmosphere. Additionally, high-resolution X-ray spectroscopy (with XRISM and NewAthena) will enable direct measurements of turbulence and cooling rates in both aligned and misaligned objects identified by AXIS. As these are key diagnostics of feedback efficiency, the combination of high-resolution X-ray spectroscopy with high-resolution X-ray imaging will reveal the role of jet reorientation in achieving more isotropic and effective feedback.

**Special Requirements:** None

*14. Multiwavelength Insights into Perseus, Virgo, and Centaurus Clusters with AXIS*

**Science Area: Galaxy clusters, intracluster medium, multiphase gas, radio jets, non-thermal particles**

**First Author:** Benjamin Vigneron (Université de Montréal)

**Co-authors:** Julie Hlavacek-Larrondo (Université de Montréal), Priyanka Chakraborty (Center for Astrophysics | Harvard & Smithsonian), Esra Bulbul (MPE), Congyao Zhang (Masaryk University, The University of Chicago), Francesco Ubertosi (Universitá di Bologna)

**Abstract:** We propose AXIS observations of three nearby galaxy clusters to investigate the multiphase environments surrounding Brightest Cluster Galaxies (BCGs) and the interactions between radio jets and the Intracluster Medium (ICM). These systems feature extensive filamentary structures within a cooling phenomenon, from hot ICM to cold molecular gas, regulated by Active Galactic Nuclei (AGN) feedback.

Despite extensive multi-wavelength studies, questions remain about the kinematics correlation between gas phases and processes such as metal entrainment, shock generation, and turbulent mixing at jet-ICM interfaces. AXIS's arcsecond spatial resolution and moderate spectral resolution will enable detailed mapping of X-ray emission at filamentary scales and characterization of thermodynamic states across jet-ICM boundaries.

Our targets, NGC 1275 (Perseus), NGC 4696 (Centaurus), and M87 (Virgo), exhibit different jet powers and environmental conditions, providing an ideal comparative study. By combining AXIS observations with existing multi-wavelength data, we will characterize the physical conditions across different temperature regimes, quantify metal transport processes, and identify shock structures, thereby advancing our understanding of AGN feedback cycles and the multiphase nature of galaxy clusters.

**Science:** *Multiphase Nature of the Intracluster Medium.* Brightest Cluster Galaxies (BCGs) are the most massive galaxies in the Universe and typically lie close to the spatial and gravitational center of their host galaxy cluster ([364], [124]). With stellar masses exceeding the Milky Way's by an order of magnitude ([552]), these colossal systems challenge our understanding of galaxy formation and evolution. Despite lacking gaseous disks, $\sim 10-15\%$ of BCGs exhibit prodigious optical line emission with H$\alpha$ luminosities exceeding $10^{42}$ erg/s ([109]). This optical emission can manifest in blobs and filaments around BCGs, often reaching staggering sizes of more than 50 kpc (see [338], [290]).

The environment surrounding BCGs reveals a complex, multiphase ecosystem. The intracluster medium (ICM), the hot ionized gas permeating the cluster, is gradually cooling from X-ray temperatures ($10^7-10^8$K, [107]) down to cold molecular gas temperatures ($10-10^3$K, [440]). This multiphase environment, however, is largely influenced by the activity of the central supermassive black hole (SMBH) through its radio jets, which entail shocks, turbulence, and mixing in a complex feedback cycle ([175,176], [548]).

Multi-wavelength observations of nearby galaxy clusters have allowed us to better grasp the spatial correlation between optical, radio, infrared, and X-ray emissions associated with BCGs and their surrounding ICM (see [146], [145], [182]). Nevertheless, critical questions persist regarding the kinematic relationships between different gas phases of this environment across multiple spatial scales. Namely, establishing kinematics and potential surface brightness correlations through dedicated high-spatial and spectral resolution multi-wavelength observation campaigns would allow a breakthrough in our comprehension of the formation paradigm of this complex multiphase environment by revealing markers of chaotic cold accretion within the ICM through shared kinematics and surface brightness properties between soft X-ray and optical filaments. Understanding these interactions, particularly as they relate to central AGN activity, requires instruments capable of unprecedented spatial and spectral resolution in the X-ray regime.

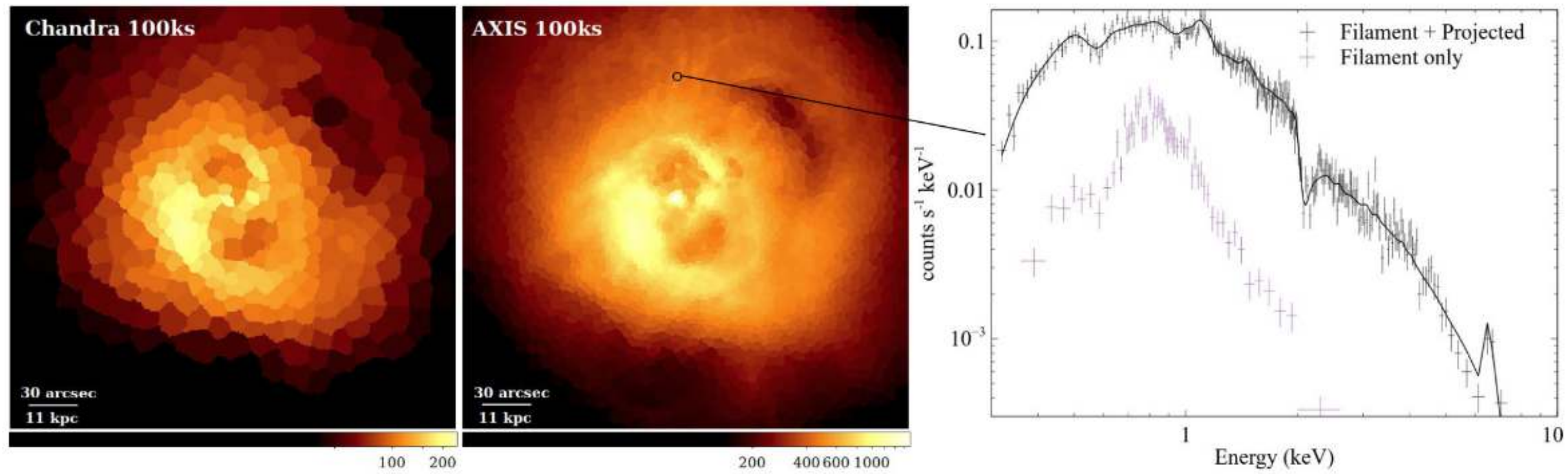

**Figure 20.** Image taken from [436]. **Left**: Chandra X-ray image of the Perseus cluster (100 ks exposure from launch), spatially binned to achieve 7500 counts per region. The color scale indicates counts per square arcsecond. **Center**: Simulated AXIS image of the same region in the Perseus cluster with identical exposure time (100 ks) and count threshold (7500 counts per region), demonstrating AXIS's superior spatial resolution. **Right**: Simulated AXIS spectra extracted from the 2.5 arcsecond radius region highlighted in the center panel. AXIS achieves a tenfold reduction in spatial bin size compared to Chandra. The black curve shows the total observed spectrum and model, while the purple curve isolates the intrinsic emission from the soft X-ray filament (with projected emission subtracted using an adjacent off-filament region).

AXIS's exceptional capabilities, and specifically its arcsecond spatial resolution combined with moderate spectral resolution, present a unique opportunity to map X-ray emission from the ICM at filamentary scales in nearby galaxy clusters. Recent studies suggest intriguing surface brightness correlations between X-ray features and optical filamentary nebulae ([367]), but AXIS will enable the first comprehensive mapping of these structures on sub-arcsecond scales. By combining AXIS observations with existing multi-wavelength data (optical, radio, and infrared), we propose to investigate the multiphase nature of the ICM across several temperature regimes. This coordinated analysis will significantly deepen our understanding of the complex feedback mechanisms operating within galaxy clusters.

*Interaction between Jets and ICM.* A critical aspect of our proposed investigation is also to examine the complex interactions between radio jets and the ICM across our proposed targets. These powerful jets, emanating from the central AGN, play a fundamental role in transporting energy and metals throughout the cluster environment ([449]). Previous observations and simulations have provided tantalizing glimpses of metal entrainment along jet paths, shock fronts at jet termination points, and turbulent mixing layers at jet-ICM interfaces ([216], [257], [343]). However, the detailed physics of these interactions—including energy dissipation mechanisms, metal transport processes, and the development of instabilities remains poorly constrained due to insufficient spatial and spectral resolution in existing X-ray data. AXIS's unprecedented arcsecond spatial resolution will enable us to map temperature and metallicity variations across jet-ICM boundaries with precision that is currently impossible with existing instruments. These observations will reveal the thermodynamic state of a gas experiencing jet interactions, quantify the spatial extent of metal enrichment through jet entrainment, and characterize shock structures by their distinctive spectral signatures. We aim to compare these features across the targets identified below, which span different jet powers and environmental conditions, such as AGN accretion rates and activity, ICM cavity sizes and localization, as well as kinematics and correlations with other phases. These innate differences provide a broader sample, allowing us to develop a comprehensive picture of jet-ICM coupling mechanisms that shape the evolution of galaxy clusters.

*Galaxy Cluster Targets.* We have identified three exceptional nearby galaxy cluster targets that maximize the scientific return from AXIS observations:

- **The Perseus Cluster of Galaxies ($z = 0.0183$):**

NGC 1275 is the BCG of the nearby Perseus Cluster and exhibits one of the most extended optical filamentary nebulae ever discovered ($50 \times 80$ kpc). By the AXIS's launch, an exceptional multi-wavelength dataset will be available for this target. High-spectral-resolution SITELLE observations of the entire optical filamentary structure offer detailed spectroscopy at a spatial scale of 0.32" ([128], [544]). Cold molecular gas has been detected throughout the optical filaments via IRAM and SMA radio observations ([440], [439], [280]), while ALMA and VLA observations revealed the jets and inner radio activity of the AGN ([352], [183]). Additionally, upcoming JWST observations (PI: Hlavacek-Larrondo, program number 7033) utilizing NIRSpec and NIRCam will provide unprecedented insights into the cooler and coronal temperature gas phases. This comprehensive dataset, combined with AXIS's capabilities, will enable a complete characterization of the multiphase medium surrounding NGC 1275.

The Perseus cluster also hosts a complex radio mini-halo at its core [181] and a recently discovered giant radio halo [535], providing a unique laboratory for studying particle acceleration and the interaction between thermal and non-thermal plasma in the ICM. The mini-halo is associated with the AGN 3C 84, located in NGC 1275. Mini-halos are typically found in relaxed, cool-core clusters and are thought to be powered by turbulence from gas sloshing within the ICM [188]. The Perseus mini-halo extends over 300 kpc and exhibits a complex morphology with numerous filaments and edges that trace underlying X-ray gas structures [181]. Its radio emission is asymmetric, with radial filaments extending eastward and northeastward over 150–170 kpc. The brightness profile follows an exponential decay with a scale radius of $30.4 \pm 0.2$ kpc. Spectral index mapping between 144 MHz and 1.5 GHz reveals a relatively uniform spectral index of $-0.9$ to $-1.1$, with slight flattening near NGC 1272. High-resolution VLA imaging shows that the mini-halo is bounded by cold fronts, consistent with gas sloshing driven by off-axis minor mergers [559]. Feedback from 3C 84 likely contributes to turbulence and re-acceleration of relativistic electrons, driving synchrotron emission [181,183,418]. A larger, more diffuse giant radio halo has been identified with LOFAR at 120–168 MHz [535]. Extending up to 1.1 Mpc, it envelops the mini-halo and follows the X-ray morphology of the ICM. The halo's circular shape is punctuated by bright patches and filaments, indicating a turbulent origin. The spectral index across the halo is steeper than that of the mini-halo, ranging from $-1.2$ to $-1.4$, consistent with ageing and radiative losses of relativistic electrons. The discovery of a giant halo in a cool-core cluster like Perseus challenges the conventional view that such halos are exclusive to merging clusters.

- **The Centaurus Cluster of Galaxies ($z = 0.0104$):**

NGC 4696 is the BCG of the nearby Centaurus Cluster and exhibits an extended filamentary nebula of ionised gas ($\sim 9 \times 7$ kpc) displaying a peculiar swirling structure as well as large amounts of dust ([154], [110]). This galaxy and its environment have been studied extensively at various wavelengths, together with the optically emitting filaments surrounding NGC 4696, mostly through Integral Field Unit (IFU) observations with WiFeS and VIMOS ([156], [78]). Infrared observations of the BCG environment were also obtained with the Herschel instrument ([340]), and additional JWST NIRSpec observations are planned to be obtained before AXIS' launch (PI: Hlavaceck-Larrondo, program number 5354), thus offering in-depth information on the central AGN accretion flow. On the other hand, VLA radio observations have revealed AGN jets and bubbles actively reshaping the ICM ([495]), while cold molecular gas within filaments has been detected using Spitzer ([246]). The combination of these diverse datasets with new AXIS observations will allow for a dedicated investigation of the multiphase properties of the ICM in this dynamic environment.

- **The Virgo Cluster of Galaxies ($z \sim 0.00428$):**

As one of the most thoroughly studied galaxies in the local Universe, M87 features a large ($>$1 arcmin) emission-line nebula curving around its AGN-powered radio jets, as observed with the VLA ([86]). This system exhibits strong spatial correlations between filaments and structures in the hot X-ray emitting gas ([589]), making it an ideal target for AXIS observations. Rich optical dataset from MUSE/VLT ([369]) provides detailed spectroscopic mapping of the emission-line nebula, while comprehensive infrared observations from Herschel ([24], [35]) reveal the dust and cooler gas components. Furthermore, JWST/NIRSpec & NIRCam observations were obtained to observe the central AGN on parsec scales (PI: Walsh, program number: 2228) and to perform wide-field imaging of the cluster's galaxies (PI: Tully, program number: 3055), allowing us to spatially compare different regions on small and large scales with future AXIS observations. As the closest and most detailed view of BCG/ICM interactions available, M87 offers an unparalleled opportunity to study feedback processes at the highest possible spatial resolution.

The incredible spatial resolution of AXIS will therefore be instrumental in resolving the complex interplay between different gas phases surrounding these BCGs, potentially revolutionizing our understanding of feedback processes in galaxy clusters.

**Exposure time (ks):** 100 ks per target for a total of 300 ks

**Observing description:** To investigate the multiphase nature of the ICM and the impact of AGN feedback in shaping and influencing the surrounding ICM, we propose deep AXIS observations of three nearby galaxy clusters: Perseus, Centaurus, and Virgo. These systems are ideal laboratories for studying the interaction between AGN-driven outflows and the surrounding hot, warm, and cold gas phases, leveraging extensive multi-wavelength datasets available prior to the AXIS mission.

We have selected NGC 1275 (Perseus Cluster), NGC 4696 (Centaurus Cluster), and M83 (Virgo Cluster) as our initial primary targets based on their well-characterized multiphase structures and their extensive coverage across optical, infrared, radio, and X-ray wavelengths. Each system exhibits large-scale filamentary nebulae that correlate with AGN-inflated X-ray cavities generated through the AGN jets, providing a direct window into the kinematics and thermodynamics of gas cooling, heating, and mixing processes in the ICM.

The Perseus Cluster (NGC 1275, z = 0.0183) hosts the most extended optical filamentary nebula known, spanning 50 × 80 kpc, with strong spatial correlations between its X-ray, optical, and cold molecular gas components. Additionally, previous SITELLE, IRAM, ALMA, and VLA observations provide flux and kinematics maps of the ionized and molecular gas phases. AXIS will thus allow unprecedented spatially resolved X-ray spectroscopy of the ICM–filament interface, revealing the thermodynamic structure of the cooling gas. Moreover, thanks to AXIS's stable PSF at large radii and its wide field of view, a detailed point-to-point analysis between the X-ray and radio emissions can also be performed out to the radius of the radio halo. This will, in turn, allow us to better comprehend the interplay between large-scale emission structures within galaxy clusters.

The Centaurus Cluster (NGC 4696, z = 0.0104) features an extended ionized nebula ($\sim 9 \times 7$ kpc) with a peculiar swirling morphology, linked to AGN activity. IFU observations from WiFeS and VIMOS provide detailed optical kinematics, while Spitzer and Herschel data trace molecular and infrared components. AXIS will therefore probe the small-scale X-ray structure of the ICM, enabling comparisons between the optical filaments and their X-ray counterparts.

Finally, the Virgo Cluster (M87, z = 0.004283) is among the most extensively studied galaxies in the local Universe, with a large ($>$1 arcmin) emission-line nebula that curves around AGN-driven radio jets. Optical (MUSE), infrared (Herschel), and radio (VLA) observations reveal a complex multi-wavelength picture of feedback-driven interactions. Consequently, AXIS observations will provide crucial insights into the formation and evolution of X-ray filaments, complementing existing multi-wavelength data.

Our goal will be to obtain deep AXIS X-ray imaging and spectroscopy for each of these targets, enabling a detailed study of the spatially resolved X-ray emission from the ICM at filamentary scales. The high spatial and moderate spectral resolution of AXIS will allow us to:

- **Map the thermodynamic properties** (temperature, density and pressure) of the X-ray-emitting gas surrounding the BCGs on filamentary scales.
- **Quantify metal transport processes** and **identify shock structures** from jets generated by the AGN interacting with the ICM.
- **Compare X-ray emission with existing optical, infrared, and radio datasets**, establishing the first comprehensive view of multiphase gas interactions at sub-arcsecond resolution.

Preliminary simulations focusing on the Perseus Cluster ([436]) demonstrate that a relatively modest exposure time of 100 ks will substantially improve imaging and detection of ICM emission at subarcsecond scales. As shown in Figure 20, we compare a Chandra image of the Perseus Cluster (obtained using spatial binning with 7500 counts per region) with a simulated AXIS image using similar binning parameters. AXIS capabilities enable a remarkable tenfold reduction in spatial bin size compared to Chandra.

Additionally, the simulated spectra shown on the right side of Figure 20 correspond to a 2.5 arcsecond radius region highlighted in the central panel. The total spectrum (black) clearly differentiates itself from the X-ray filament emission (purple), demonstrating AXIS's ability to extract soft X-ray filament signatures from the overall ICM emission —a critical feat for our science goals. Moreover, upcoming Chandra Legacy surveys and previous deep Chandra observations (PI: Fabian) of our targets will unfortunately be insufficient to achieve our desired goals, since the soft X-ray response of the Space Telescope has degraded significantly over time, thus preventing a detailed analysis of this critical link between the ICM emission and cooler gas phases.

Consequently, AXIS observations will provide an unprecedented window into the small-scale interplay between AGN feedback and the cooling ICM, offering crucial constraints for theoretical models of gas cooling, AGN outflows, and the regulation of baryons in cluster environments.

**Joint Observations and synergies with other observatories in the 2030s:** By 2026, JWST observations will be available for the Centaurus and Perseus clusters. Currently, SITELLE or MUSE data are already accessible for Perseus, Virgo, and Centaurus. Additionally, ALMA observations are available, although limited to the central disk in Perseus, with low detection levels for Virgo.

**Special Requirements:** None

*15. Emission-line filaments surrounding the brightest central galaxies*

**Science Area: ISM, ICM, charge exchange**

**First Author:** Liyi Gu, SRON Space Research Organisation Netherlands

**Abstract:** The brightest central galaxies (BCGs) in galaxy clusters are often surrounded by extensive networks of emission-line filaments, yet their formation and excitation mechanisms remain uncertain. The classical cooling flow model, which predicts the radiative condensation of intracluster gas, is challenged by XMM-Newton RGS observations that reveal a suppression of cooling below one-third of the ambient cluster temperature. This raises a key question: if the ICM is not cooling efficiently, what is the origin of the warm, optical emission-line filaments?

One scenario proposes that the filaments originate from the uplift of cold gas by AGN-driven outflows and jets, with ionization and excitation occurring through AGN kinematics and charge exchange interactions with the surrounding hot ICM. This scenario could be rigorously tested with AXIS, whose high sensitivity and spatial resolution will allow for separation of filaments from the ambient ICM. Additionally, by measuring the strength of charge exchange emission, AXIS could constrain the total mass and energetics associated with AGN-driven uplift.

The second scenario suggests that filaments form via local thermal instabilities, resulting in a stratified, multi-phase structure where emission is primarily driven by thermal or photoionization processes. Charge exchange emission is suppressed in a multi-phase structure.

Future high-resolution X-ray observations with AXIS, combined with optical and UV integral field spectroscopy, will provide critical constraints on the physical conditions and emission mechanisms of these filaments. Additionally, targeted RGS or XRISM (if the gate valve opens) spectroscopy could help differentiate between charge exchange and thermal or photoionization processes, offering new insights into filament formation and its connection with AGN feedback in BCGs.

**Science:** **Emission mechanism of the filaments**

Subtracting the surrounding thermal component from the filament spectrum using AXIS will allow us to obtain a high-quality, uncontaminated spectrum of the filaments.

By comparing this spectrum to different emission models – thermal collisional, photoionization, and charge exchange components – we can determine the dominant excitation processes at play. As shown in Figure 21, a high-quality CCD spectrum such as AXIS can readily distinguish between these processes. This is because charge exchange spectra are typically line-dominant, whereas thermal and photoionization models include continua and exhibit different characteristic transitions. If the spectrum is consistent with thermal emission, it suggests that the filaments are embedded in a multi-phase medium that is still experiencing residual cooling. A spectral signature of photoionization would indicate that radiation from an AGN or young stars is playing a key role in exciting the gas. Alternatively, the detection of strong charge exchange lines would suggest that interactions between the cold filaments and the surrounding hot ICM are the primary driver of the observed emission.

**Heating and cooling of the filaments**

For a thermal filament, it is straightforward to compare the observed temperature with model predictions based on rapid cooling. By constructing a spatial temperature map of the filament, we can search for regions with additional heating.

For photoionized filaments, the ionization state of the emission lines can provide valuable insights into the luminosity and type of the ionizing source.

For a charge-exchanging filament, the total intensity of the emission lines offers a rough estimate of the total mass within the filament. Additionally, the line ratio between highly excited transitions and the low-lying Rydberg series can provide useful constraints on the interaction speed between the cold filament gas and the hot ICM.

**Metallicity of the filaments**

It is straightforward to compare the metallicity of the filament with that of the surrounding ICM to investigate its origin. If the filament shares a similar metallicity with the ICM, it may have condensed out of the hot phase through cooling or thermal instabilities. Conversely, if the filament exhibits an enhanced metallicity, it could indicate enrichment by stellar or AGN-driven outflows from the BCG.

**Comparing with other observations**

Comparing AXIS results with XRISM velocity measurements of the surrounding ICM will provide insights into the dynamics of the filaments, helping to distinguish between uplift and instability formation scenarios.

In addition, comparison with multi-wavelength data – H$\alpha$, CO, [O III], and other relevant lines – will allow for a more comprehensive determination of the formation mechanism.

**Exposure time (ks):** three clusters, about 300 ks per cluster

**Observing description:** To investigate the nature of cold gas filaments in the cores of galaxy clusters, we propose deep AXIS observations of three nearby clusters: Perseus, Centaurus, and Abell 1795. The filaments in the Perseus cluster exhibit remarkable linearity; some extend over 6 kpc in length while being only about 70 pc wide. This morphology suggests they are supported by magnetic fields that stabilize them against tidal shear and prevent dissipation into the surrounding hotter ICM [148]. Similar bright X-ray filaments are also observed extending from the cores in both the Centaurus cluster and Abell 1795.

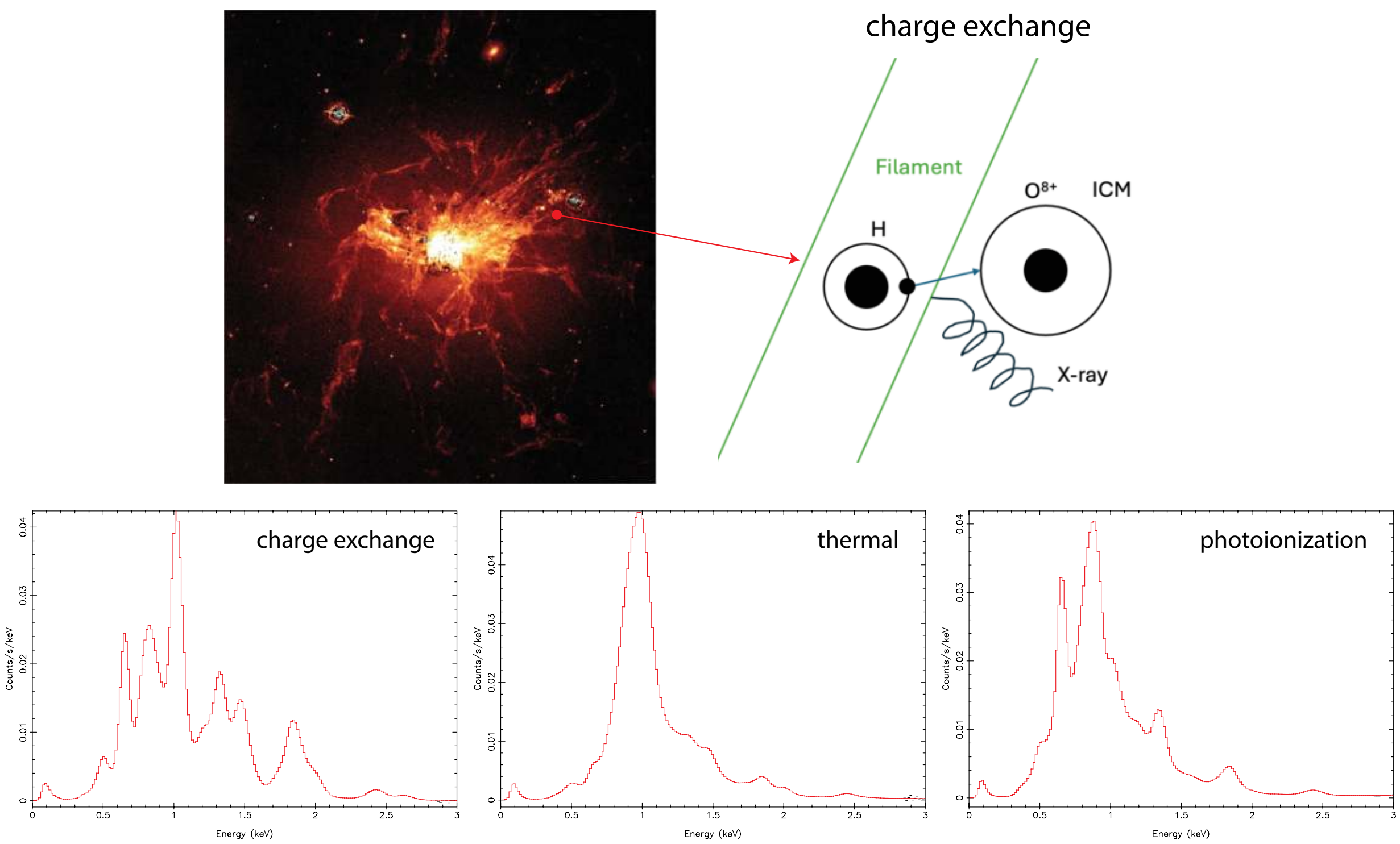

**Figure 21.** Filaments in the Perseus cluster and the comparison of underlying spectral models. (upper) H$\alpha$ image of the Perseus cluster core [148]. A schematic on the right illustrates the possible charge exchange process occurring at the interface between cold filaments and the surrounding hot ICM. (lower) Representative spectral models for charge exchange, thermal, and photoionized emission, each convolved with the *Chandra* CCD response. The X-ray flux is set to observed values in the Perseus cluster [560]. The spectral features are clearly different at CCD resolution, allowing for separation of the underlying emission mechanisms with AXIS.

These three clusters have previously been studied using Chandra ACIS data [560]. Their spectra are predominantly soft, and the soft X-ray flux of each major filament in Perseus has been quantified. However, the sensitivity of existing Chandra observations is insufficient to definitively distinguish whether the filament emission is primarily driven by charge exchange, thermal processes, or photoionization.

The primary goal is to obtain deep AXIS spectra of the filaments to:

- Robustly identify the emission mechanism of the filaments. If charge exchange is favored, this could lead to a definitive detection of charge exchange emission in clusters.
- Determine the metallicity of the filaments, which can provide insights into their role in transporting metals from galaxies to the ICM.
- Compare the X-ray properties with optical and infrared data. If charge exchange processes are occurring, the AXIS spectra could independently constrain the collision velocity, allowing for meaningful comparisons with results derived from longer wavelengths.

Practically, this would require approximately 300 ks per cluster to surpass the depth of archival Chandra data significantly. This estimate is based on a rough calculation assuming a factor of 7 increase in the effective area at 1 keV with AXIS compared to Chandra. It also accounts for the total exposure time of existing archival Chandra data prior to ACIS becoming nearly insensitive below 1 keV due to contamination.

**Joint Observations and synergies with other observatories in the 2030s:** Tracers of the multiphase gas from ELTs and ALMA for the three clusters will be useful in constraining the charge exchange model.

**Special Requirements:** None

### d. Galaxy cluster mergers

*16. The ICM at sharp focus: cold fronts, plasma depletion layers, connection with radio filaments*

**Science Area: Galaxy clusters, intracluster medium, magnetic fields, particle acceleration**

**First Author:** F. Gastaldello (INAF-IASF Milano)

**Co-authors:** A. Botteon (INAF-IRA Bologna), S. Ghizzardi (INAF-IASF Milano), J. ZuHone (CfA), I. Bartalucci (INAF-IASF Milano), M. Balboni (University of Bologna), A. Bonafede (University of Bologna), N. Biava (INAF-IRA Bologna, TLS), Maxim Markevitch (NASA GSFC)

**Abstract:** The arcsec resolution and large effective area of AXIS allow the plasma physics of the ICM to be probed with unprecedented detail. Phenomena only hinted at by Chandra can be used to probe the conduction and viscosity of the ICM high $\beta$ plasma and its magnetic structure, in conjunction with the features revealed in synchrotron emission by current and modern radio telescopes. In particular, in this project we aim to reveal the science cases that AXIS can tackle for what concerns: i) plasma depletion layers (PDL) in shear flows associated with cold fronts and sloshing motions ii) the ICM properties that can be probed by tails of radio galaxies or thin filaments. We will show detailed simulations for the PDL in the cluster A1068.

**Science:** The structure we observe today on the largest scales of the cosmos originates from tiny density perturbations that were left over after the Big Bang. Under the influence of gravity, small overdense clumps of matter have merged over time, leading to a web-like structure of galaxies, galaxy groups, galaxy clusters, and large-scale cosmic filaments, spanning the observable Universe [54]. This succession of mergers injects kinetic energy into the newly formed structures, which is eventually dissipated into heat. This is why most of the normal, baryonic matter is today in the form of a diffuse plasma with temperatures reaching millions to hundreds of millions of degrees, filling the space around and between galaxies in the cosmic web [e.g., 315]. However, many of the properties that characterize this plasma - viscosity, heat conductivity, and the interplay between particles and magnetic fields - are still unknown and need to be characterized observationally because of the large uncertainty of the theoretical estimates.

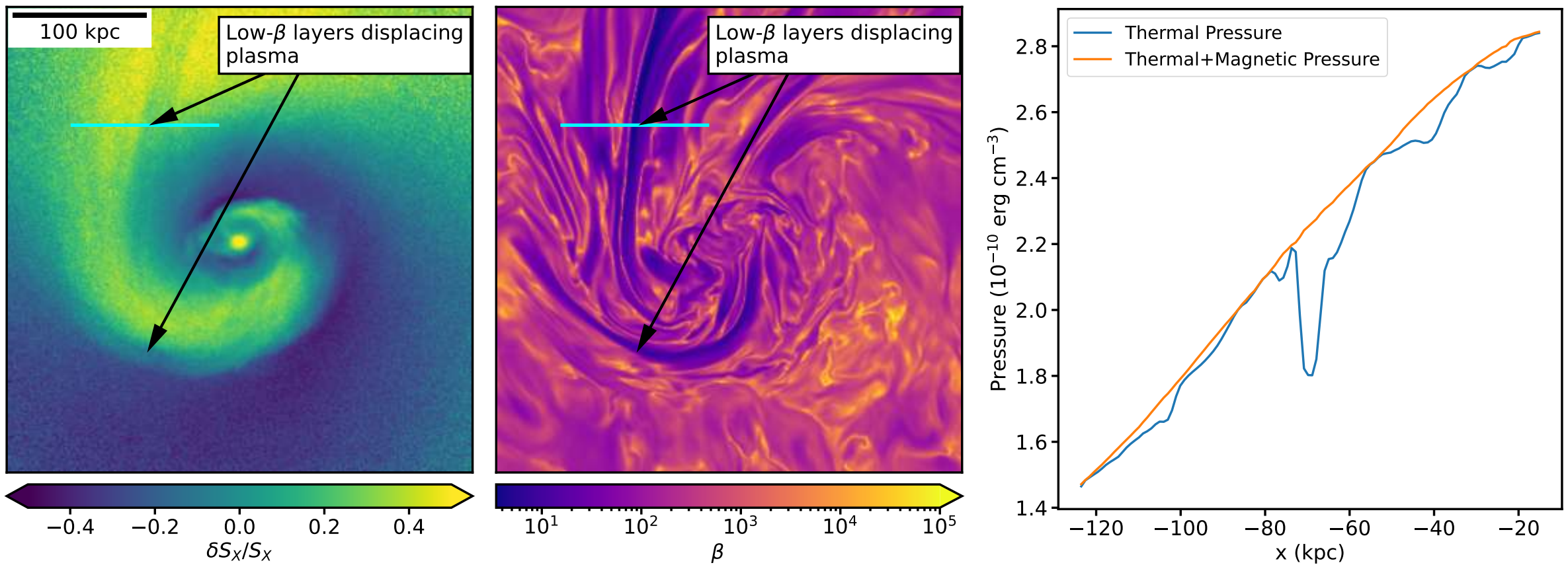

**Figure 22.** left: SB residuals projection from an AXIS simulation from a large cool-core cluster with gas sloshing. center: slice through plasma beta parameter. sloshing motions produce strong magnetic fields through shear amplification, which can amplify fields down to $\beta \sim 3$ near cold fronts and push plasma out, revealing plasma depletion layers in X-ray SB which can be detected with AXIS.

Key measurements can be performed with a high-resolution and high-effective-area instrument such as AXIS, building on the legacy of Chandra discoveries of surface brightness edges [312,616] and sharp features in the plasma distribution of the ICM, and obtaining better physical insights.

Cold fronts are the interfaces between a cool, dense gas cloud and a hot, tenuous ambient medium. The cool gas is the remnant of an infalling subcluster or gas from the core that has been displaced by sloshing. The significantly increased sensitivity of AXIS with respect to Chandra will enable us to display the full array of constraints on plasma microphysics that the cold front can afford by systematically studying these interfaces in an azimuthally resolved fashion, both in their thermodynamic and metal abundance properties. Comparison with radio features, such as edges in mini-radio halos or discontinuities in radio galaxy tails [58,97] will be enlightening.

Plasma depletion layers (PDL) forming in magnetic layers amplified by the velocity field at sloshing cold fronts are the latest key sharp feature in the ICM discovered by Chandra [313, and references therein]. They were initially discovered outside of Earth's magnetopause and subsequently found to be a common feature of planetary magnetospheres. A PDL is a region of reduced plasma density and enhanced magnetic field strength that can form in the magnetosheath (the shocked solar wind) adjacent to the magnetopause boundary [618]. It results from the pileup of interplanetary magnetic field (IMF) lines against the magnetospheric obstacle, and the associated partial evacuation of plasma away from the region along the direction of the piled-up magnetic field. In close analogy to the effect described above, PDLs have been recognized in MHD simulations of the ICM when the magnetic field in the ICM is stretched and amplified by a plasma flow with subsonic but super-Alfvénic velocity, as indeed in the proximity of sloshing cold fronts [131,293,614]. The magnetic pressure then rises and the gas is squeezed out of the region to keep pressure equilibrium. If enough gas is squeezed out of the region, a deficit of X-ray emission is visible, as in the case of A2142 [568] and of A1068 [36], see Fig.22. PDLs offer the possibility to disentangle the effects of magnetic fields and viscosity on cold front stability, and also to observe the structure of the magnetic field in thermal emission, providing an analogue to what can be inferred in the radio band from the synchrotron emission of relativistic particles.

Filamentary structures are being observed in an increasing number of synchrotron sources, and they trace the places where to point AXIS. For example, in the case of Abell 2657 [60], a bifurcated radio arc and thinner strands in the outer radio arcs have been interpreted as a radio bubble shredded by gas sloshing [151,615], similar to other striking features found in other environments down to the scale of the large population of radio filaments in the Galactic center [see 591, for similarities]. These features are poorly matched to existing X-ray observations. With the advent of radio facilities that offer increasing sensitivity and spatial resolution, the number of such features is expected to grow enormously. The corresponding increase in X-ray sensitivity at high angular resolution, provided by AXIS, is necessary to make the required connections.

**Exposure time (ks):** 50ks on A1068

**Observing description:** A selection of targets of interest for the scientific cases described in this proposal is listed below.

- Cold fronts: Abell 2142, Abell 3667, Abell 3376, Abell 1775
- PDLs: Abell 2142, Abell 520, Virgo, Perseus, Abell 1068
- radio filaments: Abell 2657, Abell 194 (see [432])

We used here to highlight the key capabilities of AXIS a simulation of the cluster A1068 at z=0.139. This is the highest redshift PDL in a sloshing core known to date. The simulation is based on the 27 ks Chandra observation (ObsID: 1652) analyzed and discussed in [36] where the evidence of a PDL is reported in a sloshing cool core with multiple cold fronts (previously recognized by [309] selecting this target for deeper Chandra observations), see Fig.23. We used the spatial distribution of the X-ray emission and the global spectrum (modeled with a `phabs*apec` model in `XSPEC`) obtained with this data to simulate an *AXIS* observation of 50 ks using `SIXTE`. The input flux of the model in the 0.5–7.0 keV band was $1.2 \times 10^{-11}$ erg s cm$^{-2}$ s$^{-1}$. The result of the simulation –a count image in 0.5–7.0 keV band– is shown in Figure 23 (bottom panel). A comparison of the SB profiles across the cold front and the PDL is also shown, highlighting the exquisite AXIS statistics. Finally, we extracted spectra from the PDL and compared the AXIS spectra (in black) with those of Chandra: the 7000 counts with 50ks AXIS (to be compared with the 500 from the 27ks Chandra observation) allow us to constrain the density and the temperature at the 2% and 4% level, respectively. This will allow us for the first time to characterize the physical properties of these sharp features.

**Joint Observations and synergies with other observatories in the 2030s:** Radio facilities with high sensitivity and spatial resolution: MeerKAT, LOFAR 2.0, SKA

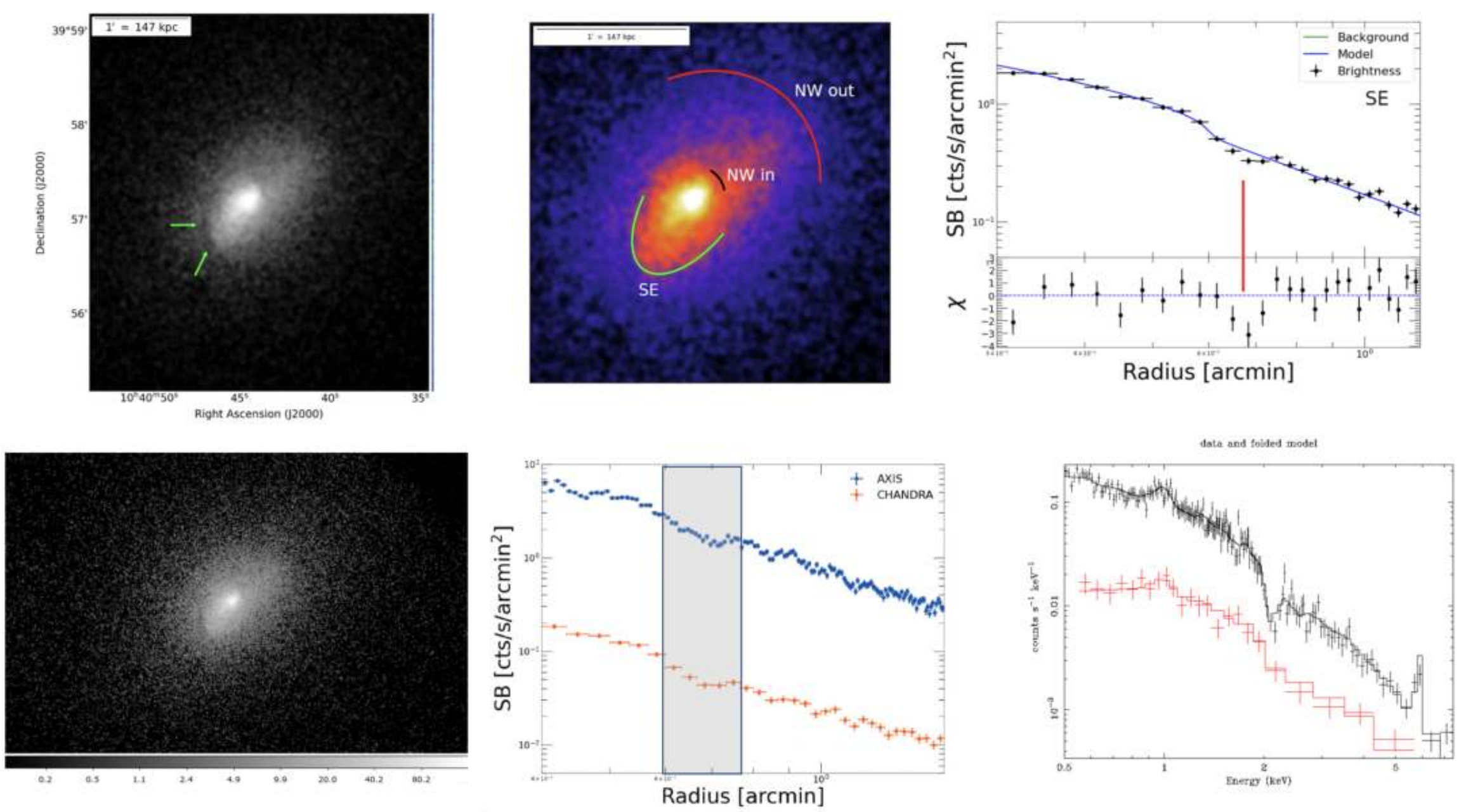

**Figure 23.** The AXIS case for a 50 ks observation of the A1068 cluster. *Top panel:* In the left panel the A1068 Chandra image in the 0.5-7 keV band with green arrows indicating the position of the PDL. The central panel indicates the position of the cold fronts. The right panel shows the best-fitting broken power-law model (blue line) with associated residuals of the SB discontinuity in the SE. The red line indicates the position of the PDL. Figure taken from [36]. *Bottom panel:* In the left panel the 50 ks AXIS image of A1068 in the 0.5-7 keV band. In the central panel, the comparison of the Chandra and AXIS SB profiles across the cold front and the PDL (highlighted by the shaded box), with the exquisite AXIS statistics allowing a 1 arcsec binning. In the right panel, the comparison of the Chandra (red) and AXIS (black) spectra extracted from the PDL, i.e., an elliptical sector with the same angular extension as in the Chandra profile and 6 arcsec wide, corresponding to the Chandra density depression.

**Special Requirements:** A stable PSF over the field of view will clearly allow a better characterization of sharp features of the ICM, also at large radii for local clusters. This must be matched to a good knowledge of NXB.

*17. X-ray and radio surface brightness edges in the ICM*

**Science Area: galaxy clusters, intracluster medium, non-thermal phenomena**

**First Author:** Andrea Botteon (INAF-IRA)

**Co-authors:** Marco Balboni (University of Bologna), Fabio Gastaldello (INAF-IASF Milano), Simona Giacintucci (NRL), Myriam Gitti (University of Bologna), Wonki Lee (Yonsei University), Maxim Markevitch (NASA GSFC), Kamlesh Rajpurohit (CfA), Timothy Shimwell (ASTRON), Francesco Ubertosi (University of Bologna), Reinout van Weeren (Leiden University), John ZuHone (CfA)

**Abstract:** *Chandra's* superb angular resolution has played a crucial role in detecting numerous shocks and cold fronts in the intracluster medium (ICM) of both relaxed and merging clusters. These fronts trace density jumps that appear as sharp edges in the X-ray surface brightness distribution. High-fidelity radio observations of extended cluster sources (radio halos and mini halos) with Square Kilometre Array (SKA) pathfinders and precursor instruments have revealed that diffuse radio emission is rich in substructure and often exhibits surface brightness discontinuities that correlate with those observed in the X-ray-emitting gas. This suggests a strong interplay between thermal and nonthermal components in the ICM. Radio observations have now become highly efficient in detecting discontinuities in the ICM, often outperforming X-ray observations, which generally require very long exposures to identify these edges. This trend is expected to become even more pronounced with the advent of the SKA. Fortunately, future observations with *AXIS* will help bridge this gap. Thanks to its high-angular resolution, which remains approximately constant across its wide field-of-view and its large collecting area, *AXIS* will efficiently study the connection between thermal and nonthermal components in clusters and investigate the nature of radio/X-ray surface brightness edges in the ICM.

**Science:** Galaxy clusters are the most massive collapsed structures in the Universe. Their total mass is dominated by dark matter ($\sim$80%), which forms deep gravitational potential wells where baryonic matter ($\sim$20%) virializes. The majority of the baryonic matter in clusters resides in the ICM, a hot ($kT \sim 2 - 10$ keV) and tenuous ($n_{\rm e} \sim 10^{-3} - 10^{-4}$ cm$^{-3}$) plasma that emits primarily via thermal bremsstrahlung in the X-ray band. In many galaxy clusters, radio observations have revealed the presence of diffuse synchrotron sources, indicating the existence of nonthermal electrons and magnetic fields mixed within the ICM [see 533, for a review]. These extended radio sources allow us to probe particle acceleration mechanisms in dilute plasmas and study the intricate interplay between thermal and nonthermal components in the ICM. Among the extended diffuse sources that can be found in clusters – the so-called "radio halos" – a distinction is commonly made between *mini* and *giant* halos, which is not solely based on their size, as their names might suggest. Mini halos are typically confined within the central cooling region of *relaxed* clusters ($r < 0.2 r_{500}$[1]), while giant halos extend above the Mpc-scale (even $r > r_{500}$) and are associated with *merging* clusters. Although the radio-emitting particles in both types of halos likely originate via turbulent particle reacceleration, the underlying physical processes driving their formation differ. Turbulence in mini halos is believed to be induced by the central active galactic nucleus or by sloshing motions in the cool core region, while in giant halos it is driven by energetic cluster mergers. Alternative processes, such as hadronic collisions or mechanisms involving shocks (which can reaccelerate particles and adiabatically compress nonthermal plasma), may also contribute. However, their role is likely subdominant or confined to specific regions within the halos [see 70, for a review].

---

[1] The $r_{500}$ radius encloses a mean overdensity of 500 with respect to the critical density at the cluster redshift.

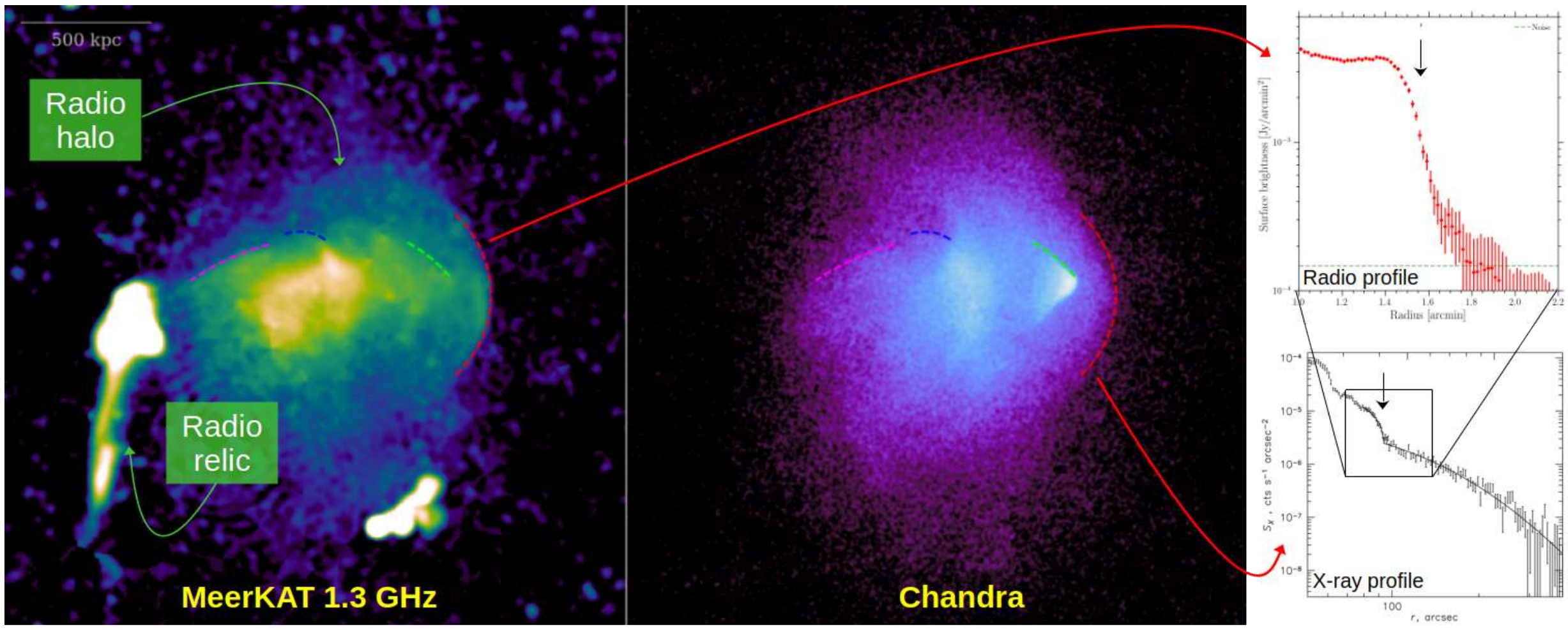

**Figure 24.** MeerKAT 1.28 GHz image (*left*) and deep, exposure-corrected, *Chandra* 500 ks image in the 0.5–2.0 keV band (*center*) of the Bullet cluster where discrete sources were cosmetically removed. The four dashed arcs mark the position of the radio edges detected in the MeerKAT image. These are co-located with shocks and/or cold fronts in the *Chandra* image. Panels on the *right* show the radio (*top*) and X-ray (*bottom*) surface brightness profiles extracted across the prominent bow shock in the west (red arc). Adapted from [57,307].

Both relaxed and merging clusters are important laboratories for studying several aspects of plasma astrophysics. In particular, the sub-arcsecond resolution of *Chandra* has enabled detailed investigations of previously unseen structures, such as shocks and cold fronts [see 312, for a review]. These features appear as sharp surface brightness discontinuities in X-ray observations but differ in the nature of their temperature and pressure jumps across the front. Shocks represent pressure discontinuities where gas is compressed and heated in the downstream (post-shock) region, resulting in higher temperatures compared to the upstream (pre-shock) region. In contrast, cold fronts exhibit the opposite temperature variation, with the denser, cooler gas moving through a hotter, less dense medium, while maintaining approximate pressure equilibrium across the boundary. Thanks to its exceptional spatial resolution, *Chandra* has played a crucial role in linking mini halo boundaries to sloshing cold fronts [e.g. 187,189,318] and in associating the edges of some giant halos with shock fronts [e.g. 308,310,311]. These discoveries provide key observational evidence connecting diffuse radio emission with the dynamical state of the ICM and the mechanisms responsible for particle acceleration and magnetic field amplification in galaxy clusters.

Except for a few cases, both mini and giant halos have been historically observed as sources with smooth and regular morphologies that broadly follow the X-ray-emitting gas. However, highly sensitive observations with modern radio interferometers –precursors and pathfinders of the SKA– are now providing a new picture. In recent years, the number of detected substructures in radio halos, such as prominent and previously unknown radio surface brightness discontinuities and filamentary structures, has been rapidly increasing [36,49,50,57,59,61,183,190,196,219,258,405,534,535]. A textbook example is the Bullet cluster, reported in Figure 24. Despite the Bullet cluster being well known as a highly disturbed system with a prominent shock co-located with the western boundary of the radio halo [310,462], new MeerKAT observations at 1.28 GHz from the MeerKAT Galaxy Cluster Legacy Sample [258,466] have revealed even more complexity. In addition to the previously reported edge coincident with the western shock, three additional radio surface brightness discontinuities have been discovered within the halo [57], as shown in Figure 24. Remarkably, all four radio edges are co-located with X-ray discontinuities identified in the deep 500 ks *Chandra* observation of the cluster, and their surface brightness profiles show similar shapes.

The investigation of the nature of these radio/X-ray edges is currently ongoing. In X-rays, a projection of a 3D spherical discontinuity is generally adopted to describe the underlying density profile for the surface brightness jumps associated with shocks and cold fronts. This model is motivated by the fact that these edges mark contrasts in the thermal electron density $n_e$ and that the X-ray emissivity in the hot ($\gtrsim$2.5 keV) ICM is $j_x \propto n_e^2 (kT)^{1/2}$, which is very weakly dependent of the gas temperature in the soft X-ray band [e.g. 139]. In radio, the synchrotron emissivity at frequency $\nu$ for a simple power-law spectrum is $j_r \propto n_{CRe} B^{(\delta+1)/2} \nu^{-(\delta-1)/2}$, thus it depends on the magnetic field strength $B$, density of the emitting nonthermal electrons $n_{CRe}$, and slope of the electron energy distribution $\delta$ (which is related to the observed synchrotron spectrum via $\delta = 2\alpha + 1$). Therefore, determining what causes the jump is not trivial due to the degeneracy between $n_{CRe}$ and $B$. However, by combining radio and X-ray observations of these edges, it is possible to investigate their nature under simple assumptions (e.g., assuming that the thermal and nonthermal electron density contrasts are the same or that $B$ is adiabatically compressed).

While MeerKAT observations with integration times of ∼6-10 h appear sufficient to detect radio edges [57], detecting the same discontinuities in X-rays with *Chandra* typically requires exposures $\gtrsim$100 ks (≈28 h). Even longer X-ray exposures are often necessary to accurately measure the temperature jump across the fronts. These measurements are crucial for distinguishing between shocks and cold fronts and for revealing the processes responsible for enhancing radio synchrotron emission in their downstream regions. This highlights a major observational bottleneck: while modern radio observations are now very efficient in detecting ICM discontinuities, the ability to interpret these features and to robustly determine the underlying physical processes critically depends on complementary X-ray measurements of the thermal gas properties across the same regions. Currently, such joint radio/X-ray analyses are limited to a few clusters with exceptionally deep *Chandra* exposures, making the need for a next-generation mission like *AXIS*, increasingly urgent.

Thanks to its large collecting area and high angular resolution, *AXIS* will enable an efficient study of these edges in combination with radio interferometers currently in the completion phase (e.g., LOFAR2.0, MeerKAT+, and SKA), thereby mitigating the need for prohibitively long exposure times in X-rays. Compared to *Chandra*, *AXIS* will significantly enhance this type of analysis, also thanks to its wide field of view, which will allow the detection of edges throughout the entire X-ray-emitting region without the severe point spread function (PSF) degradation that affects *Chandra* at large off-axis distances. This will elevate the study of radio/X-ray discontinuities in the ICM to the next level, enabling us to conduct systematic studies of these edges and determine the origin of these structures, thereby disentangling the roles of shocks, turbulence, and magnetic field amplification. This is essential for understanding how energy is dissipated during cluster formation and for unveiling the processes that govern the microphysics of the ICM.

**Exposure time (ks):** 100 ks

**Observing description:** The Bullet cluster (RA 06:58:26.88, DEC -55:58:21) is a prime example to showcase the improvements that *AXIS* will bring over *Chandra*. It is a highly disturbed system at $z = 0.296$, featuring multiple X-ray edges, which has already been observed with a deep *Chandra* mosaic, totaling 500 ks of on-target time. This is one of the longest *Chandra* observations ever performed on a merging cluster. In Figure 25 (left panel) we show an image in the 0.5–2.0 keV band obtained with an archival 100 ks observation (ObsID: 5356). We used the spatial distribution of the X-ray emission and the global spectrum (modeled with a `phabs*apec` model in `XSPEC`) obtained with this data to simulate an *AXIS* observation with the same duration using `SIXTE`. The input flux of the model in the 0.5–2.0 keV band was $5.1 \times 10^{-12}$ erg s cm$^{-2}$ s$^{-1}$. The result of the simulation –a count image in 0.5–2.0 keV band– is shown in Figure 25 (right panel). The side-to-side comparison of the two images (displayed with a matched colorscale), demonstrates the significant improvement in count statistics that *AXIS* will provide. We estimate that *AXIS* will collect $\sim$6 times more counts than a *Chandra* observation of the same duration, exceeding even the total number of counts collected in the deep 500 ks *Chandra* mosaic (Figure 24). Importantly, *AXIS* will provide an approximately constant PSF across the entire field-of-view where the thermal cluster emission is detected. This represents a significant advantage in the characterization of the surface brightness jumps with respect to *Chandra*, which requires edges to be positioned at the aimpoint to optimize its PSF. Since our simulation is based on a single *Chandra* ObsID with the aimpoint at the bow shock, the simulated *AXIS* image underestimates the full capability of the instrument in detecting sharp features throughout the cluster.

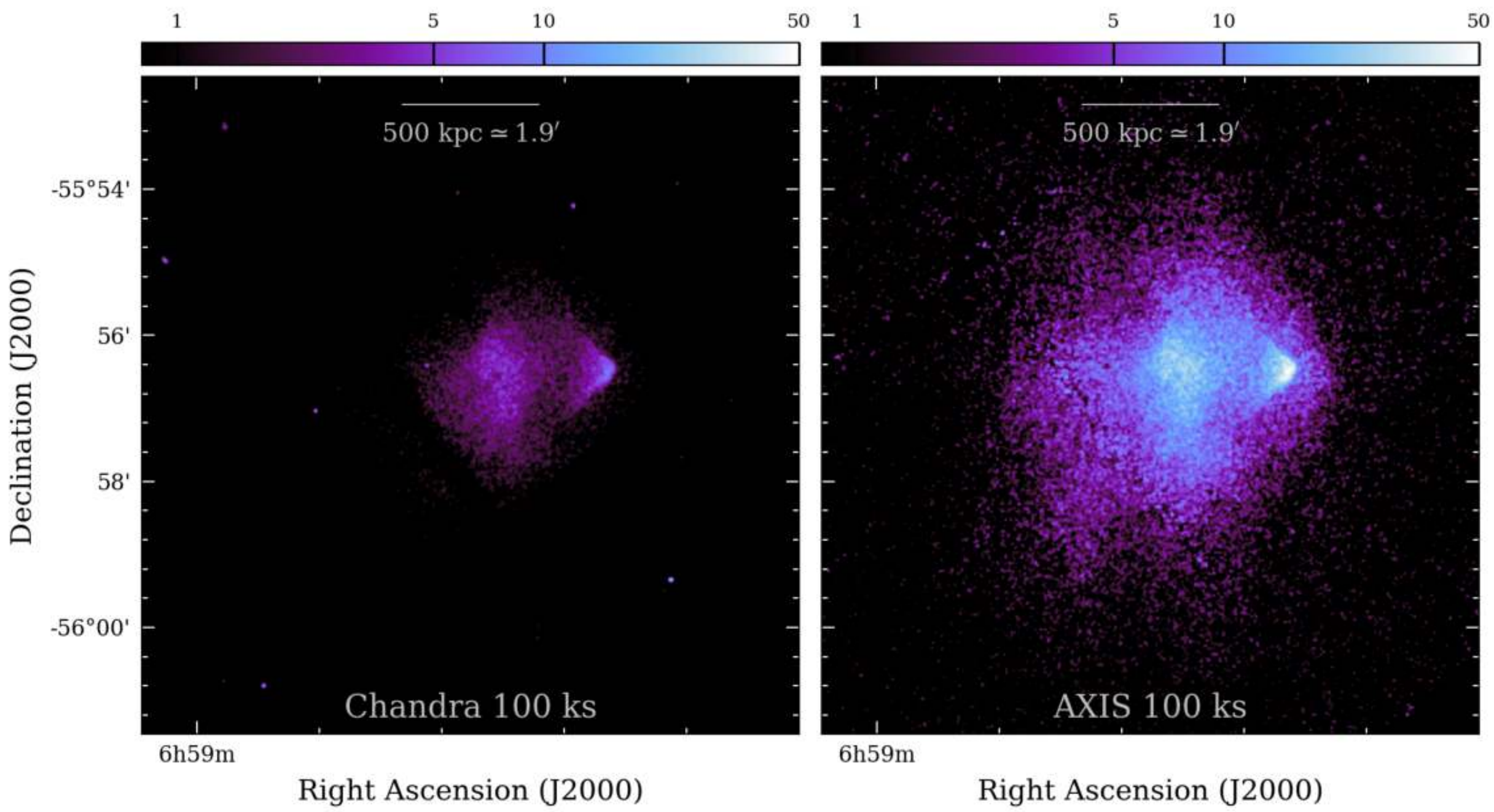

**Figure 25.** *Chandra* archival (*left*) and *AXIS* simulated (*right*) observations of the Bullet cluster. Both images have an exposure time of 100 ks and cover the energy range 0.5–2.0 keV. The images have units of counts and have a matched colorscale.

**Joint Observations and synergies with other observatories in the 2030s:** Radio surface brightness discontinuities in mini and giant halos have been reported primarily with LOFAR, VLA, and MeerKAT. Currently, both MeerKAT and LOFAR are undergoing instrumental upgrades (MeerKAT+ and LOFAR2.0) that will enhance their sensitivity and resolution, while in the 2030s, SKA will become fully operational. The synergy between *AXIS* and these interferometers will enable the joint modeling of radio and X-ray edges in the ICM.

**Special Requirements:** Low and well-understood background

*18. The structure of cluster merger shocks*

**Science Area: Galaxy clusters, intracluster medium**

**First Author:** H. R. Russell (University of Nottingham)

**Co-authors:** J. ZuHone (CfA), F. Gastaldello (INAF/IASF-Milano), S. Randall (Center for Astrophysics | Harvard & Smithsonian), E. Bulbul (MPE), C. Zhang (Masaryk), A. Gill (MIT), A. Tumer (GSFC, UMBC), A. Sarkar (MIT), M. Calzadilla (CfA)

**Abstract:** AXIS's high spatial resolution and large increase in effective area will reveal the structure of cluster merger shocks in unprecedented detail. By mapping the gas properties on arcsec scales across the faint, pre- and highly-structured postshock regions, AXIS will explore new plasma microphysics and demonstrate their unknown but potentially crucial impact on heat transport on macro scales. We propose a 1 Ms AXIS observation of the galaxy cluster merger Abell 2146 to measure the electron-ion thermal equilibration timescale and map the strong, normal, and weaker, inclined shock fronts generated as a multitude of gas flows converge or diverge in the merger's wake.

**Science:** Massive galaxy clusters form at the largest nodes in the cosmic web, where they regularly merge with smaller clusters and groups (for a review, see [312]). In these mergers, the constituent galaxies and dark matter halos are essentially collisionless particles (e.g., [105]) and travel ahead of the spectacular collision of hot atmospheres, which behave like fluids. *Chandra* X-ray observations of hot atmospheres reveal shock fronts, sharp edges associated with cold fronts, large-scale turbulent eddies, and tails of gas stripped from cool cores by ram pressure. Vast shock fronts are typically $\sim$ 500 kpc across with Mach numbers $M \sim 1-3$. They are a primary heat source for the intracluster medium (ICM), dissipating energies $> 10^{63}$ erg and raising the gas temperature in accordance with the newly-formed, deeper gravitational potential (e.g., [421]). Shocks and turbulence generated by the merger can also accelerate particles to relativistic speeds. The resulting synchrotron emission produces large-scale radio haloes and relics (e.g., [533]).

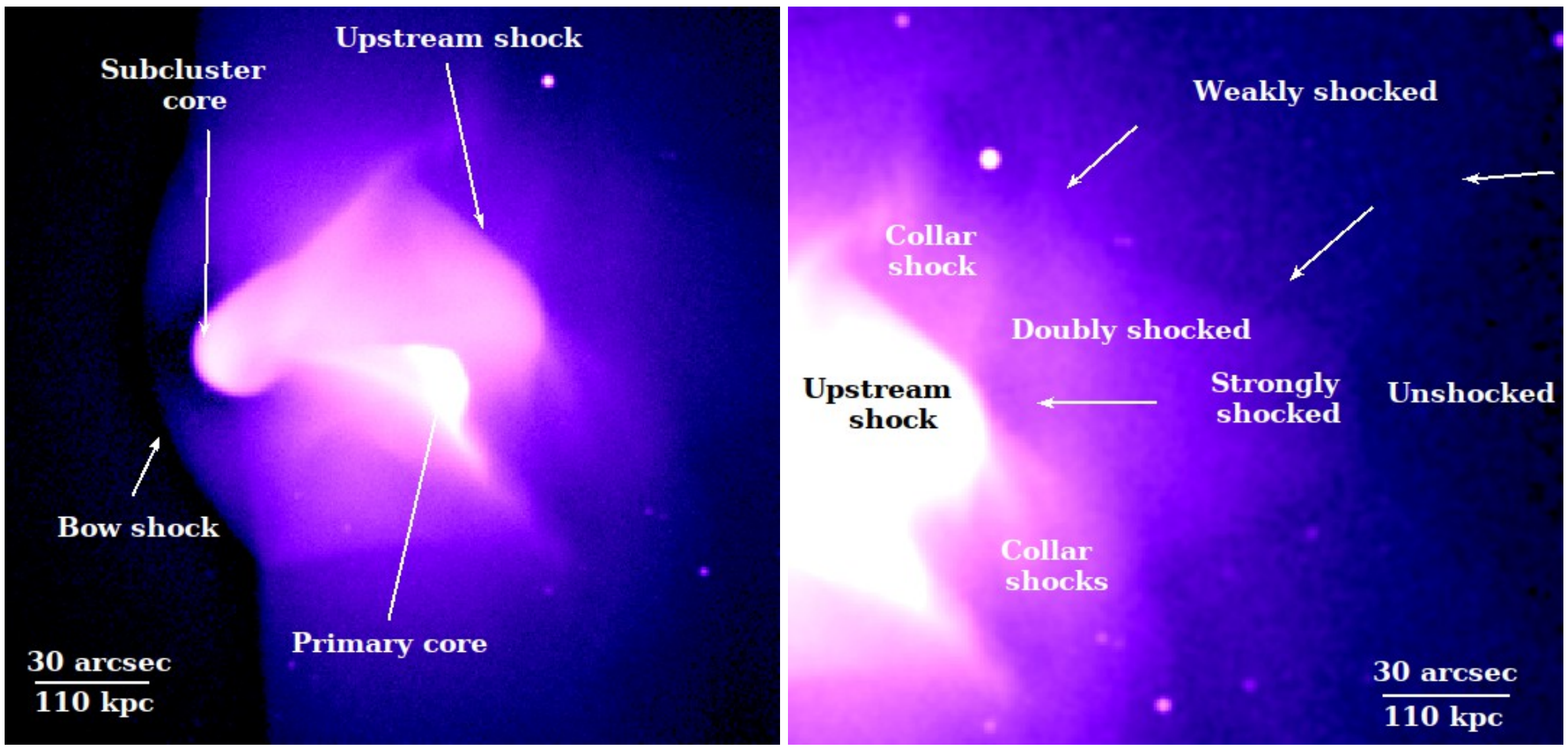

**Figure 26.** Left: Simulation of a 1 Ms AXIS observation of galaxy cluster merger Abell 2146, based on hydrodynamical simulations from [90]. Right: Focus on the upstream shock and the complex preshock structure.

Detections of shock fronts, with sharp, unambiguous jumps in both density and temperature, are rare. This is primarily an observational shortcoming. Shocks are easiest to detect shortly after first pericenter passage, when they are close to the bright center of the cluster, and when the merger axis is oriented close to the plane of the sky. Only a handful are known, and only three clusters host bright enough and strong enough bow shocks for studies of their structure: the Bullet cluster ([307]), A520 ([567]), and A2146 ([435]). Hydrodynamical simulations of these mergers ([90]), and new radio observations of halos with LOFAR and SKA pathfinders (e.g., [466]), hint at a far more complex picture with a whole series of strong, normal and weaker, inclined shock fronts generated as a multitude of gas flows converge or diverge in the merger (Fig. 26).

In the low densities of the ICM, merger shocks are likely to be collisionless and propagate via plasma-wave interactions between the ICM particles and the cluster's magnetic field ([307]). Collisionless shocks occur over a wide range of spatial scales in astrophysics, from accretion shocks at the intersection of massive cosmic structure filaments to supernova remnants and solar wind shocks (e.g., [306]). Ions are heated in a narrow shock layer with a width of order the Larmor radius. Electrons likely remain significantly cooler, equilibrating with the ions through collisions, unless an additional electron heating process is present. Only clusters provide the opportunity to map the electron equilibration process in a single observation that is unaffected by systematic errors from cross-calibration. Cluster shock fronts can produce a straightforward benchmark and provide an important guideline for more refined models in heliospheric shocks and SNRs. This is a hugely challenging observation for *Chandra*, and current results are in conflict ([307,435,567]).

AXIS's high spatial resolution and large increase in effective area will reveal merger shock structure in unprecedented detail and routinely measure postshock equilibration timescales to explore new plasma microphysics and their impact on macro-scale heat transport (Fig. 27).

**Exposure time (ks):** 1 Ms observation of the merging galaxy cluster Abell 2146

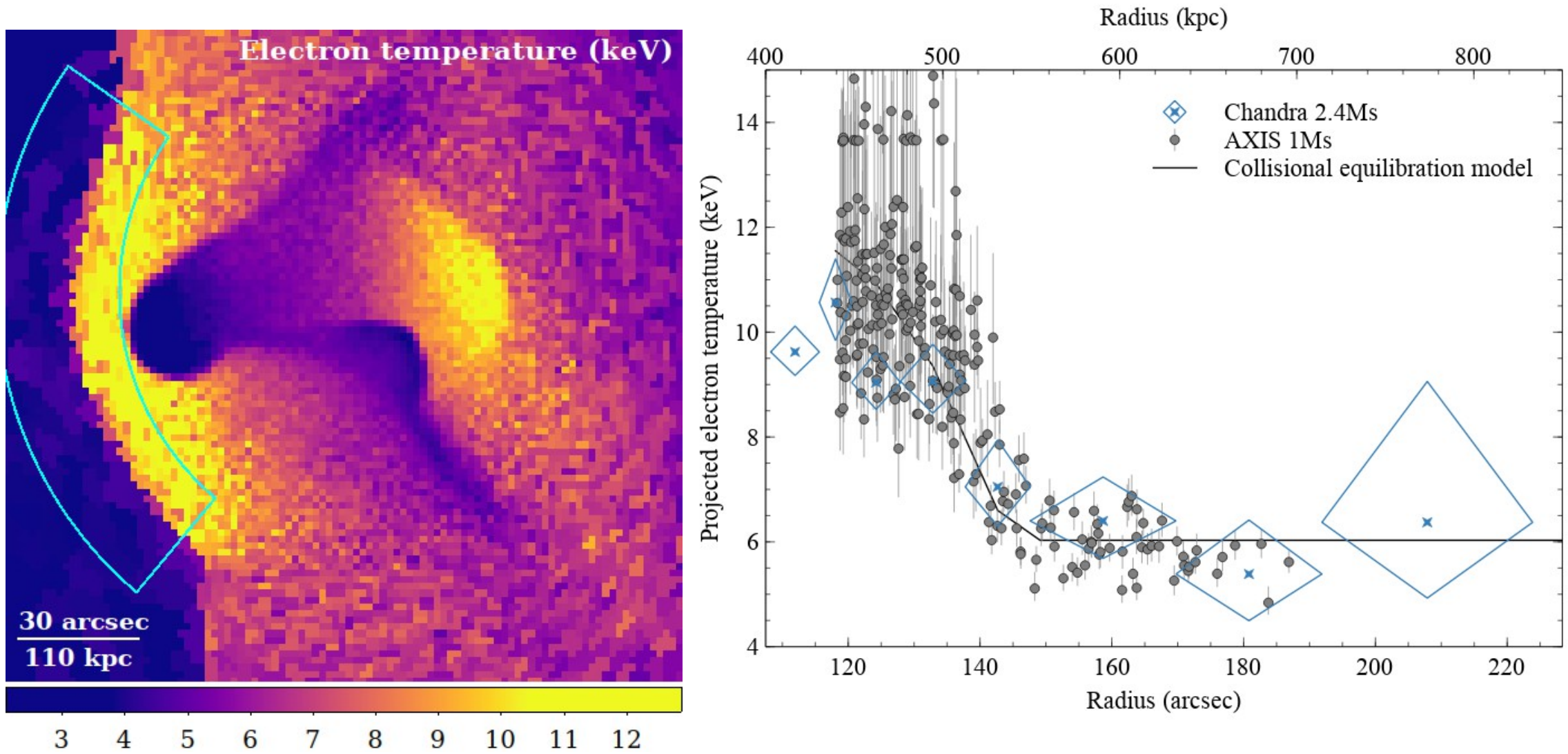

**Figure 27.** Left: Simulated electron temperature map for a 1 Ms AXIS observation of Abell 2146. Right: Simulated electron temperature profile across the bow shock extracted from the sector shown left. The simulated profile is compared to results from a 2 Ms *Chandra* observation [435] and the theoretical expectation from collisional equilibration of electrons and ions in the postshock region.

**Observing description:** Deep *Chandra* observations of Abell 2146 ($z = 0.23$, 3.6 kpc/arcsec) reveal the supersonic passage of a subcluster through the center of a primary cluster roughly 0.1 Gyr ago, producing two large shock fronts ahead of each cluster core [434]. A detailed dynamical analysis of the galaxies tightly constrained the merger geometry and crucially determined that the angle between the merger axis and the plane of the sky is $13 - 19$ deg [578]. Each shock front is over 400 kpc in length but appears remarkably narrow over this distance. Modest asymmetries in the merger, such as the offset location of the primary cluster core, are due to a non-zero impact parameter. Hydrodynamical simulations show that the cluster cores likely passed $\sim 100$ kpc from each other at closest approach [90].

From the electron temperature profile behind the bow shock, [435] ruled out rapid thermal equilibration of the electrons and the shock-heated ions at the $6\sigma$ level. The observed temperature profile favours slower collisional equilibration, but systematic uncertainties dominate measurement of the slope. The drop off in shock Mach number with angle around the front does not follow the theoretical expectation of $\cos\theta$, where $\theta$ is the angle between the velocity of the subcluster's core and the normal to the shock front). This indicates that the shock velocity is not steady and the pre-shock medium is not uniform. We require high-resolution maps of the gas temperature on a few arcsec scales to sample non-uniformity in the pre-shock medium and variations in shock velocity. Additionally, we must measure the post-shock electron temperature and density on scales ranging from a few arcsec to 10 arcseco. The hot atmosphere at larger distances behind the shock was heated by the shock a few $\times 10^7$ yr to $10^8$ yr ago, or roughly around core passage. The pre-shock conditions were likely different, as evidenced by the rapid flattening of the post-shock electron temperature profile. The changing state of the pre-shock gas necessitates detailed mapping of the gas properties on a few arcsec scales on either side of the shock front to ensure accurate measurement of the equilibration timescales.

The upstream shock forms when material stripped from the subcluster's core is swept upstream and collides with remaining infalling material. Hydrodynamical simulations demonstrate that this produces complex gas flows and a series of shocks that rapidly evolve [90]. Fig. 26 shows the main features produced as unshocked material falls in from the right-hand side and, under the gravitational pull of the primary cluster, converges towards the merger axis. The proposed AXIS observation will detect, for the first time, the resulting series of strong, normal, and weaker inclined shock fronts.

Narrow collisionless shocks will appear broader in projection if local gas motions warp their smooth shapes. Both shock widths in Abell 2146 are consistent with collisionless shocks blurred by local gas motions of $290 \pm 30$ km s$^{-1}$.

Diffusion and conduction are sedtrongly suppressed across the leading edge of the dense, cool subcluster core. The cool, multiphase gas at $0.5 - 2$ keV is separated from the shock-heated ICM by a cold front less than 2 kpc wide. The subcluster's cool core is disintegrating under ram pressure, producing a long tail of stripped material mixing into the primary cluster's ICM. The ICM transport processes clearly dictate the morphology and evolution of these spectacular structures observed in Abell 2146.

**Joint Observations and synergies with other observatories in the 2030s:** Existing VLA observations reveal an extended radio halo and a relic at each shock front in A2146 ([217]). We request ngVLA observations at $1 - 2$ GHz to study the detailed halo and relic structure, map variations in spectral index and compare features with the X-ray maps to investigate particle acceleration by shocks and turbulence. Future radio observatories (LOFAR2.0, SKA, as well as ngVLA) will map the spatial and spectral details of relativistic electrons at the same spatial scales that AXIS can deliver for thermal plasma. In combination with future advances in particle-in-cell simulations, we anticipate a step change in our understanding of energy dissipation in large-scale structure.

**Special Requirements:** Low and well-understood background

## e. The Circumgalactic Medium and Connections to the Cosmic Web

*19. Spiral-rich and dynamically young groups*

**Science Area:** Galaxy groups, intra-group medium, galaxies, multiphase gas

**First Author:** Ewan O'Sullivan (Center for Astrophysics | Harvard & Smithsonian)

**Co-authors:** Anna Wolter (INAF Osservatorio Astronomico di Brera), Weiguang Cui (University Autónoma de Madrid)

**Abstract:** Low mass, spiral-rich galaxy groups are typically rich in cold atomic and molecular gas, but poor in hot gas and thus X-ray faint [113,606]. They are at the boundary of technical feasibility for current X-ray observatories, with deep exposures capable of tracing the hot gas available for only a handful of cases. In these systems, we find disturbed gas morphologies which suggest the groups are dynamically young, with galaxy interactions playing an important role in the formation of the intra-group medium. Stephan's Quintet is the most famous example, with the $\sim$900 km s$^{-1}$ collision between an intruder galaxy and a tidal HI filament producing a ridge of shock-heated $\sim 10^7$ K plasma apparently cooled primarily via line emission from warm $H_2$. HCG 16 provides another interesting case, with interacting spiral members embedded in a $\sim$425 kpc HI bridge, at least part of which is cospatial with hot X-ray emitting gas, perhaps ejected from the galaxies by starburst superwinds. These groups offer windows on a range of physical processes related to group and galaxy evolution, including shocks and cooling in highly multiphase media, interactions between galaxy winds and the intra-group medium, the impact of galaxy-galaxy interactions, and the long-term survivability of cold gas drawn out of galaxies by tidal forces. *AXIS*'s combination of exceptional soft-band sensitivity, excellent spatial resolution and a large field of view is ideally suited to investigate these systems, and we propose deep (at least 100 ks) observations of these two exemplars as a first step in understanding the hot gas content of low-mass groups.

**Science:** Galaxy groups are arguably the most important environment for the study of galaxy evolution. Around 55% of galaxies reside in groups, which also host the majority of Baryonic matter [137] and dark matter [112]. Furthermore, the group environment brings those galaxies into close proximity at low relative velocities, driving tidal interactions and mergers. Groups are a highly diverse class, ranging from lower-mass, dynamically-young systems dominated by spirals and with most of their gas in the form of HI or $H_2$ in or near the galaxies (e.g., the Local Group) through to high-mass "miniature clusters" with evolved galaxy populations embedded within an extended hot, X-ray emitting intra-group medium (IGrM). The presence of a hot IGrM appears to be linked to galaxy evolution; groups with elliptical members are significantly more X-ray luminous than spiral-only systems [348] and all known X-ray bright groups ($L_X > 10^{42}$ erg s$^{-1}$) include at least one elliptical.

Our understanding of the processes that drive the transformation of groups as they grow and evolve has significant gaps. It is widely accepted that mergers between spirals can produce ellipticals, though multiple cycles of merging may be required. The fate of their cold gas component and the origin of the hot IGrM are less clear. Studies of HI in compact groups show that a large fraction of the cold gas can be tidally stripped during galaxy interactions, producing tails and filaments outside the galaxies, or even a cold diffuse IGrM [245,262,542]. In more massive groups with significant hot halos in place, gas may also be drawn out of galaxies by ram-pressure stripping [425]. In galaxy clusters we expect much of the hot gas to be accreted directly into the halo, heated by gravitationally-driven shocks, but in groups, the links between galaxy evolution and gas content suggest that additional processes are in play.

Only a handful of spiral-rich groups undergoing transformation are known [e.g., 493,514]. The two best-studied examples are HCG 16 and HCG 92, more commonly known as Stephan's Quintet. HCG 16 ($z$=0.013, RA=02$^h$09$^m$31$^s$, Dec.=-10°09′31″) is the more dynamically youthful of the two, consisting of four main spiral galaxies (NGC 833, 835, 838 & 839) embedded in a giant (30′/425 kpc HI filament which links them to a fifth (NGC 848, see Fig. 28) and extends 70 kpc beyond it [229]. The HI had likely already been stripped from the main spirals before being drawn out into a filament as the fifth galaxy passed them [247]. Two of the main spirals (NGC 833 and NGC 835) are interacting, and all four host starbursts and/or AGN are thought to have been triggered by tidal interactions [199,376,520]. Based on ∼160 ks of *Chandra* observations, [377] confirmed the presence of a bridge of hot gas linking the galaxies and apparently cospatial with the HI filament. This is clearly not a relaxed halo, and while they estimated that 20-40% of the hot gas may have originated from the starburst winds of the galaxies, the overall hot gas mass is similar to the HI deficiency of the group (the amount of HI missing compared to the mass expected from the galaxy population). This raises the possibility that at least part of the HI has been heated into the X-ray phase.

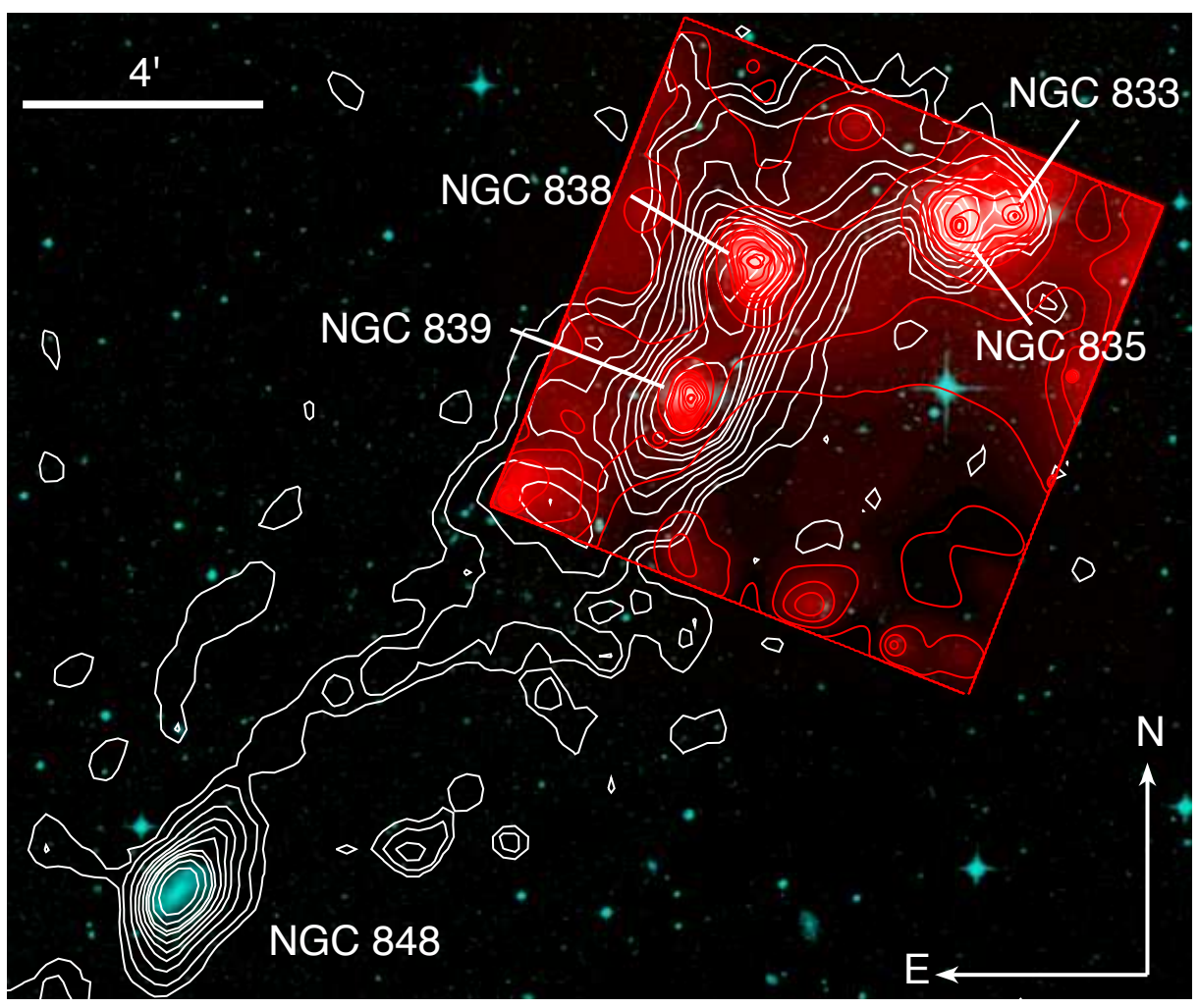

**Figure 28.** Adaptively smoothed *Chandra* 0.5-2 keV (red) and DSS optical (cyan) image of HCG 16, overlaid with VLA HI contours. The *Chandra* data trace emission from the four main galaxies and the bridge of diffuse hot gas linking them, but cannot determine whether it extends the full length of the HI filament.

Stephan's Quintet (hereafter SQ, $z$=0.0215, RA=22$^h$35$^m$57$^s$, Dec.=+33°57′36″) provides an example of how such heating can take place. The group has several members but is centered on three large galaxies (NGC 7317, NGC 7318a, NGC 7319) with similar recession velocities, with multiple disturbed structures (gas and stellar tails, tidal dwarf galaxies) providing evidence of past interactions [227,415,486]. Much of the HI and molecular gas appears to have been stripped from the galaxies, eventually being drawn out into a ∼180 kpc tidal filament. Then, ∼5 Myr ago, a high-velocity infalling spiral galaxy (NGC 7318b) collided with this filament at ∼900 km s$^{-1}$, driving a strong shock through the cold gas in a ∼35 kpc section. This heated the diffuse low-density gas in the filament to ∼1 keV, and reaccelerated relativistic particles to produce a clumpy ridge of radio and X-ray emission tracing the shock zone (see Fig. 29). Smaller X-ray filaments extend from this ridge to connect to the bar and nucleus of NGC 7319, while the northern end of the ridge encloses a tidal dwarf galaxy (Starburst A in Fig. 29); clearly, the original cold gas structure was morphologically complex.

Despite the presence of X-ray and relativistic plasmas, the shock is also rich in dust and molecular gas, the latter detected via CO, $H_2$, [CII], and even $H_2O$ emission [e.g., 17,18,106,138,204]. The luminosity of warm $H_2$ in the shock actually exceeds the X-ray luminosity by a factor of 3 [106], making $H_2$ line emission *the dominant cooling mechanism in the shock region*. It is suggested that while the diffuse component of the tidal filament was shock-heated, dense molecular clouds were able to survive the shock, rapidly radiating away energy via line emission. The hot and cold phases are now in close contact, with the energy transferred from the hot phase to the molecular clouds via a turbulent cascade which acts to cool the X-ray plasma [204]. This may explain the counterintuitive finding that the temperature in the core of the shock (∼0.6 keV) is cooler than that of the surrounding X-ray emission (∼1 keV) [375]. This cooling mechanism has implications beyond SQ; similar conditions are thought to occur in many galaxy collisions and mergers, in AGN jet/gas interactions within galaxy disks [267,365], and perhaps even in the filamentary nebulae at the center of galaxy cluster cooling flows. SQ also hosts multi-phase gas on larger scales. As well as a ∼150 kpc radius X-ray halo which traces the diffuse stellar halo [160,515,516], the shock zone is surrounded by diffuse radio emission whose polarization properties indicate an ordered magnetic field on scales of at least 70 kpc [357], while low-resolution FAST mapping suggests a low-density HI halo extending up to 600 kpc [94,585].

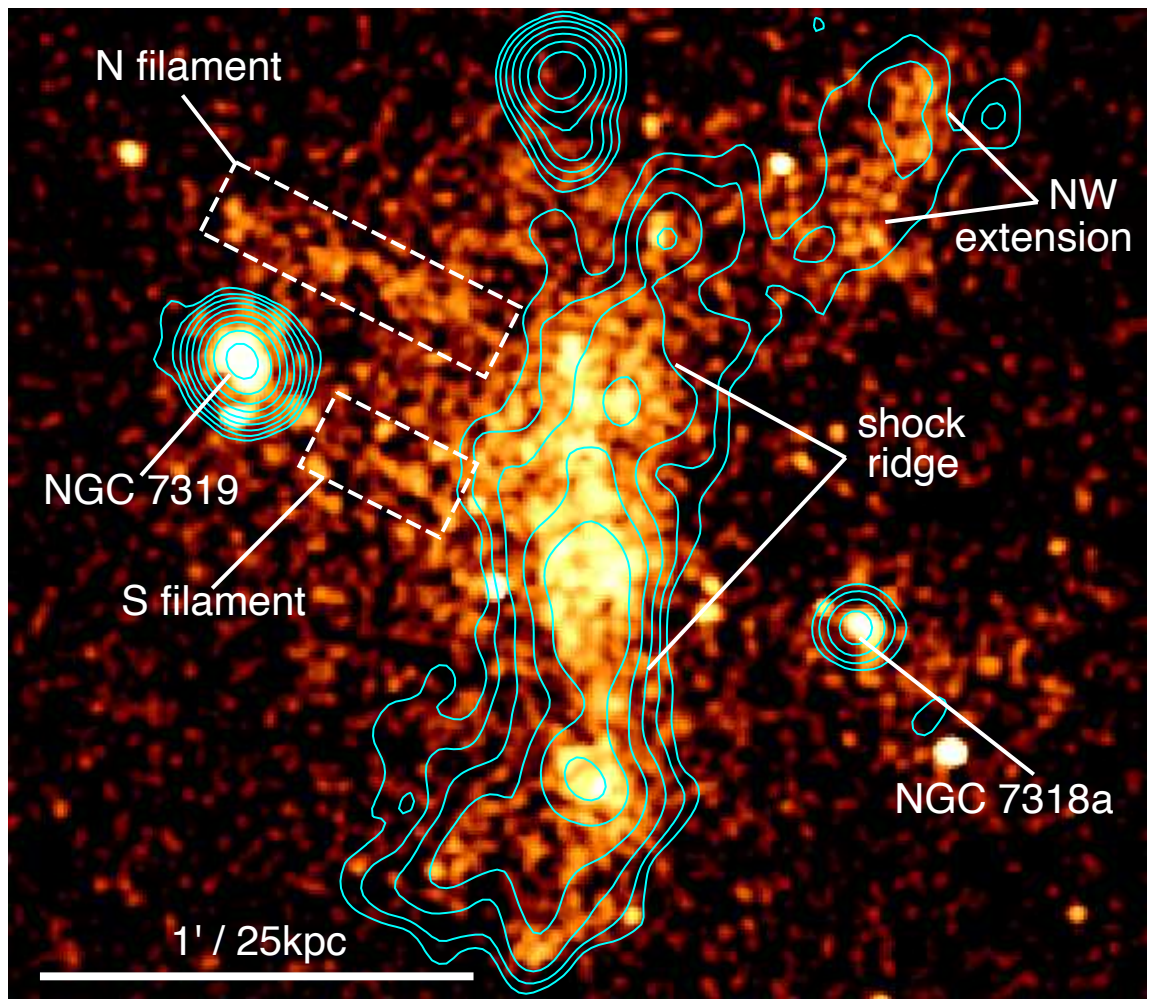

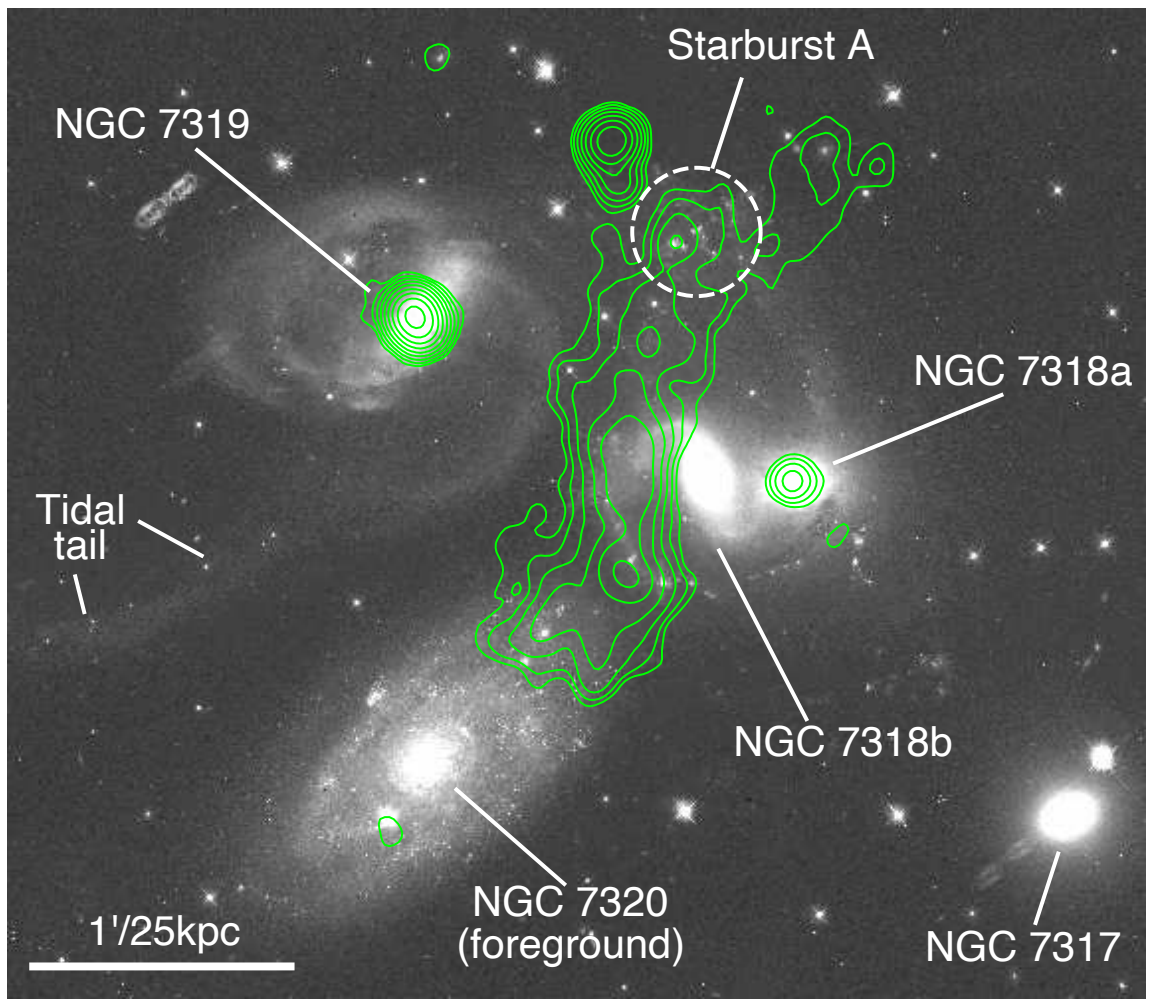

**Figure 29.** *Left*: *Chandra* 0.3-2 keV image with 6′ (2.5 kpc) resolution VLA 1.4 GHz contours overlaid, starting at 3×r.m.s. (180 $\mu$Jy) and increasing in steps of factor 2. *Right*: *HST* WFC3 F606w image of SQ with the same VLA 1.4 GHz contours overlaid.

These two groups, as exemplars of their class, offer windows on a range of physical processes: shocks and cooling in highly multiphase media; the build-up of the IGrM; the survivability of cold gas in the hot IGrM outside galaxies; interactions between galaxy winds and the IGrM; and the impact of galaxy-galaxy and galaxy-IGrM interactions. Unfortunately, at present, only a handful of such systems are known, though additional examples will likely be identified as deeper X-ray and radio surveys become available. Their hot gas components of HCG 16 and SQ are both faint and highly complex, with surface brightness structures down to arcsecond scales, and at present we have only very limited information on the variation of gas properties in these structures because even the available early-mission *Chandra* data lacks the sensitivity for spectral mapping. The HI filament in HCG 16 extends off the S3 chip in its *Chandra* data (see Fig. 28) and the *XMM* observation is affected by straylight from a nearby bright source. Only AXIS has the combination of high spatial resolution, sensitivity, and field of view necessary to thoroughly explore these systems.

**Exposure time (ks):** 200

**Observing description:** We base our feasibility estimates for *AXIS* on the available *Chandra* and *XMM* observations, which provide information on the X-ray foreground and background emission as well as the range of sources in the groups themselves. Spectral simulations suggest that *AXIS* will provide almost an order of magnitude improvement in count rates over *Chandra* for the soft diffuse emission, with ∼100 ks observations providing >15,000 source counts (0.3-2 keV) from the known section of the HCG 16 gas bridge, >70,000 counts from diffuse emission structures in the inner regions of SQ (26,000 counts in the shock region) plus an additional ∼40,000 from the surrounding halo.

The primary goals of these observations would be to understand the physical state of the hot gas in the two groups, and thus constrain its origin. As a first step, we would *determine the extent, morphology and physical properties of the hot gas component.* In HCG 16 we would be able to determine whether the hot gas extends along the whole HI filament (hinting at a shock origin) or is only localized around the galaxies (suggesting formation via winds). We would also be able to perform spectral analysis of subregions of the bridge for the first time, tracing temperature and abundance structure. In SQ, the morphology is better understood, but *AXIS* would enable the determination of gas properties in the numerous substructures (filaments, NW extension, tidal tails, etc.) for the first time. In both systems, deep imaging opens a significant discovery space, as additional substructures are likely to become visible.

In SQ, we would *test our understanding of the shock and cooling physics*. An *AXIS* observation would support mapping of the gas temperature and density on scales of a few arcseconds in the shock zone, for comparison with other gas phases, as well as the radio emission, dust, and star formation regions. If the turbulent cascade can cool the X-ray gas efficiently, we should see low temperatures and high densities in regions with strong $H_2$ emission (as traced by *JWST*) [19]. Conversely, if shock driven turbulence in the warm gas, visible from H$\alpha$ and CO line widths (traced by WEAVE and SITELLE IFU data, ALMA/ACA) becomes excessive, the connection between the hot an cold phases may be severed, suppressing cooling and star formation [19,21,129,138]. The *Chandra* and JVLA+MeerKAT radio continuum data already show hints of asymmetry in the shock, and this may be connected to the angle of the disk of the intruder galaxy, NGC 7318b, during the collision with the HI filament. Identifying regions with minimal impact from line cooling will allow a clean constraint on the shock strength and obliquity, as well as investigation of its impact on the pre-existing hot halo of the group.

In addition, *AXIS* would provide high-quality data on *the member galaxies and their environmental interactions*. We would expect to obtain 15,000-35,000 source counts (0.3-10 keV) from each of the four main galaxies in HCG 16, as well as imaging NGC 848 at arcsecond resolution for the first time. Tracing the galaxy winds in HCG 16 will let us determine whether gas is being stripped from them by their motion relative to the HI filament (as the radio data suggest). In SQ, we could expect $\sim$4,500 counts from the nucleus of the bright Seyfert galaxy NGC 7319 and may be able to find X-ray structures associated with the radio jet of NGC 7319, which appears to be interacting with the galaxy disk [392]. We also note that a single *AXIS* pointing will cover most of the area around SQ in which low-resolution FAST HI mapping has identified a halo of gas-rich minor galaxies. Given the narrow off-axis PSF, we would expect to detect X-ray emission from these, allowing us to measure their hot gas content and star formation rates.

**Joint Observations and synergies with other observatories in the 2030s:** These systems are rich in multiphase gas and are thus excellent targets for new facilities. SQ, in particular, has an exceptional multiwavelength data set and has been a first-light/early-release target for numerous observatories (e.g., *JWST* MIRI and NIRCAM, the WEAVE IFU on WHT). Both systems would benefit from additional *JWST* and ALMA mapping to trace the warm and cold molecular gas, as well as expanded ground-based IFU observations to cover the ionized phase. Currently, VLA, GMRT, and MeerKAT have provided the best radio continuum and HI maps of the groups; however, LOFAR 2.0, SKA, and potentially ngVLA will offer even greater sensitivity. This will be particularly useful for tracing the unique, large-scale, ordered magnetic field structure in SQ, which is likely linked to the collisional shock.

**Special Requirements:** None

*20. Ram pressure stripping in clusters and groups*

**Science Area: galaxy clusters, galaxy groups, intracluster medium, star-forming galaxies**

**First Author:** Alessandro Ignesti (INAF Osservatorio di Padova)

**Co-authors:** Yuanyuan Su (University of Kentucky), Myriam Gitti (Università di Bologna)

**Abstract:**In clusters and groups, ram pressure stripping (RPS) induced by the intracluster medium (ICM) or intergroup medium (IGM) can be a crucial evolutionary driver for galaxies, as it effectively quenches their star formation by removing the interstellar medium (ISM). Investigating the RPS physics is essential to understanding how galaxies evolve in dense environments. In this context, X-ray observations play a crucial role, as they allow us to measure the environmental plasma's thermal properties, study the RPS-induced mixing between the ISM and ICM/IGM, and detect AGN activity triggered by gas inflows induced by ram pressure. These studies can also probe the complex physics of the ICM/IGM-ISM mixing, providing insights into the fundamental laws governing the interplay between astrophysical plasmas. AXIS, due to its large FOV, will be able to study multiple galaxies in the same cluster/group with a single observation, thus greatly increasing the number of galaxies with high-angular-resolution X-ray data required for these studies.

**Science:** The balance between the gas inflows and removals drives the baryon cycles in galaxies, and thus their evolution. While the inflows sustain star formation and allow galaxies to grow in mass, gas removal can lead to the quenching of star formation. Galaxies can lose their gas via internal or external processes. The latter, generally known as environmental effects, are crucial in shaping galaxy properties in clusters and groups [55,125,553]. They can be divided into two main categories, those driven by gravitational forces between galaxies, such as mergers and tidal interaction [28] or harassment due to fast encounters [344], and those resulting from the hydrodynamical interaction between the galaxies and the environmental plasma, either the intracluster medium (ICM) or the intragroup medium (IGM). This category includes thermal evaporation [323], viscous stripping [361], and ram pressure stripping [RPS, 205]. The latter is an external pressure exerted by the environmental plasma, which can overcome the stellar disk binding force and strip the ISM components from the galaxy [205]. The gas loss induced by RPS can effectively quench the star formation in the stellar disk [55, for a review], thus making it an important quenching pathway for satellite galaxies [e.g., 506,551,554,570]. Although RPS is expected to be the strongest during the galaxies' first infall in clusters and groups, especially those following the most radial orbits [42], optical studies revealed that almost all galaxies in clusters undergo an RPS event within their lifetime [553]. RPS can play a role also in low-mass clusters such as Fornax [458], and groups [425]. In the optical/UV band, RPS galaxies are characterized by extraplanar, unilateral debris extending beyond their stellar disks, and striking tails of ionised gas traced by the H$\alpha$ emission.

X-ray studies of clusters and groups can provide crucial insights into the physics of RPS. To begin with, X-ray observations represent the only way to characterize the environmental plasma's thermal properties, such as density and temperature, which define the external pressure affecting galaxies. Secondly, X-ray observations enable us to investigate a range of physical processes triggered by RPS in infalling galaxies. In the soft X-ray band, RPS galaxies show extended emission, typically following the trail of stripped material detected at other wavelengths (Figure 44, left panel) with a typical surface brightness of $10^{-17} - 10^{-16}$ erg s$^{-1}$ cm$^{-2}$ [e.g., 29,56,77,120,399,411,489–491,603]. Spectral analysis revealed that the average temperature of the extraplanar emission is typically in the 1-2 keV range, which is in between the ISM temperature and the one of ICM/IGM, possibly indicating that the X-ray emission comes from the mixing layer formed at the interface between the stripped ISM and the surrounding plasma. RPS galaxies are expected to be visible also in the hard X-ray band. On the one hand, RPS can trigger gas inflows toward the galaxies' centers [6], resulting in a larger-than-usual fraction of AGN activity [391,398]. On the other hand, RPS galaxies have been shown to host a larger-than-usual number of ultra-luminous X-ray sources in their stripped tails, although the physical connection between the two phenomena is still unclear [399].

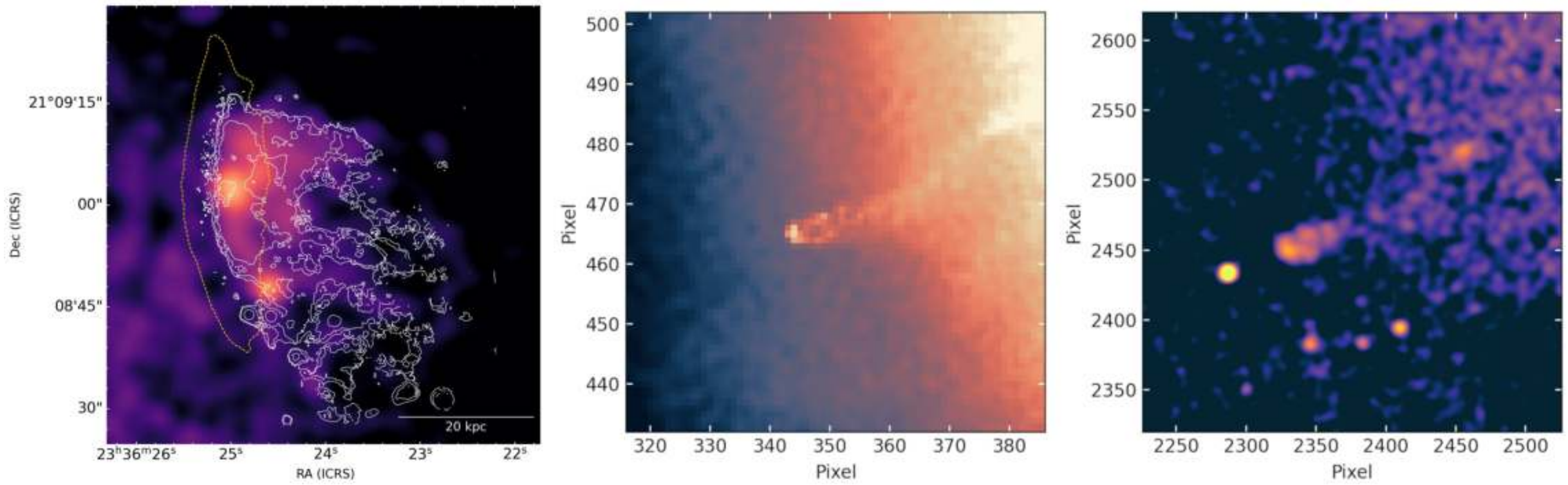

**Figure 30.** Left: *Chandra* image of a ram pressure stripped galaxy from Poggianti et al. [399]. The gold and silver contours indicate, respectively, the stellar disk and the emission-only H$\alpha$ emission. Center: image of an RPS satellite in a cluster in the IllustrisTNG cosmological simulation. Right: Simulated AXIS image of the RPS satellite in the center panel with an exposure time of 20 ks. The cluster is placed at $z = 0.1$. Particle and astrophysical backgrounds are included.

Hence, we propose two possible lines of research that can be addressed by new, deep, high-angular-resolution X-ray observations.

- **Fate of the stripped ISM:** A currently open question in this field of study is the nature of the process that allows the stripped ISM to resist the thermal conduction from the hot, external plasma and, instead, to collapse into cold gas and form new stars outside of the stellar disk [e.g., 180,197,459]. Understanding this process can shed new light on the more vast astrophysical problems of the mixing between hot and cold plasmas. New key insights, such as temperature and metallicity distribution, can emerge from the study of the mixing layer formed on the stripped ISM left behind by the RPS galaxies, which new, high-angular resolution X-ray observations can probe. Combined optical and X-ray studies of RPS galaxies have revealed a spatial correlation between the optical emission, typically the H$\alpha$ line, and the thermal X-ray emission [77,399,491]. This result suggests that the X-ray emitting plasma originates from the stripped ISM, which maintains a higher density than the surrounding ICM, hence a higher X-ray brightness. The same correlation is also observed for the cold plasma filaments observed at the center of relaxed galaxy clusters [367], thus suggesting similarities in the origin of these structures. Furthermore, MUSE observations revealed that the stripped ISM velocity structure can be influenced by the surrounding ICM turbulence [231,278], providing complementary evidence of the interplay between the ISM and ICM and further indication of turbulent mixing occurring between the two plasmas. Additional evidence of a complex interplay between the plasmas comes from the study of the optical spectrum of the stripped material. MUSE observations revealed that the farthest part of the stripped tail can show LINER-like optical line ratios induced by the photoionizing emission emitted by the surrounding plasma and the mixing layer [77]. Therefore, resolved spectral studies of these mixing layers will map their temperature and metallicity, addressing under which conditions the ICM-ISM mixing can lead to the survival of the stripped ISM.
- **The role of infalling satellites in enriching the peripheries of clusters and groups with metals:** The metal-rich ISM stripped from the infalling satellite galaxies will eventually evaporate in the surrounding plasma, thus increasing the local metallicity. Observations and numerical simulations suggest that the satellite enrichment contribution to the total cluster metal content is negligible for $z < 2$ [321]. However, their local contribution remains unknown, as a systematic study of metallicity in the trails of RPS galaxies is lacking. This question could be addressed by conducting a survey of the ICM/IGM metallicity in the surroundings of galaxies currently ongoing RPS, i.e., with tails of stripped material observable at any wavelength, and fully stripped galaxies, in which the ISM has been fully dissipated in the ICM/IGM.

The study of ram pressure in clusters and groups is currently limited by the scarcity of X-ray data with high angular resolution, which is necessary to conduct resolved studies and explore physical scales below 10 kpc. For this reason, the Advanced X-ray Imaging Satellite (AXIS) represents the only viable option in the near future for advancing this field of study.

**Exposure time (ks):** 100 ks

**Observing description:** We propose to observe the near RPS galaxy JO206 ($z = 0.0511$, RA 21 13 47.41 DEC +02 28 34.4) as a case study to test the AXIS's capability to pursue the aforementioned science goals. This target provides comprehensive coverage of ancillary observations, spanning a range from low-frequency radio continuum to UV. Currently, it lacks deep, high-resolution X-ray observations capable of detecting the extraplanar emission. Based on the archival *Chandra* data (project 23610079, PI Ignesti) in which the extrapalanar emission is undetected, we estimate an upper limit of the $2 \times 10^{-14}$ erg s$^{-1}$ cm$^{-2}$ in the 0.5-2 keV band. Based on the expected AXIS sensitivity, an exposure of 100 ks would be able to reach a sensitivity of $\sim 10^{-16}$ erg s$^{-1}$ cm$^{-2}$, which would represent an improvement of two orders of magnitude. Therefore, it will guarantee the detection of the extraplanar emission with a physical resolution of $\sim 1.5$ kpc, which will permit resolved morphological and spectroscopic studies of its properties and a comparison with other multiwavelength observations.

We further note that the combination of high angular resolution across the large FOV and the high effective area at low energies offered by AXIS entails that the proposed science case can be explored even without pointed observations on RPS galaxies. The great advantage of these studies, hence, lies in the synergy with other cluster and group observation programs. Therefore, we argue that the study of RPS galaxies can become an auxiliary science goal for cluster and group observations, which can reach a sensitivity of $\sim 10^{-16}$ erg cm$^{-2}$ s$^{-1}$, thus enhancing the scientific output of future AXIS projects. We test the capability of AXIS in exploring ram pressure stripping galaxies in every system at $z < 0.1$ by using SIXTE to predict how AXIS would observe an infalling satellite in a simulated TNG cluster at $z = 0.1$. With a modest exposure time of 20 ks, AXIS could recover the extraplanar emission with a resolution of $\sim 3$ kpc. Moreover, the serendipitous detection of extraplanar, soft X-ray emission in a spiral galaxy can strongly indicate ongoing ram pressure. Thus, large FOV X-ray observations targeting clusters and groups can help to expand the sample of known RPS galaxies and provide the targets for follow-up studies at other wavelengths.

**Joint Observations and synergies with other observatories in the 2030s:** The study of RPS galaxies is part of the proposed SKA science topics for the Extragalactic continuum working group ('Exploring the physics of ram pressure stripping in clusters and groups with radio continuum observations', PI Ignesti). SKA observations will map both the neutral and the nonthermal ISM phases in RPS galaxies; thus, they will greatly synergize with AXIS to study how the stripped ISM can evolve in the ICM.

**Special Requirements:** None

*21. Interplay of the microphysics and the global properties in the weakly collisional medium around galaxies*

**Science Area:** plasma physics, intracluster medium, diffuse magnetized medium around galaxies

**First Author:** P. P. Choudhury (University of Oxford)

**Co-authors:** H. R. Russell (Nottingham), J. A. ZuHone (CfA)

**Abstract:** Detailed investigations of the intracluster and circumgalactic medium can take a leap forward using AXIS due to the higher sensitivity, spatial resolution, and field of view. The dialogue between global dynamics (e.g., mergers, shocks, accretion, ram-pressure stripping, AGN/SN feedback, etc.) and how it transpires into microphysical processes at the gyroscales of the energetically weak magnetic field is poorly understood in both theory and observations. The combination of higher spatial resolution and spectral mapping opens up the possibility of mapping the persistence of temperature and density contrasts across a wider range of length scales and amplitudes. This can answer key questions in both galaxy formation and plasma physics as following: *1. Are weak shocks sustained in the medium, and can we identify the equation of state (isentropic versus isobaric)? 2. Do strong temperature contrasts exist, and can we model suppression factors for energy transport? 3. Is it possible to calculate the power spectra across length scales for the temperature contrasts, and what is the implication for the viscous cut-off? 4. Do large-scale temperature contrasts exist, and do these map the magnetic fields (potential for synergistic radio observations)?*

**Science:** Galaxy formation is governed by a delicate balance between cosmological accretion, radiative processes, state of ionization, and last but not least, the black hole co-evolution that regulates the circumgalactic (CGM) and intracluster medium (ICM). AGN feedback plays a crucial role in preventing runaway cooling; however, the efficiency of energy transport through thermal conduction, viscosity, cosmic ray propagation, and turbulence remains an open question. Magnetic fields further shape these processes, influencing the distribution of heat and momentum across the hot plasma. Mergers and large-scale structure formation generate shocks, cold fronts, and turbulent fluctuations, which trace the interaction between astrophysical plasma and cosmic magnetic fields. Understanding these structures is essential for constraining AGN-driven heating, suppressing conduction, exploring alternative modes of energy transport, and coupling cosmic rays to the ICM, to name a few.

With its superior spatial resolution across the wide field of view and increased sensitivity, AXIS will enable high-precision mapping of these thermodynamic properties. By resolving weak shocks, density fluctuations, and large-scale temperature contrasts, AXIS will provide key insights into the physics of galaxy formation, bridging observations with theoretical models of plasma transport, feedback, and cosmic magnetism. The following four specific scientific cases are proposed, with consideration given to their implications for galaxy formation physics.

- **Probing the Equation of State in weak shocks:** Chandra's sharp imaging has been crucial in identifying and characterizing shock fronts, cold fronts, and other discontinuities in galaxy clusters and the respective morphologies (Fabian et al. 152, Young et al. 589). There have been works on edge detection techniques using Chandra and XMM-Newton (e.g., Walker et al. 562). AXIS's high spatial resolution and effective area will enable detailed mapping of density and temperature variations associated with weak shocks in the ICM (at higher SNR). AXIS's increased effective area (by $\sim 10$ times) will also allow for observations of the CGM. However, due to the CGM's lower surface brightness, a longer exposure time (by a factor of $3-4$) or stacking will be required to obtain the same sensitivity of existing ICM observations (e.g., Fabian et al. 145). By precisely measuring these variations, we can determine whether the shocks are isentropic (compressive) or isobaric (pressure-balanced), revealing the underlying equation of state. This will provide crucial insights into the thermodynamics of the hot plasma and the mechanisms that sustain weak shocks.

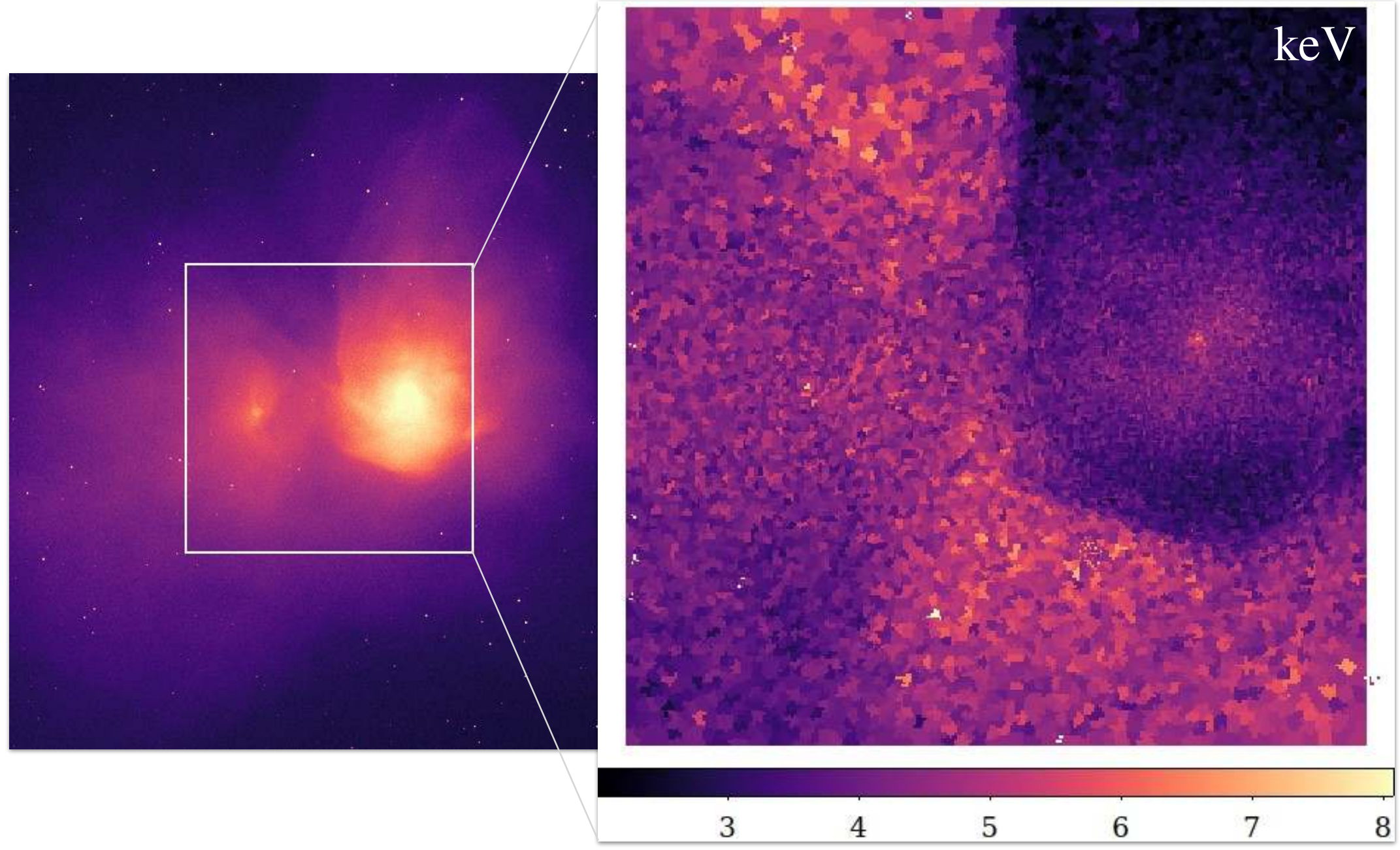

**Figure 31.** Mock AXIS observation (500 ks) for a $10^{15}\,M_{\odot}$ merging cluster from the suite of The Next Generation (TNG) cosmological simulations (left panel). The image is 20 arcmin on a side or 1.2 Mpc. A temperature map for the region within the white square is shown in the right panel. A wide range of scales (limited by the lowest resolution of the simulations) in temperature contrasts appear in the diffuse plasma of the cluster outskirts.

Understanding the equation of state will also constrain the transport properties of the plasma, since acoustic waves can propagate and weak shocks can be sustained only under certain plasma conditions. Further, sustenance of compressive fluctuations underpins the role of isentropic acoustic waves as energy carriers from AGNs in galaxy clusters.

- **Quantifying energy transport suppression in temperature contrasts :** The persistence of strong temperature contrasts in the ICM challenges classical models of thermal conduction. Using Chandra's high spatial resolution, it has been deduced that the cold fronts may have widths below the Coulomb mean free path (Ascasibar & Markevitch 22, **?** , Richard-Laferrière et al. 419). Electron gyroscale numerical simulations suggest that, in collisionless plasmas, energy transport is suppressed by a factor related to the ratio of thermal to magnetic energy (plasma $\beta$), compared to predictions based on free-streaming electrons (e.g., Komarov et al. 260, Roberg-Clark et al. 424). A milder suppression may arise in other physical conditions (e.g., Komarov et al. 261). AXIS will allow us to map these contrasts in unprecedented detail (due to the overall sensitivity improvement), quantifying the suppression factors for energy transport. By comparing observational data with theoretical models, we can determine the mechanisms that inhibit thermal conduction, potentially constraining the role of the magnetic field. The ratio of the length scales of a sharp temperature contrast along gravity to across it may indicate anisotropic propagation and the presence of thermally unstable internal gravity waves, which can be a mechanism to sustain large-amplitude temperature contrasts (Choudhury & Reynolds 99).

- **Investigating viscous cut-offs through temperature contrast power spectra:** Deep Chandra observations of the Coma cluster enabled comparing power spectra of density fluctuations with those of numerical simulations of clusters using classical Spitzer viscosity (Zhuravleva et al. 608). This revealed that the effective viscosity is orders of magnitude below what is expected, either due to anisotropic momentum transport or particle scattering off gyroscale fluctuations. AXIS's ability to map density/temperature contrasts across a wide range of length scales (combination of high spatial resolution, wider FoV and uniform PSF across it) will enable the calculation of their power spectra accurately (e.g., small-scale temperature structures are revealed in the diffuse regions of merging TNG clusters in Fig. 31). By calculating the slopes of these spectra, we can search for evidence of viscous cut-offs, which provide constraints on the momentum transport. This can be compared to the refined transport models of galaxy clusters, which consider details of scattering by microinstabilities to understand the exact microphysical processes at play.
- **Mapping magnetic fields through large-scale temperature contrasts:** Large-scale temperature contrasts in the ICM and CGM, induced by mergers, may trace the geometry of coherent magnetic fields of comparable scale. This is because the magnetic field is expected to be frozen in the fluid flow. AXIS's superior FoV and uniform PSF across it will help to capture these large-scale features very well. In addition, the higher sensitivity may capture such features in the CGMs with comparable or longer exposure times (by a factor of $\sim$3, due to the low surface brightness of the CGM). This will enable us to determine the relationship between temperature structures and magnetic field configurations (for example, see 22). Synergistic radio (e.g., Bonafede et al. 48) and X-ray observations will provide the spatial correlation of magnetic field and temperature features indicating magnetic draping, and can inform on the amplitude and geometry. Such observation will leverage current theoretical and computational efforts using MHD and particle-in-cell simulations to understand the role of magnetic field in preventing temperature phase mixing and therefore sustaining sharp density/temperature contrasts, which may contribute to star formation or black hole growth in due course.

**Exposure time (ks):** 950 ks total

**Observing description:** Our goals require measurements of temperature and surface brightness fluctuations on the scale of the mean free path. This can be spatially resolved in the nearest galaxy clusters, including Perseus, M87, and Coma. For Perseus and M87, excluding the central few arcmin, which are dominated by radio bubble structures, the mean free path is $1-3$ arcsec. The Coma cluster is hotter, and the mean free path is larger at $\sim$ 5 arcsec. We require $\sim$ 3000 source counts per region on these scales in Perseus and Coma to determine the temperature to $\sim$ 10%. The lower temperature of M87 provides improved temperature diagnostics from Fe L emission, and therefore, we need fewer counts ($\sim$ 1000) for a similar constraint. This will also be sufficient to measure surface brightness fluctuations with a few % level accuracy. With sharp spatial resolution across the field of view, AXIS will map the temperature and density structure on the required scales across these cluster cores.

- The Perseus cluster ($z \approx 0.018$) is the prototype for AGN-driven feedback in cool core clusters that can host a variety of signatures for the equation of state. For 2 arcsec regions, we require 340 ks in the weak shock regions previously diagnosed. For CGM studies, synergy with eRosita (e.g., Fig. 3 in Zhang et al. 605) or tSZ (e.g., Das et al. 117) could be informative to decide on the brightest target.
- M87 is the closest galaxy cluster ($z \approx 0.0044$) and provides the sharpest view of shock fronts, cool gas flows along the radio lobes and cavities. For 2 arcsec regions, we require 190 ks to map fluctuations beyond the bright core and the cool arms to the fainter regions beyond the shock front.

- Coma ($z \approx 0.023$) is well studied for turbulence using Chandra and represents an important example of cluster core structure in the absence of current feedback. The merger dynamics may trace magnetic field morphology, as revealed by LOFAR (e.g., Bonafede et al. 49) and SKA (e.g., Vazza et al. 538). Given the larger mean free path, we propose to use 5 arcsec regions and require 420 ks.

**Joint Observations and synergies with other observatories in the 2030s:** Next generation radio observatories such as SKA, which trace radio lobes and halos and reveal the magnetic field morphology.

**Special Requirements:** None

*22. Brightest Cluster Galaxy Corona and galaxy cluster co-evolution*

**Science Area: Galaxy Clusters, intracluster medium, interstellar medium, active galactic nuclei**

**First Author:** Ayşegül Tümer (University of Maryland Baltimore County, NASA/GSFC)

**Co-authors:** Massimo Gaspari (University of Modena and Reggio Emilia)

**Abstract:** Brightest cluster galaxies (BCGs) play a critical role in the evolution of galaxy clusters. To date, an evolutionary link between non-cool core (NCC) and cool core (CC) clusters remains poorly established. BCG coronae —mini versions of cool cores— are good candidates to unravel this relation.

**Science:** The intracluster medium (ICM) is a hot, optically thin plasma ($\sim$$10^7$–$10^8$ K) that fills the space between galaxies in clusters and emits primarily in X-rays via thermal bremsstrahlung [444]. It comprises $\sim$12% of cluster mass and is often near hydrostatic equilibrium in relaxed clusters. However, many clusters show disturbed X-ray morphologies, indicating ongoing or past mergers. ICM thermodynamic properties (temperature, pressure, entropy) are sensitive to such dynamical activity. X-ray emission leads to plasma cooling, which is proportional to the square of the gas density [393]. In dense regions, this implies short cooling times, potentially shorter than the cluster age [147,316].

Brightest cluster galaxies (BCGs), located at cluster centers, play a key role in regulating ICM cooling via feedback from central supermassive black holes [487]. A self-regulated active galactic nuclei (AGN) feedback loop is the leading mechanism proposed to counteract cooling [103,431,549], although supernovae and thermal conduction are also considered.

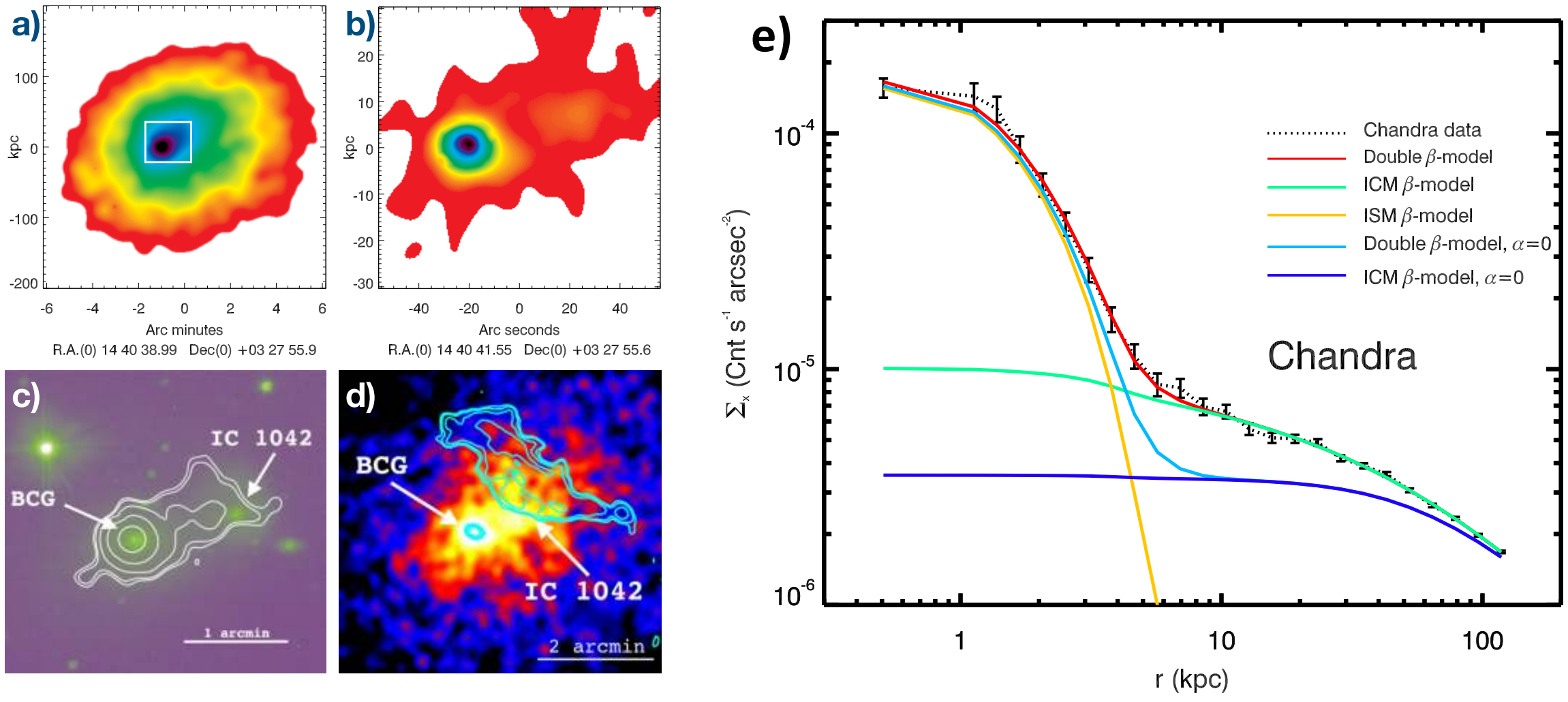

**Figure 32.** (a) *XMM-Newton* $SB_x$ map (0.5-2.5 keV). White box indicates the extent of *Chandra* $SB_x$ map shown in panel (b). (c) RTT 150 optical image overlaid by the extrapolated *Chandra* X-ray contours. (d) *Chandra* wavelet image overlaid by GMRT 325 MHz contours, where contour levels correspond to [1, 1.4, 3, 9]$\times$ mJy beam$^{-1}$. (e) *Chandra* surface brightness fit of the double $\beta$-model in the 0.5 - 2.5 keV energy band in the circular region with $r$ = 4″ for MKW 08. Projected surface brightness (*dotted black line*); the double $\beta$-model (*red*); the $\beta$-model corresponding to the ICM (*green*); the $\beta$-model corresponding to the ISM (*yellow*); the double $\beta$-model with $\alpha$ parameter is set to zero (*light blue*); the $\beta$-model of the ICM with $\alpha$ parameter is set to zero (*dark blue*) [519]

Clusters form hierarchically through mergers of smaller halos, reaching masses of $\sim 10^{14}$–$10^{15}$ $M_{\odot}$, with merger energies up to $10^{65}$ erg, second only to the Big Bang. These events also influence the cosmic web, making clusters key probes of large-scale structure.

Despite their diversity, clusters exhibit self-similarity when scaled by mass, allowing statistical studies. Currently, clusters are categorized as cool-core (CC) or non-cool-core (NCC), but classification varies depending on criteria such as central temperature drop, cooling time, or mass deposition rate [225], introducing systematic uncertainties into both cluster and cosmological studies.

A third category, BCG coronae or mini-cool cores, has been proposed [487]. These compact, ISM-origin structures resemble CCs but on smaller scales and may survive strong ICM stripping, AGN feedback, and rapid cooling, offering a promising solution to current classification challenges. BCGs reside near the center of galaxy clusters and are the most luminous and the most massive galaxies within the cluster. They host a supermassive black hole (SMBH), and through the AGN feeding and feedback, there is an intimate link between the properties of the BCGs and their host clusters [177]. BCG coronae have a typical radius of $r < 4$ kpc. These coronae can be identified by their prominent iron L-shell bump, and they exhibit extended soft X-ray emission. The temperature values of this sample of embedded coronae vary in the range $0.3 < kT < 1.7$ keV and the abundance of the coronal gas was found to be $Z \sim 0.8$ $Z_{\odot}$. Various clusters which were previously identified as NCC proved to have coronae with luminous radio-loud AGNs [487]. This picture suggests that the distinction between CC and NCC clusters is more complex and not yet well understood. X-ray instruments with high sensitivity and high spatial resolution are required to study the co-evolution of the AGN, ISM, and ICM within the BCG of NCC types of clusters to identify parameters with which galaxy cluster self-similarity relations can be categorized. This is only possible by constraining the thermodynamics and spatial scales of the multiphase gas within the BCG as well as the surrounding environment. At the BCG corona spatial scale, only local coronae can be studied in depth due to the lack of spatial resolution and sensitivity of the current X-ray telescopes. AXIS will be able to determine how cluster cores evolve in relation to their surrounding environment and the activity from their central AGN. Studying whether the BCG coronae remain isothermal in hostile environments of NCCs will help understand their potential to become large cool cores in the future of their evolution, or be completely destroyed, turning into NCCs. The mini-cool cores, hence, are one of the most elusive yet essential pieces in understanding the large-scale structure evolution of the universe.

The most in detail X-ray study of a BCG coronae is achieved for NGC 5718 (z = 0.027) in the galaxy group MKW08 using Chandra and XMM-Newton data revealing the size, $r \simeq 3$ kpc; temperature, $kT \simeq 1$ keV; 0.5–2.5 keV luminosity, $L_{X,} = 1.32 \times 10^{41}$ erg s$^{-1}$, and mass, $M_{gas} \simeq 6.17 \times 10^{7}$ $M_{\odot}$). A prominent emission of AGN from the BCG ($L_X = 3.04 \times 10^{40}$ erg s$^{-1}$) points to possible scenarios that the group has undergone a merger, currently disrupting its once larger CC, or is in the preliminary stages of CC formation. However, within 3 kpc, spectra points to multiple emission features attributed to AGN, ISM, and ICM, yet spatially unresolved. Moreover, having undergone violent dynamical activities, many BCGs are accompanied by unique features, such as X-ray tails, whose origins are not well understood. Fig. 32 shows Photon images of MKW08 and surface brightness profile indicating the multicomponent emission. Although the galaxy group is nearby, the superior resolution of Chandra is not able to distinguish contributions from multiple emission sources at kpc scales.

- AXIS will enable the extraction of surface brightness (SB) profiles of galaxy clusters thanks to its large effective area, wide FOV, and superior and off-axis independent PSF.
- We will be able to disentangle the spectral features of multiple structures within the FOV by detecting various structures (cavities, filaments, jets..) and be able to constrain the properties of the coronal gas, intracluster medium and AGN.

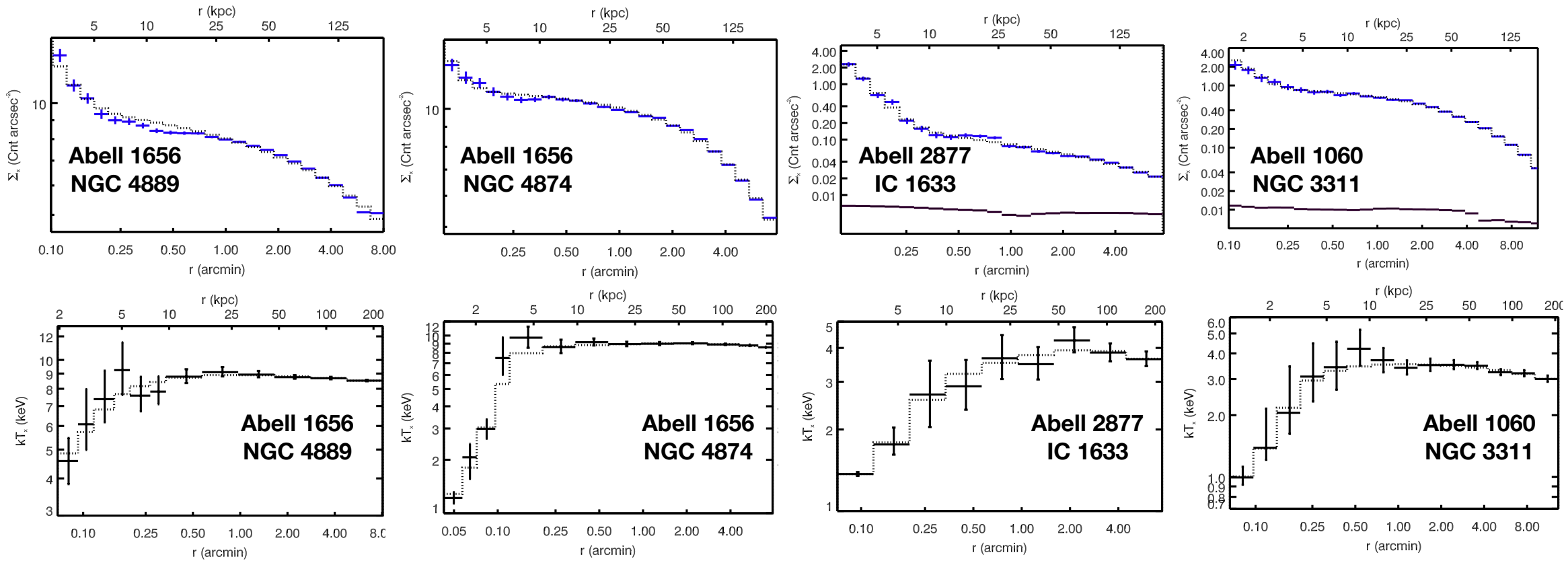

**Figure 33.** Preliminary results on *Chandra* radial surface brightness profiles for the 0.3-2.5 keV and temperature profiles of clusters from the sample and their BCGs.

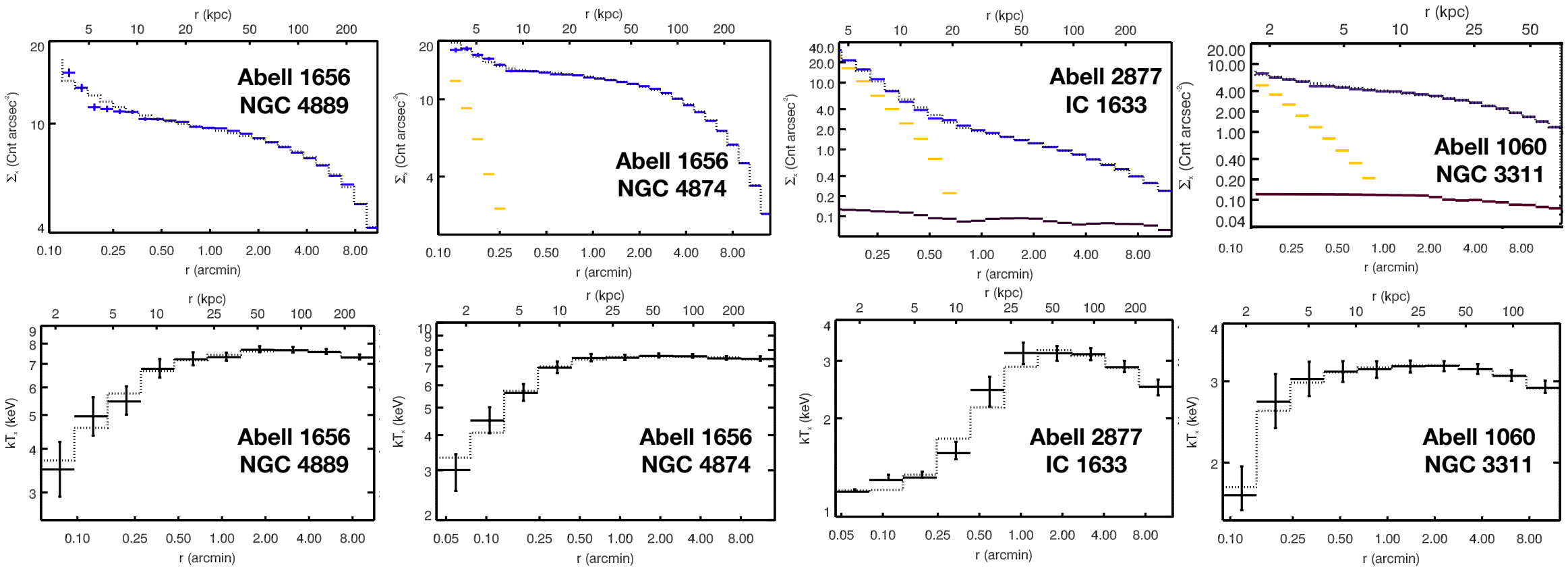

**Figure 34.** Preliminary results on *XMM-Newton* radial surface brightness profiles for the 0.3-2.5 keV and temperature profiles of clusters from the sample and their BCGs.

- Thanks to the high spatial resolution of AXIS, we will be able to study the evolution of the BCG coronae at various redshifts.

**Exposure time (ks):** 500 ks

**Observing description:** We aim to sample the known BCG coronae with AXIS. The radial surface brightness profiles of cluster/group of galaxies will be fitted with an analytical function called the modified double $\beta$-model [545]. The first $\beta$-model is used to fit the ICM surface brightness, where the second is to describe the surface brightness of the ISM of the BCG. The surface brightness model will be used to pinpoint the radius at which the emission from the coronal gas meets the emission from the ICM beta model (Fig. 32).

Once a first estimation of the BCG coronae radii is established, I will use various spectral models to find the best model that describes the emission components within this radius (e.g.;apec, powerlaw). The spectral analysis will also be used to assess the thermodynamic properties of the surrounding ICM at various radii scaled by $R_{500}$. The size of the corona, temperature, and density of the ICM at different radii, the power law index of the AGN emission, and redshift will be compared for all BCG coronae in the

sample to assess correlations. To achieve the campaign, we will observe galaxy clusters/groups, including MKW08, Abell 1656, Abell 1060, Abell 2877, and Abell 400, centered on the BCGs, for 100 ks each.

**Joint Observations and synergies with other observatories in the 2030s:** Multiwavelength observations to trace multiphase cool gas nebulae (ALMA, ELTs). Next generation radio observatories to identify jet activity (ngVLA, SKA, LOFAR2.0).

**Special Requirements:** None

*23. Thermo- and chemo-dynamics in the ICM: from the core to the outskirts, from galaxy groups to massive clusters.*

**Science Area:** Galaxy clusters, galaxy groups, ICM, metal abundance

**First Author:** S. Ghizzardi (INAF/IASF-Milano)

**Co-authors:** M. Rossetti(1), S. De Grandi(2), G.Riva(1), F. Gastaldello(1), L. Lovisari(1), S. Molendi(1) (with affiliations: (1) INAF/IASF-Milano; (2) Osservatorio Astronomico di Brera)

**Abstract:** The thermodynamic and chemical properties of the intracluster medium (ICM) provide critical insights into gravitational processes and non-gravitational feedback mechanisms, as well as into the cosmic enrichment history. However, precise measurements of these properties in cluster outskirts remain challenging due to low surface brightness and background contamination. Leveraging AXIS's capabilities, we aim to measure temperature and iron abundances up to $R_{500}$ for a sample of systems across a broad mass range, from galaxy group scale to massive clusters. This will help test self-similarity deviations and shed light on open issues like the "Fe Conundrum".

**Science:** Galaxy clusters hold a pivotal role in both cosmology and astrophysics, offering a unique window into the formation and evolution of large-scale structure. Within a hierarchical framework of structure formation, galaxy groups and clusters grow primarily through the gravitational accretion of matter along the filaments of the large-scale cosmic web. This process occurs mainly via subsequent mergers, which heat the intracluster medium (ICM) and the intragroup medium (IGrM) to X-ray-emitting temperatures.

Given the scale-free nature of gravitational collapse, clusters are expected to behave self-similarly, with their thermodynamical properties tightly correlating with the total gravitational mass [249]. However, deviations from these relations do exist and clearly indicate the presence of non-gravitational processes, such as cooling, feedback from active galactic nuclei (AGN) or supernovae, bulk motions, and turbulence induced by accretion events. These processes introduce complexities that lead to departures from the simple, self-similar behavior [see, e.g., 54,177,192,402,403,547]. While the thermodynamic properties of the ICM reveal the energetic processes at play, the chemical abundances offer a complementary perspective, tracing the enrichment history of the Universe. Since clusters are the largest virialized systems, they are representative of the Universe as a whole, and the metal content in them retains an imprint of the cosmic enrichment history. Metals originate within stars, before being expelled from galaxies into the ICM through supernovae (SN) explosions and stellar winds. For this reason, the enrichment history of galaxy clusters is intimately related to the integrated star formation history of the cluster. The abundance and spatial distribution of metals in the ICM is strictly connected to its dynamical history and the feedback mechanisms of active galactic nuclei (AGNs), outflows or jets [e.g. 38,155,174,518], which inject and spread the metal content in the surrounding environment. Studying both the thermodynamic and chemical properties of the ICM and IGrM is therefore essential, as they represent two sides of the same coin: each reveals crucial information about the physical processes governing the formation and evolution of cosmic structures.

**From the core to the outskirts**

Pushing measurements of both thermodynamic properties and chemical abundances in the outskirts of clusters is essential to gain deeper insights into cosmic history. The reason is twofold. First, the outskirts encompass by far the largest volume of galaxy clusters, making them a crucial region for understanding the overall structure and composition of these systems. Second, they serve as direct witnesses to freshly accreted material [558] and so preserve a record of cluster formation and growth.

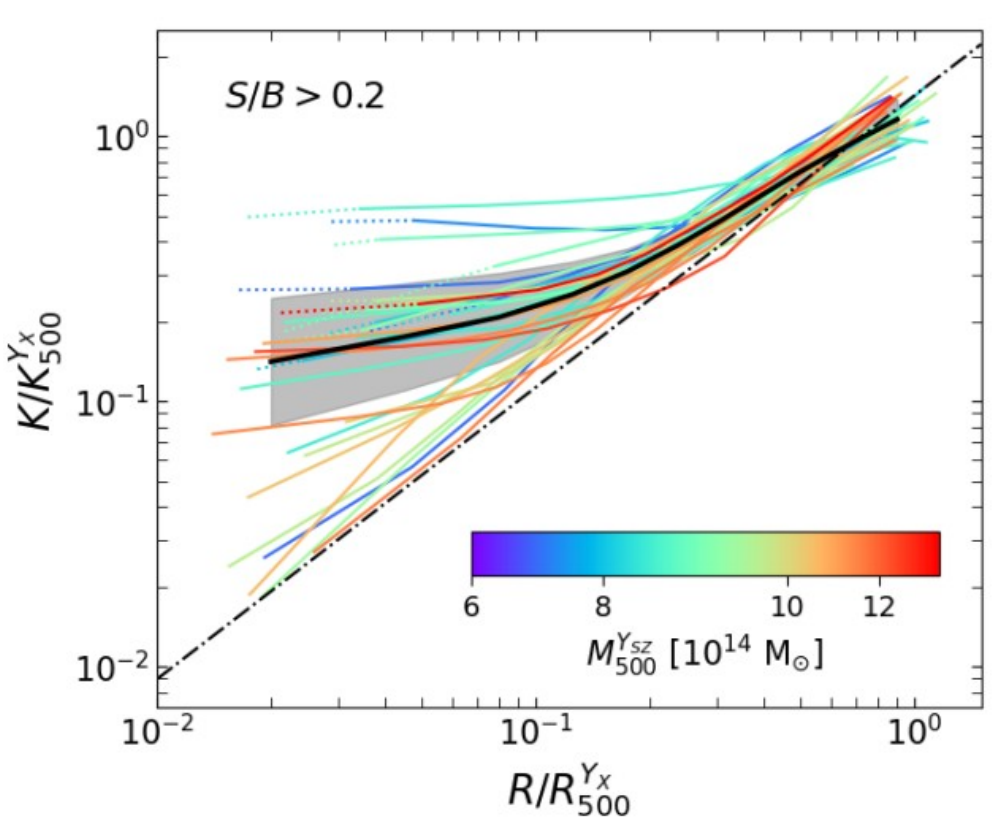

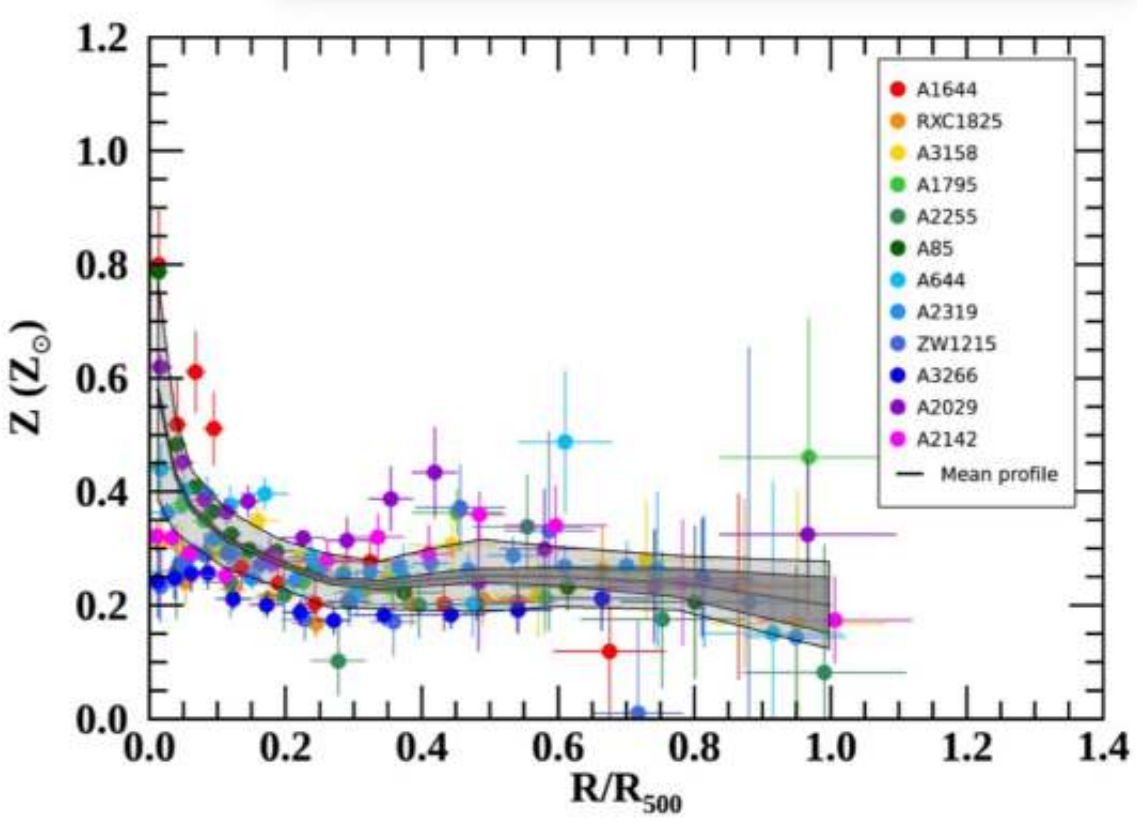

**Figure 35.** *(Left panel:)* Entropy profiles of HIGHMz clusters [423], with their median profile (solid black) and the measured intrinsic dispersion (gray shaded area). Profiles are color-coded according to their mass. The black dash-dotted line is the prediction from non-radiative simulations [549]; *(Right panel:)* Iron abundance profiles for X-COP clusters [186] with average profile (thick line) overlaid. The dark and light shaded areas indicate, respectively, the statistical error and the total scatter.

Of course, obtaining robust measurements of the ICM's thermodynamic properties and chemical abundances in low-surface-brightness regions is challenging. XMM-Newton and Chandra observations have allowed us to explore the thermodynamic profiles of bright and massive clusters up to $R_{500}$. Indeed, recent results [430] have shown that we can finally keep systematic errors on temperature profiles below 10% up to $R_{500}$, where the source intensity is only 20% of the background, after 20 years of efforts characterizing the XMM-Newton background.

Thanks to the lower particle background level, Suzaku explored the outskirts of a dozen galaxy clusters, but provided controversial results, with some studies showing an inversion of the entropy profiles. Indeed, Riva et al. [423] showed that systematics in temperature measurements beyond $R_{500}$ may induce a similar behavior in the entropy profiles measured by XMM-Newton and that steadily increasing entropy profiles can be derived once this error is taken into account (see Fig. 35, left panel). It is important to point out that while the combination of SZ observations with X-ray imaging has provided a new method to derive thermodynamic properties beyond $R_{500}$ [e.g. 185], its application has been limited so far to massive clusters resolved by current millimeter telescopes.

For what concerns the chemical enrichment, Ghizzardi et al. [186] presented the first iron abundance profiles extending out to $R_{500}$ for a representative sample of massive clusters (Fig. 35, right panel). This result has been achieved exploiting the data of the XMM Cluster Outskirt Project (X-COP) sample [132]. However, measurements of iron abundances at these radii is still challenging, especially so for lower mass systems. Recent pilot studies of some intermediate mass clusters ($M_{500} = 2 - 4 \times 10^{14}$ M$_\odot$) reached $\sim 0.8 R_{500}$, (Riva 2024 PhD Thesis; Riva et al *in prep*), but most of the systems below $M_{500} \sim 4 \times 10^{14}$ M$_\odot$, down to the group size, typically can be characterized up to $\sim 0.5 R_{500}$.

**From massive clusters to galaxy groups**

Both for thermodynamic studies and for chemical characterization, it is necessary to explore all the mass scales from groups to massive systems. For the most massive clusters, the dominant physical process shaping the thermodynamic properties of the ICM is gravity, with global properties obeying self-similar relations and entropy profiles closely following the predictions of semi-analytic models from the cores [549]. At lower masses, non-gravitational processes start to have a growing impact, with significant deviations from self-similarity and entropy profiles showing an excess at all radii (see Fig. 36). Recently, Riva et al. [423] compared the entropy profiles of heterogeneous samples at different masses and estimated the

deviation from the self-similar mass scale to minimize the scatter in the entropy profiles: such deviations are larger in the cores (where ongoing cooling and AGN feedback have a large impact) but are significant also at $R_{500}$. Molendi et al. (2025, *in press*) could reproduce these results by proposing a modification to the self-similar models, assuming the decoupling of the accretion of baryons and dark matter during the formation phase, at a mass scale of $1 \times 10^{13}$ $M_{\odot}$.

The above results are obtained by comparing heterogeneous samples in the literature, which are derived from analyses with different assumptions. In the framework of the CHEX-MATE project [96] we intend to perform a homogeneous analysis of the thermodynamic profiles of clusters covering a decade in mass from $2 \times 10^{14}$ $M_{\odot}$ to $2 \times 10^{15}$ $M_{\odot}$ at $z < 0.2$. This will eliminate the scatter induced by the analysis methods, which can recover the true one, caused by deviations from self-similarity. Still, it would be essential to extend this analysis to the group scale, where deviations are expected to be larger [e.g., 134]. Moreover, in this regime, the ongoing feedback from the central AGN may also significantly affect the shape of the entropy profile out to $0.2 - 0.3 R_{500}$ (see discussion in Molendi et al. 2025, *in press*).

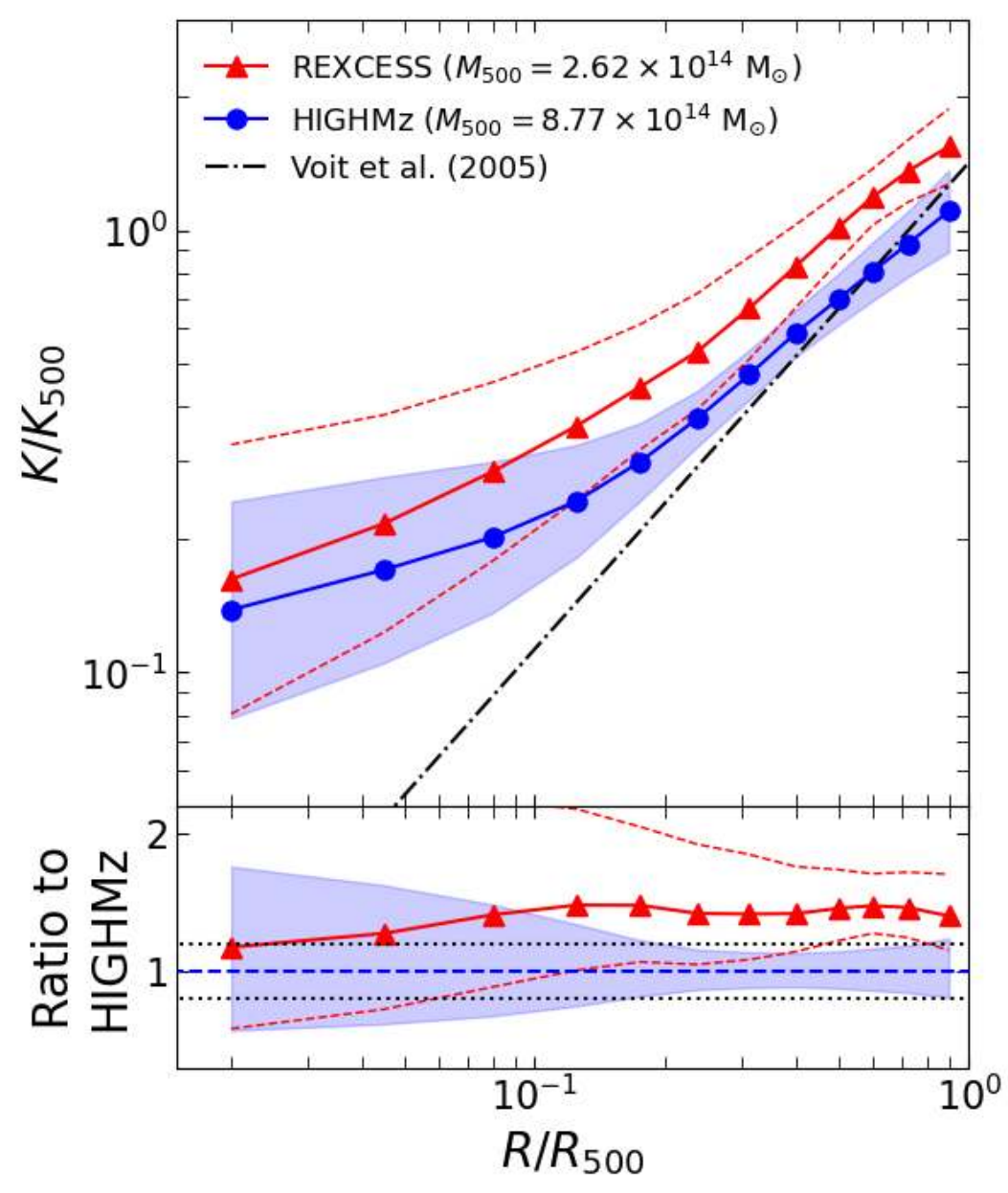

**Figure 36.** Comparison of the median entropy profile for the HIGHMz sample [423] with the REXCESS sample [46]. *(top panel)*: median entropy profiles for the two samples, with their intrinsic scatter; *(bottom panel)*: ratio to the HIGHMz median profile. The shaded area in both panels is the intrinsic scatter of the HIGHMz clusters, while the horizontal dotted lines indicate a 20% variation. The black dash-dotted line is the prediction from non-radiative simulations [549].

Studying iron abundances across the entire range of dynamical masses is crucial for understanding the mechanisms driving cluster enrichment. An important open debate concerns the different behavior of iron profiles in the outskirts. Whereas there is striking evidence that massive systems exhibit a flat profile beyond $0.3 R_{500}$, at about $\sim 0.4$ $Z_{\odot}$, [186,530,576, see also Fig. 35, right panel] the case of galaxy groups remains less clear. Some studies suggest a similar flattening trend in these lower-mass systems [e.g. 285,329], while other studies report steadily declining radial abundance profiles in groups that decrease uninterruptedly down to $\sim 0.1 - 0.2$ $Z_{\odot}$ [e.g. 409,410,488], thus making the general picture more uncertain at this mass scale. Beyond tracing iron distribution, by combining X-ray and optical measurements, we can conduct a comprehensive census of both the stellar mass and the total iron mass, including the amount of iron spread into the ICM and the fraction still locked into stars and galaxies. Comparing these estimates with the iron mass expected from supernovae (both SNIa and SNcc) allows us to assess the efficiency of metal production in clusters. This is quantified by the effective iron yield $\mathcal{Y}_{\rm Fe}$ which measures the efficiency of the stars in producing the iron amount detected in the ICM/IGrM.

Figure 37 shows iron yields derived for the X-COP cluster sample [black triangles; 186], compared with predicted values from theoretical models and measured SN explosion rates [169,416, yellow and gold band respectively]. Massive X-COP systems feature a discrepancy by a factor $3 - 7$ with respect to theoretical predictions. This inconsistency is known as the "Fe Conundrum". It seems to be absent at the scale of galaxy groups (green squares from Renzini & Andreon [416]), suggesting that this discrepancy is unique to massive systems. At the intermediate mass scale, data remain limited. Only two measurements

(blue and red dots) are currently available from the pilot study by Riva (2024 PhD thesis; Riva et al *in prep*), leaving this mass range substantially unexplored, despite its critical role in bridging the gap between massive clusters and galaxy groups.

The reason why the total iron amount in massive systems so largely exceeds the predicted value is far from understood, and the different behavior claimed on the galaxy group scale exacerbates even more the "Fe conundrum", making more puzzling the whole picture. New insights into this controversial topic have been provided by the recent work of Molendi et al. [342], who developed a simple model capable of predicting several observational properties in the group and massive cluster regimes. However, robust measurements across the entire scale are needed to validate the model further. Investigating the differences across various mass scales and resolving discrepancies with theoretical expectations is critical for understanding the mechanisms driving cluster enrichment.

**Exposure time (ks):** 2Ms

**Observing description:** AXIS provides the ideal capabilities to address these challenges. Its effective area, wide field of view, and high spatial resolution which keeps constant through the field of view, make it uniquely suited to get robust measures of thermodynamical quantities and iron abundance up to $R_{500}$ for all the mass scales — from the most massive clusters down to galaxy groups. A critical factor in obtaining robust measurements in regions of extremely low surface brightness is, of course, the accurate characterization of all background components at low levels. Concerning the particle background, which will be the dominant contribution in the hard band, we will need to characterize it at better than 5% level to keep systematics below 10%. A low level of the Cosmic X-ray Background will be ensured by the exquisite AXIS resolution, which will resolve 90% of the point sources of the field. However, to assess the level of the CXB residual flux and of the Galactic foregrounds, we will need a reference region free from cluster emission. This requirement translates to selecting clusters with $R_{200}$ smaller than $\sim 10'$ (and correspondingly $R_{500}$ below $\sim 6.5'$), allowing us to exploit the outermost part of the field of view ($10' - 12'$) for background calibration.

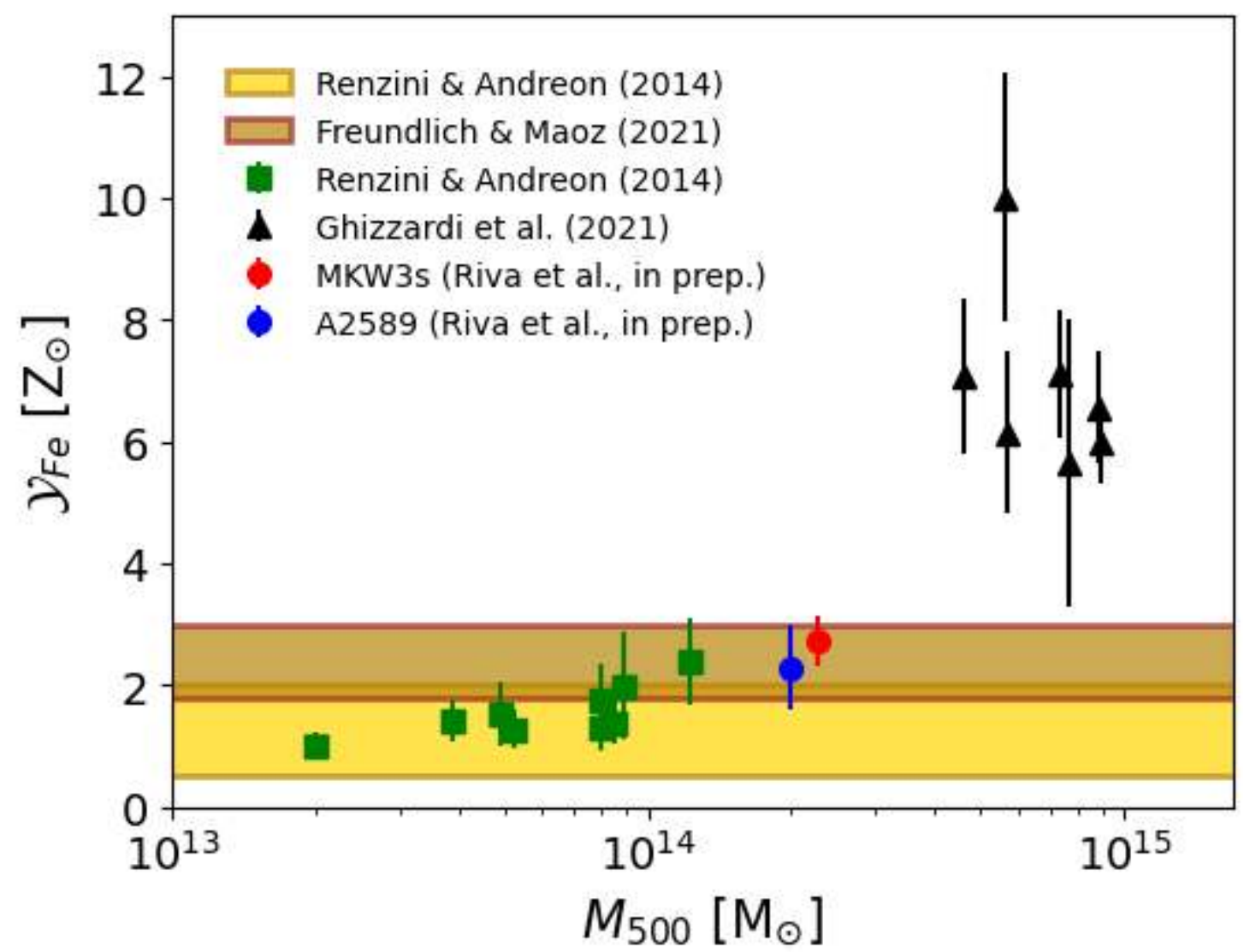

**Figure 37.** Iron yields vs $M_{500}$. Green squares and black triangles are measurements for galaxy groups [416] and massive clusters [186], respectively. The red dot and the blue dot are iron yields for two intermediate-mass clusters, MKW3s and A2589 (Riva PhD Thesis; Riva et al *in prep*). Yellow and gold regions represent theoretical expectations from supernovae.

We started from the *eRASS* cosmological catalogue of galaxy clusters [72] and selected systems having $R_{500}$ between $6'$ and $6.5'$. This criterion selects over 180 objects with masses in the $5 \times 10^{13}$ and $1 \times 10^{15}$ M$_\odot$ range. This provides a notable starting pool, from which we will select about 20 objects uniformly distributed across the mass range $5 \times 10^{13}$ and $1 \times 10^{15}$ M$_\odot$. To quantify the feasibility of measuring temperature and iron abundances in the cluster outskirts, we performed spectral simulations using XSPEC for three representative systems of different masses: one massive object ($M_{500} = 9.8 \times 10^{14}$ M$_\odot$, $z = 0.226$,

$T = 6.2$ keV), one intermediate-mass galaxy cluster ($M_{500} = 2.6 \times 10^{14}$ M$_\odot$, $z = 0.135$, $T = 2.8$ keV), and one galaxy group ($M_{500} = 5.1 \times 10^{13}$ M$_\odot$, $z = 0.075$ , $T = 1.1$ keV). Simulated spectra have been generated for an annular region with radii $5' - 6.5'$ (corresponding approximately to $\sim 0.75 - 1\ R_{500}$) and assuming an exposure time of 100 ksec, to test the possibility of measuring temperature and iron abundances in the outskirts.

Simulated spectra included 3 main components:

- *Cluster component:* phabs*apec: For each cluster we rescaled the flux provided in the *eRASS* catalogue (within $R_{500}$), to the annulus $5' - 6.5'$.
- *NXB component:* We adopted the NXB model provided by the AXIS collaboration for the L2 orbit, rescaled on the annulus area
- *Sky component:* The sky component is mainly due to three contributions: the Galactic Halo, the Local Hot Bubble and the Cosmic X-ray background. We refer to the modelization of these components obtained by the CHEX-MATE collaboration [430] XMM-Newton observations. We rescaled the CXB contribution considering the higher spatial resolving capability of AXIS. We assumed that the resolved fraction of the CXB with 100 ksec of observing time is 90%.

We fitted synthetic simulated spectra, and in all the cases we measured temperatures at a precision level of $2\% - 4\%$ and iron abundances at a precision level of $17\% - 20\%$.

As already mentioned above, while the spatial resolution and effective area of AXIS enable excellent resolution of a significant fraction of the CXB, the primary limiting factor remains the NXB. In our simulations, we fixed its contribution. However, in reality, this component is highly variable. To assess the effect of an inaccurately characterized NXB, we ran simulations in which its contribution was intentionally misestimated. We refitted the synthetic spectra while assuming NXB normalizations offset by $+5\%$, $-5\%$, $+10\%$, and $-10\%$ from the "true" simulated values, mimicking uncertainties in background characterization. These variations have a very low impact on the statistical errors and thus on the precision of our measurements. On the contrary, they induce significant biases on the measured values. Variations of 5% in the normalization of the NXB shifts values for temperatures and iron abundances of $\sim$ 10% with respect the “true” values; a 10% variation of the NXB normalization induces biases on temperature of the order of $\sim 20 - 25\%$ and biases as high as 35% for iron abundances measures. Despite the high statistics and the high precision of measures, a poor characterization of the NXB would severely compromise the quality of measurement, making them highly inaccurate. We definitely need to characterize the NXB at a level of 5% or better to fully harness the high potentialities of AXIS and avoid undesired systematic biases.

**Joint Observations and synergies with other observatories in the 2030s:** Large optical surveys revealing the galaxy population of these clusters and cluster masses through weak lensing (e.g., Rubin, Roman, ELTs). Radio observations tracing AGN activity and merger history via radio halos (SKA and LOFAR2.0).

**Special Requirements:** The NXB is required to be characterized at a level of 5% or better to fulfill our scientific goals.

## *24. Chemical Enrichment in the Outskirts of Galaxy Groups*

**Science Area: galaxy groups, intracluster medium, IGM enrichment, cosmic enrichment history**

**First Author:** Arnab Sarkar (MIT Kavli Institute for Astrophysics and Space Research & University of Arkansas, Fayetteville)

**Co-authors:** Yuanyuan Su (University of Kentucky), Scott W. Randall (Center for Astrophysics | Harvard & Smithsonian)

**Abstract:**
Understanding the distribution of metals in the ICM/IGrM is critical to unlocking the history of baryonic feedback in the Universe. While massive galaxy clusters, with their deep potential wells, retain much of their enriched material produced by the stars inside their virial radius, the situation in lower-mass systems is far more dynamic–and far less understood. Galaxy groups, due to their shallower gravitational potentials, are highly susceptible to metal loss through powerful non-gravitational processes, including supernova-driven winds and AGN feedback [179]. These mechanisms can eject significant amounts of enriched gas, dramatically altering the metal budget of the system [286]. Previous studies suggest a mass dependence in metal retention among groups, with iron mass-to-light ratios (IMLRs) within $0.5R_{200}$ increasing with mass. Yet, despite their importance, the outskirts of galaxy groups remain largely unexplored due to their low surface brightness, the impact of contaminating X-ray background and foreground emission, and instrumental background. This proposed study aims to bridge the gap. Leveraging the superb arcsec-resolution and exceptional soft X-ray sensitivity, AXIS is uniquely positioned to trace metal abundances out to—and beyond—the virial radius in these systems, providing definitive insights into the chemical evolution of the low-mass end of the halo population. This is particularly important for galaxy groups, since they are expected to have earlier formation epochs compared to rich clusters, and can thus help better discriminate between early- and late-stage chemical enrichment models.

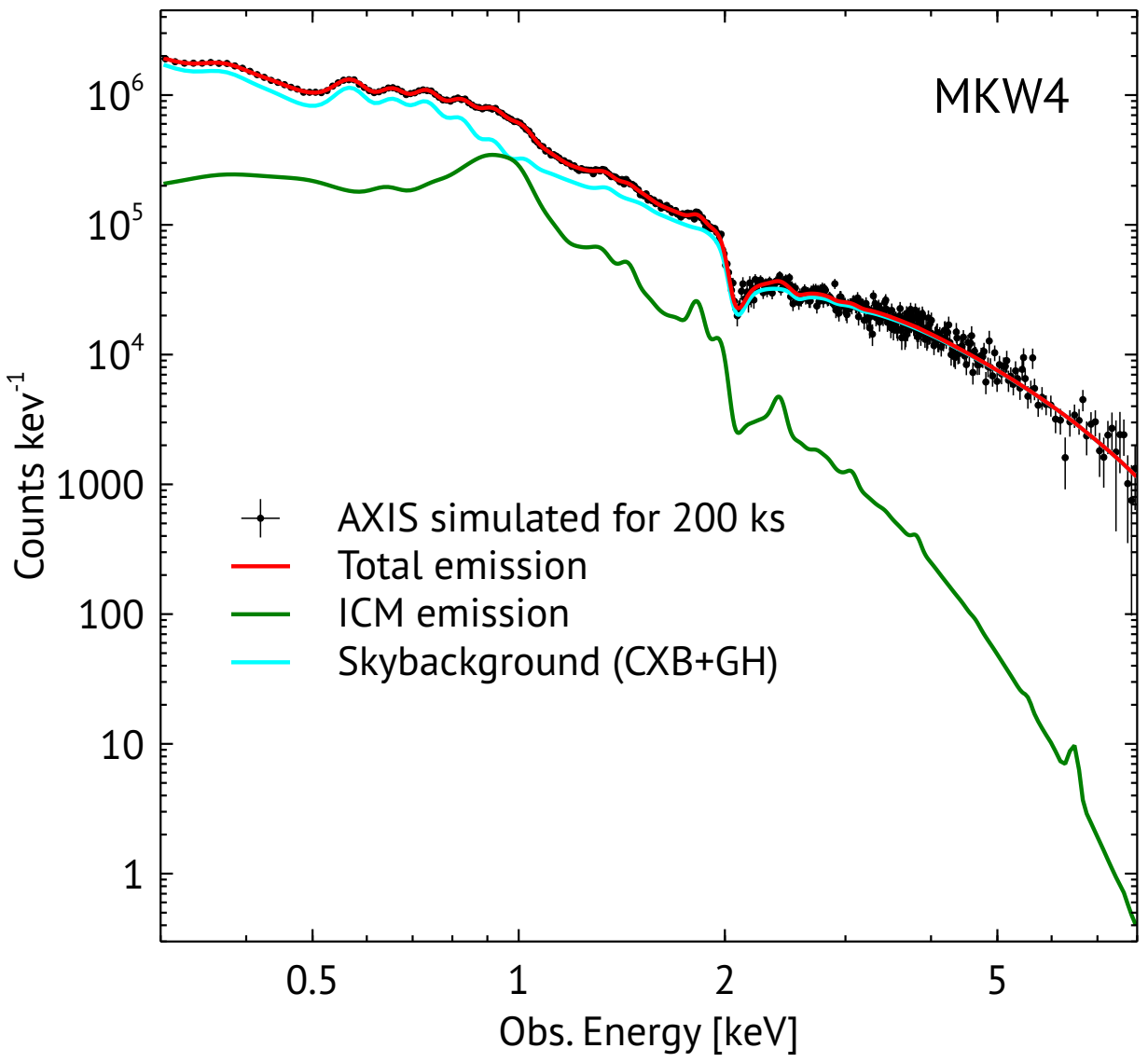

**Figure 38.** AXIS 200 ks Simulated spectrum for MKW4 at $R_{200}$. Black data points represent the simulated data. Green and Cyan show the ICM emission and total X-ray sky-background components, respectively. Sky-background includes contributions from Cosmic X-ray Background and Galactic Halo foreground emission. We adopted Sky-background components from the Suzaku study of MKW4 [445,446]. ICM emission component was modeled with an absorbped `VAPEC`. ICM temperature, chemical abundances were adopted from Sarkar et al. [446].

**Science:** Precisely measuring the radial abundance profiles out to $R_{200}$ is essential for tracing the chemical enrichment history of the groups. Our primary science goals are –

- **Constraining enrichment epoch:** Simulations predict that the bulk of metal enrichment occurred during cosmic noon ($z \sim 2$–3), driven by intense stellar activity and powerful outflows from supermassive black holes [37]. These early feedback processes are thought to have ejected and distributed metals well beyond their host galaxies, before large-scale structures fully assembled. This

"early enrichment" scenario results in a uniform metal distribution at the outskirts of clusters and groups, largely independent of halo mass. AXIS is uniquely positioned to test this scenario. With its unprecedented sensitivity, AXIS will directly measure Fe abundances beyond $R_{500}$ in galaxy groups, where the surface brightness is lowest and current measurements are lacking precision. Detecting a significantly lower Fe abundance in group outskirts, compared to the canonical $0.3Z_{\odot}$ observed in clusters, would challenge the early enrichment model paradigm.

- **AGN feedback:** AGN are among the most powerful engines of baryonic feedback in galaxy groups, redistributing metals within IGrM. In these lower-mass systems, the shallower gravitational potential allows AGN-driven outflows—fueled by supermassive black holes in the brightest group galaxies (BGGs)—to uplift enriched gas from group cores and transport it to larger radii by buoyantly rising bubbles and jets, leading to an abundance enhancement along the wakes [407]. AXIS, with its superb arcsec-angular resolution, stable off-axis PSF, and exceptional sensitivity, will provide detailed 2D maps of metal abundances out to larger radii, allowing us to directly trace how AGN feedback reshapes the chemical enrichment at the group scale. AXIS will precisely measure the Fe yield, which is the ratio of total Fe mass released by stars to the total stellar mass formed for a given stellar population, at $\gtrsim R_{500}$ [416]. A significantly large Fe yield at larger radii implies that AGN activity in group cores has efficiently transported metals into the outskirts.
- **Relative contributions of SNe Ia and SNcc:** Over the past few decades, advances in stellar nucleosynthesis have clarified the origins of various elements in the ICM/IGrM. Lighter elements such as oxygen (O), magnesium (Mg), and neon (Ne) are primarily synthesized by massive stars and released into the ICM through core-collapse supernovae (SNcc) [359]. In contrast, heavier elements like argon (Ar), calcium (Ca), iron (Fe), and nickel (Ni) are mainly produced by Type Ia supernovae (SNe Ia) [233]. Elements such as silicon (Si) and sulfur (S) are generated in comparable proportions by both SNcc and SNe Ia. As a result, the relative abundance of O, Mg, and Si–when compared to Fe–encodes the imprint of both SNe Ia and SNcc, offering a direct diagnostic of their respective contributions to chemical enrichment in the IGrM. AXIS will precisely measure these abundance ratios at the outskirts of groups, where previous measurements suffer from large uncertainties exceeding 25% ($1\sigma$). As shown in Figure 39, AXIS will dramatically reduce these uncertainties, enabling robust comparisons with theoretical nucleosynthetic yield models. This will enable us to disentangle the contributions of different supernova types and constrain the physical processes driving their explosions.

**Exposure time (ks):** 200 ks for each target, totalling 1 Msec.

**Observing description:** Obvious examples of bright, nearby groups that could be targeted for this study are MKW4, Antlia, ESO3060170, NGC5044 and NGC5813 (where, for NGC5813, the long-term energy injection rate into the IGM is known, due to the unique morphology and properties of this group). These groups are nearby and bright, making them ideal candidates for resolving substructures with AXIS and mapping chemical abundances of different elements with sufficient photon counts. Also, these groups have been extensively studied using deep Suzaku and Chandra observations. Figure 38 shows the 200 ks simulated spectrum for MKW4 at $R_{200}$. As illustrated in Figure 39, a 200 ks exposure for each of five selected targets, totaling 1 Msec, would be required to constrain the X/Fe abundance ratio to within 10% uncertainties ($1\sigma$ confidence level) at the outskirts of the groups. A possible extension to "higher redshift" (non-local) targets should be investigated, especially SZ-selected targets.

**Joint Observations and synergies with other observatories in the 2030s:** The combination of AXIS's high spatial resolution and NewAthena's advanced spectroscopic capabilities will provide a comprehensive

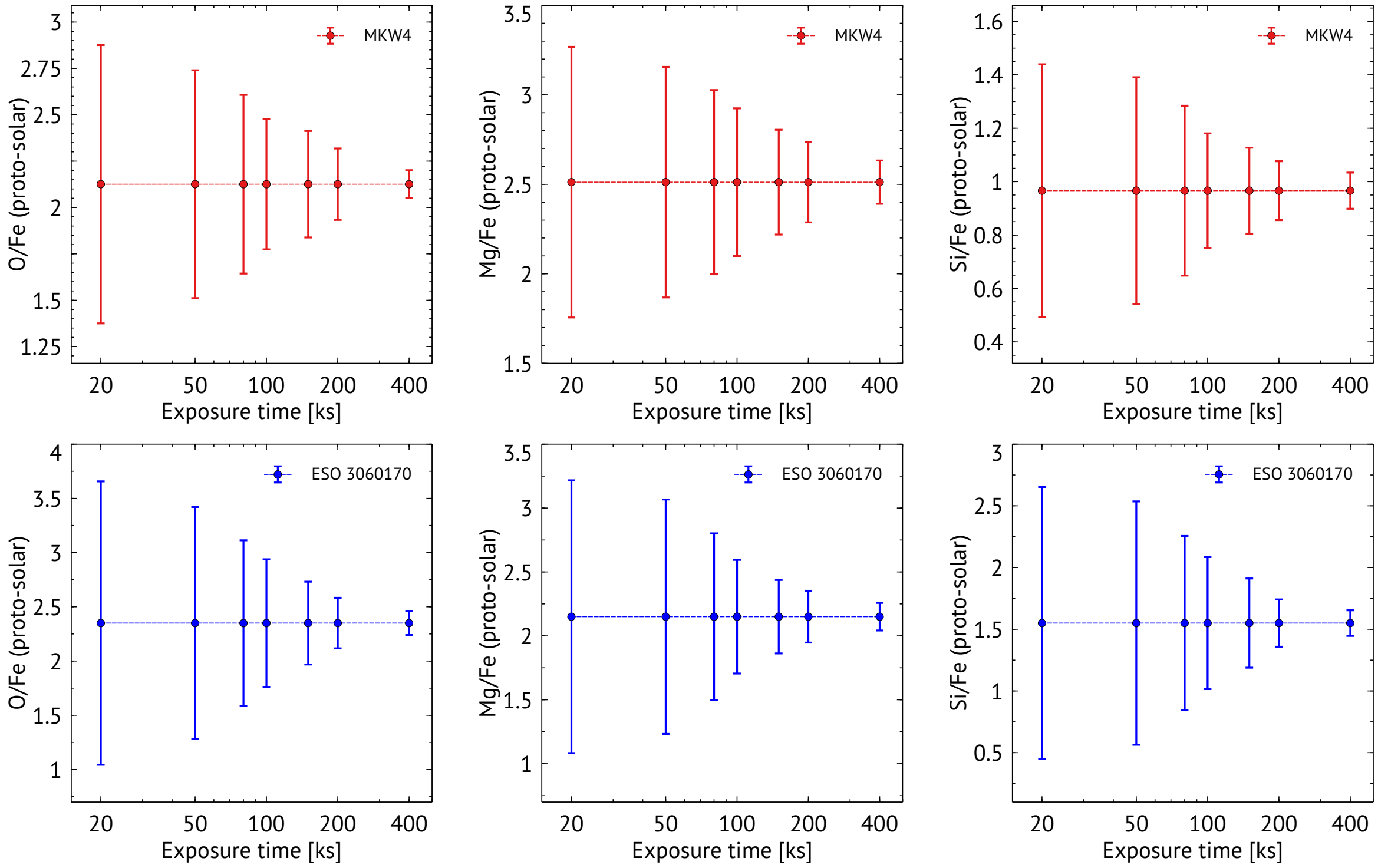

**Figure 39.** Simulated AXIS constraints on the abundance ratios O/Fe, Mg/Fe, and Si/Fe at $R_{200}$ for two galaxy groups: MKW4 (top panels) and ESO 3060170 (bottom panels), as a function of exposure time. The input abundance ratios are based on joint Suzaku and Chandra measurements from Sarkar et al. [446]. These simulations demonstrate that a minimum exposure of 200 ks per target is required to achieve better than 10% precision ($1\sigma$) on X/Fe ratios in group outskirts—enabling a robust decomposition of SNcc and SNe Ia contributions to IGrM enrichment.

view of chemical enrichment in the largely understudied galaxy groups. NewAthena will measure detailed abundance patterns, while AXIS will spatially map their distributions. This synergy will offer valuable insights into the roles of supernovae, AGN feedback, and mergers in enriching the intragroup medium and shaping galaxy group evolution.

**Special Requirements:** Low and well-understood background.

*25. An AXIS-ELT/MOSAIC joint survey of a common cosmological volume before the cosmic noon*

**Science Area:** galaxies, AGN, CGM/IGM

**First Author:** Jiang-Tao Li (Purple Mountain Observatory)

**Co-authors:** Feige Wang, Rui Huang, Jinyi Yang, Sean Johnson, Joel Bregman (University of Michigan)

**Abstract:** When AXIS is launched in the 2030s, numerous multi-wavelength cosmological surveys with unprecedented depth will be available for studying AGN, galaxies, galaxy clusters, and large-scale structures. Among these, the 39m ELT will conduct a deep spectroscopic survey using its multi-object spectrograph, MOSAIC, focusing on probing the IGM, CGM, and dark matter at $z \sim 3$, before the cosmic noon. This survey will include spectroscopic observations of $\sim 9,000$ galaxies, along with spatially resolved spectroscopy of $\sim 480$ objects over a sky area of $\sim 0.5$ deg$^2$. We propose an AXIS survey of a cosmological field overlapping with the ELT/MOSAIC survey. A joint analysis of optical and X-ray data will enable a statistical study of AGN with precise redshift and SMBH mass measurements, an estimation of the contributions from stellar X-ray sources and the hot CGM through stacking X-ray images of different galaxy subsamples, and the identification of X-ray counterparts of galaxy clusters and other large-scale structures. This combined AXIS/MOSAIC survey will provide the first in-depth and statistically meaningful view of various types of X-ray sources before the cosmic noon, enriched by high-quality multi-wavelength complementary data.

**Science:** AXIS will offer unprecedented capabilities for surveying AGN and other X-ray sources before the cosmic noon ($z \gtrsim 2$; e.g., [305]). The scientific impact of these deep X-ray surveys will be significantly enhanced when complemented by multi-wavelength observations. Such observations will enable precise redshift measurements and spectroscopic identification of X-ray source counterparts, facilitate the discovery of large-scale structures such as galaxy clusters and cosmic webs, and support joint studies of multi-phase gaseous structures.

The 39m Extremely Large Telescope (ELT) will be the world's largest general-purpose ground-based telescope. Its Multi-Object Spectrograph, MOSAIC, will operate with Ground Layer Adaptive Optics (GLAO) to deliver high image quality across the full field of view (FOV) of the ELT ($\sim$ 40 arcmin$^2$). MOSAIC will offer two observation modes: the Multi-Object Spectroscopy (MOS) mode, which can simultaneously observe up to 140 objects using fiber-fed spectrographs, and the multi-IFU (mIFU) mode, which integrates these fibers into six 2.2″ integral field units (IFUs). MOSAIC will be one of the ELT's second-light instruments, with its first light expected around 2035.

The ELT/MOSAIC "Inventory of Matter" Science Working Group (SWG2) has defined a Science Reference Program (SRP) covering a sky area of $\sim 0.5$ deg$^2$, utilizing $2 \times 50$ MOSAIC patrol fields (each of the 50 fields observed twice). The program aims to obtain medium-resolution ($R \sim 5,000$) optical spectra in MOS mode for $\sim 9,000$ Lyman Break Galaxies (LBGs) and quasars, reaching an $r$-band magnitude limit of $< 26$ at $2.5 < z < 3.1$, with a signal-to-noise ratio of S/N $> 3$ (for an early version of this program, see [236]). Additionally, parallel near-IR multi-IFU (mIFU) observations will be conducted with high spectral resolution ($R \sim 19,000$) for $\sim 480$ selected galaxies with stellar masses in the range $M_* = 10^{8.5-10.5}$M$_\odot$ within the same FOV.

The proposed AXIS and MOSAIC surveys within a common cosmological volume will provide the deepest X-ray/optical catalog of AGN and galaxies in the pre-cosmic noon era at $z \sim 3$. In particular, the MOSAIC survey will include high-quality optical spectroscopy of sub-$L^\star$ and more massive galaxies, enabling studies of galaxy clusters, large-scale structures such as cosmic webs, and gaseous components, including the circumgalactic and intergalactic medium (CGM and IGM). Specifically, this survey will allow us to explore the following key scientific questions:

- *Statistics of X-ray bright AGN.* Most individually detectable point-like X-ray sources at high redshift are AGN (e.g., [288]). X-ray observations uniquely probe the accretion disk coronae of supermassive black holes (SMBHs). We will first cross-identify these detected X-ray sources with optical spectra from the MOSAIC/MOS survey to determine their redshifts and SMBH masses (via the widths of specific emission lines; e.g., [543]). These measurements will also allow us to estimate their bolometric luminosities and Eddington ratios. We will then conduct statistical analyses using the measured X-ray and optical properties of these AGN (e.g., [275]). These analyses will include, but are not limited to: constructing the AGN X-ray luminosity function (LF; e.g., Fig. 40); comparing AGN across different redshift ranges in terms of their rest-frame X-ray luminosity, optical-to-X-ray spectral index ($\alpha_{\rm OX}$), and other parameters such as bolometric luminosity, SMBH mass, and Eddington ratio; stacking AGN in different subgroups to determine their X-ray spectral shape, often characterized by the X-ray photon index ($\Gamma$); and exploring the potential of using AGN X-ray properties, specifically the $\alpha_{\rm OX} - L_{\rm UV}$ relation, as a standard candle to constrain cosmological models (e.g., [289]).
- *Stellar X-ray sources.* Most stellar X-ray sources at $z \sim 3$ will not be individually detectable, except for a few ultra-luminous X-ray sources (ULXs) with luminosities of $L_{\rm X} \gtrsim 10^{40}$ ergs s$^{-1}$ (e.g., [157,232]). Even for these ULXs, the angular resolution of AXIS is often insufficient to determine whether they are offset from the galactic center, and the number of collected X-ray photons is likely too low to distinguish them from AGN spectroscopically. In such cases, optical spectra from MOSAIC will be critical for identification, particularly in low-mass galaxies without an AGN. We will search for X-ray-luminous non-AGN stellar sources and constrain their occurrence rate. Additionally, MOSAIC observations will provide key galaxy properties, including redshifts, stellar masses, and star formation rates (SFRs). We will then stack X-ray data from different galaxy subgroups to estimate the integrated contribution from hot gas and individually faint stellar X-ray sources (typically have $L_{\rm X} \lesssim 10^{39}$ ergs s$^{-1}$; e.g., [223]). The latter often dominate the stellar disk and bulge of galaxies (e.g., [274,276]). Quantifying the contribution of stellar X-ray sources in different galaxy types is crucial not only for studying the star formation history of galaxies but also for understanding the reionization and thermal balance of the IGM (e.g., [240,248]).
- *Stacking analysis of the extended hot CGM.* The MOSAIC survey is designed to simultaneously probe the IGM, CGM, and dark matter (via rotation curves) using both absorption and emission lines. Key gas tracers include Ly$\alpha$ and various metal lines in the rest-frame UV band, which trace gas across a broad temperature range of $T < 10^6$ K. Spatially resolved X-ray observations provide complementary information on the hotter CGM at $T \gtrsim 10^6$ K surrounding starburst or massive galaxies (e.g., [274,276]). By stacking X-ray images of different galaxy subsamples from AXIS, as proposed above, we will search for extended X-ray emission beyond the stellar disk and bulge. With its high angular resolution across the entire FOV, AXIS is optimized to remove contamination from AGN-scattered photons, enabling cleaner detection of extended hot gas emission compared to existing X-ray telescopes such as Chandra, XMM-Newton, and eROSITA (e.g., [273,604]). This advantage is particularly significant for high-$z$ galaxies with small angular sizes ($1'' \sim 8$ kpc at $z \sim 3$). A comprehensive view of the multi-phase CGM, integrating rest-frame UV and X-ray data, will further constrain the baryon budget across different galaxy subsamples (e.g., [65,256,575]).
- *The first virialized galaxy clusters and other large-scale structures.* With measured galaxy redshifts down to dwarf galaxies with $M_* \sim 10^{8.5}$ M$_\odot$, we can identify galaxy overdensities (such as cosmic webs) in both spatial and redshift domains and further determine gravitationally bound systems (galaxy groups and clusters) by measuring their velocity dispersions. The MOSAIC survey is designed to exclude known massive superclusters to avoid biases from environmental effects

(e.g., the supercluster in the COSMOS field; [111]). However, the survey area spans a comoving physical scale of $\gtrsim$ 60 cMpc, significantly larger than an individual galaxy cluster. This allows us to identify numerous new galaxy groups, clusters (Fig. 40), and larger-scale structures. The pre-cosmic noon epoch at $z \sim 3$ marks the formation of the first massive galaxy clusters, where the hot intra-cluster medium (ICM) begins to virialize (e.g., [511,569]). We will compare the X-ray properties of AXIS-detected galaxy clusters to established X-ray scaling relations that link hot ICM properties (luminosity, temperature, metallicity, etc.) to the depth of the gravitational potential (described by halo mass, stellar mass, velocity dispersion, etc.) of low-$z$ objects (e.g., [400]). Such comparisons will provide insight into the virialization processes of the most massive gravitationally bound systems in the Universe. Furthermore, extended hot gas emission from cosmic webs is expected and has been indicated by Sunyaev-Zel'dovich (SZ) effect observations (e.g., [119]). We will also search for diffuse X-ray emission associated with these large-scale structures.

**Exposure time (ks):** We estimate the required exposure time based on the point-source detection limit and survey area. We request a 0.5-2 keV flux detection limit (observational frame; assuming $\sim$ 5 net counts for a robust detection) comparable to that of the $\sim$ 7 Ms Chandra Deep Field South (CDF-S) survey [288], i.e., $F_{0.5-2\ \mathrm{keV}} \sim 4.6 \times 10^{-18}$ ergs s$^{-1}$ cm$^{-2}$ over $\sim$ 1 arcmin$^2$ from the optical axis. This detection limit is similar to that of the central region ($<$ 20% of the total observed area) of the $\sim$ 7 Ms CDF-S survey [288]. The corresponding average detection limit over the entire FOV is $\sim 5.8 \times 10^{-18}$ ergs s$^{-1}$ cm$^{-2}$. The AXIS exposure time required per tile in the mosaicked field to reach these detection limits is $\sim$ 800 ks [80]. To cover the full $\sim$ 0.5 deg$^2$ area of the MOSAIC survey, we will need a $2 \times 2$ tiling pattern, requiring a total exposure time of $\sim$ 3.2 Ms.

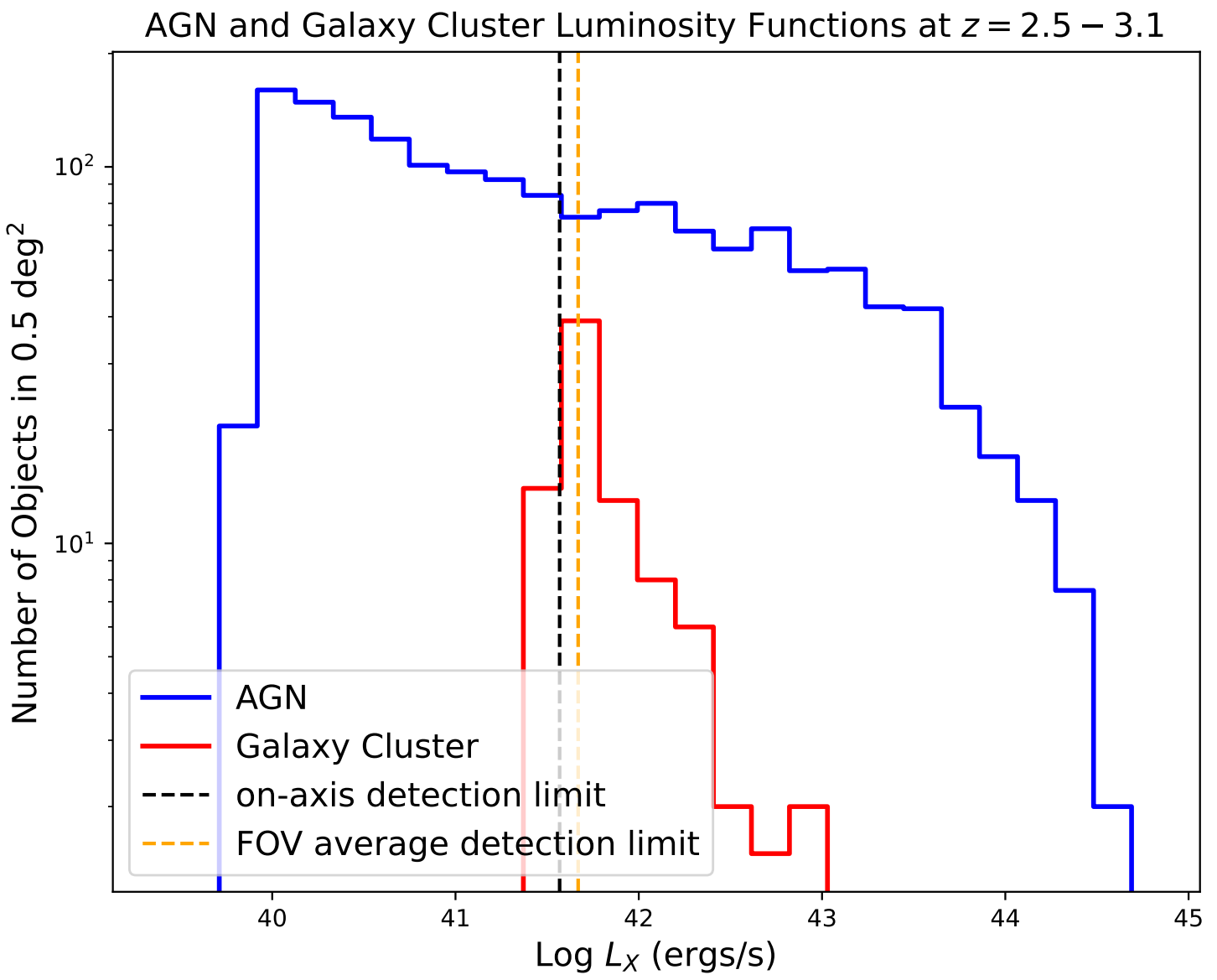

**Figure 40.** The AGN and galaxy cluster luminosity functions in a 0.5 deg$^2$ sky area and $\Delta z \approx 0.6$ redshift bin at $z \sim 3$, scaled from the mock catalog of [305]. The two dashed lines mark the detection limit of the proposed AXIS survey on-axis or averaged over the entire FOV, respectively.

At $z \sim 3$, the requested detection limit corresponds to luminosities of $L_{0.5-2\ \mathrm{keV}} \sim 3.7 \times 10^{41}$ ergs s$^{-1}$ and $4.7 \times 10^{41}$ ergs s$^{-1}$ on axis or averaged over the entire FOV, respectively. The central regions of each tile will be sufficiently deep to detect the brightest ULXs, sometimes referred to as hyperluminous X-ray sources (HLXs) if $L_X \gtrsim 10^{42}$ ergs s$^{-1}$ (e.g., [157,232]). For the AGN statistical study, we anticipate detecting $\sim$ 680 AGN within a field of view of $\approx$ 0.5 deg$^2$ and a redshift bin of $\Delta z \approx 0.6$ at $z \sim 3$, above the designed detection limit of the AXIS survey (Fig. 40; [305]), in addition to numerous foreground sources.

Except for AGN, the number of other X-ray sources at $z \sim 3$ detectable by the proposed AXIS survey remains highly uncertain due to their expected low abundance [305]. For galaxies and galaxy clusters, we will primarily study them through stacking X-ray images. Using the mock galaxy cluster X-ray catalog from [305], we estimate that $\sim$ 70 galaxy clusters with X-ray luminosities above the FOV averaged detection limit could be present within the joint MOSAIC/AXIS survey area and redshift bin (Fig. 40).

This sample size makes it feasible to divide the galaxy clusters into subgroups and measure the luminosity, temperature, and extent of the ICM from stacked images. For the CGM around galaxies, adopting scaling relations between diffuse X-ray luminosity and stellar mass or SFR [277], we estimate that the total stacked X-ray signal from the $\sim 9,000$ galaxies spectroscopically confirmed in the MOSAIC survey will yield a few hundred net counts from the hot CGM. The stacked X-ray image will thus help confirm the presence of extended diffuse hot gas emission from galaxies and provide a rough estimate of its contribution to the baryon budget [65,273]. Stellar X-ray sources are typically brighter and more compact than the hot CGM, making them easier to detect. However, to avoid contamination, we will exclude galaxies hosting X-ray-bright AGN from our analysis. The detectable extent of the hot CGM strongly depends on its spatial distribution, the surrounding environment, and the level of background fluctuations. Based on our experience with stacking analyses of nearby galaxies [273], we can typically trace the hot CGM out to galactocentric radii of $\sim (20-50)$ kpc. In this regime, the high angular resolution of AXIS is crucial for separating diffuse emission from bright, point-like sources. Although the instrument background tends to increase in lower orbits, this has a limited impact on CGM detection, as the angular scale of the detectable emission spans only a few arcseconds, allowing for reliable local background estimation.

In the above justification, we focus on the pre-cosmic noon era at $z \sim 3$ to align with the primary redshift range probed by the ELT/MOSAIC survey of the IGM, CGM, and dark matter. This approach enables a systematic joint X-ray and optical spectroscopic study at what is likely the highest redshift that can be examined with sufficient statistical robustness. Of course, various types of X-ray sources and their counterparts can be more effectively studied at lower redshifts using the proposed AXIS survey in combination with other multi-wavelength data (e.g., [44,198,266,451]).

**Observing description:** We request the proposed AXIS observations to cover approximately the same sky area as the planned ELT/MOSAIC survey. The specific field has not yet been finalized—candidates include the COSMOS field—and will be refined based on forthcoming multi-wavelength surveys over the next $\sim 10$ years, such as those from the Vera C. Rubin Observatory [e.g., 44], the Nancy Grace Roman Space Telescope [e.g., 451], and the Chinese Space Station Telescope (CSST; Gong et al. 198). Additionally, AXIS may carry out wide or deep field surveys that could already fulfill our scientific objectives. The central goal of the joint AXIS–ELT/MOSAIC program is to enable unprecedentedly deep optical characterization of X-ray sources and their environments over a contiguous area that samples diverse cosmic environments and potentially encompasses large-scale structures. All justifications provided above are based on the area and redshift coverage of the MOSAIC survey, rather than any fixed sky location. Several potential cosmological deep fields with extensive multi-wavelength data—such as the Euclid deep fields [140,141] and the LSST deep drilling fields [63,266]—could serve as viable targets for similar joint MOSAIC/AXIS observations.

**Joint Observations and synergies with other observatories in the 2030s:** Approximately $\sim 60$ nights of ELT/MOSAIC observations are proposed as part of the SRP [236], scheduled to commence after 2035. By the time this program is conducted, numerous multi-wavelength imaging and low-resolution spectroscopic surveys (e.g., [44,198,266,451]) will likely be available, facilitating the selection of targets for follow-up multi-object spectroscopic observations. Combined with the proposed AXIS survey, these datasets will enhance full-wavelength coverage by incorporating observations from ALMA, JWST, LSST, Euclid, Roman, and ngVLA, etc., and possibly also time-domain analyses (e.g., [63]). These analyses will be more feasible for studying brighter and lower-redshift objects.

**Special Requirements:** Each field will be observed in 5–10 separate exposures, spaced 1–2 months apart. This cadence enables the study of source variability and ensures coverage across detector gaps or field

boundaries. The temporal spacing is primarily designed to provide a basic characterization of AGN variability.

*26. Connections to the cosmic web*

**Science Area:** Galaxy clusters, cosmic web, WHIM filaments

**First Authors:** Ka-Wah Wong (SUNY Brockport), Stephen Walker (The University of Alabama in Huntsville)

**Co-authors:** Mohammad Mirakhor (The University of Alabama in Huntsville)

**Abstract:** The cosmic web contains a large fraction of the Universe's baryons, primarily in the form of low-density, intergalactic medium (IGM) or Warm-Hot Intergalactic Medium (WHIM) at temperatures of $10^5$–$10^7$ K. Detecting this gas in emission has remained challenging due to its faintness and diffuse nature. AXIS, with its high spatial resolution, large effective area, and stable background, is uniquely suited to detect and characterize both the WHIM and the circumgalactic medium (CGM). We propose deep observations of 12 galaxy clusters in the redshift range $z = 0.1$–$0.3$, aimed at mapping the faint X-ray emission near the virial radius and tracing filaments across the cosmic web. These data will enable measurements of the density and temperature of diffuse gas, detection of infalling clumps and substructures, and mapping of the thermodynamic state of cluster outskirts. Comparing hydrostatic mass estimates from X-ray data with gravitational lensing measurements will constrain non-thermal pressure and feedback processes. This program will test predictions from cosmological simulations and shed light on how baryons cycle between galaxies and large-scale structures.

**Science:** The cosmic web forms the backbone of gas flows on cosmological scales. In the local universe, about half of all baryons are expected to reside in the WHIM, with temperatures in the range $10^5 < T < 10^7$ K. This hot intergalactic medium is believed to be the main reservoir of the "missing baryons" in the local universe and the ultimate depository of metals and entropy produced by galaxies over cosmic time. The hotter phase of the WHIM ($10^6 < T < 10^7$ K) is detectable in the soft X-ray band [see 414,557 for a review], but due to its very low density, it is extremely faint in emission (see Fig. 13 in [436]). Only instruments with a combination of large effective area and low, stable background—such as AXIS—will be capable of detecting this emission.

AXIS will open a new window into the WHIM by probing the theoretically predicted $10^6$–$10^7$ K IGM, which is expected to dominate the baryonic budget at low redshifts. By combining X-ray emission mapping with X-ray and UV absorption line studies from future missions [420], AXIS will enable, for the first time, a complete census of baryons and metals in the local universe. Cosmological simulations predict specific spatial distributions of WHIM and metal enrichment patterns, influenced by processes such as AGN feedback and gravitational collapse. AXIS will directly test these predictions by mapping the WHIM filaments, the cluster outskirts, and the CGM around galaxies.

These observations address key open questions in cosmology and galaxy formation:

- How is matter funneled into the most massive knots of the cosmic web?
- How and when does accreted matter mix with the intracluster medium (ICM)?
- How does matter cycle between galaxies and large-scale structures?
- What is the interplay between feedback and gravitational collapse on cosmic scales?

AXIS's contributions align with the broader Cosmic Ecosystems theme identified in the ASTRO2020 Decadal Survey, which seeks to link observations and modeling of stars, galaxies, and the gas and energetic processes that govern their formation, evolution, and destinies by mapping the circumgalactic and intergalactic medium in emission. It will directly support the Priority Science Area: Unveiling the Hidden Drivers of Galaxy Growth, aiming to revolutionize our understanding of how diffuse cosmic web gas feeds galaxies and triggers star formation.

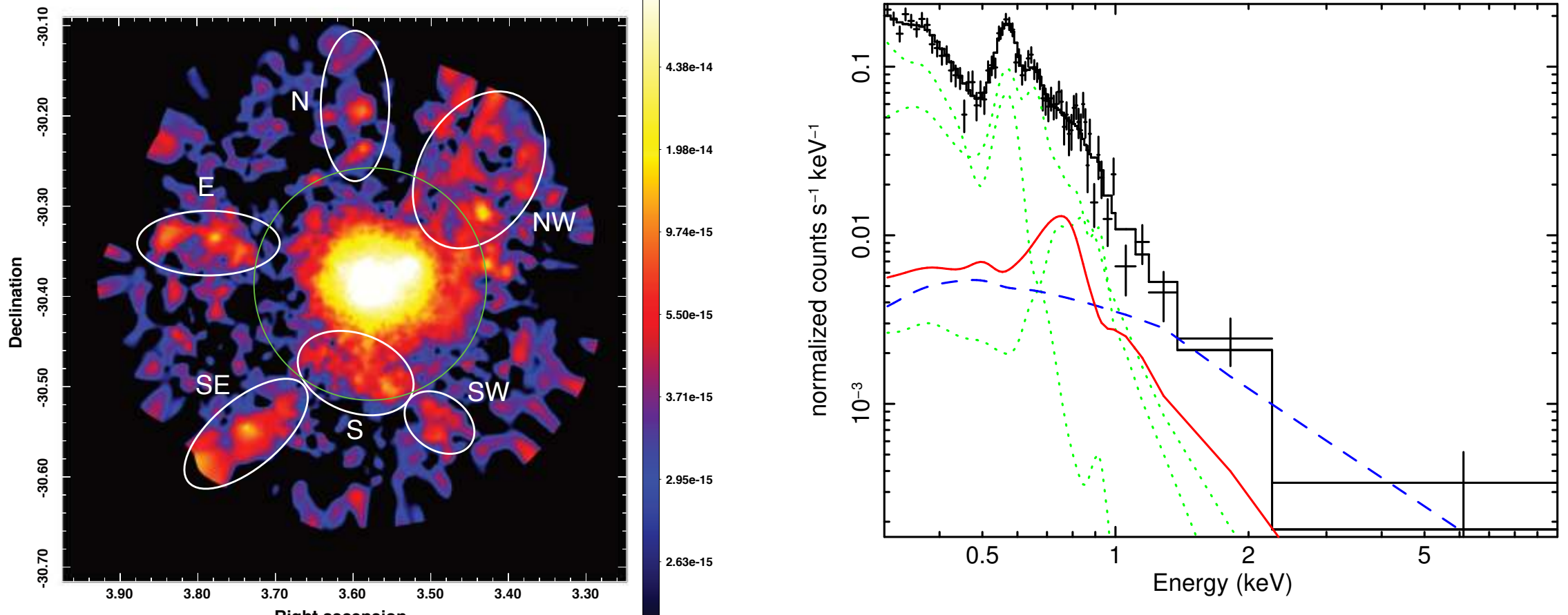

**Figure 41.** *Left:* XMM-Newton image of the massive galaxy cluster Abell 2744 in the 0.5–1.2 keV band [135]. The green circle indicates the virial radius of about 2 Mpc. Detected filaments are marked by white ellipses. *Right:* Simulated AXIS spectrum of a 1 keV filament with typical surface brightness near the virial radius, seven times fainter than that detected in Abell 2744 (solid red line). For a 100 ks exposure and a region size comparable to the SW filament of Abell 2744, approximately 600 photon counts will be detected in the 0.3–10 keV band. The green dotted lines represent the GXB, modeled with three thermal components. The blue dashed line shows the CXB. NXB is included in the simulation but not shown here.

In addition to detecting diffuse emission from the WHIM, AXIS will make precise measurements of gas density and temperature in cluster outskirts, enabling accurate hydrostatic mass estimates. Simulations predict that non-thermal pressure support increases at large radii, reaching 10–30% of the total pressure, and provides additional support against gravity. Comparing hydrostatic masses from AXIS with gravitational lensing masses from Euclid and Roman will constrain the level of non-thermal pressure support in these low-density regions [see 558 for a review], offering insights into the thermodynamic state of the ICM and its connection to large-scale structure.

One of AXIS's key technical advantages is its high spatial resolution, paired with a large effective area and a low and stable background—a combination unmatched by current and past missions such as Chandra, XMM-Newton, XRISM, and Suzaku. While XMM-Newton and Suzaku have detected X-ray emission in cluster outskirts and along filaments, the dominant source of uncertainty in those measurements comes from unresolved point sources, which can be confused with truly diffuse, extended emission. AXIS will overcome this limitation by resolving and removing small-scale gas clumps and detecting diffuse emission out to about twice the virial radius, $2R_{200}$ ($\sim$2 Mpc for low-mass clusters and up to 5 Mpc for the most massive). These gas clumps are believed to be infalling substructures, and simulations show that clumping increases dramatically in the $(1–2)R_{200}$ region before the clumps are stripped by the ICM [350,426,537]. If not resolved, clumping biases gas density and temperature profiles, leading to systematic errors in hydrostatic mass estimates [469,508].

Resolving these clumps is essential not only for robust cluster mass modeling, but also for understanding how galaxies lose their CGM through ram-pressure stripping in the cluster environment [87]. These faint clumps—believed to contribute significantly to total gas clumping—remain unresolved with current X-ray telescopes, and their physical properties are largely unexplored. AXIS will also allow studies of non-equilibrium ionization and electron-ion equilibration in cluster outskirts [5,15,23,583,584].

To advance our understanding of the cosmic web and the baryon cycle in large-scale structures, AXIS will target the following key physical measurements and observables:

- Density and temperature of IGM filaments and their connections to galaxy cluster outskirts, targeting systems in the redshift range $z = 0.1$–$0.3$.
- Detection and characterization of faint gas clumps and infalling substructures in cluster outskirts by measuring their gas density and temperature ($kT$).

These measurements are critical for constraining the thermal and dynamical state of the low-density plasma in the cosmic web and for testing cosmological simulations that model baryon evolution and feedback processes across cosmic time.

**Exposure time (ks):** 12 clusters, about 400 ks per cluster, totaling 4.8 Ms

We plan to observe 12 galaxy clusters in the redshift range $z = 0.1$–$0.3$. Galaxy maps will be used to identify the directions of large-scale filaments, which will then be followed up with AXIS observations. Each cluster will require approximately four pointings to cover the filaments, with each pointing having an exposure time of $\sim$100 ks. This results in a total exposure time of 400 ks $\times$ 12 clusters $\approx$ 4.8 Ms. Potential target clusters include Abell 2744 ($z = 0.308$), Abell 3016/3017 ($z = 0.220$), Abell 98 ($z = 1.04$), as well as other clusters with known filamentary structures identified by existing X-ray missions or the eROSITA survey.

**Observing description:** We propose to observe a small sample of massive clusters to measure the hot gas properties around the virial radius ($R_{200}$) and the surrounding filaments. Massive clusters are chosen because they are more likely to host denser accreting filaments. The clusters will be chosen in the redshift range $z = 0.1$–$0.3$, such that approximately four pointings will be sufficient to cover the filaments around the virial radius [436].

An example is Abell 2744, a massive cluster with a mass of $2 \times 10^{15}\, M_\odot$ located at $z = 0.3$, where several filaments have been detected with XMM-Newton (left panel in Fig. 41; [135]). The cluster is currently undergoing a merger involving at least four distinct components, supporting the hypothesis that Abell 2744 may be an active node in the cosmic web. While X-ray emission has been detected with XMM-Newton, a major source of uncertainty arises from unresolved point sources that can be confused with diffuse emission. With our proposed AXIS observations, approximately 95% of unrelated point sources will be resolved and removed, enabling a robust determination of the nature of the diffuse emission.

Our feasibility analysis is based on the criterion that we can detect cluster emission near the virial radius, where the surface brightness is approximately $10^{-16}$ erg s$^{-1}$ cm$^{-2}$ arcmin$^{-2}$ in the 0.3–2.0 keV band. This is about seven times fainter than the filaments detected in Abell 2744, ensuring that we are capable of detecting even fainter filaments around clusters.

Using `Xspec`, we simulated X-ray spectra of a filament region with a size similar to the SW filament shown in the left panel of Fig. 41, but with a surface brightness seven times fainter. The filament was modeled using an `APEC` thermal plasma model absorbed by the Galactic column density. We assumed a temperature of 1 keV—at the lower end of values detected with XMM-Newton—and a metallicity of 0.2 Solar. The simulation also included the Galactic X-ray background (GXB), which dominates the emission below 1 keV, as well as the unresolved Cosmic X-ray Background (CXB) and the Non-X-ray Background (NXB). The simulated spectrum is shown in the right panel of Fig. 41.

With an exposure time of 100 ks, we find that approximately 600 counts will be detected in the 0.3–10 keV range, allowing the temperature and normalization to be measured with uncertainties of about 10% and 20%, respectively. This is sufficient to determine the gas density with an uncertainty of approximately 10%.

We also assessed the systematic uncertainties due to the dominant non-X-ray background (NXB). Varying the NXB level by 5% changes the measured temperature by less than 1%, owing to the distinctive spectral shape of the 1 keV thermal emission. The normalization changes by about 10%, which is smaller than or comparable to the statistical uncertainty. This result also justifies the chosen exposure time of 100 ks, beyond which systematic uncertainties would begin to dominate.

The hotter phase of the filament, at around 2 keV as observed with XMM-Newton, can also be detected. Our simulations confirm that both the temperature and normalization can be measured.

**Joint Observations and synergies with other observatories in the 2030s:** Comparison with gravitational lensing estimates from Euclid and Roman will enable measurements of the non-thermal pressure components in the ICM as a function of radius, helping to constrain the dominant physical processes in the low-density plasma.

In the coming two decades, projects such as CMB-S4 [1], the Simons Observatory [498], and LiteBIRD [209] are expected to significantly advance our understanding of massive halos through observations of the Sunyaev–Zel'dovich (SZ) effect.

On the radio side, the leading long-term facility is the Square Kilometre Array (SKA). Its forecasted Phase 2 sensitivity improvements below $\lesssim$200 MHz, expected beyond 2030, will greatly enhance the search for low surface brightness radio structures in the cosmic web.

According to recent numerical simulations, a threefold improvement in sensitivity is expected to enable the systematic detection of radio emission associated with accretion shocks in the cluster outskirts, as well as the imaging of the radio-brightest regions of the larger cosmic web.

**Special Requirements:** Systematic uncertainties in the NXB should remain below about 5%.

*27. Inflows and outflows from galaxy clusters near and beyond the virial radius*

**Science Area:** Galaxy clusters, intracluster medium

**First Author:** Congyao Zhang (Masaryk University, The University of Chicago)

**Abstract:** AXIS's large effective area and high spatial resolution across the field of view will revolutionize our ability to detect and characterize sharp gaseous structures (e.g., shocks, contact discontinuities) in the cluster outskirts, $R_{500}$, and potentially beyond. A low and well-understood detector background will be key and, if realized, could allow AXIS to study the large, relatively unknown volumes towards the edges of clusters that reveal their assembly history.

**Science:** Galaxy clusters are assembled through cosmic inflows, including both smooth accretion and occasional mergers of other halo structures (e.g., galaxies, groups). The smoothly accreted gaseous baryons are shock-heated while crossing borders between the hot ICM and cold IGM, and pile up in the cluster outskirts. Mergers, on the contrary, penetrate the ICM, stir the gas cores, and further power Mpc-scale outflows in the form of runaway merger shocks [596,597]. The steep gas density radial profiles in the cluster outskirts ($\rho \propto r^{-3}$ or steeper) allow the runaway merger shocks to survive for an extended time, forming a "shock habitable zone" outside $R_{500}$ [597]. On the other hand, such a steep gas density profile results in an extremely faint X-ray signal, i.e., the X-ray surface brightness decreases as $\propto \int_{\rm LOS} \rho^2 \sim r^{-5}$ or steeper [291].

Fig. 42 illustrates our current theoretical understanding of the gaseous discontinuities (i.e., shocks and contact discontinuities) formed in the cluster outskirts. Bow shocks gradually detach from the infalling subhalos that drive them, typically occurring between $R_{500}$ and $R_{200}$. They transform into runaway merger shocks, propagating all the way to the cluster peripheries. The Coma cluster and its interaction with the infalling group NGC 4839 showcase a textbook example of such processes[102,292]. The long-lived runaway merger shock eventually overtakes the accretion shock, shaping a new boundary of the ICM [595,600] and leaving a Mpc-scale contact discontinuity near the cluster virial radius [594]. Promising candidates for the virial contact discontinuity have been identified in the Perseus cluster [561].

Thanks to its large effective area, AXIS will significantly enhance our detection of shock fronts and contact discontinuities outside $R_{500}$, providing key information on the large-scale inflows and outflows that trace the cluster's assembly history. The capability of AXIS is illustrated in Fig. 43. A major difficulty of probing signals from the cluster outskirts is the faint ICM surface brightness relative to the background, including the Galactic foreground dominating the soft band ($\lesssim 2$ keV), non-X-ray background (NXB; dominating the hard band, e.g., $\gtrsim 2$ keV), and cosmic X-ray background (CXB). The CXB can be largely removed thanks to AXIS's high angular resolution. Fig. 43 shows that the ICM surface brightness near $R_{500}$ is close to the total AXIS background (incl. 5% CXB) around $E = 1$ keV. A careful modeling of the background, especially the NXB, is essential for extracting the ICM signals and measuring their gas properties. It is worth noting that a good knowledge of the background, with uncertainties $\lesssim 10\%$, is required to probe the ICM near $R_{200}$.

Here are several major science goals of probing discontinuities outside $R_{500}$ with AXIS:

- Characterizing electron-ion non-equilibrium in the cluster outskirts: the collisionless nature of runaway merger shocks makes them an ideal target for probing the presence of electron-ion non-equilibrium and constraining the plasma physics of the high-$\beta$ ICM. Given the typical merger rate of galaxy clusters, AXIS will enable the identification of a large number of merger shocks in the cluster outskirts. Synergizing radio surveys with AXIS follow-ups will enable the connection of X-ray shocks with radio relics, which is essential for understanding the mechanism of electron re-acceleration powered by shocks with moderate Mach numbers.

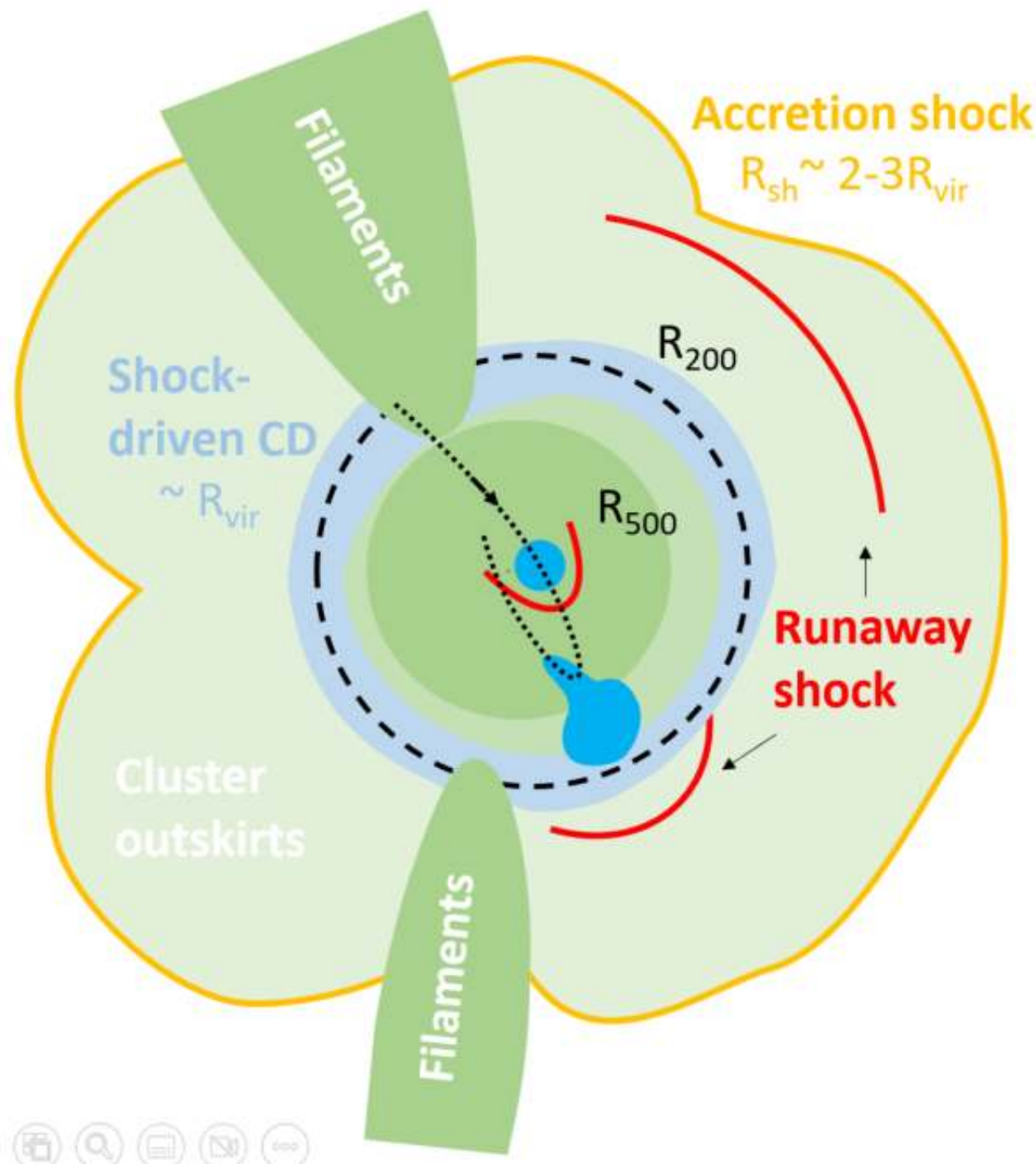

**Figure 42.** A sketch summarizing gaseous discontinuities formed in the cluster outskirts through mergers and accretion. The radius $R_{500}$, $R_{200}$, and external shock front $R_{sh}$ are indicated by the inner green region, dashed black line, and yellow boundary, respectively. An infalling subcluster (blue) moves in the main cluster along the trajectory shown as the dotted black line. The red arcs illustrate merger shocks, which experience a transition from bow shocks to runaway shocks [597]. The latter is associated with radio relics in the cluster peripheries. The long-lived runaway shocks eventually overtake the accretion shock and shape the new boundary of the ICM (accretion shocks). In this process, Mpc-scale contact discontinuities (CD) are formed near the virial radius $R_{vir}$ (blue annulus in the sketch; see 162]. AXIS will dramatically advance our ability to measure sharp gaseous structures (e.g., runaway shocks, contact discontinuities) in the cluster outskirts, which is essential to understanding the assembly history of galaxy clusters.

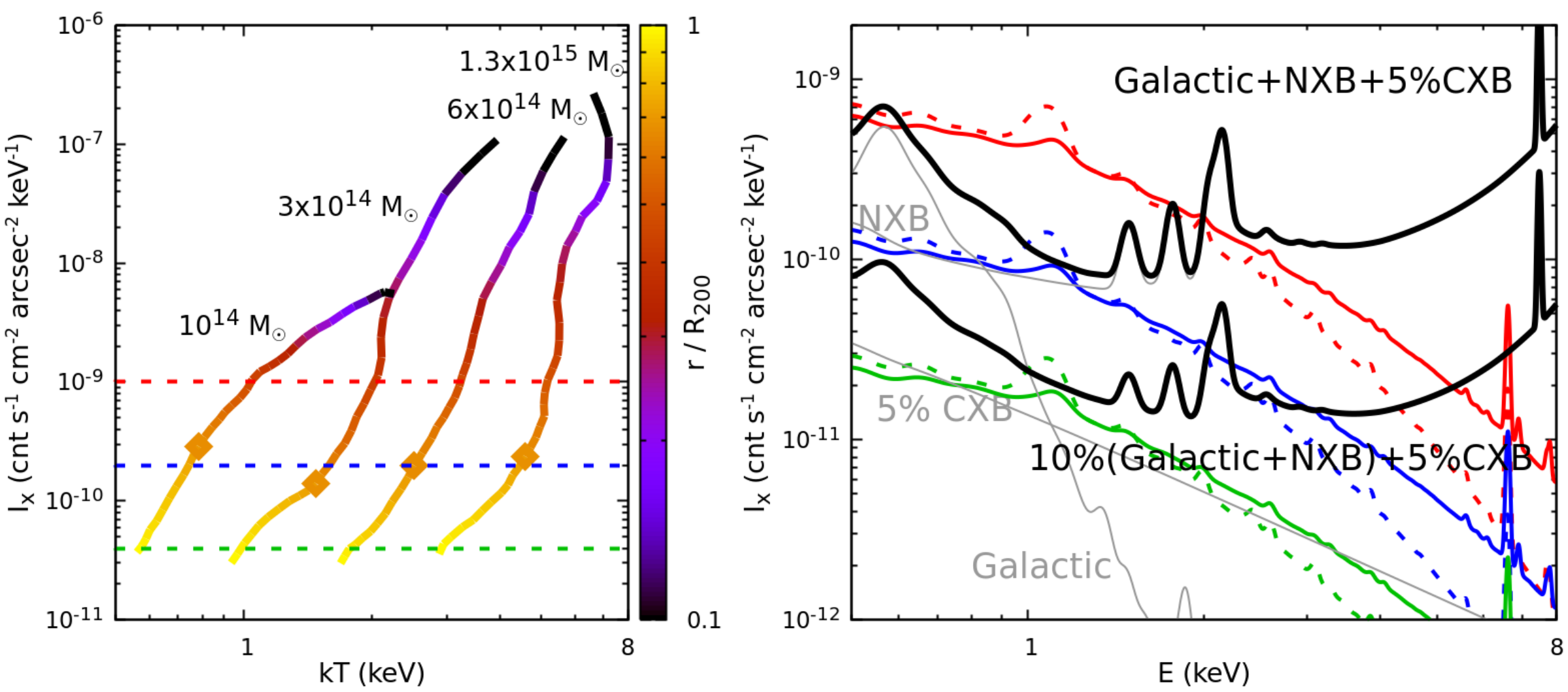

**Figure 43.** Left Panel: X-ray surface brightness (0.5 − 8 keV) vs. gas temperature as a function of the radius. It showcases four galaxy clusters with different halo masses selected from the TNG-300 cosmological simulation [see 601, and references therein for more details]. The line color encodes the radius in units of the virial radius $R_{200}$. The diamond marks the radial position of $R_{500}$. Right panel: a comparison of the ICM spectra (color lines) with the AXIS background (thick black lines), including the Galactic foreground, NXB, and 5% CXB components (thin grey lines). The ICM spectra are models resembling the signals from the regions near $0.4R_{200}$ (red), $0.65R_{200} \simeq R_{500}$ (blue), and $R_{200}$ (green), corresponding to the horizontal dashed lines marked in the left panel. The solid and dotted colored lines represent gas temperatures of 2 and 4 keV, respectively.

- Measuring the ICM gas density radial profiles up to at least $R_{200}$: The evolution of runaway shocks largely depends on the steepness of the gas density radial profiles in the diffuse ICM [597]. AXIS will provide an unprecedented sample of these profiles out to $\gtrsim R_{200}$ by carefully excluding contributions from cold gas clumps. This is key to understanding the fate of runaway shocks, the evolution of merger-driven outflows, and the extent to which these outflows redistribute matter and energy in the cluster outskirts.
- Probing accretion shocks: Large-scale filaments penetrate the ICM due to their high, smooth mass accretion rate [555,600,610]. Near the tip of these penetrating filaments (often around the virial radius), accretion shocks with a concave shape are formed. This region is the only place where we have the opportunity to probe accretion shocks —- high Mach number, collisionless shocks that separate the ICM and IGM —- even with next-generation X-ray telescopes.
- Searching for Mpc-scale contact discontinuities formed by shock collisions in the cluster outskirts: These structures provide direct evidence of merger-driven outflows and their role in breaking the self-similarity of the ICM in the cluster outskirts. By combining X-ray observations with the Sunyaev-Zel'dovich effect, we can constrain how electrons are heated by collisionless shocks [600].

**Exposure time (ks):** 800 ks

**Observing description:** Perseus would be a priority target, where we will thoroughly search for shocks, cold fronts, and other X-ray excesses associated with the past merger events, such as those that have been identified in *Chandra* X-ray (a cold front at 730 kpc, half the virial radius; [559]) and weak lensing (a subhalo at $\simeq$ 500 kpc; [228]) measurements. We propose 10 AXIS pointings ($4 \times 50$ ks and $6 \times 100$ ks) to cover the radial range up to $\simeq 700 - 800$ kpc. Coma would be another suitable nearby target that allows us to study detailed merger-driven gaseous structures (e.g., slingshot tails, bridges associated with radio diffuse emissions), while massive clusters at higher redshifts could fit better within the AXIS FOV.

**Joint Observations and synergies with other observatories in the 2030s:** Multi-wavelength observations will be essential to maximize our scientific achievements. Combining SZ (e.g., CMB-S4) and X-ray images will extend our detections of the ICM to larger radii and provide constraint power on the clumpiness of gas distributions. The detection of shocks and X-ray fluctuations will be crucial for interpreting the faint, diffuse radio structures (e.g., halos, relics) probed by the ongoing and upcoming high-resolution, low-frequency radio observations (e.g., LOFAR, SKA). HUBS, with its capability for high-resolution X-ray spectroscopy, will play a significant complementary role in mapping the ICM.

**Special Requirements:** Low and well-understood background. $< 10\%$ systematic uncertainty of the NXB is necessary for robust detections of the ICM signals near $R_{200}$.

*28. The dynamical state of intermediate to high redshift clusters*

**Science Area:** phenomenologies in merging clusters

**First Author:** Renato Dupke (National Observatory, University of Michigan)

**Co-authors:** Vinicius Bessa (National Observatory), Yolanda Jimenez Teja (Instituto de Astrofisica de Andalucia), Rebeca Batalha (CEA-Saclay), Weiguang Cui (University Autonoma de Madrid)

**Abstract:** By the end of the decade, next-generation observatories will push the astrophysics and cosmology studies with galaxy clusters into the systematics-limited regime, allowing the detection of an unprecedented number of high-z clusters and groups within a wide mass range for which weak lensing data and accurate membership will be available, allowing us to probe in detail the evolution of clusters and groups of galaxies through earlier epochs as well as refine their use for cosmological goals. In particular, out-of-equilibrium phenomena in the intracluster medium (ICM) not only present a challenge on their own but also introduce significant departures from the assumptions of hydrostatic equilibrium and sphericity, biasing cluster mass estimates and increasing the uncertainties in cosmological constraints. The specifications of AXIS will enable the detailed characterization of global and spatially resolved ICM morphological and thermodynamic properties for an unprecedentedly large sample of intermediate- and high-z clusters, which is crucial to understanding the physical nature of the local phenomenology and improving the sample purity of relaxed systems suitable for cosmology studies.

**Science:** Understanding the different phenomena in dynamically active clusters is very important to evaluate the systematics in measuring the cluster mass & baryon fraction through X-rays for high-z clusters and to improve the purity of samples raised for cosmology. X-ray observed/derived configuration of the distribution of thermodynamic and chemical quantities/parameters provides invaluable information to determine the evolutionary history of clusters and of larger scale structures such as cosmic filaments near the massive cluster-nodes. There are many proxies, with different levels of accuracy, used to estimate the degree of dynamical activity in clusters, including looking for substructures in phase space [126,136], geometric indicators based on surface brightness asymmetries [207,387,590,593], the presence of radio halos [84,161], and, more recently, the distribution of the intracluster light fraction (ICL fraction) across different optical bands [243]. The latter has a natural mutual synergy with the high-resolution spatially resolved X-ray analysis from AXIS. The ICL fraction based solely on optical data can find the markers of cluster merging (Fig. 40-Right), but the best characterization of mergers comes from X-rays, and the complementary analysis between the two techniques will help to understand how the merging markers of the ICL fraction are created. On the other hand, the ICL also traces the merger as a collisionless but highly structured medium, providing additional important information about the merger's configuration history [e.g. 244].

All in all, these methods are calibrated or backed up by the analysis of spatially resolved X-ray spectra at lower redshifts, which provide the quintessential diagnostic for a cluster's activity level, and then applied to higher-z clusters, where detailed X-ray observations are very expensive with current instruments. Naturally, several limitations are associated with this approach, including the adoption of universal thresholds to distinguish relaxed from perturbed clusters in parameter spaces for higher-z systems, where many astrophysical processes are not yet well-constrained. Furthermore, there are uncertainties when considering projection effects, which can hide substructures and other signatures of mergers, while the low S/N regime in high-z systems limits the thermodynamic characterization of the ICM through spectral modeling.

The science capabilities of AXIS, accompanied by excellent effective area and near homogeneous point spread function (**PSF**), will allow for the detection of sharp surface brightness features in the plasma

distribution, and enough statistics for spatially resolved spectroscopic measurements of thermodynamic and chemical parameters in the ICM for galaxy clusters across a wide range of redshifts.

**Exposure time (ks):** 200 ks

**Observing description**: Abell 2744 (the Pandora cluster) is a prime target, an extremely active node of the cosmic web ([135]) passing through a very violent multiple cluster merger, with many sub-clumps (>4) detected through lensing and X-ray analyses ([3,237,328,332,379]) with various phenomena that are rarely seen together in other clusters, including a gas-poor Dark Matter (**DM**) cluster-size clump (called dark core) and a (DM+galaxy)-poor "ghost" subsystem (called "interloper"), thrown ahead possibly by a gravitational slingshot effect and maybe associated to a shock front (**SF**) and radio relic [389]. Interestingly, the "thrown out" gas (with no DM or BCGs) shows indications of a cold front (**CF**) [e.g. 89,237,253], which creates several open questions regarding the stability of CFs in the absence of magnetic fields far away from clusters cores.

AXIS will be uniquely able to address this plethora of questions and also tie them to the large-scale structures at the cluster's outskirts. Furthermore, the characterization of these features at intermediate redshifts will help us to identify them in high redshift clusters such as in SPT-CLJ0615-5746 ($z = 0.972$), originally classified as relaxed by the Symmetry-Peakiness-Alignment criteria [302,304], having a relatively regular morphology as seen from a (240ks Chandra observation), but later found to have strong indications of being in a strong merger from a re-analysis of the Chandra data [242].

Currently, more than 2 Ms of observations with Chandra's ACIS-I have confirmed and refined previous analyses, finding three CFs and two SFs in the system so far [89]. With effective areas of 4200 $cm^2$ (at 1 keV) and 830 $cm^2$ (at 6 keV) AXIS would be easily double the statistics necessary for that analysis with a $\sim$ 200ks observation (using the effective areas of Chandra at cycle 27). More importantly, AXIS PSF stability across the detector, from 1.25″ at the optical axis to an average of 1.5″ over the field of view (**FoV**), would enhance the scientific results extraordinarily, when compared to Chandra's almost an order of magnitude PSF increase over the extension of the cluster, which covers roughly one ACIS-I CCD. The ability to resolve the different regions in the cluster's outskirts so finely will allow the analysis of thermodynamic parameters across surface brightness discontinuities throughout different regions of the cluster with unprecedented accuracy. This is crucial to disentangle the physics of this most complex cluster merger. Furthermore, [135] was able to detect several putative inflowing (cold) material through cosmic filaments in the very outskirts of the cluster with a (effective) 87-97ks XMM observation. The larger FoV of AXIS will allow us to zero in on the properties of the incoming material to the cluster, without having to deal with the complex methods to remove the known causes of intrinsic uncertainties of X-ray background and off-axis XMM PSF. With more than twice the effective area of XMM, low background, and stable PSF, AXIS will be able to characterize the filaments associated with cluster galaxies, determining the thermodynamic parameters, mass, baryon fraction, and metal abundances with a precision by a factor of >2, from 5% to 25%.

Chemical enrichment is a powerful tracer of dynamical activity in the outskirts of galaxy clusters. The chemical composition of the ICM encodes essential information about the assembly and enrichment history of clusters. In particular, abundance ratios provide insights into the relative contributions of core-collapse and Type Ia supernovae, while also tracing the mixing and transport processes within the hot plasma [31,186,215,330,470]. However, no existing stellar nucleosynthesis model successfully reproduces all observed abundance ratios. This discrepancy may be due to systematic uncertainties in spectral modeling, which depend on different atomic databases and plasma codes [e.g. 331], as well as spatially varying enrichment processes across different cluster regions. Addressing these open questions requires both spatially resolved spectroscopy and high sensitivity, key capabilities uniquely provided by AXIS.

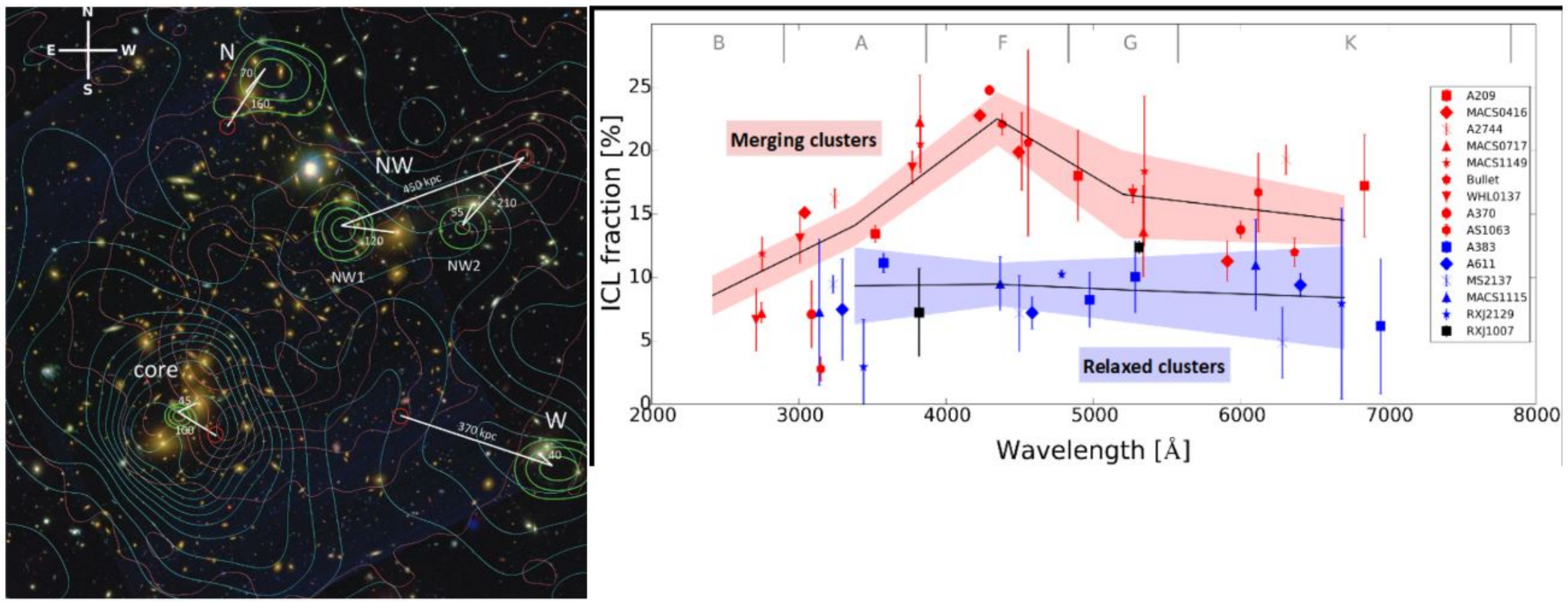

**Figure 44. Left:** The field of Abell 2744, with different phenomena. The false-color background is provided by HST/ACS (the two pointings can be identified by the higher, blue background noise level), VLT, and Subaru images on a field size of 240" × 240" ($\sim$ 1.1 Mpc on a side). Overlaid in cyan are the surface mass density contours, most concentrated in the 'core' area, and in magenta, the more evenly distributed X-ray luminosity contours. The peak positions of the core, N, NW, and W clumps are indicated by the green likelihood contours. Contours are 86%, 61%, and 37% of the peak likelihood for each clump. The small red circles indicate the positions of local overdensities in the gas distribution, corresponding to each individual dark matter clump. The white rulers show the separation between dark matter peaks and the bright clusters of galaxies and local gas peaks [332]. **Right:** ICL fraction rest-frame color distribution for merging (red) and relaxed (blue) clusters studied in [243]. Black lines indicate the error-weighted mean of each main-sequence spectral-type subsample, and colored shadowed areas indicate the mean of the errors. Gray vertical lines at the top of the figure split the wavelength range expected for the peak emission for an average main-sequence star, labeled with gray letters. Although the distribution for relaxed clusters is mostly flat, that of merging clusters shows an excess in the region corresponding to the emission peaks of late A- and early F-type stars. We also present the case for the extremely relaxed fossil group RXJ1007, represented by black squares, which is consistent with an extremely relaxed system ([130]).

The clusters' outskirts are especially valuable in this context, as they remain sensitive to recent accretion and the infall of preprocessed material from surrounding filaments and galaxy groups. This material may already carry chemical signatures from prior enrichment episodes, enabling us to test whether metals were injected into the ICM in situ or were accreted after prior processing. AXIS will enable the detection and characterization of fine-scale abundance structures in clusters at intermediate to high redshifts. Its combination of high spatial resolution and large effective area will allow precise measurements of elemental abundances across the ICM, including in low surface brightness regions in clusters' outskirts, where inflowing material from filaments can be detected.

**Joint Observations and synergies with other observatories in the 2030s:** The LSST will create the largest uniform sample of clusters, using high-precision lensing measurements to constrain their mass profiles and distribution extremely well.

JWST deep cluster surveys, such as UNCOVER, MegaScience [485], VENUS [170], with full filter coverage, will provide detailed observations of the ICL, which will be a key ingredient, together with high-resolution X-ray observations, to disentangle the cluster merger history.

J-PAS IFU coverage of clusters and groups of galaxies up to z$\sim$1 will provide photospectra for over 400 M galaxies in the northern hemisphere, with a precision of 0.3% photo-z for about a quarter of them,

allowing the determination of galaxy membership and also their SF properties among other spectral characteristics[32][52].

The LSST will create the largest uniform sample of clusters, using high-precision lensing measurements to constrain their mass profiles and distribution extremely well.

**Special Requirements:]** None

## f. Galaxy and Cluster Formation

### *29. Spatially resolved X-ray study of strongly lensed HyLIRGs*

**Science Area:** high-z galaxies, starburst, AGN feedback, high-mass X-ray binary population

**First Author:** Q. Daniel Wang (University of Massachusetts, Amherst)

**Co-authors:** Carlos Garcia Diaz and Min S. Yun (University of Massachusetts, Amherst), Nicholas Foo and Patrick S. Kamieneski (Arizona State University), Kevin C. Harrington (National Astronomical Observatory of Japan), Eric F. Jimenez-Andrade (Universidad Nacional Autónoma de México), Daizhong Liu (Purple Mountain Observatory), Belén Alcalde Pampliega (ESO), Brenda L. Frye (University of Arizona), James D. Lowenthal (Smith College), Massimo Pascale (University of California, Berkeley), Amit Vishwas (Cornell University), and Dazhi Zhou (University of British Columbia)

**Abstract:** Hyperluminous infrared galaxies (HyLIRGs) are the most extreme dusty star-forming galaxies (DSFGs) observed in the early Universe and remain poorly understood. AXIS observations of strongly lensed HyLIRGs at cosmic noon will enable us to resolve their Einstein rings and separate the AGN and non-AGN contributions with minimum dust-obscuration effect. Such a measurement will allow us to determine the population of high-mass X-ray binaries (HMXBs) and to quantify the effect of AGN on the IR emission and the coevolution of AGN and extreme star formation. Existing deep XMM-Newton observations indicate the presence of AGN and an enhanced specific HMXB population (relative to the star formation rate) in HyLIRGs. This enhancement may be due to the effective dynamical production of HMXBs in compact stellar clusters, which tend to be formed in extreme star formation conditions. Therefore, the AXIS observations will provide deep insights into high-energy astrophysical processes in extreme star-forming galaxies in the high-redshift Universe.

**Science:** High-energy astrophysical phenomena are believed to play a central role in regulating galaxy formation and evolution. Such phenomena are mostly associated with star formation (SF) and massive

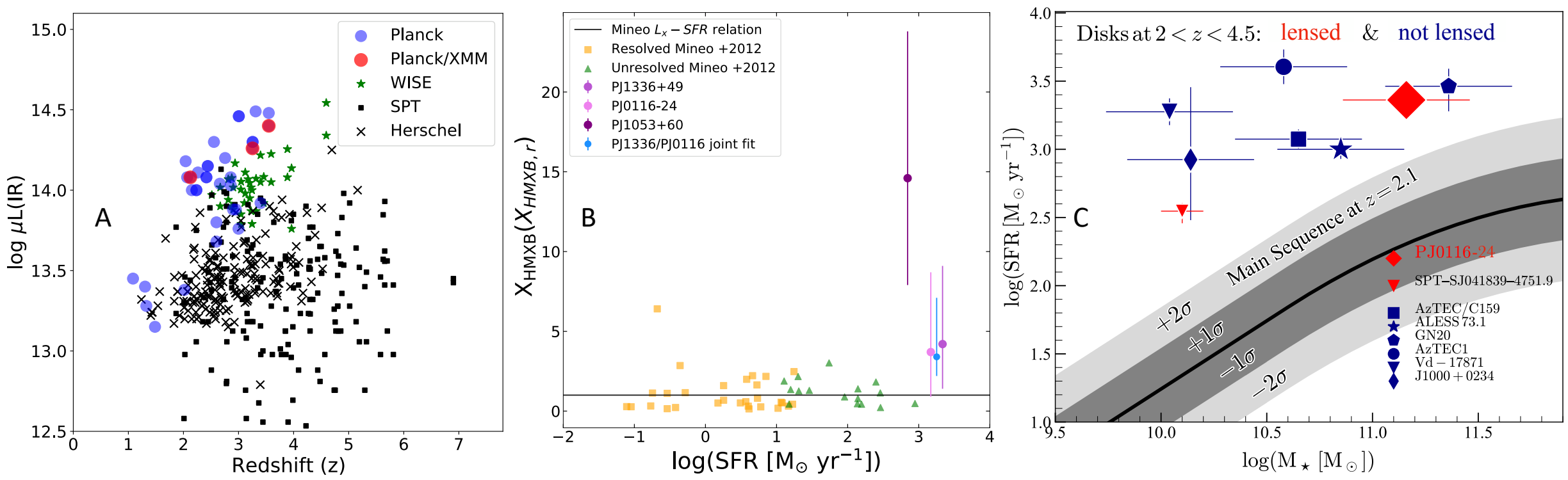

**Figure 45.** *Illustration of existing studies of HyLIRGs [566]:* **(A)** *Apparent IR (8 − 1000 μm) luminosity vs. redshift plot of DSFGs, showing the PASSAGES sources, along with other sources as marked.* **(B)** *The XMM-Newton* $X_{HMXB} = L_x/\mathrm{SFR}$ *measurements for the three HyLIRGs, compared with the data for individual galaxies included in the study by [336], normalized to their best-fit* $L_X$*−SFR relation for those local galaxies with resolved X-ray emission.* **(C)** *Comparison of one of the XMM-Newton targets (PJ0116-24) with a population of high-z DSFGs known to have gaseous disks, but with SFRs significantly above the galaxy main sequence at the redshift range indicated. The AXIS will allow for a unique spatially resolved X-ray study of HyLIRGs.*

black hole (MBH) accretion, which are largely coordinated processes, especially during the early rapid coevolution stage. However, it remains unclear how this regulation works and whether key scaling relations derived locally are applicable to high-z extreme star-forming galaxies evolving under very different conditions. This is largely due to the limited sensitivity and resolving power of existing X-ray observations, a primary tool for studying high-energy phenomena.

**HyLIRGs as laboratories for high-energy astrophysics**

As the most extreme version of dusty star-forming galaxies (DSFGs), HyLIRGs are identified by their large rest frame infrared (IR) luminosity $\gtrsim 10^{13}$ L$_\odot$, corresponding to a star formation rate (SFR) $\gtrsim 10^3$ M$_\odot$ yr$^{-1}$, and observed primarily at $z > 1$ [e.g., Fig. 45A; 566]. There are no analogs to their extreme luminosity, large inferred gas mass (up to $10^{11} M_\odot$), and a gas mass fraction (40-80%) in the local Universe. Therefore, a detailed multiwavelength investigation of HyLIRGs is the only way to probe the underlying physical processes. However, this has been challenging because HyLIRGs are dusty and distant, generally faint in bands other than the far-infrared, and rarely well-resolved.

**X-ray HyLIRGs through strong gravitational lenses**

Strong lensing allows us to break observational limitations by magnifying HyLIRGs by a factor of $\mu > 2$, making such observations possible [e.g., 251, and references therein]. The effectiveness of this approach has recently been demonstrated in a unique large XMM-Newton observing program [566], in which three strongly lensed HyLIRGs were observed with a total exposure of 530 ks (Fig. 45). They were selected from the Planck All-Sky Survey to Analyze Gravitationally Lensed Extreme Starbursts (PASSAGES), which provides a sample of 30 gravitationally lensed DSFGs, representing the brightest IR galaxies observed in the Universe [e.g., 251, and references therein]. Extensive multiwavelength observations have been taken for this sample of DSFGs (e.g., from JVLA, ALMA, SMA, LMT, JWST, HST, and/or Gemini). Detailed lens modeling further shows that most of these DSFGs are magnified by $\mu = 2 - 28$. They are mostly intrinsically luminous [$L_{IR} = (0.2 - 6) \times 10^{13}$ L$_\odot$], rivaling the luminosities of even the brightest unlensed objects known at their respective redshifts. The X-ray luminosity of PJ1336 + 49 (z = 3.254) appears significantly higher than the non-AGN luminosity expected from the local $X_{HMXB}$ [336,566]. However, its X-ray spectrum is consistent with those of high-mass X-ray binaries, especially the so-called ultraluminous X-ray sources. This, together with the lack of a point-like radio or near-mid-IR component, indicates the non-AGN origin of the X-ray emission. The well-detected radio continuum forms an Einstein ring of radius $\sim 1.2''$ [251,566], which is unresolved at the XMM-Newton resolution and can hardly be separated from the lensing galaxy in both the IR and X-ray bands (even with AXIS). In contrast, the other two (PJ1053+60 and PJ0116-24) are marginally resolved. However, the X-ray emission of PJ1053+60 (z=3.549) is strongly contaminated by a foreground AGN identified later, which is a few arcseconds off the lensed image. Therefore, we propose to observe PJ1053+60 and PJ0116-24 (z=2.12) with AXIS, which will allow us to fully resolve their X-ray emission.

AXIS observations of these two strongly lensed HyLIRGs will allow us to resolve their X-ray emission far beyond the reach of unlensed high-z galaxies. This unique X-ray observing capability, in synergy with the largely available high-resolution multiwavelength data, will enable us to probe the role of high-energy activities in regulating these extreme ecosystems, providing key insights into the evolution of similar galaxies in general and testing the dynamical production scenario of HMXBs. As such, this study will break new scientific ground in high-energy extragalactic astronomy by observing X-ray emission from high-z DSFGs in a spatially resolved manner.

Specifically, the strong lensing magnification will allow us to 1) conduct a spatially resolved spectral study of the X-ray emission of the two HyLIRGs; 2) examine the co-evolution of extreme star formation and AGN and the latter's contribution to the extreme IR luminosities of the galaxies; and 3) determine the overall population of HMXBs, constraining their formation and evolution in extremely star-forming

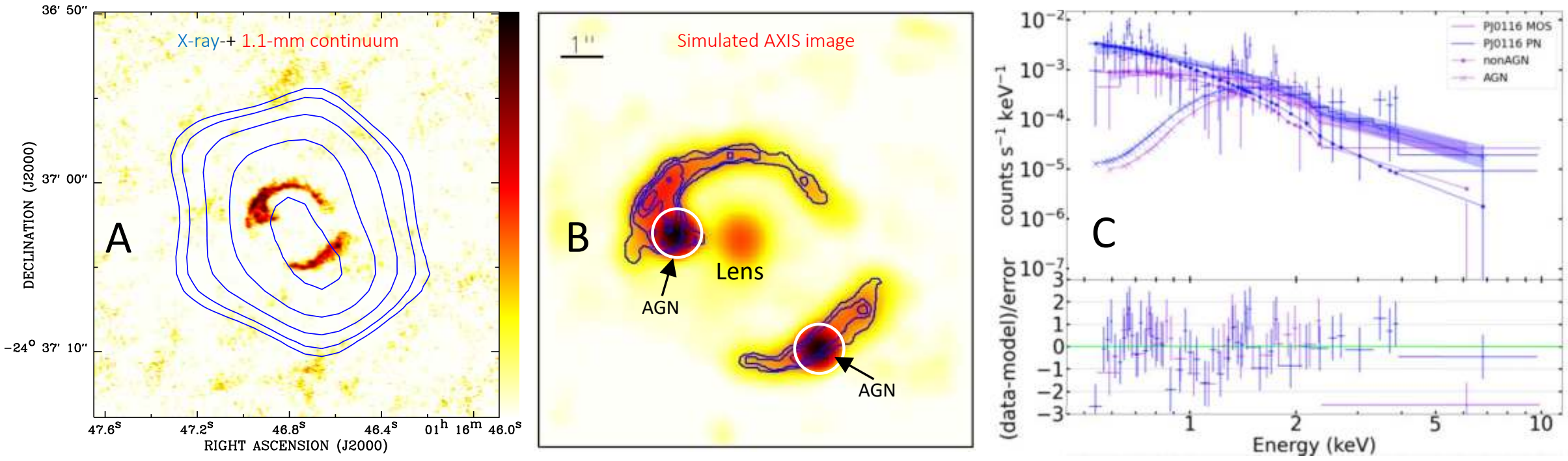

**Figure 46.** *X-ray properties of our targets PJ0116-24:* **(A)** *XMM-Newton 0.3-7 keV intensity contours (resolution $\sim 6.6''$ FWHM) overlaid on the ALMA 1.1-mm continuum intensity image (resolution $\sim 0.2''$);* **(B)** *The continuum intensity contours overlaid on a mock 0.5-7 keV map, which can potentially be obtained with an AXIS observation, together with the circles around the AGN images, each of which contains the 60% PSF energy;* **(C)** *XMM-Newton spectra of the HyLIRG fitted with a double power law with the two components also separately plotted (see [566] for details). The AXIS observation will spatially test this decomposition in a spatially resolved fashion.*

environments, which will have strong implications for understanding the formation process of gravitational wave sources, the ionization and heating of the interstellar medium, the role of HMXBs in heating the intergalactic medium, and using the X-ray luminosity of a DSFG as an SFR tracer.

In short, the AXIS observations will allow us to explore new frontiers in extragalactic high-energy astrophysics by spatially resolving X-ray emission from high-z galaxies, taking advantage of the mission's combined capability of superb sensitivity and spatial resolution, the magnifying power of strong gravitational lensing, and the exceptional intrinsic IR brightness of our target HyLIRGs.

**Exposure time (ks):** As a pilot project, we propose 100 ks each for two selected HyLIRGs, PJ0116-24 and PJ1053+60. The two targets have comparable X-ray fluxes, excluding the contribution from the foreground AGN projected in the vicinity of the latter.

**Observing description:** We study the feasibility based on various mock data and request an exposure of 100 ks for each target to obtain a total of $\sim 10^3$ net counts (in the 0.5-8 keV or rest-frame 1.6-25 keV band, where dust-obscuration or X-ray absorption is minimal), estimated from the existing XMM-Newton spectrum of PJ0116-24, as an example (Fig. 46C; [566]). About 55% of the counts are expected to be from the AGN, with the remainder from the non-AGN component of this source. Fig. 46B shows a mock X-ray intensity image that includes contributions from both the foreground lensing elliptical galaxy at $z_{GL} = 0.555$ and the sky background, as well as the non-AGN component from the lensed PJ0116-24. The latter component is mocked from the existing *ALMA* continuum image of the galaxy (Fig. 46A), assuming that the dust emission intensity follows the surface SFR and the currently best estimated $X_{HMXB}$ [566]. The lensing galaxy should produce $\sim 10\%$ counts, estimated from a real Chandra observation of a representative elliptical galaxy at z=0.0156 (NGC 1600) and with a mass consistent with the lens mass, taking into account the difference in redshift. The lensing galaxy contribution is concentrated in the central $1''$ radius and is also much softer than the expected non-AGN (or HMXB) emission. The radial distribution of the contribution can be reasonably modeled, either empirically or using the standard $\beta$ model. Therefore, the small lensing galaxy contamination of the lensed DSFG emission (covering an effective area of only $\sim 10$ arcsec$^2$ $\sim 3''$ from the lensing galaxy center) can be well estimated (to an accuracy of $\sim 10\%$) and thus accounted for in the data analysis. A similar accuracy can also be obtained to remove the AGN contamination, as shown in Fig. 46B. The mock spectral analysis also shows that the spectral shape of the

non-AGN component can be reasonably well characterized (e.g., to an accuracy better than $\sim 20\%$ for the power-law index at the 90% confidence level.

Ideally, one would like to expand the project to obtain a statistically significant sample of HyLIRGs. They may be selected from the remaining PASSAGES DSFGs, although without existing X-ray measurements. This would enable us to determine how consistent the $X_{HMXB}$ factor may be or its dispersion among HyLIRGs. We could then further address such questions as what the AGN occupation probability is and how the HMXB population may change with extreme galaxy properties (e.g., specific star formation rate) in such galaxies, providing badly needed constraints on the role of high-energy astrophysical phenomena in regulating galaxy formation and evolution at cosmic noon.

**Joint Observations and Synergies with other observatories in the 2030s:** New observatories, such as Euclid, Roman, LSST, ng-VLA, and SKA, as well as JWST, will dramatically improve our ability to study strongly lensed, distant DSFGs. These improvements will be achieved through an unprecedented scale and depth of surveys, as well as multiwavelength capabilities. The expected number of lensed DSFGs will be $> 10^2\times$ higher than the current known number. JWST observations will enable us to discover faint, dusty lensed galaxies and uncover deeply embedded AGNs through deep-infrared imaging and color selection. Near- and mid-infrared integral field unit observations will allow us to map the dust attenuation and the distribution of intrinsic stellar mass. Multiwavelength observations will also greatly improve measurements of foreground lensing galaxies, lens modeling, magnification determination, and source-plane imaging of lensed galaxies. Radio observations with high-resolution imaging and spectroscopy capabilities will enable us to penetrate dust, resolve star formation distribution, and reveal gas dynamics in lensed systems. These capabilities will complement AXIS X-ray observations, helping us to better understand the interplay among the different galactic components in the ecosystems of lensed DSFGs.

**Special Requirements:** None

*30. Ionizing Radiation from Extremely Metal-Poor and Early Universe Galaxies*

**Science Area:** low-metallicity galaxies, stellar evolution, early universe epoch of heating

**First Author:** Bret Lehmer, University of Arkansas

**Co-authors:** Mihoko Yukita (Johns Hopkins), Antara Basu-Zych (UMD Baltimore County)

**Abstract:** Recent investigations have shown that low-metallicity star-forming galaxies (SFGs) show signatures of hard ionizing spectra (based primarily on the presence of nebular He II 4686), with a rising fraction of SFGs showing these signatures with decreasing metallicity and increasing redshift. Modern stellar population synthesis models struggle to reproduce these emission lines, leading to the imperative questions: (1) what are the important sources of ionization in low-metallicity/high-redshift galaxies?; (2) what are their detailed intrinsic ionizing spectra?; and (3) what is their impact on the epochs of heating and reionization in the early Universe? Among the leading candidates of the hard ionizing spectra are X-ray binaries and hot gas. Due to the faintness of high-redshift galaxy targets, along with the fact that redshifting of low-energy photons limits constraints on key contributions from the ionizing spectrum, the best sources to study in X-rays are nearby extremely metal-poor galaxies (XMPs) that share similar properties to high-redshift galaxies. The high-spatial resolution and low-energy response of *AXIS* make for the most ideal combination of properties for studying XMPs in detail and directly constraining the emission from the most promising candidates for the hard ionizing radiation in low-metallicity galaxies throughout the Universe.

**Science:** Observations of star-forming galaxies (SFGs) have shown that the X-ray luminosity, $L_X$, not only scales with the total star-formation rate [SFR; e.g., 336,337,406], but that $L_X$/SFR significantly increases towards lower metallicity [e.g., 68,123,270]. Such a metallicity dependence is expected to arise primarily from binary evolutionary effects, in which stellar-wind efficiencies and angular-momentum losses are higher at high metallicities, causing binary systems to lose more mass and widen their orbits more. Low-metallicity binary populations therefore have more tightly bound orbits and suffer less wind mass loss over their lifetimes, producing heavier black holes and yielding more luminous X-ray sources [e.g., 168]. An additional metallicity dependence is expected from hot gas emission associated with stellar feedback processes, as the ionizing photon flux of a collisional plasma is higher at low metallicity for a given temperature [e.g., 373] and the emergent X-ray flux is higher for a given hydrogen column density due to the lack of metal absorption features [e.g., 173].

The metallicity dependence of the X-ray emission within SFGs may have critical implications for key astrophysical problems. For instance, high-redshift galaxies ($z \gtrsim 5$) and nearby extremely metal poor galaxies (XMPs; $<0.2\ Z_\odot$) often show signatures of hard ionization from extreme emission lines (e.g., C IV, He II, O III, Ne V, and C III) that require intrinsic fluxes of photons with $E \gtrsim 50$ eV well in excess of those produced by stellar evolution models [e.g., `Starburst99`, `BPASS`, and `BEAGLE`; 370,464,476,527]. Additionally, cosmological simulations that track the spin temperature evolution of the very early universe ($z \gtrsim 8$) have implicated X-ray emission as potentially the main source of heating the neutral intergalactic medium, prior to reionization [see, e.g., 211,381,388] and early results from HERA already require an elevated X-ray/SFR ratio at $z \approx 8$ [213].

The X-ray properties of high-redshift galaxies, which are dominated by X-ray binaries (XRBs), hot gas, and sometimes active galactic nuclei (AGN), unfortunately remain inaccessible: the X-ray fluxes of even the most star-forming active galaxies at $z \approx 3$ are typically $\lesssim$few $\times 10^{-18}$ ergs cm$^{-2}$ s$^{-1}$, well below the detection limits of the deepest *Chandra* surveys and the detectable rest-frame energies from such sources ($\gtrsim$1 keV) is too high to probe low-temperature hot gas and absorption associated with the key emergent ionizing flux. With *AXIS* it will be possible to both (1) constrain the detailed intrinsic and emergent spectral

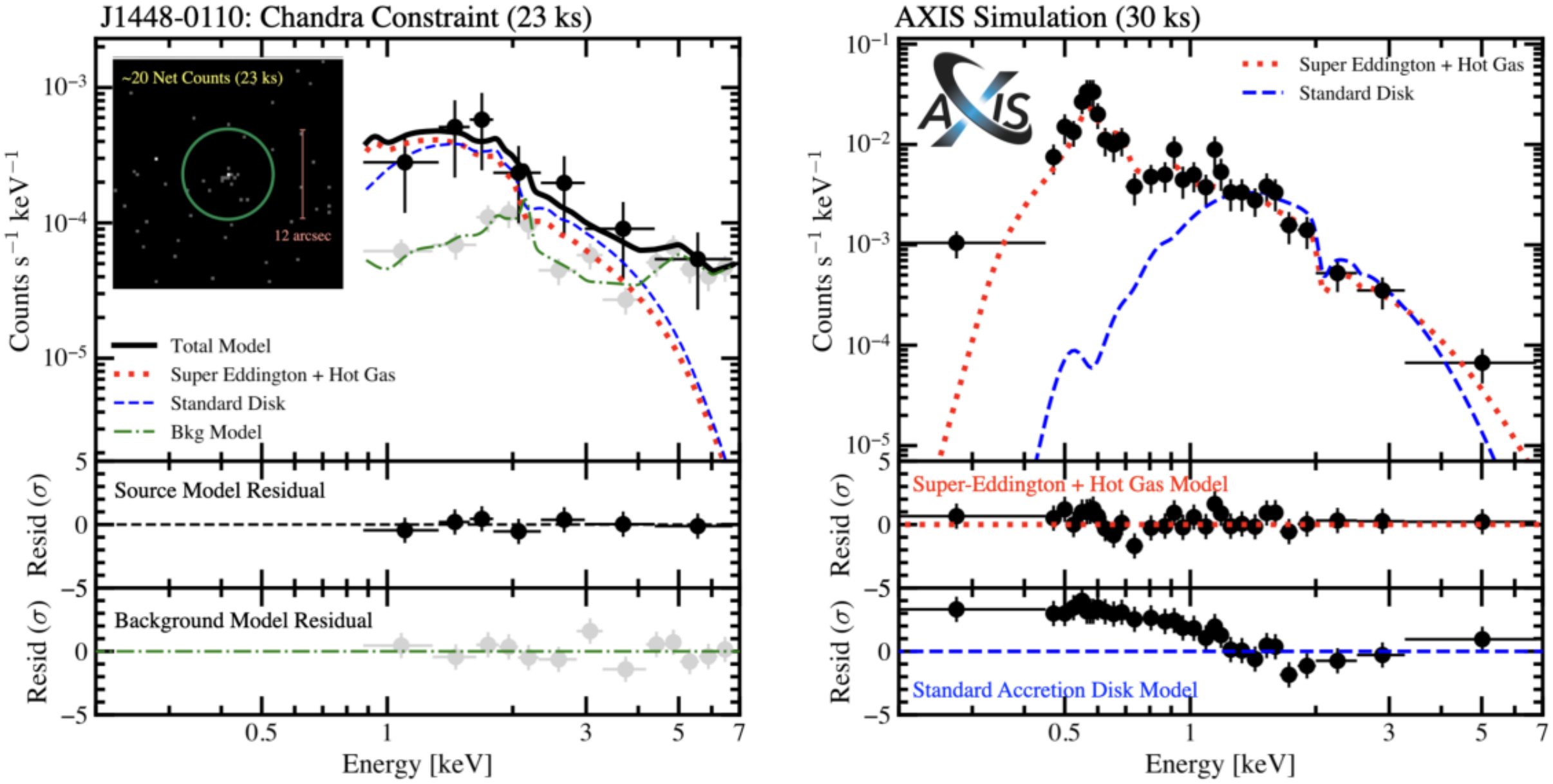

**Figure 47.** *Due to their relative rarity, extremely metal-poor galaxies (XMPs) are distant and faint. Current observations of XMPs with Chandra (left) provide only weak spectral constraints with degenerate solutions. The data can be modeled equivalently as either a standard accretion disk model (blue dashed) or a more physically motivated advection-dominated accretion disk (super Eddington) with hot gas (red dotted). Due to the much softer response and higher effective area, a nearly equivalent AXIS exposure (right) would easily break this degeneracy, and clearly constrain the hot gas component and absorption.* **It is clear that the range of viable models shown here produce a vast range of emergent and intrinsic ionizing photons in the extreme metal poor regime. *AXIS* constraints are needed to establish the relevance of these photons for producing extreme emission lines observed in low-metallicity and high-redshift galaxies and X-ray heating of the high-redshift IGM.**

contributions from XRBs and hot gas, including the critical low-energy X-ray regime, for relatively-rare, yet nearby ($D \approx$ 100-300 Mpc) XMPs and (2) detect direct X-ray emission from populations of galaxies at $z \approx$ 3–10. The combination of these two capabilities will provide powerful insights into the ionization problem and nature of the role of X-ray emission in heating the early Universe. Here, we highlight how *AXIS* will provide significantly improved constraints on the nature of extremely metal-poor galaxies.

Samples of XMPs can be selected by identifying local SFGs that have key properties comparable to those of high-redshift galaxies. Starting with ∼25,000 SDSS DR16 compact SFGs from Izotov et al. [234], we can down-select for XMPs that are non-AGN with properties similar to $z \approx$ 3–10 main-sequence galaxies studied by *JWST* [e.g., 114]: SFR $> 1\ M_\odot\ \mathrm{yr}^{-1}$ and metallicities between $Z = 0.05$–$0.3\ Z_\odot$ (i.e., 12+log(O/H) = 7.4–8.2) IR WISE colors $W_1 - W_2 < 0.8$, $m_{\rm FUV} < 20$. Such a downselect reveals several 10s of objects at $D \lesssim 200$ Mpc that represent the best targets for studying with X-ray observatories. These XMPs are observed to have a broad range of nebular emission features, including a large fraction of sources with enhanced He II $\lambda$4686 over that predicted by star formation alone. A subsample of these objects has been targeted with moderately deep (20–50 ks) *Chandra* observations to obtain detections of the galaxies.

In Figure 47, we show the case of J1448−0110, which has a 23 ks depth archival *Chandra* observation (*left panel*). In total ∼20 net counts are detected, which provides only a very weak spectral constraint, primarily at energies $\gtrsim$1 keV. J1448−0110 provides a median example of distant XMP constraints currently available in X-rays. The spectrum can be ambiguously modeled with very basic models (e.g., `POWERLAW`), or somewhat more physically-motivated absorbed accretion disk models (*blue-dashed curves*), or models that include components that have been observed in other galaxies along with super-Eddington accretion

models that are more appropriate for ULXs that are present in these XMPs (*red-dotted curves*). The ambiguity in spectral modeling translates to large uncertainties in both the emergent low-energy X-ray flux ($E < 1$ keV), which is important for providing ionizing radiation to the IGM, and also in the intrinsic ionizing radiation required to power extreme emission lines.

To illustrate potential improvements with *AXIS*, we performed simulations of the two model scenarios that are ambiguous in *Chandra* fits: i.e., the absorbed standard accretion disk (e.g., `TBABS*DISKBB`; *blue-dashed curves*) and hot gas plus super-Eddington (e.g., $\texttt{TBABS}_{\rm gas}$`*APEC + `$\texttt{TBABS}_{\rm ULX}$`*DISKPBB`) scenario. In our *AXIS* simulations, we adopted the latter model as being more realistic, as it captures components that we know exist in nearby galaxies. The right panel of Figure 47 shows that an *AXIS* exposure of 30 ks (comparable to the existing *Chandra* exposure) would detect ∼300 counts, and easily distinguish between these scenarios (see residuals in bottom panels).

**Exposure time (ks):** 500 ks

**Observing description:** The above demonstrates the power of *AXIS* for constraining the nature of the hard ionizing spectra observed in XMPs and high-redshift galaxies. With the relatively short *AXIS* exposures required, it would be possible to survey a diversity of XMPs. A program to observe with *AXIS*, at ≈20–30 ks depth, ≈20 XMPs (total of ∼500 ks) that span a range of these properties would provide transformative understanding of ionizing radiation in a diversity of extremely low metallicity galaxies throughout the Universe.

Such a program would allow for the detection and characterization of scaling relations between galaxy properties (e.g., star-formation rate) and hot gas and XRB emission in the extremely low-metallicity regime. The program would provide unique insights into how extremely low-metallicity stars explode and provide feedback to their ISMs (as traced by the hot gas properties), the fraction of low-energy photons that escape from their host galaxies in this regime, and the potential impact of the escaping X-ray photons on heating the early IGM. Furthermore, this program would allow for direct testing of how intrinsic ionizing emission from the X-ray emitting components impacts the state of the ISMs of XMPs by determining how X-ray spectral properties vary with optical/UV nebular emission-line probes of spectral hardness (e.g., the intensity of C IV $\lambda$1550, He II $\lambda$1640, O III] $\lambda$1666, C III] $\lambda$1908, and He II $\lambda$4686).

**Joint Observations and synergies with other observatories in the 2030s:** The above program would have various synergies with current or future observatories that may be operating alongside *AXIS*. High-redshift H I 21 cm emission probes (e.g., HERA) will provide insights into the evolution of the IGM temperature at $z \gtrsim 8$, which is expected to be impacted heavily by the escaping X-ray radiation from extremely metal-poor galaxies. In conjunction with *AXIS* observations, the detailed UV morphologies and spectral properties of the XMP sample will be probed directly by the forthcoming *UVEX* mission [e.g., 265]. These observations will enable highly sensitive and direct probes of the detailed intrinsic hardness of the ionizing spectrum, based on the detailed emission-line portfolio. Finally, the connection between nearby XMPs and high-redshift galaxies can be made through direct detections of high-redshift galaxy population X-ray emission from *AXIS* deep surveys and probes of high-redshift galaxy UV spectra from *JWST*.

**Special Requirements:** None

*31. Evolution with redshift of low-mass systems scaling relations*

**Science Area:** Galaxy groups, intra-group medium

**First Author:** Lorenzo Lovisari (INAF-IASF Milano)

**Co-authors:** Fabio Gastaldello (INAF-IASF Milano); Massimo Gaspari (U. of Modena & Reggio Emilia); Dominique Eckert (University of Geneve)

**Abstract:** Galaxy groups are remarkable astrophysical laboratories that provide unique insights into the physics of galaxy evolution and structure formation, the role of feedback processes, the history of metal enrichment, the nature of dark matter, and the physics of hot, diffuse, magnetized plasmas. It is well established that nearby groups (and clusters) do not appear as expected in simple gravity-only models (e.g., their gas entropy shows a significant excess to that achievable by pure gravitational collapse), and extra energy input is required to explain the observed properties. However, the timing and mechanism by which this energy is injected into the intragroup medium (IGrM) remain a matter of debate. For instance, hydro simulations suggest that stellar feedback from SNe and feedback from SMBH predict very similar radial and integrated properties at low redshifts but differ strongly at higher redshifts. Thus, analyzing high-redshift galaxy groups is crucial to disentangle these scenarios. Therefore, studying the evolution of the scaling properties for a large sample of systems at high redshift, where the impact of non-gravitational processes is expected to be more pronounced and distinguishable, can provide key insights into the formation and evolution of groups and clusters. Beyond $z > 0.5$, groups (and clusters) span from a few tens of arcseconds to a few arcminutes; i.e., they are extended but small and faint objects. The high spatial resolution and large effective area of AXIS are crucial not only for detecting and resolving these systems, but also for accurately measuring their temperature and luminosity profiles. These detailed profiles are essential for constraining scaling relations and for understanding the thermodynamic evolution of the intragroup medium at high redshift.

**Science:** The study of galaxy groups offers a unique window into the processes shaping the evolution of cosmic structures. Their properties reflect their dynamic history, including the complex interplay between gravitational collapse, cooling processes, feedback from active galactic nuclei (AGN), and the cosmic environment. While non-gravitational processes also affect massive clusters, their relative impact is significantly greater in low-mass systems, where shallower potential wells make the IGrM more susceptible

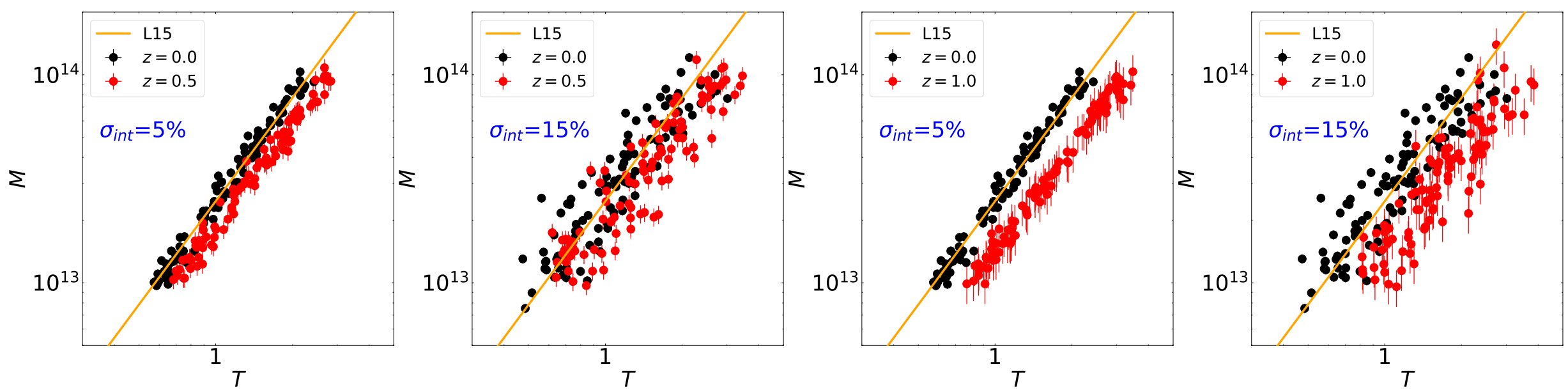

**Figure 48.** Visualization of the self-similar evolution of the M-T relation. In every panel, the black points represent a group sample at $z \sim 0$ while the red points are drawn from a population of systems at $z \sim 0.5$ (*left* and *center-left* panels) and $z \sim 1.0$ (*center-right* and *right* panels). The *left* and *center-right* panels refer to the case of a small intrinsic scatter (i.e., ∼5%) while the *center-left* and *right* panels refer to the case of a large intrinsic scatter (i.e., ∼15%). Every sample consists of 100 groups and is drawn from the M-T relation measured in [286].

to heating, cooling, and dynamical disturbances. As such, understanding the physical properties of groups provides key insights into the growth of large-scale structures and the interplay between visible and dark matter in galaxy formation and evolution. However, our understanding of radiative cooling, of the energy injected by supernovae SNe and AGNs into the IGrM, and of the stochastic, transient perturbations caused by mergers remains limited. This is due to a combination of observational challenges, including the low surface brightness and small angular size of groups, as well as the complex interplay of physical processes at these mass scales. As a result, their impact on the properties of scaling relations (i.e., shape, scatter, and evolution) remains poorly constrained, which limits the use of low-mass systems in cosmological studies.

The scaling relations between the integrated properties can be used to test for departures from the self-similar expectations (e.g., how properties scale with mass and redshift in a purely gravitational scenario) and to estimate their masses using easily measured X-ray properties (e.g., luminosity and temperature). The key predictions of the self-similar model are that the slopes of the observed scaling relations follow the self-similar predictions (i.e., objects of different sizes are the scaled versions of each other). The model also predicts that the X-ray scaling relations are redshift-dependent (e.g., [284], and references therein), reflecting the decrease in time of the mean density of the Universe. The evolution of the X-ray scaling relations is expected to be influenced by non-gravitational processes, whose relative importance increases with cosmic time as feedback mechanisms (e.g., from AGNs and SNe) inject energy into the intragroup medium and counteract radiative cooling, particularly in lower-mass systems where gravitational heating is less dominant. Unfortunately, due to their fainter and cooler nature, and despite groups being more common than clusters, it is more difficult to detect them over the background, especially at higher redshifts. As a consequence of the big challenges to detect large and representative samples of groups beyond the local Universe, the literature on this subject is very limited and the results are very uncertain. So far, the few studies (i.e., [12,239,380,457,529]) that focused on such a topic did not find convincing evidence for the evolution of the X-ray properties of galaxy groups beyond the self-similar expectations, possibly due to limited sample sizes, low signal-to-noise data, and the difficulty in resolving group-scale structures at high redshift. A characterization of such an evolution is one of the goals of the next generation of X-ray instruments, such as AXIS.

The observed cluster properties are stochastically related to the "true" properties by the intrinsic scatter and the measurement uncertainties (see, e.g., Fig. 6 in [284]). That means that the ability to constrain the evolution of the scaling relations strongly depends on the amplitudes of both intrinsic scatter and statistical errors (assuming that the selection function is known). While the nature sets the first (and is currently unknown, especially at the group scale where the measurement errors are large and the scatter cannot be properly constrained), the second depends on the quality of the data (which in turn depends on factors such as effective area of the detector, exposure time on targets, level and knowledge of the background, etc.). Key factors for constraining the evolution of the scaling relations are also the size of the sample and the average redshift of the systems. Qualitatively speaking, the higher the redshift of the systems in the sample, the easier it is to detect an evolution of the relations even in the presence of a higher intrinsic scatter. For a given scatter, the larger the statistical uncertainties, the larger the sample size needs to be to properly constrain the normalization (and so the evolution) of the scaling relations. A visualization of the M-T relation is shown in Figure 1 (similar plots can be easily obtained for any other relation). When the true scatter is small, one can disentangle evolution even at intermediate redshifts (e.g., at $z \sim$0.5) without stringent requirements on the uncertainty measurements. In contrast, if the scatter is large, you can either build a larger sample to obtain precise measurements or focus on high-$z$ systems.

**Exposure time (ks):** $\sim$ 2 Ms

**Observing description:** Given a combination of intrinsic scatter and statistical uncertainties, we estimated the number of systems required at different redshifts to constrain the evolution of the M–T relation and

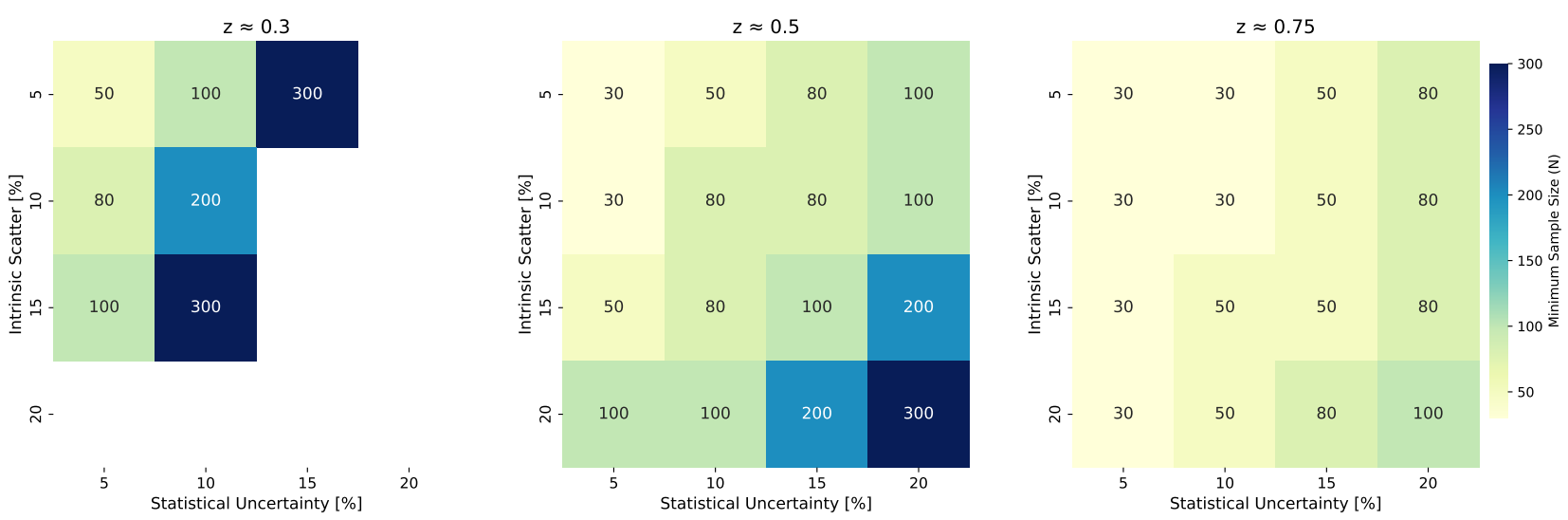

**Figure 49.** Minimum sample size required to detect evolution in the M-T normalization parameter $\alpha$ at $\geq 1\sigma$ significance, as a function of intrinsic scatter and statistical uncertainty, for redshifts $z \approx 0.3$, 0.5, and 0.75. Each panel displays the smallest sample size that achieves the required significance among the tested configurations. Due to the limited number of discrete sample sizes explored, the grids are not fully populated, and finer sampling in parameter space (e.g., intermediate sample sizes) would help refine the detection thresholds.

test its consistency (or lack thereof) with self-similar predictions. The simulations are performed as follows. We used the best-fit relations from [286] as a reference for the local galaxy groups. For each combination of intrinsic scatter (i.e., 5, 10, 15, 20%; while the scatter for groups is unknown the scatter of the M-T relation for galaxy clusters is usually estimated to be 10-15% – we adopted a wider range to account for the possibility that group scatter may be smaller, similar, or larger) and statistical uncertainties (i.e., 5, 10, 15, 20%) we simulated samples of systems with M=[1-10]$\times 10^{13} M_{\odot}$, different sizes (i.e., N varying from 30 to 300), and different average redshifts (i.e., 0.3, 0.5, 0.75) accounting for the redshift evolution. Then, we fitted the data points with LIRA [456] and estimate how well with each combination we were able to disentangle self-similar evolution and no evolution in the scaling relations. We discarded all the combinations that did not allow for disentangling the evolution with a reasonable number of systems. In practice, given that the evolution of the scaling relations manifests as a change in normalization, for each combination, we measured the normalization and its associated uncertainty, and retained only those cases where the difference between the measured and input normalizations exceeded $1\sigma$, ensuring that the evolution could be detected.

For each redshift and each mass M=[1-10]$\times 10^{13} M_{\odot}$ we measured the exposure time required to obtain the requested statistical uncertainties (i.e., 5, 10, 15, 20%). The group spectra were simulated with an APEC model for which the normalizations were set to match the fluxes of the systems calculated using the L-M relation. For instance, if the intrinsic scatter is $\sigma_{int} \sim$10%, a sample of 50-80 groups at $z \sim$0.75 is sufficient to measure the evolution if you can estimate the groups' properties (i.e., the temperature and mass) at a $\sim$10-15% level. That can be done with average observations of $\sim$25 ks (a bit longer for smaller systems and shorter for the more massive ones).

**Joint Observations and synergies with other observatories in the 2030s:** By the early 2030s, dedicated surveys (e.g., Vera Rubin Observatory and Euclid in the Optical/Infrared, eROSITA in X-rays, and several "Stage 3" ground-based mm-wave observatories) will increase the number of known groups and clusters out to high redshift. While eROSITA will be limited to $z \lessapprox$1, the SZ-effect surveys are expected to provide robust catalogs out to at least $z \sim$1.5, and observatories like Euclid or Vera Rubin will hunt candidates up to about $z \sim$2. There is also a good chance that a few tens of the first groups at $z >$2 will be serendipitously discovered by AXIS during its mission lifetime. Although such distant systems will be invaluable for probing the early evolution of group-scale halos, they are not the primary targets for feasibility studies presented here, as their detailed characterization will require significantly longer

exposure times. Nonetheless, even a single well-characterized system at these redshifts could provide unique insights into the thermodynamic state and formation history of the earliest collapsed structures in the universe. The identification and redshift estimation can be performed through Euclid and V. Rubin multi-color surveys.

**Special Requirements:** Sensitivity to low surface brightness; low detector background; large effective area

*32. Witnessing the birth of cool cores*

**Science Area:** Galaxy clusters, intracluster medium, AGN feedback

**First Author:** M. Rossetti (INAF - IASF Milano)

**Co-authors:** I. Bartalucci, F. Gastaldello, S. Ghizzardi, G. Riva (INAF - IASF Milano), S. De Grandi (INAF - OA Brera), Y. Su (U. Kentucky)

**Abstract:** High-redshift ($z > 1$) galaxy clusters are crucial for understanding the assembly of the large-scale structure and the intricate interplay between baryonic physics, galaxy evolution, and AGN feedback during the epoch of cluster formation. Recent studies have revealed remarkably similar properties between distant cool cores and their local counterparts, implying early formation and a surprisingly stable AGN regulation over a timescale of approximately 10 Gyr. However, these important results are based on challenging analyses with current X-ray telescopes, relying heavily on numerous assumptions due to limited photon statistics, making it crucial to verify them with more direct and robust measurements. This proposal leverages AXIS's superior capabilities to address this. We will conduct snapshot observations of SZ-selected clusters at $z = 1 - 2$ to determine global properties and identify CC clusters. Subsequently, deeper AXIS follow-ups will target those clusters, enabling detailed temperature and metal abundance profiles. These observations, made possible by AXIS's larger effective area, will test the hypothesis of early CC formation and its stable properties. AXIS is uniquely positioned among current and future X-ray missions to resolve and characterize CCs at high redshift, providing unprecedented insights into their formation and evolution.

**Science:** Galaxy clusters at high redshift ($z > 1$) serve as powerful probes of the early Universe, offering critical insights into the formation and evolution of large-scale structures, the baryon cycle, and the development of galaxies and of the intracluster medium (ICM). These early epochs, approximately 1–2 Gyr after cluster collapse, correspond to the peak of cosmic star formation and AGN activity. Consequently, studying clusters during this period can provide crucial insight into when and how the delicate balance between ICM cooling and AGN feedback was established.

The number of known clusters at $z > 1$ has dramatically increased in the last decade, primarily through Sunyaev-Zel'dovich (SZ) effect and near-infrared surveys. We now recognize massive, virialized systems up to $z \simeq 2$, with their ICM already enriched with metals [303]. Recent studies ([e.g. 67,76,218,322]) have revealed that some of these high-redshift clusters exhibit cool core (CC) properties, such as peaked density profiles, suggesting early CC formation ($z > 1.5$). Interestingly, while the surrounding cluster environment grows self-similarly in size and mass, the CC properties (density, size, mass) appear to remain relatively constant [320,322,433]. This implies that AGN feedback effectively preserves and regulates the core properties over timescales of $\sim 10$ Gyr.

However, these important implications are based on X-ray observations yielding only $\sim 1000$ counts, from which only global spectral information and density profiles can be measured. These counts are insufficient to accurately extract profiles of thermodynamic quantities (temperature, entropy, cooling time) and metal abundance, which are essential for studying CC formation and evolution. This limitation underscores the challenge of obtaining detailed X-ray spectroscopic information at high redshift. Achieving these crucial measurements for $z > 1$ systems is extremely demanding in terms of exposure time for *Chandra*, the only current facility with the necessary angular resolution to resolve CCs, with typical size of $\sim 80$ kpc corresponding to $\sim 9$ arcsec at $z = 1 - 2$, and disentangle them from point sources in distant clusters. For example, even the 170 ks *Chandra* observation of the distant ($z = 1.7$) and massive cool core SpARCS 104922.6+564032.5 [218] yielded only $\sim 150$ net counts.

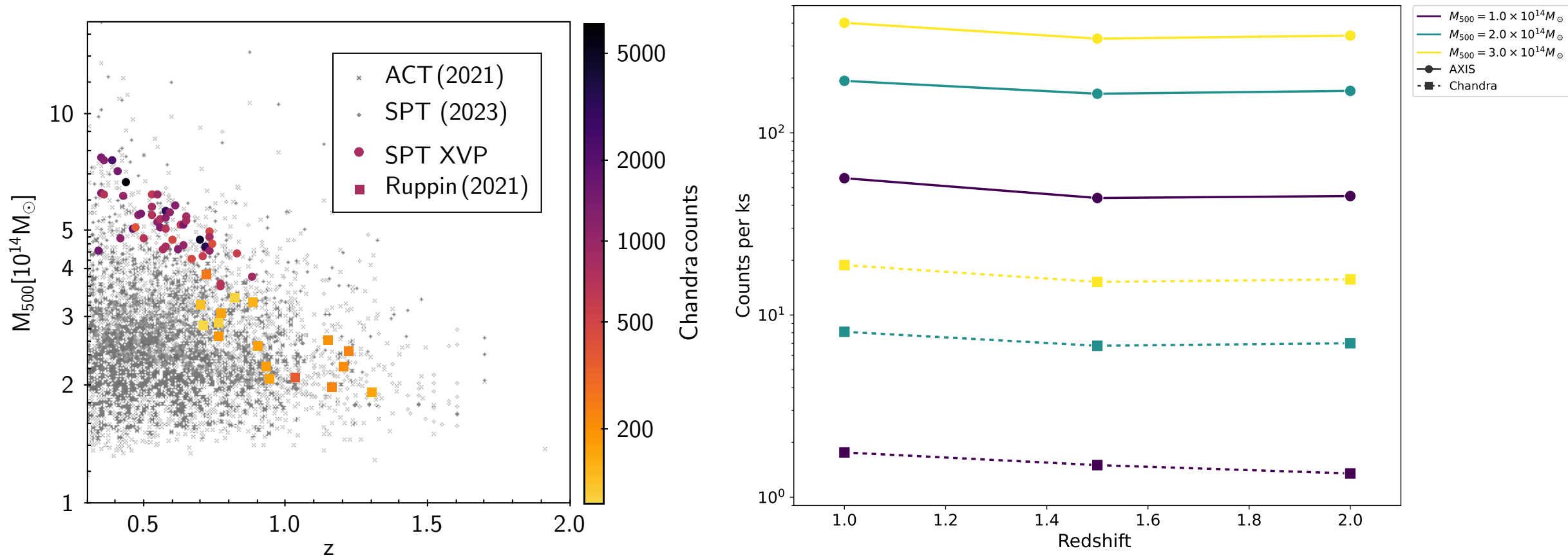

**Figure 50.** *Left:* Figure adapted from [433]. We show in grey the distribution of SPT and ACT clusters in the mass-redshift plane, focusing at z>0.3. Coloured circles and squares highlight the objects with *Chandra* follow-up, from the SPT-XVP [320] and [433] samples, respectively, colour-coded by the number of *Chandra* counts. *Right:* Cost in terms of counts per ks of AXIS observations with respect to *Chandra* ones for the high redshift clusters described in this project. We plot with different colors the values for the masses ($M_{500} = 1, 2, 3 \times 10^{14} M_\odot$) and show with circles and filled lines the AXIS counts (in the 0.5-10 keV energy range) ans with squares and dashed lines the *Chandra* ones (ACIS-S in the typical 0.7-7 keV energy range).

Despite these limitations, *Chandra* data with ∼1000 counts, such as those in the "SPT XVP" sample, have yielded insights into the distribution of thermodynamic properties [e.g. 320,442] for clusters of $M > 3 \times 10^{14} M_\odot$ and $0.3 < z < 1$ (Fig. 50). Ruppin et al. [433] explored even less massive and higher redshift objects using *Chandra* observations with less than 200 counts (Fig. 50), combined with SZ data from SPT. However, the inherent low-count statistics require relying heavily on assumptions, such as the shape of the temperature profile and hydrostatic equilibrium. While these assumptions have been studied in the local Universe, their universal validity, especially at high redshift, is uncertain and may introduce biases. Direct temperature profile measurements in a few radial bins are needed to verify these assumptions and study their impact. Unfortunately, such observations for $z > 1$ CCs are prohibitively expensive for *Chandra*: the expected count-rate for a $M \sim 2 \times 10^{14} M_\odot$ cluster at $z = 1 - 2$ is less than 10 counts per ks (Fig. 50), requiring more than 1 Ms on a single target to measure a temperature profile.

Within the framework of the AXIS Science program, we propose observing a sample of CCs in the high-redshift Universe. While these observations would require prohibitive *Chandra* exposures only for individual targets (less than 10 counts per ks for $M < 3 \times 10^{14} M_\odot$, see Fig. 50), the significantly larger effective area of AXIS will enable routine observations with much shorter exposure times for representative samples of clusters. AXIS will be the only future mission with the required angular resolution to perform detailed studies of the CC properties at high redshift. At z>1.5, even the *NewAthena* goal resolution of $9''$ will not resolve regions smaller than ∼80 kpc.

**Exposure time (ks):** 1.9 Ms (400 ks for snapshot observations of the whole sample + 1.5 Ms for the deeper observations of the CC candidates).

**Observing description:** The current and next generation of SZ surveys (Simons Observatory, CMB-S4) will significantly populate the mass-redshift plane within the crucial redshift range of $z = 1{-}2$, now poorly explored (Fig. 50) revealing new clusters with masses of $M_{500} = (1{-}3) \times 10^{14} M_\odot$. As previously discussed, this epoch is pivotal for the formation of cluster cool cores, and this mass range corresponds to

the progenitors of the massive clusters observed in the local Universe [433]. Consequently, this sample is ideally suited for studying the evolutionary trends of cluster properties.

Follow-up campaigns of SZ-selected clusters within this mass-redshift range will be essential for addressing a variety of astrophysical and cosmological questions, and AXIS's capabilities make this follow-up feasible. Snapshot observations of 3-6 ks will be sufficient to collect $\sim$1000 counts for clusters with $M_{500} = (2{-}3) \times 10^{14} M_{\odot}$ at $z = 1.5{-}2$ (Fig. 50), based on the MCXC $L - M$ relation [394] and assuming self-similar evolution. Slightly longer, but still manageable, observations ($\sim$20 ks) will extend these measurements down to $M_{500} = 10^{14} M_{\odot}$ (Fig. 50). This level of statistics will enable the extraction of density profiles, the measurement of global cluster properties (luminosity, temperature, gas mass), and the determination of morphological parameters, assessing the cluster dynamical state and the presence of a cool core. We estimate that with 400 ks we will be able to follow up a sample of about 35 clusters (20 with $M_{500} = (2{-}3) \times 10^{14} M_{\odot}$ and 15 with $M_{500} = (1{-}2) \times 10^{14} M_{\odot}$).

If the fraction of cool cores, which has been shown to remain relatively constant up to $z = 1$ [433], persists at $z \simeq$ 1-2, we expect 30% of clusters to exhibit CC signatures (about 12 clusters). These CC clusters will be targeted for deeper follow-up observations to achieve a sufficient number of counts (10,000-15,000) for extracting temperature and metal abundance profiles in 5-10 radial annuli. We estimate that relatively short observations (40-90 ks) will be sufficient to reach the goal of 15,000 counts for each cluster with $M_{500} \geq 2 \times 10^{14} M_{\odot}$ in the full redshift range $z \simeq 1 - 2$, while exposures of approximately 200 ks will be needed AXIS to gather 10,000 counts down to masses of $M_{500} = 10^{14} M_{\odot}$ at $z > 1.5$ (Fig. 50). We estimate the total exposure time for the deeper observations of candidate CC to be 1.5 Ms.

Our feasibility estimates are based on the local luminosity-mass ($L - M$) scaling relation in MCXC [394], assuming self-similar evolution with redshift. We converted the expected luminosity into fluxes and subsequently into counts in the $0.5 - 10$ keV band, assuming temperature as derived from the masses with the $M - T$ scaling relation of [20]. For the counts conversion, we used the AXIS response files and on-axis auxiliary response function (ARF), given the compact nature of these high-redshift targets. Our estimates of the required counts are based on the analysis of *Chandra* observations of the "SPT XVP", *Planck* PSZ1 and MACS samples (e.g. [428,429])

**Joint Observations and synergies with other observatories in the 2030s:** Future SZ surveys performed by Simons Observatory and CMB-S4 will be ideal to identify an unbiased population of target clusters for this project.

**Special Requirements:** sensitivity to low surface brightness; low and stable detector background; large Effective Area, good PSF

*33. Probing the physics of a pure cooling flow in a z=1.7 massive cluster of galaxies*

**Science Area:** High-redshift galaxy clusters, intracluster medium, cooling flows, starburst, AGN feedback, intracluster light

**First Author:** Julie Hlavacek-Larrondo (University of Montréal)

**Co-authors:** Benjamin Vigneron, Marine Prunier, Esra Bulbul, Michael Calzadilla

**Abstract:** SpARCS104922.6+564032.5 (SpARCS1049) is a remarkably high-redshift galaxy cluster at $z = 1.709$, making it one of the most distant known cool-core clusters. Identified in 2015 with 27 spectroscopically confirmed members, SpARCS1049 has a total mass of $M_{200} = 3.5 \pm 1.2 \times 10^{14}\, M_{\odot}$, placing it among the most massive clusters at $z > 1.5$. Deep 170 ks *Chandra* observations have revealed a compact X-ray surface brightness peak ($n_{e,0} \sim 0.07\,\mathrm{cm}^{-3}$), consistent with a well-defined cool core. However, SpARCS1049 is unique because it exhibits an intense starburst spanning 60 kpc (SFR $= 860 \pm 130\, M_{\odot}\,\mathrm{yr}^{-1}$) fueled by a large molecular gas reservoir, suggesting that large-scale runaway cooling is directly driving star formation. Notably, there is an absence of radio-loud AGN feedback, making SpARCS1049 a rare and strong candidate of a system undergoing pure cooling flow without AGN regulation.

AXIS (Advanced X-ray Imaging Satellite) will revolutionize the study of SpARCS1049's intracluster medium (ICM) and star formation processes. Its superior spatial resolution and soft X-ray sensitivity will enable precise mapping of the temperature and metallicity gradients, providing insights into early metal enrichment and cooling flow dynamics (i.e., temperature distribution in the core and mass deposition rate measurements as a function of temperature). AXIS will also search for subtle signs of AGN feedback, such as weak cavities or shocks, which could identify past episodes of jet activity. Furthermore, AXIS's ability to resolve the ICM density and entropy profiles beyond 50 kpc will clarify whether SpARCS1049 is still assembling, as indicated by discrepancies in weak-lensing and velocity dispersion estimates. A 100 ks AXIS exposure will yield a temperature estimate within 2% precision and an abundance constraint within 18%—a level of accuracy currently unattainable with *Chandra*. SpARCS1049 represents a rare and valuable laboratory for studying the evolution of high-redshift clusters, and AXIS will provide the sensitivity and resolution necessary to unravel their complex thermodynamic and feedback mechanisms.

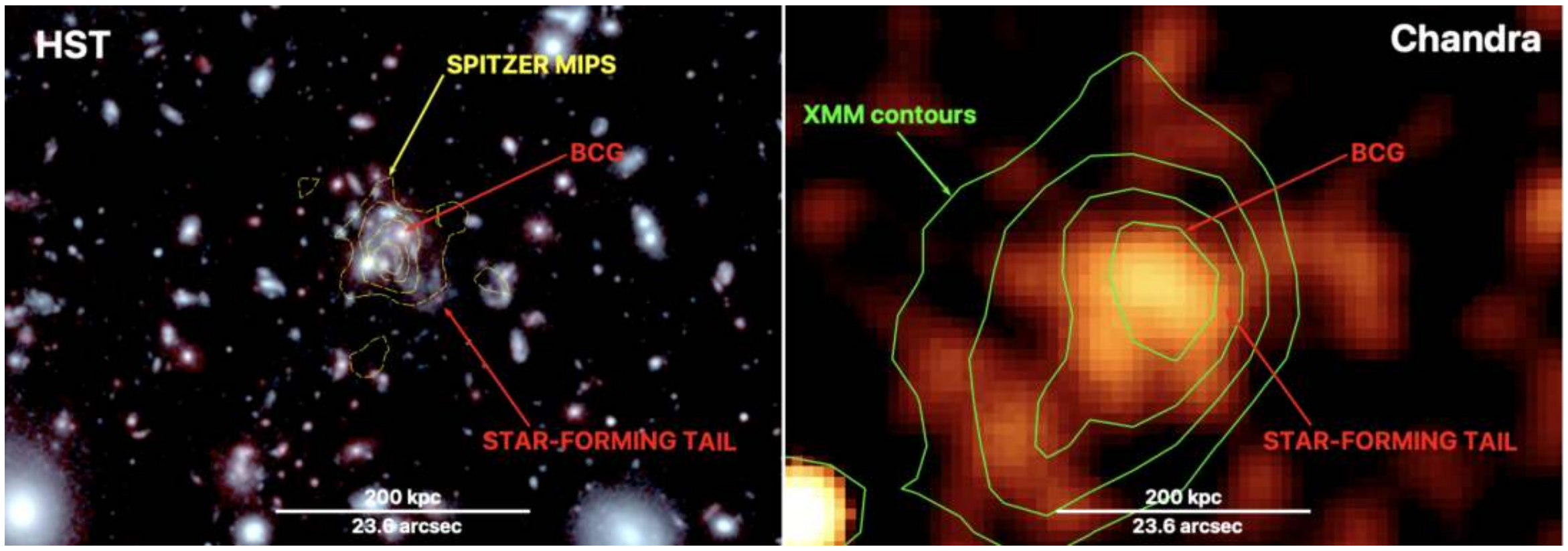

**Figure 51. LEFT:** Hubble Space Telescope images of SpARCS1049 at $z = 1.709$ using the F160W, F150W and F814W filters. The extreme starburst of $860 \pm 130\ \mathrm{M}_{\odot}\mathrm{yr}^{-1}$ is highlighted by the **yellow contours** which trace the Spitzer MIPS 24 micron emission [572]. The starburst also coincides with a large molecular gas reservoir of $1.1 \pm 0.1 \times 10^{11}\ \mathrm{M}_{\odot}$ [574]. **RIGHT:** Exposure-corrected, background-subtracted $0.5 - 7.0$ keV Chandra X-ray image of SpARCS1049, totalling 170 ks. In **green contours**, we show the XMM-Newton observations (100 ks of the 526 ks scheduled), which start at $3\sigma_{\mathrm{rms}}$ and $\sigma_{\mathrm{rms}}$ is the standard deviation of the background.

**Science: Introduction: High-Redshift Galaxy Clusters as Cosmic Laboratories**
The last decade has seen a tremendous leap in our understanding of cluster astrophysics, driven by surveys that have uncovered thousands of clusters at high redshift. Large-scale surveys such as SpARCS and SPT have been instrumental in identifying some of the most distant clusters known today [e.g., 536,573]. Despite these advances, clusters at $1.5 < z < 2$ remain poorly understood due to the significant observational challenges posed by their distance [e.g. 8,298,301]. This redshift range represents a crucial phase in structure formation, during which proto-clusters transition into fully virialized systems.

One of the key open questions in high-$z$ cluster astrophysics concerns the metal enrichment of the intracluster medium (ICM), as well as the occurrence of cooling flows and the timescale of the feedback cycle from the BCG. Observations at lower redshifts have revealed that the ICM exhibits remarkably uniform metallicities [469,576], suggesting that enrichment processes occurred early in the universe. However, direct observational confirmation of this scenario at $z > 1.5$ is still scarce. The first deep XMM-Newton study of a $z \sim 1.7$ cluster, SPT-CLJ0459-4947, provided the first metallicity measurements beyond the cluster center [300], supporting early enrichment models. The study of additional high-$z$ clusters is critical for refining these models and placing further constraints on the timing of ICM enrichment.

**SpARCS1049: A Unique High-Redshift Cool-Core Cluster**
SpARCS104922.6+564032.5 ($z = 1.709$; hereafter SpARCS1049) is a massive, optically rich galaxy cluster at $z = 1.709$, first identified in 2015 with 27 spectroscopically confirmed member galaxies [572]. Weak-lensing analyses estimate its total mass to be $M_{200} = 3.5 \pm 1.2 \times 10^{14} M_{\odot}$ [164], placing it among the most massive clusters known at $z > 1.5$.

In 2020, deep *Chandra* observations (170 ks) [218] uncovered a compact X-ray surface brightness peak, indicating a well-defined cool core with a central density of $n_{e,0} \sim 0.07$ cm$^{-3}$ (see Fig. 1). Unlike typical cool-core clusters, SpARCS1049 also hosts an intense starburst with a star formation rate (SFR) of $860 \pm 130 M_{\odot}$ yr$^{-1}$, located ~25 kpc south of the brightest cluster galaxy (BCG) [572]. The alignment of the coolest detectable ICM X-ray gas (0.7–1.0 keV) with the starburst, as observed by *Chandra*, suggests that large-scale, runaway cooling is directly fueling star formation [218]. In contrast to low-$z$ clusters where AGN feedback prevents excessive cooling, SpARCS1049's AGN shows no significant radio activity or jet structures [517], making it a rare case of AGN feedback suppression at high redshift.

Spitzer 3.6 $\mu$m data also indicate a cluster mass of $M_{500} = 3.8 \pm 1.2 \times 10^{14} M_{\odot}$ [572], consistent with HST weak-lensing estimates [164]. However, velocity dispersion measurements ($446 \pm 80$ km s$^{-1}$) suggest a significantly lower virial mass of $1 \pm 0.4 \times 10^{14} M_{\odot}$ [164], hinting that SpARCS1049 may be in the process of virialization. The cluster's outer ICM deviates from self-similar expectations beyond 50 kpc (see Fig. 2), suggesting that the outer halo is still assembling [317]. These properties position SpARCS1049 as a key system for studying cluster formation and the transition from proto-cluster to virialized state.

The cluster is scheduled to be observed deeply (526 ks) with XMM-Newton, providing insights into the ICM and its thermodynamic state beyond what was possible with *Chandra*. We show the first 100 ks of this dataset in Fig. 1. Essentially, the XMM-Newton data will allow us to obtain thermodynamic profiles and extend the X-ray surface brightness and density profiles beyond $r_{500}$ to assess deviations from self-similarity and cluster growth at high redshift. However, due to XMM's poor point spread function (PSF), we will only be able to obtain 2–3 data points for the thermodynamic profile, limiting our ability to resolve the thermodynamic properties, particularly the spatial connection between the soft X-ray gas and the starburst. AXIS, with its superior angular resolution and sensitivity, will overcome these limitations, providing a more detailed and spatially resolved view of the cooling processes and feedback mechanisms that shape SpARCS1049.

**Scientific Goals with AXIS**

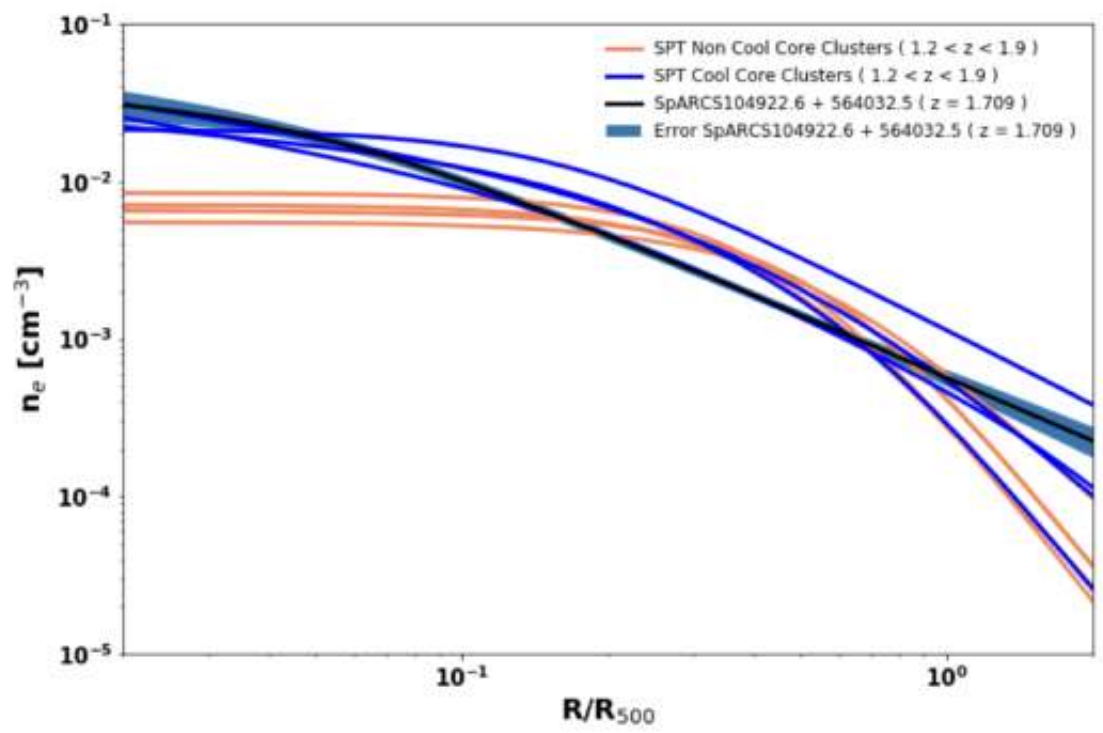

**Figure 52.** Profiles of eight $1.2 < z < 1.9$ SPT clusters of galaxies that have Chandra data (see [322] and [218] for details). The top 4 are cool core clusters as defined by their central electron density value exceeding $n_{e,0} = 0.015$ cm$^{-3}$. This figure shows that SpARCS1049 has an over-dense core (i.e., a cool core). Indeed, it has a highly peaked central density of $n_{e,0} \sim 0.07$cm$^{-3}$, making it one of the strongest cool core clusters known at high-$z$. This plot also shows that the density profile beyond 50 kpc follows a different slope, implying that the outer parts of the cluster, usually driven by self-similar processes, may not be well established in this cluster. Profiles were determined using the standard methods described in [320].

AXIS, with its high spatial resolution and unique soft X-ray sensitivity, is uniquely suited to studying the thermodynamics of SpARCS1049's ICM and understanding the role of AGN feedback in cluster evolution. The key science objectives are:

- **ICM Cooling and Metal Enrichment:** By mapping the temperature and metallicity gradients, AXIS will determine whether the ICM is enriched uniformly, as observed in low-redshift clusters [469,576].
- **AGN Feedback Constraints:** The lack of radio-loud AGN activity suggests weak or absent feedback. AXIS will search for indirect evidence of feedback, such as X-ray cavities or diffuse shock features.
- **Cluster Assembly and Virialization:** Weak-lensing mass measurements and the deprojected density profile as seen with *Chandra* indicate that SpARCS1049 is likely still assembling [164]. AXIS could also identify the presence of X-ray-emitting halos in the process of merging (similar to XDCP0044 at z$\sim$1.6) through morphology and/or different temperatures. AXIS will probe the density and entropy profiles of the ICM to assess whether they deviate from self-similar predictions for virialized clusters.

**Exposure time (ks):** 100 ks

**Observing description:** To simulate an X-ray spectrum of the high-redshift galaxy cluster SpARCS1049, we generated a synthetic *AXIS* spectrum using XSPEC. The cluster parameters were based on an observed 2 to 10 keV X-ray luminosity of $\sim 4.3 \times 10^{44}$ erg/s as observed with *Chandra*, corresponding to a flux of $\sim 2.18 \times 10^{-14}$ erg/s/cm$^2$, and a temperature of $\sim 5.71$ keV. The redshift was fixed to $z = 1.7$ and the abundance was first fixed at $0.3 Z_{\odot}$.

The simulated spectrum was created using the `phabs*apec` model in XSPEC. The Galactic hydrogen column density was set to $5.99 \times 10^{19}$ cm$^{-2}$. The cluster was assumed to have a physical diameter of $50''$, corresponding to a projected area of approximately 0.545 arcmin$^2$.

To generate the spectrum, the energy range was defined between 0.05 and 20 keV, with a binning of 19950 linear intervals. A response matrix file (RMF) and an ancillary response file (ARF) were assigned to model the instrument response. The simulation used the *fakeit* routine to create synthetic data, incorporating an exposure time of 500 ks, followed by an additional simulation with an exposure time of 100 ks.

The model parameters were then fitted to the simulated spectrum. The hydrogen column density was kept fixed, while the abundance parameter was allowed to vary. The *fit* routine was executed to determine the best-fit parameters. For a 100 ks exposure, we obtain a best-fit temperature of $5.13 \pm 0.14$ keV and an abundance of $0.291 \pm 0.055$.

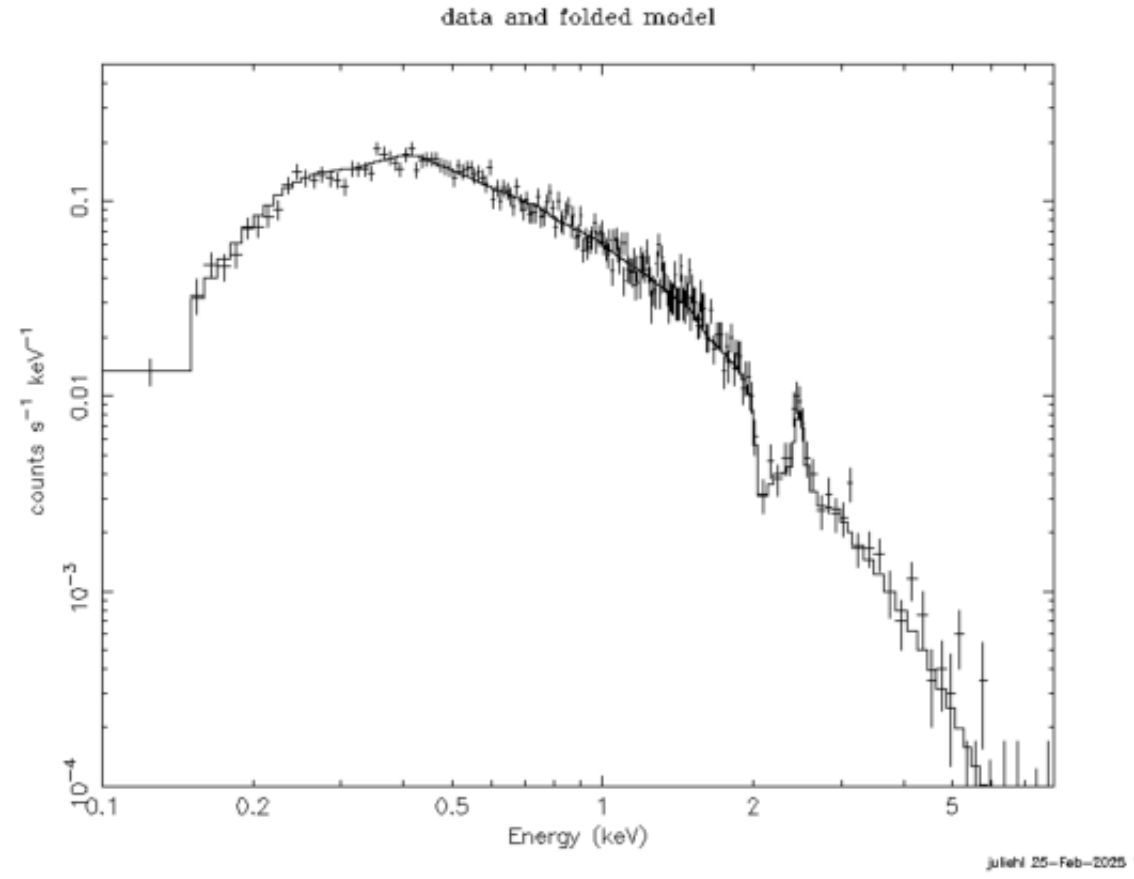

**Figure 53.** Simulated AXIS spectrum of SpARCS1049 with a 100 ks exposure. The total number of counts is ∼13,000, compared to ∼ 400 obtained with *Chandra* in 170 ks in Hlavacek-Larrondo et al. [218]. In just 100 ks with AXIS, we can estimate the temperature to 2% and the abundance to 18%, the latter of which was not possible even with 170 ks of *Chandra* observations. Hence, we could begin to probe the physics of this high-redshift cluster with a precision similar to that currently achieved for low-redshift clusters with *Chandra*.

The simulation demonstrated that with just 100 ks of *AXIS* exposure, we can estimate the temperature with a precision of 2% and constrain the abundance to approximately 18% accuracy. This level of precision was unattainable with the previous 170 ks of actual *Chandra* observations. The results suggest that deeper observations with next-generation X-ray missions, such as AXIS, will allow us to probe high-redshift cluster physics with a precision comparable to that currently achievable for low-redshift clusters.

**Joint Observations and synergies with other observatories in the 2030s:** The cluster has been extensively mapped across multiple wavelengths, providing a comprehensive dataset to study its properties. HST imaging has resolved its galaxy population and weak-lensing mass distribution, while Spitzer observations have traced its infrared emission, revealing an extreme central starburst. JWST observations are also being proposed. VLA data have provided constraints on AGN activity and molecular gas reservoirs, and JVLA CO (1-0) observations have confirmed the presence of a massive cold gas reservoir fueling star formation. Deep *Chandra* X-ray imaging (170 ks) has identified a compact cool core and confirmed the presence of a runaway cooling flow, while upcoming deep XMM-Newton observations (526 ks) will probe the thermodynamic structure of the intracluster medium. Additionally, the cluster has been observed with LOFAR, both using Dutch and international baselines, providing low-frequency radio data to study diffuse emission and potential feedback mechanisms and will be a prime target for ngVLA and SKA. These datasets, spanning optical, infrared, radio, and X-ray wavelengths, offer a unique opportunity to study the interplay between cooling, star formation, and feedback in a high-redshift cluster. However, the limited spatial resolution of current X-ray data constrains our ability to resolve the thermodynamic properties, a challenge AXIS will overcome with its superior imaging capabilities.

**Special Requirements:** None

*34. AGN populations and nascent intracluster medium in protoclusters*

**Science Area:** Galaxy clusters, protoclusters, galaxy evolution, intracluster medium, AGN population

**First Author:** Paolo Tozzi (INAF - Osservatorio Astrofisico di Arcetri)

**Co-authors:** Yuanyuan Su (U. Kentucky), Scott W. Randall (Center for Astrophysics | Harvard & Smithsonian), Arnab Sarkar (MIT), Erik B. Monson (Penn State), Roberto Gilli (INAF-OAS), Stefano Marchesi (University of Bologna, Italy), Fabio Vito (INAF-OAS), Marika Lepore (INAF-OAA), Massimo Gaspari (UNIMORE)

**Abstract:** The transition from protoclusters to massive virialized clusters occurs at cosmic noon, a crucial epoch for galaxy evolution. Based on our current understanding, the protocluster environment is expected to enhance both nuclear and star-formation activity compared to the field. During this transition, the diffuse baryons are heated by gravitational processes and nuclear feedback to end up in the hot intra-cluster medium (ICM), which is, in turn, chemically enriched by star formation products spread by feedback processes. The complex interplay of these phenomena challenges our ability to observe and understand them. High-resolution X-ray observations are the most effective approach for investigating most of the observables associated with such energetic processes, namely: nonthermal emission from nuclear activity, low-mass and high-mass X-ray binaries emission. in starburst events, inverse Compton from relativistic jets, and thermal bremsstrahlung from the ICM. Despite the high scientific returns, X-ray follow-up observations of high-redshift protoclusters have been challenging and resource-intensive for existing facilities. With its high angular resolution and large effective area, AXIS will provide the opportunity to systematically investigate the evolution of galaxies and diffuse baryons during the critical transition from protoclusters to clusters.

**Science:** The cosmic noon ($2 < z < 3$) is a crucial epoch for galaxy formation and evolution: the star formation rate in the Universe peaks at $z \geq 2$, similarly to the population of active galactic nuclei (AGN), which also corresponds to the peak in the growth rate of supermassive black holes (SMBHs). This epoch also marks the transition from protoclusters to massive virialized clusters. Protoclusters are large-scale, high-density environments with a total mass comparable to that of local massive clusters, but still far from virialization. They are expected to have a major influence on galaxy evolution, thanks to gravitational processes coupled with the larger availability of diffuse gas at higher densities and a high merger rate, which enhance star formation (SF) and AGN duty cycle in protocluster members. In addition, according to the hierarchical model of structure formation, a few halos within the protoclusters may already be virialized, heating the diffuse baryons and illuminating the proto-intracluster medium (proto-ICM) in the X-ray band. The proto-ICM is, in turn, chemically enriched by SF products spread by feedback processes. Eventually, the cumulative feedback effects and the dynamical processes associated with virialization effectively quench nuclear activity and star formation, resulting in the relatively quiescent galaxy population observed in massive clusters at $z < 1$ and in the local Universe.

Many physical processes coincide with the evolution of the protocluster galaxy population, including the cycle of mechanical AGN feedback and feeding events, the consumption and accretion of cold gas (e.g., via chaotic cold accretion), galaxy interactions, and the formation of the hot ICM through accretion shocks, among others. The complex interplay of physical processes shaping protocluster galaxy evolution remains challenging to fully observe and model. However, many of these mechanisms are highly energetic and produce X-ray signatures that can be detected and characterized with high-resolution imaging and CCD-like spectral resolution.

However, despite the high scientific returns expected, current X-ray facilities lack the necessary sensitivity and spatial resolution to systematically study these phenomena in high-redshift protoclusters.

AXIS will overcome these limitations, allowing for a more in-depth study of key X-ray observables in protoclusters, including:

- **Nuclear activity** in member galaxies (X-ray luminosity and intrinsic absorption);
- **Proto-ICM thermodynamics** (temperature profiles and mass distribution);
- **Proto-ICM chemical enrichment** (iron abundance);
- **Inverse Compton emission** from relativistic electrons in radio jets;
- **X-ray emission from low-mass and high-mass X-ray binaries** in vigorously star-forming galaxies.

By combining these X-ray diagnostics with multi-wavelength data, AXIS will address critical science questions, such as

- the environmental (density) dependence of nuclear and SF activity in protocluster galaxies;
- the triggering mechanism of X-ray AGN and SF in member galaxies (e.g., consumption of cold gas vs. galaxy interactions);
- the powering mechanism of Lyman Alpha Emitting sources;
- the formation of the first group-size virialized halos;
- the presence of cool cores and potential (quenched) cooling flows in the proto-ICM;
- the impact of mechanical/radio AGN feedback on the hot, warm, and cold baryons surrounding active galaxies [175];
- the formation of the brightest cluster galaxies;
- the duty cycle of AGN feedback and tied SMBH weather [176,177].

X-ray imaging and spectroscopy of cluster progenitors with AXIS can provide unique leverage to understand the co-evolution of galaxies, black holes, and large-scale structures at cosmic noon, opening a novel and rich scientific window into the most hectic era in cosmic history.

**Exposure time (ks):** For a sample of six protoclusters, we would require a total exposure time of 1.16 Ms.

Our immediate goal is to detect the AGN and proto-ICM emission in protoclusters. We consider a circular extraction region with a radius of 2 arcsec for unresolved sources (AGN) across the entire AXIS FoV and a circular region with a radius of 12 arcsec for the thermal ICM emission (which is considered at the aimpoint since the largest halo is presumably at the center of the observed field). A robust detection is set at 10 net counts for AGN and 30 net counts for diffuse emission. We note that, for the same source count rate, the S/N as a function of exposure time is lower for the extended sources, since the background is much larger. We adopted the L2 background as in the file axis_nxb_FOV_10Msec_20250210.pha, which resulted in a count rate of 1.02 cts/s/arcmin$^2$ in the 0.5-10 keV band. This corresponds to a background count rate of $3.6 \times 10^{-5}$ and $1.24 \times 10^{-3}$ for the extraction regions of unresolved and extended sources, respectively.

In Figure 54, we show the expected net count vs exposure time for AGN with intrinsic $L_{2-10\rm keV} = 10^{43}$ erg/s in the range $2 < z < 3$ and $10^{22} < N_H < 5 \times 10^{23}$ cm$^{-2}$ (left) and ICM in the range $2 < z < 3$ and temperatures in the range $1.2 < kT < 2$ keV with rest-frame luminosities of $5 \times 10^{43}$ erg/s and $10^{44}$ erg/s, respectively (right). The lowest horizontal lines correspond to our detection criteria. We note that these criteria are conservative and are based on the results of one of the deepest X-ray studies on protoclusters, the Spiderweb [271,510,511]. We also note that intrinsic absorption is a key parameter (at a given intrinsic luminosity) for the detection of Compton-thin AGNs, while the presence of Compton-thick AGNs is

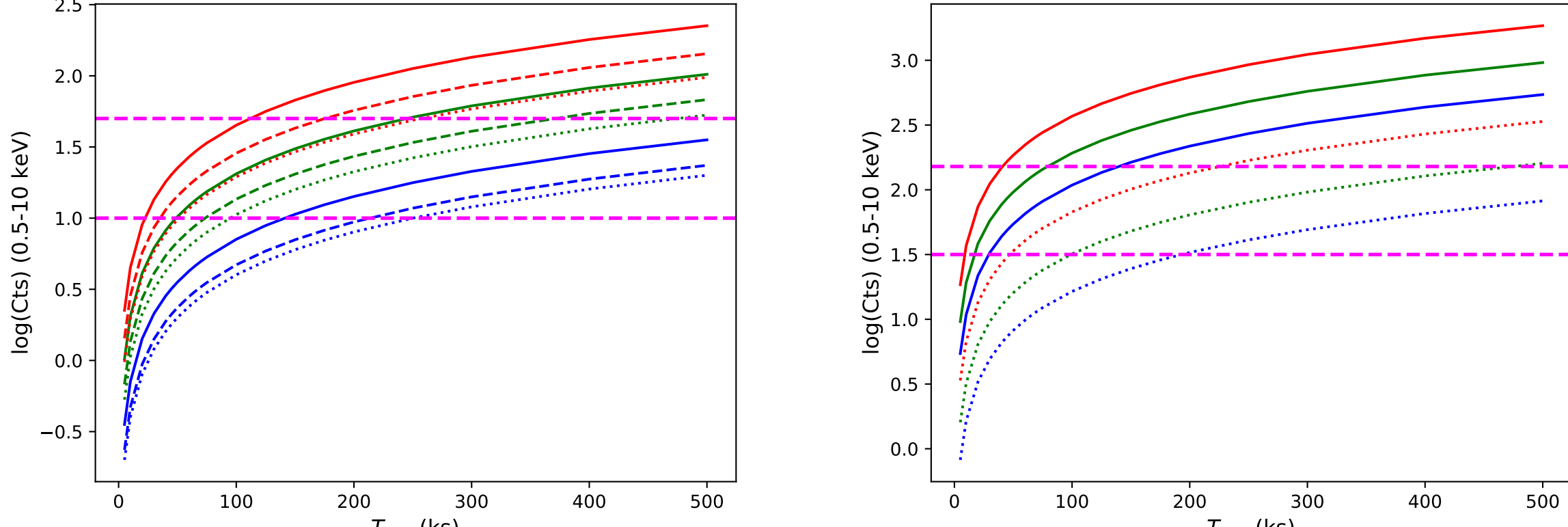

**Figure 54.** *Left:* Expected net counts for AGN as a function of exposure time. Solid lines: AGN with $L_{2-10\text{keV}} = 10^{43}$ erg/s at z = 2, with intrinsic absorption $N_H = 10^{22}$ cm−2 (red), $N_H = 10^{23}$ cm−2 (green), and $N_H = 5 \times 10^{23}$ cm−2 (blue). The count rates are computed for the average field of view (FoV). Dashed lines: the same AGN at z = 2.5. Dotted lines: the same AGN at z = 3. Horizontal dashed lines mark the thresholds at 10 and 50 net counts. *Right:* expected net counts for ICM as a function of exposure time. Solid lines: thermal emission with kT = 2 keV and $L_{0.5-10\text{keV}} = 10^{44}$ erg/s at z = 2 (red), z = 2.5 (green) and z = 3 (blue). Dotted lines: thermal emission with kT = 1.2 keV and $L_{0.5-10\text{keV}} = 5 \times 10^{43}$ erg/s at z = 2 (red), z = 2.5 (green) and z = 3 (blue). The count rates are computed at the aim point. Horizontal dashed lines mark the thresholds at 30 and 150 net counts.

expected to be more challenging and complex to model. Still, when a large number (a few hundred) of confirmed protocluster members are available, stacking techniques can be used to bring the average detection limit below our threshold, enabling the study of severely obscured sources. Regarding thermal emission, the key parameter is temperature. Due to the dramatic K-correction effect, temperatures below 1 keV are hardly visible at $z > 2$. It is worth exploring the trade-off between effective area and background below 0.5 keV to test the possible detection of $< 1$keV ICM at high redshifts. Finally, based on preliminary spectral simulations, spectral analysis is possible starting from several net counts roughly equal to 5 times larger than the detection thresholds (therefore, 50 net counts for AGN and 150 for ICM, see the upper dashed lines in Figure 54).

**Observing description:** In the next 5-10 years, we expect a large number of protoclusters to be identified thanks to wide-angle surveys from space missions, such as Euclid, and surveys from the ground, including those with LSST. We note that it is crucial to identify and characterize overdensities over a few physical Mpc (3-5 arcmin at least) to select protoclusters. Therefore, a suitable protocluster sample to be observed with AXIS will be optimally and homogeneously selected from future catalogs.

As an example, we list six protoclusters selected using different methods. The first three targets in Table 3 are selected as overdensities around radio-loud galaxies. The other three are selected as mass overdensities. Exposure times are computed from Figure 54 for the detection threshold according to their redshift. We note that the requirement for AGN detection is always higher than that for ICM detection. The primary reason is that we conservatively set the minimum AGN intrinsic luminosity a factor of 5 or 10 below that of the ICM, considering that we included strongly absorbed AGN. We also note that, due to the strong K-correction, the required exposure time for ICM detection increases with redshift faster than for AGN.

**Table 3.** Test sample of protoclusters suitable for this science case. (1): Object ID; (2): Right ascension; (3): Declination; (4) Redshift; (5): Galactic hydrogen column density in units of $10^{20}$ cm$^{-2}$; (6): Galaxy overdensity; (7) Reference to the discovery paper; (8), (9): Exposure time in ks for AXIS observations for AGN and ICM.

| Protoclusters with radio beacon | | | | | | | | |
|---|---|---|---|---|---|---|---|---|
| ID | RA | Dec | z | $N_{HGal}$ | $\delta_g$ | Ref. | $t_{exp}$ AGN | $t_{exp}$ ICM |
| MRC0156-252 | 29:38:21.9 | -24:59:32.1 | 2.02 | 1.10 | $17.8 \pm 2.4^a$ | [171] | 130 | 40 |
| USS1707+105 | 17:10:06.5 | 10:31:06.2 | 2.35 | 6.18 | $12 \pm 2^b$ | [208] | 200 | 100 |
| MRC2104-242 | 21:06:58.2 | -24:05:11.3 | 2.49 | 4.16 | $14.0 \pm 2.1^a$ | [171] | 200 | 100 |
| Protoclusters without dominant galaxy | | | | | | | | |
| ID | RA | Dec | z | $N_{HGal}$ | $\delta_g$ | Ref. | $t_{exp}$ AGN | $t_{exp}$ ICM |
| BOSS1244 | 12:43:31.2 | +35:55:12.0 | 2.23 | 1.27 | $22.9 \pm 4.9^c$ | [461] | 180 | 80 |
| HS1700+643 | 17:00:55.00 | +64:11:26.7 | 2.30 | 2.11 | $6.9 \pm 2.1^d$ | [478] | 200 | 100 |
| SSA22 | 22:17:34.0 | +00:05:01 | 3.09 | 4.23 | $7.6 \pm 1.4^e$ | [477] | 250 | 180 |

$^a$Density of IRAC sources from [171], $^b$Density of H$\alpha$ emitters from [208], $^c$Density of H$\alpha$ emitters from [461], $^d$Density of IRAC sources from [478], $^e$Density of Lyman Break Galaxies from [507]

For the test sample of six protoclusters listed in Table 3, a complete follow-up would require 1.16 Ms. We stress that the depth of each observation would reach the same depth of the Spiderweb Galaxy survey (700 ks with Chandra AO20), despite a 2× lower angular resolution. However, the useful FoV of 450 arcmin$^2$ would be $\sim 6\times$ larger than in Chandra-Spiderweb, where the used FOV was $\sim$ 80 arcmin$^2$. This is due also to the negligible variation of the AXIS PSF, which is characterized by a $HPD < 2$ arcsec across the entire FoV. Note that the 24 arcmin diameter of the AXIS FOV corresponds to 11-12 physical Mpc at $z = 2 - 3$.

**Joint Observations and synergies with other observatories in the 2030s:** NIRCam photometry and other current and upcoming NIR missions can probe the rest-frame optical at $z = 2$ and $z = 3$, providing a possible window into the emission of AGN host galaxies. X-ray-to-IR SED fitting with tools like XCIGALE, Lightning, and AGNFITTER, among others, can provide a census of the properties of AGN hosts by simultaneously modeling both AGN and host emission. Simultaneous modeling of the AGN and host becomes much easier with sensitive X-ray observations, which constrain the luminosity of the AGN component.

With JWST/NIRSpec spectroscopy, the redshifts of individual galaxies can be measured to confirm protocluster members down to a low stellar mass, allowing one to measure the AGN fraction as a function of $M_*$. JWST's dust observations combined with AXIS's measurements of hot gas provide a comprehensive view of the multi-phase gas in protoclusters. JWST identifies member galaxies and their interactions, allowing one to associate interactions with the triggering of nuclear activity. Together, they also explore the impact of AGN and supernova feedback, with JWST tracing their influence on SF and AXIS mapping their effects on the surrounding hot gas and chemical enrichment. This synergy provides a transformative understanding of protocluster formation and evolution.

Other key facilities from the ground with strong synergies are

- ALMA: ALMA identifies the most rapidly star-forming protocluster populations, enabling constraint of their AGN fraction with AXIS, and possibly enabling the study of their XRB populations. ALMA also enables studies of molecular gas kinematics, relating the growth of bulges and bars in rapidly growing protocluster galaxies to the growth of SMBHs, as probed by AXIS.
- Future ELTs: Narrowband surveys and next-gen integral field spectroscopy can identify Ly-$\alpha$ nebulae and structures in the ICM. AXIS's survey capability can place AGN in the context of large-scale Ly-$\alpha$ structure.

**Special Requirements:** None

## g. Cosmology

*35. Cosmology with Dynamically Relaxed Clusters of Galaxies*

**Science Area:** Galaxy clusters, intracluster medium, cosmology

**First Author:** Adam Mantz (KIPAC)

**Co-authors:** Steven W. Allen (KIPAC, Stanford, SLAC), R. Glenn Morris (KIPAC, SLAC), Anthony M. Flores (KIPAC, Stanford, Rutgers), Abby Pan (KIPAC, Stanford), Taweewat Somboonpanyakul (Chulalongkorn University), Haley Stueber (KIPAC, Stanford)

**Abstract:** Observations of clusters of galaxies provide multiple, powerful probes of cosmology [8,54,547], and have played a key role in establishing the current "concordance" model of cosmology, in which the mass-energy budget of the Universe is dominated by dark matter and dark energy, with the latter being consistent with a cosmological constant [11,105,297,546,579]. The most dynamically relaxed, massive clusters play a special role in this work. The impact of projection on determinations of the intracluster medium (ICM) thermodynamic structure is minimized by the relative morphological and geometric simplicity presented by relaxed clusters. At the same time, their proximity to hydrodynamic equilibrium makes it possible to precisely reconstruct the 3D mass profile, including dark matter, with only small biases due to non-thermal support expected [16,351].

The gold standard for assessing the dynamical state of a cluster is the morphology of the ICM, which is most sensitively probed by high angular resolution X-ray observations. For this reason, historically, Chandra has played a key role in both identifying and exploiting the most massive, relaxed clusters, despite having less effective area than most contemporaneous observatories. Various measures of cluster morphology are in use, the most effective of which broadly aim to quantify the sharpness of the surface brightness (or ICM density) peak in the cluster center and the symmetry of the brightness distribution on larger scales [302,341,443]. At intermediate and high redshifts, the ability to resolve steep central temperature and density gradients in relaxed clusters on scales of a few arcseconds and smaller is especially important.

Here we review the key cosmological tests enabled by X-ray observations of relaxed clusters, and consider the potential role of AXIS in identifying and exploiting new systems of this type, noting that both high angular resolution and large collecting area are critical to advancing the field. We find that an investment of time comparable to what has been accomplished with Chandra would achieve order-of-magnitude improvements in dark energy constraints, including in evolving models, and make possible percent-level constraints on the Hubble constant from clusters, independent of the distance ladder and the baryon acoustic scale.

### Cosmology Enabled by Relaxed Clusters

X-ray observations of samples of relaxed clusters have provided key constraints on the mean matter density, $\Omega_\mathrm{m}$, and the cosmic expansion, through measurements of the gas mass fraction, $f_\mathrm{gas}$ [10,301,304,579]. Physically, $f_\mathrm{gas}(r,z) = \Upsilon(r,x)\,\Omega_\mathrm{b}/\Omega_\mathrm{m}$, where $\Upsilon(r,z)$ accounts for the baryonic depletion of the ICM and the fraction of baryons in the X-ray emitting phase. For massive clusters with deep gravitational potentials (in practice defined as those with average temperatures $> 5\,\mathrm{keV}$), $\Upsilon(r,z)$ is predicted to have very small intrinsic scatter and little or no evolution for radii outside the cluster center, making $f_\mathrm{gas}$ at these radii a standard quantity. In detail, the $f_\mathrm{gas}$ values we measure under the assumption of a reference cosmological model depend on the true cosmology as [10]

$$f_{\rm gas}^{\rm ref}(z) = K(z)\,\Upsilon(z)\left(\frac{\Omega_{\rm b}}{\Omega_{\rm m}}\right)\left(\frac{d^{\rm ref}(z)}{d(z)}\right)^{3/2}, \quad (1)$$

where $K(z)$ encodes any systematic bias in the $f_{\rm gas}$ measurements, whether due to astrophysics (e.g. non-thermal pressure) or instrument calibration. The normalization of the $f_{\rm gas}^{\rm ref}(z)$ curve, in practice precisely determined at low redshifts, thus constrains $\Omega_{\rm m}$, given external priors on $H_0$ and $\Omega_{\rm b}h^2$ [579], while the last term is sensitive to the expansion history generally, and thus to dark energy. Notably, current results on dark energy from the redshift-dependent signal (Fig 55a) are limited by the sparsity of the data at higher redshifts ($z \gtrsim 0.6$), and not by systematic uncertainties [304]. The measured intrinsic scatter is exceptionally small: 4.3% in $f_{\rm gas}$, equivalent to 2.9% in distance [304] (compare to 4.6% uncertainty in distance from individual supernovae Ia [455]). Consequently, there is ample opportunity to increase the sample size and/or improve the precision of existing measurements before systematic uncertainties result in diminishing returns.

Given the ongoing controversy regarding the value of the Hubble constant, it is worth noting that constraints from the shape of $f_{\rm gas}(z)$ are independent of $h$. Constraints from its normalization are weakly dependent, as $h^{1/2}$ given the usual practice of adopting an external prior on $\Omega_{\rm b}h^2$. Conversely, the normalization of $f_{\rm gas}(z)$ can be combined with constraints on the cosmic baryon fraction from the Cosmic Microwave Background (CMB; essentially uncorrelated with the direct CMB constraint on $h$) to yield an independent estimate of the Hubble constant [304] (Fig 55b).

Another measure of cosmic distance can be obtained by combining X-ray observations with measurements of the Sunyaev-Zel'dovich (SZ) effect towards a cluster [39,51,453,468,565]. SZ measurements probe the Compton $y$ parameter, which is proportional to the line-of-sight integral of the electron pressure, $y \propto \int d\ell\, n_{\rm e}(\ell)T(\ell)$. The same quantity can be independently inferred from X-ray measurements of the density and temperature, given a reference cosmological model. Analogously to Eq. 1, we can write the relationship between these values as

$$y_{\rm X-ray}^{\rm ref} = y_{\rm SZ}\left[\frac{d^{\rm ref}(z)}{d(z)}\right]^{1/2}, \quad (2)$$

where equality of the Comptonization from the two observations (probing the same gas) places constraints on $d(z)$. The relatively weak dependence on distance above means that this technique is primarily useful for constraining the normalization of $d(z)$ (i.e., the Hubble constant), making it highly complementary to the $f_{\rm gas}$ test. While the requirement that both X-ray and SZ data probe the same ICM pressure does not directly depend on hydrostatic equilibrium, the simple morphologies of relaxed clusters make them the preferred targets for this work as well, minimizing systematic uncertainties due to asymmetry and projection effects. The challenge of independently verifying the calibration of X-ray temperature measurements limits current constraints on $h$ from this technique; however, promising routes to improving this calibration could lead to percent-level constraints on the timescale of a decade. Discussed further by [565], strategies include a dedicated, CubeSat-scale X-ray source for direct, in-situ calibration, as well as measurements of relativistic corrections to the SZ effect, which provide an independent constraint on the ICM temperature.

The ability to derive constraints on the enclosed mass as a function of 3D radius in relaxed clusters from X-ray data (in contrast to strong and weak lensing, which are less directly related to the 3-dimensional potential) provides an additional cosmological probe. Within the cold dark matter paradigm, simulations predict that the density profiles of relaxed halos, on all observable scales, can be well approximated by a simple, universal profile, parametrized by total mass and concentration [73,172,353]. The concentration is expected to be weakly anti-correlated with mass at low redshifts, gradually decreasing with redshift and eventually trending to be constant with mass at $z \gtrsim 3$. While the mass dependence of concentration has

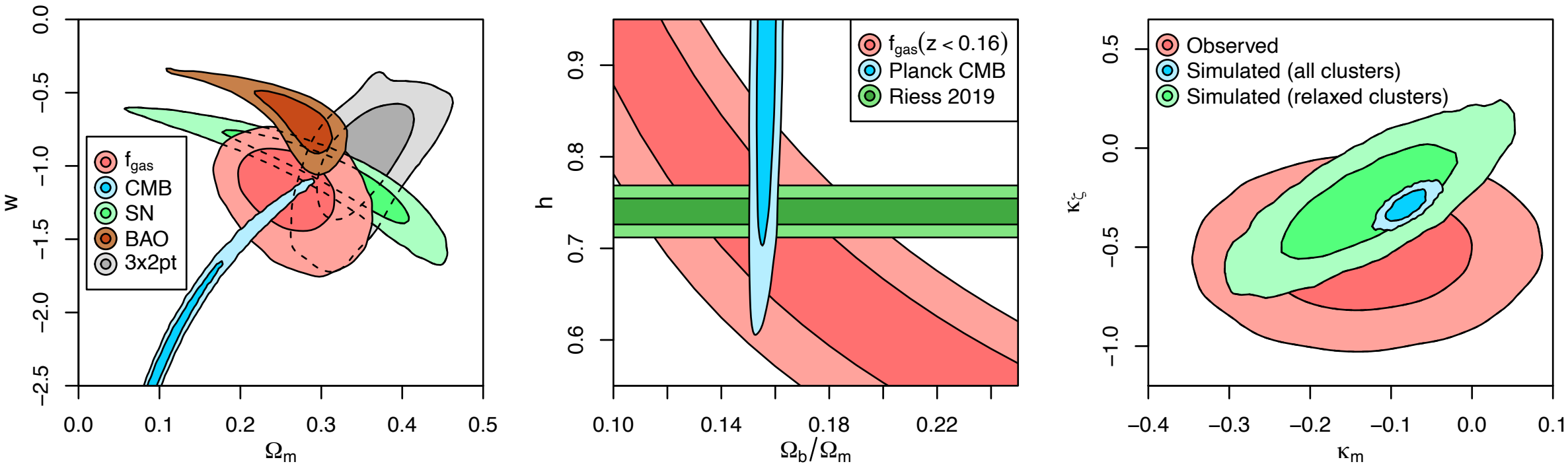

**Figure 55.** From [116,304]. Left: Constraints on flat, constant-$w$ cosmological models from cluster $f_{\rm gas}$, *Planck* CMB [396,397], Pantheon supernovae [455], baryon acoustic oscillations [7], and galaxy clustering and lensing [2] data. Center: Constraints on the Hubble parameter and the cosmic baryon fraction from $f_{\rm gas}$ data at $z < 0.16$ only, *Planck* CMB data [396,397], and the Cepheid distance ladder [422]. The *Planck* analysis assumes a flat, constant-$w$ model, while the $f_{\rm gas}$ and distance-ladder constraints are essentially independent of the cosmological model. Right: Constraints on power-law slopes of NFW concentration with mass ($\kappa_m$) and $1+z$ ($\kappa_\zeta$) from observed relaxed clusters and hydrodynamical simulations. The latter include a selection of relaxed clusters accomplished using the same X-ray morphology criteria as the observed clusters. The consistency shown here is a powerful test of ΛCDM.

been confirmed for some years [299,452,528,545], evolution with redshift was only recently detected and remains relatively poorly constrained with current samples of relaxed clusters [116] (Fig 55c). Additional constraints can be obtained from the observed shapes of halos, as probed by the ellipticity of their X-ray emission as a function of radius [460].

**Identifying Relaxed Clusters**

The best current probe of the dynamical state of a galaxy cluster is the morphology of its ICM, as revealed by high spatial resolution X-ray imaging. The high sensitivity of X-ray emissivity to the ICM density can reveal a variety of astrophysical features, in particular shock and cold fronts [312], gas sloshing [22], and the cool, dense cores found in some clusters [147]. The most important morphological features historically used in both qualitative and quantitative support of a relaxed dynamical state can broadly be described as [47,74,238,302,341,363,408,443]:

1. Presence of a centrally located cool core – while this feature is frequently associated with localized gas disturbances due to feedback processes associated with AGN [146,325], observationally, their presence anti-correlates with large-scale, disruptive merger activity [75,212,230,302,335,427,471].
2. Elliptical symmetry on intermediate-to-large scales, reflecting an absence of cold fronts, large-scale sloshing or other merger-driven substructure.

In particular, the Symmetry-Peakiness-Alignment (SPA) metrics implementing these features [302] have been used to identify a highly relaxed sample used in recent cosmological work [116,299,301,304,565].

An eventual goal in this area is to validate and refine all such metrics against a "ground truth" established by robust hydrodynamic simulations, which span a range of cluster masses and redshifts, subgrid physics, cosmologies, and observing conditions. While such tests remain beyond the state of the art, it is encouraging that mock X-ray images of clusters from The Three Hundred Project simulations [116], identified as relaxed by the SPA algorithm, have, on average, had more time to recover from their last major merger than others. Along with empirical evidence such as the reduced intrinsic scatter in $f_{\rm gas}$

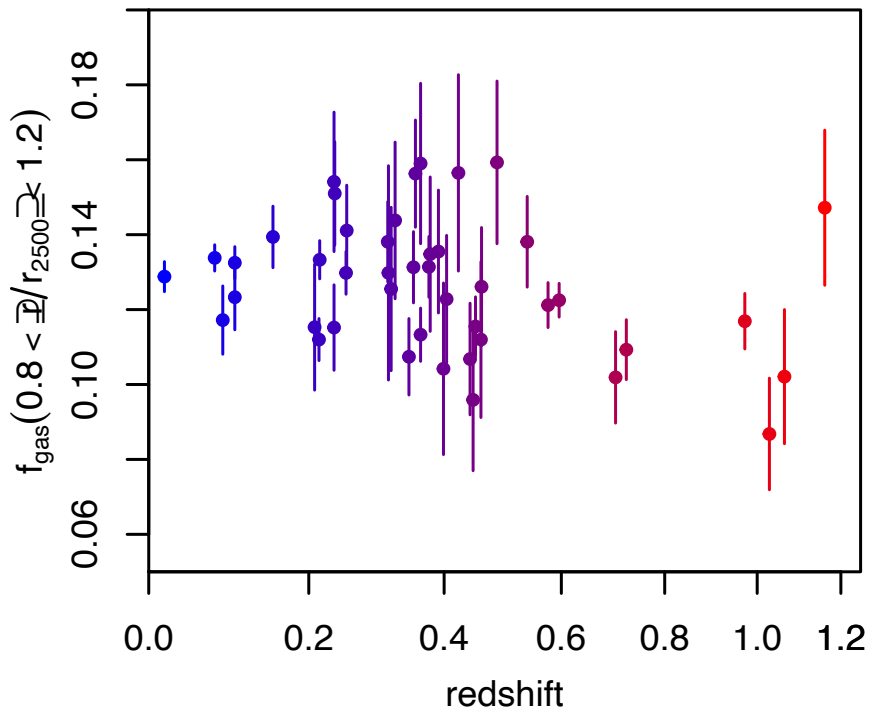

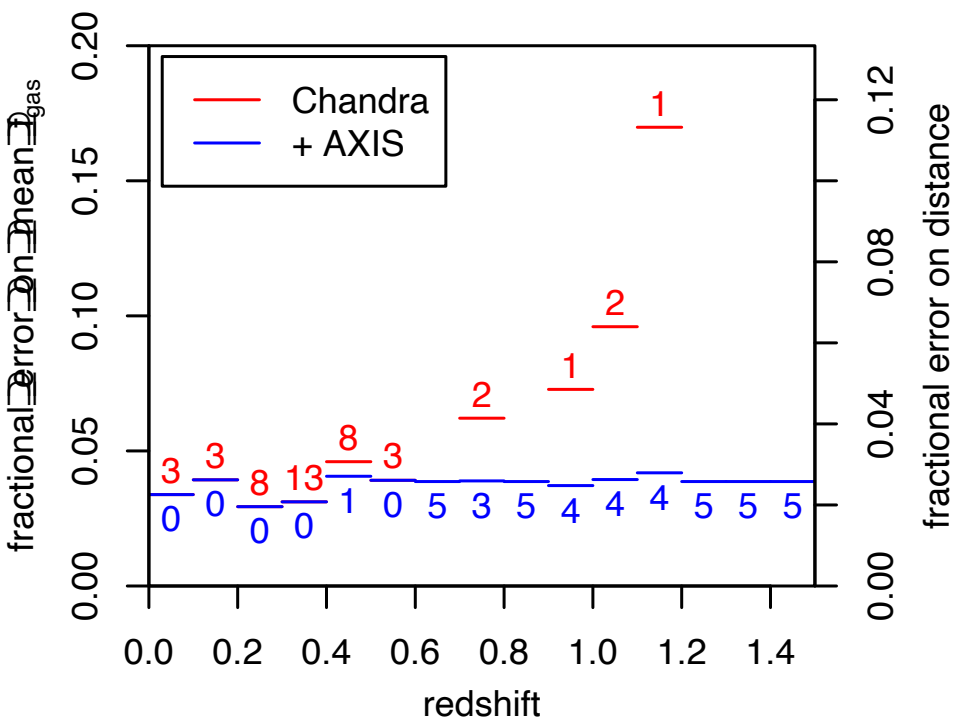

**Figure 56.** Left: Present constraints on $f_{gas}(z)$ from relaxed clusters [304], leading to the cosmological parameter constraints in Fig. 55. Right: The fractional error on the mean $f_{gas}$ value, and corresponding uncertainty in cosmic distance, in redshift bins of $\Delta z = 0.1$ from the current data are shown with red lines., with the number of clusters contributing to each bin given above each one. Blue numbers show the number of additional clusters in each bin, with $f_{gas}$ individually measured to 7.5%, we consider for our projected future constraints. The resulting mean $f_{gas}$ (distance) uncertainty in each bin (blue lines) is approximately uniform at 4% (2.5%).

for SPA-selected relaxed clusters compared with by-eye selection [301], this work supports the notion that X-ray morphological indicators grounded in physical intuition are adept at selecting dynamically relaxed clusters. Arcsecond angular resolution such as AXIS provides is a requirement for robustly measuring features like those described above, both due to the relatively small scales and sharp features involved, but also to prevent contaminating emission from point-like sources from washing out the ICM morphological signal. At the redshifts of greatest interest for cosmology, where clusters subtend relatively smaller angles, this consideration takes on even greater importance.

**Projections for Future $f_{gas}$ Data**

The impact of an enlarged sample of relaxed clusters of dark energy constraints from $f_{gas}$ in particular was studied by [9,301]. We have produced projected constraints following their procedure, updating the current data set to which new cluster observations are added to reflect the present state of the field. Fig. 56a shows this current data set [304], which becomes relatively sparse at $z \gtrsim 0.6$. Noticing that the existing data constrain the mean $f_{gas}$ in redshift bins approximately uniformly at the $\sim 4\%$ level (corresponding to $\sim 2.5\%$ in distance) out to $z = 0.6$ (Fig. 56b), we consider a future sample where AXIS observations have been used to identify and follow up relaxed clusters to achieve the same precision out to $z = 1.5$. Assuming the new observations are designed to provide a precision of 7.5% on individual $f_{gas}$ measurements, the number of new clusters this entails in each redshift bin (41 in total) is shown in blue in Fig. 56b, for comparison with the current sample (red). Note that, so long as the individual measurement precisions are larger than the intrinsic scatter, the projected results depend only on how much exposure time is spent on clusters at a given redshift, not on how the time is divided among those clusters (i.e. their individual precisions) in detail.

The projected constraints (blue shading) are compared with current results (red) in Fig. 57. These forecasts assume some improvements in simulation-based priors on the depletion parameter which remain forthcoming (see [9,301]), but that this assumption is largely important only for the $\Omega_m$ constraints shown; for the dark energy parameters, the improvement is driven by the expansion of the data set at intermediate-to-high redshifts. For each of the models in Fig. 57, the area of the parameter constraints shown shrink by a factor of 8–9 compared with current results. We note that no attempt to optimize the

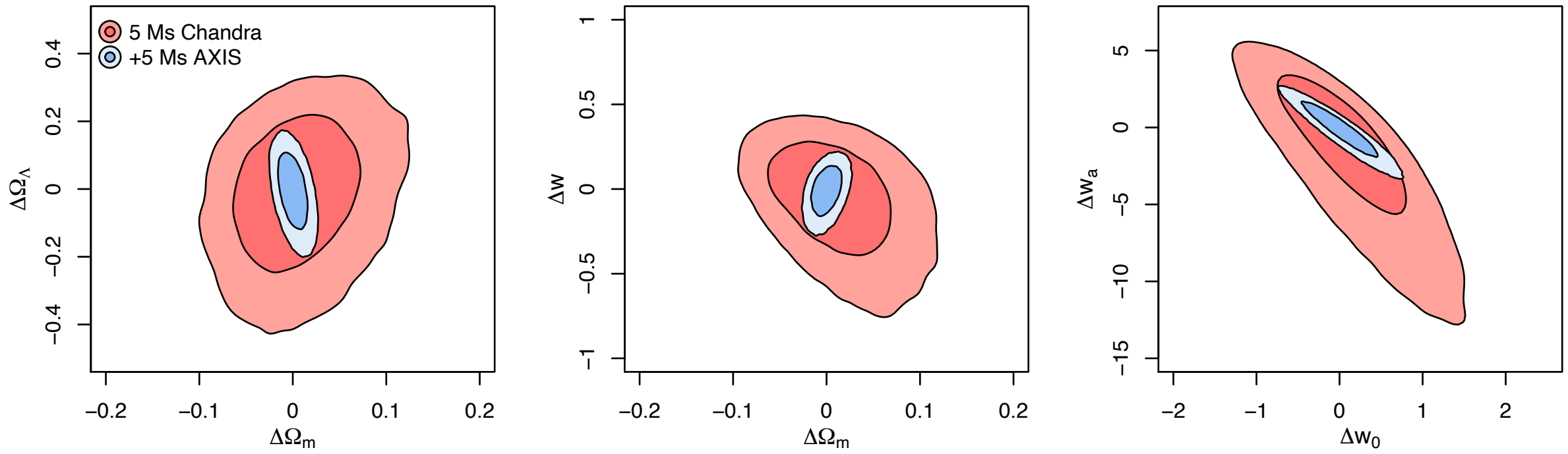

**Figure 57.** Constraints on non-flat ΛCDM (left), flat constant-$w$ (center) and flat, evolving-$w$ models from current $f_{\rm gas}$ data (red [304]), and forecasts including 41 additional relaxed clusters with redshifts and precision shown in Fig. 56b. The axes show the difference in parameter values from the fiducial model for which new observations are simulated.

redshift distribution or relative exposure times of the new clusters for any particular model has been attempted here; this exercise is only intended to provide a broad idea of the scale of possible improvement.

**Projections for Future X-ray+SZ Data**
To the extent that we are only interested in estimating $h$ from the combination of X-ray and SZ data, the fractional precision of future constraints can be simply estimated as $2(\sigma_{\rm sys}^2 + (\sigma_{\rm int}^2 + \sigma_{\rm stat}^2)/N)^{1/2}$, where $N$ is the number of clusters observed, $\sigma_{\rm int}$ is the intrinsic scatter in the measured ratio $y^{\rm ref}_{\rm X-ray}/y_{\rm SZ}$, $\sigma_{\rm stat}$ is the measurement error per cluster, and $\sigma_{\rm sys}$ is the systematic error budget, which is dominated by uncertainty in the absolute calibration of X-ray temperature measurements [9,565]. Given a path to reduce $\sigma_{\rm sys}$ to $\sim 0.01$ by calibrating X-ray temperatures to those measured through the relativistic SZ effect, the systematic floor for $h$ is 2%. If we adopt $\sigma_{\rm int} \sim 0.13$ from current data [565] and require a comparable $\sigma_{\rm stat} = 0.15$, a 2.5% constraint on $h$ would require $\sim 700$ cluster observations (which could include data acquired for the $f_{\rm gas}$ project). While this sample is considerably larger than the one considered in the previous section, the required number of relaxed clusters should be available at redshifts $< 0.45$, even if we restrict new clusters to $z > 0.3$ and adopt a conservatively low relaxed cluster fraction of 5% [302]. The combined constraint on $h$ from this method and the low-redshift $f_{\rm gas}$ data plus baryon fraction from the CMB (assuming improved simulations of the depletion parameter, as in the previous section) would achieve a precision of 1.5%, an invaluable complement to other cosmological estimates.

**Observing Description:** The best targets generally for cosmological work using massive clusters at intermediate-to-high redshifts (for the $f_{\rm gas}$ test) will be selected from the upcoming mm and optical/IR surveys, most notably by the Simons Observatory, CMB-S4, Rubin/LSST, *Euclid* and *Roman*. While these data will vastly expand the catalog of known clusters at such redshifts, they do not directly provide a selection of *relaxed* clusters. However, the combination of galaxy photometry and low-resolution SZ data (sufficient to measure the center of the ICM, and supplemented by eROSITA X-ray information where possible) can effectively identify the most promising *candidate* relaxed clusters, with a purity of $\sim$ 40–60% [82] compared with $< 10\%$ odds of choosing a relaxed cluster at random [302]. Low-frequency radio data from LOFAR and SKA may provide additional refinements, e.g. by revealing the presence of radio halos or relics. At lower redshifts (for the X-ray+SZ test) such a pre-selection is already in principle possible using existing data.

A future AXIS observing program for relaxed cluster cosmology would thus consist of two phases: short observations of relaxed cluster candidates identified from existing data to provide a bona fide relaxed

sample, followed by deeper observations of that sample to measure the ICM density and temperature as a function of radius, from which the cosmologically relevant signals can be extracted. The short observations require only $\sim 1000$ counts to robustly probe a cluster's dynamical state, compared with more than an order of magnitude more for precise measurements of temperature and mass; assuming a pre-selection of relaxed candidates with 50% purity, the first phase thus represents a small fraction of the total time investment.

Based on the empirical relationship between *Chandra* exposure time, cluster redshift and the resulting $f_{\rm gas}$ constraints from [9], and the increased effective area of AXIS compared with *Chandra*, the fiducial set of new observations described in Sec. 35 would require $\sim 5$ Ms. As noted, the redshift distribution and target $f_{\rm gas}$ precision as a function of redshift have not been optimized for any particular cosmological model for these projections; such an exercise is beyond our scope here, but could plausibly reduce the total exposure requirements for a given improvement in cosmological parameters.

In contrast to $f_{\rm gas}$, the constraints on $h$ from combining X-ray and SZ data can be accomplished using clusters at any redshift, and are thus most efficiently obtained from relatively low redshift clusters. The intrinsic scatter in this probe is relatively higher, however. The strategy for this test would thus naturally involve observing to shallower depth a larger sample of lower redshift clusters selected from SZ surveys (since the intrinsic scatter also limits the utility of more precise individual measurements). For a fiducial sample of 700 relatively low-redshift clusters measured with $\sigma_{\rm stat} = 0.15$ (Sec. 35), the required exposure time is only 1–2 ks per cluster with AXIS, and $< 1.5$ Ms in total. We note that the target of $\sigma_{\rm stat} = 0.15$ represents a few thousand counts per cluster, the sort of observations that have been widely used for "basic characterization" of clusters with Chandra and XMM-Newton, such that much of the required data could naturally become available through more general efforts to survey various cluster populations. This is particularly true if the requirement of dynamical relaxation can be loosened for the X-ray+SZ cosmology project, although the effect that this would have on the intrinsic scatter of the observable has yet to be studied in detail.

**Joint Observations and synergies with other observatories in the 2030s:** Simons Observatory, CMB-S4, Rubin, Euclid, Roman, LOFAR, SKA (as discussed above).

**Special Requirements:** Low and well-understood background

**Acknowledgments:** We thank everyone who has contributed to the development of the AXIS Probe mission.

## References:

1. Abazajian, K., Addison, G., Adshead, P., et al. 2019, arXiv e-prints, arXiv:1907.04473
2. Abbott, T. M. C., Aguena, M., Alarcon, A., et al. 2022, Ph. Rev. D, 105, 023520
3. Abriola, D., Della Pergola, D., Lombardi, M., et al. 2024, A&A, 684, A193
4. Adamcová, B., Svoboda, J., Kyritsis, E., et al. 2024, A&A, 691, A27
5. Akamatsu, H., Hoshino, A., Ishisaki, Y., et al. 2011, PASJ, 63, S1019
6. Akerman, N., Tonnesen, S., Poggianti, B. M., Smith, R., & Marasco, A. 2023, ApJ, 948, 18
7. Alam, S., Aubert, M., Avila, S., et al. 2021, Ph. Rev. D, 103, 083533
8. Allen, S. W., Evrard, A. E., & Mantz, A. B. 2011, ARA&A, 49, 409
9. Allen, S. W., Mantz, A. B., Morris, R. G., et al. 2013, arXiv:1307.8152, arXiv:1307.8152 [astro-ph.CO]
10. Allen, S. W., Rapetti, D. A., Schmidt, R. W., et al. 2008, MNRAS, 383, 879
11. Allen, S. W., Schmidt, R. W., Ebeling, H., Fabian, A. C., & van Speybroeck, L. 2004, MNRAS, 353, 457
12. Alshino, A., Ponman, T., Pacaud, F., & Pierre, M. 2010, MNRAS, 407, 2543
13. Amorín, R. O., Rodríguez-Henríquez, M., Fernández, V., et al. 2024, A&A, 682, L25
14. Amorín, R. O., et al. 2010, ApJL, 715, L128
15. Andreon, S., Moretti, A., Böhringer, H., & Castagna, F. 2023, MNRAS, 519, 2366
16. Ansarifard, S., Rasia, E., Biffi, V., et al. 2020, A&A, 634, A113
17. Appleton, P. N., Guillard, P., Boulanger, F., et al. 2013, ApJ, 777, 66
18. Appleton, P. N., Guillard, P., Togi, A., et al. 2017, ApJ, 836, 76
19. Appleton, P. N., Guillard, P., Emonts, B., et al. 2023, ApJ, 951, 104
20. Arnaud, M., Pointecouteau, E., & Pratt, G. W. 2005, A&A, 441, 893
21. Arnaudova, M. I., Das, S., Smith, D. J. B., et al. 2024, MNRAS, 535, 2269
22. Ascasibar, Y., & Markevitch, M. 2006, ApJ, 650, 102
23. Avestruz, C., Nagai, D., Lau, E. T., & Nelson, K. 2015, ApJ, 808, 176
24. Baes, M., Clemens, M., Xilouris, E., et al. 2010, Astronomy & Astrophysics, 518, L53
25. Baldassare, V. F., Geha, M., & Greene, J. 2020, ApJ, 896, 10
26. Baldassare, V. F., Reines, A. E., Gallo, E., & Greene, J. E. 2015, ApJ, 809, L14
27. Barkana, R., & Loeb, A. 2001, Physics Reports, 349, 125
28. Barnes, J. E., & Hernquist, L. 1992, ARA&A, 30, 705
29. Bartolini, C., Ignesti, A., Gitti, M., et al. 2022, ApJ, 936, 74
30. Basu-Zych, A. R., et al. 2013, ApJ, 774, 152
31. Batalha, R. M., Dupke, R. A., & Jiménez-Teja, Y. 2022, The Astrophysical Journal Supplement Series, 262, 27
32. Benitez, N., Dupke, R., Moles, M., et al. 2014, arXiv e-prints, arXiv:1403.5237
33. Berdina, L., & Tsvetkova, V. S. 2018, arXiv e-prints, arXiv:1801.05650
34. Best, P. N., von der Linden, A., Kauffmann, G., Heckman, T. M., & Kaiser, C. R. 2007, MNRAS, 379, 894
35. Bianchi, S., Giovanardi, C., Smith, M., et al. 2017, Astronomy & Astrophysics, 597, A130
36. Biava, N., Bonafede, A., Gastaldello, F., et al. 2024, A&A, 686, A82
37. Biffi, V., Mernier, F., & Medvedev, P. 2018, Space Science Reviews, 214, 123
38. Biffi, V., Planelles, S., Borgani, S., et al. 2018, MNRAS, 476, 2689
39. Birkinshaw, M., Hughes, J. P., & Arnaud, K. A. 1991, ApJ, 379, 466
40. Bîrzan, L., Rafferty, D. A., McNamara, B. R., Wise, M. W., & Nulsen, P. E. J. 2004, ApJ, 607, 800
41. Bîrzan, L., Rafferty, D. A., Nulsen, P. E. J., et al. 2012, MNRAS, 427, 3468
42. Biviano, A., Poggianti, B. M., Jaffé, Y., et al. 2024, ApJ, 965, 117
43. Blanton, E. L., Randall, S. W., Clarke, T. E., et al. 2011, ApJ, 737, 99
44. Blum, B., Digel, S. W., Drlica-Wagner, A., et al. 2022, arXiv e-prints, arXiv:2203.07220
45. Bogdán, Á., & Gilfanov, M. 2008, MNRAS, 388, 56

46. Böhringer, H., Schuecker, P., Pratt, G. W., et al. 2007, A&A, 469, 363
47. Böhringer, H., Pratt, G. W., Arnaud, M., et al. 2010, A&A, 514, A32
48. Bonafede, A., Brüggen, M., Rafferty, D., et al. 2018, MNRAS, 478, 2927
49. Bonafede, A., Brunetti, G., Rudnick, L., et al. 2022, ApJ, 933, 218
50. Bonafede, A., Gitti, M., La Bella, N., et al. 2023, A&A, 680, A5
51. Bonamente, M., Joy, M. K., LaRoque, S. J., et al. 2006, ApJ, 647, 25
52. Bonoli, S., Marín-Franch, A., Varela, J., et al. 2021, A&A, 653, A31
53. Boorman, P. G., Svoboda, J., Stern, D., et al. 2025, arXiv e-prints, arXiv:2505.08885
54. Borgani, S., & Kravtsov, A. 2011, Advanced Science Letters, 4, 204
55. Boselli, A., Fossati, M., & Sun, M. 2022, A&A Rev., 30, 3
56. Boselli, A., Cuillandre, J. C., Fossati, M., et al. 2016, A&A, 587, A68
57. Botteon, A., Markevitch, M., van Weeren, R. J., Brunetti, G., & Shimwell, T. W. 2023, A&A, 674, A53
58. Botteon, A., Giacintucci, S., Gastaldello, F., et al. 2021, A&A, 649, A37
59. Botteon, A., van Weeren, R. J., Brunetti, G., et al. 2022, Sci. Adv., 8, eabq7623
60. Botteon, A., Gastaldello, F., ZuHone, J. A., et al. 2024, MNRAS, 527, 919
61. Botteon, A., van Weeren, R. J., Eckert, D., et al. 2024, A&A, 690, A222
62. Bourne, M. A., & Yang, H.-Y. K. 2023, Galaxies, 11, 73
63. Brandt, W. N., Ni, Q., Yang, G., et al. 2018, arXiv e-prints, arXiv:1811.06542
64. Bregman, J. N. 1980, ApJ, 236, 577
65. Bregman, J. N., Hodges-Kluck, E., Qu, Z., et al. 2022, ApJ, 928, 14
66. Brighenti, F., & Mathews, W. G. 2006, ApJ, 643, 120
67. Brodwin, M., McDonald, M., Gonzalez, A. H., et al. 2016, ApJ, 817, 122
68. Brorby, M., Kaaret, P., et al. 2016, MNRAS, 457, 4081
69. Brüggen, M., Heinz, S., Roediger, E., Ruszkowski, M., & Simionescu, A. 2007, MNRAS, 380, L67
70. Brunetti, G., & Jones, T. W. 2014, Int. J. Mod. Phys. D, 23, 30007
71. Brunker, S., et al. 2020, ApJ, 898, 68
72. Bulbul, E., Liu, A., Kluge, M., et al. 2024, A&A, 685, A106
73. Bullock, J. S., Kolatt, T. S., Sigad, Y., et al. 2001, MNRAS, 321, 559
74. Buote, D. A., & Tsai, J. C. 1995, ApJ, 452, 522
75. Burns, J. O., Hallman, E. J., Gantner, B., Motl, P. M., & Norman, M. L. 2008, ApJ, 675, 1125
76. Calzadilla, M. S., Bleem, L. E., McDonald, M., et al. 2023, ApJ, 947, 44
77. Campitiello, M. G., Ignesti, A., Gitti, M., et al. 2021, ApJ, 911, 144
78. Canning, R., Fabian, A., Johnstone, R., et al. 2011, Monthly Notices of the Royal Astronomical Society, 417, 3080
79. Cao, X. 2000, A&A, 355, 44
80. Cappelluti, N., Foord, A., Marchesi, S., et al. 2024, Universe, 10, 276
81. Cardamone, C., et al. 2008, MNRAS, 399, 1191
82. Casas, M. C., Putnam, K., Mantz, A. B., Allen, S. W., & Somboonpanyakul, T. 2024, ApJ, 967, 14
83. Cashman, F. H., Fox, A. J., Savage, B. D., et al. 2021, ApJ, 923, L11
84. Cassano, R., Ettori, S., Giacintucci, S., et al. 2010, ApJ, 721, L82
85. Cavagnolo, K. W., McNamara, B. R., Wise, M. W., et al. 2011, ApJ, 732, 71
86. Cayatte, V., Van Gorkom, J., Balkowski, C., & Kotanyi, C. 1990, Astronomical Journal (ISSN 0004-6256), vol. 100, Sept. 1990, p. 604-634., 100, 604
87. Cen, R., Pop, A. R., & Bahcall, N. A. 2014, Proceedings of the National Academy of Science, 111, 7914
88. Chadayammuri, U., Bogdán, Á., Oppenheimer, B. D., et al. 2022, ApJ, 936, L15
89. Chadayammuri, U., Bogdán, Á., Schellenberger, G., & ZuHone, J. 2024, arXiv e-prints, arXiv:2407.03142
90. Chadayammuri, U., ZuHone, J., Nulsen, P., et al. 2022, MNRAS, 509, 1201
91. Chartas, G., Cappi, M., Vignali, C., et al. 2023, in AAS/High Energy Astrophysics Division, Vol. 20, AAS/High Energy Astrophysics Division, 300.01
92. Chartas, G., Cappi, M., Vignali, C., et al. 2021, ApJ, 920, 24

93. Chen, B., Dai, X., Kochanek, C. S., et al. 2011, ApJ, 740, L34
94. Cheng, C., Xu, C. K., Appleton, P. N., et al. 2023, ApJ, 954, 74
95. Chevalier, R. A., & Clegg, A. W. 1985, Nature, 317, 44
96. CHEX-MATE Collaboration, Arnaud, M., Ettori, S., et al. 2021, A&A, 650, A104
97. Chibueze, J. O., Sakemi, H., Ohmura, T., et al. 2021, Nature, 593, 47
98. Chisholm, J., Tremonti, C., & Leitherer, C. 2018, MNRAS, 481, 1690
99. Choudhury, P. P., & Reynolds, C. S. 2025, MNRAS, 537, 3194
100. Choudhury, T. R., & Ferrara, A. 2006, MNRAS, 371, L55
101. Churazov, E., Forman, W., Jones, C., & Böhringer, H. 2003, ApJ, 590, 225
102. Churazov, E., Khabibullin, I., Lyskova, N., Sunyaev, R., & Bykov, A. M. 2021, A&A, 651, A41
103. Churazov, E., Sunyaev, R., Forman, W., & Böhringer, H. 2002, MNRAS, 332, 729
104. Cielo, S., Babul, A., Antonuccio-Delogu, V., Silk, J., & Volonteri, M. 2018, A&A, 617, A58
105. Clowe, D., Bradač, M., Gonzalez, A. H., et al. 2006, ApJ, 648, L109
106. Cluver, M. E., Appleton, P. N., Boulanger, F., et al. 2010, ApJ, 710, 248
107. Collaboration, H. 2016, Nature, 535, 117
108. Cooper, J. L., Bicknell, G. V., Sutherland, R. S., & Bland-Hawthorn, J. 2009, ApJ, 703, 330
109. Crawford, C., Allen, S., Ebeling, H., Edge, A., & Fabian, A. 1999, Monthly Notices of the Royal Astronomical Society, 306, 857
110. Crawford, C. S., Hatch, N., Fabian, A., & Sanders, J. 2005, Monthly Notices of the Royal Astronomical Society, 363, 216
111. Cucciati, O., Lemaux, B. C., Zamorani, G., et al. 2018, A&A, 619, A49
112. Cui, W. 2024, arXiv e-prints, arXiv:2406.03829
113. Cui, W., Jennings, F., Dave, R., Babul, A., & Gozaliasl, G. 2024, MNRAS, 534, 1247
114. Curti, M., Maiolino, R., Curtis-Lake, E., et al. 2024, A&A, 684, A75
115. Dai, X., Chartas, G., Agol, E., Bautz, M. W., & Garmire, G. P. 2003, ApJ, 589, 100
116. Darragh-Ford, E., Mantz, A. B., Rasia, E., et al. 2023, MNRAS, 521, 790
117. Das, S., Chiang, Y.-K., & Mathur, S. 2023, ApJ, 951, 125
118. Dauser, T., Falkner, S., Lorenz, M., et al. 2019, A&A, 630, A66
119. de Graaff, A., Cai, Y.-C., Heymans, C., & Peacock, J. A. 2019, A&A, 624, A48
120. De Grandi, S., Eckert, D., Molendi, S., et al. 2016, A&A, 592, A154
121. Domínguez-Fernández, P., ZuHone, J., Weinberger, R., et al. 2024, ApJ, 977, 221
122. Donahue, M., & Voit, G. M. 2022, Phys. Rep., 973, 1
123. Douna, V. M., Pellizza, L. J., Mirabel, I. F., & Pedrosa, S. E. 2015, A&A, 579, A44
124. Dressler, A. 1978, Astrophysical Journal, Part 1, vol. 223, Aug. 1, 1978, p. 765-775, 777, 779-787., 223, 765
125. Dressler, A. 1980, ApJ, 236, 351
126. Dressler, A., & Shectman, S. A. 1988, AJ, 95, 985
127. Dressler, A., et al. 2015, ApJ, 806, 19
128. Drissen, L., Martin, T., Rousseau-Nepton, L., et al. 2019, Monthly Notices of the Royal Astronomical Society, 485, 3930
129. Duarte Puertas, S., Iglesias-Páramo, J., Vilchez, J. M., et al. 2019, A&A, 629, A102
130. Dupke, R. A., Jimenez-Teja, Y., Su, Y., et al. 2022, ApJ, 936, 59
131. Dursi, L. J., & Pfrommer, C. 2008, ApJ, 677, 993
132. Eckert, D., Ettori, S., Pointecouteau, E., et al. 2017, Astronomische Nachrichten, 338, 293
133. Eckert, D., Gaspari, M., Gastaldello, F., Le Brun, A. M. C., & O'Sullivan, E. 2021, Universe, 7, 142
134. Eckert, D., Gastaldello, F., O'Sullivan, E., et al. 2024, Galaxies, 12, 24
135. Eckert, D., Jauzac, M., Shan, H., et al. 2015, Nature, 528, 105
136. Einasto, M., Vennik, J., Nurmi, P., et al. 2012, A&A, 540, A123
137. Eke, V. R., Frenk, C. S., Baugh, C. M., et al. 2004, MNRAS, 355, 769
138. Emonts, B. H. C., Appleton, P. N., Lisenfeld, U., et al. 2025, ApJ, 978, 111

139. Ettori, S. 2000, MNRAS, 311, 313
140. Euclid Collaboration, Moneti, A., McCracken, H. J., et al. 2022, A&A, 658, A126
141. Euclid Collaboration, Zalesky, L., McPartland, C. J. R., et al. 2024, arXiv e-prints, arXiv:2408.05296
142. Everett, J. E., Zweibel, E. G., Benjamin, R. A., et al. 2008, ApJ, 674, 258
143. Fabbiano, G. 1989, ARA&A, 27, 87
144. Fabbiano, G., Krauss, M., Zezas, A., Rots, A., & Neff, S. 2003, ApJ, 598, 272
145. Fabian, A., Sanders, J., Allen, S., et al. 2003, Monthly Notices of the Royal Astronomical Society, 344, L43
146. Fabian, A. C. 2012, ARA&A, 50, 455
147. Fabian, A. C., Crawford, C. S., Edge, A. C., & Mushotzky, R. F. 1994, MNRAS, 267, 779
148. Fabian, A. C., Johnstone, R. M., Sanders, J. S., et al. 2008, Nature, 454, 968
149. Fabian, A. C., Sanders, J. S., Allen, S. W., et al. 2003, MNRAS, 344, L43
150. Fabian, A. C., Sanders, J. S., Taylor, G. B., et al. 2006, MNRAS, 366, 417
151. Fabian, A. C., ZuHone, J. A., & Walker, S. A. 2022, MNRAS, 510, 4000
152. Fabian, A. C., Sanders, J. S., Ettori, S., et al. 2000, MNRAS, 318, L65
153. Fabian, A. C., Sanders, J. S., Allen, S. W., et al. 2011, MNRAS, 418, 2154
154. Fabian, A. C., Walker, S., Russell, H., et al. 2016, Monthly Notices of the Royal Astronomical Society, 461, 922
155. Fabjan, D., Borgani, S., Tornatore, L., et al. 2010, MNRAS, 401, 1670
156. Farage, C., McGregor, P., Dopita, M., & Bicknell, G. 2010, The Astrophysical Journal, 724, 267
157. Farrell, S. A., Webb, N. A., Barret, D., Godet, O., & Rodrigues, J. M. 2009, Nature, 460, 73
158. Faucher-Giguère, C.-A., & Quataert, E. 2012, MNRAS, 425, 605
159. Fedorova, E. V., Zhdanov, V. I., Vignali, C., & Palumbo, G. G. C. 2008, A&A, 490, 989
160. Fedotov, K., Gallagher, S. C., Konstantopoulos, I. S., et al. 2011, AJ, 142, 42
161. Feretti, L., Giovannini, G., Govoni, F., & Murgia, M. 2012, A&A Rev., 20, 54
162. Feruglio, C., Fiore, F., Carniani, S., et al. 2015, A&A, 583, A99
163. Fielding, D., Quataert, E., Martizzi, D., & Faucher-Giguère, C.-A. 2017, MNRAS, 470, L39
164. Finner, K., James Jee, M., Webb, T., et al. 2020, ApJ, 893, 10
165. Flury, S. R., Jaskot, A. E., Ferguson, H. C., et al. 2022, ApJS, 260, 1
166. Forman, W., Churazov, E., Jones, C., et al. 2017, ApJ, 844, 122
167. Forman, W., Kellogg, E., Gursky, H., Tananbaum, H., & Giacconi, R. 1972, ApJ, 178, 309
168. Fragos, T., Lehmer, B. D., Naoz, S., Zezas, A., & Basu-Zych, A. 2013, ApJ, 776, L31
169. Freundlich, J., & Maoz, D. 2021, MNRAS, 502, 5882
170. Fujimoto, S., Coe, D., Abdurro'uf, A., et al. 2025, Vast Exploration for Nascent, Unexplored Sources (VENUS), JWST Proposal. Cycle 4, ID. #6882
171. Galametz, A., Stern, D., De Breuck, C., et al. 2012, The Astrophysical Journal, 749, 169
172. Gao, L., Navarro, J. F., Cole, S., et al. 2008, MNRAS, 387, 536
173. Garofali, K., Lehmer, B. D., Basu-Zych, A., et al. 2020, ApJ, 903, 79
174. Gaspari, M., Melioli, C., Brighenti, F., & D'Ercole, A. 2011, MNRAS, 411, 349
175. Gaspari, M., Ruszkowski, M., & Sharma, P. 2012, ApJ, 746, 94
176. Gaspari, M., Temi, P., & Brighenti, F. 2017, MNRAS, 466, 677
177. Gaspari, M., Tombesi, F., & Cappi, M. 2020, Nature Astronomy, 4, 10
178. Gaspari, M., McDonald, M., Hamer, S. L., et al. 2018, ApJ, 854, 167
179. Gastaldello, F., Simionescu, A., Mernier, F., et al. 2021, Universe, 7, 208
180. Ge, C., Luo, R., Sun, M., et al. 2021, MNRAS, 505, 4702
181. Gendron-Marsolais, M., Hlavacek-Larrondo, J., van Weeren, R. J., et al. 2017, MNRAS, 469, 3872
182. Gendron-Marsolais, M., Hlavacek-Larrondo, J., Martin, T., et al. 2018, MNRAS: Letters, 479, L28
183. Gendron-Marsolais, M.-L., Hlavacek-Larrondo, J., van Weeren, R. J., et al. 2020, MNRAS, 499, 5791
184. Genzel, R., Eisenhauer, F., & Gillessen, S. 2010, Reviews of Modern Physics, 82, 3121
185. Ghirardini, V., Bulbul, E., Kraft, R., et al. 2021, ApJ, 910, 14
186. Ghizzardi, S., Molendi, S., van der Burg, R., et al. 2021, A&A, 646, A92

187. Giacintucci, S., Markevitch, M., Brunetti, G., et al. 2014, ApJ, 795, 73
188. Giacintucci, S., Markevitch, M., Cassano, R., et al. 2019, ApJ, 880, 70
189. Giacintucci, S., Markevitch, M., Venturi, T., et al. 2014, ApJ, 781, 9
190. Giacintucci, S., Venturi, T., Markevitch, M., et al. 2024, ApJ, 961, 133
191. Gilbertson, W., et al. 2022, ApJ, 926, 28
192. Giodini, S., Lovisari, L., Pointecouteau, E., et al. 2013, Space Science Reviews, 177, 247
193. Gitti, M., Brighenti, F., & McNamara, B. R. 2012, Advances in Astronomy, 2012, 950641
194. Gitti, M., Feretti, L., & Schindler, S. 2006, A&A, 448, 853
195. Gitti, M., Nulsen, P. E. J., David, L. P., McNamara, B. R., & Wise, M. W. 2011, ApJ, 732, 13
196. Gitti, M., Bonafede, A., Brighenti, F., et al. 2025, arXiv e-prints, arXiv:2503.13735
197. Giunchi, E., Gullieuszik, M., Poggianti, B. M., et al. 2023, ApJ, 949, 72
198. Gong, Y., Miao, H., Zhou, X., et al. 2025, arXiv e-prints, arXiv:2501.15023
199. González-Martín, O., Masegosa, J., Márquez, I., Guerrero, M. A., & Dultzin-Hacyan, D. 2006, A&A, 460, 45
200. Gordon, S. M., Kirshner, R. P., Long, K. S., et al. 1998, ApJS, 117, 89
201. Grandi, P., Torresi, E., Macconi, D., Boccardi, B., & Capetti, A. 2021, ApJ, 911, 17
202. Grimm, H.-J., Gilfanov, M., & Sunyaev, R. 2003, MNRAS, 339, 793
203. Gronke, M., & Oh, S. P. 2020, MNRAS, 492, 1970
204. Guillard, P., Boulanger, F., Pineau des Forets, G., & Appleton, P. N. 2009, A&A, 502, 515
205. Gunn, J. E., & Gott, III, J. R. 1972, ApJ, 176, 1
206. Gupta, A., Mathur, S., Kingsbury, J., Das, S., & Krongold, Y. 2023, Nature Astronomy, 7, 799
207. Hashimoto, Y., Böhringer, H., Henry, J. P., Hasinger, G., & Szokoly, G. 2007, A&A, 467, 485
208. Hatch, N. A., Kurk, J. D., Pentericci, L., et al. 2011, MNRAS, 415, 2993
209. Hazumi, M., Ade, P. A. R., Adler, A., et al. 2020, in Society of Photo-Optical Instrumentation Engineers (SPIE) Conference Series, Vol. 11443, Space Telescopes and Instrumentation 2020: Optical, Infrared, and Millimeter Wave, ed. M. Lystrup & M. D. Perrin, 114432F
210. Heinz, S., & Churazov, E. 2005, ApJ, 634, L141
211. Heneka, C., & Mesinger, A. 2020, MNRAS, 496, 581
212. Henning, J. W., Gantner, B., Burns, J. O., & Hallman, E. J. 2009, ApJ, 697, 1597
213. HERA Collaboration, Abdurashidova, Z., Adams, T., et al. 2023, ApJ, 945, 124
214. Hirschauer, A. S., Crouzet, N., Habel, N., et al. 2024, AJ, 168, 23
215. Hitomi Collaboration, Aharonian, F., Akamatsu, H., et al. 2017, Nature, 551, 478
216. Hlavacek-Larrondo, J., Fabian, A., Sanders, J., & Taylor, G. 2011, Monthly Notices of the Royal Astronomical Society, 415, 3520
217. Hlavacek-Larrondo, J., Gendron-Marsolais, M. L., Fecteau-Beaucage, D., et al. 2018, MNRAS, 475, 2743
218. Hlavacek-Larrondo, J., Rhea, C. L., Webb, T., et al. 2020, ApJ, 898, L50
219. Hoang, D. N., Zhang, X., Stuardi, C., et al. 2021, A&A, 656, A154
220. Hogan, M. T., Edge, A. C., Hlavacek-Larrondo, J., et al. 2015, MNRAS, 453, 1201
221. Hogan, M. T., Edge, A. C., Geach, J. E., et al. 2015, MNRAS, 453, 1223
222. Hopkins, P. F., Hernquist, L., Cox, T. J., et al. 2006, ApJS, 163, 1
223. Huang, R., Li, J.-T., Cui, W., et al. 2023, ApJS, 268, 36
224. Huchra, J., Gorenstein, M., Kent, S., et al. 1985, AJ, 90, 691
225. Hudson, D. S., Mittal, R., Reiprich, T. H., et al. 2010, A&A, 513, A37
226. Huško, F., Lacey, C. G., Schaye, J., Schaller, M., & Nobels, F. S. J. 2022, MNRAS, 516, 3750
227. Hwang, J.-S., Struck, C., Renaud, F., & Appleton, P. N. 2012, MNRAS, 419, 1780
228. HyeongHan, K., Jee, M. J., Lee, W., et al. 2025, Nature Astronomy, arXiv:2405.00115 [astro-ph.GA]
229. Ianjamasimanana, R., Verdes-Montenegro, L., Sorgho, A., et al. 2025, arXiv e-prints, arXiv:2502.09691
230. Ichinohe, Y., Werner, N., Simionescu, A., et al. 2015, MNRAS, 448, 2971
231. Ignesti, A., Brunetti, G., Gullieuszik, M., et al. 2024, ApJ, 977, 219
232. Israel, G. L., Belfiore, A., Stella, L., et al. 2017, Science, 355, 817

233. Iwamoto, K., Brachwitz, F., Nomoto, K., et al. 1999, ApJS, 125, 439
234. Izotov, Y. I., Guseva, N. G., Fricke, K. J., et al. 2021, A&A, 646, A138
235. Izotov, Y. I., Orlitová, I., Schaerer, D., et al. 2016, Nature, 529, 178
236. Japelj, J., Laigle, C., Puech, M., et al. 2019, A&A, 632, A94
237. Jauzac, M., Eckert, D., Schwinn, J., et al. 2016, MNRAS, 463, 3876
238. Jeltema, T. E., Canizares, C. R., Bautz, M. W., & Buote, D. A. 2005, ApJ, 624, 606
239. Jeltema, T. E., Mulchaey, J. S., Lubin, L. M., Rosati, P., & Böhringer, H. 2006, ApJ, 649, 649
240. Jeon, M., Pawlik, A. H., Bromm, V., & Milosavljević, M. 2014, MNRAS, 440, 3778
241. Jiang, T., et al. 2019, ApJ, 872, 145
242. Jiménez-Teja, Y., Dupke, R. A., Lopes, P. A. A., & Vílchez, J. M. 2023, A&A, 676, A39
243. Jiménez-Teja, Y., Dupke, R., Benítez, N., et al. 2018, ApJ, 857, 79
244. Jiménez-Teja, Y., Román, J., HyeongHan, K., et al. 2025, A&A, 694, A216
245. Johnson, K. E., Hibbard, J. E., Gallagher, S. C., et al. 2007, AJ, 134, 1522
246. Johnstone, R. M., Hatch, N., Ferland, G. J., et al. 2007, MNRAS, 382, 1246
247. Jones, M. G., Verdes-Montenegro, L., Damas-Segovia, A., et al. 2019, A&A, 632, A78
248. Justham, S., & Schawinski, K. 2012, MNRAS, 423, 1641
249. Kaiser, N. 1986, MNRAS, 222, 323
250. Kalberla, P. M. W., et al. 2005, Astronomy Astrophysics, 440, 775
251. Kamieneski, P. S., Yun, M. S., Harrington, K. C., et al. 2024, ApJ, 961, 2
252. Kataoka, J., Sofue, Y., Inoue, Y., et al. 2018, Galaxies, 6, 27
253. Kempner, J. C., & David, L. P. 2004, MNRAS, 349, 385
254. Kim, C.-G., & Ostriker, E. C. 2018, ApJ, 853, 173
255. Kim, C.-G., Ostriker, E. C., & Raileanu, R. 2017, ApJ, 834, 25
256. Kim, D.-W., & Fabbiano, G. 2024, ApJ, 976, 112
257. Kirkpatrick, C., McNamara, B., & Cavagnolo, K. 2011, The Astrophysical Journal Letters, 731, L23
258. Knowles, K., Cotton, W. D., Rudnick, L., et al. 2022, A&A, 657, A56
259. Kochanek, C. S., Keeton, C. R., & McLeod, B. A. 2001, ApJ, 547, 50
260. Komarov, S., Schekochihin, A. A., Churazov, E., & Spitkovsky, A. 2018, Journal of Plasma Physics, 84, 905840305
261. Komarov, S. V., Churazov, E. M., Kunz, M. W., & Schekochihin, A. A. 2016, MNRAS, 460, 467
262. Konstantopoulos, I. S., Gallagher, S. C., Fedotov, K., et al. 2010, ApJ, 723, 197
263. Kouroumpatzakis, K., & Svoboda, J. 2025, A&A, 696, A133
264. Kovlakas, K., Zezas, A., Andrews, J. J., et al. 2021, MNRAS, 506, 1896
265. Kulkarni, S. R., Harrison, F. A., Grefenstette, B. W., et al. 2021, arXiv e-prints, arXiv:2111.15608
266. Lacy, M., Surace, J. A., Farrah, D., et al. 2021, MNRAS, 501, 892
267. Lanz, L., Ogle, P. M., Evans, D., et al. 2015, ApJ, 801, 17
268. Lanzuisi, G., Matzeu, G., Baldini, P., et al. 2024, A&A, 689, A247
269. Lehmer, B., et al. 2010, ApJ, 724, 559
270. Lehmer, B. D., Monson, E. B., Eufrasio, R. T., et al. 2024, ApJ, 977, 189
271. Lepore, M., Di Mascolo, L., Tozzi, P., et al. 2024, A&A, 682, A186
272. Levy, R. C., Bolatto, A. D., Leroy, A. K., et al. 2022, ApJ, 935, 19
273. Li, J.-T., Bregman, J. N., Wang, Q. D., Crain, R. A., & Anderson, M. E. 2018, ApJ, 855, L24
274. Li, J.-T., Bregman, J. N., Wang, Q. D., et al. 2017, ApJS, 233, 20
275. Li, J.-T., Wang, F., Yang, J., et al. 2021, MNRAS, 504, 2767
276. Li, J.-T., & Wang, Q. D. 2013, MNRAS, 428, 2085
277. —. 2013, MNRAS, 435, 3071
278. Li, Y., Luo, R., Fossati, M., Sun, M., & Jáchym, P. 2023, MNRAS, 521, 4785
279. Li, Z., Jones, C., Forman, W. R., et al. 2011, ApJ, 730, 84
280. Lim, J., Ao, Y., et al. 2008, The Astrophysical Journal, 672, 252
281. Liu, W., Sun, M., Nulsen, P., et al. 2019, MNRAS, 484, 3376

282. Lopez, L. A., Mathur, S., Nguyen, D. D., Thompson, T. A., & Olivier, G. M. 2020, ApJ, 904, 152
283. Lopez, S., Lopez, L. A., Nguyen, D. D., et al. 2023, ApJ, 942, 108
284. Lovisari, L., & Maughan, B. J. 2022, in Handbook of X-ray and Gamma-ray Astrophysics, ed. C. Bambi & A. Sangangelo, 65
285. Lovisari, L., & Reiprich, T. H. 2019, MNRAS, 483, 540
286. Lovisari, L., Reiprich, T. H., & Schellenberger, G. 2015, A&A, 573, A118
287. Lucero, D. M., Carignan, C., Elson, E. C., et al. 2015, arXiv e-prints, arXiv:1504.04082
288. Luo, B., Brandt, W. N., Xue, Y. Q., et al. 2017, ApJS, 228, 2
289. Lusso, E., Risaliti, G., Nardini, E., et al. 2020, A&A, 642, A150
290. Lynds, R. 1970, Astrophysical Journal Letters, 159, L151-L154 (1970)., 159
291. Lyskova, N., Churazov, E., Khabibullin, I. I., et al. 2023, MNRAS, 525, 898
292. Lyskova, N., Churazov, E., Zhang, C., et al. 2019, MNRAS, 485, 2922
293. Lyutikov, M. 2006, MNRAS, 373, 73
294. Mac Low, M.-M., McCray, R., & Norman, M. L. 1989, ApJ, 337, 141
295. Madau, P., & Dickinson, M. 2014, ARA&A, 52, 415
296. Makarov, D., Prugniel, P., Terekhova, N., Courtois, H., & Vauglin, I. 2014, A&A, 570, A13
297. Mantz, A., Allen, S. W., Rapetti, D., & Ebeling, H. 2010, MNRAS, 406, 1759
298. Mantz, A., Allen, S. W., Battaglia, N., et al. 2019, Bulletin of the American Astronomical Society, 51, 279
299. Mantz, A. B., Allen, S. W., & Morris, R. G. 2016, MNRAS, 462, 681
300. Mantz, A. B., Allen, S. W., Morris, R. G., et al. 2020, Monthly Notices of the Royal Astronomical Society, 496, 1554
301. —. 2014, MNRAS, 440, 2077
302. —. 2015, MNRAS, 449, 199
303. Mantz, A. B., Abdulla, Z., Allen, S. W., et al. 2018, A&A, 620, A2
304. Mantz, A. B., Morris, R. G., Allen, S. W., et al. 2022, MNRAS, 510, 131
305. Marchesi, S., Gilli, R., Lanzuisi, G., et al. 2020, A&A, 642, A184
306. Marcowith, A., Bret, A., Bykov, A., et al. 2016, Reports on Progress in Physics, 79, 046901
307. Markevitch, M. 2006, in ESA Special Publication, Vol. 604, The X-ray Universe 2005, ed. A. Wilson, 723
308. Markevitch, M. 2010, arXiv e-prints, 1010.3660
309. Markevitch, M. 2021, Revealing layers of enhanced magnetic field in ICM via X-ray imaging, Chandra Proposal ID #23800485
310. Markevitch, M., Gonzalez, A. H., David, L. P., et al. 2002, ApJ, Letters, 567, L27
311. Markevitch, M., Govoni, F., Brunetti, G., & Jerius, D. 2005, ApJ, 627, 733
312. Markevitch, M., & Vikhlinin, A. 2007, Physics Rep., 443, 1
313. Markevitch, M., Bulbul, E., Churazov, E., et al. 2019, Bulletin of the American Astronomical Society, 51, 569
314. Martizzi, D., Quataert, E., Faucher-Giguère, C.-A., & Fielding, D. 2019, MNRAS, 483, 2465
315. Martizzi, D., Vogelsberger, M., Artale, M. C., et al. 2019, MNRAS, 486, 3766
316. Mathews, W. G., & Bregman, J. N. 1978, ApJ, 224, 308
317. Maughan, B. J., Giles, P. A., Randall, S. W., Jones, C., & Forman, W. R. 2012, MNRAS, 421, 1583
318. Mazzotta, P., & Giacintucci, S. 2008, ApJ, Letters, 675, L9
319. McConnell, N. J., Ma, C.-P., Gebhardt, K., et al. 2011, Nature, 480, 215
320. McDonald, M., Benson, B. A., Vikhlinin, A., et al. 2013, ApJ, 774, 23
321. McDonald, M., Bulbul, E., de Haan, T., et al. 2016, ApJ, 826, 124
322. McDonald, M., Allen, S. W., Bayliss, M., et al. 2017, ApJ, 843, 28
323. McKee, C. F., & Cowie, L. L. 1977, ApJ, 215, 213
324. McKee, C. F., & Ostriker, J. P. 1977, ApJ, 218, 148
325. McNamara, B. R., & Nulsen, P. E. J. 2007, ARA&A, 45, 117
326. —. 2012, New Journal of Physics, 14, 055023
327. McQuinn, K. B. W., Mitchell, N. P., & Skillman, E. D. 2015, ApJS, 218, 29
328. Medezinski, E., Umetsu, K., Okabe, N., et al. 2016, ApJ, 817, 24

329. Mernier, F., de Plaa, J., Kaastra, J. S., et al. 2017, A&A, 603, A80
330. Mernier, F., Biffi, V., Yamaguchi, H., et al. 2018, Space Science Reviews, 214, 129
331. Mernier, F., Werner, N., Lakhchaura, K., et al. 2020, Astronomische Nachrichten, 341, 203
332. Merten, J., Coe, D., Dupke, R., et al. 2011, MNRAS, 417, 333
333. Mezcua, M. 2017, International Journal of Modern Physics D, 26, 1730021
334. Mezcua, M., Civano, F., Fabbiano, G., Miyaji, T., & Marchesi, S. 2016, ApJ, 817, 20
335. Million, E. T., Allen, S. W., Werner, N., & Taylor, G. B. 2010, MNRAS, 405, 1624
336. Mineo, S., Gilfanov, M., & Sunyaev, R. 2012, MNRAS, 419, 2095
337. —. 2012, MNRAS, 426, 1870
338. Minkowski, R. 1957, in Symposium-International Astronomical Union, Vol. 4, Cambridge University Press, 107
339. Mitsuishi, I., Yamasaki, N. Y., & Takei, Y. 2011, ApJ, 742, L31
340. Mittal, R., O'Dea, C. P., Ferland, G., et al. 2011, Monthly Notices of the Royal Astronomical Society, 418, 2386
341. Mohr, J. J., Fabricant, D. G., & Geller, M. J. 1993, ApJ, 413, 492
342. Molendi, S., Ghizzardi, S., De Grandi, S., et al. 2024, A&A, 685, A88
343. Moll, R., Schindler, S., Domainko, W., et al. 2007, Astronomy & Astrophysics, 463, 513
344. Moore, B., Katz, N., Lake, G., Dressler, A., & Oemler, A. 1996, Nature, 379, 613
345. Moster, B. P., Somerville, R. S., Maulbetsch, C., et al. 2010, ApJ, 710, 903
346. Mou, G., Yuan, F., Bu, D., Sun, M., & Su, M. 2014, ApJ, 790, 109
347. Muñoz, J. A., Kochanek, C. S., Fohlmeister, J., et al. 2022, ApJ, 937, 34
348. Mulchaey, J. S., Davis, D. S., Mushotzky, R. F., & Burstein, D. 2003, ApJS, 145, 39
349. Muratov, A. L., Kereš, D., Faucher-Giguère, C.-A., et al. 2017, MNRAS, 468, 4170
350. Nagai, D., & Lau, E. T. 2011, ApJ, 731, L10
351. Nagai, D., Vikhlinin, A., & Kravtsov, A. V. 2007, ApJ, 655, 98
352. Nagai, H., Onishi, K., Kawakatu, N., et al. 2019, The Astrophysical Journal, 883, 193
353. Navarro, J. F., Frenk, C. S., & White, S. D. M. 1997, ApJ, 490, 493
354. Nelson, D., Springel, V., Pillepich, A., et al. 2019, Computational Astrophysics and Cosmology, 6, 2
355. Nguyen, D. D., & Thompson, T. A. 2021, MNRAS, 508, 5310
356. —. 2022, ApJ, 935, L24
357. Nikiel-Wroczyński, B., Soida, M., Urbanik, M., Beck, R., & Bomans, D. J. 2013, MNRAS, 435, 149
358. Nogueras-Lara, F., Schödel, R., Gallego-Calvente, A. T., et al. 2020, Nature Astronomy, 4, 377
359. Nomoto, K., Tominaga, N., Umeda, H., Kobayashi, C., & Maeda, K. 2006, *Nucl. Phys. A*, 777, 424
360. Norman, C. A., & Ikeuchi, S. 1989, ApJ, 345, 372
361. Nulsen, P. E. J. 1986, MNRAS, 221, 377
362. Nulsen, P. E. J., & McNamara, B. R. 2013, Astronomische Nachrichten, 334, 386
363. Nurgaliev, D., McDonald, M., Benson, B. A., et al. 2013, ApJ, 779, 112
364. Oemler Jr, A. 1976, Astrophysical Journal, Vol. 209, p. 693-709, 209, 693
365. Ogle, P. M., Lanz, L., & Appleton, P. N. 2014, ApJ, 788, L33
366. Oguri, M. 2006, MNRAS, 367, 1241
367. Olivares, V., Picquenot, A., Su, Y., et al. 2025, Nature Astronomy, arXiv:2501.01902 [astro-ph.GA]
368. Olivares, V., Su, Y., Nulsen, P., et al. 2022, MNRAS, 516, L101
369. Olivares, V., Salome, P., Combes, F., et al. 2019, Astronomy & Astrophysics, 631, A22
370. Olivier, G. M., Berg, D. A., Chisholm, J., et al. 2021, arXiv e-prints, arXiv:2109.06725
371. Oppenheimer, B. D., Bogdán, Á., Crain, R. A., et al. 2020, ApJ, 893, L24
372. Orr, M. E., Hayward, C. C., Hopkins, P. F., et al. 2018, MNRAS, 478, 3653
373. Oskinova, L. M., & Schaerer, D. 2022, A&A, 661, A67
374. O'Sullivan, E., Giacintucci, S., David, L. P., et al. 2011, ApJ, 735, 11
375. O'Sullivan, E., Giacintucci, S., Vrtilek, J. M., Raychaudhury, S., & David, L. P. 2009, ApJ, 701, 1560
376. O'Sullivan, E., Zezas, A., Vrtilek, J. M., et al. 2014, ApJ, 793, 73
377. O'Sullivan, E., Vrtilek, J. M., David, L. P., et al. 2014, ApJ, 793, 74

378. Ott, J., Walter, F., & Brinks, E. 2005, MNRAS, 358, 1423
379. Owers, M. S., Randall, S. W., Nulsen, P. E. J., et al. 2011, ApJ, 728, 27
380. Pacaud, F., Pierre, M., Adami, C., et al. 2007, MNRAS, 382, 1289
381. Pacucci, F., Mesinger, A., Mineo, S., & Ferrara, A. 2014, MNRAS, 443, 678
382. Panagoulia, E. K., Fabian, A. C., Sanders, J. S., & Hlavacek-Larrondo, J. 2014, MNRAS, 444, 1236
383. Pandya, V., Fielding, D. B., Anglés-Alcázar, D., et al. 2021, MNRAS, 508, 2979
384. Pannuti, T. G. 2001, PASP, 113, 1438
385. Pannuti, T. G., Duric, N., Lacey, C. K., et al. 2000, ApJ, 544, 780
386. Pannuti, T. G., Schlegel, E. M., & Lacey, C. K. 2007, AJ, 133, 1361
387. Parekh, V., van der Heyden, K., Ferrari, C., Angus, G., & Holwerda, B. 2015, A&A, 575, A127
388. Park, J., Mesinger, A., Greig, B., & Gillet, N. 2019, MNRAS, 484, 933
389. Pearce, C. J. J., van Weeren, R. J., Andrade-Santos, F., et al. 2017, ApJ, 845, 81
390. Pearson, T. J., & Readhead, A. C. S. 1988, ApJ, 328, 114
391. Peluso, G., Vulcani, B., Poggianti, B. M., et al. 2022, ApJ, 927, 130
392. Pereira-Santaella, M., Álvarez-Márquez, J., García-Bernete, I., et al. 2022, A&A, 665, L11
393. Peterson, J. R., & Fabian, A. C. 2006, Phys. Rep., 427, 1
394. Piffaretti, R., Arnaud, M., Pratt, G. W., Pointecouteau, E., & Melin, J. B. 2011, A&A, 534, A109
395. Pillepich, A., Nelson, D., Truong, N., et al. 2021, MNRAS, 508, 4667
396. Planck Collaboration, Aghanim, N., Akrami, Y., et al. 2020, A&A, 641, A5
397. —. 2020, A&A, 641, A8
398. Poggianti, B. M., Jaffé, Y. L., Moretti, A., et al. 2017, Nature, 548, 304
399. Poggianti, B. M., Ignesti, A., Gitti, M., et al. 2019, ApJ, 887, 155
400. Ponman, T. J., Cannon, D. B., & Navarro, J. F. 1999, Nature, 397, 135
401. Porraz Barrera, N., Lopez, S., Lopez, L. A., et al. 2024, ApJ, 968, 54
402. Pratt, G. W., Croston, J. H., Arnaud, M., & Böhringer, H. 2009, A&A, 498, 361
403. Pratt, G. W., Arnaud, M., Piffaretti, R., et al. 2010, A&A, 511, A85
404. Predehl, P., Sunyaev, R. A., Becker, W., et al. 2020, Nature, 588, 227
405. Rajpurohit, K., Osinga, E., Brienza, M., et al. 2023, A&A, 669, A1
406. Ranalli, P., Comastri, A., & Setti, G. 2003, A&A, 399, 39
407. Randall, S. W., Nulsen, P. E. J., Jones, C., et al. 2015, ApJ, 805, 112
408. Rasia, E., Meneghetti, M., & Ettori, S. 2013, The Astronomical Review, 8, 010000
409. Rasmussen, J., & Ponman, T. J. 2007, MNRAS, 380, 1554
410. —. 2009, MNRAS, 399, 239
411. Rasmussen, J., Ponman, T. J., & Mulchaey, J. S. 2006, MNRAS, 370, 453
412. Refsdal, S. 1964, MNRAS, 128, 295
413. Reines, A. E., Greene, J. E., & Geha, M. 2013, ApJ, 775, 116
414. Reiprich, T. H., Basu, K., Ettori, S., et al. 2013, Space Science Reviews, 177, 195
415. Renaud, F., Appleton, P. N., & Xu, C. K. 2010, ApJ, 724, 80
416. Renzini, A., & Andreon, S. 2014, MNRAS, 444, 3581
417. Rhoads, J. E., Wold, I. G. B., Harish, S., et al. 2023, ApJ, 942, L14
418. Richard-Laferrière, A., Hlavacek-Larrondo, J., Nemmen, R. S., et al. 2020, MNRAS, 499, 2934
419. Richard-Laferrière, A., Russell, H. R., Fabian, A. C., et al. 2023, MNRAS, 526, 6205
420. Richter, P., Paerels, F. B. S., & Kaastra, J. S. 2008, Space Science Reviews, 134, 25
421. Ricker, P. M., & Sarazin, C. L. 2001, ApJ, 561, 621
422. Riess, A. G., Casertano, S., Yuan, W., Macri, L. M., & Scolnic, D. 2019, ApJ, 876, 85
423. Riva, G., Pratt, G. W., Rossetti, M., et al. 2024, A&A, 691, A340
424. Roberg-Clark, G. T., Drake, J. F., Reynolds, C. S., & Swisdak, M. 2018, Phys. Rev. Lett., 120, 035101
425. Roberts, I. D., van Weeren, R. J., McGee, S. L., et al. 2021, A&A, 652, A153
426. Roncarelli, M., Ettori, S., Borgani, S., et al. 2013, MNRAS, 432, 3030

427. Rossetti, M., Eckert, D., Cavalleri, B. M., et al. 2011, A&A, 532, A123
428. Rossetti, M., Gastaldello, F., Eckert, D., et al. 2017, MNRAS, 468, 1917
429. Rossetti, M., Gastaldello, F., Ferioli, G., et al. 2016, MNRAS, 457, 4515
430. Rossetti, M., Eckert, D., Gastaldello, F., et al. 2024, A&A, 686, A68
431. Roychowdhury, S., Ruszkowski, M., Nath, B. B., & Begelman, M. C. 2004, ApJ, 615, 681
432. Rudnick, L., Brüggen, M., Brunetti, G., et al. 2022, ApJ, 935, 168
433. Ruppin, F., McDonald, M., Bleem, L. E., et al. 2021, ApJ, 918, 43
434. Russell, H. R., Sanders, J. S., Fabian, A. C., et al. 2010, MNRAS, 406, 1721
435. Russell, H. R., Nulsen, P. E. J., Caprioli, D., et al. 2022, MNRAS, 514, 1477
436. Russell, H. R., Lopez, L. A., Allen, S. W., et al. 2024, Universe, 10, 273
437. Ruszkowski, M., & Pfrommer, C. 2023, A&A Rev., 31, 4
438. Sacchi, A., Bogdán, Á., & Truong, N. 2025, ApJ, 983, 178
439. Salomé, P., Combes, F., Revaz, Y., et al. 2008, Astronomy & Astrophysics, 484, 317
440. Salomé, P., Combes, F., Edge, A. C., et al. 2006, Astronomy & Astrophysics, 454, 437
441. Sanders, J. S., & Fabian, A. C. 2007, MNRAS, 381, 1381
442. Sanders, J. S., Fabian, A. C., Russell, H. R., & Walker, S. A. 2018, MNRAS, 474, 1065
443. Santos, J. S., Rosati, P., Tozzi, P., et al. 2008, A&A, 483, 35
444. Sarazin, C. L. 1986, Reviews of Modern Physics, 58, 1
445. Sarkar, A., Su, Y., Randall, S., et al. 2021, MNRAS, 501, 3767
446. Sarkar, A., Su, Y., Truong, N., et al. 2022, MNRAS, 516, 3068
447. Sarkar, K. C., Nath, B. B., & Sharma, P. 2015, MNRAS, 453, 3827
448. Schellenberger, G., David, L. P., Vrtilek, J., et al. 2021, ApJ, 906, 16
449. Schindler, S., & Diaferio, A. 2008, Space Science Reviews, 134, 363
450. Schlegel, E. M. 1994, ApJ, 434, 523
451. Schlieder, J. E., Barclay, T., Barnes, A., et al. 2024, in Society of Photo-Optical Instrumentation Engineers (SPIE) Conference Series, Vol. 13092, Space Telescopes and Instrumentation 2024: Optical, Infrared, and Millimeter Wave, ed. L. E. Coyle, S. Matsuura, & M. D. Perrin, 130920S
452. Schmidt, R. W., & Allen, S. W. 2007, MNRAS, 379, 209
453. Schmidt, R. W., Allen, S. W., & Fabian, A. C. 2004, MNRAS, 352, 1413
454. Schneider, E. E., Ostriker, E. C., Robertson, B. E., & Thompson, T. A. 2020, ApJ, 895, 43
455. Scolnic, D. M., Jones, D. O., Rest, A., et al. 2018, ApJ, 859, 101
456. Sereno, M. 2016, LIRA: LInear Regression in Astronomy, Astrophysics Source Code Library, record ascl:1602.006
457. Sereno, M., Umetsu, K., Ettori, S., et al. 2020, MNRAS, 492, 4528
458. Serra, P., Maccagni, F. M., Kleiner, D., et al. 2023, A&A, 673, A146
459. Serra, P., Oosterloo, T. A., Kamphuis, P., et al. 2024, A&A, 690, A4
460. Shen, X., Brinckmann, T., Rapetti, D., et al. 2022, MNRAS, 516, 1302
461. Shi, D. D., Cai, Z., Fan, X., et al. 2021, ApJ, 915, 32
462. Shimwell, T. W., Brown, S., Feain, I. J., et al. 2014, MNRAS, 440, 2901
463. Shin, J., Woo, J.-H., & Mulchaey, J. S. 2016, ApJS, 227, 31
464. Shirazi, M., & Brinchmann, J. 2012, MNRAS, 421, 1043
465. Shopbell, P. L., & Bland-Hawthorn, J. 1998, ApJ, 493, 129
466. Sikhosana, S. P., Knowles, K., Hilton, M., Moodley, K., & Murgia, M. 2023, MNRAS, 518, 4595
467. Silk, J., & Rees, M. J. 1998, A&A, 331, L1
468. Silk, J., & White, S. D. M. 1978, ApJ, 226, L103
469. Simionescu, A., Allen, S. W., Mantz, A., et al. 2011, Science, 331, 1576
470. Simionescu, A., Nakashima, S., Yamaguchi, H., et al. 2019, MNRAS, 483, 1701
471. Skory, S., Hallman, E., Burns, J. O., et al. 2013, ApJ, 763, 38
472. Smith, R. N., Tombesi, F., Veilleux, S., Lohfink, A. M., & Luminari, A. 2019, ApJ, 887, 69
473. Spilker, J. S., Champagne, J. B., Fan, X., et al. 2025, arXiv e-prints, arXiv:2502.05283

474. Spitzer, Jr., L. 1956, ApJ, 124, 20
475. Stark, D. P., et al. 2011, ApJL, 728, L2
476. Stark, D. P., Ellis, R. S., Charlot, S., et al. 2017, MNRAS, 464, 469
477. Steidel, C. C., Adelberger, K. L., Dickinson, M., et al. 1998, The Astrophysical Journal, 492, 428
478. Steidel, C. C., Adelberger, K. L., Shapley, A. E., et al. 2005, ApJ, 626, 44
479. Strickland, D. K., & Heckman, T. M. 2007, ApJ, 658, 258
480. —. 2009, ApJ, 697, 2030
481. Strickland, D. K., Heckman, T. M., Colbert, E. J. M., Hoopes, C. G., & Weaver, K. A. 2004, ApJ, 606, 829
482. Strickland, D. K., Heckman, T. M., Weaver, K. A., Hoopes, C. G., & Dahlem, M. 2002, ApJ, 568, 689
483. Sturm, E., González-Alfonso, E., Veilleux, S., et al. 2011, ApJ, 733, L16
484. Su, M., Slatyer, T. R., & Finkbeiner, D. P. 2010, ApJ, 724, 1044
485. Suess, K. A., Weaver, J. R., Price, S. H., et al. 2024, ApJ, 976, 101
486. Sulentic, J. W., Rosado, M., Dultzin-Hacyan, D., et al. 2001, AJ, 122, 2993
487. Sun, M. 2009, ApJ, 704, 1586
488. —. 2012, New Journal of Physics, 14, 045004
489. Sun, M., Donahue, M., Roediger, E., et al. 2010, ApJ, 708, 946
490. Sun, M., Jones, C., Forman, W., et al. 2006, ApJ, 637, L81
491. Sun, M., Ge, C., Luo, R., et al. 2021, Nature Astronomy, 6, 270
492. Svoboda, J., et al. 2019, ApJ, 880, 144
493. Tamburri, S., Trinchieri, G., Wolter, A., et al. 2012, A&A, 541, A28
494. Tang, X., & Churazov, E. 2017, MNRAS, 468, 3516
495. Taylor, G., Fabian, A., & Allen, S. 2002, Monthly Notices of the Royal Astronomical Society, 334, 769
496. Tewes, M., Courbin, F., Meylan, G., et al. 2013, A&A, 556, A22
497. The HI4PI Collaboration, Ben Bekhti, N., Floer, L., et al. 2016, Astronomy Astrophysics, 594, A116
498. The Simons Observatory Collaboration, Abitbol, M., Abril-Cabezas, I., et al. 2025, arXiv e-prints, arXiv:2503.00636
499. Thompson, T. A., Quataert, E., Zhang, D., & Weinberg, D. H. 2016, MNRAS, 455, 1830
500. Timmerman, R., van Weeren, R. J., Botteon, A., et al. 2022, A&A, 668, A65
501. Tombesi, F., Cappi, M., Reeves, J. N., & Braito, V. 2012, MNRAS, 422, L1
502. Tombesi, F., Cappi, M., Reeves, J. N., et al. 2011, ApJ, 742, 44
503. —. 2010, A&A, 521, A57
504. Tombesi, F., Meléndez, M., Veilleux, S., et al. 2015, Nature, 519, 436
505. Tombesi, F., Veilleux, S., Meléndez, M., et al. 2017, ApJ, 850, 151
506. Tonnesen, S., Bryan, G. L., & van Gorkom, J. H. 2007, ApJ, 671, 1434
507. Topping, M. W., Shapley, A. E., Steidel, C. C., Naoz, S., & Primack, J. R. 2018, The Astrophysical Journal, 852, 134
508. Towler, I., Kay, S. T., & Altamura, E. 2023, MNRAS, 520, 5845
509. Tozzi, G., Cresci, G., Marasco, A., et al. 2021, A&A, 648, A99
510. Tozzi, P., Pentericci, L., Gilli, R., et al. 2022, A&A, 662, A54
511. Tozzi, P., Gilli, R., Liu, A., et al. 2022, A&A, 667, A134
512. Tremonti, C. A., Heckman, T. M., Kauffmann, G., et al. 2004, ApJ, 613, 898
513. Trenti, M., et al. 2010, ApJL, 714, L202
514. Trinchieri, G., Iovino, A., Pompei, E., et al. 2008, A&A, 484, 195
515. Trinchieri, G., Sulentic, J., Breitschwerdt, D., & Pietsch, W. 2003, A&A, 401, 173
516. Trinchieri, G., Sulentic, J., Pietsch, W., & Breitschwerdt, D. 2005, A&A, 444, 697
517. Trudeau, A., Webb, T., Hlavacek-Larrondo, J., et al. 2019, MNRAS, 487, 1210
518. Truong, N., Rasia, E., Biffi, V., et al. 2019, MNRAS, 484, 2896
519. Tümer, A., Tombesi, F., Bourdin, H., et al. 2019, A&A, 629, A82
520. Turner, M. J. L., Reeves, J. N., Ponman, T. J., et al. 2001, A&A, 365, L110
521. Ubertosi, F., Giacintucci, S., Clarke, T., et al. 2024, A&A, 691, A294
522. Ubertosi, F., Gitti, M., Brighenti, F., et al. 2021, ApJ, 923, L25

523. —. 2023, ApJ, 944, 216
524. Ubertosi, F., Giroletti, M., Gitti, M., et al. 2024, A&A, 688, A86
525. Ubertosi, F., Schellenberger, G., O'Sullivan, E., et al. 2024, ApJ, 961, 134
526. Ubertosi, F., Gong, Y., Nulsen, P., et al. 2025, A&A, 693, A171
527. Umeda, H., Ouchi, M., Nakajima, K., et al. 2022, ApJ, 930, 37
528. Umetsu, K., Zitrin, A., Gruen, D., et al. 2016, ApJ, 821, 116
529. Umetsu, K., Sereno, M., Lieu, M., et al. 2020, ApJ, 890, 148
530. Urban, O., Werner, N., Allen, S. W., Simionescu, A., & Mantz, A. 2017, MNRAS, 470, 4583
531. Urban, O., Simionescu, A., Werner, N., et al. 2014, MNRAS, 437, 3939
532. Vakulik, V., Schild, R., Dudinov, V., et al. 2006, A&A, 447, 905
533. van Weeren, R. J., de Gasperin, F., Akamatsu, H., et al. 2019, Space Sci. Rev., 215, 16
534. van Weeren, R. J., Ogrean, G., Jones, C., et al. 2017, ApJ, 835, 197
535. van Weeren, R. J., Timmerman, R., Vaidya, V., et al. 2024, A&A, 692, A12
536. Vanderlinde, K., Crawford, T. M., de Haan, T., et al. 2010, ApJ, 722, 1180
537. Vazza, F., Eckert, D., Simionescu, A., Brüggen, M., & Ettori, S. 2013, MNRAS, 429, 799
538. Vazza, F., Ferrari, C., Bonafede, A., et al. 2015, in Advancing Astrophysics with the Square Kilometre Array (AASKA14), 97
539. Veilleux, S., Cecil, G., & Bland-Hawthorn, J. 2005, ARA&A, 43, 769
540. Veilleux, S., Rupke, D. S. N., & Swaters, R. 2009, ApJ, 700, L149
541. Veilleux, S., Meléndez, M., Sturm, E., et al. 2013, ApJ, 776, 27
542. Verdes-Montenegro, L., Yun, M. S., Williams, B. A., et al. 2001, A&A, 377, 812
543. Vestergaard, M., & Peterson, B. M. 2006, ApJ, 641, 689
544. Vigneron, B., Hlavacek-Larrondo, J., Rhea, C. L., et al. 2024, The Astrophysical Journal, 962, 96
545. Vikhlinin, A., Kravtsov, A., Forman, W., et al. 2006, ApJ, 640, 691
546. Vikhlinin, A., Kravtsov, A. V., Burenin, R. A., et al. 2009, ApJ, 692, 1060
547. Voit, G. M. 2005, Reviews of Modern Physics, 77, 207
548. Voit, G. M., Bryan, G. L., O'Shea, B. W., & Donahue, M. 2015, The Astrophysical Journal Letters, 808, L30
549. Voit, G. M., Kay, S. T., & Bryan, G. L. 2005, MNRAS, 364, 909
550. Voit, G. M., Meece, G., Li, Y., et al. 2017, ApJ, 845, 80
551. Vollmer, B., Cayatte, V., Balkowski, C., & Duschl, W. J. 2001, ApJ, 561, 708
552. Von Der Linden, A., Best, P. N., Kauffmann, G., & White, S. D. 2007, Monthly Notices of the Royal Astronomical Society, 379, 867
553. Vulcani, B., Poggianti, B. M., Smith, R., et al. 2022, ApJ, 927, 91
554. Vulcani, B., Poggianti, B. M., Tonnesen, S., et al. 2020, ApJ, 899, 98
555. Vurm, I., Nevalainen, J., Hong, S. E., et al. 2023, A&A, 673, A62
556. Wagner, A. Y., Umemura, M., & Bicknell, G. V. 2013, ApJ, 763, L18
557. Walker, S., & Lau, E. 2022, in Handbook of X-ray and Gamma-ray Astrophysics, 13
558. Walker, S., Simionescu, A., Nagai, D., et al. 2019, Space Science Reviews, 215, 7
559. Walker, S. A., Hlavacek-Larrondo, J., Gendron-Marsolais, M., et al. 2017, MNRAS, 468, 2506
560. Walker, S. A., Kosec, P., Fabian, A. C., & Sanders, J. S. 2015, MNRAS, 453, 2480
561. Walker, S. A., Mirakhor, M. S., ZuHone, J., et al. 2022, ApJ, 929, 37
562. Walker, S. A., Sanders, J. S., & Fabian, A. C. 2016, MNRAS, 461, 684
563. Walter, F., Weiss, A., & Scoville, N. 2002, ApJ, 580, L21
564. Wambsganss, J., Brunner, H., Schindler, S., & Falco, E. 1999, A&A, 346, L5
565. Wan, J. T., Mantz, A. B., Sayers, J., et al. 2021, MNRAS, 504, 1062
566. Wang, Q. D., Diaz, C. G., Kamieneski, P. S., et al. 2024, MNRAS, 527, 10584
567. Wang, Q. H. S., Giacintucci, S., & Markevitch, M. 2018, ApJ, 856, 162
568. Wang, Q. H. S., & Markevitch, M. 2018, ApJ, 868, 45
569. Wang, T., Elbaz, D., Daddi, E., et al. 2016, ApJ, 828, 56

570. Watts, A. B., Cortese, L., Catinella, B., et al. 2023, arXiv e-prints, arXiv:2303.07549
571. Weaver, R., McCray, R., Castor, J., Shapiro, P., & Moore, R. 1977, ApJ, 218, 377
572. Webb, T., Noble, A., DeGroot, A., et al. 2015, ApJ, 809, 173
573. Webb, T. M. A., Muzzin, A., Noble, A., et al. 2015, ApJ, 814, 96
574. Webb, T. M. A., Lowenthal, J., Yun, M., et al. 2017, ApJ, 844, L17
575. Werk, J. K., Prochaska, J. X., Tumlinson, J., et al. 2014, ApJ, 792, 8
576. Werner, N., Urban, O., Simionescu, A., & Allen, S. W. 2013, Nature, 502, 656
577. Wertz, O., Stern, D., Krone-Martins, A., et al. 2019, A&A, 628, A17
578. White, J. A., Canning, R. E. A., King, L. J., et al. 2015, MNRAS, 453, 2718
579. White, S. D. M., Navarro, J. F., Evrard, A. E., & Frenk, C. S. 1993, Nature, 366, 429
580. Willott, C. J., Rawlings, S., Blundell, K. M., & Lacy, M. 1999, MNRAS, 309, 1017
581. Wise, M. W., McNamara, B. R., Nulsen, P. E. J., Houck, J. C., & David, L. P. 2007, ApJ, 659, 1153
582. Wittor, D., & Gaspari, M. 2020, MNRAS, 498, 4983
583. Wong, K.-W., & Sarazin, C. L. 2009, ApJ, 707, 1141
584. Wong, K.-W., Sarazin, C. L., & Ji, L. 2011, ApJ, 727, 126
585. Xu, C. K., Cheng, C., Appleton, P. N., et al. 2022, Nature, 610, 461
586. Yang, H., et al. 2017, ApJ, 847, 38
587. Yang, H. Y. K., Gaspari, M., & Marlow, C. 2019, ApJ, 871, 6
588. Yeung, M. C. H., Ponti, G., Freyberg, M. J., et al. 2024, A&A, 690, A399
589. Young, A. J., Wilson, A., & Mundell, C. 2002, The Astrophysical Journal, 579, 560
590. Yuan, Z. S., & Han, J. L. 2020, MNRAS, 497, 5485
591. Yusef-Zadeh, F., Arendt, R. G., & Wardle, M. 2022, ApJ, 939, L21
592. Yusef-Zadeh, F., Hewitt, J. W., Arendt, R. G., et al. 2009, ApJ, 702, 178
593. Zenteno, A., Hernández-Lang, D., Klein, M., et al. 2020, MNRAS, 495, 705
594. Zhang, C., Churazov, E., Dolag, K., Forman, W. R., & Zhuravleva, I. 2020, MNRAS, 498, L130
595. —. 2020, MNRAS, 494, 4539
596. Zhang, C., Churazov, E., Forman, W. R., & Jones, C. 2019, MNRAS, 482, 20
597. Zhang, C., Churazov, E., Forman, W. R., & Lyskova, N. 2019, MNRAS, 488, 5259
598. Zhang, C., Churazov, E., & Schekochihin, A. A. 2018, MNRAS, 478, 4785
599. Zhang, C., Zhuravleva, I., Gendron-Marsolais, M.-L., et al. 2022, MNRAS, 517, 616
600. Zhang, C., Zhuravleva, I., Kravtsov, A., & Churazov, E. 2021, MNRAS, 506, 839
601. Zhang, C., Zhuravleva, I., Markevitch, M., et al. 2024, MNRAS, 530, 4234
602. Zhang, D. 2018, Galaxies, 6, 114
603. Zhang, Y., Yang, X., Wang, H., et al. 2013, ApJ, 779, 160
604. Zhang, Y., Comparat, J., Ponti, G., et al. 2024, A&A, 690, A267
605. —. 2024, A&A, 690, A268
606. —. 2025, Astronomy and Astrophysics, 693, A197
607. Zheng, X., Ponti, G., Locatelli, N., et al. 2024, A&A, 689, A328
608. Zhuravleva, I., Churazov, E., Schekochihin, A. A., et al. 2019, Nature Astronomy, 3, 832
609. —. 2014, Nature, 515, 85
610. Zinger, E., Dekel, A., Birnboim, Y., Kravtsov, A., & Nagai, D. 2016, MNRAS, 461, 412
611. Zubovas, K., & King, A. 2012, ApJ, 745, L34
612. ZuHone, J., Ehlert, K., Weinberger, R., & Pfrommer, C. 2021, Galaxies, 9, 91
613. ZuHone, J. A., & Hallman, E. J. 2016, pyXSIM: Synthetic X-ray observations generator, Astrophysics Source Code Library, record ascl:1608.002
614. ZuHone, J. A., Markevitch, M., & Lee, D. 2011, ApJ, 743, 16
615. ZuHone, J. A., Markevitch, M., Weinberger, R., Nulsen, P., & Ehlert, K. 2021, ApJ, 914, 73
616. Zuhone, J. A., & Roediger, E. 2016, Journal of Plasma Physics, 82, 535820301

617. ZuHone, J. A., Vikhlinin, A., Tremblay, G. R., et al. 2023, SOXS: Simulated Observations of X-ray Sources, Astrophysics Source Code Library, record ascl:2301.024
618. Zwan, B. J., & Wolf, R. A. 1976, *J. Geophys. Res.*, 81, 1636

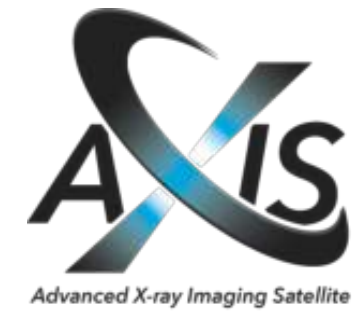

*AXIS GO Science Book*

# AXIS Active Galactic Nuclei GO Science

**T. Anguita[78], R. Arcodia[51], E. Bañados[53], R. D. Baldi[36], L. Battistini[34], A. R. Basu-Zych[56,97], F. E. Bauer[65,79], E. Behar[63], A. Belfiore[35], E. Bertola[37], H. Best[2,10], S. Bianchi[26], D. Bogensberger[6], A. Borghese[32], E. Bronzini[41,88], M. Brusa[84], E. Bulbul, A. Caccianiga[38], M. Capalbi[39], N. Cappelluti[101], D. Castro[7], G. Chartas[11], F. Civano[56], A. Comastri[41], T. Connor[7], W. Cui[9,13,44], F. Dammando[36], A. De Rosa[34], R. Decarli[41], C. Garcia Diaz[23], S. Dichiara[14], S. DiKerby[55], R. Dupke[28,61], M. Elvis[7], G. Fabbiano[7], J. Fagin[2,10], A. Falcone[14], E. Fedorova[39], G. Ferrand[21,67], N. Foo[4], A. Foord[98], B.L. Frye[87], C. Fryer[49], H. M. Günther[50], S. Gagnon[72], L. Gallo[66], M. Gaspari[102], R. Gilli[34], M. Gitti[24,41], P. Grandi[41], A. Gross[96], K.C. Harrington[57], L. Ighina[7,38], J. Irwin[85,86], A. Jana[78], F. Jennings[93], J. Jiang[105], E.F. Jimenez-Andrade[81], P.S. Kamieneski[4], E. Kammoun[6], A.K.H. Kong[58], M. Koss[28], E. Lambrides[56], G. Lanzuisi[41], B.D. Lehmer[23], J. Li[66], D.Z. Liu[66], M. I. Llave[41], J.D. Lowenthal[69], E. Lusso[94], R. Maiolino[92], J. Maithil[7], S. Marchesi[34,41,75,88], J. Matthews[104], G. Mazzolari[41,52,88], C. Mazzucchelli[80], E. T. Meyer[98,100], R. Middei[4,7,39], G. Migliori[36], M. Millon[70], A. Moretti[38], C. Morgan[76], P. Nair[73], E. Nardini[37], F. Neira[107], T. T. Q. Nguyen[69], L.M. Oskinova[64], M. O'Dowd[2,10], F. Pacucci[7,32], E. Paic[107], B. A. Pampliega[27], F. Panessa[39], M. Parvatikar[34], A. Peca[28,99], E. S. Perlman[31], P. O. Petrucci[46], R. W. Pfeifle[56,60,97], E. Piconcelli[34,39], A. Pizzetti[26], A. Del Popolo[25], A. Ptak[56], G. Pugliese[1], F. Rigamonti[34], S. W. Randall[7], C. Reynolds[99], A. Ricarte[7], C. Ricci[78], G. Risaliti[94], D. Rogantini[15], B. Rothberg[77], H. Russell[103], P. Severgnini[34], J. Singh[34], A. Sacchi[7,35], F. Salvestrini[40], A. Sarkar[23,50], E. Schwartzman[59], R. Serafinelli[39,80], Y. Shen[96], M. Signorini[30,31,34], J. Sisk-Reynes[7], D. Sluse[83], R. Soria[34], D. Stern[48], F. Tombesi[74], N. Torres-Albà[18,19], E. Torresi[41], A. Tortosa[39], P. Tozzi[37], G. Vernardos[2,10], A. Vishwas[12], F. Vito[41], D. Walton[95], Q.D. Wang[16], K. Weaver[56], D. Wilkins[60], B.J. Williams[56], K.-W. Wong[22], K. Wu[55], Y. Xu[42,47], G. Younes[8,56], A. Young[89], M.S. Yun[23], S. Zane[55], L. Zappacosta[39], F. Coti Zelati[42,47], S. Zhang[20], Y. Zhang[7], X. Zhao[5], D.Z. Zhou[90], M. Zhuang[96], F. Zou[17]**

[1] Anton Pannekoek Institute for Astronomy, University of Amsterdam, The Netherlands [2] American Museum of Natural History (AMNH), USA [3] Arizona State University, USA [4] ASI-SSDC [5] Caltech, USA [6] CEA Saclay [7] Center for Astrophysics & Smithsonian, Cambridge, MA, USA [8] Center for Space Sciences and Technology, University of Maryland Baltimore County, 1000 Hilltop Cir, Baltimore County, 21250, Maryland, USA [9] Centro de Investigación Avanzada en Física Fundamental (CIAFF), Universidad Autónoma de Madrid, Cantoblanco, E-28049 Madrid, Spain [10] City University of New York (CUNY), USA [11] College of Charleston [12] Cornell University [13] Departamento de Física Teórica, M-8, Universidad Autónoma de Madrid, Cantoblanco, E-28049, Madrid, Spain [14] Department of Astronomy and Astrophysics, The Pennsylvania State University, 525 Davey Lab, University Park, PA 16802, USA [15] Department of Astronomy and Astrophysics, University of Chicago, Chicago, IL 60637, USA [16] Department of Astronomy, University of Massachusetts, Amherst, MA 01003, USA [17] Department of Astronomy, University of Michigan, 1085 S University, Ann Arbor, MI 48109, USA [18] Department of Astronomy, University of Virginia, 530 McCormick Road, Charlottesville, VA 22904, USA [19] Department of Physics and Astronomy, Clemson University, Kinard Lab of Physics, Clemson, SC 29634, USA [20] Department of Physics and Astronomy, Michigan State University, East Lansing, MI 48823, USA [21] Department of Physics and Astronomy, University of Manitoba, Canada [22] Department of Physics, SUNY Brockport, Brockport, NY 14420, USA [23] Department of Physics, University of Arkansas, Fayetteville, AR 72701, USA [24] Dipartimento di Fisica e Astronomia (DIFA), Università di Bologna, Via Gobetti 93/2, 40129 Bologna, Italy [25] Dipartimento di Fisica e Astronomia, Università degli Studi di Catania, Italy [26] Dipartimento di Matematica e Fisica, Università degli Studi Roma Tre, Via della Vasca Navale 84, I-00146, Roma, Italy [27] ESO/ALMA, Chile [28] Eureka Scientific [29] European Southern Observatory, Santiago, Chile [30] European Space Agency (ESA), European Space Astronomy Centre, Madrid, Spain [31] Florida Institute of Technology, USA [32] Harvard University, USA [33] Institut dÉstudis

Espacials de Catalunya (IEEC), Barcelona, E-08034, Spain [34] INAF, Italy [35] INAF–Istituto di Astrofisica Spaziale e Fisica Cosmica (IASF), Milano, Italy [36] INAF–Istituto di Radioastronomia (IRA), Bologna, Italy [37] INAF–Osservatorio Astrofisico di Arcetri (OAA), Firenze, Italy [38] INAF–Osservatorio Astronomico di Brera, Milano, Italy [39] INAF–Osservatorio Astronomico di Roma (OAR), Rome, Italy [40] INAF–Osservatorio Astronomico di Trieste (OATs), Trieste, Italy [41] INAF–Osservatorio di Astrofisica e Scienza dello Spazio (OAS), Bologna, Italy [42] Institut de Planétologie et d'Astrophysique de Grenoble (IPAG), Université Grenoble Alpes, France [43] Institut fuer Astronomie und Astrophysik, Universitaet Tuebingen, Germany [44] Institute for Astronomy, University of Edinburgh, Royal Observatory, Edinburgh EH9 3HJ, United Kingdom [45] Institute of Space Sciences (ICE, CSIC), Campus UAB, Carrer de Can Magrans s/n, Barcelona, E-08193, Spain [46] Jet Propulsion Laboratory, California Institute of Technology, USA [47] Los Alamos National Laboratory, T-CNLS, Los Alamos, NM, USA [48] Massachusetts Institute of Technology (MIT) Kavli Institute for Astrophysics and Space Research, Cambridge, MA, USA [49] Massachusetts Institute of Technology (MIT), Cambridge, MA, USA [50] Max Planck Institute for Extraterrestrial Physics, Germany [51] Max Planck Institute for Astronomy (MPIA), Heidelberg, Germany [52] Michigan State University Dept. of Physics and Astronomy, USA [53] Mullard Space Science Laboratory, University College London, UK [54] NASA Goddard Space Flight Center, Greenbelt, MD, USA [55] National Astronomical Observatory of Japan [56] National Tsing Hua University, Taiwan [57] Naval Research Laboratory, USA [58] Oak Ridge Associated Universities, USA [59] Observatorio Nacional, Rua Gal. José Cristino 77, São Cristóvão, Rio de Janeiro, Brazil [60] Department of Astronomy, The Ohio State University, 140 West 18th Avenue, Columbus, OH 43210, USA [61] Physics Department Technion, Haifa, Israel [62] Potsdam University, Germany [63] Pontificia Universidad Católica de Chile (PUC), Santiago, Chile [64] Purple Mountain Observatory, Nanjing, China [65] RIKEN Center for Interdisciplinary Theoretical and Mathematical Sciences (iTHEMS), Japan [66] Saint Mary's University, Canada [67] Smith College, USA [68] Stanford University, USA [69] Stockholm University, Sweden [70] The George Washington University, USA [71] The University of Alabama, USA [72] Tor Vergata University of Rome, Italy [73] UNIBO-DIFA [74] United States Naval Academy (USNA), USA [75] United States Naval Observatory, USA [76] Universidad Andres Bello, Chile [77] Universidad Catolica, Chile [78] Universidad Diego Portales, Santiago, Chile [79] Universidad Nacional Autónoma de México (UNAM), Mexico [80] University of Alabama, USA [81] University of Alabama Dept. of Physics and Astronomy, USA [82] University of Arizona, USA [83] University of Bologna, Italy [84] University of Bristol, UK [85] University of British Columbia, Canada [86] University of California, Berkeley, USA [87] University of Cambridge, UK [88] University of Edinburgh, UK [89] University of Florence, Florence, Italy [90] University of Hertfordshire, UK [91] University of Illinois Urbana–Champaign (UIUC), USA [92] University of Maryland, Baltimore County (UMBC), Baltimore County, MD, USA [93] University of Maryland, College Park, USA [94] Department of Physics, University of Miami, Coral Gables, FL 33124, USA [95] University of Modena and Reggio Emilia, Italy [96] University of Nottingham, UK [97] University of Oxford, UK [98] University of Warwick, UK [99] Yale University, USA [100] École Polytechnique Fédérale de Lausanne (EPFL), Switzerland

---

**a. Survey Fields**

*1. Obscured AGN at EoR and Beyond - Number Densities, Seeding*

**Science Area:** Active Galactic Nuclei

**First Author: Erini Lambrides** (NPP Fellow/NASA-Goddard, erini.lambrides@nasa.gov)
**Co-authors: Andrew F. Ptak** (NASA-Goddard, andrew.f.ptak@nasa.gov ), **Francesca Civano** (NASA-Goddard, francesca.m.civano@nasa.gov )

**Abstract:**

First results from JWST have raised more questions than answers on the prevalence, growth, and impact of high-z growing SMBHs (i.e., active galactic nuclei or AGN) with recent studies suggesting a shocking level of overabundance (>100*x*) at z> 5. A critical limitation in our understanding of high-z AGN is that almost all direct observations from the ground or with JWST are of *unobscured or mildly attenuated* AGN. Before JWST, theory and observation find the majority of early AGN are obscured ($>N_H > 10^{23}$), evading the most vetted AGN selection criteria in the early Universe — yet, the JWST AGN that drive the extreme number densities at high-z are largely selected from these very same criteria. It is crucial we hunt for the most obscured AGN at the highest redshifts towards the aim of (1) cornerstone AGN demography studies in the early Universe and (2) selecting the most representative sources for early Universe AGN feedback studies. Currently, among all telescope missions in operation or planned for the future, only AXIS can efficiently select heavily obscured early Universe AGN and build the samples required to answer the most outstanding questions about early AGN growth. In particular, the AXIS deep (7Ms) survey will uncover thousands of AGN in the first half of the Universe. Simply scaling the current number densities of predicted z> 5 X-ray AGN with the predicted obscuration fractions of AGN with $N_H > 10^{23}$ cm$^{-2}$ coupled with new number density constraints from JWST, we anticipate AXIS will be able to detect the majority of moderately powerful ($L_{Bol} > 10^{43}$erg/s) obscured AGN $N_H > 10^{23}$ cm$^{-2}$) between 5 <z< 9.

**Science:**

Before JWST, theory and observation find the increasing clumpiness and ISM densities of early galaxies would heavily obscure over 80% of all AGN at $z>6$[116,130,226,429] — escaping the most common selections in rest-UV-optical data sets (e.g. Balmer line broadening, morphological compactness). Yet, the JWST AGN that drive the extreme number densities at high-z are largely selected from these very same criteria. Even recent JWST observations that have yielded a barrage of reddened $z>5$ AGN, referred to as LRDs (see previous section) may represent a distinct population. These LRDs are by definition not heavily obscured – they show evidence of their AGN nature in *rest-frame optical* emission, and those with spectroscopic follow-up, detection of the broad-line region [241].

In addition to probing representative early BH growth, obscured AGN are an excellent test-bed for AGN co-evolution studies – unlike with unobscured AGN, the host galaxy properties of obscured AGN (e.g. $M_\star$, morphology) are more accessible in regimes where the AGN emission is significantly attenuated (i.e. optical and UV); (2) most AGN, especially at $z > 2$, are heavily obscured [226]. Thus, our ability to understand the growth of most AGN at $z > 5$ is dependent on our ability to select obscured AGN. In the deepest Chandra coverage ever obtained – the 7Ms CDFS survey [360], despite the median $N_H$ $\sim 10^{23}$ cm$^{-2}$, the small spatial area and insufficient sensitivity, only a handful of powerful early obscured AGN were discovered [584].

To capture statistically representative samples of luminous obscured AGN at these epochs for AGN-Galaxy co-evolution studies and probing the total early AGN demography we need: 1) Sufficiently wide fields ($\geq 0.2\,\mathrm{deg}^2$) to capture the rare luminous sources, 2) deep enough X-ray data to identify AGN across a wide range $N_H$, and 3) high-resolution UV/optical/NIR/MIR/sub-mm imaging and UV/optical spectroscopic data to probe host galaxy and AGN properties to high-$z$ [332,333]. Decades of efforts have

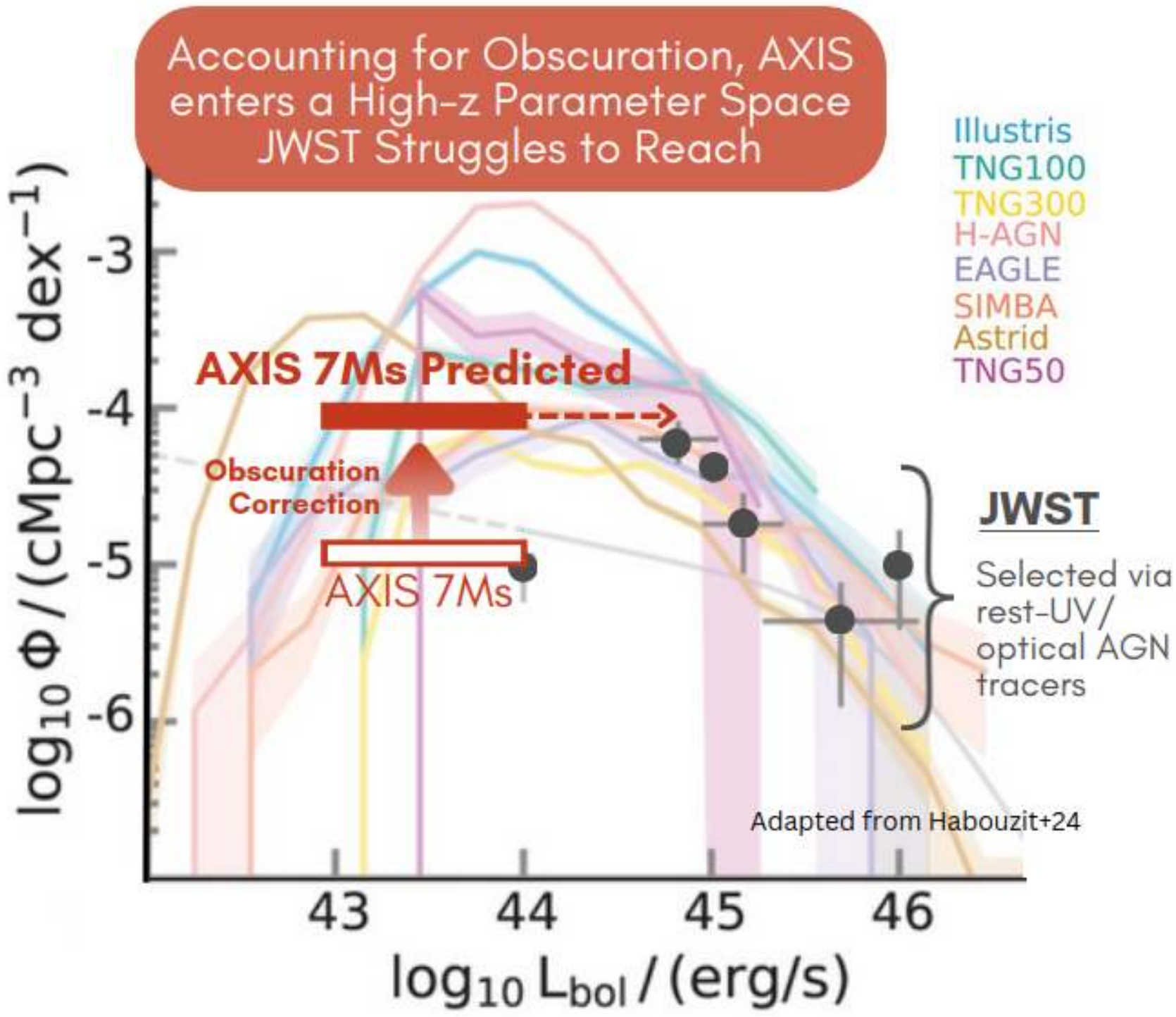

**Figure 1.** AXIS will uniquely detect the majority of AGN with $L_{Bol} > 10^{43}$ erg/s. Adapted from Fig. 2 in [253], the range of predicted total number densities from different cosmological simulations at z= 5 are shown in different colors. The black points represent the observed number density constraints from recent JWST results [241]. In the open and filled red rectangles, we show the $L_{Bol}$ range that the AXIS 7Ms Deep survey will be sensitive to at this epoch – with the open red rectangle representing the extrapolated JWST number constraints, and the filled red rectangle the correction when accounting for the full obscured population.

shown that no single energy regime alone can detect all the powerful, obscured sources at early times [262]. Thus, deep X-ray surveys in conjunction with multiwavelength datasets are *critical* in understanding which AGN we are selecting and the role their galaxy properties play in their identification.

Thus, simply by having the ability to select more representative samples of early AGN, we can probe several outstanding questions:

1. Is there significant black hole growth in most early massive galaxies?
2. Which black hole seeding models can most efficiently reproduce their number densities?
3. What is the nature of high-z AGN obscuration and how does this inform high-z gas-to-dust ratios?
4. How early can dust form in the Universe?
5. How does this impact the role most AGN play in re-ionizing the Universe?

**Selecting High-z Obscured AGN with AXIS:**

As we push AXIS to reach higher and higher redshifts, the total number of significant counts per detection decreases. Canonical X-ray rooted methods of inferring the level of obscuration such as X-ray spectral fitting become less robust, and even comparing the relative flux between softer energy bands (0.5 - 2 keV) to harder energy bands (2 -7 keV) less informative, as the rest-frame energies probed at early times (i.e. z> 7) enter the higher energy, and thus less obscuration sensitive, regimes. As was

done with deep Chandra surveys at more intermediate redshifts, identifying the rest-UV/optical AGN signatures (or lack thereof) and comparing them to the X-ray properties is an efficient means of identifying significantly obscured AGN. This is especially relevant for even high-fidelity observations with one of the current premier telescopes in existence, JWST struggles to select obscured AGN along multiple axes of investigation: 1) lack of point source in JWST/NIRCam Imaging 2) Difficulty in identifying ionization line diagnostics that are effective in the lower metallicity regimes and 3) Lack of access to reprocessed AGN emission in the rest-frame MIR.

There has been significant effort in the field to identify signatures of hidden AGN emission at $z > 5$ with JWST via hunts ionization diagnostics that theoretically should work [264] – yet they remain untested until independent lines of evidence can confirm they are indeed selecting obscured accreting SMBHs vs exotic high-z SF processes. With a 7Ms 25x25 arcmin deep field with AXIS, $\sim 40$ sources are predicted to be detected. Extrapolating the predicted obscuration fractions at $5 < z < 9$ via [226] to this fiducial sample, in 1 we show the expected detectable obscured AGN space densities (solid red line). We compare these values to predictions to simulations, and the current AGN number densities derived from JWST. Pre-JWST informed space densities are bolometric luminosity functions that overshoot predictions by at least an order of magnitude. The solid and open red boxes are rooted in pre-JWST number densities, and if indeed the bolometric and completeness of JWST-selected AGN is robust, can be treated as a lower limit of the detectable number of observed sources. While JWST may not identify the AXIS predicted sources as AGN, their host galaxies and the properties within can be determined.

**Immediate Objectives:**

We will:

- Cross-match X-ray detections in ancillary, incoming, and future deep rest-UV -optical photometric and spectroscopic data-sets (i.e, JWST, ELTs, Roman) to identify the highest-z sources.
- X-ray stack known AGN candidates and massive galaxies detected beyond $z > 6$ in any wavelength (including radio, UV, optical, NIR, sub-mm).
- For sources with significant counts (estimated > 30 counts), X-ray spectral fitting to determine the power-law slope, iron line detection, etc., level of obscuration.
- Joint X-ray spectral analysis for detected sources with $\sim 30$ counts to determine X-ray spectral properties on a population level.
- Compare the estimated $N_H$ via X-ray only methods to estimates derived from other data sets.

**Exposure time (ks):** Archival study using directed science deep fields.
**Joint Observations and synergies with other observatories in the 2030s:**
**Special Requirements:** None.

*2. AXIS-NEXUS: One Field for both SMBH Evolution and Time-domain Study*

**Science Area:** Active Galactic Nuclei, Extragalactic Survey, Time-Domain Studies
**First Author: Xiurui Zhao** (Caltech, xiurui.zhao.work@gmail.com)
**Co-authors: Yue Shen** (UIUC, shenyue@illinois.edu), **Francesca Civano** (NASA-GSFC, francesca.m.civano@nasa.gov), **Nico Cappelluti**, (University of Miami, ncappelluti@miami.edu), **Arran Gross** (UIUC, acgross@illinois.edu) **Stefano Marchesi** (University of Bologna, Italy, stefano.marchesi@unibo.it), **Mingyang Zhuang** (UIUC, mingyang@illinois.edu)

**Abstract:** The North ecliptic pole EXtragalactic Unified Survey (NEXUS) is a JWST Multi-Cycle (Cycles 3–5; 368 primary hours, PI: Y. Shen) GO Treasury spectroscopic and imaging survey centered around the North Ecliptic Pole (NEP). Over three years, the program will conduct a uniform spectroscopic and photometric survey covering 400 arcmin$^2$. Given the line sensitivity of JWST/NIRCam slitless spectroscopy, all emission-line galaxies (e.g., AGN and star-forming galaxies) with mAB $<$ 24 (mAB $<$ 30) in F444W will have reliable spectroscopic redshifts across the entire 400 arcmin$^2$ field (central 50 arcmin$^2$ field, respectively). The NEXUS field also overlaps with the ultra-deep field of Euclid, offering an unprecedented optical-to-MIR spectroscopic coverage.

We propose a deep AXIS survey of the NEXUS field, which will enable us to: 1) Match the JWST sensitivity of an extragalactic survey with X-ray for the first time. 2) Achieve the first spectroscopic redshift- and classification-complete deep X-ray contiguous survey. 3) Probe the co-evolution of SMBHs and their host galaxies using a large, well-constrained sample of AGN host galaxies with detailed morphology and stellar population properties. 4) Constrain the X-ray luminosity functions of AGN out to $z \sim 6$. 5) Constrain the obscured AGN evolution up to the Cosmic Dawn. 6) Discover and characterize a large sample of X-ray transients with well-characterized host properties. 7) Constrain the gas distribution surrounding the SMBHs by monitoring thousands of AGN in five years.

**Science:**
**Introduction:** The NEXUS survey aims to acquire deep (observed frame) IR spectra and imaging of all galaxies in a contiguous extragalactic field for the first time. The survey comprises two overlapping tiers. The wide tier, covering about 400 arcmin$^2$, will perform NIRCam/WFSS 2.4–5$\mu$m grism spectroscopy every year for three cycles. Given the line sensitivity of JWST/NIRCam slitless spectroscopy [530], **all emission-line galaxies (e.g., AGN or star-forming galaxies, SFGs) at** $m_{\rm AB} < 24$ **(in F444W) will have reliable spectroscopic redshifts in the entire NEXUS field.** The deep tier, covering about 50 arcmin$^2$, located in the center of the wide tier, will perform high-multiplexing and high-throughput NIRSpec 0.6–5.3$\mu$m MOS/PRISM spectroscopy *over three years with a 2-month cadence starting June 2025*. The deep field will provide broadband optical-to-IR high-quality spectra for more than 10,000 astronomical objects. Therefore, **all emission-line galaxies at** $m_{\rm AB} < 30$ **(in F444W) will have reliable spec-$z$ in the deep tier.** The combination of uniform and broad spectroscopy and cadenced observations of NEXUS is unique among all JWST treasury programs. In addition, the NEXUS field is located in the Euclid Ultra-Deep Field (UDF), featuring the deepest 0.9–2 $\mu$m Euclid spectroscopy among all Euclid deep fields [169]. Therefore, each $m_{\rm AB} < 22.2$ galaxy in the NEXUS field will have unprecedented broadband spectroscopic coverage in 0.9 $\mu$m–5$\mu$m. Furthermore, NEXUS features multiband optical to mid-IR imaging coverage by Euclid VIS/NISP and JWST NIRCam/parallel MIRI, enabling deep photometry covering 0.5–12 $\mu$m down to $m_{\rm AB} < 28$ in 400 arcmin$^2$ to produce reliable photo-zs. Continuum sensitivity and wavelength coverage of JWST and Euclid are plotted in Fig. 2 left.

These unique features make NEXUS the newest and most prominent field for tracing AGN evolution and performing time-domain studies. X-ray observations are essential for directly probing the intrinsic accretion power, measuring the circumnuclear material surrounding SMBHs, and investigating the

high-energy processes in AGN. AXIS will be the best X-ray telescope for these objectives due to its exceptional spatial resolution, sensitivity, and photon collecting area, all of which are crucial for detecting and characterizing faint targets in the NEXUS field.

**Immediate Objectives:**

**1) First Spec-*z* and Classification-Complete X-ray Contiguous Survey**

X-rays provide a more pure and efficient selection of the AGN populations compared with selections at other wavelengths. To measure the AGN evolution and SMBH growth history, multiple deep and wide contiguous extragalactic surveys in X-ray have been performed over the last few decades, measuring the number counts, luminosity function, and obscuration distribution of accreting SMBHs. However, the lack of reliable spec-*z* and classifications of all X-ray detected sources prevented these works from achieving a complete census of the X-ray sources, thus hindering the aforementioned scientific objectives. Indeed, 23%–59% of the X-ray detections **do not** have spec-*z* in the COSMOS [377], CDFS [360], and SERVS [428] fields, not to mention reliable classifications and black hole mass measurements.

The JWST-NEXUS survey, thanks to its uniform spectroscopic survey down to deep IR fluxes, ensures that all galaxies with $m_{AB} < 24$ will have reliable spec-*z* in a large contiguous area. The *AXIS*-NEXUS survey is designed to reach an X-ray flux limit (0.5–2 keV flux $F_{0.5-2} \sim 10^{-17}$ cm$^{-2}$) such that the vast majority (>90%) of X-ray sources will have counterparts with $m_{AB} < 24$ [360,377], and thus complete spectroscopy will already be in hand for this sample. This allows all X-ray detected targets to have spec-*z* without introducing biases against low-luminosity or type 2 AGN, making AXIS-NEXUS survey the first spec-*z* complete, contiguous, deep X-ray survey. More intriguingly, the deep optical to IR spectroscopic and imaging data would best constrain the morphology and stellar population of galaxies, allowing us to have a complete X-ray sample to probe the co-evolution of SMBH and their host galaxies.

The NEXUS sample also provides a unique, spectroscopically complete sample of *galaxies* down to faint limits within which the proposed *AXIS* observations can isolate AGN. As such, our survey will provide unprecedented constraints on the probability distribution of accretion rates in galaxies [8,428], including dwarf galaxies at lower redshifts and probing the AGN occupation of galaxies around the characteristic mass scale of the stellar mass function out to $z \sim 4$. Spectroscopic tracers not only provide robust SFRs to link SMBH and galaxy assembly over cosmic time, but also provide higher-order galaxy properties that can constrain the diverse mechanisms that fuel AGN throughout cosmic time [427,632].

**2) Least Biased X-ray Luminosity Functions of AGN**

Accurate measurements of the X-ray luminosity function (XLF) of AGN at different redshifts are instrumental in constraining AGN-galaxy co-evolution, SMBH growth history, and AGN demographics [9,86,223,574]. One challenge is the lack of spec-*z* for all targets in previous contiguous area surveys. Only 55% of the $z > 3$ AGN in Vito et al. [583] and 58% of the $z < 3$ AGN in Aird et al. [9] have spec-*z*, resulting in a large uncertainty in the redshift incompleteness correction or the utilization of photo-*z*. Thanks to the uniform and complete spec-*z* compilation of the NEXUS field, we can accurately constrain the XLF of AGN without the need for the uncertain redshift incompleteness correction (see Fig.2 for the expected XLF of AGN measured with *AXIS*-NEXUS assuming the XLF in Gilli et al. [223]).

**3) Uncover the Obscured AGN Population**

Obscured ($N_H \geq 10^{22}$ cm$^{-2}$) AGN represents a fundamental phase of SMBH growth during which most of the BH mass is accreted. Obscured AGN are thought to constitute a significant fraction (70%–90% in the local Universe and the fraction is believed to increase at higher redshifts [18,86]), which may explain the lack of X-ray detections of the Little Red Dots (likely AGN) discovered by JWST in the early Universe. However, obscured AGN are difficult to find in large populations due to significant obscuration and the lack of deep IR and X-ray surveys. In addition, in most of the previous contiguous surveys in which only optical spectra were taken, type 2 AGN with narrow emission lines at $z \gtrsim 0.5$ were easily misclassified as

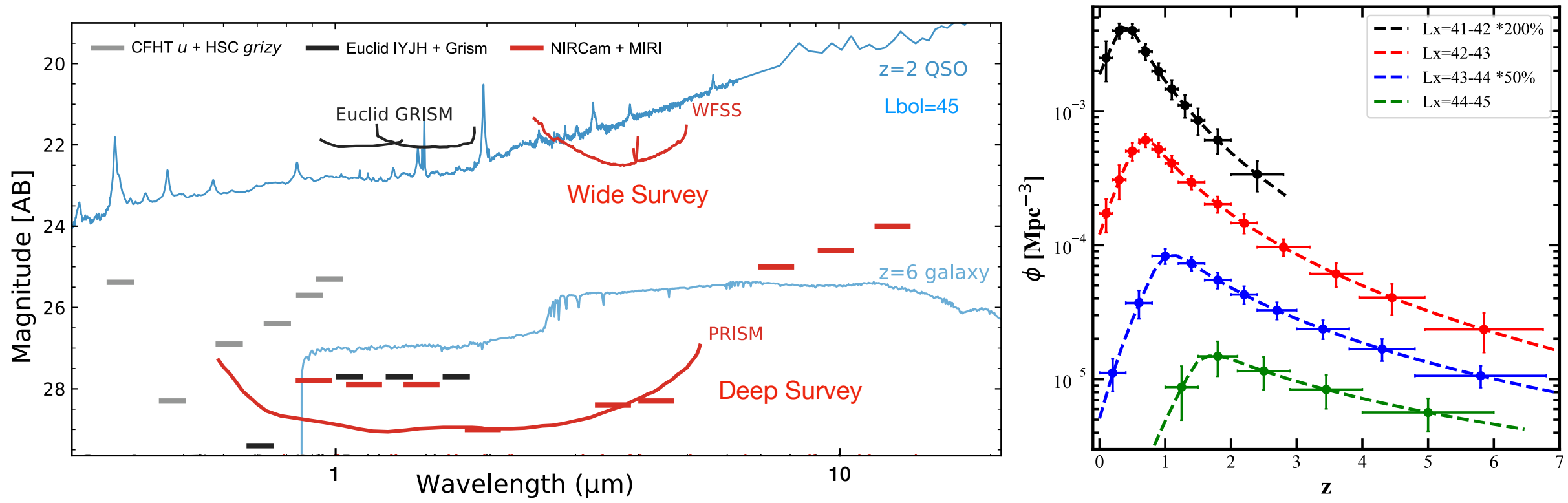

**Figure 2. Left:** Continuum sensitivity and wavelength coverage of JWST, Euclid, and ground-based telescopes [adapted from 530]. **Right:** Space density of AGN with different luminosities at different redshifts expected from the AXIS-NEXUS survey assuming Gilli et al. [223] population synthesis model.

SFGs. Therefore, the statistics and evolution across cosmic time of obscured AGN are still highly uncertain. The JWST and Euclid broadband 0.9–5 $\mu$m spectroscopic observations allow for a reliable classification of X-ray sources using different optical and IR emission lines up to $z = 6$.

We emphasize that future ngVLA observations will be significant in testing the effectiveness of radio diagnostics, such as radio-excess selection techniques [78,543] and the slope of the radio spectrum, to unveil the most obscured AGN populations.

**4) A Time Domain Field to Study X-ray Transients and Variables**

Time-domain astronomy is highlighted as one of the high-priority science themes by the 2020 Decadal Survey [422]. The NEXUS field is located in the continuous view zone (CVZ) of multiple other space missions as well, e.g., JWST, Euclid [169], WISE [613], SPHEREx [158], Roman [600], and The NEO Surveyor, enabling multiwavelength monitoring of the NEXUS field with existing or upcoming data. JWST will survey NEXUS with a two-month cadence, while Euclid will survey the field with a monthly cadence, ultimately making it the deepest Euclid field across the entire sky (with 165 times more exposure than the typical wide survey). From the ground, the Wide Field Survey Telescope [WFST, 599] and the Young Supernova Experiment on the Pan-STARRS telescope [YSE, 291] are both photometrically monitoring the NEXUS field in the optical with few-day cadences. With AXIS, we propose to perform the entire desired 1 Ms exposure in 50 epochs, each with 20 ks, making the AXIS-NEXUS a machine for discovering X-ray transients and variables. We highlight below a selection of representative targets of interest.

**AGN** exhibit variability across all wavelengths on timescales ranging from minutes to years. NEXUS is an ideal survey for investigating the circumnuclear environment of accreting SMBHs in AGN, which remains poorly understood. The long-term, deep X-ray observations of AXIS-NEXUS will enable a large-scale monitoring of both intra- and inter-observation variability of thousands of AGN across minutes to years. This is crucial for probing the gas and dust distribution surrounding the SMBHs by monitoring their line-of-sight obscuration variability as well as discovering both transient and changing-look AGN. **Tidal Disruption Events (TDEs)** are highly energetic, multiwavelength events that occur when the tidal forces of an SMBH disrupt a star at the center of a galaxy. To date, approximately 200 TDEs have been detected [136], mostly in the optical. TDEs provide an excellent sample for studying the evolution of black hole accretion disks and coronae. In a six-month *eROSITA* survey, 13 TDEs were detected at redshifts up to $z \sim 0.6$ [518]. We note that the X-ray detected TDE rate by eROSITA is about 10 times lower than expected in optical [518]. Therefore, the monitoring of the NEXUS field will provide significant clues of the X-ray and non-X-ray TDEs (host galaxy properties, time delay, intrinsic X-ray weakness, etc). This

will be crucial to understanding the tidal disruption processes of stars around the SMBHs. Other X-ray transients, though rare, are also valuable targets of the X-ray NEXUS survey, e.g., Fast X-ray Transients (FXTs), Gamma-Ray Bursts (GRBs), SNe, Fast Radio Bursts (FRBs), and Ultraluminous X-ray sources (ULXs).

**Exposure time (ks):** 1 Ms

**Observing description:** The survey includes a 1 Ms exposure of the NEXUS field (one pointing of AXIS can cover the entire NEXUS field), reaching a 0.5–2 keV sensitivity of $\sim 10^{-17}$ erg cm$^{-2}$ $^{-1}$, so that all JWST spectroscopically detected AGN in the field will have X-ray counterparts at 3-$\sigma$ detection level. The entire 1 Ms exposure is suggested to be taken in 50 epochs with 20 ks (corresponding to $\sim 2 \times 10^{-16}$ erg cm$^{-2}$ $^{-1}$ at 0.5–2 keV) in each epoch in **five years**. The intervals between epochs are suggested to be sampled from days, weeks, months, to years.

**Joint Observations and synergies with other observatories in the 2030s:** The survey has great synergies with:
JWST. The NEXUS field has been approved for three cycles of JWST observations. Further JWST observations, taken simultaneously with AXIS, will extend the NIR monitoring baseline to more than a decade.
Euclid. The NEXUS field is in the ultra-deep field of Euclid. Further observations simultaneous with AXIS will allows simultaneous X-ray and optical/NIR monitoring of AGN in more than a decade.
UVEX. The AGN accretion variability can be directly traced through UV emissions using UVEX, thanks to its large effective area. With UVEX, we are able to trace the X-ray and UV variability of AGN simultaneously and probe the connection of the corona and accretion rate.
NEWAthena. NEWAthena will provide an unprecedented photon-collecting area and spectral resolution, making AXIS and NEW Athena highly complementary.
ngVLA. Recent studies have shown that the corona produces both X-ray and sub-mm/radio emissions. Therefore, simultaneous monitoring of the NEXUS field with ngVLA and AXIS allows the best constraints on the physics of AGN corona.

**Special Requirements:** Monitoring

*3. Characterizing heavily obscured SMBH accretion*

**Science Area:** Active Galactic Nuclei, Variability studies.

**First Author:** Stefano Marchesi, University of Bologna, Italy, stefano.marchesi@unibo.it
**Co-authors:** Nuria Torres-Alba (nuria@virgina.edu); Xiurui Zhao (xiurui.zhao.work@gmail.com); Francesca Civano (francesca.m.civano@nasa.gov); Massimo Gaspari (massimo.gaspari@unimore.it), Andrealuna Pizzetti (ESO/ALMA, Chile).

**Abstract:** Spatially resolved, multi-epoch studies of nearby supermassive black holes (SMBHs) accreting matter in very dense, obscured environments can provide important insights on the geometry of the obscuring medium, as well as on the processes (possibly interconnected) that cause both the accretion and the obscuration. AXIS, with its excellent angular resolution combined with its large effective area up to 10 keV, represents the ideal instrument to perform this type of analysis. In particular, we present a GO program to target with AXIS a sample of nearby Seyfert 2 galaxies that have been determined to be heavily obscured thanks to current-generation X-ray facilities.

**Science:**

Obscuration in active galactic nuclei (AGNs) has been studied largely over the electromagnetic spectrum, from the optical [e.g., 339,537], to the infrared [e.g., 279,424], and to the X-rays [e.g., 223,495]. It is commonly accepted that the obscuration is caused by a "dusty torus", i.e., a distribution of molecular gas and dust located at ∼1–10 pc from the accreting supermassive black hole (SMBH). While the existence of this obscuring material is universally accepted, its geometric distribution and chemical composition remain a matter of debate. Several works reported observational evidence favoring a "clumpy torus" scenario, where the obscuring material is distributed in clumps formed by optically thick clouds [e.g., 89,279,380,424,501]. This claim was recently strengthened by several works based on *Chandra* imaging, which found extended hard (E>3 keV) X-ray emission in the so-called "cross-cone" region at the kpc scale in nearby ($z$ <0.03), heavily obscured AGN. The cross-cone region corresponds to the area perpendicular to the ionization cones observed in the optical, and as such is aligned to the obscuring torus. Thus, the observation of hard X-ray emission in this area is strong evidence for the clumpy nature of the torus, allowing these photons to interact with the host galaxy medium on ∼kpc scales [e.g., 174,175,572]. Finally, recent theoretical/numerical models of accretion onto SMBHs also predict a highly clumpy and chaotic multiphase medium, in particular within $r$<100 pc of the AGN [see 213, for a review].

If the obscuring torus is indeed inhomogeneous, one would expect to observe significant variability in the torus line-of-sight column (l.o.s.) density ($N_{H,l.o.s.}$) and even, in some cases, a "changing look" scenario, i.e., a transition from a Compton-thick (CT-) state (where $N_{H,l.o.s.}>10^{24}$ cm$^{-2}$) to a Compton-thin one. This transition should occur in a period as short as a day and as long as several months, assuming a typical range of obscuring clouds filling factors, velocities, and distances from the accreting BH [e.g., 424]. Initially, multi-epoch X-ray studies have primarily focused on unobscured Sy 1 galaxies and Compton-thin sources with $\log N_{H,l.o.s.}\sim$22–23 [387]. Although such works detected significant $N_{H,l.o.s.}$ variability and used it to study the torus structure, unobscured and mildly obscured AGN are not ideal targets for studying the torus's average properties. In these sources, the 0.5–70 keV emission is dominated by the primary hot corona emission. In heavily obscured AGN, instead, the main component is the absorbed reprocessed component, whose strength is linked to the torus column density and covering factor, and can be clearly observed.

As a direct consequence of this line of reasoning, recent work has focused on the analysis of multi-epoch X-ray observations of nearby Seyfert 2 galaxies selected in the all-sky *Swift*–BAT surveys, and having a line of sight (l.o.s.) column density $\log N_{H,l.o.s.}$>23 [380,474,475,561,562]. Notably, all the targets in these works have been observed at least once, and often more than once, with the *NuSTAR* telescope, meaning that their spectral shape is well characterized up to ∼50 keV. Consequently, the average

properties of the obscuring medium (such as its average column density and its covering factor), which are not expected to vary over time-scales of years, are already well constrained, and can be be used in synergy with new AXIS observations to obtain broadband X-ray spectra with excellent count statistic in the 0.3–50 keV energy range.

With this proposal, we thus aim to target four 20 ks AXIS observations of 10 known l.o.s. column density–variable, heavily obscured (i.e., with $\log N_{H,l.o.s.}$>23) AGN in the nearby Universe ($z$<0.05), all of which have been observed with *NuSTAR* at least once [the sources are selected from the works of 380,474,475,561,562]. With the proposed monitoring campaign, we will measure all the significant X-ray spectral properties of the proposed targets four times throughout one AXIS cycle. This number of observations is motivated by the fact that it has been shown [562] that by observing an obscured AGN at least four times the probability of measuring significant l.o.s. column density variability is ∼95 %. With each 20 ks observation, we will constrain the torus line of sight column density and the AGN 2–10 keV intrinsic luminosity with uncertainties <10 %. Such an accuracy can be achieved only with >60 ks *Chandra* observations, and ∼ 30 ks XMM-*Newton* observations, that however lack the angular resolution required to perform the spatially resolved analysis described below.

Once the monitoring is over, we will use the information on the variability timescales to measure multiple properties of the obscuring material, such as the distance between the obscuring clouds and the accreting supermassive black hole (SMBH; $d \propto {t_{100}}^2 \Delta N_{\rm H}^{-2}$, where $t_{100}$ is the variability time in units of 100 ks, and $\Delta N_{\rm H}$ the difference in column density between two observations; see, e.g., [65,501]), the velocity, number, size and density of the clouds, and the overall geometry of the torus (see, e.g., [328,387]). By combining the available archival observations with those proposed in this monitoring, we will be able to sample a broad range of values for all these parameters. For example, we will measure distances between the SMBH and the obscuring clouds as small as hundreds of Schwarzschild radii and as large as $\sim 10^3$ pc, i.e., from within the broad line region to outside the obscuring torus. We will also be able to measure the size of clouds as small as $R \sim 10^{12}$ cm and as large as $R \sim 10^{14}$ cm. For both $d$ and $R$, the reported numbers cover the whole range of observed values for the parameters (see, e.g., [244]). We will also use the l.o.s. column density and intrinsic luminosity measurements to test different feeding-feedback scenarios, such as those proposed within the Chaotic Cold Accretion (CCA) framework [213]. We will also test existing models of hydromagnetic disk-wind (e.g. [203]), which we will use to simulate obscured X-ray spectra matching the real AXIS and *NuSTAR* ones, to constrain the global wind geometry and its connection to the $N_{\rm H,l.o.s.}$ variability over <1 month time-scales. Specifically, such a physically motivated wind model will allow us to determine the opening angle of the disk wind, the gas density distribution along the line of sight, and its density at the base of the innermost wind radius near the disk, among other parameters. In doing so, we will also link the measured total column density of the obscuring medium to a global mass outflow rate $\dot{M} \propto nr^2v$ with kinetic power $\dot{E} \propto \dot{M}v^2$.

Following the approach discussed in previous works with *Chandra* on a subsample of nearby Seyfert 2 galaxies different from those proposed here [e.g., 174,175,572], we will also combine the proposed AXIS observations in a single one, and search for extended emission at energies above 3 keV. Where detected, the 3–10 keV extended emission would support the non-homogeneity of the obscuring material surrounding the accreting SMBH. In particular, we will search for hard X-ray emission in the so-called "cross-cone" region, which is the region perpendicular to the ionization cones and aligned with the torus. In a homogeneous torus scenario, hard X-ray photons should not be detected in these regions; therefore, a significant detection of emission above 3 keV (i.e., necessarily connected to AGN activity) in the cross-cone area would be an independent proof of the obscuring clumpiness, which only AXIS could identify.

**Exposure time (ks):** 800 ks (10 sources monitored with 4 20 ks AXIS observations)

**Observing description:** We plan to monitor the proposed targets with four 20 ks AXIS observations, with scheduling constraints that will allow us to sample a range of variability timescales (i.e., $t$∼1–10–100 days)

corresponding to SMBH-cloud distances varying from less than 0.1 pc (i.e., in proximity of the broad line region) to 10–20 pc (i.e., within the obscuring torus). Based on the average fluxes of the targets in our sample, 20 ks of AXIS observation will ensure the detection of at least 1,000 net counts in the 0.5–10 keV band. This will allow us to put constraints on the l.o.s. column density and on the intrinsic AGN luminosity with accuracy <10 %. We report in Figure 3 an example of a simulated 20 ks AXIS spectrum of NGC 1358, one of the targets in our sample, jointly fitted with one of the real, 50 ks *NuSTAR* spectra available for this target. The AXIS spectrum is simulated assuming a l.o.s. column density $\log(N_{H,l.o.s.,cm^{-2}})$ = 24, and a 2–10 keV intrinsic luminosity $\log(L_{2-10keV,cgs})$ = 43. In the inset, we report the confidence contours for $N_{H,l.o.s.}$ in the AXIS observations as a function of the power-law photon index. As can be seen, $N_{H,l.o.s.,cm^{-2}}$ is constrained with accuracy <10 % at the 99 % confidence level, as mentioned. We note that $N_{H,l.o.s.}$ has been computed independently for the AXIS and *NuSTAR* observations, since the *NuSTAR* observation had a best-fit $N_{H,l.o.s.}$ = 8 $\times 10^{23}$ cm$^{-2}$. Consequently, the excellent constraints on $N_{H,l.o.s.,cm^{-2}}$ are due to the AXIS spectral quality, combined with the constraints on the reprocessed emission provided by the existing *NuSTAR* data. Finally, the stacked 80 ks AXIS image that will be obtained at the end of the monitoring will instead be used to search for extended emission in the 3–10 keV band.

**Joint Observations and synergies with other observatories in the 2030s:** As mentioned in the scientific justification, the proposed observations will build on the legacy of years of observations with *NuSTAR*, which could in principle be operational when AXIS is launched; if that is the case, we will ask for simultaneous observations with the two telescopes. We note, however, that these are not strictly needed to achieve the planned goals. Besides *NuSTAR*, complementary analysis on the properties of the obscuring medium can be performed using ALMA and JWST [see, e.g., 381] on NGC 424, a nearby Seyfert 2 galaxy hosting a Compton-thick AGN [378]. By combining AXIS, ALMA, and JWST, we will achieve unprecedented angular resolution and sensitivity in three different bands, thereby allowing us to obtain a comprehensive characterization of the complex environment surrounding accreting supermassive black holes.

**Special Requirements:** Monitoring.

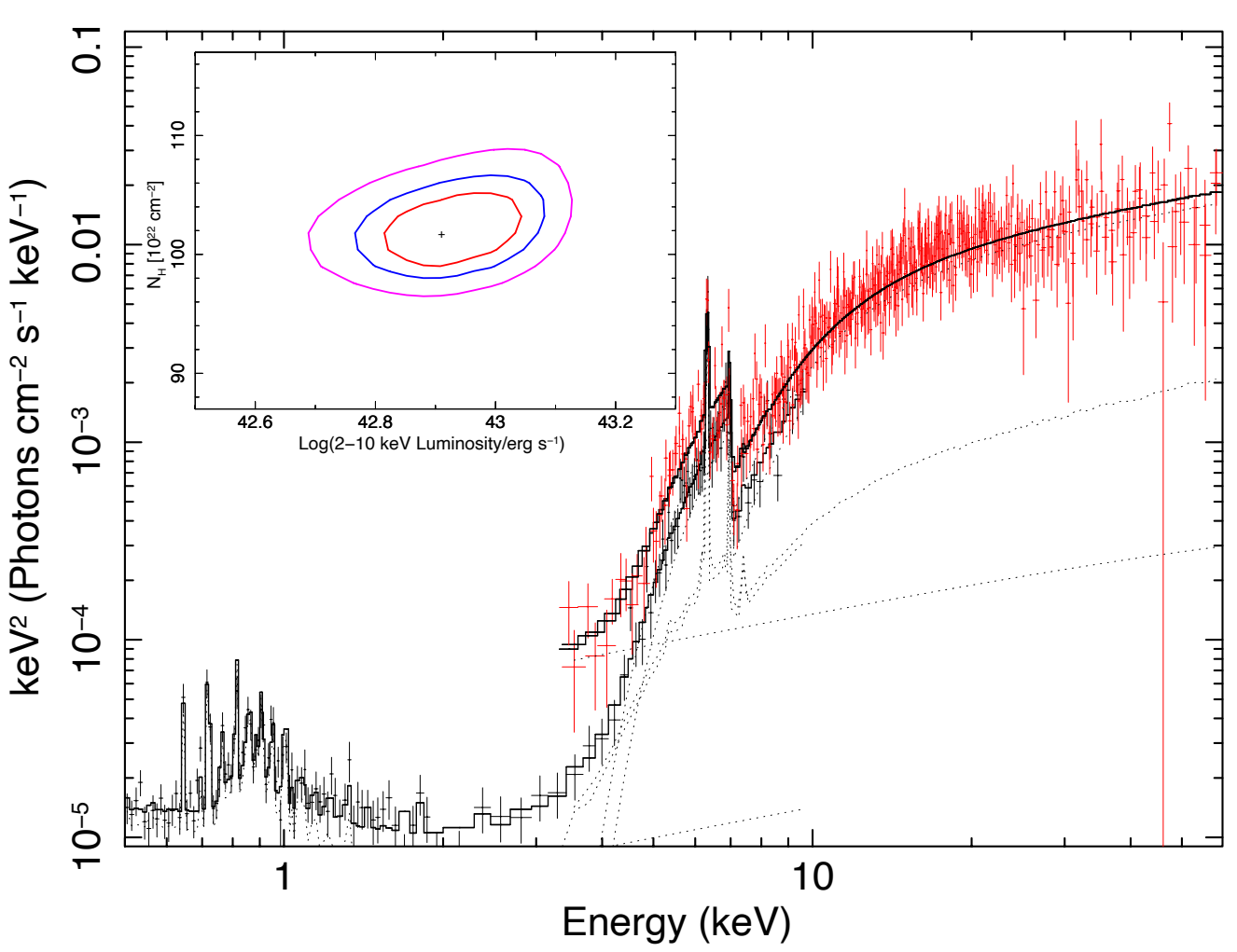

**Figure 3.** Simulated AXIS 20 ks spectrum of NGC 1358, jointly fitted with a real, 50 ks *NuSTAR* spectrum (both *NuSTAR* cameras are included). In the inset, we report the confidence contours for the l.o.s. column density as a function of the 2–10 keV luminosity, as measured in the AXIS observation. As can be seen, both parameters are accurately constrained, thus allowing us to precisely measure variability (both intrinsic and due to column density variations) over time scales as short as days.

*4. The accretion physics of QSOs at the Cosmic Dawn*

**Science Area:**AGN
**First Author:** Fabio Vito (INAF-OAS Bologna, fabio.vito@inaf.it)
**Co-authors:** Franz Erik Bauer (PUC), Andrea Comastri (INAF-OAS Bologna), Stefano Marchesi (UNIBO-DIFA), Enrico Piconcelli (INAF-OAR), Alessia Tortosa (INAF-OAR), Luca Zappacosta (INAF-OAR), Francesco Salvestrini (INAF-OATs)
**[Abstract: ]** Recent X-ray observational campaigns on high-redshift luminous QSOs revealed that the typical photon index Γ of their X-ray power-law emission steepens significantly at $z \gtrsim 6$. Possible explanations include the cooling of the hot corona due to peculiarly soft UV seed-photon fields, in turn due to the truncation of the inner accretion disk as a consequence of the launching of nuclear winds, and the presence of low-energy cut-offs related to the temperature and optical depth of the corona. These scenarios can hardly be proven with current X-ray facilities, as accurate measurements of Γ for high-redshift QSOs require time-consuming observations with Chandra and XMM-Newton even for the brightest targets. We propose to use AXIS to conduct a survey of a large sample of 18 $z \gtrsim 6$ QSOs, measuring their photon indices with an accuracy of 10% through individual exposures of 10–50 ks. The targets will be selected among known $z > 6$ QSOs and objects that will be discovered with, e.g., Euclid and Roman. The sample will span a wide range in BH mass ($10^7 - 10^{10}$ M$_\odot$) and bolometric luminosity ($\log L = 46 - 48$), allowing us to study for the first time the dependence of the steepening of Γ on these parameters. We will also observe 6 bright QSOs at $z > 6$ (5–30 ks each) for which previous X-ray observations indicate steep photon indices to constrain the possible presence of a $\approx 20$ keV cutoff in each individual target. The results will unveil key details of the accretion physics of high-redshift QSOs.
**Science:** The discovery of hundreds of luminous ($L_{bol} > 10^{47}$ erg/s) QSOs powered by already massive ($M_{BH} = 10^9 - 10^{10} M_{sun}$) SMBHs at $z \gtrsim 6$, i.e. only $< 1$ Gyr after the Big Bang, allows us to collect invaluable insights into the processes involved in the formation and early growth of the first BHs. Multiwavelength campaigns revealed that the typical spectral energy distribution of high-redshift QSOs is overall similar to that of luminous QSOs at later cosmic times, implying a lack of strong evolution of the SMBH accretion physics [e.g., 185]. This is at odds with the need for extremely fast, efficient, and possibly super-Eddington accretion required to form $10^9$ M$_{sun}$ SMBHs in a few 100s million years. Few relevant differences exist [e.g., 67], and carry key information on possibly peculiar accretion conditions and geometries that might have favored the fast growth of SMBHs at the Cosmic Dawn.

The photon index Γ, i.e. the slope of the X-ray primary power-law emission, carries information about the physical state of the hot corona, such as temperature and density, the UV photon field incoming from the accretion disk, and possibly the Eddington ratio of the QSO [e.g., 499]. Thus, Γ is an observational proxy for the accretion physics within a few gravitational radii from the SMBH, and a relatively easy quantity to measure in the X-ray spectra of unobscured QSOs.

The typical value of Γ for luminosity-matched samples of optically selected bright QSOs is remarkably constant at all cosmic times, suggesting that the typical QSO accretion state is largely time independent. Only recently, X-ray observations of increasingly large samples of high-redshift QSOs found a significant steepening of Γ at $z \gtrsim 6$ ([587,596,624]; Fig.4, left). This result is one of the few observational properties of high-redshift QSOs that sets them apart from their lower-redshift counterparts and has been possible to obtain only through time-consuming programs with Chandra and XMM-Newton, targeting each QSO with observations lasting tens to hundreds of kiloseconds.

Possible physical explanations for the Γ steepening at high redshift (see [567,624] for extended discussions) include:

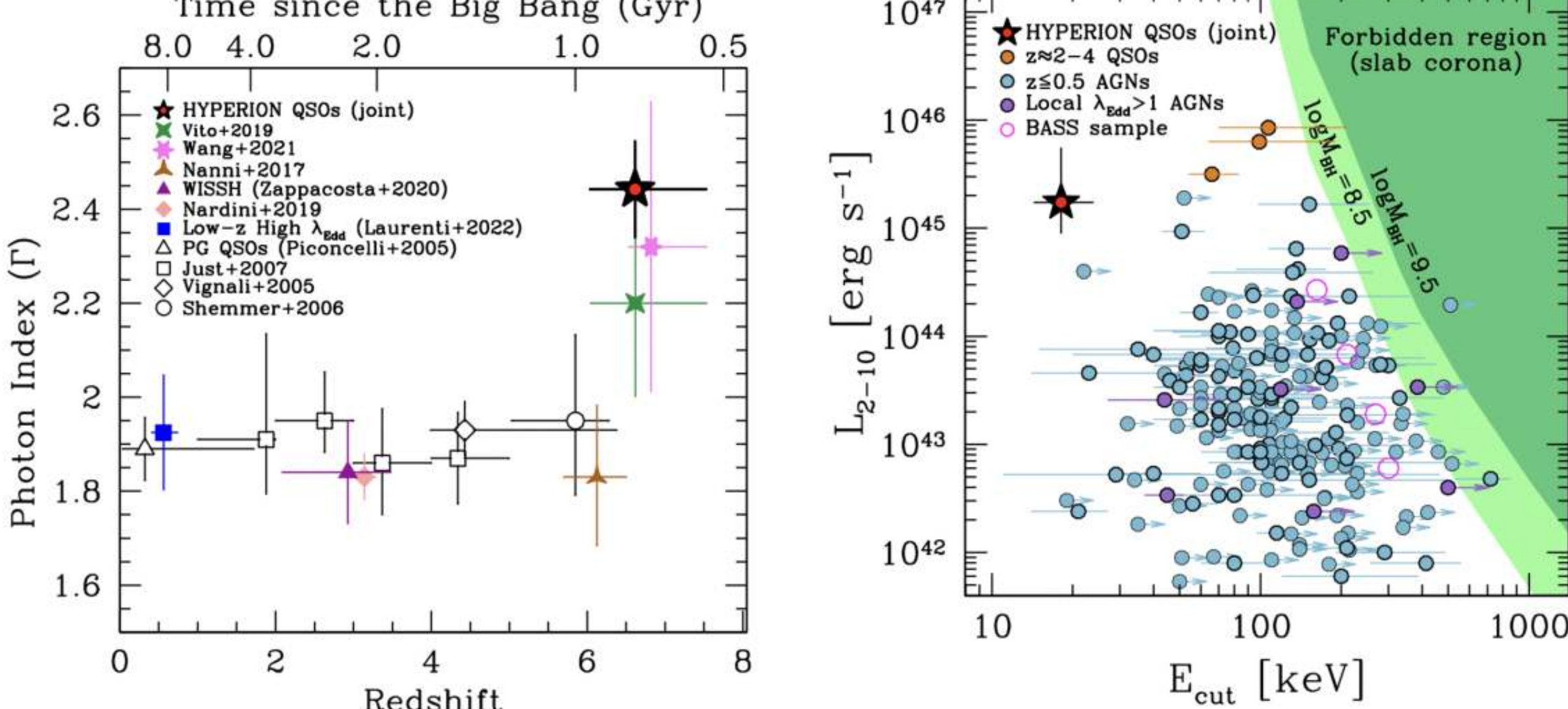

**Figure 4.** Left panel: average effective photon index of optically selected, luminous QSOs as a function of redshift. QSOs at Cosmic Dawn are characterized by significantly steeper photon indices than their lower redshift counterparts. Right panel: average X-ray luminosity versus cut-off energy for $z > 6$ QSOs (red star), compared with other samples of luminous QSOs. The observed steep photon index of high-redshift QSOs might be caused by the presence of a $\approx 20$ keV cut-off. From [624].

- A low-energy cut-off of the primary power-law (e.g., cool coronae). [624] found that a rest-frame 20–30 keV cut-off would mimic the observed strong steepening of $\Gamma$ at $z > 6$. Such energy values for the cut-off have been rarely observed in QSOs at lower redshift (Fig. 4, right).
- High Eddington ratios producing geometrically thick accretion disks and cooler coronae. This scenario might be linked to the launching of fast winds from the inner regions, and would be consistent with the larger incidence and strength of nuclear winds observed in $z > 6$ QSOs (e.g., [67]). [567] found a clear relation between $\Gamma$ and the blueshift of the C IV emission line, which is not observed for QSOs at lower redshifts (Fig. 5).

Probing these scenarios is beyond the capabilities of Chandra and XMM-Newton, as it requires significantly increasing the $z \gtrsim 6.5$ sample of QSOs with sensitive X-ray coverage and obtaining X-ray spectra of at least some of such objects to perform relatively accurate spectral analysis. Currently, this has been possible only for a handful of QSOs and only among the most luminous ones, or using stacking techniques to derive average values. AXIS will provide the sensitivity needed to detect faint point sources with tens of net counts in relatively short exposures and meet those requirements for individual QSOs spanning larger ranges of luminosity and redshift.

We propose an AXIS survey of a sample of 18 optically selected QSOs at $z = 6 - 8$ to probe the strong evolution of $\Gamma$ at such redshifts. We will more than double the number of individual QSOs at $z > 6$ for which $\Gamma$ is constrained at the 10% level, and, for the first time, extend this investigation down to logLx$\approx$ 44.5 and up to $z \approx 8$. We will probe any dependence of $\Gamma$ on X-ray and bolometric luminosity, redshift, BH mass, incidence, and strength of nuclear winds. We will also observe an additional 6 luminous QSOs included in previous X-ray investigations ([587,596]) that constrained their photon indices to be $\Gamma > 2.2$. We request exposure times suitable to constrain their individual cut-off energies $E_c$ at the 50%

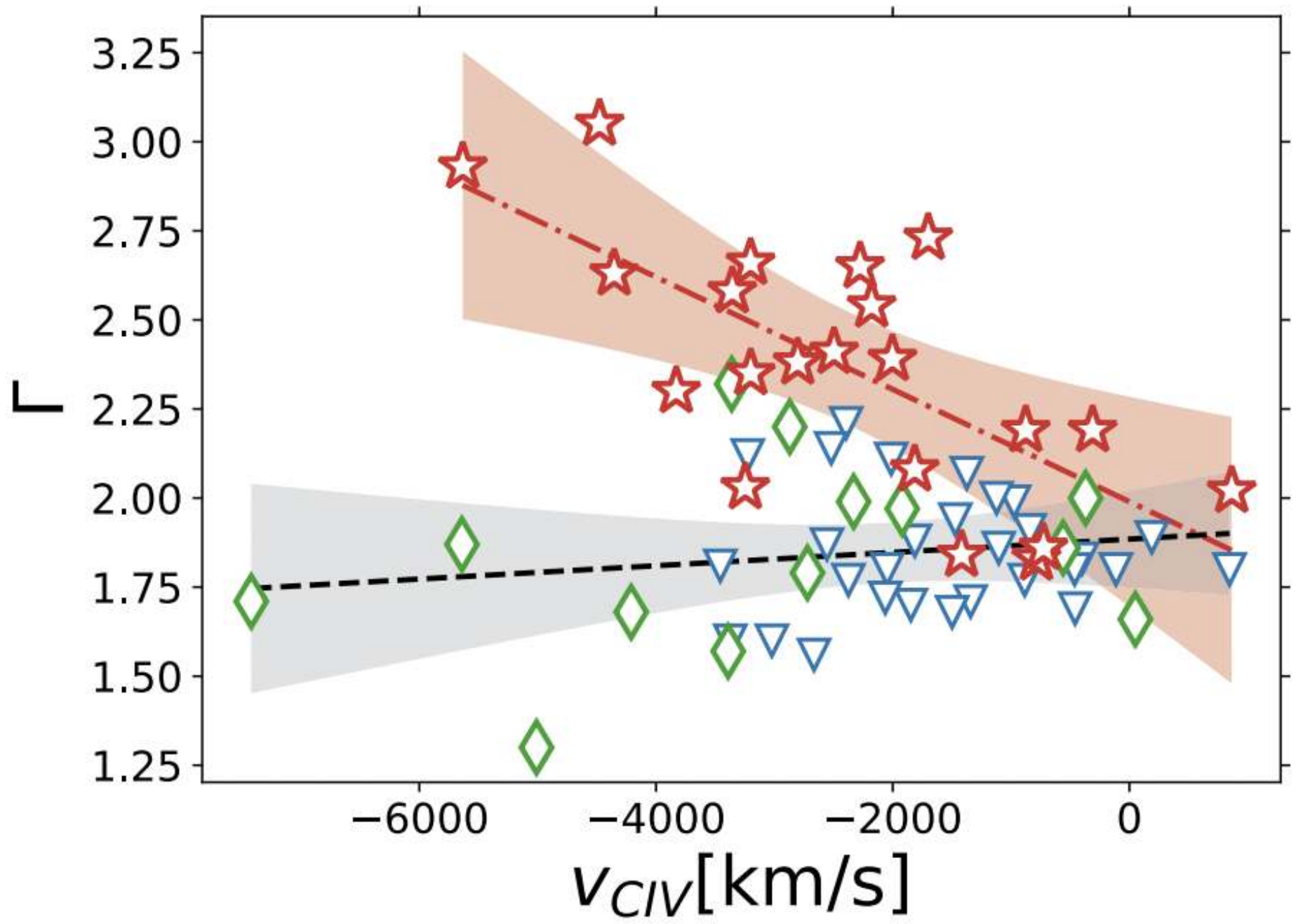

**Figure 5.** Photon index vs. C IV blueshift for a sample of luminous QSOs at $z > 6$ (red stars) and at Cosmic noon (blue and green diamonds). The anti-correlation seen for high-redshift sources indicates a strong interplay between coronal properties and disk wind launching which is not seen at lower redshift. From [567].

level, assuming an intrinsic $E_{\rm c} = 20$ keV, such that we will probe whether the observed photon index steepening at high redshift is due to the presence of cutoffs at such energies.
**[Exposure time (ks): ]** 601 ks
**Observing description:**

We consider three redshift bins $z = 6 - 6.5$, $6.5 - 7$, and $7 - 8$, each one split into a low-luminosity ($\log L_{\rm bol} = 46 - 47$) and a high-luminosity ($\log L_{\rm bol} > 47$) bins, for a total of 6 subsamples. We aim to observe three QSOs for each subsample, for a total of 18 targets (Tab. 1). Exposure times were computed to detect 60 net counts per target, assuming simple powerlaw emission with $\Gamma = 2$, and Galactic absorption ($5 \times 10^{20}\,{\rm cm}^{-2}$). This model was renormalized to produce $\log L_X = 44.5$ and 45 for the two luminosity bins (consistent with known Kbol corrections), at the redshift corresponding to the center of each redshift bin. Our spectral simulations with XSPEC using the latest AXIS response files indicate that with such a number of counts, we can retrieve the input $\Gamma$ value with 10% uncertainties. The requested exposure times are not strongly dependent on the assumed value of $\Gamma$, within a reasonable range. We will select the targets among known high-redshift QSOs [e.g., 185] and QSOs that will be identified in future years by, e.g., Roman and Euclid space telescopes (e.g., [168]). We will give priority to objects for which high-quality rest-frame UV/optical spectra exist and thus have accurate measurements of BH mass and nuclear wind properties, if present.

Moreover, we will observe six QSOs previously covered with sensitive *Chandra* and XMM-*Newton* observations ([587,596]) to constrain the possible presence of a $\approx 20$ keV cutoff. The targets are selected to be bright ($F_{0.5-2{\rm keV}} > 10^{-15}\,{\rm erg\,cm^{-2}\,s^{-1}}$) and characterized by a steep powerlaw ($\Gamma > 2.2$ at the $> 1\sigma$ confidence) in previous runs of spectral analysis. We performed XSPEC simulations assuming cut-off powerlaw spectra ($\Gamma = 2$, $E_c = 20$ keV) and normalizing them to the known soft band fluxes. We estimated

**Table 1.** Target summary and exposure times

| ID | $z$ bin | $\log L_{bol}$ bin | Exposure Time (ks) |
|---|---|---|---|
| $z6 - lowL$ | 6–6.5 | 46–47 | $3 \times 30$ ks $= 90$ ks |
| $z6 - highL$ | 6–6.5 | $> 47$ | $3 \times 10$ ks $= 30$ ks |
| $z6.5 - lowL$ | 6.5–7 | 46–47 | $3 \times 40$ ks $= 120$ ks |
| $z6.5 - highL$ | 6.5–7 | $> 47$ | $3 \times 12$ ks $= 36$ ks |
| $z7 - lowL$ | 7–8 | 46–47 | $3 \times 50$ ks $= 150$ ks |
| $z7 - highL$ | 7–8 | $> 47$ | $3 \times 15$ ks $= 45$ ks |
| ATLAS J029–36 | 6.03 | 47.4 | 30 ks |
| SDSS J0100+2802 | 6.30 | 48.2 | 5 ks |
| PSO J338+29 | 6.67 | 46.8 | 20 ks |
| VDES J0020–3653 | 6.83 | 47.2 | 25 ks |
| ULAS J1120+0641 | 7.09 | 47.3 | 25 ks |
| ULAS J1342+0928 | 6.54 | 47.2 | 25 ks |

the requested exposure times required to obtain best-fit $E_c \approx 20 \pm 10$ keV (Tab. 1). The results of the observations of this bright QSO subsample will be used to confirm whether the presence of a $\approx 20$ keV cutoff is driving the apparent steepening of the photon index, thus helping us to interpret the outcomes of the main sample.

**Joint Observations and synergies with other observatories in the 2030s:** The proposed project has strong synergies with, e.g., JWST, Roman, VLT, ELT, which will be used to discover and characterize the target QSOs. Rest-frame optical/UV spectroscopic observations are required, in particular, to estimate the BH masses of the target QSOs and the C IV blueshift to better investigate the connection between photon index steepening and the presence of disk winds.

**Special Requirements:** none.

*5. Unveil the origin of X-ray and radio weakness in JWST-discovered AGN*

**Science Area:** AGN

**First Author: Giovanni Mazzolari** (Max Planck Institute for Extraterrestrial Physics, INAF-OAS, Università di Bologna , gmazzolari@mpe.mpg.de) **Co-authors: Matilde Signorini** (ESA; INAF Firenze, Italy), **Alessandro Peca** (Eureka Scientific, USA; Yale University, USA), **Stefano Marchesi** (INAF Bologna, Italy), **Roberto Gilli** (INAF Bologna, Italy), **Fabio Vito** (INAF Bologna, Italy), **Marcella Brusa** (Università di Bologna, Italy), **Giorgio Lanzuisi** (INAF Bologna, Italy), **Andrea Comastri** (INAF Bologna, Italy), **Roberto Maiolino** (Cambridge University, UK), **Guido Risaliti** (Università di Firenze, Italy), **Elena Bertola** (INAF Arcetri, Italy), **Jiachen Jiang** (University of Warwick, UK), **Francesco Salvestrini** (INAF Trieste, Italy), **Enrico Piconcelli** (INAF Roma, Italy), **Fabio Pacucci** (Harvard University, US), **Luca Zappacosta** (INAF Roma, Italy)

**Abstract:**
The Advanced X-ray Imaging Satellite (AXIS) will be crucial in uncovering the nature of the X-ray weakness observed in the abundant population of JWST-selected AGN at high redshift. The vast majority of these sources are undetected in the current deepest X-ray surveys, implying X-ray luminosities that are 2–3 dex below expectations. Furthermore, these sources remain almost undetected at radio wavelengths, possibly indicating a weakness also at radio frequencies. Viable explanations of the observed X-ray weakness include Compton-thick absorption from a dense, dust-free medium near the black hole or super-Eddington accretion, which produces softer, beamed X-ray emission. These scenarios can also explain the possible radio weakness of these sources.
Here, we demonstrate that future AXIS deep X-ray observations will not only be able to detect the X-ray emission of these objects but also to distinguish between the two aforementioned scenarios. Additionally, AXIS observations will be complemented by future radio surveys performed by the Square Kilometer Array Observatory (SKAO), which, with sensitivities reaching the few tens of nJy, will further probe the physical mechanisms governing their accretion and emission properties.

**Science:**
The *James Webb Space Telescope* (JWST) has discovered a large population of Active Galactic Nuclei (AGN) in the early Universe. Both photometric and spectroscopic AGN selections based on JWST data revealed AGN densities at $z \sim 5$ that are $\sim 1$ dex larger compared to previous predictions [240,309,368,393,396,520,573]. Interestingly, a large fraction of these AGN revealed a significant lack of X-ray emission, even from the stack in the deepest X-ray fields (like CDFS or CDFN), with luminosity upper limits ($L_{2-10\mathrm{keV}} \lesssim$ few $\times\, 10^{41}$erg/s) that are 2-3 dex lower than expected from their $H\alpha$, optical or bolometric luminosities [17,331,370,622]. This observed X-ray weakness was found in samples of both type-1 and type-2 AGN at high redshift, and also in JWST-selected sources at intermediate redshift [$z \sim 2$ 293]. At the same time, works studying the radio properties of JWST photometrically and spectroscopically selected AGN did not reveal any detection down to very faint radio fluxes ($\sim$ 100nJy from the stack) and across multiple radio frequencies [230,397,457]. These upper limits correspond to average rest-frame radio luminosities lower than $L_{5\mathrm{GHz}} \leq 10^{39}\mathrm{erg\ s^{-1}}$ that are, at most, compatible with a radio-quiet AGN nature, possibly further suggesting that they might also be characterized by a radio weakness [397].
Different scenarios have been proposed to justify both the observed X-ray and radio weakness of the JWST selected AGN, including Compton-thick obscuration from an extremely dense and dust-free medium close to the central black hole [17,370,622], super Eddington accretion [305,331,438] or the lack/quench of the X-ray corona [421,623]. *AXIS*, in synergy with the future *Square Kilometer Array Observatory* (*SKA*), is the only facility that can answer to the compelling questions about the nature and the physical processes

characterizing these sources.

In the first scenario, the observed X-ray weakness would be due to the presence of an extremely dense gas distribution on the scale of the broad-line region (BLR). Such a Compton thick gas distribution has to be almost dust-free, to allow the BLR emission to be detected in X-ray weak type-1 AGN, and nearly spherical (with a covering factor $\sim 4\pi$) to justify the very high fraction of X-ray weak AGN detected at high-z with JWST spectroscopy ($\sim 98-99\%$ of the sources). The BLR clouds themselves are optimal candidates for being responsible for this absorption, given their characteristic high column densities [504]. This scenario is supported by the detection of $H\alpha$ equivalent width (EW) larger than in standard, optically/UV selected AGN/quasars and by the presence of absorption features in the broad Balmer emission line in a significant fraction of JWST selected type-1 AGN [148,285,293,595]. [397] also suggested that such a dense gas distribution can determine a large enough free-free opacity coefficient to almost completely absorb radio emission.

The second possible scenario explaining the observed X-ray weakness would instead be related to episodes of super-Eddington accretion producing a much softer and beamed X-ray emission. Investigating this scenario, [305,365,438] showed that the "slim disk" originating from super Eddington accretion can generate a much softer X-ray spectrum and that the orientation effect connected with the beamed emission can explain the very low fraction of X-ray detected objects at high-z. An indication of softer X-ray spectra when accretion occurs close to the Eddington limit also comes from the X-ray spectral analysis of X-ray detected $z > 6$ quasars of the HYPERION sample [567,624]. Using a theoretical model of super-Eddington accretion, [331] was able to reproduce the observed dearth of X-ray emission and also the faint ultraviolet emission lines of these objects, like CIV and HeII. Using general relativity magneto-hydro dynamic (GRMHD) simulations, [438] showed that in the super Eddington scenario, the innermost part of the accretion disk is characterized by much lower temperatures compared to the "standard" accretion disk, as well as by lower energy particle jets originating from accretion onto the SMBH. Both of these properties can also lead to reduced synchrotron radio emission in these sources.

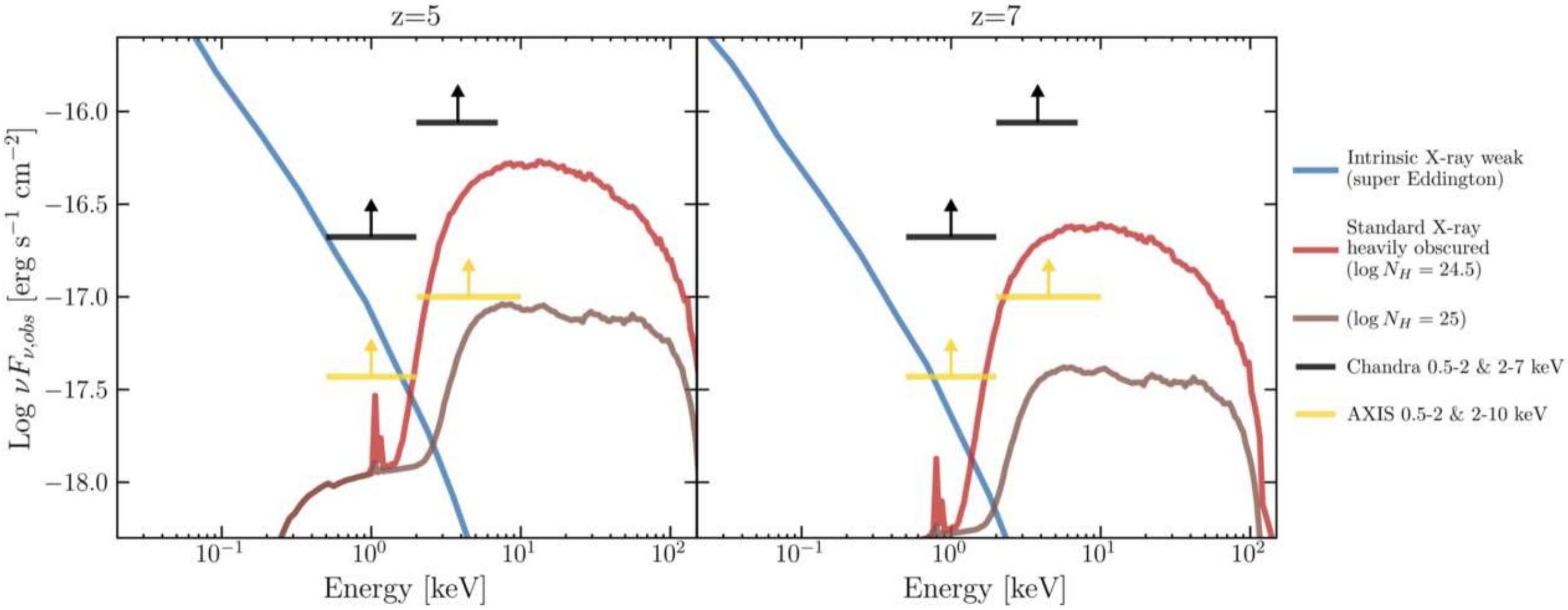

**Figure 6.** Predicted X-ray SEDs of an AGN at $z = 5$ (left) and $z = 7$ (right) with $\log L_{bol} = 44.7$ erg s$^{-1}$, the average luminosity of JWST-selected X-ray weak AGN. Blue line shows the X-ray spectrum of an intrinsic X-ray weak AGN accreting at super-Eddington, while red and brown represent a "standard" AGN obscured by $\log N_H \sim 24.5$ and 25 cm$^{-2}$. Black lines mark the 0.5–2 keV and 2–7 keV *Chandra* 7 Ms CDFS limits (20% completeness); while golden lines show the 0.5–2 keV and 2–10 keV limits for the AXIS deep survey.

In Fig. 6 we show the SED in the X-ray band derived by considering the two scenarios mentioned above. In particular, we simulated the X-ray spectra considering an object at $z = 5$ and at $z = 7$ with $\log L_{bol} = 44.7$ erg s$^{-1}$, which corresponds to the average redshift and bolometric luminosity of the sample of JWST-selected, X-ray weak AGN. The red and brown spectra correspond to the case of heavy obscuration and are generated using XSPEC and an ad-hoc obscured AGN spectral model obtained using the Monte Carlo radiative transfer code BORUS [40]. In this model, the gas is assumed to be uniformly distributed, and the line-of-sight obscuring medium can have a different column density than the average column density of the torus to account for clouds passing in/out of the line of sight. In Fig. 6 we considered two different values of $\log N_{H,los} = \log N_{H,tor} = 24.5, 25$ cm$^{-2}$. Instead, the X-ray spectrum in the case of an intrinsic X-ray weakness due to super-Eddington accretion is taken directly from the SED model presented in [438], which was derived using GRMHD simulations of super-Eddington accretion. This spectrum is extremely steep, determining a much softer X-ray emission compared to the obscured scenario. In the same figure, we also plot the flux limits in the 0.5-2 keV and 2-7 keV bands at the 20% completeness level for the CDFS [351] and for the Deep Survey that will be performed by *AXIS* (both covering $\sim 450$ arcmin$^{-2}$). Clearly, while both the intrinsically weak X-ray and the heavily obscured SEDs remain below the detection limits of *Chandra*, they are expected to be both detected at the *AXIS* deep survey sensitivities. Furthermore, thanks to their drastically different slopes, *AXIS* observations will not only detect these objects but also unveil the true nature of the observed X-ray weakness. Given that *AXIS* is expected to detect sources characterized by both these X-ray spectral shapes, in case of non-detection, the only remaining scenario would be the lack (or a significant suppression) of the overall X-ray emission due, for example, to a much less efficient (quenched, or absence of) X-ray corona [22].

On the radio side, the great improvement represented by the Square Kilometer Array Observatory (SKAO) by the end of this decade will allow achieving rms at 1.4GHz of the order of 50 nJy (in the Ultra Deep survey) [481]. SKAO observations are expected to clearly detect the radio emission produced by the accretion onto these objects if they are "standard" RQ AGN, following the well-established luminosity correlations between the luminosity in the radio band and in the X-rays, optical, or [OIII] emission [444]. In the case of intrinsically weak X-ray emission (with $\log L_X \sim 41.5$, at $z \sim 5$), SKAO observations are expected to still marginally detect their radio emission, provided that the fundamental plane relations or the direct $L_X - L_{radio}$ correlations used to predict the radio luminosities are still valid at these redshifts. Therefore, the synergies of AXIS and SKAO observations will be fundamental and decisive in shedding light on the physical properties characterizing the accretion on these early SMBH.

**Exposure time (ks): Archival study using directed science deep fields.**

**Observing description:**

The scientific goals of this proposal will be achieved thanks to the data collected in the deep survey that *AXIS* will perform on well-known extragalactic fields (already covered by *JWST* imaging and spectroscopy and where most of the *JWST*-selected AGN reside). There is the possibility of achieving similar results by taking ad-hoc pointed observations of lensed sources in the galaxy cluster fields already observed by *JWST* (like the Abell cluster), where the lensing effects can further strengthen the capabilities of *AXIS* requiring shorter exposure times to retrieve comparable sensitivities. However, due to the depth of the required observations, it is expected that, in any case, the targets that this proposal will investigate will depend on the fields observed by the *AXIS* surveys.

**Joint Observations and synergies with other observatories in the 2030s:** SKAO and JWST as clearly expressed in the main body.

**Special Requirements:** none

*6. X-ray to UV relation and application to cosmology*

**Science Area:**
**First Author:** Guido Risaliti, University of Florence, Italy guido.risaliti@unifi.it
**Co-authors:** Franz E. Bauer (Universidad Catolica, Chile), Francesca Civano (NASA-GSFC), Elisabeta Lusso (University of Florence), Riccardo Middei (CfA | Harvard & Smithsonian), Emanuele Nardini (INAF), Andrea Sacchi (CfA | Harvard & Smithsonian), Matilde Signorini (ESA).
**Abstract:** The non-linear X-ray to UV luminosity relation is a powerful tool for measuring quasar distances at redshifts of 2 to 7, where no other cosmological standard candle is available. To date, this method has been applied to approximately 2,000 sources with SDSS optical spectra and serendipitous XMM-Newton and/or Chandra observations. AXIS will revolutionize this study: within just a few years, it will serendipitously observe thousands of SDSS quasars, providing high S/N spectra for each. Even larger samples will emerge from upcoming quasar catalogs, such as those from Rubin and Euclid. The AXIS quasar sample will be transformative for cosmological distance measurements in two key ways: 1. It will produce a high-precision Hubble diagram of quasars at high redshifts, enabling powerful cosmological tests. 2. It will rigorously validate the method by testing the redshift independence of the X-ray to UV relation at the percent level.
**Science:**

In the past few years, it has been proved that quasars can be turned into *standard candles* by using the non-linear relation between their monochromatic luminosities at 2500 Å ($L_{UV}$) and 2 keV ($L_X$). A log-linear law reproduces this relation (known for more than 40 years): $\log(L_X)=\gamma\log(L_{UV})+\beta$, with $\gamma\sim0.6$, $\beta\sim8$, and a dispersion of ∼0.4 dex [363]. In principle, such a relation can be written in a cosmology-independent form through fluxes, which can then be used to obtain an independent measurement of the quasar distances [502,503] Yet, until a few years ago the large scatter of the relation has always deterred any attempts to use it for cosmological purposes. The situation has now radically changed:
*(1)* The redshift-independence of the relation has been tested through an analysis of its slope $\gamma$ in small redshift intervals, confirming that $\gamma$ is constant from $z\sim0.3$ to $z\sim6$. The normalisation of the Hubble Diagram is obtained against that of supernovae Ia (SNe Ia), analogously to what is done to calibrate SNe Ia using Cepheids. The perfect match of the shape of the Hubble diagram between quasars and SNe Ia in the common redshift range is a strong confirmation of the reliability of the method.
*(2)* We cross-correlated the latest SDSS quasar catalog with the XMM-Newton and Chandra point source catalogs to obtain a sample of more than 10,000 quasars with both UV and X-ray high-quality spectra. This improvement in sample statistics enabled us to apply stringent filters, thereby avoiding X-ray absorption, dust extinction, and systematic effects resulting from the flux limits of the parent samples. As a result, we obtained an unbiased quasar sample optimized for a detailed study of both the origin of the dispersion of the relation and possible systematics and biases. Once objects affected by dust and/or gas absorption, or with a poor quality X-ray spectrum (due to calibration issues or high background flares) are removed from the sample, the observed dispersion drops to $<0.24$ dex.
*(3)* A more detailed analysis of a few tens of sources with the highest quality X-ray observations in the sample showed that the intrinsic dispersion of the X-ray-to-UV relationship must be lower than 0.1 dex. This result suggests that the physical mechanism responsible for the observed relation is universal for all quasars at all redshifts and luminosities.
*(4)* The Hubble diagram obtained with the sample of quasars mentioned above shows a $\sim4\sigma$ discrepancy with the standard ΛCDM model at redshifts z>1.5.

The results summarized above demonstrate the significant relevance of this study to both the physics of quasars and cosmology, while also highlighting its current limitations. The high-quality observations show that it is possible to obtain strong cosmological constraints and new insights in quasar physics;

on the other hand, the relatively low quality of the X-ray observations is the "bottleneck" of the project: the sample with a dispersion of the order of 0.1 dex consists of just a few tens of sources, way too few for any cosmological application, while the total sample of more than 2,000 quasars provides distances with uncertainties of about a factor of 2 in individual sources. Such large uncertainties not only limit the precision of cosmological tests but also make it difficult to rule out the presence of hidden systematics that may affect the reliability of the results.

To overcome these limitations, we need to build large quasar samples with X-ray spectral observations of the same quality as the best 2-3% of the currently available Chandra or XMM-Newton observations. This is precisely what AXIS will be able to achieve effortlessly, only through serendipitous observations of quasars with available rest-frame optical/UV spectral observations.

It is expected that new large quasar samples will become available in the next few years from instruments such as DESI, Euclid, Rubin, and UVEX. However, here we only assume the availability of the latest SDSS quasar catalog (DR16, Lyke et al. 364). Even if this is a very conservative assumption, it is more than enough to illustrate the transformative contribution by AXIS to this field.

In Fig. 7 we show a spectrum from a pointed XMM-Newton observation of a z$\sim$3 bright quasar, and a simulated spectrum obtained from a 10 ks AXIS off-axis serendipitous observation of one of the faintest sources in the DR16 quasar catalog. The XMM-Newton spectrum is from the "golden" sample with a dispersion lower than 0.1 dex. This simulation shows that every SDSS quasar serendipitously falling into an AXIS field of view will have a good enough X-ray spectrum to provide a distance estimate with an uncertainty of the order of 0.1 dex. To predict an AXIS Hubble diagram of quasars, we consider that

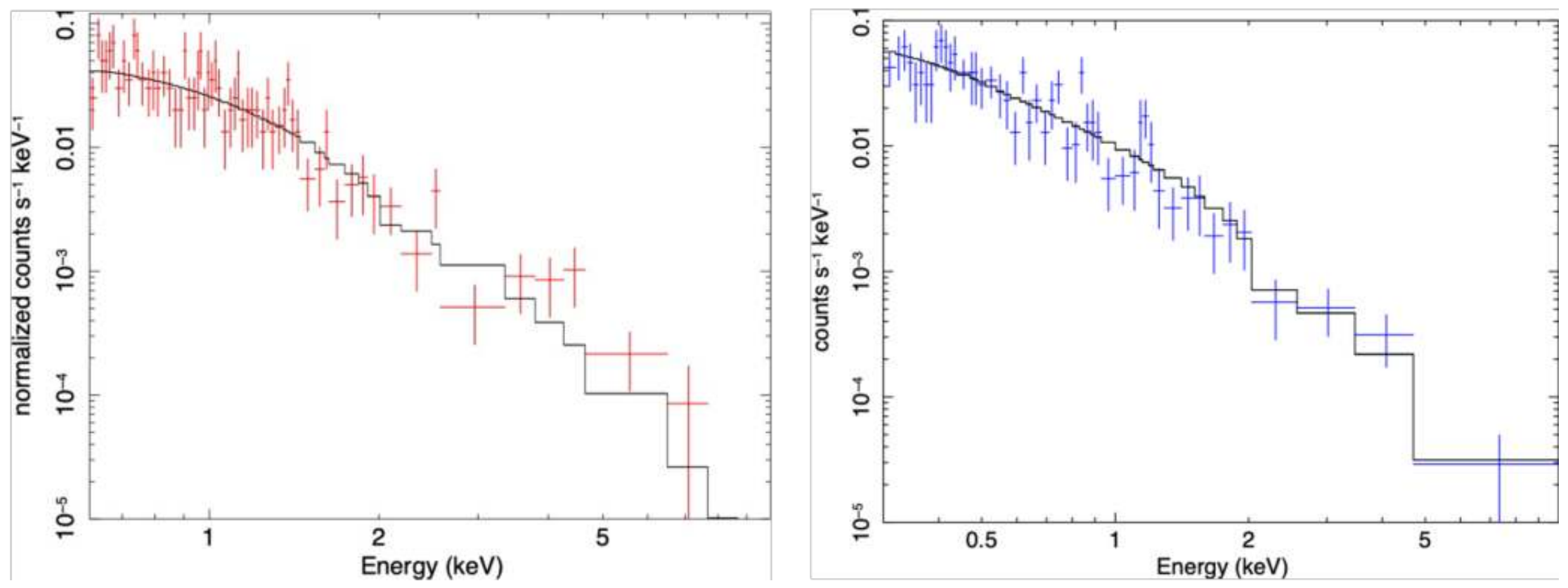

**Figure 7.** Left: Spectrum from a pointed XMM-Newton observation of a z$\sim$3 bright quasar. Right: Simulated AXIS spectrum for an off-axis exposure of 10 ks of a "worst case" SDSS DR16 quasar with a 0.5-10 keV flux of $7\times10^{-15}$ erg cm$^{-2}$ s$^{-1}$.

the SDSS DR16 covers approximately 1/3 of the sky; therefore, a random AXIS observation has a 33% chance to cover an SDSS field. The exact area covered by AXIS observations will depend of the observing strategy. However, if we conservatively assume an observed area of 120 deg/year, and considering that the density of SDSS quasars is $\sim$50/deg$^2$, we expect $\sim$2,000 quasars serendipitously observed every year. In Fig.8 we show the predicted X-ray to UV relation for three small redshift intervals, and the expected Hubble diagram of quasars after 3 years of observations. Such data quality will rule out the presence of systematics and provide the most precise Hubble diagram ever produced up to redshift z$\sim$5.

**Exposure time (ks):** Archival deep and wide surveys.

**Observing description:** All the observations will be serendipitous, with no need for dedicated observing time.

**Joint Observations and synergies with other observatories in the 2030s:**

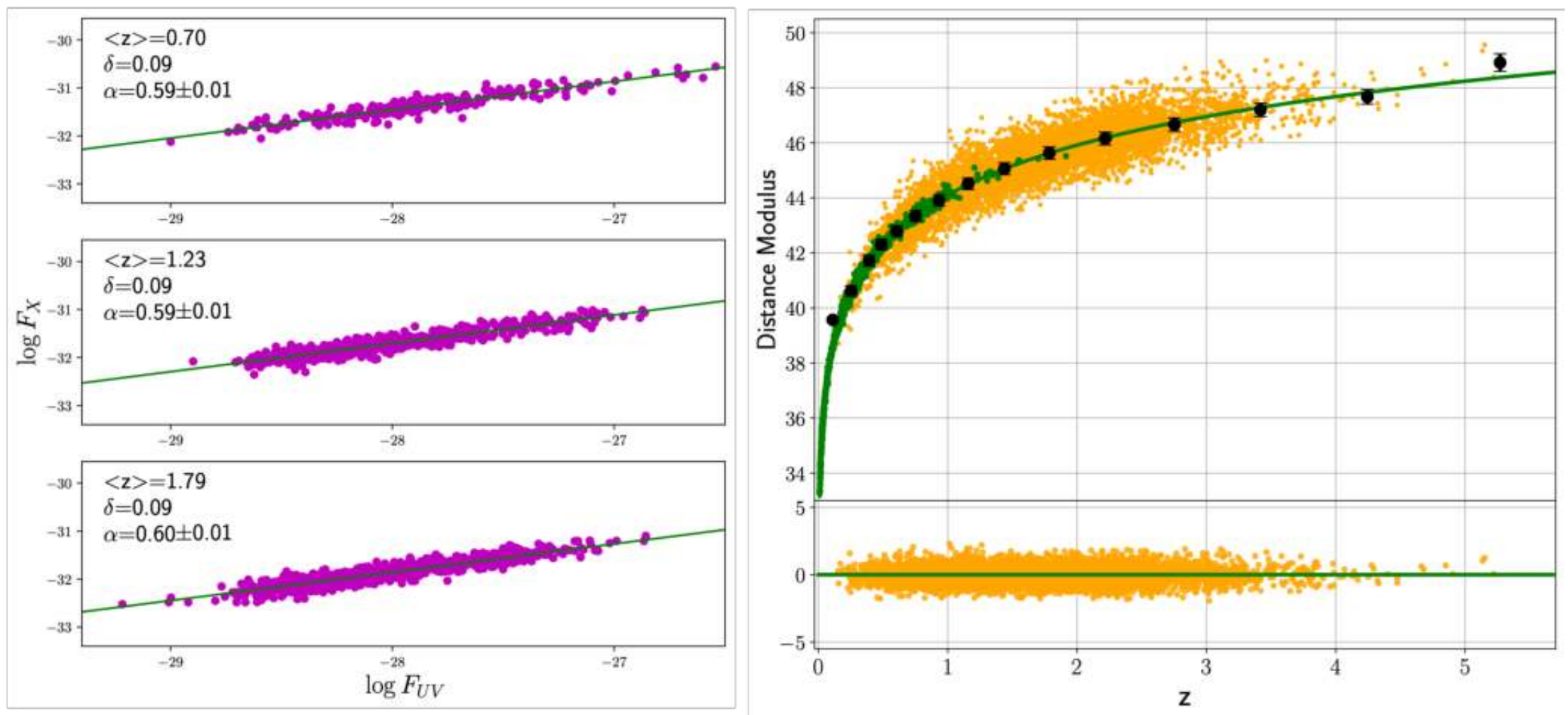

**Figure 8.** Left: examples of the X-ray to UV relation in small redshift intervals. Right: Hubble Diagram of supernovae (green points) and quasars (orange points) after 3 years of AXIS observations. The black points are quasar averages in small redshift bins.

**Special Requirements:** None

*7. SMBH spin survey with the AXIS deep field survey*

**Science Area: Supermassive black holes**

**Authors:** **Júlia Sisk Reynés** (Center for Astrophysics | Harvard & Smithsonian): julia.sisk_reynes@cfa.harvard.edu) **Co-authors:** **Jiachen Jiang** (University of Warwick): Jiachen.Jiang@warwick.ac.uk); **Dan Wilkins** (The Ohio State University): wilkins.401@osu.edu), **Joanna (Asia) Piotrowska** (Caltech): joannapk@caltech.edu, **Andrew Young** (University of Bristol): Andy.Young@bristol.ac.uk, **Dominic Walton** (University of Hertfordshire): d.walton4@herts.ac.uk, **Jaya Maithil** (Center for Astrophysics | Harvard & Smithsonian): jaya.maithil@cfa.harvard.edu, **Angelo Ricarte** (Center for Astrophysics | Harvard & Smithsonian): angelo.ricarte@cfa.harvard.edu,

**Abstract:**

The spin distribution of supermassive black holes (SMBHs) encodes information of their growth over cosmic times via merger-driven vs. accretion-driven scenarios. The key questions we will tackle are: what is the spin of moderately-accreting ($\lambda_{\rm Edd} = 0.1 - 0.3$) SMBHs in type-1 AGN across $z = 0 - 2.5$ as revealed by *AXIS*? What are the implications for our understanding of SMBHs over cosmic times via merger-driven vs. coherent-accretion driven scenarios? How is the growth of SMBHs linked to their environment, as suggested by black hole spin?

**[Science]:**

The no-hair theorem of General Relativity states that, at a fundamental level, astrophysical black holes are defined by two quantities: mass ($M_{\rm BH}$) and angular momentum. The angular momentum is usually quantified via the dimensionless spin parameter, $a^* = cJ/GM^2_{\rm BH}$, where $c$ is the speed of light in vacuum, $G$ is the gravitational constant, $J$ is the angular momentum, and $M_{\rm BH}$ is the black hole mass. For a Kerr black hole, $a^*$ can take any value $\in \pm 0.998$, where a positive or negative sign indicates the black hole is rotating in the same direction or in a counter direction with respect to the accretion flow, respectively. Furthermore, spin is a compelling indicator of recent growth. Semi-analytic and numerical models of cosmic structure formation predict that coherent accretion onto SMBHs will increase the observed SMBH spin magnitude ($a^*$), whereas black hole mergers will, on average, decrease the observed SMBH spin magnitude. The exact relationship between spin and the SMBH growth history will depend on understanding the physics of different accretion modes. Most notably, jets powered by spin via the Blandford-Znajek mechanism process will remove angular momentum, relating to epochs in which the 'quasar mode of AGN feedback' is operating.

Reflection of X-rays of the innermost regions of the accretion disk probe the extreme environment just outside the event horizon. If the inner disk truncates at the innermost stable circular orbit (ISCO) —itself a function of $a^*$— then the X-ray reflection spectrum enables a measurement of $a^*$ from determining the ISCO radius. X-ray reflection spectroscopy models have been used to describe the reflection spectra of $\sim$ 60 AGN with masses $M_{\rm BH} \geq 10^{5-9} M_\odot$ [41,372,373,488]. Interpreting the observed distribution of SMBH spins, as determined by X-ray reflection spectroscopy, is, however, challenging due to a combination of factors. First, these studies are based on flux-limited samples. Second, some of the current bounds are only lower limits. Third, some of the current spin estimates have been obtained based on the reflection interpretation of the soft excess. Fourth, most spin measurements exist for $z < 0.1$.

We propose performing an SMBH spin survey by modeling the relativistically broadened Fe-K$\alpha$ features in AGN up to $z \sim 2.5$ detected by the *AXIS* Deep Survey (5-Ms). In the absence of strong gravitational lensing and microlensing, only *AXIS* –among current and foreseeable future missions– will be able to reliably detect relativistically broadened Fe-K$\alpha$ signatures up to these cosmologically interesting redshifts with its collecting area and broadband coverage. A systematic study of SMBH spin versus redshift would offer a new opportunity for calibrating subgrid physics models in cosmological simulations [see discussions in 473]. Redshift-resolved information on spin parameter trends in SMBH populations

provides an additional dimension for comparison against cosmological simulations, which is beyond the capabilities of current observations.

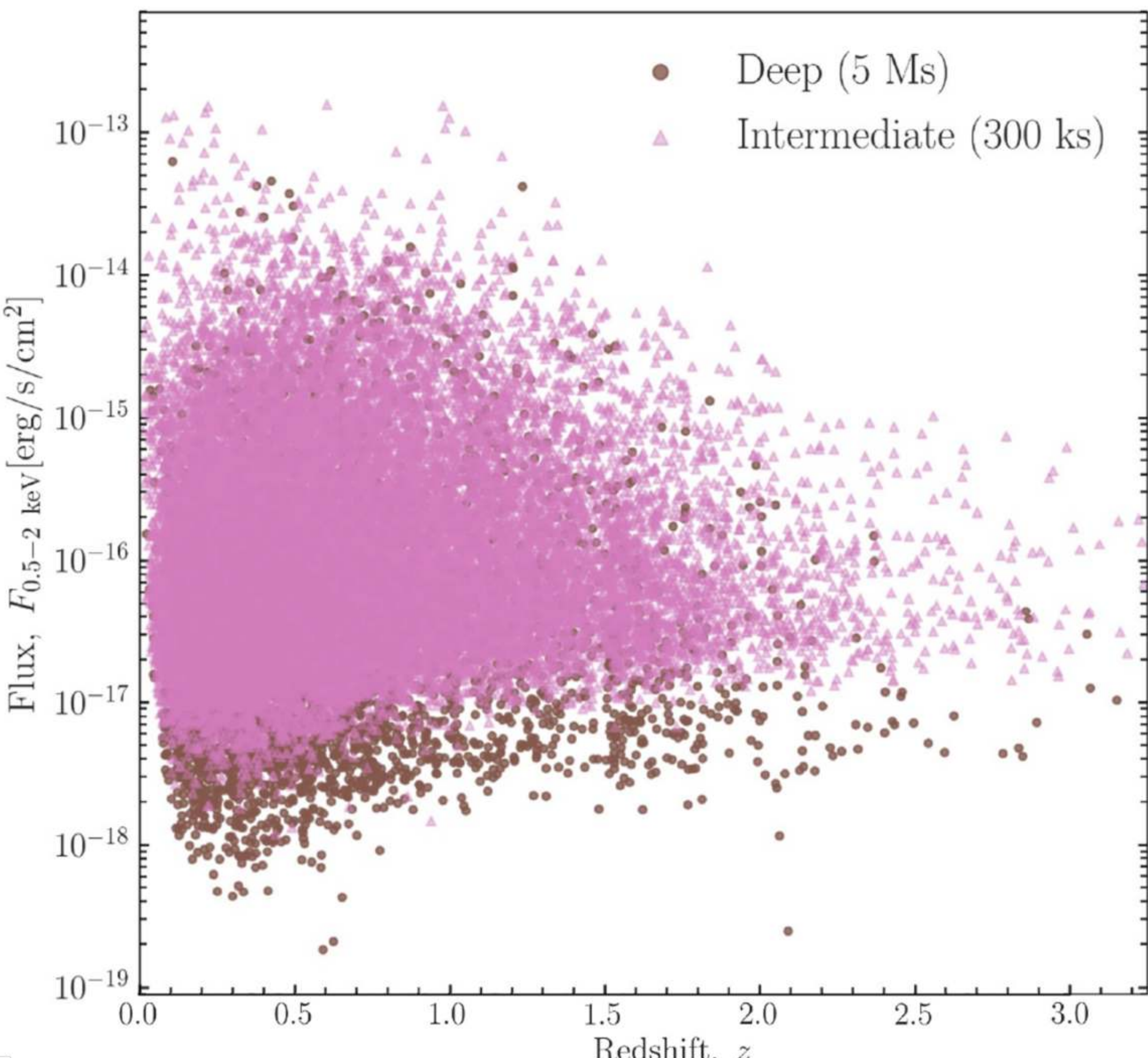

**Figure 9.** Flux and count capabilities expected from the *AXIS* Intermediate and Deep Surveys ($t$ = 300 ks, 5 Ms, respectively). Taken from Marchesi et al. [379] in early 2023.

X-ray reflection spectrosocopy

In the presence of high-quality spectral data and in the absence of spectral degeneracies induced by warm absorbers and/or ultra-fast outflows, X-ray reflection spectroscopy models can be fit to observed data to predict the reflection spectrum from the hot X-ray corona. These models predict the X-ray reflection spectrum as coronal emission is reflected from the inner accretion disk by taking into account the relativistic effects of Doppler shifts and gravitational redshifts. The most obvious signature of coronal emission reprocessed in the ionized accretion disk is the appearance of the Fe-$K\alpha$ emission line, which has a rest frame energy of $6.4 - 6.97$ keV depending on the ionization state of the iron in the disk. For a wide

range of ionization states, the reflection spectrum from the inner disk also contains a forest of soft X-ray emission, which can be broadened into a pseudo-continuum soft excess.

When used to describe the observed features in the Fe-K band of moderately accreting AGN, reflection spectroscopy can be used to locate the inner edge of the accretion disk. If the disk truncates at the innermost stable circular orbit (as expected for moderately accreting AGN), reflection models can be used to describe the observed features to subsequently estimate several fundamental parameters of the system: $a^*$, and the iron abundance and inclination of the inner accretion disk. In general, depending on the quality of the data, one may expect a degeneracy between the best-fit inclination and spin values [539]. In contrast, the best-fit metallicity value of the inner disk may reach a few times the Solar value [488]. In addition, [182] report on the degeneracy between the extent or scale height of the corona and the inferred spin. The latter suggests the existence of non-negligible systematics in current X-ray reflection spectroscopy models. While access to the Compton hump is critical to break some of these degeneracies (e.g. leveraging to the hard-energy coverage by *NuSTAR* for local AGN), we will thoroughly explore the parameter space of the reflection models when measuring black hole spins.

How will we estimate SMBH spin from X-ray reflection in the *AXIS* Deep Survey?

We will simulate spectra of typical type-1 AGN at a range of observed fluxes down to the flux sensitivity of the *AXIS* Deep Survey, for each source redshift (Figure 9). Each of these simulated spectra will be generated for different model parameters, including: the photon index $\Gamma$, the inclination $Inc$, and the black hole spin ($a^* \in [-0.998, +0.998]$). For each simulated spectrum, we will robustly assess to what extent the fundamental model parameters ($a^*$, and inner disk iron abundance and inclination) can be recovered based on modeling the relativistically blurred iron line with RELXILL provided a phenomenological irradiation profile. This assessment will allow us to compile an error bar in $a^*$ (for a given $a^*$) and assess its evolution with the redshift of the source. As an example, Figure 10 shows the total number of photon counts vs. redshift needed to reliably model the relativistically-blurred Fe-K$\alpha$ emission features in simulated *AXIS* spectra assuming a maximally-spinning black hole ($a^* = +0.998$). The colorbar shows the uncertainty in the inferred spin when all the 1 000 simulated spectra (featuring $a^* = +0.998$) are fitted simultaneously with the RELXILL model. For a given redshift $z$, all spectra were simulated to reproduce an unabsorbed luminosity of $L_{0.5-2} = (1-2) \times 10^{41}$ erg/s, corresponding to a flux of $\sim 10^{-14}$ erg/s/cm$^2$ at redshift $z = 2$. Each simulated spectrum is drawn from a total number of photon counts $\in 10^{4-6}$. The number indicated by each vertical line is the average spin best-fit value inferred from fitting all simulated spectra at a given $z$.

As the redshift increases, we will be able to constrain the continuum above the Fe K$\alpha$ line and start being sensitive to the Compton hump. Having access to fitting the Compton reflection hump allows reducing systematic uncertainties in the broad iron line parameters, ultimately shrinking the statistical error bars in the inferred spin magnitude value. At $z = 2.5$, $E = 8$ kev corresponds to the rest-frame $E = 30$ keV –within the realm of the Compton hump.

An important caveat is that Figure 9 does *not* capture other model degeneracies that could affect the X-ray continuum. At present, the simulation setup does not account for realistic spectral complexity beyond fitting the relativistically-broadened Fe-K$\alpha$ line, thus not accounting for: 1) multiple reflection components, 2) ionized emission lines, and 3) ionized mass outflows. In the case of 1), multiple reflection components may appear due to coronal reprocessing in the innermost regions of the disk, as well as cold, slowly moving matter surrounding the vicinity of the AGN. The current simulation setup can be readily extended to consider the latter, e.g., by simulating spectra with a single or both reflection components and by comparing the goodness-of-fit statistics of a narrow-line-only vs. relativistic-reflection-only vs. relativistic+cold reflection models. A similar implementation could readily address 2) by also implementing the search for

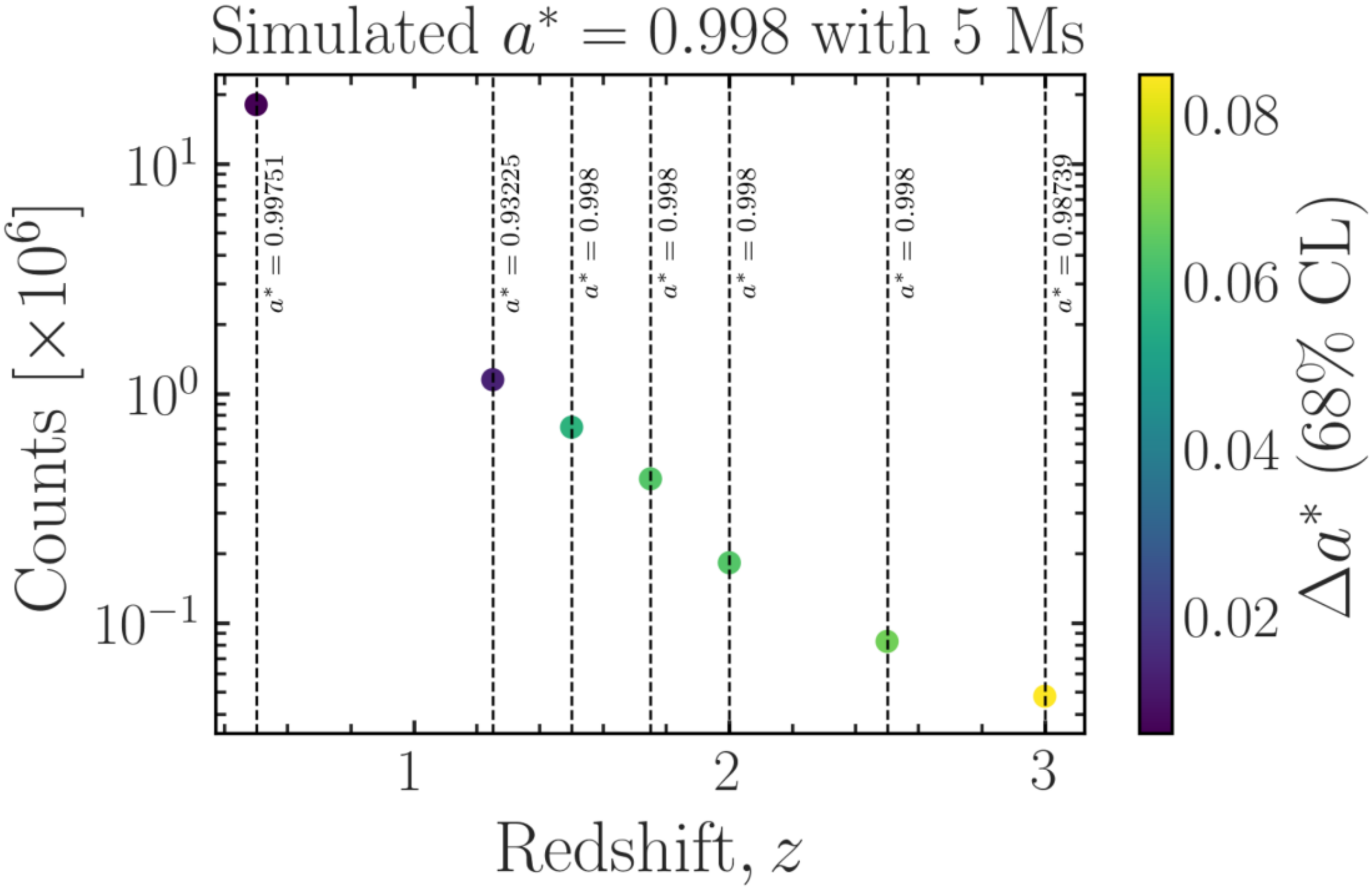

**Figure 10.** Recovery of the black hole spin magnitude when 1 000 mock *AXIS* spectra from the 5 Ms Deep Survey (each simulated with the RELXILL model for $a^* = +0.998$) are fitted with a RELXILL where the black hole spin magnitude is fitted for. The total counts shown in the vertical axis are inferred from the simulated spectra properties in the 0.5-8.0 keV band. Each simulated spectrum assumes an exposure of 5 Ms and used *AXIS*' average field-of-view ARF available in early 2023. The simulated spectra have been assigned fluxes to correspond to an unabsorbed luminosity of $L_{0.5-2} = (1-2) \times 10^{41}$ erg/s. This corresponds to a flux of $\sim 10^{-14}$ erg/s/cm$^2$ at redshift z = 2.

Gaussian line scans [313]. However, addressing 3) would pose a nontrivial exercise given the severe spectral degeneracies between ionized mass outflows and relativistic reflection features. [448] demonstrated that fitting a single process emission model (either relativistic reflection or ionized outflow) to a hybrid line profile due to both components results in large systematic biases in the estimates of key parameters, including the SMBH spin $a^*$, and subsequently discussed various strategies to mitigate this effect. These strategies include having access to harder X-ray energies to cover the Compton hump, and high-energy data that covers the Compton hump. Machine learning techniques may also be implemented to identify the presence of a given component in the spectrum [447].

To improve signal-to-noise statistics in the Fe-K$\alpha$ band for high redshifts, we will perform Bayesian hierarchical analysis to assess the reliability of parameter inference (for $a^*$, inner disk iron abundance, and inner disk inclination) in our simulations. This will allow us to identify which model parameters $a^*$ are degenerate with, as a function of the redshift, and to assess to what extent the true $a^*$ is recovered. These spectra will incorporate a more complex spectral model as described above to explore potential parameter degeneracies introduced by additional components fully. In addition, we will consider stacking broadened spectra from simulated, isolated point sources.

We will extend the work done within, e.g., the *Suzaku* spin survey [594]. Knowledge on the redshift of all sources is required. For those AGN whose black hole spins will be estimated by the proposed *AXIS* spin survey and which will have a mass estimate available ($M_{\rm BH}$, e.g. from reverberation mapping or single-epoch H$\beta$, Mg II, C IV line measurements), *AXIS* will help populate the observed SMBH spin distribution drawn from X-ray reflection further. These may help uncover SMBHs with $M_{\rm BH} > 10^8 M_\odot$ which are currently relatively rare in this distribution, and are crucially needed to test state-of-the-art models of cosmic structure formation. A rigorous statistical analysis of the observed distribution is required to test models for accretion-driven vs. merger-driven growth.

Finally, we note that the proposed survey will enable a population-level analysis assessing how the spin varies with the Eddington fraction, ionization parameter, disk density, emissivity profile/corona geometry, reflection fraction, and inner disk iron abundance. Currently, such an exploration is limited to the heterogeneous sample of $\sim 50-70$ massive black hole spins in the literature, which have all been studied by different groups, and which are all drawn from flux-limited samples.

**Observing description:**

- When *AXIS* launches, the SMBH spin survey will require the spectroscopic redshifts of all AGN whose spectra will be analyzed be known, unless these are extracted from spectral fitting. In addition, to populate the spin vs. mass plane, SMBH mass estimates also need to be known. However, we are leveraging the FTO Deep Survey, and thus likely targeting well-studied sky regions (e.g., GOODS-N, GOODS-S already have extensive multiwavelength coverage, including reliable optical data). In addition, a catalog of X-ray properties and optical/NIR counterparts will be provided for the *AXIS* Deep Survey.

**Exposure time (ks):** Archival deep.
**Joint Observations and synergies with other observatories in the 2030s:**

The proposed spin survey has the potential to complement other major surveys, such as *Euclid*, *Rubin*, and *SKA*, in addressing the following questions about the *jet-spin connection*: i.e., how are jets launched? How can relativistic jets get launched via the extraction of the SMBH's rotational energy? What is the connection between the host galaxy properties and the spin of the central BH (AGN feedback)? What are the spins of SMBHs in merging galaxies?

Some of the topics this science case will relate to are:

- *Spin-AGN feedback connection*: See [112] for observational studies connecting spins of SMBHs with molecular gas tracers used for star formation.
- *Spin-AGN feedback connection:* See [75] for theoretical work on the role of BH spins in radiative AGN feedback. See [161,553] for references to models driving jet feedback from extracting rotational (spin) energy from the SMBH.
- *Jet-spin connection:* See [446] for a study of GRMHD simulations that suggests how higher BH spins launch more powerful jets.
- *Jet-spin connection:* See [281] for such a discussion in the context of observations of the radio-loud galaxy Cen A from observations taken by the Event Horizon Telescope.

**Special Requirements:**

Spectra cannot be (significantly) piled up.

*8. Coronal properties of high-z quasars*

**Science Area:** Active Galactic Nuclei
**First Author: Elias Kammoun** (Caltech; ekammoun@caltech.edu)
**Co-authors: Xiurui Zhao** (Caltech, xiurui.zhao.work@gmail.com); **Dom Walton** (University of hertfordshire, d.walton4@herts.ac.uk); **Pierre-Olivier Petrucci** (Institute of Planetology and Astrophysics of Grenoble, pierre-olivier.petrucci@univ-grenoble-alpes.fr); **Stefano Bianchi** (Università degli Studi Roma Tre, stefano.bianchi@uniroma3.it); **Enrico Piconcelli** (INAF-Osservatorio Astronomico di Roma, enrico.piconcelli@inaf.it); **Alessia Tortosa** (INAF-Osservatorio Astronomico di Roma, alessia.tortosa@inaf.it); **Luca Zappacosta** (INAF-Osservatorio Astronomico di Roma, luca.zappacosta@inaf.it)
**Abstract.** The hard X-ray emission in active galactic nuclei (AGN) is believed to originate from a region of hot and trans-relativistic electrons known as the X-ray corona. This emission, typically described by a power law with a photon index (Γ) and a high-energy cutoff ($E_{\rm cut}$), is indicative of Comptonization by thermal electrons. Within this framework, the $\Gamma - E_{\rm cut}$ plane can be mapped onto the temperature–optical depth plane, assuming a specific coronal geometry. Despite various hypotheses regarding the origin, geometry, and composition of the corona, our understanding of this fundamental component remains incomplete. At high cosmological redshift, the observed energy cutoff will be closer to the observed range, allowing a better determination of the coronal properties. We propose to use surveys and targeted observations of high-redshift quasars to constrain the coronal properties of these sources. This will allow us to study the evolution of the X-ray corona's properties with redshift and their dependence on quasars' properties.
**Science.** A physically compact X-ray corona is expected to be radiatively compact, meaning that the ratio of X-ray luminosity to corona radius ($L_{\rm X}/R_{\rm c}$) is large. Any increase in source luminosity for a given size leads to electron-positron pair production rather than heating the coronal plasma. This pair production increases the number of particles sharing the energy, thereby acting as a thermostat that regulates the coronal temperature as a function of compactness [$\ell \propto L_{\rm X}/R_{\rm c}$; 248]. Current measurements enabled by *NuSTAR*, with its unprecedented sensitivity above 10 keV, show that AGN coronae are broadly consistent with this scenario [see e.g., 58,180,181,563]. However, these constraints are currently limited to bright, nearby sources ($z < 0.1$) with moderate luminosities of the order of $L_{2-10\,{\rm keV}} \sim 10^{42-44}\,{\rm erg\,s^{-1}}$. For such sources, the allowed range of coronal temperatures without violating the pair-production limit remains relatively broad.

In contrast, high-redshift ($z > 1$) radio-quiet quasars - typically with luminosities exceeding $10^{45}$, erg, s$^{-1}$ - offer an ideal opportunity to constrain high-energy cutoffs. In these objects, the limited range of allowed $E_{\rm cut}$ values, required to satisfy the pair-production limit, combined with cosmologically redshifting $E_{\rm cut}$ into the observable band, makes them powerful probes of coronal properties at high luminosity. Consequently, they provide a crucial test of the runaway pair production theory. To date, high-energy cutoffs have been successfully measured for only a handful of $z > 1$ quasars [e.g., 58,294,295,334,384]. We propose using AXIS to extend these measurements to a larger sample of high-redshift quasars. Observing high-z QSOs, where coronal properties can be tightly constrained, is crucial for investigating the potential evolution of X-ray coronae with accretion parameters, such as the accretion rate and SMBH mass.

Understanding the X-ray coronal properties of high-redshift quasars is pivotal for addressing fundamental questions about the growth of supermassive black holes and the evolution of their accretion processes. This study will provide essential diagnostics of the physical conditions in the corona, leveraging these measurements to further our understanding of both microphysical processes near black holes and the macro-evolution of galaxies in the early Universe. The main key scientific objectives are listed below:

1. **Tracing Black Hole Growth:**
    - High-redshift observations offer a unique window into the early Universe, allowing us to study active galactic nuclei (AGN) when they were in their formative stages.
    - Comparing the coronal properties of high-$z$ QSOs with those of local AGN will help trace the evolutionary path of supermassive black holes.
2. **Investigating the Corona Physics:**
    - The *photon index* characterizes the slope of the X-ray spectrum, providing insights into the energy distribution of the coronal electrons and the efficiency of Comptonization.
    - The *high-energy cutoff* is directly related to the temperature of the electrons in the corona. Measuring this cutoff allows us to infer the maximum energy of the electron distribution and thus the physical conditions (e.g., electron temperature, optical depth) within the corona, constrain the pair-production thermostat model, and shed light on the electron distribution in the X-ray corona.
3. **Testing Accretion Models:**
    - Observational constraints on the photon index and high-energy cutoff serve as stringent tests for theoretical models, such as thermal Comptonization, that describe the energy dissipation near the black hole.
    - Any evolution in these parameters over cosmic time may indicate changes in the fundamental accretion physics and energy dissipation mechanisms.
4. **Implications for Galaxy Evolution:**
    - AGN feedback, mediated through energetic X-ray emissions, plays a critical role in regulating star formation and the thermodynamics of the interstellar medium.
    - This study will contribute to a broader understanding of how AGN influence galaxy evolution and the overall structure formation in the Universe.

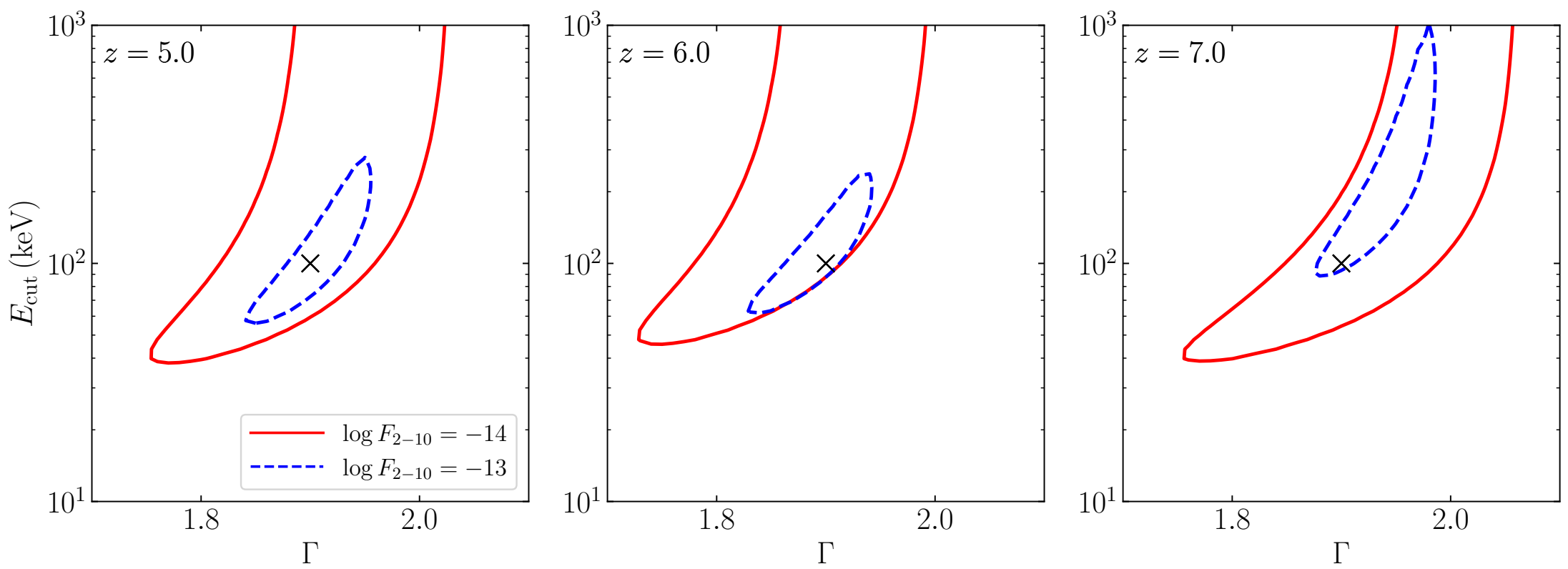

**Figure 11.** Cutoff energy vs photon index contours obtained by simulating 150 ks of *AXIS* observations of QSOs at $z = 5, 6$, and 7 (left to right) and flux levels of $\log F_{2-10\,\rm keV}/\rm cgs = -14$ and $-13$ (solid and dashed contours, respectively). The input parameters are shown as black crosses.

The proposed study of high-$z$ QSOs not only aims to deepen our understanding of accretion physics near supermassive black holes but also to inform models of cosmic structure formation. By providing robust measurements of the X-ray coronal properties at high redshift, we expect to bridge the gap between local Universe studies and the conditions prevailing in the early cosmos, thereby offering crucial insights into the evolution of AGN and their host galaxies.

**Plan.** We plan to use AGN spectra from the *AXIS* Deep Survey, with known spectroscopic redshifts, to measure the coronal properties of these sources. We will employ both phenomenological and physical models to characterize the X-ray spectra of these sources. We will then study the evolution of these properties with redshift and the dependency of various parameters such as BH mass and accretion rate. To demonstrate the capabilities of *AXIS* and the validity of our approach, we simulated a set of spectra assuming $\Gamma = 1.9$, $E_{\rm cut} = 100\,{\rm keV}$ (which can be considered as a pessimistic value - we expect high-luminosity quasars to have a lower cutoff energy), and $2 - 10\,{\rm keV}$ fluxes of $10^{-13}$ and $10^{-14}$ cgs. We considered sources at $z = 5, 6$, and 7, assuming an exposure time of 150 ks only. The constraints on $\Gamma$ and $E_{\rm cut}$ are shown in Fig. 11.

**Exposure time (ks):** Archival Deep
**Special Requirements:** N/A.

*9. Compton-Thick and Compton-Thin AGN evolution up to $z \sim 6$ (and beyond) with AXIS Deep and Wide Surveys*

**Science Area:** Active Galactic Nuclei

**First Author:** Alessandro Peca, Eureka Scientific, US; Yale University - Department of Physics, US; peca.alessandro@gmail.com

**Co-authors:** Matilde Signorini (ESA; INAF Firenze, Italy), Giovanni Mazzolari (INAF Bologna, Italy), Enrico Piconcelli (INAF Bologna, Italy), Francesco Salvestrini (INAF Trieste, Italy), Andrealuna Pizzetti (ESO/ALMA, Chile)

**Abstract:** Active Galactic Nuclei (AGNs) play a fundamental role in galaxy evolution, yet their observed properties are significantly affected by obscuration from both their circumnuclear material and their host galaxies. While local studies ($z \lesssim 0.5$) have provided constraints on the fraction of Compton-thick (CK, $N_H > 10^{24}$ cm$^{-2}$) AGNs, the redshift evolution of this class of AGNs remains largely uncertain. For Compton-thin (CN, $10^{22} < N_H < 10^{24}$ cm$^{-2}$) AGNs, we can reach higher redshifts up to $z \sim 4$–5, but with large uncertainties and for a limited range of luminosity. Current X-ray surveys lack the sensitivity and angular resolution required to effectively detect and characterize the populations of obscured AGNs beyond $z \sim 0.5$. AXIS, with its unprecedented sensitivity and resolution, will enable a direct measurement of the evolution of obscuration from both AGNs and their host galaxies across cosmic time. Using the AXIS deep and intermediate surveys, we will compute the intrinsic fraction of obscured AGNs up to $z \sim 6$ (and beyond for a specific range of luminosity and $N_H$), analyze their distribution as a function of redshift and luminosity, and explore how host galaxy properties influence AGN obscuration. Our study will leverage synergies with ALMA, JWST, Roman, Hubble, and Euclid to characterize the large-scale host galaxy environment and its contribution to obscuration. Combined with AXIS measurements of the total line-of-sight column density from X-rays, this will allow us to disentangle the contributions from nuclear and host-scale material, providing a comprehensive view of AGN obscuration across different physical scales.

**Science:**

**Scientific Background** Obscuration in Active Galactic Nuclei (AGNs) is fundamental to understanding the growth of supermassive black holes (SMBHs) and their co-evolution with host galaxies [e.g., 262,266,570]. Type 2 AGNs can be broadly classified into Compton-thick (CT, $N_{\rm H} > 10^{24}$ cm$^{-2}$) and Compton-thin (CN, $10^{22} < N_{\rm H} < 10^{24}$ cm$^{-2}$) categories, both playing crucial roles in shaping the cosmic X-ray background (CXB). While observations from hard X-ray ($\gtrsim$10 keV) facilities such as *NuSTAR* and *Swift*/BAT have provided valuable constraints on the local fraction of CK AGNs, estimated to be ∼30–50% for $z \lesssim 0.5$ [e.g., 19,496], these estimates remain limited primarily to low redshifts due to sensitivity constraints. Nevertheless, their advantage lies in providing a relatively unbiased view of obscuration, as photons above 10 keV are less affected by absorption, enabling robust detection and characterization of heavily obscured CK AGNs. In contrast, softer X-ray telescopes like *Chandra* and *XMM-Newton* excel in detecting CN AGNs because CN material has a comparatively modest effect on photons in the ∼ 0.5–10 keV energy range. Benefiting also from significantly enhanced sensitivities, and aided by the fact that at high redshift the observed spectra are redshifted into softer energies, these telescopes have successfully extended the detection and characterization of CN AGNs out to higher redshifts ($z \sim 4$–5), though these observations still suffer from substantial uncertainties and limited luminosity range coverage [e.g., 10,454,479,585].

The energy range of AXIS (0.2–10 keV) is ideal for detecting redshifted hard X-rays from distant obscured AGNs. With unprecedented sensitivity and angular resolution, AXIS will significantly surpass current facilities, enabling the detection of faint AGNs and resolving their emission even in dense, crowded cosmic environments. Additionally, the stability of AXIS's point spread function (PSF) ensures high-quality spectral and spatial data across the full field of view, facilitating precise measurements of AGN obscuration

properties. These capabilities also enable the direct determination of redshifts from X-ray spectra alone, particularly for obscured AGNs where optical/NIR methods may be unreliable [452,453].
***Why is studying the evolution of obscuration important?*** Understanding the evolution of obscuration in AGNs is critical because it directly informs models of SMBH growth and galaxy evolution [e.g., 266,570]. Currently, X-ray luminosity functions used to model AGN populations primarily rely on extrapolations from low-redshift data, limiting their accuracy at higher redshifts and for a broad range of luminosities [e.g., 10,19,454,585]. By directly measuring the intrinsic fractions of obscured AGNs up to $z \sim 6$ and beyond, AXIS will substantially enhance existing models and reduce uncertainties across a wide luminosity range. The improved obscuration constraints will enable a detailed investigation of how obscuration evolves as a function of cosmic time and AGN luminosity. Moreover, the origins of AGN obscuration are debated—whether primarily originating from a compact nuclear torus or from large-scale host galaxy structures, such as the interstellar medium (ISM). Observational studies suggest that at higher redshifts, obscuration could be increasingly dominated by the gas-rich, dusty ISM within actively star-forming galaxies [e.g., 20,226]. By integrating AXIS X-ray observations with multiwavelength data from ALMA (for molecular gas and dust content), JWST, Roman, and Euclid (for detailed mapping of dust extinction and stellar populations), our analysis will compare the total line-of-sight obscuration measured in the X-rays with independent ISM estimates of gas and dust, allowing us to disentangle nuclear-scale obscuration from that occurring on host galaxy scales. This comparison will clarify the respective contributions and interactions between AGNs and their host galaxies across cosmic time. Additionally, recent JWST discoveries of X-ray weak AGN candidates, such as the so-called "little red dots," have led to the hypothesis that many of these sources may be heavily obscured AGNs. AXIS's superior sensitivity will enable precise measurements of the line-of-sight column density ($N_H$), helping to distinguish between intrinsic faintness and extreme obscuration. This capability is essential for confirming the nature of these elusive sources and for building a complete census of AGN activity at early cosmic epochs.
**Goals of this proposal and expected impact:** With this proposal, we aim to deliver the most robust measurements of obscured AGN fractions to date, both CK and CN, extending these constraints up to $z \sim 6$ (and beyond for a specific range of luminosity and $N_H$, see Figure 12) across different luminosity bins. Furthermore, we will thoroughly examine the cosmic evolution of AGN obscuration, linking it explicitly to AGN luminosity, the host galaxy's ISM, and possibly other host galaxy properties. By deriving the $N_H$ distribution as a function of redshift and luminosity and directly connecting it to the X-ray luminosity function (XLF), we will significantly refine current population synthesis models. Crucially, our analysis will disentangle nuclear torus obscuration from host galaxy ISM obscuration, providing unprecedented clarity regarding the multiple scales of obscuration affecting AGNs. This study will establish a foundational knowledge base for future AGN research, informing population synthesis and CXB modeling efforts for decades to come. Additionally, the results will maximize the scientific returns of future, coordinated follow-up studies with the NewAthena mission, drawing upon historical analogies of highly productive synergies between Chandra and XMM-Newton. Moreover, the detailed multiwavelength approach outlined here—leveraging ALMA's unique capabilities to quantify dust content, alongside JWST, Roman, Hubble, and Euclid to measure the host=galaxy sizes—will enrich the broader astrophysical community by significantly advancing our understanding of AGN-galaxy co-evolution.

**Feasibility, Plan, and Expected Outcomes** Data Sources: AXIS deep and intermediate surveys
Our approach utilizes AXIS's planned deep and intermediate surveys, complemented by sophisticated X-ray spectral fitting, statistical bias correction, and multiwavelength integration from publicly available archives (ALMA, JWST, Roman, and Euclid). Dedicated simulations provided by the AXIS team, utilizing the latest AXIS response matrices, confirm AXIS's ability to accurately measure obscuration up to $N_{\rm H} \sim 10^{25}\,{\rm cm}^{-2}$, and to identify significant samples of obscured AGNs across luminosity and redshift bins.

In brief, we will perform detailed X-ray spectral analysis of the detected sources in the AXIS surveys. For sources without spectroscopic redshift identifications from optical spectroscopy, we will estimate redshifts directly from the X-ray spectra [e.g., 452,453]. From this spectral analysis, we will derive the number of obscured AGNs, which will then be corrected for observational biases [e.g., 454,535]. These bias-corrected numbers will be used to compute the intrinsic fraction of obscured AGNs (both CN and CK) across different luminosity bins (see Figure 12), as well as their respective X-ray luminosity functions. We will then integrate the resulting luminosity functions to quantify the contribution of obscured AGNs to the growth of SMBHs and their role in the cosmic X-ray background. Dedicated simulations provided by the AXIS team show predictions for the AXIS deep and intermediate surveys, indicating that both Compton-thick (CK) and Compton-thin (CN) AGNs will be detected at high redshifts and across various luminosity bins in sufficient numbers to meet the proposed scientific goals. In Figure 12, we present the predicted fractions of CK and CN AGNs as a function of redshift, categorized by luminosity bins (2–10 keV). Using the AXIS deep survey, we will effectively trace the CN fraction up to $z \sim 4$ for both high luminosity ($10^{44}$–$10^{46}$ erg s$^{-1}$) and low luminosity ($10^{40}$–$10^{42}$ erg s$^{-1}$) sources, and up to $z \sim 8$ for intermediate luminosities ($10^{42}$–$10^{44}$ erg s$^{-1}$). For CK sources, high luminosities will be traced up to $z \sim 4$, intermediate luminosities up to $z \sim 6$, and lower luminosities up to $z \sim 1$. The AXIS intermediate survey, in contrast, will greatly enhance the detection of high-luminosity AGNs up to $z \sim 6$, leveraging larger cosmic volumes necessary to sample these rare, luminous objects adequately. However, this gain comes at the cost of reduced sensitivity to lower luminosity sources, which require the deeper observations of the AXIS deep survey for sufficient sampling. These complementary characteristics underline the strategic synergy between the two surveys.

We will then use available ALMA data to estimate the gas and dust content of these sources. This will be done using CO emission lines where available, or through SED fitting using the Rayleigh-Jeans tail of the dust continuum, [e.g., 522,523]. The dust content will be converted into gas mass using appropriate dust-to-gas conversion factors [e.g., 142,226]. We will further combine ALMA, optical, and NIR high-resolution imaging to constrain the physical extent of the host galaxy. From this, we will derive line-of-sight $N_H$ estimates attributable to the ISM [see e.g., 226]. These ISM-derived values will then be compared to the total X-ray-derived $N_H$ to infer the contribution from nuclear-scale (torus) obscuration. Collectively, this multi-facility strategy ensures our ability to robustly address key open questions about obscured AGNs and their evolution throughout cosmic history.

In summary, this proposal will provide:

- Precise, bias-corrected measurements of the intrinsic obscured AGN fractions and their XLFs, both for a broad range of luminosity, up to high redshifts ($z \sim 6$);
- A direct comparison between nuclear and host-galaxy scale obscuration components, enabling deeper insight into the structure and physical conditions of AGNs and their surrounding environments;
- Comprehensive insights into the cosmic evolution of AGN obscuration and its connection to host galaxy evolution, by comparing AGN and host galaxy properties as a byproduct of the multiband data used.

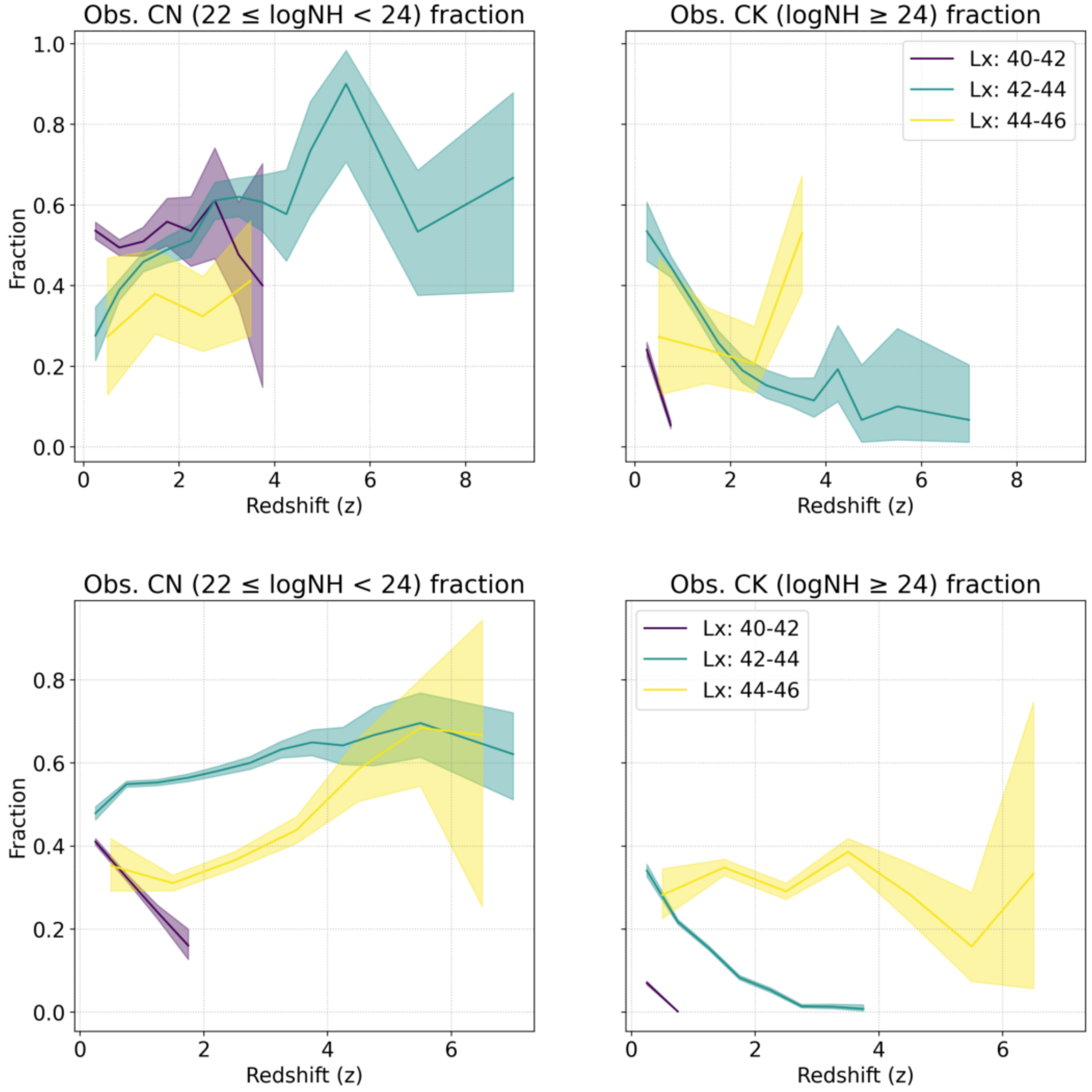

**Figure 12.** Observed fractions of Compton-thin (CN) AGNs (left) and Compton-thick (CK) AGNs (right) as a function of redshift, determined from the combined results of AXIS deep (top) and intermediate (bottom) surveys. Each panel is divided into 2–10 keV luminosity bins: yellow denotes $10^{44}$–$10^{46}$ erg/s, green to $10^{42}$–$10^{44}$ erg/s, and purple $10^{40}$–$10^{42}$ erg/s. For each bin, at least three objects were required to ensure statistical reliability. These results demonstrate the potential to refine and extend estimates of the obscured AGN fraction across a broad luminosity range and to higher redshifts, beyond current constraints.

**Exposure time (ks):** 7Ms (AXIS Deep); 335ks (AXIS intermediate)
**Observing description:**
Archival proposal for using the planned AXIS intermediate and deep surveys.
**Joint Observations and synergies with other observatories in the 2030s:** JWST, ALMA, Roman, Hubble, and Euclid
**Special Requirements:** None None

**b. Mergers**

*10. A Legacy Survey of Dual AGN in Nearby Mergers ($z < 0.1$) with AXIS*

**Science Area: AGN**
**First Author: Michael Koss (mike.koss@eurekasci.com**
**Co-authors: Stefano Bianchi (Università degli Studi Roma Tre, stefano.bianchi@uniroma3.it)**
**Abstract:**

We propose a wide-field AXIS complete volume-limited all sky survey to discover and characterize dual AGN in all nearby luminous AGN ($z < 0.1$, $L_{2-10,\mathrm{keV}} > 10^{43}$ erg s$^{-1}$) from a sample of 525/3500 AGN in major mergers. Dual AGN—where both SMBHs in a merging system are actively accreting—represent critical phases in the evolution of SMBH binaries and are key to understanding how galaxy mergers drive black hole growth. By increasing the number of systems studied by more than an order of magnitude, we can reach $\sim$1% precision (Fig. 1), enabling robust comparison to simulations and high-$z$ samples [197].

AXIS's high spatial resolution and exceptional sensitivity in the X-ray band are uniquely suited to uncover dual AGN in these systems, even when heavily obscured (Compton-thick, CT, $N_{\mathrm{H}} = 10^{24}$ cm$^{-2}$), and resolve them down to physical separations of 1 kpc. This survey dramatically improves the resolution limit ( 7 kpc) of AXIS dual AGN surveys in the more distant universe, including the AXIS PI-led medium survey. Our program thus serves as a crucial low-redshift complement, providing nearby analogs of the luminous dual AGN systems found at high redshift.

This legacy survey will establish the most complete census of dual AGN in the local universe and set key benchmarks for theoretical models of galaxy mergers and black hole pairing. It will directly inform predictions for gravitational wave signals from SMBH binaries observable by LISA and pulsar timing arrays and synergize with ongoing and future multiwavelength efforts, including JWST, ALMA, ngVLA/SKA, Euclid, Roman, and the next generation of extremely large telescopes.

**Science:**

Galaxy mergers are a key driver of black hole growth and AGN activity, as predicted by simulations [44,152]. Both SMBHs can accrete simultaneously as dual AGN on kpc scales during these events, eventually forming gravitational-wave-emitting binaries. The discovery of a buried dual AGN in NGC 6240 [311] catalyzed efforts to find similar systems, leading to hundreds of candidates across deep fields and higher redshifts [140,514]. These systems offer a unique window into SMBH pairing and merger-triggered accretion. They are critical for predicting the low-frequency gravitational wave background probed by LISA and pulsar-timing arrays.

We propose a general observer AXIS survey to identify and characterize dual AGN in all nearby ($z < 0.1$), luminous ($L_{2-10} > 10^{43}$ erg/s) AGN undergoing late-stage major mergers (mass ratio $\lesssim$6:1, separation <30 kpc), selected from Swift-BAT, eROSITA, and other all-sky catalogs (525/3500). AXIS's exceptional sensitivity (6–10× Chandra) and broad energy range (0.2–10 keV) make it uniquely capable of detecting even Compton-thick companions missed in previous surveys.

Our survey will deliver the most precise measurement of the local dual AGN fraction (to <1% uncertainty), a critical benchmark for models of SMBH growth. Current low-$z$ estimates ( 5%; [314]) are in tension with simulations predicting much lower rates ($\sim$0.1–1%; [197]. Resolving this discrepancy requires a large, volume-limited sample like ours.

These rare systems are largely absent from deep fields (Fig. 2) — only two such luminous mergers appear in the AXIS COSMOS medium field — making our targeted survey essential. AXIS's $\sim$1.5$''$ PSF resolves dual nuclei down to $\sim$1 kpc at $z < 0.1$, well-matched to the separation scales where theory predicts peak dual activity [580]. Even sub-kpc pairs can be identified through multiwavelength data (e.g., radio cores).

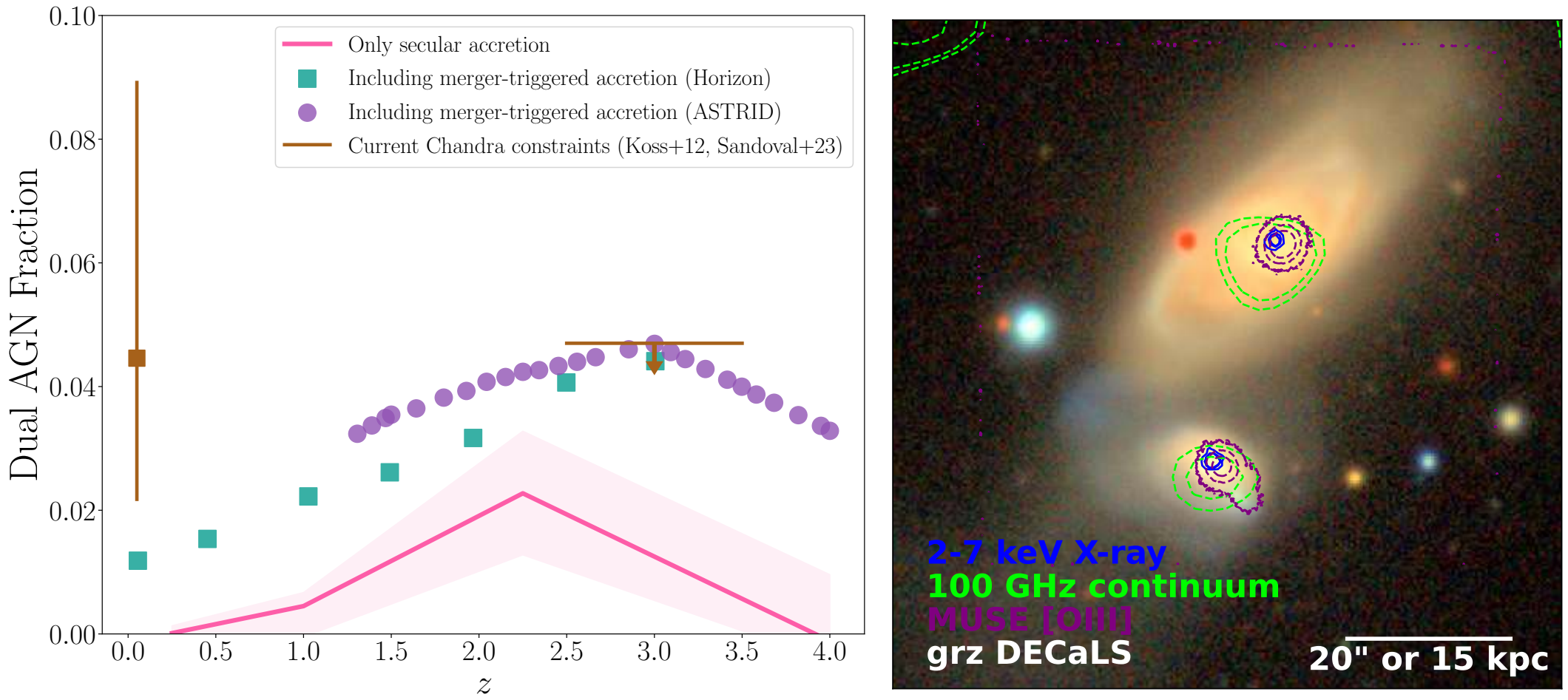

**Figure 13. Left: Dual AGN Fraction vs. Redshift:** Current constraints (brown square; [314]) suggest a $\sim$5% dual AGN fraction at $z < 0.05$, but with large uncertainty. Simulations (purple/green curves; [197]) predict much lower values ($\lesssim$1%). Early JWST results at $z \sim$ 1–3 (orange point; [464]) suggest higher dual fractions, echoing local observations. Our AXIS survey (blue area) will improve precision at $z < 0.1$ to $\sim$1%, anchoring dual AGN evolution at low $z$. A significantly higher fraction (potentially 5˘10% as implied by initial Chandra observations) would be a clear deviation from theoretical predictions, indicating that mergers play a more pronounced role in SMBH fueling than currently modeled. Our result will provide an essential anchor at $z \approx 0$ for the evolution of dual AGN fraction out to $z < 4$, which AXIS and other missions will measure for the first time. **Right:** Mul. tiwavelength view of a luminous merger (BAT 193, $z = 0.05$), combining [O,III] from MUSE (green), 2–7 keV X-rays from AXIS (blue), and 100 GHz ALMA continuum (red) over a *grz* image. Such data, available for many targets, allows us to identify AGN and study host properties in detail.

Crucially, AXIS overcomes Chandra's limitations at high energies. For example, a Compton-thick AGN with $L_{2-10} \sim 10^{42.4}$ erg/s at $z = 0.1$ would yield only $\sim$3 counts in 100 ks with Chandra, but 150–200 counts with AXIS (or 5$\sigma$ detections in 10–20 ks), enabling a near-complete census of obscured SMBHs.

Our sample spans a broad range of merger stages and AGN properties (e.g., luminosity, obscuration, environment), enabling us to investigate how the incidence and properties of dual AGN evolve during coalescence. We will test whether late-stage mergers preferentially host Compton-thick AGN [498] and whether dual activity is more common at high luminosities ($L_X > 5 \times 10^{43}$ erg/s), where merger-driven accretion is expected to dominate [267,571]. By increasing by a factor of 14 the number of luminous duals studied to date [318], AXIS will uniquely constrain how SMBH fueling, obscuration, and symmetry depend on merger dynamics. Finally, we will combine AXIS data with optical/IR spectra, radio imaging, and high-resolution morphologies to investigate the physical conditions that enable dual accretion. This rich multiwavelength view will clarify the role of mergers in triggering AGN and shape predictions for dual AGN at higher redshifts.

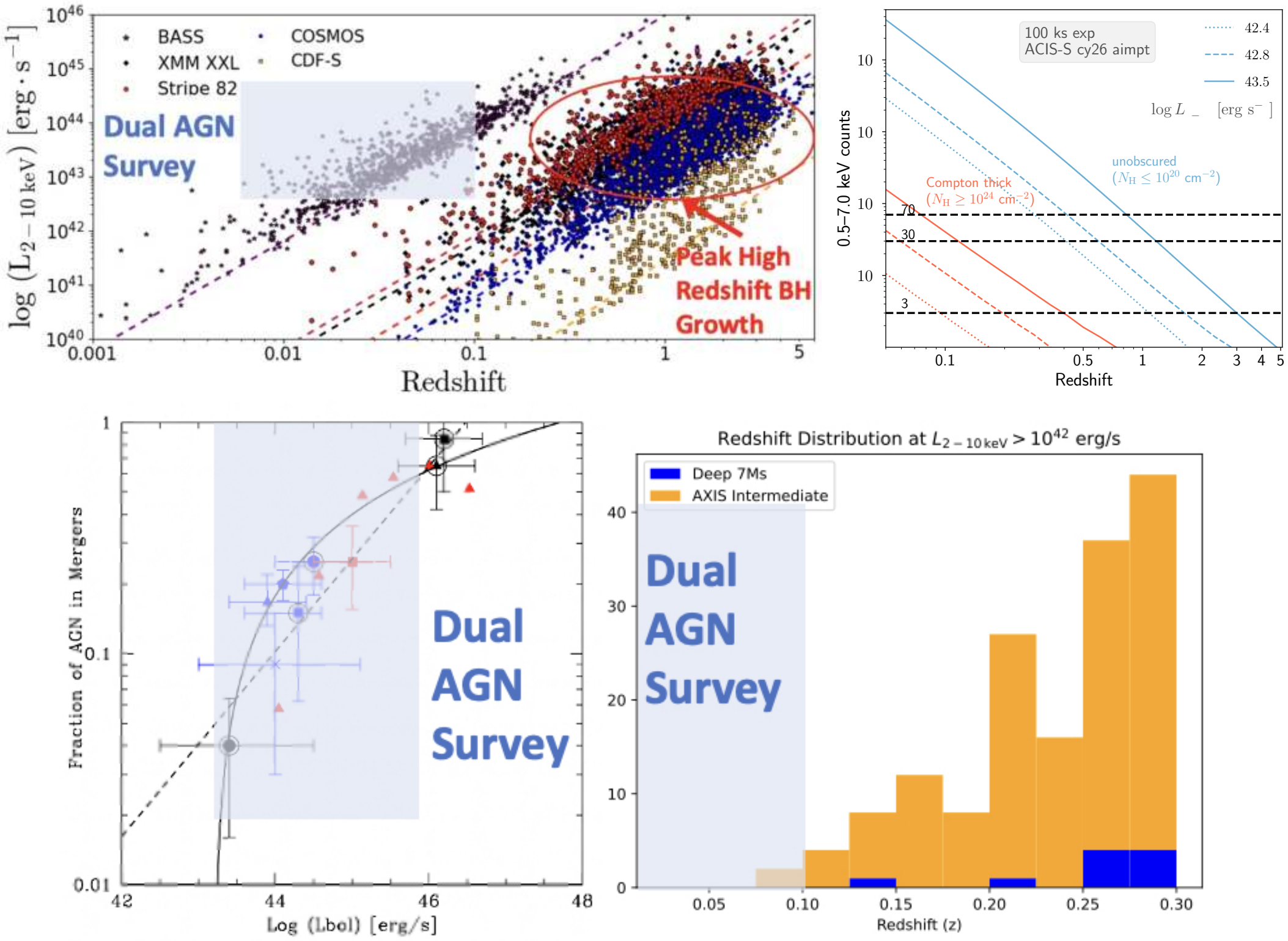

**Figure 14. Left:** 2–10 keV luminosities of the *Swift*/BAT AGN sample (blue) compared to high-$z$ X-ray surveys [316]. These local AGN ($z < 0.1$) serve as high-resolution analogs of peak SMBH growth at $z \sim 1$–2 [318], providing crucial context for dual AGN found by AXIS, Chandra, JWST, Gaia, Euclid, and ngVLA. **Right:** Chandra detection limits for obscured AGN in 100 ks: unobscured AGN (blue) remain visible at $z > 1$, but CT AGN (red) fall below detectability even at $z \sim 0.1$. **Bottom Left:** Predicted merger-driven AGN fraction vs. luminosity [571]. Our survey (blue shading) probes a poorly studied range just below the quasar regime. **Bottom Right:** Histogram of AGN with $L_{2-10} > 10^{43}$ erg/s at $z < 0.1$ from AXIS deep (blue) and medium (orange) surveys. These wide-field surveys capture few such sources—on average, $\sim$2 AGN and no mergers—underscoring the need for targeted observations to explore luminous local dual AGN fully.

**Key Science Questions:**

1. **Dual AGN demographics:** What fraction of local luminous AGN are in mergers with dual accreting SMBHs, and how does this depend on AGN luminosity, environment, and SMBH mass? We aim for $\sim$1% precision (Fig. 1), enabling robust comparison to simulations and high-$z$ samples [197].
2. **Obscuration vs. merger stage:** Do late-stage mergers harbor more Compton-thick AGN? By measuring $N_H$ and luminosity for each nucleus, we'll test whether dual AGN are more obscured than singles and whether obscuration peaks at advanced stages.
3. **Merger characteristics:** Do dual AGN prefer specific merger configurations (e.g., major mergers, gas-rich hosts)? We'll compare duals to single AGN to isolate key drivers of simultaneous SMBH fueling.
4. **Multiwavelength comparison:** How do X-ray-selected duals compare to optical/IR-selected ones? With BASS spectra and radio data, we'll assess selection biases and improve diagnostics in dusty starbursts.

5. **Implications for GW sources:** Do observed dual AGN match simulation predictions [66]? A higher local incidence would imply more common or longer-lived pre-binary phases, refining LISA/PTA merger rate forecasts.

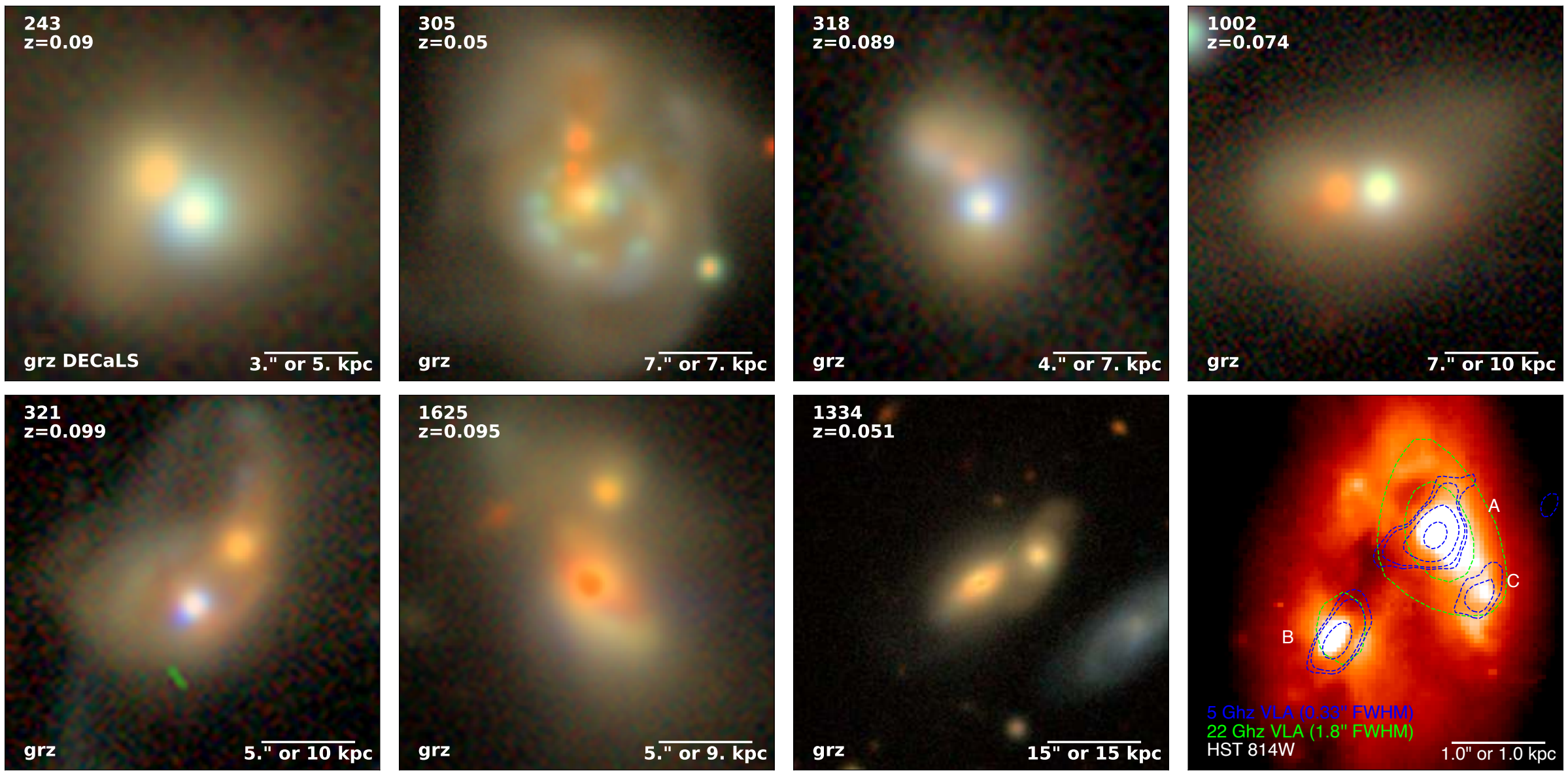

**Figure 15. Nearby Merging Galaxy Examples:** Color images ($g/r/z$) of seven $z < 0.1$ luminous AGN mergers not previously observed with Chandra. All are confirmed major mergers showing dual stellar nuclei and tidal features. AXIS may see blended emission for the closest pairs (e.g., top-left, $1.5'' \sim 2.6$ kpc), but extended structure and multiwavelength data enable separation. The sample spans diverse merger stages, which is ideal for studying the occurrence of dual AGN. Many show ionized gas (blue) aligned with soft X-rays and LINER/composite line ratios—prime AXIS targets. *Bottom Right:* A confirmed dual AGN (from Chandra), demonstrating the value of combining high-res X-ray, AO, and radio imaging to isolate dual SMBHs.

**[Exposure time:] 1070 ks** To ensure detection of Compton-thick (CT) AGN—the most challenging case—we assume $N_{\rm H} = 10^{24}$ cm$^{-2}$, $\Gamma = 1.9$ (intrinsic), and an edge-on torus model (`MYTORUS`, 90° inclination; [630]), with a 1% scattered fraction [497]. Using `fakeit`, we simulate counts for AXIS. Our goal is to observe AGN with $L_{2-10\,keV\,int} = 10^{43}$ erg/s, and detect a CT AGN within at least a factor of 10 in luminosity ($L_{2-10\,keV\,int} = 10^{42}$ erg/s) to match the sensitivity of AXIS moderate field, but up to Compton-thick AGN. We request an initial observation to reach 10 counts, where any sources reaching this threshold would be observed to get 30 counts to estimate the intrinsic luminosity within a factor of 10 based on our fakeit simulations. This leads to an exposure of 2.2 ks on average or 220 ks to observe 100 duals, with a small number of additional exposures for the sources at the threshold of 10-30 counts in the secondary.

**Observing description:** For target selection, we will observe AGN at $z < 0.1$, assuming the population synthesis models of [224,575], there are 3500 AGN at $z < 0.1$ at $L_{2-10\,keV\,int} = 10^{43}$ erg/s and $N_{\rm H} < 10^{24}$ cm$^{-2}$ at $> 10$ deg outside of the Galactic plane, of which we expect 15% to be in major mergers [315] leading to 525 AGN. Of these, 37/525 of the brightest mergers (from the Swift BAT sample [318]) have already been observed with Chandra (or XMM in a few widely separated sources). Observing the remaining 488 and increasing the sample size by $\sim$14 and the precision by a factor of 4 in the dual AGN fraction would require 1.07 Ms.

**[Joint Observations and synergies:]**

**GMT, TMT and ELT**: will deliver optical and near-IR imaging and spectroscopy at resolutions of 0.01–0.1$''$ (with adaptive optics) with instruments (e.g., GMTIFS, TMT-IRIS, ELT-MICADO/HARMONI) that can spatially resolve the two nuclei in even the closest dual AGN in our sample and obtain individual spectra. This will enable dynamical measurements (separate stellar velocity dispersions or gas kinematics for each nucleus, improving SMBH mass estimates) and detailed ionization diagnostics (e.g., confirming both nuclei as AGN via spatially separated emission-line ratios).

**JWST**: JWST's mid-IR spectroscopy (with MIRI at 7–12 $\mu$m) can pinpoint deeply buried AGN hot dust emission, even when optical or X-ray might be obscured. JWST's 0.1–0.7$''$ resolution in the near/mid-IR allows it to resolve dual nuclei in many of our targets. For example, JWST NIRCam or MIRI can directly image the two nuclei and measure their relative mid-IR luminosities – a powerful independent indicator of accretion (since an AGN will heat dust to emit strongly at mid-IR). JWST spectroscopy (NIRSpec, MIRI LRS) can also detect high-ionization lines ([Ne,V], etc.) from each nucleus, confirming AGN activity. A combined AXIS+JWST approach is essential for Compton-thick cases where X-rays see only a reflection: JWST can reveal the primary AGN through its thermal dust signature. JWST and AXIS will provide a complete energy budget of each AGN (unabsorbed IR and absorbed X-ray emission) for a complete SED analysis.

**Roman Space Telescope**: Roman's wide-field near-IR imaging (starting $\sim$2027) will complement our study by identifying many new close mergers using the high-quality NIR imaging in the High Latitude Wide Area Survey that could find rare dual AGN candidates (e.g., double-nucleus infrared galaxies) that escape smaller surveys. Roman's time-domain surveys could also detect any variability-induced offsets (via photocenter wobble) hinting at unresolved binary AGN in our systems [110].

**ngVLA and SKA**: High-resolution radio observations are a crucial counterpart to X-rays for finding AGN through heavy dust and gas. The ngVLA will operate at frequencies up to tens of GHz with sub-arcsecond to milliarcsecond resolution and extreme sensitivity. We will use ngVLA to observe each merger at high radio frequencies (e.g., 10–100 GHz). At these frequencies, star formation-related synchrotron emission is diminished, and a compact, flat-spectrum radio core (or a tiny jet) reveals an AGN [541,542]. A detected compact core in the radio spectrum strongly confirms the presence of an AGN. Conversely, if an X-ray-faint nucleus shows an intense 15–100 GHz core, that suggests a Compton-thick AGN (radio-bright but X-ray suppressed) [335]; similarly, SKA (especially SKA-VLBI or SKA-mid Band5 at 15 GHz) can provide complementary coverage for the southern targets.

**ALMA:** ALMA's high-resolution (0.1$''$ or better) imaging of molecular gas (CO lines) in our targets can map the gas distribution and kinematics feeding the dual AGN. Many of our targets already have ALMA CO or continuum data. In synergy, ALMA/ngVLA data can correlate the presence of dual AGN with massive gas inflows or nuclear disks in each nucleus, providing a physical explanation for why some mergers ignite two AGN.

**Special Requirements:**

**Bright Source and Pile-up Mitigation**: A few AGN ($F_{2-10} \sim 10^{-11}$ erg cm$^{-2}$ s$^{-1}$) will require 1/8 subarray readout or custom frame times to avoid pile-up.

**Orientation Constraints**: Roll angle restrictions may be needed to minimize readout streaks.

**PSF Knowledge and Astrometric Alignment**: Accurate PSF modeling and astrometric alignment ($\sim$0.2″) will distinguish close ($< 1''$) nuclei. Optical/IR positions from AO or HST/JWST will guide dual source identification.

### *11. Dual AGN in galaxy mergers up to cosmic noon with AXIS*

**Science Area:** AGN
**First Author:** Alessandra De Rosa, alessandra.derosa@inaf.it
**Co-authors:** Paola Severgnini, paola.severgnini@inaf.it, Stefano Bianchi stefano.bianchi@uniroma3.it, Manali Parvatikar manali.parvatikar@inaf.it, Lorenzo Battistini lorenzo.battistini@inaf.it, Jasbir Singh jasbir.singh@inaf.it, Fabio Rigamonti fabio.rigamonti@inaf.it, Roberto Serafinelli roberto.serafinelli@mail.udp.cl
**Abstract:** AGN pairs play a crucial role in understanding the growth of supermassive black holes (SMBHs) during galaxy mergers and serve as natural precursors to gravitational waves (GWs). X-ray observations are essential for confirming dual AGN candidates, examining their accretion properties, and linking them to those of the surrounding environment. However, X-ray investigations have lagged behind optical studies. The current limitation is related to the number of systems observed in X-rays compared to the large number of dual AGN in optical datasets. Better sampling and expanding the projected separation and redshift parameter space are key to shedding light on how nuclear activity is triggered in dual AGN and the role of the environment. X-ray studies provide these answers. Chandra's sensitivity allows us to detect and spatially resolve close systems (separations $<8$ kpc) up to redshifts $z\sim1$. However, a comprehensive X-ray census and detailed study of dual AGN at higher redshifts is still lacking. AXIS, with its stable point spread function (PSF) at increasing off-axis angles and its exceptional sensitivity in the soft X-ray domain, plays a crucial role in enabling the detailed study of weak dual AGN at redshifts beyond $z=2$. This X-ray sampling of the high-redshift population will provide key insights into the dynamical evolution and mass growth of SMBHs as they move toward coalescence. By comparing these observational results with simulations, it will be possible to evaluate the effects of selection biases in optical versus X-ray surveys and investigate the potential role of the nuclear and large-scale environment in triggering AGN activity.
**Science:** The detection and characterization of dual SMBHs in merging and interacting systems are fundamental to understanding their formation and accretion history across cosmic ages [591]. Dual SMBHs are the natural consequence of galaxy mergers, during which tidal and/or gravitational torques drive gas to the nucleus. However, it is still unclear how and when SMBHs become active during their long journey in the galaxy-merger remnant down to the very late phase of SMBH binary coalescence [16]. Dual AGN lose angular momentum through distinct physical processes at different stages of separation at kiloparsec scale and serve as precursors to binary AGN, where the SMBHs are gravitationally bound (see [140] for a review on dual and binary AGN). These systems are the primary focus of both upcoming space-based missions (such as LISA, [16]) and ongoing ground-based experiments (like PTA, [6]).

One of the most spectacular and unexpected discoveries of JWST was the large fraction of galaxy mergers and dual AGN ($\sim$30%, [464]) at high redshift ($z>3$). Using one of the largest high-redshift Chandra catalogs, no evidence for dual AGN has been found at $z=2.5$–$3.5$ [514]. *The main limitation is the small number of sources observed at the highest spatial resolution with sufficient signal-to-noise ratio for detection.*

Low-$z$ dual AGN have been studied in X-rays, although more rarely for separations below 30–40 kpc, showing evidence for higher obscuration at the latest stages of the merger [139,246,307,470,500], as expected from simulations [72]. Since galaxy mergers are a complex process that strongly affects gas kinematics, the mixing of metals, star-formation efficiency, and the activation of central SMBHs, this process can also be responsible for the observed obscuration. It remains unclear how this obscuration evolves throughout the merger phase and whether the absorbing gas is located on a nuclear or galactic scale. Whether obscuration, AGN activity, and galaxy interaction are related is still debated as no enhancement of merger fraction in AGN samples has been found if properly compared with control samples of inactive galaxies [117]. *These conflicting results may be a consequence of different sample selection criteria (such as the*

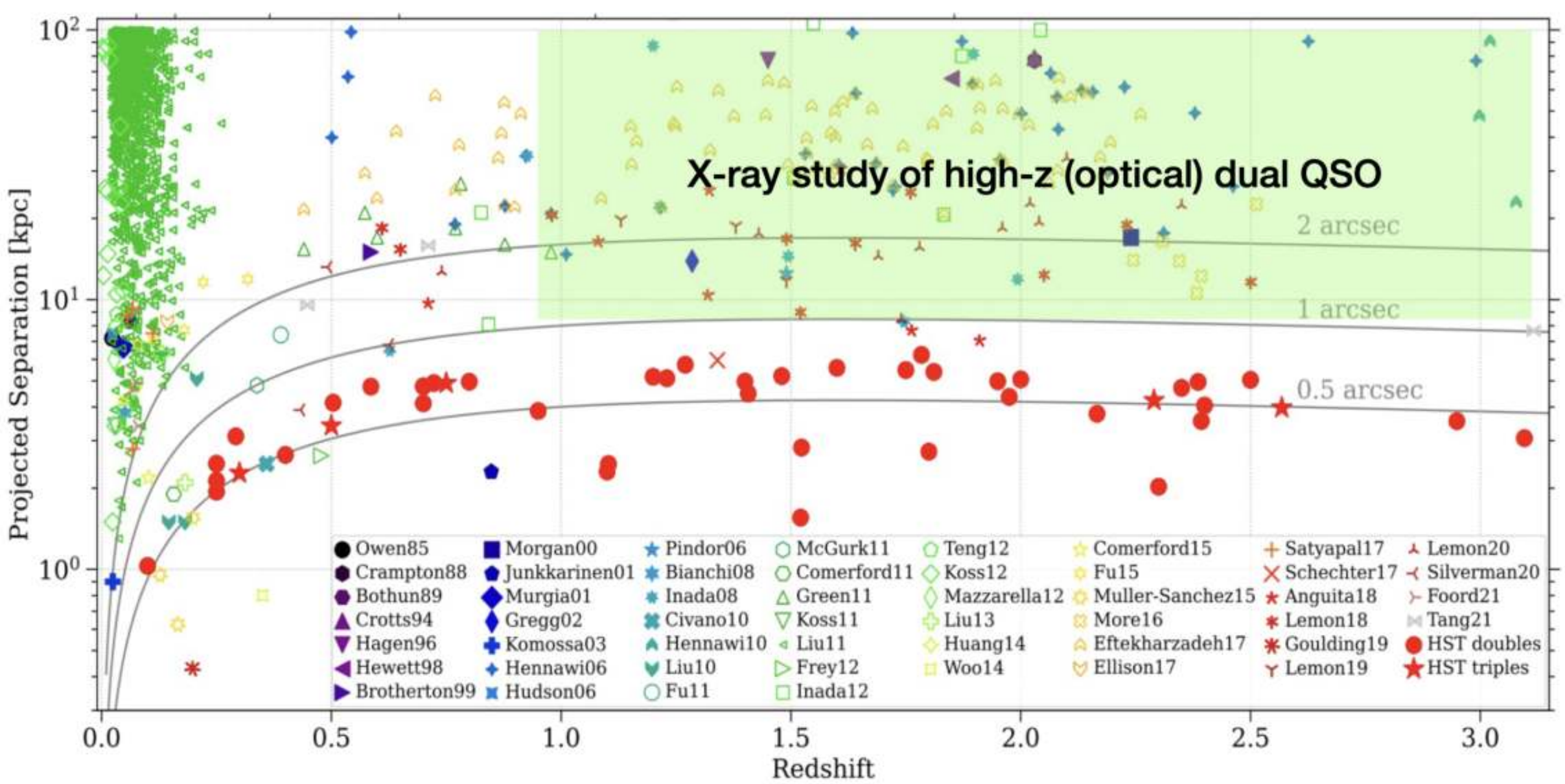

**Figure 16.** Collection of optically classified dual AGN and candidates in literature (adapted from [113,374]). Low-$z$ regimes have been extensively studied in X-rays, whereas this AXIS program focuses on the higher-redshift range (z=1–3), which remains largely unexplored. A key limitation is the high sensitivity and angular resolution that only AXIS can achieve at high redshifts. *The unique combination of the exquisite AXIS PSF and high sensitivity in the soft energy domain will allow us to investigate a large, statistically robust sample of dual AGN at high redshift (green shaded area) in their 2–10 keV rest-frame.*

*merger stage of interacting galaxies and/or AGN luminosity) and observational biases (e.g., nuclear obscuration and AGN variability).*

The investigation of high-$z$ dual AGN at the intermediate stages of mergers (kpc distances) demands significantly enhanced angular resolutions to resolve the two nuclei through imaging and high sensitivity directly. Currently, only a handful of dual AGN with such low pair separations are known at $z>0.5$ [113,164,269,374,514] (see Fig. 16). Although the star formation and BH accretion rate densities peak at $z$=1–3, little is known about dual systems at these redshifts [582], mainly due to the limitation of telescope sensitivity and PSF. *Furthermore, the X-ray domain is still largely unexplored for dual AGN at these redshifts.*

Considering the angular separations needed to resolve the nuclei at high-z (at z=1, 12 kpc corresponds to an angular separation of 1.5 arcsec), and the redshift range, AXIS PSF, together with superior sensitivity in the soft X-ray energy domain with respect to Chandra, will provide unique observations for probing nuclear processes down to weak fluxes (associated with an intrinsically low level of activity and/or strong obscuration).

*B) Immediate Objectives:*

The accretion properties of black holes (such as $\dot{M}$ and $L_{Edd}$) and the characteristics of the nuclear environment (including obscuration) can only be effectively probed through X-ray observations of AGN. However, despite this, it has recently been found that less than 1% of dual-AGN systems are detected and studied in X-rays, primarily at low redshifts, while over 80% of the detected dual systems have been identified through optical spectroscopy [469].

Although most X-ray studies have relied on archival data from serendipitous detections at low-$z$, they have demonstrated the enormous power of high-energy sampling in this field [139,141,196,314,500,514] to confirm dual BH activity, investigate the accretion properties in mergers, and study how the SMBH

environment is modified by a companion AGN. The current limitation arises from the relatively small number of systems observed in X-rays compared to the large population of dual AGN identified in optical datasets. *No X-ray study has yet systematically quantified the physical characteristics, such as bolometric luminosity or accretion rate, of dual AGN at z>1, leaving a gap in our understanding of their evolution with respect to isolated systems.* Therefore, it is essential to improve sampling and broaden the range of projected separations and redshift parameter space.

We propose to observe all the closest optical dual AGN from optical selection that AXIS can spatially resolve (corresponding to a projected separation $r_p$=12–100 kpc) and $z = 1 - -3$. These data will provide the first statistically robust sample of intermediate- to late-stage high-redshift dual AGN. Combined with the information gained from low-$z$ studies, this programme will allow us to answer the long-standing question: *do mergers enhance SMBH nuclear activity?*
Through X-rays, we will address key open issues and achieve the following main objectives:

- *How do dual AGN evolve across cosmic time?* Through the first statistically rigorous sample, extracted from a complete optical selection, we will study high-$z$ AGN in optically identified dual systems. Using X-ray and optical spectra of this homogeneous sample, we will derive observational parameters such as $L_{bol}$, $M_{gal}$, $L_{Edd}$, and E(B-V) which will be compared with the trends observed in dual AGN at lower redshifts. These studies will enable us to verify and constrain possible models of dual AGN evolution.
- *How do emission properties differ between high-z dual AGN and isolated AGN?* We will compare the results from this sample with those obtained for isolated AGN (matched in redshift) to investigate differences in their physical properties (such as luminosity, accretion rate, absorption) and the relative importance of dual AGN in the fueling and growth processes of SMBHs. This study is crucial in shedding light on possible triggering mechanisms.
- *What is the relation between SMBH growth and environment (galaxy/nuclear gas)?* We will measure the AGN obscuration tracking the $N_H$ as a function of $r_p$ and $z$ for the sources proposed here, as well as those from the literature. Additionally, thanks to the optical spectra and SED fitting, we can evaluate the column density derived from E(B - V) as estimated from both the Narrow and Broad Line Region (NLR, BLR), and compare it with the $N_H$ measured from the X-ray analysis. This will allow us to locate the gas and dust in the AGN environment (NLR, BLR, or torus).

AGN pairs represent the precursors of binary SMBHs. The redshift evolution of dual AGN as a function of their separation $r_p$ will offer crucial insights into the dynamics of SMBH binaries. Knowledge of properties in dual AGN ($L_{Edd}$, BH mass ratio, and gas properties, as derived in our program) allows us to refine models of SMBH binary evolution, which directly impacts estimates of GW event rates detectable by future observatories, such as LISA.

**Exposure time (ks):** 500 ks. 50 ks each system optically confirmed as dual AGN at high-z. A preliminary sample selection included $> 10$ systems.

**Observing description:** The selected sample. High-energy investigations of optically-selected AGN pairs are key to derive their accretion properties, free from significant bias towards heavily obscured sources up to absorption column densities of $\sim 10^{24}$ cm$^{-2}$. Thus, we considered AGN pairs optically selected that AXIS can spatially resolve (separation >1.5" , see Fig. 16). To estimate the bolometric luminosity of our targets, we adopted the monochromatic luminosity at a rest-frame wavelength of 3000 Å from [508], when this information was available in publicly accessible spectra. In the cases where these data were not available for one of the AGN in the pair, we rescaled the monochromatic luminosity to that of the

companion AGN using the SDSS magnitudes in the filter closest to 3000 Å (i-mag $z<1.7$ and z-mag for $z>1.7$). We then applied the bolometric corrections from [162] to estimate the 2–10 keV luminosity. Finally, we established a threshold for the X-ray flux based on both the need to build a statistically robust sample and the maximum exposure available in a GO programme. The final sample consists of >10 targets (with angular separation above 1.5", rp=10–100 kpc).

Estimates of the exposure time

Here we ask for 50 ksec net exposure for each system in the sample selected in the previous section (> 10 pairs) which will allow us to collect more than 300 net counts for each nucleus of the pair. As described above, we adopted the monochromatic luminosity at a rest-frame wavelength of 3000 Å from [508] to estimate the bolometric luminosity of our targets. We then used the bolometric corrections from [162] to calculate the X-ray luminosity. Given that the selected sources are optically classified as NL and BL QSO, we simulated the case of an absorbed power-law ($\Gamma$=2, $N_H$=$10^{23}$ $cm^{-2}$) for the weakest and average-flux spectra. With 300 net counts we will be able to measure the luminosity, photon index, and $N_H$ with 5, 10, and 30% accuracy, respectively (at 90% c.l.). It is worth mentioning that, for the redshift selection (1–3), the AXIS effective area will offer the best opportunity to measure the accretion properties in the 2–10 keV rest frame, including the Fe K line emission region.

**Joint Observations and synergies with other observatories in the 2030s:** GW observatories, Optical

**Special Requirements:** None

*12. Understanding the Selection Function of Spatially Resolved Mid-IR-Selected Dual AGNs*

**Science Area:** AGN

**First Author:** Ryan W. Pfeifle (University of Maryland, Baltimore County; NASA Goddard Space Flight Center; Oak Ridge Associated Universities; ryan.w.pfeifle@nasa.gov, rpfeifle@gmu.edu)
**Co-authors:** Kimberly Weaver (NASA Goddard Space Flight Center), Emma Schwartzman (Naval Research Laboratory), Barry Rothberg (United States Naval Observatory, George Mason University)

**Abstract:**

Dual active galactic nuclei (dual AGNs) are predicted to be key tracers of supermassive black hole (SMBH) growth induced during galaxy interactions and mergers, and their accretion histories across the merger sequence set our expectations for the properties of gravitationally bound (and later inspiralling) SMBH binaries. Dual AGN searches and follow-up analyses have primarily focused on dual AGN candidates selected via optical spectroscopy and/or in later-stage galaxy mergers (separations $< 10\,\mathrm{kpc}$), potentially biasing against dust-obscured, heavily obscured, and optically elusive dual AGN populations. All-sky surveys from the Wide-Field Infrared Survey Explorer (*WISE*) have now enabled the selection of spatially-resolved dual mid-IR AGNs across a variety of merger phases, with nuclear separations on the order of $\sim 5 - 100\,\mathrm{kpc}$. Recently assembled samples of spatially resolved mid-IR dual AGN represent prime systems for studying powerful, dusty dual AGNs and their hosts. With advances in ground-based optical spectroscopic fiber surveys along with ground-based follow-up, such mid-IR samples of dual AGNs can now be identified and studied across the optical, mid-IR, and radio. AXIS, with its 1.5″ on-axis PSF and large effective area in the $2 - 10\,\mathrm{keV}$ passband, will usher in a more complete multiwavelength picture of the activity, environments, and selection function for these dual AGN by providing constraints on their X-ray properties; many of these dual AGNs are too heavily obscured or X-ray faint for current facilities to characterize properly. AXIS observations of this mid-IR dual AGN sample will directly complement (a) prior and current work with *Chandra* focusing on optically- and hard X-ray-selected dual AGN samples, and (2) future work with AXIS focusing on optically-selected dual AGNs. Only through a complete, multiwavelength picture of dual AGNs – identified through a variety of selection methodologies – will we be able to understand the observational selection function, environments, behavior, evolution, and potential importance of dual AGNs across the merger sequence.

**Science:**

Over two decades have passed since the discovery of the scaling relations between the mass of supermassive black holes (SMBHs) and the properties of their massive host galaxies [193,215,250], yet the relative contributions of different physical processes to the establishment of these scaling relations has so far remained a topic of contention. Secular processes (such as bar instabilities [e.g., 209] or cold accretion flows in the early universe [e.g. 151,337]) as well as galaxy merger-induced SMBH and galaxy growth [167] have long been discussed as potential contributors to these scaling relationships; the literature is particularly divided over the relative importance of galaxy mergers for SMBH growth, with decades of work providing conflicting results potentially due to different systematic approaches and biases [see 167, and references therein]. Cases of dual accreting SMBHs, known as dual active galactic nuclei (dual AGNs), are predicted to trace significant merger-driven SMBH mass growth across cosmic time, as secular processes alone are not expected to explain the relative fraction of dual AGNs at any given redshift based on cosmological simulations [110,197]. As dual AGNs are also predicted to evolve into gravitationally bound binary SMBHs, the environments and growth history of dual AGNs directly relate to the properties of binary SMBHs within this hierarchical growth paradigm. These SMBH pairs are expected to contribute to the gravitational wave background as continuous wave sources once they reside within the sub-pc binary regime [and will be detectable by pulsar timing arrays, PTAs, 88], and the eventual inspiral and merger of these massive binary SMBHs will be detectable by future space-based gravitational wave

detectors, such as the Laser Interferometer Space Antenna [LISA, 16]. Thus, dual AGNs may offer a unique and important perspective for a variety of astrophysical topics, including the relative importance of galaxy mergers for inducing SMBH growth and potentially contributing to the establishment of galaxy-SMBH scaling relations, SMBH mass and accretion histories, the properties of SMBH binaries, contributions to the gravitational wave background, and hierarchical SMBH growth across cosmic time.

To gain a statistically complete understanding of dual AGN environments and SMBH accretion activity within these merger-induced systems, it is essential to have a diverse sample drawn using various selection strategies to mitigate biases inherent to any single wavelength. Significant efforts have already been conducted to study dual AGNs in the local universe using samples drawn from Sloan Digital Sky Survey (SDSS) optical imaging and spectroscopy [e.g., 261,354,355,597], and a variety of follow-up works have focused on the optical spectroscopic, radio, and X-ray properties of these optically-selected candidates [e.g., 118,141,202,268,352,353,414]. Such studies have yielded hundreds of candidates in the local universe and dozens of confirmed dual AGN cases [469]. Additional, smaller samples have been identified using hard X-ray pre-selection [314], radio pre-selection [628], selection based on distinct radio sources [200], mid-IR color pre-selection [470,517], and very recently spatially resolved mid-IR colors ([45], Pfeifle et al., in prep.) All-sky mid-IR color selection with the Wide-Field Infrared Survey Explorer (WISE, [614]) is particularly useful, as the mid-IR traces dust-reprocessed AGN emission and is relatively insensitive to line-of-sight obscuration [546], making it possible to select obscured and unobscured dual AGNs exhibiting a variety of BPT optical classes as well as optically elusive dual AGNs [e.g., 317,470]. In addition, close angular pairs of mid-IR dual AGNs can be resolved down to separations of $\sim 6''$ in *WISE* imaging at 3.4 $\mu$m, 4.6 $\mu$m, and 12 $\mu$m, enabling the selection of dual AGNs across the majority of the merger sequence with separations $\sim 5 - 100\,\mathrm{kpc}$. **We propose an AXIS GO program to observe a new sample of spatially resolved mid-IR dual AGNs recently identified in ongoing galaxy mergers in the nearby universe ($z < 0.4$, Pfeifle et al., in prep.)**

Through an AXIS GO program, we will provide complete X-ray coverage for this novel, statistically significant sample of mid-IR dual AGNs. This will be one of the first samples of dual AGNs with statistically complete X-ray coverage. We aim to accomplish the following objectives:

- **Detection of X-ray counterparts to dual mid-IR AGNs.** X-ray AGN counterpart confirmation will rely upon a luminosity threshold (e.g., $L_{2-10\,\mathrm{keV}} \geq 10^{42}\,\mathrm{erg\,s^{-1}}$) and/or the spectral properties of the sources. This will yield the occupation fraction of dual and single X-ray AGNs within this population
- **Constraints on the X-ray spectroscopic properties of individual sources within these dual AGNs.** With its 1.5″ on-axis PSF, AXIS will provide spatially resolved spectra of distinct AGNs within each pair. We will place direct constraints on the X-ray spectroscopic properties (when sufficient counts are detected) of individual AGNs, such as the shape of the intrinsic power laws, the presence of Fe K$\alpha$ emission, the line-of-sight column densities ($N_{\mathrm{H}}$), and the unabsorbed $2 - 10\,\mathrm{keV}$ luminosities. Though highly accurate $N_{\mathrm{H}}$ constraints require hard X-ray ($> 10\,\mathrm{keV}$) observations to access the Compton hump beyond 20 keV [e.g. 79], modest constraints for modestly obscured and even heavily obscured dual AGNs can be obtained in the $2 - 10\,\mathrm{keV}$ band [especially when Fe K$\alpha$ is detected, 311,470,471]. For low signal-to-noise detections, indirect diagnostics (e.g., comparison of AXIS detections against the infrared or optical [e.g., 470]) will be used to infer loose constraints on column densities and provide priors for basic spectroscopic fitting using robust Bayesian fitting codes (e.g., Bayesian X-ray Analysis, BXA, [87]).
- **Constrain relationships between these dual AGNs and their host environments.** We will compare the X-ray spectroscopic properties (luminosity, $N_{\mathrm{H}}$, etc.) of these dual AGNs against host merger morphological stages and nuclear pair separations, host star formation rates, etc., to test for correlations between the activity and environments of these dual AGNs and the properties of their

large-scale host environments. Prior works have suggested that optically- and hard X-ray-selected dual AGN samples show anti-correlations between nuclear separation and luminosity [268,314], and potentially between separation and column density [though this has been called into question recently, 141,246]; we will directly test these correlations for mid-IR-selected dual AGNs. Folding in merger morphological information during these tests will be particularly important, as nuclear separation alone is a degenerate parameter (different merger stages and orbital configurations can be found at the same separation) and may lead to ambiguous results for these relationships.

- **Multiwavelength Dual AGN Characterizations.** We will use the X-ray information obtained with AXIS and archival 3.5 GHz radio information available through the VLA All Sky Survey (VLASS, spatial resolution $\sim 2.5''$) to estimate SMBH masses for large numbers of obscured and unobscured dual AGNs via the Fundamental Plane [249]. By extension, we will estimate accurate accretion rates and Eddington ratios for a significant number of these mid-IR dual AGNs, enabling comparisons with optical and hard X-ray selected samples from the literature. As mentioned above, when sources are detected with low significance/counts, indirect diagnostics (for example, $L_{2-10\,\mathrm{keV}}$ vs. $L_{12\,\mu\mathrm{m}}$) will be used to constrain parameters such as column density indirectly. The optical BPT classes of these dual AGNs, obtained via archival spectroscopic surveys (SDSS, DESI, LAMOST) and follow-up observations (Gemini, Keck) will be compared against the derived column densities, X-ray luminosities, and accretion rates. We will constrain the fractions of unobscured, obscured, and Compton-thick dual AGNs that fall into any one given BPT class/optical type.
- **Constrain the multiwavelength selection function for mid-IR dual AGNs.** Exhaustive multiwavelength follow-up observations and studies are currently being performed to ascertain the mid-IR, radio, optical morphological, and optical spectroscopic properties of these new mid-IR dual AGNs (Pfeifle et al., in prep.). These studies will provide the occupation fractions of radio and optical AGNs among mid-IR dual AGNs. With AXIS, we will determine the X-ray AGN occupation fraction in this sample, and we will study how this occupation fraction compares to the fractions of AGNs detected at other wavelengths as well as how these fractions compare to the host environments. With information spanning the optical, mid-IR, radio, and now the X-rays, we will have a virtually complete understanding of the selection function for the population of mid-IR dual AGNs. The analysis of the multiwavelength occupation fraction dovetails with the objective above focusing on multiwavelength source characterization.

**Exposure time (ks):** 430 ks on-source time
**Observing description:**

**Sample selection:** The sample of mid-IR dual AGNs was drawn from the *WISE* all-sky mid-IR survey, and full details of the sample selection and curation process are described in Pfeifle et al. (in prep). Briefly: pairs of mid-IR AGNs were drawn from the AllWISE point source catalog, where each source satisfied the mid-IR color cut $W1 - W2 > 0.8$ [546, with W1 magnitude limit $W1 < 15$], and pairs were required to have angular separations of $6'' \lesssim \theta < 60''$. Each AGN in a given pair was required to exhibit distinct W1 and W2 magnitudes, and AGNs were required to have high signal-to-noise (S/N$>$ 10) detections in both W1 and W2. Objects affected by image artifacts or other photometry quality issues were removed. The sample was limited to objects within the DeCaLS footprint and were morphologically classified using the DeCaLS optical imaging. Only mid-IR AGN pairs in galaxy mergers and merger candidates were retained, where galaxy mergers were identified based on clear morphological features, such as tidal tails, bridges, shells, and debris fields. Archival and follow-up optical spectroscopy have provided redshift coverage for a large fraction of this sample, aiding in the removal of projected pairs and providing BPT optical class information. Currently, this sample consists of $\sim 130$ dual AGNs and dual AGN candidates, the majority of which are too faint to be observed with *Chandra* and *XMM-Newton*. For this AXIS program, we focus on

dual mid-IR AGNs with separations of $< 20''$ and those without Chandra observations (78). Dual AGNs with separations of $> 20''$ are better suited for NewAthena observing programs (HEW $\sim 9''$).

**AXIS Exposure Times:** To estimate the predicted X-ray fluxes of these dual mid-IR AGNs, we rely upon the known correlation between the observed X-ray $2-10$ keV luminosity and the mid-IR 12 $\mu$m luminosity [24]. We use the W3 luminosity of each nucleus as a proxy for the 12 $\mu$m luminosity, and infer the $2-10$ keV flux of each nucleus. Next, we use PIMMS to estimate the expected (observed) AXIS count rate using the intrinsic $2-10$ keV flux, and we assume a power law index of 1.8 and provide the redshift of the merging system; given the potential for high obscuring columns [72], we adopt an extragalactic column density of $5 \times 10^{23}$ cm$^{-2}$ in addition to the expected Galactic absorption. Using these count rates, we estimate the exposure time to detect $\sim 80-100$ counts in the fainter nucleus in a given dual mid-IR AGN system. $\sim 80-100$ counts will provide sufficient S/N to fit the spectra directly and modestly constrain the AGN X-ray properties; such a threshold will also allow the detection of X-ray AGNs even if the AGNs are fainter in the X-ray band by $\sim 0.3-0.5$ dex. If the AGNs are brighter than predicted in the X-rays, it will allow for more complex spectroscopic fitting. The average predicted count rate per nucleus is 4.7e-02 counts/s, yielding average exposure times of $\sim 5500$ ks per system (not including overheads). This observation program could not be feasibly carried out with Chandra: a Chandra observing program would require exposure times $3\times -4\times$ longer than those proposed here (and would result in a program time of 1.3-1.7 Ms for only 78 systems).

**AXIS Simulations:** To demonstrate the feasibility of our observation plan and efficacy of our exposure time estimates, we use SOXS [633] and SIXTE [135] to simulate AXIS observations for a small number of dual mid-IR AGNs within this sample. SOXS was used to generate the SIMPUT files, where we adopt point source models for the AGNs, separations matching our example objects (5.7$''$, 7.9$''$, 12.2$''$, and 16.4$''$), and we adopt a simple power law model with normalization set to reproduce the predicted X-ray fluxes for the nuclei. SIXTE was then used to generate the AXIS $2-10$ keV imaging using the SIMPUT files from SOXS; Figure 17 demonstrates the simulated AXIS $2-10$ keV observations for these four representative targets. In each case, both nuclear X-ray point sources are clearly detected above the background and spatially resolved. Figure 17 also depicts the same observations carried out with Chandra, demonstrating that Chandra would not detect enough counts for reliable spectroscopic constrains for many of the AGNs in the sample.

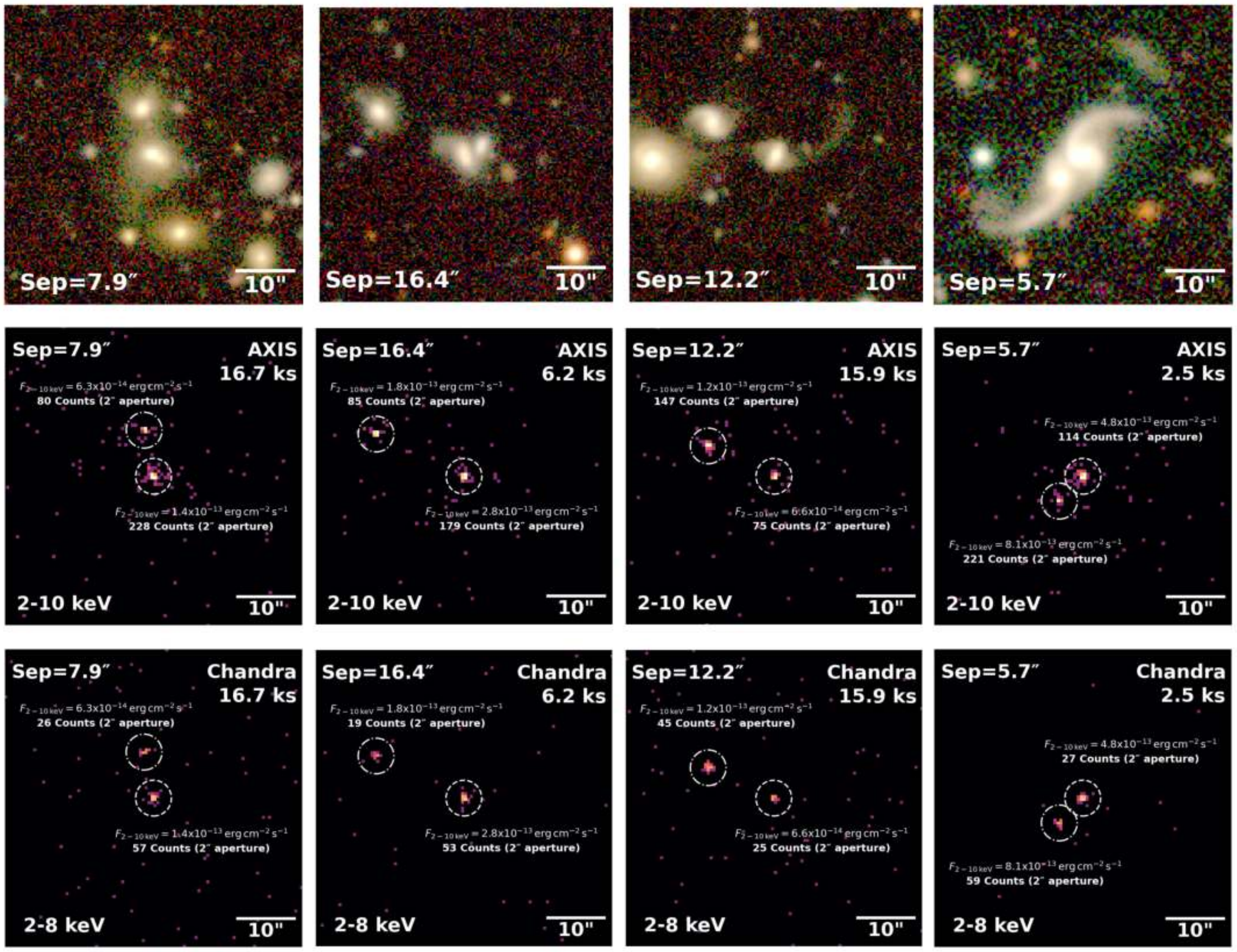

**Figure 17. AXIS simulations of mid-IR dual AGNs.** This figure illustrates AXIS's capability of detecting dual AGNs in a representative subset of mid-IR dual AGNs in this sample. Top row: DeCaLS tricolor imaging of four mid-IR dual AGN candidates with separations $1.5'' < \theta < 20''$. Middle row: AXIS $2-10$ keV simulations for each of these four dual AGN candidates, with exposure times designed to detect the weaker of the two AGNs with $\sim 80-100$ counts to enable direct spectroscopic fitting. Bottom row: analogous Chandra $2-8$ keV simulations using the same exposure time as that adopted for AXIS. Each panel that shows X-ray imaging includes the separation of the two AGN candidates, a 10″ scale bar, the adopted exposure time, and the energy band. Additionally, each X-ray panel includes the adopted AGN fluxes and the detected counts in 2″ apertures in the simulated imaging. These AXIS and Chandra simulations demonstrate that AXIS will be able to detect the X-ray dual AGNs with modest exposure times, whereas Chandra is not able to detect both AGNs with the required number of counts; a Chandra observational program would require exposure times $\sim 3-4\times$ longer.

**Joint Observations and synergies with other observatories in the 2030s:**

- **Radio Synergies:** VLA, VLBA, SKA, ngVLA, ALMA
- **Infrared Synergies:** *Euclid* (archival), *Nancy Grace Roman*, *JWST*, *WISE* (archival)
- **X-ray Synergies:** *XMM-Newton* (archival), *NewAthena*
- **Optical Spectroscopic Synergies:** SDSS, LAMOST, DESI
- **Optical Imaging Synergies:** DES/DeCaLS, SDSS, Pan-STARRS

As described above, this AXIS observational program will leverage synergies with a wide variety of multiwavelength telescopes, surveys, and archives. Prior and ongoing spectroscopic surveys (e.g., SDSS, DESI, LAMOST) and pointed, follow-up observations from facilities such as Keck, Gemini, and

SOAR will provide complementary optical spectroscopic information necessary for determining optical BPT classes, black hole masses (for cases of broad emission lines), approximating narrow line region sizes, and searching for evidence of feedback. This sample was of course built using the WISE archive, demonstrating the clear synergy between AXIS and archival infrared surveys. The current VLA All Sky Survey (VLASS) provides coverage for a significant fraction of this sample, and will play an important role in understanding the radio properties, black hole masses, and radio AGN fraction in this sample; the ngVLA and SKA may also provide valuable imaging sufficiently deep to probe faint radio AGNs in these systems. Archival near-IR observations from Euclid and archival (and potentially pointed) observations with Nancy Grace Roman and JWST will provide unprecedented views of the near-IR and mid-IR emission properties of these dual AGNs and their host galaxies, and these facilities will be particularly useful for studying the most heavily dust obscured and gas absorbed dual AGN candidates. ALMA will be used to probe the cold molecular gas in the nearest and brightest mid-IR dual AGNs, and the constraints drawn from those observations will be compared against the direct line-of-sight column densities derived via the AXIS spectra. This AXIS program targets explicitly only $< 20''$ dual AGNs, while NewAthena will provide the X-ray coverage of more widely separated dual AGNs; together, these two X-ray facilities will provide complete X-ray coverage of this sample.
**Special Requirements:** None

*13. A Legacy Survey of Confirmed and Candidate Dual AGNs: AXIS Science with the Big Multi-AGN Catalog*

**Science Area:** AGN
**First Author:** Ryan W. Pfeifle (University of Maryland, Baltimore County; NASA Goddard Space Flight Center; Oak Ridge Associated Universities, ryan.w.pfeifle@nasa.gov, rpfeifle@gmu.edu)
**Co-authors:** Kimberly Weaver (NASA Goddard Space Flight Center), Emma Schwartzman (Naval Research Laboratory), Barry Rothberg (United States Naval Observatory, George Mason University)
**Abstract:**

Dual active galactic nuclei (dual AGNs) are predicted to be clear tracers of merger-triggered supermassive black hole (SMBH) growth throughout cosmic time. As galaxy mergers offer one avenue for the co-evolution of both the SMBHs and their hosts and establishment of known mass and light scaling relations, and since the pairing rate, accretion histories, environments, and mass evolution of dual AGNs place observational constraints on our expectations for the pairing of gravitationally bound binary SMBHs, substantial efforts have been put forth to identify and analyze dual AGNs over the last five decades. A literature complete and living repository for multi-AGNs and candidates, the Big Multi-AGN Catalog, was recently released and now offers the opportunity for archival and new follow-up studies of the known dual AGN population drawn from a variety of selection methods. Due to observational constraints, X-ray observations have been conducted for only a relatively small proportion of all known dual AGNs and candidates, thereby limiting our understanding of the X-ray and multiwavelength properties among currently identified dual AGNs. With its arcsecond on-axis PSF and superior sensitivity at $2-10$ keV, AXIS is the prime facility to carry out a legacy X-ray science program of dual AGNs and candidates from the literature. This legacy science program will serve two purposes: (a) provide critically important X-ray spectroscopic information for confirmed dual AGNs currently lacking sufficient X-ray coverage, and (b) provide X-ray coverage for curated lists of dual AGN candidates and thereby confirm further cases of X-ray dual AGNs. Dual AGN candidates ruled out from these X-ray observations will nonetheless benefit other galaxy merger studies focusing on, for example, star formation, which require X-ray coverage for large statistical samples. This legacy program will complement multiwavelength archival studies of this literature complete sample, enabling a detailed exploration of the selection function for dual AGNs in the literature. Furthermore, this program will directly test the efficacy of prior selection strategies, allowing us to better understand how the X-ray properties of dual AGNs differ as a function of selection methodology, and ultimately narrow the current list of potential dual AGN candidates.

**Science:**

Since the discovery of the first dual active galactic nuclei (dual AGNs) in the 1970s [see 469, and references therein], a substantial amount of work has been dedicated to the identification of dual AGNs in galaxy mergers [e.g., 113,200,200,201,314,355,517,628]. Theoretical predictions suggest dual AGNs are sites of vigorous merger-induced supermassive black hole (SMBH) growth [e.g., 72,110], with dual AGN fractions directly tracing merger-induced SMBH growth across the history of the universe [110,197]. Dual AGNs may therefore represent an important phase of merger-driven SMBH growth, potentially aiding in the construction of SMBH-galaxy scaling relations [193,215,250], as well as forming the foundation for the properties of gravitationally bound binary SMBHs once the dual SMBHs have sunk to the centers of their merging hosts [50,110]. Thus, the activity of dual AGNs effects not just our expectations for galactic structure, star formation, and the correlation with the host mass, but also affects our expectations for the properties of continuous wave sources and the in-spiral and gravitational wave event rate for bound SMBH binaries [88]. Despite their potential importance and the vast amount of work dedicated to their identification and study, the relative importance of dual AGNs has yet to be observationally ascertained.

Several thousand candidate dual AGNs have been published in the literature over the last five decades. Yet, only a small fraction of total candidates have been confirmed as bona fide dual accreting

SMBHs [469]. Substantial fractions of these candidates have been drawn from optical spectroscopic or imaging surveys [e.g. 261,355,597], while smaller fractions of candidates have been drawn via the hard X-rays [314], radio imaging [200,628], or mid-IR imaging [45,470,517], among other methods [see 469]. Multiwavelength information is often required for unambiguous confirmation of dual AGNs, particularly for dual AGNs with close angular separations [118,119,201,202,352,414] and/or separations otherwise below the resolution limit of the facility used for selection [113,314,470,517,628]. Medium- and high-resolution X-ray observations (via *XMM-Newton* and *Chandra*, respectively) have been used with great success to confirm dual AGNs in the past [61,119,245,268,311,314,352,470], though only a relatively small number of dual AGNs have been confirmed via the X-rays due to the observational demands of such observations. Nonetheless, X-ray observations are one of the cleanest ways to identify bona fide dual AGNs (and thus represent important observational diagnostics for these elusive systems), thanks to the ubiquitous nature of X-ray emission in AGNs and the ability of X-ray emission to penetrate relatively high column densities without significant attenuation (up to a few times $10^{23}\,\mathrm{cm}^{-2}$). Even at higher column densities up to and exceeding $N_{\mathrm{H}} = 10^{24}\,\mathrm{cm}^{-2}$ where the softer X-ray continuum in the 2-10 keV band is substantially depleted, dual AGNs can still be detected and studied [e.g., 61,470,471, etc.]. Furthermore, X-ray spectroscopy allows for direct constraints on the intrinsic accretion power of the central engines and properties of the SMBH environment such as the line-of-sight absorption [which can be directly compared against predictions from simulations, e.g., 72]. X-rays are thus one of the best tools we can use to identify and study further cases of dual AGNs among the vast numbers of dual AGN candidates from literature.

Recently, the first literature-complete catalog of all multi-AGNs from literature was assembled and published[1], which includes dual AGNs, binary AGNs, recoiling AGNs, and N-tuple AGN systems drawn from $\sim 600$ articles in the literature from 1970-2020 [469]. This archival repository now offers the opportunity for large-scale archival analyses and multiwavelength follow-up observations to confirm and better characterize larger samples of dual AGNs. **We propose a large-scale AXIS GO survey of dual AGN candidates drawn from the literature between 1970-2020 to increase the number of confirmed dual AGNs, characterize the X-ray spectroscopic properties of previously known dual AGNs and new dual AGNs confirmed via this program, and quantify the multiwavelength selection function for the known population of dual AGNs and candidates.** AXIS will boast a superior $2-10\,\mathrm{keV}$ effective area to both *XMM-Newton* and *Chandra*, with a on-axis spatial resolution ($1.5''$) similar to the on-axis resolution of *Chandra* ($0.5''$). Whereas the vast majority of dual candidates could not have been observed previously with *XMM-Newton* or *Chandra* due to prohibitively high exposure times, the sensitivity afforded by AXIS will enable large-scale X-ray coverage that will result in the detection of substantial numbers of dual AGNs in a fraction of the time. **Our AXIS GO program will acquire significant $2-10\,\mathrm{keV}$ X-ray coverage for large pools of candidate and confirmed dual AGNs from the literature, providing the means to accomplish the following objectives and planned analyses:**

- **Confirmation of new bona fide dual AGNs and identification of counterparts to known dual AGNs.** The observed X-ray $2-10\,\mathrm{keV}$ luminosities ($L_{2-10\,\mathrm{keV}} \geq 10^{42}\,\mathrm{erg\,s^{-1}}$) and/or spectroscopic properties of the detected X-ray sources will be used to determine the presence of X-ray AGN counterparts. This will yield the occupation fraction of dual and single X-ray AGNs within the populations of confirmed and candidate dual AGNs, independent of the original selection method.
- **Spectroscopic constraints on the X-ray properties of confirmed and candidate dual AGNs.** For the chosen sample of dual AGNs and candidates from the literature (see observation plan below), AXIS will spatially resolve the individual AGNs in each given pair. We will use the spatially resolved

[1] https://github.com/thatastroguy/thebigmac

spectra to place direct constraints on spectroscopic properties (when sufficient counts are detected) of distinct AGNs such as the shape of the intrinsic power laws, the presence of Fe K$\alpha$ emission, and the line-of-sight column densities ($N_{\rm H}$), and unobscured luminosities. Though hard ($> 10$ keV) X-ray observations are required for highly accurate constraints on $N_{\rm H}$ [e.g., 79], modest constraints for even heavily obscured dual AGNs can be obtained in the $2 - 10$ keV band [especially when Fe K$\alpha$ is detected, 470]. For low S/N detections, we will use multiwavelength diagnostics to infer source properties and provide informative priors for spectroscopic fitting via the Bayesian X-ray Analysis software [BXA, 87], which is well suited for fitting in the low count regime.

- **Multiwavelength Dual AGN Characterizations.** Studies using ongoing surveys and archival datasets will soon provide vast amounts of multiwavelength information for dual AGNs and candidates within the Big MAC. We will leverage the synergy between archival multiwavelength datasets and the AXIS X-ray imaging to estimate a range of AGN properties in these systems. For example, we will use the Fundamental Plane [249] to estimate SMBH masses in dual AGNs exhibiting significant AGN-driven X-ray and radio emission. Using the detected X-ray emission and SMBH masses derived via the Fundamental Plane or broad optical emission lines detected in archival surveys (e.g., SDSS, DESI, LAMOST) or targeted spectra (e.g., Keck, Gemini, etc.), we will further estimate accretion rates and Eddington Ratios for large numbers of dual AGNs; these mass and accretion rate constraints will enable statistically robust comparisons to predictions from cosmological simulations [e.g., 110]. Other properties, such as $N_{\rm H}$, can be inferred via comparisons to the optical and mid-IR in cases of low S/N detections (as mentioned above). The fractions of unobscured, obscured, and Compton thick dual AGNs will be compared against the infrared colors and luminosities and the optical BPT classes [37] to understand what fraction/populations of obscured/unobscured dual AGNs are selectable in the mid-IR and optical, and conversely what fraction of obscured/unobscured AGNs are optically-elusive [e.g., 317,470].
- **Constrain relationships between dual AGNs and their host environments.** We will leverage the superior statistical power (relative to prior, smaller samples) afforded by these new AXIS observations to test potential relationships between dual AGNs and their host environments. Prior works have suggested that optically- and hard X-ray-selected dual AGN samples show anti-correlations between nuclear separation and luminosity [268,314], and potentially between separation and column density [141,246]. We will compare the X-ray luminosities and $N_{\rm H}$ of confirmed and candidate dual AGN against the nuclear pair separation to test these tentative correlations (with and without taking into account the original selection strategy of different populations to see if the selection method has any effect). Furthermore, we will compare the X-ray properties with the host merger morphological stage and level of disturbance, the number of companion galaxies, host star formation rates, and other factors to search for any heretofore unknown correlations between the properties of the dual AGNs and their host environments. Comparisons against host morphology – in addition to nuclear separation – are critical, as pair separation alone is a degenerate property; a variety of merger stages and orbital configurations can be observed at a given separation. While we have particularly emphasized the hard energy $2 - 10$ keV band for our science case, the soft X-ray band ($0.3 - 2$ keV) will play a crucial role in constraining properties such as the star formation rates and properties of the intergalactic medium around and between the AGNs.
- **Constrain the multiwavelength selection function of confirmed dual AGN populations.** Ongoing and planned archival studies will soon provide the mid-IR, optical, and radio AGN fractions among confirmed and candidate dual AGNs within the Big MAC through SDSS, DESI, LAMOST, *WISE*, and VLASS data products (private communication). With AXIS, in concert with future NewAthena snapshot programs, we will provide the most statistically significant constraint on the X-ray AGN occupation fraction among confirmed and candidate dual AGNs from the literature. We will compare

this X-ray occupation fraction against the AGN fractions obtained at mid-IR, optical, and radio wavelengths, and use the available multiwavelength coverage to provide a rigorous examination of selection effects among the known dual AGN population. In particular, special attention will be paid to the original selection strategy employed in the literature (and recorded within the Big MAC) for a given dual AGN or set of dual AGNs when studying the AGN fractions at different wavelengths. As dual AGNs have historically been challenging to select, confirm, and characterize, this analysis will provide important insights into the potential selection biases affecting our current samples.

These AXIS observations will serve as a treasury program for the field of dual AGNs and galaxy mergers, providing X-ray coverage across a large sample of objects with a large range in X-ray luminosities and column densities, host masses, black hole masses, nuclear separation, and merger morphology. In particular, this program will increase the X-ray coverage of the candidate dual AGN population (with pair separations $1.5'' < \theta < 20''$) by a factor of $\sim 6-10\times$, lending superior statistical power to future X-ray studies of dual AGNs. Finally, this treasury program will apply to a variety of topics related to galaxy mergers and AGNs, including but not limited to AGN fractions and properties in mergers, obscured AGN populations, AGN feedback, star formation rates in galaxy mergers, and even studies of X-ray binary populations (for the nearest galaxy mergers).

**Exposure time (ks):** 2.9 Ms **(Spread across the nominal mission lifetime)**

**Observing description:**

**Sample Selection:** We began with the full Big MAC DR1 and first limited the sample to only systems classified as confirmed (rank=1) or likely candidate (rank=0.5) dual AGNs; we discarded binary AGN candidates and recoiling AGN candidates. To ensure AXIS resolves the two AGNs or AGN candidates, systems with separations $< 1.5''$ were discarded, and we also removed any candidates where a second nucleus has yet to be confirmed (e.g., the vast majority of candidates selected via double-peaked optical spectroscopic emission lines, [469]). We then narrowed the catalog further to dual AGNs and candidates with angular separations $< 20''$. While AXIS could resolve these systems, systems with large angular separations are better suited as targets for NewAthena X-ray programs (expected HEW $\sim 9''$). Finally, due to the difficulty in identifying and studying the host galaxies of dual AGNs at high redshifts (and because late-stage mergers become inaccessible by $z = 0.5$, i.e., separations $r_p \lesssim 9-10$ kpc cannot be resolved) and to limit prohibitively high exposure times in this program, we limit the redshift of the sample to $z \leq 0.1$. Higher redshift dual AGNs and candidates will be studied in separate observing programs. These requirements yield a sample size of $\sim 190$ dual AGNs and candidates from the literature with separations $1.5'' < \theta < 20''$.

**AXIS Time Estimations:** For this program, we require an X-ray AGN to exhibit either (1) an absorbed (observed) $2-10$ keV luminosity of $\geq 10^{42}$ erg s$^{-1}$, or (2) exhibit clear spectroscopic signatures of an AGN (i.e., Fe K$\alpha$ emission, an intrinsic $2-10$ keV luminosity $> 10^{42}$ erg s$^{-1}$ after correct for intrinsic absorption, etc.). We therefore aim to detect X-ray AGNs down to a luminosity limit of $5\times10^{40}$ erg s$^{-1}$, which equates to $\sim 2.3\times10^{-13}$ erg cm$^{-2}$ s$^{-1}$ at $z = 0.01$, $\sim 8.5\times10^{-15}$ erg cm$^{-2}$ s$^{-1}$ at $z = 0.05$, and $\sim 1.9\times10^{-15}$ erg cm$^{-2}$ s$^{-1}$ at $z = 0.1$. To determine the exposure times required for observations of the Big MAC systems, we used Webpimms to determine the expected AXIS $2-10$ keV point source count rate for each flux limit and redshift noted above; these count rate limits were in turn used to estimate the exposure times required to detect a $2-10$ keV point source with a significance of $3\sigma$ at the flux limit of a given observation. As mentioned above, even heavily absorbed and Compton-thick AGNs can be identified via $2-10$ keV X-ray imaging. Our adopted luminosity limit intrinsically takes into account the potential for high absorbing columns; assuming a simple power law continuum, our luminosity limit allows for the detection of AGNs with *intrinsic* $2-10$ keV luminosities of $10^{42}$ erg s$^{-1}$ even when absorbed by a column density of $10^{24}$ cm$^{-2}$. While our exposure estimates and luminosity limit are based on the hard X-ray

$2-10$ keV band, our observations will also provide critical information in the soft energy $0.3-2$ keV band that will be used to constrain star formation rates and study the properties of the intergalactic medium in these systems.

**AXIS Simulations:** As a sanity check, we also simulated mock AXIS observations using SOXS [633] and SIXTE [135] for point source pairs at the separations of four representative dual AGN candidates from the sample. In these simulations, one AGN has a luminosity consistent with the luminosity limit of the survey ($5 \times 10^{40}$ erg cm$^{-2}$ s$^{-1}$) while the other has a luminosity 5$\times$ higher (an arbitrary choice that helps to differentiate between the two point sources in the imaging). As these simulations rely solely on the observed $2-10$ keV flux, we employ simple point source models and simple power-law models in SOXS to generate the SIMPUT files. We then use SIXTE and the AXIS response files to generate the AXIS simulated imaging. To compare the performance of AXIS against Chandra, we also generate analogous Chandra simulations using the same SIMPUT files and exposure times. Figure 18 demonstrates these four simulated cases, emphasizing our ability to detect dual AGNs close to the flux limit of the AXIS imaging. Figure 18 also emphasizes that Chandra would not be able to accomplish the goals of this survey at the same luminosity limit. As an additional check, we also generated AXIS simulations with the same exposure depths but with "optimal" $2-10$ keV fluxes for the AGNs (where the AGN luminosities are an order of magnitude higher than in the fiducial luminosities mentioned above); in such cases, the depth of the AXIS simulations will provide sufficient counts for complex spectral fitting ($\sim$hundreds of counts, see Figure 19).

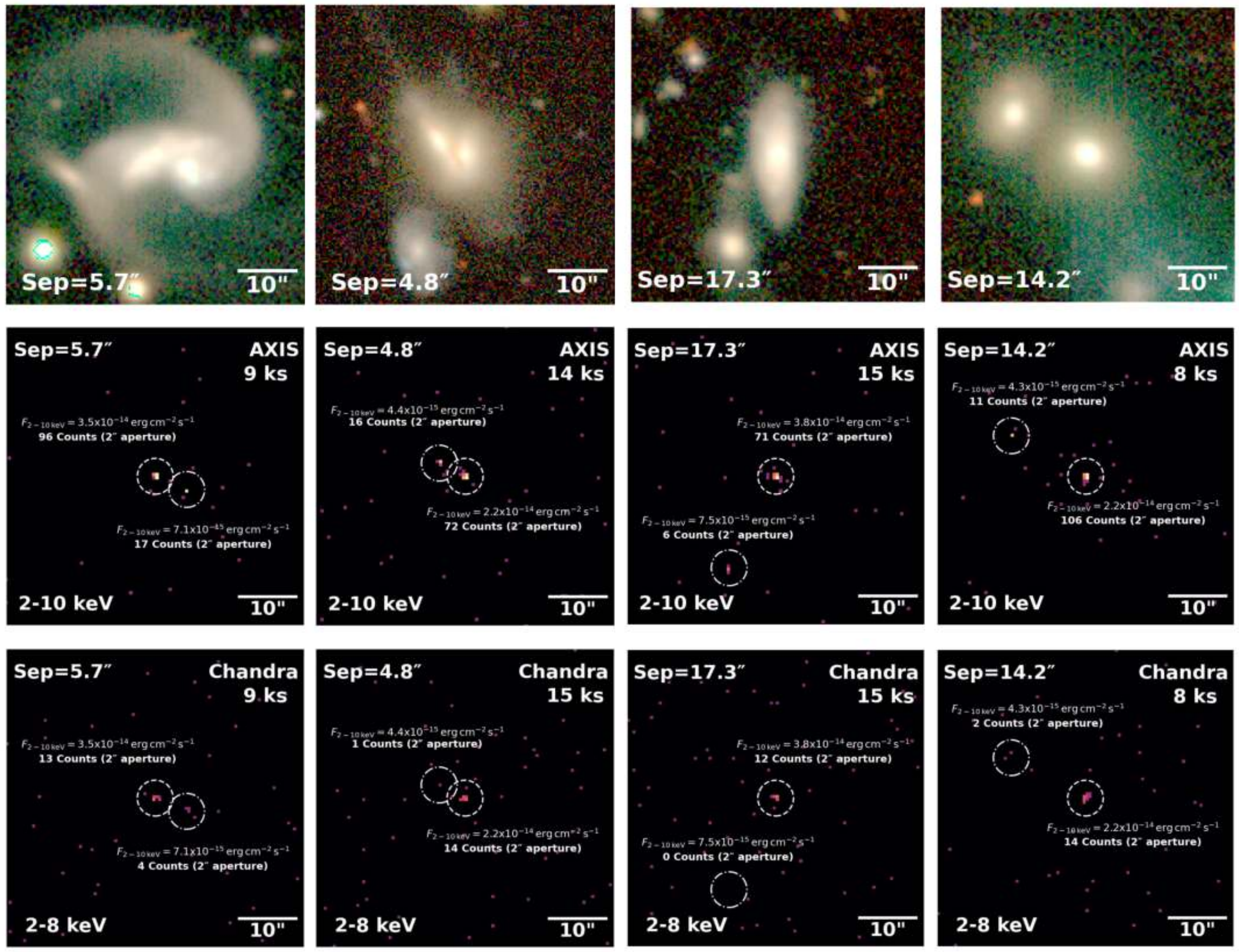

**Figure 18. AXIS simulations of dual AGNs.** This figure illustrates AXIS's capability of detecting dual AGNs down to the adopted luminosity limit in this observing program. Top row: DeCaLS tricolor imaging of four dual AGN candidates from the Big MAC with separations $1.5'' < \theta < 20''$. Middle row: AXIS $2-10$ keV simulations for each of these four dual AGN candidates, with exposure times designed to detect AGNs with $\sim 3\sigma$ significance at the luminosity limit of the observing program. Bottom row: analogous Chandra $2-8$ keV simulations of these dual AGN candidates using the same exposure time as that adopted for AXIS. Each panel that shows X-ray imaging includes the separation of the two AGN candidates, a 10″ scale bar, the adopted exposure time, and the energy band. Additionally, each X-ray panel includes the adopted AGN fluxes and the detected counts in 2″ apertures in the simulated imaging. To aid the eye, these simulations are designed such that the weaker of the two AGNs has a luminosity $5 \times 10^{40}$ erg cm$^{-2}$ s$^{-1}$ (the luminosity limit of the survey) and the brighter AGN has a luminosity $5\times$ higher (an arbitrary choice to make the two AGNs clearly differentiable). These AXIS and Chandra simulations demonstrate that AXIS will generally be able to detect the X-ray dual AGNs in these candidates down to the luminosity limit of the observing program, whereas Chandra is not able to detect both AGNs with $3\sigma$ significance; a Chandra observational program would require exposure times $\sim 3-4\times$ longer.

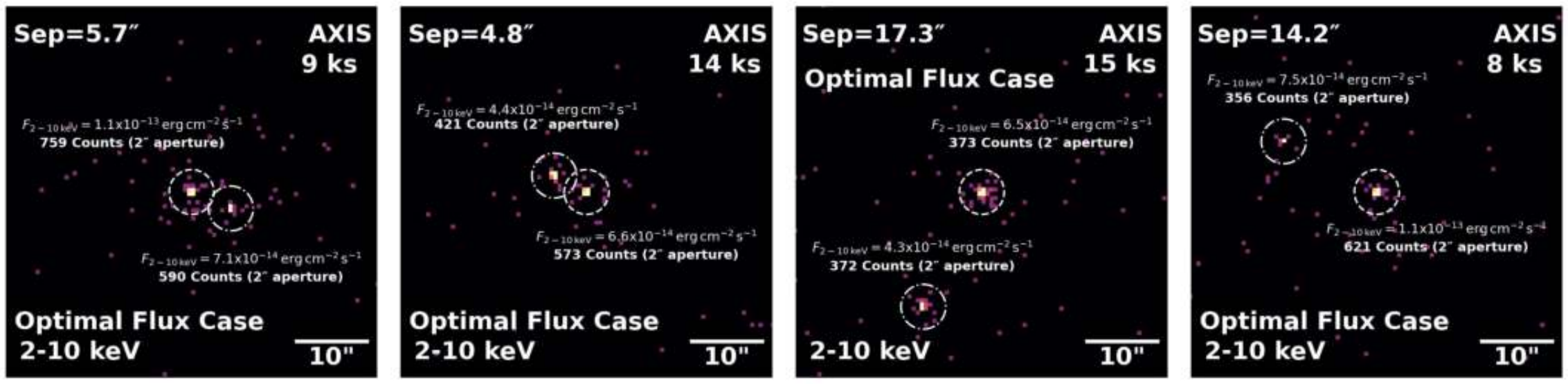

**Figure 19. AXIS simulations of Big MAC dual AGNs (optimal flux case).** This figure is similar to Figure 18, but rather than showing fluxes close to the luminosity limit of the survey, these AXIS simulations demonstrate the "optimal" case where the AGNs are at least an order of magnitude above the adopted luminosity limit. Each panel that shows X-ray imaging notes the separation of the two AGN candidates, a 10″ scale bar, the adopted exposure time, and the energy band. Additionally, each X-ray panel includes the "optimal" AGN fluxes and the detected counts in 2″ apertures in the simulated imaging. In an "optimal" scenario, the X-ray AGNs will be much brighter than the survey's limiting luminosity, allowing for complex spectroscopic fitting of the sources. This complex spectroscopic fitting would still not be possible with Chandra even under an "optimal" source flux scenario without exposure times $3\times -4\times$ higher.

**Joint Observations and synergies with other observatories in the 2030s:**

- **Radio Synergies:** VLA, VLBA, SKA, ngVLA, ALMA
- **Infrared Synergies:** *Euclid* (archival), *Nancy Grace Roman*, *JWST*, *WISE* (archival)
- **X-ray Synergies:** *XMM-Newton* (archival), *NewAthena*
- **Optical Spectroscopic Synergies:** SDSS, LAMOST, DESI
- **Optical Imaging Synergies:** DES/DeCaLS, SDSS, Pan-STARRS

This AXIS treasury program will leverage synergies with a wide variety of multiwavelength telescopes, surveys, and archives. Ground-based spectroscopic surveys (e.g., SDSS, DESI, LAMOST) and pointed, follow-up observations from facilities such as SOAR, Gemini, Keck, etc., will provide complementary optical spectroscopic information necessary for determining optical BPT classes, black hole masses (for cases of broad emission lines), approximating narrow line region sizes, searching for and studying AGN feedback, and comparing the general optical and X-ray properties of this sample. Vast archives of multiwavelength imaging, including data from WISE, 2MASS, GALEX, SDSS, Pan-STARRS, and the VLA All Sky Survey, will provide valuable multiwavelength information on these dual AGNs and candidates, offering additional lines of evidence to determine the true nature of dual AGNs. Ground-based optical imaging, for example, will continue to play an important role in understanding the optical morphological properties and stages of the host galaxies. As another example, the current VLA All Sky Survey (VLASS) provides coverage for a significant fraction of the sky and will be used to understand the radio properties and radio AGN fraction among the known population of dual AGN candidates. Future radio facilities, such as the ngVLA and SKA, will usher in the necessary sensitivity to detect faint radio AGNs across these archival samples of dual AGN candidates. Leveraging AXIS X-ray observations with radio imaging, we will also place constraints on black hole masses via the Fundamental Plane for a statistically large number of dual AGNs for the first time. Archival near-IR observations from Euclid and archival (and potentially pointed) observations with Nancy Grace Roman and JWST will provide unprecedented views of the near-IR and mid-IR emission properties of these dual AGNs and their host galaxies. These facilities will be particularly useful for studying the most heavily dust-obscured and gas-absorbed dual AGN candidates (as host dust is an excellent tracer of obscured, powerful AGNs). ALMA will be used to probe the cold molecular gas in the nearest and brightest dual AGNs and candidates, and the constraints

drawn from those observations will be compared against the direct line-of-sight column densities derived directly or indirectly via AXIS and other multiwavelength data. This AXIS program specifically targets only $< 20''$ dual AGNs, while NewAthena will provide the X-ray coverage of more largely separated dual AGNs; together, these two X-ray facilities will provide statistically significant X-ray coverage of the known population of dual AGN candidates.
**Special Requirements:** None

*14. X-ray detection of AGN in high-redshift quasar companions*

**Science Area:** AGN
**First Author:** Thomas Connor (Center for Astrophysics | Harvard & Smithsonian, thomas.connor@cfa.harvard.edu)
**Co-authors:** Eduardo Bañados (Max-Planck-Institut für Astronomie), Roberto Decarli (INAF—OAS Bologna), Chiara Mazzucchelli (Universidad Diego Portales), Daniel Stern (Jet Propulsion Laboratory, California Institute of Technology)
**Abstract:** Recent sub-mm observations of high-redshift ($z \gtrsim 6$) quasars have revealed a surprising fact: many of the quasar hosts have companion galaxies located at the same redshift, within several to tens of kpc of separation, at a rate far above expectations from the field. Notably, both hosts and companions are rapidly forming stars (SFR $> 100 M_\odot$, as measured by [C II]), and the companion galaxies are comparably-massive or even *more massive* than their adjacent quasar hosts. Yet, heretofore, almost all attempts to detect AGN signatures in these companion galaxies have failed. These non-detections point to one of two paths: either the companions do not host massive, accreting black holes—implying a rapid disconnect of galaxy–black hole scaling relations—or their black holes are heavily obscured by thick screens of galactic obscuration. We propose deep observations with *AXIS* of three of the most promising high-redshift quasar systems with known, massive companions. With the unparalleled imaging capabilities of *AXIS*, we will either detect the long-sought companion black holes or set robust constraints on the roles of environment alone in early black hole growth and evolution.

**Exposure time (ks):** 310 ks

**Joint Observations and synergies with other observatories in the 2030s:** ALMA will remain the best tool for identifying companion galaxies for further observation. *JWST* enables emission line diagnostics for any galaxy with detectable NIR emission. The role of mergers will be potentially quantifiable with next-generation gravitational wave observatories.

**Science:**
The brightest X-ray beacons of the early universe are quasars, highly-luminous ($L_X \gtrsim 10^{44}$ erg s$^{-1}$) active galactic nuclei (AGN) powered by supermassive black holes (SMBHs) with masses of $M_{\rm BH} > 10^9$ $M_\odot$. *Chandra* and *XMM-Newton* have characterized these sources in the Epoch of Reionization to the most distant redshifts of detection ([26]), to deep levels of obscuration ([618]), and in statistically meaningful numbers ([587,624]). X-ray observations of these sources have revealed powerful ([125,271]), variability ([382,590]), and, at slightly lower redshifts, evidence for super-Eddington accretion ([550]). High-energy observatories are therefore a powerful tool into understanding the processes that drive the growth and evolution of these first cosmic titans.

At the heart of this matter is a fundamental question: how do black holes, formed from seeds limited to low masses by the constraints of physics, grow to such massive scales in such short times? Past work has identified SMBHs that have masses of $10^{10}$ M$_\odot$ less than 900 Myr after the Big Bang ([122,165,615]), $10^9$ M$_\odot$ only 700 Myr after the Big Bang ([601,616]), and even $10^7$ $M_\odot$ a mere 450 Myr after the Big Bang ([74]). In contrast, models for typical, Eddington-limited growth, with standard assumptions, cap the growth rate of black holes at one decade of mass every $\sim$115 Myr. Rectifying such massive observed SMBHs with such slow growth requires massive seeds, which raise their own theoretical challenges. And, while individual quasar detections provide constraints on what parameter space black hole growth must be capable of obtaining, they do not, *a priori*, break degeneracies between massive seeds and faster-than-Eddington growth. For that, secondary indicators are necessary.

One parameter capable of constraining seeding mechanisms is the overall abundance of SMBHs across the first billion years of the Universe; specifically, any mechanism that produces a seed black hole with finely limited parameters should be expected to yield a far smaller distribution of black holes than more forgiving models. Direct collapse black hole models, for instance, are capable of producing massive ($M > 10^4$ $M_\odot$) seeds, but require a bright, nearby source of Lyman–Werner photons to dissociate $H_2$ molecules that would too effectively cool the atomic gas and trigger star formation ([59]).

The one variable that needs controlled against in this situation is environment: SMBHs are believed to trace the largest overdensities in the early Universe, and so, being located in statistical outliers, they are more likely to be statistical outliers themselves. Luckily, recent work has opened up this possibility—[143] serendipitously identified four companion galaxies to $z > 6$ quasars, located within $|\Delta v| < 600$ km s$^{-1}$ and less than 100 kpc in projection of the quasar host galaxies, using ALMA [C II] measurements. With measured star formation rates exceeding 100 $M_\odot$ yr$^{-1}$ and with dust masses comparable to *or above* those of the quasar host galaxies, yet lacking quasars detectable in NIR imaging ([398]), these galaxies represent as ideal of control populations as could be desired for testing the diversity of AGN growth in the early Universe.

Follow-up ALMA observations have significantly expanded on this work, detecting additional companion galaxies and providing a more detailed characterization of those already observed. X-ray follow-up, however, has been less fruitful. While [586] reported a detection of a $z = 6.515$ quasar companion with *Chandra*, further observations by [588] ruled out the original report to a strong significance. Likewise, in a 140 ks *Chandra* observation of a $z = 6.59$ quasar with a $\sim 3 \times 10^{10}$ M$_\odot$ host, [124] reported no X-ray emission from the $\sim 4 \times 10^{10}$ $M_\odot$ companion only 8 kpc away (Figure 20). Indeed, despite hundreds of ks of *Chandra* observations dedicated to the hunt, the only candidate detection of X-rays in one of these companion galaxies comes from [123], who were only able to reduce the probability of a background fluctuation to $P = 0.021$ in their report of X-rays from a merging galaxy at $z = 6.23$.

There are three pathways to explaining this dearth of non-detections of X-ray sources from galaxies perfectly matched in size and environment to hosts of massive quasars: companion galaxies host SMBHs that are actively accreting, but the X-ray, optical, and NIR signatures are shrouded by thick intragalactic obscuration; companion galaxies host SMBHs that are only weakly accreting, and their X-ray emission.

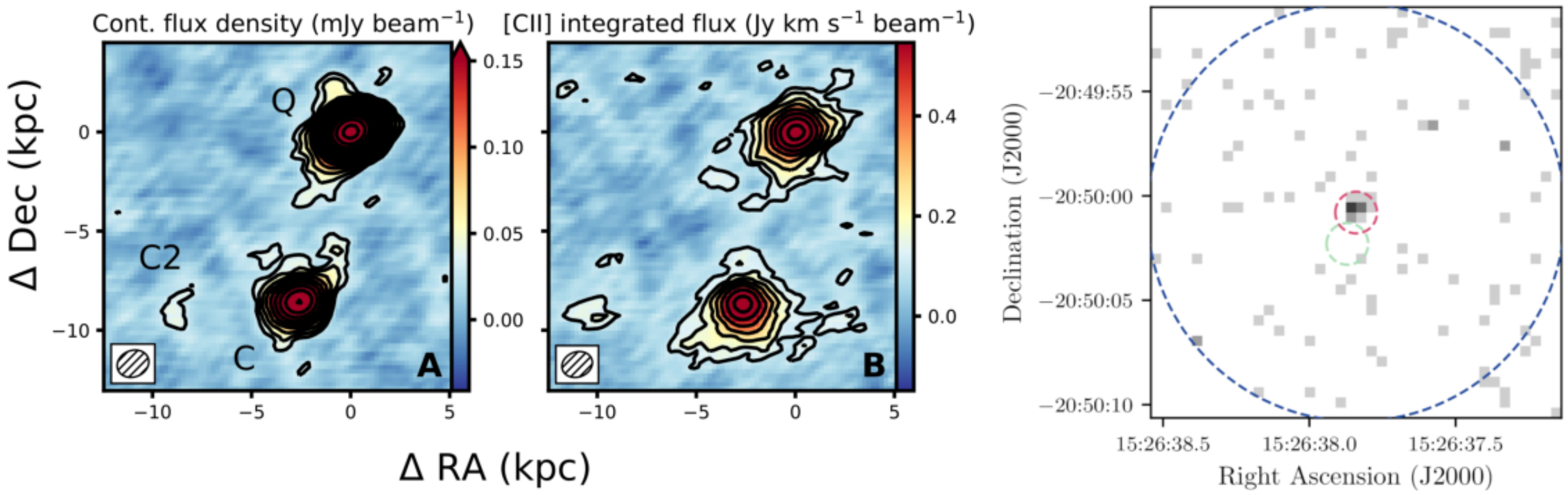

**Figure 20. Left**: ALMA [C II] and continuum observations of the $z = 6.59$ quasar and companion system PJ231−20, as presented in [423]. **Right**: 140 ks ACIS-S 0.5–7.0 keV observation of the same quasar system, showing no X-ray photons coincident with the companion in the *Chandra* Cycle 20 observations ([124]).

is thus weaker than that of the quasars; or the companions do not host SMBHs of comparable mass so that not even weak X-ray emission should be seen. Each possibility can be constrained by *AXIS*, and each would point to a different path for SMBH growth.

- **Obscured AGN**: Recent works have pointed toward a large fraction of AGN in the early Universe being obscured ([27,288]), potentially by their host galaxies [227]. Such a scenario would strongly align with non-detections: if each powerful AGN has only a low probability of being unobscured, the odds would multiply for two such galaxies. Detection of a powerful-but-obscured AGN in these companions would allow for a significantly larger population of AGN in the early Universe, such that the extreme outliers would be more common in number.
- **Sub-Eddington AGN**: One of the common assumptions concerning high-redshift AGN is that they are, when observed, accreting at the Eddington limit—both as a matter of convenience as well as to mitigate concerns about rapid growth. However, most AGN are sub-Eddington in the local Universe (e.g., [144]), and a large (if smaller) fraction of high-redshift AGN should also be expected to be in a more quiescent state. Faint X-ray fluxes—supported by insights from spectral modeling ([83])—would reveal the conditions between bursts of accretion, providing some of the best insights into the duty cycle needed to grow the first SMBHs.
- **Less massive black holes**: A long-standing pillar of black hole science is the relation between the stellar mass of a galaxy and its black hole ([193]); no X-ray detections of these sources would imply that this relation is not in effect at these redshifts. This is already being seen with black holes far more massive than expected ([441])—non-detections would confirm this effect independent of the role of local environment. Such stochastic growth of black holes independent of their galaxy properties follows similar results showing the disconnect between galaxy growth and star formation ([303,304]).

We propose a GO program with *AXIS* to observe the three most promising candidate high-redshift quasar companion galaxies to search for X-ray emission. With *AXIS*'s fantastic angular resolution and large effective area, we will either 1) detect obscured emission, pointing toward quasars being a unique sub-population that has evolved past obscuration, 2) see the faint X-ray emission of sub-Eddington accretion, revealing a population of slow growers, or 3) detect no X-ray emission, painting a picture in which galaxy–black hole growth is decoupled, possibly implicating the extreme SMBHs as stochastic deviations. While this experiment has the potential for a non-detection, we note that *a non-detection is still scientifically informative*. Furthermore, even with companion non-detections, these observations will still yield deep insights into high-redshift quasars, helping to build *AXIS*'s first sample of $z \gtrsim 6$ AGNs.

**Table 2.** High-z companion targets

| Target | RA (J2000) | Dec (J2000) | z | Separation (kpc) | $M_{Quasar\ Host}$ ($\times 10^{10}\ M_\odot$) | $M_{Companion}$ ($\times 10^{10}\ M_\odot$) | *Chandra* (ks) | *AXIS* (ks) |
|---|---|---|---|---|---|---|---|---|
| J0842+1218 | 08:42:29.2 | +12:18:52.5 | 6.07 | 47 | 1.5–4.8 | 0.8–2.5 | 29 | 100 |
| PJ231−20 | 15:26:37.8 | −20:50:00.7 | 6.59 | 9.1 | 2.0–6.2 | 2.7–8.4 | 140 | 130 |
| J2100−1715 | 21:00:54.6 | −17:15:22.5 | 6.08 | 61 | 1.3–4.1 | 4.1–14 | 82 | 80 |

**Table note:** Values from [423]. Separation is projected distance. *Chandra* is duration of archival observation.

**Observing description:**
We propose targeting three high-redshift quasars (listed in Table 2) with known, ALMA-detected companion galaxies that are comparable in mass and star-formation rate to the quasar hosts. All separated by $\lesssim 10''$ from the quasar hosts, these require the arcsecond-scale angular resolution of *AXIS*. All targets have also been observed with *Chandra*, where no detections were reported. As such, *AXIS*'s significant increase in effective area, particularly at soft energies compared to post-2015 ACIS, is vital.

We simulate our sources using `XSPEC`, as described below, with the most up-to-date *AXIS* response files and the L2 non-X-ray background. Detection probabilities are based on Poisson evaluations of expected source and background counts, where we assume a background rate in a $2''$ source aperture of 0.00754 (0.3–2.0) and 0.01338 (2.0–8.0) ct/ks. Final exposure times in Table 2 are the minimum needed to achieve all objectives. Per the science justification, there are three possibilities we are sensitive to:

- **Heavily-obscured AGN**: Based on [227], we assume that the AGN in these companion galaxies are obscured by material at the same redshift with column densities of $N_H = 10^{25}$ cm$^{-2}$. Luckily, given the redshift of these sources, we would expect relatively little dimming above 1–2 keV ([224]) in the observed frame. For this goal, we set the target of detecting an AGN comparable in luminosity to the quasar, but with thick obscuration, modeled as `phabs` $\times$ `zphabs` $\times$ `pow`, with $\Gamma = 2.2$ (e.g., [624]) and $L_{X,pow} = 7 \times 10^{44}$ erg s$^{-1}$ and $N_{H,z} = 10^{25}$ cm$^{-2}$. With our exposures, we will detect these sources to a probability of $P = 0.999$ against background fluctuations.
- **Faint AGN**: The three quasars have already been observed with *Chandra*, and so we know their X-ray luminosities ($L_{X,Q}$). From works at slightly-lower redshift, we would expect a peak in the Eddington rate of $\sim 0.1\lambda_{\rm Edd}$ ([7]). We therefore set a target of $0.1 L_{X,Q}$ to characterize faint AGN. Assuming a power-law model with $\Gamma = 1.9$ (being sub-Eddington; cf. [83]), we will be able to detect 30 photons from these sources.
- **No X-ray Source**: We propose to observe three quasars for a shorter time instead of one for a much longer time to exploit the multiplicative effects of probability. A $1\sigma$ lower outlier should occur with probability $P = 31.73\%/2$, such that we can expect at least one target to be within the $1\sigma$ expectations to 99.5% probability. As such, by targeting the $-1\sigma$ limit of expectations, we can marginalize over sample variability to ensure three non-detections is a robust indicator of no X-ray sources. Adopting black hole masses of $M_\bullet = 10^7\ M_\odot$ (at the low end of expectations for companion masses; [551]), Eddington ratios of $\lambda_{\rm Edd} = 1$, and bolometric corrections of $K_X = 10$ ([162]), and assuming $\Gamma = 1.9$ (under the assumption that the observed high-redshift $\langle\Gamma\rangle \sim 2.2$ is for extreme quasars, this is a more typical AGN value, e.g., [442]), we expect to detect enough counts to reject a background fluctuation to $P = 0.99$ for these sources. Even with individual non-detections, the stacked sample will probe to $L_X = 8 \times 10^{43}$ erg s$^{-1}$, assuming no obscuration, based on the same $P = 0.99$ value.

**Special Requirements:** None.

**c. Feedback**

*15. X-ray emission from kpc-scale AGN Jets*

**Science Area:** AGN
**First Author:** Eric S. Perlman (Florida Institute of Technology, eperlman@fit.edu)
**Co-authors:** (with affiliations) Eileen T. Meyer (UMBC, meyer@umbc.edu)
**Abstract:** Jets are common in AGN, being seen in about 10% of AGN. They emerge relativistically from deep in the central engine near the supermassive black hole and carry mass and energy from the innermost nuclear regions out through the host galaxy and into the surrounding cluster, finally terminating hundreds of kiloparsecs from the black hole. The energies carried by these outflows are tremendous and can transform a galaxy and the surrounding cluster. Despite the transformation of our knowledge about X-ray jets brought about by *Chandra*, which was basically responsible for discovering that X-ray emission from jets exists in many objects, we know very little about the mechanisms by which it is generated. This is because of the faintness of X-ray jets. We propose *AXIS* imaging spectroscopy of a sample of 20-40 jetted AGN of all types, chosen using *Chandra* results. This will finally allow us to uncover their physics, revealing a treasure trove of structural details and X-ray spectra of each X-ray emitting component, distinguishing between emission and particle acceleration mechanisms, and testing the balance between synchrotron and Comptonization processes in jets in the X-rays. This will also tell us to what extent jets are highly beamed on kiloparsec scales, and which objects are, thus giving us information about jet kinematics and deceleration over large scales, where currently none exists.
**Science:**

Jets are a common feature of AGN, likely created deep in the central engine by a coupling between the black hole's accretion process and the magnetic field. AGN jets are highly relativistic outflows, which accelerate to Lorentz factors as high as 20-50 and carry energy and mass from the sub-parsec nuclear regions out through the host galaxy on kiloparsec scales, finally terminating in lobes on scales of hundreds to thousands of kiloparsecs. The energy carried by these powerful outflows can be comparable to that of the AGN as well as the galaxy in which the AGN sits. They can transform a galaxy and the surrounding cluster (see e.g., [255]), rapidly quenching star formation in galaxies where a (likely merger related) nuclear infall of material triggered both massive starbursts as well as the AGN activity, and counteracting the formation of cooling flows in galaxy clusters, where the tremendous power of the outflows is released as p dV work.

The last thirty years has seen a dramatic explosion of our knowledge regarding AGN jets. The launch of the *Chandra X-ray Observatory* and *Hubble Space Telescope* led to the discovery of X-ray and optical emission from many dozens of AGN radio jets, and we now know that such emissions are a common feature of the class. Much remains unknown about jets, as illustrated by the discussion in our Astro2020 White Paper [458]. For example, the X-ray emission from jets may arise via synchrotron emission from a high-energy population of electrons, but that population may be either the high-energy tail of the particles that create the emission seen in the optical and radio (in some objects), or a largely separate population (in others). Other processes may also be involved; however, most notably, inverse-Comptonization, either of Cosmic Microwave Background photons by the jet's relativistic particles or of the radio photons created within the jet. Another poorly constrained issue in jets is the scales over which they maintain relativistic velocities and where they decelerate. While *Hubble* and *Chandra* have seen relativistic velocities in jet components on kiloparsec scales, the evidence for deceleration on these scales remains scant.

The X-ray emission in jets is faint. While *Chandra* has resulted in a revolution in our knowledge of these objects, jet components are typically seen to have fewer than 100 photons detected, even in deep images. As a result, the X-ray structure of jets is not well known – do a few bright components typically

dominate it, or is X-ray emission at lower levels common? The same can be said for the shape of X-ray spectrum and the overall morphology of the spectral energy distribution in the UV-to X-ray region. The massive increase in X-ray sensitivity from *AXIS* will enable us to refine our understanding of jets in the X-rays.

We propose to obtain *AXIS* imaging spectroscopy of a sample of 20-40 jetted AGN of all types, over a range of redshifts. Our sample is drawn from *Chandra* observations and includes objects of all types, such as FR I and FR II radio galaxies, quasars, and BL Lac objects. This sample size will allow *AXIS* to observe statistically significant numbers of each object class, enabling us to sample well what nature truly provides. A smaller sample size may result in us missing an important part of the puzzle, including the range of parameter space over which the IC/CMB mechanism can account for the X-ray emission, or the range of parameter space for X-ray particle acceleration or the energy budget of the jet as a whole. Because of the massive increase in sensitivity possible with AXIS, images comparable in depth to hundreds of ks with *Chandra* will be achievable in about a tenth of the time. An example is seen in Figure 21, the deep *Chandra* image of 3C 273, the only quasar jet for which we currently have sufficient exposure to see both knots and extended structure, allowing us to obtain X-ray spectral indices for all jet components.

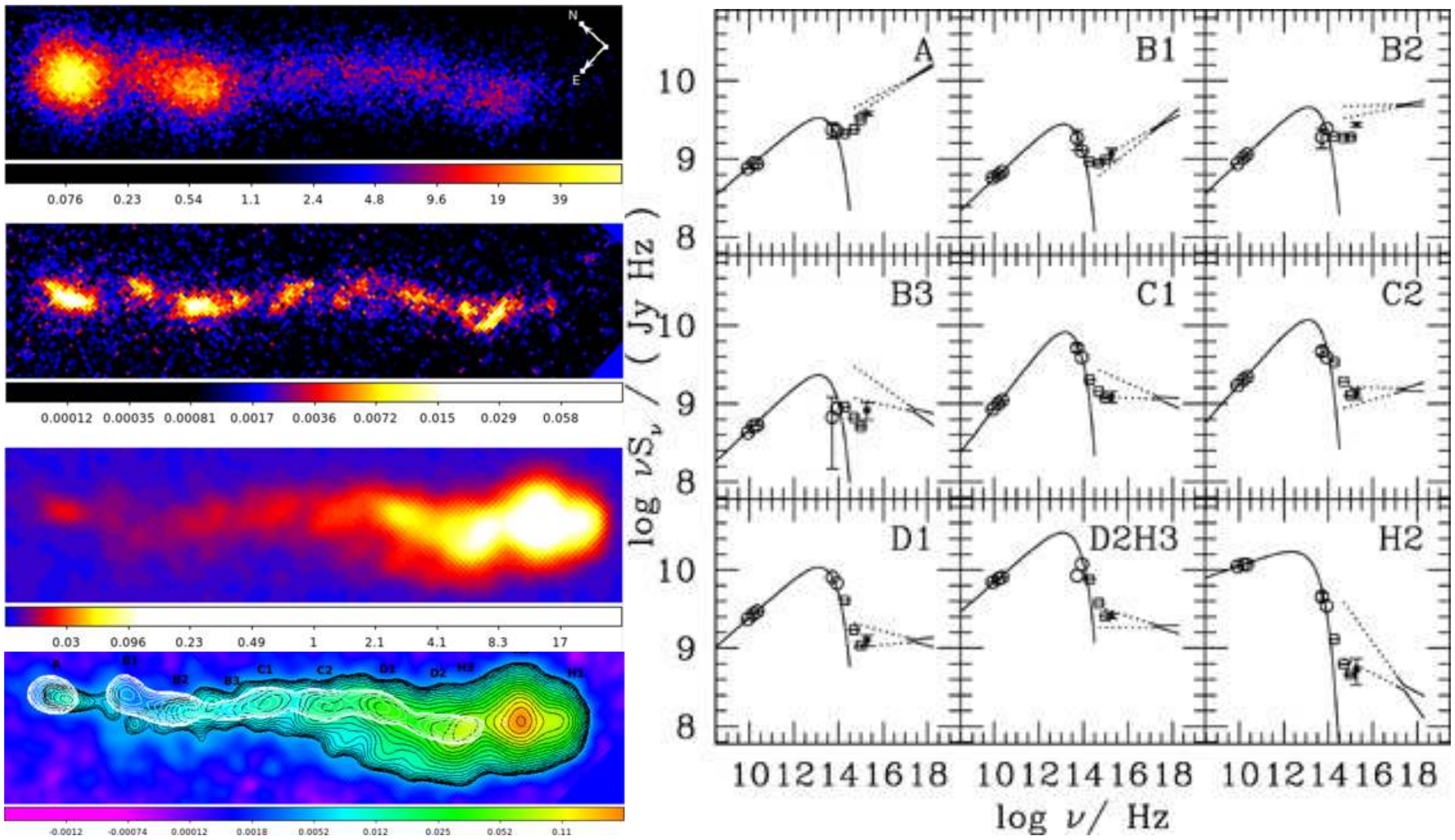

**Figure 21.** At left, the jet of 3C 273, as seen in multiple different wavebands [283,284,376]: At top: The combined *Chandra* image of the jet, representing about 200 ks of integration. Second from top: With HST/ACS/SBC, in the F150W band, i.e., wavelength 150 nm. Third from top: In the radio, as seen with the JVLA. At bottom: The radio image overplotted with contours from a deconvolved *Chandra* model. At right: the spectral energy distribution of different jet components. Several things are notable. Firstly, there are significant differences between the radio, optical/UV, and X-ray morphologies of the jet, with the X-ray emission originating from a much narrower region than the radio. Second, the UV and X-ray emission represents a different spectral component than seen at lower (radio and optical) photon energies. In 3C 273, this component is attributed to a high-energy, locally accelerated particle population occupying a small fraction of the jet cross-section; however, in other sources, it may originate from different emission mechanisms. This interpretation is critically dependent on knowing not only the shape of the broadband SED but also the X-ray spectral index.

will reveal a rich new treasure trove of morphological and spectral detail. This will allow us to break ground in several areas.

- **X-ray morphology of jets.** We will be able to detect both X-ray emitting knots and fainter, inter-knot regions in the objects in our sample. Right now, in most X-ray jets, we only see the very brightest knots, so our knowledge of the X-ray emission processes in jets is likely incomplete. Different processes may be at work in fainter regions, particularly between knots, where particle acceleration is likely to be weaker. If the X-ray emission from these regions is also synchrotron in nature, as it is in 3C 273 (Figure 21), it should occupy a much narrower region of the jet cross-section and/or be more localized to knot regions. In this case, differences in the knot locations would also indicate differences in the localization of the high-energy particle population. In this case, we would also expect the X-ray emitting regions to be highly polarized at lower frequencies, as has been observed with HST polarimetry [94,459–462].
- **X-ray spectra and SEDs for synchrotron-emitting jets.** Obtaining X-ray spectra for X-ray emitting jet components is critical to understanding their emission mechanisms. As seen in Fig. 21, without the X-ray spectral information, we would not know that the X-ray and UV emissions in 3C 273 lie on a different emission and particle component than the radio and optical emissions. Currently, there is knot-by-knot X-ray spectral information for only three jets: Cen A [232,256], M87 [461], and 3C 273 [283,284,376], and this has required *Chandra* exposure times of hundreds of kiloseconds. In those objects, the X-ray spectra inconjuction with *HST* multiband images and polarimetry allow us to obtain the particle energy distributions in each jet component, measure the need for particle acceleration and the response of the jet magnetic field, and along with the X-ray morphology localize the X-ray emission within the jet. This has allowed those authors to discuss topics such as distributed particle acceleration, magnetic reconnection and reconfinement, spine-sheath models, and the like.
- **X-ray spectra and SEDs for IC/CMB X-ray jets.** If on the other hand the emission is due to inverse-Comptonized CMB emission, first suggested in 2001[98] we expect the morphology of the X-ray emission to be much more similar to the low frequency radio emission, with an X-ray spectral index much more similar to that seen in the low-frequency radio since that particle population would scatter it. It would also be unpolarized in the optical. These *AXIS* observations, in conjunction with future observations, will disentangle this mystery, and will be able to tell us which objects can be explained by the IC/CMB hypothesis, and whether these are the most Compton dominated objects and hence most highly beamed objects seen at GeV energies[409], or only the highest redshift. IC-CMB jets, because they are the most highly beamed, also define the upper edge of the range of jet energy budgets.
- **X-ray variability: the tightest constraints.** We will also use the far better statistics in these images to search for X-ray variability on timescales of weeks or longer in the knots for objects observed multiple times. This was found for several X-ray jets by Chandra, e.g., M87[259,463], Pic A[257], Cen A[544], and not only cements their emission as high-energy synchrotron radiation, but indicates that the particle energy distribution in jets reach multi-TeV energies and is subject to the local dynamics in the particle acceleration regions where they are emitted[410], which must be at most light-months in size. Variability on shorter timescales is not expected given the resolved nature of most jet components and the physical sizes implied. Our research, up to date, has barely scratched the surface of this subject, but the statistics will improve massively with *AXIS*.

**Exposure time (ks):**$\sim$ 1**Ms**
**Observing description:** ( Our target list of 40 AGN has been drawn from *Chandra* results and represents the jets for which the highest-quality results can be obtained. This is simple imaging spectroscopy, which can be performed simultaneously with other programs for these objects. Experience with *Chandra* leads

us to conclude that 25 ks integration with *AXIS* will allow us to obtain 50-200 counts per knot on typical jet region as well as significant amounts of interknot emission, as was seen in jets where such deep *Chandra* imaging exists, for example M87 [259,459,461,463], Cen A [232,256] and 3C 273 [283,284,376]. Those observations needed much longer integration times with *Chandra*, but we can obtain better results with *AXIS* in about a tenth of the observing time.

- Quasars: 3C 273, PKS 0637-752, PKS 1030-357, PKS 0920-397, 3C 279, 3C 9, 1045-188, PKS 1127-145 , PKS 1136-135, 4C +49.22, PKS 1202-272, 4C 19.44, PKS 1510-089, 1745+624, PKS 2101-49
- FR II radio galaxies: Pic A, 3C 111, 3C 390.3, 3C 227, 3C 219
- FR I radio galaxies: M87, Cen A, 3C 120, 3C 15, NGC 6251, 3C 66B, M84, NGC 4261, 3C 465, Cen B, 3C 129, 3C 31, NGC 315, 3C 31
- BL Lacs: PKS 0521-36, 3C 371, OJ 287, S5 2007+777, PKS 2201+044, AP Lib

Typical observing time is 25 ks per object, based on experience with *Chandra*. This will allow us to trace out faint components in each jet, including discrete knots as well as extended emission. Most knots will be detected with tens to hundreds of counts apiece, which combined with the high sensitivity and much broader bandpass of *AXIS* will allow constraints on the X-ray spectrum that are sufficient for our aims.

**Joint Observations and synergies with other observatories in the 2030s:** ngVLA, ALMA, JWST, various ground-based telescopes.

**Special Requirements:** pileup, coordinated observations, monitoring, TOO.

*16. A test bed for jet evolution and disc-outflow connection in jetted Narrow-Line Seyfert 1 galaxies*

**Science Area:**
**First Author:** Filippo D'Ammando (INAF, dammando@ira.inaf.it)
**Co-authors:** Luigi Gallo (Saint Mary's University)

**Abstract:** The discovery by the *Fermi Gamma-ray Space Telescope* of variable $\gamma$-ray emission from a few radio-loud narrow-line Seyfert 1 galaxies (NLSy1) revealed the presence of an emerging third class of AGN with powerful relativistic jets. This has raised important questions about the conditions required for jet formation and launching, the disc-jet connection, and possible outflows in these sources.

X-ray emission is one of the most intriguing aspects of this class of sources. The spectra are steep above 2 keV, which is similar to FSRQs rather than radio-quiet NLSy1s, indicating that non-thermal jet emission dominates the spectra. However, the majority of sources for which good-quality spectra are available also show a soft excess at low energies, most plausibly related to the accretion disc, which makes them different from typical blazars. A tentative detection of a narrow Fe line at 6.4 keV has been reported for one source. We want to exploit the unique combination of spatial and spectroscopic resolution in X-rays of the AXIS satellite for:

- studying the connection between the jet and accretion flow (starting from soft excess and Iron emission lines);

- searching for intrinsic absorption and winds/outflows, investigating the jet/wind interaction;

- understanding the origin of feedback and its relation with the AGN activity, in particular with the relativistic jets, in these sources.

This investigation will be performed on a sample of (radio-loud) jetted NLSy1 galaxies not limited only to current $\gamma$-ray-emitting NLSy1.

**Science Justification**

Active Galactic Nuclei (AGN) are the most luminous persistent sources of high-energy radiation in the Universe. However, only a small percentage of AGNs are radio-loud, and this characteristic is commonly ascribed to the presence of relativistic jets, roughly perpendicular to the accretion disc. Accretion of gas onto the supermassive black hole (SMBH) is thought to power these collimated jets. However, the nature of the coupling between the accretion disc and the jet remains one of the outstanding open questions in high-energy astrophysics [e.g. 69,405]. Certainly, relativistic jets are the most extreme expression of the power that can be generated by a SMBH in the center of an AGN, with a large fraction of the power emitted in $\gamma$ rays. Before the launch of the *Fermi Gamma-ray Space Telescope* satellite, only two classes of AGN were known to generate these structures and thus to emit up to the $\gamma$-ray band: blazars and radio galaxies, both hosted in giant elliptical galaxies [71]. The first 16 years of observations by *Fermi*-LAT confirmed that blazars are the most numerous class of identified sources in the extragalactic $\gamma$-ray sky; however, *Fermi*-LAT observations revealed narrow-line Seyfert 1 galaxies (NLSy1) as a new class of $\gamma$-ray-emitting AGN with blazar-like properties [3]. It is a very small class, consisting of only about a dozen sources to date [11,131].

NLSy1s are a class of AGN identified by Osterbrock & Pogge [435] and characterized by narrow permitted lines (FWHM (H$\beta$) $< 2000$ km s$^{-1}$), [OIII]/H$\beta < 3$, and a bump due to Fe II [e.g. 476]. In X-rays they exhibit strong X-ray variability, steep spectra [$\Gamma_X > 2$ 242], relatively high luminosity, and substantial soft X-ray excess [76]. These observational characteristics indicate systems with smaller masses of the central black hole ($10^6 - 10^8 M_\odot$) compared to those in blazars and radio galaxies, as well as higher accretion rates (close to or above the Eddington limit). Moreover, some of the strongest candidates of objects with relativistic broad lines and reverberation lags are NLSy1s [e.g. 138,183,594]. The inner few gravitational radii of NLSy1 are exposed, allowing us to study the X-ray corona and the inner accretion disc. In many of these, the corona is associated with a collimated outflow and is interpreted as the base

of an aborted jet [e.g. 604,605]. Many NLSy1 also exhibit strong wind features that could be launched from the inner region [e.g. 449]. Only a small fraction of NLSy1 ($<7\%$) are associated with jetted emission and classified as radio-loud [312], and objects with very high values of radio-loudness ($R > 100$) are even more sparse ($\sim$2.5%). At radio frequencies, NLSy1s display a compact structure with strong and variable emission. The brightness temperatures are well above the inverse Compton limit [e.g., 620]. They also show a flat radio spectrum, which, together with powerful emission and variability, suggests the presence of a relativistic jet in some of the NLSy1s. This was confirmed by the detection by *Fermi*-LAT of persistent $\gamma$-ray emission in some of them.

The $\gamma$-ray observations, as well as the multiwavelength properties, provide clear evidence for the existence of powerful jets close to the line of sight. For example, some of the sources exhibit large-amplitude $\gamma$-ray flares with apparent isotropic $\gamma$-ray luminosities of $\sim 10^{48}$ erg s$^{-1}$. This is comparable to flares in bright flat spectrum radio quasars (FSRQs), indicating that NLSy1s host relativistic jets as powerful as in those sources [see e.g. 131, and the reference therein]. Given that NLSy1s are usually associated with spiral galaxies, low-mass black holes, and high accretion rates [e.g. 146], they do not fit in with the typical paradigm of systems that host powerful relativistic jets [e.g. 388]. This makes the $\gamma$-ray NLSy1s and strong radio-loud NLSy1s very interesting in terms of understanding the conditions required for jet formation, for studying the disc-jet connection, and the jet-wind interaction. Those sources are the best candidates to perform those studies.

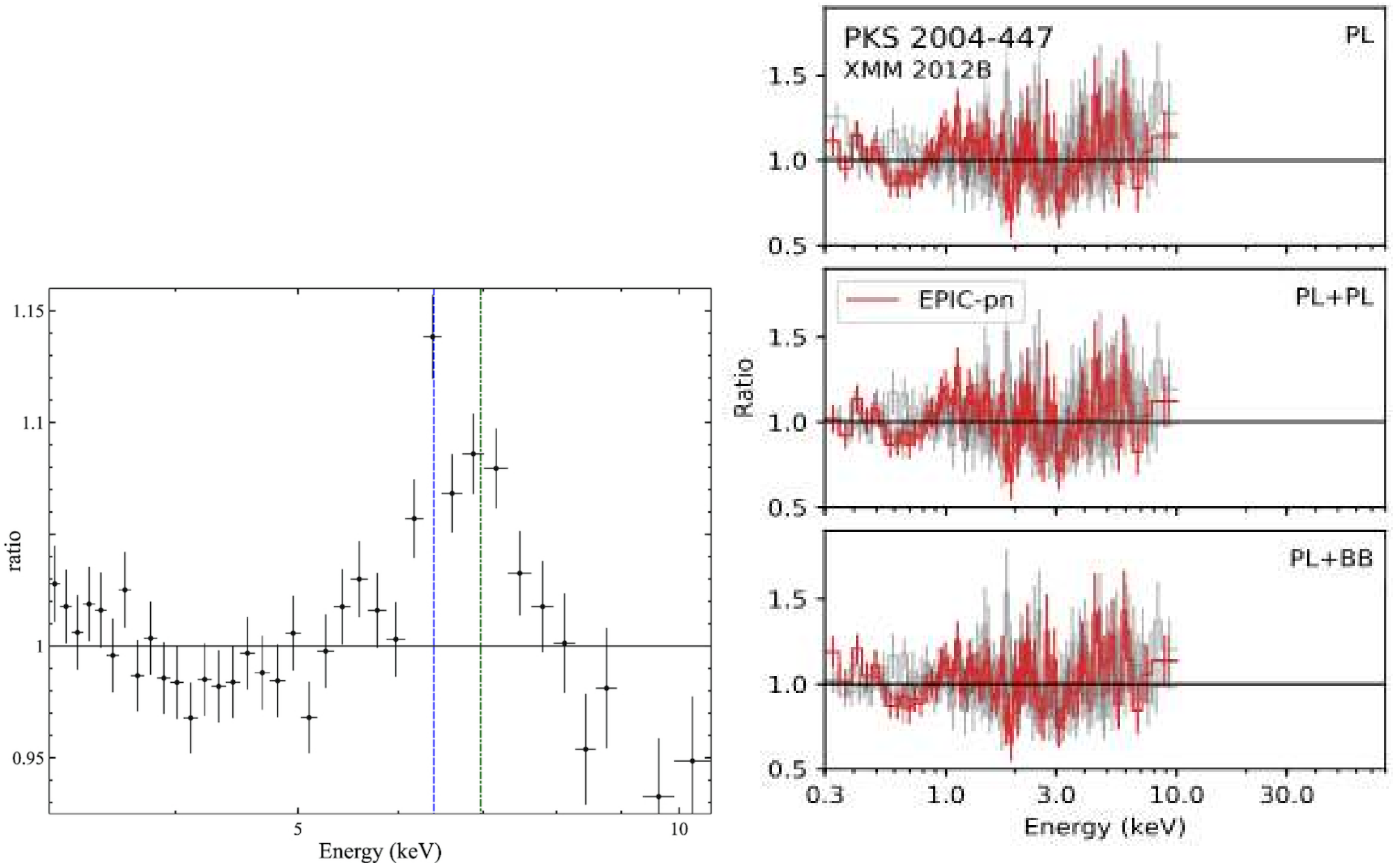

**Figure 22.** *Left*: The ratio of EPIC-pn to a simple power law for 1H 0323+342; energies are in the rest frame of the source. The blue line represents the position of neutral iron at 6.4 keV, while the green line shows that of hydrogen-like Fe XXVI at 6.97 keV. We find an equivalent width of 175 $\pm$ 40 eV for the broad Iron line. Adapted from [416]. *Right:* EPIC-pn spectra of PKS 2004-447 collected in different periods fitted with a single power-law, two power-laws, and a power-law and a black body model. In red, the 2012B observation is highlighted. Adapted from [443].

**The X-ray spectra of $\gamma$-ray-emitting NLSy1**

X-ray emission is one of the most intriguing aspects of $\gamma$-ray-emitting NLSy1. The spectra are hard above 2 keV ($\Gamma \sim 1.2 - 1.7$), which is similar to FSRQs rather than radio-quiet NLSy1s and shows that jet emission dominates the spectra. However, during low-activity states of the jet, the thermal features can become evident in their spectra [e.g., 132]. A narrow Fe line at 6.4 keV has been reported for a single source, 1H 0323+342 (Fig. 22, keft panel), which has the softest X-ray spectrum of the population [132,324].

**The soft X-ray excess**

The majority of sources for which good-quality spectra are available also exhibit a soft excess at low energies, distinguishing them from typical blazars. A plausible explanation for the spectra is that the underlying Seyfert emission, originating from the corona and accretion disc, has a noticeable contribution at low energies. The soft X-ray excess is a ubiquitous yet poorly understood feature in the X-ray spectra of AGN in general, but it is often exceptionally strong in radio-quiet NLSy1s (compared to regular broad-line Seyferts), making it plausible that it would be detectable in the $\gamma$-ray emitting sources even though the jet emission is strong. In agreement with this, analyses of XMM-Newton observations of several $\gamma$-ray NLSy1s have shown that the spectra can be well described by models that include reflection from the accretion disc and/or Comptonisation in a warm corona in addition to jet emission [e.g. 133,324,336]. If the former is the origin of the soft excess, it implies that the accretion disc extends very close to the SMBH suggesting a high spinning black hole with implications also for the jet formation and development. If a warm corona produces the soft excess, it requires an optically thick vertically extended structure above the disk where a large part of the accretion power should be released, and this can have an implication on the accretion flow stability and production of outflows.

Another possibility is that the soft X-ray emission originates from the jet itself. For example, a spectrum with a soft excess will be observed if the tail of the synchrotron emission extends into the X-ray range. It has also been suggested that the bulk Comptonisation by a blob of plasma traveling along the jet could produce an excess at soft X-rays [99]. However, this feature should be transient, in apparent contradiction with the fact that it is observed in the majority of sources. To discriminate between different models for the soft excess and study the disc-jet connection in $\gamma$-ray NLSy1s, it is essential to have detailed X-ray spectra, such as those available with AXIS. In this context, the improved AXIS sensitivity in the soft X-ray band, combined with excellent background control, can significantly enhance the characterization of features in soft X-ray observations and disentangle the different scenarios proposed for the origin of the soft X-ray excess.

**Outflows: WA and UFO**

Outflows in AGNs are one of the fundamental mechanisms by which the central SMBH interacts with its host galaxy. Outflows usually come in two forms: warm absorbers (WA) and ultra-fast outflows (UFO). The absorption features of ionized materials with a typical temperature of approximately $10^5$ K and an ionization state of $\xi$ = 10 -1000 erg cm s$^{-1}$ are often shown in the soft X-ray observations of radio-quiet Seyfert 1 AGNs [e.g. 163]. These materials are estimated to be located within or just outside the broad line region (BLR). Some might be even closer to the nucleus, e.g., at the distance of the accretion disc, or further out in the host galaxy [e.g. 150]. They are 'warm' relative to the more ionized and hotter plasma in the innermost accretion region just outside the SMBH. Why are WAs so rare in radio-loud AGNs, and how is this connected with a dominant jet emission? The coexistence of jets and winds is contentious, as it is usually associated with different accretion modes (radio and quasar modes, respectively).

WAs are found in almost 50% of radio-quiet Seyfert 1 AGNs [418]. The most prominent features include O VII, O VIII and Fe xvii–xviii absorption lines between 0.5-0.8 keV in the rest frame. Absorption lines are sometimes found to be blueshifted by a velocity of a few hundred to a few thousand km s$^{-1}$ [402,

e.g.]. WA are produced at sub-pc to pc-scale and can provide information about the connection between the accretion disc, BLR, and winds at a larger scale. Compared to radio-quiet AGNs, WA detection is relatively rare in radio-loud AGNs, where significant radio jet emission exists in the data [195,392, e.g.]. In cases of blazars, the jet is aligned with the line of sight, and this can mask the absorption features of WAs [e.g. 149]. Others propose that radiation-driven outflows may form in the BLR of radio-loud quasars, assuming that powerful quasars are accreting closer to the Eddington limit than their lower luminosity counterparts [e.g. 489]. The absorbers are thus too ionized to show absorption features. However, no decisive conclusions have been drawn for the absence of WAs in the majority of radio-loud AGNs. The current positive detection of WAs is limited to a small sample of bright radio-loud AGNs. In this context, jetted NLSy1 are ideal targets. Some jetted NLSy1s show tentative evidence of WAs in CCD-resolution data, in particular the $\gamma$-ray-emitting NLSy1 PKS 2004-447 [443], but archival high-resolution data do not have enough S/N to confirm the existence of WAs in these sources (Fig. 22, right panel).

However, UFOs are characterized by slightly relativistic velocities (0.1-0.3 $c$) usually inferred from the detection of blue-shifted Iron K absorption lines in the 6-9 keV energy band, and typically high ionization states. The high velocity of these features implies that they are produced near the accretion disc, where radiation pressure and magnetic fields can produce such a level of acceleration. UFOs are believed to play an important role in AGN feedback, but not only the launching mechanisms but also the connection with WAs and jets is still debated. UFOs are detected in a significant fraction of radio-loud AGN [e.g. 558], suggesting that the presence of relativistic jets does not preclude the existence of such features. However, no UFOs have been detected from a $\gamma$-ray-emitting NLSy1 so far. AXIS observations can provide the first detection of a UFO in a $\gamma$-ray-emitting NLSy1.

**Detection and origin of Fe emission line**

In radio-loud AGN, it is not so common to observe the Fe emission line, because the beamed jet component weakens the line strength. The hint of detection of a Fe emission line in 1H 0323+342, stacking 7 XMM-Newton EPIC-pn observations performed during 2015-2018 (see Fig. 22, left panel), is exciting and consistent with the classical nature of NLSy1. It opens the possibility of investigating the origin of this emission line in a $\gamma$-ray-emitting NLSy1. This line has been interpreted as a relativistic broadened Fe K$\alpha$ line [416]. However, the disc inclination obtained by the analysis of *XMM*-Newton data (60°-70°) is inconsistent with the luminosity and flaring activity observed in $\gamma$-rays. The reflection may originate within a funnel geometry, as seen in jetted TDEs like Swift J1644, resulting in a steeper inclination measurement. Alternatively, the disk can be warped or torn, and the angle measured by the jet and Fe K$\alpha$ naturally did not coincide at the moment. Moreover, although the blurred reflection fit is statistically acceptable, there are other possible explanations, like a mixture of narrow emission components. New AXIS observations will be fundamental in confirming and disentangling the nature of this Fe line in 1H 0323+342 and potentially in other $\gamma$-ray-emitting NLSy1s, such as PKS 2004−447.

Potentially, different topics can be studied with AXIS for jetted NLSy1:

- intrinsic absorption, soft X-ray excess, and Fe line
- wind and outflows
- jet-wind interaction
- disc-jet connection
- the relation between feedback and jet activity

AXIS will be the best X-ray telescope for investigating the relationship between two different modes of outflows in radio-loud AGN, wind and jet, due to its combination of excellent spatial resolution, sensitivity, and photon collecting area, which is crucial for characterizing the X-ray spectra of jetted NLSy1. In fact, the on-axis effective area of AXIS is close to or higher than that of *XMM-Newton*/EPIC cameras, and

significantly higher than that of *Swift*-XRT, *NuSTAR* and *Chandra*-ACIS. In addition, AXIS will have better spectral resolution compared to other X-ray satellites within the range of energies where these features are located, and this will enable us to measure these features more accurately, even during periods of moderate jet dominance. Finally, by extending the sensitivity range below 0.5 keV, AXIS will allow us to refine the accretion models and better understand the interplay between outflows, discs, and jets.

In addition to confirmed $\gamma$-ray-emitting NLSy1 we would like to conduct a similar investigation for jetted radio-loud NLSy1 not showing strong evidence of $\gamma$-ray activity yet for a direct comparison.

The proposed project has strong synergies with several facilities, operating from radio to VHE (see below). This program can be integrated with other AXIS observational campaigns on the same targets.

**Exposure time (ks):** 250 ks.

75 ks for the two $\gamma$-ray-emitting NLSy1, 50 ks for the other two radio-loud NLSy1. Based on previous *XMM-Newton* observations of the sources, and taking into account the flux variability observed in these sources, the observation time is estimated to detect the different features discussed above at least at 3-$\sigma$ level.

**Observing description:**

Sources to be investigated

- 1H 0323+342
- PKS 2004-447
- Mrk 1239
- IRAS 17020+4544

Mrk 1239 and IRAS 17020+4544 can be changed to other two jetted NLSy1 with similar features and X-ray flux observed for other GO programs.

**Joint Observations and synergies with other observatories in the 2030s:** SKA, ngVLA, ALMA, mm-VLBI, VLBA, Vera Rubin, Nancy Grace Roman, UVEX, JWST, IXPE, XRISM, XMM-Newton, NuSTAR, NewAthena, Fermi-LAT, CTAO

**[Archival data:]** Wise (IR), SDSS, LAMOST, DESI (optical), XMM-Newton (X-rays)

**Special Requirements:** : None

*17. Fast SMBH growth in the densest regions of the z>2 universe*

**Science Area:** AGN

**First Author:** Fabio Vito (INAF-OAS Bologna, fabio.vito@inaf.it)
**Co-authors:** Stefano Marchesi (UNIBO-DIFA), Monica Isla Llave (INAF-OAS Bologna), Andrea Comastri (INAF-OAS Bologna), Paolo Tozzi (INAF-OAA), ...

**Abstract:** The Mpc-scale environment is one of the main drivers of galaxy evolution, but its impact on the supermassive black hole (SMBH) growth and evolution has been largely overlooked. A few studies have investigated the AGN content of $z > 2$ protoclusters, which are the densest regions of the high-redshift universe, where any environmental effect can be most strongly and clearly appreciated. They generally found enhanced AGN activity with respect to the field environment at similar redshifts and their descendant structures, i.e., galaxy clusters. The incidence of luminous ($L_X > 10^{45}\,\mathrm{erg\,s^{-1}}$) AGN appears particularly high in overdense regions, resulting in remarkably flat X-ray luminosity functions in the high-luminosity regime. Still, the small number of protoclusters for which the AGN population has been studied in detail, mainly via deep (i.e., 100–700 ks) Chandra observations, limits considerably our understanding of the physical link between environment ans SMBH growth. We propose to survey a large sample of $z = 2 - 4$ protoclusters with AXIS to characterize their AGN content and probe, with high statistical power, the influence of an overdense environment on SMBH growth at high redshift. Our immediate goals are 1) to constrain the AGN fraction in protoclusters, as a function of distance from the structure centers and host galaxy properties such as stellar mass, 2) to derive reliably the AGN X-ray luminosity function in protoclusters down to $Lx \approx 10^{43}\,\mathrm{erg\,s^{-1}}$, 3) to compare the outcomes with the expectations from the field environment and measure the AGN enhancement in protoclusters with unprecedented accuracy and robustness.

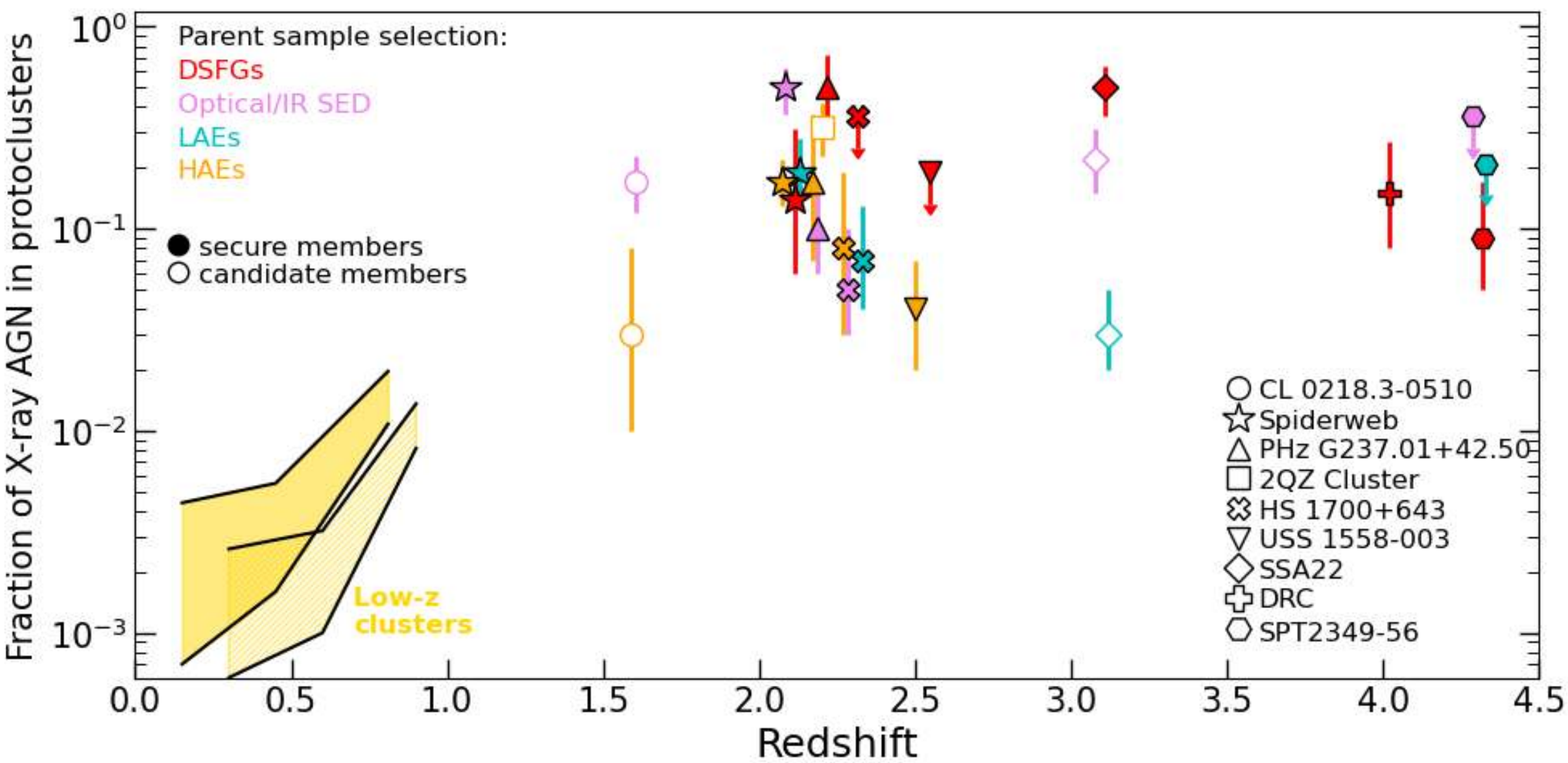

**Figure 23.** Fraction of X-ray selected AGN in a sample of $z > 2$ protoclusters as a function of the redshift. Different symbols and colors correspond to different structures and selection methods used for the parent galaxy population, as reported in the figure. Filled and empty symbols correspond to spectroscopically identified or candidate members of the protocluster, respectively. The AGN fraction in overdense regions at high redshift is significantly higher than in the field environment and low-redshift galaxy clusters (yellow stripes). From [589].

**Science:** The impact of the large-scale environment on galaxy evolution is clearly visible in $z = 0 - 2$ galaxy clusters. Their centers are dominated by massive and nearly quiescent galaxies that host the most massive SMBHs known in the local universe (e.g., [48,608]). Therefore, their progenitors, i.e. protoclusters at $z > 2$ (e.g., [436]) must be characterized by efficient galaxy and SMBH growth, which is sustained by the continuous infalling of gas from the forming cosmic web and the high rates of galaxy interactions and mergers (e.g., [263,277**?** ]. The radiative and mechanical feedback produced by AGN is then expected to play a fundamental role in regulating, and eventually hindering, further galaxy and SMBH growth during the gravitational collapse, shaping the observable properties of the resulting galaxy clusters (e.g. [177,213,225]).

While several works proved the impact of protocluster environemnt on star-formation (e.g., [114, 166,340]), only a limited number of studies focused on the ongoing SMBH growth in protoclusters (e.g., [569,589] and references therein). They generally found an enhanced incidence of AGN with respect to local galaxy clusters and the field environment at similar redshifts, suggesting that SMBH growth is favored in the densest regions of the high-redshift universe. Moreover, circumstantial evidence indicates that such enhancement is more substantial for luminous ($\log L_x > 1e45\,\mathrm{erg\,s^{-1}}$) AGN (Fig. 24), which are the likely progenitors of the $10^{10}\,\mathrm{M_\odot}$ BHs that redside in $z = 0$ clusters. Such fast accreting SMBHs will likely strongly affect the evolution of the entire structures in which they are placed via powerful AGN feedback, thus shaping the observable properties of local galaxy clusters. Still, these results are based on a small number of $z = 2 - 4$ structures which were covered by long-exposure (up to 700ks) Chandra observations.

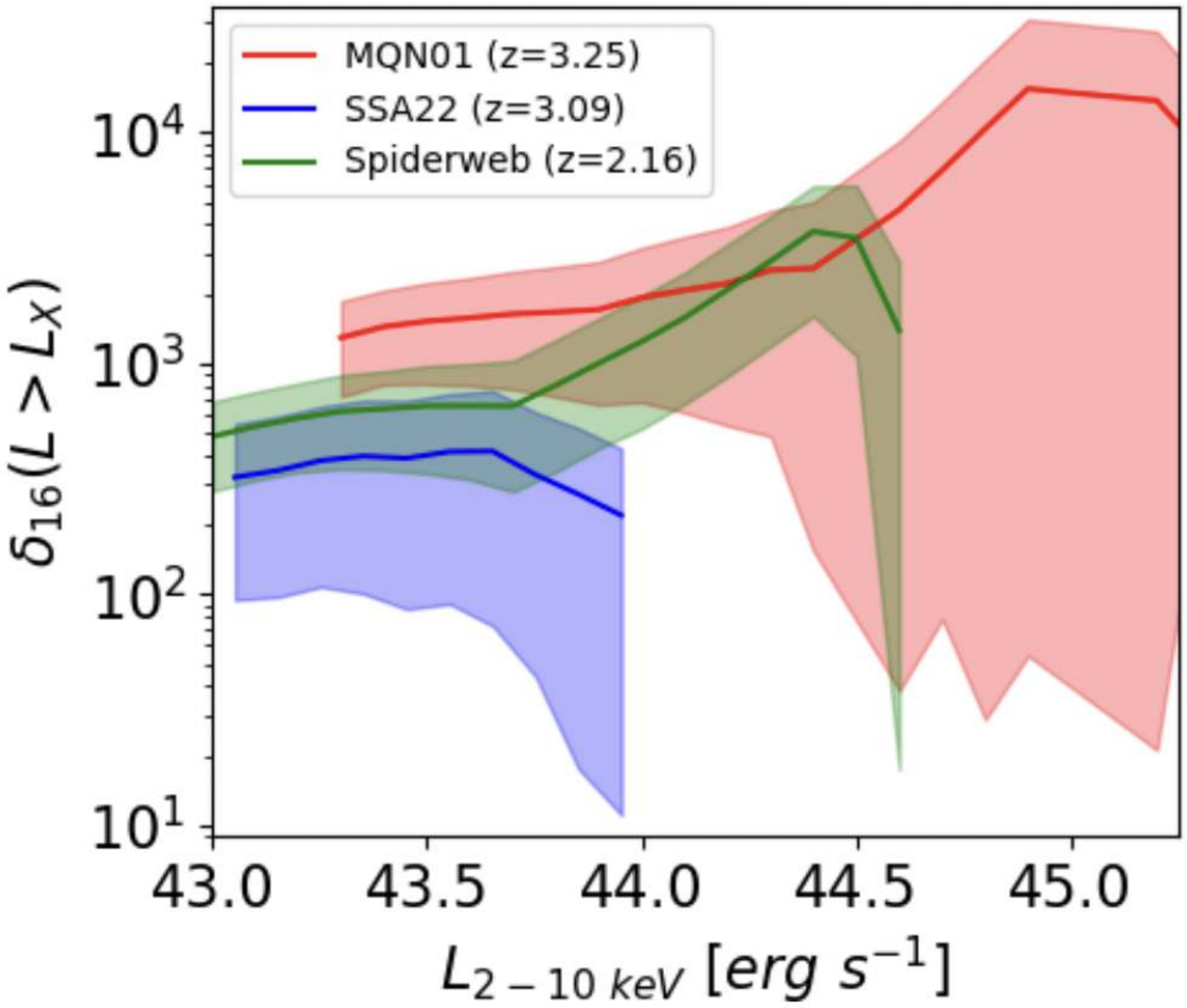

**Figure 24.** Cumulative AGN overdensity (i.e., fraction of the space density of AGN with luminosity $> L_X$ in protoclusters and the field environment) as a function of luminosity, for three protoclusters at $z \approx 2.2 - 3.3$. The AGN enhancement in protoclusters is tentatively luminosity dependent, with more powerful objects being more enhanced. From 2025A&A...694A.165T.

We propose to perform an AXIS survey of a statistically significant sample of 12 protoclusters at $z = 2 - 4$ 1) to constrain the AGN incidence in high-redshift protoclusters as a function of redshift, structure mass and overdensity, distance from the structure center, 2) to measure the AGN XLF in protoclusters and its possible cosmic evolution, 3) to compare the results with expectations from the field environment, using control samples of galaxies and AGN included in the AXIS deep and medium surveys. This program will more than double the number of protoclusters for which such an investigation is possible (e.g., Fig. 23), allowing us to control for quantities such as the protocluster selection methods and the overdensity levels of these structures.

**Exposure time (ks):** 435ks

**Observing description:**

We aim to observe 3 protoclusters in 4 bins of redshift over the range z=2-4, for a total number of 12 targets. This will increase the current number of protoclusters in that redshift range with sensitive X-ray coverage by a factor of $gtrsim$2, allowing us to explore dependencies with secondary quantities, such as redshift, structure overdensity, and selection method. The targets will be collected from already known protoclusters, and structures that will be discovered in the next 5–10 years with state-of-the-art and future facilities such as SPT, Euclid, Rubin, Subaru-PSF, and Roman. We will prioritize protoclusters with the largest number of members, for which the environmental effects are expected to be more clearly visible and that will return the highest number of AGN, as well as those with the most complete multiwavelength coverage, to enable characterization of the host galaxies via, e.g., spectral energy distribution fitting.

We aim to detect AGN in each protocluster down to log$L_X \approx 44$, even if obscured by Compton-thick material. The same observations will detect AGN down to log$L_X \approx 43$ if mildly obscured (log$N_H$=22). This is required to sample with high completeness the AGN population that contributes the most to the total accretion power in protoclusters, and obtain meaningful constraints on the AGN fraction and XLF. We used XSPEC spectral simulations, assuming the MYtorus model (ref) with the most up-to-date AXIS response files to compute the exposure time required to detect 5 net counts in the $0.3 - 10$ keV band at the desired luminosity limits. We request exposure times of 27 ks, 34 ks, 40 ks, and 44 ks for each protocluster in the $z = 2 - 2.5$, $2.5 - 3$, $3 - 3.5$, and $3.5 - 4$ redshift bins, respectively. Therefore, the total requested exposure time is 435 ks. Remarkably, that exposure time is required by Chandra to observe at most 2–3 protoclusters down to similar sensitivity.

**Joint Observations and synergies with other observatories in the 2030s:** Joint observations with optical, IR, and millimeter facilities would support this program by identifying and spectroscopically confirming protocluster galaxy members. Synergies are foreseeable with all of the state-of-the-art and future observatories working at such wavelengths, e.g., ALMA, JWST, Roman, ELT, VLT, SKA.

**Special Requirements:** None.

*18. Relativistic jets from quasars in the early Universe*

**Science Area: AGN, Jets, Quasars, High-redshift**
**First Author:** Luca Ighina (CfA, Harvard & Smithsonian)
**Co-authors:** Alessandro Caccianiga (INAF-Brera), Thomas Connor (CfA, Harvard & Smithsonian), Eileen T. Meyer (University of Maryland), Alberto Moretti(INAF-Brera), Eric Perlman (Florida Institute of Technology)

**Abstract:**
With this GO program, we aim at studying the X-ray properties of high-$z$ ($z \gtrsim 6$) jetted quasars. Given its unique combination of collective area and angular resolution, AXIS is the ideal telescope for studying the emission and physical properties of both the core and extended jet components in these systems. By targeting the still unexplored $z > 6$ jetted quasar population, we can put the strongest constraints on the observed redshift evolution of their high-energy properties. These observations will be crucial in understanding the impact of relativistic jets in the context of supermassive black hole formation and growth, as well as AGN-galaxy feedback in the early Universe.

**Science:**

## Scientific Background

The study of quasars at $z > 6$ is one of the most important ways to investigate the evolution of supermassive black holes (SMBHs) and their host galaxies in the primordial Universe. The observation of very massive BH ($\sim 10^{8-9}$ $M_{\odot}$; e.g. [601]) at high redshift challenges our understanding of BH formation and evolution in the early Universe (e.g. [593]). While the presence of an already massive seed BH ($\sim 10^{4-6}$ $M_{\odot}$; [51]) or an accretion rate above the Eddington limit (e.g. [361]) is typically required, radio/jetted quasars can also play a crucial role in providing a large mass build-up in such a short time. These systems are characterized by the presence of two, bipolar jets of relativistic material launched close to the central BH and extending up to Mpc-scales (e.g. [389]).

The presence of jets can enhance the accretion onto the central SMBH by converting part of the accreting material into jet kinetic power instead of producing radiation (e.g. [290]), which would limit the accretion of additional material. Since relativistic jets can carry away a substantial amount of kinetic power but very little mass, **jetted quasars can, in fact, accrete more material compared to QSOs with a similar luminosity but without relativistic jets (e.g. [121])**. The recent discovery of many high-$z$ blazars (i.e., quasars with a jet oriented close to our line of sight; e.g. [55]) implies the existence of a large population of jetted quasars at the same redshifts (e.g. [519]). If confirmed, the rapid increase of the fraction of jetted quasars in the most massive systems at $z > 6$ would support the jet-enhanced accretion scenario (e.g. [222]). At the same time, relativistic jets are often invoked as a major cause of galaxy feedback to explain the large number of massive quenched galaxies at $z > 3$, especially after the multiple recent discoveries with JWST (e.g. [96]). Indeed, the large mechanical energy of jets can remove and/or heat the gas in the interstellar medium and, consequently, hamper star formation.

The electrons within the jets interact with the CMB photons through an Inverse Compton process (IC/CMB), producing X-ray radiation. Although in the local Universe the amount of radiation produced by this process is still debated (see e.g. [408]), due to the CMB energy density increase, $\propto (1+z)^4$ (e.g. [521]), this interaction dominates the X-ray emission produced by extended regions of the jets (several kilo-parsecs away from the core, see e.g. [271]). Therefore, by modeling the radio+X-ray spectral energy distribution (SED) of kpc-scale relativistic jets can be used to estimate the power they can carry and release in the surrounding medium even at high redshift (e.g. [612]) independently of other methods (such as the

detection of cavities, for example, e.g. [179]). However, resolving both the radio and X-ray emission of kpc-scale jets at high-$z$ is extremely challenging, and only four X-ray jets have been detected at $z > 4$ so far ([401]).

In addition, recent works found an evolution of the core X-ray emission of jetted quasars as a function of redshift, with $z > 4$ sources typically having a stronger and softer X-ray emission compared to $z < 2$ radio-bright quasars (see [634]). The origin of this evolution is still unclear. While the IC/CMB interaction (on scales not resolved even with *Chandra*) could, in principle, result in an increment of the overall X-ray luminosity, the steepening of the average spectral index might imply be that most of the very high-$z$ jetted quasars are accreting faster than their low-$z$ counterparts. Indeed, for very high-accretion rates, potentially above the Eddington limit, we expect the X-ray radiation produced by the accretion process to be steep ($\Gamma_X \gtrsim 2.5$; see e.g. [439]). Once again, the number of radio quasars at $z > 6$ with dedicated X-ray observations is still too limited and often not deep enough to constrain the photon index accurately ([634]) to perform detailed statistical analyses on their high-energy properties.

In this context, **the observation of $z > 6$ radio quasars is extremely valuable since they provide the strongest constraints on any redshift evolution.** In particular, X-ray observations play a crucial role in understanding the impact of relativistic jets on the host galaxy and the accretion process. With this proposal, we aim to target all $z > 6$ jetted quasars currently known to constrain both their core and extended emission.

## Aim of the observations

We propose AXIS observations of $z > 6$ radio-bright, jetted quasars to constrain their X-ray emission on multiple scales. The goals we expect to achieve can be summarized as follows:

- **Constrain the X-ray emission in high-$z$ jetted systems**. Several studies have recently reported an increase in overall X-ray emission in radio-bright, jetted quasars as a function of redshift. This trend can either be associated with an evolution of the emission produced by the relativistic jets (IC/CMB; e.g. [631]) or with different accretion conditions at these high redshifts ([121]). The characterization of this evolution is crucial for understanding the role of relativistic jets in the observed properties of $z > 6$ SMBHs.
- **Identify blazars in the primordial Universe**. Blazars, that is, quasars with jets oriented close to the line of sight, are characterized by a strong and hard X-ray emission and can therefore be identified with X-ray observations. In particular, at $z > 6$ AXIS is sensitive to the $\gtrsim$2–70 keV radiation, where we expect the jet to dominate the energy output with respect to X-ray corona. By modeling their high-energy emission, we can estimate critical physical parameters, such as the viewing angle and bulk Lorentz factor ($\Gamma$; see Fig. 25 and [221]). From a well-defined sample of blazars confirmed through X-ray observations we can then estimate the total number of jetted quasars with similar properties and at the same redshift, but with a jet oriented in different direction: $N_{\rm tot} \approx N_{\rm blazar} \times 2\Gamma^2$ ([222]). This estimate is crucial to constrain the number density and relative abundance of jetted quasars in the early Universe.
- **Characterize the high-energy emission of high-$z$ relativistic jets at kpc scales.** The angular resolution and the sensitivity of AXIS make it a unique telescope to resolve and study the extended high-energy emission from relativistic jets. At these redshifts and scales (kpc) the X-ray emission is dominated by the IC/CMB-jet interaction (see Fig. 25 and [271]). Modeling the multiwavelength spectral energy distribution of these systems will enable us to estimate the power relativistic jets carry up to very large scales and that they can release in the surrounding medium at high-$z$ (e.g. [612]). Moreover, we can also extrapolate the expected X-ray radiation from IC/CMB in similar, lower-redshift jets, where the origin of the high-energy emission is still debated (e.g., [80]). Moreover,

as a bonus result, the large FoV and the relatively constant angular resolution across it will allow for the serendipitously detection of extended X-ray also at lower redshift (see e.g. [536] for an example at $z \sim 2$).

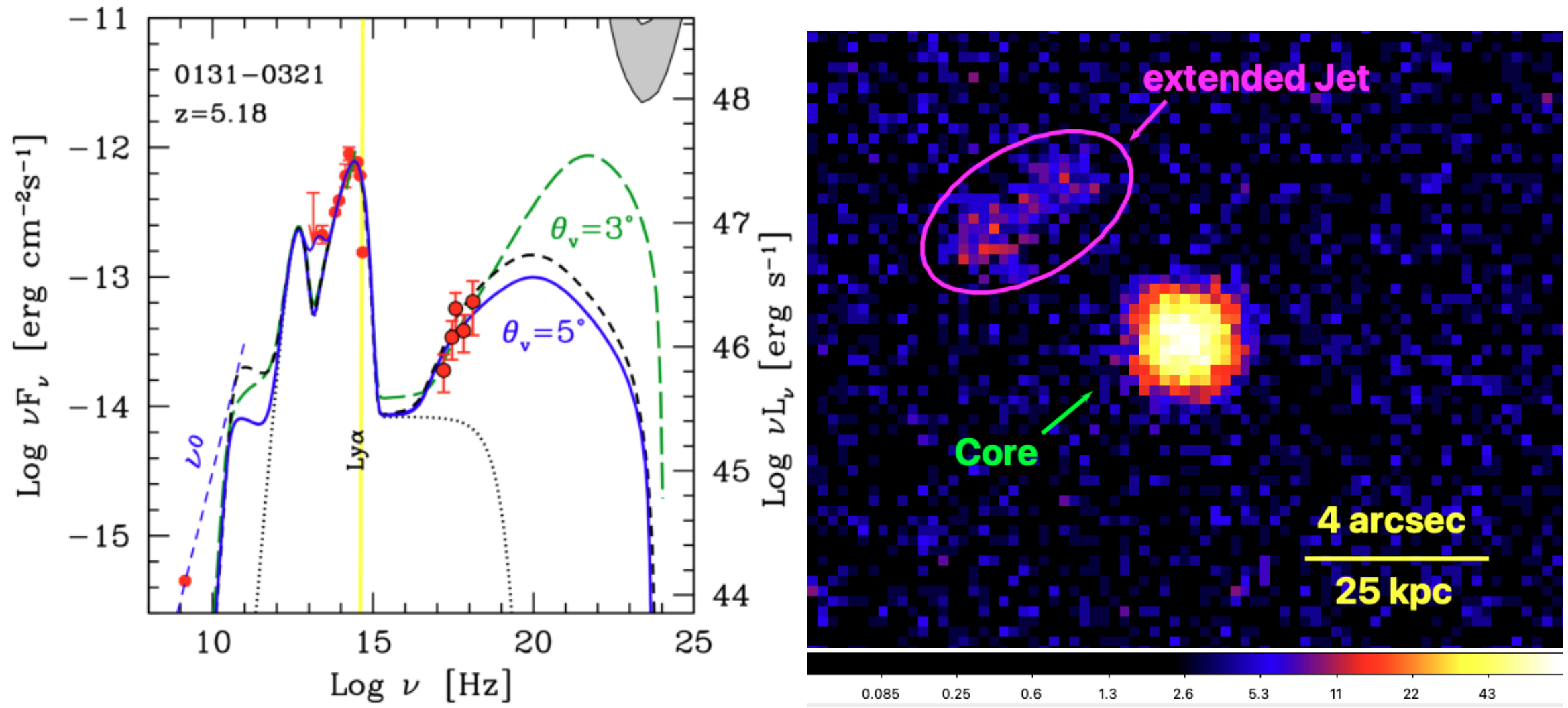

**Figure 25. Left panel:** Example of the multiwavelength SED of a $z = 5.18$ quasar identified as a blazar through the modeling of its X-ray emission. Adapted from [220]. **Right panel:** AXIS simulation of a 100 ksec exposure of the only $z > 6$ extended X-ray jet currently known (see [271]). As is clear from the image, AXIS can easily detect extended components in high-$z$ jetted systems, with about 400 net photons. As a reference, this is a factor $\sim 20$ larger than the available *Chandra* observations.

**Exposure time (ks):** 1MS, 60-80 ksec per target.
**Observing description:** We are targeting all the $z > 6$ jetted quasars, which, to date, are $\sim 12$.
**Joint Observations and synergies with other observatories in the 2030s:** VLA, ngVLA, MeerKAT, SKA, JWST, EUCLID
**Special Requirements:** None

*19. X-ray bubbles and AGN energy*

**Science Area:**

**First Author:** Weiguang Cui (Universidad Autónoma de Madrid, weiguang.cui@uam.es)
**Co-authors:** Fred Jennings (University of Edinburgh)

**Abstract:**

How galaxy quenching is a key question in current galaxy formation models. Active Galactic Nuclei (AGN) feedback is believed to play a crucial role in regulating star formation within galaxies by injecting energy into the intracluster medium (ICM) through jets, thereby creating buoyant X-ray cavities. However, due to the uncertainty in AGN energy output and the unknown mechanisms behind the launching of AGN winds, it is challenging to model this feedback properly in numerical simulations. Therefore, these cavities, as tracers of AGN feedback, are key to understanding the physical model behind them. However, their detailed morphology, energetics, and interaction with the ICM remain poorly constrained. The Advanced X-ray Imaging Satellite (AXIS), with its unprecedented combination of high angular resolution 1.5'' on-axis, large effective area, and wide field of view, offers a transformative opportunity to study X-ray cavities in galaxy clusters with unmatched precision. We propose a survey of a sample of around 15 nearby (z < 0.1) galaxy clusters to map cavity systems, measure their thermodynamic properties, and quantify their impact on ICM heating. This study will resolve the physics of jet-CGM interactions, constrain AGN feedback efficiency, and refine models for baryon processes.

**Science:**

AGN-driven jets inflate cavities in the hot CGM, redistributing energy and suppressing runaway cooling. Cavity sizes, ages, and surrounding pressure profiles encode the energy budget of feedback, yet current data lack the resolution and sensitivity to probe cavity edges, faint structures, or older, faded cavities at larger radii.

The key aims we are trying to achieve with this proposal are:

- to estimate the AGN energy from detected X-ray cavities.

  Following the calculation from [282], we can estimate the total energy of each detected X-ray cavity. Furthermore, with the AXIS's high-sensitivity spectra measurement, we can map out the temperature structure around the cavity. Using its line broadening, we will be able to get its LOS velocity information. Both of which will provide accurate results and help with the last item.
- to estimate the AGN frequency through these detections in different galaxy groups and clusters.

  With the identification of multiple cavity events, through the measurement of $t_{buoyang}$, we can estimate the jet event frequency, which allows us to investigate the AGN duty cycle in detail.
- to track the dynamic evolution of cavities, and understand their disruption/mixing with the CGM/ICM.

  Through the cavity distance to the galaxy/BH center, we can have a rough time evolution, which will allow us to build a complete picture from its launch, to dissolve into the CGM fully. With the spectra measurement, we can calculate the Mach number, which helps us to measure the dissipation rate.

AXIS's high spatial resolution is excellent for these tasks because it can resolve fine structures (e.g., cavity rims, filaments) and detect smaller/older cavities, while its large effective area enables robust spectral mapping of temperature and pressure gradients.

**Observing description:**

To achieve the goals outlined in this proposal, we will conduct high-resolution X-ray imaging and spectroscopy of a sample of nearby AGN and galaxy clusters known to harbor X-ray cavities. Observations will target the central regions of these systems, focusing on identifying the cavity boundaries, analyzing

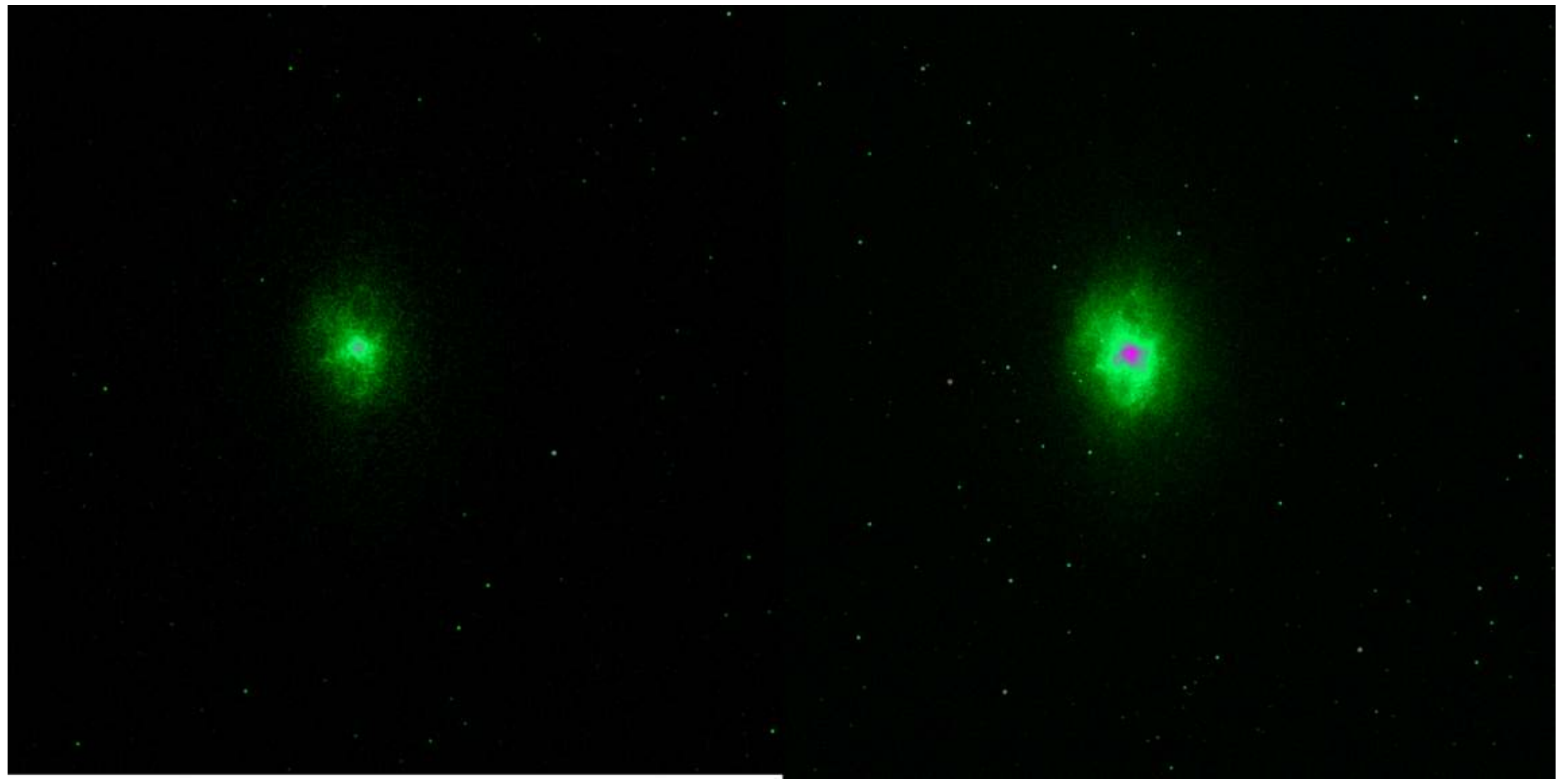

**Figure 26.** Mock AXIS observations of a galaxy group of mass, $M_{500} = 10^{13.8} M_{\odot}$, selected from the HYENAS project [128], at z=0.1 with clear X-ray cavities. Using both 10k exposure time (left) and 100k (right), we can clearly see the features of X-ray cavities in the simulated galaxy groups due to their recent AGN feedback.

the X-ray emission from the surrounding hot gas, and measuring the spectral properties of the cavity regions.

As shown in the Figure 26, a 10ks exposure time is enough to detect the X-ray cavities in massive galaxy groups at z=0.1. However, a longer exposure time (100ks shown on the right panel of Figure 26) is required to detect these small cavities in the central region of the galaxy group/cluster.

**Joint Observations and synergies with other observatories in the 2030s:**

**Exposure time (ks):** 1000.
**Special Requirements:** Nonw

*20. Accretion and ejection in radio galaxies from low to high regimes: radio-to-VHE synergies*

**Science Area: Active Galactic Nuclei**

**First Author:** Giulia Migliori (INAF-IRA Bologna; giulia.migliori@inaf.it), Eleonora Torresi (INAF-OAS Bologna; eleonora.torresi@inaf.it), Ranieri D. Baldi (INAF-IRA Bologna; ranieri.baldi@inaf.it)
**Co-authors:** Stefano Marchesi (University of Bologna; INAF-OAS Bologna), Ettore Bronzini (University of Bologna; INAF-OAS Bologna), Paola Grandi (INAF-OAS Bologna)

**Abstract:** Over the past two decades, our understanding of the connection between accretion and ejection phenomena in jetted extragalactic radio sources has undergone a profound transformation. Two populations, named low-excitation (LERGs) and high-excitation radio galaxies (HERGs), associated with radiatively inefficient and efficient accretion systems, respectively, have been primarily investigated in relation with their jet characteristics, and there is now growing evidence that the launch of a relativistic jet is intimately related to the accretion process at work in the central engine. We propose using AXIS to investigate the connection between accretion and ejection in powerful radio galaxies (HERGs), leveraging its superior X-ray sensitivity and spatial resolution. Indeed, X-ray observations are crucial for probing the innermost accretion regions and studying jet-disk interactions. Another major recent breakthrough has come from the latest generation of HE-to-VHE observatories, which have established new classes of gamma-ray-emitting radio-loud AGN, including young radio sources and low-power radio sources. The latter are particularly interesting because they are the most numerous population in the local universe, and one of them has even been detected at $\sim$20 TeV by LHAASO, challenging our current understanding of radiative processes and particle acceleration mechanisms, relevant in both low and high regimes of accretion, even at low regimes. To achieve a comprehensive understanding of the accretion-ejection link, a multiwavelength approach is essential. In the coming years, we will benefit from the synergy of wide-field, highly sensitive instruments such as SKA, LSST, and CTAO. In this context, AXIS, with its unique combination of angular resolution and sensitivity, will play a crucial role in advancing our exploration of the X-ray band emitting central engines to probe the accretion disk and jet base. In this GO, we will focus on the AXIS observations of targets in the local Universe to unveil the accretion-ejection physics involved in high and low-power radio galaxies in light of their possible forthcoming detections at very high energies.

**Science:**

Radio galaxies represent unique laboratories for studying phenomena related to the accretion of the SMBH, as well as purely non-thermal processes associated with jet emission. The large inclination angle of their jets relative to the observer's l.o.s. ($\theta$ >10 deg) facilitates the observation of radiative signatures of both outflowing and accreting matter at comparable levels [300,577]. In addition to the well-known radio classification into edge-darkened FRI and edge-brightened FRII [186], radio galaxies are also divided into efficiently-accreting high-excitation radio galaxies (HERGs) and inefficiently-accreting low-excitation radio galaxies (LERGs), based on their optical spectra [90,260,278,329], which represent the excitation level of the gas in the narrow-line region and thus reflect the efficiency of the accretion flow. Recently, a new class of radio sources, named FR0s, has been identified Baldi et al. [35]. FR0s are compact, low-power LERGs and may represent the ordinary picture of the radio-loud AGN phenomenology in the local Universe [32].

In this scientific case, we focus on the physics of the jet and its relationship with accretion in radio galaxies, aiming to demonstrate the role of AXIS in characterizing the ejection (in synergy with the GeV/TeV bands) and in probing the accretion/ejection connection (in synergy with the otpical and radio band). AXIS's higher sensitivity to soft and hard X-rays, along with its larger effective area compared to both XMM-Newton and Chandra, and its spatial resolution similar to the on-axis resolution of Chandra, can provide a more stringent view of accretion and ejection mechanisms in radio galaxies.

*1. AXIS and the study of low-power AGN*

AXIS will play a crucial role in the study of low-power radio-loud AGN. With this term we refer to radio galaxies with low accretion rates and bolometric outputs ($L_{X-ray} < 10^{42}$ erg s$^{-1}$), which generally show compact jets and accrete inefficiently (LERG). As such, this definition includes CSOs, FR0s, faint FRIs, and other classes. Interest in these objects arises from their large numbers in the local Universe; for instance, FR0s constitute the bulk (80%) of the radio-loud population at z<0.05 [34,36]. Moreover, these objects are particularly intriguing because, thanks to the Fermi satellite, we know they can emit in the gamma-ray band — examples include PKS1718-649 [413], NGC3894 [483], and Tol1326-379 [236]. In recent years, significant effort has been dedicated to understanding the acceleration mechanism(s) responsible for emission at these energies. The picture has become even more complex following LHAASO's detection of the low-power radio galaxy NGC 4278 at 20 TeV [92], opening up new interpretative scenarios. A powerful approach to study these sources is the construction and modeling of their multi-frequency spectral energy distributions (SED). In particular, the X-ray band, which lies between the synchrotron and inverse Compton regimes (in a leptonic scenario), is crucial for distinguishing between different models. Additionally, studying the SED of low-power AGN will allow us to make predictions about their detectability at very-high energies and potentially estimate their contribution to the gamma/TeV extragalactic background, as has already been done for FR0s [545]. These sources are generally very nearby; therefore, a Chandra-like resolution, such as that of AXIS, is essential for separating genuine AGN emission from that of galactic components. The latter becomes particularly relevant in nearby sources, and the only way to properly account for it is through spatially resolved studies, like the one conducted by [456] on NGC 4278.

For statistical population studies, AXIS will revolutionize the view of low-power AGN with X-ray (2-10 keV) luminosity $< 10^{42}$ erg s$^{-1}$ in the local Universe, z<0.3, with the following proposed surveys:

- Deep Field: a single pointing with an exposure of 7 Ms (area 450 arcmin$^2$) will detect $\sim$ 120 low-power AGN;
- Intermediate Field: Mosaic-mode observations of $\sim$ 335 ks on a field of 2×2 deg$^2$ will detect $\sim$2500 low-power AGN.

More than 10% of such an X-ray AGN population, i.e., $\sim$300 sources, is expected to be consistent with being low-power *radio-loud* AGN. The characterization of the X-ray properties of this population, combined with observations from forthcoming large-area, high-sensitivity radio and optical/IR surveys (e.g., SKA, ngVLA, LSST, Roman), will shed light on the accretion-ejection mechanism operating at the center of radio-loud AGN. In addition, based on the Fermi catalog, we expect that at least a few percent of these targets will also be detected in the GeV band, providing a multiband study of their SED for a more complete understanding of the jet launching and acceleration/deceleration processes across a broad range of frequencies.

*2. AXIS and the accretion/ejection link*

X-rays are the only type of radiation capable of penetrating down to the AGN central engine, enabling the estimation of key parameters such as the spectral shape of the continuum and the characterization of reprocessing structures, including the iron line and the reflection (in case of efficiently accreting systems – HERGs). X-rays have also consistently played a crucial role in the multi-wavelength (MW) context. For example, in two well-known sources, e.g., 3C 111 and 3C 120 [107,108,390], dips observed in the X-ray flux followed by the ejection of superluminal knots in VLBI images have demonstrated a clear connection between the radiative state near the black hole (probed by X-rays) and events occurring in the jet (seen in radio maps), as typically observed in black hole X-ray binaries. 3C 111 and 3C 120 remain, to date, the only sources where the jet-disk connection has been observed. Interestingly, they are also two

gamma-ray sources detected by the Fermi satellite during its first months of operations [4]. For 3C 111, [237] discovered that the gamma-ray flare also occurred after the X-ray dip and coincided with the ejection of the radio knot, further demonstrating the strong connection between these different energy bands. During the X-ray dip, the Suzaku spectrum of 3C 111 showed an absorption line attributed to a fast disk wind [39], which disappeared once the X-ray flux increased again. If confirmed, this could explain the transient nature of disk winds in radio galaxies as a signature of an ongoing disk-jet perturbation. With this in mind, we propose monitoring radio galaxies similar to 3C 111 and 3C 120 with AXIS—specifically, face-on HERGs (also known as Broad Line Radio Galaxies, BLRGs) that have been detected by the Fermi satellite to date (about 15 sources). The X-ray light curve will provide insight into the temporal evolution of the X-ray flux, highlighting the presence of potential dips, while the spectra will provide information on the spectral shape of the continuum and the iron line. The detection of the iron line is a crucial signature indicating that X-rays are produced in the accretion disk regime. Finally, we will synergistically use high-resolution radio maps to monitor the jet's behavior alongside X-ray data, as well as data from the Fermi survey. This will allow us to test the jet-disk coupling in other sources by leveraging the capabilities of AXIS.

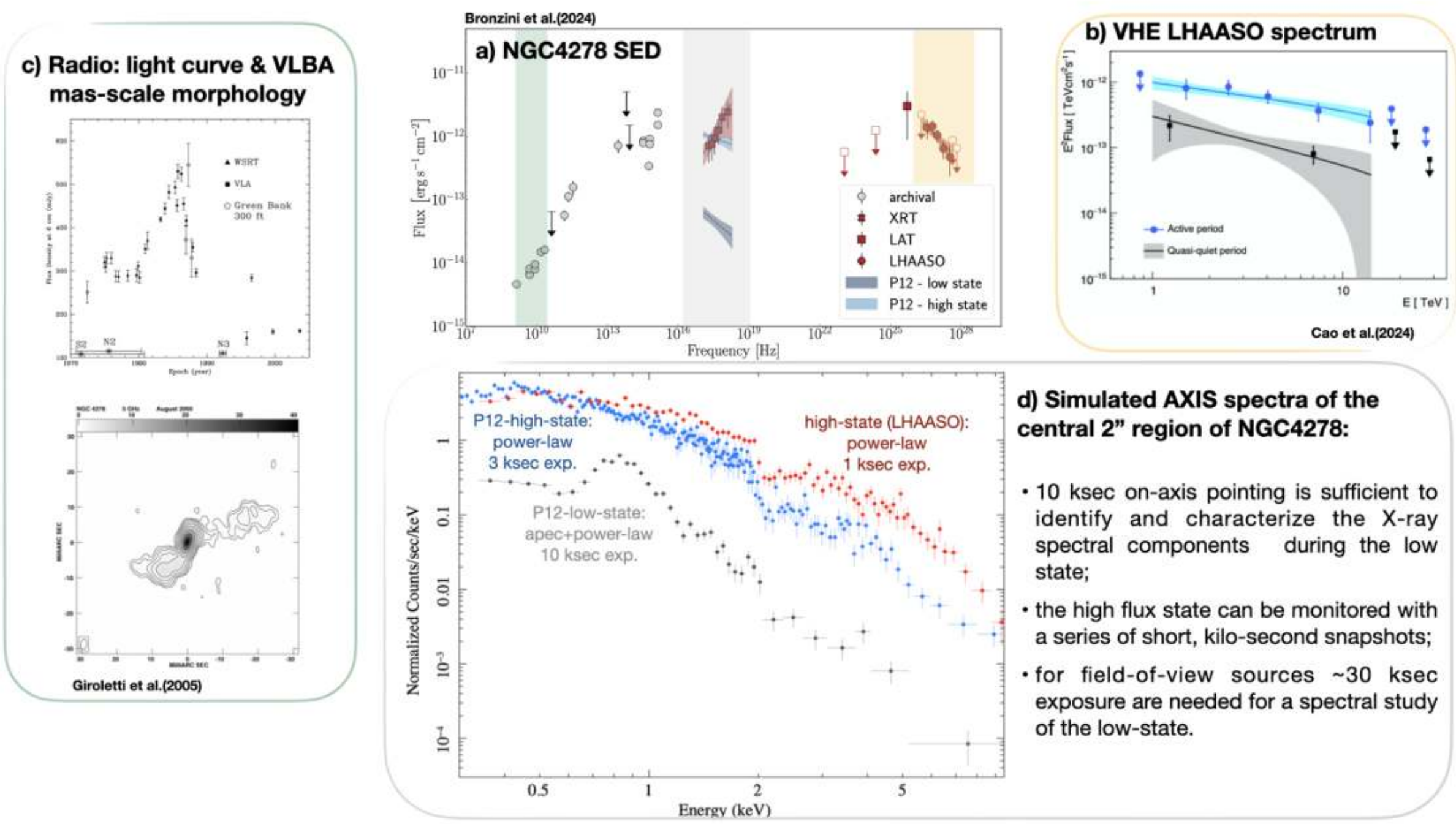

**Figure 27.** Simulated AXIS spectra for the low-power AGN NGC4278. Panels **(a), (b), (c)** are adapted from [85], [92],[228], respectively.

**Exposure time (ks):** Archival Deep + 1-10 ksec per target on-axis, 30 ksec off-axis.

**Observing description:**

*Case study 1* – To estimate the necessary exposure time we consider the case of a nearby low-power radio AGN (e.g., NGC4278, $D_L$=16.4 Mpc, $M_{BH}$=3×10$^8$ $M_\odot$) with a detection in the HE-to-VHE regime where X-ray radiation is likely to trace the inefficient accretion emission in the low state, and the emergence of the non-thermal, jet emission in the high state. An X-ray monitoring is key to constraining the processes producing the high-energy emission and the shape of the electron energy distribution (hence indicating the acceleration mechanism). Additional information can be obtained through radio (VLBI) observations,

which can probe the parsec-scale jet morphology and track the ejection of new jet components. Our simulations are shown in Fig. 27. Pointed observations will enable efficient spectral characterization and monitoring. Relevant for the surveys, a modest (factor 3) increase in the exposure time will be sufficient to achieve similar results for off-axis targets.

*Case study 2* – Simulations of BLRGs (3C 111-like) show that 10 ksec on-axis(/30 ksec off-axis) exposures are needed to constrain the spectral model of the continuum and significantly detect spectral features such as the iron line.

**Joint Observations and synergies with other observatories in the 2030s:** SKA, eMERLIN, ngVLA, ALMA, mm-VLBI, LSST, Roman, IXPE, XRISM, XMM, NewAthena, Fermi, CTAO.

**Special Requirements:** None

*21. X-ray winds and outflows from active galactic nuclei*

**Science Area:** Active Galactic Nuclei

**First Author: Roberto Serafinelli** (Universidad Diego Portales; roberto.serafinelli@mail.udp.cl)
**Co-authors: Alessia Tortosa** (INAF-OAR; alessia.tortosa@inaf.it), **Francesco Tombesi** (Università di Roma Tor Vergata; francesco.tombesi@roma2.infn.it), **Stefano Bianchi** (Università Roma Tre; stefano.bianchi@uniroma3.it), **Pierre-Olivier Petrucci** (Université Grenoble Alpes; pierre-olivier.petrucci@univ-grenoble-alpes.fr), **Yerong Xu** (ICE-CSIC; yerong.xu@ice.csic.es)

**Abstract:**

AXIS can be used for deep observations of a selected AGN sample with known warm absorbers (WA) and ultra-fast outflows (UFO) or to search for such features. Both are tracers of ionized absorbers in the surrounding environment of AGN, characterized by absorption features at $\sim 1$ keV in the case of WAs and at $7-8$ keV in the case of UFOs. AXIS will allow measurements of absorption features from highly ionized elements (e.g., Fe, O, Ne), providing direct constraints on the ionization state, column density, and velocity structure of these outflows. The soft X-ray sensitivity of AXIS enables more accurate probing of warm absorbers than previous missions, aiding in understanding their connection to the AGN environment. With an effective area of 4200 cm$^2$ at 1 keV (on-axis), about $\sim 3$ times larger than XMM-Newton, AXIS is particularly powerful for studying warm absorbers. This will open the window for fainter, higher-redshift AGN, too faint to be studied in detail with the currently available telescopes. Its effective area at 6 keV is comparable to XMM-Newton, enabling UFO studies at similar levels to those observed with XMM.

By leveraging the unique capabilities of AXIS, we will investigate the variability of these outflows on multiple timescales, exploring their response to changes in AGN luminosity and the physical conditions that drive their evolution. Combining AXIS data with multiwavelength observations (e.g., JWST, VLT, ELT, LBT) will help study the connection between X-ray outflows and ionized winds in the optical and infrared, providing insights into the role of AGN-driven winds in galaxy evolution.

AXIS will provide crucial constraints on the ionization and kinematic structure of these winds. With a five-year prime mission in low-Earth orbit, AXIS will push the boundaries of AGN outflow science, uncovering the physical drivers behind powerful winds in AGN, such as distinguishing between magnetic and radiative driving at launch.

**Science:**

**Warm Absorbers and Ultra-Fast Outflows:** Ionized outflows in active galactic nuclei (AGNs) are crucial for understanding black hole accretion and feedback. These winds, detected in X-ray spectra as blue-shifted absorption features, come in two primary forms: warm absorbers [WA, e.g., 73] and ultra-fast outflows [UFOs, e.g., 559]. Warm absorbers, present in $\sim 50\%$ of Seyfert 1 galaxies [e.g., 327], have velocities of a few hundred to a few thousand km s$^{-1}$ and are characterized by absorption from highly ionized species such as O VII, O VIII, and Fe L or Fe M complex absorption features. They are thought to originate at sub-parsec to parsec scales, providing insight into the connection between the accretion disk, the broad-line region, and larger-scale AGN-driven winds.

UFOs, on the other hand, are detected through highly blueshifted Fe XXV and Fe XXVI absorption lines at velocities of $v \gtrsim 0.1c$. Their extreme speeds suggest an origin close to the accretion disk, where radiation pressure and magnetic fields can accelerate gas to relativistic velocities. UFOs are believed to play a fundamental role in AGN feedback, potentially regulating star formation and black hole growth by injecting energy into the interstellar medium. However, many open questions remain regarding their launching mechanisms, variability, and connection to lower-velocity outflows, such as warm absorbers.

- **Tracking the variability of Warm Absorbers:** The ability to measure ionization state ($U$) and column density ($N_{\rm H}$) with only 1 ks AXIS exposures would represent a breakthrough in time-resolved studies of AGN outflows. As demonstrated in our simulation for NGC 4051 (Fig. 28, top left panel), AXIS

high sensitivity in the spectral energy band $E = 0.3 - 2$ keV, where warm absorbers are prominent, allows precise tracking of these parameters on short timescales, enabling robust constraints on the physical properties of warm absorbers and ultra-fast outflows. This capability far exceeds that of currently-operating X-ray telescopes such as *Chandra* and *XMM-Newton* by a factor of 10 and 3, respectively. A time-resolved approach, similar to that of [322], who used a 100-ks *XMM-Newton* observation to monitor absorber variability in response to AGN luminosity changes, could provide valuable information about the variability of warm absorbers, particularly their location. However, their study was limited by the need for coarser time binning due to lower effective area and spectral resolution.

With AXIS, a 100-ks observation of NGC 4051 could be divided into 100 independent 1-ks snapshots, providing unprecedented temporal resolution in tracking the response of ionized gas to continuum variations. This will allow us to measure recombination timescales with high accuracy, directly constraining the electron density and location of the absorbing gas. Compared to previous short-timescale studies that relied on a handful of flux states, AXIS enables continuous monitoring of outflow properties, resolving the dynamic interplay between AGN radiation and ionized gas structures. By leveraging this capability, we will determine whether outflows maintain a stable structure or undergo rapid changes, providing key insights into their role in AGN feedback.

- **Probing Entrained Ultra-Fast Outflows:** Entrained ultra-fast outflows [E-UFOs, e.g., 524] represent a newly identified phase of AGN winds, where ultra-fast outflows interact with the surrounding interstellar medium (ISM), sweeping up and accelerating ambient gas. E-UFOs are typically observed at $E \sim 1$ keV and exhibit velocities comparable to classical UFOs ($v_{\mathrm{out}}/c \sim 0.1 - 0.2c$) but with significantly lower ionization and column densities, suggesting a complex multiphase structure. These outflows provide crucial insight into how AGN winds couple with their host galaxies; however, current X-ray missions struggle to precisely characterize their properties in a large sample of AGN due to limited spectral resolution and sensitivity in the soft X-ray band.

  To assess AXIS capabilities in detecting and characterizing E-UFOs, we performed simulations in which we injected an E-UFO component into the spectrum of NGC 4051. Our 10*ks* AXIS simulations demonstrate that the velocity of the E-UFO can be constrained to within 0.001*c* (see Fig. 28, top right panel), an order of magnitude improvement over *XMM-Newton* estimates. This precision enables robust measurements of the wind's kinematics, breaking degeneracies in spectral modeling and providing direct constraints on its acceleration mechanisms. Additionally, the superior soft X-ray sensitivity of AXIS allows for the detection of lower column density absorbers down to $N_{\mathrm{H}} \sim 10^{19-20}$ cm$^{-2}$, offering a more complete picture of the multiphase structure of outflows.

  By leveraging AXIS's high effective area, we will systematically investigate E-UFOs across a well-selected AGN sample, tracking their variability on short timescales. This will determine whether E-UFOs are transient or persistent features and whether their properties evolve in response to changes in AGN luminosity. Such an approach will enable a direct comparison with theoretical models of AGN wind-ISM interactions, shedding light on the efficiency of AGN feedback in regulating black hole growth and star formation. With AXIS, we will achieve an unprecedented level of detail in characterizing E-UFOs, providing a critical step toward understanding their role in shaping the evolution of AGN and their host galaxies.

- **Probing Ultra-Fast Outflows for local AGN:** While AXIS significantly improves soft X-ray sensitivity, its effective area in the 6˘10 keV band is comparable to XMM-Newton. At 6 keV, AXIS provides 830 cm$^2$ (on-axis) and 570 cm$^2$ (FoV-average), ensuring that UFOs remain detectable at the same level as XMM for local sources. This enables robust measurements of key Fe XXV–XXVI absorption features, which are crucial for constraining UFO velocities and ionization states.

  Fig. 28, lower left panel, presents residuals from a UFO injected into a 100 ks AXIS spectrum of NGC

4051, demonstrating that AXIS can detect these features with the same fidelity as XMM-Newton. This means that we can continue to detect UFOs in local sources in the Fe K band, maintaining sensitivity comparable to that of XMM-Newton in the 6˘10 keV range. While AXIS does not significantly exceed the effective area of XMM in this band, its real added value lies in ensuring continuity of high-energy spectral observations at a time when XMM-Newton is nearing the end of its operational lifetime and no other comparable facility will be available until NewAthena launches in 2037. AXIS will uniquely complement the next generation of ground- and space-based observatories across the electromagnetic spectrum (e.g., JWST, ALMA, VLT, ELT, SKA), many of which are expected to be fully operational in the 2030s. Its ability to obtain simultaneous, high-resolution X-ray spectra in the Fe K band will be critical for linking the inner, fast-moving regions traced by UFOs to larger-scale outflows observed in the optical, infrared, submillimeter, and radio wavelengths. This connection is crucial for understanding the complete energetics, acceleration mechanisms, and multiphase structure of AGN-driven winds.

With XMM-Newton nearing the end of its operational lifetime, XRISM can only observe the brightest AGN, and NewAthena is still far from its predicted launch date, AXIS will take the lead in high-energy UFO studies. It will ensure continuity in Fe K-band observations, maintaining our ability to study AGN feedback and outflow physics, and especially their variability, in the coming decades.

- **Probing Ultra-Fast Outflows at Cosmic Noon:** AXIS will also open new frontiers in the study of UFOs at Cosmic Noon ($z \sim 1-4$), the peak of AGN and galaxy co-evolution. Detecting high-velocity outflows in these distant quasars is critical for understanding SMBH feedback in the early Universe. At these redshifts the Fe XXV and XXVI will be around 1 to 3 keV [e.g., 103], depending on the redshift of the source. Therefore AXIS effective area will be much larger than both XMM and especially Chandra, with whom these kind of sources are best observed, since they are often found in gravitationally lensed systems [e.g., 106].

  We simulated a 50 ks spectrum of a bright AGN at $z = 2.2$ with a flux of $5 \times 10^{-13}$ erg cm$^{-2}$ s$^{-1}$, representative of very luminous or gravitationally lensed quasars. We injected a UFO with $\log U = 2.5$ and $v_{\rm out} = 0.2$, typical of more powerful high-redshift quasars. The resulting contour plot (Fig. 28, bottom right panel) shows that fitting the UFO with a Gaussian line yields closed $3\sigma$ confidence level contours, demonstrating the ability of AXIS to detect and characterize UFOs at these epochs.

  This capability will enable the first systematic studies of Fe K UFOs in high-redshift quasars, which are currently limited to sparse samples, helping to determine whether powerful outflows were common during early SMBH growth. AXIS will thus compare the outflow physics and incidence between local AGN and the feedback processes shaping galaxies at Cosmic Noon.

**Exposure time (ks):** 250 ks.
100 ks per local AGN (2 sources), 50 ks for bright $z \sim 2$ AGNs.

**Observing description:**

For warm absorber variability studies, the following two sources are particularly well-suited (but not the only ones), as they also intermittently host ultrafast outflows.

- NGC 4051, Narrow-Line Seyfert 1, 100ks, $z = 0.00234$
- Ark 564, Narrow-Line Seyfert 1, 100ks, $z = 0.0247$

Their varying ionized absorption properties make them ideal candidates for investigating the physical conditions driving warm absorber variability and the potential connection between WAs and UFOs.
A 100 ks observation with AXIS will provide a high signal-to-noise spectrum, enabling both the detection of

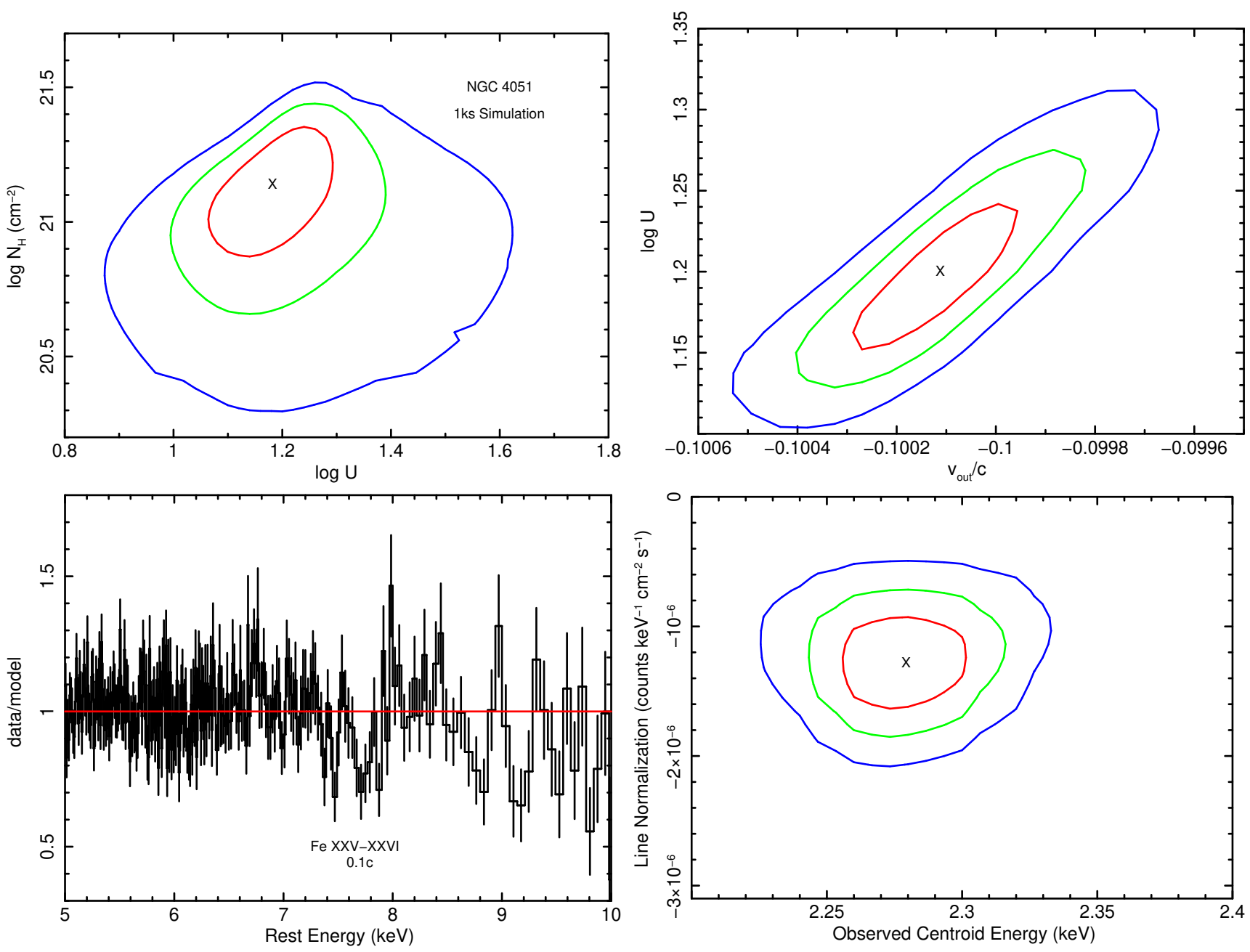

**Figure 28.** Top Left. Contour plot between $\log U$ and $\log N_{\rm H}$ in a 1 ks AXIS observation of NGC 4051. The same exposure with XMM would result in unconstrained parameters. The red, green and blue lines represent the 68% (1$\sigma$), 95% (2$\sigma$) and 99.7% (3$\sigma$) Top Right. Contour plot between $\log U$ and the outflow velocity $v_{\rm out}$ for a 10 ks observation of NGC 4051, when a E-UFO with velocity $v \sim 0.1c$ is injected. Bottom Left. Data-to-model ratio of a 100ks observation of NGC 4051 when a UFO is injected and the spectrum is fitted with a simple power law. Bottom Right. Contour plot between the normalization of the line and centroid energy when a UFO with $\log U = 2.5$ and $v_{\rm out} = 0.2c$ is fitted with an absorption Gaussian line.

a UFO, if present, and a detailed time resolved analysis of warm absorber variability. By capturing spectral changes throughout the observation, we can probe the response of the ionized absorber to fluctuations in AGN luminosity and determine key properties, such as changes in ionization state, density, and recombination timescales. This will help establish whether WAs and UFOs arise from the same outflowing gas or represent distinct components of AGN feedback.
Additionally, combining these observations with multi-epoch data from XMM-Newton and future missions will allow us to track the long-term evolution of these absorbers, further constraining their physical origin and connection to the AGN accretion process. More sources can also be proposed in future AXIS Announcements of Opportunity, expanding the sample and enabling broader statistical studies of AGN outflows across different luminosities and accretion states.
The best source for the observation of UFOs at Cosmic Noon is

- APM 08279+5255, quasar, $z = 3.911$

This quasar is particularly notable for hosting ultra-fast outflows (UFOs) with velocities reaching up to $0.6 - 0.7c$, among the fastest observed in any active galactic nucleus (AGN). These outflows have been detected through absorption features in X-ray spectra, indicating highly ionized winds originating from the accretion disk surrounding the central supermassive black hole. The presence of such high-velocity outflows suggests that magnetic acceleration mechanisms may be at play, as purely radiative driving mechanisms are limited to velocities below $v_{\rm out} = 0.3 - 0.5c$
Observations with VLT, ELT, or other state-of-the-art optical facilities will allow us to connect X-ray ionized absorbers with optically observed large-scale outflows. ALMA will provide the connection with molecular outflows
**Joint Observations and synergies with other observatories in the 2030s:** VLT, ELT, ALMA, JWST
**Special Requirements:** None.

*22. Distinguishing AGN jet vs. "central engine" spectral components*

**Science Area: Active Galactic Nuclei**
**First Author:** Elena Fedorova (INAF-OAR, olena.fedorova@inaf.it)
**Co-authors:** Milvia Capalbi (INAF-OAR, milvia.capalbi@inaf.it); Antonio del Popolo (Dipartimento di Fisica e Astronomia, University of Catania, antonino.delpopolo@unict.it)
**Abstract:** Distinguishing the spectral components of the "jet base" and the "central engine" in the X-ray to $\gamma$-ray spectra of radio loud AGN can provide us with the possibility to investigate the nature of its central regions in a more detailed way and to test the models of its structure. Here, we describe the proposed recipe for separating the "jet base" and "central engine" components in the AGN spectrum. It enables us to estimate individually the time variations of the "central engine" (i.e., accretion disc and corona) and synchrotron-self-Compton (SSC) or Inverse Compton (IC) jet emission in an RL AGN based on the radio, X-ray, and $\gamma$-ray observational data. We utilized the connections between the synchrotron radio- and X-ray SSC/IC jet spectra, their photon indices, and the dependence between the nuclear continuum and Fe-K iron fluorescent line emission near 6.4 keV to separate the nuclear and jet base contributions to the total X-ray continuum. We demonstrated the efficiency of this method on multi-mission datasets collected by XMM-Newton, Swift, Suzaku, and INTEGRAL for 3C 111 and 3C 120, and propose a sample of AGN for which the use of this method could also be promising. The spectral resolution of AXIS in the 5-7 keV range (where the Fe-K lines are typically located) appears promising for applying the second recipe at the same or even higher quality level as XMM-Newton/EPIC and Suzaku/XIS.
**Science:** The fact that active galactic nuclei (AGN) can manifest themselves as radio loud (RL) or radio quiet (RQ) is clearly known since 1989 when the criterion of radio loudness was formulated in [301]. This criterion is based on the radio-to-optical brightness relation $R = L_R/L_V$ where $L_R$ is the AGN luminosity in radio frequencies (usually at 5 GHz) and $L_V$ is its luminosity in the blue optical light band (near 689 THz) $R \leq 1$ means that the object is RQ; $R \geq 10$ corresponds to an RL AGN. $R = 1 - 10$ can be referred to as radio moderate (RM). The last class with strong variations of radio loudness from typically RL to RQ values and unstable jets becoming more common in the last decade is radio transitional (RT) [383]. RT AGN have variable spectral shape similarly to X-ray binaries [190,191]. Both objects of our investigation, 3C 111 and 3C 120, likely belong to this class rather than to RL one [356].

The modern schemes of the RL/RQ dichotomy and jet formation encompass various scenarios, depending on the hypothetical origin of the magnetic field and its localization (central or peripheral). The central BH spin, as well as other characteristics of the "central engine", significantly affect the profiles of emission lines formed in its vicinity. This opens a possibility to test such schemes of AGN "central engine" structure as the spin or gap paradigm (the model explaining the jet presence in terms of higher values of black hole spin in comparison with RQ AGN [81,210] or retrograde accretion disk [211]), and the "magnetic" scheme considering the central supermassive object as the source of (electro)magnetic field launching jets (especially for the case of "exotic" object such as wormhole/"magnetic tunnel"[134,298,299] or naked singularity [252] as its origin) using the observational data in X-rays concerning the fluorescent lines like Fe-K (near 6.4 keV) and Ni-K (near 7 keV). These lines can be found within the range of 5-8 keV, where the on-axis effective area of AXIS is comparable to or higher than that of the XMM-Newton/EPIC PN camera, and significantly higher than that of Swift/XRT, NuSTAR, and Chandra/ACIS. This makes AXIS a worthy successor for these missions, continuing the investigation of this kind.

Here we add the recent observational data by SWIFT and INTEGRAL to our previous investigation of 3C 111 and 3C 120 published in [188,189], applying the two recipes of distinguishing between the spectral components induced by jet base and accretion disk/corona proposed in [189]. 3C 111 is supposed to be an S1 type RL AGN [515] nested in the nearby elliptical broad-line radio galaxy (BLRG) at redshift z = 0.0485 [525] and the central BH mass around $2\times 10^8$ $M_\odot$ in [109]. 3C 120 is a radio transitional (RT) AGN [356]

hosted in the nearby lenticular/disc S0 BLRG at the redshift z = 0.0336 [338], with the central BH virial near 5.7 $\pm$ 2.7$\cdot 10^7$ $M_\odot$ [480]. The jet's composition is likely to be $e^- - e^+$ pair-dominated [626], with the particle dominating over the magnetic field at the jet base [625].

There are two main competing models to explain the variability of RT AGNs like 3C 120 and 3C 111: "disk-jet cycle" [356] with the jet-forming zone situated in the innermost region of AGN and "plasmoid magnetic reconnection" [531,534] with the magnetically turbulent jet-forming regions at $\approx 10^3 R_g$ away from the center. In [188] we compared the results of the spectral fitting for 3C 120 with these models and found that the jet-forming zone in 3C 120 is more plausibly situated in the central region.

**Two Ways to Distinguish between Nuclear and Jet Base Spectral Components:**Jet base of subparsec size is composed mainly of ultrarelativistic $e^- - e^+$ (leptonic) or $e^- - p$ (hadronic) plasma. Being accelerated at shock fronts inside the jet, it emits synchrotron radiation visible at a wide wavelength range from radio to UV (in some cases even to X-rays). Accelerated plasma particles are distributed over energies after power-law dependence $N(E) \propto E^{-\gamma_e}$ [326], generating the self-absorbed synchrotron emission with the spectral energy distribution $F_\nu \propto (\nu/\nu_0)^{5/2} \cdot [1 - \exp(-(\nu/\nu_0)^{-\frac{\gamma_e - 1}{2}})]$, where $\nu_0 > \nu$ is the break frequency at which an optical depth is $\tau = 1$.

For the cases $\nu >> \nu_0$ (transparent) and $\nu << \nu_0$ (absorbed), this formula can be approximated with the simple power-law with slope $\alpha = (\gamma_e - 1)/2 = \Gamma + 1$ and $\alpha = 1/2$ where $\Gamma$ is the photon index.

As was shown in [120], the jet X-ray self-Comptonized synchrotron (SSC) or inverse Compton (IC) spectra have exactly the same photon indices as the jet synchrotron emission. Thus, as the first way, we adopt the same power-law photon index value to fit the X-ray SSC/IC and radio synchrotron emission spectra of the jet base. Therefore, we use the frozen values of the photon index $\Gamma^{fr}$ for the jet base component of our X-ray spectral model, which are equal to those obtained from the radio spectrum. The Planck radio spectrum of 3C 111 was fitted with a simple power-law model with the photon index $\Gamma = 1.66 \pm 0.06$, and that of 3C 120 with the photon index $\Gamma = 1.24 \pm 0.05$.

The second proposed method is to utilize the energy fluxes in Fe K lines formed in the accretion disk of an AGN, which are thus correlated with the primary nuclear continuum of the "central engine". For the simplest situation, we presumed a linear dependence of the Fe-K$\alpha$ line flux and the primary nuclear emission flux: $F^{nucl}_{5-7} = \kappa F^{line}_{5-7}$.

**Fitting the Spectra and Separation of the Jet Base Contribution:** The model for 3–300 keV X-ray spectra of 3C 111 and 3C 120 consists of such main components: the disk/corona continuum emission represented by *pexrav* for 3C 111 [367] and *xillver-comp* model for 3C 120 [204], reflected and mildly absorbed by neutral medium, represented by the *tbabs* model including the Galactic absorption as the lowest possible value of the column density $N_H$ (of 7.4 $\times 10^{21}$ cm$^{-2}$ for 3C 111 [560] and 1.9 $\times 10^{21}$ cm$^{-2}$ for 3C 120 [607]; Fe K$\alpha$ line emission near 6.4 keV; SSC/IC emission of the jet base (broken power-law). Intercalibration constants were used in the simultaneous modeling of XMM-Newton/EPIC, Suzaku/XIS or SWIFT/XRT with INTEGRAL/ISGRI or Swift/BAT spectra, with values in the range 0.8–1.5. The errors, lower and upper limits indicated below correspond to a confidence level 90% ($\Delta\chi^2$ = 2.71). Swift/BAT and INTEGRAL spectra were treated within observational time intervals that overlapped with shorter intervals of XMM-Newton/EPIC, Suzaku/XIS, or Swift/XRT observations; they were processed together to trace the variability details. All the observational data on 3C 111, collected by XMM-Newton, Suzaku, INTEGRAL, and Swift, were divided into 15 observational periods; 12 of them were described earlier in [189]. Here we processed the last three Swift observations and consider them together with the previous ones. The fitting has been performed with frozen value $\Gamma_{fr}$ = 1.66 of the jet base power-law SSC/IC emission. During the last three observations of 3C 111, variations in both the jet and "central engine" spectra were detected. Spectrum of the jet followed the broken power-law pattern during first two of them, with two breaks with

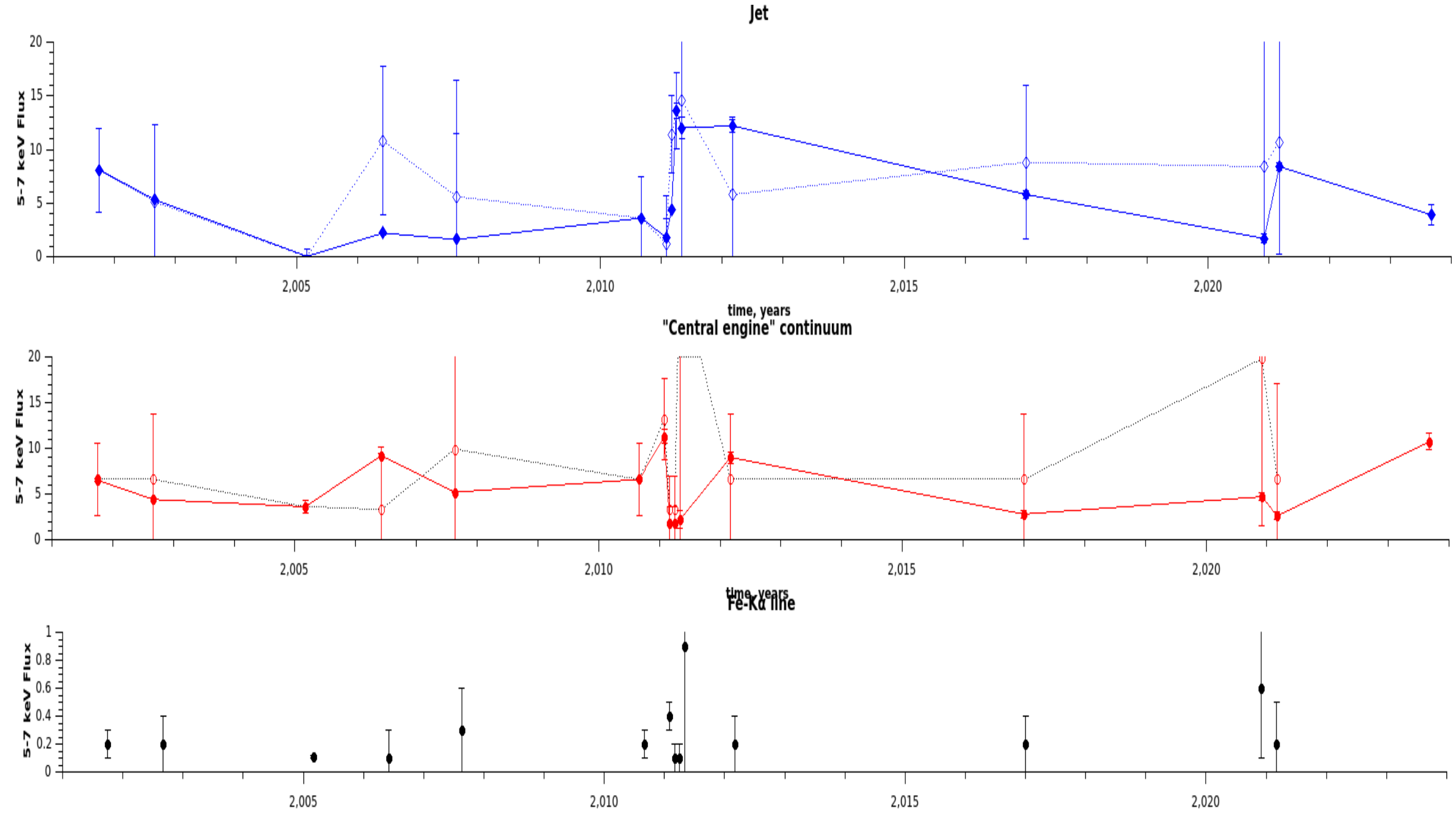

**Figure 29.** 5-7 keV flux curves of the jet base (above), "central engine" (middle), and Fe-K line (below) in 3C 111. Solid lines + filled symbols: model fluxes; dashed lines + empty symbols: fluxes estimated using the "central engine" continuum-Fe-K line flux dependence.

similar values (8.0±2.1 and $8.1^{+2.6}_{-1.4}$ keV for lower break and $16.6^{+8.4}_{-3.4}$ keV and 19.8±3.4 keV for the upper), with similar negative photon index between them (-2.8±1.6 for the first observation and -2.2±1.5 for the second) and quite steep above the second one (1.86±1.0 and 2.19±0.7). During the third observation, the jet spectrum followed the single power-law pattern. Photon index and high-energy cut-off in the spectrum of the "central engine" were varying as well (photon index - from 1.4±0.4 and $1.0^{+0.3}_{-0.2}$ during the first and third observations to 2.2±0.7 during the second one; cut-off from $6.2^{-4.2}_{-2.1}$ keV during the first observation to $59^{+50}_{-28}$ keV during the third and $233^{+\infty}_{-34}$ during the second one). Fe-K$\alpha$ line at 6.4 keV was detected during the first two observations only, with an equivalent width of 0.27±0.08 keV during the first and 0.08±0.02 keV during the second observation. We also estimated the jet and "central engine" 5-7 keV emission fluxes, using both the first and the second methods described above. We used the value of $\kappa = 33 \pm 3$ obtained in [189] as the new data appear to be too jet-dominated to estimate it from them. The resulting flux curves for the Fe.K$\alpha$ line, jet, and "central engine" calculated within the model and utilizing this second method are shown in Fig. 29.

The observational data on 3C 120 (both new and previously published) collected by XMM-Newton, Suzaku, INTEGRAL, and Swift were divided into 23 observational periods, 20 of which were described in [188]. The last three Swift observations are processed here and considered together with the previous ones. The fitting has been performed with the frozen value of the lower-knee photon index of jet SSC/IC emission $\Gamma_{fr}$ = 1.24.

The only break at $181^{+\infty}_{-101}$ keV (first observation) and $40^{+\infty}_{-24}$ keV (second observation) was found in the jet spectra for the first two observations; the breaks at 8.4±0.7 keV and 15±1 keV with the negative photon index (-1.1±0.4) segment between them were found for the third one. The upper photon indices were at the levels of -2.5±2.0, $0.6^{+0.7}_{-0.4}$ and $1.61^{+0.20}_{-0.16}$.

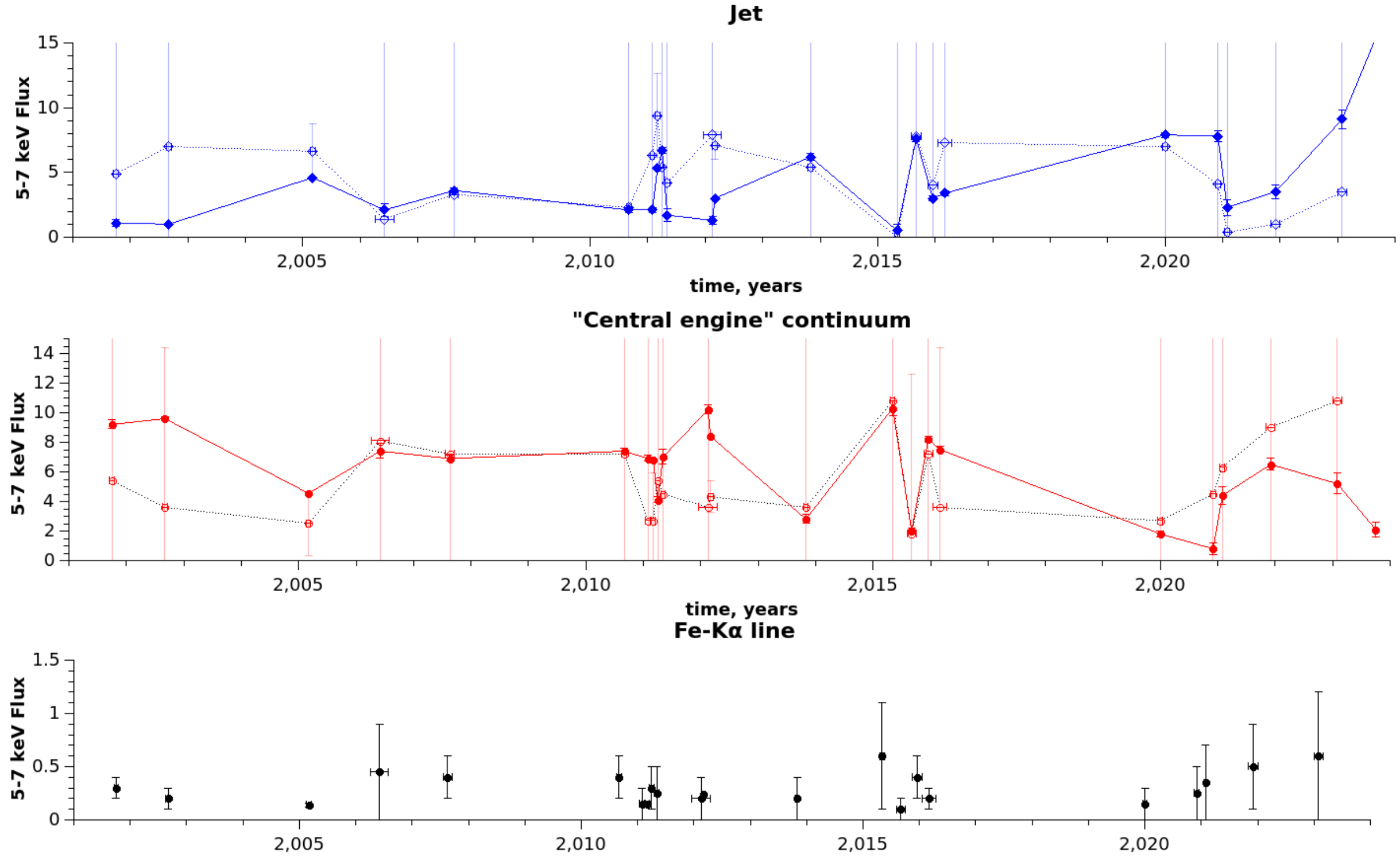

**Figure 30.** 5-7 keV flux curves of the jet base (above), "central engine" (middle), and Fe-K line (below) in 3C 120. Solid lines + filled symbols: model fluxes; dashed lines + empty symbols: fluxes estimated using the "central engine" continuum - Fe-K line flux dependence.

The variations of neither photon index ($\approx$ 2-2.5) and high-energy cut-off (defined only for the second observation at the level of 53$\pm$45), nor ionization potential (>1.5) and plasma electronic temperature (<2.6 keV) were not detected. The emission line Fe-K$\alpha$ was clearly detected during the first two observations with similar parameters (at energies 6.39$\pm$0.15 keV and 6.35$\pm$0.3 keV). During the third observation, the absorption line at 6.27$\pm$0.16 keV with a line strength of 0.14$\pm$0.10 was detected.

We have also estimated the fluxes in 5-7 keV from the jet and "central engine" components in the same way as for 3C 111, using the value of $\kappa = 18^{+90}_{-15}$ obtained in [188]. The resulting flux curves for the jet and "central engine" estimated using the spectral model and the method based on the Fe-K lines and the "central engine" continuum flux dependence are shown in Fig. 30.

**Exposure time (ks):** In our analysis, we used the XMM-Newton/EPIC, Suzaku/XIS, SWIFT (BAT and XRT), and INTEGRAL (ISGRI) datasets covering the period 2000-2024. The observation LOG of the data collected before January 2019 for 3C 111 and before February 2022 for 3C 120 is shown in [188,189]. The Swift/XRT and INTEGRAL/ISGRI data obtained during the period 2021/01-2024/08 (for 3C 111) and 2022/07-2024/08 (for 3C 120) were processed in the same mode as in [189]. The observation LOG of the data collected before February 2022 is shown in [188]. The Swift (XRT and BAT) data obtained during the period 2022/07-2024/08 (data revolutions 36369, 37594, 89848, and 96109) and INTEGRAL/ISGRI (spacecraft revolutions 2531-2562, 2598-2609, 2669-2700) were processed here in the same way as for 3C 111.

**Observing description:** As we can see in the Fig.29, 30, both the methods used here gave compatible flux values within the error bars in different spectral components; however, the error bars of the second method often reach more than 50% levels, especially in the SWIFT/XRT observations. The better quality of

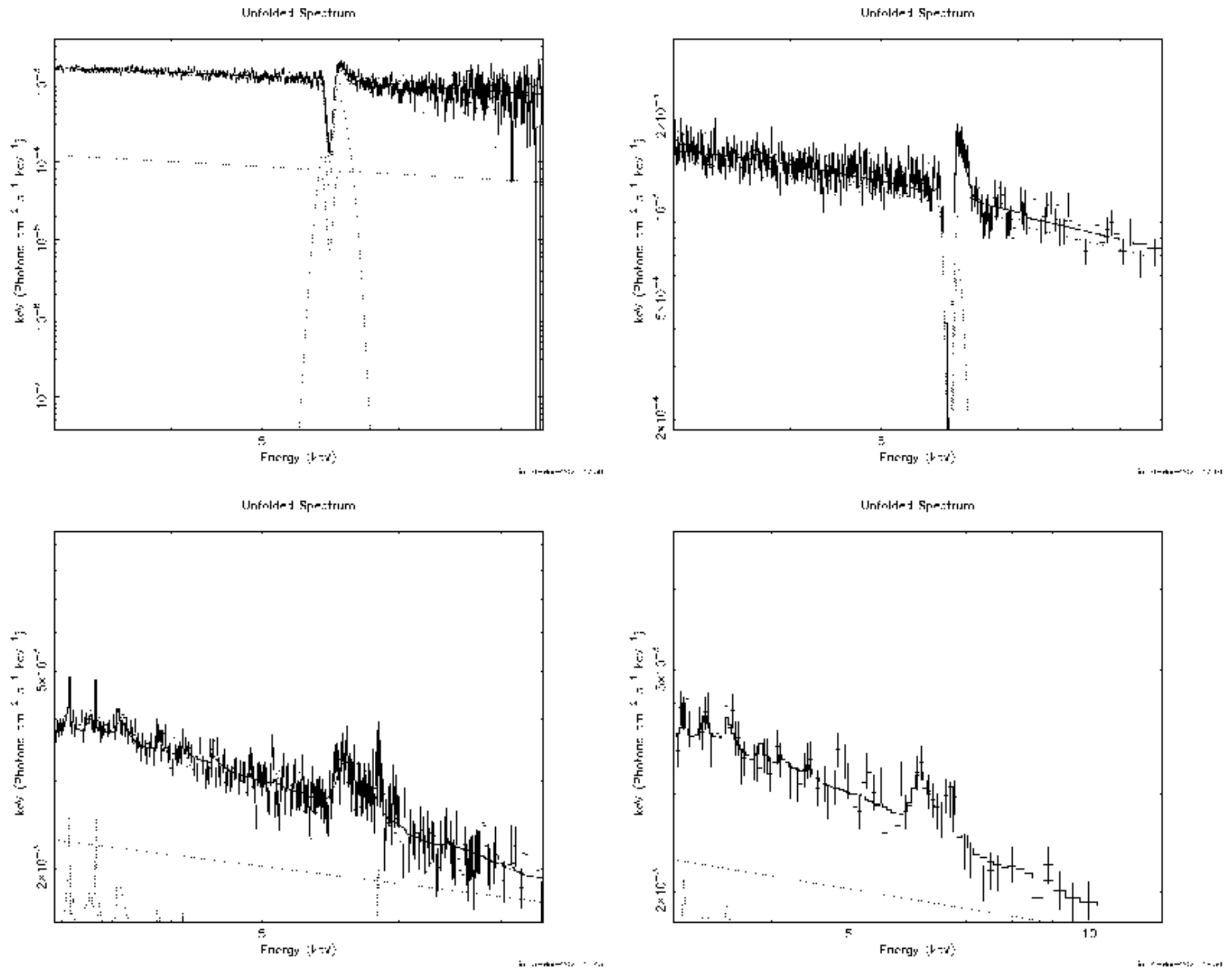

**Figure 31.** AXIS simulated spectrum of 3C 111 (upper left) and 3C 120 (lower left); Suzaku/XIS spectrum of 3C 111 (upper right) during August 2008; XMM-Newton/EPIC PN spectrum of 3C 120 (lower right) during September 2002.

the results from the second method was achieved using XMM-Newton/EPIC and Suzaku/XIS. In Fig. 31, we compare the Suzaku/XIS and XMM-Newton/EPIC spectra with the simulated AXIS spectra based on their spectral models.

The spectral resolution of AXIS near 6 keV (within the range of energies where Fe-K lines are usually located) is 5 eV (in comparison with: XMM-Newton/EPIC PN 15 eV, Suzaku/XIS 2 eV, Swift/XRT 50 eV), giving the possibility to measure the line flux (and thus the fluxes in spectral continuum components) at a higher level of accuracy and to investigate line profiles in a more detailed way, even during the period of moderate jet dominance.

**Perspectives for other AGN:** 3C 111 and 3C 120 are definitely not unique objects whose structure can be studied with the two methods described here. Taking into account that at radio range we can see the lower knee of synchrotron emission or, in some cases, also a plateau or some kind of transitional area, the first of these methods (i.e., equal photon indices of synchrotron and SSC/IC emission of a jet) is applicable in such situations:

- in X-rays we see the lower knee of the SSC/IC emission of the jet only, with $\Gamma = \Gamma_0$;
- the lower knee and the plateau or transitional area of SSC/IC emission of the jet are visible in X-rays: than as well as for the synchrotron emission, we have $\Gamma_0$ for the lower knee, free parameter of the break energy $E_b$ and free or frozen to the value of $\Gamma_1$ (if it is visible in radio synchrotron emission) transitional photon index);

**Table 3.** Perspective AGN to be investigated with the two proposed recipes.

| Target name | Type | redshift | Description |
|---|---|---|---|
| 3C 111 | RT S1 | 0.049 | RT AGN with intriguing variability; Fe-K line profiles as a clue |
| 3C 120 | RT S1 | 0.033 | to understand the the "central engine" |
| 3C 138 | CSS S1.5 | 0.759 | CSS with dip between Synchrotron and SSC within 0.04-20 keV |
| 3C 216 | HPQ S0 | 0.67 | blazar with the syncrothron-SSC dip at 0.04-40 Kev |
| 3C 273 | LPQ FSRQ | 0.158 | 0.01-10 keV syncrotron-SSC dip, broad Fe-K lines detected |
| 3C 279 | HPQ FSRQ | 0.536 | strong variability, emission of the controversial nature above 400 KeV |
| 3C 286 | CSS QSO | 0.85 | synchrotron-SSC dip at 0.04-400 keV, bright AD visible in UV |
| 3C 309.1 | CSS QSO | 0.904 | synchrotron-SSC dip at 0.04-40 KeV bright AD visible in UV |
| 3C 345 | HPQ FSRQ | 0.533 | synchrotron-SSC dip in 0.04-40 KeV, Fe-K lines detected |
| 3C 380 | S1.5 LPQ | 0.692 | synchrotron-SSC dip at 0.04-400 KeV, bright AD visible in UV |
| 3C 390.3 | BLRG S1 | 0.056 | broad Fe-K lines detected |
| 3C 454.3 | BL LAC | 0.859 | double BH candidate, strong variability synchrotron-SSC dip in 0.04-40 keV |
| Mrk 1501 | RL S1 | 0.089 | flaring jet |
| 4C 15.05 | FSRQ | 0.8336 | synchrotron-SSC dip at 0.4-40 KeV |
| 8C 1849+670 | Bl LAC | 0.657 | hybrid SSC/IC blazar |
| NGC 7213 | RM LINER | 0.006 | no-accretion disk candidate |
| OJ 287 | BL LAC | 0.306 | double BH candidate, strong variability, synchrotron-SSC dip at 4-40 KeV |

- both the lower knee of SSC/IC and the upper knee of synchrotron emission of the jet are seen in X-rays, where the "synchrotron-SSC dip" is situated. In such a case, we can use $\Gamma_0$ for the lower knee of SSC / IC, and the photon index of the upper knee synchrotron as a free fitting parameter or frozen to the value obtained from the fitting of high-energy $\gamma$-ray spectrum (Fermi/LAT of higher).

In the third case, when the synchrotron-SSC dip is situated within the area covering from soft X-rays to soft $\gamma$-rays, both methods are applicable even for blazars, as within the dip we can see the emission of the "central engine" of an AGN, including the Fe-K lines. Based on these criteria, we have compiled a list of AGN of various types for which the methods of spectral component distinguishing can provide some insight (see Table 3).

The high enough spectral resolution provided by AXIS will enable us not only to apply both methods of the spectral component distinguishing more effectively, but also to analyze the shape of relativistic line profiles (like Fe-K and Ni-K) as a clue to understand the nature of the AGN "central engine" and a central massive object nested there.

**Exposure time (ks):** 500 ks.

**Joint Observations and synergies with other observatories in the 2030s:** XMM-Newton, SWIFT, ATHENA + Fermi, CTA + radio band missions (MWA, VLA, VLBI)

**Special Requirements:** None

**d. IMBHs**

*23. Search for intermediate mass black holes in GCs, UCDs, late-type spirals and dwarfs*

**Science Area: intermediate mass black holes; nuclear black holes; co-evolution of black holes and galaxies**

**First Author:** (with affiliation, email) Roberto Soria (INAF, roberto.soria@inaf.it)
**Co-authors:** (with affiliations) Riccardo Arcodia (MIT), Andrea Sacchi (CfA | Harvard & Smithsonian), Albert Kong (NTHU, Taiwan)

**Abstract:**
i) Context. Determining the occupation fraction and mass distribution of nuclear black holes (BHs) in low-mass galaxies in the local universe, as a function of galaxy type, size, and history, provides key tests for models of high-redshift BH seeding and quasar growth, and for predictions of gravitational wave emission resulting from galaxy mergers (hence, predictions of what space-based observatories such as *LISA*, *TAIJI* and *TianQin* will detect). Finding a point-like nuclear X-ray source in a low-mass galaxy is a strong hint that the galaxy contains a nuclear BH; combining the X-ray properties with optical, IR, and radio data, we constrain the black hole mass and its accretion properties.

ii) Objectives. First, we aim to constrain the nuclear BH occupation fraction in low-mass galaxies, and their mass and Eddington ratio distribution. In particular, we want to explore the intermediate mass range ($\sim$1,000–$10^4$ $M_\odot$). Second, we will improve our understanding of the scaling relations between stellar mass and nuclear BH mass in low-mass galaxies of different types and evolutionary histories, including galaxies that have lost a substantial fraction of their original stars via tidal stripping, and in the most massive globular clusters. Third, we will constrain the rates of two classes of transient phenomena (both characterized by soft, thermal X-ray spectra): quasi-periodic eruptions and tidal disruption events on intermediate mass black holes (IMBHs).

iii) Strategy. *AXIS*'s large field of view, high sensitivity, and excellent PSF ($<2''$ even at $12'$ off-axis) are key features that make this mission ideal for surveys of galaxies in clusters and compact groups. In particular, we will be able to detect nuclear BHs as faint as $\approx 10^{37}$ erg s$^{-1}$ in Virgo Cluster dwarfs with only a 10-ks observation. Thus, *AXIS* will reveal the presence of faint nuclear BHs that have so far escaped our detection and hampered our knowledge of galaxy/black hole scaling relations. Moreover, *AXIS*'s sensitivity peaks in the soft X-ray band. This makes it ideal for a search of IMBHs in the high/soft state ($L_X \sim 10^{41}$–$10^{43}$ erg s$^{-1}$) in galaxies and globular clusters at least as far as $z = 0.3$ in 10 ks. It will also hugely expand the volume open to searches of tidal disruption events in dwarfs and globular clusters. Finally, *AXIS*'s sharp PSF will be ideal for the search and follow-up study of multiwavelength counterparts.

**Science:**
Over the last decade, intermediate mass black holes (IMBHs) in the mass range $10^3 \lesssim M_{\rm IMBH}/M_\odot \lesssim 10^5$ have become a hot topic of research in several contexts. In the early universe, IMBHs are a possible starting point for the growth of nuclear supermassive black holes (SMBHs) at redshift $\sim$6–12 [25,369,593,617]. In the local universe, IMBHs are predicted to be found in the nuclei of late-type disk and dwarf galaxies, according to an extrapolation of galaxy scaling relations to small stellar masses [115]. The debate on the presence of an IMBH in the core of the most massive globular clusters and ultracompact dwarfs remains unsettled, and has important implications on the origin of those stellar populations—either formed in situ or from tidal stripping of accreted and disrupted satellite dwarfs [239]. IMBHs may be floating in the halo of massive galaxies, perhaps still inside a tightly bound stellar cluster, also as a result of gravitational recoil in mergers [231]. IMBHs crossing the accretion disk of a more massive nuclear SMBH have been invoked as one possible scenario for sharp, recurrent soft X-ray flares in AGN (quasi-periodic eruptions;

[101,199,348]). Moreover, the forthcoming suite of space-based gravitational wave detectors such as *LISA* will be mostly sensitive to timescales associated with IMBH-IMBH mergers [16].

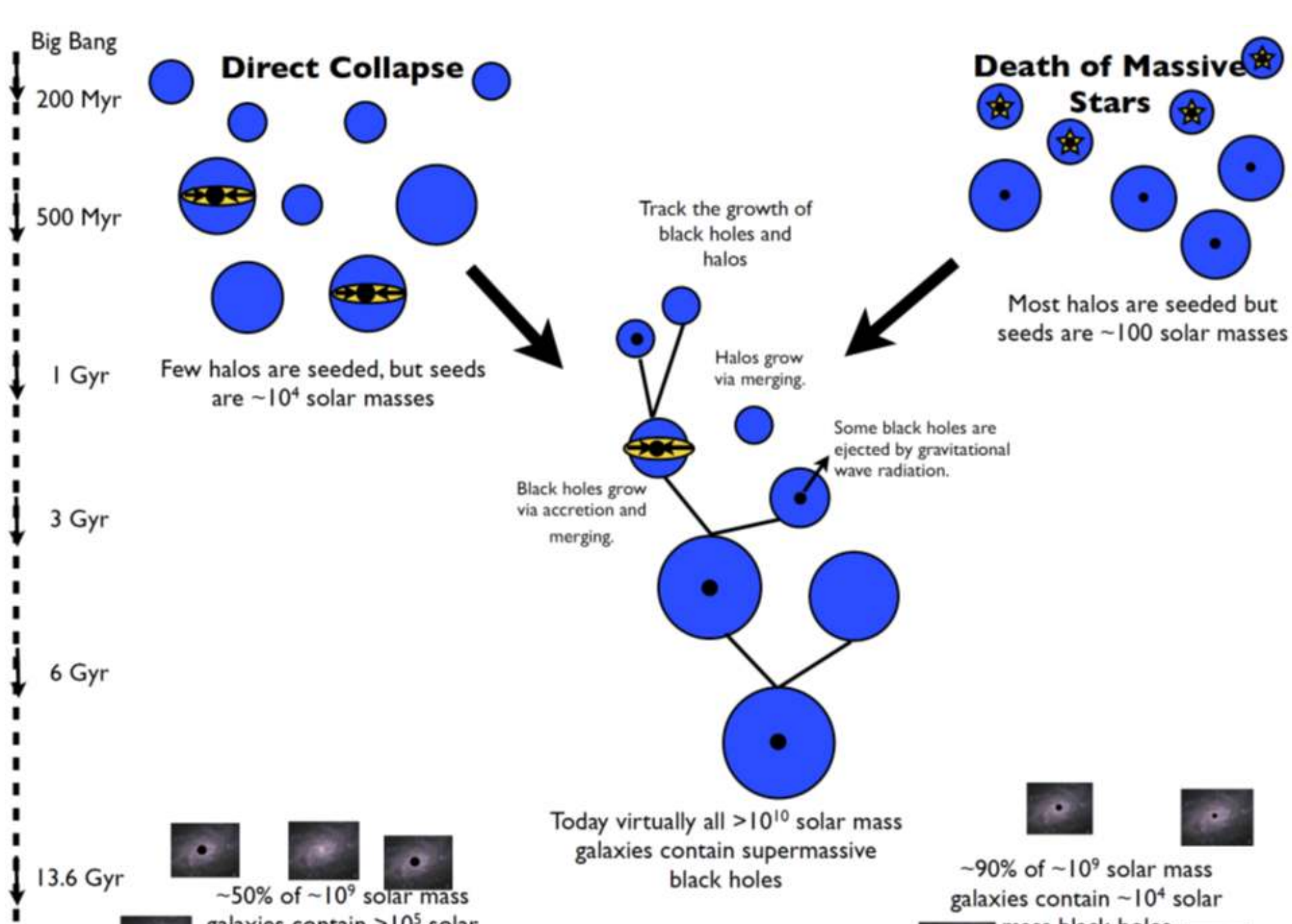

**Figure 32.** Determining the fraction of (small) galaxies in the local universe that contain a nuclear IMBH will enable us to distinguish between different scenarios of nuclear IMBH formation (initial seed mass) and growth (via gas accretion and/or mergers). From [238].

Despite all the contexts in which IMBHs are expected to be found, direct evidence of their existence and location from electromagnetic observations remains scarce and/or controversial. With AXIS, we will search for and investigate IMBHs in X-ray active states. We want to use the X-ray and multiband information to distinguish IMBHs from supermassive BHs on the one hand, and stellar-mass ultraluminous X-ray sources on the other. We will structure our study in four parts.

- *Nuclear IMBHs in the high/soft state.* Theoretical models and observational studies show [487] that for a well defined range of accretion rates, between ≈0.02–1 times the Eddington accretion rate, the accretion flow is dominated by a radiatively efficient, optically thick accretion disk, without substantial contributions from a hotter, optically thin, radiatively inefficient corona. This accretion regime is generally referred to as the high/soft state, or thermal dominant state. The X-ray spectrum in this state is dominated by a thermal component, with a luminosity $L_{\rm disk} \approx 0.1\dot{M}c^2 \approx 4\pi R_{\rm in}^2 \sigma T_{\rm in}^4$, where $R_{\rm in}$ is approximately the radius of the innermost stable circular orbit (inner radius of the disk) and $T_{\rm in}$ is the peak color temperature in the inner disk [527]; $kT_{\rm in} \approx 230(\dot{m}/M_4)^{1/4}$ eV [157,323], where $\dot{m}$ is the scaled accretion rate in Eddington units and $M_4$ is the BH mass in units of $10^4 M_\odot$. Therefore, point-like nuclear X-ray sources with apparent luminosities $L_{\rm X} \sim 10^{41}$–$10^{43}$ erg s$^{-1}$, a soft thermal spectrum, and peak disk-blackbody temperatures $kT_{\rm in} \approx 0.2$–0.3 keV are strong IMBH

candidates. Nuclear sources in the same luminosity range are very common (low-luminosity AGN) but only those powered by an IMBHs should be in the high/soft state.

There are approximately two dozen known nuclear sources with soft, likely thermal X-ray spectra [513]. Some are extremely variable, by more than an order of magnitude on timescales of few years, which is very unusual for AGN. They are consistent with accretion disk emission from an IMBH in the high/soft state, but alternative scenarios are also possible. For example, when a nuclear BH is fed by a tidally disrupted star (tidal disruption event), the X-ray emission from the transient accretion flow may also be soft and thermal, without the presence of a hot corona, regardless of nuclear BH mass [218,493]. With *AXIS*, we will substantially increase the sample of thermal AGN, measure their temperature and luminosity distribution, and monitor the X-ray luminosity evolution over several years. We will finally determine whether a soft thermal spectrum is the signature of a nuclear IMBH, and distinguish between steady accretion from large-scale gas inflows and sporadic accretion from tidal disruption events. Once we have built a reliable and statistically significant sample of galaxies with a nuclear IMBH, we will use those galaxies to calibrate and extend the scaling relations between BH mass and stellar mass of the host galaxy, for different morphological types [115,234].

- *Off-nuclear IMBHs in the high/soft state.* A small number of point-like X-ray sources with $L_X \sim 10^{41}$–$10^{43}$ erg s$^{-1}$, a soft thermal spectrum consistent with disk-blackbody emission, and peak temperatures $kT_{\rm in} \approx 0.2$–$0.3$ keV have also been found outside galaxy nuclei, in the outskirts of early-type galaxies [102,187,287,345]. The combination of luminosity, temperature, and spectral state suggests IMBH accretors, for the same reasons outlined above. There are at least three reasons for the off-nuclear location of such IMBHs. They may be formed in situ inside massive globular clusters, via core collapse; or they may have been nuclear IMBHs of accreted and disrupted satellite galaxies, that have not sunk into the nuclear region of the larger galaxy; or they may have been ejected from the nuclear region by recoil processes during a nuclear BH-BH merger. The sample of X-ray-selected off-nuclear IMBH candidates is still too small to draw any conclusions on either their origin or their feeding process; looking at a much larger volume of space, *AXIS* will enable us to answer those questions.

- *Nuclear IMBHs in the low/hard state* Although IMBHs in the high/soft state are easier to discover and identify, a much larger number of IMBHs are expected to be in the low/hard state, at luminosities $L_X < 10^{41}$ erg s$^{-1}$. Their power-law X-ray spectra and photon index will be indistinguishable from those of SMBHs in the same luminosity range (low-luminosity AGN). Thus, our search for this subsample of IMBHs will be based on an optical selection. We will identify subclasses of galaxies that are more likely to host a nuclear IMBH, based on their morphology, stellar mass, and diagnostic optical line emission. With *AXIS*, we will then search for point-like, nuclear X-ray sources in those galaxies, modeling the occupation fraction and Eddington ratio distribution. We will estimate (at a statistical level) the likelihood that the nuclear sources are IMBHs rather than, for example, stellar-mass X-ray binaries. We will also search for core radio emission, which further constrains the nuclear BH mass via fundamental plane relations.

- *Quasi periodic eruptions* Quasi-periodic eruptions (QPEs) are sharp, recurrent soft X-ray transients from nuclear BHs, repeating on timescales $T_{\rm rec} \sim 2.5$–$100$ hr and lasting only few 1000 s [21,101,199,348]. Their peak luminosities ($L_{\rm peak} \sim 10^{42}$–$10^{44}$ erg s$^{-1}$) and blackbody-like spectra with temperatures $\approx 50$–$250$ eV largely overlap the parameter space of IMBHs in the high/soft state described above. However, QPE emission is more likely to have a completely different origin. Current models suggest that QPEs occur when an orbiting lower-mass companion crosses the accretion disk of a nuclear

SMBH. Each collision shocks the disk gas to temperatures $T \sim 10^6$ K, leading to a repeated pattern of soft X-ray flares. For the collider, typical observed collisional cross sections $\sim R_\odot$ are consistent with either a low-mass star or an IMBH. Intriguingly, this phenomenon appears to occur only in nuclear BHs that have become active after a recent tidal disruption event, rather than in persistent AGN. Understanding the origin of QPEs will constrain models of nuclear BH growth and gravitational wave emission from extreme-ratio inspirals. So far, only a dozen nuclear BHs have been found to show QPEs, based on archival (mostly serendipitous) *XMM-Newton* and *Chandra* data, and on the relatively shallow eROSITA survey. A wide-field survey with *AXIS* will be crucial to shed light on the origin of QPEs.

**Exposure time (ks):** 300.

**Observing description:**

The nuclear regions of late-type disk galaxies and star-forming dwarfs (those where IMBHs are most likely to lurk) are complex environments, with at least four possible sources of X-ray emission:

— direct, point-like emission from the nuclear IMBH, either in the high/soft state (disk-blackbody spectrum) or low/hard state (power-law spectrum);

— a point-like soft excess, emitted by hot plasma photoionized by the nuclear BH;

— extended, soft, thermal-plasma emission from a nuclear star-forming region, powered by supernova remnants and stellar winds;

— power-law-like emission from high mass X-ray binaries or ultraluminous X-ray sources in the nuclear region, slightly offset from the optical/radio nucleus.

The broad point-spread functions of *XMM-Newton*/EPIC and eROSITA cannot spatially resolve those components, hindering our search for nuclear IMBHs. *Chandra* can resolve the nuclear source, but its low and steadily decreasing sensitivity (especially in the soft band) and small field of view limit its use for a wide-field, deep search of new candidates. Instead, *AXIS* provides an ideal combination of sensitivity and spatial resolution.

With *AXIS*, we can detect IMBHs in the high/soft state ($L_X \sim 10^{41}$–$10^{43}$ erg s$^{-1}$) and select the most promising candidates (based on hardness ratios and multiband counterparts) at least as far as $z = 0.3$ in 10 ks. Specifically, we estimate $\approx$30 ct in 10 ks for a $10^{41}$ erg s$^{-1}$ source at redshift 0.3. Deeper follow-up observations ($t_{\rm exp} \sim 100$ ks) will provide enough counts for spectral modeling. The planned Intermediate and Wide surveys will constrain the volume density of such IMBHs in the local universe. In addition, *AXIS* will make it feasible to map rich nearby galaxy clusters, such as the Coma Cluster, with a reasonably small number of pointings ($\sim$10–20), thanks to its uniformly good PSF across the field of view. Conversely, for fainter nuclear IMBHs in the low/hard state, *AXIS* can reach $L_X \sim 10^{37}$ erg s$^{-1}$ at the distance of the Virgo Cluster.

In summary, i) we will make use of the planned Deep and Intermediate surveys, and support the development of a wide-area (tens of sq. deg) survey together with other science cases and working groups (*e.g.*, TDAMM); ii) we will make use of all cumulative data acquired in the archives; iii) we will perform dedicated follow-up on the most promising targets (IMBH candidates in the high soft state, nuclear sources with QPEs, fainter nuclear sources in galaxies that are strong candidate IMBH hosts, off-nuclear IMBH candidates, etc).

We do not propose any specific targets at this time. The proposed science case will exploit the survey programs of *AXIS*, which will be completed as part of its core science. In particular, we will benefit from AWES. Furthermore, given *AXIS*'s wide field of view, coupled with its flat and narrow PSF, we will be able to search for IMBHs that might be serendipitously imaged while observing other targets.

**Joint Observations and synergies with other observatories in the 2030s:** LISA, SKA

**Special Requirements:** None.

**e. Variability**

*24. X-ray variability*

**Science Area:** Active Galactic Nuclei

**First Author: Alessia Tortosa** (INAF-OAR;)
**Co-authors: Matilde Signorini** (European Space Agency; matilde.signorini@esa.int) **Stefano Bianchi** (Università Roma Tre; stefano.bianchi@uniroma3.it), **Pierre-Olivier Petrucci** (Université Grenoble Alpes; pierre-olivier.petrucci@univ-grenoble-alpes.fr) **Riccardo Middei** (ASI-SSDC; riccardo.middei@ssdc.asi.it) **Claudio Ricci** (Universidad Diego Portales; claudio.ricci@mail.udp.cl) **Roberto Serafinelli** (Universidad Diego Portales; roberto.serafinelli@mail.udp.cl) **Eric S. Perlman** Florida Institute of Technology

**Abstract:** Active Galactic Nuclei (AGN) are characterized by variability across a broad range of timescales and wavelengths, providing key insights into their emission properties. In the X-ray band, variability spans both short ($< 1000$ secs) and long timescales (years), offering a window into the innermost regions of AGNs. With this GO program, we aim at exploiting AXIS's fast readout times, high time resolution, and deep-field capabilities to investigate rapid X-ray fluctuations and study the underlying processes driving AGN variability. By resolving short-term changes in the accretion disk-corona system, AXIS will enable testing of AGN variability models, while its high-resolution imaging and low instrumental background will facilitate the separation of AGN emission from surrounding sources. This is essential for understanding variability, especially in fainter AGN populations. Moreover, AXIS's time-domain surveys will provide crucial data on long-term AGN variability, offering insights into changes in accretion rates and black hole feedback. In addition, AXIS observations will enable the detection of rare events, such as changing-look AGNs and tidal disruption events (TDEs). With improved photon statistics and the ability to track variability over months to years, AXIS will contribute to a comprehensive understanding of AGN growth, feedback mechanisms, and their connection to galaxy evolution. Comparisons with multiwavelength observations from JWST, LSST, and SKA will deepen our understanding of how AGN variability influences star formation and cosmic feedback, while expanding the scope of variability studies to high-redshift quasars.

**Science:** Variability is a distinctive feature shared by all classes of AGN, occurring over a wide range of timescales and amplitudes across all the wavelengths [400,576]. In the X-ray band, variability is observed on both short ($< 1000$ sec, [400,579]) and long timescales (years, [276,399,516]) giving insight into the innermost regions of the AGN. Thus, its study can help us understand the emission properties of AGNs and better characterize the growing population of extremely variable AGNs identified in the optical and X-ray regimes.

One method used to study the temporal structure of the variations is the power spectral density (PSD) analysis. If the temporal frequency is $\nu = 1/t$, where $t$ is the time, the observed power spectrum is generally modeled as a power-law of the form: $P_\nu \propto \nu^\alpha$. For short timescales (high frequencies) $\alpha \sim -2$, while for long timescales (low frequencies) $\alpha \sim -1$. The PSD break timescales, $T_B$, can be obtained by fitting a broken power law to the observed PSD. This parameter has been found to be positively correlated with the black hole mass. However, NLSy1 galaxies, which typically accrete at very high Eddington ratios ($L_{\rm bol}/L_{\rm Edd} = \lambda_{\rm Edd}$), display a different behavior, with their break timescales being shorter for a given black home mass. One possible explanation is that the break timescales could also depend on a second parameter, such as the accretion rate or the black hole spin. A few objects have quasi-periodic oscillations, which indicate processes in the accretion disk.

AXIS will have fast readout times and high time resolution, making it ideal for detecting rapid X-ray fluctuations in AGN. This will help in probing short-term changes in the accretion disk-corona system and test models of AGN variability. AXIS's deep-field and time-domain surveys will enable more

comprehensive statistical studies of long-term AGN variability, allowing for the tracking of changes in accretion rates and black hole feedback. Additionally, AXIS, with its angular resolution, will provide high-resolution imaging with sub-arcsecond spatial resolution, comparable to *Chandra* but over a larger field of view. This helps separate AGN emission from surrounding sources (e.g., host galaxy, nearby X-ray binaries), which is crucial for studying variability. With its low instrumental background, AXIS will allow the detection of weaker variability signals and fainter AGN populations.

Some of the results we expect to achieve with this AXIS GO proposal:

- **Rapid X-ray Variability: probing the inner accretion flow.** X-ray emission from AGN originates near the event horizon, where extreme gravity dominates. Rapid changes in brightness provide direct insight into the physics of the corona and disk. With its time resolution ($\sim$ seconds), AXIS will detect fast flickering of the corona, revealing its structure and heating mechanisms. Moreover, studying the energy-dependent variability, it will be possible to determine whether changes originate from fluctuations in the disk, coronal emission, or relativistic effects. AXIS's better photon statistics will capture detailed light curves for fainter AGN, improving reverberation mapping studies;
- **Long-Term X-ray Variability: tracing AGN growth and feedback.** AGN exhibit variability over months to years, reflecting changes in accretion rates, disk instabilities, and interactions with the host galaxy. Frequent monitoring of key AGN with repeated deep-field observations will enable the tracking of accretion rate fluctuations in hundreds of AGN. This type of study will enable us to detect AGN shutdown and reactivation, identifying AGN switching states and capturing the moment when black holes turn on or off. Additionally, it will be possible to connect X-ray variability to multiwavelength studies by comparing AXIS data with JWST, LSST, and SKA, which will reveal how AGN variability affects star formation and feedback.
- **Catch rare transient events like changing-look AGN and TDEs.** AXIS's combination of high sensitivity, rapid readout, and low background makes it an excellent tool for detecting and characterizing transient X-ray phenomena. Changing-look AGN—active galaxies that dramatically switch between Type 1 (bright, unobscured) and Type 2 (faint, obscured) states—can provide crucial information about accretion state transitions, disk instabilities, or sudden changes in obscuration. AXIS will detect these changes in real-time, allowing follow-up studies across multiple wavelengths to understand the physical mechanisms driving them.
- **Expand variability studies to high-redshift quasars.** Current X-ray variability studies are mostly limited to lower-redshift AGN due to sensitivity constraints. AXIS, with its deep-field capabilities and high time resolution, will enable variability studies of high-redshift quasars ($z > 4-6$). This will allow us to investigate how the variability properties of quasars evolve over cosmic time, shedding light on early black hole growth, accretion physics, and feedback mechanisms. By detecting faint, distant AGN and tracking their variability, AXIS will help distinguish between different accretion models at high redshift. For instance, it can test whether early quasars exhibit stronger variability due to less stable accretion flows or different disk-corona properties compared to their lower-redshift counterparts. This will be crucial for understanding the co-evolution of supermassive black holes and their host galaxies in the early Universe.
- **Search for periodicities and quasi-periodicities.** Oscillations, whether periodic or quasi-periodic, originate in the accretion disk that circles the central supermassive black hole and represent a variety of mechanisms that can exist within the disk materials. These can include mechanical oscillations or corrugations, warps due to external stresses or even in spacetime, orbiting hotspots and/or overdensities within the disk, which can accelerate or heat the material, a binary black hole, as well as a variety of other possibilities. They are usually modulated either at the orbital frequency or at the Lense-Thirring frequency (if the BH is spinning). In AGN, QPOs are much more difficult to find due to the faintness and low count-rate of the source, as well as the long time scale of the variations (both

**Table 4.** Example Targets

| Target | Type | Exposure Time (ks) | Observing Strategy |
|---|---|---|---|
| NGC 4151 | Type 1 | 20 | Short-term variability |
| MCG-6-30-15 | Type 1 | 20 | Short-term variability |
| 1H 0707-495 | NLSy1 | 20 | Rapid variability & reverberation |
| IRAS 13224-3809 | NLSy1 | 20 | Rapid variability & reverberation |
| Mrk 1018 | CL AGN | 100 | Multi-epoch monitoring |
| NGC 1566 | CL AGN | 100 | Multi-epoch monitoring |
| PKS 2155-304 | Blazars | 20 | High-cadence & rapid flares |
| Mrk 421 | Blazars | 20 | High-cadence & rapid flares |
| Fairall 9 | High luminous QSO | 100 | Reverberation studies |
| PG 1211+143 | High luminous QSO | 100 | Reverberation studies |
| SDSS J1148+5251 | High-z QSO | 50 | Variability evolution |
| ULAS J1120+0641 | High-z QSO | 50 | Variability evolution |

the orbital and Lense-Thirring frequencies are proportional to mass), than they are in X-ray binaries. As a result, only very few have been found, but there are many unconfirmed claims.

- **Excess variance studies.** Accurately determining the AGN's power spectra can be difficult, since it requires high-quality data, long exposures, and sometimes monitoring campaigns, to extend time coverage that adequately covers relevant PSD frequency ranges that include potential breaks. Given such difficulties, it is common practice to quantify the X-ray variability of AGNs in terms of the so-called normalised excess variance ($\sigma^2_{\rm NXS}$, Nandra et al. 417). Although it does not contain the same amount of information as the PSD, the normalised excess variance can be used to confirm the PSD results in large samples of AGN, and it also allows the discovery of new correlations between the X-ray variability amplitude and other AGN's physical parameters.

**Exposure time (ks):** Total exposure time is 620 ks. This program can be integrated with other AXIS observational campaigns.

**Observing description:** The proposed observations will target a sample of 12 AGN and QSOs, selected based on their known X-ray variability properties and their potential for providing new constraints on accretion processes. For short-term variability, we will observe bright AGN with continuous exposures of $\sim 20$ ks per target.

For long-term variability (months to years), we propose multi-epoch observations with individual exposures of $\sim 10 - 20$ ks, revisiting targets at intervals of weeks to months. For high-redshift AGN ($z > 6$) deep exposures ($\sim 50$ ks) will be employed to maximize sensitivity to variability trends. Targets summary is shown in Tab 6.

Moreover, we estimated how many AGN with a good excess variance measure, i.e., without upper limits, would be observed by AXIS. To this end, we need unobscured AGN with at least 40 counts in the 2–10 keV range and 10, 20, and 40 ks long observations. We simulated the observations assuming power law spectra with Galactic absorption $N_h \sim 2 \times 10^{20}\,\rm cm^{-2}$, photon index $\Gamma = 1.7, 1.8, 1.9$, and the on-axis response and the Field Of View average response. The minimum 2–10 keV flux needed to have 40 counts in the 2–10 keV energy band are $5 \times 10^{-15}$, $1.25 \times 10^{-14}$, $2 \times 10^{-14}$ erg cm$^{-2}$ s$^{-1}$ for 10, 20 and 40 ks long observations, respectively (see Fig.33). Looking at the mock catalog from [**?** ], among the unobscured sources, 7841 objects have a flux of at least $1.25 \times 10^{-14}$ erg cm$^{-2}$ s$^{-1}$ in the 2–10 keV energy band, which is the limit considering the average FoV response for 20 ks long observations. Thus, if AXIS observes 100 square degrees with at least 20 ks per exposure, we would get 7841 objects, with a redshift distribution that peaks around 1 and a tail up to redshift 4. This will bring a substantial decrease of the dispersion of the $\sigma^2_{\rm NXS}$ vs. $M_{\rm BH}$ and the $\sigma^2_{\rm NXS}$ vs. $\lambda_{\rm Edd}$ relations. Additionally, since the X-ray variability could potentially also be used as a tool to estimate intrinsic luminosities, AGN may be used as standard candles to test cosmological

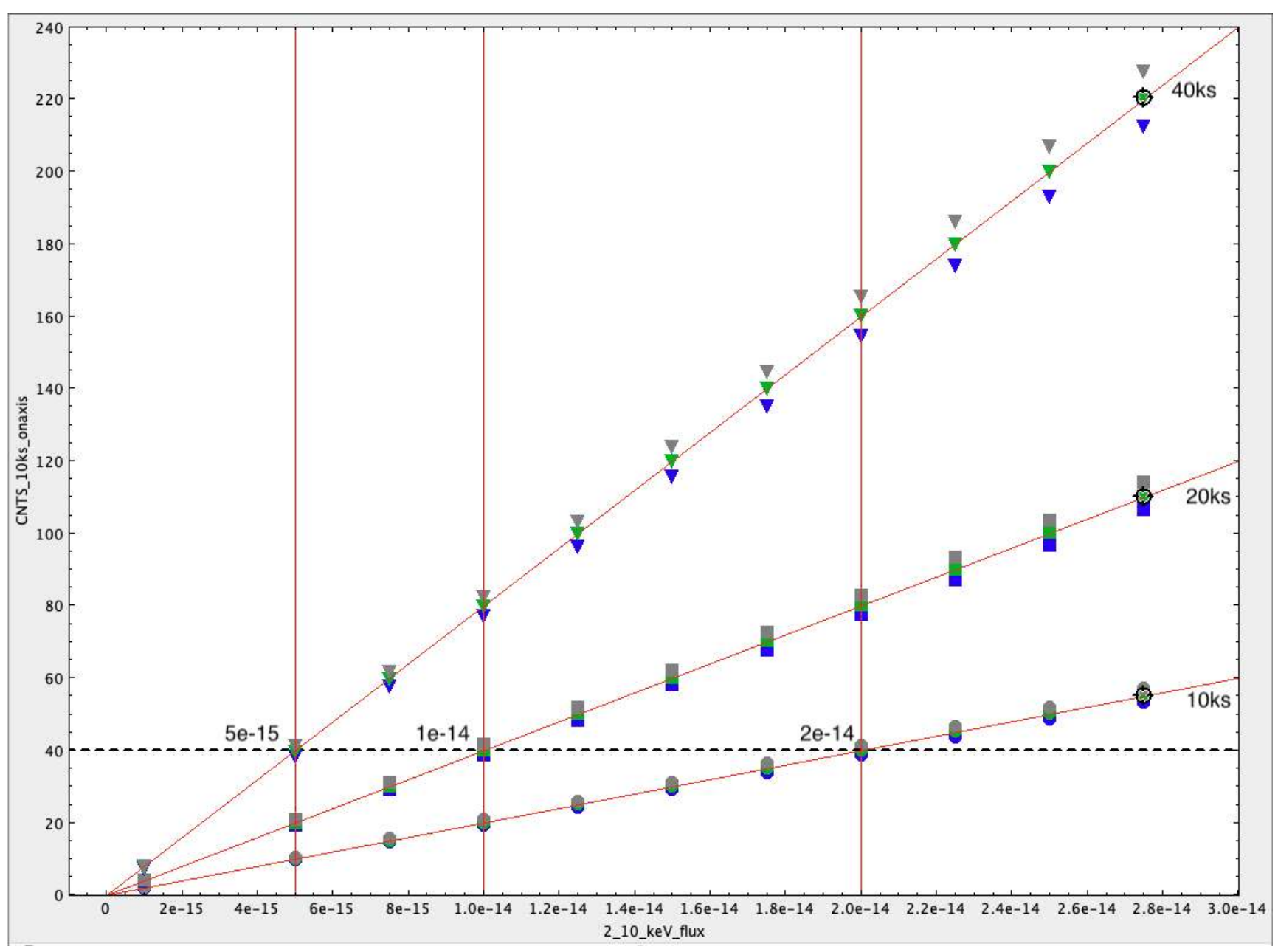

**Figure 33.** Counts vs 2–10 ks flux trend for on-axis sources observed with different exposure times : 10 ks (cirles), 20 ks (squares), 40 ks (triangles).

models [325], incorporating broad emission line properties alongside variability data. With AXIS, it will be possible to test the possible cosmological implementation up to $z \sim 4$.

**Joint Observations and synergies with other observatories in the 2030s:** JWST, LSST, SKA

**Special Requirements:**

## f. Accretion

### *25. X-ray Microlensing as a window into quasar structure and black hole spin evolution*

**Science Area: AGN, Quasar Structure, SMBH spin, Accretion Disk Properties**
**First Author: George Chartas** (College of Charleston, chartasg@cofc.edu)
**Co-authors:** Elena Bertola (INAF – Osservatorio Astrofisco di Arcetri), Christopher Morgan (USNA), Joshua Fagin (CUNY, AMNH), Eric Paic (EPFL), Favio Neira (EPFL), Henry Best (CUNY, AMNH), Timo Anguita (Universidad Andres Bello), Martin Millon (Stanford University), Matthew O'Dowd (CUNY, AMNH), Dominique Sluse (Universite de Liege), Georgios Vernardos (CUNY, AMNH),
**Abstract:** Current X-ray observations and simulations show that gravitational lensing can be used to infer the structure near the event horizons of black holes, constrain the dynamics and evolution of black-hole accretion, test general relativity in the strong-gravity regime, and place constraints on the evolution of the spin parameter of quasars. These science goals currently cannot be achieved in a statistically large sample of $z = 0.5 - 5$ lensed quasars due to the limited capabilities of current X-ray telescopes and the relatively low number ($\sim 400$) of known lensed quasars. The latter limitation will be resolved with the multiband and wide-field photometric optical survey of LSST which is expected to lead to the discovery of $> 4{,}000$ additional gravitationally lensed systems. These science goals can be reached with AXIS having a spatial resolution of $\sim 1.5$ arcsec to resolve the lensed images and a collecting area of $\sim$0.5 m$^2$ at 1 keV.

**Structure and evolution of AGN environments**

Kochanek ([310]) described a microlensing light-curve method that compares the variability amplitudes in the X-ray and optical of individual lensed images to place tight constraints on the structure of the accretion disk and the X-ray emitting regions. An application of this microlensing light-curve method to *Chandra* observations of several bright lensed quasars reveal that the X-ray emitting corona is very compact, $\sim 6 - 50 r_g$ ($r_g = GM_{BH}/c^2$) [e.g., 105,478]. The compactness of the hot corona has been independently confirmed through X-ray reverberation mapping studies of nearby Seyfert galaxies [e.g., 91,297].

The microlensing light-curve method is currently limited to the relatively small number of known gravitationally lensed systems and the sensitivity limitations of *Chandra*. The former limitation will soon be lifted with the aid of multiband and wide-field photometric optical surveys of the Rubin Observatory and *EUCLID*. The Large Synoptic Survey Telescope *LSST* survey ($\sim$ 2025−2035) alone is expected to lead to the discovery of >4,000 additional gravitationally lensed quasars (e.g., *LSST* Science Book; [432]). In Fig. 34 we show the angular separation of lensed quasars expected to be discovered by *LSST*. A high-resolution and high-throughput X-ray mission (FWHM $\sim$ 1.5arcsec), however, will be required to resolve a large fraction of the newly discovered *LSST* lensed quasars. The application of the microlensing light-curve method to the *LSST* lensed quasars over a range of quasar redshifts, black hole masses, radio loudness values, and Eddington ratios will constrain the dependence of the sizes of X-ray emitting regions over this parameter space.

The application of the microlensing light-curve method to a high signal-to-noise X-ray spectrum of a lensed quasar will also constrain the sizes of X-ray emitting regions of quasars, ranging from the hot corona and inner accretion flow to the molecular and dusty torus. Isolating these emission regions is possible in a high S/N X-ray spectrum because the microlensing light-curve method can be applied to select energy bands over which these emission components dominate. *LSST* will monitor the fluxes of images of gravitationally lensed quasars over its ten-year survey period and can provide robust triggers for X-ray high-resolution, high-throughput missions to cover single caustic crossing events.

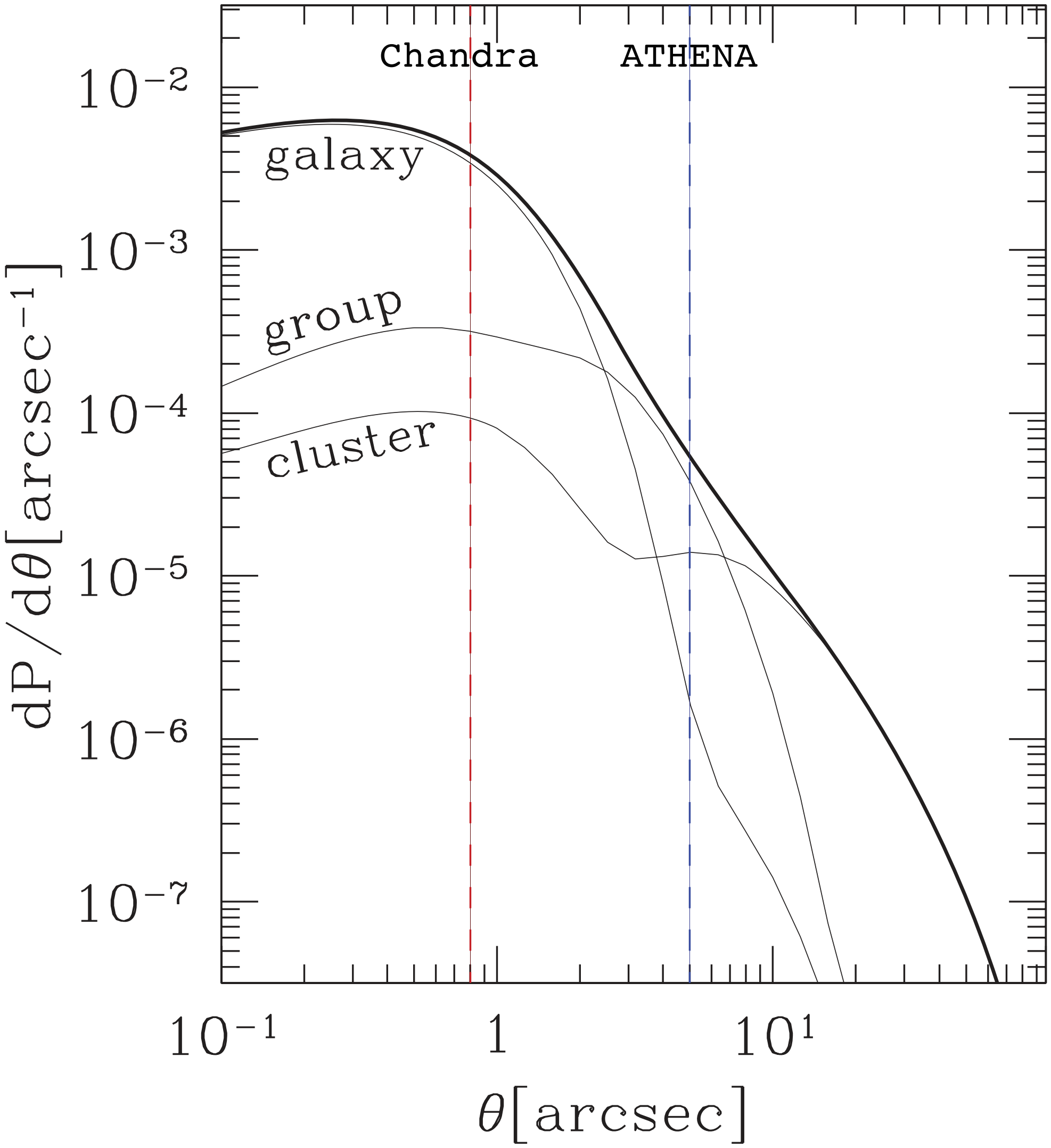

**Figure 34.** The distribution of lens image separations for three scales: galaxy, group, and cluster-scales, predicted by a halo model (Oguri 2006). The thick line shows the total distribution. The vertical lines represent the half-power diameters of Chandra and Athena.

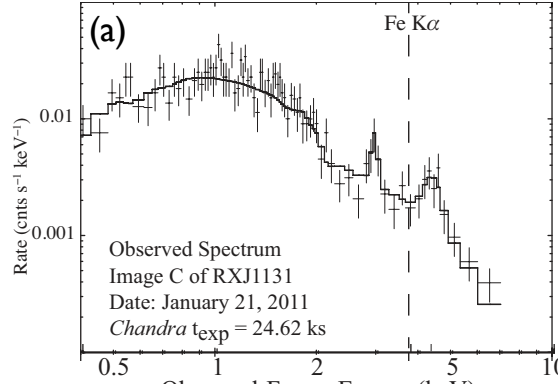

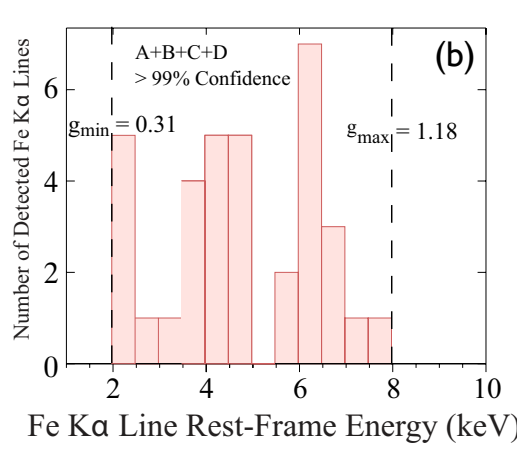

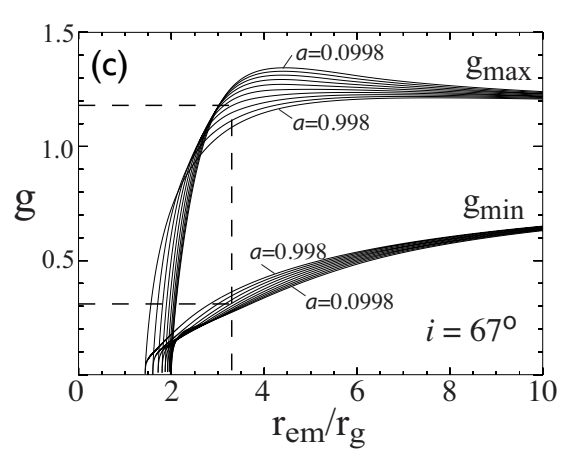

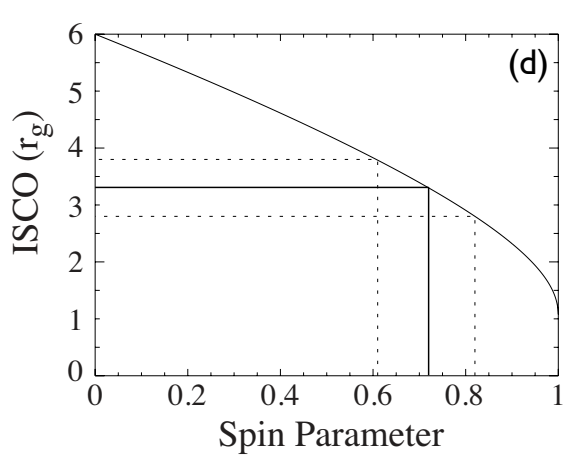

**Figure 35.** (a) Spectrum of image C showing shifted Fe lines due to microlensing. (b) Distribution of the Fe K$\alpha$ line energies for all images and all observations of RXJ1131. (c) Extremal shifts of the Fe K$\alpha$ line energy for spin values ranging between 0.098 and 0.998 in increments of 0.1. Horizontal lines represent the observed values of $g = E_{\rm obs}/E_{\rm rest}$ of the most redshifted and blueshifted Fe K$\alpha$ lines. (d) The inner radius of the accretion disk is constrained to be $r_{\rm ISCO}$ = 3.3 $\pm$ 0.5 $r_{\rm g}$ that corresponds to a spin parameter of $\alpha$ = 0.72 $\pm$ 0.10.

**Measuring SMBH spin and its evolution**

The mechanisms leading to the growth of supermassive black holes can be studied by constraining the evolution of the SMBH spins of lensed quasars with redshifts in the range of $0.5 - 5$. Numerical simulations suggest that the evolution of the spin of supermassive black holes depends on the accretion rate, the mode of accretion, and mergers. Specifically, in the high redshift universe, numerical simulations predict that accretion rates onto the central SMBHs of galaxies are high, and so the spin evolution is dominated by the high accretion. At low redshifts, galaxies are in general more massive and gas-poor, and SMBH$-$SMBH mergers are predicted to lead to the decrease of SMBH spins [e.g., 160]. Numerical simulations by Volenteri et al. [592] also find that the typical spin and radiative efficiency of SMBHs decrease with cosmic time.

We plan to apply the $g$-distribution microlensing method [e.g., 104] to X-ray observations of lensed quasars to measure their spin and ISCO sizes over the $z = 0.5 - 5$ redshift range and thus strongly constrain numerical models of black hole growth [592]. Quasar microlensing produces energy shifts of the Fe K line emitted from the accretion disk [104,320,321]. These shifts originate from general and special relativistic effects when the magnification caustic selectively magnifies emission very close to the event horizon of the black hole. The distribution of Fe K$\alpha$ energy-shifts ($g$-distribution) for $z$=0.658 quasar RX J1131$-$1231 is shown in Fig.35, where $g = E_{obs}/E_{rest}$ is the fractional energy shift of the iron line. By comparing the distribution of energy shifts (g-distribution method) to simulated ones, we constrain the ISCO, spin, and inclination angle of distant quasars (see Fig.35). To study the evolution of ISCO and spin of quasars over the $z$ = 0.5 $-$ 5 redshift range, we propose to monitor a sample of $\sim$10 X-ray bright quadruply lensed quasars over this redshift range. For each quasar, we would build a $g$-distribution to infer the ISCO and spin.

**Exposure time (ks):** Two targets will be triggered per observing cycle. The total proposed exposure time per cycle is 300 ksec.

**Observing description:** To select the best targets for studying AGN microlensing, we began with a list of 306 known lensed quasars, including those detected in recent Gaia surveys. We found that 156/306 of the lensed quasars in our list fall in the area of the sky (with German data rights) covered in the four passes of the eROSITA survey (eRASS:4). By using a positional match within 10 arcsec ($\sim$2$\sigma$ of the eROSITA positional uncertainty), we found 120/156 detections in eRASS:4. From this target, we filtered for lensed quasars with image separations greater than 1.5 arcsec and with combined 2 $-$10 keV fluxes exceeding 1 $\times$ $10^{-13}$ erg s$^{-1}$ cm$^{-2}$. This filtering resulted in 18 lensed quasars (to be updated).

For this proposal, we focus on measuring the energy shifts of the microlensed accretion disk lines and flux changes of the continuum during a single microlensing caustic crossing. The data collected from these observations will address the science goals of this proposal.

We propose to trigger a monitoring program of 15 $\times$10 ks observations of 2 objects per AXIS cycle from our sample list of 18 lensed quasars (to be updated), with the detection of the onset of a microlensing event in any of the AXIS resolved images of the lensed quasars in our sample. The AXIS triggers will be provided by a neuro-network tool developed by the LSST strong lensing team ([184]). The cadence of the monitoring observations of each target will be estimated beforehand and depend on the gravitational radius $r_g$ and the estimated caustic crossing speed.

**[Joint Observations and synergies with other observatories in the 2030s: LSST, Athena]**

**Special Requirements:** Monitoring, ToO.

*26. X-raying the Narrow Line Region in AGN*

**Science Area:** Circumnuclear matter in AGN
**First Author: Stefano Bianchi** (Università degli Studi Roma Tre, stefano.bianchi@uniroma3.it)
**Co-authors: Riccardo Middei** (Center for Astrophysics, Harvard & Smithsonian, INAF-OAR, SSDC-ASI), **Jiachen Jiang** (University of Warwick) **Giuseppina Fabbiano** (Center for Astrophysics, Harvard & Smithsonian), **Martin Elvis** (Center for Astrophysics, Harvard & Smithsonian).
**Abstract:** The spatial and kinematic overlap between soft X-ray emission and the Narrow Line Region (NLR) in obscured AGN provides key insights into the ionized gas structure and AGN feedback on galaxy scales. High-resolution X-ray spectroscopy with *Chandra* and XMM-*Newton* has shown that soft X-ray emission in these regions is dominated by photoionized gas, with evidence indicating that radiation pressure compression (RPC) is the primary mechanism shaping the NLR's density structure. However, existing X-ray observations are often photon-starved, limiting our ability to spatially resolve and spectroscopically characterize the NLR across a large AGN sample. AXIS, with its significantly larger effective area and excellent soft X-ray sensitivity, will transform our study of the X-ray NLR. Although its angular resolution is slightly broader than *Chandra*'s, AXIS compensates with a superior photon collection, allowing precise measurements of faint extended emission structures that remain elusive with current observatories. Moreover, its fine pixel sampling minimizes the need for sub-pixel reconstruction techniques, allowing more efficient and reliable imaging of the NLR. With these advantages, AXIS will provide unprecedented constraints on the density and ionization structure of the X-ray NLR, directly testing RPC predictions and alternative confinement mechanisms such as thermal pressure or magnetic fields. By resolving the extent and morphology of X-ray-emitting gas in a statistically significant AGN sample, AXIS will quantify the outflow energetics and their role in AGN feedback. Its capability to detect low surface brightness emission with high signal-to-noise ratio will enable a comprehensive X-ray census of the NLR, revolutionizing our understanding of AGN-driven photoionization and its impact on galaxy evolution.
**Science:** Understanding the structure and physical conditions of the ionized gas in the vicinity of AGN is essential for unveiling the processes that regulate the co-evolution of supermassive black holes and their host galaxies. In particular, the NLR of obscured AGN, extending from tens to thousands of parsecs, acts as a key interface between the nuclear engine and the larger-scale interstellar medium. High-resolution optical and infrared studies have revealed complex morphologies and ionization structures in the NLR, often aligned with radio outflows or dust features. Meanwhile, integral-field spectroscopy has mapped the kinematics of ionized gas across large AGN samples.

High-resolution X-ray observations have significantly advanced our understanding of the NLR in AGN. *Chandra* imaging has revealed soft X-ray emission extended on kiloparsec scales, often co-spatial with [OIII] emission, indicating a shared photoionized origin [e.g. 62]. Grating spectroscopy with XMM-*Newton*/RGS and *Chandra*/LETG and HETG has shown that these regions are dominated by narrow emission lines from highly ionized species, and radiative recombination continua, consistent with photoionization rather than thermal processes [243]. The presence of strong recombination features and line ratios further supports the dominance of photoionized gas. A growing body of evidence suggests that radiation pressure compression (RPC) plays a key role in shaping the density structure of the NLR, offering a self-consistent explanation for its observed stratification and ionization structure [46,63,548].

However, X-ray observations have lagged behind in terms of spatial and spectral coverage, hampered by the modest sensitivity and photon collection of current facilities. This is particularly striking when compared to the wealth of high-resolution optical, infrared, and radio data now available, which has enabled detailed morphological and kinematic studies of the NLR and host galaxy environment. The gap in X-ray coverage has limited our ability to fully assess the impact of AGN photoionization and

outflows, especially in the band where the AGN radiation field most directly influences the gas conditions. Additionally, soft X-ray data are crucial for disentangling the contribution of photoionized gas from that of collisionally ionized plasma, which can arise from star formation activity and shock-heated regions associated with AGN-driven or galactic-scale outflows. Properly resolving and characterizing these components is essential for a complete understanding of the physical processes at work in the circumnuclear environment.

The proposed AXIS program aims to systematically characterize the soft X-ray emission in a complete, flux-limited sample of X-ray obscured Seyfert galaxies. These objects, selected to ensure that the bright central continuum is strongly suppressed in the soft X-rays, are ideally suited to reveal the extended photoionized emission that traces the NLR. The sample is already well characterized at optical, infrared, and radio wavelengths, enabling a robust multiwavelength analysis of the spatial and spectral properties of the circumnuclear gas. With AXIS's high throughput and optimized soft X-ray response, exposure times can be significantly reduced compared to *Chandra* for equivalent signal-to-noise, allowing a large number of sources to be observed with moderate integration. This efficiency gain will enable us to construct a statistically significant dataset that covers a broad range of AGN luminosities, host galaxy types, and environmental conditions. It also compensates for the lower spatial resolution of AXIS compared to Chandra, enabling the detection of lower surface brightness features and dramatically expanding the number of sources that can be observed. Moreover, its fine pixel sampling minimizes the need for sub-pixel reconstruction techniques, resulting in more efficient and robust imaging of extended emission. This enhances the fidelity of spatially-resolved structures in the NLR, even at modest exposure times, and strengthens AXIS's capability to recover detailed morphology in synergy with high-resolution multiwavelength datasets.

| Name | RA (J2000) | Dec (J2000) | z | Distance (Mpc) | log $N_H$ |
|---|---|---|---|---|---|
| MRK573 | 25.9907 | 2.3498 | 0.0172 | 71.80 | >24.5 [60] |
| NGC835 | 32.3525 | -10.1358 | 0.0128 | 55 | 23.66 [47] |
| NGC1068 | 40.6696 | -0.0133 | 0.0038 | 16 | >25 [47] |
| NGC1167 | 45.4265 | 35.2057 | 0.01653 | 72 | 21.50 [12] |
| NGC1386 | 54.1924 | -35.9994 | 0.003 | 12 | >24.5 [47] |
| ESO428-G014 | 109.1301 | -29.3247 | 0.0057 | 23 | 23.3 [194] |
| NGC3081 | 149.8730 | -22.8263 | 0.0081 | 25 | 23.8 [495] |
| NGC3079 | 150.4908 | 55.6798 | 0.0037 | 16 | 24.56 [495] |
| NGC3393 | 162.0977 | -25.1621 | 0.0125 | 61 | 24.4 [495] |
| NGC4102 | 181.5963 | 52.7109 | 0.0028 | 17 | 24.14 [495] |
| NGC4945 | 196.3645 | -49.4682 | 0.0019 | 4 | 24.8 [495] |
| NGC5135 | 201.4336 | -29.8337 | 0.0137 | 83 | >24 [495] |
| NGC5252 | 204.5665 | 4.5426 | 0.0231 | 112 | 22.7 [93] |
| NGC5643 | 218.1699 | -44.1746 | 0.004 | 16 | 25 [495] |
| NGC5728 | 220.5997 | -17.2532 | 0.0093 | 43 | 24.14 [495] |
| ESO137-G034 | 248.8088 | -58.08 | 0.009 | 42 | 24.36 [495] |
| IC5063 | 313.0098 | -57.0688 | 0.0114 | 52 | 23.56 [495] |
| NGC7212 | 331.7554 | 10.2311 | 0.0266 | 120 | >22.9 [47] |

**Table 5.** The proposed sample. All information taken from NED, except log $N_H$.

Heavily obscured AGN have two key benefits for studying the extended X-ray emission. First, according to the Unified Model, we observe these AGN from the side - through the torus - so the outflow cones are oriented nearly in the plane of the sky, enhancing their apparent size. Second, the obscuring gas acts like a natural coronagraph, dimming the bright central source and making surrounding faint X-ray features easier to detect. In light of this we used the Swift/BAT BASS catalog (12–150 keV) of obscured

AGN [319], complemented by a similar selection based on [OIII] luminosity [504] to extract the brightest heavily obscured AGN, all with column densities log$N_H$ >23.3. In Table 5 we report the list of sources to be observed, ordered by RA. All sources of our sample have extensive multiwavelength coverage. Cross-comparison with existing narrow-band imaging, IFU data, and high-resolution radio maps will enable a comprehensive analysis of the interplay between AGN radiation, outflows, and host galaxy structure. This coordinated approach will provide a coherent picture of the excitation mechanisms at work in the NLR, test models of gas confinement, and constrain the energetics and incidence of AGN feedback on kiloparsec scales.

**Exposure time (ks):** The observing strategy involves 10 ks snapshot observations of 18 heavily-obscured Seyfert galaxies, for a total request of 180 ks.

**Observing description:** We simulated AXIS observations of four representative targets from our sample, shown in Fig. 36. These images were performed using the SIXTE software package, assuming 10 ks exposures and based on archival *Chandra* data, using the observed soft X-ray morphologies, spectral shapes, and fluxes. Despite the shorter exposure time and lower angular resolution compared to Chandra, the AXIS simulations clearly recover the extended soft X-ray emission. The resulting images demonstrate that AXIS will not only detect but also characterize the diffuse photoionized structures across a wide range of targets, confirming the feasibility and scientific promise of this program.

**Joint Observations and synergies with other observatories in the 2030s:** NewAthena, all major multi-wavelength facilities with high spatial resolution.

**Special Requirements:** None

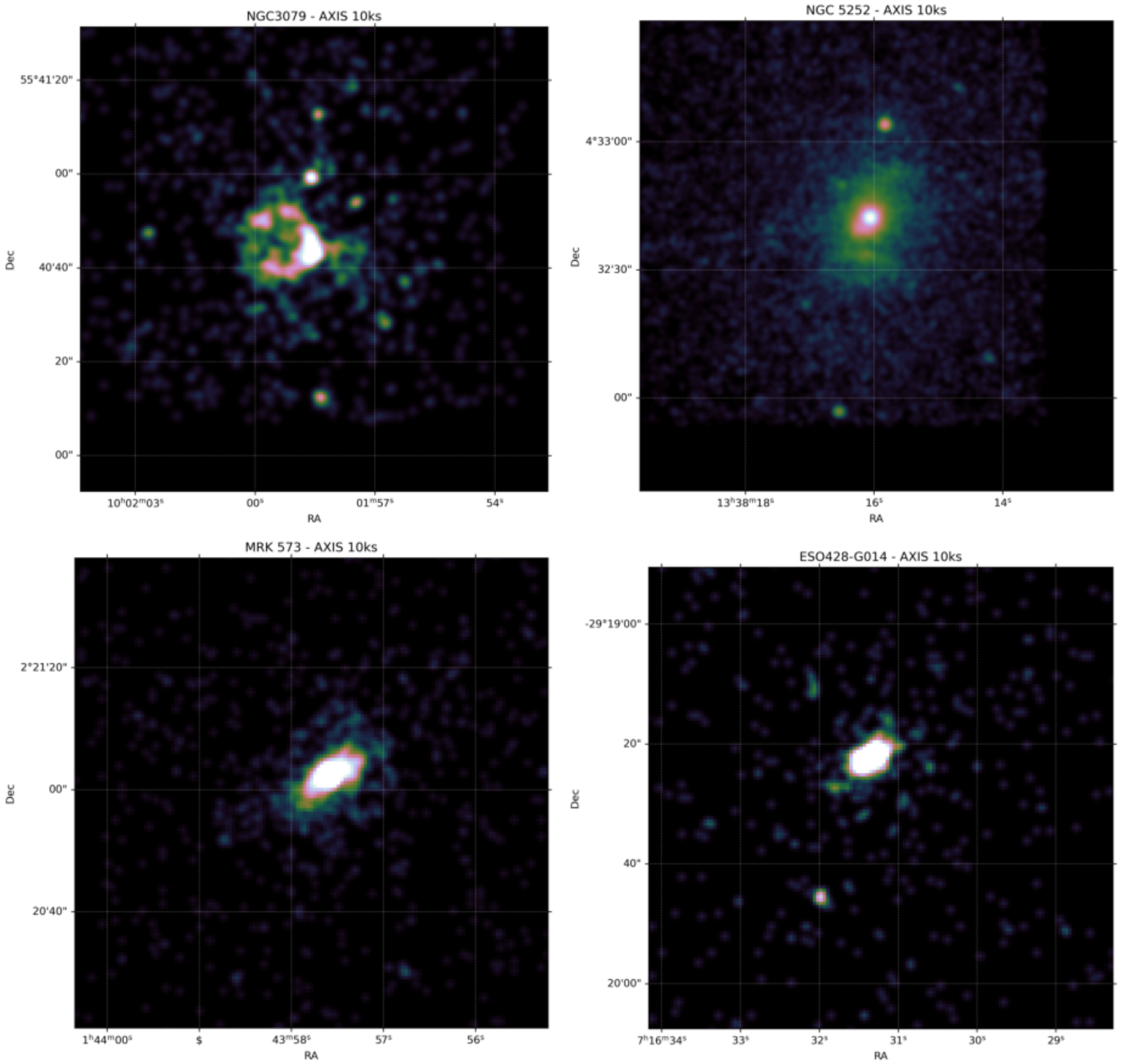

**Figure 36.** Simulated AXIS images for representative targets of the proposed sample of obscured Seyfert galaxies. The extended soft X-ray emission (0.5-2.5 keV) is well resolved with only 10ks also for the fainter sources.

*27. X-ray emission from Super-Eddington accreting AGN*

**Science Area:** Active Galactic Nuclei

**First Author: Alessia Tortosa** (INAF-OAR; alessia.tortosa@inaf.it)
**Co-authors: Stefano Bianchi** (Università Roma Tre; stefano.bianchi@uniroma3.it), **Pierre-Olivier Petrucci** (Université Grenoble Alpes; pierre-olivier.petrucci@univ-grenoble-alpes.fr) **Riccardo Middei** (ASI-SSDC; riccardo.middei@ssdc.asi.it) **Enrico Piconcelli** (INAF-OAR; enrico.piconcelli@inaf.it) **Claudio Ricci** (Universidad Diego Portales; claudio.ricci@mail.udp.cl) **Roberto Serafinelli** (Universidad Diego Portales; roberto.serafinelli@mail.udp.cl) **Luca Zappacosta** (INAF-OAR; luca.zappacosta@inaf.it)

**Abstract:** AGN accreting above their Eddington limit are so far poorly understood, and it is still largely debated what the physical properties of the accretion flow from the direct relativistic environment of the highly accreting supermassive black holes (SMBHs) are. Accretion above the Eddington limit is invoked as a key ingredient for the launching of powerful outflows and winds capable of controlling the growth and evolution of the AGN and the host galaxy, and to be the mechanism allowing the SMBHs we observe in most massive galaxies to reach masses of $M_{\rm BH} \sim 10^6 - 10^9 M_{\rm BH}$. We want to exploit the unique combination of spatial and spectroscopic resolution in X-rays of the AXIS satellite to study the X-ray emission of the Super Eddington Accreting Massive Black Holes (SEAMBHs) sample [159,598], the best sample of bona-fide super-Eddington sources available, which contains exclusively objects with black hole masses estimated via reverberation mapping, to investigate the innermost regions of these extreme sources. AXIS will provide unprecedented constraints on the spectral parameters of AGN in this poorly sampled regime which will be crucial to shed light on the X-ray emitting region of extremely high accreting AGN, on the powerful ionised X-ray outflows and winds emanating from AGN accreting in the super- and hyper-Eddington regime and on the feedback mechanism between the AGN and host-galaxy. AXIS, with its significantly larger effective area and excellent soft X-ray sensitivity, will allow us to connect nuclear outflows to galactic-scale outflows for the first time in highly accreting sources, thanks to synergies with multiwavelength facilities (e.g., VLT, JWST, ALMA). All this is pivotal to help our understanding on the fast growth of the first SMBHs in the early Universe. Indeed, highly accreting AGNs, particularly at low redshift, offer a benchmark for understanding the rapid growth of the first SMBHs in the early Universe, which remains a mystery to be solved. Moreover, being local, the study of these targets provides a perfect testbed to offer key insights into the super-Eddington systems powering the growth of SMBHs in the early Universe, as well as to explain the X-ray weakness of infant BHs and Little Red Dots.

**Science:** The main scientific questions driving this study are:

- How do AGN accrete?
- How does the change of accretion flow in SMBHs affect the X-rays emitting region?
- How does this affect the interaction between the AGN and its host galaxy?
- What is the driving mechanism of powerful outflows? Are nuclear AGN-driven UFOs and the galaxy-scale outflows connected?
- How did SMBHs grow to billion-solar-mass in less than 1 Gyr?

To tackle these questions, this study will focus on:

i) X-ray spectral characterization of hyper- and super-Eddington accreting AGNs. Thanks to the high sensitivity of AXIS, the proposed study will gather unprecedented quality spectra (each with $> 2100$ net-counts in the 2–10 keV energy range). With the proposed program, we will measure the photon-index of the primary X-ray power-law ($\Gamma$). This will allow to shed light on evolution of the $\Gamma - \lambda_{\rm Edd}$ relations in this extreme accretion regime. Indeed, while this relation has been widely explored in the sub-Eddington regime (e.g. [84,350,568]), there is a clear lack of statistics in the super-Eddington regime (see left panel of Fig 37). Moreover, we will investigate the origin of the soft excess at this extremely high accretion regime.

ii) Variability analysis of the targets. The AXIS observations will provide high-quality X-ray light

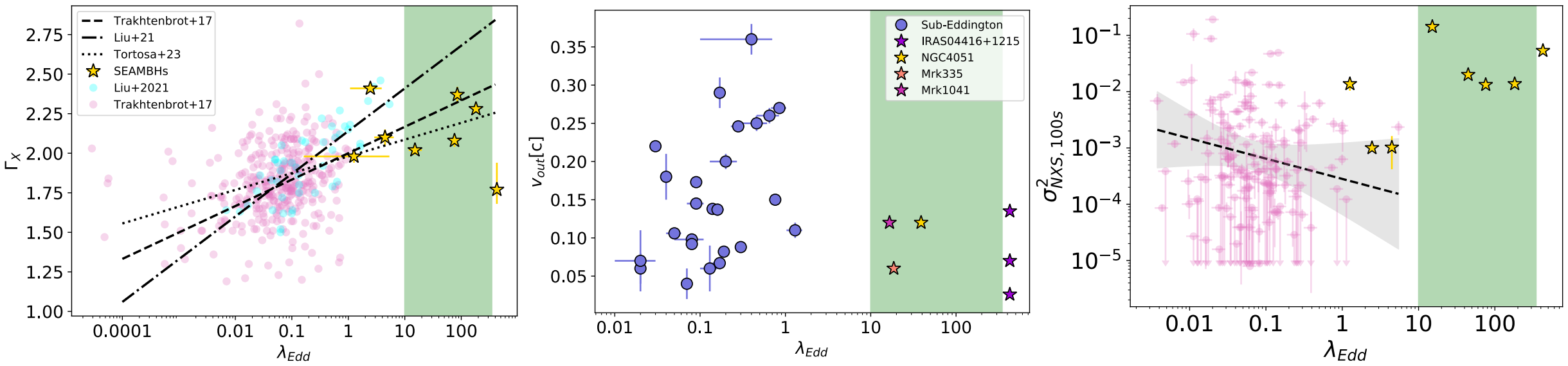

**Figure 37.** Left Panel: X-ray power-law photon index ($\Gamma$) vs. Eddington Ratio ($\lambda_{\rm Edd}$) relation for the sample of SEAMBHs sources (yellow stars, dotted line, [159]) together with the values for different sub-Eddington from [568] (BASS sample, pink dots, dashed line) and from [350] (cyan dots, dotted-dashed line). Central Panel: $\lambda_{\rm Edd}$ vs. velocity of the outflows ($v_{\rm out}$) taken from the literature for a sample of sources accreting in the sub-Eddington regime (blue dots, from [359] and references therein) and for super-Eddington sources (stars: IRAS 04416+1215 from [564], NGC 4051 from [559], Mrk 335 and Mrk 1041 from [272]); Right Panel: $\lambda_{\rm Edd}$ vs. excess variance ($\sigma^2_{\rm NXS}$) relations for the BASS sample (pink dots, [565]) compared with the values obtained for the SEAMBHs from [566] (yellow stars). The relations are obtained from the $0.2-10$ keV light curves binned with 100 s. **In each panel, the green region represents the region of each parameter space we will populate with the proposed program.**

curve data that will allow us to study the temporal structure of the variations of the sources, as well as the relation between the excess variance ($\sigma^2_{NXS}$; [581]), a quantity used to describe the variability amplitude of a light curve) and the physical properties of the sources (i.e. $M_{\rm BH}$, $L_x$, $\lambda_{\rm Edd}$). The relation between $\sigma^s_{NXS}$ and the BH mass is a well known property of AGN [477,564], with larger $M_{\rm BH}$ corresponding to lower $\sigma^2_{NXS}$. It has been shown that the $M_{\rm BH}$-$\sigma^2_{NXS}$ relation flattens at lower $M_{\rm BH}$ suggesting changes in accretion processes or variability mechanisms and this flattening is stronger for objects with high accretion than for objects with low accretion [357]. This study offers the opportunity to extend the $M_{\rm BH}$-$\sigma^2_{NXS}$ correlation at higher Eddington ratio.

iii) Full characterization of the ionized absorbers in these super-Eddington sources exploiting the high sensitivity in the 0.3–2 keV X-ray range and the low background of AXIS. We will be able to confirm the presence of X-ray winds in these targets, measuring with high accuracy the parameters of the ionized absorbers, i.e., ionization parameter, column density, and velocity of the wind, using detailed photoionization models such as XSTAR and WINE [358]. Thus, we will have a complete picture of the nuclear outflows produced in the poorly sampled super-Eddington sources in the local Universe, which will help to probe the inner accretion flow structure, to determine whether super-Eddington AGN sustain their high accretion rates via radiation pressure, magnetic driving, or other mechanisms.

**Exposure time (ks):** Total exposure time is 200 ks. This program can be integrated with other AXIS observational campaigns.

**Observing description:**

- IRAS 04416+1215: z=0.089, $\log(\lambda_{\rm Edd}) = 2.67$, exposure $t_{\rm exp} = 20$ ks;
- SDSS J080101.41+184840.7: z=0.139, $\log(\lambda_{\rm Edd}) = 2.33$, $t_{\rm exp} = 20$ ks;
- SDSS J093922.89+370943.9; z=0.186, $\log(\lambda_{\rm Edd}) = 2.35$, $t_{\rm exp} = 20$ ks;
- Mrk 493: z=0.031, $\log(\lambda_{\rm Edd}) = 2.17$, $t_{\rm exp} = 20$ ks;
- IRASF 12397+3333: z=0.043, $\log(\lambda_{\rm Edd}) = 2.03$, $t_{\rm exp} = 20$ ks;
- Mrk 142: z=0.044, $\log(\lambda_{\rm Edd}) = 1.91$, $t_{\rm exp} = 20$ ks;
- PG 2130+099: z=0.062, $\log(\lambda_{\rm Edd}) = 1.49$, $t_{\rm exp} = 20$ ks;
- Mrk 382: z=0.034, $\log(\lambda_{\rm Edd}) = 0.81$, $t_{\rm exp} = 20$ ks;
- PG 0026+129: z=0.142, $\log(\lambda_{\rm Edd}) = 0.45$, $t_{\rm exp} = 20$ ks;

- PG 0953+414: z=0.234, $\log(\lambda_{\rm Edd}) = 0.39$, $t_{\rm exp} = 20\,{\rm ks}$;

**Joint Observations and synergies with other observatories in the 2030s:** NewAthena, VLT, ELT, LSST, VLA, VLBA, Euclid, Roman, JWST, ALMA.
**Special Requirements:** None

*28. Constrain the origin of the soft X-ray excess*

**Science Area: AGN, Quasar Structure, SMBH spin, Accretion Disk Properties**
**First Author: Pierre-Olivier Petrucci** (Institute of Planetology and Astrophysics of Grenoble, Grenoble, France, pierre-olivier.petrucci@univ-grenoble-alpes.fr)
**Co-authors:** Ehud Behar (Physics Department Technion, Haifa, Israel), Stefano Bianchi Università degli Studi Roma Tre, Roma, Italy)

**Abstract:**

A majority of Active Galactic Nuclei (AGN) show the presence of a soft X-ray emission, below 2 keV, in excess with respect to extrapolation of the 2-10 keV power law. The origin of this soft X-ray excess is a long-standing issue in our understanding of the AGN X-ray emission. Different models have been proposed to explain the soft X-ray excess but presently, only two appear viable: either blurred ionized reflection in the accretion disk, or thermal Comptonization in an optically thick ($\tau \gg 1$) and warm (kT$\sim$1 keV) plasma (the so-called warm corona). The reality can also be a mix of the two, with their relative importance varying from object to object. The high sensitivity of AXIS in the soft band will allow a detailed spectral and timing study of the soft X-ray excess component, which will allow us to test its origin 1) by constraining the spectral properties of the soft X-ray excess in a broad range of X-ray luminosity, and, especially at low luminosity 2) by studying the soft X-ray excess fast variability and time lags with respect to the high-energy emission

**Science:** The soft X-ray excess is a ubiquitous yet poorly understood feature in the X-ray spectra of active galactic nuclei (see e.g. [64,229] and references therein). This excess, observed as an additional component above the extrapolation of the hard X-ray power-law in the 0.1–2 keV range, challenges our understanding of AGN physics. Despite extensive observational and theoretical studies, its exact origin remains debated, with leading explanations including (see Fig. 38) thermal Comptonization in a warm corona (e.g., [129,157,366,465] and relativistically blurred and ionized reflection from the accretion disk (e.g., [127,286]).

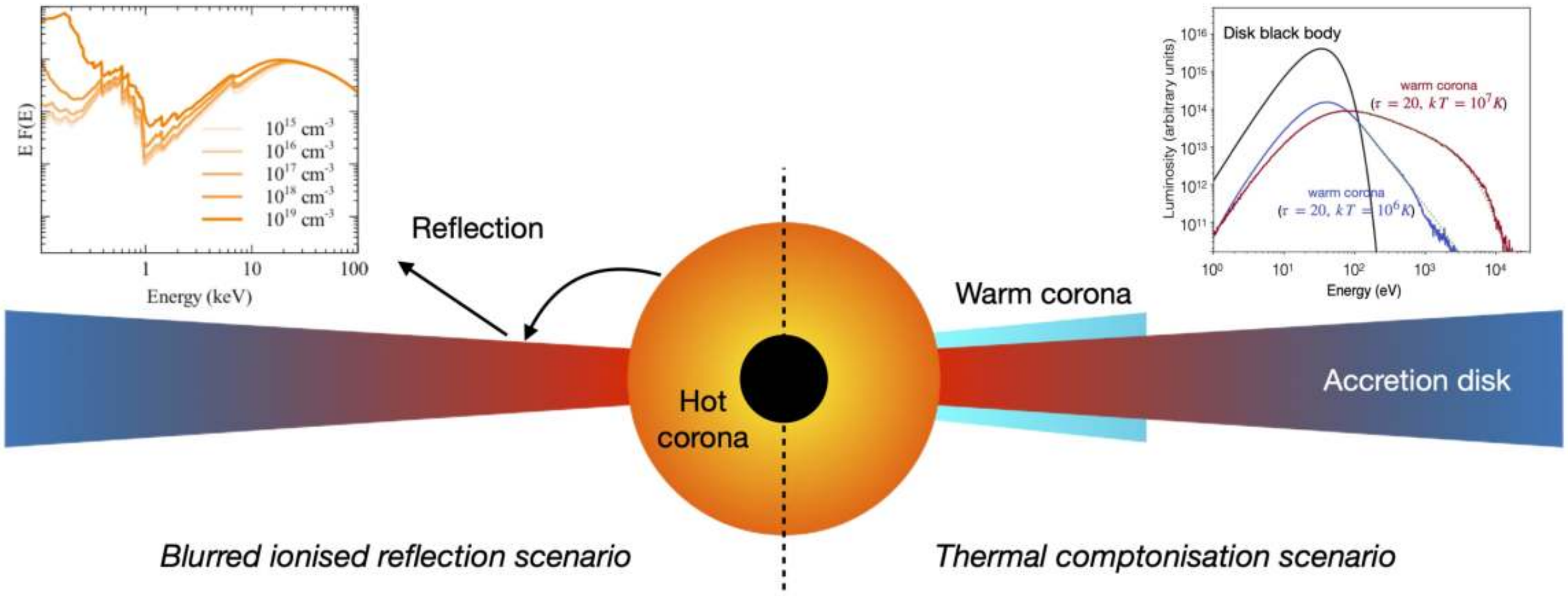

**Figure 38.** Sketchs of the two leading scenarii for the soft X-ray excess: Left: relativistically blurred and ionized reflection. The reflection spectra for different disk density in the inset are from [286]. Right: warm corona. The spectra for different warm corona temperature (but the same $\tau = 20$) are from [467].

Depending on its true origin, the soft X-ray excess can have several physical consequences for our understanding of the accretion physics in AGN. For instance, theoretical works suggested that the warm corona could be a powerful, extended, and optically thick plasma covering an almost

non-dissipative accretion disk, i.e., all the accretion power would be released mainly in the warm corona (e.g., [38,466,467,505]. If true, this result would be at odds with the commonly accepted behavior of standard optically thick, geometrically thin accretion flows, where the gravitational power is believed to be released in the deeper layers. The consequences would be important, and would have a direct impact on our understanding of the accretion disk vertical equilibrium, the expected spectral and reflected emission from such accretion flow, its capacity of producing outflows/jets. In this respect, if the warm corona is a permanent feature in AGN, then it could affect the growth of supermassive black holes and the regulation of the host galaxies by the AGN through feedback. On the other hand, if blurred reflection is at the origin of the soft excess (e.g. [205,206]), it has to originate from very close to the event horizon. This would imply that the accretion disk extends very close to the black hole, suggesting a highly spinning black hole. This would also favor a lamp post-type of geometry for the primary X-ray source. Interestingly, the soft excess also appears to evolve with luminosity and Eddington ratio (e.g., [296] and references therin), suggesting that its physical properties may be different in different accretion states.

AXIS will offer several advantages for studying the soft X-ray excess in active galactic nuclei (AGN): The high sensitivity of AXIS in the soft band will allow a detailed spectral and timing study of the soft X-ray excess component, which will allow us to test its origin:

- The improved sensitivity in the 0.3–2 keV soft X-ray range, combined with a very good control of the background, will significantly improve the signal-to-noise ratio for soft X-ray observations, either spectrally but also from a timing point of view. Indeed, the shortest variabilities (ks or less) are expected in the reflection scenario compared to the warm corona one. Also, in the reflection scenario, we also expect soft X-ray excess fast variability correlated, or with short lags, with the high-energy emission
- This will also be crucial for detecting faint excess components in AGN spectra and to probe the Soft Excess in a large range of luminosities (especially at low luminosity), black hole masses, accretion rates and redshifts (naturally limited however to $z \leq 1$ due to the Galactic absorption that will limit the soft X-ray excess detection). Comparing its properties in Seyfert galaxies, quasars, and low-luminosity AGN can reveal dependencies on these parameters.
- Very broad band SED, combining AXIS with UV facilities would be very useful to provide the AGN inner accretion flow, linking the accretion disk evolution (in UV) with the (warm and hot) coronae (see e.g. [296]).

**Exposure time (ks):** 300 ks spread between a sample of 20-30 high and low luminosity radio quiet AGN.
**Observing description:** This program can be integrated with other AXIS observational campaigns, the soft X-ray excess being a spectral components among the other spectral components of AGN. Targets with signatures of relativistic reflection will be of prime importance to test the reflection scenario. Low luminosity AGN are also very interesting targets, poorly observed up to now, because they require rather long exposure time with the current instruments (Chandra or XMM). A sample of AGN of different masses and luminosities (like the one of [192]) would be, eventualy, a great legacy of AXIS.
**Joint Observations and synergies with other observatories in the 2030s:**
**Special Requirements:** None

*29. High-Cadence Monitoring of M31* and the Nuclear Region of M31*

**Science Area:** Accretion

**First Author:** Stephen DiKerby (Michigan State University Dept. of Physics and Astronomy, dikerbys@msu.edu)

**Co-authors:** Shuo Zhang (Michigan State University Dept. of Physics and Astronomy), Jimmy Irwin (University of Alabama Dept. of Physics and Astronomy, Eureka Scientific)

**Abstract:**

Every large galaxy has at its heart a supermassive black hole (SMBH), related in observationally concrete but theoretically mysterious ways to the galaxy as a whole. In the past decade, two SMBHs in particular (M87* and Sgr A*) have been studied not only as unseen gravitational forces at the heart of galactic cores but also as explicitly resolved objects in their own right. Beyond these two SMBHs, few other systems can be explored at sub-Bondi radius scales, among them M31*, the SMBH at the core of the Andromeda galaxy (M31). As the closest SMBH outside of our own galaxy, M31* is a unique laboratory for investigating non-AGN SMBH in a field spiral galaxy. The X-ray emission from M31* can only be studied with arcsecond-resolution X-ray observations to differentiate the emission from M31* from the flux from three other nearby nuclear sources.

Over the past 20 years, Chandra has conducted sporadic and uneven monitoring of the nuclear region of M31. The 20-year light curve for M31* ([153,343]) shows periods of quiescence with occasional flares, but the exact nature, frequency, and duration of these flares are not well constrained. With its $\sim$arcsecond spatial resolution, impressive effective area, and wide energy band, AXIS is uniquely suited to pursue a more comprehensive monitoring campaign of M31* and the surrounding M31 nuclear region. With monitoring exposures as brief as 1.5 ks once per week, AXIS can construct a detailed light curve for M31*, identifying several new X-ray flares per year and conclusively solving the question of the origin of X-ray flares from our nearest neighboring SMBH.

**Science:**

Sagittarius A* and M87*, the two SMBHs directly imaged by the Event Horizon Telescope [170,172], border opposite ends of SMBH luminosity and mass. Sgr A* has a mass of $4 \times 10^6 M_\odot$ [219,406] and does not power an AGN, instead resting in a quiescent state with occasional radio, IR, or X-rays flares [28,217,629]. M87*, on the other hand, has a mass over three orders of magnitude greater ($6 \times 10^9 M_\odot$ [214]) and powers an AGN blasting a jet deep into intergalactic space [156,532]. Resolving these systems at the finest scales has returned substantial advances in testing relativity and in the study of accretion in SMBH systems, and continued monitoring of these systems has illuminated the link between accretion at the smallest scales and high-energy radiation [13].

Few other SMBHs can be directly imaged at scales similar to the Bondi radius ($R_B \approx 2GM/v_s^2$ for $M$ the mass of the object and $v$ the characteristic speed or sound speed of the ambient medium), a characteristic scale of accretion onto a SMBH. One of the less-studied systems among them is M31*, the SMBH of the Andromeda galaxy, with a mass of $\approx 10^8 M_\odot$[57] and an angular Bondi radius of $\sim 5''$ or $\sim 15$ pc. Being $\approx 100\times$ more distant than Sgr A*, M31*'s event horizon is not directly observable, but the immediate accretion environment around it can be observed with X-ray observations with arcsecond spatial resolution.

The observation of M31* is complicated by the crowded inner nucleus of the bulge of M31. There are three other X-ray point sources within $2''$ of the SMBH [207,344], which are all likely X-ray binaries with substantial X-ray variability. Measuring the X-ray flux of M31* requires approximately arcseconds resolution imaging capabilities currently only accessible by Chandra-HRC or -ACIS (with sub-pixel reconstruction) or in the future with AXIS, as the overlapping PSFs of M31* and the three other nuclear sources require 2D image modeling at arcsecond scales to disentangle the emission from each object.

In a recent publication [153], expanding the dataset and analysis in [343], we demonstrated how persistent monitoring of the nuclear region of M31 by Chandra can investigate the X-ray emission from M31* despite the challenges of resolving the X-ray emission from the SMBH system. In 2011, [343] examined the 2001-2010 light curve of M31* via 2D spatial reconstruction, finding a dramatic X-ray flare in 2006 that catapulted the quiescent $0.5 - 8.0$ keV X-ray flux of M31* from $< 1$ counts/ks to over 70 counts/ks. Since that 2006 flare, we showed in [153] that M31* has remained in a persistent high state to the present day, with elevated flux of $\approx 5$ counts/ks. We also detected another X-ray flare in 2013, demonstrating that M31* has an X-ray duty cycle of $\approx 1\%$ and likely has had other flares unobserved with Chandra in the intervening years.

Interestingly, the 2013 and 2006 M31* flares do not exhibit substantially different hardness ratios compared to M31* in its post-2006 status quo. Indeed – when fitted with a thermal plasma or power law model – the 2013 flare shows X-ray emission much more similar to Sgr A* in a quiescent, non-flaring state than with Sgr in a flaring state. This surprising result suggests that the transition between non-flaring and flaring states occurs via different mechanisms or pathways for M31* compared to Sgr A*. The identification of additional X-ray flares from M31*, of which only two are presently known, is a prerequisite for understanding the mechanism of these flares, and AXIS is the only future X-ray probe with the capability to discover and investigate these flares.

Figure 39 shows a 1.5 ks AXIS observation of the nuclear region of M31 simulated with the `sixte` package based on the typical fluxes of each source within a few arcseconds of M31*, compared to a Chandra-HRC observation with the same duration. In both Chandra and AXIS, the four nuclear sources are so close that direct source detection using a routine like `ximage` would not yield distinct detections; instead, 2D image reconstruction is necessary to model the fluxes of each source simultaneously. Despite having a slightly coarser on-axis PSF compared to Chandra-HRC, AXIS's substantially larger effective area enables it to measure the flux of each nuclear source with precision on the order of increased photon statistics alone.

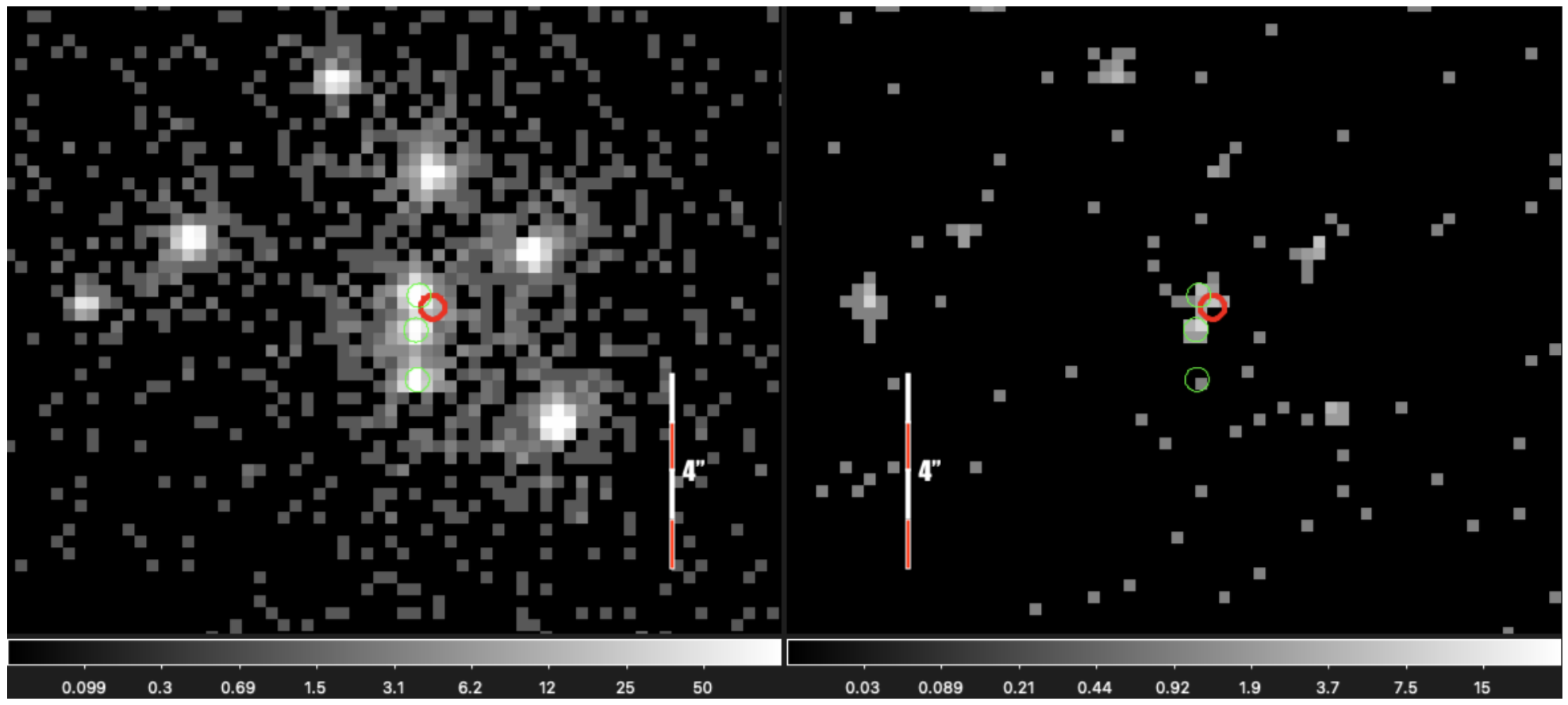

**Figure 39.** A simulated 1.5 ks AXIS observation (left) of the nuclear region of M31 compared to a 1.5 ks Chandra-HRC observation. M31* is highlighted in red in both observations, along with the three other nuclear sources (green circles) that contribute contaminant photons.

AXIS's unique capabilities, in particular its fine PSF and its substantial effective area, will enable several lines of research in the nuclear region of M31 with only brief monitoring observations at regular intervals. AXIS observations will achieve several unique science goals in substantially less exposure time compared to Chandra. These science returns include:

- The monitoring of the X-ray variability of M31* and the detection of X-ray flares thought to occur with an $\approx$ 1% duty cycle, allowing for the construction of a conclusive theoretical model for the accretion and energetic processes at M31*
- The construction of light curves for the other three nuclear sources (S1, SSS, N1) for fortuitous observations where all three are in low states, like the 2013 flare, allowing spectral modeling of M31* without contamination by other sources of photons
- Monitoring over twenty other X-ray sources in the bulge of M31 with a single pointing, building a long-running survey of X-ray binaries in M31 as a fortuitous secondary science benefit.

By extending the X-ray light curve of M31*, AXIS will test competing emission models for the environment immediately surrounding M31*. Previous observations with Chandra did not accumulate sufficient counts to strongly determine whether a non-thermal or thermal emission model is appropriate for M31*, nor are the hardness ratios modeled in [343] and [153] detailed enough to establish whether the flares detected in 2006 represent transitions from one emission mechanism to another. AXIS's substantially larger effective area means that the summed exposure of this campaign of observations will enable strong discrimination between non-thermal and thermal models for X-ray emission from M31*.

Furthermore, a more regular and comprehensive light curve for M31* will constrain the variability timescale of the X-ray emission from the SMBH, constraining the size of the emission region. The detection by AXIS of another event like the 2013 flare noted in [153] in particular will allow for much more detailed spectral modeling with little contamination from the other nuclear sources, and a direct test of whether M31*'s X-ray emission is thermal or non-thermal in origin.
**Exposure time (ks):** 80 ks/year. 1.5 ks weekly
**Observing description:**

- Primary Target: M31* at (J2000) RA = 00:42:44.302, Dec = +41:16:08.173
- Exposure: 1.5 ks, weekly, except when sun-constrained between March and May of each year

**Joint Observations and synergies with other observatories in the 2030s:** The "double nucleus" of M31 has also been extensively characterized in optical and infrared light, so it would be feasible to conduct a joint monitoring campaign with current and future space-based observatories like JWST or SPHEREx to constrain possible joint IR-X-ray variability in M31*.
**Special Requirements:** None

*30. Resolving Hot Gas inside Bondi Radius of SMBHs*

**Science Area:** Accretion

**First Author:** Ka-Wah Wong (SUNY Brockport, kwong@brockport.edu)
**Co-authors:** Helen Russell (University of Nottingham, helen.russell@nottingham.ac.uk), Jimmy Irwin (University of Alabama, jairwin@ua.edu)

**Abstract:**

Hot gas around a supermassive black hole (SMBH) should be captured within the gravitational "sphere of influence", characterized by the Bondi radius. Deep Chandra observations have spatially resolved the Bondi radii of five nearby SMBHs that are believed to be accreting in hot accretion mode. In contrast to earlier hot-accretion models that predicted a steep temperature increase within the Bondi radius, none of the resolved temperature profiles exhibit such an increase. The temperature inside the Bondi radius appears to be complex, indicative of a multi-temperature phase of hot gas with a cooler component at about 0.2–0.3 keV. The density profiles within the Bondi regions are shallow, suggesting the presence of strong outflows. These findings may be explained by recent realistic numerical simulations, which indicate that large-scale accretion within the Bondi radius can be chaotic, with cooler gas raining down in some directions and hotter gas outflowing in others. With an angular resolution similar to Chandra and a significantly larger collecting area, AXIS will collect enough photons to map the emerging accretion flow within and around the "sphere of influence" of a large sample of active galactic nuclei (AGNs). AXIS will reveal transitions in the inflow that ultimately fuel the AGN, as well as outflows that provide feedback to the environment.

**Science:**

Supermassive black holes (SMBHs) are ubiquitous in the centers of galaxies [265]. Accretion processes determine the growth of SMBHs and feedback mechanisms play a critical role in the evolution of their host galaxies [178,404]. Understanding how SMBHs are fueled by their surroundings and how matter and energy are transported back into the environment is crucial for comprehending the cosmic ecosystem.

Immersed in the diffused hot gas of the host galaxy, the gravitational potential of the black hole dominates the thermal energy of the hot gas inside the "sphere of influence," which is defined by the characteristic Bondi radius, $R_B = 2GM_{\rm BH}/c_s^2$, where $M_{\rm BH}$ is the mass of the black hole and $c_s$ is the sound speed of the gas at a distance far away from the black hole [77]. Although the Bondi accretion model is an oversimplification, it serves as a baseline for comparing theoretical and observational accretion models. The inferred Bondi accretion rate is often used in studies linking accretion and AGN feedback [e.g., 15,512] and in cosmological simulations [e.g., 472]. More importantly, directly studying the hot gas inside and around the Bondi radius allows us to probe how the gas is captured and accreted onto the black hole [see, e.g., 29,68,82,208,251,330,434,485,611].

Most SMBHs are low-luminosity AGNs (LLAGNs), radiating well below the Eddington limit ($L_{\rm Edd}$) [265]. While luminous AGNs typically radiate at $\sim$10% $L_{\rm Edd}$, LLAGNs are much fainter, with many emitting at $\sim 10^{-5} L_{\rm Edd}$ or even lower at $\sim 10^{-8} L_{\rm Edd}$. Bright AGNs are thought to accrete via a classical optically thick, geometrically thin accretion disk, whereas LLAGNs are believed to accrete via a hot, optically thin, geometrically thick flow, sometimes referred to as a radiatively inefficient accretion flow (RIAF).

For nearby low-luminosity active galactic nuclei (LLAGNs) such as Sgr A$^*$ and NGC 3115, hot gas density within $R_B$ can be determined through X-ray observations, allowing us to estimate the available material to fuel the supermassive black holes (SMBHs) within their "sphere of influence". The Bondi accretion rate is given by $\dot{M}_B = 4\pi\lambda R_B^2 \rho c_s$, where $\rho$ is the gas density near the Bondi radius and $\lambda$ is the adiabatic index of the hot gas [29,77,609]. However, the observed X-ray luminosity of these LLAGNs is

orders of magnitude smaller than what would be expected from the Bondi accretion rate with a ∼10% radiative efficiency. Thus, it is not simply that they are starved for gas to explain such low luminosity.

Extensive theoretical work has been conducted to understand why LLAGNs are under-luminous. One explanation is that during hot accretion, most of the energy is carried by ions and is advected into the black hole before it can be radiated (advection-dominated accretion flows or ADAFs; [270,420,486]). Another possibility is that much of the gas never reaches the black hole, instead circulating in convective eddies (convective-dominated accretion flows or CDAFs; [5,419,484]) or being expelled in strong outflows (advection-dominated inflow-outflow solutions or ADIOS; [49,70]). These models can collectively be classified as radiatively inefficient accretion flow (RIAF) models.

Simulations of hot accretion flows confirm the presence of significant outflows, which prevent most of the inflowing material from reaching the black hole [274,341,549]. Numerical hydrodynamic simulations have been conducted to model hot accretion flows, which generally reproduce analytic self-similar solutions during the time-averaged "steady-state" flow in the simulations [274,549]. These simulations confirm that strong winds or outflows can be driven during accretion [341,549]. More realistic magnetohydrodynamic simulations have been conducted, including magnetically arrested disk (MAD) models and other magnetic field-dominated accretion flows that can drive powerful relativistic jets [273,275,403,554]. Additionally, the roles of cooling and thermal conduction have been explored within these models [212,528,529]. magnetohydrodynamic (MHD) simulations suggest that strong magnetic fields may play a key role in regulating accretion, sometimes forming magnetically arrested disks (MADs) that drive powerful relativistic jets [273,275,403,554]. The roles of cooling, conduction, and other effects have also been explored [212,528,529]. A more comprehensive review of these hot accretion models can be found in [619].

Observationally, measuring the temperature and density of hot gas inside and around the Bondi radius provides a powerful test for these models. While all RIAF models predict that gas temperature increases toward the black hole, the density profile varies significantly between models. Yet, observational constraints have proven to be challenging due to the small angular scale of the Bondi radius of black holes. Even for the closest SMBHs, the angular sizes of the Bondi radii are on the order of a few arcseconds or less (see Figure 40 below).

With *Chandra*'s sub-arcsecond resolution, several Bondi regions have now been resolved, including in SgrA$^*$, NGC3115, and M87 [29,507,509,510,609]. These observations provide some of the first direct constraints on the hot gas structure in LLAGNs. For example, NGC 3115 has a Bondi radius of ∼2.4–4.8 arcseconds, making it one of the best targets for studying hot accretion [609,610]. Surprisingly, instead of a temperature increase, the gas appears to be multiphase, with components at ∼1 keV and ∼0.3 keV. This suggests that some of the gas is cooling and feeding a rotating disk rather than free-falling onto the black hole.

Similar findings have emerged from studies of M87, which has one of the largest Bondi radii (∼5 arcsec) and is among the most gas-rich LLAGNs [509,510]. M87 shows no temperature increase within $R_B$, and its density profile suggests significant outflows. These results challenge the idea that Bondi or ADAF accretion models are fully applicable. Instead, they indicate that inflows and outflows coexist, creating a complex accretion structure.

Additional evidence for multi-temperature gas comes from recent *Chandra* observations of NGC1600 and M84, both of which have resolved Bondi regions [43,507]. In NGC1600, a clear multi-temperature phase is seen, while in M84, no such evidence is found—perhaps due to uniform gas temperature or the presence of even cooler gas below 0.2 keV that is undetectable with *Chandra*. In any case, multi-temperature gas appears to be a common feature inside Bondi radii [611].

These results suggest that classical hot accretion models may need revision, particularly regarding temperature and density structures. Cooling and inflow/outflow interactions likely play a larger role

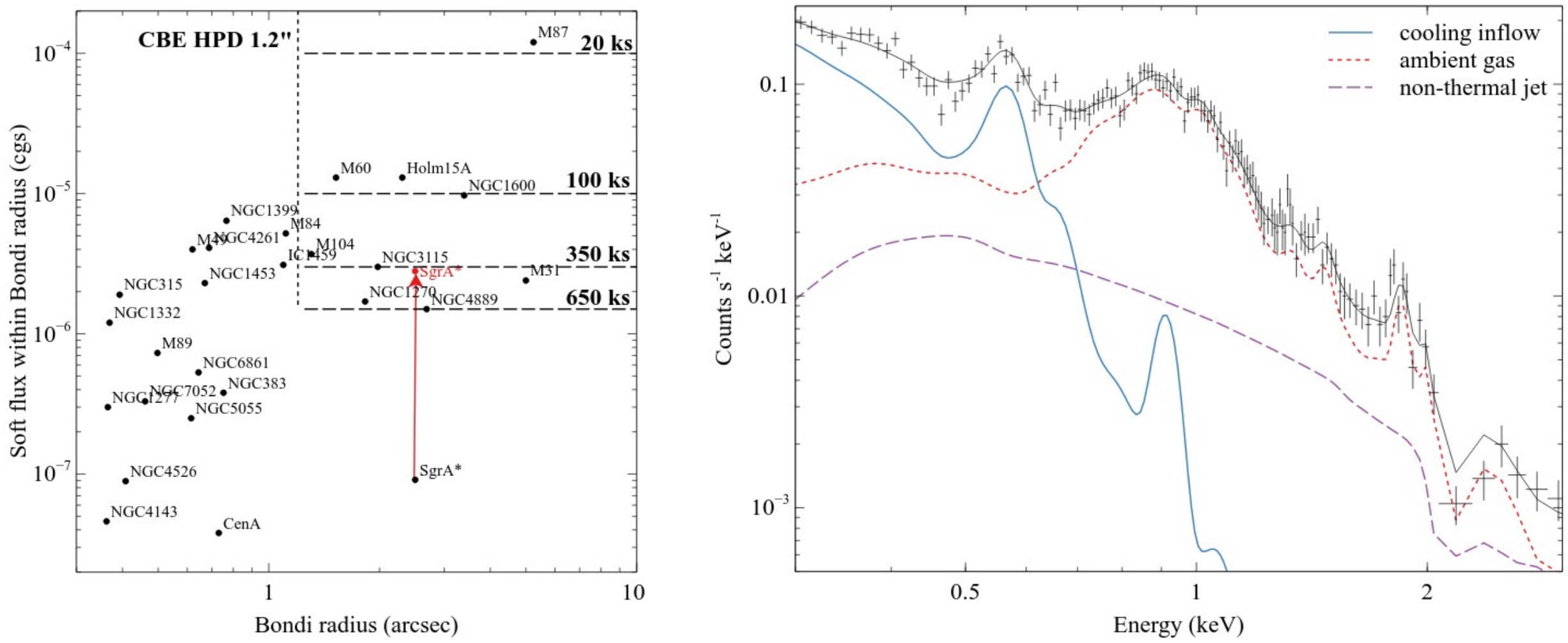

**Figure 40.** (**Left**) Distribution of the sizes of the Bondi radius[2] for a sample of SMBHs [611]. The vertical line separates samples that can be resolved with AXIS's current best estimate HPD of 1.2 arcsec, illustrating its ability to measure the Bondi radius in a significant number of sources. Exposure times required for at least 3000 counts in the soft band 0.5–1 keV, except Sgr A* where the estimation was based on hard flux (highlighted in red) due to high absorption in soft X-rays. The total number of counts across the full energy range would be significantly higher. (**Right**) Simulated AXIS spectrum of M87 with 100 ks exposure for a region with the size of AXIS's HPD with a total of ∼10,000 photon counts in 0.3–10 keV (including both thermal gas emission and non-thermal emission from the AGN and jet) [611].

than previously thought. However, deeper X-ray observations are required to confirm these findings and explore whether even cooler gas phases exist.

With AXIS's higher sensitivity to soft X-rays, even cooler gas within Bondi radii will be detected, providing a clearer picture of accretion and feedback in LLAGNs [491,611]. By extending the sensitivity range below 0.5 keV, AXIS will enable us to refine accretion models and gain a deeper understanding of the interplay between inflows, outflows, and black hole fueling.

**Exposure time (ks):** 1,100 ks

M87, NGC 1600, Holm 15A, M60: 100 ks each

M31, NGC 3115: 350 ks each

**Observing description:**

Current studies with *Chandra* are limited by photon statistics. With the current baseline spatial resolution of 1.5 arcsec and the current best estimate of 1.2 arcsec (Half Power Diameter, HPD), AXIS will have a resolution very similar to *Chandra* but with a significantly larger photon-collecting area. This will enable us to resolve and map the hot accretion flow in and around the Bondi radius of approximately 10 nearby AGNs (Figure 40), revealing how gas transitions from inflow to fuel the AGN and how outflows along the jet axis regulate accretion.

With the planned exposures (Figure 40), AXIS will collect at least 3,000 counts in the soft band (0.5–1 keV) within $R_B$, providing density and temperature measurements with less than 10% uncertainty. The

---

[2] It should be noted that the concept of Bondi radius is not well defined in a realistic system due to the multi-temperature nature of the hot gas and the nontrivial outer boundary condition. Nevertheless, we can still estimate the size of the Bondi radius observationally to characterize the rough size of the accretion region. However, the sizes shown here are only approximate, as we assume the hot gas temperature to be 0.5 keV, which underestimates the Bondi radii for cooler systems.

total number of counts across the full energy range will be significantly higher. Multi-temperature gas has already been detected in three LLAGNs, providing critical constraints on accretion and feedback models. Expanding the sample with higher photon statistics will enable us to systematically investigate multi-temperature gas, which appears to be a common phenomenon. The sample includes a variety of LLAGNs, from quiescent (NGC 3115) to moderately active (NGC 1600) to highly active with a relativistic jet (M87), allowing us to study accretion and outflows in different environments and search for correlations with jet power.

The best-resolved targets, such as NGC 3115 and M87, will enable more detailed comparisons with numerical simulations that predict cooler gas in inflow regions and hotter gas in outflow regions (e.g., Figure 2 in [611]). In M87, a 100 ks exposure within its Bondi radius will yield $\sim$10,000 counts in the 0.3–10 keV range, allowing a high-resolution temperature map to study the multi-temperature gas in unprecedented detail. In contrast, *Chandra* would collect only about 170 counts in the same region with the same exposure.

Hot gas around and beyond the Bondi radius will also be mapped at the highest possible resolution, helping us understand large-scale accretion and feedback to the environment. This will also benefit LLAGNs with slightly unresolved Bondi radii, significantly increasing the sample size.

For M87 and Sgr A*, EHT has resolved plasma at event-horizon scales [171,173], but the connection between Bondi-scale inflows and SMBH feeding remains uncertain [251,289]. Numerical simulations spanning $10^3$–$10^6 r_g$ reveal structures such as turbulent torii, shocks, and filaments. The Bondi regions observed with AXIS will provide critical constraints for linking large-scale accretion with event-horizon-scale flows, improving our understanding of how SMBHs are fueled.

**Joint Observations and synergies with other observatories in the 2030s:**

**Special Requirements:**

The only target with a potential pile-up issue is M87, and only if AXIS performs worse than Chandra in this regard. Given AXIS's effective area is 5–10 times larger than Chandra's, the frame time would need to be at least 10 times shorter than Chandra's 0.4 s frame time in 1/8 subarray mode–i.e., less than 40 ms. With the current design including a potential window mode with a 10 ms frame time, this should be sufficient to mitigate pile-up for this target.

*31. Shedding light on fundamental physics (including dark matter) with AXIS*

**Science Area:** AGN/Dark matter/Fundamental Physics

**First Author: Júlia Sisk Reynés**, Center for Astrophysics | Harvard& Smithsonian, julia.sisk_reynes@cfa.harvard.edu

**Co-authors: Thong T. Q. Nguyen** (Stockholm University): thong.nguyen@fysik.su.se ; **James Matthews** (University of Oxford): james.matthews@physics.ox.ac.uk

**Abstract:**

We highlight some possible avenues to probing fundamental physics –specifically, in the area of physics beyond the Standard Model searches– using *AXIS*. We discuss several applications related to the study of Weakly-Interacting Slim Particles (WISPs), including axion-like particles, axion dark matter, and the sterile neutrino. We also introduce possible searches for other dark matter candidates: the dark photon, $B-L$ vector dark matter, and primordial black holes. *AXIS*'s large collecting area, PDF stability off-axis, and spatial resolution make it an excellent observatory to probe an array of theoretical models beyond the Standard Model of particle physics.

**Science:**

Axion-like particles

The Standard Model of particle physics is the current leading theoretical framework describing the fundamental particles and their interactions. This framework is extremely adept at describing a wide range of phenomena in nature, as well as many experimental results. Nevertheless, the Standard Model is challenged by several so-called 'tensions', including: the lack of understanding of the dark matter and dark energy components in the universe; and the so-called 'strong charge-parity (CP) problem' – the absence of CP violation by the strong nuclear force, which clashes with the predictions of the theory.

Historically, the first axion model –the quantum chromodynamics or QCD axion– emerged as the leading solution to the strong CP problem through the so-called Peccei-Quinn mechanism [455]. This mechanism consists of adding a $U(1)$ symmetry in the Standard Model –known as the Peccei-Quinn symmetry–, which leads to a new particle –the QCD axion– when this symmetry undergoes spontaneous symmetry breaking [431,602,603]. The QCD axion is characterized by two inversely proportional fundamental quantities: its mass, $m_{\rm a}$, and its decay constant, $f_{\rm a}$. These quantities are usually quoted in units of electronvolts and Gigaelectronvolts, respectively, and for the QCD axion, $f_{\rm a}$ is the energy scale at which the Peccei-Quinn symmetry breaks spontaneously. In general, one may think of $f_{\rm a}$ as a combination of $g_{a,j}$ factors, where each $g_{a,j}$ quantifies the coupling strength of the QCD axion to the $j$-th particle in the Standard Model – e.g., photons –, quoted in units of inverse Gigaelectronvolts. In addition to solving the strong CP problem, the QCD axion is a leading dark matter candidate (see [2,154,482] for early theoretical works; and [100,433] for recent review on searches for QCD axion dark matter). Currently, most laboratory-based efforts aimed at detecting the QCD axion probe the $10^{0-2}$ $\mu$eV mass range. This mass range is the 'sweet spot' to search for QCD axions that simultaneously solve the strong CP problem and would comprise the entirety of dark matter in the universe. This range in masses is driven by the possible production mechanisms for QCD axions in the Early Universe. Fig. 16 of Ref. [100,433] illustrates current bounds on $g_{\rm afl}$ for such axions ['Dark matter (direct detection)'] compared to the bounds on generic axions that need *not* comprise dark matter, inferred from both astrophysics ['Astrophysics ($\gamma \rightarrow a$)'] and independent laboratory-based facilities ['Experiments ($\gamma \rightarrow a$)'].

Axion-like particles (which we refer to as 'axions', for simplicity) are model-independent generalizations of the QCD axion and compelling dark matter candidates. The driving motivation to

search for such axions is that, albeit no longer solving the strong CP problem, they are ubiquitous in many theories extending beyond the Standard Model, such as type-IIB string theories. One of the key observables predicted by these theories is the ensemble prediction for $g_{a\gamma}$ –the coupling strength of axions to photons– for a given axion mass, $m_a$ [23,137,145,216,552]. These ensemble predictions examine a range of possible values for $g_{\mathrm{a}\gamma}$ across various string theory compactifications and model assumptions.

Axions will plausibly be *indirectly* detected through their interaction with photons, triggered as photons traverse an intervening magnetic field. For a given axion mass $m_{\mathrm{a}}$, this induces *energy-dependent* spectral distortions in the incident photon spectrum and is characterized by: the coupling strength $g_{a\gamma}$, and the structure of this external magnetic field. For reference, Fig. 41 shows the energy-dependent 'survival probability' curve for X-ray photons interacting with the intracluster medium (ICM) in two massive cool-core clusters – CL1821+643 (orange) and Perseus (blue) – for axion parameters $m_{\mathrm{a}} = 10^{-13}$ eV, $g_{\mathrm{a}\gamma} = 10^{-12}$ GeV$^{-1}$. Notably, the amplitude of the axion signal increases as the energy increases, whereas the frequency of the distortions decreases as the energy increases. Note that the energy-dependent 'survival probability' curves in Fig. 41 were inferred numerically by solving the so-called 'wave equation' for photons traveling through an external magnetic field as described in Sec. 3 of [538]. This approach maps the ICM magnetic field into coherent cells in the range 3.5 – 10 kpc – that is, the range of coherence lengths expected for Perseus – each within which the magnetic field orientation is set independently and randomly. The wave equation is solved in each box, using the solution for cell $i$ as the input state for box $i+1$ for coherence cells located $r = 10$ kpc $-$ 1.5 Mpc to the cluster center.

Therefore, we can search for axion-induced distortions by analyzing the spectra of astrophysical sources (e.g., a cluster-hosted AGN) interacting with intervening astrophysical magnetic fields (e.g., the intracluster medium, or ICM, of the host cluster), provided a suitable model for the magnetic field is available. If these spectral distortions are not detected, one can instead set upper bounds on $g_{a\gamma}$ over a range of axion masses. Figure 42 shows the tightest bounds set to date on light axions that need not comprise dark matter ($m_a < 10^{-12}$ eV), inferred from high-resolution X-ray spectroscopy of the cluster-hosted AGN H1821+643 [538]. Figure 42 also shows projected bounds on such axions from simulated observations of NGC 1275 –the central engine of the Perseus cluster– for both *AXIS* and *Athena* [540].

The current tightest bounds on light axions to date exclude all photon-axion coupling values $g_{a\gamma} > 6.3 \times 10^{-13}$ GeV$^{-1}$ at 99.7% confidence. These were inferred from *Chandra* Transmission Grating observations of the extraordinarily luminous quasar H1821+643 [538] (exposure: 570 ks). This quasar is hosted by CL 1821+643, a massive cooling-core cluster at redshift $z = 0.3$ [511]. The high-quality data permitted an extraction of the intrinsic quasar emission free from that of the cluster, thanks to the dispersion along the grating arrays (which led to residuals at the 2.5% level). Ref. [538] found no evidence for axion-induced distortions in the quasar spectrum. The posteriors on axions were inferred through a Bayesian analysis that compared the goodness-of-fit statistics of an astrophysical-only model (typical of a type-I AGN) with a library of models that took into account the multiplicative effect of photon-axion interconversion along the cluster line of sight. All (energy-dependent) photon-axion models in this library were calculated for a given $(m_a, g_{a\gamma})$ and a set of realistic realizations of the ICM magnetic field in CL 1821+643.

Broadly, we expect studies employing observations of cluster-hosted AGN with current X-ray observatories to be limited by S/N statistics, and therefore to be limited to the brightest cluster-hosted AGN observed with the longest available archival observations. With its significant improvement in collecting area compared to *Chandra*, Ref [540] showed that *AXIS* will improve on the current tightest bounds on light axions by a factor of 3 with a simulated observation of NGC1275 (at the 95.5% level for an exposure of 200 ks) based on the non-detection of spectral distortions due to photon-axion interconversion as the AGN emission interacts with the magnetized ICM of the host cluster (Perseus). These limits were inferred by using a similar Bayesian approach to Ref. [538] using a realistic library of photon-axion

interconversion models in the Perseus cluster (see also Ref. [490]). Note that there are several caveats with this projection: 1) The effects of systematic uncertainties were not considered. The importance of assessing these systematic uncertainties is illustrated in Figure 42 and was discussed in detail by [540] in the context of *NewAthena*; 2) The area response file ($\mathtt{ARF}_0$) and response matrix files used to simulate an *AXIS* observation of NGC1275 were those available in early 2023; and 3) Ref. [540] did not consider the possible influence of photon pileup in the simulated *AXIS* spectra, assuming that *AXIS* will have bright-source modes permitting accurate CCD spectroscopy of sources with fluxes at least as high as a few milliCrab. To address caveat 2), to first order, the projected bounds on $g_{a\gamma}$ shown in Fig. 42 can be rescaled to those expected for an updated collecting area by multiplying such bounds ($g_{a\gamma}$ for a given mass $m_a$) by a rescaling quantity $f$, where $f = \mathtt{ARF}_{\rm new}/\mathtt{ARF}_0$.

Searches for unidentified X-ray lines

Another prospect for using *AXIS* to constrain fundamental physics is to search for unidentified X-ray lines in the direction of the Milky Way halo, for instance, to search for decay signatures of sterile neutrino dark matter particles with masses in the keV range. Sterile neutrinos are well-motivated extensions to the Standard Model as they would naturally explain the observed sum of the active neutrino masses [the three Standard Model neutrinos; 1,450]. In addition, sterile neutrinos are compelling dark matter candidates [155,533]. If sterile neutrinos make up the entirety of dark matter, they may decay into photons and active neutrino species ($\chi \rightarrow \gamma + \nu$ and $\chi \rightarrow \gamma + \gamma$). These photons may be detected through distinct spectral features in X-ray astrophysical data. A well-known example in the literature is the 3.5 keV line expected from the decay of a $m_\chi \sim 7$ keV sterile neutrino decaying into two X-ray photons (see discussions and references in [147], highlighting the importance of background modeling in dark matter decay searches through stacked observations for unidentified X-ray lines).

Ref. [198] used archival data from *XMM-Newton* –specifically, observations taken across the full sky– to search for evidence of the decay of dark matter particles in the $m_\chi = 6 - 15$ keV range within the Milky Way's ambient halo. With its improved angular resolution, collecting area, and PSF stability off-axis, we expect that *AXIS* observations will improve on these limits.

Dark photons and $B - L$ vector dark matter

The existence of dark photons and $B - L$ vector dark matter arises from extending the theory of the Standard Model with extra $U(1)$ symmetries [176]. These hypothetical particles can interact with other SM particles through kinetic mixing with photons, or baryon-minus-lepton numbers ($B - L$) coupling. These two interactions are the subject of string theory and string compactification studies in the literature [95,233].

These vector particles are compelling dark matter candidates and may arise as the consequences of inflation [235] or freeze-in [254] mechanisms in the Early Universe. They can decay into final states containing Standard Model particles [426]. For masses $< 2m_e \sim 1$ MeV (where $m_e$ is the electron rest mass), these vector dark matter candidates decay into a distinct three-photon final state. This decay, often referred to as the 'dark photon trident' [346,347], produces photons with energies on the order of keV, with a unique continuous X-ray spectrum that peaks at an energy which is approximately one third of the vector dark matter particle mass, rendering these particles detectable by *AXIS* and enabling the exploration of the sub-MeV parameter space for dark photons and $B - L$ vector dark matter [425].

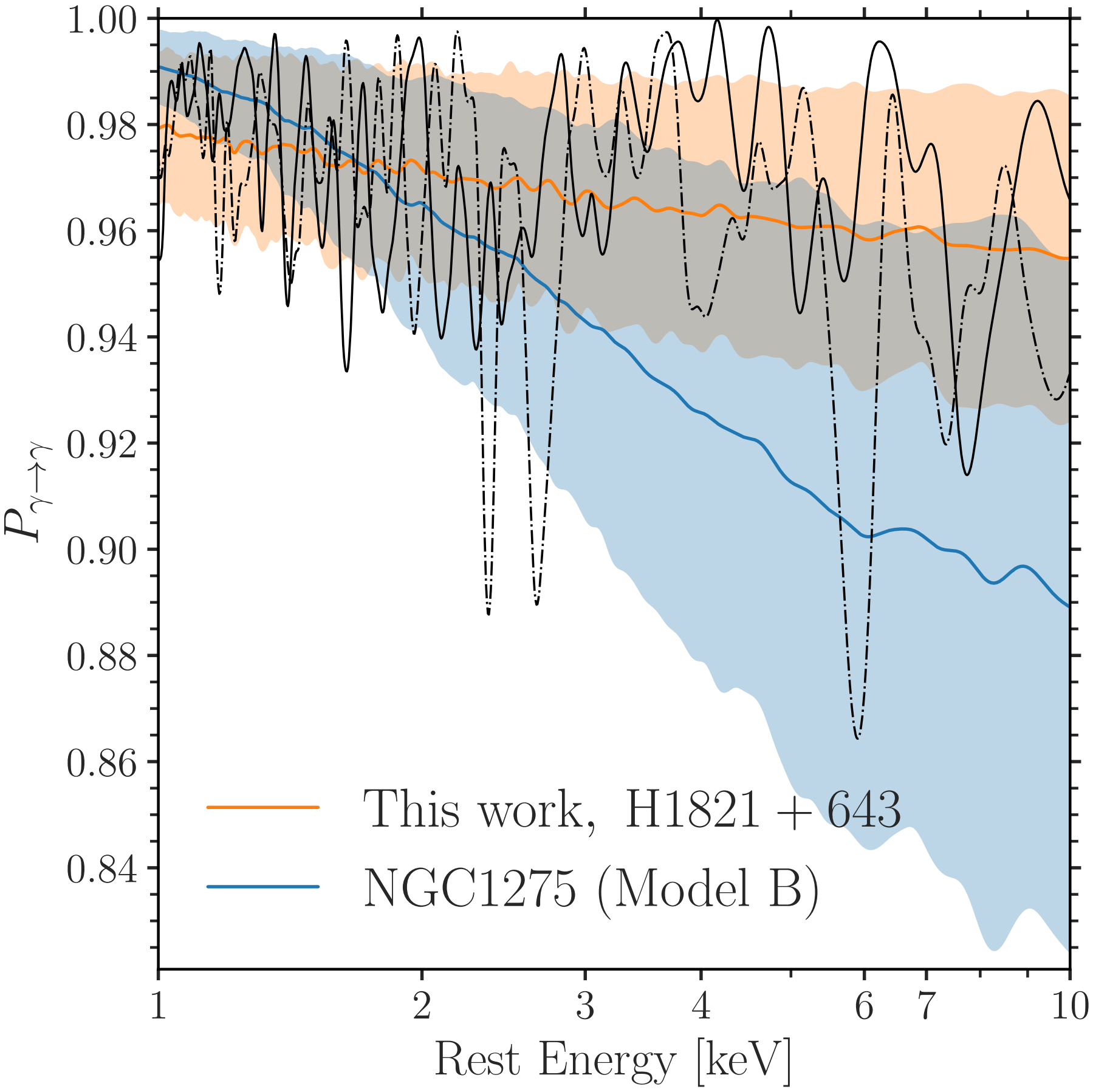

**Figure 41.** Mean photon survival probability as a function of energy computed for X-ray photons traveling from the inner 10 kpc to 1.5 Mpc in the cool-core clusters CL1821+643 (orange) and Perseus (blue). [blue, Model B of 490]. The shaded regions represent the standard deviations of 500 curves in each energy bin, calculated using 1000 logarithmic bins between $1-10$ keV. The black solid and dot-dashed lines show survival probability curves for two individual realizations of the magnetic field. All calculations are performed for ALP parameters $g_{\rm afl} = 10^{-12}\,\mathrm{GeV}^{-1}$, $m_{\rm a} = 10^{-13}$eV. Figure taken from [538].

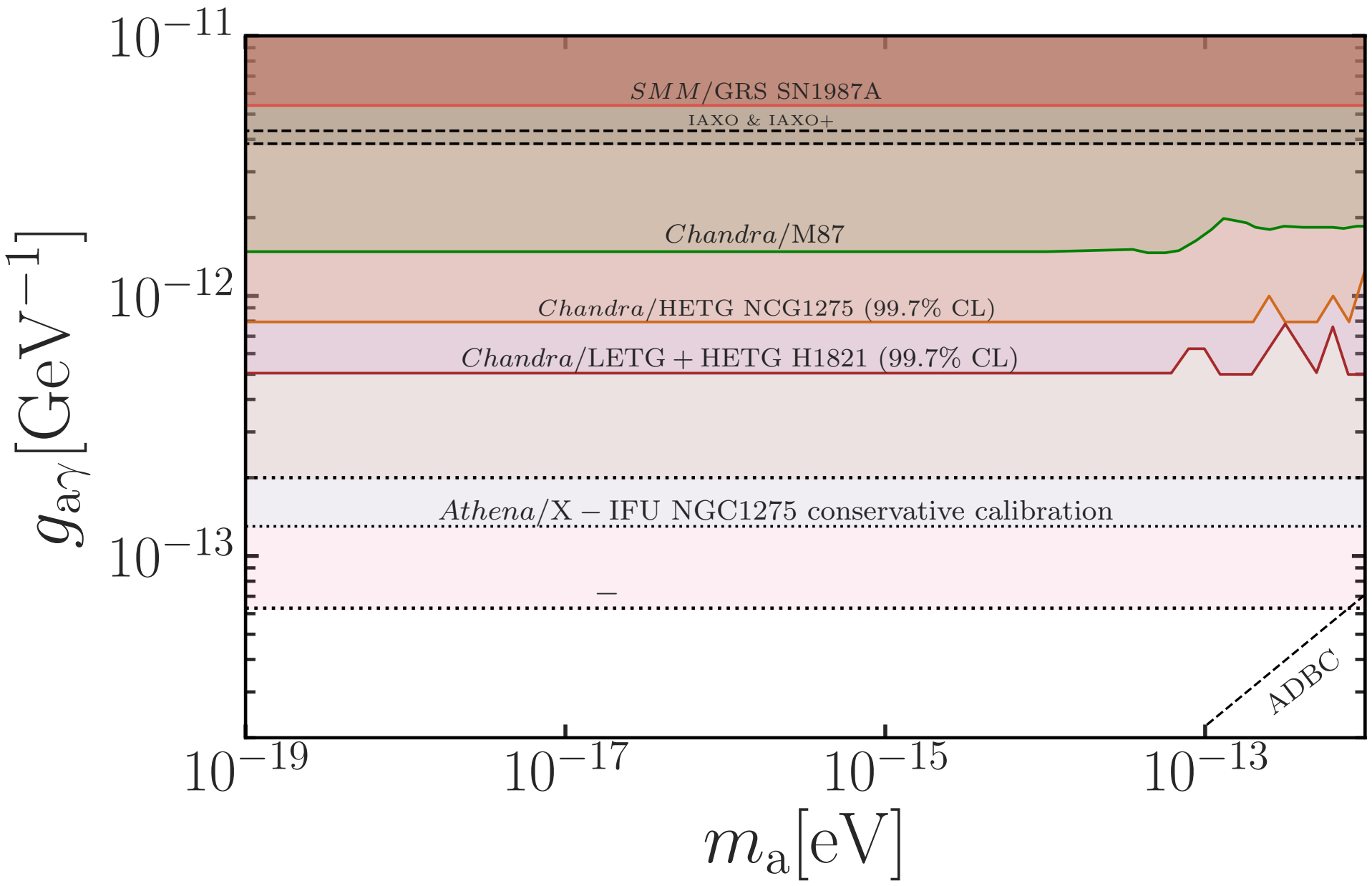

**Figure 42.** Projected bounds on axion-like particles from a 200-ks simulated observation of NGC1275 (the Perseus cluster's central engine) with *AXIS* (using on-axis ARF and RMF from early 2023) and *Athena*. The shaded area above each exclusion curve shows the all values of the photon-axion coupling constant $g_{a\gamma}$ (vertical axis) that are excluded for a given axion mass (horizontal axis). Solid lines show current exclusion limits from: *Chandra* observations of two cluster-hosted AGN: M87* [391] and H1821+643 [538], and from the SN1987A event [451] (see Ref. [375] for a recent reanalysis). Projected bounds from the future birefringent cavity experiment *ADBC* (dashed black line) are shown. *ABDC* will constrain axion-like particle dark matter [349]. Figure taken from [540].

Primordial black hole dark matter

As an alternative to particle dark matter, primordial black holes (PBHs) have been widely discussed and considered as dark matter candidates in terms of both their production mechanisms and detection strategies [97]. Inflation [437], reheating [14] and cosmological phase transition [395] models can generate PBHs. As dark matter candidates, PBHs exhibit a broad mass range spanning $10^{15}$–$10^{52}$ grams. Numerous detection strategies –including gravitational waves, microlensing, gamma-ray, and cosmic-ray observations– are being employed to explore this parameter space and to determine the fraction of the total dark matter relic abundance that PBHs may constitute.

Primordial black holes (PBHs) with masses ranging from $10^{15}$–$10^{17}$ grams provide a compelling window for X-ray searches [362]. In this mass range, PBHs can evaporate and subsequently produce high-energy cosmic rays. These high-energy cosmic rays can subsequently undergo Inverse Compton Scattering (ICS) processes with low-energy photons associated with the cosmic microwave background, as well as with optical and UV photon emissions from stars. These ICS processes can produce X-ray emission with a continuous spectrum that depends on the PBH mass and spin, falling into the observable energy range to *AXIS*. Utilizing *AXIS* data with the diffuse astrophysical X-ray background will not only constrain the possibility of PBHs constituting a fraction of dark matter, but will also provide insight on the spin of these compact objects.

**Exposure time (ks):** Archival Deep.

**Observing description:**

- All results will be guaranteed by observations that *AXIS* is expected to take, for example:
- Axion-like particles: unobscured, bright cluster-hosted AGN with well-known spectra (e.g., NGC1275, M87*, H1821+643, Hydra); $t_{\rm clean}$ = few $\times$ 100 ks. The longer the exposure, the more sensitive the axion limits will be.
- Search for unidentified X-ray lines: from observations of point sources in the direction of the Milky Way's ambient halo.
- Dark photons, $B - L$ vector dark matter, primordial black holes: data from the *AXIS* deep survey, in some cases, by stacking spectra of point sources in the direction of the Milky Way's halo.

**Joint Observations and synergies with other observatories in the 2030s:**

There are no obvious possible synergies or joint observations except for the case where high-quality spectra of point sources in the direction of the Milky Way halo taken by different facilities (e.g. *Chandra*, *XMM-Newton*, *INTEGRAL*) may be stacked to constrain dark photons, $B - L$ vector dark matter, and primordial black holes.

**Special Requirements:**

- Pileup: In relation to axion projections from *AXIS*: Ref. [540] did not consider the possible influence of photon pileup in the simulated spectra. This study also assumed that *AXIS* will have bright-source modes permitting accurate CCD spectroscopy of sources with fluxes at least as high as a few milliCrab.

## *32. Changing state AGNs*

**Science Area: AGN, Accretion**

**First Author:** Arghajit Jana (Universidad Diego Portales, Santiago, Chile, arghajit.jana@mail.udp.cl)
**Co-authors:** Claudio Ricci (Universidad Diego Portales; claudio.ricci@mail.udp.cl)

**Abstract:**

Changing State AGNs (CSAGNs) are a rare class of active galactic nuclei that undergo dramatic transitions between Type 1 and Type 2 states over months to years, exhibiting significant variability in their optical, UV, and X-ray emission [e.g., 430,494]. These transitions are thought to be driven by changes in the accretion rate, triggered either by disk instabilities [e.g., 547,556] or external perturbations, such as tidal disruption events [e.g., 407]. The transitions are linked to changes in the geometry of the inner accretion flow, the properties of the corona, disk-corona coupling, and the presence or suppression of relativistic jets [280,342,411]. *AXIS*, with its exceptional angular resolution, deep X-ray sensitivity, and time-domain capabilities, will enable precise monitoring of spectral state transitions and reverberation mapping to probe disk-corona interactions. By tracking the evolution of the X-ray corona across different AGN states, *AXIS* will provide unprecedented insights into the dynamics of accretion flows, coronal physics, and the connection between AGNs and stellar-mass black hole accretion systems [506].

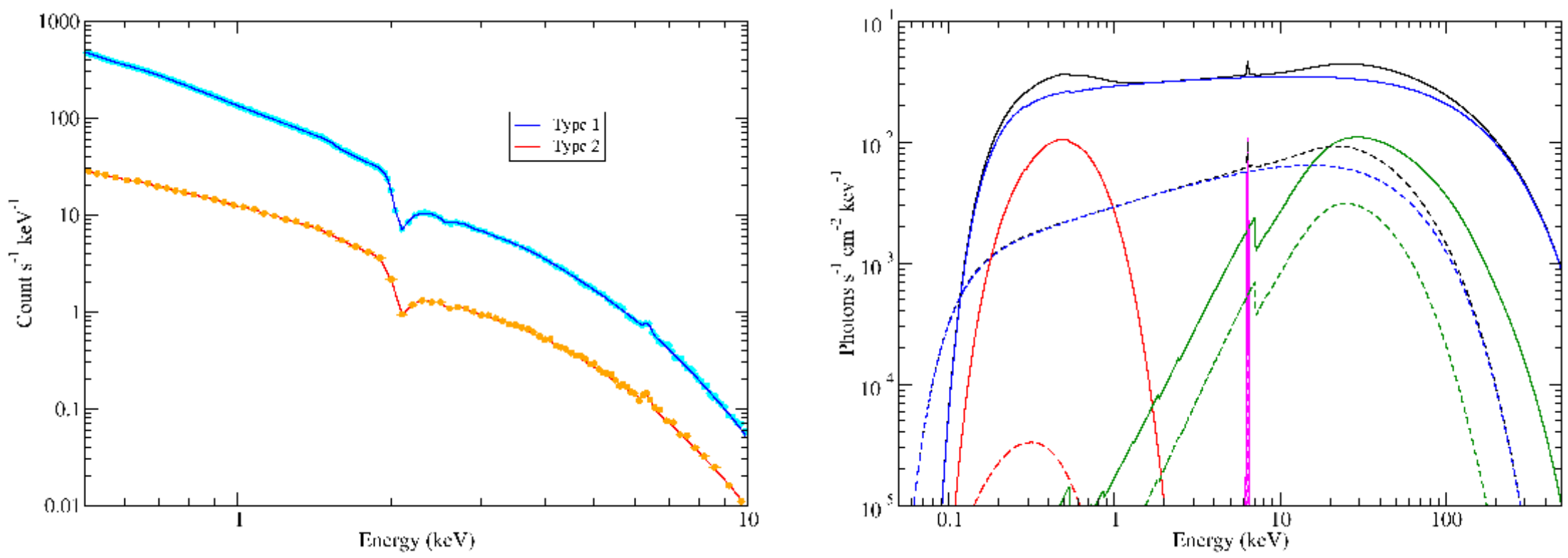

**Figure 43.** Left panel: Simulated *AXIS* spectra of changing-state AGN NGC 1566 in type 1 (blue) and type 2 state (orange). Right panel: best-fitted theoretical model for corresponding spectra. The solid and dashed lines represent different spectral components from type 1 and type 2 states, respectively. The black, blue, red, magenta, and green lines represent the total spectra, continuum, soft-excess, iron line, and reprocessed emission, respectively.

**Science:** CSAGNs are a subset of AGNs that exhibit dramatic spectral and flux variability over months to years, transitioning between Type 1 and Type 2 states [e.g., 494]. These transitions challenge our understanding of the structure and variability of the inner accretion flow. They also provide an opportunity to probe the physics of accretion and feedback processes on timescales much shorter than those associated with typical AGN evolution. While optical and X-ray studies have provided valuable insights into these extreme transformations, a dedicated observational campaign with AXIS can revolutionize our comprehension of the physical mechanisms driving these changes.

Recent studies have suggested that variations in accretion rates drive intrinsic changes in AGN emission, leading to the disappearance or re-emergence of broad emission lines [494,547]. The soft X-ray excess is believed to be responsible for ionizing the BLR clouds [430]. In the type 1 state, SE flux is generally found to be high, and decreases in the type 2 state. Hence, the warm corona, which is believed to produce the SE, also appears to evolve with the accretion rate. Figure 43 shows the *AXIS*-simulated spectra of

a CSAGN NGC 1566 in type 1 and type 2 state in the left panel. The right panel of Figure 43 shows the theoretical best-fitted model, which clearly shows the change of different spectral components in type 1 and type 2 states.

The CS transitions are found to occur around Eddington ratio ($\lambda_{\rm Edd}$), $\lambda_{\rm Edd} \sim 0.01 - 0.02$. At this $\lambda_{\rm Edd}$, inner accretion geometry is also believed to change. At $\lambda_{\rm Edd}$ ($> 0.01$), the inner accretion flow is believed to consist of a hot corona along with a standard accretion disk that extends up to ISCO. At $\lambda_{\rm Edd}$ ($< 0.01$), the inner accretion disk is believed to be replaced by an advection dominated accretion flow (ADAF). The warm corona is linked with the inner part of the accretion disk. In type 2 state, the inner accretion disk is replaced by ADAF. This naturally weakens the warm corona, hence the SE diminishes. Interestingly, the black hole X-ray binaries (BHXBs) also show soft↔hard state transitions, with transition $\lambda_{\rm Edd} \sim 0.01$. This links between the accretion properties of the CSAGNs and BHXBs.

High-cadence monitoring and deep observations with *AXIS* will allow us to measure the response of the coronal emission to changes in the accretion disk luminosity, shedding light on the timescales and physical processes involved. Monitoring CSAGNs will allow us to understand the geometry of the inner accretion flow, as well as the dynamics of the warm corona. Understanding these processes will also improve constraints on the coupling between the accretion disk and the corona, which is a fundamental aspect of AGN physics.

Addition of the variable accretion rate, in some cases, variable obscuration could also led to changing-look transition [627]. In such scenarios, transient winds or outflows from the outer BLR or the torus may obscure the BLR, triggering the CL transition. Continuous monitoring with *AXIS* could provide valuable insights into these processes.

The synergy between AXIS and multi-wavelength observatories, such as LSST and ALMA, will be crucial in constructing a comprehensive picture of CSAGN evolution. *AXIS* observations of soft X-ray emission can be compared with IR variability to constrain dust reverberation, while simultaneous UV and optical data will provide essential diagnostics of accretion state transitions. In particular, time-domain studies combining LSST's optical light curves with *AXIS* X-ray monitoring will help establish causal links between different emission components, revealing how energy is redistributed within the AGN. This multiwavelength approach will bridge the gap between the small-scale physics of AGNs and their impact on the host galaxy environment.

The main goals of this study are:

- **Understanding the geometry of the inner accretion flow.**
- **Exploring the nature of the warm corona, and its role for CS transitions.**
- **Understanding the link between the black hole X-ray binaries and the CSAGNs.**

**Exposure time (ks):** Around $\sim$ 300ks per source.
**Observing description:**

The targets will be selected from the alert given by LSST or other all-sky monitors.

For one source, when it goes into outburst (showing CS event), it needs to be monitored for at least 2 years, with a cadence of $\sim 1 - 3$ months.

**Joint Observations and synergies with other observatories in the 2030s:** LSST and ALMA.
**Special Requirements:** Monitoring, TOO.

*33. X-ray and radio coupling in variable radio quiet AGN*

**Science Area: AGN, jets, corona, jets**
**First Author: Francesca Panessa** (INAF/IAPS, francesca.panessa@inaf.it)
**Co-authors:** Ranieri D. Baldi, Ari Laor, Ehud Behar, Ian McHardy, Pierre-Olivier Petrucci, Marcello Giroletti, Gabriele Bruni, Claudio Ricci

**Abstract:**

Recent studies have revealed a persistent radio excess in radio-quiet AGN (RQ AGN), with flat-spectrum emission in the range $\sim 50 - 230$ GHz that closely correlates with X-ray luminosity ($L_{\rm radio}/L_{\rm X-ray} \sim 10^{-5}$). High-resolution and micro-lensing observations identify this emission as arising from extremely compact regions, likely on sub-parsec scales, suggesting a common origin with the X-ray–emitting corona. Intra-day variability in both X-ray and millimeter bands provides compelling evidence for such compact emission regions, with variability timescales directly constraining source sizes via light-travel arguments.

AXIS, with its exceptionally high sensitivity combined with an optimal angular resolution and sub-hour temporal resolution, will revolutionize our ability to study the innermost structures of AGN. By isolating the hard X-ray coronal component and capturing variability on timescales ranging from minutes to hours, AXIS will directly probe the dynamic processes governing coronal heating and emission. This will enable robust tests of whether millimeter and X-ray emission arise from the same compact region, such as a magnetically confined corona or the base of a weak jet.

When combined with high-resolution, high-cadence monitoring from next-generation radio/mm facilities (e.g., ALMA, ngVLA, SKA), AXIS will enable simultaneous time-domain studies of RQ AGN cores, resolving structures down to tens of gravitational radii. This synergy will place powerful constraints on the size, structure, and physical origin of the emitting regions, providing transformative insights into the accretion-ejection mechanisms that operate in the majority of AGN across cosmic time.

**Science:**

Most of the properties of matter and light close to the innermost regions near supermassive black holes (SMBH) are still unknown. Coronal X-ray emission produced in the hot, tenuous corona surrounding SMBHs at the centers of galaxies is generated by the inverse Compton scattering of photons from the accretion disk by hot electrons in the corona. Unveiling the physical origin of the X-ray corona is a key element in understanding the accretion-ejection process, SMBH feedback, and galaxy evolution.

The almost ubiquitous presence of compact radio cores in AGN has led to the suggestion that the corona might be heated and confined by the same magnetic fields responsible for producing the radio jet [52,306]. This hypothesis is supported in X-ray BH Binaries, where a correlation between the corona and the jet is found. Indeed, persistent radio jets are only observed in 'hard' accretion states, during which the coronal continuum dominates the X-ray spectrum [190,385]. In bright AGN and quasars, vertical magnetic fields could efficiently power coronae without necessarily launching a strong jet [42]. Overall, the connection between the hot corona and jet emission remains poorly understood.

The majority of AGN nuclei are radio-quiet (RQ AGN), i.e., their radio power is significantly below $10^{24}$ W/Hz at 1.4 GHz. Radio-Loud (RL) AGN are bright radio sources generally separated by the RQ AGN population according to the radio-loudness parameter defined by Kellermann et al. [302] [3]. In RL AGN, the origin of the radio emission is attributed to synchrotron emission from powerful jets extending from the nucleus to kiloparsec (kpc) and sometimes megaparsec (Mpc) scales. In RQ AGN, radio emission

---

[3] $R = L_R/L_V$ where $L_R$ is the AGN radio luminosity at 5 GHz and $L_V$ is the optical luminosity in the B band , see also Terashima & Wilson [557]

is likely a combination of different origins, such as nuclear star-forming regions, winds/outflows, low power jets, and/or the corona (see **?** ).

The investigation of the radio morphology and spectral slope at different angular resolutions and frequencies provides important clues in defining the emitting radio source in RQ AGN. However, in the majority of cases, this observational evidence is not sufficient to discriminate between the different possible origins. Comparing X-ray and radio flux variations is an optimal strategy for understanding the various possible emission processes.

X-ray variability is stochastic and occurs at different time-scales [386], possibly due to the response of the accretion flow and corona to intrinsic fluctuations [415,578]. Radio variability was almost totally unexplored in RQ AGN so far, until recently, when it has been tested in a few selected AGN (e.g., Baldi et al. 33, Behar et al. 53, Panessa et al. 445, Petrucci et al. 468, Williams et al. 606).

The simultaneity of the radio and X-ray monitoring is fundamental to addressing their possible common origin. Indeed, correlated and lagged variability can provide strong support for a possible physical connection. Specifically, if the radio emission derives from a synchrotron jet, we should observe an X-ray/radio correlation, similar to those observed in XRBs [126]. Assuming a scaling with BH mass, a jet origin would predict the low-frequency radio to lag behind the X-ray photons in RQ AGNs by tens of days, such as those found by Bell et al. [54], i.e., a 24d radio lag in the LINER NGC 7213. No correlated emission was observed in the radio/X-ray light curves in NGC 4051 [292]. Inter-day temporal correlation between radio and X-ray light curves was also investigated in local Seyfert galaxies, with no sign of significant correlated variability [111,445].

Since synchrotron absorption decreases strongly with increasing frequency, it is expected that at high frequencies above 50 GHz, emission will come from smaller regions and will be more variable. It was found that the nearby Seyfert galaxy NGC 7469 varies at 95 GHz (confidence level 99.98%) on a timescale of days (Baldi et al. 33; see Fig. 1 below), whereas it appears to be steady for years at 5 GHz. The 95 GHz variability time and amplitude are comparable, but somewhat less than those in the X-rays. To further test the alleged connection with the X-ray source, simultaneous monitoring of both radio and X-ray emissions is necessary.

Radio emission in the mm-band correlates tightly with X-ray luminosity ($L_{\rm radio}/L_{\rm X-ray} \sim 10^{-5}$).

Indeed, this relation resembles the Gudel-Bentz relation seen in coronally active stars, where the Nuepert effect is observed, where $L_{\rm radio} \propto dLx/dt$, which may also be present in AGN [335]. The underlying physics is a reconnection event, where the magnetic field in a specific region is converted into an electric field, which accelerates electrons. Relativistic electrons produce synchrotron radio emission, but they also heat the corona, increasing X-ray emission. So, generally, the radio, in particular the mm emission which comes from the most compact scales, is an indicator of the coronal heating process, while the X-rays indicate the coronal Compton cooling process.

High-resolution observations at milliarcsecond scales show that this emission originates from extremely compact cores (typically a few to tens of pc). To probe smaller regions, this study utilizes intra-day variability as a tool, relying on the fact that variability timescales limit source size due to light travel time. The variability of X-ray and radio emission, associated with the corona in RQ AGN, is indeed found to be correlated, indicating a common physical origin. AXIS, with its high sensitivity, will allow us to study the AGN core emission by investigating the coronal hard X-ray component. A synergy with high-resolution radio arrays will revolutionize the temporal analysis of X-ray/radio light curves, providing new constraints on the accretion-ejection mechanisms ongoing in radio-quiet AGN.
**Exposure time (ks):** 180 ks.
**Observing description:**

**The sample:**

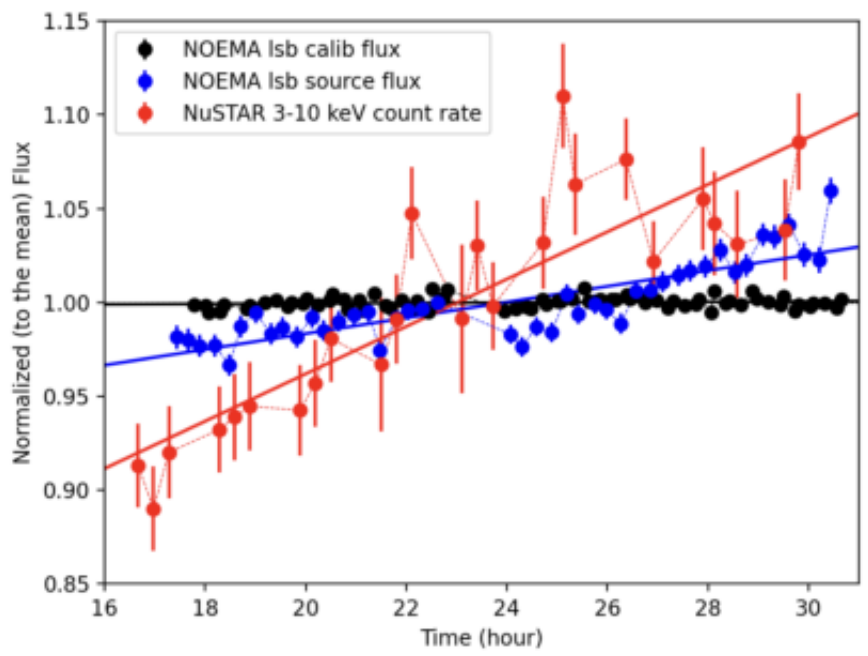

**Figure 44.** MCG+08-11-11 radio/X-ray variability [468].

We have selected a sample of local (< 200 Mpc) Seyfert galaxies known to be variable in their radio emission. In particular, some of them have been tested in the mm frequency range.

Indeed, NGC 7469 was known for a significant variability at 95 GHz, by a factor of two within four to five days [33]. Later, Behar et al. [53] provided the simultaneous mm/X-ray monitoring of NGC 7469 lasting ∼70 days in X-rays, of which ∼ 50 days of contemporaneous mm and X-ray observations with a cadence of one to two day. In [468], a simultaneous mm/X-ray campaign of an RQ AGN was carried out for MCG +08-11-11. The 3–10 keV X-ray XMM-Newton light curve and the 100 GHz NOEMA light curve were studied on timescales of 14 hours. Both fluxes showed slight increases, but no correlated variability was found.

The most recent result has been obtained for IC 4329A Shablovinskaya et al. [526], with a joint ALMA (∼ 100 GHz) and X-ray (NICER/XMM-Newton/Swift) campaign. No significant correlation has been found on timescales of less than 20 days; however, the mm flux density has varied significantly (by a factor of 3) within four days, exceeding the simultaneous X-ray variability and showing the largest mm variations ever detected in RQ AGN over daily timescales. This suggests a synchrotron origin for the mm radiation in a source of ∼ 1 light day size, most likely coincident with an X-ray corona. Similarly, as reported by [412] for the RQ AGN GRS 1734-292 with ALMA observations at 100 GHz.

The sparse number of observing campaigns does not allow for drawing any stable conclusions on the nature of mm variability in RQ AGN and its connection with the X-ray emission.

**The monitoring campaign:**

Here, we propose to observe the selected targets with AXIS to detect X-ray variability in correlation with the radio variability. At the lowest X-ray flux of the sample, i.e., $2\times10^{-11}$ cgs, we will detect 7000 counts in 1 ks observation. Observing each source for 20 ks will allow us to test different variability timescales within one single observation (intra-hours, intra-day), constraining the emitting size region down to a few $R_{sh}$. In addition, the arcsecond spatial resolution of AXIS will be crucial in isolating the nuclear emission from possible contaminating emissions (such as star-forming emission, ULXs, and XRBs).

The AXIS observations will be combined with high-resolution, intra-hour, and intra-day monitoring from next-generation radio/mm facilities (e.g., ALMA, ngVLA, SKA) to cross-correlate the X-ray and radio fluxes. In particular, with 10 min observation on source with ngVLA at 100 GHz (with a spatial resolution of ∼ 1.3 mas) it will be possible to reach 25 microJy. Similarly, with ALMA at 100 GHz, ∼ 26 microJy/beam will be reached with the same exposure on scales of ∼ 10 mas. This will allow a short-time sampling of the mm light curves to be compared with the AXIS light curves.

**Joint Observations and synergies with other observatories in the 2030s:** ALMA, ngVLA, SKA

**Special Requirements:** None

**Table 6.** Example Targets: 1) Name; 2) Seyfert type; 3) Luminosity distance (Mpc); 4) Scale (pc/arcsec); 5) Black hole mass in solar masses; 6) 2-10 keV nuclear flux density (erg $s^{-1}$ $cm^{-2}$).

| Target (1) | Type (2) | Dist (3) | scale (4) | BH Mass (5) | F(2-10 keV) (6) |
|---|---|---|---|---|---|
| NGC 4151 | Type 1 | 18.4 | 88.5 | $8.9\times10^7$ | $1.49\times10^{-10}$ |
| MCG +08-11-11 | Type 1 | 92.1 | 429 | $2.8\times10^7$ | $4.52\times10^{-11}$ |
| NGC 7213 | Type 1 | 22.8 | 110 | $8.9\times10^7$ | $2.30\times10^{-11}$ |
| NGC 4051 | Type 1 | 13.7 | 65.9 | $1.7\times10^6$ | $1.70\times10^{-11}$ |
| Mrk 110 | Type 1 | 163 | 736 | $4.8\times10^7$ | $2.83\times10^{-11}$ |
| Ark 564 | Type 1 | 106 | 491 | $2.6\times10^6$ | $2.42\times10^{-11}$ |
| NGC 7469 | Type 1 | 67.2 | 316 | $9.1\times10^6$ | $2.60\times10^{-11}$ |
| NGC 5548 | Type 1 | 80.1 | 375 | $7\times10^7$ | $2.90\times10^{-11}$ |
| IC 4329A | Type 1 | 75 | 356 | $6.8\times10^7$ | $1.20\times10^{-10}$ |

*34. A Time-Domain X-ray Study of Little Red Dots: Flickering AGN at Cosmic Dawn*

**Science Area:** AGN, Time-domain Studies
**First Author:** **Fabio Pacucci** (Center for Astrophysics | Harvard & Smithsonian, fabio.pacucci@cfa.harvard.edu)
**Co-authors: Nico Cappelluti** (University of Miami, ncappelluti@miami.edu), **Jiachen Jiang** (University of Warwick, Jiachen.Jiang@warwick.ac.uk)

**Abstract:**
Little Red Dots (LRDs) are compact, red JWST-discovered sources at $z \gtrsim 4$ that show broad-line AGN features and V-shaped SEDs. Remarkably, they remain weak in X-rays or undetected, even in deep surveys or stacking analyses. Two LRDs, PRIMER-COS 3866 ($z = 4.66$) and JADES 21925 ($z = 3.1$), are X-ray detected, with moderately obscured AGN. These rare detections offer the first opportunity to study X-ray variability in this emerging population at early cosmic epochs. We propose a time-domain monitoring program with AXIS to detect X-ray variability, measure the coronal variability timescale, and constrain the physical processes governing SMBH accretion at high redshift. Using ten 30 ks observations, logarithmically spaced over $\sim$ 1 year, we aim to detect variability on timescales from days to years (rest-frame). This will be the first systematic X-ray variability study of this key high-$z$ AGN population.
**Science:**

The "Little Red Dots" (LRDs) discovered by JWST [258,308,394] form a unique population of ultra-compact, red AGN candidates at $z \sim 4-8$. Their characteristic V-shaped SEDs, small effective radii ($\sim$100–300 pc, see, e.g., Baggen et al. 30, Guia et al. 247), and broad Balmer lines [241,309] point to the presence of actively accreting SMBHs. Despite this, nearly all LRDs are undetected in the deepest X-ray surveys [17,371,440,621].

Two exceptional LRDs break this pattern: PRIMER-COS 3866 at $z = 4.66$ and JADES 21925 at $z = 3.1$, which are clearly detected in deep Chandra fields [309]. Their obscuration-corrected $2-10$ keV luminosities are $\log L_{2-10} = 44.7$ and 43.7 erg s$^{-1}$, respectively, corresponding to $\log L_{\rm bol} \sim 46-45$, respectively, under standard bolometric corrections [162]. Inferred black hole masses are $\sim 10^7$–$10^8\, M_\odot$, consistent with Eddington-limited accretion. These two systems, the only X-ray-detected members of a sample of over 300 LRDs [309], offer a rare opportunity to probe AGN variability and corona physics at high redshift.

Recent variability studies with JWST have shown that LRDs are surprisingly quiescent in the rest-frame UV and optical [555]. Even spectroscopically confirmed broad-line LRDs show little photometric variability across multiple epochs. This UV quietness may result from host galaxy dilution, obscuration, or limited time sampling. However, X-rays are less affected by these issues and directly trace the innermost corona [492].

We propose to use AXIS to search for X-ray variability in the two known X-ray LRDs and constrain the variability timescale associated with their coronae. We base our predictions on the physically motivated model from Ishibashi & Courvoisier [276], where the characteristic X-ray variability timescale $\tau_X$ is associated with the heating and cooling times of hot coronal electrons. It scales with black hole mass and accretion rate as:

$$\tau_X \sim 5 \left( \frac{M_{\rm BH}}{10^8\, M_\odot} \right)^2 \left( \frac{1\, M_\odot\, {\rm yr}^{-1}}{\dot{M}} \right) {\rm days}\,, \tag{1}$$

where the numerical coefficient accounts for typical electron cooling and Compton heating conditions. For LRDs with $M_{\rm BH} \sim 10^8\,M_\odot$ and $\dot{M} \sim 1-2\,M_\odot\,yr^{-1}$, this gives:

$$\tau_X \sim 1\text{–}5\,\text{days}\,(\text{rest frame}), \tag{2}$$

or $\sim 5-25$ days in the observed frame at $z \sim 4$. This characteristic timescale $\tau_X$ has also been linked to the break observed in AGN X-ray power spectra, which marks a transition between slow, long-timescale variability and more rapid, damped fluctuations [276]. Variability in X-rays is often described by a power spectral density that steepens at a characteristic break frequency $\nu_X = 1/\tau_X$. At low frequencies (long timescales, $\Delta t > \tau_X$), variability power typically follows a flicker noise spectrum with slope $\alpha \sim -1$. At high frequencies (short timescales, $\Delta t < \tau_X$), the variability is suppressed, and the PSD steepens to $\alpha \sim -2$, characteristic of red noise. This break reflects the inability of the corona to respond to rapid fluctuations on timescales shorter than $\tau_X$. This effect has broadly been interpreted as a signature of the Compton cooling and heating timescale of the hot plasma [276].

To detect such variability, we propose 10 epochs spaced logarithmically over a 1-year period. This cadence is designed to cover the full range of variability timescales expected from the corona, from just above the characteristic electron heating time $\tau_X$ (a few days in the rest frame) up to year-scale secular fluctuations. While $\tau_X$ represents the physical threshold below which variability is damped, longer baselines are needed to fully map the structure function and capture rare, high-amplitude transient events. The logarithmic spacing allows efficient sampling of both fast coronal fluctuations and slower accretion instabilities. Each 30 ks observation will yield $\sim 200$ counts per epoch per source, based on AXIS response simulations for sources with $F_{2-10} \sim 10^{-15}\,\text{erg}\,\text{s}^{-1}\,\text{cm}^{-2}$, calculated via WebPIMMS. Variability amplitudes of 10–20% will be easily detectable with AXIS. In contrast, Chandra, due to a smaller effective area, will not be able to detect variability at any significant confidence level (see Fig. 45). A non-detection will rule out typical coronal variability and support either jet-dominated or corona-quenched emission mechanisms [440].

In this physical framework, coronal variability is expected to be very sensitive to electron density fluctuations. Luminosity fluctuations, as a function of changes in electron density $n_e$, are given by:

$$\frac{\Delta L_X}{L_X} \sim \frac{9}{7}\frac{\Delta n_e}{n_e}\,. \tag{3}$$

Hence, even small changes in electron density can produce 10–30% luminosity variability [276].

Beyond secular variability, this program will also enable the detection of rare, short-lived high-energy transients:

- **Coronal eclipses**, caused by obscuring gas clumps;
- **Mini-outbursts**, due to magnetic or thermal instabilities;
- **Tidal disruption events (TDEs)**, a signature of early black hole growth, as recently suggested by [56].

This will be the first systematic X-ray variability study of this key AGN population, and is graphically summarized in Figure 45. The experiment is possible only with AXIS's large effective area, high angular resolution, and flexible scheduling time. If variability is detected, it will provide direct evidence of coronal processes in SMBHs hosted in compact galaxies at high redshift. If absent, it would point toward alternative accretion modes or extreme obscuration, or absence of AGN in the LRDs altogether (see, e.g., Baggen et al. 31).

**Exposure time (ks):**
300 ks total.

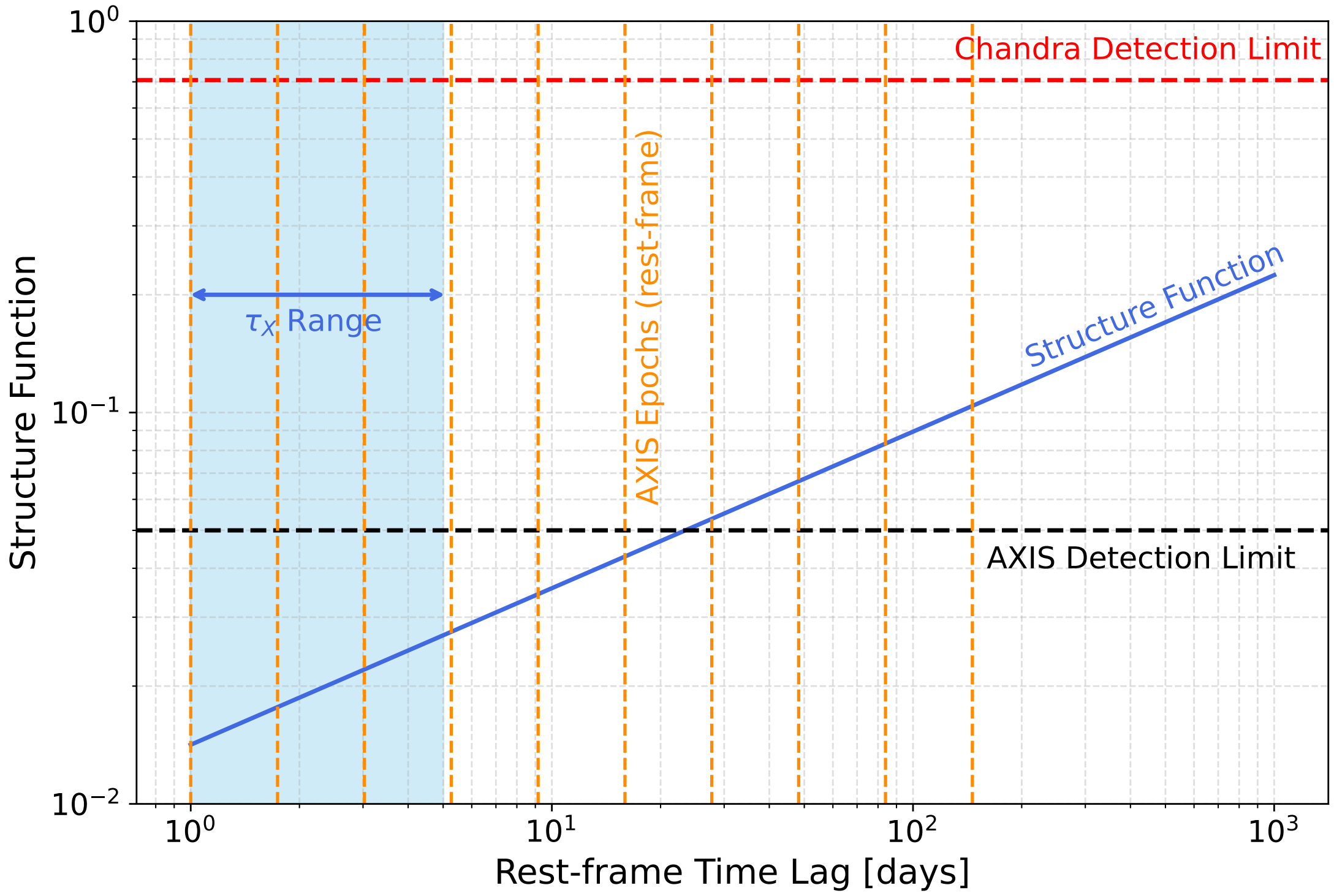

**Figure 45.** Predicted structure function of X-ray variability for an LRD with $M_{\rm BH} = 10^8\, M_\odot$. The blue curve shows the expected root-mean-square fractional variability as a function of rest-frame time lag. The shaded blue region indicates the theoretical suppression regime due to electron heating/cooling ($\tau_X = 1 - 5$ days, see main text for details). AXIS observation epochs (orange lines) are spaced logarithmically and probe both above and inside the suppressed regime. The black dashed line shows AXIS variability detection threshold; Chandra's threshold (red) is shown for comparison and lies far above any expected signal. AXIS can detect the predicted variability at rest-frame lags $\gtrsim$20 days, providing a direct test of coronal physics in high-redshift AGN.

10 epochs of 30 ks per source, for 2 sources (PRIMER-COS 3866 and JADES 21925). These fields overlap with widely studied survey regions and may be included in future AXIS PI-led surveys. In that case, this GO program will leverage those datasets.

**Observing description:**
We propose 10 observations of each source, each 30 ks in duration, spaced logarithmically over a 1-year period. This spacing ensures sensitivity to both rapid variability near the expected $\tau_X$ timescale (days to weeks) and longer-term variability associated with secular accretion instabilities. Each observation uses standard AXIS full-field imaging in the $0.5 - 7$ keV band. No roll-angle or orientation constraints are required.

**Joint Observations and synergies with other observatories in the 2030s:**
This program complements JWST spectroscopy (e.g., CEERS, JADES), rest-UV variability searches with Roman, and deep radio follow-up with ngVLA. AXIS provides the only feasible X-ray time-domain capability to measure coronal variability in the early Universe.

**Special Requirements:**
Monitoring (logarithmic cadence); no special orientation or scheduling constraints.

1. Abazajian, K. N., Acero, M. A., Agarwalla, S. K., et al. 2012, arXiv e-prints, arXiv:1204.5379
2. Abbott, L. F., & Sikivie, P. 1983, Physics Letters B, 120, 133
3. Abdo, A. A., Ackermann, M., Ajello, M., et al. 2009, ApJ, 707, L142
4. —. 2010, ApJ, 720, 912
5. Abramowicz, M. A., Igumenshchev, I. V., Quataert, E., & Narayan, R. 2002, ApJ, 565, 1101
6. Agazie, G., Anumarlapudi, A., Archibald, A. M., et al. 2023, ApJ, 951, L8
7. Aird, J., Coil, A. L., & Georgakakis, A. 2018, MNRAS, 474, 1225
8. —. 2019, MNRAS, 484, 4360
9. Aird, J., Coil, A. L., Georgakakis, A., et al. 2015, MNRAS, 451, 1892
10. —. 2015, MNRAS, 451, 1892
11. Ajello, M., Baldini, L., Ballet, J., et al. 2022, ApJS, 263, 24
12. Akylas, A., & Georgantopoulos, I. 2009, A&A, 500, 999
13. Algaba, J. C., Baloković, M., Chandra, S., et al. 2024, A&A, 692, A140
14. Allahverdi, R., Brandenberger, R., Cyr-Racine, F.-Y., & Mazumdar, A. 2010, Annual Review of Nuclear and Particle Science, 60, 27
15. Allen, S. W., Dunn, R. J. H., Fabian, A. C., Taylor, G. B., & Reynolds, C. S. 2006, MNRAS, 372, 21
16. Amaro-Seoane, P., Andrews, J., Arca Sedda, M., et al. 2023, Living Reviews in Relativity, 26, 2
17. Ananna, T. T., Bogdán, Á., Kovács, O. E., Natarajan, P., & Hickox, R. C. 2024, ApJ, 969, L18
18. Ananna, T. T., Treister, E., Urry, C. M., et al. 2019, ApJ, 871, 240
19. —. 2019, ApJ, 871, 240
20. Andonie, C., Alexander, D. M., Greenwell, C., et al. 2024, MNRAS, 527, L144
21. Arcodia, R., Merloni, A., Nandra, K., et al. 2021, Nature, 592, 704
22. Arcodia, R., Merloni, A., Comparat, J., et al. 2024, A&A, 681, A97
23. Arvanitaki, A., Dimopoulos, S., Dubovsky, S., Kaloper, N., & March-Russell, J. 2010, Ph. Rev. D, 81, 123530
24. Asmus, D., Gandhi, P., Hönig, S. F., Smette, A., & Duschl, W. J. 2015, MNRAS, 454, 766
25. Bañados, E., Venemans, B. P., Mazzucchelli, C., et al. 2018, Nature, 553, 473
26. Bañados, E., Connor, T., Stern, D., et al. 2018, ApJ, 856, L25
27. Bañados, E., Momjian, E., Connor, T., et al. 2025, Nature Astronomy, 9, 293
28. Baganoff, F. K., Maeda, Y., Morris, M., et al. 2003, ApJ, 591, 891
29. —. 2003, ApJ, 591, 891
30. Baggen, J. F. W., van Dokkum, P., Labbé, I., et al. 2023, ApJ, 955, L12
31. Baggen, J. F. W., van Dokkum, P., Brammer, G., et al. 2024, ApJ, 977, L13
32. Baldi, R. D. 2023, A&A Rev., 31, 3
33. Baldi, R. D., Behar, E., Laor, A., & Horesh, A. 2015, MNRAS, 454, 4277
34. Baldi, R. D., Capetti, A., & Giovannini, G. 2015, A&A, 576, A38
35. —. 2016, Astronomische Nachrichten, 337, 114
36. Baldi, R. D., Capetti, A., & Massaro, F. 2018, A&A, 609, A1
37. Baldwin, J. A., Phillips, M. M., & Terlevich, R. 1981, PASP, 93, 5
38. Ballantyne, D. R. 2020, MNRAS, 491, 3553
39. Ballo, L., Braito, V., Reeves, J. N., Sambruna, R. M., & Tombesi, F. 2011, MNRAS, 418, 2367
40. Baloković, M., García, J. A., & Cabral, S. E. 2019, Research Notes of the American Astronomical Society, 3, 173
41. Bambi, C., Brenneman, L. W., Dauser, T., et al. 2021, Space Science Reviews, 217, 65
42. Bambic, C. J., Quataert, E., & Kunz, M. W. 2024, MNRAS, 527, 2895
43. Bambic, C. J., Russell, H. R., Reynolds, C. S., et al. 2023, MNRAS, 522, 4374
44. Barnes, J. E., & Hernquist, L. E. 1991, ApJ, 370, L65
45. Barrows, R. S., Comerford, J. M., Stern, D., & Assef, R. J. 2023, ApJ, 951, 92
46. Baskin, A., Laor, A., & Stern, J. 2014, MNRAS, 438, 604

47. Bassani, L., Dadina, M., Maiolino, R., et al. 1999, ApJS, 121, 473
48. Bassini, L., Rasia, E., Borgani, S., et al. 2019, A&A, 630, A144
49. Begelman, M. C. 2012, MNRAS, 420, 2912
50. Begelman, M. C., Blandford, R. D., & Rees, M. J. 1980, Nature, 287, 307
51. Begelman et al., M. C. 2006, MNRAS, 370, 289
52. Behar, E., Vogel, S., Baldi, R. D., Smith, K. L., & Mushotzky, R. F. 2018, MNRAS, 478, 399
53. Behar, E., Kaspi, S., Paubert, G., et al. 2020, MNRAS, 491, 3523
54. Bell, M. E., Tzioumis, T., Uttley, P., et al. 2011, MNRAS, 411, 402
55. Belladitta, S., Moretti, A., Caccianiga, A., et al. 2020, A&A, 635, L7
56. Bellovary, J. 2025, ApJ, 984, L55
57. Bender, R., Kormendy, J., Bower, G., et al. 2005, ApJ, 631, 280
58. Bertola, E., Vignali, C., Lanzuisi, G., et al. 2022, A&A, 662, A98
59. Bhowmick, A. K., Blecha, L., Torrey, P., et al. 2022, MNRAS, 510, 177
60. Bianchi, S., Chiaberge, M., Evans, D. A., et al. 2010, MNRAS, 405, 553
61. Bianchi, S., Chiaberge, M., Piconcelli, E., Guainazzi, M., & Matt, G. 2008, MNRAS, 386, 105
62. Bianchi, S., Guainazzi, M., & Chiaberge, M. 2006, A&A, 448, 499
63. Bianchi, S., Guainazzi, M., Laor, A., Stern, J., & Behar, E. 2019, MNRAS, 485, 416
64. Bianchi, S., Guainazzi, M., Matt, G., Fonseca Bonilla, N., & Ponti, G. 2009, A&A, 495, 421
65. Bianchi, S., Piconcelli, E., Chiaberge, M., et al. 2009, ApJ, 695, 781
66. Bird, S., Ni, Y., Di Matteo, T., et al. 2022, MNRAS, 512, 3703
67. Bischetti, M., Feruglio, C., D'Odorico, V., et al. 2022, Nature, 605, 244
68. Blandford, R., & Globus, N. 2022, MNRAS, 514, 5141
69. Blandford, R. D. 2001, Progress of Theoretical Physics Supplement, 143, 182
70. Blandford, R. D., & Begelman, M. C. 1999, MNRAS, 303, L1
71. Blandford, R. D., & Rees, M. J. 1978, *Phys. Scr.*, 17, 265
72. Blecha, L., Snyder, G. F., Satyapal, S., & Ellison, S. L. 2018, MNRAS, 478, 3056
73. Blustin, A. J., Page, M. J., Fuerst, S. V., Branduardi-Raymont, G., & Ashton, C. E. 2005, A&A, 431, 111
74. Bogdán, Á., Goulding, A. D., Natarajan, P., et al. 2024, Nature Astronomy, 8, 126
75. Bollati, F., Lupi, A., Dotti, M., & Haardt, F. 2024, A&A, 690, A194
76. Boller, T., Brandt, W. N., & Fink, H. 1996, A&A, 305, 53
77. Bondi, H. 1952, MNRAS, 112, 195
78. Bonzini, M., Padovani, P., Mainieri, V., et al. 2013, MNRAS, 436, 3759
79. Boorman, P. G., Gandhi, P., Buchner, J., et al. 2025, ApJ, 978, 118
80. Breiding, P., Meyer, E. T., Georganopoulos, M., et al. 2023, MNRAS, 518, 3222
81. Brenneman, L., Miller, J., Nantra, P., et al. 2009, in astro2010: The Astronomy and Astrophysics Decadal Survey, Vol. 2010, 26
82. Brighenti, F., & Mathews, W. G. 1999, ApJ, 527, L89
83. Brightman, M., Silverman, J. D., Mainieri, V., et al. 2013, MNRAS, 433, 2485
84. —. 2013, MNRAS, 433, 2485
85. Bronzini, E., Grandi, P., Torresi, E., & Buson, S. 2024, ApJ, 977, L16
86. Buchner, J., Georgakakis, A., Nandra, K., et al. 2015, ApJ, 802, 89
87. —. 2015, ApJ, 802, 89
88. Burke-Spolaor, S., Taylor, S. R., Charisi, M., et al. 2019, A&A Rev., 27, 5
89. Burtscher, L., Meisenheimer, K., Tristram, K. R. W., et al. 2013, A&A, 558, A149
90. Buttiglione, S., Capetti, A., Celotti, A., et al. 2010, A&A, 509, A6
91. Cackett, E. M., Zoghbi, A., Reynolds, C., et al. 2014, MNRAS, 438, 2980
92. Cao, Z., Aharonian, F., Axikegu, et al. 2024, ApJ, 971, L45
93. Cappi, M., Mihara, T., Matsuoka, M., et al. 1996, ApJ, 456, 141
94. Cara, M., Perlman, E. S., Uchiyama, Y., et al. 2013, ApJ, 773, 186

95. Carenza, P., Ferreira, T., & Nguyen, T. T. Q. 2025, arXiv e-prints, arXiv:2507.08932
96. Carnall et al., A. C. 2023, MNRAS, 520, 3974
97. Carr, B. J., Kohri, K., Sendouda, Y., & Yokoyama, J. 2016, Ph. Rev. D, 94, 044029
98. Celotti, A., Ghisellini, G., & Chiaberge, M. 2001, MNRAS, 321, L1
99. Celotti, A., Ghisellini, G., & Fabian, A. C. 2007, MNRAS, 375, 417
100. Chadha-Day, F., Ellis, J., & Marsh, D. J. E. 2022, Science Advances, 8, eabj3618
101. Chakraborty, J., Kara, E., Arcodia, R., et al. 2025, arXiv e-prints, arXiv:2503.19013
102. Chang, Y.-C., Soria, R., Kong, A. K. H., et al. 2025, arXiv e-prints, arXiv:2503.00904
103. Chartas, G., Brandt, W. N., Gallagher, S. C., & Garmire, G. P. 2002, ApJ, 579, 169
104. Chartas, G., Krawczynski, H., Zalesky, L., et al. 2017, ApJ, 837, 26
105. Chartas, G., Rhea, C., Kochanek, C., et al. 2016, Astronomische Nachrichten, 337, 356
106. Chartas, G., Cappi, M., Vignali, C., et al. 2021, ApJ, 920, 24
107. Chatterjee, R., Marscher, A. P., Jorstad, S. G., et al. 2009, ApJ, 704, 1689
108. —. 2011, ApJ, 734, 43
109. —. 2011, ApJ, 734, 43
110. Chen, N., Di Matteo, T., Ni, Y., et al. 2023, MNRAS, 522, 1895
111. Chen, S., Laor, A., & Behar, E. 2022, MNRAS, 515, 1723
112. Chen, Y., Gu, Q., Fan, J., et al. 2024, MNRAS, 534, 2134
113. Chen, Y.-C., Hwang, H.-C., Shen, Y., et al. 2022, ApJ, 925, 162
114. Chiang, Y.-K., Overzier, R. A., Gebhardt, K., & Henriques, B. 2017, ApJ, 844, L23
115. Chilingarian, I. V., Katkov, I. Y., Zolotukhin, I. Y., et al. 2018, ApJ, 863, 1
116. Circosta, C., Vignali, C., Gilli, R., et al. 2019, A&A, 623, A172
117. Cisternas, M., Jahnke, K., Inskip, K. J., et al. 2011, ApJ, 726, 57
118. Comerford, J. M., Gerke, B. F., Stern, D., et al. 2012, ApJ, 753, 42
119. Comerford, J. M., Pooley, D., Barrows, R. S., et al. 2015, ApJ, 806, 219
120. Condon, J. J., & Ransom, S. M. 2016, Essential Radio Astronomy
121. Connor, T., Bañados, E., Cappelluti, N., & Foord, A. 2024, Universe, 10, 227
122. Connor, T., Stern, D., Bañados, E., & Mazzucchelli, C. 2021, ApJ, 922, L24
123. Connor, T., Bañados, E., Stern, D., et al. 2019, ApJ, 887, 171
124. Connor, T., Bañados, E., Mazzucchelli, C., et al. 2020, ApJ, 900, 189
125. Connor, T., Bañados, E., Stern, D., et al. 2021, ApJ, 911, 120
126. Corbel, S., Coriat, M., Brocksopp, C., et al. 2013, MNRAS, 428, 2500
127. Crummy, J., Fabian, A. C., Gallo, L., & Ross, R. R. 2006, MNRAS, 365, 1067
128. Cui, W., Jennings, F., Dave, R., Babul, A., & Gozaliasl, G. 2024, Monthly Notices of the Royal Astronomical Society, 534, 1247
129. Czerny, B., & Elvis, M. 1987, ApJ, 321, 305
130. D'Amato, Q., Gilli, R., Vignali, C., et al. 2020, A&A, 636, A37
131. D'Ammando, F. 2019, Galaxies, 7, 87
132. —. 2020, MNRAS, 496, 2213
133. D'Ammando, F., Larsson, J., Orienti, M., et al. 2014, MNRAS, 438, 3521
134. Damour, T., & Solodukhin, S. N. 2007, Ph. Rev. D, 76, 024016
135. Dauser, T., Falkner, S., Lorenz, M., et al. 2019, A&A, 630, A66
136. De Colle, F., & Lu, W. 2020, New A Rev., 89, 101538
137. De la Torre Luque, P., Carenza, P., & Nguyen, T. T. Q. 2025, arXiv e-prints, arXiv:2507.01962
138. De Marco, B., Ponti, G., Cappi, M., et al. 2013, MNRAS, 431, 2441
139. De Rosa, A., Vignali, C., Husemann, B., et al. 2018, MNRAS, 480, 1639
140. De Rosa, A., Vignali, C., Bogdanović, T., et al. 2019, New A Rev., 86, 101525
141. De Rosa, A., Vignali, C., Severgnini, P., et al. 2023, MNRAS, 519, 5149
142. Decarli, R., Walter, F., Aravena, M., et al. 2016, ApJ, 833, 69

143. Decarli, R., Walter, F., Venemans, B. P., et al. 2017, Nature, 545, 457
144. Delvecchio, I., Daddi, E., Aird, J., et al. 2020, ApJ, 892, 17
145. Demirtas, M., Gendler, N., Long, C., McAllister, L., & Moritz, J. 2023, Journal of High Energy Physics, 2023, 92
146. Deo, R. P., Crenshaw, D. M., & Kraemer, S. B. 2006, AJ, 132, 321
147. Dessert, C., Foster, J. W., Park, Y., & Safdi, B. R. 2024, ApJ, 964, 185
148. D'Eugenio, F., Maiolino, R., Perna, M., et al. 2025, arXiv e-prints, arXiv:2503.11752
149. Di Gesu, L., & Costantini, E. 2016, A&A, 594, A88
150. Di Gesu, L., Costantini, E., Arav, N., et al. 2013, A&A, 556, A94
151. Di Matteo, T., Khandai, N., DeGraf, C., et al. 2012, ApJ, 745, L29
152. Di Matteo, T., Springel, V., & Hernquist, L. 2005, Nature, 433, 604
153. DiKerby, S., Zhang, S., & Irwin, J. 2025, arXiv e-prints, arXiv:2502.01365
154. Dine, M., & Fischler, W. 1983, Physics Letters B, 120, 137
155. Dodelson, S., & Widrow, L. M. 1994, Ph. Rev. L, 72, 17
156. Doeleman, S. S., Fish, V. L., Schenck, D. E., et al. 2012, Science, 338, 355
157. Done, C., Davis, S. W., Jin, C., Blaes, O., & Ward, M. 2012, MNRAS, 420, 1848
158. Doré, O., Werner, M. W., Ashby, M. L. N., et al. 2018, arXiv e-prints, arXiv:1805.05489
159. Du, P., Hu, C., Lu, K.-X., et al. 2014, ApJ, 782, 45
160. Dubois, Y., Volonteri, M., & Silk, J. 2014, MNRAS, 440, 1590
161. Dubois, Y., Beckmann, R., Bournaud, F., et al. 2021, A&A, 651, A109
162. Duras, F., Bongiorno, A., Ricci, F., et al. 2020, A&A, 636, A73
163. Ebrero, J., Kaastra, J. S., Kriss, G. A., et al. 2016, A&A, 587, A129
164. Eftekharzadeh, S., Myers, A. D., Hennawi, J. F., et al. 2017, MNRAS, 468, 77
165. Eilers, A.-C., Simcoe, R. A., Yue, M., et al. 2023, ApJ, 950, 68
166. Elbaz, D., Daddi, E., Le Borgne, D., et al. 2007, A&A, 468, 33
167. Ellison, S. L., Viswanathan, A., Patton, D. R., et al. 2019, MNRAS, 487, 2491
168. Euclid Collaboration, Barnett, R., Warren, S. J., et al. 2019, A&A, 631, A85
169. Euclid Collaboration, Mellier, Y., Abdurro'uf, et al. 2024, arXiv e-prints, arXiv:2405.13491
170. Event Horizon Telescope Collaboration, Akiyama, K., Alberdi, A., et al. 2019, ApJ, 875, L1
171. —. 2019, ApJ, 875, L1
172. —. 2022, ApJ, 930, L12
173. —. 2022, ApJ, 930, L12
174. Fabbiano, G., Elvis, M., Paggi, A., et al. 2017, ApJ, 842, L4
175. Fabbiano, G., Paggi, A., Karovska, M., et al. 2018, ApJ, 855, 131
176. Fabbrichesi, M., Gabrielli, E., & Lanfranchi, G. 2020, arXiv e-prints, arXiv:2005.01515
177. Fabian, A. C. 2012, ARA&A, 50, 455
178. —. 2012, ARA&A, 50, 455
179. Fabian, A. C., Celotti, A., Blundell, K. M., Kassim, N. E., & Perley, R. A. 2002, MNRAS, 331, 369
180. Fabian, A. C., Lohfink, A., Belmont, R., Malzac, J., & Coppi, P. 2017, MNRAS, 467, 2566
181. Fabian, A. C., Lohfink, A., Kara, E., et al. 2015, MNRAS, 451, 4375
182. Fabian, A. C., Parker, M. L., Wilkins, D. R., et al. 2014, MNRAS, 439, 2307
183. Fabian, A. C., Zoghbi, A., Ross, R. R., et al. 2009, Nature, 459, 540
184. Fagin, J., Paic, E., Neira, F., et al. 2025, ApJ, 981, 61
185. Fan, X., Bañados, E., & Simcoe, R. A. 2023, ARA&A, 61, 373
186. Fanaroff, B. L., & Riley, J. M. 1974, MNRAS, 167, 31P
187. Farrell, S. A., Webb, N. A., Barret, D., Godet, O., & Rodrigues, J. M. 2009, Nature, 460, 73
188. Fedorova, E., & Del Popolo, A. 2023, Universe, 9, 212
189. Fedorova, E., Hnatyk, B. I., Zhdanov, V. I., & Del Popolo, A. 2020, Universe, 6, 219
190. Fender, R. P., Belloni, T. M., & Gallo, E. 2004, MNRAS, 355, 1105
191. Fender, R. P., Homan, J., & Belloni, T. M. 2009, MNRAS, 396, 1370

192. Fernández-Ontiveros, J. A., & Muñoz-Darias, T. 2021, MNRAS, 504, 5726
193. Ferrarese, L., & Merritt, D. 2000, ApJ, 539, L9
194. Feruglio, C., Fabbiano, G., Bischetti, M., et al. 2020, ApJ, 890, 29
195. Fiore, F., Elvis, M., Mathur, S., Wilkes, B. J., & McDowell, J. C. 1993, ApJ, 415, 129
196. Foord, A., Gültekin, K., Nevin, R., et al. 2020, ApJ, 892, 29
197. Foord, A., Cappelluti, N., Liu, T., et al. 2024, Universe, 10, 237
198. Foster, J. W., Kongsore, M., Dessert, C., et al. 2021, Ph. Rev. L, 127, 051101
199. Franchini, A., Bonetti, M., Lupi, A., et al. 2023, A&A, 675, A100
200. Fu, H., Myers, A. D., Djorgovski, S. G., et al. 2015, ApJ, 799, 72
201. Fu, H., Yan, L., Myers, A. D., et al. 2012, ApJ, 745, 67
202. Fu, H., Zhang, Z.-Y., Assef, R. J., et al. 2011, ApJ, 740, L44
203. Fukumura, K., Mehdipour, M., Behar, E., et al. 2024, ApJ, 968, 70
204. García, J., Dauser, T., Reynolds, C. S., et al. 2013, ApJ, 768, 146
205. García, J. A., Fabian, A. C., Kallman, T. R., et al. 2016, MNRAS, 462, 751
206. García, J. A., Kara, E., Walton, D., et al. 2019, ApJ, 871, 88
207. Garcia, M. R., Murray, S. S., Primini, F. A., et al. 2000, ApJ, 537, L23
208. Garcia, M. R., Hextall, R., Baganoff, F. K., et al. 2010, ApJ, 710, 755
209. García-Burillo, S., Combes, F., Schinnerer, E., Boone, F., & Hunt, L. K. 2005, A&A, 441, 1011
210. Garofalo, D. 2009, in American Astronomical Society Meeting Abstracts, Vol. 214, American Astronomical Society Meeting Abstracts #214, 416.10
211. Garofalo, D. 2013, Advances in Astronomy, 2013, 213105
212. Gaspari, M., Brighenti, F., & Temi, P. 2015, A&A, 579, A62
213. Gaspari, M., Tombesi, F., & Cappi, M. 2020, Nature Astronomy, 4, 10
214. Gebhardt, K., Adams, J., Richstone, D., et al. 2011, ApJ, 729, 119
215. Gebhardt, K., Bender, R., Bower, G., et al. 2000, ApJ, 539, L13
216. Gendler, N., Marsh, D. J. E., McAllister, L., & Moritz, J. 2024, JCAP, 2024, 071
217. Genzel, R., Schödel, R., Ott, T., et al. 2003, Nature, 425, 934
218. Gezari, S. 2021, ARA&A, 59, 21
219. Ghez, A. M., Klein, B. L., Morris, M., & Becklin, E. E. 1998, ApJ, 509, 678
220. Ghisellini, G., Tagliaferri, G., Sbarrato, T., & Gehrels, N. 2015, MNRAS, 450, L34
221. Ghisellini, G., & Tavecchio, F. 2009, MNRAS, 397, 985
222. Ghisellini et al., G. 2013, MNRAS, 432, 2818
223. Gilli, R., Comastri, A., & Hasinger, G. 2007, A&A, 463, 79
224. —. 2007, A&A, 463, 79
225. Gilli, R., Mignoli, M., Peca, A., et al. 2019, A&A, 632, A26
226. Gilli, R., Norman, C., Calura, F., et al. 2022, A&A, 666, A17
227. —. 2022, A&A, 666, A17
228. Giroletti, M., Taylor, G. B., & Giovannini, G. 2005, ApJ, 622, 178
229. Gliozzi, M., & Williams, J. K. 2020, MNRAS, 491, 532
230. Gloudemans, A. J., Duncan, K. J., Eilers, A.-C., et al. 2025, arXiv e-prints, arXiv:2501.04912
231. González, J. A., Hannam, M., Sperhake, U., Brügmann, B., & Husa, S. 2007, Ph. Rev. L, 98, 231101
232. Goodger, J. L., Hardcastle, M. J., Croston, J. H., et al. 2010, ApJ, 708, 675
233. Goodsell, M., Jaeckel, J., Redondo, J., & Ringwald, A. 2009, Journal of High Energy Physics, 2009, 027
234. Graham, A. W., Chilingarian, I. V., Nguyen, D. D., et al. 2025, arXiv e-prints, arXiv:2503.10958
235. Graham, P. W., Mardon, J., & Rajendran, S. 2016, Ph. Rev. D, 93, 103520
236. Grandi, P., Capetti, A., & Baldi, R. D. 2016, MNRAS, 457, 2
237. Grandi, P., Torresi, E., & Stanghellini, C. 2012, ApJ, 751, L3
238. Greene, J. E. 2012, Nature Communications, 3, 1304
239. Greene, J. E., Strader, J., & Ho, L. C. 2020, ARA&A, 58, 257

240. Greene, J. E., Labbe, I., Goulding, A. D., et al. 2023, arXiv e-prints, arXiv:2309.05714
241. —. 2024, ApJ, 964, 39
242. Grupe, D., Komossa, S., Leighly, K. M., & Page, K. L. 2010, ApJS, 187, 64
243. Guainazzi, M., & Bianchi, S. 2007, MNRAS, 374, 1290
244. Guainazzi, M., La Parola, V., Miniutti, G., Segreto, A., & Longinotti, A. L. 2012, A&A, 547, A31
245. Guainazzi, M., Piconcelli, E., Jiménez-Bailón, E., & Matt, G. 2005, A&A, 429, L9
246. Guainazzi, M., De Rosa, A., Bianchi, S., et al. 2021, MNRAS, 504, 393
247. Guia, C. A., Pacucci, F., & Kocevski, D. D. 2024, Research Notes of the American Astronomical Society, 8, 207
248. Guilbert, P. W., Fabian, A. C., & Rees, M. J. 1983, MNRAS, 205, 593
249. Gültekin, K., King, A. L., Cackett, E. M., et al. 2019, ApJ, 871, 80
250. Gültekin, K., Richstone, D. O., Gebhardt, K., et al. 2009, ApJ, 698, 198
251. Guo, M., Stone, J. M., Kim, C.-G., & Quataert, E. 2023, ApJ, 946, 26
252. Gyulchev, G., Kunz, J., Nedkova, P., Vetsov, T., & Yazadjiev, S. 2020, European Physical Journal C, 80, 1017
253. Habouzit, M., & Department of Astronomy, Chemin d'Ecogia, U. o. G. 2025, MNRAS, 537, 2323
254. Hall, L. J., Jedamzik, K., March-Russell, J., & West, S. M. 2010, Journal of High Energy Physics, 2010, 80
255. Hardcastle, M. J., & Croston, J. H. 2020, New A Rev., 88, 101539
256. Hardcastle, M. J., Kraft, R. P., Sivakoff, G. R., et al. 2007, ApJ, 670, L81
257. Hardcastle, M. J., Lenc, E., Birkinshaw, M., et al. 2016, MNRAS, 455, 3526
258. Harikane, Y., Zhang, Y., Nakajima, K., et al. 2023, ApJ, 959, 39
259. Harris, D. E., Cheung, C. C., Stawarz, Ł., Biretta, J. A., & Perlman, E. S. 2009, ApJ, 699, 305
260. Heckman, T. M., & Best, P. N. 2014, ARA&A, 52, 589
261. Hennawi, J. F., Strauss, M. A., Oguri, M., et al. 2006, AJ, 131, 1
262. Hickox, R. C., & Alexander, D. M. 2018, ARA&A, 56, 625
263. Hill, R., Chapman, S., Scott, D., et al. 2020, MNRAS, 495, 3124
264. Hirschmann, M., Charlot, S., Feltre, A., et al. 2023, MNRAS, 526, 3610
265. Ho, L. C. 2008, ARA&A, 46, 475
266. Hopkins, P. F., & Hernquist, L. 2009, ApJ, 694, 599
267. —. 2009, ApJ, 694, 599
268. Hou, M., Li, Z., & Liu, X. 2020, ApJ, 900, 79
269. Hwang, H.-C., Shen, Y., Zakamska, N., & Liu, X. 2020, ApJ, 888, 73
270. Ichimaru, S. 1977, ApJ, 214, 840
271. Ighina, L., Moretti, A., Tavecchio, F., et al. 2022, A&A, 659, A93
272. Igo, Z., Parker, M. L., Matzeu, G. A., et al. 2020, MNRAS, 493, 1088
273. Igumenshchev, I. V. 2008, ApJ, 677, 317
274. Igumenshchev, I. V., Abramowicz, M. A., & Narayan, R. 2000, ApJ, 537, L27
275. Igumenshchev, I. V., Narayan, R., & Abramowicz, M. A. 2003, ApJ, 592, 1042
276. Ishibashi, W., & Courvoisier, T. J. L. 2009, A&A, 504, 61
277. Ivison, R. J., Biggs, A. D., Bremer, M., Arumugam, V., & Dunne, L. 2020, MNRAS, 496, 4358
278. Jackson, N., & Rawlings, S. 1997, MNRAS, 286, 241
279. Jaffe, W., Meisenheimer, K., Röttgering, H. J. A., et al. 2004, Nature, 429, 47
280. Jana, A., Ricci, C., Temple, M. J., et al. 2025, A&A, 693, A35
281. Janssen, M., Falcke, H., Kadler, M., et al. 2021, Nature Astronomy, 5, 1017
282. Jennings, F. J., Babul, A., Davé, R., Cui, W., & Rennehan, D. 2025, Monthly Notices of the Royal Astronomical Society, 536, 145
283. Jester, S., Harris, D. E., Marshall, H. L., & Meisenheimer, K. 2006, ApJ, 648, 900
284. Jester, S., Meisenheimer, K., Martel, A. R., Perlman, E. S., & Sparks, W. B. 2007, MNRAS, 380, 828
285. Ji, X., Maiolino, R., Übler, H., et al. 2025, arXiv e-prints, arXiv:2501.13082
286. Jiang, J., Fabian, A. C., Dauser, T., et al. 2019, MNRAS, 489, 3436
287. Jin, C. C., Li, D. Y., Jiang, N., et al. 2025, arXiv e-prints, arXiv:2501.09580

288. Johnson, J. L., & Upton Sanderbeck, P. R. 2022, ApJ, 934, 58
289. Johnson, M. D., Akiyama, K., Blackburn, L., et al. 2023, Galaxies, 11, 61
290. Jolley, E. J. D., & Kuncic, Z. 2008, MNRAS, 386, 989
291. Jones, D. O., Foley, R. J., Narayan, G., et al. 2021, ApJ, 908, 143
292. Jones, S., McHardy, I., Moss, D., et al. 2011, MNRAS, 412, 2641
293. Juodžbalis, I., Ji, X., Maiolino, R., et al. 2024, arXiv e-prints, arXiv:2407.08643
294. Kammoun, E. S., Risaliti, G., Stern, D., et al. 2017, MNRAS, 465, 1665
295. Kammoun, E. S., Igo, Z., Miller, J. M., et al. 2023, MNRAS, 522, 5217
296. Kang, J.-L., Done, C., Hagen, S., et al. 2025, MNRAS, 538, 121
297. Kara, E., Alston, W. N., Fabian, A. C., et al. 2016, MNRAS, 462, 511
298. Kardashev, N. S., Novikov, I. D., & Shatskii, A. A. 2006, Astronomy Reports, 50, 601
299. Kardashev, N. S., Novikov, I. D., & Shatskiy, A. A. 2007, International Journal of Modern Physics D, 16, 909
300. Kataoka, J., Stawarz, Ł., Takahashi, Y., et al. 2011, ApJ, 740, 29
301. Kellermann, K. I., Sramek, R., Schmidt, M., Shaffer, D. B., & Green, R. 1989, AJ, 98, 1195
302. —. 1989, AJ, 98, 1195
303. Kelson, D. D., & Abramson, L. E. 2022, Research Notes of the American Astronomical Society, 6, 162
304. Kelson, D. D., Abramson, L. E., Benson, A. J., et al. 2020, MNRAS, 494, 2628
305. King, A. 2024, MNRAS, 531, 550
306. King, A. L., Lohfink, A., & Kara, E. 2017, ApJ, 835, 226
307. Kocevski, D. D., Brightman, M., Nandra, K., et al. 2015, ApJ, 814, 104
308. Kocevski, D. D., Onoue, M., Inayoshi, K., et al. 2023, ApJ, 954, L4
309. Kocevski, D. D., Finkelstein, S. L., Barro, G., et al. 2024, arXiv e-prints, arXiv:2404.03576
310. Kochanek, C. S. 2004, ApJ, 605, 58
311. Komossa, S., Burwitz, V., Hasinger, G., et al. 2003, ApJ, 582, L15
312. Komossa, S., Voges, W., Xu, D., et al. 2006, AJ, 132, 531
313. Kosec, P., Pinto, C., Fabian, A. C., & Walton, D. J. 2018, MNRAS, 473, 5680
314. Koss, M., Mushotzky, R., Treister, E., et al. 2012, ApJ, 746, L22
315. Koss, M., Mushotzky, R., Veilleux, S., & Winter, L. 2010, ApJ, 716, L125
316. Koss, M., Trakhtenbrot, B., Ricci, C., et al. 2017, ApJ, 850, 74
317. Koss, M. J., Blecha, L., Bernhard, P., et al. 2018, Nature, 563, 214
318. Koss, M. J., Trakhtenbrot, B., Ricci, C., et al. 2022, ApJS, 261, 1
319. —. 2022, arXiv e-prints, arXiv:2207.12428
320. Krawczynski, H., & Chartas, G. 2017, ApJ, 843, 118
321. Krawczynski, H., Chartas, G., & Kislat, F. 2019, ApJ, 870, 125
322. Krongold, Y., Nicastro, F., Elvis, M., et al. 2007, ApJ, 659, 1022
323. Kubota, A., Tanaka, Y., Makishima, K., et al. 1998, PASJ, 50, 667
324. Kynoch, D., Landt, H., Ward, M. J., et al. 2018, MNRAS, 475, 404
325. La Franca, F., Bianchi, S., Ponti, G., Branchini, E., & Matt, G. 2014, ApJ, 787, L12
326. La Mura, G., Busetto, G., Ciroi, S., et al. 2017, European Physical Journal D, 71, 95
327. Laha, S., Guainazzi, M., Dewangan, G. C., Chakravorty, S., & Kembhavi, A. K. 2014, MNRAS, 441, 2613
328. Laha, S., Markowitz, A. G., Krumpe, M., et al. 2020, ApJ, 897, 66
329. Laing, R. A., Jenkins, C. R., Wall, J. V., & Unger, S. W. 1994, in Astronomical Society of the Pacific Conference Series, Vol. 54, The Physics of Active Galaxies, ed. G. V. Bicknell, M. A. Dopita, & P. J. Quinn, 201
330. Lalakos, A., Gottlieb, O., Kaaz, N., et al. 2022, ApJ, 936, L5
331. Lambrides, E., Garofali, K., Larson, R., et al. 2024, arXiv e-prints, arXiv:2409.13047
332. Lambrides, E. L., Chiaberge, M., Heckman, T., et al. 2020, ApJ, 897, 160
333. Lanzuisi, G., Civano, F., Marchesi, S., et al. 2018, MNRAS, 480, 2578
334. Lanzuisi, G., Gilli, R., Cappi, M., et al. 2019, ApJ, 875, L20
335. Laor, A., & Behar, E. 2008, MNRAS, 390, 847

336. Larsson, J., D'Ammando, F., Falocco, S., et al. 2018, MNRAS, 476, 43
337. Latif, M. A., Whalen, D. J., Khochfar, S., Herrington, N. P., & Woods, T. E. 2022, Nature, 607, 48
338. Lavaux, G., & Hudson, M. J. 2011, MNRAS, 416, 2840
339. Lawrence, A. 1991, MNRAS, 252, 586
340. Lemaux, B. C., Cucciati, O., Le Fèvre, O., et al. 2022, A&A, 662, A33
341. Li, J., Ostriker, J., & Sunyaev, R. 2013, ApJ, 767, 105
342. Li, R., Ricci, C., Ho, L. C., et al. 2024, ApJ, 975, 140
343. Li, Z., Garcia, M. R., Forman, W. R., et al. 2011, ApJ, 728, L10
344. Li, Z., Wang, Q. D., & Wakker, B. P. 2009, MNRAS, 397, 148
345. Lin, D., Strader, J., Carrasco, E. R., et al. 2018, Nature Astronomy, 2, 656
346. Linden, T., Nguyen, T. T. Q., & Tait, T. M. P. 2024, arXiv e-prints, arXiv:2402.01839
347. —. 2024, arXiv e-prints, arXiv:2406.19445
348. Linial, I., & Metzger, B. D. 2023, ApJ, 957, 34
349. Liu, H., Elwood, B. D., Evans, M., & Thaler, J. 2019, Ph. Rev. D, 100, 023548
350. Liu, H., Luo, B., Brandt, W. N., et al. 2021, On the Observational Difference Between the Accretion Disk-Corona Connections among Super- and Sub-Eddington Accreting Active Galactic Nuclei, arXiv:2102.02832 [astro-ph.GA]
351. Liu, T., Tozzi, P., Wang, J.-X., et al. 2017, ApJS, 232, 8
352. Liu, X., Civano, F., Shen, Y., et al. 2013, ApJ, 762, 110
353. Liu, X., Greene, J. E., Shen, Y., & Strauss, M. A. 2010, ApJ, 715, L30
354. Liu, X., Shen, Y., Strauss, M. A., & Greene, J. E. 2010, ApJ, 708, 427
355. Liu, X., Shen, Y., Strauss, M. A., & Hao, L. 2011, ApJ, 737, 101
356. Lohfink, A. M., Reynolds, C. S., Jorstad, S. G., et al. 2013, ApJ, 772, 83
357. Ludlam, R. M., Cackett, E. M., Gültekin, K., et al. 2015, MNRAS, 447, 2112
358. Luminari, A., Piconcelli, E., Tombesi, F., et al. 2018, A&A, 619, A149
359. Luminari, A., Tombesi, F., Piconcelli, E., et al. 2020, A&A, 633, A55
360. Luo, B., Brandt, W. N., Xue, Y. Q., et al. 2017, ApJS, 228, 2
361. Lupi et al., A. 2024, A&A, 686, A256
362. Luque, P. D. l. T., Koechler, J., & Balaji, S. 2024, Ph. Rev. D, 110, 123022
363. Lusso, E., Comastri, A., Vignali, C., et al. 2010, A&A, 512, A34
364. Lyke, B. W., Higley, A. N., McLane, J. N., et al. 2020, ApJS, 250, 8
365. Madau, P., & Haardt, F. 2024, arXiv e-prints, arXiv:2410.00417
366. Magdziarz, P., Blaes, O. M., Zdziarski, A. A., Johnson, W. N., & Smith, D. A. 1998, MNRAS, 301, 179
367. Magdziarz, P., & Zdziarski, A. A. 1995, MNRAS, 273, 837
368. Maiolino, R., Scholtz, J., Curtis-Lake, E., et al. 2023, arXiv e-prints, arXiv:2308.01230
369. —. 2024, A&A, 691, A145
370. Maiolino, R., Risaliti, G., Signorini, M., et al. 2024, arXiv e-prints, arXiv:2405.00504
371. —. 2025, MNRAS, 538, 1921
372. Mallick, L., Fabian, A. C., García, J. A., et al. 2022, MNRAS, 513, 4361
373. Mallick, L., Pinto, C., Tomsick, J., et al. 2025, arXiv e-prints, arXiv:2501.15380
374. Mannucci, F., Scialpi, M., Ciurlo, A., et al. 2023, A&A, 680, A53
375. Manzari, C. A., Park, Y., Safdi, B. R., & Savoray, I. 2024, Ph. Rev. L, 133, 211002
376. Marchenko, V., Harris, D. E., Ostrowski, M., et al. 2017, ApJ, 844, 11
377. Marchesi, S., Lanzuisi, G., Civano, F., et al. 2016, ApJ, 830, 100
378. Marchesi, S., Ajello, M., Zhao, X., et al. 2019, ApJ, 872, 8
379. Marchesi, S., Gilli, R., Lanzuisi, G., et al. 2020, A&A, 642, A184
380. Marchesi, S., Zhao, X., Torres-Albà, N., et al. 2022, ApJ, 935, 114
381. Marconcini, C., Feltre, A., Lamperti, I., et al. 2025, arXiv e-prints, arXiv:2503.21921
382. Marcotulli, L., Connor, T., Bañados, E., et al. 2025, ApJ, 979, L6
383. Marecki, A., & Swoboda, B. 2011, A&A, 525, A6

384. Marinucci, A., Vietri, G., Piconcelli, E., et al. 2022, A&A, 666, A169
385. Markoff, S., Nowak, M. A., & Wilms, J. 2005, ApJ, 635, 1203
386. Markowitz, A., Edelson, R., & Vaughan, S. 2003, ApJ, 598, 935
387. Markowitz, A. G., Krumpe, M., & Nikutta, R. 2014, MNRAS, 439, 1403
388. Marscher, A. P. 2009, arXiv e-prints, arXiv:0909.2576
389. Marscher, A. P., & Jorstad, S. G. 2011, ApJ, 729, 26
390. Marscher, A. P., Jorstad, S. G., Gómez, J.-L., et al. 2002, Nature, 417, 625
391. Marsh, M. C. D., Russell, H. R., Fabian, A. C., et al. 2017, JCAP, 2017, 036
392. Mathur, S., Wilkes, B., Elvis, M., & Fiore, F. 1994, ApJ, 434, 493
393. Matthee, J., Naidu, R. P., Brammer, G., et al. 2023, arXiv e-prints, arXiv:2306.05448
394. —. 2024, ApJ, 963, 129
395. Mazumdar, A., & White, G. 2019, Reports on Progress in Physics, 82, 076901
396. Mazzolari, G., Scholtz, J., Maiolino, R., et al. 2024, arXiv e-prints, arXiv:2408.15615
397. Mazzolari, G., Gilli, R., Maiolino, R., et al. 2024, arXiv e-prints, arXiv:2412.04224
398. Mazzucchelli, C., Decarli, R., Farina, E. P., et al. 2019, ApJ, 881, 163
399. McHardy, I. M. 2001, in Astronomical Society of the Pacific Conference Series, Vol. 224, Probing the Physics of Active Galactic Nuclei, ed. B. M. Peterson, R. W. Pogge, & R. S. Polidan, 205
400. McHardy, I. M., Papadakis, I. E., Uttley, P., Page, M. J., & Mason, K. O. 2004, MNRAS, 348, 783
401. McKeough et al., K. 2016, ApJ, 833, 123
402. McKernan, B., Yaqoob, T., & Reynolds, C. S. 2007, MNRAS, 379, 1359
403. McKinney, J. C., Tchekhovskoy, A., & Blandford, R. D. 2012, MNRAS, 423, 3083
404. McNamara, B. R., & Nulsen, P. E. J. 2007, ARA&A, 45, 117
405. Meier, D. L. 2003, New A Rev., 47, 667
406. Melia, F., & Falcke, H. 2001, ARA&A, 39, 309
407. Merloni, A., Dwelly, T., Salvato, M., et al. 2015, MNRAS, 452, 69
408. Meyer, E. T., Georganopoulos, M., Sparks, W. B., et al. 2015, ApJ, 805, 154
409. Meyer, E. T., Iyer, A. R., Reddy, K., et al. 2019, ApJ, 883, L2
410. Meyer, E. T., Shaik, A., Tang, Y., et al. 2023, Nature Astronomy, 7, 967
411. Meyer, E. T., Laha, S., Shuvo, O. I., et al. 2025, ApJ, 979, L2
412. Michiyama, T., Inoue, Y., Doi, A., et al. 2024, ApJ, 965, 68
413. Migliori, G., Siemiginowska, A., Sobolewska, M., et al. 2016, ApJ, 821, L31
414. Müller-Sánchez, F., Comerford, J. M., Nevin, R., et al. 2015, ApJ, 813, 103
415. Mummery, A., & Turner, S. G. D. 2024, MNRAS, 530, 4730
416. Mundo, S. A., Kara, E., Cackett, E. M., et al. 2020, MNRAS, 496, 2922
417. Nandra, K., George, I. M., Mushotzky, R. F., Turner, T. J., & Yaqoob, T. 1997, ApJ, 476, 70
418. Nandra, K., & Pounds, K. A. 1994, MNRAS, 268, 405
419. Narayan, R., Igumenshchev, I. V., & Abramowicz, M. A. 2000, ApJ, 539, 798
420. Narayan, R., & Yi, I. 1994, ApJ, 428, L13
421. Nardini, E., Lusso, E., Risaliti, G., et al. 2019, A&A, 632, A109
422. National Academies of Sciences. 2021, Pathways to Discovery in Astronomy and Astrophysics for the 2020s
423. Neeleman, M., Bañados, E., Walter, F., et al. 2019, ApJ, 882, 10
424. Nenkova, M., Sirocky, M. M., Nikutta, R., Ivezić, Ž., & Elitzur, M. 2008, ApJ, 685, 160
425. Nguyen, T. T. Q., De la Torre Luque, P., John, I., et al. 2025, arXiv e-prints, arXiv:2507.13432
426. Nguyen, T. T. Q., John, I., Linden, T., & Tait, T. M. P. 2024, arXiv e-prints, arXiv:2412.00180
427. Ni, Q., Aird, J., Merloni, A., et al. 2023, MNRAS, 524, 4778
428. Ni, Q., Brandt, W. N., Chen, C.-T., et al. 2021, ApJS, 256, 21
429. Ni, Y., Di Matteo, T., Gilli, R., et al. 2020, MNRAS, 495, 2135
430. Noda, H., & Done, C. 2018, MNRAS, 480, 3898
431. Noether, E. 1971, Transport Theory and Statistical Physics, 1, 186

432. Oguri, M., & Marshall, P. J. 2010, MNRAS, 405, 2579
433. O'Hare, C. A. J. 2024, arXiv e-prints, arXiv:2403.17697
434. Olivares, H. R., Mościbrodzka, M. A., & Porth, O. 2023, A&A, 678, A141
435. Osterbrock, D. E., & Pogge, R. W. 1985, ApJ, 297, 166
436. Overzier, R. A. 2016, A&A Rev., 24, 14
437. Özsoy, O., & Tasinato, G. 2023, Universe, 9, 203
438. Pacucci, F., & Narayan, R. 2024, arXiv e-prints, arXiv:2407.15915
439. —. 2024, ApJ, 976, 96
440. —. 2024, ApJ, 976, 96
441. Pacucci, F., Nguyen, B., Carniani, S., Maiolino, R., & Fan, X. 2023, ApJ, 957, L3
442. Pal, I., Anju, A., Sreehari, H., et al. 2024, ApJ, 976, 145
443. Paliya, V. S., Parker, M. L., Jiang, J., et al. 2019, ApJ, 872, 169
444. Panessa, F., Baldi, R. D., Laor, A., et al. 2019, Nature Astronomy, 3, 387
445. Panessa, F., Pérez-Torres, M., Hernández-García, L., et al. 2022, MNRAS, 510, 718
446. Parfrey, K., Philippov, A., & Cerutti, B. 2019, Phys. Rev. Lett., 122, 035101
447. Parker, M. L., Lieu, M., & Matzeu, G. A. 2022, MNRAS, 514, 4061
448. Parker, M. L., Matzeu, G. A., Matthews, J. H., et al. 2022, MNRAS, 513, 551
449. Parker, M. L., Pinto, C., Fabian, A. C., et al. 2017, Nature, 543, 83
450. Particle Data Group, Zyla, P. A., Barnett, R. M., et al. 2020, Progress of Theoretical and Experimental Physics, 2020, 083C01
451. Payez, A., Evoli, C., Fischer, T., et al. 2015, JCAP, 2015, 006
452. Peca, A., Cappelluti, N., Marchesi, S., Hodges-Kluck, E., & Foord, A. 2023, arXiv e-prints, arXiv:2311.08449
453. Peca, A., Vignali, C., Gilli, R., et al. 2021, ApJ, 906, 90
454. Peca, A., Cappelluti, N., Urry, C. M., et al. 2023, ApJ, 943, 162
455. Peccei, R. D., & Quinn, H. R. 1977, Ph. Rev. L, 38, 1440
456. Pellegrini, S., Wang, J., Fabbiano, G., et al. 2012, ApJ, 758, 94
457. Perger, K., Fogasy, J., Frey, S., & Gabányi, K. É. 2025, A&A, 693, L2
458. Perlman, E., Meyer, E., Eilek, J., et al. 2019, Bulletin of the American Astronomical Society, 51, 59
459. Perlman, E. S., Biretta, J. A., Zhou, F., Sparks, W. B., & Macchetto, F. D. 1999, AJ, 117, 2185
460. Perlman, E. S., Clautice, D., Avachat, S., et al. 2020, Galaxies, 8, 71
461. Perlman, E. S., & Wilson, A. S. 2005, ApJ, 627, 140
462. Perlman, E. S., Padgett, C. A., Georganopoulos, M., et al. 2006, ApJ, 651, 735
463. Perlman, E. S., Adams, S. C., Cara, M., et al. 2011, ApJ, 743, 119
464. Perna, M., Arribas, S., Lamperti, I., et al. 2023, arXiv e-prints, arXiv:2310.03067
465. Petrucci, P. O., Ursini, F., De Rosa, A., et al. 2018, A&A, 611, A59
466. Petrucci, P. O., Paltani, S., Malzac, J., et al. 2013, A&A, 549, A73
467. Petrucci, P. O., Gronkiewicz, D., Rozanska, A., et al. 2020, A&A, 634, A85
468. Petrucci, P. O., Piétu, V., Behar, E., et al. 2023, A&A, 678, L4
469. Pfeifle, R. W., Weaver, K. A., Secrest, N. J., Rothberg, B., & Patton, D. R. 2024, arXiv e-prints, arXiv:2411.12799
470. Pfeifle, R. W., Satyapal, S., Secrest, N. J., et al. 2019, ApJ, 875, 117
471. Piconcelli, E., Vignali, C., Bianchi, S., et al. 2010, ApJ, 722, L147
472. Pillepich, A., Springel, V., Nelson, D., et al. 2018, MNRAS, 473, 4077
473. Piotrowska, J. M., García, J. A., Walton, D. J., et al. 2024, Frontiers in Astronomy and Space Sciences, 11, 1324796
474. Pizzetti, A., Torres-Albà, N., Marchesi, S., et al. 2022, ApJ, 936, 149
475. —. 2025, ApJ, 979, 170
476. Pogge, R. W. 2000, New A Rev., 44, 381
477. Ponti, G., Papadakis, I., Bianchi, S., et al. 2012, A&A, 542, A83
478. Pooley, D., Rappaport, S., Blackburne, J., et al. 2009, ApJ, 697, 1892
479. Pouliasis, E., Ruiz, A., Georgantopoulos, I., et al. 2024, A&A, 685, A97

480. Pozo Nuñez, F., Ramolla, M., Westhues, C., et al. 2012, A&A, 545, A84
481. Prandoni, I., & Seymour, N. 2015, in Advancing Astrophysics with the Square Kilometre Array (AASKA14), 67
482. Preskill, J., Wise, M. B., & Wilczek, F. 1983, Physics Letters B, 120, 127
483. Principe, G., Migliori, G., Johnson, T. J., et al. 2020, A&A, 635, A185
484. Quataert, E., & Gruzinov, A. 2000, ApJ, 539, 809
485. Quataert, E., & Narayan, R. 2000, ApJ, 528, 236
486. Rees, M. J., Begelman, M. C., Blandford, R. D., & Phinney, E. S. 1982, Nature, 295, 17
487. Remillard, R. A., & McClintock, J. E. 2006, ARA&A, 44, 49
488. Reynolds, C. S. 2021, ARA&A, 59, 117
489. Reynolds, C. S., & Fabian, A. C. 1995, MNRAS, 273, 1167
490. Reynolds, C. S., Marsh, M. C. D., Russell, H. R., et al. 2020, ApJ, 890, 59
491. Reynolds, C. S., Kara, E. A., Mushotzky, R. F., et al. 2023, arXiv e-prints, arXiv:2311.00780
492. —. 2023, 12678, 126781E
493. Ricci, C., & Trakhtenbrot, B. 2023, Nature Astronomy, 7, 1282
494. —. 2023, Nature Astronomy, 7, 1282
495. Ricci, C., Ueda, Y., Koss, M. J., et al. 2015, ApJ, 815, L13
496. Ricci, C., Trakhtenbrot, B., Koss, M. J., et al. 2017, ApJS, 233, 17
497. —. 2017, ApJS, 233, 17
498. Ricci, C., Bauer, F. E., Treister, E., et al. 2017, MNRAS, 468, 1273
499. Ricci, C., Ho, L. C., Fabian, A. C., et al. 2018, MNRAS, 480, 1819
500. Ricci, C., Privon, G. C., Pfeifle, R. W., et al. 2021, MNRAS, 506, 5935
501. Risaliti, G., Elvis, M., Fabbiano, G., et al. 2007, ApJ, 659, L111
502. Risaliti, G., & Lusso, E. 2015, ApJ, 815, 33
503. —. 2019, Nature Astronomy, 3, 272
504. Risaliti, G., Maiolino, R., & Salvati, M. 1999, ApJ, 522, 157
505. Różańska, A., Malzac, J., Belmont, R., Czerny, B., & Petrucci, P. O. 2015, A&A, 580, A77
506. Ruan, J. J., Anderson, S. F., Eracleous, M., et al. 2019, ApJ, 883, 76
507. Runge, J., & Walker, S. A. 2021, MNRAS, 502, 5487
508. Runnoe, J. C., Brotherton, M. S., & Shang, Z. 2012, MNRAS, 422, 478
509. Russell, H. R., Fabian, A. C., McNamara, B. R., & Broderick, A. E. 2015, MNRAS, 451, 588
510. Russell, H. R., Fabian, A. C., McNamara, B. R., et al. 2018, MNRAS, 477, 3583
511. Russell, H. R., Fabian, A. C., Sanders, J. S., et al. 2010, MNRAS, 402, 1561
512. Russell, H. R., McNamara, B. R., Edge, A. C., et al. 2013, MNRAS, 432, 530
513. Sacchi, A., Risaliti, G., & Miniutti, G. 2023, A&A, 671, A33
514. Sandoval, B., Foord, A., Allen, S. W., et al. 2024, ApJ, 974, 121
515. Sargent, W. L. W. 1977, ApJ, 212, L105
516. Sartori, L. F., Schawinski, K., Trakhtenbrot, B., et al. 2018, Monthly Notices of the Royal Astronomical Society: Letters, 476, L34
517. Satyapal, S., Secrest, N. J., Ricci, C., et al. 2017, ApJ, 848, 126
518. Sazonov, S., Gilfanov, M., Medvedev, P., et al. 2021, MNRAS, 508, 3820
519. Sbarrato, T., Ghisellini, G., Tagliaferri, G., et al. 2022, A&A, 663, A147
520. Scholtz, J., Maiolino, R., D'Eugenio, F., et al. 2023, arXiv e-prints, arXiv:2311.18731
521. Schwartz et al., D. 2019, Astronomische Nachrichten, 340, 30
522. Scoville, N., Sheth, K., Aussel, H., et al. 2016, ApJ, 820, 83
523. Scoville, N., Faisst, A., Weaver, J., et al. 2023, ApJ, 943, 82
524. Serafinelli, R., Tombesi, F., Vagnetti, F., et al. 2019, A&A, 627, A121
525. Sguera, V., Bassani, L., Malizia, A., et al. 2005, A&A, 430, 107
526. Shablovinskaya, E., Ricci, C., Chang, C. S., et al. 2024, A&A, 690, A232
527. Shakura, N. I., & Sunyaev, R. A. 1973, A&A, 24, 337

528. Shcherbakov, R. V., & Baganoff, F. K. 2010, ApJ, 716, 504
529. Shcherbakov, R. V., Wong, K.-W., Irwin, J. A., & Reynolds, C. S. 2014, ApJ, 782, 103
530. Shen, Y., Zhuang, M.-Y., Li, J., et al. 2024, arXiv e-prints, arXiv:2408.12713
531. Shende, M. B., Subramanian, P., & Sachdeva, N. 2019, ApJ, 877, 130
532. Sheridan, C., Neilsen, J., Haggard, D., Markoff, S., & Nowak, M. 2024, in American Astronomical Society Meeting Abstracts, Vol. 243, American Astronomical Society Meeting Abstracts, 110.03
533. Shi, X., & Fuller, G. M. 1999, Ph. Rev. L, 82, 2832
534. Shukla, A., & Mannheim, K. 2020, Nature Communications, 11, 4176
535. Signorini, M., Marchesi, S., Gilli, R., et al. 2023, A&A, 676, A49
536. Simionescu, A., Stawarz, Ł., Ichinohe, Y., et al. 2016, ApJ, 816, L15
537. Simpson, C. 2005, MNRAS, 360, 565
538. Sisk-Reynés, J., Matthews, J. H., Reynolds, C. S., et al. 2022, MNRAS, 510, 1264
539. Sisk-Reynés, J., Reynolds, C. S., Matthews, J. H., & Smith, R. N. 2022, MNRAS, 514, 2568
540. Sisk-Reynés, J., Reynolds, C. S., Parker, M. L., Matthews, J. H., & Marsh, M. C. D. 2023, ApJ, 951, 5
541. Smith, K. L., Mushotzky, R. F., Vogel, S., Shimizu, T. T., & Miller, N. 2016, ApJ, 832, 163
542. Smith, K. L., Mushotzky, R. F., Koss, M., et al. 2020, MNRAS, 492, 4216
543. Smolčić, V., Novak, M., Bondi, M., et al. 2017, A&A, 602, A1
544. Snios, B., Wykes, S., Nulsen, P. E. J., et al. 2019, ApJ, 871, 248
545. Stecker, F. W., Shrader, C. R., & Malkan, M. A. 2019, ApJ, 879, 68
546. Stern, D., Assef, R. J., Benford, D. J., et al. 2012, ApJ, 753, 30
547. Stern, D., McKernan, B., Graham, M. J., et al. 2018, ApJ, 864, 27
548. Stern, J., Laor, A., & Baskin, A. 2014, MNRAS, 438, 901
549. Stone, J. M., Pringle, J. E., & Begelman, M. C. 1999, MNRAS, 310, 1002
550. Suh, H., Scharwächter, J., Farina, E. P., et al. 2025, Nature Astronomy, 9, 271
551. Sun, Y., Rieke, G. H., Lyu, J., et al. 2025, arXiv e-prints, arXiv:2503.03675
552. Svrcek, P., & Witten, E. 2006, Journal of High Energy Physics, 2006, 051
553. Talbot, R. Y., Sijacki, D., & Bourne, M. A. 2024, MNRAS, 528, 5432
554. Tchekhovskoy, A., & McKinney, J. C. 2012, MNRAS, 423, L55
555. Tee, W. L., Fan, X., Wang, F., & Yang, J. 2025, ApJ, 983, L26
556. Temple, M. J., Ricci, C., Koss, M. J., et al. 2023, MNRAS, 518, 2938
557. Terashima, Y., & Wilson, A. S. 2003, ApJ, 583, 145
558. Tombesi, F., Cappi, M., Reeves, J. N., et al. 2011, ApJ, 742, 44
559. —. 2010, A&A, 521, A57
560. Tombesi, F., Reynolds, C. S., Mushotzky, R. F., & Behar, E. 2017, ApJ, 842, 64
561. Torres-Albà, N., Marchesi, S., Zhao, X., et al. 2023, A&A, 678, A154
562. Torres-Albà, N., Hu, Z., Cox, I., et al. 2025, ApJ, 981, 91
563. Tortosa, A., Bianchi, S., Marinucci, A., Matt, G., & Petrucci, P. O. 2018, A&A, 614, A37
564. Tortosa, A., Ricci, C., Tombesi, F., et al. 2022, MNRAS, 509, 3599
565. Tortosa, A., Ricci, C., Arévalo, P., et al. 2023, MNRAS, 526, 1687
566. Tortosa, A., Ricci, C., Ho, L. C., et al. 2023, MNRAS, 519, 6267
567. Tortosa, A., Zappacosta, L., Piconcelli, E., et al. 2024, A&A, 691, A235
568. Trakhtenbrot, B., Ricci, C., Koss, M. J., et al. 2017, MNRAS, 470, 800
569. Travascio, A., Cantalupo, S., Tozzi, P., et al. 2025, A&A, 694, A165
570. Treister, E., Schawinski, K., Urry, C. M., & Simmons, B. D. 2012, ApJ, 758, L39
571. —. 2012, ApJ, 758, L39
572. Trindade Falcao, A., Fabbiano, G., Elvis, M., Paggi, A., & Maksym, W. P. 2023, ApJ, 950, 143
573. Übler, H., Maiolino, R., Curtis-Lake, E., et al. 2023, A&A, 677, A145
574. Ueda, Y., Akiyama, M., Hasinger, G., Miyaji, T., & Watson, M. G. 2014, ApJ, 786, 104
575. —. 2014, ApJ, 786, 104

576. Ulrich, M.-H., Maraschi, L., & Urry, C. M. 1997, ARA&A, 35, 445
577. Urry, C. M., & Padovani, P. 1995, PASP, 107, 803
578. Uttley, P., & Malzac, J. 2025, MNRAS, 536, 3284
579. Uttley, P., & McHardy, I. M. 2005, MNRAS, 363, 586
580. Van Wassenhove, S., Volonteri, M., Mayer, L., et al. 2012, ApJ, 748, L7
581. Vaughan, S., Reeves, J., Warwick, R., & Edelson, R. 1999, Monthly Notices of the Royal Astronomical Society, 309, 113
582. Vignali, C., Piconcelli, E., Perna, M., et al. 2018, MNRAS, 477, 780
583. Vito, F., Gilli, R., Vignali, C., et al. 2014, MNRAS, 445, 3557
584. Vito, F., Brandt, W. N., Yang, G., et al. 2018, MNRAS, 473, 2378
585. —. 2018, MNRAS, 473, 2378
586. Vito, F., Brandt, W. N., Bauer, F. E., et al. 2019, A&A, 628, L6
587. —. 2019, A&A, 630, A118
588. Vito, F., Brandt, W. N., Ricci, F., et al. 2021, A&A, 649, A133
589. Vito, F., Brandt, W. N., Comastri, A., et al. 2024, A&A, 689, A130
590. —. 2025, A&A, 694, L16
591. Volonteri, M., Haardt, F., & Madau, P. 2003, ApJ, 582, 559
592. Volonteri, M., Sikora, M., Lasota, J. P., & Merloni, A. 2013, ApJ, 775, 94
593. Volonteri et al., M. 2021, Nature Reviews Physics, 3, 732
594. Walton, D. J., Nardini, E., Fabian, A. C., Gallo, L. C., & Reis, R. C. 2013, MNRAS, 428, 2901
595. Wang, B., de Graaff, A., Davies, R. L., et al. 2024, arXiv e-prints, arXiv:2403.02304
596. Wang, F., Fan, X., Yang, J., et al. 2021, ApJ, 908, 53
597. Wang, J.-M., Chen, Y.-M., Hu, C., et al. 2009, ApJ, 705, L76
598. Wang, J.-M., Du, P., Hu, C., et al. 2014, ApJ, 793, 108
599. Wang, T., Liu, G., Cai, Z., et al. 2023, Science China Physics, Mechanics, and Astronomy, 66, 109512
600. Wang, Y., Zhai, Z., Alavi, A., et al. 2022, ApJ, 928, 1
601. Wang et al., F. 2021, ApJ, 907, L1
602. Weinberg, S. 1978, Ph. Rev. L, 40, 223
603. Wilczek, F. 1978, Ph. Rev. L, 40, 279
604. Wilkins, D. R., & Gallo, L. C. 2015, MNRAS, 449, 129
605. Wilkins, D. R., Gallo, L. C., Silva, C. V., et al. 2017, MNRAS, 471, 4436
606. Williams, D. R. A., Baldi, R. D., McHardy, I. M., et al. 2020, MNRAS, 495, 3079
607. Willingale, R., Starling, R. L. C., Beardmore, A. P., Tanvir, N. R., & O'Brien, P. T. 2013, MNRAS, 431, 394
608. Willis, J. P., Canning, R. E. A., Noordeh, E. S., et al. 2020, Nature, 577, 39
609. Wong, K.-W., Irwin, J. A., Shcherbakov, R. V., et al. 2014, ApJ, 780, 9
610. Wong, K.-W., Irwin, J. A., Yukita, M., et al. 2011, ApJ, 736, L23
611. Wong, K.-W., Russell, H. R., Irwin, J. A., Cappelluti, N., & Foord, A. 2024, Universe, 10, 278
612. Worrall et al., D. M. 2020, MNRAS, 497, 988
613. Wright, E. L., Eisenhardt, P. R. M., Mainzer, A. K., et al. 2010, AJ, 140, 1868
614. —. 2010, AJ, 140, 1868
615. Wu, X.-B., Wang, F., Fan, X., et al. 2015, Nature, 518, 512
616. Yang, J., Wang, F., Fan, X., et al. 2020, ApJ, 897, L14
617. —. 2021, ApJ, 923, 262
618. Yang, J., Fan, X., Wang, F., et al. 2022, ApJ, 924, L25
619. Yuan, F., & Narayan, R. 2014, ARA&A, 52, 529
620. Yuan, W., Zhou, H. Y., Komossa, S., et al. 2008, ApJ, 685, 801
621. Yue, M., Eilers, A.-C., Ananna, T. T., et al. 2024, ApJ, 974, L26
622. —. 2024, arXiv e-prints, arXiv:2404.13290
623. Zappacosta, L., Piconcelli, E., Giustini, M., et al. 2020, A&A, 635, L5

624. Zappacosta, L., Piconcelli, E., Fiore, F., et al. 2023, A&A, 678, A201
625. Zargaryan, D., Gasparyan, S., Baghmanyan, V., & Sahakyan, N. 2017, A&A, 608, A37
626. Zdziarski, A. A., & Egron, E. 2022, ApJ, 935, L4
627. Zeltyn, G., Trakhtenbrot, B., Eracleous, M., et al. 2022, ApJ, 939, L16
628. Zhang, Y.-W., Huang, Y., Bai, J.-M., et al. 2021, AJ, 162, 276
629. Zhao, J.-H., Young, K. H., Herrnstein, R. M., et al. 2003, ApJ, 586, L29
630. Zhao, X., Marchesi, S., Ajello, M., et al. 2021, A&A, 650, A57
631. Zhu et al., S. F. 2019, MNRAS, 482, 2016
632. Zou, F., Yu, Z., Brandt, W. N., et al. 2024, ApJ, 964, 183
633. ZuHone, J. A., Vikhlinin, A., Tremblay, G. R., et al. 2023, SOXS: Simulated Observations of X-ray Sources, Astrophysics Source Code Library, record ascl:2301.024
634. Zuo et al., Z. 2024, MNRAS, 530, 360

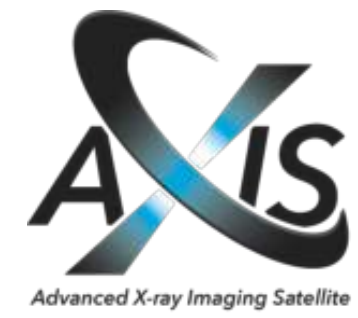

*AXIS White Paper*

# AXIS TDAMM GO Science

**The AXIS Time-Domain and Multi-Messenger Science Working Group: Nafisa Aftab[1], Igor Andreoni[2], Riccardo Arcodia[3], Maria Cristina Baglio[4], Arash Bahramian[5], Franz E. Bauer[6], Andy Beardmore[7], Paz Beniamini[8], Tamara Bogdanovic[9], David Bogensberger[10], Matteo Bonetti[11], Enrico Bozzo[12], Marica Branchesi[13], Sergio Campana[4], Fiamma Capitanio[14], Jonathan Carney[2], S. Bradley Cenko[15,†], Joheen Chakraborty[3], Lia Corrales[10], Francesco Coti Zelati[16], Filippo D'Ammando[17], Alessandro Di Marco[14], Simone Dichiara[18], Steven Dillmann[19], Joseph Durbak[20], Hannah Dykaar[21], Alessia Franchini[11], Chris Fryer[22], Suvi Gezari[20], Margherita Giustini[23], Andrea Gnarini[24], Daryl Haggard[21,†], Sebastian Heinz[25], Wynn Ho[26], Christopher Irwin[27], Nazma Islam[15], Wynn Jacobson-Galan[28], Chetana Jain[29], Amruta Jaodand[30], Peter Jonker[31], Erin Kara[3], Manish Kumar[1], Alexander Kutyrev[15], Fabio La Monaca[14], Tingting Liu[32], Daniele Bjørn Malesani[33], Julie Malewicz[9], Antonios Manousakis[34], Tatsuya Matsumoto[27], Giovanni Miniutti[23], Sachindra Naik[35], Michela Negro[36], Mason Ng[21], Scott C. Noble[15], Brendan O'Connor[37], Dheeraj Pasham[3], Biswajit Paul[1], Aaron Pearlman[21], Maria Petropoulou[38], Tsvi Piran[39], Pragati Pradhan[40], Giovanna Pugliese[41], Jonathan Quirola-Vásquez[31], Lauren Rhodes[21], Roberto Ricci[42], Ketan Rikame[1], Patrizia Romano[4], Alicia Rouco Escorial[43], Geoffrey Ryan[44], Takanori Sakamoto[45], Jeremy D. Schnittman[15], Alberto Sesana[11], Rahul Sharma[46], Robert Stein[20], Nial Tanvir[7], Ira Thorpe[15], Andrea Tiengo[46], Eleonora Troja[42], Hendrik van Eerten[47], Georgios Vasilopoulos[38], Susanna D. Vergani[48], Zorawar Wadiasingh[20], Muskan Yadav[42], Yuhan Yao[49], George Younes[50], and Bing Zhang[51]**

[1] Raman Research Institute [2] UNC Chapel Hill [3] MIT [4] INAF-OAB [5] Curtin Institute of Radio Astronomy [6] Universidad de Tarapacá [7] University of Leicester [8] Open University of Israel [9] Georgia Institute of Technology [10] University of Michigan [11] Universitá degli Studi di Milano-Bicocca [12] INTEGRAL Science Data Centre [13] Gran Sasso Science Institute [14] INAF-IAPS [15] NASA Goddard Space Flight Center [16] ICE-CSIC [17] INAF-IRA [18] Penn State [19] Stanford University [20] University of Maryland, College Park [21] McGill University [22] Los Alamos National Laboratory [23] Centro de Astrobiologia - INTA-CSIC [24] Universitá Roma Tre [25] University of Wisconsin - Madison [26] Haverford College [27] University of Tokyo [28] California Institute of Technology [29] University of Delhi [30] Center for Astrophysics | Harvard & Smithsonian [31] Radboud University [32] Georgia State University [33] Niels Bohr Institute [34] SAASST - University of Sharjah [35] PRL Ahmedabad [36] Louisiana State University [37] Carnegie Mellon University [38] National and Kapodistrian University of Athens [39] Hebrew University of Jerusalem [40] Embry-Riddle Aeronautical University [41] University of Amsterdam [42] Universitá degli Studi di Roma Tor Vergata [43] ESA [44] Perimeter Institute [45] Aoyama Gakuin University [46] IUSS Pavia [47] University of Bath [48] CNRS - Observatoire de Paris [49] University of California, Berkeley [50] George Washington University [51] University of Nevada, Las Vegas

† AXIS TDAMM Science Working Group Co-Lead

---

## a. Multi-Messenger Astronomy and Cross-Facility Synergies

### *1. Accurate localization of high-redshift short GRBs with AXIS*

**First Author:** S. Dichiara (Penn State University, sbd5667@psu.edu)

**Co-authors:** E. Troja (U. Rome), B. O'Connor (CMU), P. Beniamini (ARCO), Chris Fryer (LANL), A. Kutyrev (NASA/UMD), J. Durbak (UMD), T. Sakamoto (AGU)

**Abstract:** Mergers of two neutron stars (NSs) are loud sources of gravitational waves (GWs), progenitors of gamma-ray bursts (GRBs), and the only confirmed cosmic source of heavy r-process metals. However, their poorly constrained age distribution does not allow us to understand their role in the chemical evolution of our universe. To understand whether NS mergers are the primary cosmic source of heavy elements, we need to establish their frequency in the early universe. Due to the limited sensitivity of GW detectors, which can only detect these mergers in the nearby Universe, observations of distant short GRBs open a unique window into the population of high-redshift NS mergers. These observations are our best tool to constrain the rate of NS mergers in the early universe.

The Advanced X-ray Imaging Satellite (AXIS) will play a pivotal role in this effort by delivering accurate X-ray localizations for a large sample of short GRBs, overcoming the limitations of optically driven searches. AXIS's capabilities will significantly enhance our ability to precisely localize GRB afterglows, identify their host galaxies, and investigate their environment across various cosmic epochs. The combination of AXIS's rapid response and astrometric precision will enable us to build an unbiased sample of neutron star (NS) mergers at cosmological distances, providing critical constraints on their formation channels and age distribution.

**Science:** Neutron star (NS) mergers are the leading progenitors of short GRBs, and recent observations suggest that they may be the primary source of r-process elements in the Universe [71]. The kilonova associated with the first gravitational wave signal from a binary neutron star merger (GW170817 [4]), linked to the short GRB170817A [5], was detected across UV, optical, and near-IR bands [e.g., 64,111,167]. Additionally, radio and X-ray emissions began rising at the kilonova site several days after the GW trigger [e.g., 84,210]. These observations provided, for the first time, detailed data on r-process element production in the ejecta and off-axis emissions from a GRB jet. The discovery of GW170817A marked a pivotal moment in modern astrophysics, highlighting the importance of NS mergers in understanding the origin of heavy elements. About half of the elements heavier than iron are formed through rapid neutron capture (r-process) reactions [36,39], yet the astrophysical site(s) of these processes remain elusive, posing one of the thorniest problems in nuclear astrophysics [51].

Historically, the primary focus of GRB research was to establish the connection between short GRBs and NS mergers. As a result, most observational resources were directed toward closer events ($z < 1$), where afterglows, kilonovae, and host galaxies could be better characterized through optical observations. However, with the connection between NS mergers and GRBs now firmly established, short GRBs are increasingly used as probes of NS mergers in the distant universe.

The redshift distribution of short GRBs provides critical insights into the age of their stellar progenitors [13,24,138,229], helping to estimate their contribution to cosmic chemical enrichment. Initial findings based on the first three short GRB localizations (GRB 050509B, GRB 050709, and GRB 050724), at $z \approx 0.2$, suggested that they occur at significantly lower redshifts than long GRBs (with a mean redshift of 2.5 for long GRBs) [108]. However, with a larger sample of events, it is now clear that the redshift distribution of short GRBs is significantly broader, ranging from $z=0.1$ to $>2$. The redshift distribution of short GRBs is a

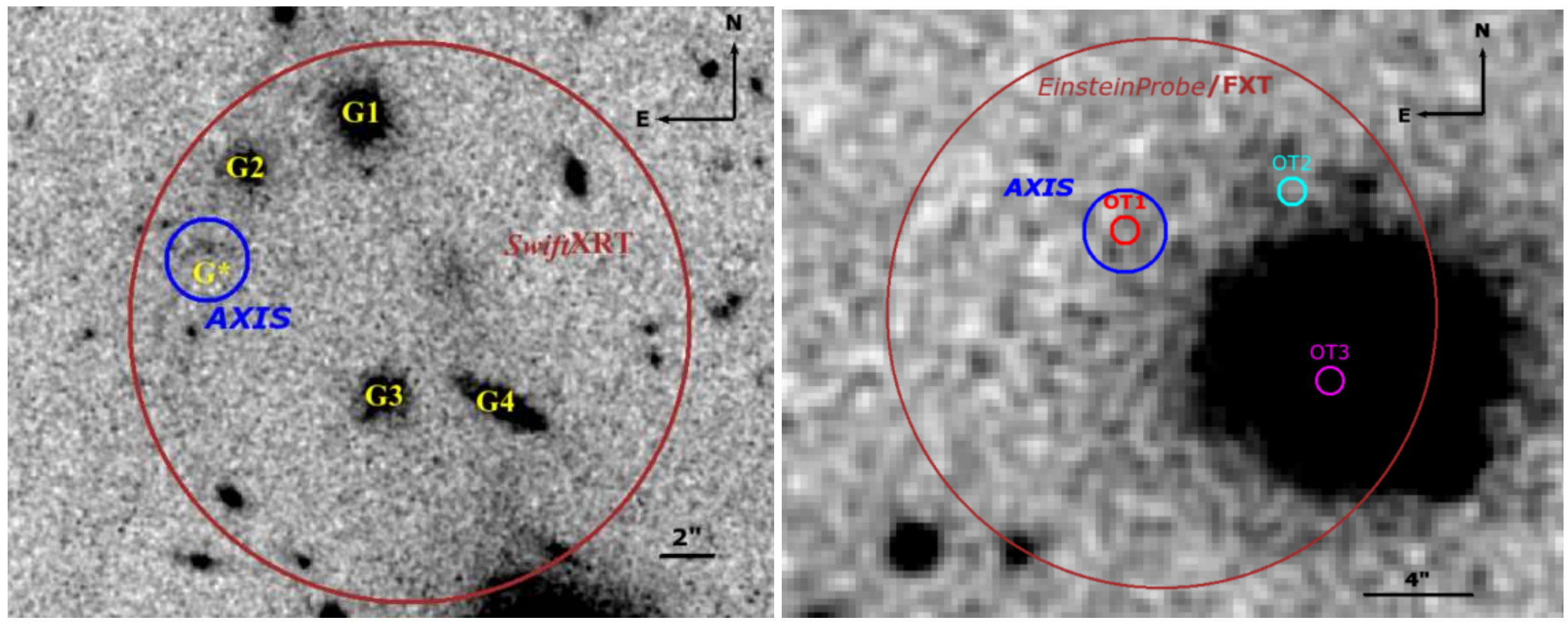

Figure 1 Left: The *Swift*/XRT error region of the short GRB 230906A with multiple candidate host galaxies. The precise localization provided by AXIS will enable us to pinpoint the host galaxy. Right: Example of the Einstein Probe (EP) transient EP250207b. Three different optical transients were identified within the EP error region (10 arcsec radius). AXIS would unambiguously discriminate between multiple candidates.

powerful discriminant between different progenitor models and formation channels. However, it remains uncertain due to the limited number of known events at z> 1.

An accurate GRB localization is essential for the robust identification of its distant host galaxy, which typically requires an optical or NIR position. However, detecting short GRBs is particularly challenging because their optical afterglows are faint. Additionally, at higher redshifts, absorption by intervening material - such as neutral hydrogen in the intergalactic medium - can further diminish the observed brightness in the optical band, requiring sensitive coverage at near-IR wavelengths. The rate of short GRB localizations is significantly higher at X-ray energies, as demonstrated by the Neil Gehrels Swift Observatory and its high rate ( 70%) of X-ray afterglow detections. However, when the afterglow is only detected in X-rays, with a localization uncertainty of 3-10 arcseconds, the chance of incorrect host galaxy identification becomes much higher, as seen in Figure 1. Multiple galaxies are identified within a typical XRT error circle, and probabilistic arguments tend to favor the association with brighter, lower redshift hosts. These observational biases prevent the identification of high-redshift short GRBs. For this reason, several observing programs with the Chandra X-ray Observatory have been dedicated to refining the X-ray localization of short GRB afterglows to sub-arcsecond accuracy, leading to the identification of GRB111117A at z=2.21 [191,195], one of the most distant short GRB known so far.

However, due to the typical Chandra's response time of 48 hours or longer, the number of well-localized X-ray afterglows remains limited to a handful of bursts.

AXIS is uniquely positioned to close this observational gap. With its rapid Target-of-Opportunity (ToO) response and exceptional X-ray sensitivity, AXIS can detect over 90% of short GRB afterglows with a brief 1 ks exposure (Figure 2). Its superior angular resolution enables precise localization of the X-ray transient, which is crucial for unambiguously identifying the GRB counterpart and its host galaxy (see Figure 1).

As shown in Figure 2, many X-ray afterglows are long-lived and remain detectable for several days (cyan). However, these bright, long-lasting afterglows are more likely to have an optical and radio counterpart. In contrast, most X-ray afterglows fade at a faster rate, with X-ray fluxes dropping below $10^{-13}$ erg cm$^{-2}$ s$^{-1}$ within a few hours post-burst (dark blue). These events critically need rapid-response

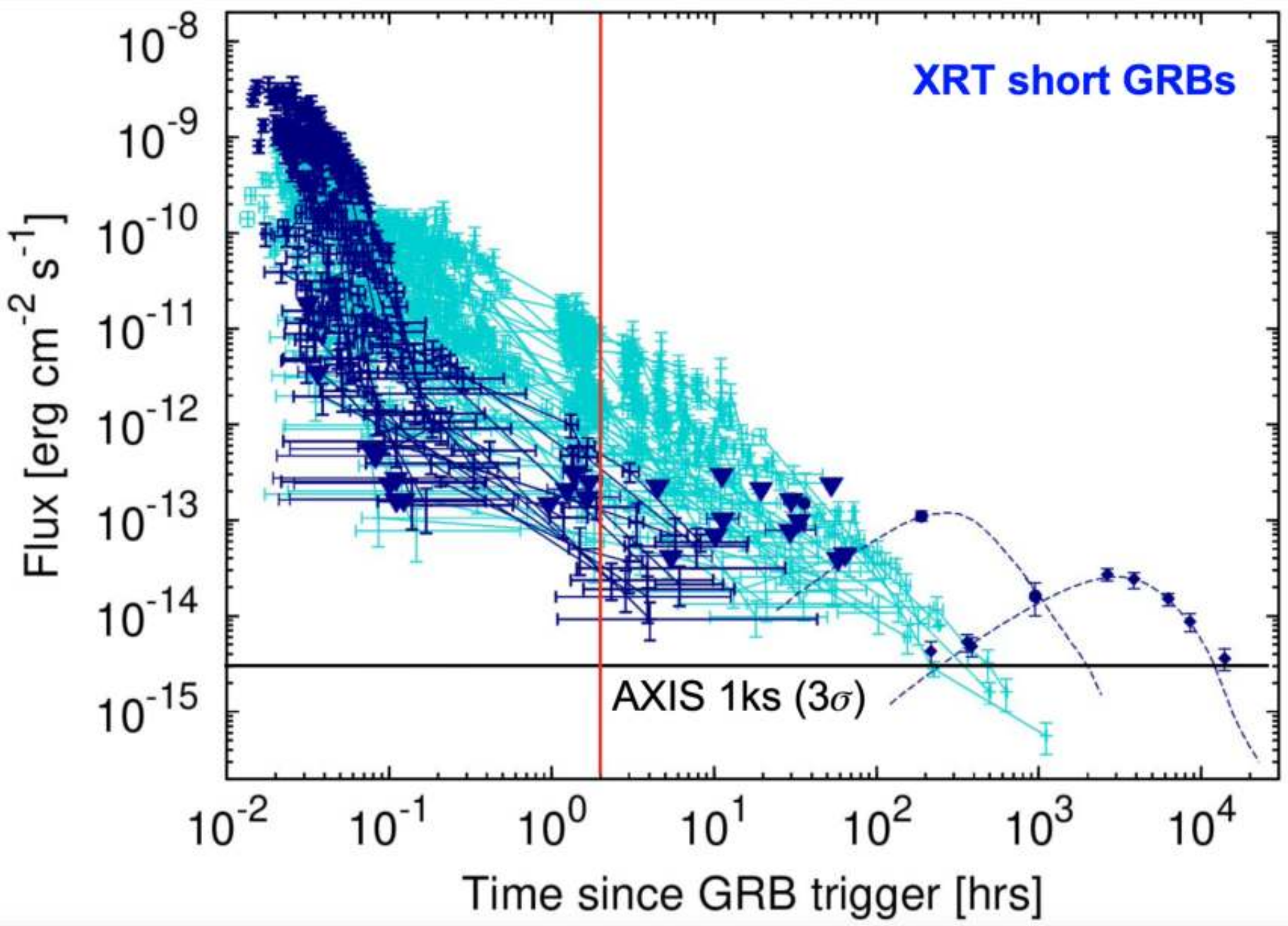

Figure 2 *Swift*/XRT light curves of short GRBs divided into two classes: 1) short-lived X-ray afterglow (dark-blue) and 2) long-lived X-ray afterglow (cyan). In addition, we show the X-ray afterglows of GRB150101B and GW170817 [210,211], two off-axis afterglows (blue dashed lines). The black solid line shows the AXIS sensitivity for a 1 ks observation (at 3$\sigma$ level). The red vertical line is the 2-hour AXIS response time.

observations, ideally within a few hours since the burst, to detect the X-ray counterpart and achieve arcsecond localization. The primary challenge for Chandra observations was the slow response time - typically 2-5 days - by which point the afterglow had faded below the detection threshold. AXIS's faster response and higher sensitivity would significantly improve detection rates and enable secure identifications of high-redshift NS mergers.

**Exposure time (ks):** $\sim$1 ks for ToO for observations starting within 2-4 hours since the GRB trigger. With 10 ks per year.

**Observing description:** Target summary (names): Newly detected short GRBs. Rate: 10 events/yr
Specify critical AXIS Capabilities:

- Good angular resolution
- High sensitivity (down to few $\times 10^{-15}$ erg cm$^{-2}$ s$^{-1}$ in 1 ks)
- Rapid response time (<4 hour ToO capability)

**Joint Observations and synergies with other observatories in the 2030s:** JWST, Roman, LIGO A+, Cosmic Explorer, Einstein Telescope

**Special requirements:** ToO $<$ 4 hours.

*2. Characterization of Orphan and Off-axis Afterglows from Neutron Star Mergers*

**First Author:** Brendan O'Connor (Carnegie Mellon University, boconno2@andrew.cmu.edu)

**Co-authors:** Eleonora Troja (University of Rome, Tor Vergata), Simone Dichiara (Pennsylvania State University), Geoffrey Ryan (Perimeter Institute), Paz Beniamini (Open University)

**Abstract:** The merger of two neutron stars is capable of launching a relativistic jet that creates a shock wave producing non-thermal synchrotron radiation across the entire electromagnetic spectrum. These jets, and their "afterglow" emission, are commonly found following short-duration gamma-ray bursts (GRBs), as was the case for the first neutron star merger detected in gravitational waves (GWs), known as GW170817. However, both upcoming optical and near-infrared wide-field surveys (e.g., Roman, LSST) and upcoming GW observing runs will uncover a hidden population of off-axis neutron star mergers with their jets oriented away from Earth, such that we do not observe the prompt short GRB emission. Target of Opportunity (ToO) observations with AXIS can reveal the afterglow emission produced by the off-axis jet, and constrain the viewing angle, energetics, collimation, and angular structure of their outflows. These mergers are a signpost of heavy element production, and their collimation is a key property required to understand their intrinsic rate within the Universe and their true influence in depositing heavy elements across cosmic time.

**Science:** The 2017 detection of the binary neutron star (BNS) merger GW170817 in gravitational waves (GWs) robustly linked the production of heavy elements, relativistic jets, and short gamma-ray bursts (GRBs), solidifying their previously hypothesized progenitor [4,5]. The BNS merger, and its subsequent remnant, launches a highly collimated relativistic jet that propagates into the surrounding environment – the interstellar medium (ISM) – producing an external shock that radiates non-thermal synchrotron radiation from radio to optical to X-ray wavelengths. This synchrotron radiation is referred to as the "afterglow", and characterizing its behavior allows for inferences of the physical properties of the jet and the environment.

The significant majority of cosmological GRBs are viewed directly down the core ("on-axis") of the relativistic jet, whereas GW170817, due to its close proximity to Earth (40 Mpc), was detected at a large viewing angle of $\sim$20 degrees from the jet's core ("off-axis"). An off-axis afterglow displays a very different lightcurve shape compared to the fading on-axis afterglows of cosmological GRBs (Figure 3; see, e.g., [25,26,188,189]). Instead, the off-axis viewing angle leads to a rising lightcurve shape that peaks once the jet has decelerated enough that the photons from the jet's core are no longer beamed away from the observer. Therefore, the power-law slope of the rising lightcurve, time of the afterglow lightcurve peak, and subsequent temporal decay can be used to measure the observer's viewing angle, angular structure of the jet's energy and Lorentz factor, and the jet's half-opening angle. As only a single event has been robustly characterized in this way (Figure 2), extending these observations to a population of events is crucial for understanding the relativistic jets of GRBs and their jet-launching mechanisms, which produce the observed angular structure. However, such lightcurves are difficult to detect and identify at large distances from Earth, and are likely limited to a few hundred Mpc from Earth – though this depends on the exact orientation and jet structure (Figure 3; see, e.g., [25,26,188,189]).

An avenue to identify these objects comes from another outflow launched by the cataclysmic BNS merger. The rapid decompression of neutron-rich material ejected by the merger leads to the production of heavy elements through the rapid neutron capture process (r-process). The subsequent decay of these heavy elements heats the ejecta, leading to a luminous ultraviolet, optical, and infrared transient, referred to as the kilonova, that evolves rapidly on timescales of a few days. As the ejecta is moving at slower speeds, it does not suffer from the same relativistic beaming effects that apply to the afterglow light. Instead, kilonovae are expected to be quasi-isotropic, though their exact dependence on viewing angle is still debated. Regardless, kilonovae are expected to be readily detectable by wide-field ultraviolet, optical, and infrared surveys such as ULTRASAT, Rubin Observatory, and the Nancy Grace Roman Space

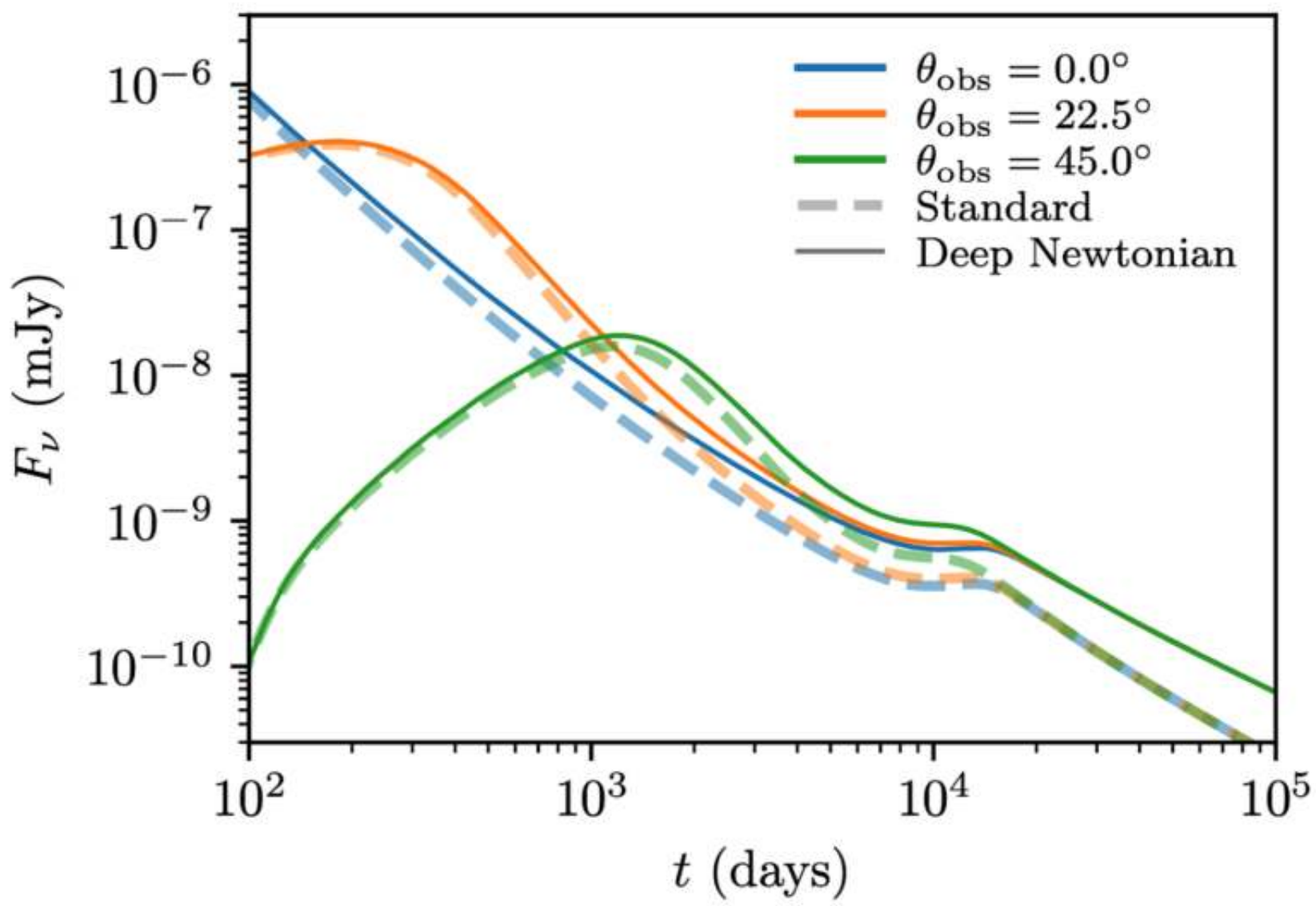

Figure 3 Impact of viewing angle on the observed afterglow lightcurve shape [189]. Off-axis viewing angles shift the peak time later and reduce the peak flux. Adapted from Ryan et al. 2025 [189].

Telescope, among others. The detection and identification of a kilonova marks the presence of a neutron star merger, and, therefore, very likely, the presence of a relativistic jet with an off-axis viewing angle.

A further alternative way to discover the afterglow of relativistic jets launched by BNS mergers is to monitor "orphan" afterglows discovered by wide-field surveys, such as Rubin at optical wavelengths, Roman at near-infrared wavelengths, and the Square Kilometer Array (SKA) at radio wavelengths. The past decade has seen a rise in the discovery of non-gamma-ray triggered afterglows by surveys such as the Zwicky Transient Facility (ZTF). The next decade of wide-field surveys is likely to increase this discovery rate.

Therefore, we propose that Target of Opportunity (ToO) observations with AXIS will enable critical studies of relativistic jets, utilizing wide-field surveys to pinpoint the subarcsecond locations of candidate BNS mergers. AXIS will be used to detect and characterize the afterglow emission produced by relativistic outflows launched by neutron star mergers to:

1. Determine the presence of a relativistic jet
2. Constrain the angular structure and geometry of the relativistic outflow
3. Constrain the fraction of neutron star mergers that launch successful relativistic jets
4. Probe the nature of the central engine (Figure 4)

These observations are directly connected to the electromagnetic follow-up of GW events (e.g., Figure 4) and allow us to expand the sample of events that have well characterized kilonovae and afterglows, beyond the current handful of events known to date (e.g., GW170817, GRB 211211A, GRB 230307A). This follow-up campaign with AXIS will provide crucial information regarding the production of relativistic jets in BNS mergers (Figure 5). The rapid ToO response and incredible sensitivity of AXIS make it well-suited to carry out this critical science in the 2030s (Figure 5).

**Exposure time (ks):** A short 6 ks exposure per epoch should be sufficient to reach $5 \times 10^{-16}$ erg cm$^2$ s$^{-1}$ with 5 observations total for 30 ks.

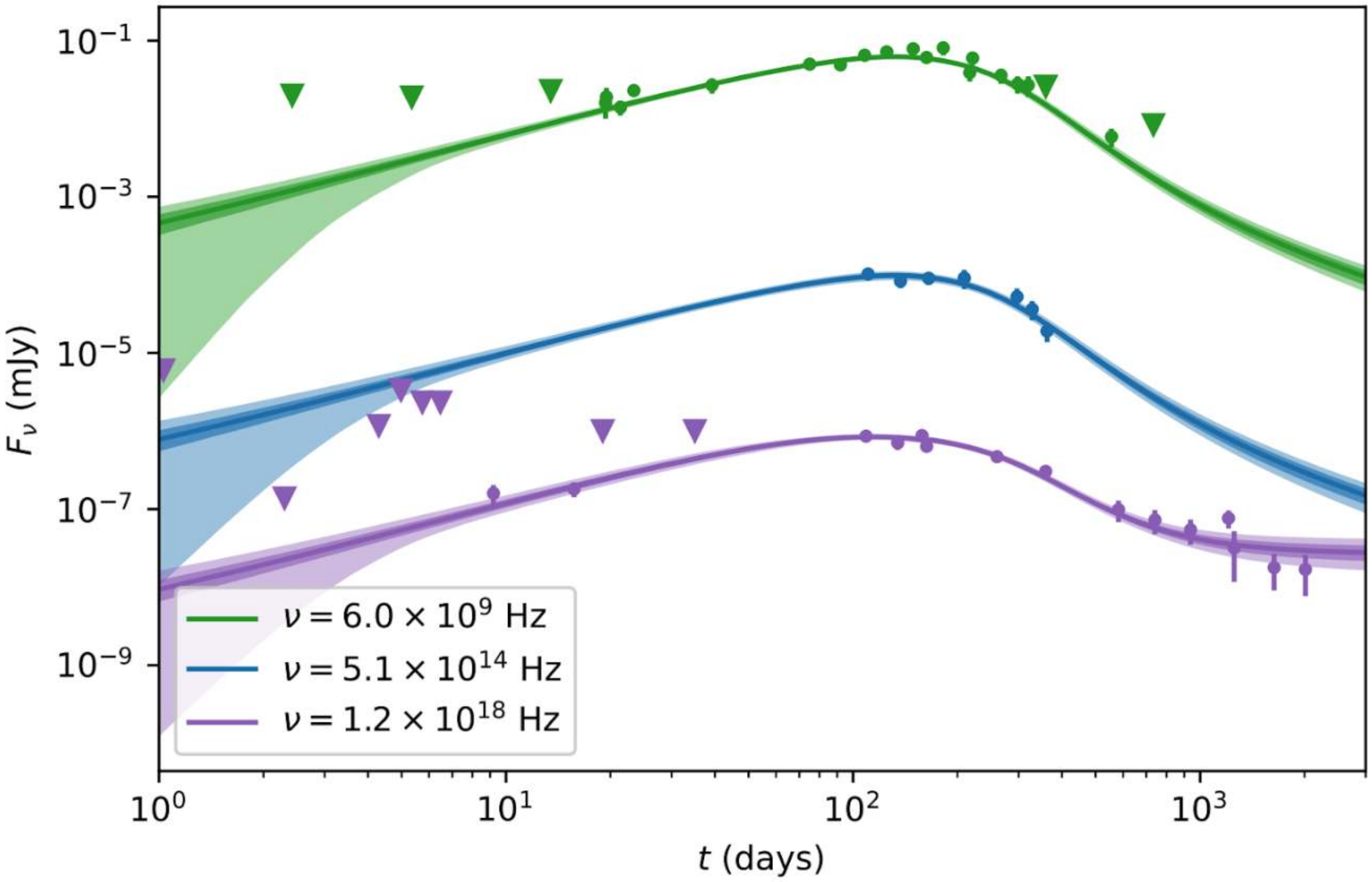

Figure 4 Observed afterglow lightcurves for GW170817 [189] at radio (green), optical (blue), and X-ray (purple) wavelengths. Late-time X-ray observations with Chandra revealed the need for an additional constant luminosity component, potentially a kilonova afterglow or long-lived central engine activity. Future observations with AXIS can reveal the long-term behavior of this X-ray source and whether this flattening is observed in future off-axis afterglows. Adapted from [189].

**Observing description:** We will employ an observing strategy wherein five observations occur with temporal log-spacing to probe the onset of the afterglow and characterize its temporal behavior (rise, peak, and fall), which will allow for robust constraints on the viewing angle and collimation of the source. Initial observations should be conducted between one day and one week after the discovery of a kilonova candidate. The observations will be approximately 6 ks per epoch to produce the required depth.

The angular resolution and low background of AXIS are critical for these studies, enabling sensitive observations that extend the probable distance horizon for off-axis afterglows. The angular resolution is crucial for robustly linking the X-ray source to its kilonova counterpart and avoiding contamination from other X-ray sources.

**Joint Observations and synergies with other observatories in the 2030s:** A variety of important facilities have strong synergies with these observations that can increase their science return. First, the discovery of candidate kilonovae by Roman, LSST, ULTRASAT, or UVEX will serve as a filler pool of targets for this program. Late-time observations at multiple wavelengths are required to fully characterize the off-axis afterglow and constrain the characteristic synchrotron break frequencies (and their evolution). To this end, sensitive optical and infrared observations with JWST and ELT, as well as sensitive radio observations with ngVLA and SKA, will be critical.

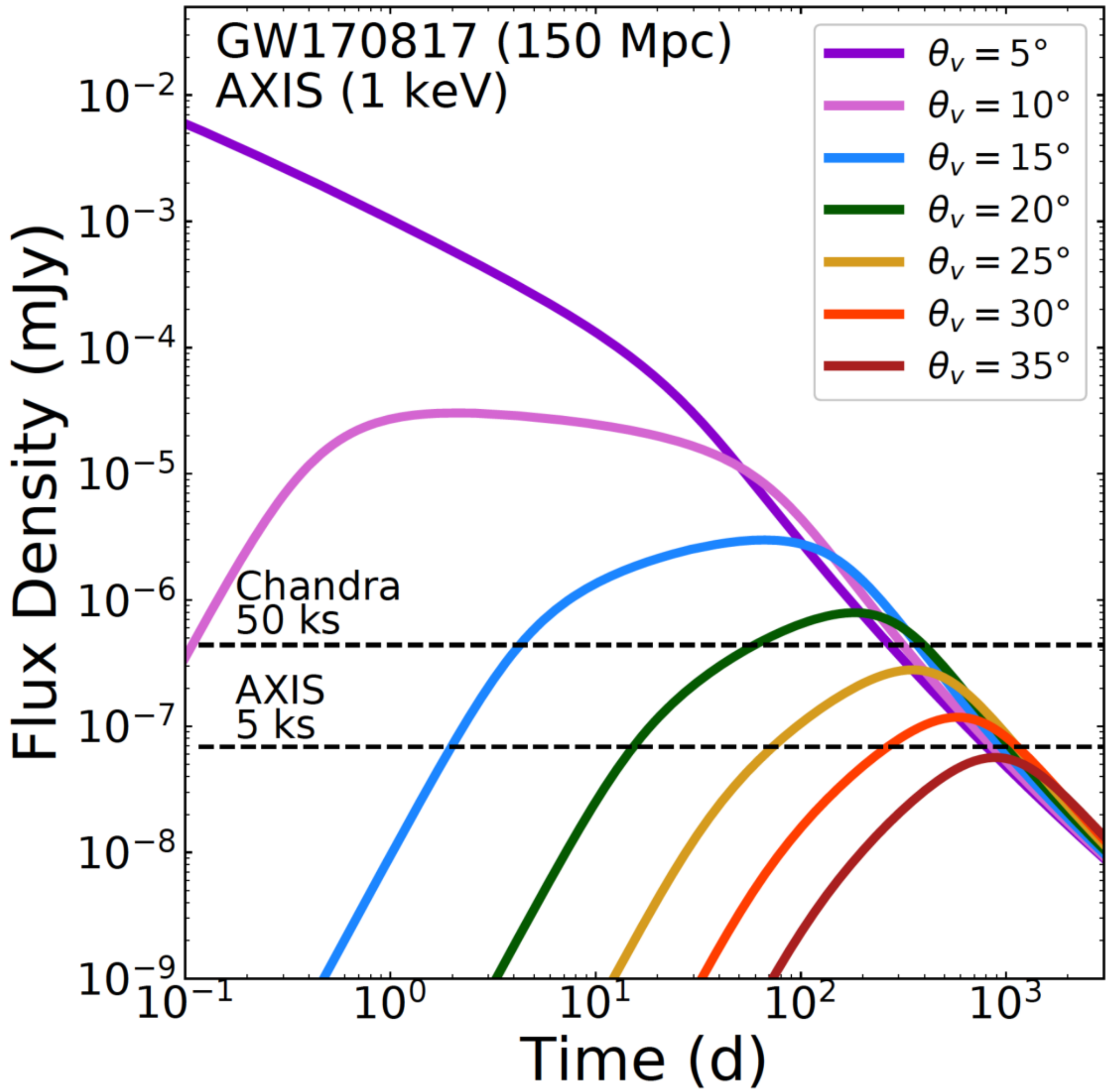

Figure 5 GW170817-like X-ray afterglows for a variety of viewing angles versus the Chandra and AXIS sensitivity. AXIS performs quite well and can detect even very far-off-axis afterglows at 150 Mpc. Afterglow lightcurves were computed using the `afterglowpy` software [188].

**Special Requirements:** An initial 1-day ToO response will be sufficient. This will probe the potential on-axis afterglow case. If the afterglow is found to be off-axis, then the following observations can occur temporally with log-spacing, preferably out to a few hundred days to probe the likely peak time (Figure 5).

*3. Searching for late-time X-ray flares from neutron star mergers*

**First Author:** Eleonora Troja (University of Rome Tor Vergata - Italy, eleonora.troja@uniroma2.it)

**Co-authors:** M. Yadav (University of Rome Tor Vergata), B. O'Connor (Carnegie Mellon University), H. vanEerten (University of Bath), B. Zhang (University of Nevada - Las Vegas), T. Piran (Racah Institute of Physics), G. Ryan (Perimeter Institute), R. Ricci (University of Rome Tor Vergata), T. Matsumoto (University of Tokyo)

**Abstract:** Late-time X-ray flares are a predicted but yet-to-be-confirmed signature of neutron star (NS) mergers, arising from the interaction between the sub-relativistic merger ejecta and the surrounding interstellar medium. Observations of gamma-ray bursts (GRBs), gravitational wave (GW) sources, and their kilonovae unambiguously show that a large mass of fast-moving ejecta is released during or right after the merger. The progressive deceleration of this ejecta by the ambient medium will inevitably lead to the production of nearly isotropic non-thermal emission, resulting in a faint but potentially detectable X-ray signal.

To search for these elusive X-ray flares, we propose using the Advanced X-ray Imaging Satellite (AXIS) to conduct X-ray follow-up observations of nearby, well-localized NS mergers on timescales of months to years after the event. AXIS, with its high sensitivity and angular resolution, is uniquely suited to systematically search for faint X-ray emission from NS mergers and distinguish it from the host galaxy background. The discovery and characterization of this new type of X-ray signal would provide an unprecedented insight into merger dynamics, ejecta properties, and the nature of the merger remnant, thus providing novel constraints on the behavior of ultra-dense matter in extreme conditions.

**Science:** Mergers of binary NSs serve as powerful astrophysical laboratories for studying the formation of highly relativistic outflows [179,187], the production of heavy r-process elements [51], and the behavior of hot and dense matter [6,161]. The detection of gravitational waves (GWs) from GW170817 and its accompanying electromagnetic signals firmly established NS mergers as progenitors of short gamma-ray bursts (GRBs) and kilonovae [4,5]. Observations of kilonovae in GRBs [175,209] demonstrated that a substantial fraction of the NS mass is dynamically ejected during the merger or expelled through post-merger winds, providing both heavy r-process nucleosynthesis sites and a reservoir of fast-moving ejecta. As this ejecta expands and interacts with the ambient interstellar medium (ISM), it should gradually decelerate and power non-thermal synchrotron radiation [153]. This late-time emission, visible from radio to X-ray wavelengths [112], represents a long-predicted but yet undetected signature of NS mergers [60,182,194]. It is expected to be nearly isotropic, unlike the beamed prompt GRB and afterglow, and may peak months to years after the merger (Figure 6). Its brightness is determined by the merger environment and, critically, by the nature of the central compact object (magnetar or BH). At the same time, its temporal evolution encodes the stratification of the ejecta's velocity. Detecting this signal would fill a missing piece in our understanding of merger dynamics, particle acceleration, and the nature of the merger remnant.

Numerical simulations of binary NS mergers show that, while the bulk of the ejecta coast at a sub-relativistic velocity ($\beta_0 \approx 0.2\, c$), a less massive tail of material may expand with substantially higher velocities. This latter fast-moving component does not contribute significantly to the optical/nIR kilonova emission, and can only be probed through its non-thermal afterglow radiation visible as a late-time flare. The energy distribution of this ejecta can be approximated by a simple power-law $E(>\beta) \propto (\beta/\beta_0)^{-k}$, where $k$ is the velocity index [94]. The value of $k$ depends on the merger dynamics and the NS EoS: a shallow distribution (smaller $k$) is expected when a fast, shock-heated tail of ejecta is launched near the remnant's bounce phase, thus pointing to the formation of a massive NS remnant; conversely, a steep distribution (larger $k$) would be consistent with the prompt collapse to BH, suppressing the component of shock-driven ejecta.

Figure 6 shows that the value of $k$ shapes the onset of late-time X-ray emission: the steeper the velocity profile, the steeper the flare rise. The observed temporal profile would thus allow us to constrain the energy distribution of the merger ejecta and probe the properties of matter right after the merger, at

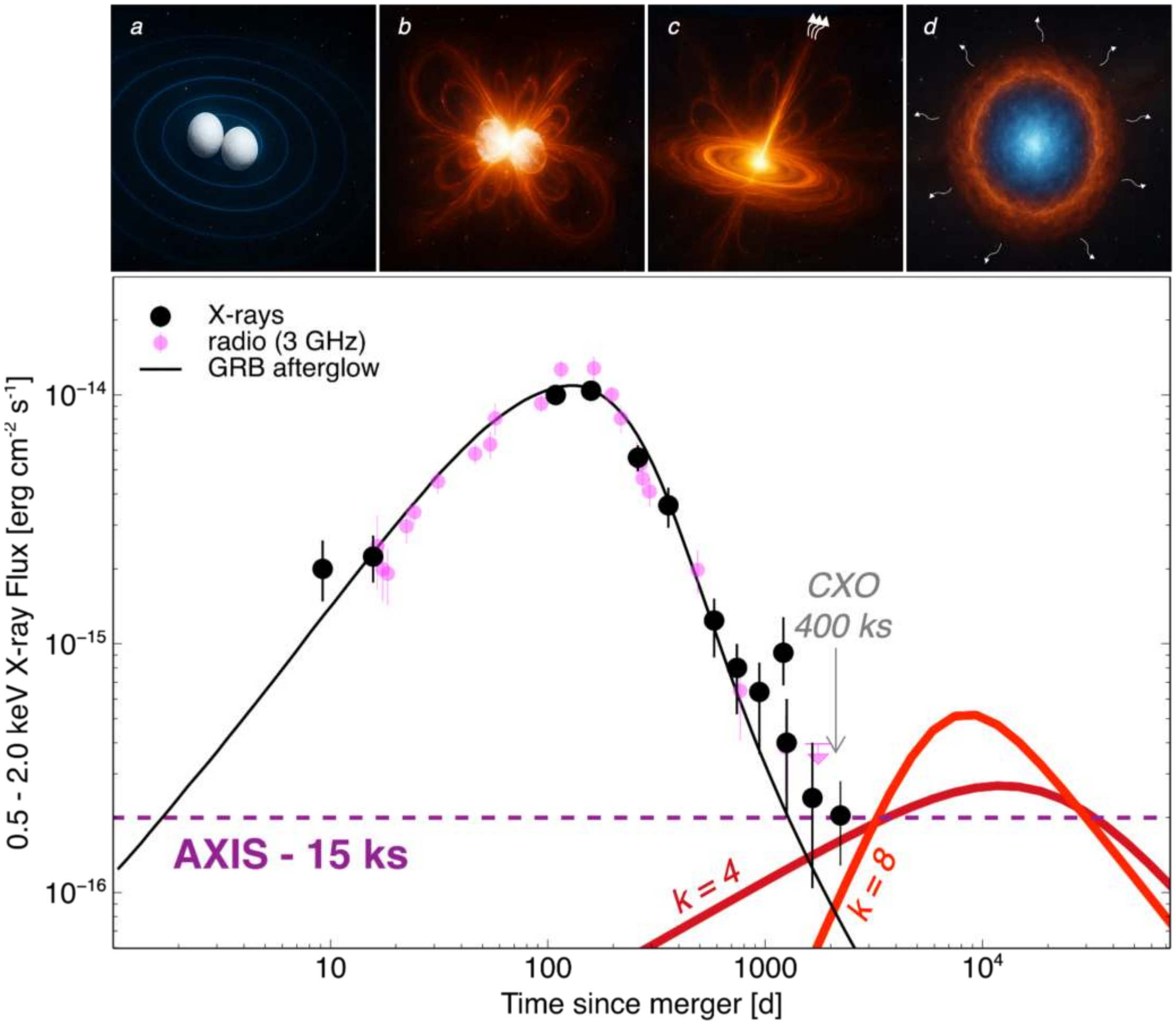

Figure 6 *Top panels (a-d):* different stages of a binary NS coalescence. After the inspiral (a) and merger (b) phase, the merger remnant launches a collimated jet (c), producing the GRB and afterglow emission. Months–years later, the sub-relativistic ejecta plows into the ambient medium, generating a nearly isotropic X-ray flare (d). *Bottom panel:* X-ray (black) and 3 GHz radio (magenta) light curves of GW170817 at 40 Mpc compared to an off-axis GRB afterglow model (black line) [188,189], followed by the predicted late-time X-ray flare (red thick lines) for different velocity indices $k$. The horizontal dashed line marks the *AXIS* 0.5–2 keV sensitivity in a 15 ks exposure.

temperatures and densities higher than the pre-merger phase constrained by the LIGO/Virgo limits on tidal deformability [6]. For context, late-time *Chandra* and VLA observations of GW170817 could only place mild constraints on the value of $k$ and have yet to reveal any late-time flare [20,212].

The main challenge is that these late-time signals are intrinsically faint, and, to date, only a single NS merger has been securely localized in the nearby Universe, where such emission is within reach. Moreover, some of these mergers occur close to their galaxy's center, and most X-ray facilities, except for the *Chandra* X-ray Observatory (CXO), lack sufficient angular resolution to cleanly separate the X-ray flare from the background galaxy's light. Thanks to its combination of high sensitivity and sharp imaging, the Advanced X-ray Imaging Satellite (AXIS) will open a new discovery space by enabling the first systematic late-time monitoring of NS mergers.

This capability is especially timely, as detection rates are expected to rise. In the next decade, planned upgrades to the LVK network are predicted to increase the number of nearby GW events from one every few years to several per year [31]. Moreover, the advent of the Vera Rubin Observatory (VRO) and the Nancy Grace Roman Space Telescope may drive the discoveries of new nearby kilonovae, independently of their gamma-ray or GW signals [45]. Looking ahead, third-generation GW observatories such as the Einstein Telescope (ET) and Cosmic Explorer (CE) will extend reach and localization precision even further, turning nearby GW events from rare occurrences into a steady stream [96]. Together, these advances will substantially expand the sample of nearby NS mergers that can be targeted with AXIS, enabling robust population studies rather than single-event case work.

**Exposure time (ks):** 150 ks $yr^{-1}$

**Observing description:** We propose to target with regular (≈1-2 yr) cadence 10 NS mergers within 150 Mpc, selected based on their progenitor (e.g., NS-NS and NS-BH), environment (prioritizing higher density), and kilonova properties. We envision, on average, a 15 ks exposure per target to reach a $3\sigma$ on-axis sensitivity of $\approx 2 \times 10^{-16}$ erg $cm^2$ $s^{-1}$ (0.5-2.0 keV), assuming a power-law spectrum with photon index $\Gamma$=2 and $N_H$=$5\times10^{20}$ $cm^{-2}$. The exact exposure time would be adjusted depending on the source's distance.

**Critical AXIS Capabilities:** effective area and low background to detect weak X-ray sources; angular resolution to disentangle the target from the galaxy's background light. These capabilities make AXIS uniquely suited to isolate and characterize faint X-ray emission in complex galactic environments.

**Joint Observations and synergies with other observatories in the 2030s:** AXIS observations would enhance discoveries from LVK, VRO/LSST, Roman, ET, and CE, deepen our understanding of kilonovae observed with HST, JWST, ULTRASAT and UVEX, while providing essential X-ray coverage to joint radio monitoring with the ngVLA and SKA.

*4. Multi-messenger science synergies with future and next-generation GW detectors*

**First Author:** Susanna D. Vergani (CNRS, Paris Observatory PSL - France susanna.vergani@obspm.fr)

**Co-authors:** Marica Branchesi (Gran Sasso Science Institute); Wynn Ho (Haverford College)

**Abstract:** AXIS encompasses four characteristics required for MM studies: a large field-of-view, high sensitivity and spatial resolution, rapid Target of Opportunity Observations, and rapid alert dissemination. With such characteristics, AXIS will be able to respond to different challenges. AXIS will be suitable to perform MM observations in the context of LISA. Indeed, X-rays have been identified as the best spectral domain for detecting the EM signatures of massive BH mergers that LISA will detect, and AXIS specifications cover the requirements predicted by simulations of such EM counterparts. AXIS will also be a key facility for MM studies of the EM counterparts of GW BNS and BHNS events detected by the ground-based Next Generation Interferometers, as well as GRB and afterglow emission, and possibly KN remnants. AXIS will be able to rapidly react to GW alerts to start as soon as possible the search for on- or off-axis afterglow; thanks to its sensitivity and field of view, it will be able to cover the large sky error region of GW events, detect transient counterparts, and work in synergy with the Vera Rubin Observatory and other facilities observing at different wavelengths from space and ground to identify the GW counterparts among several detected transients correctly. Finally, AXIS observations could prove crucial for studying isolated neutron stars and neutron stars in non-merging systems. Many of these sources are candidates for producing detectable persistent GWs, and many are only bright in X-rays and

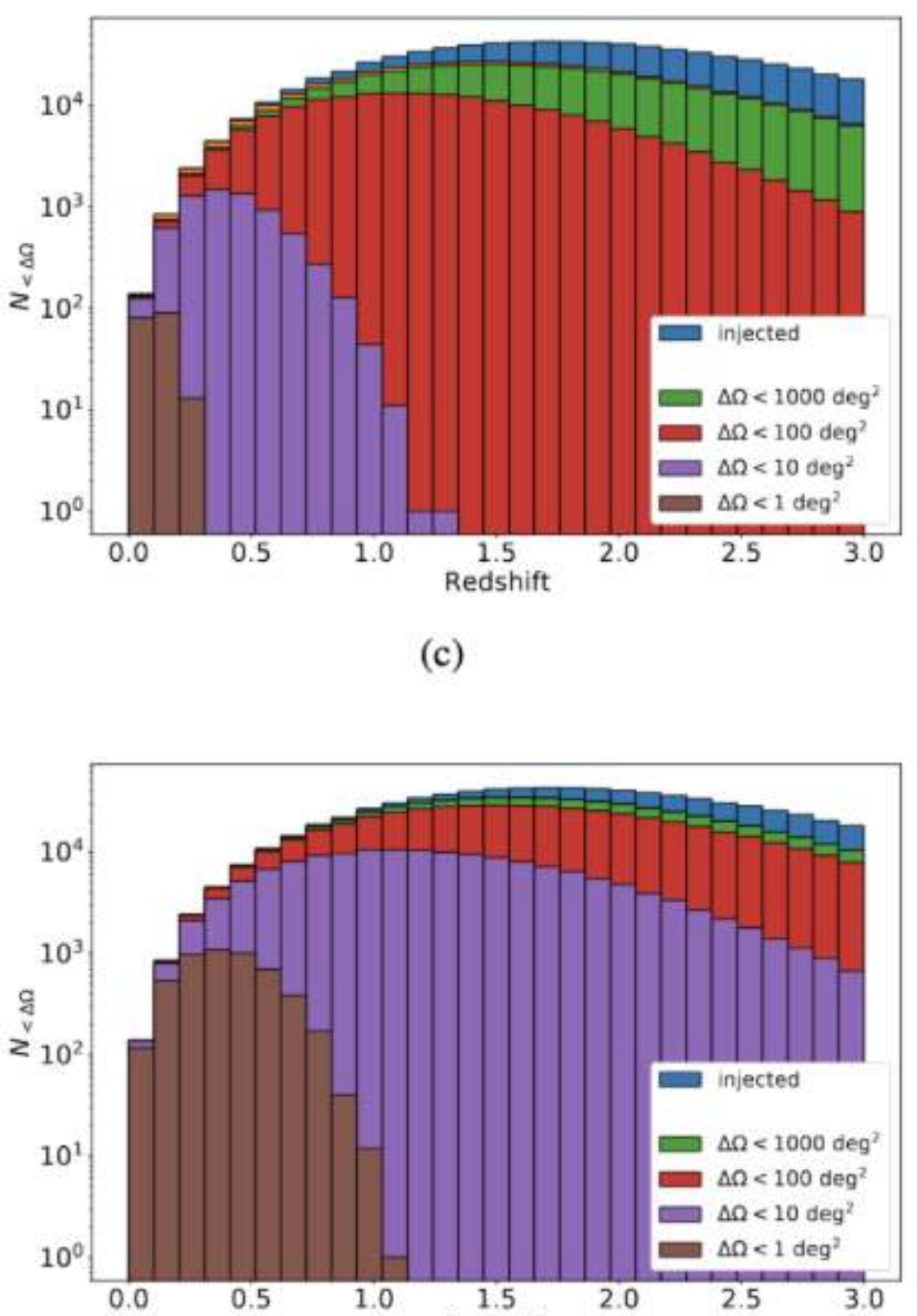

Figure 7 Simulation of number of detections (per year) of BNS by ET and CE in ET+CE (top panel) and ET+2CE (bottom panel) configurations. The injected population of events is in blue, whereas the other colors correspond to the different precisions of the sky localizations of the GW events. Adapted from Ronchini et al. 2022 [185].

are in crowded fields. AXIS could identify new GW candidates, as well as confirm sources detected first by GWs.

**Science:** AXIS is needed to detect, identify, and characterize electromagnetic (EM) counterparts to gravitational wave (GW) sources since ground- and space-based GW interferometers of the future will revolutionize the field of multi-messenger (MM) astrophysics. These new GW detectors will transform the field, which is currently observationally constrained to the local Universe and a small number of events, into a domain that explores cosmic history, encompassing hundreds of thousands of objects (Figure 7). This paradigm shift necessitates the development of new observational facilities and strategies to detect, identify, and characterize the EM counterparts of GW events.

Simulations of EM counterparts in the context of the LISA mission (Figure 8), combining state-of-the-art astrophysical models for galaxy formation and evolution, as well as the black hole characteristics (accreting gas, spin, etc.), show that some LISA-detectable black hole mergers are sufficiently bright to be detected with sensitive X-ray instruments.

Simulations performed in the context of the next generation ground-based GW detectors, Einstein Telescope (ET) or Cosmic Explorer (CE), explored the characteristics of the counterparts of binary neutron star mergers (BNS) and black hole-neutron star mergers (BHNS) detected by those interferometers. The

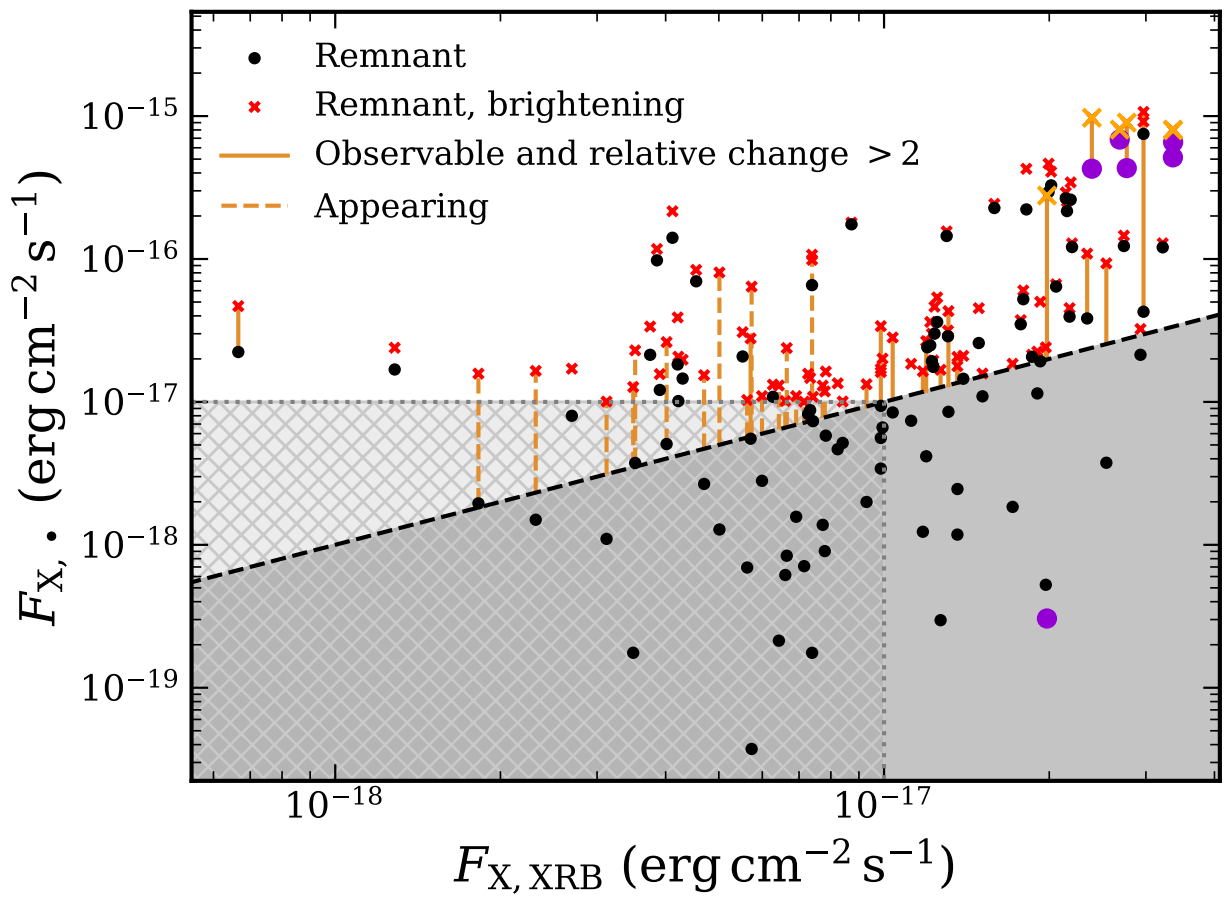

Figure 8 Red crosses and black dots connected by orange lines show the brightening in X-rays of LISA detectable mergers from galaxy simulations that would be detectable by sensitive X-ray instruments. The red dashed horizontal line corresponds to the sensitivity of AXIS with a few $10^4$ s exposure. Adapted from [55].

results show that X-ray observations are important players in the detection of the gamma-ray burst (GRB) afterglow counterparts of GW events, both for on- and off-axis GRBs (Figure 9). Furthermore, next-generation GW interferometers will not be able to provide a complete parameter estimation analysis for all sources they detect, but they will utilize outside alerts to select the GW signals for full analysis. AXIS will be able to independently detect short GRBs and/or their afterglows and alert GW interferometers.

The difficulties in EM counterpart observations and detections primarily reside in the expected faintness, transient nature, and large error regions of GW events. AXIS encompasses the four characteristics required for multi-messenger studies: a large field of view, high sensitivity and spatial resolution, rapid Target of Opportunity Observations, and rapid alert dissemination. AXIS is crucial to achieving the goal of multi-messenger exploration of the Universe.

**Exposure time (ks):** Ranging from a few ks for on-axis afterglows to a few tens of ks for off-axis ones and for EM counterparts of massive BH mergers, to be repeated to tile the GW error regions (strategy to be defined).

**Observing description:** The observing strategy will depend on the GW estimated parameters and error regions, as well as on the pre-merger information and possible counterparts detected at other wavelengths.

**Joint Observations and synergies with other observatories in the 2030s:** Einstein Telescope, Cosmic Explorer, Vera Rubin Observatory, SKA

**Special requirements:** TOO (< few hrs), TAM.

### *5. Exploring the synergy between high-energy neutrinos and X-rays with AXIS*

**First Authors:** Maria Petropoulou (National and Kapodistrian University of Athens, mpetropo@phys.uoa.gr) and Robert Stein (University of Maryland, rdstein@umd.edu)
**Co-authors:** Giovanna Pugliese (Anton Pannekoek Institute of Astronomy, University of Amsterdam, pugliese@astroduo.org), Filippo D'Ammando (INAF-Institute for Radioastronomy, dammando@ira.inaf.it), Takanori Sakamoto (Aoyama Gakuin University), Georgios Vasilopoulos (National and Kapodistrian University of Athens, gevas@phys.uoa.gr)

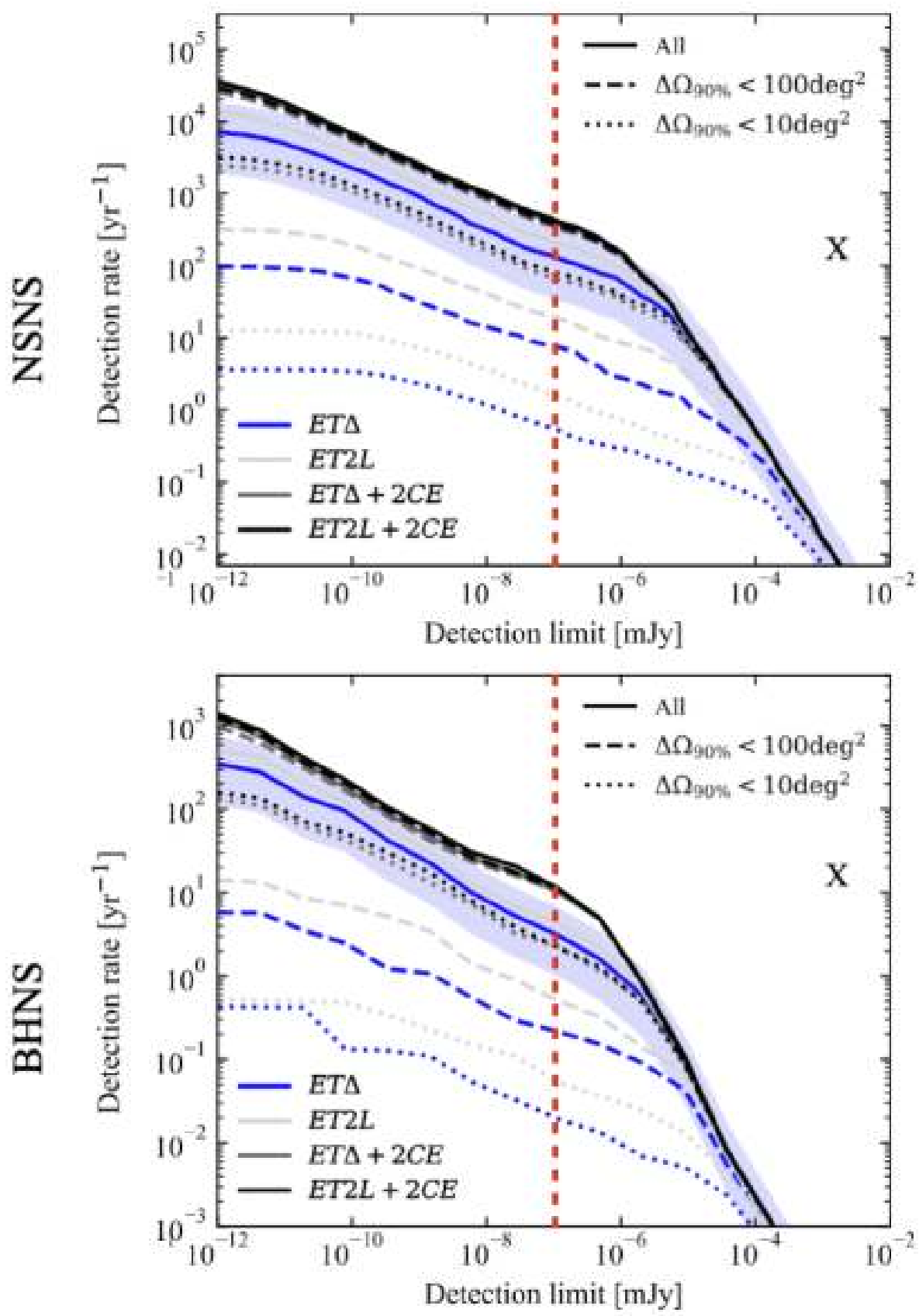

Figure 9 Simulations of detectable rates as a function of X-ray detection limits of GRB afterglows as counterparts of GW BNS (upper panel) and BHNS (lower panel) events detected by ET and CE. The vertical red dashed line corresponds to AXIS sensitivity in $\sim 10^4$ s of exposure. Adapted from Colombo et al. 2025 [48].

**Abstract:** High-energy astrophysical neutrinos (TeV–PeV) provide a unique window into the most extreme cosmic environments, including AGN cores and jets, GRBs, and TDEs. Produced by interactions between relativistic nuclei and radiation or matter, neutrinos are crucial for identifying astrophysical particle accelerators and probing extreme physics. Their weak interaction with matter allows them to travel cosmic distances unimpeded. Despite their discovery by IceCube in 2013, their origins remain largely unknown. Neutrino follow-up programs across the electromagnetic spectrum have identified candidate sources, primarily jetted AGNs and TDEs. Notably, all have been detected in X-rays. Theoretical models also suggest a strong connection between X-ray photons and neutrino production, making X-ray observations essential for neutrino astrophysics. As IceCube upgrades and new observatories (KM3NeT, P-One, Baikal-GVD) turn online to form a global network providing improved sensitivity over the whole sky, and localizations of neutrinos to $< 1$ sq. deg., AXIS will advance neutrino science through two complementary approaches: (1) rapid automated tiling of neutrino localization regions to identify counterparts and (2) long-term monitoring of candidate sources. With a 0.125 sq. deg. field of view and enhanced sensitivity (60 times more sensitive than Swift), AXIS can efficiently cover a substantial fraction of these regions. To maximize its impact, we recommend mission software capable of handling multi-pointing (tiling) observation requests. **AXIS's high-cadence follow-up of neutrino alerts will enable timely identification**

**and spectral characterization of potential neutrino counterparts, solidifying its role in multi-messenger astrophysics.**

**Science:** High-energy neutrinos are produced when relativistic nuclei interact with radiation or matter. Often referred to as the "ghost particles" of the Universe due to their weak interaction with matter and tiny masses, neutrinos can traverse cosmic distances unimpeded. High-energy astrophysical neutrinos are one of the most powerful messengers, offering a direct window into the most extreme astrophysical phenomena. IceCube first discovered a high-energy astrophysical neutrino flux in 2013 [97]. We now know that some of these neutrinos originate within our own Galaxy [100], but the vast majority are extragalactic. Further IceCube observations have identified the nearby Seyfert galaxy NGC 1068 as the most significant individual source of high-energy neutrinos [99]. However, much of the flux remains unexplained. To aid in the identification of transient or variable neutrino sources, follow-up programs that tile the degree-scale localization area of neutrinos are operated by telescopes throughout the electromagnetic spectrum. These include radio (Effelsberg, OVRO), IR (WINTER), optical (ASAS-SN, DECam, Pan-STARRs1, Tomo-e-Gozen and ZTF), UV/X-ray (Swift), and gamma-rays (Fermi, HESS, MAGIC, and VERITAS). These programs have led to the successful identification of candidate counterparts for individual neutrinos, including the flaring blazar TXS 0506+056 [98] and tidal disruption events [178,203]. In many respects, X-ray emission is the most crucial wavelength for understanding astrophysical neutrino sources. The very high-energy (VHE, $E_\gamma \gg 100$ GeV) gamma rays produced alongside high-energy neutrinos are typically absorbed long before they reach Earth, either through direct attenuation in the source or through scattering over long cosmic distances. Stacking searches for neutrinos from gamma-ray selected AGN samples have placed firm upper limits on their contribution to the diffuse neutrino flux [e.g. 1,139]. Moreover, there is no strong correlation of high-energy neutrinos with GeV gamma-ray flares from blazars [e.g. 1,73,223]. Therefore, gamma rays are not a good tracer of neutrino emission. However, the production of TeV – PeV neutrinos via the p$\gamma$ interaction requires X-ray photons to serve as a target. Additionally, the luminosity of the electromagnetic cascade initiated by the attenuation of VHE photons produced in hadronic processes can be related to the X-ray source luminosity [151]. These make searches at X-ray wavelengths especially powerful for identifying neutrino sources. Note that all candidate neutrino sources identified thus far have been detected at X-ray wavelengths.

With AXIS, we envisage two complementary approaches for advancing neutrino science:

1. rapid automated tiling of neutrino localizations to find candidate X-ray counterparts and,
2. targeted longer-term X-ray monitoring of candidate counterparts identified by AXIS or other surveys.

AXIS is the only instrument capable of performing this task. There are multiple classes of sources that we will study with AXIS, as shown in Figure 10. We outline below indicative science cases.

Active Galactic Nuclei and Blazars

The first evidence of a high-energy neutrino association with an astrophysical source was obtained with the blazar TXS 0506+056 [98], supported by multiple follow-up observations conducted throughout a broad energy range. Notably, Swift-XRT observed the blazar $\sim$3 hr after the neutrino alert and detected significant X-ray variability within 10 days of the event [113]. X-ray observations in the 0.1–10 keV range played a crucial role in constraining theoretical models and providing a physical interpretation of this first multi-messenger flare [72,113,151], see also Figure 11 (right panel). More recently, IceCube reported compelling evidence of high-energy neutrinos from the Seyfert galaxy NGC 1068 [99]. The reported neutrino luminosity is comparable to the intrinsic broadband X-ray flux from its nucleus, suggesting a physical connection of X-ray photons and TeV neutrinos [150]. Providing a theoretical basis for the

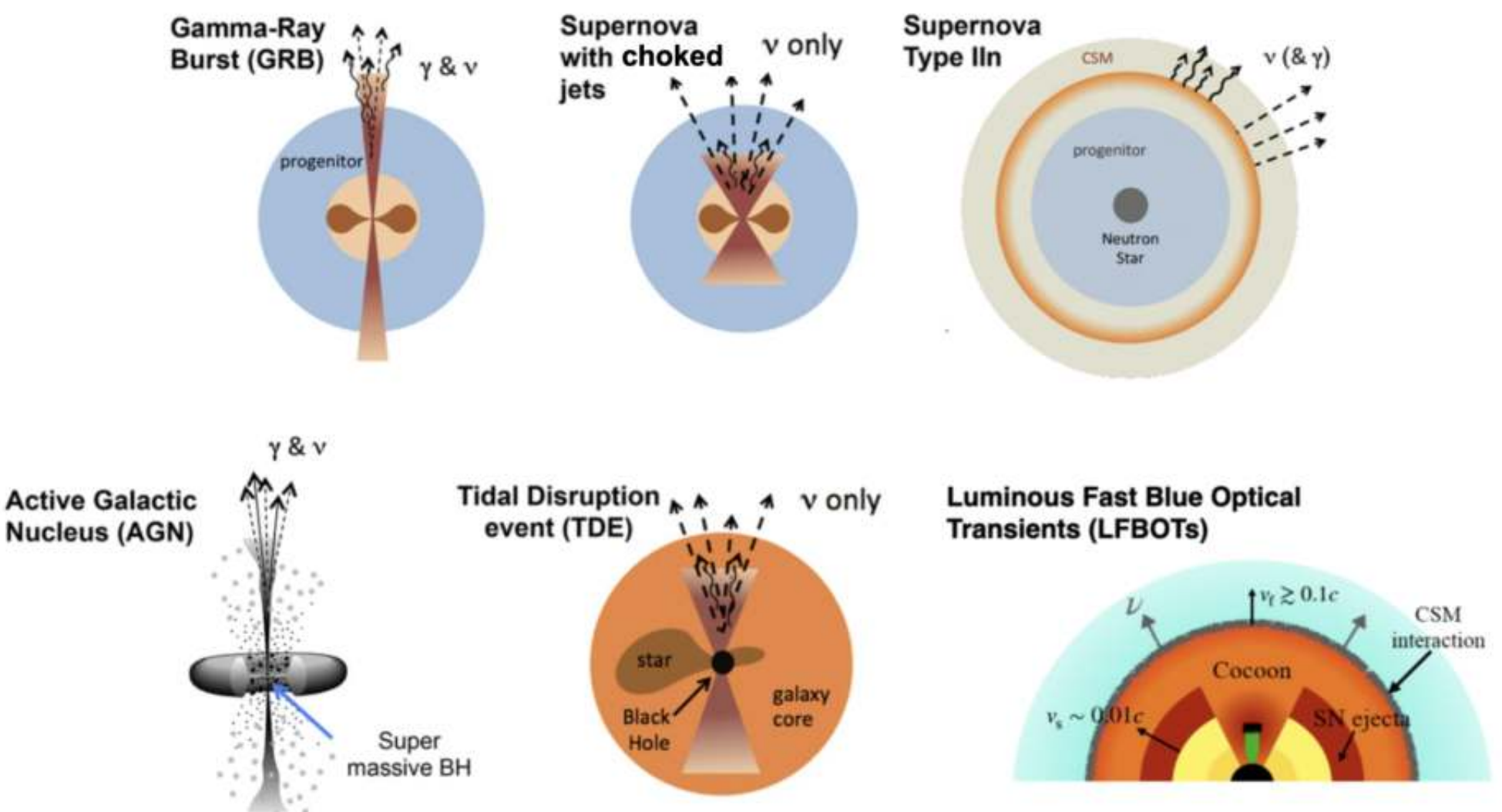

Figure 10 Time-varying neutrino source candidates targeted by our program (adapted from Refs. [22],[79]).

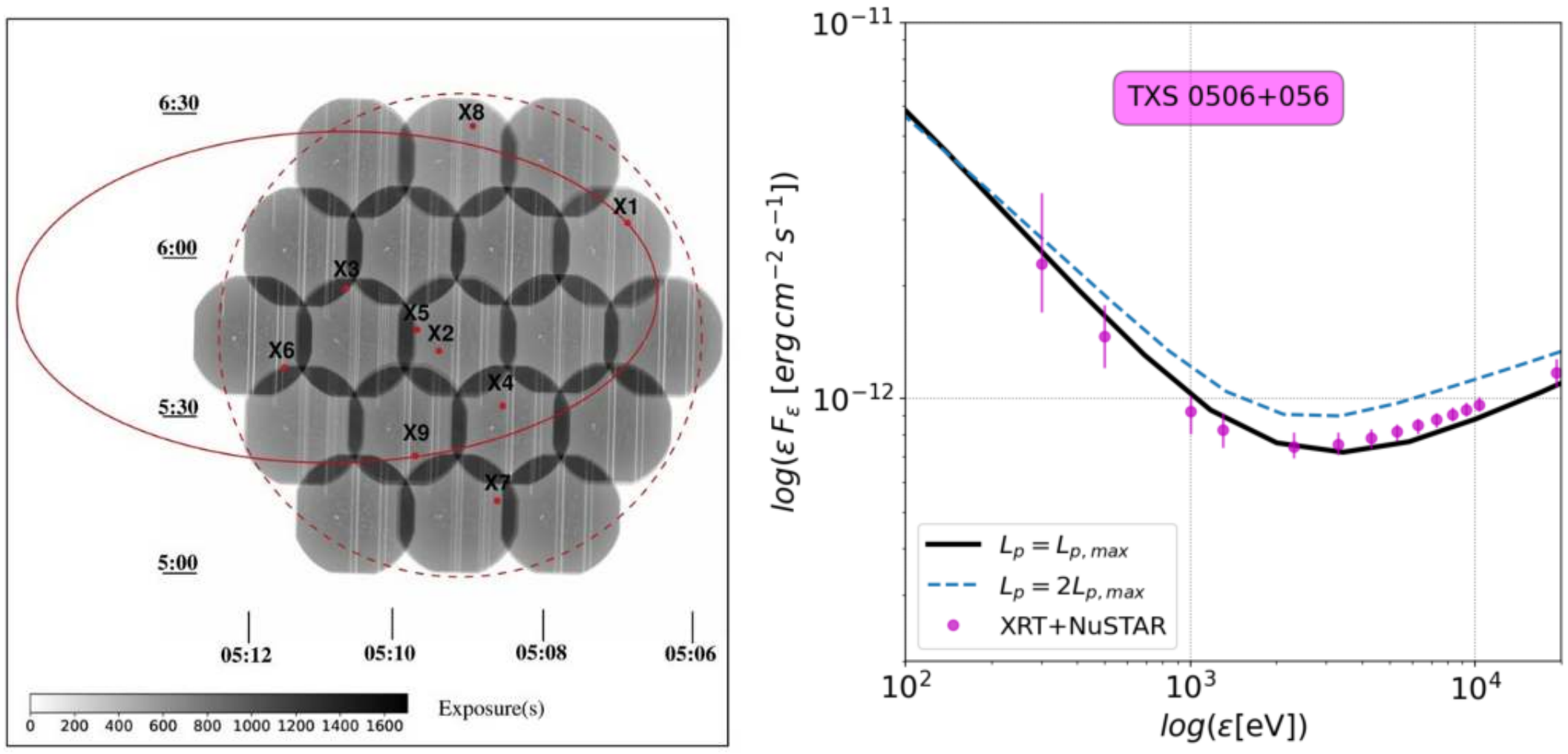

Figure 11 *Left panel:* Follow-up tiling observations performed with Swift-XRT in search of X-ray counterparts (red points) to the high-energy neutrino event IC 170922A whose 90% containment region is shown by the red solid ellipse [98]. The typical exposure time for each field is 0.8 ksec (see colorbar). Figure adopted from Ref. [113]. *Right panel:* Flux points from a joint fit to XRT and NuSTAR data comprised of 8 and 1 observations, respectively [113]. The data reveal the presence of a trough around 2 keV, which has been crucial for constraining neutrino predictions. A leptohadronic model (dashed line) with twice the neutrino luminosity of the baseline model (solid line) can be excluded.

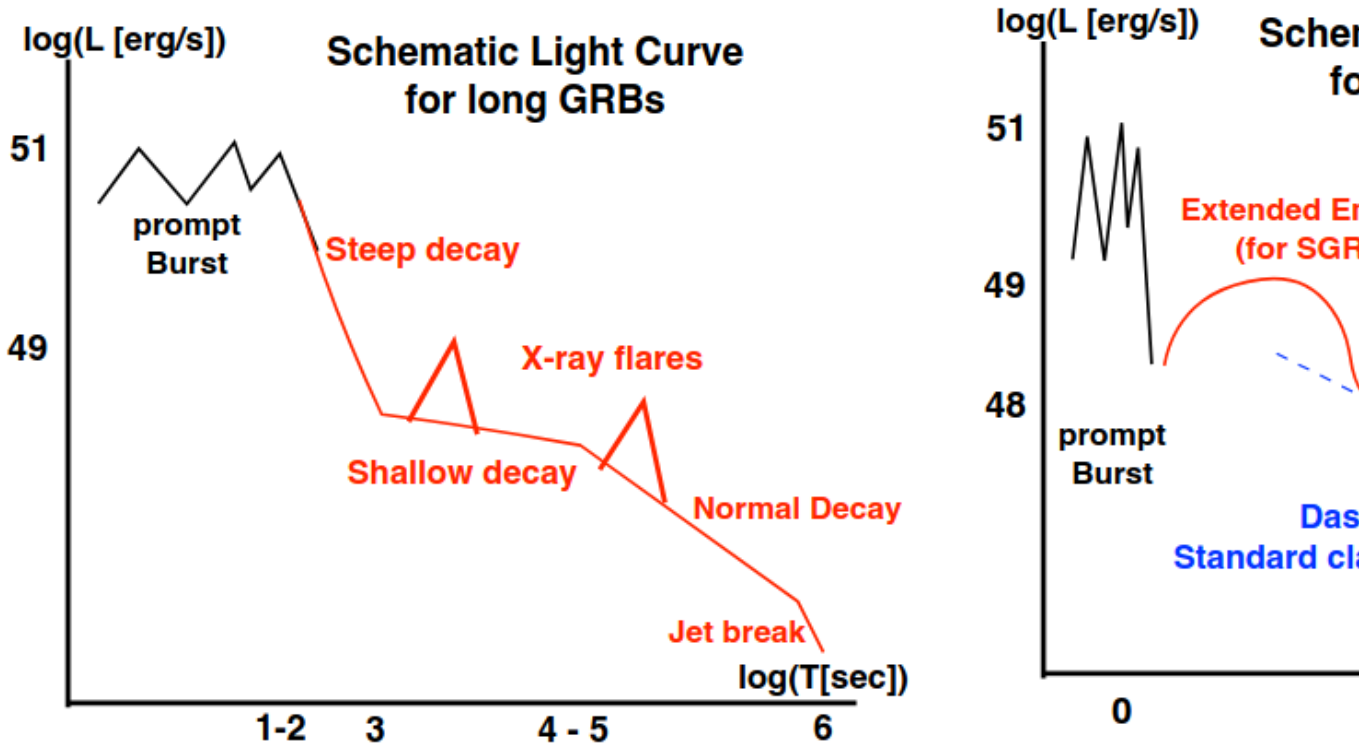

Figure 12 Sketches of typical X-ray light curves, for both long (left image) and short (right image) GRBs. Both classes of GRBs exhibit X-ray plateaus and/or X-ray flares, but with different temporal distribution and duration. High-energy neutrinos are expected, according to different models, during the early afterglow phase or X-ray flares and plateaus. Figure adapted from Ref. [116].

X-ray/neutrino connection in AGN coronae has gained significant momentum in recent years [see, e.g., 67,148].

AXIS' enhanced sensitivity, 60 times greater than Swift, will enable continuous monitoring of AGN with short exposure times, providing a detailed understanding of their X-ray baseline emission. Furthermore, AXIS's rapid response time ($<$ 2 hours, comparable to Swift) will be ideal for high-cadence follow-up observations of candidate sources in the direction of neutrino alerts, enabling timely identification and spectral characterization of potential neutrino counterparts.

Gamma-Ray Bursts

The search for neutrinos in association with Gamma-Ray Bursts (GRBs) has been an active field for almost 30 years [129]. Theoretical models proposed that the GRB prompt and afterglow phases could be related to the acceleration of ultra-high-energy cosmic rays (UHECRs) and neutrino production [80,166, 169,172,215,227]. Despite multiple campaigns that have been carried out by the IceCube Collaboration to identify the counterpart of high-energy neutrinos within the region of GRBs coincident in time with the IceCube detection, no GRB has been associated with neutrinos up to today [2,3,129]. However, theoretical studies suggest that the X-ray flares and X-ray plateaus observed in several GRB afterglow light curves could be related to internal dissipation mechanisms and late-time engine activity and consequently to neutrino production, both in long- and short-duration GRBs, as shown in Figure 40 [116,121,226]. Furthermore, theoretical studies suggest that non-standard GRBs, such as low-luminosity GRBs (LLGRBs) and choked GRBs, could be prime candidates for cosmic-ray acceleration and high-energy neutrino production [e.g. 142,149,186,196].

In general, the X-ray emission from GRBs can be fast-fading (within 12-24 hours after the trigger), but in some cases, such as GRB 170817A/GW 170817, it can last for several days [28,53,159]. In either case, thorough monitoring of the initial X-ray emission and its temporal evolution are essential for modeling the data and comparing different theories, requiring prompt slewing/observations within a few seconds after the initial prompt emission.
As neutrino sensitivity improves with IceCube upgrades and the continued development of KM3NeT, AXIS can play a fundamental role in the rapid follow-up of neutrino alerts and the identification of GRB counterparts. More specifically, based on the AXIS exposure time and Table 1, and taking into consideration

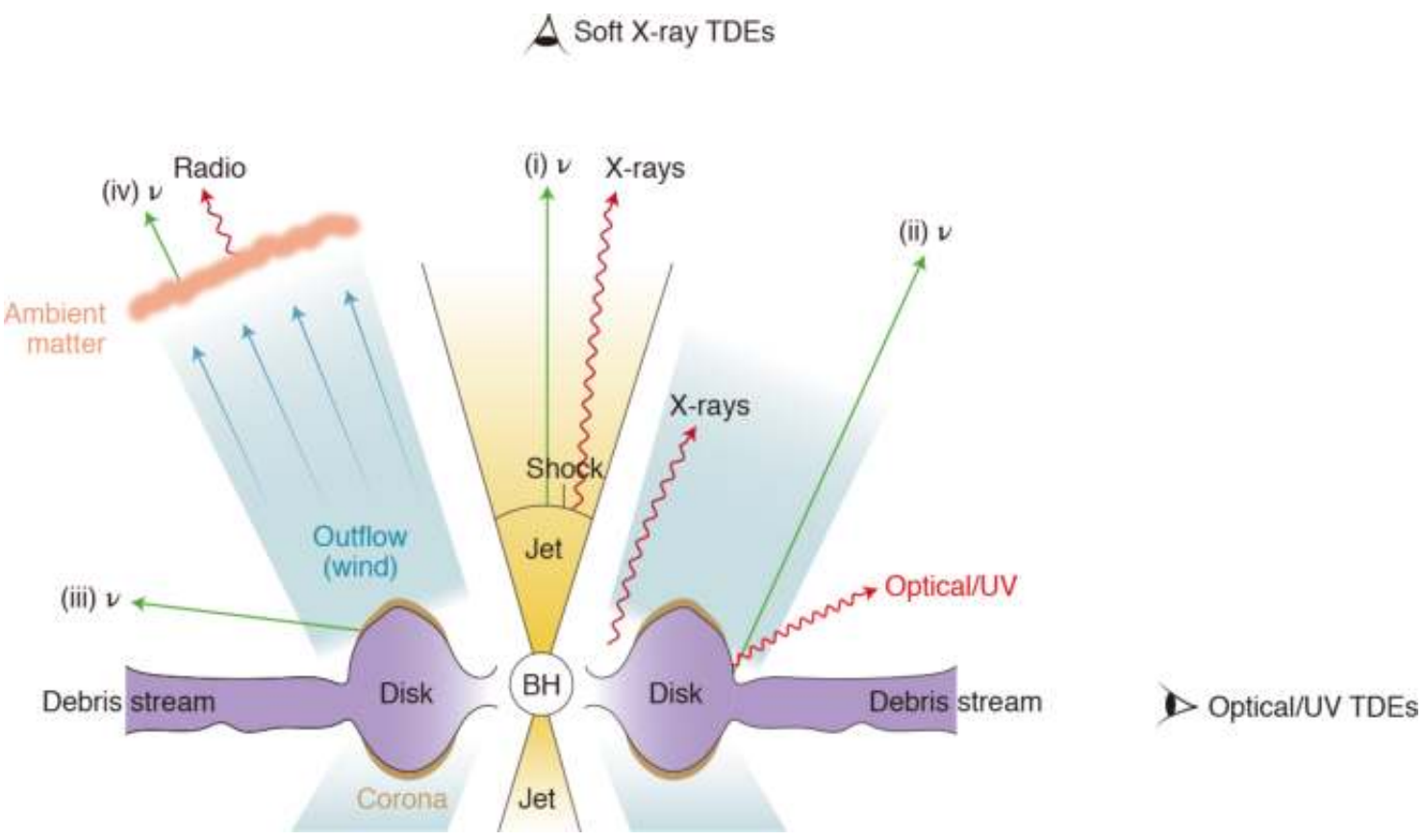

Figure 13 Illustration of the four main neutrino production sites in TDEs, from [87]. For some viewing angles, TDEs would only be detectable through neutrinos and X-rays.

that the distribution of the X-ray emission photon index peaks at $\Gamma < 1.5$, and an average X-ray flux ($0.3 - 10$ KeV) between $10^{-12}$ and $10^{-11}$ erg cm$^{-2}$ s$^{-1}$ [133], we got an estimate for the exposure time of less than 10 s and a count rate of less than 3, both for ONAXIS and FOVAVG.

Another critical aspect to consider is the role that neutrinos play as time-domain multi-messenger signals in an astro-particle context, where the synergy between AXIS and future facilities such as CTAO, ELT, and Roman is a key element in unveiling the physics behind acceleration processes in GRB jets.

Tidal Disruption Events

When a star passes close to supermassive black holes, the strength of tidal forces can shred the star, generating luminous emission across the electromagnetic spectrum. These 'Tidal Disruption Events' (TDEs) are now routinely detected by all-sky surveys, most notably in the optical [86]. Some optical TDEs exhibit X-ray emission [see e.g. 86], although complementary searches at X-ray wavelengths have revealed a broader sample of TDEs which lack optical counterparts [192]. One possible explanation is that the discrepancy is primarily due to a viewing angle effect [52], as illustrated in Figure 13. A handful of TDEs even launch relativistic jets, identified by luminous and highly beamed emission [30]. It was theorized that TDEs should emit neutrinos via $p\gamma$ interactions, either in relativistic jets, winds, or through corona (Figure 13; see [87] for a recent review). Optical follow-up campaigns have already identified two candidate neutrino-TDEs [178,203], with both sources detected in X-rays. AXIS's sensitivity will be capable of identifying any X-ray-bright TDEs, regardless of whether they are accompanied by optical emission.

**Observing description:** By the time of AXIS's launch, we expect there will be several neutrinos localized to $< 1$ sq. deg. each month. We envisage that AXIS can directly tile a substantial fraction of these neutrino alert localizations directly with its 0.125 sq. deg. field of view. As outlined in the previous section, multiple source classes are potential neutrino emitters, and we do not know in advance from which source class the detected neutrino was produced. **Therefore, we propose a source-agnostic strategy that can discover X-ray counterparts from a variety of possible source classes** shown in Figure 10. We envision the following workflow:

1. A current or future neutrino observatory detects a neutrino, and an alert is issued using the standard machine-readable format.
2. The neutrino alert is automatically evaluated with a pre-defined AXIS trigger algorithm, based on the sky area and probability that the neutrino is astrophysical in origin.

3. For neutrino alerts that meet our trigger criteria, we will determine a series of up to 8 tilings (1 sq. deg.) using the reported alert localization.
4. A ToO will be submitted within one minute of neutrino detection. We expect observations to begin in less than 1 hour (a priority for GRBs and jetted TDEs that fade rapidly).
5. Each individual tile will consist of one 1 ks snapshot.
6. We will identify new or flaring sources coincident with the neutrino.
7. We will cross-match the detected AXIS sources with source catalogs at different wavelengths. In particular, we will use eROSITA to identify known X-ray sources [141].
8. Spectral hardness, cross-matches, and contextual information will be used for classification.
9. We will promptly report these candidates to the community, enabling multi-wavelength follow-up.

In Table 1, we present the expected count rates in the 0.3-10 keV range with AXIS for power-law spectra with different indices (hard to soft) and integrated energy fluxes relevant to AGN jets, cores, GRBs, or TDEs.

| Flux (0.3-10 keV) | $\Gamma = 1.5$ | | $\Gamma = 2.0$ | | $\Gamma = 2.5$ | |
|---|---|---|---|---|---|---|
| (erg cm$^{-2}$ s$^{-1}$) | ONAXIS | FOVAVG | ONAXIS | FOVAVG | ONAXIS | FOVAVG |
| $5 \times 10^{-14}$ | $3.894 \cdot 10^{-2}$ | $3.303 \cdot 10^{-2}$ | $6.415 \cdot 10^{-2}$ | $5.478 \cdot 10^{-2}$ | $9.507 \cdot 10^{-2}$ | $8.150 \cdot 10^{-2}$ |
| $1 \times 10^{-13}$ | $7.787 \cdot 10^{-2}$ | $6.606 \cdot 10^{-2}$ | $1.283 \cdot 10^{-1}$ | $1.096 \cdot 10^{-1}$ | $1.901 \cdot 10^{-1}$ | $1.630 \cdot 10^{-1}$ |
| $5 \times 10^{-13}$ | $3.894 \cdot 10^{-1}$ | $3.303 \cdot 10^{-1}$ | $6.415 \cdot 10^{-1}$ | $5.478 \cdot 10^{-1}$ | $9.507 \cdot 10^{-1}$ | $8.150 \cdot 10^{-1}$ |
| $1 \times 10^{-12}$ | $7.787 \cdot 10^{-1}$ | $6.606 \cdot 10^{-1}$ | 1.283 | 1.096 | 1.901 | 1.630 |
| $5 \times 10^{-12}$ | 3.894 | 3.303 | 6.415 | 5.478 | 9.507 | 8.150 |

Table 1 Predicted count rates (counts/s) for AXIS for power-law spectra with different photon indices ($\Gamma$) and integrated fluxes (0.3–10 keV) assuming an absorption $N_H = 10^{21}$ cm$^{-2}$. The power-law spectra considered may apply to AGN cores, jets, TDEs, or GRBs.

*Neutrino Triggers*

Given that these alerts are future ToOs, we do not have an a priori target list. We envisage triggering on well-localized ($< 1$ sq. deg.) neutrinos with a high probability of being astrophysical (this information will be provided by the neutrino alert issued by the observatory). Based on the recent rate of suitable 'Gold' IceCube neutrino alerts, we expect 10 triggers per year and assume a further 10 triggers from other neutrino telescopes that will be in operation by the AXIS launch. With 20 triggers per year, we would be able to place meaningful constraints on the origin of astrophysical neutrinos within a couple of years of AXIS operations [see e.g. 204, for results from a similar optical program].

*Tiling of Neutrino Localization*

We will tile the localization, which is typically elliptical (see left panel in Figure 11), to maximize the enclosed probability. Bivariate Gaussians can approximate neutrino localizations, and we will select tilings that prioritize the innermost/most likely region. While we would consider up to 8 pointings for a given neutrino (corresponding to 1 sq. deg.), we assume that the area of a typical neutrino alert can be covered with 4 pointings. An illustration of a similar X-ray tiling, as done by Swift-XRT, is shown in Figure 11.

*AXIS Observations*

For blazars with TXS-like X-ray fluxes and photon indices ($F_X \sim 10^{-12}$ erg cm$^{-2}$ s$^{-1}$, $\Gamma \sim 2 - 2.5$), AXIS will detect approximately 1280-1900 (1096 - 1600) photons in a 1 ks ONAXIS (FOVAVG) observation (see Table 1), providing superb photon statistics for spectral analysis. Even in the case of a hard spectrum ($\Gamma = 1.5$) and a flux of $5 \times 10^{-14}$ erg cm$^{-2}$ s$^{-1}$ AXIS will be able to detect at high significance ($> 5\sigma$)

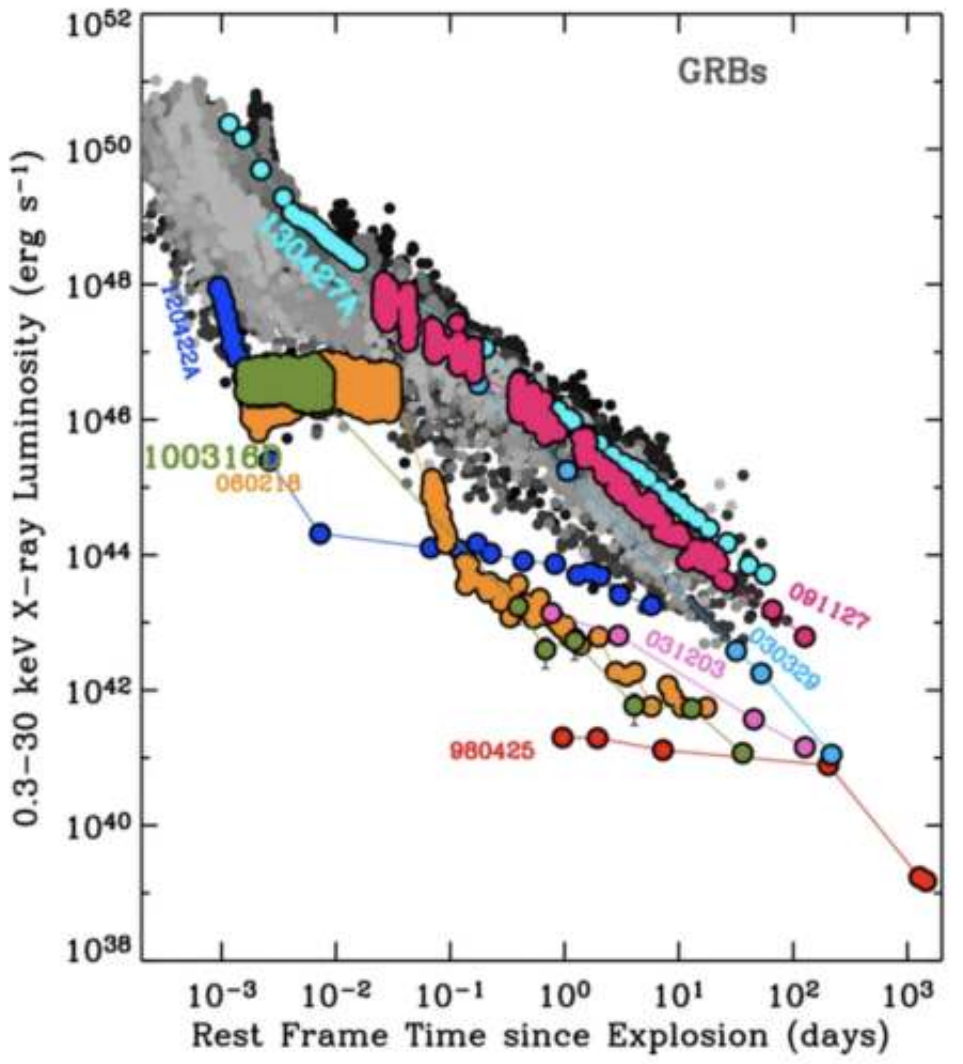

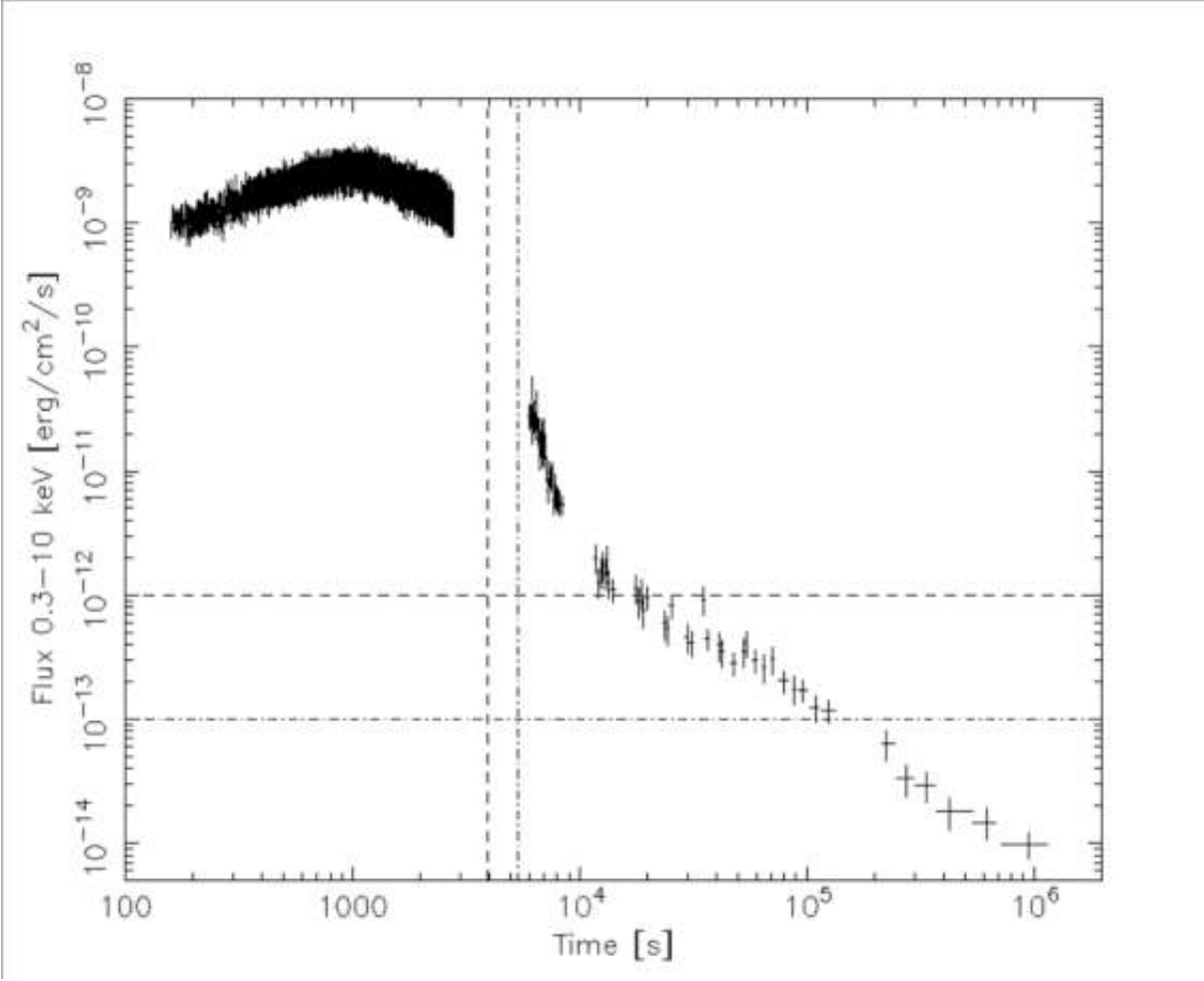

Figure 14 *Left panel:* X-ray luminosity light curves of LLGRBs (colored points) comparing with known redshift long GRBs shown in grey [134]. *Right panel:* XRT flux light curve of the famous low-luminosity GRB 060218. Vertical lines indicate the completion time of a tiling AXIS observation starting after 1 hr of the neutrino alert for a flux of $10^{-12}$ erg cm$^{-2}$ $s^{-1}$ (dotted) and $10^{-13}$ erg cm$^{-2}$ s$^{-1}$ (dash dotted).

other candidates with respect to the brightest source in the field in 1 ks. Tiling observations for the search for X-ray counterparts to the high-energy neutrino event IC 170922A were performed with Swift, with an average exposure of 0.8 ksec, and 8 X-ray candidate sources were detected (see Figure 11). A similar strategy with AXIS would reveal more sources, as fainter counterparts would be detectable. The scan of the neutrino alert area with AXIS would be completed in approximately 4 ks, assuming a $<$ 1 ks observation per tiling.

For fluxes relevant to LLGRBs, $F_X \sim 10^{-13}$ erg cm$^{-2}$ s$^{-1}$ and $\Gamma = 2$, AXIS will provide a $5\sigma$ detection in approximately 0.23 ks (using FOVAVG rates in Table 1). If the AXIS observation starts 1 hr after the neutrino alert, and assuming a slewing time of 30 s for each maneuver of the spacecraft, the AXIS observation will be completed in 5.3 ks after the neutrino alert. Figure 14 (right panel) illustrates the expected completion times of an AXIS observation, assuming $10^{-12}$ and $10^{-13}$ erg cm$^{-2}$ s$^{-1}$, compared to the Swift XRT light curve of one of the famous LL GRB 060218. In summary, we will observe each tile for 1 ks, using an average of 4 tilings per neutrino. This will be sufficient to detect a typical LLGRB at $z = 2$ in less than 1 hour after the burst, and a typical TDE at $z = 0.3$.

**Exposure time (ks):** We expect about 20 neutrino alerts per year, and each neutrino alert will be observed for 1 ks, with 4 fields per neutrino on average. In total, this amounts to 80 ks per year.

**Joint Observations and synergies with other observatories in the 2030s:** IceCube is the only operating detector publishing real-time neutrino alerts, with typical 90% localizations of $\sim$ 1 square degrees. IceCube will soon be joined by new water-based neutrino observatories that are under construction or will soon be completed (e.g., KM3NeT, P-One, Baikal-GVD), for which even more precise angular resolution is expected. By the time of AXIS's launch, we expect there will be several neutrinos localized to $<$ 1 sq. deg. each month. A similar workflow to the one described in the previous section will be operated by other observatories, for example, the Vera C. Rubin Observatory in the optical [15]. This unified community approach provides substantial synergy, enabling efficient candidate characterization through multi-wavelength analysis. All source populations in Figure 10 are expected also to have some degree of optical/NIR emission, and

combining the data will enable us to quickly distinguish between potential sources, such as supernovae, TDEs, and AGN.

**Special requirements:** AXIS observations should start within $< 2$ hr to identify rapidly decaying X-ray transients in the neutrino event localization area. Mission software capable of handling requests for observations of multiple adjacent pointings should be developed.

*6. Probing the X-ray spectra of massive black hole binaries with AXIS*

**First Author**: Tingting Liu (Georgia State University, tliu20@gsu.edu)
**Co-authors**: Julie Malewicz (Georgia Institute of Technology), Tamara Bogdanovic (Georgia Institute of Technology)
**Abstract:** Supermassive black hole binaries (SMBHBs) and massive black hole binaries (MBHBs) are promising sources of multi-messenger signals. They can be detected through gravitational waves (GWs) using pulsar timing arrays (PTAs) and the Laser Interferometer Space Antenna (LISA) and are observable across the electromagnetic (EM) spectrum as binary active galactic nuclei (AGN). Theoretical studies of these binary AGN predict a range of EM observables, including distinctive X-ray spectra. We will use AXIS to observe an SMBHB/binary AGN candidate and search for X-ray spectral features, such as excess soft X-ray emission and anomalous iron K $\alpha$ line profiles, which could indicate the presence of minidisks in an SMBHB system.
**Science:**

Massive black hole binaries (MBHBs) with masses between $\sim 10^6$–$10^9$ $M_\odot$ are thought to form in galaxy mergers. A series of physical processes – dynamical friction and interactions with stars and gas – bring the pair of massive black holes (MBHs) to sub-parsec separations, where gravitational wave (GW) emission begins to dominate the binary's evolution. Consequently, these systems are strong GW sources for pulsar timing arrays (PTAs) and the Laser Interferometer Space Antenna (LISA).

PTAs use the precise radio timing of millisecond pulsars – extremely stable cosmic clocks – to detect deviations in the pulse arrival times caused by nanohertz GWs. Multiple PTA collaborations have already reported evidence for a stochastic GW background [10,62,143,176,220], consistent with the ensemble signal from a population of SMBHBs (e.g., [9,63]), and the loudest binaries may be individually resolved as "continuous wave" sources (e.g., [23,115]). These PTA sources are expected to be massive ($> 10^9 M_\odot$), have high mass ratios ($q > 0.1$), and locate at relatively low redshifts ($z < 1$; e.g., [9]).

LISA, adopted in 2024 and expected to launch in $\sim$2035, consists of a constellation of three spacecraft exchanging laser beams while trailing the Earth's orbit around the Sun. It measures changes in the distances between test masses inside the spacecraft via laser interferometry and is sensitive to millihertz GWs [12,49]. During its four-year primary mission, LISA is expected to detect up to hundreds of MBHBs (e.g., [21,117]). These binaries are typically less massive ($\sim 10^6 M_\odot$) but can be detected out to high redshifts ($z \sim 10$).

Since galaxy mergers can funnel substantial amounts of gas to the nucleus of the system, MBHBs can accrete and radiate as (binary) active galactic nuclei (AGN), making them promising multi-messenger sources and unique laboratories for probing accretion physics in binary black hole systems. The prospects for the multi-messenger studies of these sources depend on our ability to uniquely identify binary AGN and distinguish their electromagnetic (EM) emission from that of a single SMBH. Over the past decade, numerous theoretical studies have sought to understand accretion onto SMBHBs and the distinctive EM signatures these systems may produce (see, e.g., [82] for a recent review). These signatures mainly include (1) periodic modulation of the source flux on the binary orbital timescale (e.g., [56,65,130,158]) and (2) unusual spectral features in the UV/optical and X-ray bands (e.g., [66,184]).

In the second category, SMBHB EM signatures include enhanced hard X-ray emission due to accretion streams striking the accretion disks attached to each black hole (the "minidisks") [66,184], double broad iron K$\alpha$ lines resulting from the line-of-sight motion of the two minidisks [198], and excess soft X-ray emission lines originating from the minidisks [131]. AXIS is well-suited to probe the latter two types of signatures thanks to its high sensitivity. These signatures are expected to be particularly prominent for SMBHBs detectable by PTAs, as their larger orbital separations and longer orbital periods allow for time-resolved spectroscopy. For MBHBs detectable by LISA, the AXIS wide field of view is advantageous for rapid source localization: LISA can only constrain the position of a $10^6 M_\odot$ MBHB within $\sim 10$ deg$^2$ approximately one week before merger [132,168]. Identifying the source *before* the merger requires efficient tiling of this error region. Given the short orbital periods ($\sim$ hours) and the short evolutionary timescales ($\sim$ weeks) of the LISA sources, only orbit-averaged spectroscopy will be possible, and the reliable identification of MBHB EM signatures must first be "rehearsed" on SMBHBs detectable by PTAs, which evolve on much longer timescales. Here, we propose to investigate the X-ray spectra of PTA-detectable SMBHBs with AXIS. Specifically, we will search for excess soft X-ray emission at $\sim 1$ keV and anomalous iron K $\alpha$ line features at $\sim 6$ keV which are both indicative of minidisks accreting from a circumbinary disk.

Numerical simulations of SMBHBs indicate that minidisks accrete through a pair of streams that connect to the circumbinary cavity wall. The total accretion onto the SMBHB, which powers the total EM output of the binary AGN, is divided between the two minidisks, and the ratio of accretion rates depends on the binary mass ratio and favors the less massive black hole (the "secondary", e.g., [66]). When the binary is relatively widely separated, as is the case for PTA-detectable SMBHBs, the minidisks are persistent. They can be considered "mini" versions of geometrically thin and optically thick standard accretion disks. Since X-rays probe the inner regions of the SMBHB system that are dominated by the minidisks and/or their coronae, the observed X-ray spectrum of the binary AGN is a composite of the two minidisk spectra.

Recent work by [131] considered these effects – the different accretion rates between the two minidisks and the binary orbital motion – and computed the reflection spectra for a range of binary mass ratios and total accretion rates. They find that (1) the SMBHB exhibits complex soft X-ray emission lines because of their strong dependence on the ionization levels of the minidisks and (2) the source exhibits an abnormal or variable iron K $\alpha$ line profile that results in unconstrained or variable spin parameter estimates.

To assess the detectability of the soft X-ray feature with AXIS, we simulate a 100-ks observation at three different orbital phases. The whole reflection spectrum at one of these phases is shown in the upper panel of Figure 15 for illustration. Next, we fit the mock observation at each epoch using a single AGN spectral model, mimicking a spectral analysis performed under the assumption that the source is a "regular" AGN. The redshift and absorbing column density are fixed at their injected values. As can be seen in the lower set of panels, the complex emission lines at soft energies (between 0.2 and 1 keV) cannot be captured by a single AGN model at any epoch, resulting in significant residuals (i.e., large $\chi^2/dof$ values).

Additionally, there are strong residuals at around 6 keV for $f = 90°$, due to the poor fit to the iron K $\alpha$ line at this orbital phase. In fact, the fit is so poor that the uncertainty in the spin parameter cannot be determined, and the best-fit value is near-maximal retrograde. We summarize the fit results in Table 2.

**Exposure time (ks): 100 ks**

**Observing description:**

We will use AXIS to follow up on possible SMBHB detections reported by PTAs. Following the observation of the stochastic GW background, measurements of background anisotropy (indicative of an astrophysical origin of the background) and even individual source detections are possible in the near future. Time-to-detection calculations suggest that the first continuous-wave (i.e., single-source) detection is likely by $\sim 2030$, which is well aligned with the AXIS mission timeline. Given the large localization

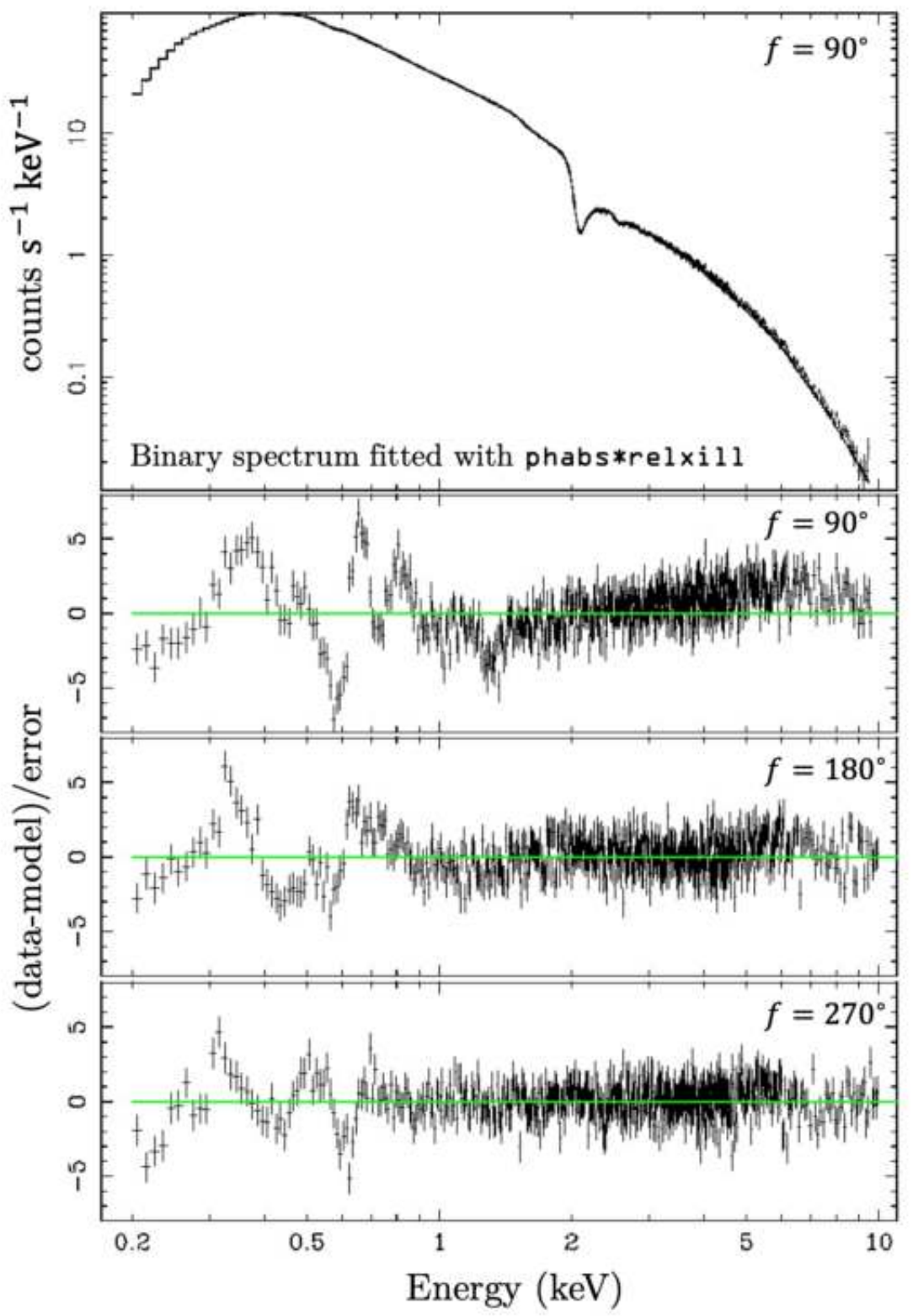

Figure 15 A simulated 100 ks AXIS observation of an SMBHB ($z = 0.1$, $M_{\rm tot} = 10^9 M_\odot$, mass ratio $q = 0.2$, Eddington ratio $\lambda_{\rm tot} = 0.1$, spin of each black hole $a = 0.99$). $f = 90°$ ($f = 270°$) corresponds to the orbital phase where the secondary black hole is moving towards (away from) the observer. We adopt a Galactic absorption of $N_{\rm H} = 3 \times 10^{20}$ cm$^{-2}$. The complex excess soft X-ray features at $\sim 1$ keV, originating from the minidisks, cannot be adequately fit by a single AGN spectral model (`phabs*relxill`), highlighting the potential of using AXIS to identify distinctive X-ray spectral signatures of SMBHBs.

Table 2. Fit results of the mock observations

| $f$ | $\chi^2$ / d.o.f. | $a$ | $\Gamma$ | $\log \xi$ | $A_{\rm Fe}$ | $f_{\rm refl}$ |
|---|---|---|---|---|---|---|
| 90 | 2,104 / 933 | $-0.998$ | 2.00 | 2.97 | 0.57 | 0.33 |
| 180 | 1,343 / 973 | $0.998^{+0.000p}_{-0.037}$ | $1.98 \pm 0.00$ | $2.88^{+0.01}_{-0.02}$ | $0.79^{+0.03}_{-0.02}$ | $0.56 \pm 0.02$ |
| 270 | 1,169 / 973 | $0.998^{+0.000p}_{-0.019}$ | $1.97 \pm 0.00$ | $2.94 \pm 0.02$ | $0.73^{+0.03}_{-0.01}$ | $0.62^{+0.02}_{-0.03}$ |

Note. — $f-$ orbital phase; $\chi^2$ / d.o.f. $-$ goodness of fit; $a-$ spin of each black hole; $\Gamma-$ photon index; $\log \xi-$ ionization parameter; $A_{\rm Fe}-$ iron abundance; $f_{\rm refl}-$ reflection fraction. Uncertainties correspond to a 90% confidence range on the fit parameter. A "p" in an error bar indicates that the error range is pegged at the upper or lower limit of the parameter.

regions associated with PTA detections, X-ray follow-up is advantageous when searching for the EM counterpart, due to the significantly lower number of sources compared to the optical. Furthermore, X-rays are less susceptible to obscuration, which is likely in SMBHB systems formed in gas-rich galaxy mergers.

With a 100 ks observation, AXIS can detect the predicted soft X-ray signature with high confidence and reliably identify the true EM counterpart to a GW detection. This observation will also serve as a pilot study for LISA (which is expected to launch in $\sim$ 2035, also well matched to the AXIS timeline), since the LISA-detectable MBHBs are expected to exhibit similar X-ray signatures. The strong synergies between AXIS and PTAs/LISA will open a new chapter in astrophysics – the multi-messenger detection of massive black hole binaries, approximately fifteen years after the first multi-messenger detection of a binary neutron star merger.
**Joint Observations and synergies with other observatories in the 2030s:** Pulsar timing arrays, LISA.

*7. AXIS–UVEX Synergies*

**First Author:** Suvi Gezari (University of Maryland, suvi@umd.edu)

**Abstract:** UVEX, the Medium Explorer NASA Mission scheduled for launch in 2030 (with a baseline mission duration of 2 years and potential for an extended mission), has numerous scientific synergies with AXIS. The concurrent operation of these two telescopes, with complementary capabilities in imaging and spectroscopy in the UV and X-ray ranges, respectively, would enable unique TDAMM science. UVEX will provide imaging in the FUV and NUV bands across the entire sky, with 10 epochs of varying cadences spread over the 2-year mission. It also offers long-slit FUV and NUV TOO spectroscopy with a response time of $<$ 6 hours and a spectral resolution of R $>$ 1000. Below, we outline science cases for joint UVEX and AXIS observations during a potential UVEX extended mission.
**Science:** Many energetic transients are bright simultaneously in the X-rays and UV. AXIS discovery in the X-rays with rapid response ($<$ 6 hours) UVEX follow-up UV spectroscopy, and UVEX discovery in the FUV and NUV imaging with rapid response ($<$ 2 hours) AXIS follow-up X-ray imaging, will enable new joint UV+X-ray studies of the earliest phases of a supernova, of rapidly evolving transients like LFBOTs and kilonovae, as well as monitoring observations of long-lived transients such as tidal disruption events (TDEs) and changing-look AGN.

1. UVEX rapid response ($<$ 6 hours) UV spectroscopy of AXIS-discovered X-ray transients from shock breakout in core-collapse SNe would enable the earliest spectra of a supernova explosion, and would probe the mass-loss history and stellar structure of the SN progenitor. For nearby SNe ($<$ 250 Mpc), UVEX can provide a detailed spectral sequence from 115 nm to 265 nm, which probes the structure and chemical composition of the progenitor star and its pre-explosion mass-loss history.
2. AXIS X-ray follow-up of UVEX discovered tidal disruption events would probe the joint UV/X-ray evolution of a large sample of TDEs, and probe the relative timing and evolution of the two emission components, tracing the inner accretion disk (X-rays) and the circularizing stellar debris streams and/or associated outflows (UV). Joint UV/X-ray observations of TDEs at late times can be used to measure the inner and outer radius of the newly formed TDE accretion disk, and infer the mass of the central black hole from its accretion disk luminosity. UVEX will have the FUV and NUV per-epoch sensitivity ($m_{lim}$= 24.5 mag) to detect and identify thousands of TDEs as UV transients out to $z = 3$. Similarly, AXIS can probe soft X-ray emission typical of TDEs ($L_X > 10^{42}$ erg/s) out to $z = 3$.
3. UVEX rapid response ($<$ 6 hours) UV spectroscopy of jetted TDEs discovered by AXIS in the X-rays would enable the joint study of the newly formed jet and its associated TDE accretion flow. UV absorption and emission features would trace the kinematics and viewing angle of the TDE accretion disk and its associated outflows, and probe the conditions necessary for jet formation.

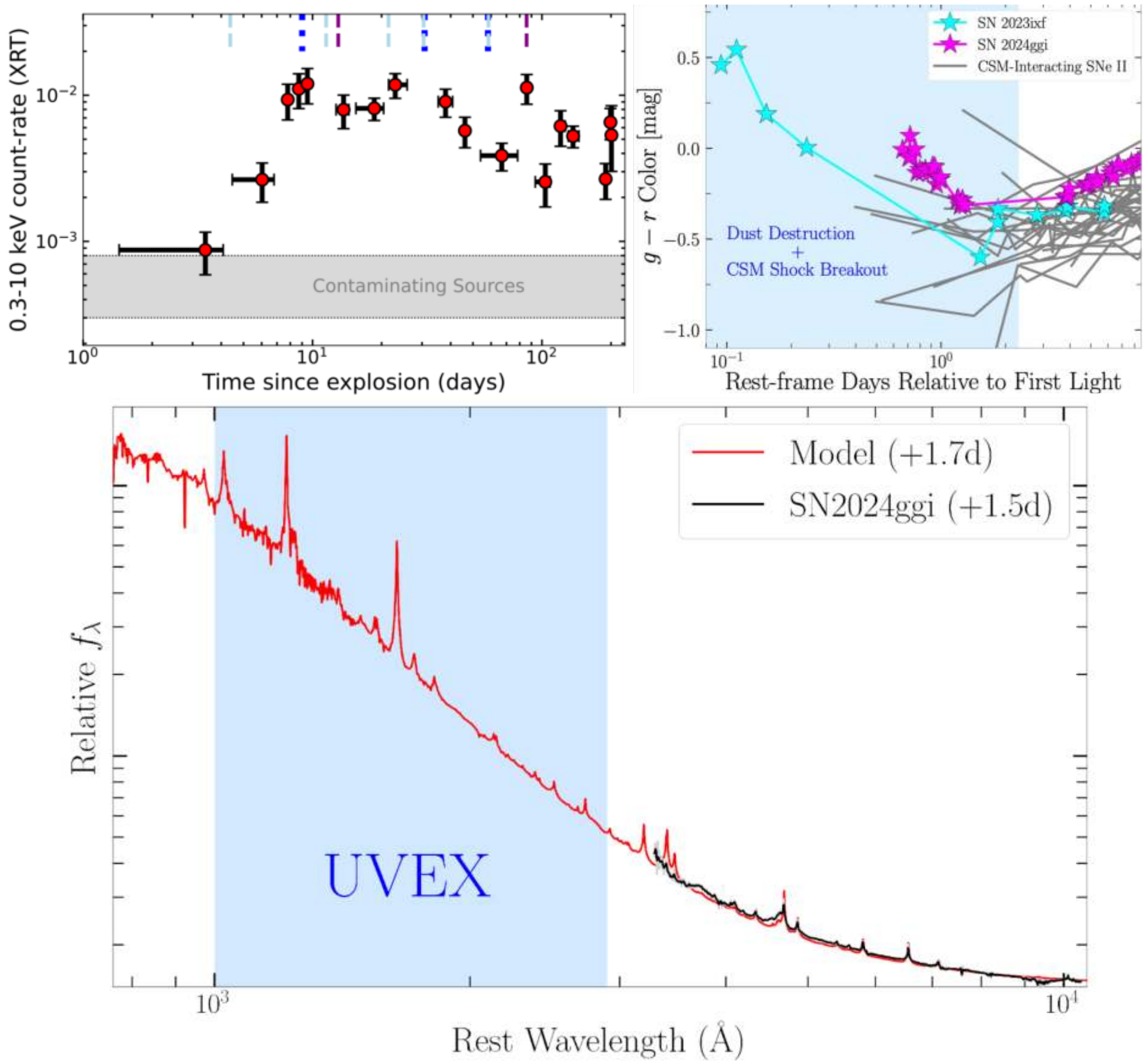

Figure 16 **AXIS rapid-response follow-up X-ray imaging of UVEX transient discoveries, and UVEX rapid-response UV spectroscopy of AXIS transient discoveries, will open up a new parameter space for prompt characterization of energetic transients.** *Upper left*: Rising X-ray emission from shock breakout (SBO) in dense circumstellar material (CSM) as observed in the nearby type II SN 2023ixf. Adapted from [155]. *Upper right*: Early-time color evolution showing the extended SBO phase in young SNe II as the fast-moving ejecta collide with dense CSM, creating a dramatic rise in temperature as the shock wave breaks out. Adapted from [104,105]. *Lower*: Very early-time spectrum of type II SN 2024ggi (black) showing high-ionization spectral lines from ejecta interaction with confined CSM. In red, we show a well-matched non-LTE radiative transfer model created using the `CMFGEN` code [54,91], which demonstrates that UV wavelengths carry the majority of the flux and spectral line information at early times. The UVEX wavelength coverage is shown in blue.

4. AXIS X-ray follow-up of gravitational wave sources, in coordination with UVEX UV follow-up, and ground-based optical and space-based NIR follow-up, would provide a comprehensive panchromatic view of the earliest phases of a kilonova.
5. AXIS X-ray follow-up of UVEX discovered changing-look AGN could be used for multi-band reverberation mapping.
6. AXIS rapid response ($< 2$ hours) X-ray follow-up of UVEX discovered LFBOTs would probe the central engine, potentially powering these events.

**Exposure time (ks):** Various

**Observing description:** Various targets: Infant SNe, LFBOTs, kilonovae, TDEs, changing-look AGN

**Joint Observations and synergies with other observatories in the 2030s:** UVEX

**Special requirements:** Rapid response TOO capability

**b. Extreme Transients and Compact Object Outbursts**

*8. Probing wandering black holes with off-nuclear tidal disruption events*

**First Author:** Yuhan Yao (UC Berkeley, yuhanyao@berkeley.edu)

**Abstract:** The hierarchical merger-driven process of galaxy assembly naturally predicts the existence of a significant number of off-nuclear massive black holes (MBHs). Some of them are intermediate-mass black holes (IMBHs) that come from the disruption of satellite galaxies, and some of them are MBH pairs/binaries and recoiling MBHs, which represent the progenitors and outcomes of massive black hole binaries that are the primary sources for the Pulsar Timing Array and the upcoming LISA Observatory. Tidal disruption events (TDEs) are produced when a star wanders too close to a MBH to be disrupted, and they provide a unique probe of wandering MBHs. AXIS, with its unprecedented sub-arcsecond angular resolution across the entire $\sim$0.1 deg$^2$ field of view, its deep X-ray surveys, and the rapid ToO capability, is well-suited to identify and characterize a sample of offset TDEs in the X-rays.

**Science:** Off-nuclear MBHs can arise through three channels:

1. Long dynamical friction (DF) timescales. Because nearly every big galaxy harbors a central MBH, galaxy mergers lead to the formation of MBH pairs. On $\sim$kpc scales, MBHs are brought together by DF. In certain cases, such as minor mergers in massive galaxies or when the secondary MBH undergoes rapid complete tidal stripping [57,114], the DF timescale exceeds the Hubble time, meaning the secondary MBH remains offset. Cosmological simulations show that those wanderers dominate the MBH population at the low-mass end [180].
2. Slingshot kick during 3-body interaction. Since the timescale of a MBH pair/binary is long ($\sim$Gyr), a third MBH can enter the system in a subsequent galaxy merger. In such cases, close triple interactions will eject the least massive black hole [92,190], giving it a "slingshot" kick and producing an offset wandering MBH.
3. Gravitational wave (GW) kick during MBH coalescence. GW radiation drives MBH binaries to coalescence, making them the primary sources for PTA and LISA. In some cases, the GWs carry enough linear momentum to impart a substantial kick to the newly merged MBH, creating a recoiling MBH at off-nuclear positions.

The demographics of offset MBHs provide key information on the formation and evolution of ultra-compact dwarf galaxies, intermediate-mass black holes (IMBHs), and GW sources. Direct evidence of offset MBHs comes from dynamical mass measurements. Using this method, eight MBHs have been detected at the centers of stripped nuclei residing in the halos of their host galaxies ([83,197], and references therein). However, this technique targets nuclear star clusters and is therefore restricted to local galaxies within $\sim$20 Mpc and mainly probes wanderers formed via the first channel.

A more widely adopted approach for detecting offset MBHs involves searches for dual or binary active galactic nuclei (AGN) and offset AGN. However, these methods face substantial selection effects, as AGN only probe MBHs that are actively accreting [29].

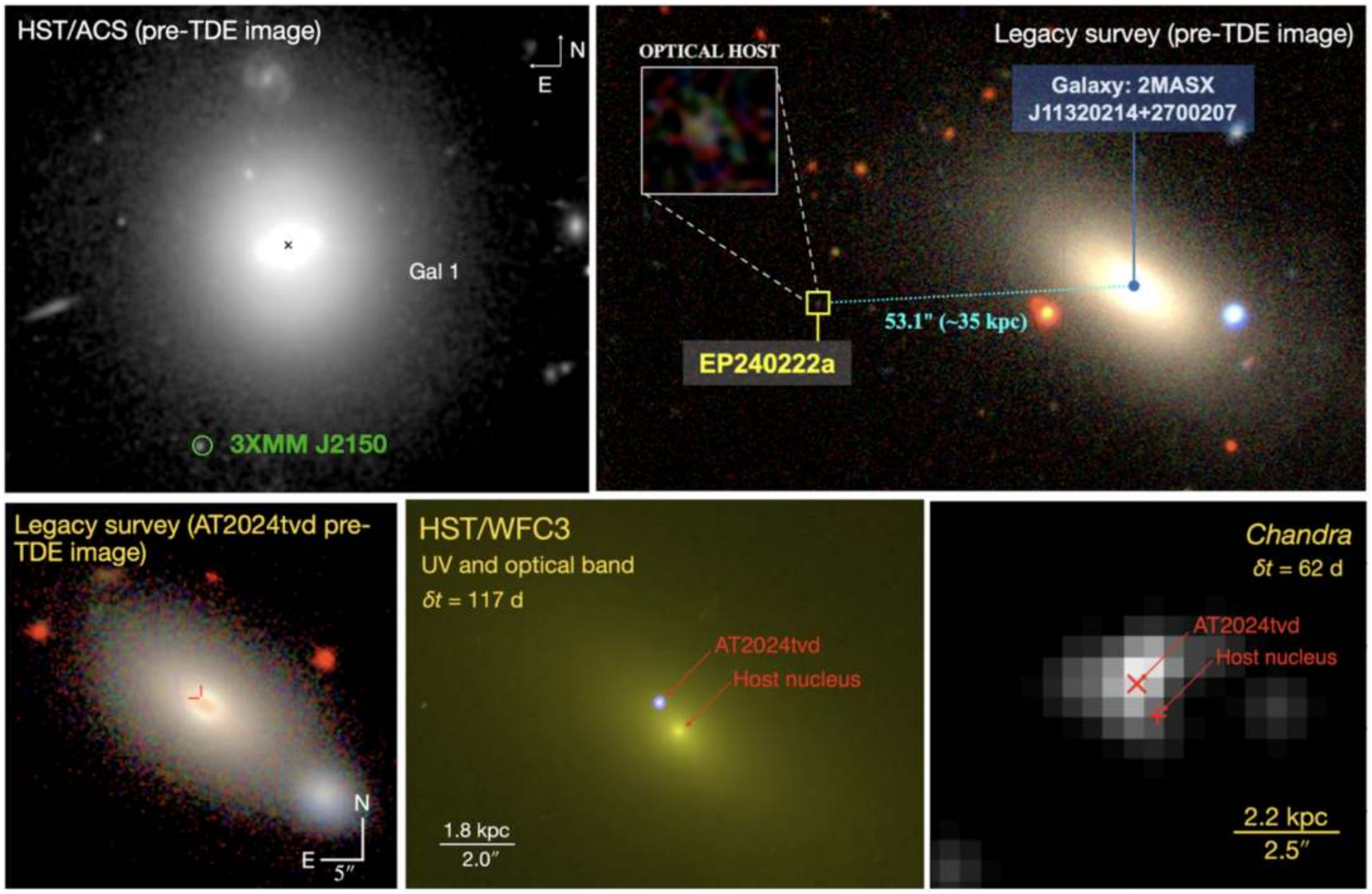

Figure 17 **AXIS will provide sufficient angular resolution and sensitivity to uncover the elusive population of off-nuclear TDEs.** *Upper*: Pre-TDE optical images of 3XMM J2150 and EP240222a show stripped nuclei at the locations of the TDEs [109,124]. *Lower*: AT2024tvd is offset from its galaxy nucleus by 0.8 kpc but not associated with any massive cluster. Chandra observations are key to associating the location of the X-ray transient with the optical transient [222].

In contrast, tidal disruption events (TDEs) are produced when a star wanders close enough to a MBH to be disrupted, and they occur across all types of galaxies [136,192,221]. Therefore, off-nuclear TDEs offer a unique pathway to probe MBHs irrespective of the state of merger-driven accretion [181].

To date, three offset TDEs have been identified. In all cases, Chandra observations played a crucial role in confirming the TDE nature and the off-nuclear locations.

- 3XMM J2150, identified by searching for soft X-ray flares in the XMM-Newton archives [124,125], exhibited no optical counterpart. Archival Chandra observations associated the transient with a marginally resolved source 12.5 kpc from the center of its parent galaxy in pre-flare HST imaging ([124]; Figure 17, upper left panel).
- EP240222a [109], initially detected by the wide-field X-ray telescope aboard the Einstein Probe X-ray satellite, had only arcmin-scale localization. Prompt follow-up with Chandra refined its position to sub-arcsec accuracy, pinpointing its origin in a stripped nucleus 35 kpc from the center of its parent galaxy (Figure 17, upper-right panel).
- AT2024tvd [222], discovered by the Zwicky Transient Facility (ZTF) optical survey, is merely 0.8 kpc from the center of its host galaxy. Archival optical imaging shows that it is within the Galactic bulge, placing an upper limit of $M_{\rm star} < 10^{7.6}$ $M_\odot$ on any possible massive cluster at its location. While its initial TDE classification was based on broad hydrogen lines and bright UV emission—features common in TDEs—similar characteristics can occasionally appear in rare massive star explosions [76,81]. A rising soft X-ray source near the optical transient supports the TDE interpretation, as

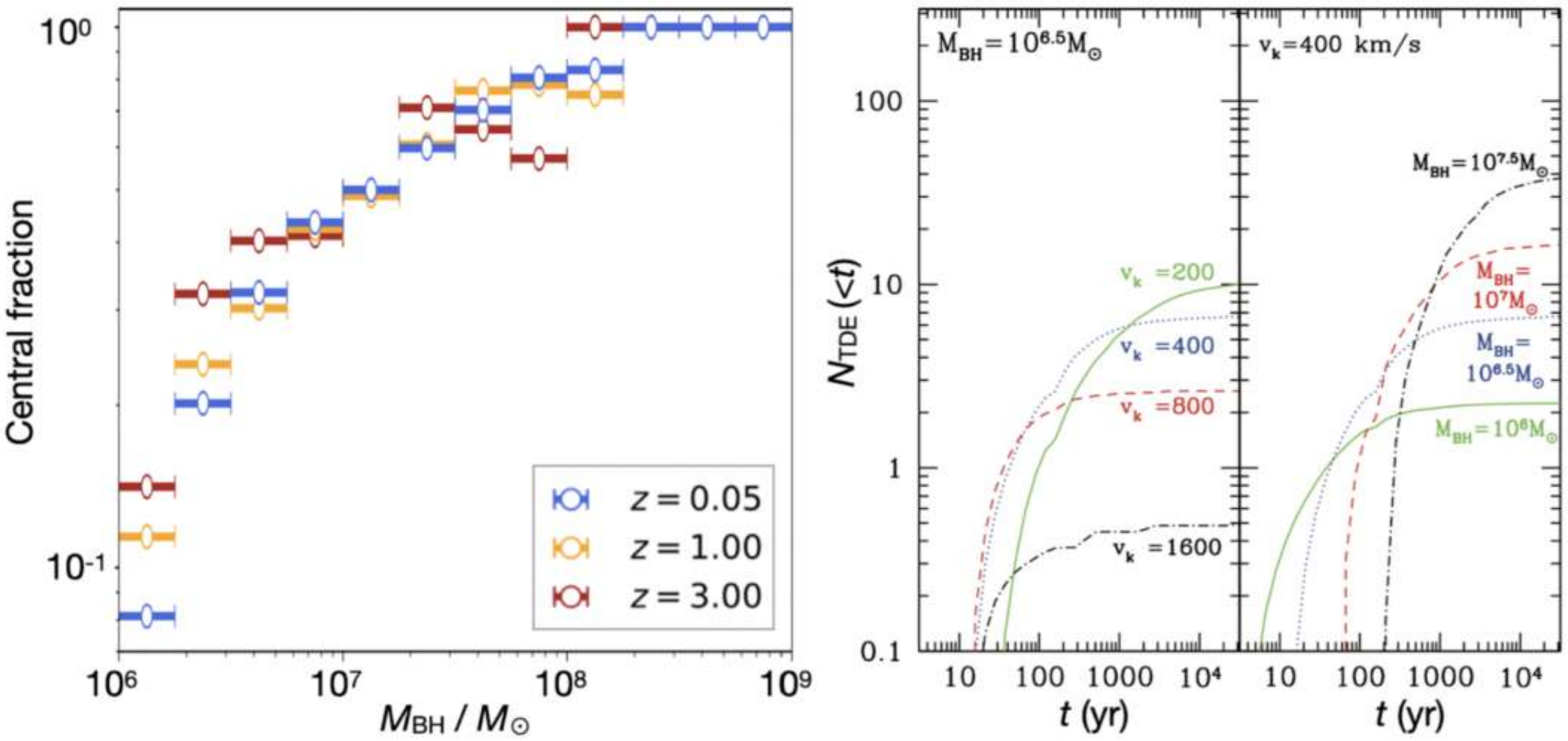

Figure 18 **AXIS observations of TDEs will allow us to probe fundamental galaxy evolution properties that are exceedingly challenging via other means.** *Left*: The fraction of MBHs of a given mass that are centrally located in the ROMULUS25 cosmological simulation [181]. This fraction drops far below unity at low masses, meaning that the typical low-mass MBH is a wanderer. If these wandering MBHs can retain an unresolved stellar component cluster, they may dominate the TDE rate in their mass bins. *Right*: expected number of TDEs in less than t years since a MBH merger, for varying kick velocities and masses of the merger remnant [205]. Since the MBH of a TDE can be constrained by fitting to the X-ray spectra [218] or the UV/optical light curve [147], the rate of offset TDEs powered by GW recoiling MBHs (i.e., channel 3) probes GW kick velocities.

such soft X-ray spectra are expected from accretion in TDEs but not in supernovae (SNe). However, without precise X-ray localization, an alternative scenario could not be excluded: AT2024tvd might be a hydrogen-rich SN powered by interaction, while the X-ray transient could be an unrelated, optically faint, X-ray-bright TDE coincidentally occurring in the same galaxy. A Chandra DDT observation resolved this ambiguity by confirming the X-ray transient's spatial coincidence with AT2024tvd, definitively establishing its TDE nature (Figure 17, lower panels).

Among the $\sim$200 known TDEs, 3XMM J2150 and EP240222a remain the only two associated with IMBHs ($M_{BH} < 10^5$ M$_{\odot}$). Their offset IMBHs are consistent with the formation mechanism described in channel 1. The MBH in AT2024tvd, with a much smaller offset of 0.8 kpc, could have originated through either channel 1 or channel 2. Channel 3 can be ruled out, as bright radio emission is detected from both the TDE and the host galaxy nucleus, indicating the presence of an AGN at the galactic center [222].

Systematically identifying a sample of offset TDEs opens avenues for addressing key questions in astrophysics that AGN-based studies cannot fully answer, such as assessing the role of galaxy mergers in seeding off-nuclear wanderers (Figure 18, left panel) and constraining GW kick velocities across the galaxy population (Figure 18, right panel).

With current and upcoming wide-field optical time-domain surveys such as ZTF (2018–), LS4 (2025–; [144]), and LSST (2025–; [103]), the Roman Space Telescope (2027–), the number of off-nuclear TDE candidates is expected to grow significantly. These surveys provide real-time public alerts that enable rapid identification and follow-up observations.

However, to firmly establish a transient as an off-nuclear TDE, we must confirm that its X-ray emission is spatially offset from the nucleus of the host galaxy. Chandra can do this for the nearest events at $z \lesssim 0.05$.

Compared to Chandra, AXIS has significantly better sensitivity, a wider field of view, and a more uniform off-axis PSF. Therefore, AXIS probes a larger volume, and for each observation, it has sufficient sources for registration, reducing systematic astrometric uncertainty. We anticipate that AXIS will play a crucial role in confirming and characterizing the off-nuclear TDE candidates.

**Exposure time (ks):** 1-10 for each trigger with 100 ks total.

**Observing description:** Targets will be triggered by ToO programs. Exposure is about 1-10 ks depending on the xdistance: 1 ks for $z = 0.05$; 10 ks for $z = 0.15$. Angular resolution is the most critical.

**Joint Observations and synergies with other observatories in the 2030s:** LSST, Roman.

**Special requirements:** A response time of within one week.

*9. Jetted Tidal Disruption Events*

**First Author:** Igor Andreoni (UNC Chapel Hill, igor.andreoni@unc.edu)

**Co-authors:** Jonathan Carney (University of North Carolina), Brad Cenko (NASA/GSFC), Hannah Dykaar (McGill), Daryl Haggard (McGill), Brendan O'Connor (Carnegie Mellon University), Dheeraj Pasham (MIT), Lauren Rhodes (McGill), Robert Stein (University of Maryland), Yuhan Yao (UC Berkeley)

**Abstract:** When a star strays too close to a supermassive black hole, the gravitational forces cause the star to be pulled apart, and some of the material is lost into space, but about half of the stellar mass accretes onto the black hole. These tidal disruption events (TDEs) are typically observed in the X-ray, optical, and radio bands. A rare and exceptionally energetic subset of TDEs exists, which occur when, during the stellar destruction and subsequent accretion of debris onto the black hole, collimated streams of matter are expelled at velocities close to the speed of light. Jetted TDEs provide a unique opportunity for astronomers to investigate the processes of accretion and jet formation out to cosmological distances. AXIS will be key to unveiling a large population of these rare transients thanks to its unique depth and angular resolution capabilities. First, AXIS will pinpoint the locations of promising candidates identified by X-ray and gamma-ray observatories, providing crucial information such as confident host galaxy associations and spatial coincidences with the galaxy nucleus. Second, AXIS monitoring of the X-ray light curve will enable observations of the accretion mode transition from a super-Eddington to a sub-Eddington state, which can be used as an independent constraint on the black hole mass. AXIS will therefore be transformative for the study of massive black holes and accretion across cosmic time.

**Science:** The tidal disruption and subsequent accretion of a star that wanders too close to a massive black hole can illuminate the electromagnetic spectrum. These phenomena provide rare opportunities to study massive black holes in distant galaxies that would otherwise be dormant. Tidal disruption events (TDEs) were first proposed by theorists but have now been established as a class of transients observed primarily in X-ray, radio, and optical bands. For a recent review, see e.g. [74].

In some rare instances, TDEs can launch jets of matter at velocities approaching the speed of light. If the jet is aligned with our line of sight, its overall brightness is Doppler boosted by several orders of magnitude. As such, relativistic or "jetted" TDEs can be found at cosmological distances ($z > 1$), as opposed to the more common types of thermal TDEs, which are typically found in the nearby universe. They are instrumental in inferring, for instance, the black hole mass, magnetic field, and spin. However, only a handful of jetted TDEs have been discovered to date. It was estimated that only $\sim$1% of TDEs launch jets [14,206] and only a small fraction of those will be favorably aligned with the line of sight.

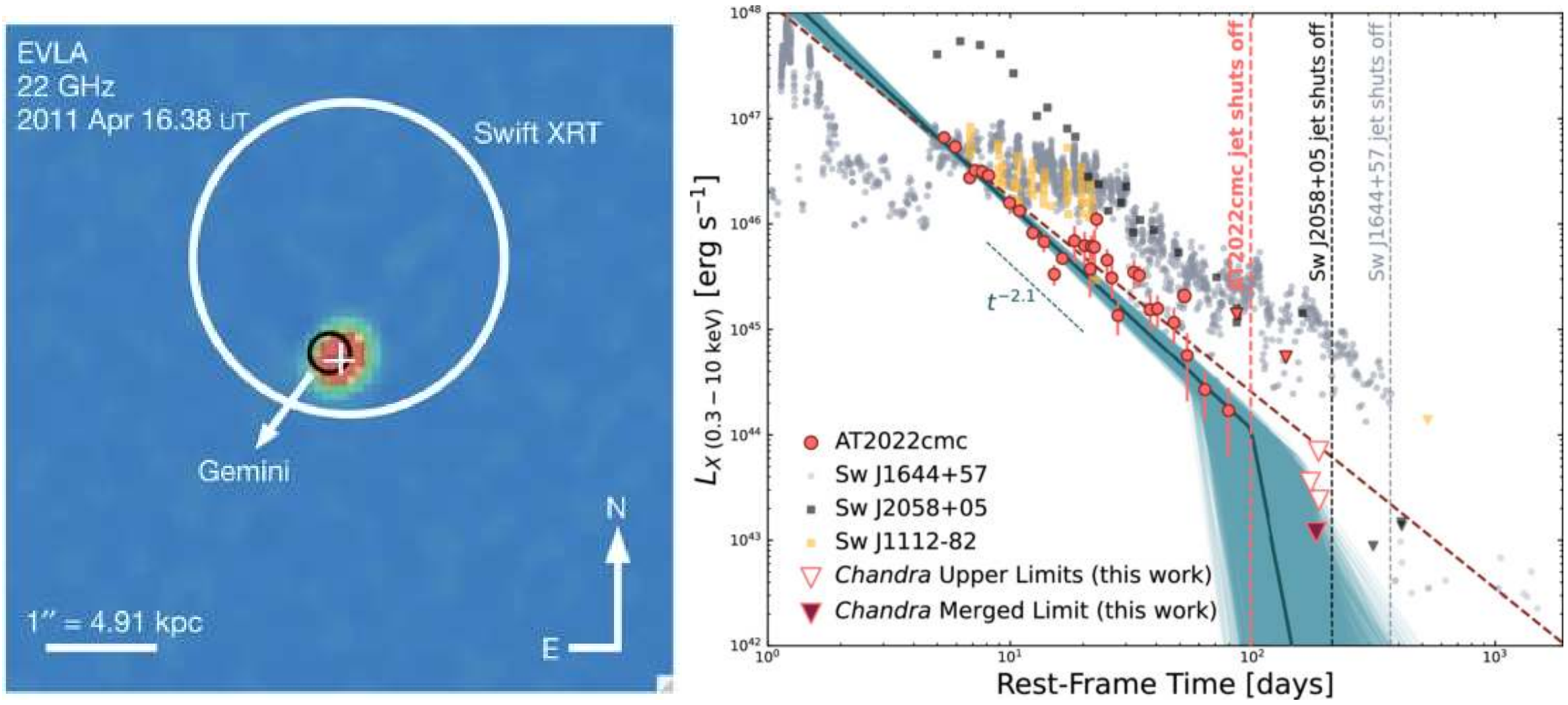

Figure 19 *Left:* Localization of the jetted TDE Swift J1644+57. The Swift/XRT instrument found a bright source, which was then associated with the center of a galaxy by deep multi-wavelength follow-up observations with the EVLA (figure from [225]). AXIS will be able to rapidly identify the position of the transient with sub-arcsecond precision within hours of discovery. *Right:* The jet shut-off of the optically-discovered jetted TDE AT2022cmc observed by Chandra, which enabled the black hole mass to be constrained to $\lesssim$ a few $10^5$ solar masses (figure from [61]).

Sensitive X-ray and gamma-ray wide-field monitors can detect jetted TDE candidates in the form of long-duration transients, with a much more extended and brighter emission than typical gamma-ray bursts. However, their ability to localize the transient typically ranges from a few arcseconds to arcminutes. An optical or radio transient counterpart must typically be identified to make a confident association between the X-ray source and its host galaxy (Figure 19, left). On the other hand, AXIS will be able to observe the source within a few hours of the initial detection and promptly pinpoint the source to a specific galaxy. This will enable the accurate measurement of the distance and, therefore, the energetics of the event. An enormous isotropic equivalent luminosity, combined with the co-location of the relativistic X-ray transient with the nucleus of the host galaxy, will be the smoking gun enabling the prompt identification of a rare jetted TDE.

Following the initial characterization, we propose using AXIS to monitor jetted TDEs regularly. Past observations of these phenomena suggest that the light curves are typically flat and "bursty" in the first few days/weeks from the event, then they fade following a power-law until the X-ray flux drops suddenly (Figure 19, right). The sensitivity of AXIS, whose effective area is $\sim$ 5-10x that of Chandra, will be key to determining when this occurs. It has been suggested [224] that the X-ray shut-off marks the accretion mode transition from a super-Eddington to a sub-Eddington state, which can be used to constrain the black hole mass (e.g., [61,165,207]) independently of black hole mass-bulge luminosity relations.

In conclusion, AXIS will be transformational for jetted TDE science. The combination of sensitivity and excellent angular resolution will enable us to rapidly ($\sim$hours) identify jetted TDEs that would otherwise go unnoticed out to very high redshifts. AXIS could detect the brightest emission from Sw J1644+57 beyond $z \sim 12$, and the X-ray light curve shut-off out to $z \sim 1.2$ with $\approx$10 ks of exposure time or $z \sim 2$ with 100 ks of exposure time, assuming a luminosity of $L_X \sim 10^{43}$ erg s$^{-1}$ after the drop (Figure 19, right).

**Exposure time (ks):** 1 ks for the initial follow-up and localization, 10-100 ks for the long-term monitoring of confirmed jetted TDEs.

**Observing description:** Observations will be carried out via TOO; therefore, the target coordinates and fluxes are not known at this time. Promising targets for AXIS follow-up will be found by gamma-ray or X-ray wide-field monitors. AXIS observations of each target will be performed in two phases:

1. Rapid TOO, possibly within a few hours from trigger. These observations will likely require short exposure times of $\sim$1 ks, and will be critical to localize the X-ray transient precisely. This is particularly important, especially when an optical counterpart cannot be found due to sun constraints or because it is dimmed by dust (for instance, Sw J1644+57 showed a dense circumnuclear medium and high extinction, which would make the detection of an optical signal challenging at large distances). AXIS localization will enable host association, determination of the position of the transient within the host, distance measurement via spectroscopic or photometric redshift, and therefore a rapid estimate of the X-ray luminosity. These early measurements will prompt extensive multi-wavelength follow-up with other facilities.
2. Long-term monitoring to determine the overall trend of the light curve and the time at which the flux drops off (see above). The exposure time can be adjusted based on previous measurements, likely several ks in the first months, and then increased up to $\sim$10 ks. When the observations result in a non-detection, a deep exposure from $\sim$10 ks up to 100 ks may be required to determine the amplitude of the light curve change. We suggest monitoring the light curve evolution with AXIS every 2-3 weeks, unless other telescopes are available and the source is bright enough until the shut-off time.

**Joint Observations and synergies with other observatories in the 2030s:** Synergistic observations will be of paramount importance to get a multi-wavelength description of new jetted TDEs.

**Special requirements:** ToO $\lesssim$ 2 hr to prompt further multi-wavelength follow-up Monitoring with a cadence of 2-3 weeks to find the X-ray light curve drop-off time.

*10. Quasi-periodic eruptions - Rates and Discoveries*

**First Author:** Riccardo Arcodia (MIT, rarcodia@mit.edu)

**Co-authors:** David Bogensberger (UMich), Matteo Bonetti (UniMib), Joheen Chakraborty (MIT), Alessia Franchini (Uni. Zuerich), Margherita Giustini (CAB), Erin Kara (MIT), Giovanni Miniutti (CAB), Scott C. Noble (NASA/GSFC), Jeremy D. Schnittman (NASA/GSFC), Alberto Sesana (UniMib), Ira Thorpe (NASA/GSFC)

**Abstract:** Quasi-periodic eruptions (QPEs) are repeated soft X-ray flares from the nuclei of galaxies. They are thought to be the electromagnetic counterpart, or precursor, of extreme mass ratio inspirals (EMRIs). These peculiar flares have so far manifested only in the soft X-rays; thus, there is currently no other channel to discover them. Here, we discuss the prospects of discovering QPEs with AXIS with a dedicated wide-area survey, given the current estimate of QPE rates from eROSITA. While current X-ray telescopes may find more individual sources through targeted follow-up observations, only dedicated blind and systematic X-ray wide-area surveys for QPEs allow us to estimate their intrinsic volumetric rates. Thus, performing a wide-area survey with AXIS is the only prospect of refining the current uncertain QPE rates before or during LISA. If their association with extreme mass ratio inspirals is confirmed, this would be the only data-driven constraint on the expected number of LISA EMRI events.

**Science:** We adopted the QPE rate inferred from the eROSITA QPEs [16], extrapolating the luminosity function (constrained down to $\log(L_X) \sim 41.7$ by eROSITA) down to a peak $\log(L_X) \sim 40.5$ (breaking the single power-law artificially to a constant value below $\log(L_X) \sim 41.5$). We chose this lower value of $\log(L_X) \sim 40.5$ since it's observable by AXIS within 2 ks at distances compatible with known QPEs, without going to lower values where the contrast with star formation may reach unity. To infer the number of QPEs discovered by AXIS, we sample this extended luminosity function in bins of 0.6 dex spanning peak $\log(L_X) \sim 40.5$-45. We integrate the luminosity function in each bin and multiply it by the maximum volume spanned by the central luminosity of the bin, set by the exposure of the survey snapshot and by AXIS's sensitivity. QPEs are repeating transients, and to identify them (i.e., distinguish them from any constant soft X-ray source, or a variable but non-repeating soft X-ray source), one needs to observe the same galaxy at least twice, ideally more. In fact, a single (reasonably short) few-ks snapshot cannot resolve individual flares in most cases, as eruptions may last up to one day or more [90]. Thus, QPEs or QPE candidates will be identified by significant high-amplitude variability between visits. This strategy implies that an ideal QPE survey will observe the same regions of the sky multiple times, and that the sensitivity required to detect and identify a QPE source is not only the instrumental sensitivity, but also that of a highly significant detection above background. We adopt a $5\sigma$ sensitivity for soft thermal sources, given a specific exposure, as the relevant sensitivity for QPE peak detection, allowing us to identify significant differences between a QPE detection and a non-detection across multiple visits.

To design a QPE survey, we note that there are disadvantages in having a single deep exposure rather than many shallower ones. For instance, this is due to the above argument regarding the repeating nature of QPEs, as well as their duty cycle of $\approx$10% (estimated by taking the near-peak duration as the average duration between 50% peak flux points). Thus, most of the QPE sources in single deep (e.g., 10 ks) exposures will not be identified unless QPEs last $< 10$ ks (only true for $\sim$half of the known QPEs so far) and are entirely within the observation (which occurs $\sim$ 10% of the time). Instead, while a single 1 ks exposure yields a $N_{QPE}$ / deg$^2$ which is less than 1/10 of the 10 ks numbers, repeating 10 times over the same area will cancel out the duty cycle and guarantee that most/all QPE sources in the volume will be identified, even if the volume is smaller. Furthermore, a wide-area survey made of several short exposures, rather than single deeper ones, may also serve other science cases for TDAMM (e.g., TDEs and others, as well as variable AGN). Finally, we note that it is not worthwhile to perform more snapshots per location than $\sim$10, otherwise one will identify more eruptions of the same source rather than discovering new QPEs in the volume. Hence, we suggest here performing a wide-area survey that visits the same regions of the sky several times, with the total area (and thus exposure) determining the number of QPE discoveries.

Figure 20 shows the number of QPEs discovered and identified (accounting for the sampling probability) as a function of area for two possible surveys: one corresponds to $10 \times 1$ ks-long visits (green line), the other to $5 \times 2$ ks-long visits (blue line and $1\sigma$ contours). The latter (through the deeper exposure) finds slightly more sources per total exposure, even if penalized by half the sampling attempts, although this gain is small.

The total survey exposure also increases with the $x$-axis, as approx $\sim 8 \times t_{\rm snapshot}$ / deg$^2$, using a 0.125 deg$^2$ FoV. Quantitatively, in order to find $\sim$20 QPEs with a $10 \times 1$ ks survey, the total exposure would be $\sim$7.5 Ms over $\sim$94 deg$^2$, while for a $5 \times 2$ ks survey the exposure would be lower, $\sim$7 Ms over $\sim$87 deg$^2$. As they are very similar, we invite the broader TDAMM community to consider their ideal depth and cadence to ensure this effort serves the widest range of science cases. The $5 \times 2$ ks ($10 \times 1$ ks) survey would return to the same spot every $\sim$16d ($\sim$9d), which is, for instance a great cadence to find and identify X-ray TDEs and our proposed survey would discover up to tens [78,192]. Finally, the location of this survey can be chosen to complement existing and planned multi-wavelength coverage, serving as a legacy AXIS field.

**Exposure time (ks):** Depends on final survey design, of order 7 Ms.

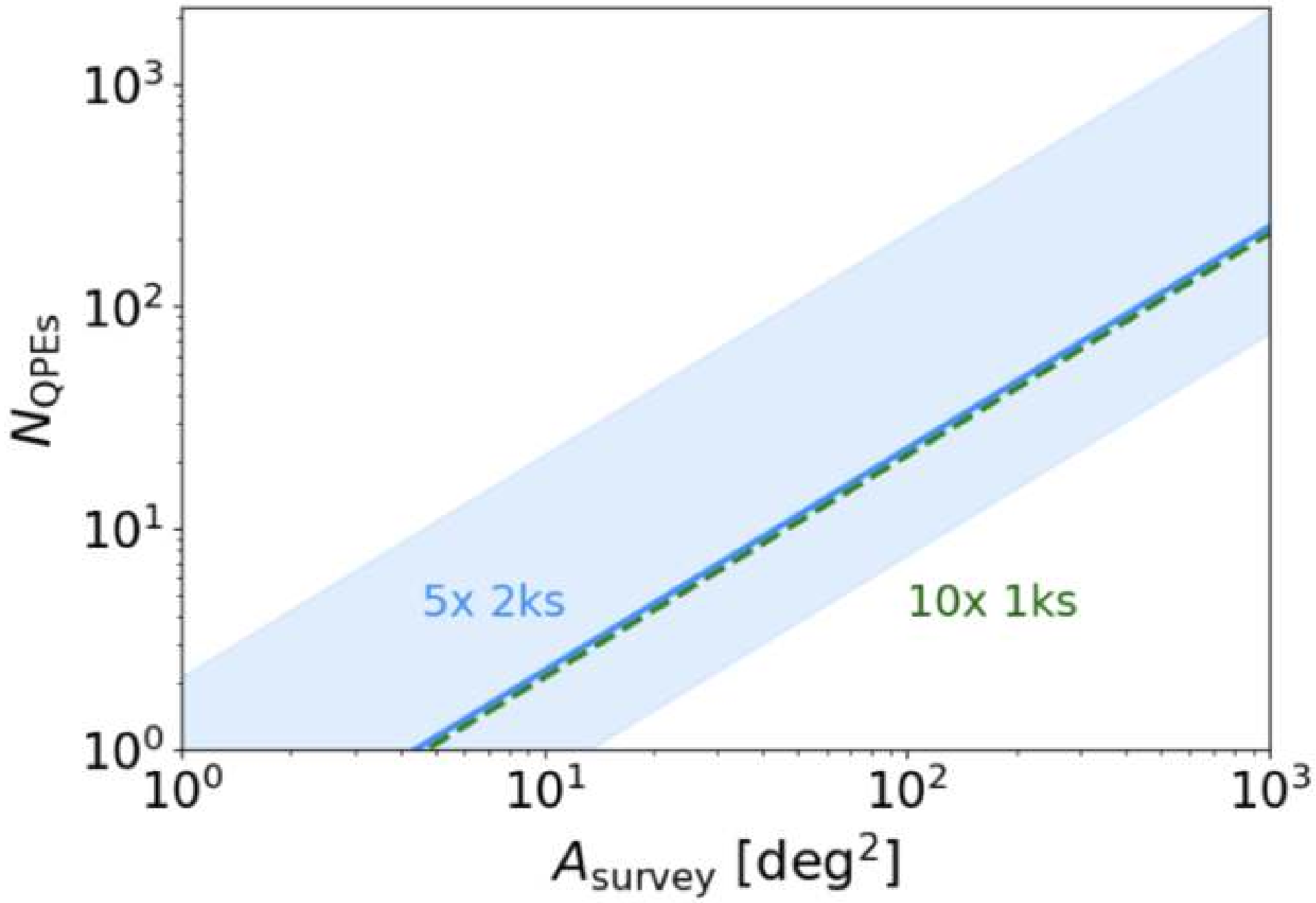

Figure 20 **AXIS has sufficient sensitivity and field-of-view to conduct a transformative survey to discover new QPEs.** A cadenced survey (e.g., $10 \times 1$ ks) over $\sim$94 deg$^2$ would uncover a sample of $\sim 20$ new QPEs, while also enabling discovery of other time-domain sources and serving as an X-ray complement to Rubin Deep-Drilling Fields and Roman survey fields.

**Observing description:** Raster scan to cover the full survey area, repeating several times (see text).

**Joint Observations and synergies with other observatories in the 2030s:** LISA, Rubin, Roman.

**Special requirements:** Rapid slew and settle for efficient observations.

*11. Quasi-periodic eruptions – Monitoring and follow-up*

**First Author:** Joheen Chakraborty (MIT, joheen@mit.edu)

**Co-authors:** Riccardo Arcodia (MIT), David Bogensberger (UMich), Matteo Bonetti (UniMib), Alessia Franchini (Uni. Zuerich), Margherita Giustini (CAB), Erin Kara (MIT), Giovanni Miniutti (CAB), Scott C. Noble (NASA/GSFC), Jeremy D. Schnittman (NASA/GSFC), Alberto Sesana (UniMib), Ira Thorpe (NASA/GSFC)

**Abstract:** Quasi-periodic eruptions (QPEs) are recurring soft X-ray flares from the nuclei of some nearby low-mass galaxies. They are thought to be the electromagnetic counterparts of extreme-mass ratio inspirals (EMRIs). As QPEs are an X-ray-only phenomenon thus far, next-generation X-ray facilities are necessary to extend our reach toward deeper model constraints, wider population studies, and a more complete understanding of their behavior. Here, we discuss how the unique capabilities of AXIS will be well-suited for monitoring programs of QPEs, enabling new physical tests of their models via X-ray timing and spectroscopy. AXIS will also be a primary discovery engine for QPEs. By taking advantage of the now-established QPE/TDE connection, dedicated X-ray monitoring of Rubin-discovered TDEs may be expected to discover $\sim$10 new QPEs per year.

**Science:**

*TDE Follow-Up:* At least some QPEs occur in the aftermath of TDEs discovered in X-ray [40,145] and optical surveys [42,157,173], with an estimated $9^{+9}_{-5}$% of optical TDEs showing QPEs within 5 years [42]. AXIS provides an ideal X-ray synergy for the upcoming suite of time-domain surveys. For instance, Rubin is expected to accelerate the TDE discovery rate up to $\sim$1000 per year [38], providing a rich parent sample for dedicated long-term monitoring and QPE discovery. With a $\sim$20 ks observation, AXIS will reach a sensitivity of $\sim 10^{-16}\,\mathrm{erg\,cm^{-2}\,s^{-1}}$, extending the reach of QPE detectability (for an assumed peak luminosity of $10^{43}\,\mathrm{erg\,s^{-1}}$) to $z \sim 3$. This is well-matched to the volume probed by Rubin TDEs as well as LISA-detectable EMRIs, and a factor $>$200,000$\times$ the volume currently accessible by X-ray observatories. By designating a small percentage of all TDEs for AXIS long-term follow-up – say 100 TDEs/year – the expected yield would be $\sim$9 new QPEs in known TDEs per year. This efficacy may be enhanced by selecting for specific TDE properties, and the coming years will likely reveal which sub-types are particularly efficient in forming QPEs (e.g., recent work suggests optical/IR coronal line emission to be associated with QPEs at an enhanced rate [42,164]. This will quickly result in an appreciable sample of QPEs in known TDEs suitable for a systematic, population-level study to determine any relationships between the precursor TDE properties and emergent QPE flares.

*Short-term Timing:* The L2 orbit will enable long, uninterrupted exposures spanning several hundred ks, which is a totally new capability compared to current X-ray telescopes. This allows the detection of dozens of consecutive bursts, without any ambiguity on missed eruptions, providing an ideal dataset for QPE timing analysis. The QPE timing method, especially for short-period sources/low-mass SMBHs/higher-eccentricity orbiters, has the prospect to provide dynamical SMBH mass and spin estimates given a sufficiently dense and long-baseline monitoring program [146,230]. Moreover, recent detections of super-periodic modulation in QPE timings have suggested rapid disk precession and/or sub-milliparsec SMBH binaries as additional components which may affect the orbiter-SMBH system [41,70,146]. Thus far, we have been limited to relatively short timing baselines, making an unambiguous interpretation difficult; with the AXIS capability of mapping several consecutive super-periods, we may uniquely disentangle each of the oscillatory modes affecting QPE timings. With an assumed AXIS observation of 400 ks uninterrupted, compared to the current XMM capability of 100 ks, we find the measurement error on EMRI eccentricity reduces by 60%, and the error on disk precession timescale reduces by 82%.

*Long-term Timing:* The extended timing baselines achievable by AXIS will also be key in measuring orbital decay in QPEs driven by gravitational-wave emission and dissipative orbiter-disk interactions [17,127]. The significant quasi-periodicity in QPEs means timing programs are confronted with an ambiguity in uniquely assigning burst numbers to determine any secular drift. Long baselines also enable the disentangling of whether apparent orbital changes are monotonic or oscillatory, thereby differentiating the effects of inspiral and precession. This problem is best solved by continuous viewing windows spanning as long as possible; for a handful of stable QPE sources, long-baseline AXIS exposures can yield well-constrained $\dot{P}$ measurements. These measurements can then jointly constrain the properties of the orbiter and the accretion disk (e.g., its surface density profile), providing a new probe of the SMBH environment. This measurement will also be relevant for the characteristic EMRI lifetime in the presence of environmental gas, which remains an uncertainty for LISA rate estimates.

*Spectroscopy:* The soft X-ray sensitivity and spectral resolution of AXIS will also be well-suited for deep studies of the QPE emission region. Current X-ray facilities have the SNR to detect only the continuum blackbody emission in most QPEs; there are a few exceptions for QPEs with exceptional brightness/long timescales (ZTF19acnskyy [43,90]) or the most extensive X-ray coverage (GSN 069 [119,145]), in which outflow-like features have been robustly detected upon the continuum. The physical parameters and

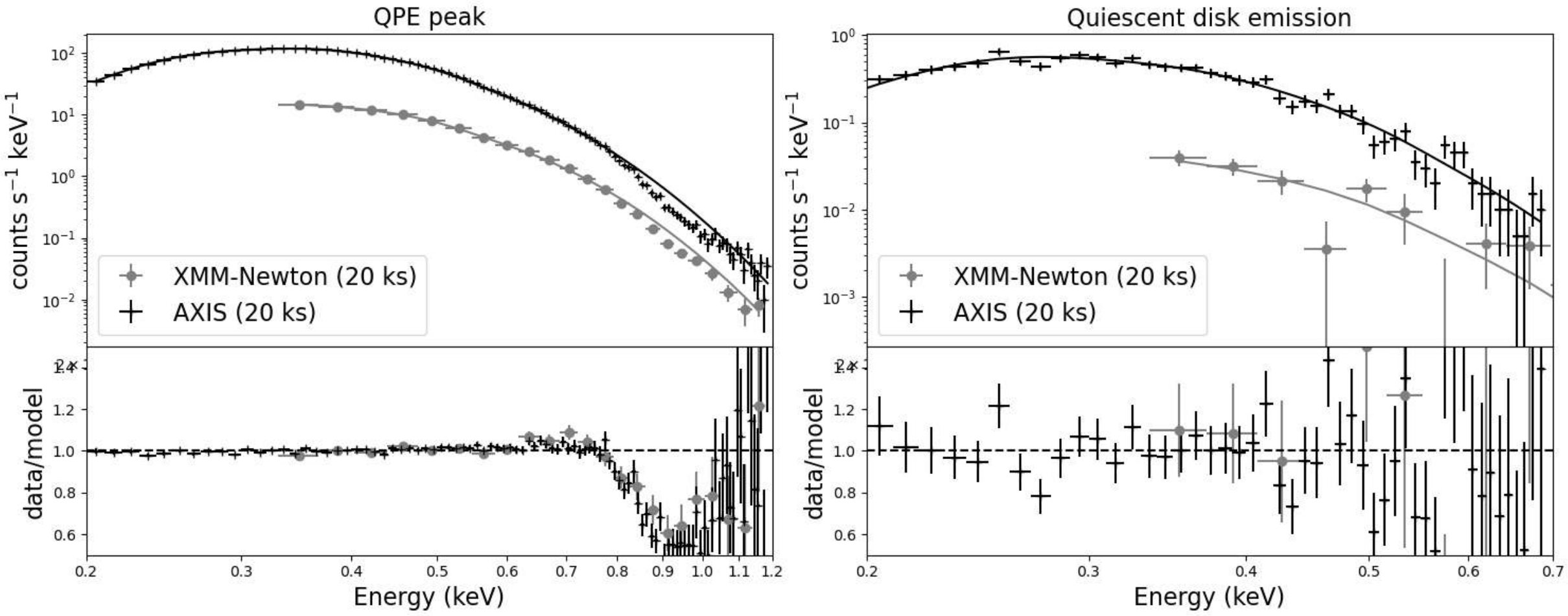

Figure 21 **Left:** QPE peak spectrum of ZTF19acnskyy observed with XMM (gray squares) and a simulated 20 ks AXIS exposure (black crosses). With its superior effective area and count rate, AXIS will achieve orders-of-magnitude higher SNR, enabling time-resolved study of outflow signatures and detection of discrete spectral features currently out of reach for current generation observatories. **Right:** quiescence emission (between QPEs) indicating a compact, bare accretion disk. The XMM data are drawn from [90].

time-variability of these spectral features will provide the best constraints on the evolution of the ejecta powering the QPE emission, diagnosing its bulk velocity, density, and ionization state. These properties can distinguish between models for the QPE emission, which is currently thought to be shock-driven, but still faces significant uncertainty in whether the energetics and timescales can be explained by the theoretical models of their accretion disks. The superior effective area of AXIS will significantly enhance the SNR of QPE spectra, allowing for deeper, time-resolved looks at the physical properties of the emitting material (Fig. 21).

*Quiescent Emission:* AXIS will also enable the study of the quiescent emission between the QPE flares, which is a key component in developing a complete picture of the eruptions. All viable QPE models invoke the accretion disk in some capacity, such as powering QPEs directly via disk instabilities [163], suggesting the collisions drive enhanced mass-transfer in the disk which are observed as QPEs [128], or proposing that QPEs are powered by shock-heated material ejected from the disk during star-disk collisions [126,214]. The properties of the direct accretion disk emission may thus distinguish between models via, e.g., whether oscillations in the quiescent emission are detected on timescales comparable to the eruptions, whether secular evolution is detected in the disk temperature/luminosity concordant with the flare evolution, and whether the radial scale inferred from the accretion disk is compatible with an orbiter-disk collisions powering the bursts. However, in most QPEs, the quiescence is too faint to be detectable by current-generation X-ray facilities, as it is typically 10–50$\times$ fainter than the peak. With its 70–100$\times$ better sensitivity compared to Swift/NICER, AXIS will extend our reach to the accretion disk in all known QPEs (Fig. 21).

**Exposure time (ks): (a)** 20 ks per TDE to constrain QPE emergence; **(b)** depends on source-specific properties, likely $\mathcal{O}(100 - 1000\ \mathrm{ks})$ per source with 2 Ms for the total survey.

**Observing description: (a)** Follow-up of a parent sample of $\mathcal{O}(100)$ Rubin optical TDEs to constrain QPE-TDE connection; **(b)** monitoring of known QPE sources to measure long-term evolution, spectral properties, and timing modulation effects

**Joint Observations and synergies with other observatories in the 2030s:** LISA, Rubin, Roman

**Special requirements: (a-b)** Repeated monitoring of known TDEs and QPEs out to late times

**c. Neutron Star and Black Hole X-ray Binary Physics**

*12. Transient behavior of neutron star X-ray binaries*

**First Author**: Pragati Pradhan (Embry-Riddle Aeronautical University, Prescott, AZ, pradhanp@erau.edu)

**Co-authors:** Enrico Bozzo (ISDC, University of Geneva), Patrizia Romano (INAF–Osservatorio Astronomico di Brera, IT), Alicia Rouco Escorial (ESA/ESAC), Biswajit Paul (Raman Research Institute, Bangalore, India), Antonios Manousakis (SAASST & UoSharjah, UAE)

**Abstract:** With its exceptional sensitivity, high spatial resolution, and flexible scheduling, AXIS will transform our understanding of transient high-mass X-ray binaries (HMXBs), including supergiant fast X-ray transients (SFXTs), Be X-ray binaries (BeXRBs), and symbiotic X-ray binaries (SyXRBs). By capturing rapid outbursts, quiescent states, and state transitions, AXIS will illuminate the physics of wind-fed accretion, magnetospheric gating, and disk instabilities. It will track flaring evolution in SFXTs, map circumstellar disk structures in BeXRBs, and characterize accretion variability in SyXRBs. Additionally, AXIS will detect and monitor faint and transient X-ray sources, including low-luminosity XRBs along the Galactic plane and previously uncharacterized populations in the local group. Its multi-epoch observations will probe long-term variability, enabling the identification of dormant or weakly accreting compact objects. AXIS will also play a crucial role in identifying electromagnetic counterparts to gravitational wave sources, particularly short-period XRBs that will serve as bright LISA targets. These observations will provide key constraints on mass transfer mechanisms, outburst recurrence, and the influence of compact objects on their binary environments, advancing our understanding of accretion physics and compact object evolution. The extreme variability of these HMXBs is evident in studies of the Milky Way (MW) X-ray binaries, as well as the Large Magellanic Cloud (LMC) and Small Magellanic Cloud (SMC), which have also revealed population-level trends shaped by metallicity and age. AXIS will provide new insights into massive star formation under conditions similar to those in higher-redshift galaxies.

**Science:**

High-mass X-ray binaries are broadly classified based on the nature of their companion star. Be X-ray binaries consist of a neutron star orbiting a Be star, which is typically surrounded by a circumstellar 'decretion' disk. The neutron star accretes material from this disk, leading to distinct outburst behavior. The most common are Type I outbursts, which occur periodically at periastron due to enhanced accretion and last for a few days to weeks. In contrast, Type II outbursts are irregular, significantly more luminous (by an order of magnitude or more), and can persist for several weeks. See Fig. 22 for an example of two BeXRBs. These giant outbursts may be tied to changes in the structure of the Be star's disk, although the precise mechanisms remain debated. BeXRBs exhibit remarkable variability, yet the cause of the drastic differences in aperiodic X-ray behavior among systems remains poorly understood. Supergiant X-ray binaries (SGXBs) feature a neutron star accreting from the dense stellar wind of an OB supergiant. Within this category, supergiant fast X-ray transients) represent a distinct subclass, characterized by sporadic, few hours-long outbursts that can reach peak luminosities comparable to classical SGXBs and spend long intervals of time in faint quiescent states with luminosities down to $10^{31}$ erg/s. Discovered in 2005 with INTEGRAL, SFXTs share many properties with classical SGXBs, including similar companion stars and orbital period distributions. However, the SFXTs' total dynamic range between outbursts and quiescence usually achieves factors of 105-106, to be compared with the typical factors of 20-100 of the classical

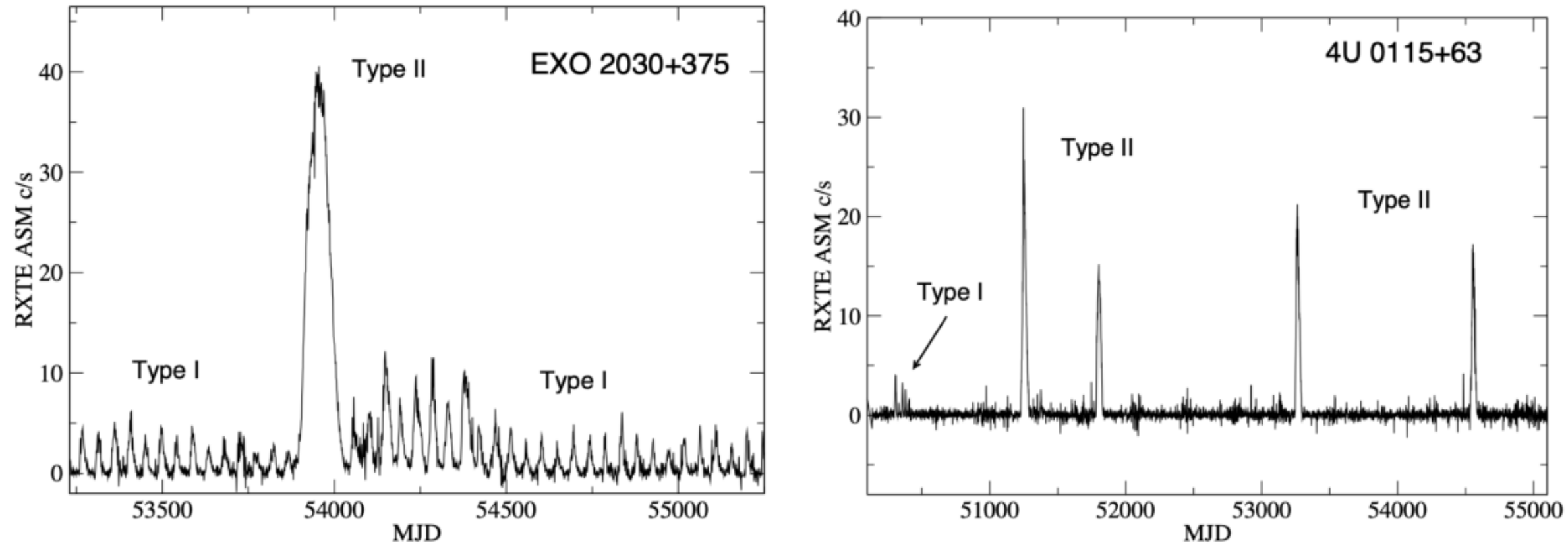

Figure 22 Examples for Type I and Type II bursts for two BeXRBs. Notice the regularity and magnitude of Type I versus Type II outbursts. Image from [177]

systems. See Fig. 23 for lightcurves of IGR J17544-2619, the SFXT prototype, and Vela X-1, the classical SGXB prototype.

The outbursts and flares of wind-fed SGXBs, including classical systems and SFXTs, are widely thought to be triggered by the accretion of dense clumps from the wind of OB supergiants (being inherently inhomogeneous) onto neutron stars, combined with the effect of magnetic and centrifugal gating associated with the spin and magnetic field of the compact object. As of today, there is no convincing evidence of a dichotomy in the properties of massive star winds between classical systems and SFXTs [170], and thus it is likely that the more extreme behavior of SFXTs is related to the magnetic and spin properties of the NS.

Observations have shown that, in both classical and SFXTs, the stellar wind clumps partially obscure the compact objects when passing close to them. This leads to remarkable variations in absorption column density during X-ray observations [32,174] while variations in the shape of the X-ray continuum emission are generally interpreted as being due to the switch between different accretion modes. These are regulated by the different physical processes at work, which can enhance or depress the instantaneous mass accretion rate by exploiting the (possible) NS fast rotation and/or its (possibly) strong magnetic field. Accurate measurements of the absorption column densities and continuum shape across flares and outbursts of classical SGXBs and SFXTs are thus the key to understanding the physics of wind-fed systems, and more generally utilize these objects to directly probe in situ the properties of massive star winds. There is a widely spread interest in understanding better the characteristics of clumps in massive star winds, as on these clumps depend on the effective mass loss rate of massive stars, which in turn affects, e.g., the chemical evolution of Galaxies, that of the Universe on a larger scale, as well as the production of progenitors of specific gravitational wave sources. So far, the literature has largely shown that the most effective machine to probe the spectral variation across flares and outbursts of SGXBs is XMM-Newton, thanks to its unique combination of large effective area in the soft X-ray domain, the sensitivity to faint objects, and the good energy resolution (probing changes, e.g., in the centroid energy of iron lines).

Symbiotic X-ray binaries (SyXRBs) represent a distinct and less common category, consisting of a neutron star accreting from the wind of an evolved, late-type giant. These systems differ significantly from both BeXRBs and SGXBs, as their donor stars are much cooler (M or K spectral types), leading to different wind emission mechanisms (likely dust-driven, rather than line-driven) and distinct accretion regimes. SyXRBs exhibit variability spanning a wide range of timescales, from short-term flickering to long-term outbursts lasting years. The nature of this variability is not yet fully understood, but it is likely influenced by the complex interaction between the neutron star and the slow, dense stellar wind

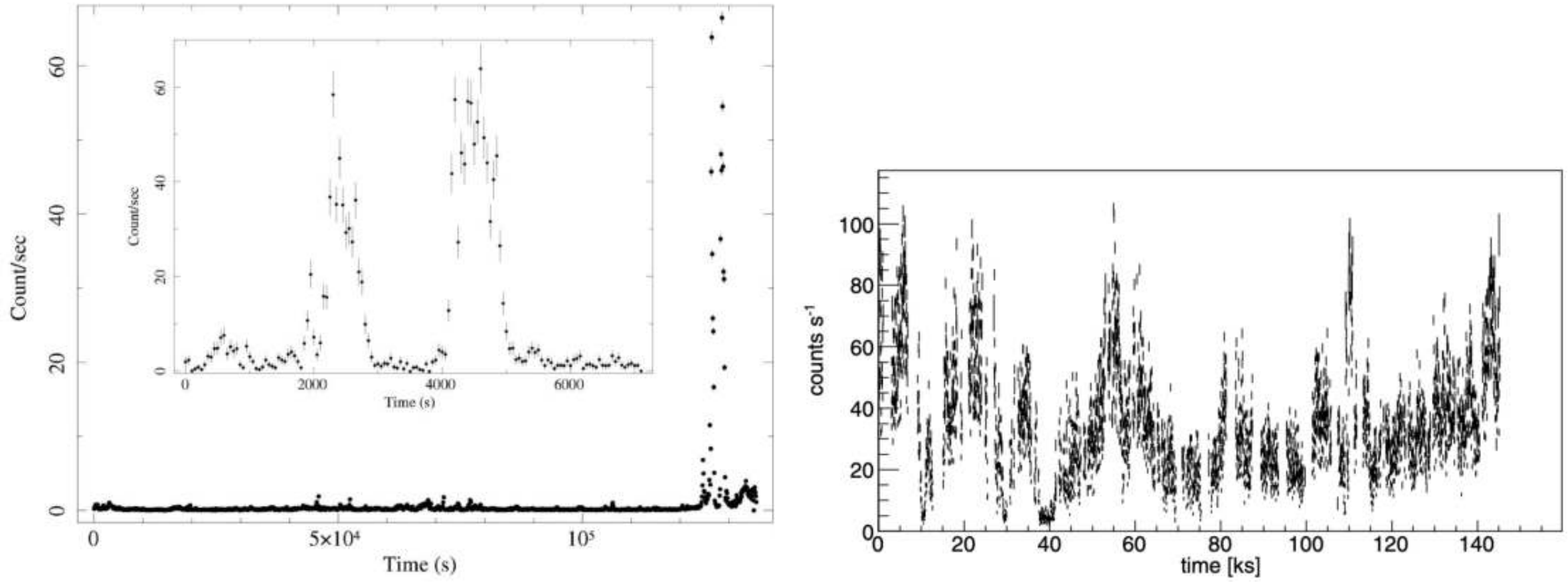

Figure 23 Pointed *XMM* observations of the erratic transient SFXT IGR J17544–2619 [33] on the left, and a classical SGXB Vela X-1 with *Suzaku*/XIS on the right [160]. Note the long period of quiescence and dynamic range of luminosity on the left versus the persistent (yet variable) behaviour of Vela X-1 on the right. The start times for both are arbitrary.

of the giant companion. Some SyXRBs, such as GX 1+4, exhibit strong pulsations and torque reversals, suggesting that wind properties and magnetospheric interactions play a crucial role in regulating their accretion. See Fig. 24 for X-ray variability of SyXRB, GX 1+4.

AXIS offers key capabilities that will significantly enhance our understanding of HMXBs, particularly in variability studies. Given its larger area, fine energy resolution, low background, and fast repointing capabilities, AXIS could take the investigation of wind-fed systems to the next level. It will indeed be possible to follow in detail even the faintest flares and outbursts from classical SGXBs to SFXTs, performing a time-resolved spectroscopy with the required accuracy down to the quiescent level. This could potentially reveal the yet poorly understood triggering conditions for the onset of magnetic and centrifugal barriers, which will inevitably be the cause of the drop in luminosities of the SFXTs, often below $10^{31}$–$10^{32}$ erg/s. The winds in SyXRBs, along with their own mechanisms for triggering the different variability of these sources, could also be similarly explored. Furthermore, high sensitivity surveys at low flux could allow us to widen the number of SGXBs known, as such classes remain largely unexplored with only a few tens of systems known (several being candidates and not yet confirmed). AXIS's superb spatial resolution could also facilitate the association of faint new sources discovered with Optical/IR counterparts, which is a necessary step to identify obscured OB supergiant and firmly establish the classification of newly discovered wildly variable X-ray sources. AXIS's rapid response time will make it easy to track details of the onset and development of BeXRB normal (as well as giant) outbursts. This will allow us to better constrain the accretion dynamics and accretion/decretion disk formation processes. It will also allow us to disentangle the open question about whether these systems enter the propeller regime immediately after the giant outbursts or not, and if this regime exists when it occurs.

By surveying nearby galaxies, AXIS can also identify extra-Galactic BeXRBs and SGXBs, further expanding on the samples of known objects in these classes and refining our understanding of the diverse variability in HMXBs.

**Exposure time (ks):** The sources are generally bright ($L_X \sim 10^{31} - 10^{38}$ erg/s) for SFXTs or BeXRBs and therefore a typical exposure of $\sim 40$ ks per pointing will be enough to constrain the spectral parameters to an accuracy of a few percent. Depending on the state in which it is captured, we can also perform time-resolved spectroscopy and pulse profile analysis for pulsating sources, among other techniques.

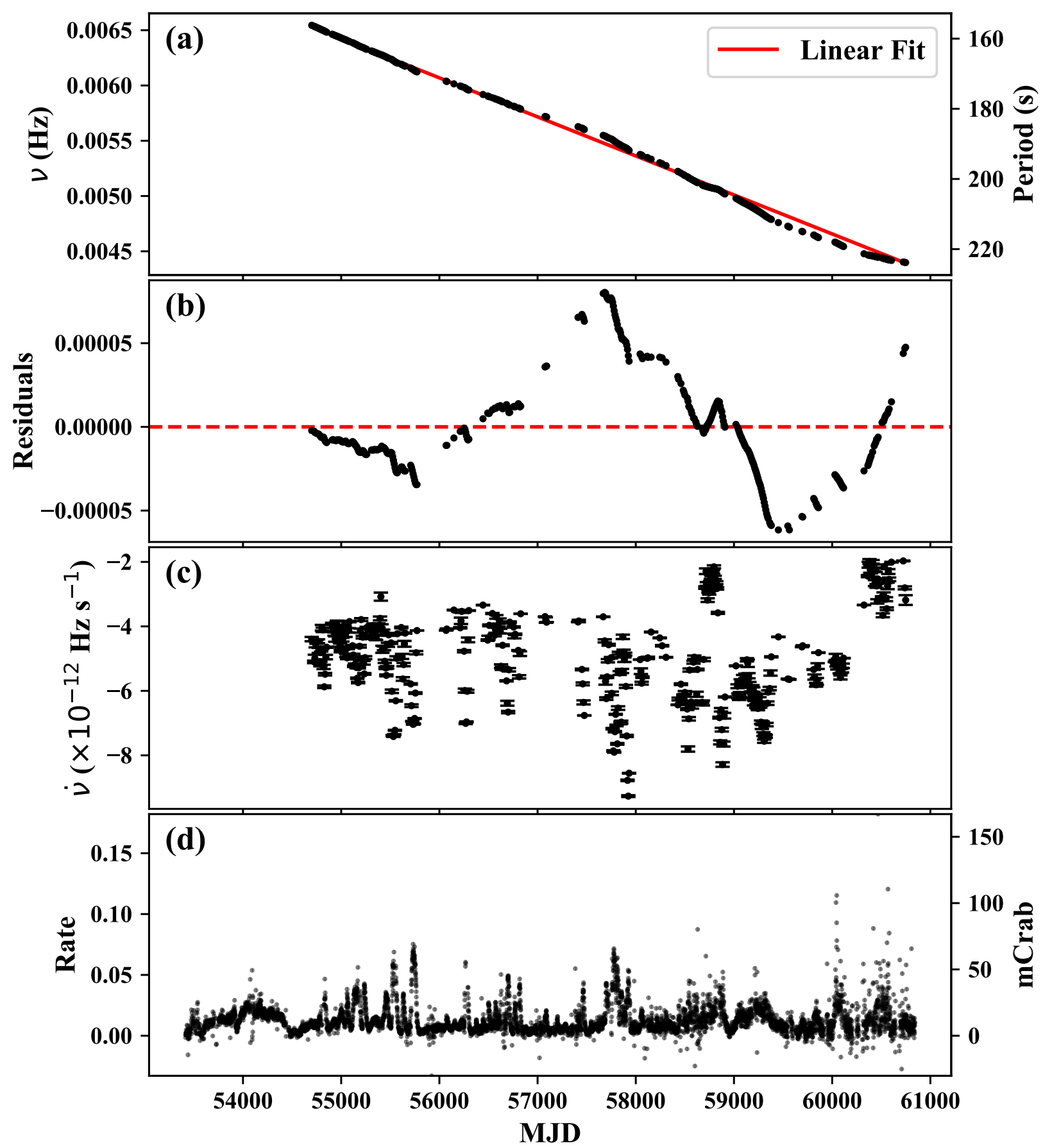

Figure 24 Top: (a) Fermi/GBM monitoring of the spin frequency of SyXRB, GX 1+4 fit with a linear fit with slope as -3.53$\times$ $10^{-7}$ Hz d$^{-1}$ (or, $\dot{P} \sim$ -1.90$\times$ $10^{-7}$ ss$^{-1}$), (b) Residuals that show deviations from a linear trend. (c) Frequency derivative $\dot{\nu}$ (proxy for torque of accreted matter) derived from Fermi GBM observations (d) Swift/BAT light curves binned over 5 days. Note the erratic aperiodic variability in the source. Figure adapted from Sextro et al., in prep.

Historical data show that in a 40-ks-long observation, we typically expect 3-6 flares for SFXTs, several moderate flares from classical SGXBs, and coverage over many of the variability timescales displayed by the SyXRBs.

**Observing description:** The observing strategy consists of three main components:

(i) **(>) 40 ks-long pointed observations:** these data will allow us to detect and study different kinds of X-ray variability in classical SGXBs, SFXTs, and SyXRBs. The primary objective is to utilize the exceptional sensitivity, broad soft X-ray coverage, good energy resolution, and large effective area of AXIS to perform time-resolved spectroscopy across flares and short-lived outbursts, thereby studying the triggering mechanisms and their coupling with various types of stellar winds. Einstein Probe also expands the sample of variable HMXBs beyond the brightest sources, allowing AXIS to study less luminous systems that were previously inaccessible. This will enable us to perform detailed studies, also in the quiescent or low-luminosity regimes of these sources, which are currently limited by the sensitivity of existing instruments.

(ii) **Target of Opportunity (ToO) Observations:** If AXIS responds to transient HMXBs within a matter of hours to a few days, we can probe BeXRB normal/giant outbursts, ensuring detailed spectral and timing coverage of their evolution. The detection of new sources by the SFXT outbursts typically lasts a few hours ($<$ 2 hours), and a swift response time is needed to detect SFXT outbursts.

(iii) **Long-Term Variability Studies:** AXIS will leverage its Galactic Plane Survey and other systematic monitoring programs to investigate the long-term evolution of HMXBs. By capturing variability over extended periods, AXIS will refine our understanding of accretion dynamics, disk evolution, and duty cycles in different HMXB populations. This approach will help identify new transient sources, track secular changes in known systems, and provide a comprehensive view of the diverse behaviors exhibited by HMXBs. Apart from higher sensitivity, soft X-ray coverage, good energy resolution, and large effective area, the wider FoV will help discover more transients (e.g., SFXTs) during the surveys.

**Joint Observations and synergies with other observatories in the 2030s:**

**New Transient Detections with Einstein Probe** The Einstein Probe is rapidly identifying new X-ray transients (e.g., ATel 17068), with a positional accuracy of $\sim$ 5 arc-min. AXIS will provide high-resolution spatial information and simultaneously measure the X-ray spectra of new sources, distinguishing between neutron star/black hole binaries or CVs. By extending studies beyond the brightest, well-known HMXBs, AXIS will help refine the population statistics.

**Polarimetric Measurements with IXPE** AXIS and IXPE will offer complementary views of HMXBs, particularly those with strong magnetic fields such as BeXRBs and supergiant systems. IXPE's polarimetric capabilities will probe the emission process and geometry [68].

**UV Emission from Massive Stars (with UVEX)** Since many HMXBs have massive donor stars with strong outflows, AXIS and UVEX observations will help determine whether UV flares precede or lag behind X-rays. The planned observing strategy for UVEX is to cover all-sky (e.g., arxiv 2111.15608 ). This will provide UV coverage of HMXBs, tracking changes in the UV emission driven by stellar winds from the companion stars.

**NIR for Dusty Sources (with ELT)** The Extremely Large Telescope (ELT) will provide near-infrared (NIR) follow-up of HMXB outbursts, particularly in highly obscured or dusty environments like SyXRBs and IGR J16318-4848 [69]. This is crucial for understanding the stellar wind environment processes of highly reddened systems and embedded HMXBs. By leveraging these multiwavelength and polarimetric capabilities, AXIS will significantly enhance our understanding of HMXB accretion physics, magnetic field interactions, and variability across the electromagnetic spectrum.

### *13. Catching outbursting black hole and neutron star X-ray binaries with AXIS*

**First Author**: Maria Cristina Baglio (INAF-OAB, maria.baglio@inaf.it)

**Co-authors:** Amruta Jaodand (SAO, Harvard & Smithsonian), Francesco Coti Zelati (ICE-CSIC, IEEC), Fiamma Capitanio (INAF-IAPS), Arash Bahramian (Curtin Institute of Radio Astronomy), Andrea Gnarini (Università degli Studi Roma Tre), Fabio La Monaca (INAF-IAPS), Alessandro Di Marco (INAF-IAPS), Mason Ng (McGill University), Sergio Campana (INAF-OAB)

**Abstract:** X-ray binaries (XRBs) are some of the brightest sources in the X-ray sky. They usually appear in the sky as they sporadically brighten by a factor of up to $10^6$ in luminosity, and they are typically detected thanks to regular observations with X-ray all-sky monitors. These periods of enhanced activity are known as outbursts, and they correspond to an increase in the mass accretion rate onto the compact object (a black hole or a neutron star) from the companion star, which transfers mass via an accretion disc. Outbursts can last from weeks up to several months; once they come to an end, systems transition back to quiescence, becoming faint (with X-ray luminosities $< 10^{33}$ erg s$^{-1}$) and feeding at low accretion

rates. At these very low luminosities, only the most sensitive, state-of-the-art X-ray satellites can detect an XRB. The very first phases of XRB outbursts remain poorly studied due to the limited sensitivity of current X-ray monitoring instruments; yet, identifying their onset is key to understanding their triggers. Optical monitoring has proven to be a powerful tool in this context, as it is predicted that outbursts should first be observed in the optical band before manifesting in X-rays. Therefore, accurately measuring the optical-to-X-ray delay is essential to understanding the physics of the outburst onset. AXIS will provide an unprecedented opportunity to study the earliest moments of XRB outbursts. Its rapid response time for Target of Opportunity observations will be crucial to observe XRBs soon after the first optical detection. AXIS's exceptional sensitivity, $\sim 20$ times better than Swift/XRT, allows reaching a flux limit of $\sim 10^{-14}\,\mathrm{erg\,cm^{-2}\,s^{-1}}$ even in short 10 ks exposures, enabling precise monitoring of the outburst's early evolution with unprecedented accuracy. By precisely measuring the optical-to-X-ray delay, AXIS will provide critical constraints on the speed at which heating waves propagate through the accretion disc, helping us distinguish between different outburst triggering mechanisms. In addition, the excellent angular resolution of AXIS (1.5″ for on-axis observations) will help resolve XRBs in crowded regions.

**Science:** X-ray binaries (XRBs) are the brightest sources in the transient X-ray sky. They normally appear in the sky as they brighten by a factor of up to $10^6$ in luminosity, and they are detected thanks to observations performed with X-ray all-sky monitors. These periods of enhanced activity are known as outbursts and correspond to an increase in the mass accretion rate onto the compact object (a black hole -BH- or a neutron star -NS-) from the companion star, which transfers mass via an accretion disc. Outbursts last from weeks up to months; once they end, systems transition back to quiescence, being faint (with X-ray luminosities $< 10^{33}\,\mathrm{erg\,s^{-1}}$) and feeding at low rates. At these low luminosities, only the most sensitive X-ray satellites (e.g. Chandra, XMM-Newton), can detect an XRB. In XRBs, the X-ray emission is produced close to the compact object, towards the inner regions of the accretion disc. In quiescence, the inner regions of the disc are colder than in outburst, allowing us to detect its emission at lower frequencies in the ultraviolet (UV) and optical/near infrared (OIR) bands. A large fraction of quiescent XRBs (qXRBs) are detectable in the OIR using ground-based telescopes (e.g. [231]). For the brightest qXRBs, continuous OIR monitoring programs are in place using robotic telescopes. This monitoring has made it possible to observe evidence for emission from the residual accretion disc in quiescence, in addition to emission coming from the companion star. As the new outburst gets closer, activity at OIR frequencies can be observed in the form of flaring (e.g. IGR J00291+5934; [18]); in addition, a gradual long-term increase in the OIR flux can be observed in the months preceding the onset of the outburst, showing the slow build-up of the accretion disc during quiescence (see e.g., Cen X-4 and V404-Cyg; [19,27]; Fig. 25). The Disc Instability Model predicts this behavior [DIM; 85,123], i.e., the most widely accepted scenario to explain the mechanism responsible for triggering an outburst (Fig. 25). According to the DIM, an instability is driven by the ionization state of hydrogen in the disc. If all the hydrogen in the disc is ionized, the system is considered to be stable. However, if the mass accretion rate or temperature becomes low enough to allow for the recombination of hydrogen, a thermal-viscous instability can occur in the disk, oscillating between a hot, ionized state (outbursts) and a cold, recombined state (quiescence). When the system is in quiescence, the cold accretion disk accumulates mass until a critical density, and the temperature rises until the hydrogen ionization temperature is reached at a specific radius (ignition point). At the ignition point, a heating front is generated ([140,200]), and the outburst is triggered, giving rise to the observed high X-ray luminosity. Then the outburst starts to decay and the disk is depleted, bringing the system back to quiescence [123]. To probe the DIM, it is crucial to observe the very first phases of the outburst. However, it is normally the case that outbursts are detected only when the X-ray flux rises above the all-sky monitors' detection threshold. Therefore, the initial stages of the outburst rise are missed. Thus, the only chance to study the trigger of an outburst is through observations in the OIR (especially since the DIM predicts that the

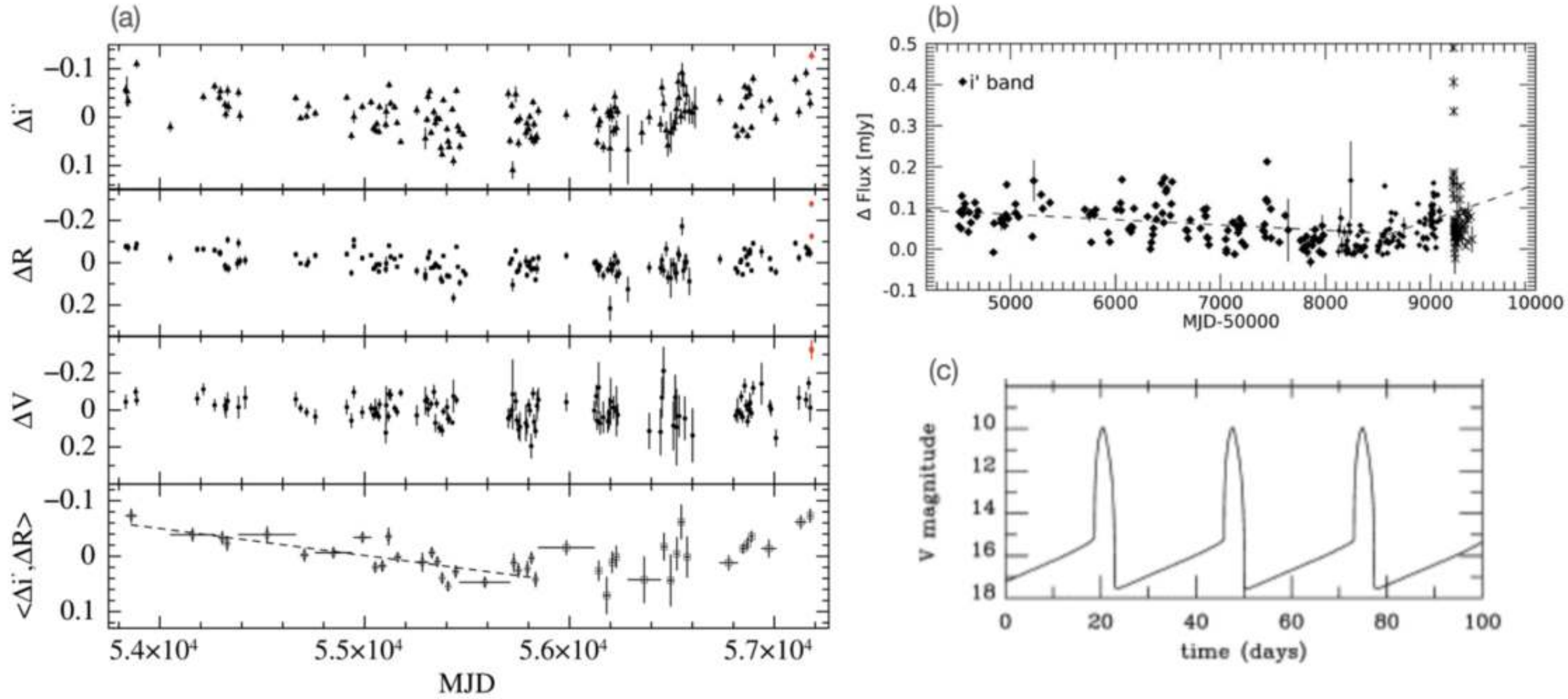

Figure 25 *(a)*: Downward trend of optical fluxes followed by an upward trend for the XRB V404 Cyg [27]. The increasing fluxes were explained as the gradual buildup of the accretion disc around the compact object that preceded the 2015 outburst of V404 Cyg. *(b)*: Downward trend of optical fluxes followed by an upward trend for the XRB Centaurus X-4 [19]. The "misfired outburst" is visible at the end of the optical light curve. *(c)*: The increasing trend of fluxes preceding the onset of a new outburst is predicted by the DIM [123], especially when irradiation is taken into account.

optical will start brightening before the X-rays; [123]). This makes continuous OIR monitoring of XRBs using robotic telescopes crucial, particularly in measuring the OIR-to-X-ray delay of the rise to outburst —a very important parameter that demonstrates how the propagation of the heating front in the accretion disc occurs. To address this challenge, continuous optical monitoring of XRBs is essential, as provided by the X-ray Binary Early Warning System (XBNEWS), which offers real-time light curves of XRBs using Las Cumbres Observatory (LCO) telescopes.

This approach proved effective during the 2019 outburst of the neutron star XRB SAX J1808.4−3658, for which an X-ray-to-optical delay of four days was measured using XBNEWS in conjunction with NICER and Swift [77]. These observations provide critical insights into the optical-to-X-ray delay during the onset of XRB outbursts, shedding light on the DIM that regulates these transitions. Additionally, long-term LCO observations of the neutron star qXRB Cen X-4 allowed the prediction of an outburst several months in advance. This event, characterized by weak activity, was identified as a "misfired outburst" [19]. Another illustrative case is the black-hole low-mass X-ray binary Swift J1753.5−0127. After remaining in quiescence since 2018, XB-NEWS detected a significant optical re-brightening in 2023 [11]. However, a Swift/XRT observation on the same day revealed no detectable X-ray emission, with an upper limit to the flux of $1.8 \times 10^{-13}$ erg cm$^{-2}$ s$^{-1}$ in the 0.5–10 keV range. Swift/XRT first detected the source when the unabsorbed 0.5-10 keV flux reached $1.7 \times 10^{-12}$ erg cm$^{-2}$ s$^{-1}$, approximately a week after the initial optical detection. However, Chandra detected a weak X-ray source at the position of Swift J1753.5−0127 just three days after the first optical detection, at a flux level of $6 \times 10^{-14}$ erg cm$^{-2}$ s$^{-1}$ in a $\sim 6$ ks observation ([93]; Fig. 26). This made it possible to measure the X-ray-to-optical delay of the outburst onset with much higher precision than what Swift/XRT could achieve, thanks to the instrument's better sensitivity.

AXIS will provide an unprecedented opportunity to study the earliest moments of XRB outbursts. Its rapid response time for Target of Opportunity (ToO) observations will be crucial for capturing XRBs

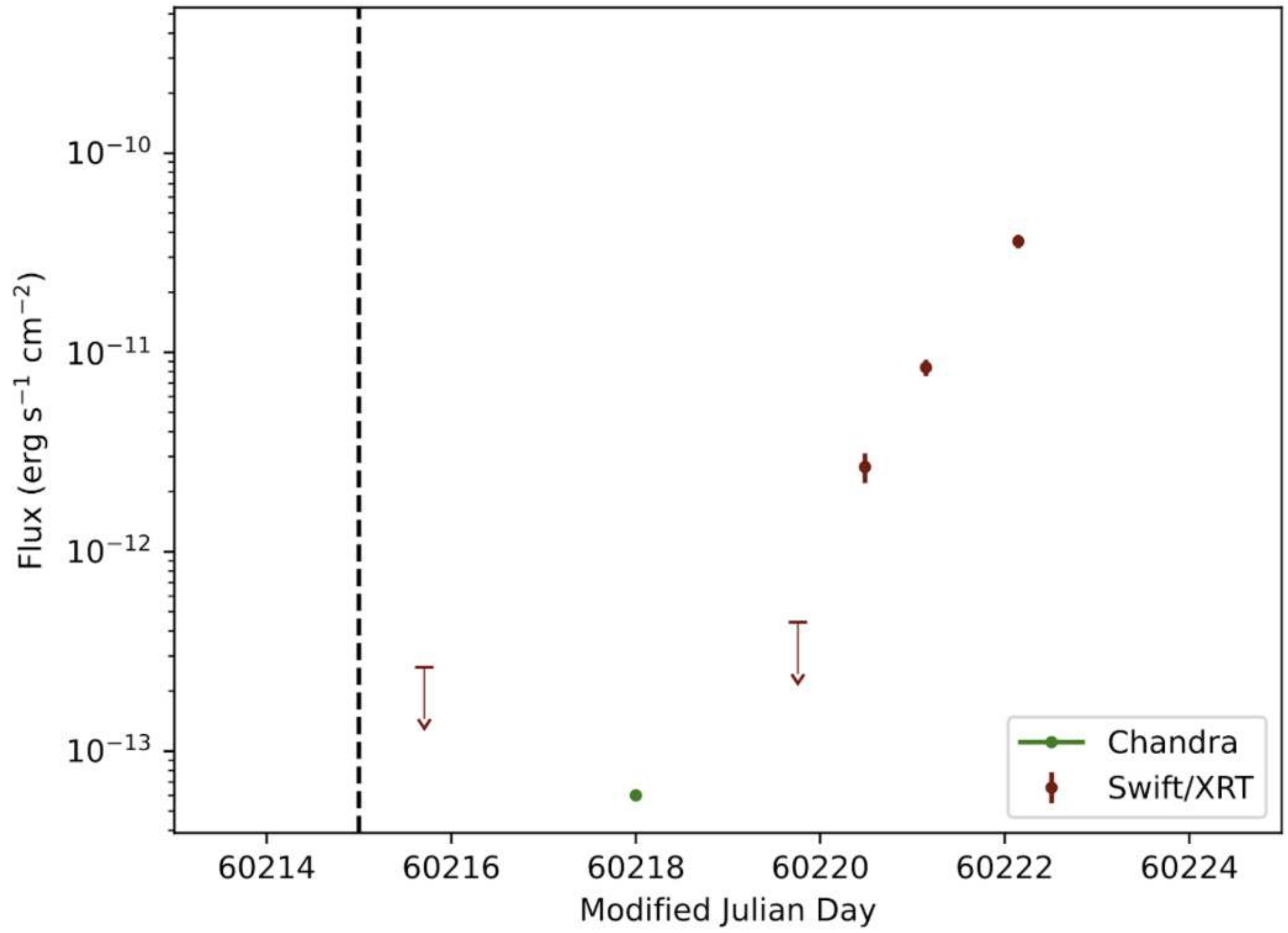

Figure 26 Early onset of the 2023 outburst of the BH XRB Swift J1753.5-0127 as observed by Swift/XRT and Chandra. With a dashed line, the time of the first optical detection of the rise to outburst is reported. Data are taken from [11,93].

soon after their first optical detection. The case of Swift J1753.5-0127 demonstrates that AXIS, with its competitive sensitivity which can reach $\sim 10^{-15}$erg s$^{-1}$ cm$^{-2}$ in 2 ks exposures, will be capable of detecting the X-ray flux increase within just 1-3 days of the first optical detection, significantly improving our ability to constrain the optical-to-X-ray delay (Fig. 26). We simulated an AXIS spectrum for a 5-ks exposure assuming an absorbed power-law model, with an NH of $10^{21}$ cm$^{-2}$, a photon index of 2, and an observed flux of $6 \times 10^{-14}$ erg s$^{-1}$ cm$^{-2}$ in the 0.3–10 keV energy band. Fitting the simulated spectrum in the 0.3–3 keV band using the Cash statistics shows that it is possible to achieve a precision on the photon index of about 10%, resulting in a Cash statistic value of 252 for 266 degrees of freedom (d.o.f.). Conversely, fitting the same spectrum with a blackbody model yields a higher Cash statistic value of 282 (266 d.o.f). Figure 27 shows the simulated spectrum (rebinned only for plotting purposes) with the best-fitting absorbed power-law model overlaid. The middle and bottom panels display the residuals for the best-fitting PL and BB models, respectively.

This simulation demonstrates that AXIS will be capable of characterizing the X-ray spectrum of the source in the early stages of an outburst—something not achievable with previous X-ray observatories. This will allow for an unambiguous determination of the source's state. By precisely measuring the optical-to-X-ray delay, AXIS will place critical constraints on the propagation speed of heating waves in the accretion disk, helping to distinguish between different outburst triggering mechanisms. Additionally, AXIS's exceptional angular resolution (1.5″ for on-axis observations) will enable the resolution of XRBs in crowded regions.

**Exposure time (ks):** 100 ks/year

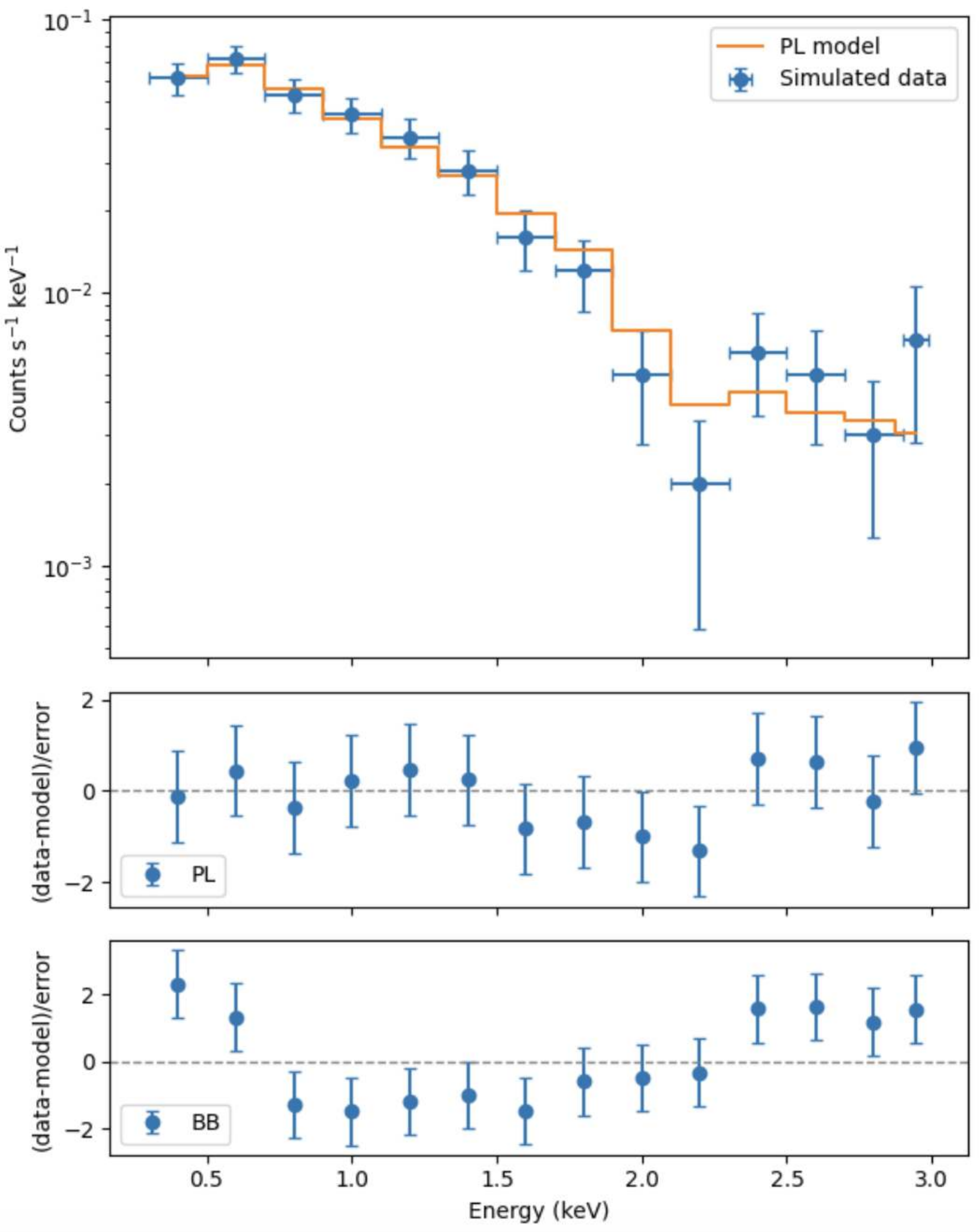

Figure 27 Top: AXIS spectrum simulated for an XRB in the earliest stages of its outburst, assuming a 5-ks on-source exposure and an observed flux of 6E-14 cgs in the 0.3–10 keV band (see text for details). The best-fit absorbed power-law model is highlighted in orange. Middle: post-fit residuals relative to this model. Bottom: post-fit residuals relative to the best-fit absorbed blackbody model.

**Observing description:** Our program aims to rapidly respond to alerts from XBNEWS and other optical transient monitoring services, enabling timely follow-up observations with AXIS. We can expect to have one outburst/year that is first detected in the optical thanks to such monitoring services. As soon as a new outburst is reported, we will trigger AXIS observations with exposure times between 5–10 ks, optimized to reach a sensitivity threshold of at least $10^{-15}\,\mathrm{erg\,s^{-1}\,cm^{-2}}$. The primary goal is to secure an X-ray detection as soon as possible after the optical trigger. If the sensitivity allows, we will also characterize the early-stage spectrum and follow its evolution with subsequent observations. We have demonstrated, through simulations based on the fluxes of Swift J1753.5$-$0127 (initial flux $6 \times 10^{-14}$, erg, cm$^{-2}$, s$^{-1}$), that even a 5 ks exposure is sufficient to achieve this. The specific exposure duration will be adjusted based on the known absorption properties of the source. For transients with well-characterized absorption, we will tailor the observation time accordingly. However, in cases where the absorption is unknown—such as for newly discovered transients—we will adopt a conservative default exposure of 10 ks to ensure adequate sensitivity. Observations will be repeated daily until independent detections from other all-sky X-ray monitors confirm the outburst. This strategy will maximize the chances of capturing the early X-ray evolution of new transients while efficiently utilizing AXIS observing time.

**Joint Observations and synergies with other observatories in the 2030s:** Our AXIS observing strategy will necessitate real-time alerts from transient optical monitoring networks such as XBNEWS and other optical surveys, enabling rapid follow-up of new X-ray transients. These joint efforts will maximize our ability to detect and characterize the very early stages of outbursts of BHXBs and NS-LMXBs. In addition, the new SOXS spectrograph at the New Technology Telescope, which will start operations in the second half of 2025, will observe bright XRB outbursts at their early stages to determine the composition of the outer accretion disc at the time of the outburst trigger. Synergies with upcoming radio, optical, and gamma-ray observatories in the 2030s will be essential for a multi-wavelength understanding of the first stages of outbursts. Joint observations with AXIS and the next-generation enhanced X-ray Timing and Polarimetry [eXTP; 228] mission will enable polarimetric studies of these phases, which will give information on the geometrical properties of the inner accretion flow in LMXBs at the beginning of outbursts. The upcoming Ultraviolet Explorer [UVEX; 120] can provide simultaneous UV coverage of the early stages of outbursts, allowing for the study of the delay in onset also at these wavebands. Similarly, the ELT will allow observing the same at NIR frequencies, which will make it possible to determine if the NIR flux increases together with the optical when an outburst starts, or if it is delayed or even anticipated. Finally, the SKA will also be able to monitor the outbursts of XRBs, making it possible to determine when relativistic jets are launched and during which phase of the outburst rise this occurs.

**Special Requirements:** Rapid ToO response (<1 d) will enable observation of the very early stages of XRB outbursts soon after the first optical detection.

*14. Reprocessing studies in X-ray binaries with eclipse observations*

**First Author**: Biswajit Paul (Raman Research Institute, C. V. Raman avenue, Sadashivanagar, Bangalore 560080, India, bpaul@rri.res.in)

**Co-authors:** Sachindra Naik (Physical Research Laboratory, Ahmedabad, India), Ketan Rikame (Raman Research Institute, Bangalore, India), Nafisa Aftab (Raman Research Institute, Bangalore, India), Pragati Pradhan (Embry-Riddle Aeronautical University, Prescott, AZ, USA), Rahul Sharma (Inter University Centre for Astronomy and Astrophysics, Pune, India), Manish Kumar (Raman Research Institute, Bangalore, India), Nazma Islam (NASA GSFC and University of Maryland Baltimore County, USA), Chetana Jain (Hansraj College, University of Delhi, India)

**Abstract:** Reprocessed X-rays are key to diagnosing the environment in X-ray binary systems. The main difficulty in the study of the reprocessed X-rays is the presence of much brighter primary radiation from the compact star, together with the reprocessed radiation. In the eclipsing X-ray binary systems, the direct X-ray emission from the compact star is blocked by the companion star during the eclipse. Therefore, during an eclipse, we observe only the reprocessed emission that contains clues about the environment of the compact object like the wind velocity, chemical composition, ionization levels, etc. XMM-Newton observations of several low mass and high mass X-ray binaries during their X-ray eclipses have already provided a wealth of data regarding the reprocessing agents in these systems and have shown remarkable differences from system to system. X-ray flares and bursts have also been observed during the eclipses of some X-ray binaries, providing further details about the multiple emission components present during the eclipses and the extent of the X-ray reprocessing regions. AXIS will enable far more detailed investigations of the X-ray reprocessing agents in the X-ray binaries using their eclipse observations.

**Science:**

**1. X-ray reprocessing:** The X-rays received from X-ray binary systems consist of many different components. The dominant components are from regions near the compact object and can be one or more of these: thermal emission from neutron star surface and boundary layer, thermal emission from accretion disk, non-thermal components like Compton up-scattered X-rays from the accretion column in case of accreting pulsars, Compton up-scattered X-rays from accretion disk corona in case of black hole or low magnetic field neutron stars, Synchrotron emission from jet, etc. Some sources have additional soft components (4U 1538-522, [8]) that originate elsewhere in the binary system (Fig. 28) and in most binary systems, a part of the dominant component is reprocessed/scattered in the environment of the compact object/binary. The scattered/reprocessed emission can have spectral characteristics that are different from the dominant component, and it is often rich in fluorescence emission lines.

**2. Supergiant High Mass X-ray Binaries and Supergiant Fast X-ray Transients:** A detailed study of X-ray reprocessing in High Mass X-ray Binary (HMXB) systems showed a tremendous range of the reprocessing efficiency, i.e., the ratio of flux in the continuum emission during eclipse to out-of-eclipse, varying in the range of $\approx$0.004 to $\approx$0.125 in different sources. The iron emission line, which is primarily produced in the wind of the companion star, is, however, much more prominent during the eclipse compared to the out-of-eclipse phase, with an equivalent width exceeding 1000 eV in several sources during the eclipse. Overall, significant differences are observed in the eclipse spectra of different HMXBs, as well as in their eclipse spectra compared to out-of-eclipse spectra [8]. Though the SFXTs show very low equivalent width of the Fe K$\alpha$ emission line compared to supergiant HMXBs during out of eclipse [170], the iron line is very prominent in the SFXTs during eclipses, often reaching 1000 eV [8].

**3. Low Mass X-ray Binaries:** The eclipses in Low Mass X-ray Binaries (LMXB) are of a much shorter duration, a few hundred seconds. Despite that, it has been possible to investigate the eclipse spectral characteristics of LMXBs by combining data from multiple eclipses [7]. Surprisingly, despite having a weaker stellar wind, the ratio of the continuum flux between the eclipse and out-of-eclipse is found to be larger in the LMXBs compared to the HMXBs. In LMXBs, the iron emission line, however, does not show as much change in the equivalent width as it does for HMXBs during the eclipses.

**4. Ultra-Luminuous X-ray Sources:** A few of the Ultra-Luminuous X-ray (ULX) sources, show X-ray eclipses and from the large duration of the eclipses, these sources are thought to have massive companion star [213]. However, due to their relatively low brightness, it has not yet been possible to conduct detailed characterizations of the eclipse spectra of the ULXs.

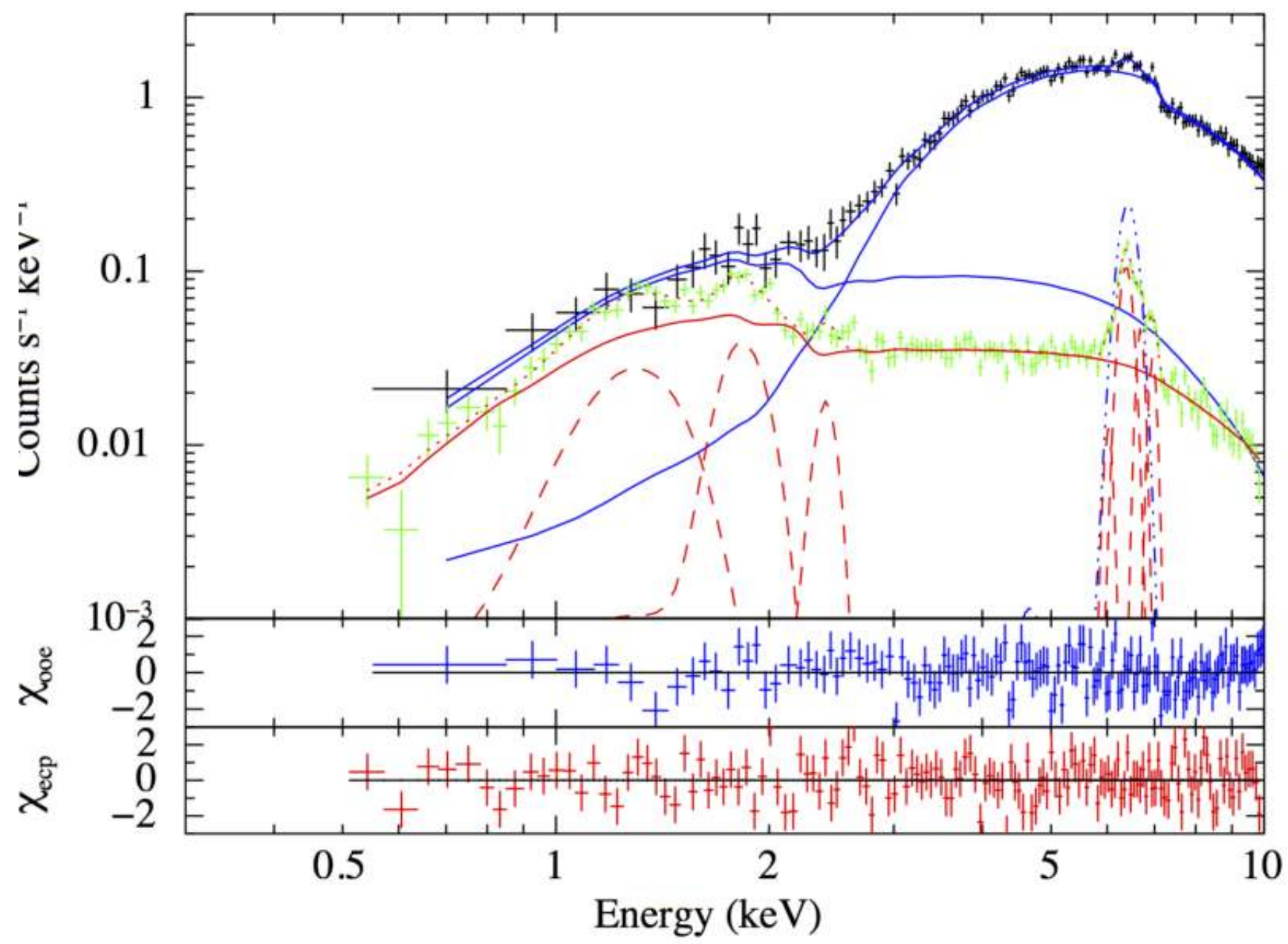

Figure 28 The eclipse (black) and out of eclipse (green) spectra of 4U 1538−522 (Aftab et al. 2019). While the hard power-law component is suppressed during the eclipse, the soft component remains unaffected, indicating a different, and much larger, origin.

**5. Flares and bursts during eclipses:** Many HMXBs are strongly variable sources, showing flares of duration from a fraction of a minute to tens of minutes. The peak luminosity of the flares can be more than 10 times the persistent luminosity, and some of these flares have recently been detected during the eclipses (Fig. 29), in the form of reprocessed emission [183]. As mentioned earlier, while the eclipse spectrum is primarily reprocessed emission, it can also contain a persistent component. A comparison of the eclipse flare spectrum against the eclipse non-flare spectrum for 4U 1700−37 showed the presence of a soft spectral component (Fig. 30), that is not produced by reprocessing. Hence, it must be produced in a large region in the binary, which is not fully masked during the eclipse of the compact star [183].

Analogous to the HMXBs, in which the flares have been seen during eclipse, thermonuclear bursts (Type-I burst) have been detected during the eclipses in LMXBs [118]. Though the direct emission is blocked, reprocessed emission, possibly from the parts of the accretion disk, the disk wind, or the ablated wind from the companion star that remain visible during the eclipse, produce the bursts during the eclipses. A few bursts have also been detected just before the eclipse's egress, with their tails visible after the eclipse's egress. These unique bursts help measure the reprocessing efficiency of these reprocessing agents and how it changes with their visibility at different orbital phases during the eclipse. The sharp rise of the bursts detected during the eclipses confirms that the X-ray emission detected from the LMXBs during their eclipses is due to scattering in the vicinity of the binary and not in the interstellar material.

**6. Partial eclipses:** There are about 15 eclipsing HMXB, eight eclipsing LMXBs and five eclipsing ULXs. Two of the HMXBs and one LMXB show a unique feature of partial eclipse [102,107,110,171]. The partially eclipsing LMXB 4U 1822-37 is understood to be an accretion disk corona source, in which the central source

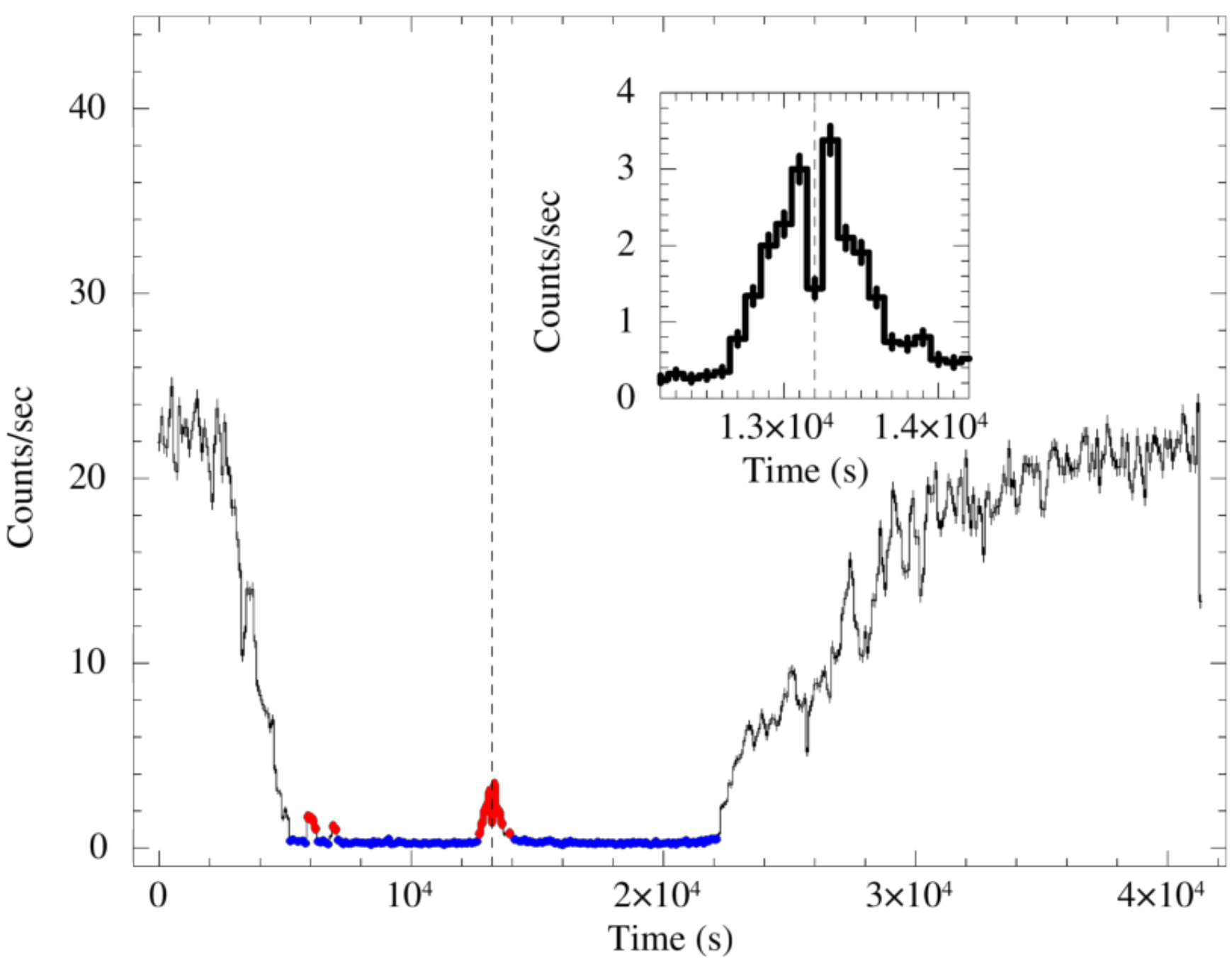

Figure 29 Eclipse in LMC X-4 showing multiple flares in XMM-Newton observation [118].

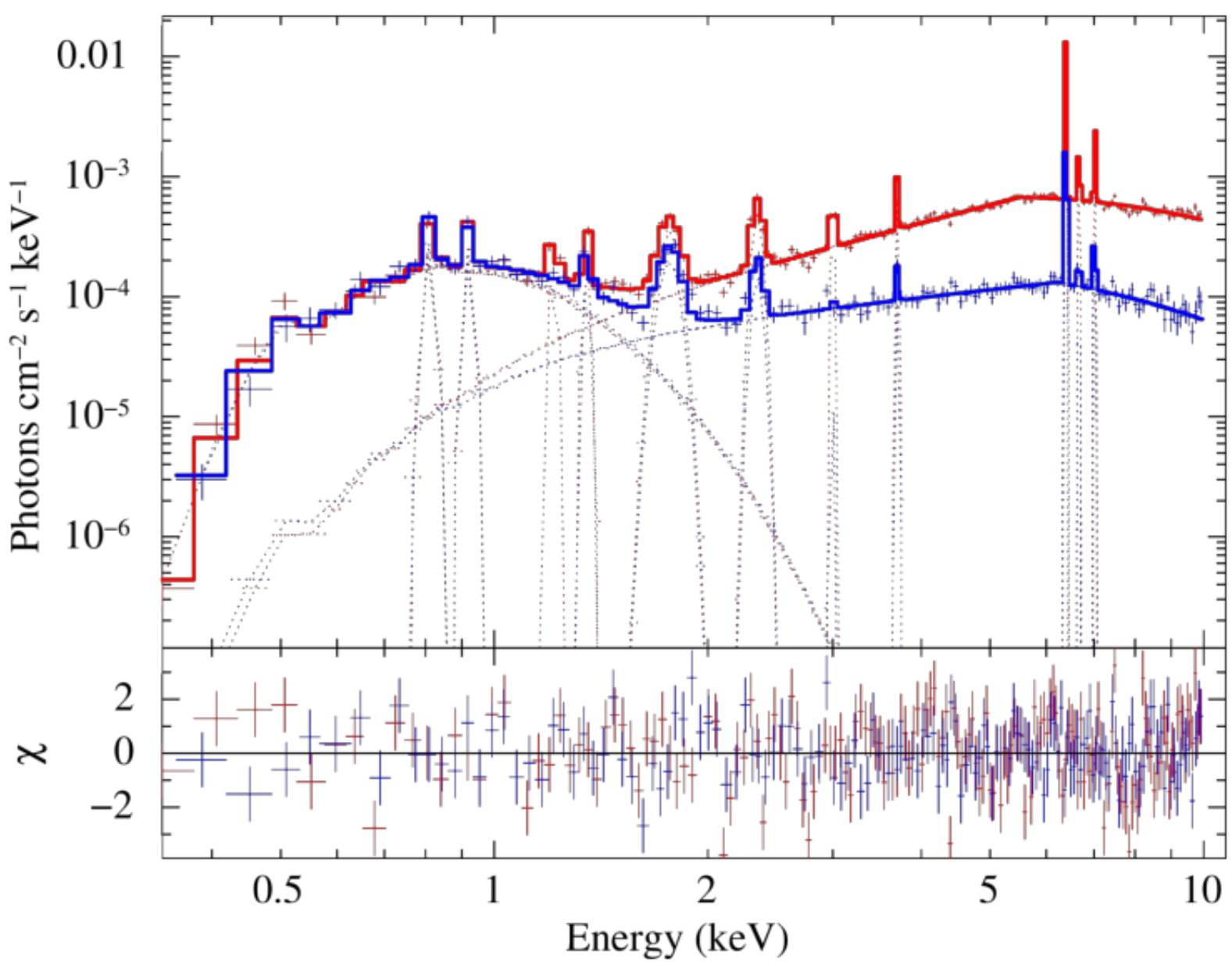

Figure 30 Flaring (in red) and non-flaring (in blue) spectrum of 4U 1700−37 during eclipse showing a persistent component that is neither reprocessed emission nor eclipsed [118].

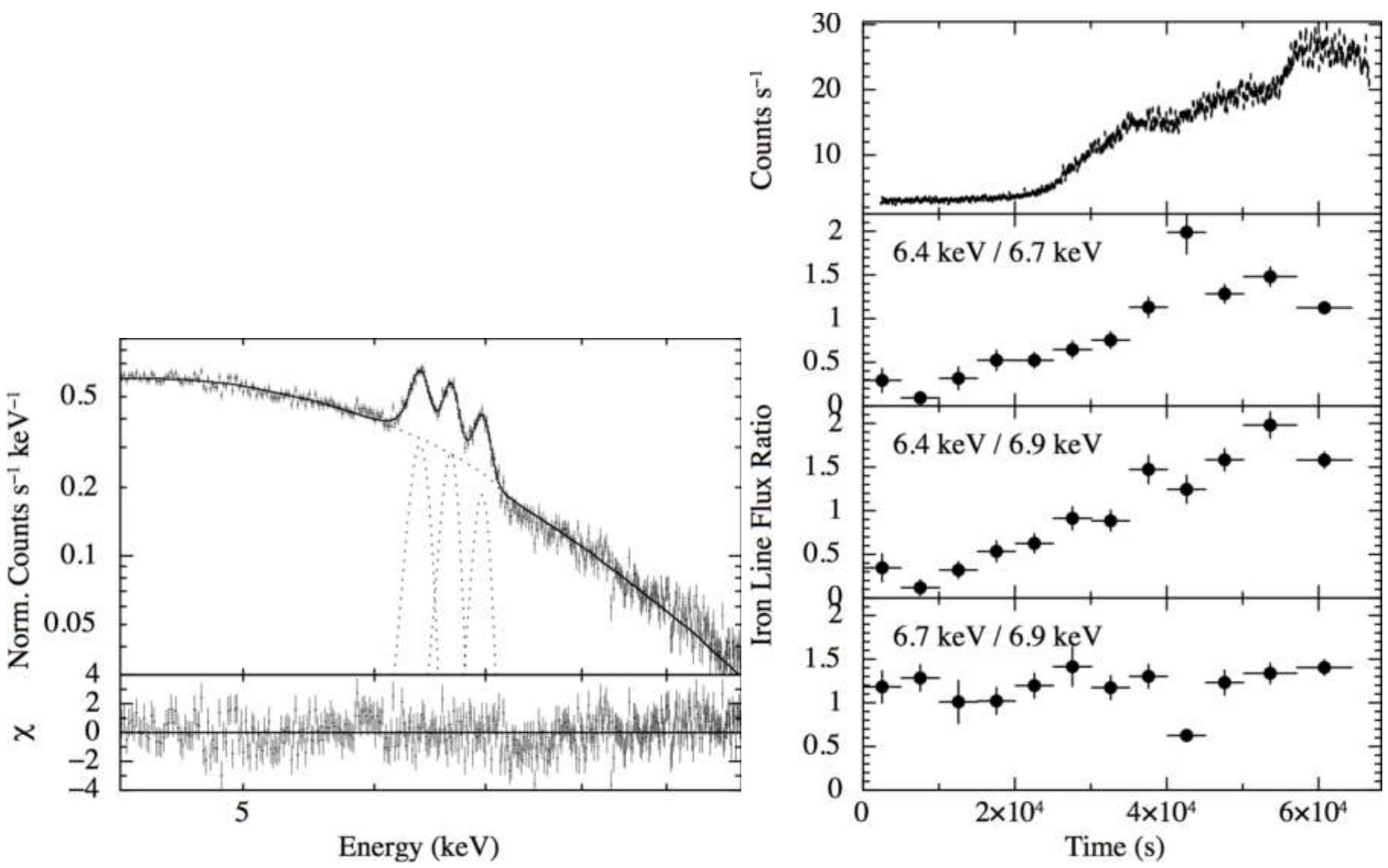

Figure 31 The X-ray spectrum of Cen X-3 with three iron emission lines (left) and time resolved spectroscopy (right) showing variable ratio of the line fluxes during the eclipse egress [152].

is blocked by a puffed up outer accretion disk, and most of the observed X-rays are from the corona, which, being an extended source can have a partial eclipse. The neutron star in 4U 1822−37 exhibits coherent pulsations with a very small pulse fraction, consistent with the scenario that most of the observed X-rays originate from the corona rather than the compact star. It is, however, difficult to explain the partial eclipse of the neutron star in the HMXB systems as the X-ray emission (with a high pulsed fraction) region is very small in size, and it is difficult to imagine a scenario in which an object with a size of 20 km can be partially eclipsed.

**7. Eclipse transitions:** Eclipse transitions in both HMXBs and LMXBs are interesting episodes and observations of eclipse transitions give useful information about the size of the the X-ray reprocessing regions. For example, Cen X-3, an accreting HMXB pulsar, exhibits three strong iron emission lines in its spectrum, corresponding to different ionization states. During the eclipse egress, the different spectral components show different timescales for the eclipse transitions (Fig. 31, [152]. The power-law continuum has a sharp transition, the neutral iron line has a slower transition, and highly ionized iron lines show a much longer eclipse transition timescale.

**AXIS observations:** Most of the X-ray reprocessing studies in X-ray binaries with eclipse observations mentioned above have been carried out using archival data. The flare and burst detections during eclipses of HMXBs and LMXBs are serendipitous discoveries. With targeted eclipse observations of HMXBs and LMXBs using AXIS, more comprehensive investigations can be carried out to understand various astrophysical processes in these systems that are only observable during eclipses or eclipse transitions. These eclipsing galactic X-ray binaries are bright (have luminosities of the order of $10^{36}$ ergs $s^{-1}$ or more) in the out-of-eclipse phase. During the eclipse, the fluxes decrease by one to two orders of magnitude. Therefore, pileup will be a concern during the out-of-eclipse observations, and that can be mitigated by carrying out the observations in Windowed Mode. For observations of eclipsing ULXs and any extra-galactic binaries, pileup will not be a concern.

**Observation durations:** Typical observation duration for each HMXB will be about 80 ks, and for LMXBs it will be about 40 ks each. All together, a detailed investigation of this information-rich topic can be done effectively in about 800 ks for 10 HMXBs and 200 ks for 5 LMXBs. For the HMXBs, these would be phase-constrained pointed observations.

### d. Stellar Explosions and Magnetar Activity

#### *15. Probing Shock Breakout and Progenitor Evolution Through X-ray Observations of Both Young and Old Supernovae*

**First Author:** Wynn Jacobson-Galán (Caltech, wynnjg@caltech.edu)

**Co-authors:** Brad Cenko (NASA/GSFC), Franz E. Bauer (UTA, MAS), Christopher Irwin (University of Tokyo)

**Abstract:** X-ray emission is a characteristic feature of supernova (SN) shock waves both during the explosion's infancy as well as later-time evolution. In the first minutes to hours of a supernova (SN), the intrinsic spectrum and duration of shock breakout (SBO) are directly linked to progenitor star mass/radius and the explosion mechanism. Additionally, luminous X-ray emission is an expected phenomenon in SNe that arise from progenitor stars enshrouded by dense circumstellar material (CSM). Below, we outline how the AXIS probe is uniquely positioned to study SN SBO and the physics of shock power in unprecedented detail.

**Science:**

**(1) Supernova SBO:** Following core-collapse of a massive star, the SN shock will "break out" of the stellar surface at a characteristic optical depth, producing luminous X-ray/UV emission – the shock breakout (SBO) spectrum and duration being intrinsically linked to progenitor star identity. As shown in Figure 1, more compact stars (e.g., Wolf Rayet, Blue Supergiant) produce brighter, shorter, and harder shock breakout emission than more extended stars (e.g., Yellow/Red Supergiants). Detection and characterization of the SBO signal using AXIS will enable the robust identification of SN progenitor identity out to larger distances, allowing for a novel approach to measuring the stripped-envelope core-collapse supernova rate within a larger volume than is accessible with ground-based, time-domain surveys.

To date, there is only one example of a confirmed X-ray SBO from a type Ib (H-stripped, He-only) progenitor (SN 2008D), likely from the explosion of a compact Wolf Rayet star [202]. Consequently, characterizing the X-ray emission from SBO in both compact and extended massive stars remains a theoretical exercise [154,216]. However, even simulations for SBO have shown that incredible amounts of diversity can exist when explosion asymmetries and 3D effects are taken into account, notably the extension of SBO duration from $\sim$minutes to $\sim$hours [75]. AXIS will revolutionize our understanding of X-ray emission during massive star SBO by both serendipitous detection/discovery of the initial SBO flash as well as rapid follow-up of SBO discovered by other UV-focused telescopes (e.g., ULTRASAT, UVEX) [120,199].

For SN progenitor systems enshrouded in dense circumstellar material (CSM), SBO X-ray emission timescales can be significantly extended as shocked gas cools from on-going SN ejecta-CSM interaction [46]. Consequently, X-ray observations of young SNe interacting with confined CSM enable us to understand the density/composition of the local circumstellar environment and, most importantly, the unknown mass-loss histories of SN progenitor stars in their final months to years before explosion – these timescales being inaccessible otherwise. It has been estimated that $> 40\%$ of type II SN progenitors (i.e., red supergiants) could be surrounded by dense CSM within a $\sim 2-3$ stellar radii [35,104]. As shown in recent studies, the

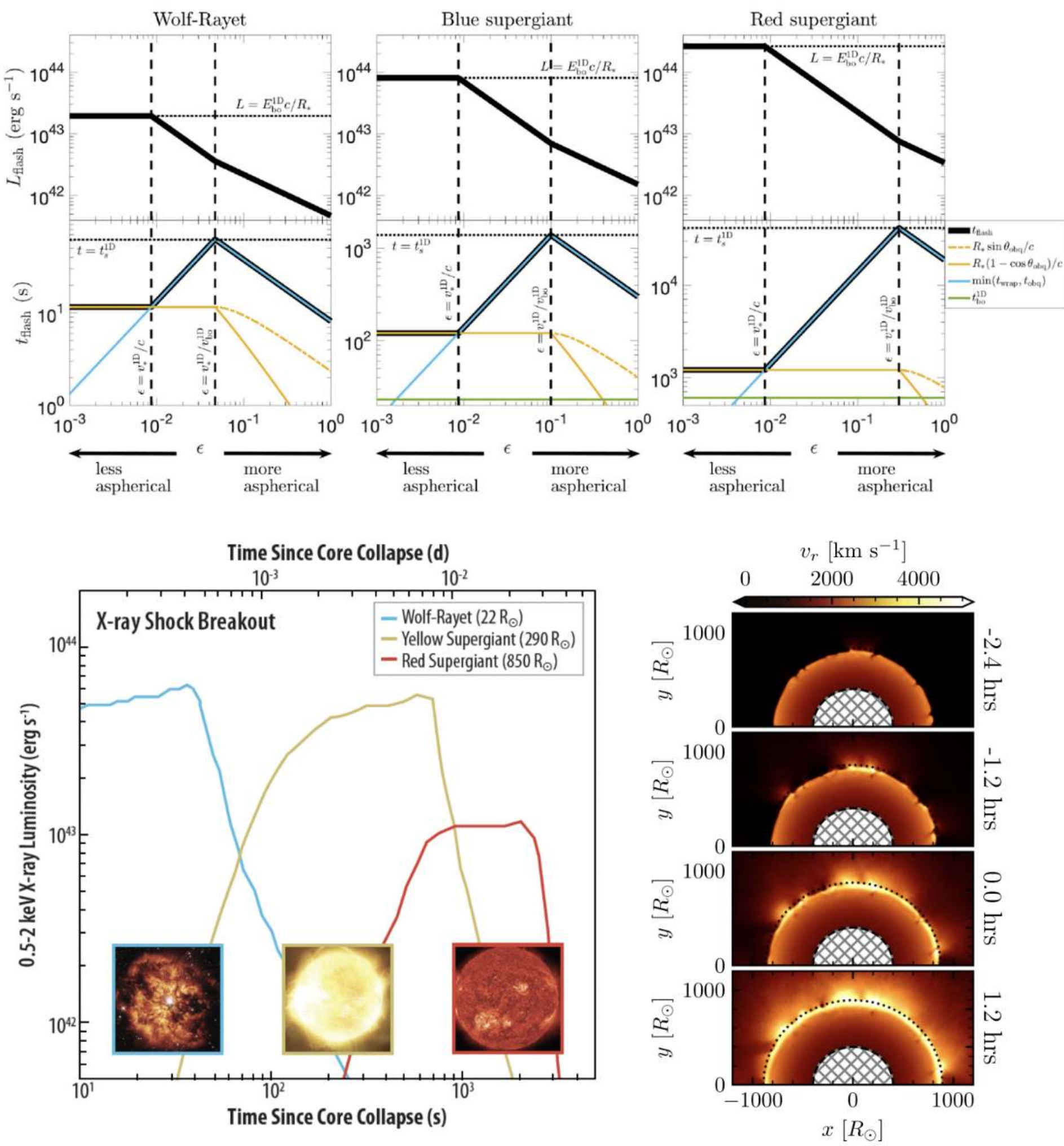

Figure 32 *Top:* Shock breakout luminosities for different supernova progenitor stars with varying levels of asymmetry. (Adapted from [101]) *Bottom Left:* X-ray emission predicted for various massive stars using theoretical 1D shock breakout models.[154] *Bottom Right:* Visualization of 3D shock breakout from a red supergiant star. (Adapted from [75]) AXIS detection of shock breakout X-ray durations, luminosities, and spectral shapes will directly constrain progenitor identity and explosion asymmetries.

SBO timescales for enshrouded red supergiants can be $\sim$1 day, with X-ray emission rising as the optical depth in the confined CSM drops with time [105,155]. Applying the expected 10 ks exposure time of the Deep Extragalactic Survey, AXIS can serendipitously detect SBO emission from all SN varieties out to $z \sim 2$.

**(2) Persistent CSM-Interaction:** After SBO, the observed X-ray emission in most SN sub-types arises from the collision of SN ejecta with distant CSM, formed during various phases of stellar evolution [44,47,135,155]. Continuous X-ray monitoring enables robust constraints to be placed on the density profile of the progenitor environment, which directly informs the identity of the progenitor star and its mass-loss mechanisms (e.g., eruptions, binary interaction).[34,59] Through its ToO capabilities and overall sensitivity, AXIS will allow observers to monitor the X-ray emission of SNe out to later phases post-explosion and

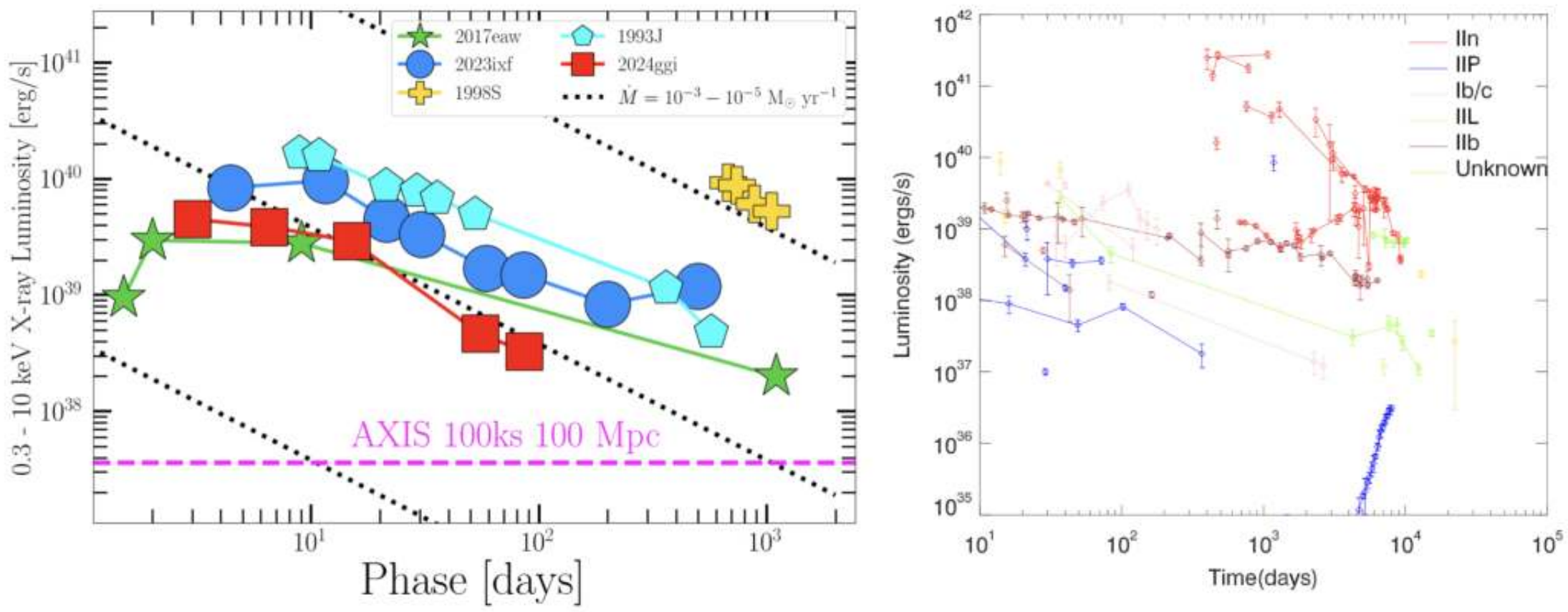

Figure 33 *Left:* X-ray light curves of nearby type II SNe that show late-time X-ray emission from on-going SN ejecta interaction with distant CSM. (Adapted from [106]) *Right:* X-ray light curves of all SN sub-types. (Adapted from [58])

with increased spectral resolution – the former probing terminal stages of stellar evolution (i.e., longer lookback times) while the latter enables robust spectral modeling to uncover essential properties of the post-shock gas (e.g., contributions from various shocks, gas temperature, shock velocities, abundances, etc). To date, the total number SNe with X-ray observations remains limited and is, at late times (>1 year post-explosion), typically reserved for very nearby events (e.g., <20 Mpc). Through both targeted GO programs and serendipitous monitoring of X-ray bright SNe in the Deep Extragalactic Survey, AXIS will significantly increase the current sample of X-ray-observed SNe.

**Target summary (names):** Infant SNe, SBO candidates, CSM-interacting SNe.

**Exposure time estimates:** 10 ks for ToO follow-up to young SNe. 100 ks for late-time monitoring of CSM-interacting SNe.

**Specify critical AXIS Capabilities:** Field of view allows for serendipitous SBO detection out to z $\sim$ 2. Rapid (< 4 hour) ToO capability for young SNe detected by UV/optical time-domain surveys. Spectral resolution and soft X-ray sensitivity for modeling the X-ray spectra of CSM-interacting SNe at $\sim$years post-explosion.

**Joint Observations and Synergies with other Observatories in 2030s:** UVEX, ULTRASAT.

**Special requirements:** ToO (< 4 hours), and monitoring (daily to monthly)

*16. Magnetar Giant flares*

**First Author:** Michela Negro (Louisiana State University, michelanegro@lsu.edu)

**Co-authors:** Zorawar Wadiasingh (University of Maryland, College Park), George Younes (George Washington University)

**Abstract**: Magnetar giant flares (MGFs) are among the most luminous high-energy transients in the nearby universe, offering unique probes of extreme magnetic dissipation, fireball dynamics, and potentially heavy-element nucleosynthesis. MGFs are characterized by an intense, short-lived MeV $\gamma$-ray spike followed by a softer, pulsating X-ray tail, with a possible late-time r-process signature. Because only three Galactic events have been firmly detected, and a handful of extragalactic candidates suggested,

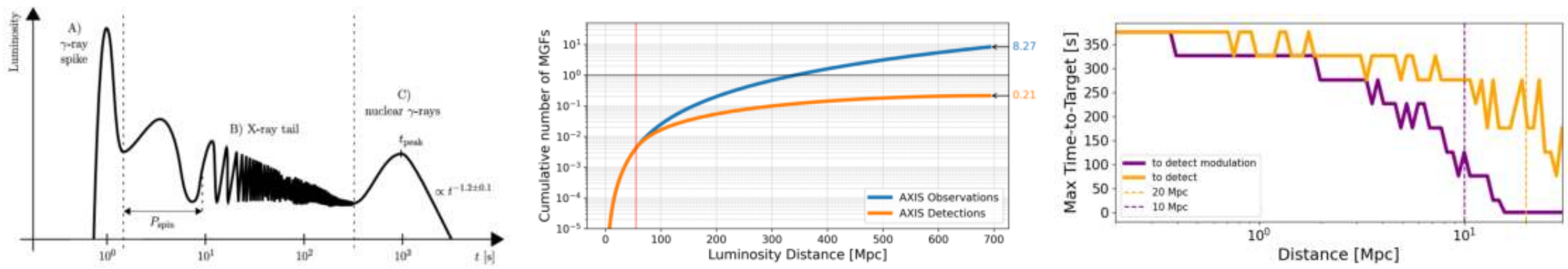

Figure 34 Figures from [156]. *Left*: Schematic light curve of a magnetar giant flare. Credits: Jakub Cehula and Anirudh Patel. *Middle*: Cumulative number of MGFs as seen by AXIS in 5-year mission as a function of the distance. *Right*: Maximum time-to-target on a magnetar giant flare as a function of distance, based on simulated AXIS observations.

constraining their rates and energetics requires sensitive X-ray and $\gamma$-ray facilities with both wide-field coverage and rapid slewing capabilities. A recent feasibility study evaluated the prospects for detecting MGFs using AXIS, with a focus on both serendipitous discoveries and rapid follow-up capabilities.

**Science**: The study by [156] examined two complementary discovery channels: (i) the serendipitous detection of the prompt $\gamma$-ray spikes directly within the AXIS field of view, and (ii) the targeted follow-up of pulsating X-ray tails in nearby galaxies. It was shown that while bright spikes could in principle be detectable out to several hundred Mpc, the combination of their intrinsic hardness and sub-second duration makes such detections unlikely over the mission lifetime. We note that other types of short transients with softer spectra (peaking at lower energies in $\nu F(\nu)$ space) could be detectable by AXIS in the prime mission serendipitously, depending on their intrinsic volumetric rates and energetics.

The softer, longer-lasting tails emerge as a much more promising pathway. Simulations indicate that AXIS could detect periodic modulations out to $\sim$5–10 Mpc and identify tails as faint X-ray transients as far as $\sim$20 Mpc, provided that rapid time-to-target response is available. These detections would represent the first extragalactic measurements of magnetar periodic signals, offering new insights into fireball physics, emission geometry, and potentially revealing quasi-periodic oscillations on millisecond timescales (which, of course, would require ms-scale time resolution).

The study also evaluated the detectability of soft X-ray line emission associated with r-process nucleosynthesis in MGFs. These signals were found to be several orders of magnitude fainter than the AXIS sensitivity limit at extragalactic distances, restricting meaningful constraints to Galactic events only. Even so, such Galactic observations would be invaluable, offering a direct test of whether MGFs contribute to heavy-element production.

**Observation strategy**: Beyond specific detectability forecasts, the study by [156] also put forward a general framework for assessing the transient reach of any instrument, combining sensitivity rescaling, volumetric rates, spectral assumptions, and mission efficiency. An important outcome of this exercise is that an optimal AXIS strategy would not rely solely on uniform sky coverage, but instead devote a fraction of its observing time to nearby, star-forming galaxies within $\lesssim$100 Mpc. In this regime, a small number of galaxies dominate the expected MGF rate, making targeted monitoring particularly advantageous. Such a strategy would maximize the chances of capturing both prompt spikes serendipitously and fainter tails (including possible orphan tails), substantially enhancing the mission's scientific return.

**Exposure time estimates**: $\sim$ 1 ks (in case of ToO); .

**Joint Observations and Synergies with other Observatories in 2030s:** MeV gamma-ray telescope (COSI or next generation gamma-ray observatory with spectroscopic capability)

**Special requirements:** ToO (<200 s); strategic non-uniform sky monitoring (targeting nearby star-forming galaxies within $\lesssim$100 Mpc).

## e. Novel X-ray Probes of the Transient Universe

### *17. X-ray Dust Tomography with AXIS*

**First Author:** Sebastian Heinz, University of Wisconsin-Madison, sheinz@wisc.edu

**Co-authors:** Andrea Tiengo, IUSS Pavia, andrea.tiengo@iusspavia.it, Lia Corrales, University of Michigan, liac@umich.edu, Andy Beardmore, University of Leicester, ab271@leicester.ac.uk

**[Abstract:]** X-ray dust scattering can be a sensitive probe of the interstellar medium. While all bright X-ray sources in the Galactic plane exhibit diffuse dust scattering halos, highly variable sources can show spectacular dust echoes in the form of rings. Studying the temporal evolution of these rings with the current generation of X-ray telescopes — Chandra, Swift, and XMM-Newton — has opened up a new field of X-ray dust tomography. Dust echoes allow us to measure geometric distances to dust clouds and X-ray sources, constrain the ISM composition (such as gas-to-dust ratios), and probe dust mineralogy and grain size distribution. High angular resolution, large effective area, large field of view, low background, and the ability to rapidly execute TOOs are crucial for pushing this field forward. Consequently, **AXIS will be ideally suited for studies of X-ray dust echoes**. Below, we discuss how the increase in echo sensitivity promised by AXIS over current telescopes will enable a several orders of magnitude increase in the number of echoes that can be studied, and how this will enable accurate and sensitive maps of the Galactic dust distribution along many sight lines. We further discuss improvements in constraining dust properties. We also address the operational requirements and the supporting infrastructure necessary for these studies to move forward.

**Science**

**Background: X-ray Dust Tomography** X-ray sources in or behind the Galactic plane are subject to dust scattering. For continuously bright sources, this leads to the formation of smooth, diffuse dust scattering halos, but for X-ray transients (like Galactic X-ray binaries [XRBs], flaring magnetars, and gamma-ray bursts [GRBs]), the small-angle scattering by grains can lead to the formation of X-ray dust scattering echoes. If an X-ray outburst is temporally short, followed by a quick return to quiescence, the echo takes the form of discrete, concentric rings that grow in radius with time since the outburst. Each ring corresponds to a concentration (cloud) of dust along the line of sight. Some such echoes that have

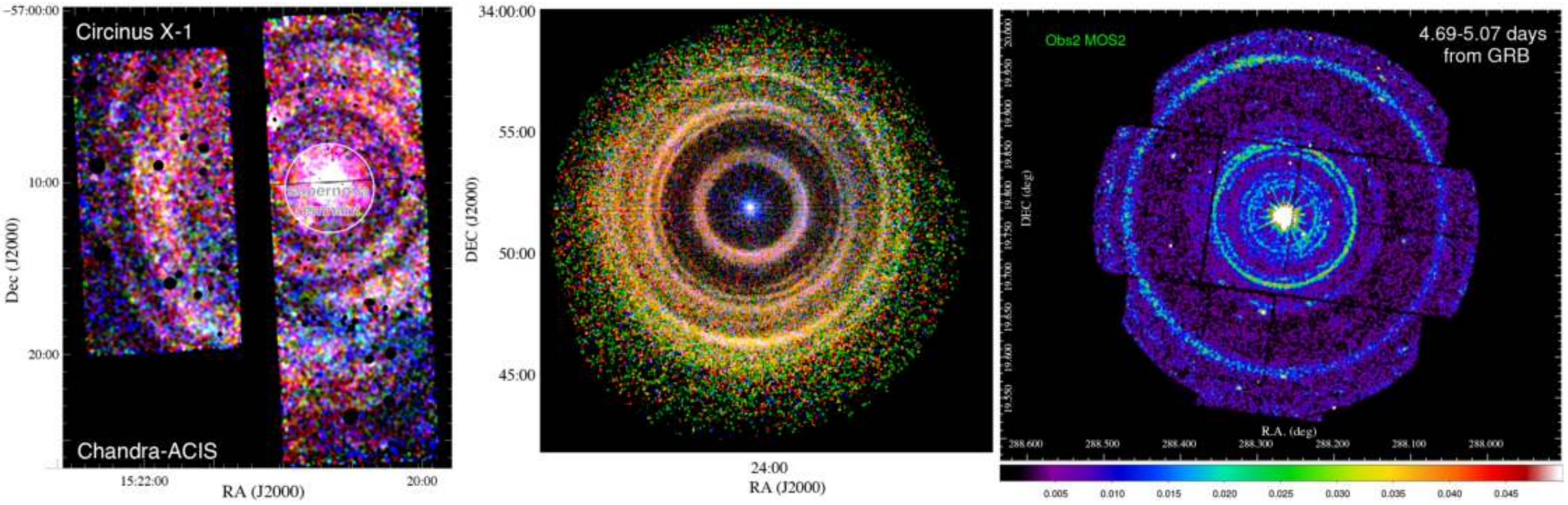

Figure 35 Examples of X-ray dust echoes from Cir X-1 [Chandra, left, 89], V404 Cyg [Swift, middle, 88], and GRB221009A [XMM, right, 208]. In each case, the echo constrains the dust column density distribution, composition, and grain size distribution along the line of sight.

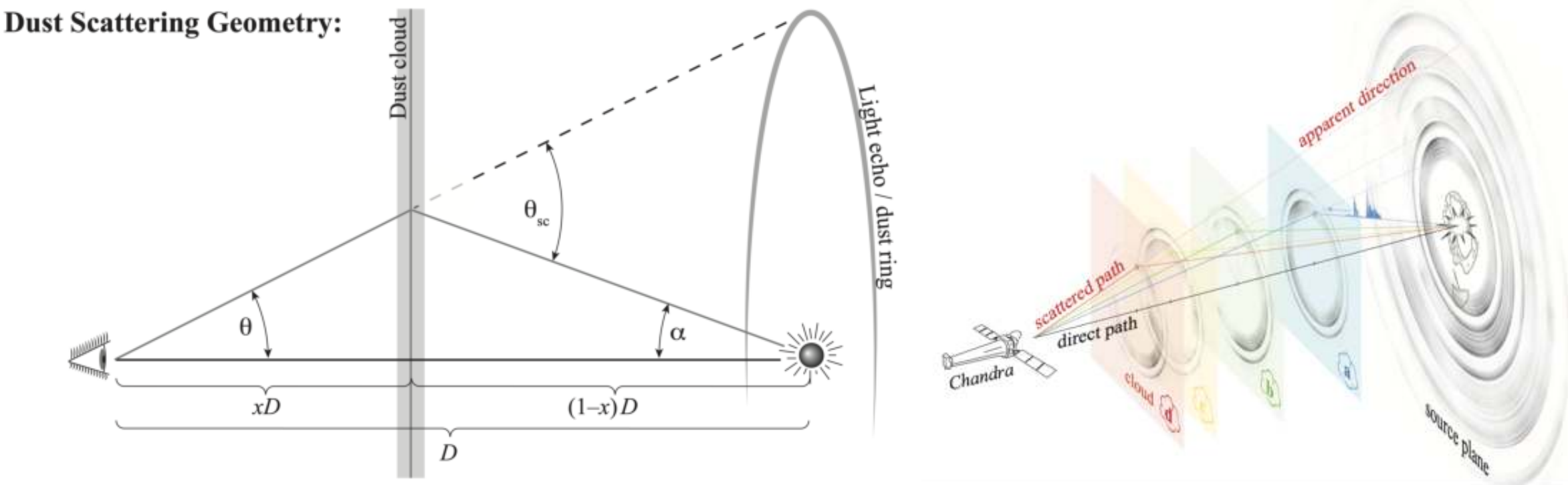

Figure 36 Left: Illustration of dust scattering geometry. The observed angle of the echo (theta) depends on the source distance $D$, the relative distance to the dust $x \equiv d/D$, and the time delay $\Delta t$. The X-ray photons are scattered by an angle $\theta_{\rm sc} \geq \theta$. Right: Illustration of how multiple concentrations of dust along the line of sight lead to the creation of complex ring echoes like those shown in Fig. 35

been observed by the current generation of imaging X-ray telescopes are shown in Fig. 35. As outlined below, the field of X-ray dust echo tomography—the study of X-ray dust echoes—enables a wide range of science, and AXIS will be the most powerful tool to advance this science in the next decade.

The simple geometry of dust scattering is illustrated in the left panel of Fig. 36. The angular size of the echo depends on the distance D to the source, the distance to the dust $d = xD$ (where $x$ is the fractional distance to the dust), and the time $\Delta t$ since the outburst as

$$\theta = \sqrt{\frac{2c\Delta t\,(1-x)}{xD}} \qquad \left[\theta(D \longrightarrow \infty) = \sqrt{\frac{2c\Delta t}{d_{\rm dust}}}\right] \tag{1}$$

where the simplifying limit $D \to \infty$ applies to the special case of light echoes from GRBs. These expressions suggest that dust echo tomography can be a powerful tool for measuring distances.

The intensity of the echo (as shown in Fig. 35) produced by a single dust cloud of equivalent Hydrogen column density $N_{\rm H}$ at distance $D$ is given by the expression [e.g., 137]

$$I_\nu = N_{\rm H} \frac{d\sigma}{d\Omega} \frac{F_\nu(-\Delta t)}{(1-x)^2} e^{-\tau_\nu} \tag{2}$$

where $F_\nu$ is the unabsorbed flux of the X-ray source a TOO time delay $\Delta t$ before the image was taken, $d\sigma/d\Omega$ is the differential dust scattering cross section per equivalent $N_{\rm H}$, and $\tau_\nu$ the total photo-electric absorption optical depth to the scattered X-rays. If the source flux is known, this expression implies that observations of dust echoes can be used to measure the dust scattering cross-section as a function of angle and energy, making X-ray dust tomography one of the most sensitive tools for constraining dust properties.

Because the dust distribution along the line of sight to the source typically contains dust grains at all distances with varying concentrations, the images of dust echoes contain numerous rings at discrete angles, as well as diffuse emission at others. A simple illustration of how the complex intensity of an image like Fig. 35 left is created can be seen in the right panel of Fig. 36. The dependence on $N_{\rm H}$ and angle in (eq. 2) therefore allows us to measure the dust column density distribution between us and the source.

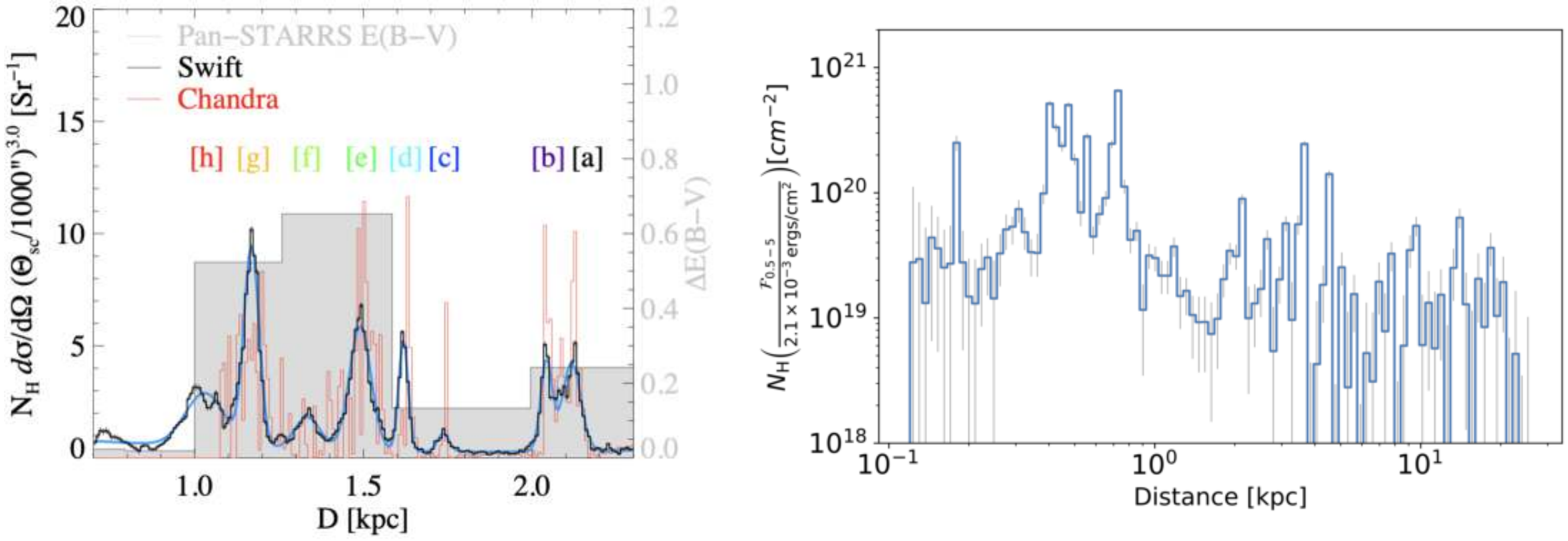

Figure 37 Left: Dust column density distribution towards V404 Cyg derived from Dust echo (observed by Swift and Chandra), demonstrating potential for mapping out the Galactic dust distribution and anchoring spiral structure, if a sufficient number of sight lines can be probed [88]. Right: Dust column density distribution towards GRB221009A from Swift [219].

The energy dependence of $d\sigma/d\Omega$ implies that, with knowledge of the source spectrum, observations of dust echoes can be used to fit spectral dust models and therefore constrain dust properties, especially when the echo can be observed at multiple times and therefore at different scattering angles.

The signal-to-noise ratio of an echo can be estimated using the expression for the flux of a single ring:

$$F_{\mathrm{ring},\nu} = \frac{2\pi c N_{\mathrm{H}}}{x\,(1-x)\,D}\frac{d\sigma}{d\Omega}\mathcal{F}_{\nu}e^{-\tau} \tag{3}$$

where $\mathcal{F} \equiv \int dt F_{\nu}$ is the fluence of the outburst creating the echo. This expression also allows us to estimate the sensitivity of AXIS for echoes from different kinds of transients, which we use below to calculate the rates of transients that AXIS will be able to follow up on. Because the occurrence of X-ray transients cannot be predicted, the field of X-ray dust tomography relies on the ability to execute TOO follow-up campaigns.

**Science Goal 1: Measure Galactic structure and dust distribution**

While it is now well established that the Milky Way is a barred spiral, the detailed structure (number and column density of spiral arms, etc.) is currently poorly constrained. Fig. 38 shows an artist's impression of what the Milky Way might look like from the North Galactic pole, but the near-far ambiguity and the limits of kinematic distance measurements (using the Galactic rotation curve) and the significant distance uncertainties of dust maps created by differential extinction beyond $\approx$ 3kpc make an accurate map of the Milky Way gas and dust distribution impossible.

For Galactic transients, (eq. 1) cannot be solved for both $x$ and $D$ independently, so absolute distance measurements for echoes from Galactic transients require a priori knowledge of either $d$ or $D$. However, in the case of GRBs, the effectively infinite distance to the source allows absolute geometric distance measurements to the Galactic foreground dust column density distribution with percent-level distance accuracy. Therefore, X-ray dust tomography can answer the fundamental question of Galactic spiral structure through a campaign of GRB follow-up observations, complemented with a TOO campaign of dust echoes from Galactic transients. Additionally, this will enable an accurate accounting of all the Galactic dust along dozens of sightlines through the plane.

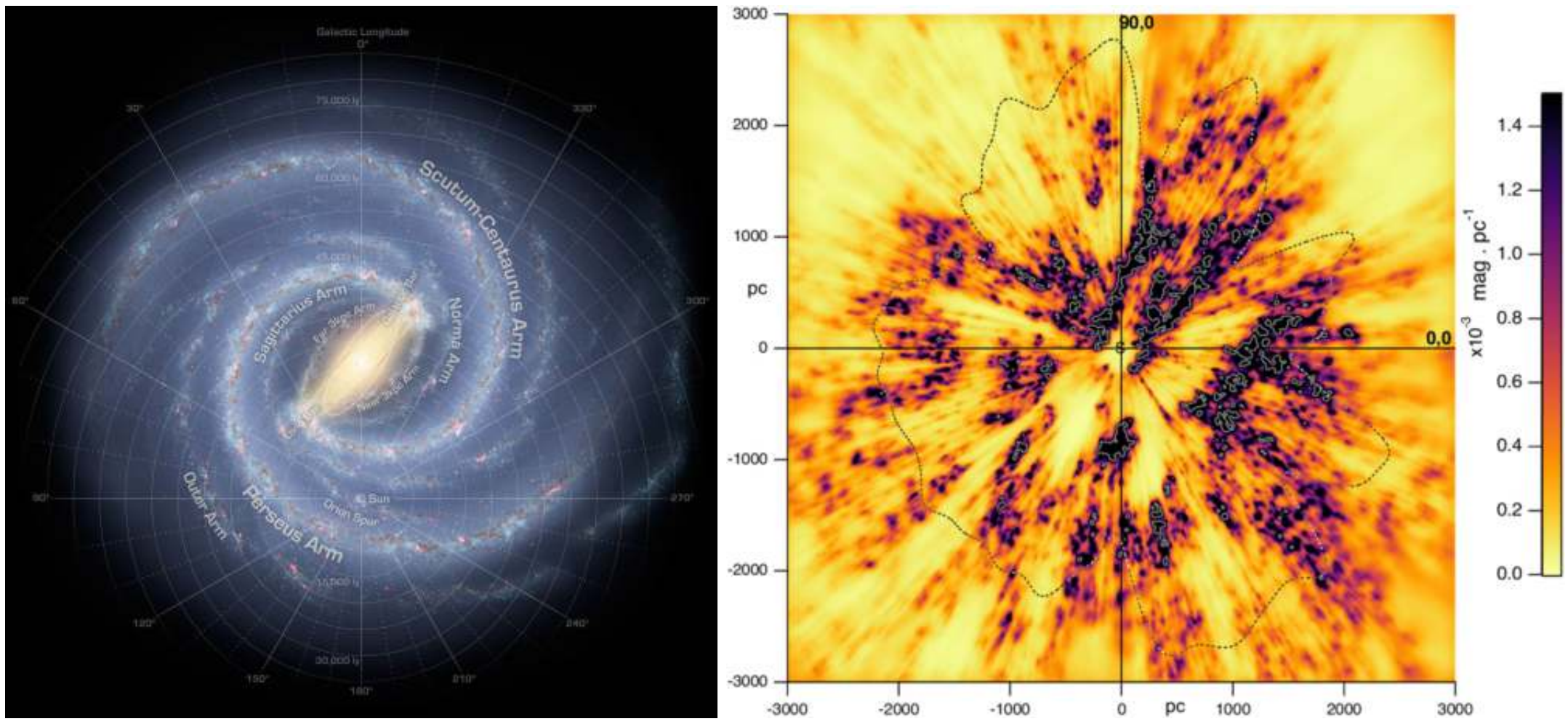

Figure 38 Left: Artist's impression of Milky Way Galaxy; spiral arm locations and numbers are subject to major uncertainties [95]. Right: Distribution of dust in Solar neighborhood [122], indicating that differential extinction dust distances are very accurate within ≈ 3kpc but carry large uncertainties beyond. Echo distances complement and anchor differential extinction distances with very high accuracy to distances across the entire Galactic plane.

These accurate dust distances from AXIS will complement and anchor the volume-filling dust distributions measured from differential extinction [193], allowing the creation of a precise 3D dust distribution through most of the Milky Way.

The combination of X-ray dust tomography with radio measurements of gas (HI, CO) will constrain the gas-to-dust ratio over an extensive range of sightlines and Galactic environments, helping to disentangle near-far distance ambiguities in kinematic distance measurements of the gas distribution. In addition, the detailed dust maps will complement the deep extinction maps we can expect from the ROMAN Galactic Plane survey [162].

**Science Goal 2: Measure the grain size distribution**

Understanding dust grain sizes has an important bearing on a wide range of astronomical fields: The grain size distribution determines dust extinction/reddening (and thus has critical observational consequences for UV-Optical telescopes) and depends strongly on the processes of dust creation and destruction, thus constraining the physics of interstellar shocks, metal enrichment by supernovae, and cold-phase dust and gas chemistry.

Because of the strong dependence of $d\sigma/d\Omega(\theta,\nu)$ on the grain size distribution $dN/da$ [e.g. 201,217], time series observations of an echo can be used to fit the scattering angle dependence of $d\sigma/d\Omega$ for different dust clouds along the line of sight and thus constrain $dN/da$. For a sample of sightlines, this will enable characterization of grain sizes (spectral shape and upper and lower grain size cutoffs) as a function of Galactic environment (informing models of grain growth and grain destruction) and constrain the variance of the Galactic dust extinction law.

**Science Goal 3: Determine dust mineralogy**

The dependence of $d\sigma/d\Omega$ on grain mineralogy will enable spectral fits of dust echoes spanning a range of dust chemistries (different contributions and chemical structures of graphites and silicates) and constrain more complex chemistries (such as the presence of ice mantles on dust grains). Understanding

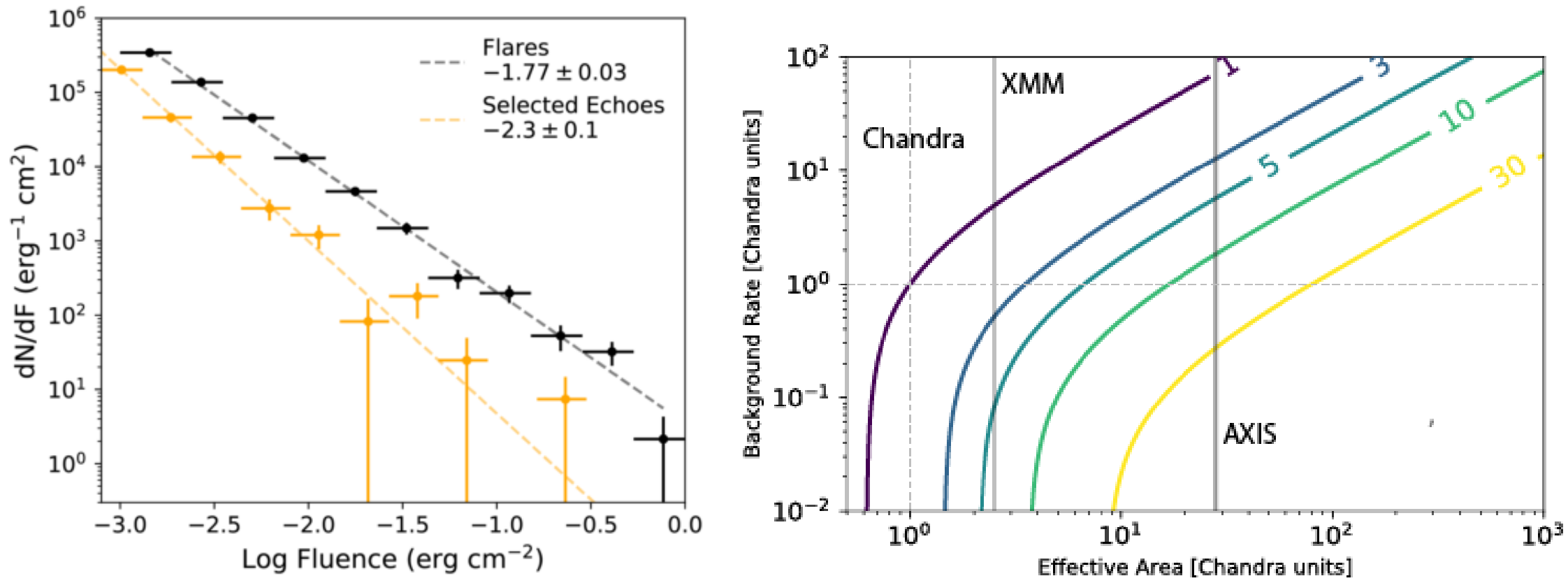

Figure 39 Left: Artist's impression of Milky Way Galaxy; spiral arm locations and numbers are subject to major uncertainties. Right: Distribution of dust in Solar neighborhood [122], indicating that differential extinction dust distances are very accurate within 3kpc but carry large uncertainties beyond. Echo distances complement and anchor differential extinction distances with very high accuracy to distances across the entire Galactic plane.

dust structure and mineralogy is key input for models of star and planet formation and, like measurements of $dN/da$, has significant consequences for modes of UV/Optical dust extinction.

**Science Goal 4: Distance determinations to Galactic X-ray transients**

In addition to measuring dust distances and concentrations, echo tomography also enables distance measurements to Galactic X-ray transients [e.g., 89]. This requires knowledge of the distance to at least one of the dust clouds along the line of sight, which can be obtained through, e.g., differential extinction distances for nearby clouds, prior echo tomography with GRBs nearby the transient, kinematic distance measurements using CO maps, or MASER measurements to individual cold clouds along the line of sight.

Distances to XRBs and transients are crucial for measuring, for example, absolute luminosities and Eddington ratios, calibrating other distance measurement techniques (such as radius expansion bursts), and situating X-ray sources within the context of Galactic structure.

**Science Campaigns:**

Below, we distinguish between Galactic and extragalactic (GRB) transients. While the scientific case for both is roughly the same, the source statistics and, therefore, the estimate of AXIS sensitivity will differ.

**[Exposure time (ks): $\approx$ 250 ksec/yr]**

**Observing description: X-ray Dust Echoes from Galactic X-ray Transients**

To enable the science potential outlined in §17, we envision a follow-up campaign for Galactic transients that satisfy simple criteria: Location behind a sufficient column of dust, a lightcurve that can be expected to create a suitably distinct echo, and a sufficiently high fluence to allow high signal-to-noise measurements of column densities, cross sections, distances, etc.

The expected sensitivity and source statistics of AXIS compared to the current generation of imaging X-ray telescopes were worked out in [50], based on flare statistics of Galactic X-ray transients observed with MAXI.

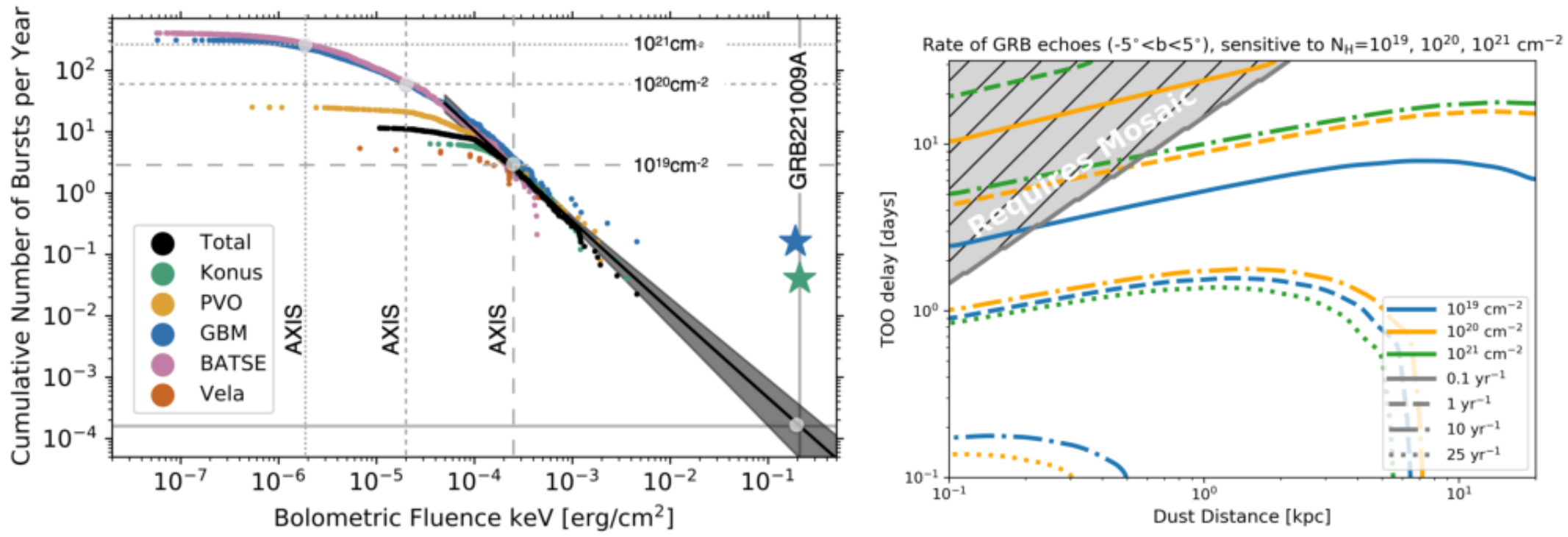

Figure 40 AXIS sensitivity to Galactic light echoes from GRBs. Left: GRB Fluence distribution [37], overplotted with AXIS sensitivity limits for different column densities. Right: Rate of GRBs within Galactic latitude of $|b| < 5^\circ$ per year for different limiting column densities per resolution element, as a function of dust distance and TOO response time. Calculations assume exposure time of 100ksec, angular resolution of 1.5”, background rates corresponding to L2 orbit. Over its mission lifetime, AXIS will be able to produce accurate geometric distance measurements to dust clouds with $N_{\rm H} \gtrsim 10^{20}$ cm$^{-2}$ along dozens of sightlines across the entire Galaxy, out to distances $d \gtrsim 20$ kpc, promising to solve long-standing questions about spiral structure, dust concentration, gas-to-dust ratios, and dust grain composition and size.

Results from the study, updated for the 2025 estimates of the effective area and background files, are shown in Fig. 39. Compared to Chandra’s sensitivity, AXIS will enable roughly an order of magnitude increase in the number of Galactic transient echoes, with an expected rate of a few per year. The comparable angular resolution will thus enable comparable science for dozens of sightlines over the nominal mission lifetime of AXIS. Importantly, Chandra observations of transient echoes are primarily limited by the severe limitations on TOO observations.

**[Exposure time (ks): 300-500 ksec/yr]**
**Observing description: Observing X-ray Dust Echoes from GRBs**

Dust echoes from GRBs have the unique advantage of eliminating the distance ambiguity between the source and the dust. This makes them ideal sources for X-ray dust tomography. Unfortunately, GRBs are generally fainter compared to the brightest Galactic X-ray transients, which has limited the number of GRB echoes that have been suitable for follow-up. The most spectacular was the echo of GRB221009A, the brightest GRB to date. Fig. 35 (right hand panel) shows the XMM image of the echo, while Fig. 37 shows the dust column density distribution towards GRB221009A, indicating that GRB echo observations can be sensitive to dust down to column densities of $N_{\rm H} \sim 10^{19}$ cm$^{-2}$ to distances of 20 kpc, i.e., across the entire galaxy.

The increased sensitivity and high angular resolution will make AXIS the ideal telescope to observe GRB dust echoes. To realize the potential for transformative science, we envision a follow-up program for GRBs within the Galactic plane (limited to Galactic latitudes of order $|b| < 5^\circ$). The GRB fluence distribution (left panel of Fig. 40) shows the column density sensitivity AXIS would have for different GRBs and the rate at which those bursts occur (that is, AXIS would be able to detect a dust concentration at a given distance down to that column density). The figure shows that observations comparable to the GRB221009A light echo can become routine with AXIS (even though GRB221009A was a once-in-$10^5$ year event when taking into account its proximity to the Galactic plane and extraordinarily high fluence.)

The sensitivity of an exposure to dust of different column densities depends on the TOO response time, the distance to the dust, and the fluence of the GRB. The right panel in Fig. 40 shows the rate at which bursts of a given column density sensitivity can be observed out to a given dust distance within $|b| < 5°$.

The dashed blue curve shows that, with a TOO response time of 1 day, AXIS will be able to observe GRBs sensitive to dust columns $N_{\rm H} \gtrsim 10^{19}\,{\rm cm}^{-2}$ about once a year. The rate of bursts sensitive to dust cloud columns of order $N_{\rm H} \gtrsim 10^{20}\,{\rm cm}^{-2}$ is approximately 10 per year. Such observations would be sufficient to measure distances to all moderate to large dust clouds and map out spiral structure with distance accuracy of order 1% to dozens of sightlines over the nominal mission lifetime, and solve the long-standing problem of spiral structure of the Milky Way. We envision this to be a key legacy program for AXIS.

**Specify critical AXIS Capabilities**

AXIS will be ideally suited for the follow-up of X-ray dust echoes. The following list of operational observatory requirements aligns closely with the AXIS specification in almost all categories. As long as telescope specifications regarding effective area, field of view, and imaging resolution remain roughly as they currently are, the main requirements posed by X-ray dust tomography revolve around operational needs like the ability to execute TOOs and the ability to either rapidly mitigate or accept bright point sources in the field of view.

- Angular resolution: Depending on the lightcurve of the transient causing the echo, dust tomography can benefit greatly from the highest possible resolution. In background-limited cases, the sensitivity to dust echo rings us inversely proportional to angular resolution. The distance accuracy achievable with dust tomography is directly proportional to resolution (see comparison of Chandra and Swift in Fig. 37). An angular resolution of 1.5" provides a distance accuracy of order 1%.
- Effective area: The diffuse signal of X-ray dust echoes requires a large effective area at energies between 1 and 4 keV. The relatively large Galactic column densities of echo sources imply peak emission around 2 keV. Figs. 5 and 6 indicate about an order-of-magnitude increase in sensitivity to Galactic transients (in terms of total numbers observable) and about 3 orders of magnitude increase vs. Swift for GRB echoes.
- FOV: The steep scattering angle dependence limits ring echoes to angular scales of $\theta \lesssim 20$ arcmin, which is very well matched to AXIS FOV. Contiguous FOV coverage will translate 1-to-1 in sensitivity.
- Background: The diffuse echo signal will benefit from the lowest possible background. We find that, while LEO would be preferable in terms of NXB, the increased observing efficiency at L2 mostly compensates for the increase in background. The sensitivity calculations made above assume the background rates published in February 2025 for an L2 orbit. The predictable time variability of the echo signal allows the CXB to be removed effectively.
- TOO requirements: As shown in Fig. 40, a response time of $\Delta t \leq 24$ hours is ideal for follow-up of GRB echoes; at minimum, a follow-up window of $\Delta t < 10$ days is required. In the case of Galactic X-ray transients, the delay times depend on source distance; a follow-up window of $\Delta t \leq 10$ days would be sufficient in most cases.
- Bright source limits: Galactic transients may re-brighten during the observation; AXIS CCDs should thus be capable of sustained observation of $\sim 1$ Crab sources. Alternatively, any risks posed by bright point sources could be mitigated by onboard logic to slew away from rapidly brightening sources in observations of Galactic transients. GRBs do not pose any requirements regarding CCD damage, given rapidly declining lightcurves.
- Scheduling or constraints: Observations of dust echoes do not require uninterrupted exposures or roll angle constraints.

- Dependence on non-AXIS triggers: X-ray dust tomography relies on triggers of X-ray transient sources from all sky monitors, both in the soft X-rays (to detect Galactic transients) and hard X-rays or gamma-rays (to detect GRBs).
- Monitoring requirements: Accurate measurements of cross sections and absolute column densities require knowledge of the transient lightcurve of the transient (which allows calculation of the intensity of the echo by convolution with column density distribution; eq. 2) and the total flux and spectrum (eq. 3). Therefore, for AXIS to enable its potential for X-ray dust tomography, one or more all-sky monitors sensitive to GRBs and the soft X-ray sky, or dedicated monitoring of the transient before the echo observation, will be required for maximum science output.

**[Joint Observations and synergies with other observatories in the 2030s:]** ROMAN/WFIRST, SKA, ALMA, ATHENA
**[Special Requirements:]** TOO ($<$ 24 hrs)

**Acknowledgments:** We thank everyone who has contributed to the development of the AXIS Probe mission concept.

## References:

1. Abbasi, R., Ackermann, M., Adams, J., et al. 2022, ApJ, 938, 38
2. —. 2022, ApJ, 939, 116
3. —. 2023, ApJ, 946, L26
4. Abbott, B. P., Abbott, R., Abbott, T. D., et al. 2017, Phys. Rev. Lett., 119, 161101
5. —. 2017, ApJ, 848, L12
6. —. 2018, Phys. Rev. Lett., 121, 161101
7. Aftab, N., & Paul, B. 2024, New A, 106, 102109
8. Aftab, N., Paul, B., & Kretschmar, P. 2019, ApJS, 243, 29
9. Agazie, G., Anumarlapudi, A., Archibald, A. M., et al. 2023, ApJ, 952, L37
10. —. 2023, ApJ, 951, L8
11. Alabarta, K., Homan, J., Russell, D. M., et al. 2023, The Astronomer's Telegram, 16262, 1
12. Amaro-Seoane, P., Audley, H., Babak, S., et al. 2017, arXiv e-prints, arXiv:1702.00786
13. Anand, N., Shahid, M., & Resmi, L. 2018, MNRAS, 481, 4332
14. Andreoni, I., Coughlin, M. W., Perley, D. A., et al. 2022, Nature, 612, 430
15. Andreoni, I., Margutti, R., Banovetz, J., et al. 2024, arXiv e-prints, arXiv:2411.04793
16. Arcodia, R., Merloni, A., Buchner, J., et al. 2024, A&A, 684, L14
17. Arcodia, R., Linial, I., Miniutti, G., et al. 2024, A&A, 690, A80
18. Baglio, M. C., Campana, S., D'Avanzo, P., et al. 2017, A&A, 600, A109
19. Baglio, M. C., Saikia, P., Russell, D. M., et al. 2022, ApJ, 930, 20
20. Balasubramanian, A., Corsi, A., Mooley, K. P., et al. 2021, ApJ, 914, L20
21. Barausse, E., Dey, K., Crisostomi, M., et al. 2023, Ph. Rev. D, 108, 103034
22. Bartos, I., & Kowalski, M. 2017, Multimessenger Astronomy
23. Bécsy, B., Cornish, N. J., & Kelley, L. Z. 2022, ApJ, 941, 119
24. Behroozi, P. S., Ramirez-Ruiz, E., & Fryer, C. L. 2014, ApJ, 792, 123
25. Beniamini, P., Gill, R., & Granot, J. 2022, MNRAS, 515, 555
26. Beniamini, P., Granot, J., & Gill, R. 2020, MNRAS, 493, 3521
27. Bernardini, F., Russell, D. M., Shaw, A. W., et al. 2016, ApJ, 818, L5
28. Bernardini, M. G., Margutti, R., Mao, J., Zaninoni, E., & Chincarini, G. 2012, A&A, 539, A3
29. Blecha, L., Sijacki, D., Kelley, L. Z., et al. 2016, MNRAS, 456, 961
30. Bloom, J. S., Giannios, D., Metzger, B. D., et al. 2011, Science, 333, 203
31. Borhanian, S., & Sathyaprakash, B. S. 2024, Ph. Rev. D, 110, 083040
32. Bozzo, E., Giunta, A., Cusumano, G., et al. 2011, A&A, 531, A130
33. Bozzo, E., Bhalerao, V., Pradhan, P., et al. 2016, A&A, 596, A16
34. Brethauer, D., Margutti, R., Milisavljevic, D., et al. 2022, ApJ, 939, 105
35. Bruch, R. J., Gal-Yam, A., Yaron, O., et al. 2023, ApJ, 952, 119
36. Burbidge, E. M., Burbidge, G. R., Fowler, W. A., & Hoyle, F. 1957, Reviews of Modern Physics, 29, 547
37. Burns, E., Svinkin, D., Fenimore, E., et al. 2023, ApJ, 946, L31
38. Bučar Bricman, K., van Velzen, S., Nicholl, M., & Gomboc, A. 2023, ApJS, 268, 13
39. Cameron, A. G. W. 1973, Space Science Reviews, 15, 121
40. Chakraborty, J., Kara, E., Masterson, M., et al. 2021, ApJ, 921, L40
41. Chakraborty, J., Arcodia, R., Kara, E., et al. 2024, ApJ, 965, 12
42. Chakraborty, J., Kara, E., Arcodia, R., et al. 2025, ApJ, 983, L39
43. Chakraborty, J., Kosec, P., Kara, E., et al. 2025, ApJ, 984, 124
44. Chandra, P., Chevalier, R. A., Chugai, N., Fransson, C., & Soderberg, A. M. 2015, ApJ, 810, 32
45. Chase, E. A., O'Connor, B., Fryer, C. L., et al. 2022, ApJ, 927, 163

46. Chevalier, R. A., & Irwin, C. M. 2011, ApJ, 729, L6
47. —. 2012, ApJ, 747, L17
48. Colombo, A., Sharan Salafia, O., Ghirlanda, G., et al. 2025, arXiv e-prints, arXiv:2503.00116
49. Colpi, M., Danzmann, K., Hewitson, M., et al. 2024, arXiv e-prints, arXiv:2402.07571
50. Corrales, L., Mills, B. S., Heinz, S., & Williger, G. M. 2019, ApJ, 874, 155
51. Cowan, J. J., Sneden, C., Lawler, J. E., et al. 2021, Reviews of Modern Physics, 93, 015002
52. Dai, L., McKinney, J. C., Roth, N., Ramirez-Ruiz, E., & Miller, M. C. 2018, ApJ, 859, L20
53. D'Avanzo, P., Campana, S., Salafia, O. S., et al. 2018, A&A, 613, L1
54. Dessart, L., Hillier, D. J., & Audit, E. 2017, A&A, 605, A83
55. Dong-Páez, C. A., Volonteri, M., Beckmann, R. S., et al. 2023, A&A, 676, A2
56. D'Orazio, D. J., Haiman, Z., & MacFadyen, A. 2013, MNRAS, 436, 2997
57. Dosopoulou, F., & Antonini, F. 2017, ApJ, 840, 31
58. Dwarkadas, V. V. 2025, Universe, 11, 161
59. Dwarkadas, V. V., & Gruszko, J. 2012, MNRAS, 419, 1515
60. Eddins, A., Lee, K.-H., Corsi, A., et al. 2023, ApJ, 948, 125
61. Eftekhari, T., Tchekhovskoy, A., Alexander, K. D., et al. 2024, ApJ, 974, 149
62. EPTA Collaboration, InPTA Collaboration, Antoniadis, J., et al. 2023, A&A, 678, A50
63. —. 2024, A&A, 685, A94
64. Evans, P. A., Cenko, S. B., Kennea, J. A., et al. 2017, Science, 358, 1565
65. Farris, B. D., Duffell, P., MacFadyen, A. I., & Haiman, Z. 2014, ApJ, 783, 134
66. —. 2015, MNRAS, 447, L80
67. Fiorillo, D. F. G., Petropoulou, M., Comisso, L., Peretti, E., & Sironi, L. 2024, ApJ, 961, L14
68. Forsblom, S. V., Poutanen, J., Tsygankov, S. S., et al. 2023, The Astrophysical Journal Letters, 947, L20
69. Fortin, F., Chaty, S., & Sander, A. 2020, The Astrophysical Journal, 894, 86
70. Franchini, A., Bonetti, M., Lupi, A., et al. 2023, A&A, 675, A100
71. Freiburghaus, C., Rosswog, S., & Thielemann, F. K. 1999, ApJ, 525, L121
72. Gao, S., Fedynitch, A., Winter, W., & Pohl, M. 2019, Nature Astronomy, 3, 88
73. Garrappa, S., Buson, S., Sinapius, J., et al. 2024, A&A, 687, A59
74. Gezari, S. 2021, ARA&A, 59, 21
75. Goldberg, J. A., Jiang, Y.-F., & Bildsten, L. 2022, ApJ, 933, 164
76. Gomez, S., Nicholl, M., Berger, E., et al. 2024, MNRAS, 535, 471
77. Goodwin, A. J., Russell, D. M., Galloway, D. K., et al. 2020, MNRAS, 498, 3429
78. Grotova, I., Rau, A., Baldini, P., et al. 2025, A&A, 697, A159
79. Guarini, E., Tamborra, I., & Margutti, R. 2022, ApJ, 935, 157
80. Guetta, D., Hooper, D., Alvarez-Mun, J., Halzen, F., & Reuveni, E. 2004, Astroparticle Physics, 20, 429
81. Gutiérrez, C. P., Mattila, S., Lundqvist, P., et al. 2024, ApJ, 977, 162
82. Gutiérrez, E. M., Combi, L., & Ryan, G. 2024, arXiv e-prints, arXiv:2405.14843
83. Häberle, M., Neumayer, N., Seth, A., et al. 2024, Nature, 631, 285
84. Hallinan, G., Corsi, A., Mooley, K. P., et al. 2017, Science, 358, 1579
85. Hameury, J. M., Knigge, C., Lasota, J. P., Hambsch, F. J., & James, R. 2020, A&A, 636, A1
86. Hammerstein, E., van Velzen, S., Gezari, S., et al. 2023, ApJ, 942, 9
87. Hayasaki, K. 2021, Nature Astronomy, 5, 436
88. Heinz, S., Corrales, L., Smith, R., et al. 2016, ApJ, 825, 15
89. Heinz, S., Burton, M., Braiding, C., et al. 2015, ApJ, 806, 265
90. Hernández-García, L., Chakraborty, J., Sánchez-Sáez, P., et al. 2025, Nature Astronomy, 9, 895
91. Hillier, D. J., & Dessart, L. 2012, MNRAS, 424, 252
92. Hoffman, L., & Loeb, A. 2007, MNRAS, 377, 957
93. Homan, J., Alabarta, K., Russell, D. M., et al. 2023, The Astronomer's Telegram, 16272, 1
94. Hotokezaka, K., & Piran, T. 2015, MNRAS, 450, 1430

95. Hurt, E.-C. K. 2013, "Artist's impression of the Milky Way"
96. Iacovelli, F., Mancarella, M., Foffa, S., & Maggiore, M. 2022, ApJ, 941, 208
97. IceCube Collaboration. 2013, Science, 342, 1242856
98. IceCube Collaboration, Aartsen, M. G., Ackermann, M., et al. 2018, Science, 361, eaat1378
99. IceCube Collaboration, Abbasi, R., Ackermann, M., et al. 2022, Science, 378, 538
100. Icecube Collaboration, Abbasi, R., Ackermann, M., et al. 2023, Science, 380, 1338
101. Irwin, C. M., Linial, I., Nakar, E., Piran, T., & Sari, R. 2021, MNRAS, 508, 5766
102. Islam, N., Maitra, C., Pradhan, P., & Paul, B. 2015, MNRAS, 446, 4148
103. Ivezić, Ž., Kahn, S. M., Tyson, J. A., et al. 2019, ApJ, 873, 111
104. Jacobson-Galán, W. V., Dessart, L., Davis, K. W., et al. 2024, ApJ, 970, 189
105. Jacobson-Galán, W. V., Davis, K. W., Kilpatrick, C. D., et al. 2024, ApJ, 972, 177
106. Jacobson-Galán, W. V., Dessart, L., Kilpatrick, C. D., et al. 2025, arXiv e-prints, arXiv:2508.11747
107. Jain, C., Paul, B., & Dutta, A. 2010, MNRAS, 409, 755
108. Jakobsson, P., Levan, A., Fynbo, J. P. U., et al. 2006, A&A, 447, 897
109. Jin, C. C., Li, D. Y., Jiang, N., et al. 2025, arXiv e-prints, arXiv:2501.09580
110. Kabiraj, S., Islam, N., & Paul, B. 2020, MNRAS, 491, 1491
111. Kasliwal, M. M., Nakar, E., Singer, L. P., et al. 2017, Science, 358, 1559
112. Kathirgamaraju, A., Giannios, D., & Beniamini, P. 2019, MNRAS, 487, 3914
113. Keivani, A., Murase, K., Petropoulou, M., et al. 2018, ApJ, 864, 84
114. Kelley, L. Z., Blecha, L., & Hernquist, L. 2017, MNRAS, 464, 3131
115. Kelley, L. Z., Blecha, L., Hernquist, L., Sesana, A., & Taylor, S. R. 2018, MNRAS, 477, 964
116. Kimura, S. S. 2022, arXiv e-prints, arXiv:2202.06480
117. Klein, A., Barausse, E., Sesana, A., et al. 2016, Ph. Rev. D, 93, 024003
118. Knight, A. H., van den Eijnden, J., Ingram, A., et al. 2025, MNRAS, 538, 2058
119. Kosec, P., Kara, E., Brenneman, L., et al. 2025, ApJ, 978, 10
120. Kulkarni, S. R., Harrison, F. A., Grefenstette, B. W., et al. 2021, arXiv e-prints, arXiv:2111.15608
121. Kumar, P., & Zhang, B. 2015, Phys. Rep., 561, 1
122. Lallement, R., Babusiaux, C., Vergely, J. L., et al. 2019, A&A, 625, A135
123. Lasota, J.-P. 2001, New A Rev., 45, 449
124. Lin, D., Strader, J., Carrasco, E. R., et al. 2018, Nature Astronomy, 2, 656
125. Lin, D., Strader, J., Romanowsky, A. J., et al. 2020, ApJ, 892, L25
126. Linial, I., & Metzger, B. D. 2024, ApJ, 973, 101
127. Linial, I., & Quataert, E. 2024, MNRAS, 527, 4317
128. Lu, W., & Quataert, E. 2023, MNRAS, 524, 6247
129. Lucarelli, F., Oganesyan, G., Montaruli, T., et al. 2023, A&A, 672, A102
130. MacFadyen, A. I., & Milosavljević, M. 2008, ApJ, 672, 83
131. Malewicz, J., Ballantyne, D. R., Bogdanović, T., Brenneman, L., & Dauser, T. 2025, arXiv e-prints, arXiv:2504.14018
132. Mangiagli, A., Klein, A., Bonetti, M., et al. 2020, Ph. Rev. D, 102, 084056
133. Margutti, R., Zaninoni, E., Bernardini, M. G., et al. 2013, MNRAS, 428, 729
134. Margutti, R., Soderberg, A. M., Wieringa, M. H., et al. 2013, ApJ, 778, 18
135. Margutti, R., Chornock, R., Metzger, B. D., et al. 2018, ApJ, 864, 45
136. Masterson, M., De, K., Panagiotou, C., et al. 2024, ApJ, 961, 211
137. Mathis, J. S., Rumpl, W., & Nordsieck, K. H. 1977, ApJ, 217, 425
138. McCarthy, K. S., Zheng, Z., & Ramirez-Ruiz, E. 2020, MNRAS, 499, 5220
139. McDonough, K., Hughes, K., Smith, D., & Vieregg, A. G. 2024, J. Cosmol. Astropart. Phys., 2024, 035
140. Menou, K., Esin, A. A., Narayan, R., et al. 1999, ApJ, 520, 276
141. Merloni, A., Lamer, G., Liu, T., et al. 2024, A&A, 682, A34
142. Mészáros, P., & Waxman, E. 2001, Phys. Rev. Lett., 87, 171102
143. Miles, M. T., Shannon, R. M., Reardon, D. J., et al. 2025, MNRAS, 536, 1489

144. Miller, A. A., Abrams, N. S., Aldering, G., et al. 2025, arXiv e-prints, arXiv:2503.14579
145. Miniutti, G., Giustini, M., Arcodia, R., et al. 2023, A&A, 670, A93
146. Miniutti, G., Franchini, A., Bonetti, M., et al. 2025, A&A, 693, A179
147. Mockler, B., Guillochon, J., & Ramirez-Ruiz, E. 2019, ApJ, 872, 151
148. Murase, K. 2022, ApJ, 941, L17
149. Murase, K., Ioka, K., Nagataki, S., & Nakamura, T. 2006, ApJ, 651, L5
150. Murase, K., Kimura, S. S., & Mészáros, P. 2020, 125, 011101
151. Murase, K., Oikonomou, F., & Petropoulou, M. 2018, ApJ, 865, 124
152. Naik, S., & Paul, B. 2012, Bulletin of the Astronomical Society of India, 40, 503
153. Nakar, E., & Piran, T. 2011, Nature, 478, 82
154. Nakar, E., & Sari, R. 2010, ApJ, 725, 904
155. Nayana, A. J., Margutti, R., Wiston, E., et al. 2025, ApJ, 985, 51
156. Negro, M., Wadiasingh, Z., Younes, G., et al. 2025, arXiv e-prints, arXiv:2509.03763
157. Nicholl, M., Pasham, D. R., Mummery, A., et al. 2024, Nature, 634, 804
158. Noble, S. C., Mundim, B. C., Nakano, H., et al. 2012, ApJ, 755, 51
159. O'Brien, P. T., Willingale, R., Osborne, J., et al. 2006, ApJ, 647, 1213
160. Odaka, H., Khangulyan, D., Tanaka, Y. T., et al. 2013, The Astrophysical Journal, 767, 70
161. Oertel, M., Hempel, M., Klähn, T., & Typel, S. 2017, Reviews of Modern Physics, 89, 015007
162. Paladini, R., Zucker, C., Benjamin, R., et al. 2023, arXiv e-prints, arXiv:2307.07642
163. Pan, X., Li, S.-L., & Cao, X. 2023, ApJ, 952, 32
164. Pasham, D. R., Coughlin, E., van Velzen, S., & Hinkle, J. 2025, arXiv e-prints, arXiv:2502.12078
165. Pasham, D. R., Cenko, S. B., Levan, A. J., et al. 2015, ApJ, 805, 68
166. Petropoulou, M., Giannios, D., & Dimitrakoudis, S. 2014, MNRAS, 445, 570
167. Pian, E., D'Avanzo, P., Benetti, S., et al. 2017, Nature, 551, 67
168. Piro, L., Colpi, M., Aird, J., et al. 2023, MNRAS, 521, 2577
169. Pitik, T., Tamborra, I., & Petropoulou, M. 2021, J. Cosmol. Astropart. Phys., 2021, 034
170. Pradhan, P., Bozzo, E., & Paul, B. 2018, A&A, 610, A50
171. Pradhan, P., Paul, B., Paul, B. C., Bozzo, E., & Belloni, T. M. 2015, MNRAS, 454, 4467
172. Pugliese, G., Falcke, H., Wang, Y. P., & Biermann, P. L. 2000, A&A, 358, 409
173. Quintin, E., Webb, N. A., Guillot, S., et al. 2023, A&A, 675, A152
174. Rampy, R. A., Smith, D. M., & Negueruela, I. 2009, The Astrophysical Journal, 707, 243
175. Rastinejad, J. C., Fong, W., Kilpatrick, C. D., Nicholl, M., & Metzger, B. D. 2025, ApJ, 979, 190
176. Reardon, D. J., Zic, A., Shannon, R. M., et al. 2023, ApJ, 951, L6
177. Reig, P. 2011, Ap&SS, 332, 1
178. Reusch, S., Stein, R., Kowalski, M., et al. 2022, Phys. Rev. Lett., 128, 221101
179. Rezzolla, L., Giacomazzo, B., Baiotti, L., et al. 2011, ApJ, 732, L6
180. Ricarte, A., Tremmel, M., Natarajan, P., & Quinn, T. 2021, ApJ, 916, L18
181. —. 2021, ApJ, 916, L18
182. Ricci, R., Troja, E., Bruni, G., et al. 2021, MNRAS, 500, 1708
183. Rikame, K., Paul, B., Sharma, R., Jithesh, V., & Paul, K. T. 2024, MNRAS, 529, 3360
184. Roedig, C., Krolik, J. H., & Miller, M. C. 2014, ApJ, 785, 115
185. Ronchini, S., Branchesi, M., Oganesyan, G., et al. 2022, A&A, 665, A97
186. Rudolph, A., Bošnjak, Ž., Palladino, A., Sadeh, I., & Winter, W. 2022, MNRAS, 511, 5823
187. Ruiz, M., Lang, R. N., Paschalidis, V., & Shapiro, S. L. 2016, ApJ, 824, L6
188. Ryan, G., van Eerten, H., Piro, L., & Troja, E. 2020, ApJ, 896, 166
189. Ryan, G., van Eerten, H., Troja, E., et al. 2024, ApJ, 975, 131
190. Ryu, T., Perna, R., Haiman, Z., Ostriker, J. P., & Stone, N. C. 2018, MNRAS, 473, 3410
191. Sakamoto, T., Troja, E., Aoki, K., et al. 2013, ApJ, 766, 41
192. Sazonov, S., Gilfanov, M., Medvedev, P., et al. 2021, MNRAS, 508, 3820

193. Schlafly, E. F., & Finkbeiner, D. P. 2011, ApJ, 737, 103
194. Schroeder, G., Margalit, B., Fong, W.-f., et al. 2020, ApJ, 902, 82
195. Selsing, J., Krühler, T., Malesani, D., et al. 2018, A&A, 616, A48
196. Senno, N., Murase, K., & Mészáros, P. 2016, Phys. Rev. D, 93, 083003
197. Seth, A. C., van den Bosch, R., Mieske, S., et al. 2014, Nature, 513, 398
198. Severgnini, P., Cicone, C., Della Ceca, R., et al. 2018, MNRAS, 479, 3804
199. Shvartzvald, Y., Waxman, E., Gal-Yam, A., et al. 2024, ApJ, 964, 74
200. Smak, J. 1984, PASP, 96, 5
201. Smith, R. K., Valencic, L. A., & Corrales, L. 2016, ApJ, 818, 143
202. Soderberg, A. M., Berger, E., Page, K. L., et al. 2008, Nature, 453, 469
203. Stein, R., van Velzen, S., Kowalski, M., et al. 2021, Nature Astronomy, 5, 510
204. Stein, R., Reusch, S., Franckowiak, A., et al. 2023, MNRAS, 521, 5046
205. Stone, N., & Loeb, A. 2011, MNRAS, 412, 75
206. Sun, H., Zhang, B., & Li, Z. 2015, ApJ, 812, 33
207. Tchekhovskoy, A., Metzger, B. D., Giannios, D., & Kelley, L. Z. 2014, MNRAS, 437, 2744
208. Tiengo, A., Pintore, F., Vaia, B., et al. 2023, ApJ, 946, L30
209. Troja, E. 2023, Universe, 9, 245
210. Troja, E., Piro, L., van Eerten, H., et al. 2017, Nature, 551, 71
211. Troja, E., Ryan, G., Piro, L., et al. 2018, Nature Communications, 9, 4089
212. Troja, E., O'Connor, B., Ryan, G., et al. 2022, MNRAS, 510, 1902
213. Urquhart, R., & Soria, R. 2016, ApJ, 831, 56
214. Vurm, I., Linial, I., & Metzger, B. D. 2025, ApJ, 983, 40
215. Waxman, E., & Bahcall, J. 1997, Phys. Rev. Lett., 78, 2292
216. Waxman, E., & Katz, B. 2017, in Handbook of Supernovae, ed. A. W. Alsabti & P. Murdin, 967
217. Weingartner, J. C., & Draine, B. T. 2001, ApJ, 548, 296
218. Wen, S., Jonker, P. G., Stone, N. C., & Zabludoff, A. I. 2021, ApJ, 918, 46
219. Williams, M. A., Kennea, J. A., Dichiara, S., et al. 2023, ApJ, 946, L24
220. Xu, H., Chen, S., Guo, Y., et al. 2023, Research in Astronomy and Astrophysics, 23, 075024
221. Yao, Y., Ravi, V., Gezari, S., et al. 2023, ApJ, 955, L6
222. Yao, Y., Chornock, R., Ward, C., et al. 2025, ApJ, 985, L48
223. Yoshida, K., Petropoulou, M., Murase, K., & Oikonomou, F. 2023, ApJ, 954, 194
224. Zauderer, B. A., Berger, E., Margutti, R., et al. 2013, ApJ, 767, 152
225. Zauderer, B. A., Berger, E., Soderberg, A. M., et al. 2011, Nature, 476, 425
226. Zhang, B., Fan, Y. Z., Dyks, J., et al. 2006, ApJ, 642, 354
227. Zhang, B., & Kumar, P. 2013, Phys. Rev. Lett., 110, 121101
228. Zhang, S.-N., Santangelo, A., Xu, Y., et al. 2022, in Society of Photo-Optical Instrumentation Engineers (SPIE) Conference Series, Vol. 12181, Space Telescopes and Instrumentation 2022: Ultraviolet to Gamma Ray, ed. J.-W. A. den Herder, S. Nikzad, & K. Nakazawa, 121811W
229. Zheng, Z., & Ramirez-Ruiz, E. 2007, ApJ, 665, 1220
230. Zhou, C., Zeng, Y., & Pan, Z. 2025, ApJ, 985, 242
231. Zurita, C., Casares, J., & Shahbaz, T. 2003, ApJ, 582, 369

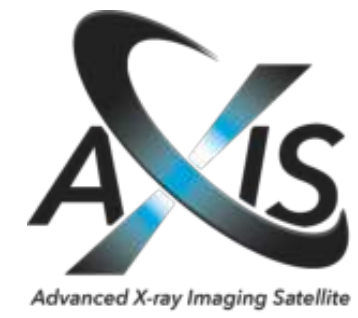

*AXIS CO-SNR GO Science Book*

# AXIS Guest Observer Community Science Cases: Compact Objects, Supernova Remnants, and Associated High-Energy Phenomena

**R. Amato[1], H. An[2], M. Bachetti[3], M. C. Baglio[4], A. Bahramian[5], S. Balman[6], A. Bamba[7], T. Bao[8], W. Becker[9], A. Belfiore[10], A. Bodaghee[11], P. Bordas[12], A. Borghese[13], G. Brunelli[14], K. Burdge[15], F. Capitanio[16], S. Casanova[17], D. Castro[18], A. Chatterjee[19], F. Coti Zelati[20], K. Dage[21], A. Decourchelle[22], N. Degenaar[23], S. DiKerby[24], N. Doerksen[25], P. Draghis[15], A. R. Escorial[13], G. Ferrand[25,26], F. Fürst[13], , F. Fraschetti[18], C. Fryer[27], S. Fujimoto[28], S. Gagnon[29], J. D. Gelfand[30], A. Gnarini[31], D. Haggard[32], J. Hare[33], C. Heinke[34], W. Ho[35], J. Homan[36], M. Imbrogno[37], N. Islam[38], G. L. Israel[1], A. Jaodand[18], O. Kargaltsev[29], D. Kirmizibayrak[39], N. Klingler[40], R. Kyer[24], B. Mac Intyre[25], A. MacMaster[25], C. Maitra[9,66], W. Majid[41], L. Mallick[25,42], S. Mandel[43], M. Mayer[44], K. Mori[43], M. Ng[32], K. Oh[24], L. Olivera-Nieto[45], M. Orio[46], D. Page[47], C. Pinto[48], G. Ponti[49], B. Posselt[50], P. Pradhan[51], M. Reynolds[52], G. A. Rodriguez-Castillo[53], S. Safi-Harb[25], C. Salvaggio[54], R. Salvaterra[10], I. Sander[25], L. R. Sandoval[55], M. Sasaki[56], A. Shporer[15], N. Sridhar[57], N. Steinle[25], C. Stringfield[43], J. Suherli[25], A. Tarana[16], Y. Terada[58], L. Townsend[59], C. Treyturik[25], N. Tsuji[60], G. Vasilopoulos[61], Z. Wadiasingh[62], D. Walton[63], A. Wolter[54], R. Wijnands[23], J. Woo[43], T. Woods[25], H. Yang[64], G. Younes[65], S. Zepf[24], S. Zhang[24], X. Zhang[12,17]**

[1] INAF-OAR, Italy [2] Chungbuk National University, South Korea [3] INAF-OACa, Italy [4] INAF-OAB, Italy [5] CIRA, Australia [6] Istanbul Univ., Turkey [7] University of Tokyo, Japan [8] INAF-Merate, Italy [9] MPE, Germany [10] INAF-IASF Milan, Italy [11] Eureka Scientific [12] University of Barcelona, Spain [13] ESA/ESAC, Spain [14] INAF Bologna, Italy [15] MIT, USA [16] INAF/IAPS, Italy [17] Institute of Nuclear Physics, Polish Academy of Science, Krakow, Poland [18] Harvard CfA [19] UPES University of Calcutta, Dehradun, India [20] ICE-CSIC, IEEC, Spain [21] Curtin, Australia [22] CEA, Saclay, France [23] University of Amsterdam, Netherlands [24] Michigan State University, USA [25] University of Manitoba, Canada [26] RIKEN, Japan [27] Los Alamos National Laboratory, USA [28] NITKC, Japan [29] George Washington U., USA [30] NYUAD, UAE [31] Università degli Studi Roma Tre, Italy [32] McGill University, Canada [33] NASA GSFC, CUA, CRESST II, USA [34] University of Alberta, Canada [35] Haverford College, USA [36] Netherlands [37] INAF, OAR, Italy [38] NASA GSFC and UMBC, USA [39] Caltech, USA [40] NASA GSFC, USA [41] JPL, USA [42] CITA, University of Toronto, Canada [43] Columbia University, USA [44] FAU Erlangen, Germany [45] MPIK, Germany [46] University of Wisconsin, Madison [47] The National Autonomous University of Mexico, Mexico [48] INAF - IASF Palermo, Italy [49] INAF-Merate, Italy [50] University of Oxford, UK [51] Embry-Riddle Aeronautical University, Arizona, USA [52] The Ohio State University, USA [53] INAF-IASF Palermo, Italy [54] INAF-OA Brera, Italy [55] The University of Texas Rio Grande Valley, USA [56] Friedrich-Alexander-University Erlangen-Nürnberg, Germany [57] Stanford University, USA [58] Saitama University, Japan [59] Southern African Large Telescope and South African Astronomical Observatory, Cape Town, South Africa [60] ICRR, University of Tokyo, Japan [61] NKUA, Greece [62] NASA GSFC/UMD/CRESST II, USA [63] University of Hertfordshire, UK [64] IRAP, France [65] NASA GSFC and UMBC, USA [66] Inter University Centre for Astronomy  Astrophysics, Pune, India

---

# Part I
# Introduction

Compact objects (CO) and Supernova Remnants (SNRs) provide nearby laboratories to probe the fate of stars and the way they impact, and are impacted by, their surrounding medium. The past five decades of multi-wavelength, especially high-energy, observations have significantly advanced our understanding of these objects, showing that they are essential to our understanding of some of the most mysterious and energetic events in the Universe, including fast radio bursts (FRBs) and gravitational wave sources. However, many questions remain unanswered. These *include*: What powers the diversity of explosive phenomena across the electromagnetic spectrum? What are the mass and spin distributions of neutron stars and stellar mass black holes? How do interacting compact binaries with white dwarfs - the electromagnetic counterparts to gravitational wave LISA sources - form and behave? Which objects inhabit the faint end of the X-ray luminosity function? How do relativistic winds impact their surroundings? How do supernova remnant shocks impact cosmic magnetism? What do neutron star kicks reveal about fundamental physics and supernova explosions?

This plethora of questions will be addressed with *AXIS* - the Advanced X-ray Imaging Satellite - a NASA Probe Mission Concept designed to be the premier high-angular resolution X-ray mission for the next decade [482,483,509]. *AXIS*, thanks to its combined (i) unprecedented imaging resolution over its full field of view, (ii) unprecedented sensitivity to faint objects due to its large effective area and low background, and (iii) rapid response capability, will provide a giant leap in discovering and identifying populations of compact objects (isolated and binaries), particularly in crowded regions such as globular clusters and the Galactic Centre, while addressing science questions and priorities of the US Decadal Survey on Astronomy and Astrophysics (Astro2020)[1]. In particular, the diversity of science cases highlighted in this Guest Observer CO-SNR Science Book focus on the Astro2020 Priority Themes: *New Messengers and New Physics* (through CO science, priority area: New Windows on the Dynamic Universe), as well as *Cosmic Ecosystems* (through SNR science, priority area: Priority Science Area: Unveiling the Hidden Drivers of Galaxy Growth).

Furthermore, as stated in the Astro2020 Decadal Survey, the *Time-Domain Multi-Messenger Astrophysics Program (TDAMM)* is a key scientific priority for the coming decade, with new capabilities for discovery on the horizon with future facilities such as the Rubin Observatory (LSST), Roman Telescope, Laser Interferometer Gravitational-wave Observatory (LIGO), Virgo, Kamioka Gravitational Wave (KAGRA), and neutrino detectors. This book's CO-SNR program focuses on a wide range of phenomena that are synergistic with time-domain, multi-wavelength, and multi-messenger science, including magnetar outbursts and their connection to FRBs, long period radio pulsars, transitional millisecond pulsars, ultra-luminous X-ray sources (ULXs), and a zoo of neutron stars and X-ray binaries, e.g., Cases 1, 4, 7 to 14, 27, 30, 34 and 35. *AXIS* will discover outflows and variability in pulsar and magnetar wind nebulae, probe accretion physics in a diversity of transient compact objects, and identify quiet black holes and neutron stars in crowded or obscured fields. A key objective is to expand the neutron star ULX population through targeted pulsation searches that exceed the current observatories' capabilities.

*AXIS* will play a pivotal role in multi-messenger astronomy, i.e., connections with other observable signals such as gravitational waves (e.g., 16, 17 and 31), cosmic rays and neutrinos (e.g., 33 and 35 to 37). *AXIS* is designed for rapid follow-up of electromagnetic counterparts to gravitational wave events detected

[1] https://nap.nationalacademies.org/catalog/26141/pathways-to-discovery-in-astronomy-and-astrophysics-for-the-2020s

by LIGO, Virgo, and KAGRA, and by future facilities such as Cosmic Explorer and Einstein Telescope. It will unveil new details in the X-ray emission from binary neutron star mergers and neutron star-black hole mergers, as well as their associated kilonova remnants. The improved sensitivity of *AXIS* will be important as future gravitational wave identifications will be made at higher redshifts than is currently possible. This CO-SNR book also highlights strong synergy with the upcoming Laser Interferometer Space Antenna (LISA), which will detect gravitational waves from ultra-compact binaries such as AM CVn systems, detached double white dwarfs, and ultracompact X-ray binaries. Many of these will be "verification binaries"—guaranteed LISA sources with known electromagnetic counterparts. *AXIS* will characterize the periods, mass transfer states, and spectral properties of these systems, aiding in the interpretation of LISA data and source classification, and enabling a suite of new multi-messenger probes of these systems.

These efforts will be strengthened through coordinated multi-wavelength programs with Roman, Rubin, JWST, SKA, CTAO, ALMA, the ngVLA, and future extremely large (optical) telescopes, see e.g., Cases 4, 6, 9, 12, 17, 22, 23 and 35. *AXIS* will help reveal the environments, progenitors, and physical mechanisms driving compact-object transients. Additionally, *AXIS* will support high-energy neutrino observatories like IceCube by conducting deep X-ray follow-up of neutrino alerts, helping to identify hidden sources such as obscured AGN, transient compact binaries, and the most powerful cosmic accelerators in our Galaxy and beyond (or 'PeVatrons', see Cases 35 and 36), driving the next generation of high-energy gamma-ray missions. In both the time-domain and multi-messenger arenas, *AXIS* provides the high-resolution, high-sensitivity X-ray view needed to connect compact objects with the most extreme events in the Universe.

This GO book is structured into the following themes. **Part II: Compact Objects** highlights a broad spectrum of science cases focused on white dwarfs, neutron stars, black holes, ultra-luminous X-ray sources, and emerging classes of transitional systems, radio transients, and other enigmatic sources. **Part III: Surveys** explores the scientific potential of *AXIS*'s efficient and powerful survey capabilities, including both Galactic and extragalactic surveys in nearby galaxies, and in synergy with the planned Galactic Plane Survey [509]. **Part IV: Diffuse Emission** addresses phenomena ranging from supernovae, novae, and kilonovae remnants, to pulsar and magnetar wind nebulae, their associated outflows and link to other enigmatic sources (such as FRBs), and the growing and intriguing class of Galactic PeVatrons – the most powerful accelerators of cosmic rays up to (or exceeding) $10^{15}$ eV energies. Several of the science cases highlighted in this book overlap with the science relevant to `TDAMM` and the `Galaxies` science working groups.

# Part II
# Compact Objects

## a. The Diversity of Isolated Neutron Stars, Pulsars, and Radio Transients

### *1. A new look at the isolated neutron star zoo with AXIS*

**First Author:** Alice Borghese (ESA/ESAC; alice.borghese@gmail.com)
**Co-authors:** Francesco Coti Zelati (ICE-CSIC,IEEC), Wynn Ho (Haverford College), Oleg Kargaltsev (GWU), Demet Kirmizibayrak (Caltech), Alicia Rouco Escorial (ESA/ESAC), Samar Safi-Harb (U. Manitoba), Janette Suherli (U. Manitoba), George Younes (NASA GSFC), GianLuca Israel (INAF OAR)

**Abstract:** The population of isolated neutron stars (NSs) counts different classes, grouped broadly according to the primary source of energy that supplies their emission. However, we have learned that these emission mechanisms are not mutually exclusive. We can distinguish four main groups. Rotation-powered pulsars (RPPs) are driven by the loss of rotational kinetic energy due to the braking caused by their magnetic fields. They are mostly detected in the radio and gamma-ray bands, but some also emit X-rays. Then, X-ray dim isolated neutron stars (XDINSs) are close-by, radio-quiet X-ray pulsars, characterized by soft thermal spectra with broad absorption features and faint optical/UV counterparts. Located near the center of young supernova remnants (SNRs), central compact objects (CCOs) emit thermal X-rays without counterparts at any other wavelength. Finally, magnetars are the most variable within the isolated NS zoo with their unpredictable bursting activity in the X-rays, gamma-rays, and radio on different timescales. They typically possess ultra-strong magnetic fields, whose instabilities and decay feed their X-ray persistent and flaring emission. Over the last few decades, theoretical studies have been proposed, highlighting the magnetic field and its evolution as key factors in explaining the puzzling diversity and evolutionary paths of the NS population. New discoveries, such as magnetar-like behaviour from RPPs, have supported these results. These discoveries have blurred the boundaries between different classes of NSs. However, we seek further observational evidence that *AXIS* can provide, due to its combination of high detection sensitivity, high angular resolution, large effective area, and low background. With *AXIS*, we will be able to *(i)* discover more INSs at the fainter end of the X-ray luminosity function to explore the evolution of X-ray emission efficiency, *(ii)* characterize the thermal emission of faint INSs to constrain magneto-thermal evolutionary models, *(iii)* search for proper motion and variability for the sources *Chandra* has been observed on a long baseline. Moreover, *AXIS* will enable an unprecedented characterization of the quiescent emission spectrum of magnetars and follow-ups of magnetar outburst decay till thousands of days since the onset to probe the state of the dense matter near the core and, ultimately, constrain the equation of state.

#### Science

*The isolated neutron star zoo*

Since their discovery, isolated neutron stars (NSs) have exhibited a remarkable variety of behaviors depending on their age, magnetic field strength, rotational dynamics, emission mechanisms, and surrounding environments. This diversity in their observational manifestations has led astronomers to classify isolated NSs into numerous categories. Below, we list the main NS groups along with their relevant properties for this proposal (for a detailed review, see, e.g., [78] and the references therein).

Rotation-powered pulsars (RPPs) represent the bulk of the NS population, counting more than 3700 of them. Powered by the loss of their rotational energy, they are mostly observed as radio or gamma-ray

pulsars. However, some of them are also detected in X-rays, with X-ray-to-radio efficiencies ranging from $10^{-5}$ to $10^{-3}$. The X-ray emission is dominated by a non-thermal contribution due to processes occurring in the magnetosphere and/or the presence of a surrounding pulsar wind nebula. In some cases, a thermal component is also present, coming from the NS cooling surface or hot spots heated by returning currents in the magnetosphere. X-ray dim isolated NSs (XDINSs) are characterised by the absence of associated supernova remnants (SNRs), pulsar wind nebulae, and coherent pulsed radio and gamma-ray emission. Among the closest known NSs, XDINSs are observed to emit primarily in soft X-rays with a nearly thermal spectrum with temperatures $kT \sim 45$–110 eV and low absorption column density values ($N_{\rm H} \sim 10^{20}\,{\rm cm}^{-2}$). Over the last few decades, broad absorption features have been discovered in the XDINS phase-averaged spectra, and in two cases, narrow phase-dependent spectral features have also been reported. The X-ray emission is pulsed at spin periods of a few seconds ($P \sim 3$–11 s) and a spin-down rate of $\dot{P} \sim 10^{-14}\,{\rm s\,s}^{-1}$, implying surface dipolar magnetic fields of $B_{\rm dip} \sim 10^{13}$ G. Central compact objects (CCOs) are soft X-ray emitters ($\sim 0.2$–5 keV), lacking a surrounding pulsar wind nebula and counterparts at any other wavelengths. While only a dozen CCOs are currently known, their locations in young SNRs suggest they may represent a significant fraction of all NSs. Their thermal spectrum is well described by the sum of two blackbodies with temperatures $kT \sim 0.2$–0.5 keV and small emitting radii spanning from 0.1 to a few km. Their emission is usually steady with a luminosity of the order of $10^{33}\,{\rm erg\,s}^{-1}$ and pulsations are detected in only three CCOs at periods in the range $\sim 0.1$–0.4 ms. For these three objects, the spin-down-measured dipole magnetic field is the smallest among all the known young isolated NSs, at odds with their relatively bright thermal emission. The magnetar census comprises about 30 sources, residing in the Galactic plane at low latitudes (except for two found in the Magellanic Clouds). They are slow rotators ($P \sim 1$–12 s) and spin down over timescales of a few thousand years ($\dot{P} \sim 10^{-13} - 10^{-11}\,{\rm s\,s}^{-1}$). These timing properties translate into $B_{\rm dip} \sim 10^{14} - 10^{15}$ G, making magnetars the strongest magnets in the Universe. The persistent X-ray luminosity, $L_{\rm X} \sim 10^{32} - 10^{36}\,{\rm erg\,s}^{-1}$, is generally larger than their spin-down luminosity, making the decay and instabilities of the super-strong magnetic field the main engine of their emission. The persistent soft X-ray (<10 keV) spectra are modeled with a thermal blackbody component ($kT \sim 0.3$–1 keV) and, in some cases, an additional component is required, either a second blackbody ($kT \sim 1$–2 keV) or a power law (photon index $\Gamma \sim 1$–2). The distinctive trait of magnetars is the transient activity on different timescales, ranging from millisecond (short bursts) to hundreds of seconds (giant flares), and month-to-year enhancement (up to three orders of magnitude) of their persistent X-ray luminosity (i.e., outbursts), which then decays back to the quiescent level in most cases. During an outburst, the X-ray spectrum undergoes an initial hardening and then slowly softens on the timescales of the luminosity relaxation.

This overall picture has become more complex with the discoveries of magnetar-like outbursts from two high-$B$ RPPs and the CCO at the center of the SNR RCW 103, three magnetars with dipolar magnetic fields within the range of those of ordinary RPPs, and radio emission from a few magnetars. These results hint at evolutionary links among the different NS classes. *AXIS* will offer unique astrophysical diagnostics to investigate how the intriguing various manifestations of NSs are related to each other.

*Objectives*

- Thanks to the combination of *AXIS* high angular resolution, low background, and large effective area, we will be able to discern the non-thermal and thermal contributions in many RPPs and characterise the faint, cold blackbody emission of some RPPs, CCOs, XDINSs, and magnetars in quiescence with unprecedented precision. These sources are particularly interesting because they can be used to constrain the NS equation of state through comparisons with theoretical cooling curves (see e.g., [468]).

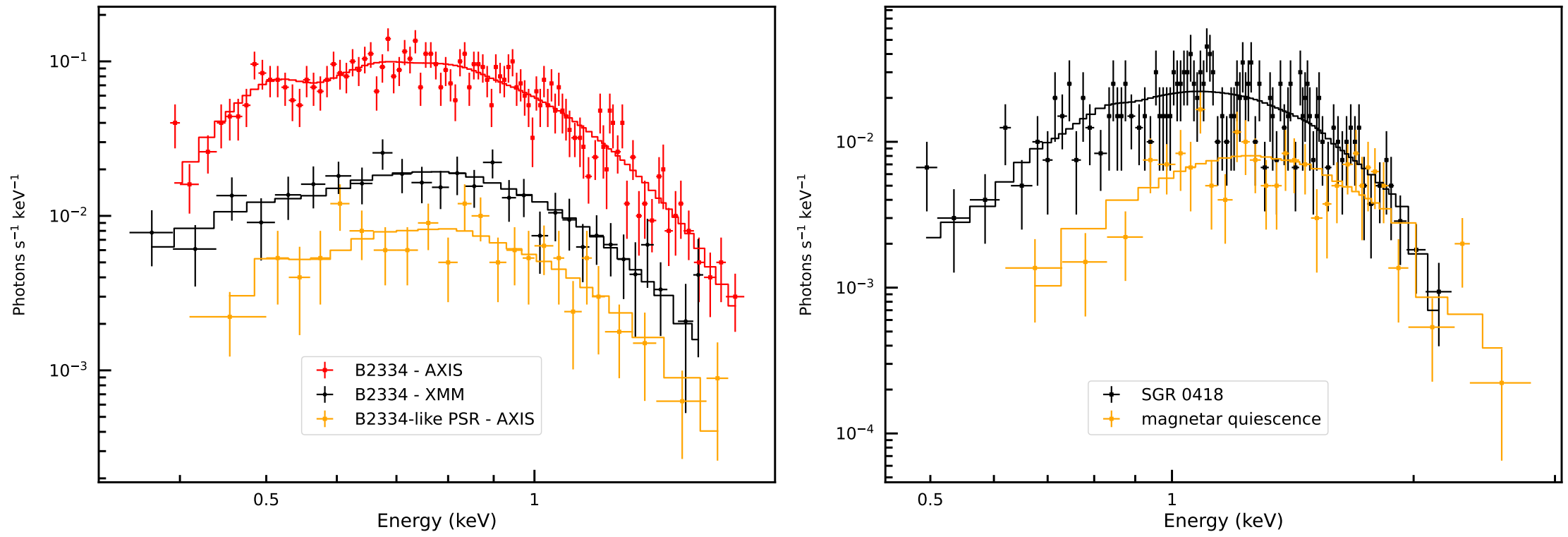

**Figure 1.** *Left*: *XMM-Newton* (black) and *AXIS* (red) spectra of the RPP PSR B2334+61, assuming the same exposure time (25 ks), modeled with a balckbody. In orange the 25-ks simulated *AXIS* spectrum for a RPP with the same temperature ($kT = 0.151$ keV) and hydrogen column density ($N_{\rm H} = 2.9 \times 10^{21}$ cm$^{-2}$) of PSR B2334+61, but with a flux lower by a factor of ten. *Right*: spectra of magnetars in quiescence simulated for a 20-ks *AXIS* observations. For SGR 0418+5729 (black), we based the simulation on a blackbody model measured by 110 ks *XMM-Newton* exposure. In orange we simulated a spectrum assuming a blackbody model with $N_{\rm H} = 10^{22}$ cm$^{-2}$, $kT = 0.3$ keV and $F_{\rm obs,0.3-10keV} \sim 5 \times 10^{-15}$ erg cm$^{-2}$ s$^{-1}$.

- The *AXIS* Galactic Plane Survey (GPS), reaching a limiting flux of $\sim 10^{-15}$ erg cm$^{-2}$ s$^{-1}$ for a source detected at $5\sigma$ confidence level, will increase the number of isolated NS candidates. Follow-up pointed observations can identify the type of isolated NSs through a pulsation search and detailed spectral analysis. *AXIS* timing resolution ($\sim$0.2 s) will allow us to detect spin periods well in the range of those of magnetars and XDINSs. Enlarging the sample of isolated NSs will enable comprehensive population synthesis studies and a deeper understanding of evolutionary tracks through magneto-thermal evolution models.
- Due to its large effective area (5–10 times larger than *Chandra*), *AXIS* is ideal to follow the evolution of magnetar(-like) outbursts until thousands of days since the onset particularly for magnetars with low quiescent luminosity (the so-called 'transient' magnetars). These sources experience larger luminosity increases during an outburst compared to magnetars with high quiescent luminosity. By enabling multi-year tracking of the soft emission decay following an outburst of faint magnetars, *AXIS* will help investigate the properties of the inner crust, where energy may be injected to power the bursting event.
- *AXIS* will allow detections of spectral features in several isolated NSs of different classes, thanks to its energy resolution and effective area. Absorption lines are common in the phase-average spectra of CCOs (e.g., 1E 1207.4–5209; [144]) and XDINSs (e.g., RX 0720.4–3125; [216]). However, their nature remains unclear; they may be due to proton cyclotron resonances or atomic transitions in a magnetized atmosphere. In magnetars, spectral features have been detected during quiescence and bursts, and in one case, a phase-dependent absorption line was reported [567]. These features are interpreted as cyclotron features, offering an additional independent gauge of the magnetic field strength. For detailed *AXIS* simulations for this goal, we refer to [509].
- Thanks to *AXIS* superb capabilities, we will be able to search for proper motion and/or variability for the sources that current X-ray satellites have observed on a long baseline. This goal can be achieved as a by-product of the GPS (with multiple visits) and through dedicated monitoring campaigns. Therefore, *AXIS* provides an excellent opportunity to extend the legacy of current X-ray observatories.

**Observing description:** To show *AXIS* contribution to characterize the faint cold thermal emission from isolated NSs, we simulated spectra of RPP PSR B2334+61, assuming a blackbody model with $N_{\rm H} = 2.9 \times 10^{21}\,{\rm cm}^{-2}$, $kT = 0.151\,{\rm keV}$ and observed 0.3–10 keV flux $F_{\rm obs,0.3-10keV} = 2 \times 10^{-14}\,{\rm erg\,cm^{-2}\,s^{-1}}$. These parameters are derived from the spectral fitting of the data extracted from the only X-ray observation available for this source, which is a ∼25-ks *XMM-Newton* pointing [363]. With a 25-ks *AXIS* observation, we will collect about 1500 photons (∼5 times more than *XMM-Newton*), allowing to obtain an estimate of the temperature with a 4% precision at $1\sigma$ confidence level (see Figure 1, left panel). Nonetheless, we can constrain $kT$ within 7% with an *AXIS* pointing of only 10 ks. Moreover, *AXIS* will be able to measure a blackbody emission as cold as the one of PSR B2334+61 with a ∼15% accuracy from a source with $F_{\rm obs,0.3-10keV} = 2 \times 10^{-15}\,{\rm erg\,cm^{-2}\,s^{-1}}$ (assuming $N_{\rm H} = 2.9 \times 10^{21}\,{\rm cm}^{-2}$ and an exposure time of 25 ks).

Regarding magnetar science, *AXIS* can follow new outbursts through a ToO program from the onset till the return to quiescence (time scales ranging from months to years). Particularly, it would be ideal to track the latest stages of the outburst decay down to lower flux values and with shorter exposures than those carried out so far (Figure 1, right panel). For example, the quiescent spectrum of the low-magnetic-field magnetar SGR 0418+5729 was constrained by merging three *XMM-Newton* observations totaling an exposure time of ∼110 ks [129]. This source is one of the magnetars with the lowest quiescent flux level known so far, $F_{\rm obs,0.3-10keV} = 10^{-14}\,{\rm erg\,cm^{-2}\,s^{-1}}$. With *AXIS*, we will be able to characterize the quiescent spectrum with a 20-ks observation, and measure $kT$ and $F_{\rm obs,0.3-10keV}$ with 4% and 7% precision, respectively, for frozen $N_{\rm H}$. Moreover, and most importantly, *AXIS* will help us to probe the quiescence of fainter magnetars. Assuming an absorbed ($N_{\rm H} = 10^{22}\,{\rm cm}^{-2}$) blackbody model with temperature $kT = 0.3\,{\rm keV}$ (typical value for magnetars in quiescence), a magnetar with $F_{\rm obs,0.3-10keV} \sim 5 \times 10^{-15}\,{\rm erg\,cm^{-2}\,s^{-1}}$ will be detected by a 20-ks *AXIS* observation. *AXIS* will open a new window by monitoring the quiescent state of magnetars to possibly look for flickering and catch the onset of a new outburst.

**Joint Observations and synergies with other observatories in the 2030s:**
Joint observations with upcoming facilities planned for the 2030s will significantly expand and complement the science achievable with *AXIS* in the study of isolated NSs. Radio-millimeter observatories such as ALMA and SKA [84] will be particularly important for follow-up monitoring of magnetar activity, enabling simultaneous multi-wavelength characterization of their transient phenomena and providing complementary constraints on their emission geometry, burst energetics, and evolution over time. Optical and infrared surveys and imaging capabilities provided by observatories like WFIRST, the Extremely Large Telescopes (ELTs), and the Vera Rubin Observatory (LSST) will allow identification and detailed study of faint nebular counterparts around CCOs and magnetars, crucial for understanding NS interactions with their surroundings. Additionally, synergies with high-energy observatories such as NewAthena [133] and the Cherenkov Telescope Array (CTAO; [111]) will deepen and extend the characterization of the emission mechanisms up to very high energies, helping to elucidate particle acceleration processes.

**[Special Requirements:]** (e.g., Monitoring (Daily, Hourly, etc), TOO (<X hrs), TAMM)
– TOO monitoring campaign to follow new magnetar outbursts. The first observation should be performed preferably within a week.
– Possible pile-up issue at the onset of a magnetar outburst.
– Monitoring campaign with biweekly observations to monitor magnetars in quiescence and other relevant isolated NSs to look for variability/proper motion.

*2. Monitoring (isolated) cooling neutron stars in complicated background regions*

**First Author:** Bettina Posselt (U. of Oxford, bettina.posselt@physics.ox.ac.uk)
**Co-authors:** Craig Heinke (U. Alberta), Wynn Ho (Haverford College), Oleg Kargaltsev (GWU)
**Abstract:** Monitoring of neutron star (NS) surface emission can provide a unique probe into the NS interior and its evolution. Observing the real-time cooling of very young (or very old) isolated neutron stars (INSs) is particularly valuable for constraining the composition, state, and physical processes in the core, which are otherwise difficult to investigate. For middle-aged isolated neutron stars, changes in surface emission provide insight into changes in magnetic field topology, internal processes such as glitches, and external interactions, including collisions with asteroids. Monitoring of pulsations provides a real-time measurement of the current (and possibly changing) spin-down of a pulsar – a crucial ingredient in NS evolution models. Some of the most promising monitoring targets reside in regions that create an intricate and dynamic X-ray backdrop from which the INS must be resolved and detected with sufficient sensitivity for spectral fits and pulsation searches. Such "backgrounds" are interesting in their own right, for instance, pulsar wind nebulae, supernova remnants, and crowded star-forming sites. Monitoring of interstellar absorption features in the thermal INS spectrum can act as a 3D probe of such environments. The excellent spatial resolution, sensitivity, and time resolution of *AXIS* will allow the monitoring of otherwise inaccessible surface emission of such INS in complex and dynamic environments. The example of constraining the cooling of the Central Compact Object (CCO) in the Cassiopeia A (Cas A) supernova remnant illustrates the need for *AXIS* and what can be achieved with monitoring observations.
**Science:** NS surface emission enables the investigation of the surface temperature distribution, surface or atmospheric composition, underlying magnetic field distribution, and gravitational redshift of the NS. Monitoring significant changes in this thermal emission reveals the processes within the NS interior, whether these involve changes in magnetic field topology or cooling processes in the core. Cooling is expected to be particularly fast when neutron stars are very young (< kyr) or very old (> a few Myr). Very old objects are rather faint and require high telescope sensitivity. Very young neutron stars reside in young supernova remnants, necessitating excellent spatial resolution to remove background contamination in the detected spectra. A striking example of reported neutron star cooling on human time scales is the CCO CXOU J232327.9+584842 in the Cas A supernova remnant (SNR). The Cas A CCO is $\sim 350\,\mathrm{yr}$ old, and the youngest INS for which surface emission is detected. The SNR is very X-ray bright and rapidly changing as it evolves. Only Chandra is currently able to separate the Cas A CCO from its parent SNR and measure a unique time sequence of temperatures, e.g., Heinke & Ho [236]. The inferred large cooling rate that early in the NS evolution indicates accelerated cooling due to neutrino emission, as one expects if neutrons have recently become superfluid in the NS core while protons were already in a superconducting state [421]. Although Chandra data allowed one to enlarge the CCO's time coverage and refine the accuracy of the deduced cooling rate (e.g.,Posselt & Pavlov 465, Shternin et al. 534, Wijngaarden et al. 614), Chandra calibration issues due to the changing ACIS optical filter contamination and pile-up in some observing modes (e.g., Posselt et al. 466) necessitate an independent confirmation. *AXIS* will be the only telescope with the spatial resolution and sensitivity to confirm the measured flux changes and achieve an even more accurate monitoring time series of the CCO, as described in the observing section below. *AXIS* resolution and sensitivity are also essential to capture the long-awaited emergence of neutron star emission in SN 1987A. Measuring the cooling of this even younger NS will be a unique opportunity to probe fundamental physics in the NS. Similarly, other CCOs can be targeted for this purpose, Ho et al. [246]. In slightly older neutron stars, *AXIS* will allow us to investigate directly the compact objects even if they are surrounded by bright pulsar wind nebulae or located in crowded regions. The excellent *AXIS* sensitivity to soft X-ray emission paired with excellent time resolution will be needed to construct detailed surface temperature maps of middle-aged pulsars, monitor for temperature changes in these

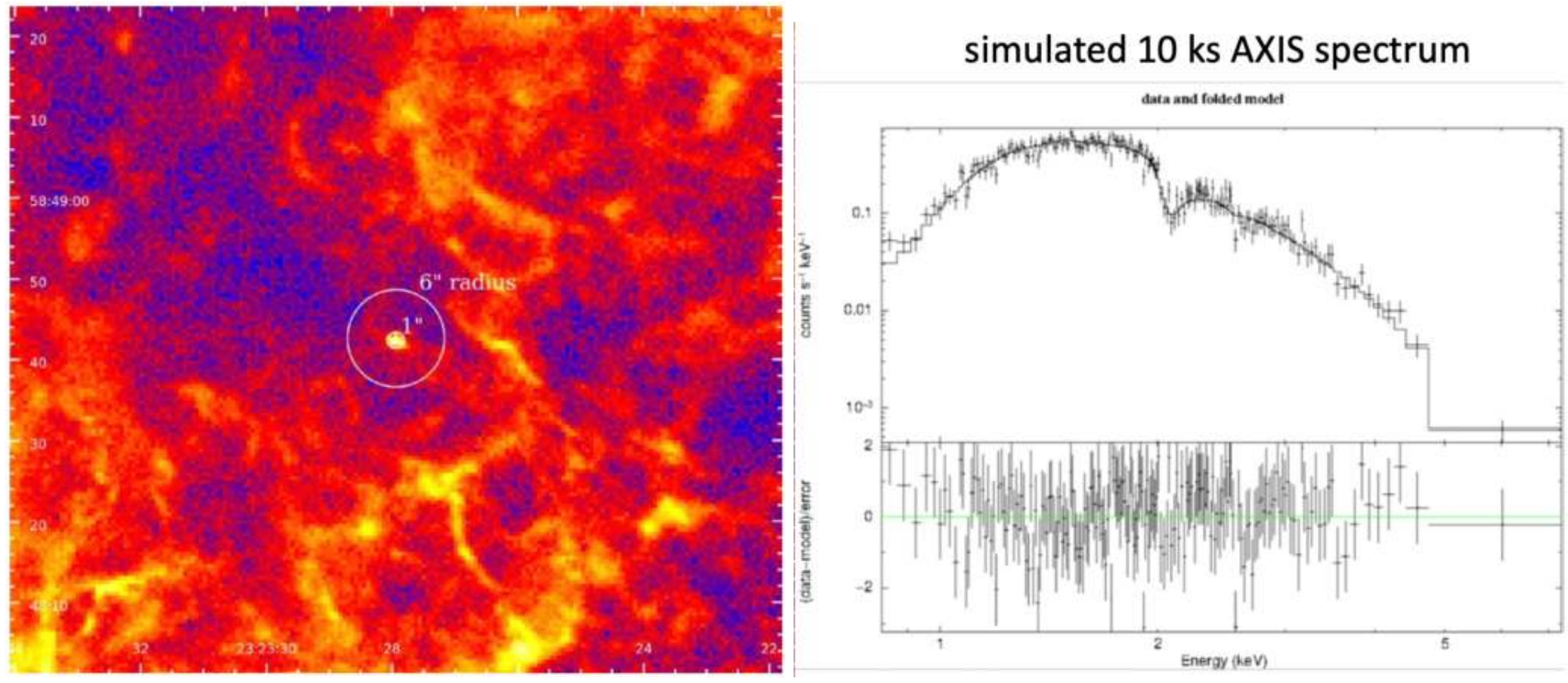

**Figure 2.** *Left*: Zoom on central portion of Cas A supernova remnant, showing how complex background requires high spatial resolution to resolve the NS. *Right*: Simulated 10 ks *AXIS* spectrum of Cas A's CCO.

maps and time-variable pulsation-phase-resolved spectral lines (caused, e.g., by the re-structuring of the near-surface magnetic field), and relate the observed variability to the neutron star's evolution. Accurately characterizing X-ray spectra of even older, faint neutron stars is crucial for disentangling thermal and non-thermal contributions and constraining the polar cap heating and thermal evolution. *AXIS*'s high effective area at low energies will enable detections and studies of many old pulsars for the first time.

In summary, *AXIS*'s capabilities are essential to monitor the thermal emission of neutron stars across the ages and reveal the physics of the neutron star surface and interior.

**[Exposure time (ks):]** dependent on target and time coverage. For the example of the Cas A CCO (see below), either 100 ks exposures if only *AXIS* observations are used, or 30 ks if the cross-calibration factor uncertainties are small enough to exploit the existing *Chandra* legacy observations.

**Observing description:** One example of an observing goal is to confirm and further constrain the cooling of the Cas A CCO. The previous Chandra observations over twenty years have shown flux changes that can be interpreted as a temperature decrease of 1% to 4% in 10 years, assuming a carbon atmosphere as the spectral model for the neutron star thermal emission (e.g., Posselt & Pavlov 465, Shternin et al. 534, Wijngaarden et al. 614). Not only is this range still large, but there also remain systematic uncertainties due to the (variable) Chandra ACIS filter contamination as well as pileup issues in the earliest observations [466]. Assuming similar spectral models as in [465,614], a 100 ks (30 ks) *AXIS* observation can constrain the temperature of the CCO with an uncertainty of $1\sigma = 0.12\,\%$ ($1\sigma = 0.21\,\%$). For three 100 ks observations spread over the nominal lifetime of *AXIS* of 5 years, *AXIS* can determine an assumed 1% temperature decrease in 10 years, $T10 = 1\%$, with an uncertainty of $1\sigma_{T10} = 0.3\,\%$.

Assuming that the three *AXIS* measurements are instead taken over a likely *actual* lifetime of 20 years – similar to current X-ray satellites – *AXIS* can determine the temperature decrease down to an uncertainty of $1\sigma_{T10} = 0.1\,\%$. Another strategy is to build on the legacy of the Chandra observations to increase the time coverage. In the following, we use only the data of two early Chandra observations (2006 and 2012) in subarray mode, which are the least affected by systematic uncertainties due to filter contamination or pileup. There will be a cross-calibration factor between Chandra and *AXIS* spectral fits, whose uncertainty must be taken into account. Here, we assume a systematic uncertainty of 1% in the cross-calibration factor, which is added quadratically to the nominal temperature fit uncertainty. Two 30 ks *AXIS* observations over a period of five (or 20) years combined with the two optimally chosen archival Chandra observations will then be able to constrain the temperature decrease in ten years down to an uncertainty level of $1\sigma = 0.5\,\%$

($1\sigma_{T10} = 0.2\,\%$). More *AXIS* observations can further increase accuracy. For six observations over 20 years, $1\sigma_{T10} = 0.1\,\%$ can be achieved. The assumed uncertainty of the cross-calibration is the dominating factor in the error budget. A careful calibration program may lower it below the assumed 1% and further increase accuracy.

**What are critical requirements to carry out INS monitoring studies with *AXIS*?**

- Critical *AXIS* capabilities for these studies are: high angular resolution, low background, sub-window capability, good (ms) timing resolution, high effective area at soft energies, pile-up mitigation strategies for objects in the field
- A stable *AXIS* performance with minimal calibration uncertainties will be crucial to achieve the best results in these monitoring programs.
- To maximize the scientific returns of *AXIS* monitoring programs with relatively modest exposure times, one should aim to exploit the decades-spanning legacies of Chandra and XMM-Newton. The uncertainties of cross-calibration factors will govern the accuracy of the results obtained with *AXIS*. A careful calibration campaign will benefit this and many other science cases.

**Joint Observations and synergies with other observatories in the 2030s:** ELTs, SKA, ALMA, NewAthena
**Special Requirements:** Monthly to yearly monitoring, pile-up mitigation strategies for bright background

*3. Catching spiders with AXIS*

**First Author:** Francesco Coti Zelati (ICE-CSIC, IEEC; cotizelati@ice.csic.es)
**Co-authors:** Hongjun An (Chungbuk National University), M. Cristina Baglio (INAF-OAB), Rebecca Kyer (Michigan State University), Amruta Jaodand (Smithsonian Astrophysical Observatory), Mason Ng (McGill University), G. Vasilopoulos (NKUA)
**Abstract**: Millisecond pulsar (MSP) "spider" binaries – redbacks (RBs), black widows (BWs), and huntsmen (HM) – are unique testbeds for studying pulsar winds, shock physics, extreme particle acceleration, and pulsar evolutionary pathways. In these systems, the pulsar wind ablates material from a low-mass companion star, forming an intrabinary shock (IBS) that dominates the high-energy emission. RBs often exhibit X-ray emission modulated at the binary orbital period (sometimes showing double peaks in the light curve), while BWs tend to be fainter and less thoroughly observed in X-rays, though they reveal similarly intriguing IBS signatures. HM, whose companions are stripped red giants, exhibit a large range of X-ray luminosity attributed to the IBS, from the $L_X \approx 10^{31}\,\mathrm{erg\,s^{-1}}$ scale of RB shock luminosity up to two orders of magnitude brighter. *AXIS*'s advanced capabilities will revolutionize spider MSP studies by enabling new insights into pulsar wind interactions and mass-loss processes within these systems. First, *AXIS* will identify new spiders by characterizing faint X-ray sources in globular clusters, including those in distant clusters expected to host numerous binary MSPs. Additionally, *AXIS* will discover and follow up on X-ray counterparts of spiders identified through targeted radio surveys of unidentified Fermi sources in the Galactic field. Second, *AXIS* will facilitate phase-resolved spectroscopy even for relatively faint binaries, particularly BW systems. Third, *AXIS* will allow deep observations of bright RB binaries. This will enable us to measure orbit-to-orbit variability in the IBS and, when combined with simultaneous optical data, probe the interplay between companion-star winds and shock formation. Finally, *AXIS*'s rapid time resolution will be critical for characterizing temporal and spectral properties of X-ray flares in bright systems. This will help reveal shock instabilities and magnetospheric reconnection events behind these phenomena, with some models predicting strong TeV emission during flares.
**Science:** Millisecond pulsar (MSP) "spider" binaries provide a unique window into high-energy astrophysics, encompassing relativistic outflows and shock acceleration, as well as the evolution of neutron stars in compact binaries. These MSPs likely result from a "recycling" process in which an old, "dead" pulsar is spun up through prolonged accretion from a low-mass companion, reinitiating particle acceleration in the pulsar's magnetosphere and reducing its surface magnetic field [11]. In spiders, the pulsar's relativistic wind interacts with the low-mass companion, stripping its outer layers via ablation [184]. This interaction produces a complex intrabinary shock (IBS) where the wind's kinetic energy is converted into non-thermal X-rays often modulated at the orbital period [253]. Such variability offers a unique testbed for particle acceleration, magnetic reconnection, and extreme magnetic field interactions.

Observations reveal two main spider classes [496,548,558]. In black widows (BWs), the companions are extremely low-mass ($<0.1\,\mathrm{M_\odot}$) and the pulsar wind wraps the shock around the companion. BWs frequently display single-peaked or irregular orbital modulations. In contrast, redbacks (RBs) host more massive ($\gtrsim 0.1\,\mathrm{M_\odot}$), often semi-degenerate companions whose stronger winds push the shock closer to the pulsar, frequently leading to double-peaked X-ray light curves. These peaks are interpreted as Doppler boosting of synchrotron emission along the shock "tail". In systems with tight orbits, Coriolis forces may shift these peaks or produce asymmetries in their amplitudes (e.g. [503]). Both BWs and RBs have been seen to experience sudden X-ray flares exceeding the quiescent flux by a factor of $>10$ and lasting from a few to tens of ks with variability timescales of 10-100 s. Such flares likely stem from rapid, sporadic magnetic reconnection within the IBS, sometimes triggering correlated optical flares [15,115]. A third emerging sub-class, "huntsmen" (HM), includes MSPs orbiting partially stripped giant companions with orbital periods of several to $\sim$10 days [547,550]; despite the giant companion and longer period, the

pulsar's wind still drives an IBS. HM provides a snapshot of the recycling process when mass transfer may pause or become inefficient, possibly due to "radio ejection" by the pulsar wind or an evolutionary bottleneck known as the "red bump" [560]. Adding to the complexity, three RBs – known as transitional MSPs – have been observed to switch, on timescales as short as weeks, between a rotation-powered state (dominated by the IBS) and an X-ray active state in which an accretion disk quenches the magnetosphere and boosts the X-ray luminosity by orders of magnitude (see Science Case 9; Papitto & de Martino 426). These transitions highlight how local plasma conditions affect shock structure and emission properties, offering unique insights into the interplay between accretion and pulsar wind processes.

In nearly all spiders, the X-ray spectrum is dominated by non-thermal IBS emission, and is typically well described by a hard power-law model with photon indices in the range $\Gamma \approx 1 - 1.5$ (e.g., [253]). This spectral hardness is consistent with magnetic reconnection accelerating electrons to high energies. Faint thermal emission from neutron star surface heating ($kT \simeq 0.15$ keV; [296]) is sometimes detected, and phase-resolved observations are key to disentangling this from the dominant variable power-law component. Population studies indicate that RBs convert pulsar spin-down power into X-ray luminosity with an efficiency of $\simeq$0.1%, roughly an order of magnitude higher than the $\simeq$0.01% seen in BWs. This disparity likely arises from the larger effective shock area provided by the stronger companion winds in RBs. However, factors such as binary inclination, mass ratio, and shock geometry also contribute [295].

Expanding the sample of spider pulsars with well-characterized X-ray properties is crucial for deepening our understanding of pulsar winds, shock dynamics, and the irradiation effects of the companion in compact binaries. In this context, *AXIS* promises to unravel the rich spider phenomenology by offering high sensitivity, low background, and sub-arcsecond resolution. *AXIS* will detect and characterize faint orbital modulations of the X-ray emission in these binaries and allow for phase-resolved spectroscopy to track subtle spectral changes, including any emerging thermal component at specific orbital phases. While in HM, IBS signatures are more challenging to detect due to their longer orbits, sufficiently long observations or multiple shorter exposures strategically spaced along the orbital phase will reliably detect modulation, if present. Additionally, by monitoring relatively bright systems over multiple consecutive orbits, *AXIS* will reveal orbit-to-orbit variability, offering insights into the dynamic evolution of the shock structure and potential episodic changes in particle acceleration and magnetic reconnection processes. *AXIS*'s rapid timing will also capture sub-orbital flares in systems like PSR J1311$-$3430 and PSR J1048+2339, where the flux can rise by factors of 5–10. Time-resolved spectroscopy will reveal the evolution of the shock structure and changes in parameters such as photon index or local absorption. Another significant advantage of *AXIS* is its large 24 arcmin field of view, which is particularly beneficial for studying MSPs in globular clusters. In fact, a single, deep *AXIS* exposure can encompass many spider MSPs located near the cluster core. This enables it to fold data from multiple orbits, unveiling modulation patterns even for those sources that are too faint for detection by current X-ray instruments. Moreover, the wide-field capability allows for coverage that extends well beyond the central regions of the cluster, potentially leading to the discovery of new spider candidates in the outskirts. In this regard, insights gained from studying the high-energy shock processes in spiders may provide crucial clues to the origin of the TeV emission detected from globular clusters such as Terzan 5 [e.g., 302].

For transitional MSPs, such as XSS J12270$-$4859, *AXIS*'s capabilities are equally critical. By comparing the IBS flux, orbital modulation, and flare patterns before transition with the X-ray emission properties after state transitions, *AXIS* can help determine how the shock emission is altered when an accretion disk interacts with the pulsar's magnetosphere. Notably, in their radio pulsar state, transitional MSPs remain observationally indistinguishable from typical RBs. Sufficiently sensitive X-ray observations might eventually reveal subtle diagnostic differences. A dedicated, multi-orbit *AXIS* campaign could establish a phase-resolved baseline that is essential for contrasting with data acquired during the X-ray active state.

This comparative analysis is crucial for understanding the mechanisms behind radio pulsar quenching and the triggers that drive the system into an active X-ray state.

In summary, *AXIS*'s combination of wide-field coverage, high sensitivity, and rapid readout will capture the full dynamical range of spider MSP behavior. By unveiling modulated X-ray emission from faint spiders, catching dramatic flares in targeted pointings, or monitoring state transitions in transitional MSPs, *AXIS* will revolutionize our understanding of pulsar wind physics, shock formation and evolution, and the interplay between rotation- and accretion-powered emission in MSPs in compact binaries.

**Exposure time (ks): 100**

**Observing description:** Figure 3 shows how an uninterrupted 100-ks AXIS observation of a globular cluster (or any spiders in the Galactic field discovered through Fermi surveys and targeted radio observations) can yield key insights into the physics of spider pulsars. The top-left panel illustrates how reliably *AXIS* will detect an orbital modulation in the X-ray light curve of a spider pulsar. Simulations were performed over a wide range of mean count rates and modulation amplitudes. The color bar represents the statistical significance with which a sinusoidal orbital variation can be distinguished from a flat (constant) model, using an F-test. Dark colors (blue and purple) imply high significance, meaning *AXIS* can robustly confirm that the observed light curve does indeed vary with orbital phase. Sources above and to the right of the white curve are likely to show detectable modulation. Thanks to *AXIS*'s low background, the exposure recovers modulations in moderately bright sources (20–30% for fluxes of a few $10^{-14}\,\mathrm{erg\,cm^{-2}\,s^{-1}}$) and in much dimmer ones when the variation is large (40–50% for fluxes as low as a few $10^{-15}\,\mathrm{erg\,cm^{-2}\,s^{-1}}$). The top-right panel displays the minimum detectable modulation amplitude for various X-ray luminosities typical of spider pulsars. Even at moderate luminosities (a few times $10^{31}$ erg/s), *AXIS* remains sensitive to modulations out to several kpc – covering much of the globular cluster population. Hence, its high angular resolution and wide FoV will enable a single observation to resolve and analyze multiple spider MSPs in crowded clusters at once, differentiating their orbital modulations even for faint sources. The bottom-left panel shows *AXIS*'s capacity to uncover orbital variability across different flux regimes using 10 phase bins. This capability enables even more refined data slicing in relatively bright sources, which can be particularly helpful in detecting features such as the double-peaked light curves sometimes observed in spider pulsars. The bottom-right panel demonstrates that a power-law model can fit the spectrum at the orbital minimum of a source with an average flux of $\simeq 7\times10^{-15}\,\mathrm{erg\,cm^{-2}\,s^{-1}}$ (0.3–10 keV) with about 25% accuracy in the photon index (15% for a fixed hydrogen absorption column density $N_{\mathrm{H}}$). This confirms *AXIS*'s ability to track potential photon index variations across orbital phases even at low flux levels, and, for brighter systems, to detect emerging thermal features in narrower phase bins.

**[Joint Observations and synergies with other observatories in the 2030s:]**

*AXIS*'s X-ray observations will be enhanced by the ability of the SKA [84] to discover many more spider pulsars and measure their spin and orbital ephemerides. These measurements will enable the folding of X-ray light curves at the binary orbital period and mapping the geometry and efficiency of pulsar winds.

Synergies with the Fermi satellite will be essential, as in the past, unidentified gamma-ray sources with pulsar-like spectra from Fermi surveys have prompted targeted radio searches that greatly expanded the known spider pulsar population (e.g., Clark et al. 120). Moreover, joint *AXIS* and gamma-ray monitoring with the upcoming CTAO [111] may reveal correlated flares (as predicted by some models), providing critical tests for models of magnetic reconnection and energy transfer in IBSs. The discovery of additional spiders also presents promising opportunities for synergy with the upcoming MeV gamma-ray mission COSI [569], as recent simulations suggest that those in ultracompact systems may exhibit a 511 keV annihilation emission line strongly modulated at the orbital period [368].

Optical observatories (including upcoming 30-meter class telescopes like the ELT, GMT, and TMT) will deliver high-cadence photometry and high-resolution spectroscopy with unprecedented sensitivity. These observations will yield precise measurements of companion radial velocities, chemical compositions,

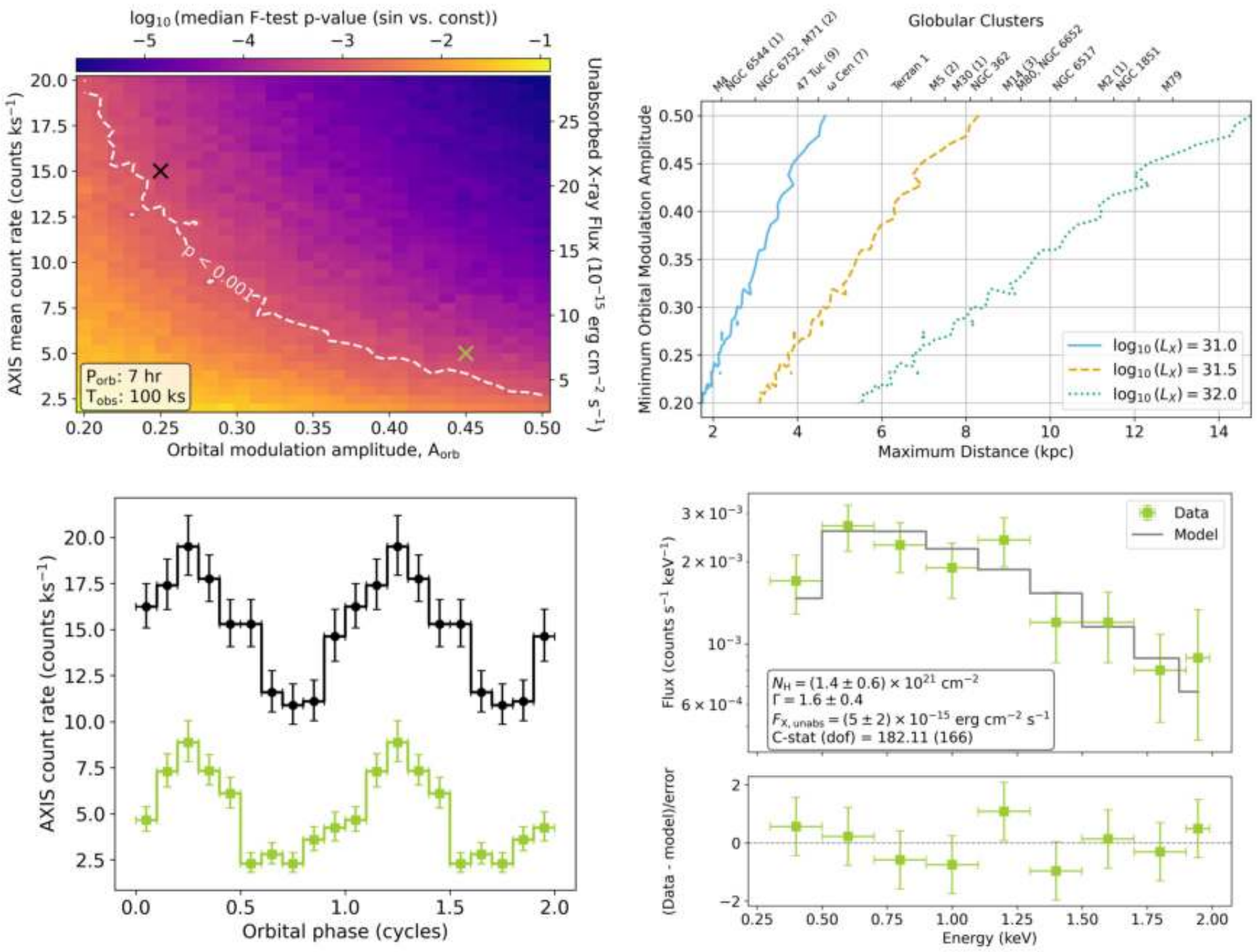

**Figure 3. Top-Left:** Detection significance of orbital modulations in simulated *AXIS* light curves of spider pulsars (100-ks exposure). The heatmap shows F-test *p*-values (assessing the improvement of a sinusoidal over a constant fit in 10 phase bins) versus mean count rate and orbital modulation amplitude (with $P_{\rm orb} = 7\,\rm hr$). Each grid point represents the result of 200 simulations. The right axis indicates the unabsorbed 0.3–10 keV flux (assuming an absorbed power-law with $N_{\rm H} = 10^{21}\,\rm cm^{-2}$ and $\Gamma = 1.5$). The white contour delineates the threshold in parameter space above and to the right of which the modulation is detected with high significance. Markers denote cases with 0.015 counts $\rm s^{-1}$ and 25% modulation (black), and 0.005 counts $\rm s^{-1}$ and 45% modulation (green). **Top-Right:** Sensitivity curves for spider pulsars showing the minimum detectable modulation amplitude versus maximum source distance for three X-ray luminosities. The top x-axis labels selected globular clusters (numbers in parentheses indicate the known spider pulsars). **Bottom-Left:** Phase-folded simulated light curves (10 phase bins, two orbital cycles shown) for the two highlighted cases. **Bottom-Right:** Spectrum at orbital minimum (phase 0.5–1) for the green-marker case, fitted over the 0.3–2 keV band using the Cash statistics. The best-fitting absorbed power-law model, along with the post-fit residuals, is also displayed.

and atmospheric dynamics, deepening our understanding of how companion stars respond to pulsar irradiation by the IBS as probed by *AXIS*. Additionally, the Vera C. Rubin Observatory (LSST) will contribute valuable time-domain optical data. Through its wide-field, deep, high-cadence imaging surveys, Rubin (LSST) will capture optical flares and brightness variations of spider companions, potentially uncovering new candidates that can be followed up by *AXIS*.

**[Special Requirements:]**

- Phase-resolved/Multi-epoch scheduling to accumulate sufficient orbital phase coverage or to cover long orbits (e.g., HM with multi-day orbital periods).
- Rapid or flexible scheduling (ToO) to catch X-ray flares and state transitions once multiband monitoring records flaring activities or indicates an imminent state transition in a spider pulsar.

*4. Constraining the X-ray emission from long-period radio transients with AXIS*

**First Author:** Jeremy Hare (NASA GSFC, CUA, CRESST II, jeremy.hare@nasa.gov)
**Co-authors:** Arash Bahramian (CIRA), Alice Borghese (ESA/ESAC), Zorawar Wadiasingh (NASA GSFC/UMD/CRESST II), Kaya Mori (Columbia University), Francesco Coti Zelati (ICE-CSIC, IEEC)
**Abstract:** Long-period radio transients (LPRTs) are a newly identified class of objects discovered in radio surveys over the past three years. To date, there are $\sim$10 Galactic sources known, and this number is rapidly growing. Currently, a dichotomy appears to exist among source classes that produce this emission. The first are synchronized white dwarf (WD) binaries, where the companion is a main sequence star, which tend to have lower radio luminosities and optical counterparts. The second class is significantly more luminous in radio and lacks any apparent optical or near-infrared (NIR) counterparts. These sources have been suggested to be neutron stars, although they exist beyond the radio death line, where rotation-powered magnetospheric coherent radio emission is expected to cease. One such source has been detected in X-rays, exhibiting pulsed emission in phase with the radio pulses during its radio-bright state. Here we propose a survey of the 10 LPRTs with the brightest radio pulses at the time of *AXIS*'s operation. Additionally, we propose a TOO observation of an LPRT that transitions to a radio-bright state. Detecting or setting deep upper limits on the X-ray emission from these sources will help us better elucidate their nature and understand the physical mechanism responsible for the emission.
**Science:** Long-period radio transients (LPRTs) are a relatively newly discovered class of bright radio sources. They have long periods ranging from about a minute up to several hours and can exhibit extraordinarily bright radio pulses with flux densities reaching tens of Jy (see e.g., [96,254,255]) that show cycle-to-cycle flux variations and sometimes even nulling behavior. Before their discovery, radio surveys and search algorithms were biased against detecting long-period sources (see, e.g., [60] for a discussion). However, since their discovery, radio survey strategies and pulse detection algorithms have been adapted to be more sensitive to these systems, leading to an increased discovery rate, with about 10 LPRTs discovered as of late 2024 (see e.g., Table 1 in [128]).

The nature of these sources and the mechanism powering their radio emission are still being debated (see e.g., [60,480]). The two most plausible scenarios are either a neutron star (NS) or a white dwarf (WD) as being responsible for the radio emission. However, both scenarios require unusual or extreme physical processes. In the NS case, many of the sources only have an upper-limit on their period derivative, but they lie near or even below the so-called "death valley" in the $P - \dot{P}$ diagram for rotation-powered pulsar emission (see e.g., Figure 4 or Figure 1 in [128]). This is the region beyond which a pulsar's spin-down power is not large enough to continue producing electron-positron pairs, responsible for the radio emission, in the magnetosphere of the pulsar; thus, radio emission should turn off [107]. This raises the question of what powers the bright, coherent radio emission in these sources if they cannot be rotationally powered. One possibility is that the radio emission is magnetically powered, similar to the X-ray emission in magnetars. This scenario is supported by the fact that one LPRT (namely PSR J0901-4046) has a measured $\dot{P}$ leading to an inferred magnetic dipole field $B \approx 10^{14}$ G [96]. Additionally, there is a young magnetar in the supernova remnant RCW 103 with a spin period of 6.67 hrs [143], suggesting that some LPRTs may be old magnetars (see [60]).

On the other hand, the WD models are motivated by the two WD pulsars discovered over the past few years, namely AR Sco and J1912-4410 [358,441]. These sources exhibit periods on the order of several minutes, with orbital periods of several hours, and emit coherent radio emission that is pulsed at the WD spin period. To date, only two LPRTs have counterparts at optical/IR wavelengths, suggesting that most are isolated objects. One issue with these models for an isolated WD is that they would require large magnetic fields (i.e., $B > 10^{10} G$) to reach the high radio luminosities ($L_R > 10^{31}$ erg s$^{-1}$) observed in some sources [60]. Recently some clarity has emerged from the two sources with optical counterparts (ILT

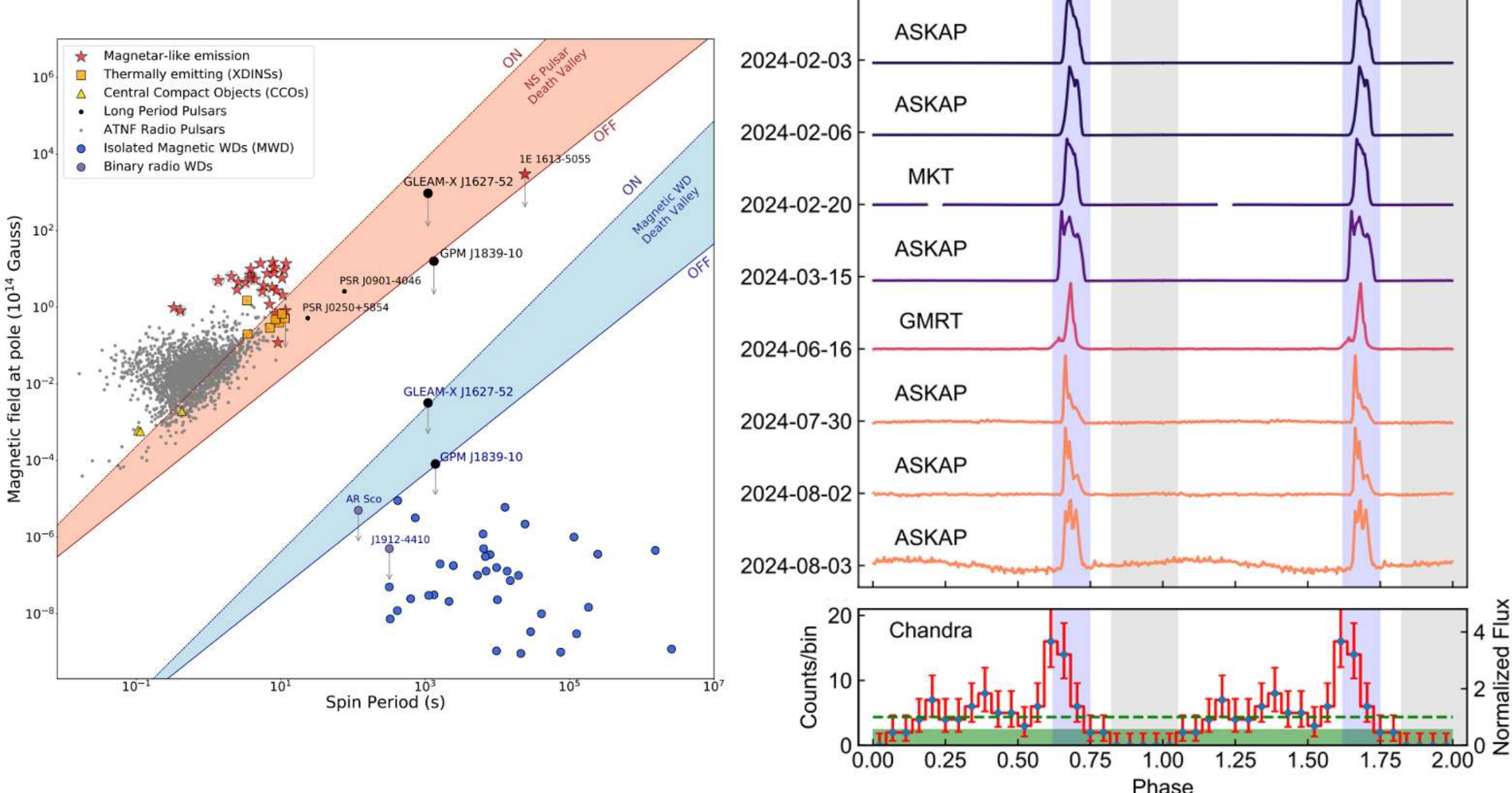

**Figure 4.** Left: P-$\dot{P}$ diagram showing the LPRTs assuming they are isolated NSs or WDs (figure adopted from [480]. Sources with arrows show the upper limit on spin-down. The red/blue-shaded regions indicate theoretical death valleys for various assumptions regarding the magnetic field structure of the source. Right: Radio and Chandra X-ray pulsations observed from ASKAP J1832-091 (figure adopted and modified from [606].)

J1101+5521 and GLEAM-X J0704-37; [147,256]), where spectroscopic radial velocity measurements showed an orbital period consistent with the radio period [147,499]. Optical spectra of both systems also showed a blue excess relative to the low-mass M-dwarf spectrum in each system, suggesting that these systems host WDs. Both sources remain undetected in X-rays, with limiting luminosities of $L_X \lesssim 10^{30}$ erg s$^{-1}$.

The emerging picture appears to suggest that the lower radio luminosity sources ($L_R \lesssim 10^{29}$ erg s$^{-1}$), except PSR J0901-4046, likely host WDs in binaries that emit coherent radio emission pulsed on the orbital period of the binary. Based on the X-ray luminosity limits, it is expected that little accretion is occurring in these systems; thus, the nature of the emission mechanism remains an open question, although some models have been proposed (see, e.g., [473]). Cataclysmic variables (CVs) typically do not have radio luminosities $L_R \gtrsim 10^{28} - 10^{29}$ erg s$^{-1}$ [52], so the radio luminous LPRTs are more likely to be NSs, though definitive proof is still lacking. Several high radio luminosity LPRTs have been observed in X-rays, with most observations resulting in only upper limits on the X-ray luminosity of $L_X < 10^{30} - 10^{32}$ erg s$^{-1}$ depending on the distance to the source and which observatory was used for follow-up (see e.g., [97,254]).

Another interesting development has come from the discovery of ASKAP J1832-091. This source has a $\sim$ 44 minute period and exhibits highly variable radio pulses with brightnesses spanning $\sim$10 Jy to a few 10s of mJy on timescales of days [606]. The brightest radio pulses imply a peak radio luminosity of $L_R = 5 \times 10^{31}$ erg s$^{-1}$. Fortunately, Chandra happened to be observing the field containing this source when it was in a radio-bright state and was able to detect it in X-rays. Interestingly, these X-rays were also strongly pulsed, with the X-ray pulsations being in phase with the radio pulsations and having a pulsed average luminosity $L_X \approx 8 \times 10^{32}$ erg s$^{-1}$ (see Figure 4 [606]). Archival X-ray observations, presumably taken when the source was in a low radio state, did not detect the source, suggesting that the X-ray and radio brightness are correlated. Unfortunately, there were not enough counts to strongly constrain the

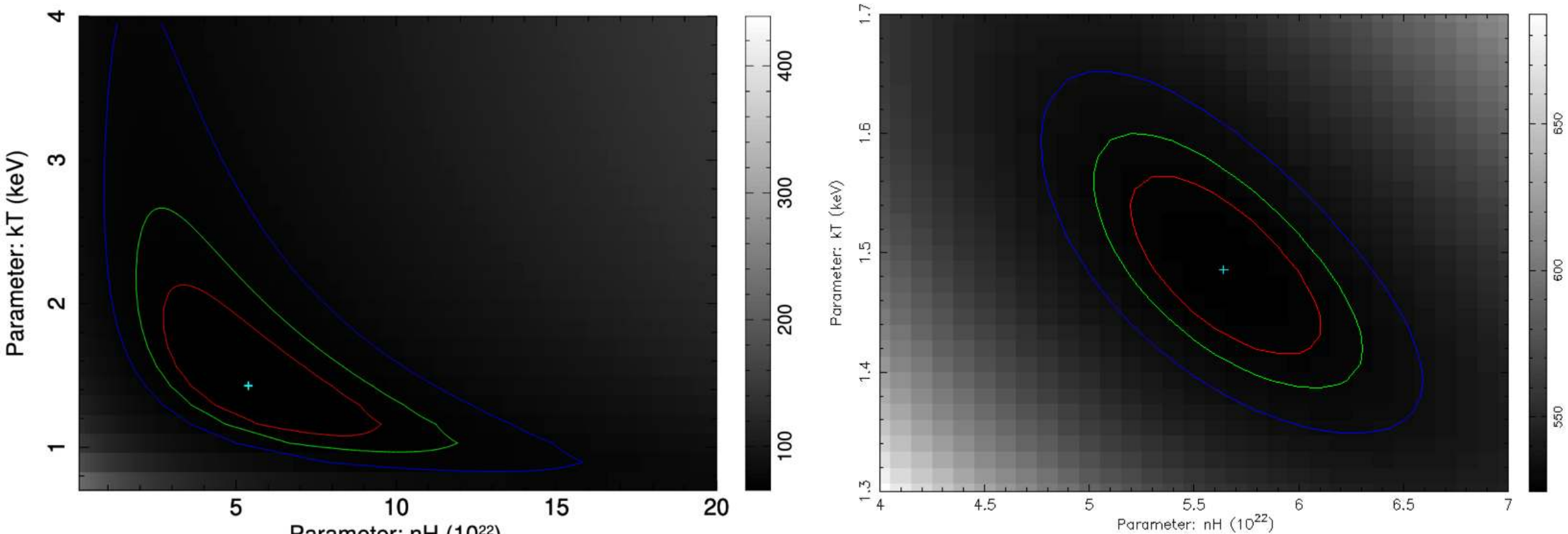

**Figure 5.** Left: Absorbing column density ($N_H$) versus temperature ($kT$) confidence contours for ASKAP J1832–091 observed by Chandra (figure adopted from [606]). Right: Simulated confidence contours from a 50 ks *AXIS* observation. Note the dramatic reduction in the scale of values spanned by the contours in both parameters. These observations will provide strong constraints on emission models from the LPRTs.

X-ray spectrum, and deep IR observations could not completely rule out the possibility of a low mass companion. However, the maximum observed radio luminosity and high X-ray luminosity favor an NS over a WD in this system, but further follow-up is needed to definitively differentiate between the two scenarios.

The LPRT landscape is rapidly evolving and likely will look very different by the time *AXIS* is launched. Already within the time of writing this proposal a new source emitting from radio to X-rays has been discovered [22]. Here we propose follow-up observations of the two WD systems to place the deepest limits on their X-ray emission, which will constrain any possible accretion occurring and help us understand the possible radio emission mechanism. We also request observations of the brightest eight persistent radio LPRTs to place the deepest constraints on their X-ray emission. Lastly, we request one TOO observation for either ASKAP J1832-091 or any other LPRT that enters a radio-bright state. We note that *AXIS*'s high angular resolution is necessary to accurately match the X-ray source positions to the radio source positions to reduce the chance coincidence probability of an unrelated X-ray source. The chance probability can be large given that many LPRTs are found in the crowded Galactic plane.

**Exposure time (ks):** 200 ks (+50 ks TOO)

**Observing description:** We propose 10 20-ks *AXIS* observations of LPRTs, including two WD systems, and the eight brightest persistent radio LPRTs discovered by the time *AXIS* launches. These observations will enable us to detect X-ray emission from LPRTs or place the most stringent limits possible on their X-ray emission. For the two WD systems, we will be able to reach limiting X-ray luminosities of $L_X \approx 5 \times 10^{27}$ erg s$^{-1}$, which will strongly constrain the accretion onto these WDs. For the other systems, we will reach a limiting luminosity of $L_X \approx 10^{28}(d/d_1)^2$ erg s$^{-1}$ where $d_1$ is an assumed distance of 1 kpc. We also request one 50-ks TOO follow-up observation for any variable or transient LPRTs that transition into a radio-bright state. The trigger criteria for this observation is any radio luminous LPRT, i.e., $L_R > 10^{30}$ erg s$^{-1}$, with a single pulse flux density $> 2$ Jy in the ASKAP band (788-1076 MHz) for 5 pulses over a 5-10 hour time period dependent on the sources period. This will ensure the source is in a bright state and not just exhibiting a short radio burst. For ASKAP J1832-091, we would collect more than 2000 counts if it reached the same X-ray brightness (i.e., $F_X \approx 10^{-13}$ erg cm$^{-2}$ s$^{-1}$) as it did in the Chandra observation. This would allow us to establish the shape of the pulse profile and to strongly constrain the source spectrum to differentiate between a NS/magnetar or WD and to test different emission mechanism models (see

Figure 5). *AXIS* simulations of this source show that a 50 ks observation is necessary to firmly distinguish between hot thermal blackbody emission (i.e., $kT \approx 1.5$ keV) and a non-thermal power-law in ASKAP J1832-091.

**Joint Observations and synergies with other observatories in the 2030s:** The SKA-low and -mid will be monitoring the southern sky, and we will use this data to inform our trigger criteria for an LPRT transitioning into a radio-bright state. DSA-2000 and ngVLA will be operational in the northern sky, likely detecting and monitoring dozens of LPRTs. LOFAR 2.0, MWA, and ASKAP will likely also still be operating in the 2030s. Existing and new radio surveys will also play a large role in discovering new LPRTs for further X-ray follow-up.

**Special Requirements:** TOO $<$ 12 hours.

**b. Accretion-Powered Neutron Stars and Black Holes**

*5. Population of high mass X-ray binaries in nearby galaxies*

**First Author:** Pragati Pradhan, Embry-Riddle Aeronautical University, pradhanp@erau.edu
**Co-authors:** Arash Bodaghee, Alicia Rouco Escorial, Breanna Binder, Chandreyee Maitra, Paul Draghis, Lee Townsend
**Abstract:** We propose 50 ks observations per galaxy of nearby galaxies selected from a sample of *eROSITA*-detected galaxies, and those that have ample multi-wavelength coverage in optical/IR. Our primary objective is to investigate population evolution by distinguishing between the number of high-mass X-ray binaries (HMXBs) and low-mass X-ray binaries (LMXBs), as well as their distribution throughout the galaxy (e.g., in the disc versus the bulge). With the excellent spatial resolution of *AXIS*, we will be able to identify point sources, enabling us to distinguish between diffuse and point emission in these galaxies.
**Science:** The latest all-sky survey in this decade was conducted by the *eROSITA* X-ray telescope that provides extensive all-sky coverage, especially for nearby galaxies out to $\sim$ 200 Mpc away. The unique strength of *eROSITA* lies in its sensitivity down to flux levels of a few $\times\ 10^{-15}$ erg s$^{-1}$ cm$^{-2}$. With the mission to conduct the most detailed X-ray All-Sky Survey to date, *eROSITA* completed 4 surveys of the X-ray sky (eRASS1-eRASS4) since its launch in 2019. The telescope is highly sensitive in the 0.2–8 keV energy range, making it an ideal instrument for capturing diffuse X-ray emission in nearby galaxies. However, the *eROSITA* photon statistics drastically worsen with increasing distance to an extended source. The FOV averaged PSF of 30 arc-seconds not only adds to source confusion at increasing distances but also hinders the identification of counterparts [367]. From the selected sample of 72 nearby galaxies, only 10 galaxies at distances less than 10 Mpc were bright for spectral analysis [311].

With *AXIS*, we aim to *provide an X-ray view of stellar population evolution*, which have good multi-wavelength coverage, are X-ray bright, and are relatively face-on, to permit ease in point source detection. *The goal of the proposal is to detect and characterize all high and low mass X-ray Binaries (XRBs) within the sample galaxies*. Some possible targets with ample multi-wavelength coverage and within 10 Mpc include NGC 4559, NGC 2903, NGC 6744, NGC 3521, NGC 5068, and NGC 4826. All of these galaxies have been imaged with HST, covering more than 90% of each galaxy. This will allow us to correlate positions from *AXIS* with objects from HST, across the optical (see, for example, Fig. 6) and sometimes including the ultraviolet or near-infrared. Given the relative proximity of these galaxies, we expect to find all high-mass, high-luminosity counterparts. The exceptions could be if the sources form a copious amount of dust or if they are embedded in a large cluster where confusion would dominate. To overcome this, we revert to IR measurements: most galaxies also have JWST observations either taken or planned. Thus, for these galaxies, we can look for mid-infrared sources that may be connected to X-ray sources, revealing not only the relative number of HMXBs to LMXBs, but also allowing us to see if there may be dust-enshrouded high-mass X-ray binaries present, e.g., analogs to the Galactic case of IGR J16318-4848 [41].

The number of X-ray binaries found in normal galaxies correlates with the Star Forming Rate (SFR) and stellar mass of the hosting galaxy. See e.g., Fig. 7 for SFR versus stellar mass. Additionally, the distribution of their X-ray luminosity functions exhibits a notable divergence between star-forming and passive galaxies. HMXBs are expected to be found in the arms of spiral galaxies and late-type systems with active star formation, while older LMXBs are found in older galaxy disks, bulges, and globular clusters. HMXBs and LMXBs also follow distinct evolutionary paths, necessitating the differentiation between these classes to understand their evolution. The specific star formation rate (sSFR), defined as the ratio of SFR to the stellar mass in the same region of a galaxy, is a proxy for assessing the dominant X-ray binary population. Regions with sSFR higher than $10^{-10}$ yr$^{-1}$ are typically dominated by HMXBs while

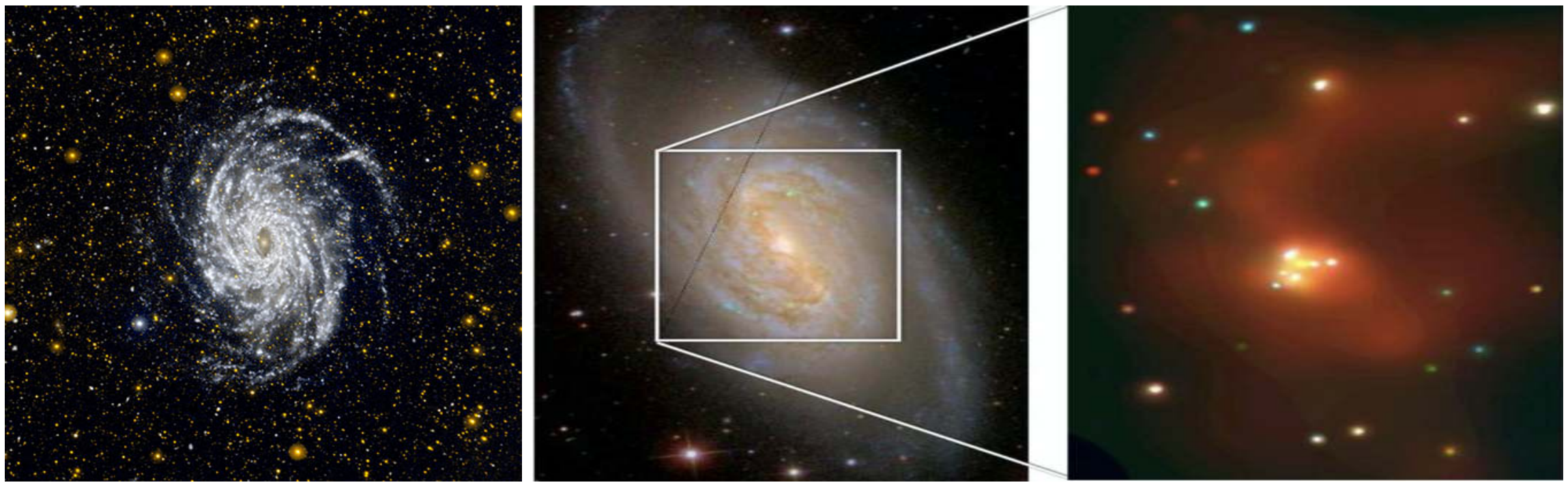

**Figure 6.** *Left: GALEX image of NGC 6744, also dubbed as 'Big Brother to the Milky Way'. Middle: The central $4' \times 6'$ region of NGC 2903 in the SDSS image. Young stars in spiral arms are traced in blue, and bright H II regions appear in green. The complex dust lane structures along the bar and spiral arms appear in a darker color. Right: The central $2' \times 3'$ NGC 2903 region of Chandra image. Red, green, and blue correspond to the energy ranges, 0.5–1.0 keV, 1.0–2.0 keV, and 2.0–8.0 keV, respectively. Figure from [637]*

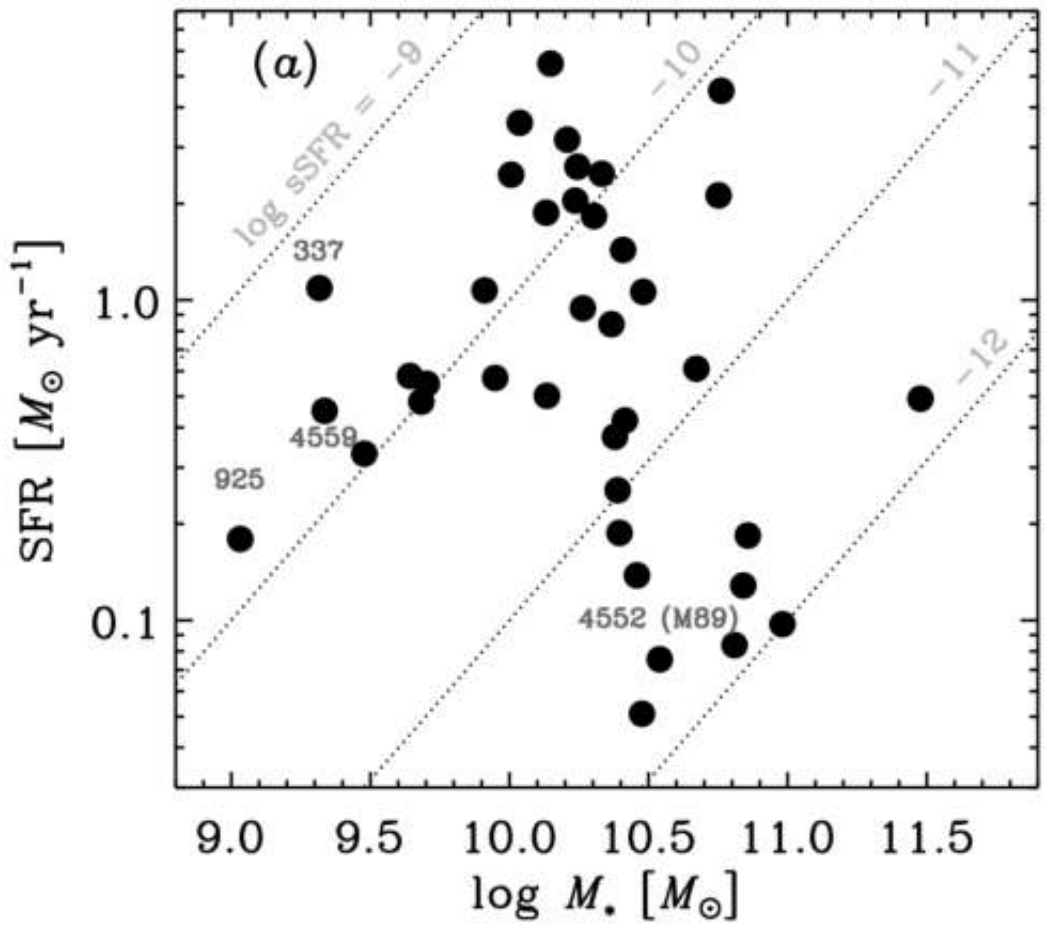

**Figure 7.** *SFR versus mass of galaxies. Figure from [324]*

sSFR below $10^{-11}$ yr$^{-1}$ suggests an LMXB-dominated environment [199]. The XRB emission can be well studied with *AXIS* since it covers an energy range greater than 3 keV above which the effective area of *eROSITA* reduces significantly. By understanding the relative numbers of HMXBs versus LMXBs, we can compare these stellar populations to binary population synthesis codes, such as BPASS [171,541]. Such comparisons will enable us to understand the importance of binary evolution in creating the population of stars in the galaxies we have chosen, which span a large range in star formation rates. Furthermore, these X-ray binaries are the progenitors of neutron star mergers [1,119], which have been shown to create the heaviest metals in the Universe. Thus, they provide excellent feedback both through their supernovae, which create the compact objects, and through future kilonovae explosions. Accurate modeling of X-ray scaling relations not only aids in comprehending X-ray emission across diverse galaxy populations, compact object formation, and stellar evolution but also informs us if the X-ray sources are consistent with stellar-mass XRBs or accreting supermassive black holes.

By December 2007, *Chandra* had already observed 383 nearby galaxies within 40 Mpc. A detailed analysis nearly a decade ago by [332], led to the detection of 17,599 independent sources, 479 of which were ULXs within 199 host galaxies. There were 18 sources detected in NGC 3521 alone, with a mean X-ray luminosity of $3\times 10^{38}$ erg/s. The updated analysis of NGC 3521 now results in 34 X-ray binaries

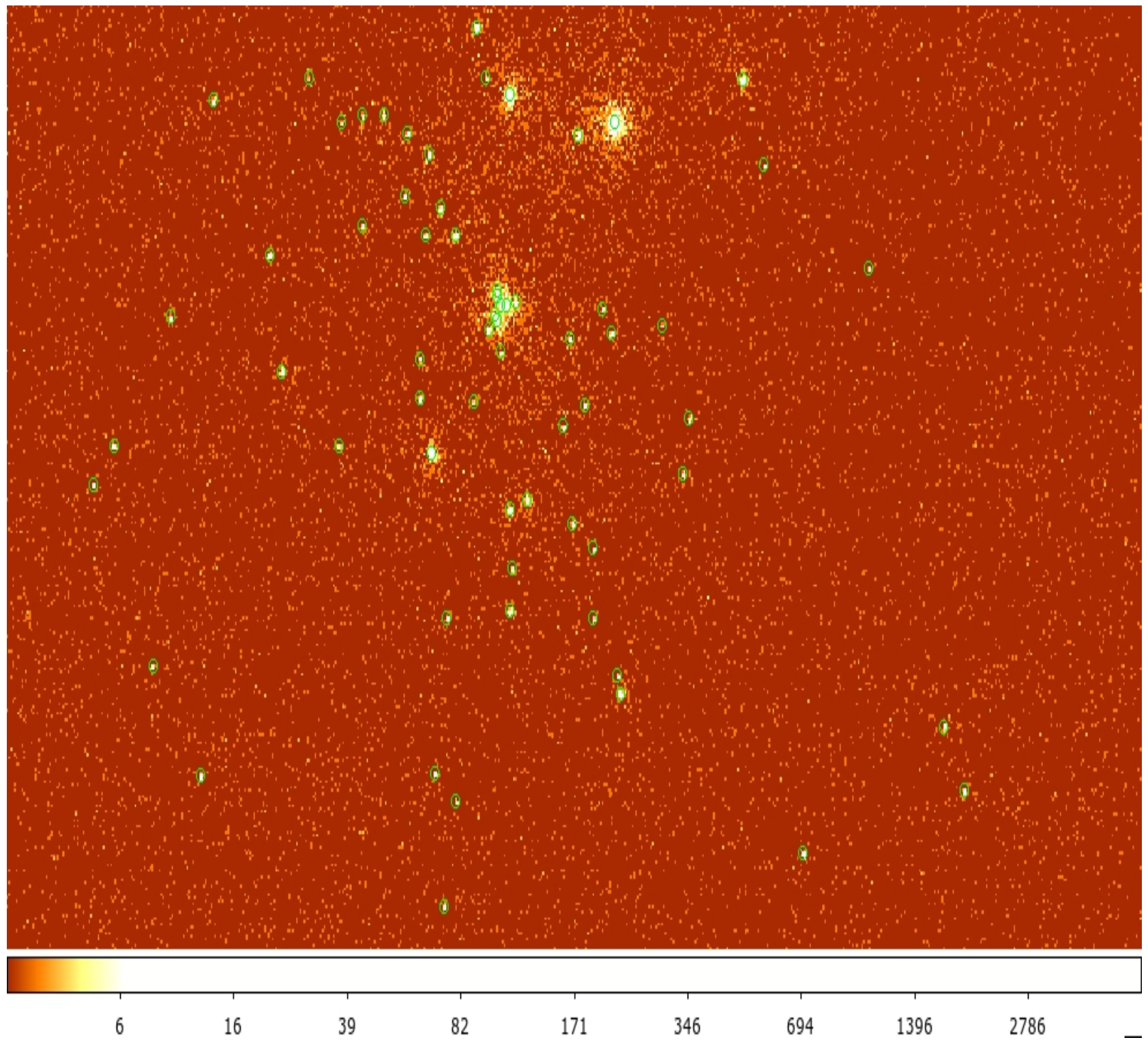

**Figure 8.** *50 ks simulation of NGC 2903 using source list from [637]*

with a mean flux of $\sim 1.5 \times 10^{-14}$ erg s$^{-1}$ cm$^{-2}$ and counts ranging from 3-200 (in 81 ks). Similarly, for NGC 4826, there are 21 X-ray binaries with a mean flux of $\sim 10^{-13.5}$ erg s$^{-1}$ cm$^{-2}$from 2-220 counts (in 26 ks). The number of point source detections in both NGC 3521 and NGC 4826 more than doubles if we relax the stringent selection criteria (details in [324]). Similarly, 92 point sources are detected in NGC 2903 [637], and Type II Be X-ray binary candidates were found very recently in NGC 6744 [585]. Furthermore, the already existing multi-wavelength observations will also help us identify the optical counterparts (e.g., star clusters and massive stars). With the PSF of *AXIS* being stable over off-axis positions, it can make more secure detections.

**[Exposure time (ks):]** 50 ks per Galaxy

**Observing description:**

We are aiming at achieving a luminosity of $10^{38}$ erg/s, typical for bright X-ray binaries. We estimate the exposure needed to achieve this for each galaxy by adapting a mean point source flux of $\sim 1.5 \times 10^{-14}$ erg s$^{-1}$ cm$^{-2}$ (taking the case of NGC 3521). We require 250 ks for a distance of 8.1 Mpc to achieve that luminosity with *Chandra*. If we assume a typical power-law of photon index 0.8, absorption $\sim 1 \times 10^{22}$ (line of sight along the galaxies is $\sim 1 \times 10^{20}$, but we take a higher value to account for local obscuration around X-ray binaries), it translates to 100 counts as a meaningful detection to be able to perform spectral analysis and period searches. *AXIS* is expected to be 5–10 times more sensitive than *Chandra* in the soft X-ray band, meaning it would need only about 1/5th the exposure time to reach similar depth (so 50 ks per Galaxy).

We used SIXTE to simulate X-ray sources expected in NGC 2903 for 50 kswith *AXIS*. The resulting image includes all point sources identified by [637], with source positions and fluxes based on their catalog (see Fig. 8). To further characterize the simulated population, we extracted and fit spectra for two representative sources, binned to 20 counts per bin. We show SrcID 13 from [637] that falls within the 90th

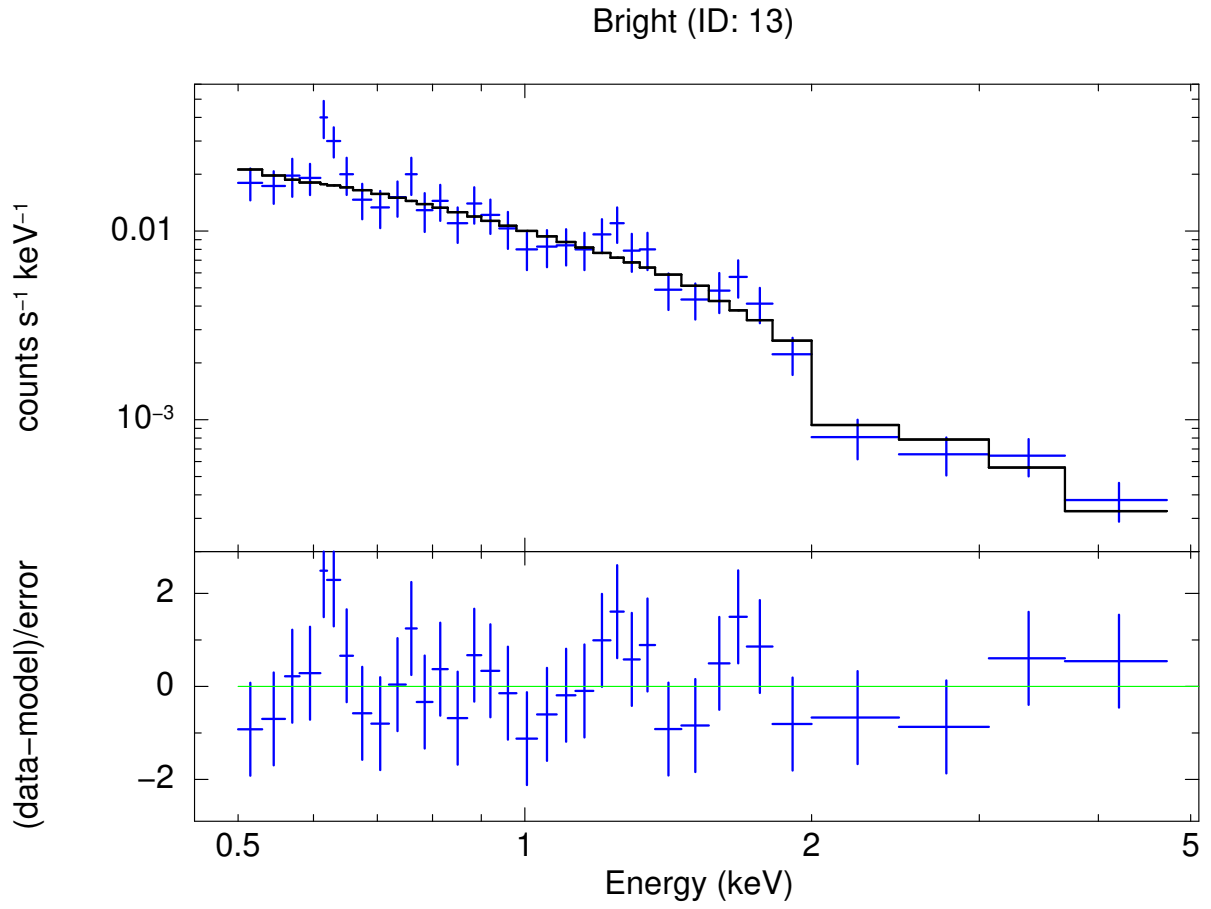

**Figure 9.** *AXIS spectrum for a bright source with AXIS.*

percentile of the most luminous sources, with an X-ray luminosity of $L = 2 \times 10^{38}$ erg/s in the 0.3–10 keV (see Fig. 9). The median luminosity of the sources in the sample is $L = 2.6 \times 10^{37}$ erg/s.

**[Joint Observations and Synergies with Other Observatories in the 2030s:]** Joint observations with JWST will enable the identification of dust-enshrouded X-ray binaries (XRBs), which may be obscured in other wavelengths. Complementary observations with HST will help identify the optical companions of these systems, providing crucial information about their stellar counterparts and evolutionary state. Together, these observations will enhance our understanding of the multi-wavelength properties of XRB populations in the target region.

**[Special Requirements:]** The observations can be conducted using a regular scanning mode, ensuring consistent coverage of the field without the need for rapid or time-sensitive follow-up.

*6. Monitoring outflows from high-mass gamma-ray binaries with AXIS*

**First Author:** Jeremy Hare (NASA GSFC, CUA, CRESST II, jeremy.hare@nasa.gov)
**Co-authors:** Hongjun An (Chungbuk National University), Arash Bodaghee, Oleg Kargaltsev (GWU), Seth Gagnon (GWU), Nazma Islam (NASA GSFC, UMBC), Arka Chatterjee (UPES University of Calcutta, India), Lee Townsend (Southern African Large Telescope and South African Astronomical Observatory, South Africa)
**Abstract:** High-mass gamma-ray binaries (HMGBs) are compact object binaries with massive companions where their broadband luminosity peaks at MeV-GeV energies. In systems hosting a neutron star, it is generally believed that the emission originates from particles accelerated at the shock between the pulsar wind and the companion wind, resulting in synchrotron and inverse Compton radiation. In systems hosting Be companion stars, a decretion disk is produced by the star, which can intersect the pulsar's orbit, given the large eccentricities of these systems. Therefore, as the pulsar orbits the Be star, it can pass through and interact with this disk-launching material out of the binary at large velocities (up to $\sim 0.1$c). Chandra observed the most prevalent example of this in the PSR B1259-63 system, where material was launched (and possibly accelerated) after periastron passages. Chandra has detected extended emission in several other systems (e.g., LS 5039, HESS J0632+057). This emission is typically relatively faint (i.e., $\gtrsim 10^{-13}$ erg cm$^{-2}$ s$^{-1}$) and observed on angular scales of a few arcseconds to arcminutes away from the binary, requiring a sensitive high angular resolution instrument to observe. Here we propose observations of several HMGBs with *AXIS* in search of extended emission. If this emission is detected, we will continue to monitor its trajectory and the evolution of its flux. This study will lead to a better understanding of the launch mechanism and evolution of this extended emission from these systems. It will also provide evidence of pulsars existing in systems where they have yet to be found.
**Science:** High mass gamma-ray binaries (HMGBs) are rare systems found in the Galaxy, with currently only about 10 known (see e.g., [112,166] for reviews). They consist of a high mass ($> 10\ M_{\odot}$) O or B-type star being orbited by either a neutron star (NS) or black hole (BH). These systems differ from the more commonly known high mass X-ray binaries (HMXBs) in that their broadband spectral energy distribution (SED) typically peaks at MeV or GeV energies, with many systems detected up to TeV energies (see e.g., [9,146,213]). For binaries hosting an NS, the high-energy emission is thought to be produced by particles accelerated in the colliding winds from the massive star and the NS. On the other hand, in binaries with a BH, the emission is thought to be from jets oriented towards the observer.

To date, there are three HMGBs hosting NSs that show radio pulsations, namely PSR B1259-63, PSR J2032+4127, and LS I +61° 303 [272,340,608], while many others, such as LS 5039 and HESS J0632+057, are suspected to also contain pulsars that remain to be detected. The lack of radio pulsations in these latter systems is not surprising. It is likely due to free-free absorption from the massive companion's wind, making these pulsations difficult to detect [641]. In fact, the pulsations in LS I +61° 303 were only recently detected by FAST in a single observation and not detected in follow-up observations at the same orbital phase, likely due to the first observation occurring during fortuitous conditions (e.g., a low density pocket in the massive star's wind [608]).

The pulsars in these systems are expected to be relatively young, with powerful pulsar winds (PW). These relativistic winds are strong enough to inhibit the massive and slow stellar wind from accreting onto the NS surface; thus, a shock forms between the PW and stellar wind. This shock accelerates particles leading to luminous X-ray emission ($L_X \approx 10^{32-34}$), which is modulated on the binary orbital period and well described by a power-law with a hard photon index (i.e., $\Gamma < 2$; see e.g., [113,352,561]). HMGBs also tend to have large eccentricities ($e > 0.5$), so the NS has a highly elliptical orbit, which means that the pulsar spends a significant amount of time at apastron. Originally, Chandra was used to search for evidence of the shocked PW at apastron and several binaries were observed and found to contain

evidence of extended emission (e.g., PSR B1259-63 [434], LS 5039 [168]), while others lacked any evidence of extended emission (e.g., HESS J0632+057 [276], PSR J2032+4127 [402]).

Here, we propose *AXIS* observations of four high-mass gamma-ray binaries at apastron to confirm and study the extended emission observed by Chandra, or to set the deepest limits possible on the lack of extended X-ray emission. This will enable us to compare and understand the parameters of binary systems that are necessary to produce extended X-ray emission versus those that do not. For sources with pulsars, it will also allow us to place constraints on the mass loss from massive stars due to stellar winds and how those winds interact with the unshocked PWs. Given the fluxes of the bright binary point sources ($F_X \approx 10^{-12} - 10^{-13}$ erg s$^{-1}$ cm$^{-2}$), faintness of the extended emission ($F_X \approx 10^{-13} - 10^{-14}$ erg s$^{-1}$ cm$^{-2}$), and small angular scales at which the extended emission is detected from the binary ($< 10''$), *AXIS* will be the only instrument suitable for this kind of study.

The most interesting case of extended emission from an HMGB is the cyclic emission from PSR B1259-63. This system hosts a young radio-detected pulsar, with a characteristic age of 330 kyr, in an elliptical 3.4-year orbit around a massive 15-30 $M_{\odot}$ Be companion star [272]. The decretion disk of the Be star is inclined to the orbit of the NS, such that the NS passes through and interacts with the decretion disk twice every orbital period. Chandra discovered extended emission when the NS was at apastron, which was initially thought to be the shocked PW [434]. Remarkably, additional follow-up observations found that the extended emission was moving away from the binary in the apastron direction (see Figure 10), allowing [278,435] to measure the projected velocity of the extended emission, which was found to be $v \approx 0.15c$. Observations following additional periastron passages allowed [226] to discover that these outflows seemed to occur every orbital period, but with some variation in their brightness and morphology. Interestingly, they also found that the clumps show a hint of acceleration in their projected motion.

The physical interpretation of this extended emission is that as the NS passes through the Be star disk, it scatters debris outward. In the periastron direction, the strong stellar winds intercept the expelled material, and it does not escape the binary to large distances [51]. However, in the apastron direction, where the pulsar spends most of its time, the PW carves out a channel in the massive stellar wind. As the pulsar passes back through the disk in the apastron direction, material from the disk gets knocked into the channel carved out by the PW, where it interacts with the relativistic PW as the pulsar heads toward apastron. The ram pressure from the PW can then accelerate this clump of disk material to the large observed velocities [51]. The X-ray emission in this scenario comes from relativistic particles accelerated at the shock between the PW and clump of disk debris, producing synchrotron emission [226,435]. The spectra of the clump exhibit hard photon indices ($\Gamma \approx 1.4$) that do not appear to show evidence of cooling, although with large errors due to the faintness of the extended emission. Unfortunately, in the past two orbital cycles, some extended emission has been seen, but it has faded before traveling to large distances from the binary, precluding a velocity measurement [228,229]. The physical cause for the apparent differences in the cycle-to-cycle brightness of the extended emission remains unclear; thus, further observations are necessary.

We propose to observe this binary twice with *AXIS*. The first observation would occur at a binary phase of 0.5, which is the earliest time a clump has been observed. In contrast, the second observation would occur at a binary phase of 0.55, where the extended emission has typically been found to be well separated from the binary. We also request short 20-ks TOO observations of the binary every 60 days if the clump is detected in the first observation. This will allow us to closely monitor the clump's motion and confirm or refute the previous evidence of accelerated motion. If the clump is as bright as seen by Chandra, we will also be able to look for evidence of spectral cooling, which was challenging to do with Chandra. These observations will allow us to understand the mechanism behind the ejections of relativistically moving plasmoids observed in the PSR B1259-63 system.

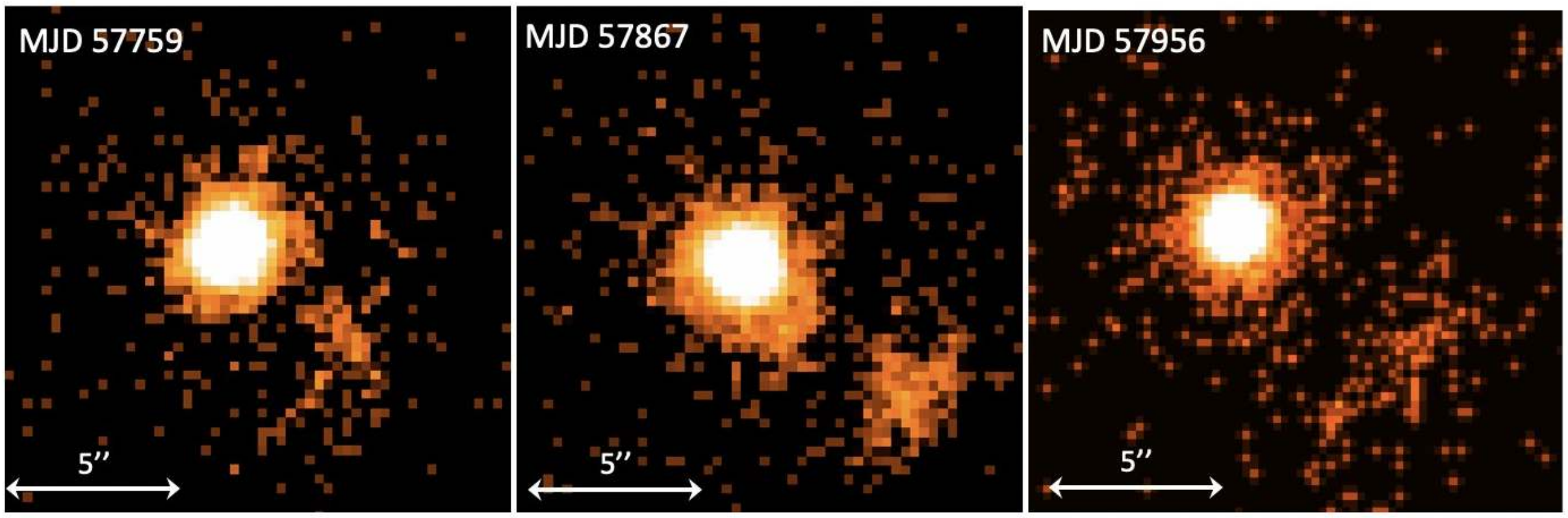

**Figure 10.** Chandra ACIS images of the extended X-ray emission launched from the high mass gamma-ray binary PSR B1259-63. The extended emission (south-west of the point source) showed signs of acceleration and reached a velocity of $\approx 0.1c$ (image adopted and modified from [226]

The only other HMGB with reported extended X-ray emission is LS 5039. This system consists of a $\sim$20 $M_\odot$ star being orbited every 3.9 days by a compact object of unknown type [385]. Recently, claims of an X-ray periodicity in this system were made, but they have been challenged (see [275,597,629]), and no radio pulsations have been detected. In [168], the authors found faint excess extended X-ray emission out to angular scales of $\sim 1'$ surrounding the binary having a flux of $F_X \approx 9 \times 10^{-14}$ at an orbital phase of $\sim$ 0.12. We request a single *AXIS* observation at the same orbital phase to confirm this extended emission and place tighter constraints on its spectrum. If the extended emission and spectral properties are confirmed, it would provide additional support for an NS in LS 5039, as it would suggest a similar physical scenario to the PSR B1259-63 system.

The two other promising candidates to host extended X-ray emission are HESS J0632+057 and LS I +61° 303. These systems both have B-type companion stars in orbits with periods of 317.3 days and 26.5 days, respectively. The longer orbital period of HESS J0632+057 makes it more similar to PSR B1259-63. Chandra observations at apastron and deep Swift-XRT observations taken across all orbital phases failed to locate any small-scale extended emission; however, the Swift-XRT image shows a hint of an X-ray shell, possibly the binary's supernova remnant, surrounding the binary [276]. Here we propose an *AXIS* observation to confirm or refute the SNR surrounding HESS J0632+057 and to determine its properties. As mentioned above, radio pulsations have been detected in LS I +61° 303, making it a suitable candidate for searching for extended X-ray emission. The source was observed by Chandra twice, with the first observation occurring near periastron passage. The second observation was taken at apastron passage, but was relatively short (only 20 ks). No extended emission was detected in either case. We propose to obtain a deeper observation of LS I +61° 303 at apastron with *AXIS*.

**Exposure time (ks):** 100 ks (+120 ks TOO)

**Observing description:** We propose 2 20 ks *AXIS* observations of PSR B1259-63, and 20 ks observations of LS 5039, HESS J0632+057, and LS I +61° 303 to search for or confirm the extended emission around the binary. If extended emission is detected around PSR B1259-63 during the first observation, we request 6 20 ks TOO observations taken roughly every 60 days over a year, which will provide us with enough measurements of the clump's position to confirm the hint of acceleration of the clump. The Chandra observations of PSR B1259-63 have spanned exposure times of 30-60 ks; therefore, a 20 ks *AXIS* exposure will allow us to collect roughly three times the number of counts as Chandra (i.e., roughly between 150-1000 counts, depending on the brightness of the extended emissions). This will be sufficient to accurately measure the clump's position to sub-arcsecond precision and to constrain its spectrum in each

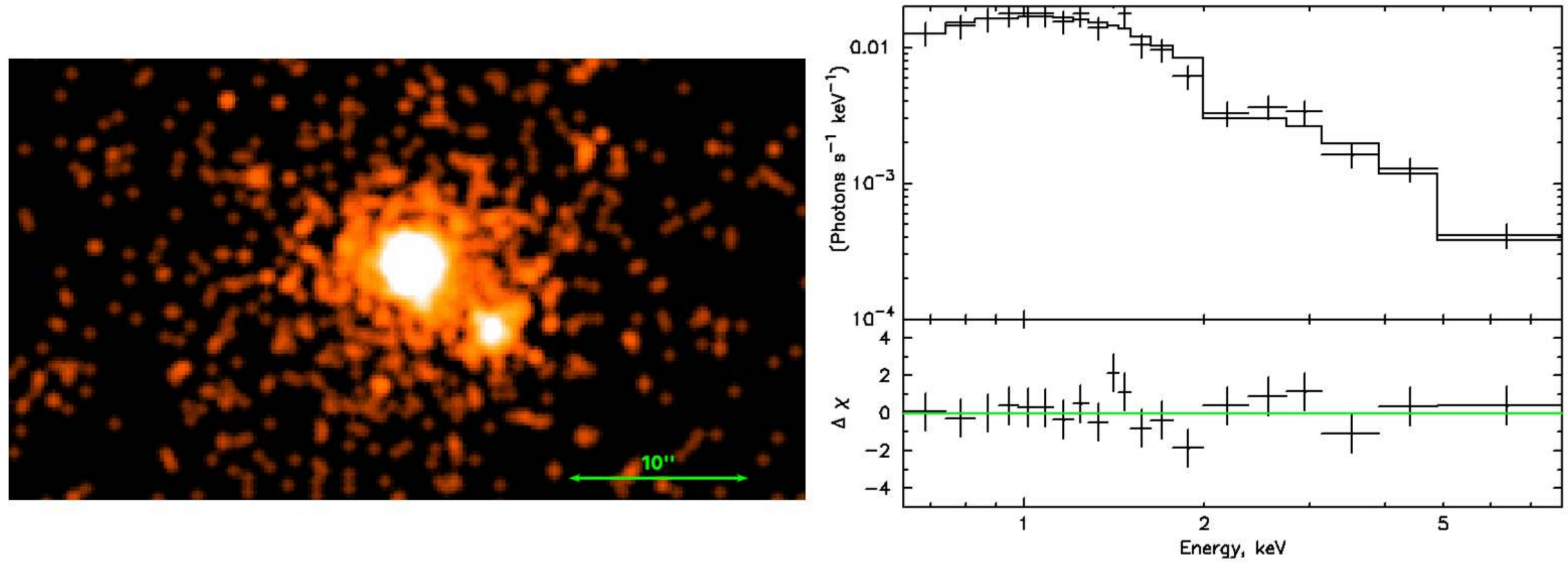

**Figure 11.** Left: AXIS simulated image (based on Chandra image from MJD 57867 shown in Figure 10) of the extended X-ray emission launched from the high mass gamma-ray binary PSR B1259-63. Right: Simulation of the X-ray spectrum extracted from the extended emission in a 20-ks X-ray observation fit with an absorbed power-law model. In 20 ks the photon index can be constrained to $\Delta\Gamma \pm 0.1$ allowing us to search for spectral evolution of the extended emission.

individual observation (see Figure 11 for a simulated *AXIS* image and spectrum from the bright clump observed by Chandra on MJD 57867). Simulations show that for an absorbed power-law model, we will be able to constrain the photon index to an accuracy of $\Delta\Gamma \pm 0.1$. For the other binaries, we request 20 ks observations at apastron. This will allow us to reach limiting fluxes of $\sim 10^{-15}$ erg s$^{-1}$ cm$^{-2}$, which is over an order of magnitude deeper than the fluxes of the extended emission detected by Chandra in PSR B1259-63 and LS 5039. We request the observations be carried out in a custom sub-array to limit the effects of pile-up due to the relatively bright emission from the point-source binaries (i.e., $F_X \approx 10^{-11} - 10^{-12}$ erg s$^{-1}$ cm$^{-2}$).

**Joint Observations and synergies with other observatories in the 2030s:** If a clump is detected in any binary, we can request SKA DDT observations for those in the southern hemisphere or VLA in the northern hemisphere to search for the X-ray emitting clump in the radio. If it is detected in radio, it will allow us to probe this extended emission on smaller angular scales and measure the broadband spectrum, from X-rays to radio wavelengths.

*7. AXIS as a sensitive thermometer in transient low-mass X-ray binaries*

**First Author:** Mark Reynolds (The Ohio State University, USA – reynolds.1362@osu.edu)
**Co-authors:** Nathalie Degenaar (University of Amsterdam, Netherlands), Jeroen Homan (Netherlands), Craig Heinke (University of Alberta, Canada), Dany Page (The National Autonomous University of Mexico, Mexico), Bettina Posselt (University of Oxford, UK), Rudy Wijnands (University of Amsterdam, Netherlands)
**Abstract:** The ultra-dense core of a neutron star is theorized to be superfluid. Constraints on the superfluid properties are essential for narrowing down the equation of state (EoS) of ultra-dense matter. Additionally, this superfluidity is a crucial element for comprehending numerous observable characteristics of neutron stars. However, we lack detailed information about the physics of the core superfluid. A measurement of the core's heat capacity can constrain its properties. The accreting neutron star in transient low-mass X-ray binaries undergoes measurable internal heating. Our primary objective is to measure the subsequent cooling of the neutron star to constrain the heat capacity of its ultra-dense core. Upon the identification of such a neutron star, *AXIS* observations will enable precise constraints on the temperature of the neutron star's crust. This measurement will provide essential information about the superfluid properties of the core and enable the exclusion of specific families of EOSs.
**Science:** *What is the equation of state of ultra-dense matter? This is one of the fundamental outstanding questions in modern physics and astrophysics. Observations with AXIS will provide key observational constraints on the temperature of neutron stars, facilitating the determination of their heat capacity and, hence, the nature of their cores.*

Neutron stars, with masses up to ∼2–2.5 $M_\odot$ and radii of ∼10–14 km, contain the densest observable form of matter in the Universe [319]. The bulk properties of this ultra-dense matter, encapsulated in the equation of state (EOS), are constrained by measurements of neutron star masses & radii, and gravitational waves from neutron star mergers (e.g., [3,83,319,373,420,490]). However, details about the microscopic properties of ultra-dense matter, such as superfluidity, do not directly follow from such studies and are challenging to obtain (e.g., [422]).

It is expected that the neutron star core, with temperatures of $\sim 10^6 - 10^8$ K and comprising about 99% of the stellar mass, is (partly) superfluid. Exactly what fraction of the core forms such a frictionless liquid plays a key role in the dynamics of neutron stars, and the long-term evolution of their temperature, magnetic field, and rotation rate (e.g., [16,66,625]). Moreover, constraints on the superfluid properties are important in the effort to narrow down the EOS of ultra-dense matter (e.g., [245]).

Obtaining observational constraints on the superfluid properties of neutron star cores is thus of paramount interest. This is, e.g., illustrated by the tremendous excitement spurred by Chandra measurements of the cooling of the young neutron star in the supernova remnant Cas A [236]. Its apparent strong temperature decay could only be explained if a large fraction of the core is superfluid (e.g., [421,535]). However, accounting for instrumental effects makes the temperature decrease much more modest than initially thought [464,614], significantly weakening the constraints on the core superfluid. Information about the core superfluid can be obtained through the heat transport coefficients, neutrino emissions, or heat capacity of neutron stars. The latter, the ability to store heat, is particularly interesting: unlike transport and emission processes, it is not determined by the uncertain, and model-dependent, description of particle collisions. A measurement of the heat capacity requires the occurrence of observable heating and cooling of a neutron star [422]. Excitingly, we encounter such exceptional circumstances in transient low-mass X-ray binaries.

Neutron stars in low-mass X-ray binaries (LMXBs) accrete matter from a ∼0.1 $M_\odot$ companion star through a disk. Transient LMXBs exhibit accretion outbursts with durations of weeks to decades that are separated by long periods of quiescence with little or no accretion occurring. In quiescence, the bare surface of the neutron star, not being outshone by the accretion disk, is directly observable. Its effective

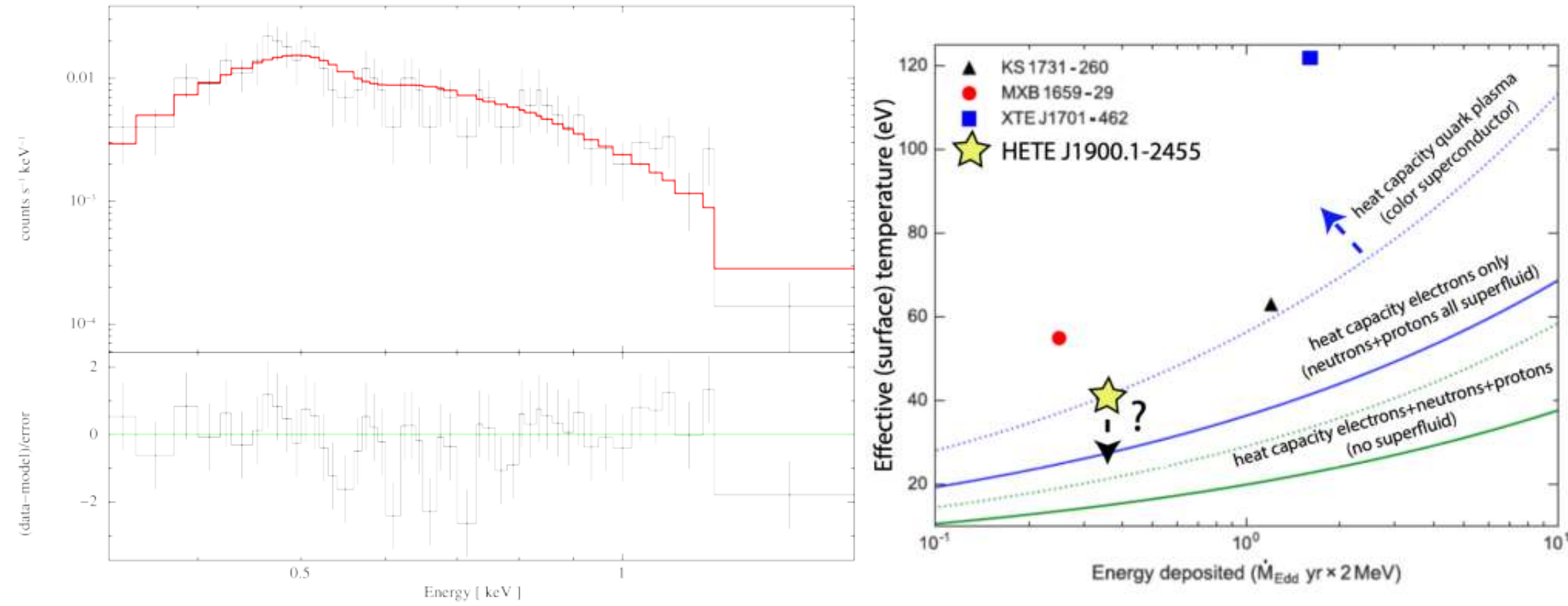

**Figure 12.** **Left:** AXIS simulation of a 50 ks observation of HETE J 1900.1–2455, with best fit model overplotted. We assume a `nsatmos` spectrum and a flux of $f_x = 2 \times 10^{-15}$ $erg\ s^{-1}\ cm^{-2}$ and fit with the same model. The measured neutron star temperature is kT = (36.3 ±2.3) eV (1$\sigma$). **Right:** Temperature versus energy deposited during accretion for 4 neutron stars (adapted from [134]). The curves indicate different heat capacities and associated superfluid properties. The temperature of HETE J1900.1–2455 measured during an observation in 2018 [148] is indicated by the (yellow star), and provides the strongest constraint. Further cooling may have subsequently occurred. With an accurate temperature measurement, we can constrain the fraction of nucleons that are paired in a superfluid. A measurement of further cooling is beyond *Chandra's* capabilities but will be straightforward for *AXIS*.

temperature can then be measured with sensitive X-ray instruments [505], and mapped onto the interior temperature [625].

During the outbursts, the neutron star crust, the ∼1-km thick layer that separates the core from the observable surface, is compressed. This causes a series of nuclear reactions that deposit ∼1.5–2 MeV of heat energy, per accreted baryon, deep in the crust [89]. An additional source of "shallow heat" is inferred for several neutron stars, but its origin is currently unknown (e.g., [155]). For sufficiently long outbursts, the heating pushes the crust out of thermal equilibrium with the core [506]. In quiescence, the accretion-heated crust is then observed to cool on a time scale of years [612]. Comparing these observations with thermal evolution simulations provides direct information about the crust structure and composition [90]. Once the crust has thoroughly cooled, the surface temperature tracks that of the dense stellar core.

[422] noted that if an accretion outburst is long and/or strong enough, the neutron star core may also rise in temperature. [134] explored this exciting possibility, showing that this allows for measuring its heat capacity and hence provides a window into constraining the superfluid properties. This is achieved by comparing the heat injected into the core, estimated from the outburst properties as observed by X-ray monitoring instruments, with the core temperature measured in quiescence with sensitive X-ray satellites, e.g., [26,148,612].

**[Exposure time (ks):] 200 ks**

**Observing description:** Based on the light curve shape and thermal evolution simulations, we can calculate the heat injected into the core during the outburst and determine when the neutron star has returned to quiescence. To demonstrate the potential for future *AXIS* observations, we consider the source HETE J1900.1-2455. The combination of its low temperature and long, well-monitored outburst made HETE J1900.1–2455 a key object providing direct and strong constraints on the core superfluid properties ([148]). Archival observations demonstrate the neutron star had fully cooled at the time of a 2018 Chandra

observation. To put firm constraints on the heat capacity, we aim to determine the temperature with a $1\sigma$ error of $\sim 5\%$.

We simulated a neutron star atmosphere model (nsatmos) with a temperature of 30–40 eV, using the known extinction ($N_H = 2 \times 10^{21}$ cm$^{-2}$) and distance (4.7 kpc; [191]), and assuming a neutron star mass and radius of 1.6 $M_\odot$ and 12 km, respectively. Since quiescent neutron star LMXBs may also exhibit power-law emission, we also simulated a set of spectra that include an additional component (pegpwrlw), that has an index of $\Gamma$ = 1.5 and contributes $\sim$ 50% to the 0.5–10 keV unabsorbed X-ray flux. Finally, we simulated a single power-law spectrum ($\Gamma$ = 1.5) yielding the same count rate. The 0.5-10 keV unabsorbed flux is $F_X \sim 2 \times 10^{-15}$ erg s$^{-1}$ cm$^{-2}$, which gives $L_X \sim 6 \times 10^{30}$ (D/4.7kpc)$^2$ erg s$^{-1}$. For these sets of spectral models, we simulated *AXIS* spectra for exposure times of 10, 25, 40, 50, and 100 ks, and determined the accuracy of the temperature measurement. We found that with 50 ks, this parameter can be measured with $\sim$ 5% uncertainty for the purely thermal model (Fig. 12, left) and with $\sim$ 8% accuracy for the thermal+powerlaw model.

Based on our spectral simulations, we require a 50 ks *AXIS* observation of a source such as HETE J1900.1–2455. This would allow us to determine the temperature of the neutron star accurately and put strong constraints on its heat capacity, if it remained at a similar brightness as during the archival Chandra observation ([148]). If this target faded, the neutron star must have a temperature $\sim$ 25 eV (a core temperature of $\sim 6 \times 10^6$ K). This would pose strong constraints on the heat capacity. Such an observation would place stringent limits on the presence of quarks in the dense core (Fig. 12, right). Measuring the NS temperature at such low values requires the capabilities of *AXIS*.

*AXIS'* combination of large effective area ($f_x \lesssim 10^{-15}$ erg s$^{-1}$ cm$^{-2}$) and superior low energy sensitivity is required to accurately constrain the temperature of the cool neutron star (kT $\sim$ 0.3 keV)

**[Joint Observations and synergies with other observatories in the 2030s:]** Multi-wavelength observations of the source decay could strengthen the constraint on the energy injected into the core; for example, UV/optical/IR and radio observations could constrain outflows present during the outburst.

**[Special Requirements:]** Initial monitoring observations of the decay to quiescence would provide constraints on the accreted mass (on a timescale of weeks to months) and identify the appropriate time for a deep pointed observation. Here, we envision 50ks of monitoring before a pointed 50ks observation. We expect 2 targets during the 5-year prime mission phase.

*8. Very faint X-ray binaries*

**First Author:** Fiamma Capitanio (INAF/IAPS, Italy, fiamma.capitanio@inaf.it )
**Co-authors:** Antonella Tarana (INAF/IAPS, Italy); Mason Ng (McGill University, Montréal, Canada), Andrea Gnarini (Università degli Studi Roma Tre, Italy), Jeremy Hare (NASA GSFC, CUA, CRESST II), Paul Draghis (MIT), Kaya Mori (Columbia University), Shifra Mandel (Columbia University), Craig Heinke (U. Alberta)

**Abstract:**

We propose to observe a sample of very faint X-ray binaries (VFXBs) with *AXIS* to gain a better understanding of their nature. VFXBs are a subclass of black hole and neutron star X-ray binaries that accrete at very low rates (below $10^{36}$ ergs/sec), a process that remains poorly understood. Thanks to its unprecedented sensitivity, *AXIS* will be able to characterize the 0.2-–10 keV spectrum of these sources, ultimately determining their accretion rate, disk parameters, reflection component, and the presence of winds or surrounding material. These observations will shed light on the nature of this class of objects and the mechanisms behind their low accretion rates. Indeed, VFXBs form an inhomogeneous group of sources that are difficult to distinguish due to their low signal-to-noise ratio. The *AXIS* GPS will identify many new VFXBs. They appear to outnumber bright X-ray binaries. The high angular resolution of *AXIS* will enable counterpart identification in other wavebands and in new and upcoming NIR/optical surveys, which can help distinguish VFXBs from other faint X-ray sources and determine their binary parameters. The nature and distribution of VFXBs are particularly interesting for studies of X-ray binary evolution and population modeling. Additionally, they could be sources of LISA-detectable gravitational waves. Since VFXBs constitute an inhomogeneous group of sources, our sample includes candidate VFXBs thought to belong to various subcategories.

**Science:** Luminous accreting low-mass X-ray binaries (with a typical flux of $\sim 10^{37}/10^{39}\,\mathrm{erg\,s^{-1}}$) have been extensively studied using many different observatories across various energy ranges, and their observational properties are becoming well known. However, in the last 20 years, the current generation of sensitive X-ray instruments (e.g., Chandra, XMM-Newton, and Swift) has discovered a growing number of X-ray binaries that exhibit peak luminosities well below the typical sensitivity of X-ray monitoring instruments ($\simeq 10^{36}\,\mathrm{erg\,s^{-1}}$). The properties of this sub-class of X-ray binaries remain poorly understood due to observational challenges posed by current instruments.

This subclass of X-ray binaries is named very-faint X-ray binaries (VFXBs) and represents an inhomogeneous group of sources. They have been observed both as transients (exhibiting days-to-weeks-long outbursts with a peak luminosity below $\sim 10^{36}\,\mathrm{erg\,s^{-1}}$) and as persistent X-ray sources (displaying a relatively steady luminosity of $\simeq 10^{34}/10^{36}\,\mathrm{erg\,s^{-1}}$).

About three dozen VFXBs are known in our Galaxy, and their number is still growing. Several of these have been established to harbor neutron stars through the detection of thermonuclear X-ray bursts [156,157,261]. However, there are also short-period black holes expected among them [28,291,342,611]. Despite these discoveries, their sub-luminous character is not fully understood.

For this reason, VFXBs provide a gateway to studying accretion processes and outflows in an underexplored mass-accretion regime.

Several explanations may account for the faint nature of these sources. Likely, more than one of the following hypotheses applies to different sub-groups of VFXBs:

- A subpopulation of VFXBs has been identified recently as short-period ultra-compact X-ray binaries (UCXBs, $P < 1.5\,\mathrm{hr}$) hosting a small or evolved donor star and consequently a small accretion disk [232,262,287].
- Another possible explanation for the faint X-ray emission of VFXBs is that the accretion flow is radiatively inefficient. Standard accretion theory predicts that in slow accretion regimes, the radiative

efficiency decreases (e.g., [399], and references therein). This should be more pronounced in black hole systems with short orbital periods (but not necessarily UCXB sources; see e.g., [291,622]). A handful of candidates have been identified among the VFXBs ([127,362,544]).

- Symbiotic systems (e.g.,[82], and references therein) could form another subclass of VFXBs ([37,532]). In these systems, the compact object (which can be a neutron star or even a white dwarf) accretes from the weak wind of the low-mass companion. This inefficient process can explain the low X-ray luminosity. A somewhat surprising aspect is that the one VFXB identified as a symbiotic X-ray binary (IGR J17445–2747; [532**? ?** ]) displays thermonuclear bursts ([366]). Symbiotic X-ray binaries are thought to host high-magnetic-field neutron stars and at least one of them hosts a giant companion star ([530]).
- Some of them may harbor neutron stars with relatively strong magnetic fields that can hinder efficient accretion onto the stellar surface (e.g., [152,232,432,610]. Several VFXBs have been identified as accreting millisecond pulsars (e.g., [12,514,515]).

However, some VFXBs require alternative explanations (e.g., [156,232]). Most known VFXBs are concentrated toward the Galactic center and the Galactic plane, where they appear to outnumber bright X-ray binaries (e.g., [149,151,390,613]).

Continued monitoring of the Galactic Centre region has revealed that many X-ray binaries that generate bright outbursts also show very faint outbursts ([149,151,156]). A fraction of VFXBs may be "normal" X-ray binaries that occasionally interrupt their quiescence to exhibit only very small outbursts. Moreover, some VFXBs were discovered only from the detection of their X-ray bursts (burst-only sources) detected by the WFCs on board the BeppoSAX satellite. The identification of this group of sources is still a matter of debate ([98,125]) The *AXIS* Galactic Plane Survey (GPS), reaching a minimum luminosity of $10^{30}$ erg s$^{-1}$ for a distance of 8 kpc, would make an important contribution by increasing the number of known VFXBs and improving our knowledge of them. With *AXIS*'s GPS, it would be possible to map out the distribution of VFXBs and compare their relative abundance with brighter X-ray binaries. Another key aspect is the spectral characterization of this subgroup of sources. With its unprecedented sensitivity, *AXIS* will be able to produce detailed spectra of these sources. These spectra could be collected both using GPS data and deep pointed observations, enabling an essential survey of VFXBs with spectra and light curves.

Moreover, to deepen our understanding of VFXBs, it is crucial to determine their binary parameters. This could be achieved through coordinated follow-up observations in the infrared and optical energy ranges. However, such studies are limited by the relatively large distances ($>$ 8 kpc) and the high absorption columns toward the Galactic center.

**[Exposure time (ks): 20 ks]**

**Observing description:**

We have selected a sample of VFXBs, some of which still have an unclear nature or a good position. Table 1 lists the names, coordinates, and main characteristics of these sources [35]. *AXIS* is expected to be 5–10 times more sensitive than Chandra in the soft X-ray band, meaning it would require only about 1/5th of the exposure time to reach a similar depth.

To illustrate *AXIS*'s capabilities, we simulated a 20 ks *AXIS* spectrum of the VFXB XTE J1719−291, a candidate NS, using XSPEC version 12.14.0. The spectral model parameters are described in [27]: $N_H = 0.33\times10^{22}$ cm$^{-2}$, $kT_{bb} = 0.33$ keV, and a power-law with $\Gamma = 1.74$. The absorbed flux in the 0.5-–10 keV band is $2.57\times10^{-12}$ erg s$^{-1}$ cm$^{-2}$, and the blackbody component contributes 30% of the total flux (0.5–10 keV). The resulting spectrum is shown in Figure 13. We verified that, with a 20 ks exposure, all parameters are well-constrained and measured with a precision better than 5%.

**(a) Persistent Sources:**

The fluxes of the VFXBs listed in Table 1 are all consistent with the tested source; therefore, we request a 20

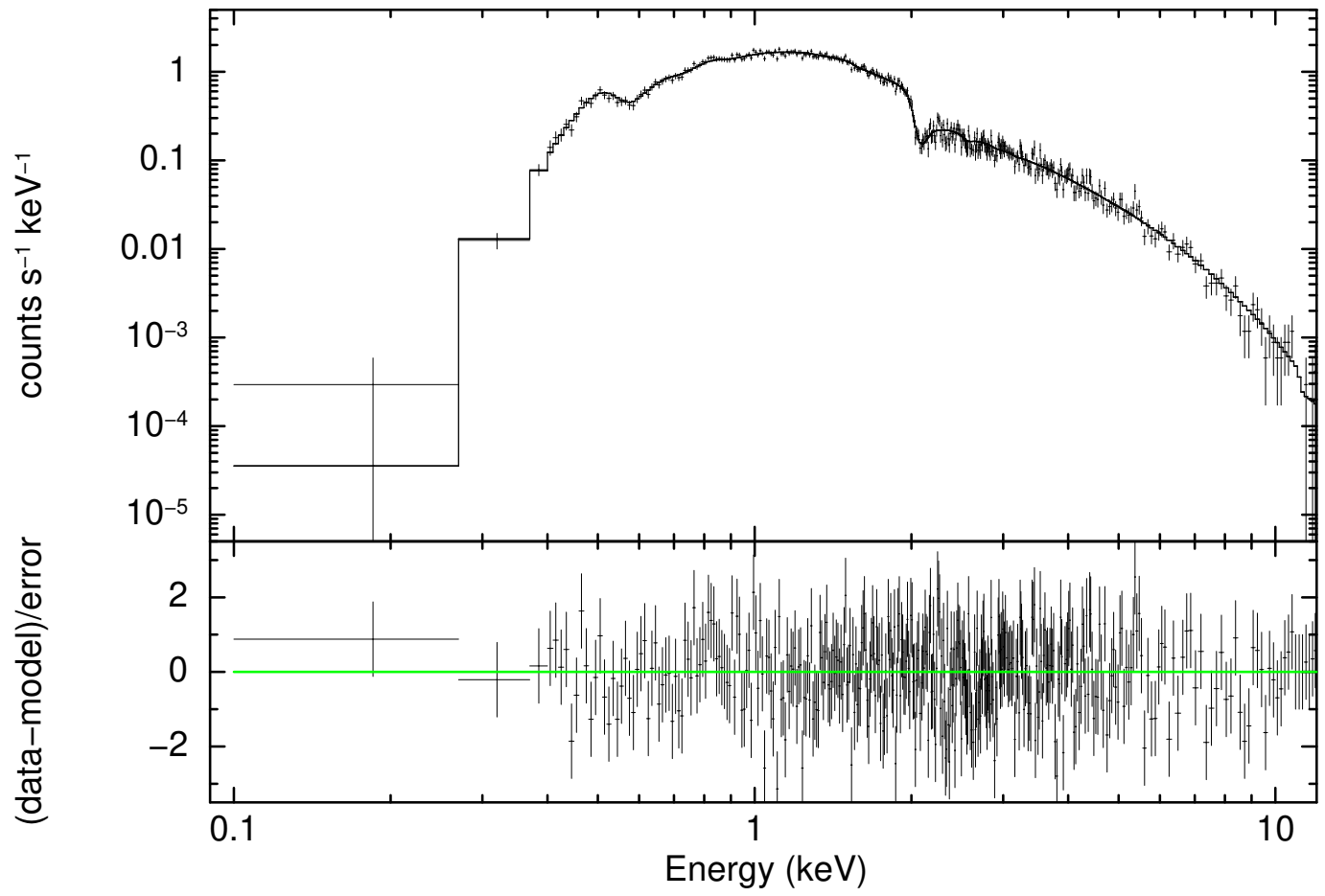

**Figure 13.** AXIS simulation of 20 ks with the model Tbabs(bb+powerlaw) performed with XSPEC for the VFXB candidate NS XTE J1719-291.

| System | Coordinates J2000 (RA, Dec) | Comments |
|---|---|---|
| IGR J17254−3257 | 17:25:25.87, -32 57 13.3 | P, NS, burster |
| SAX J1806.5−2215 | 18:06:34.08, -22 15 07.2 | QP, NS, burster |
| IGR J17494−3030 | 17:49:23.62, -30 29 59.0 | T, NS |
| XTE J1719−291 | 17:19:16.97, -29 04 10.4 | T, cNS |
| Swift J1357.2−0933 | 13:57:16.84, -09 32 38.8 | T, BH, $P_{\rm orb}$=2.6 hr |
| XTE J1728−295 | 17:28:38.97, -29 21 44.9 | T, cBH |
| XTE J1637−498 | 16:37:02.67, -49 51 40.6 | T |
| Swift J175233.9−290952 | 17:52:33.97, -29 09 52.3 | T |
| Swift J174038.1−273712 | 17:40:38.11, -27 37 12.1 | T |

**Table 1.** List of targets for VFXBs. The letters P, QP, and T refer to persistent, quasi-persistent, and transient, respectively. cBH and cNS refer to candidate BHs and candidate NSs, respectively. The indication "burster" implies that the source displays thermonuclear X-ray bursts (hence must harbor an NS). $P_{\rm orb}$ is the orbital period of the system.

ks exposure for all the persistent sources listed in the Table. In this way, we could characterize in detail the X-ray spectrum using detailed spectral models, constraining:

- the disk temperature and the inner radius for the accretion disk;
- the presence of reflection signatures;
- the optical depth and eventually the temperature of the corona for the inverse Compton component

**(b) Transient Sources:**

Furthermore, we propose monitoring transient sources during outbursts with a pointing every two days. The required exposure time depends on the flux of the outburst peak. In the case of transient sources, we could track the evolution of individual spectral components throughout the outburst. Once the source returns to the quiescent state, we request a further observation to study the source spectrum in this phase as well. In this case, too, the required exposure time depends on the source's quiescent brightness. When in quiescence, BH-XRBs are known to be less bright than NS-XRBs [98,99].

**[Joint Observations and synergies with other observatories in the 2030s:]** Joint Observations and synergies with other observatories in the 2030s: ELTs, SKA, ALMA, NewAthena, eXTP. A multi-wavelength coverage will be crucial for uncovering the nature of this subgroup of XRBs. In particular, simultaneous

eXTP observations could be essential for obtaining a broadband spectrum to constrain the electron corona temperature, which, especially for transient sources, could be too high to be determined with a spectrum extending only up to 10 keV. Additionally, radio coverage is essential for studying the accretion-ejection coupling in these accretion regimes. Simultaneous optical, infrared, and radio coverage could also be crucial for identifying the nature of the source.

**[Special Requirements:]** Some sources of the table require a ToO observation. Pointing strategy: an observation every two days.

*9. Unlocking the rapid X-ray variability of transitional millisecond pulsars*

**First Author:** Rebecca Kyer (Michigan State University, rkyer@msu.edu)
**Co-authors:** Mason Ng (McGill University), Francesco Coti Zelati (ICE-CSIC, IEEC), M. Cristina Baglio (INAF-OAB), Hongjun An (Chungbuk National University), Amruta Jaodand (Smithsonian Astrophysical Observatory)
**Abstract:** Transitional millisecond pulsars are a rare class of compact binaries composed of a low-mass stellar companion and a rapidly spinning neutron star that switch between an X-ray active state and a radio pulsar state. In their sub-luminous disk state, these systems exhibit unique X-ray variability in the form of rapid moding between high and low emission levels, in addition to occasional strong X-ray flares. This behavior has been detected in the three confirmed members of this class and is a mark of strong disk-state candidates. In the prototypical transitional system PSR J1023+0038, pulsations at the pulsar spin period have been detected in optical, UV, and X-ray observations during the X-ray high mode. More systems must be discovered and characterized to constrain models of the physical mechanism(s) behind these diverse X-ray variability signatures and unlock the full potential of transitional millisecond pulsars as unique gateways to our understanding of accretion physics. The large effective area, high time resolution, and high angular resolution of the *AXIS* probe will enable *i)* the characterization of X-ray moding in faint candidate systems in both the Galactic field and in globular clusters, *ii)* the search for pulsations at the pulsar spin period in many more transitional systems and during low modes of the brightest sources, and *iii)* the first direct X-ray observations of a state change as the accretion disk forms or dissipates.
**Science:**

Transitional millisecond pulsars (tMSPs) display unique X-ray variability during their disk state, which offers clues into complex interactions between the winds, magnetospheres, and accretion flow between a low-mass stellar companion in close orbit ($0.2 \lesssim P_{\rm orb} \lesssim 0.5$ d) around a millisecond pulsar. The tMSP disk state is described as "sub-luminous" because the typical luminosity ($L_X \sim 10^{33-34}$ erg s$^{-1}$) is several orders of magnitude fainter than the active accretion phases of other low-mass X-ray binaries hosting neutron stars. In the disk state, tMSPs display rapid moding between two distinct emission levels. Switches between the more common high mode and the $\sim 6$–7 times fainter low mode occur over only tens of seconds, and the low modes last for minutes. In addition to moding, erratic X-ray flares of a wide range of amplitudes have been detected in the disk state, with the strongest flaring episodes lasting hours. In the radio pulsar state, no signatures of an accretion disk are detected. These systems appear as ordinary "redback" spider binaries in which fainter ($L_X \sim 10^{31}$ erg s$^{-1}$) X-ray emission is produced in an intrabinary shock (see Science Case 3). Three systems have been observed to transition between these two states in just a few days, and their unique phenomenology in the disk state has led to the identification of several more candidate tMSPs. In the recycling model of millisecond pulsar formation [see 11], tMSPs seem to represent the last stages of spinning up their neutron star, with accretion turning on and off on timescales as surprisingly short as weeks. To understand these systems further, we must identify more members of this class and observe them with high-sensitivity X-ray facilities like *AXIS*.

PSR J1023+0038 is the closest tMSP at 1.37 kpc [159] as well as the only confirmed Galactic field tMSP in the disk state for the last decade. It has been the target of detailed observing campaigns and served as the template for discovering more candidate disk-state systems. X-ray observations have shown that the source occupies the high mode about 70% of the time, the low mode for about 3% of the time, and X-ray flaring occurs a few percent of the time. Low modes last tens to hundreds of seconds, with ingress and egress times shorter than about 30 seconds [33]. Simultaneous multi-wavelength observations have begun to unravel the complex interactions that occur in the disk state between the pulsar, companion, and intrabinary material, including a disk and winds. In the high mode, $\sim 12\%$ polarization of the 2–6 keV X-ray emission has been detected, aligned with a lower degree of optical polarization that suggests

a common origin in the shocked inner accretion disk [34]. Pulsations at the neutron star spin period in the X-ray, UV, and optical bands have been detected during the high and flaring modes at a pulsed luminosity too high to originate from a hot spot of accreted material on the surface of the neutron star, making PSR J1023+0038 one of the few known optical millisecond pulsars which probe an interaction between the pulsar wind and inner accretion disk [14,259]. So far, most of our theoretical understanding has been developed through the observations of PSR J1023+0038. The interaction between the pulsar wind and the accretion disk – observationally manifested in the high and low modes, and in the spin-down rate being only 27% greater than in the radio pulsar state – is believed to result from the accretion disk remaining outside the light cylinder radius most of the time [426,593]. Determining the full spectral energy distribution for more tMSPs will also allow us to anchor the emission that produces the pulsations across the entire electromagnetic spectrum, including X-rays. The capabilities of *AXIS* are necessary to search for these behaviors in more distant systems and study the full range of tMSP variability across multiple systems.

Expanding the population of tMSPs to sources that display similar properties to the tMSP disk state has uncovered sources that exhibit the same unusual X-ray moding behavior [75,130] and expanded our understanding of the range of behaviors tMSPs can display to include systems dominated by X-ray flaring [327,549], and an eclipsing system [546]. Candidate tMSPs are so far optically faint ($G > 18$) and emission from the disk often outshines the companion, making it difficult to constrain fundamental properties of each binary configuration like distance, orbital period, component masses, and inclination. Observations of the nearby PSR J1023+0038 have revealed many unique X-ray variability signatures which are inaccessible to characterize in the other tMSPs with current facilities. *AXIS* will both enable the discovery of more disk-state tMSPs and thorough characterization of their X-ray emission.

*AXIS* will enable discovery of new disk-state tMSPs: Follow-up of unassociated Fermi-LAT gamma-ray sources with X-ray and optical observations has proven effective at identifying MSPs with hydrogen-rich companions, including candidate tMSPs. Due to the several-arcminute positional uncertainties of Fermi-LAT sources, this technique is more challenging in crowded fields, such as the Galactic plane, bulge, and globular clusters, where the higher density of millisecond pulsars produces blended gamma-ray signals. The *AXIS* Galactic Plane Survey will resolve many X-ray sources in dense regions and provide an initial spectral characterization that can be used to identify new disk-state candidates. [308] estimated that about 10% of the Galactic field tMSP population within 8 kpc has been discovered so far, and the deep sensitivity of *AXIS* will enable the discovery of distant tMSPs in all parts of the sky.

The rich X-ray phenomenology of tMSPs presents the need for sensitivity at multiple timescales, from the millisecond timescale of the pulsar spin period to the days timescale of a state transition. As faint Galactic sources, the technical capabilities of *AXIS* are necessary to leverage the full population and characterize time-dependent phenomena in distant tMSPs that are inaccessible with current facilities. We next highlight how *AXIS* will be used to unlock studies of X-ray moding, pulsations, and state transitions among the current and growing population of tMSPs.

*AXIS* will constrain X-ray moding in disk-state tMSPs: Two currently known candidate tMSPs in the Galactic field are too faint to bin their X-ray light curves observed with XMM-Newton sufficiently to search for rapid mode switching [308,371]. These are ideal targets for *AXIS* due to its large effective area. In orbit around L2, *AXIS* will be able to continuously view targets in a manner ideal for simultaneous observing campaigns with radio, optical, and UV instruments, which are likely key to understanding the full picture of the complex interactions that occur during the disk state. Such campaigns have already revealed an anticorrelation between radio emission and X-ray moding levels in two of three tMSPs targeted thus far

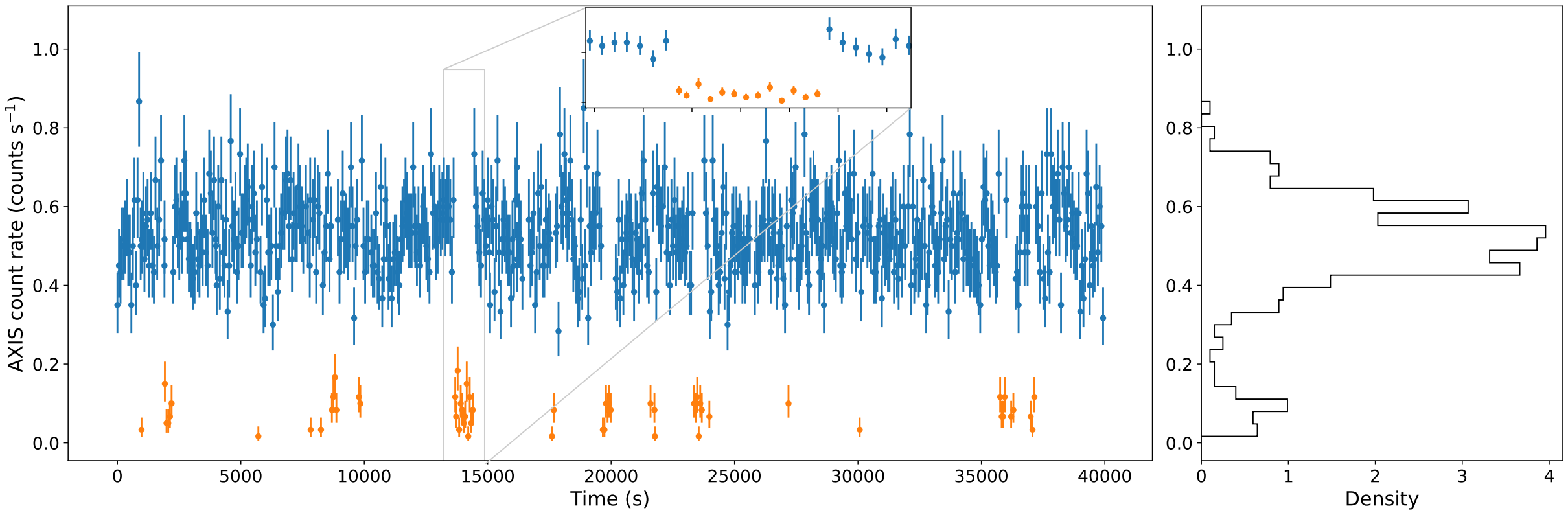

**Figure 14.** Simulated AXIS light curve for the tMSP candidate 4FGL J0639.1-8009 in its sub-luminous disk state (left), along with the corresponding count rate distribution (right). The light curve was simulated based on an unabsorbed 0.5–10 keV flux of $4 \times 10^{-13}$ erg s$^{-1}$ cm$^{-2}$, corresponding to a distance of $\approx$6-7 kpc, assuming a typical X-ray luminosity for tMSPs in the sub-luminous disk state. High and low-mode episodes are marked in blue and orange, respectively. The inset shows a zoomed-in view of the light curve around the epoch of the longest low-mode episode.

[76,211]. Due to the faintness of the radio emission in the third source targeted simultaneously with X-ray facilities, an anticorrelation with the X-ray emission could not be verified or ruled out [131].

We conducted simulations to estimate the maximum distance at which *AXIS* can detect the characteristic mode switching in the X-ray emission of tMSPs in the sub-luminous disk state. In particular, we simulated an *AXIS* light curve for the recently discovered tMSP candidate 4FGL J0639.1−8009 in its sub-luminous disk state [308]. We generated the light curve using the observed duration distributions of high and low modes, along with the time intervals during which the system switches between these modes, from the archetypal system PSR J1023+0038 [25]. We adopted a random number logic to determine whether the system is in high mode, low mode, or switching, sampling the duration for each mode and the mode switching from the distributions mentioned above. Based on the best-fitting absorbed power-law spectral model for 4FGL J0639.1−8009 derived from XMM-Newton data [308], we estimated an average *AXIS* count rate of $\simeq$0.47 ct s$^{-1}$ over the 0.3–10 keV energy band. By assuming the same high-to-low mode jump amplitude observed in PSR J1023+0038, we then estimated *AXIS* count rates of $\simeq$0.53 ct s$^{-1}$ in high mode and $\simeq$0.08 ct s$^{-1}$ in low mode. We simulated light curves that alternate randomly between high and low modes using 30-s time bins over a total uninterrupted on-source exposure time of 40 ks. For each time bin, counts were generated following a Poisson distribution, whereas count uncertainties were calculated using the method prescribed by [299].

Figure 14 shows the AXIS simulated light curve along with the corresponding count rate distribution. The results of our simulations suggest that *AXIS* will be able to detect bimodality in systems with average unabsorbed X-ray fluxes as low as a few $\times 10^{-13}$ erg s$^{-1}$ cm$^{-2}$. Assuming this flux corresponds to an average luminosity of a few $\times 10^{33}$ erg s$^{-1}$, similar to PSR J1023+0038, it implies that bimodality could be detected by *AXIS* for sources $\approx$6-7 kpc away. We performed further simulations and found that bimodality could be detected for sources as far as 8 kpc, depending on the amount of foreground absorption.

*AXIS* will search for X-ray pulsations at the neutron star spin period: X-ray pulsations at the neutron star's rotation period have been detected with a pulsed fraction of 8.5% in the high mode of PSR J1023+0038, and optical pulsations have also been detected in the flaring modes at a lower pulsed fraction than during the high mode [259,428]. In the low and flaring X-ray modes, the pulsed fraction is lower, constrained to $< 3\%$ and $< 1.3\%$ respectively. These limits are not sufficient to rule out pulsations during the low and

**Table 2.** Example targets and exposure times needed to characterize tMSPs with *AXIS*.

| Characterize... | Example Target | Distance (kpc) | Exposure Time (ks) |
|---|---|---|---|
| spectrum | New source | 1.5 | 0.2 |
| | | 5.0 | 2.2 |
| | | 8.0 | 5.7 |
| moding in faint candidate | 4FGL J0407.7−5702 | – | 40 |
| | 4FGL J0639.1−8009 | – | 40 |
| pulsations in high mode | 3FGL J1544.6−1125 | 3.8 | 5 |
| pulsations in low mode | PSR J1023+0038 | 1.37 | 7 |
| state transition | XSS J12270−4859 | 2.0 | 8 x 7 |

flaring modes entirely, as we know of several accreting millisecond X-ray pulsars with low ($\approx 1\%$) pulsed amplitudes, such as SAX J1808.4−3658 [231], HETE J1900.1−2455 [192], SAX J1748.9−2021 [433,513], MAXI J1816−195 [93], and SRGA J144459.2−604207 [403]. The large effective area of *AXIS* will enable more sensitive detections or constraints. Beyond PSR J1023+0038, *AXIS* will also enable the search for X-ray pulsations during future disk-state phases of the other confirmed systems, currently known candidate systems, and future tMSPs discovered by *AXIS*.

*AXIS* could catch rare state transitions as the accretion disk forms/dissipates: The time required to form and dissipate the disk for a state transition has not been well studied. Among the three confirmed tMSPs, a total of five transitions between the disk and pulsar states have been indirectly detected. The confirmed tMSP M28-I was detected as a radio pulsar again, just three weeks after it exhibited type I X-ray bursts in the disk state in 2013 [427,489]. The last transition of PSR J1023+0038 occurred within a window of 15 days when radio pulsations ceased to be detected [e.g., 542]. The most direct detection to date was in the Fermi-LAT light curve of XSS J12270−4859, which showed a gradual factor of 2 decrease in gamma-ray flux over about three weeks [623]. Catching these brief state transitions with *AXIS* and other facilities is essential for testing accretion physics models that aim to describe the transition mechanism.

In the disk state, the gamma-ray, X-ray, and optical emission become several times brighter than during the radio pulsar state, and the spectral features of a hot accretion disk are apparent. In the modern era of rapid alerts from monitoring surveys like ZTF and the Rubin Observatory, a sustained optical flux change of a known tMSP could promptly trigger *AXIS* to catch the formation or dissipation of an accretion disk during a state transition for the first time with Target of Opportunity scheduling. Based on the current number of transitions observed, we estimate that 1–3 transitions among the confirmed tMSPs will occur during the five-year mission of *AXIS*. In the decade when the three confirmed systems were discovered, the expected lifetimes in the radio pulsar and disk states seemed to be just a few years each. However, no transitions have been detected in the past 12–13 years, and with each newly identified candidate system that also has not yet shown a transition, the distribution of lifetimes in each state appears to be skewed toward longer timescales than initially expected. With a growing population of distant and faint disk-state candidates, the high sensitivity and flexible ToO response of *AXIS* are needed to ensure that we can study these rare events.

**[Exposure time (ks):]** 0.2–40 (see Table 2)

**Observing description:**

*AXIS* will be capable of characterizing the many X-ray signatures of tMSPs. We provide examples of targets and the corresponding minimal exposure times required for each science goal in Table 2. For discovery, the mean X-ray spectra of disk-state tMSPs can be used to identify likely candidates before scheduling long baseline observations to characterize their variability. We give the exposure times required to fit an intrinsic X-ray power law of photon index $\Gamma = 1.8$ typical of tMSPs, $N_H = 1.5 \times 10^{21}$ cm$^{-2}$ representing approximate out-of-plane absorption, and mean 0.3–10.0 keV luminosity

$L_X = 1 \times 10^{33}$ erg s$^{-1}$ at various distances with a 90% confidence interval of 0.2. As seen in Table 2, the exposure times required to characterize the mean X-ray spectra are well suited for short Target of Opportunity observations, allowing for rapid identification of new candidate tMSPs. In-plane exposure times will be strongly direction- and distance-dependent, but still feasible for characterizing distant sources. A target at 8 kpc and Galactic latitude $b = 1$ deg with expected $N_H = 8 \times 10^{21}$ cm$^{-2}$ would need 6 ks to fit the spectrum with the same precision.

To characterize X-ray moding in currently known faint candidates, we give two examples of sources with existing XMM-Newton observations. With *AXIS*, the count rates are predicted to be 2–3 times the observed XMM count rates, allowing sufficient time binning to constrain the presence of low modes lasting longer than 50 and 70 seconds in these sources, respectively. The exposure times suggested for this goal are chosen to ensure that multiple mode switches are detected even if the source is flaring during some of the observations. We recommend two targets for the search for pulsations for current disk-state tMSPs which have already shown moding, with minimal exposure times chosen to account for each source being in the high mode for about 70% of the time. For the speculative case of catching a state transition, we suggest two visits of 7 ks per week over one month to probe the accretion disk formation after a ToO has been triggered by an optical/gamma-ray brightening of XSS J12270–4859, a confirmed tMSP currently in the radio pulsar state since its last transition in 2012. Each of these individual characterization goals is an essential step to advancing our understanding of tMSPs, and the high angular resolution of *AXIS* across its field of view will allow many targets to be characterized at once in dense fields, such as globular clusters.

**[Joint Observations and synergies with other observatories in the 2030s:]**

tMSPs display variable emission across the electromagnetic spectrum, and simultaneous observations at other wavelengths are key to linking their diverse X-ray phenomenology to the physical mechanisms producing them. The first X-ray polarimetric study of a tMSP linked the polarized and pulsed X-ray emission in the high mode to shocks at the inner accretion disk in PSR J1023+0038 [34]. Joint observations with *AXIS* and the next-generation enhanced X-ray Timing and Polarimetry [eXTP; 647] mission will enable studies of the polarization degree in more distant tMSPs and during the low X-ray mode. With the upcoming Ultraviolet Explorer [UVEX; 303], simultaneous observations with *AXIS* could detect UV moding and pulsations at the neutron star rotation period in tMSPs more distant than PSR J1023+0038, to determine whether the X-ray and UV emission is in phase. In the optical, the next generation of 30-m class telescopes will unlock high-cadence, phase-resolved spectroscopic studies currently only possible for PSR J1023+0038. When combined with observations by *AXIS*, this could link the optical spectral properties to the X-ray emission for the first time. Detecting optical pulsations in more tMSPs can shed light on the nature of the synchrotron emission producing the pulsations. In the radio regime, coordinated observations with the next-generation VLA [ngVLA; 395] will extend searches for the anti-correlation between the X-ray and radio emission observed in two systems so far to the growing population of fainter candidate tMSPs.

Most candidate tMSPs and dozens of other non-accreting spider pulsars have been identified by targeting unassociated Fermi-LAT sources, and increased deep X-ray coverage of those arcminute-scale error ellipses with *AXIS* will enable further discovery. Beginning in just a few months, the Rubin Observatory will produce millions of light curves of faint optical sources in the Southern hemisphere, which will be used to identify the variable counterparts of tMSPs and other spider pulsars. In the next decade, the first radio surveys to reach $\mu$Jy sensitivity will come online, with the Deep Synoptic Array 2000 survey [DSA-2000; 224] in the Northern hemisphere and the Square Kilometre Array [SKA1; 85] covering the Southern sky. These surveys will be deep enough to be utilized to identify candidate tMSPs alongside X-ray and optical surveys. Between discovery and characterization, coordinated observations by *AXIS* and other facilities will be a vital tool to revolutionize our understanding of tMSPs in the next decade.

**[Special Requirements:]** Rapid ToO response (<1 d) will enable observation of the earliest phase of a state change as an accretion disk is forming or dissipating, as triggered by gamma-ray/optical flux changes. A time resolution of at least 0.2 ms will enable pulsation searches at the neutron star's spin period. Joint scheduling of simultaneous observations with other observatories will enable critical studies that can connect the X-ray emission to other wavelengths.

*10. Quiescent stellar mass black holes with AXIS*

**First Author:** Mark Reynolds (The Ohio State University, reynolds.1362@osu.edu)
**Co-authors**: Fiamma Capatanio (INAF/IAPS, Italy), Craig Heinke (University of Alberta, Canada), Shifra Mandel (Columbia University), Richard Plotkin (University of Nevada, Las Vegas), Jeremy Hare (NASA GSFC, CUA, CRESST II), Arka Chatterjee (UPES University of Calcutta, India)
**Abstract:**
Stellar population models predict billions of stellar mass black holes within the Milky Way. However, their detection and characterization remain significant challenges in Galactic astrophysics. *AXIS* is poised to provide a breakthrough in this area. While the average black hole may be isolated, interacting binaries have proven to be excellent systems for discovery. For instance, while the number of black holes in Low-Mass X-ray Binaries (LMXBs) is uncertain, it is estimated to be in the thousands. Observations of approximately 30 such systems with Chandra and XMM-Newton have revealed that their quiescent X-ray flux exceeds Lx $\sim 10^{30}$ erg s$^{-1}$. These systems are within the reach of *AXIS* short exposures ($\sim$5ks, reaching $F_X \sim 10^{-15}\ erg\ s^{-1}\ cm^{-2}$), out to a distance of 3 kpc or more.

*Identify a large sample of quiescent stellar mass black holes.*
The Galactic plane survey will reach detection limits approaching $10^{-15}\ erg\ s^{-1}\ cm^{2}$ for the full Galactic plane, and thus detect sources down to $L_X = 10^{30}$ erg s$^{-1}$ out to 3 kpc, and brighter sources throughout the Galaxy. This will enable the detection of a large fraction of the black hole population residing in interacting binary systems within our Galaxy. Furthermore, multi-wavelength data provided by, e.g., Rubin (LSST), Roman, and ongoing radio surveys, in conjunction with new *AXIS* observations, will allow for the identification and classification of these quiescent black holes.

*Constrain the physical properties of the quiescent accretion flow.*
The high sensitivity of *AXIS* will enable detailed X-ray spectroscopy and timing studies of the quiescent accretion flow onto stellar-mass black holes. This will allow us to constrain the properties of the accreting plasma, as well as to track the evolution of these properties over time. Furthermore, the flexible scheduling capabilities of *AXIS* will allow for coordinated simultaneous multi-wavelength campaigns, providing unprecedented insights into the broadband behavior of quiescent black hole accretion.

**Science:**
The majority of black holes in the present-day universe accrete at low fractions of their Eddington limit ($L_{Edd} \lesssim 10^{-5}$). At such low accretion rates, the inner regions of the accretion flow become geometrically thick and radiatively inefficient (e.g. [398]). An increasing fraction of the accretion power may be redistributed into a mechanical outflow (e.g., [68,174]), in the form of a radio-emitting synchrotron jet [190] and/or a wide-angle wind [636]. Most supermassive black holes accrete in an analogous state (e.g., Sgr A*, M31* through to M87*). To understand the continuum X-ray emission in low-luminosity black holes, then, is to understand the dominant mode of radiative output from black holes. *AXIS* observations of quiescent stellar mass black holes will play a key role in answering several important questions. We highlight three below.

– *What is the dominant X-ray emission mechanism in quiescence?*

Current X-ray spectra of stellar mass black holes at or below $10^{-6}$ $L_{Edd}$ are well described by a powerlaw of spectral index $\Gamma \sim 2$ [484]. If the 0.5-10.0 keV spectrum of black holes at low $\dot{m}$ is best described with a powerlaw, it must be due to synchrotron and/or synchrotron self-Comptonization (e.g. in a jet base [356]), or Comptonization in an extremely hot (e.g., 100 keV) corona that produces a very powerlaw-like spectrum below 10 keV (e.g. [398]). The emission we observe could also arise in a transition zone between an inner

Advection-Dominated Accretion Flow (ADAF) and an outer disk. The emitting gas would be cooler and produce a softer Bremsstrahlung spectrum. Indeed NuSTAR observations of V404 Cyg in quiescence suggest an electron temperature of $\sim$ 5 keV [478]. Simulations demonstrate that the temperature of the emitting plasma will be evident in the *AXIS* bandpass, e.g., a bremsstrahlung model can be statistically favored over a powerlaw in only 10ks for an input kT = 5 keV at the fluxes observed from GS1354-64 ($f_x \sim 6 \times 10^{-13}$ $erg\ s^{-1}\ cm^{-2}$). Correlated observations at IR or radio wavelengths can determine how changes in the physical properties of the accretion flow are related to the properties of the radio outflow. *AXIS* flexibility and sensitivity will facilitate detailed studies of the time-dependent behavior of the X-ray continuum (e.g., [61,257]).

*– How much mass does a black hole accrete in quiescence?*

In an ADAF the energy remains in the accreting gas rather than being radiated away. As a result, most of the energy is advected with the accretion flow, resulting in only a small percentage of the energy being radiated by the gas before it reaches the compact object. As the ADAF is radiatively inefficient, the accretion luminosity ($L_{acc} \ll$ GMM/2R) will be low [398]. In generalizing these solutions, [68] found that gas in such flows may not be bound to the black hole. These advection-dominated inflow-outflow solutions (ADIOS) predict that strong winds could be driven from black holes accreting at low $\dot{m}$.

ADAFs and ADIOS are similar in that black holes are underluminous in both circumstances, but differ in an important way: black holes can still gain mass (albeit slowly) in a standard ADAF, whereas black holes would gain very little mass were an ADIOS to hold. As most supermassive black holes in the present-day universe accrete at low fractions of their Eddington limit, it is imperative that we understand how such accretion flows work. This can be accomplished through *AXIS* observations, which will provide spectroscopic constraints. Theoretical calculations have estimated the expected strength of emission lines from an ADIOS where gas is ejected in an ionized wind, and specifically considered the case of the Galactic stellar mass black hole V404 Cyg [68,442,624]. They predict specific equivalent widths for a series of He-like Fe XXV and H-like Fe XXVI emission lines arising through collisional processes. Deep *AXIS* observations on the order of 250 ks would probe lines with EW $\simeq$ 70 eV. The detection of such emission lines would place stringent constraints on the mass accreted onto the black hole in the quiescent state.

*– How many stellar mass black holes populate the Milky Way?*

The Galaxy is expected to host $\sim$ 100 million stellar mass BHs [91,312,413], yet only 25 confirmed systems are known, with an additional 60 candidate systems of varying confidence levels awaiting further study (e.g., [126]). The GPS will detect a large fraction of the LMXBs in the Milky Way. Theoretical modelling predicts $\sim 10^4$ binaries with stellar mass black hole primaries to exist within the Galaxy (e.g., [638,639]). The *AXIS* GPS will enable the detection of these systems in quiescence, with sensitivity to systems within 3 kpc across the survey area and sensitivity to quiescent BHs beyond the GC (8 kpc) in the deeper exposures to be carried out in the bulge and GC regions. Multi-wavelength observational programs coupled with advanced numerical tools will be required to tease the systems containing BHs from the $\gtrsim 10^6$ new point sources to be revealed in the GPS. Additionally, the Roman mission will carry out detailed microlensing programs in the galactic plane during its prime mission and is expected to detect numerous isolated compact objects. These sources may present excellent opportunities for follow-up observations with the revolutionary sensitivity of *AXIS*.

In the *AXIS* prime mission, we can expect to increase the number of electro-magnetically identified black holes in the Milky Way by orders of magnitude, providing stringent constraints for future population synthesis calculations.

**[Exposure time (ks): 600 ks]**

**Observing description:** – 200ks of *AXIS* observations would be required to characterize the shape of the X-ray continuum from the accretion flow in a sample of 5–10 quiescent black holes, e.g., [455,484]. For

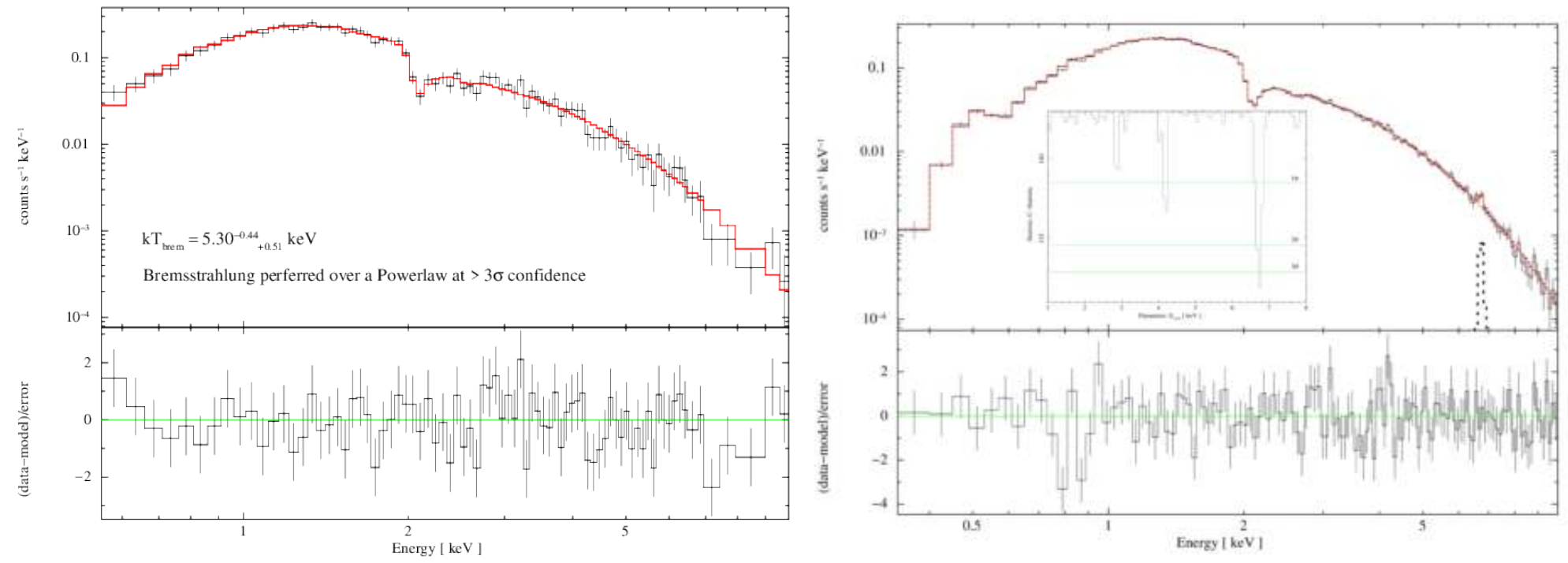

**Figure 15. Left:** Simulation of a 5 keV Bremsstrahlung model as might be expected from a quiescent black hole, e.g., V404 Cyg [478]. In 10ks, *AXIS* will constrain the spectral continuum and demonstrate spectral curvature is present. This breakthrough *AXIS* sensitivity enables time-dependent studies of the physical properties of quiescent accretion flows. **Right:** Simulated deep 250 ks observation of a system such as V404 Cyg or GS 1354-64 [478,484]. The simulation includes line emission as might be expected in an outflow (e.g., [68,442]). *AXIS* observations will place stringent constraints on the presence of such emission. Here, the Fe XXV line is measured to have $EW = 75^{+35}_{-40}$ eV (90% confidence).

brighter objects such as V404 Cyg and GS 1354-64, a 10 ks observation would reveal curvature from a 5 keV plasma as hinted at in NuSTAR observations of V404 VCyg, see Fig. 15. Coordinated multi-wavelength monitoring of sources such as V404 Cyg, A0620-00, and GS1354-64 would be designed to investigate the physics of the accretion inflow and jetted outflow.

– A 250–350 ks *AXIS* exposure will be required to statistically detect emission lines as might be expected from an outflow. The line EW are predicted to be $\sim 70$ eV (Fe XXV), but may be lower, see Fig. 15. A 350 ks observation would enable detection of 20 eV EW line emission from, e.g., GS1354-64.

– 50ks for follow-up observations of compact objects identified by the Roman microlensing campaigns (e.g., [313]) and new quiescent BHs identified during the GPS. Roman discoveries may lie in fortuitous locations. X-ray emission may originate from the interaction of these isolated BHs with diffuse gas, and this scenario should be vigorously pursued. Similarly, we might expect to discover several compelling new quiescent black holes deserving of dedicated follow-up observations.

***AXIS* and qBHs –** These sources typically exhibit fluxes $\lesssim 10^{-13}$ $erg\ s^{-1}\ cm^{-2}$. *AXIS* sensitivity in the 0.5–10.0 keV bandpass is crucial to detect and constrain curvature in the X-ray continuum. *AXIS* scheduling flexibility will revolutionize our ability to conduct multi-wavelength observing campaigns of both monitoring and ToO varieties.

**[Joint Observations and synergies with other observatories in the 2030s:]** Accretion flows from accreting black holes are fundamentally broadband in nature and require observations across the electromagnetic spectrum. We expect data from UVEX/Roman/ALMA/JVLA to play a key role in these investigations.

### c. Ultra-Luminous X-ray Sources (ULXs)

*11. Monitoring long-term ULX variability*

**First Author:** Chiara Salvaggio (INAF-OA Brera, chiara.salvaggio@inaf.it)
**Co-authors:** A. Belfiore (INAF - IASF Milan), R. Amato (INAF - OAR), G. Vasilopoulos (NKUA), M. Imbrogno (INAF - OAR), G.L. Israel (INAF-OAR), Ciro Pinto (INAF - IASF Palermo), Dominic Walton (University of Hertfordshire), A. Wolter (INAF - OA Brera), R. Salvaterra (INAF - IASF Milan), E. Ambrosi (INAF - IASF Palermo), B. A. Binder (CPP).
**Abstract:** Most ultraluminous X-ray sources (ULXs) exhibit variability in flux on different timescales, ranging from minutes to years, as well as changes in spectral shape. Some ULXs are persistent (regularly keep their luminosity beyond $10^{39}$ erg/s) others are transient (sometimes are observed below that threshold). While single ULXs are variable, X-ray luminosity functions (XLFs) seem stable, but that needs to be better tested. Diverse variability patterns have been observed among the ULX population; characterizing ULX variability is necessary for a comprehensive population study. To analyse the long-term evolution of the sources, we need multiple epochs. Thanks to its flexible pointing capability, *AXIS* will allow repeated observations of the same field. The high angular resolution, large field of view, and large effective area will enable us to disentangle clustered sources throughout the host galaxy and study a large number of sources, including new transients, with short exposures. Monitoring a sample of galaxies containing ULXs with *AXIS*, we can i) study long-term variability (months timescales), considering flux and spectral evolution, to characterize the accretion physics of the ULX population and probe the existence of a sustained super-Eddington regime; ii) compare with archival observations of the same galaxies taken with other X-ray observatories (like Chandra) to probe variability over tens of years; iii) different variability patterns may pinpoint candidates for pulsation searches, while orbital or super-orbital modulations can shed light on the binary systems; iv) push below the ULX threshold, comparing their variability to that of the upper end of the XRB population; v) explore whether the fraction of transient ULXs depends on their environment (host type, SFR, SFH, metallicity) vi) study the XLF.
**Science:** Ultraluminous X-ray sources (ULXs) are historically defined based on their luminosity, either at the peak or averaged [e.g. 273]. Among ULXs, we observe both persistent sources (with a luminosity above $10^{39}$ erg/s) and transients (some of the latter being above the threshold only for a limited amount of time) [e.g. 452]. The existence of transient sources among the ULXs implies that a source may not be classified as an ULX if observed a few times. The scenario is further complicated by the fact that most ULXs are variable in flux on timescales ranging from minutes to years, often accompanied by variability in spectral shape. Due to such variability properties, ULXs have been interpreted as X-ray binaries. Although single ULXs are variable, X-ray luminosity functions (XLFs) appear stable, but this needs to be tested with more sources [e.g. 510,643]. Diverse variability patterns have been observed among the ULX population: a characterization of ULX variability is necessary for a population study. The long-term variability (weeks-to-months timescales), considering both flux and spectral evolution, is fundamental to characterize the accretion physics of the ULX population and to probe the existence of a super-Eddington regime. Indeed, the spectral shapes in ULXs are different from those observed in the Eddington-limited Galactic X-ray binaries [e.g. 200,497]. This new spectral shape has been interpreted as a signature of an ultraluminous spectral state, which implies high accretion rates, most likely in a super-Eddington regime. For $\sim$10 sources, the super-Eddington accretion has been proved, thanks to the detection of pulsations in the X-ray emission, which means that a neutron star (NS) is contained in these ULXs [e.g. 31]. We will indicate this class of pulsating ULXs as PULXs. The super-Eddington scenario was confirmed by the detection of the long sought relativistic winds of photoionized plasma as predicted for super-Eddington radiation fields [e.g. 449].

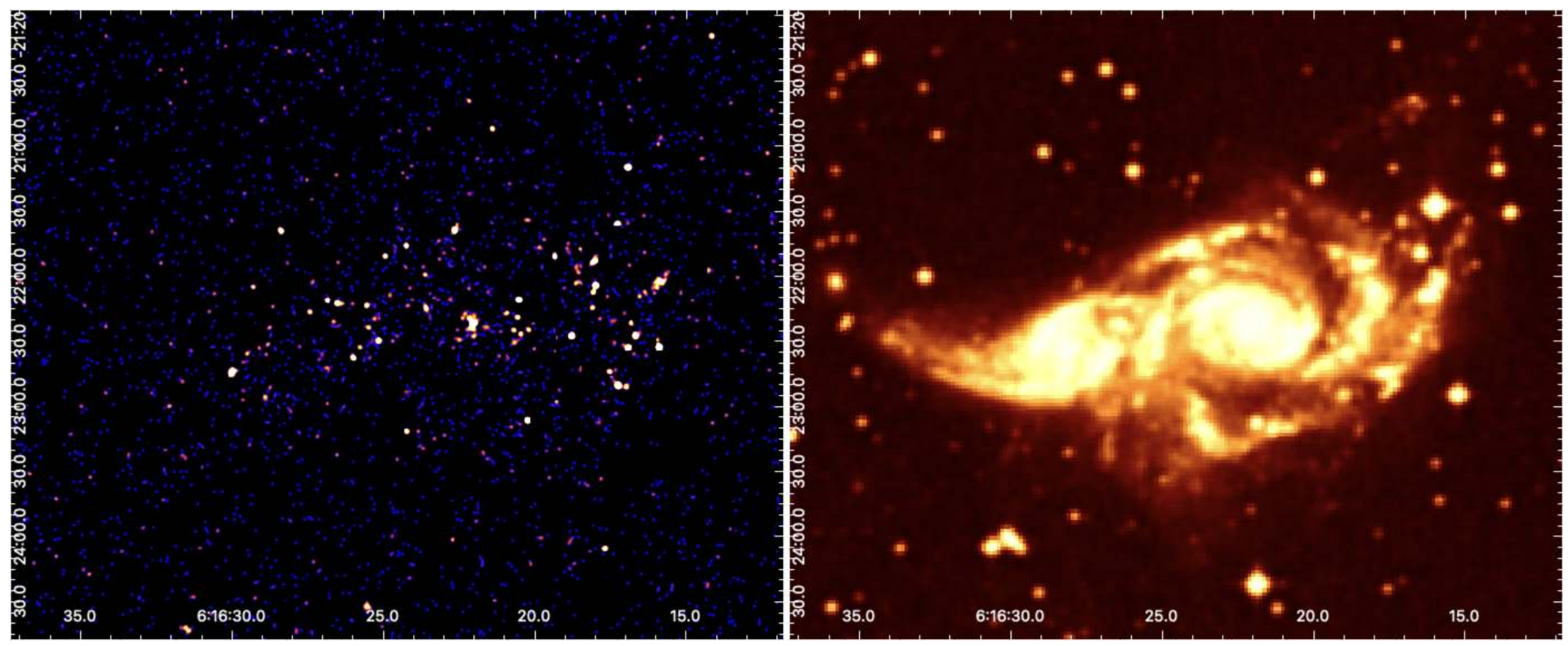

**Figure 16.** Left: Chandra smoothed image of the interacting galaxies NGC 2207/IC 2163 (Chandra obsID: 11228). Right: DSS optical image of NGC 2207/IC 2163.

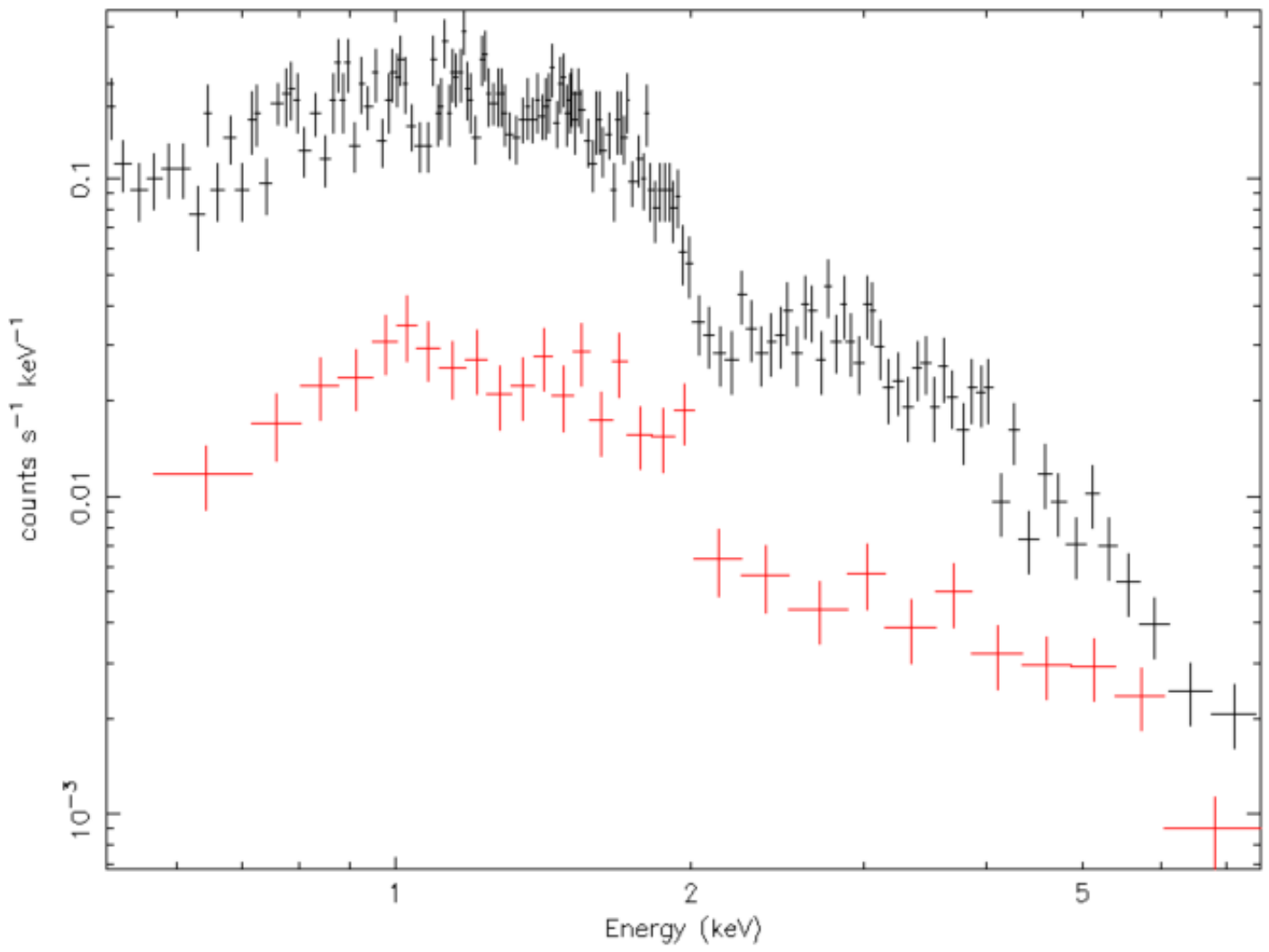

**Figure 17.** Combined spectrum of the X-ray point-like sources in the galaxies NGC 2207/IC 2163; in red: Chandra obs: 11228; in black: simulated AXIS data with *webspec*.

Most ULXs show aperiodic long-term variability. In some cases, the aperiodic variability does not show specific trends. In some sources, instead, we can identify specific variability patterns, i.e. flaring activity, flux dips, bimodal flux distribution. Flaring activity is rarely observed in ULXs [453]. Usually, the duration of a flare is on timescales of hours, which makes the detection of flaring activity difficult with the active X-ray facilities. Indeed, it is possible to observe such timescales in a deep XMM-Newton or Chandra observation, but if there are long periods without flares, it is challenging to spot the flares with a few sparse observations and thus to investigate if there is any trend in their cadence. To increment the probability of detecting flares, we need repeated visits, but with longer exposures time than the typical 1-2 ks Swift observations, not enough to catch the rise and decay phases of a flare. Flux dips are also observed in some sources, with a duration ranging from hundreds of seconds to hours. The amplitude of the flux reduction assumes a large range of values: e.g. in NGC 55 and NGC 247 ULXs a flux reduction of 80-90% is observed [543]; while in M51 ULX-7 [590],[247] the flux decrease was of 20-30%. The occurrence of dips is mainly present in the highest-flux epochs, suggesting obscuration from dense radiatively-driven winds. A bimodal flux distribution, with extended low flux epochs, which may last for a long period of time, even years, is a common property among the PULXs [e.g. 204,581]. A possible explanation is a transition in a propeller phase, which means that the strong magnetic field of an accreting NS creates a centrifugal barrier, interrupting the accretion. The ULX transits between two flux states: a high state, where accretion is active in a super-Eddington regime, and a low state, where the luminosity of the source decreases by an order of magnitude or more. Therefore, if the propeller interpretation is correct, a bimodal flux distribution may be used to identify candidate NSs. Indeed, such a flux distribution is observed also in some non-pulsating ULXs, suggesting that they may host a NS. However, the spin increase observed in some sources during low flux states (e.g., NGC300 ULX-1 [592]) would rather suggest obscuration by an optically-thick wind or a precessing accretion disk.

Sometimes, a periodic or quasi-periodic flux modulation is observed, with a period of tens of days or months [e.g. 511,591,601]. This behavior is typically observed in the PULXs and in a few other ULXs. It is still not clear whether it is a prerogative of NSs or whether it may also be found in the case of a BH accretor. These modulations are thought to be super-orbital, which is confirmed at least in the case of PULXs, for which we usually know the orbital period. Different models have been proposed in the literature to explain the periodic modulation, such as a precession of a warped accretion disk, a Lense-Thirring precession, or a precession of an NS and its magnetic dipole [e.g. 370,386]: it is still matter of study which, if any, of these models can explain the periodicities.

PULXs share common long-term variability properties: different variability patterns may pinpoint candidates for pulsation searches, while orbital or super-orbital modulations can shed light on the binary systems. Thus, it is very important to enlarge the sample of ULXs, monitored on weeks-months timescales, to search for such modulations.

With *AXIS* we will improve on the study of the long-term variability (weeks-months timescales), in terms of flux and spectral evolution, obtaining higher statistics data, in terms of number of observations, number of sources and counting statistics in the single source. In addition, we will compare the *AXIS* data with archival observations of the same galaxies taken with other X-ray observatories (like Chandra, XMM-Newton, and eROSITA) to probe variability over tens of years. The extension of the timescales over which it is possible to test variability will open the window to the individuation of evolution patterns, non-visible on shorter timescales.

In addition, thanks to the superior sensitivity of *AXIS*, we can push below the ULX threshold, comparing their variability to that of the upper end of the X-ray binary (XRB) population. We will study the XLF and its evolution among different epochs. Finally, we will test any trend among the ULX population and the host galaxy, such as the dependence of the fraction of transient ULXs on the galaxy environment (host type, SFR, SFH, metallicity).

**Observing description:** The aim of this proposal is to observe a sample of nearby galaxies rich in X-ray sources including a number of known ULXs (e.g., NGC6946, NGC5457(M101), NGC4559, M51, NGC2207). The advantage of selecting galaxies that contain several ULXs is that all of them can be monitored simultaneously.

To analyze the long-term evolution of the sources, we need multiple epochs. At the present time, the only X-ray telescope that allows such kind of observations is Swift/XRT, while the other active X-ray telescopes, e.g. those on-board XMM-Newton or Chandra, usually perform single deep pointings, which prevent us from a complete view of the ULX accretion cycle. Thanks to its flexible pointing capability, *AXIS* will allow repeated observations of the same field, with observing strategies similar to those possible with Swift/XRT. The improved angular resolution, large field of view, and larger effective area will enable us to disentangle clustered sources throughout the host galaxy and study a large number of sources, including new transients, even with short exposures. We aim to investigate the long-term flux and spectral variability of ULXs. By adjusting the observation cadence, we can explore different variability scales. We need long-term monitoring of the targets: a weekly cadence, on a yearly baseline, will allow us to test variability both on weekly and monthly timescales.

To verify the improvement in the statistics of *AXIS* compared to previous missions, we used *webspec* to simulate the X-ray population of the galaxy NGC 2207 at a distance of 36 Mpc. We used the list of point sources and the spectral fit reported in [375], for Chandra observation 11228, which has an exposure time of 13 ks (see figure 16). We simulated an *AXIS* spectrum using *webspec*, assuming the same spectral shape obtained from the analysis of the combined Chandra data, for all the point sources reported in [375]. The spectral counts for the combined spectrum, from Chandra data, are 629 in 0.5–8.0 keV, for a corresponding count-rate of $4.767\times10^{-2}$ counts $s^{-1}$. The corresponding values derived from the simulated *AXIS* spectrum, assuming the same spectral shape and exposure time, are 3770 counts and $2.9\times10^{-1}$ counts $s^{-1}$, which means an increase of a factor $\sim$six. This means that even with half the exposure time used in Chandra, we could improve the statistics by a factor of $\sim$3. This means that we can easily be sensitive to luminosity well below the ULX threshold in extragalactic environments. We estimated the *AXIS* exposure time needed to detect a source having a luminosity of $10^{36}$ and $10^{37}$ erg $s^{-1}$. We used *webpimms* to estimate the expected *AXIS* count rates. We assumed an absorbed powerlaw shape, with powerlaw index 1.8 and a background contribution of $\sim$20%. With a luminosity threshold of $10^{36}$ erg $s^{-1}$, we obtained an exposure time of 20 ks (for a galaxy distance of 5.5 Mpc, as in the case of NGC 6946) and 70 ks (in the case of NGC 2207, distance 36 Mpc), requiring a detection significance $>3\sigma$. For a luminosity of $10^{37}$ erg $s^{-1}$, we instead need 2 ks and 7 ks, respectively, for a distance of 5.5 Mpc and 36 Mpc.
**Exposure time (ks):** a total of 1000 ks for a weekly monitoring of 5/6 galaxies, lasting 1 year.
**[Joint Observations and synergies with other observatories in the 2030s:]** In cases of long-term periodicities, optical/UV telescopes (such as ELT or UVEX) can look for simultaneous modulations.
**[Special Requirements:]** (pileup, e.g., Monitoring (Daily, Hourly, etc), TOO (<X hrs), TAMM): Weekly monitoring.

*12. AXIS and the hunt for new ULX pulsars*

**First Author:** D. J. Walton (University of Hertfordshire, d.walton4@herts.ac.uk)
**Co-authors:** Matteo Bachetti (INAF-OACa), Gian Luca Israel (INAF–OAR), Matteo Imbrogno (INAF–OAR), Chiara Salvaggio (INAF-OA Brera), Andrea Belfiore (INAF-IASF Milan), Guillermo Andres Rodriguez-Castillo (INAF-IASF Palermo), Ciro Pinto (INAF-IASF Palermo), Ruben Salvaterra (INAF-IASF Milan), Demet Kirmizibayrak (Caltech), Alice Borghese (ESA/ESAC), Georgios Vasilopoulos (NKUA), Felix Fürst (ESA/ESAC), Roberta Amato (INAF-OAR)
**Abstract:** Following a series of remarkable discoveries, we now know that some of the most luminous members of the ultraluminous X-ray source (ULX) population are actually powered by highly super-Eddington neutron stars. However, much remains poorly understood about these enigmatic systems, including their overall contribution to the broader ULX population and the accretion physics that allows them to radiate at such extreme luminosities. In part, this is because only 7 confirmed ULX pulsars that show sustained super-Eddington accretion are currently known. In order to grow this sample, we propose to observe 35 ULXs (total exposure of 1.4 Ms) that have poor prior coverage from a timing perspective, but are bright enough for sensitive pulsation searches with *AXIS*. With these observations, we expect to find at least ∼7 new ULX pulsars, which would represent a significant expansion of this population.
**Science:** Ultraluminous X-ray sources are variable, off-nuclear point sources with $L_X \geq 10^{39}$ erg s$^{-1}$ (see [450] for a review). These luminosities – in excess of the Eddington limit for the 'stellar-mass' compact objects powering Galactic X-ray binaries – could imply either the presence of massive black holes (potentially the long-postulated yet elusive 'intermediate mass' black holes; IMBHs: $10^2 \lesssim M_{\rm BH} \lesssim 10^5 M_\odot$) or represent an exotic, super-Eddington accretion phase. However, even in the latter case, BH accretors have typically been assumed.

In 2014 [31] made the remarkable discovery that the ULX M82 X-2 ($L_{\rm X,peak} \sim 2 \times 10^{40}$ erg s$^{-1}$) is actually powered by a highly super-Eddington neutron star apparently exceeding its Eddington limit by a factor of ∼100, following the detection of coherent X-ray pulsations with *NuSTAR*. Six more ULX pulsars that also show ∼persistent super-Eddington accretion (or at least sustained periods at this level) have since been robustly confirmed: NGC 7793 P13 [187,269], NGC 5907 ULX1 [268], NGC 300 ULX1 [102], NGC 1313 X-2 [518], M51 ULX-7 [502] and NGC 4559 X-7 [454]. Among these, NGC 5907 ULX1 is particularly extreme in terms of its apparent luminosity, reaching $L_{\rm X,peak} \sim 10^{41}$ erg s$^{-1}$ [186], a factor of ∼ 500 above the Eddington limit! In addition, there is a small set of neutron star Be/X-ray binaries that on rare occasions briefly peak slightly above $10^{39}$ erg s$^{-1}$ during their most extreme outbursts (e.g. [575,620]), though these sources are likely somewhat distinct, as they do not exhibit the more extreme and ∼persistent levels highlighted above.

***Neutron Star ULXs:*** There is still much we do not understand about ULX pulsars. First and foremost, it is still highly uncertain what contribution they make to the overall ULX population. Although only a handful of ULXs are robustly confirmed to be pulsars, it is now speculated that such sources could actually make up a significant fraction of the broader ULX population (e.g. [294,451,602]). It is also still unclear exactly how these sources can at least appear to violate their Eddington limits to such an extreme degree. Accreting magnetized neutron stars can, in principle, reach these apparent super-Eddington luminosities through several mechanisms. High magnetic fields will collimate the accretion flow, allowing material to accrete onto the polar regions while radiation escapes from the sides of the column [53]. In addition, very large magnetic fields reduce the scattering cross section for electrons [240], reducing the radiation pressure and increasing the effective Eddington luminosity. Some authors have therefore invoked extremely magnetized (magnetar-like, i.e. $10^{14}$ G or more) neutron stars to explain the extreme luminosities observed [140,396]. In contrast, some authors suggest the magnetic fields may be as low as $10^9$ G, based on the ratio of the spin-up rate to the luminosity in ULX pulsars, which is an order of magnitude lower than

typical X-ray pulsars [290]. These authors argue that a disk truncated at a large radius, as expected for a high B-field system, would not provide the required lever arm to power the strong spin-up rates seen from ULX pulsars. Instead, 'classical' super-Eddington accretion flows (i.e., similar to those expected for black holes) are invoked outside of the pulsar magnetospheres [286]. Finally, we also note that it is difficult to explain the near-sinusoidal pulse profiles observed in the context of a highly beamed system, so large differences between the observed and intrinsic luminosities are unlikely. The nature of ULX pulsars is very much in question, since no one model can obviously explain all the observed characteristics of these extreme systems.

With so few examples currently known, identifying more ULX pulsars is still a major step in our efforts to understand both ULX demographics and the extreme accretion physics at play. Indeed, we are still at the stage where the discovery of a single new ULX pulsar would mark a major result. Beyond simply understanding the ULX population, though, both of these topics also have broader scientific significance. Accretion at extreme rates may be required to grow the $\sim 10^9\,M_\odot$ supermassive black holes now being seen at high redshift ($z \gtrsim 7$ and above; e.g. [30,347]), and ULX pulsars may offer our best local opportunity to understand this accretion regime provided their B-fields are typically low (a question we need a larger sample of ULX pulsars to answer). For the ULX demographics, although we know at least some are neutron stars, there is also still speculation that some are powered by large black holes and represent an evolutionary precursor to the $\sim 30{+}30\,M_\odot$ BH–BH mergers now being seen in gravitational waves [260,376].

***AXIS & ULX Pulsars:*** The combination of excellent imaging, large collecting area and good timing resolution means the *Advanced X-ray Imaging Satellite* (*AXIS*) is extremely well placed to significantly unearth new ULX pulsars. Here we demonstrate that with a 1.4 Ms program targeting a sample of 35 ULXs *AXIS* has the potential to $\sim$double the known population of ULXPs based on results from the latest ULX catalogues available.

**Exposure time (ks):** 1400 ks total (35 × 20–55 ks individual exposures)

**Observing description:**

The outstanding technical capabilities of *AXIS* highlighted above opens up the possibility of performing sensitive pulsation searches for ULXs that are either too faint or too confused for our current generation of X-ray observatory. We therefore propose *AXIS* observations of 35 poorly studied ULXs in order to search for X-ray pulsations and grow the small population of known ULX pulsars. These sources have been selected from the recent ULX catalogue compiled by [600], hereafter W22, based on the following criterion:

- The source is bright enough that, at its minimum typical flux 10,000 counts can be collected with *AXIS* within a $\sim$50 ks exposure (or less) allowing for sensitive pulsation searches.
- There are no previous or scheduled observations with high-timing-resolution facilities of the source that would already meet the 10,000 counts requirement.

There are a sample of 35 ULXs in the W22 catalogue that would meet these requirements. While these sources are poorly studied in terms of our knowledge of their short-term timing properties, we stress that the majority of these sources have still been observed on multiple epochs with the latest generation of X-ray observatory (*XMM-Newton*, *Swift*, *Chandra*, *eROSITA*), so their minimum typical flux estimates should be pretty robust. Based on this minimum flux, we then determine the *AXIS* exposure required to collect 10,000 counts and round this up to the nearest 5 ks, a conservative approach that results in a total request of 1.4 Ms (with the individual exposures per source ranging from 20–55 ks). Our requirement that this is achievable in a 50 ks exposure or less sets a lower limit on the minimum typical flux of our sample of $2-3\times10^{-13}$ erg cm$^{-2}$ s$^{-1}$ in the *AXIS* band.

As discussed below, observations with photon statistics at or above this level (i.e. 10,000 counts) are required to be sensitive to pulsations similar to those seen from the known ULX pulsars. To date, though,

only 25 ULXs have high-time-resolution observations in the public archive that meet this threshold (and all of these have been subject to pulsation searches). The program proposed here would therefore mark a significant expansion of this high signal-to-noise (S/N) sample. Of these 25 ULXs, 7 are now known to be pulsars, suggesting the contribution of these sources to the broader ULX population as *at least* ∼25%. However, the pulsations can be transient in these known ULX pulsars; pulsations are seen on average in ∼70% of ULX pulsar observations that meet the count threshold targeted here. As such, with this program we would expect to *detect* a minimum of 7 new ULX pulsars exhibiting *sustained* super-Eddington accretion, which would double the currently known population of ULX pulsars. We note that the sample compiled here spans the full range of luminosities seen from the known ULX pulsars, which extends up to $L_X \sim 10^{41}$ erg s$^{-1}$.

The feasibility of the *AXIS* observations proposed here is demonstrated by the previous X-ray observations of the sample in question, the characteristics of the known ULX pulsars, and simulations using the relevant *AXIS* response files. Typical pulsations from the known ULX pulsars have pulse periods of $P \sim 1$ s and pulsed fractions of $\gtrsim$5–10% in the *AXIS* band (the pulsed fraction is defined as [max−min]/[max+min], based on the maximum and minimum count rates seen across the pulse profile, and is a measure of the strength of the pulsations). In addition, ULX pulsars also often show strong, instantaneous spin derivatives of up to $\dot{P} \sim 10^{-8}$, which are determined by the combination of the accretion torque and the orbital motion of the neutron star. This means that 'accelerated' searches (i.e., searches that consider both $P$ and $\dot{P}$) are often required to detect the pulsations in these systems (e.g. [268]). Both standard Fourier theory (e.g. [322]; see Figure 18) and our simulations imply that to be able to detect pulsations similar to those outlined above with an accelerated search (using e.g. the $Z^2_2$ technique outlined by [92]), we require 10,000 counts. The expected results for a simulated 10,000-count observation of a ULX pulsar that has a pulsed fraction of ∼10% and a spin period of $P \sim 1.137$ s (based on recent observations of NGC 5907 ULX1, which typically has the lowest flux of the known ULX pulsars) is shown in Figure 19; an accelerated search clearly detects the pulsations (at $>3\sigma$ significance).

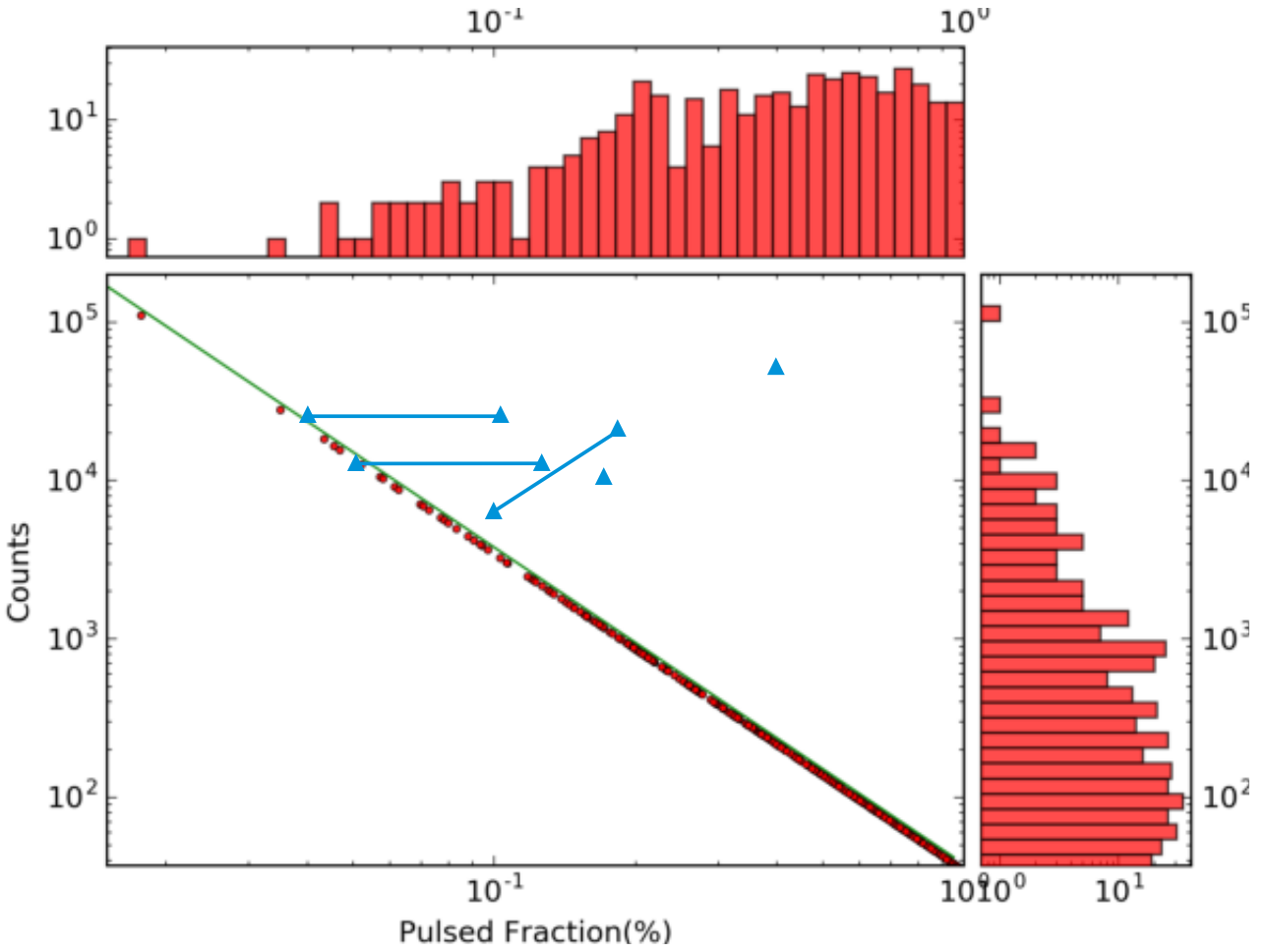

**Figure 18.** *Center:* detectable pulsed fraction vs number of counts per observation. The green line shows the predicted detection limit for pulsations (assuming a 3.5$\sigma$ detection) based on [322], while the red points show the corresponding upper limits from archival *XMM-Newton* data based on this calculation. Blue triangles show actual ULX pulsar detections (with lines joining multiple detections of the same source). The histograms show pulsed fraction limits (*top*) and counts per observation (*right*) respectively. As can be seen, $\gtrsim 10^4$ counts are required for a detection of pulsations with pulsed fractions above ∼5–10%.

In order to calculate the exposures required to record 10,000 counts, we determine the minimum typical flux seen in the archival data. *AXIS* count rates are then computed assuming a simple template for the average spectral form of a ULX pulsar in the *AXIS* bandpass (an absorbed powerlaw with $N_H = 1 \times 10^{21}$ cm$^{-2}$ and $\Gamma = 1.5$; e.g. [602]), and these rates are then used to determine the required exposures on a source-by-source basis. All of the calculated exposures are then rounded up to the nearest 5 ks. This is naturally a conservative approach, since the targets are typically seen at higher fluxes than we have assumed in our exposure calculations, so in many cases the photon statistics obtained would actually be even better than the 10,000 count threshold targeted here.

As the ULX sample compiled here has been selected based on their minimum fluxes, the proposed *AXIS* observations can be performed at any time, and observations of the sample can easily be spread over multiple years (though the individual *AXIS* exposures should ideally be continuous, owing to the strong spin-up rates seen from ULX pulsars). The timing resolution of 0.2 s *AXIS* will have in its normal observing mode should be sufficient to detect the ∼1s pulsations seen from the known ULX pulsars, so these observations should also facilitate a wealth of interesting supplementary science given the large *AXIS* field of view (e.g. the detection of new transient ULXs in these host galaxies).

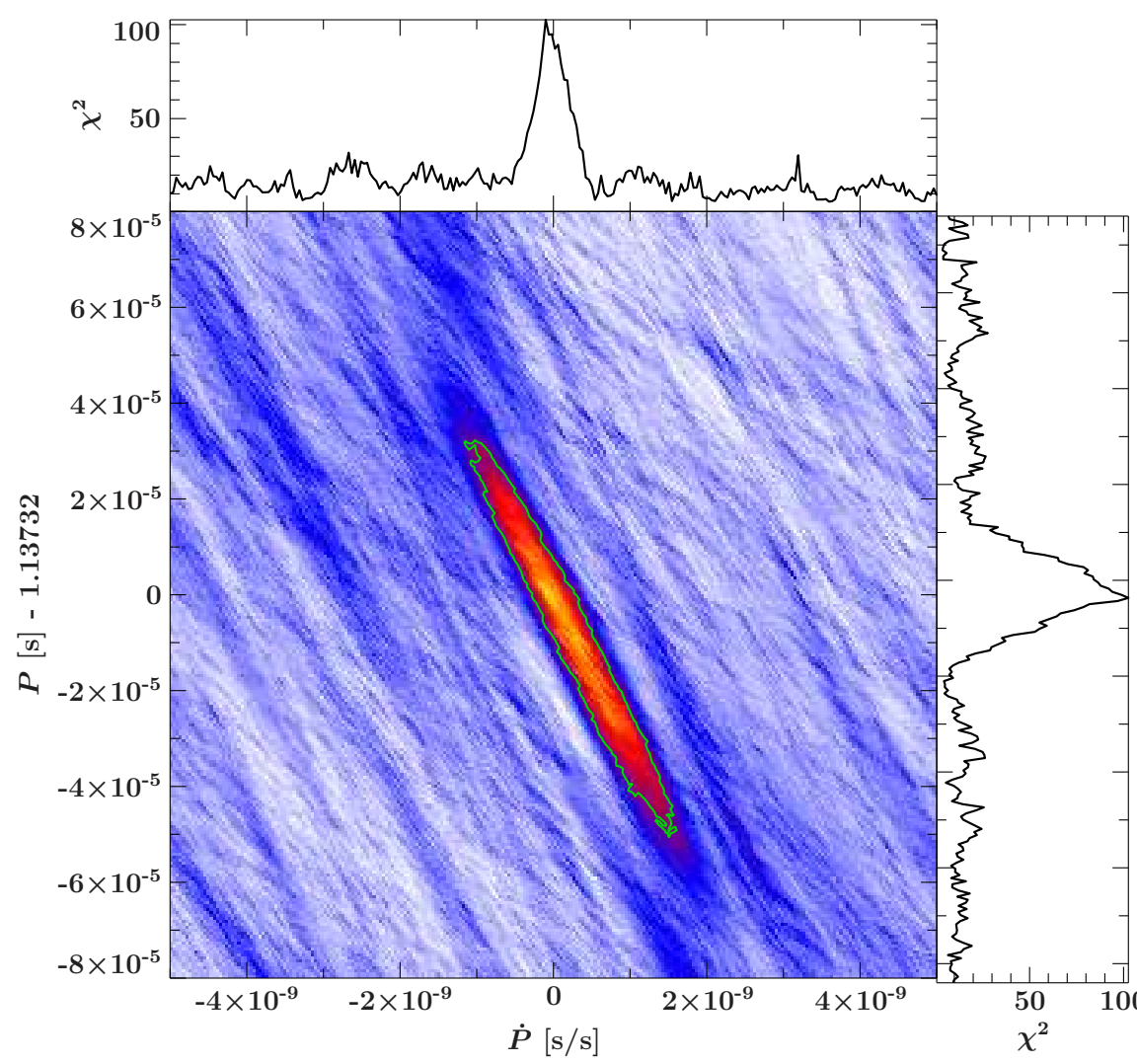

**Figure 19.** The results from a pulsation search on a simulated ULX observation with ∼10,000 counts, assuming an input pulse period of $P = 1.137$ s and a pulsed fraction of 10% (these values are based on the faintest of the ULX pulsars). The pulsations are clearly detected.

**Joint Observations and synergies with other observatories in the 2030s:**
The most relevant synergies are likely with other *AXIS* programs, e.g. programs seeking to perform general X-ray binary surveys of nearby galaxies. As such, the actual investment required in addition to any such team-led programs that will be executed during the prime mission lifetime could yet be significantly smaller than noted above.

**Special Requirements:** Pileup, timing resolution, collecting area, timing resolution, and angular resolution.

*13. The vanishing act of ULXs: nascent black holes, propeller transitions, and binary evolution*

**First Author:** Georgios Vasilopoulos (NKUA, gevas@phys.uoa.gr)
**Co-authors:** C. Maitra (MPE), Dominic Walton (University of Hertfordshire), Matteo Bachetti (INAF-OACa), Gian Luca Israel (INAF–OAR), Matteo Imbrogno (INAF–OAR), Chiara Salvaggio (INAF-OA Brera), Andrea Belfiore (INAF-IASF Milan), Guillermo Andres Rodriguez-Castillo (INAF-IASF Palermo), Ruben Salvaterra (INAF-IASF Milan), Demet Kirmizibayrak (Caltech), Roberta Amato (INAF-OAR), Alice Borghese (ESA/ESAC), Ciro Pinto (INAF - IASF Palermo), Labani Mallick (U. of Manitoba/CITA), Samar Safi-Harb (U. of Manitoba)
**Abstract:** The final stages of massive star evolution remain uncertain, particularly in interacting binary systems with compact objects like neutron stars or black holes. These systems emit bright X-rays, and the most luminous among them are known as ultra-luminous X-ray sources (ULXs). ULXs – like many X-ray binaries – can show various stages and can be quite variable, and even disappear in X-rays, showing similarities to propeller transitions in X-ray pulsars. One such source, NGC 300 ULX-1, had a brief but fascinating existence. NGC 300 ULX-1 is the only ULXP with a red supergiant (RSG), i.e., a cool star. Since 2020, the system has entered a quiescent state in both optical and X-rays. This new, never-before-seen evolutionary phase in ULXs or XRBs could be related to the last evolutionary phases of its companion, the creation of a nascent black hole or even a Thorne-Zytkow object. While CXO detected the source at a low-level X-ray Flux level in 2020, it could not provide meaningful spectral constraints. *AXIS* would be the first telescope capable of detecting NGC300 ULX-1 at its recent low flux state, offering valuable insights into the neutron star's spin evolution, accretion history, and the fate of its companion star. Similar to the test case of NGC300 ULX-1, *AXIS* would be able to follow other disappearing/fading ULXs at a level 100-1000 times lower than their peak flux, enabling investigations about the nature of the low flux state, whether that is due to obscuration, propeller effect or a binary evolutionary stages.
**Science:**

The evolution and last stages of massive stars offer an exciting field. The 'Great Dimming' of Betelgeuse caught everyone's attention, offering a plethora of scenarios to explore [e.g., 167]. Similar studies of other massive red super-giant stars have offered exciting results as well, to name a few [W60] B90, a very luminous RSG in the Large Magellanic Cloud (LMC), went through three dimming events over the past 30 years, recurring approximately every 12 years [393]. Another case is WOH G64, perhaps the most extreme RSG in the LMC given its proximity to the Humphreys-Davidson luminosity limit, which has recently shown a fast and unprecedented transition, showing indications that it is in fact an RGB Be binary very close to the common envelope phase [394]. These are examples of massive stars exhibiting extreme mass loss episodes, losing all their convective envelopes while even maintaining dusty torus around them (see Fig. 20). The question is what happens if a young and highly magnetized NS happens to stroll around a reckless star like the above? An example of such a case is NGC-300 ULX-1. In this system, a typical NS with a spin period of hundreds of seconds and a magnetic field of close to $10^{12}$ G was fed material from an evolved star for a few years, resulting in the birth and evolution of one of the closest ULXs to our Galaxy.

The X-ray activity of NGC300 ULX-1 began in 2010 with a 'supernova impostor' event, SN 2010da. After an initial bright phase, its flux quickly faded, only to reappear years later when it was reclassified as a ULX pulsar [102]. Its spin period decreased dramatically from several hundred seconds to just 16 seconds before fading again [592]. Within its bright phase, X-ray modulation could be associated with obscuration and absorption. The 15-year evolution of the system in terms of optical and X-ray activity is presented in Fig. 21 [109].

Is the system still accreting after a decade or more in the low flux state? Can we detect pulsations, and if so, what is the latest spin period? Has the system evolved into a Thorne-Zytkow object? Or was the ULX

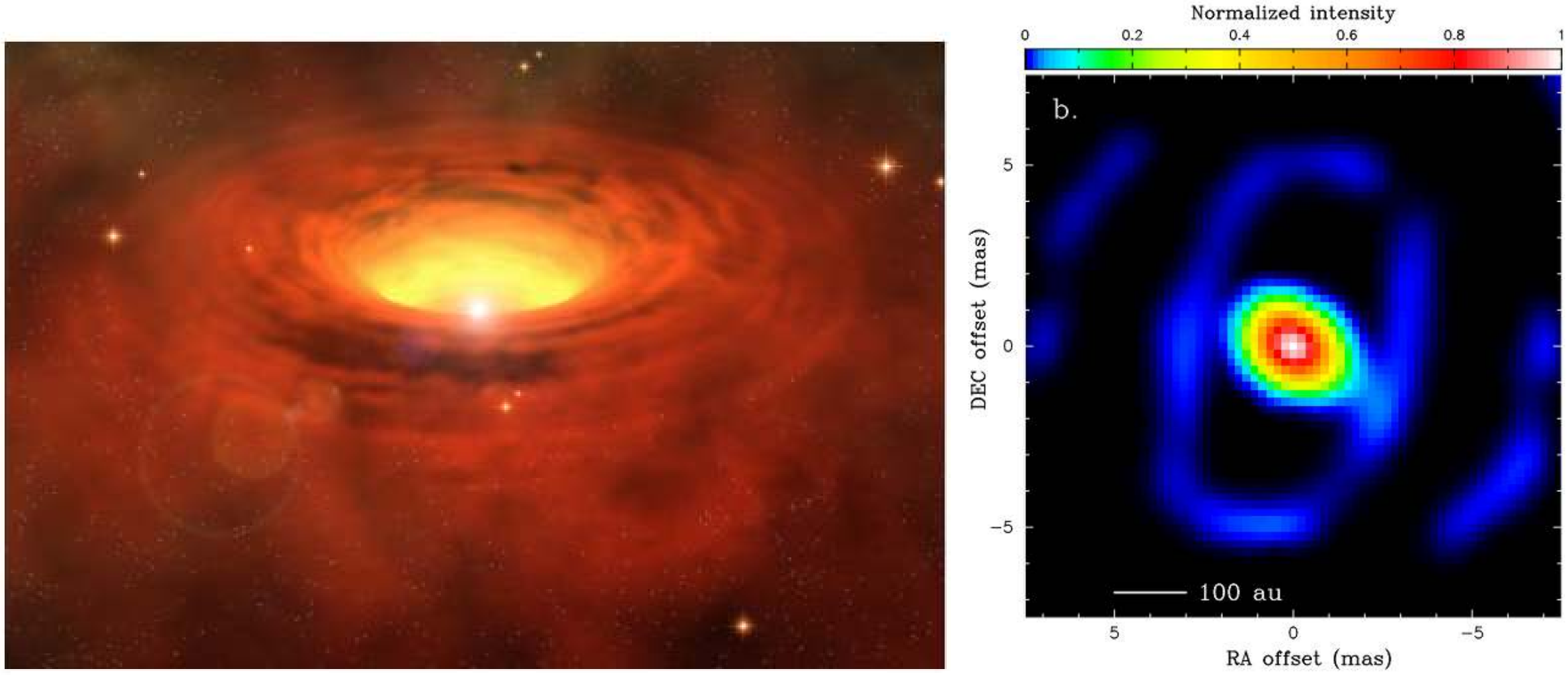

**Figure 20.** Artistic vision of WOH G64 surrounded by the dusty torus (credit: ESO). Right: Image of WOH G64 reconstructed at 2.2 $\mu$m with IR- Bis [412].

epoch related to the RSG transitioning into a yellow SG while expelling its outer shell and momentarily filling its Roche lobe?

Compared to other ULXPs, NGC300 ULX-1 has unique properties that favor the study of its spin evolution and application of accretion theory:

- Due to its large spin period $P$, the NS spin-up is much more evident compared to other ULXPs (i.e., torque $\propto \dot{P}/P^2$).
- It is the ULXP with the higher pulsed fraction (max/min pulsed flux $\sim$4). Having a pulsed fraction much larger than other ULXPs (i.e., 70% vs 10–20%), thus enabling detection of pulsations with low counts, in fact pulsations in CXO and Swift/XRT data were detected with only 100-200 counts.
- Its spin evolution is not affected by orbital modulation, as its orbital period should be larger than 2 years (size of SG companion).
- The prolonged low-flux state of is comparable to that of NGC 5907 ULX1 ($P \sim 1.1$ s), which lasted for almost 2 years (MJD 58000–58600). However, no spin period could be measured with observations obtained during a re-brightening state that lasted for about 30 days (we derived this from the analysis of public data).

*NGC300 ULX-1 is just a case study, and AXIS would be able to follow-up other transient or disappearing ULXs to ascertain their nature.* If indeed the super-Eddington phase of NGC300 ULX-1 is due to an evolutionary stage of the companion, similar studies could be performed in a number of nearby (up to 5-10 Mpc) transient ULXs. In terms of known systems, *AXIS* can be used to follow up the low state of persistent ULXP NGC 5907 ULX-1 [188], or a transient ULX discovered in NGC 1313 [575], a system where a long pulse period of $\sim$ 765.6 s has been stipulated.

*Plan B - possible reawakening:* NGC300 ULX-1 has a place of its own among ULXs as it is perhaps related to a catastrophic event. However, we still cannot confirm this scenario, and thus we should accept a scenario where the system re-brightens. This would make it perhaps the closest ULXP, and a laboratory where we can study relativistic outflows and their properties. The study of an 80 ks *XMM-Newton* observation of NGC300 ULX-1 showed evidence for a relativistic outflow, the strongest features of the absorber are iron absorption at 8–9 keV and oxygen absorption at 0.8 keV [298]. These winds are variable and the study of their variability is key to measure their clumsiness, recombination timescales and therefore density and outflow rate. *AXIS* could detect these features with 3 times less exposure time compared to

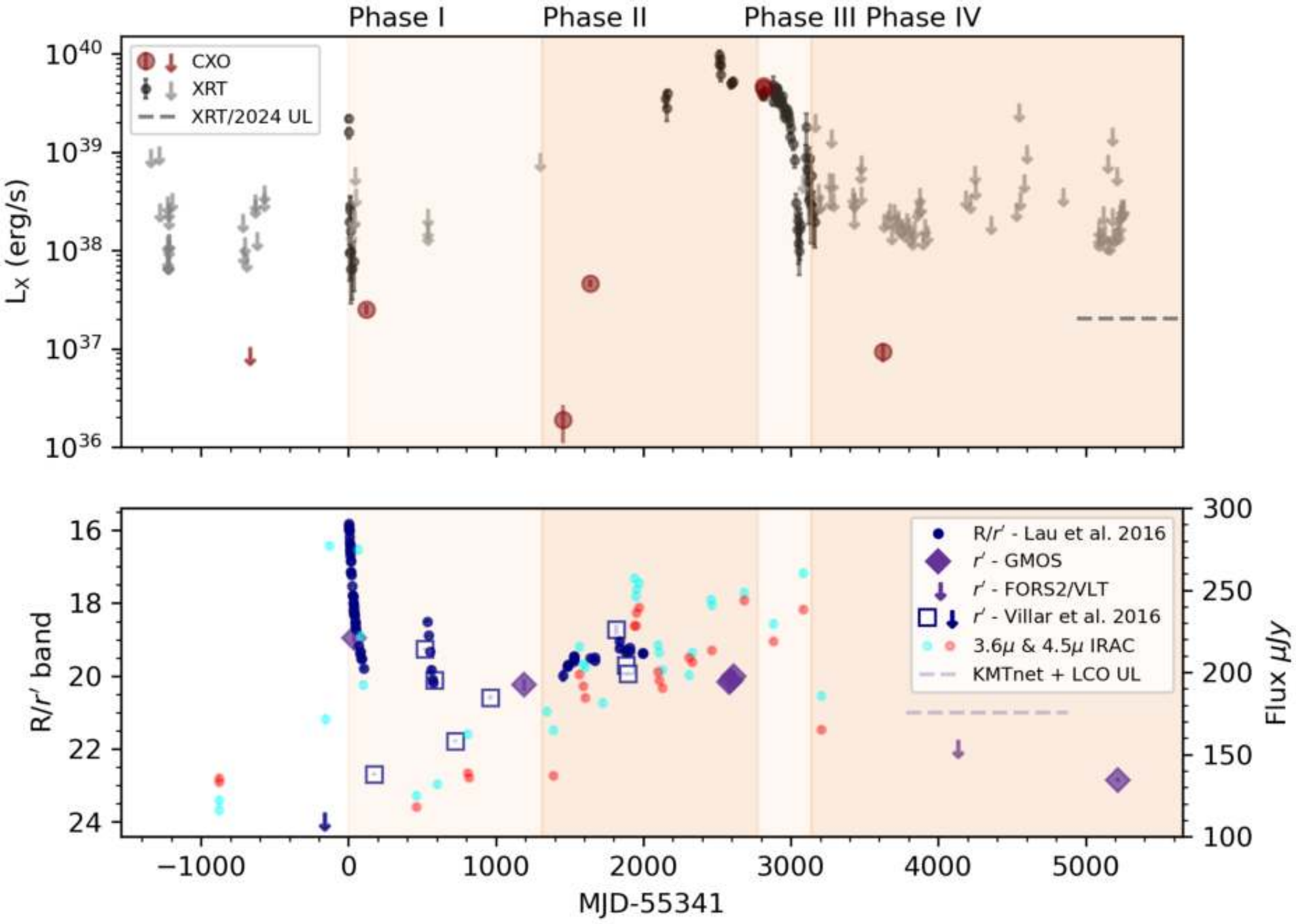

**Figure 21.** Long term X-ray light (*Upper panel*) and optical (*Lower panel*) light curve of NGC300 ULX-1. Characteristic phases following the 2010 impostor event are marked with shaded regions based on X-ray behavior. The $L_X$ is inferred from the count rates assuming constant bolometric correction, i.e. not accounting for changes in absorption. Following the impostor event, the X-ray flux drops, while in phase II the flux increases and the source breaks the ULX limit before the flux starts to gradually drop within 2018 (phase III). Within phases II and III there are strong indications that changes in intensity are mainly due to absorption (see text). Within phase IV the source is only detected with deep X-ray observations, and only upper limits were obtained by XRT snapshots or stacked observations. Optical magnitudes are taken from the literature (see text), apart from GMOS, which we analyzed. The dashed line shows an upper limit determined by the MKTnet and LCO archival data. Fig. credit: [109].

*XMM-Newton*, based on effective area and decreased background. The confirmation of the outflows and study of their variability is key to measuring their clumsiness, recombination timescales, and therefore density and outflow rate.
**[Exposure time (ks):]** 50-100 ks
**Observing description:**

The proposed research requires only a moderate amount of observation time in a specific field. Ideally, observations would be uninterrupted for an isotropic inferred Luminosity larger than $10^{37}$ erg/s. At this level, we expect $\sim$6 counts/ks, so an alternative strategy would be to obtain a series of shorter monitoring observations (5 ks each) to establish if the system is variable. Performing deeper observations if a higher flux is detected, enabling a search for pulsations. With an observation of 50 ks, *AXIS* would deliver about 300 counts if NGC300 ULX-1 is at the Flux limit of 2020 (latest CXO detection, Fig. 21), this would be sufficient for pulsation searches given the large pulsed fraction [589]. In case of re-brightening, to a state close to the maximum historic flux of the system, *AXIS* could easily detect pulsations in quite short intervals. However, with long exposures (50-100 ks) *AXIS* provides a baseline to study absorption features from outflows and even their variability. A similar strategy can be followed for other nearby known or newly discovered ULX pulsars.
**[Joint Observations and synergies with other observatories in the 2030s:]** NGC 300 ULX-1, and other similar ULXPs that may be discovered, are ideal targets for JWST. These systems represent a unique opportunity to probe eruptive mass-loss within a red supergiant-NS binary. They are likely progenitors to either a compact object binary system or elusive Thorne-Żytkow objects. Observations with NIRCam IFU and MIRI MRS would be critical in establishing the state of the donor star during eruptions or quiescent states.

*14. Too close to call: resolving two ultraluminous X-ray sources in M82 with AXIS*

**First Author:** Roberta Amato (INAF-OAR, roberta.amato@inaf.it)
**Co-authors:** Matteo Bachetti (INAF-OACa), Gian Luca Israel (INAF–OAR), Matteo Imbrogno (INAF–OAR), Chiara Salvaggio (INAF-OA Brera), Andrea Belfiore (INAF-IASF Milan), Guillermo Andres Rodriguez-Castillo (INAF-IASF Palermo), Ruben Salvaterra (INAF-IASF Milan), Demet Kirmizibayrak (Caltech), Alice Borghese (ESA/ESAC), Georgios Vasilopoulos (NKUA), Ciro Pinto (INAF-IASF Palermo), Dominic Walton (University of Hertfordshire), C. Maitra (MPE)
**Abstract:** Ultraluminous X-ray sources (ULXs) are extra-galactic, off-nuclear, point-like objects, with extreme X-ray luminosities with respect to other classes of X-ray binaries ($L_{0.5-10\ \mathrm{keV}} > 10^{39}$ erg s$^{-1}$). The discovery of pulsations from a few ULXs ultimately proved that ULXs can be powered by neutron stars accreting at super-Eddington rates, making them ideal laboratories to study extreme accretion physics. We propose to observe the galaxy M82, which hosts two of the most well-studied ULXs, X-1 and X-2, the latter being the first discovered pulsating ULX, with a spin period of 1.37 s. The two ULXs lie in the central, crowded region of M82 and are separated by just 5″. Currently, Chandra is the only X-ray mission capable of resolving them. *AXIS*, with its exceptional angular resolution and superior effective area, will clearly resolve them and provide high-quality data with much shorter exposure times than current X-ray missions. Its sub-second time resolution will allow for timing studies, extending the baseline for tracking the spin period and period derivative evolution of M82 X-2. Additionally, the nearly constant PSF of *AXIS* across a 10′ field of view will enhance source detection in M82, disentangling close X-ray sources and potentially leading to the discovery of new X-ray transients.
**Science:** Ultraluminous X-ray sources (ULXs) are a class of objects with X-ray luminosity (typically in the 0.5–10 keV range) higher than $10^{39}$ erg s$^{-1}$, that is, the Eddington limit for a typical stellar-mass black hole (BH) of 10 $M_{\odot}$. ULXs are found in galaxies up to a few hundreds Mpc away, in off-nuclear regions, hence ruling out the possibility of being active galactic nuclei. Since their discoveries in the 1980s [173,335], ULXs were often interpreted as accreting BHs of $10^2 - 10^5$ $M_{\odot}$ (intermediate-mass BHs) at or close to their Eddington limit. This paradigm swiftly shifted when coherent pulsations of the order of seconds were discovered from a few ULXs [31,102,187,268,269,502,518], ultimately proving that ULXs can be powered by neutron stars (NSs) at super-Eddington accretion rates. Pulsating ULXs (PULXs) are nowadays the only ULXs for which the nature of the compact object is certain and constitute unique laboratories to study the physics of accretion beyond the Eddington limit.

The first PULX to be discovered, and one of the most studied, is M82 X-2 [31]. It has a spin period of ∼1.37 s and a secular spin period derivative of the order of $10^{-11}$ s s$^{-1}$. M82 X-2 is the second brightest X-ray source in its host galaxy, the first being M82 X-1, at a distance of only ∼ 5″ [e.g., 87]. Both sources are located in the central region of this starburst galaxy, within high diffuse emission. M82 X-1 has not shown pulsations, and its spectral and temporal properties would point towards a BH as a compact object [88]. Isolating M82 X-2 in X-ray images from the neighboring ULX and the diffuse emission is no easy task. Currently, the only operating X-ray mission able to spatially resolve M82 X-1 and X-2 is *Chandra*, which, however, lacks the time resolution to detect the spin period of M82 X-2, preventing the study of its spin evolution over time. *XMM-Newton* and *NuSTAR* have instead been employed for timing analysis, but they are not able to spatially resolve the two ULXs, for example, for the extraction of light curves and spectra.

Thanks to its superior angular resolution and good timing resolution, *AXIS* is the first X-ray mission to enable simultaneous spectral and timing analysis of M82 X-2. Moreover, with a 5–10 times higher effective area than *Chandra*, *AXIS* will collect more photons with shorter exposures. A 10 ks exposure with *AXIS* will collect roughly 3400 photons from M82 X-2, while in comparison an archival *Chandra* observation of approximately the same length results in ∼1300 photons (Fig. 22, left and central panels; details on the simulation are given below). M82 X-2 shows a sinusoidal pulse profile, with a pulsed fraction of 5–10%

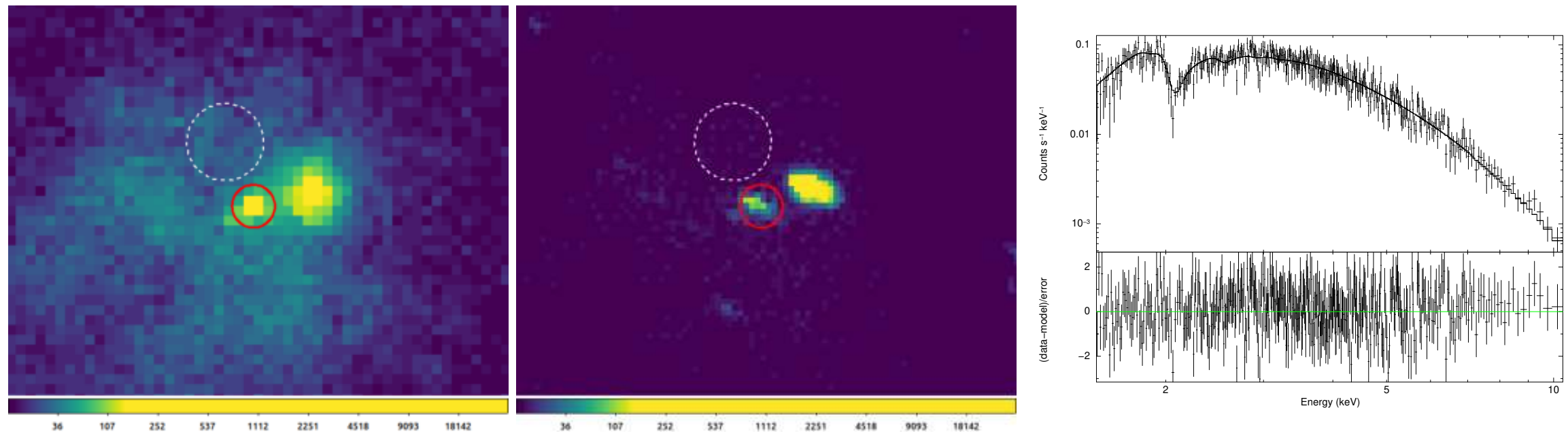

**Figure 22.** Left: 10-ks simulation of M82 X-1 and X-2 within the galaxy diffuse emission. Center: Chandra Obs.ID 17678 of M82 X-2, with an exposure of $\sim$9.3 ks. The position of M82 X-2 is marked with a red circle, while the white circle marks the area used to extract the background spectrum. Note that the pictures have the same color bar and scale. Right: Best-fit model of the 50-ks simulated *AXIS*spectrum with an absorbed cutoff power law and residual of the fit. The diffuse emission of the galaxy dominates at energies $\lesssim$ 1.5 keV.

in the energy range 2–10 keV [32]. We estimate that a $\sim$50-ks long observation would be sufficient to detect the spin period, guaranteeing the extension of the baseline to the mid 2030s to investigate the spin evolution of M82 X-2. Moreover, a high-quality statistical dataset would improve studies on the evolution of the pulse profile and enable both pulse-resolved and phase-resolved spectroscopy.

Besides M82 X-1 and X-2, a 50-ks observation will enhance the study of the X-ray population in M82. Thanks to its high sensitivity (for 50 ks of exposure, *AXIS* can reach fluxes down to $10^{-16}$ erg cm$^{-2}$ s$^{-1}$, in the 0.5–2 keV range, for on-axis sources; at the distance of M82 of 3.3 Mpc, this corresponds to an X-ray luminosity of $1 \times 10^{35}$ erg s$^{-1}$), it will be possible to detect even the faintest sources of M82. This would lead to a more comprehensive view of the entire X-ray population, with the potential to extend the X-ray luminosity function (XLF) to the lowest luminosities. Once a more complete XLF is obtained, it can be compared with the XLFs of other galaxies to investigate differences and similarities between different types of galaxies. Additionally, new serendipitous transient sources can be detected, contributing to the knowledge of the X-ray transient sky.

**[Exposure time (ks):]** 50 ks.

**Observing description:** We aim to observe the PULX M82 X-2 (RA= 148.9625, Dec=+69.679167) to: 1) spatially resolve it from the neighbor ULX M82 X-1 and extract the spectrum; 2) detect its pulsation. *AXIS* is the only X-ray mission capable of achieving both objectives. Thanks to its nominal readout rate of 5 fps, *AXIS* can detect pulsed signals down to 0.4 s, which is sufficiently accurate for the 1.37 s spin signal of M82 X-2. Assuming a perfect sinusoidal pulse profile (as all PULXs have), to effectively recover the spin signal above a detection threshold of 3.5$\sigma$, $\sim$16000 source counts are needed. To estimate the corresponding exposure time, we performed simulations with the software package SIXTE [142]. For completeness, we included in our simulations M82 X-1, X-2 and the diffuse emission of the galaxy. We based our simulation on the *Chandra* ObsID.17678, 10-ks long, analyzed in detail in [87,88], using their best-fit models for M82 X-1, X-2 and the diffuse emission. For the latter, we also employed the *Chandra* image obtained by merging all observations listed in [270], with the removal of all point-like sources. We used a 1.8$''$ circular region centered on the nominal position of M82 X-2 to extract the source's spectrum and a bigger circular region enclosing the nearby diffuse emission region to extract the background one [as done in 87] (see Fig. 22, right panel).

We found that, based on the luminosity of M82 X-2 in *Chandra* ObsID.17678, 16000 source counts can be collected with a single pointing of 50 ks. We also performed a fit of M82 X-2 spectrum with the same absorbed cutoff power law model used to simulate it. The fit returned an absorption $n_{\rm H} = 4.0 \pm 0.2$ (we used `wabs` model, though obsolete, for purposes of comparison with the results reported in the literature) and a photon index $\Gamma = 1.7 \pm 0.1$, with an accuracy on the best-fit parameters of 5% ($\chi^2$/dof=387.54/399, 1.5–10 keV band, see Fig.22, right panel). We fixed the high-energy cutoff to the value used to simulate the spectrum, since the lack of data at higher energies does not allow us to constrain it. For the purpose of comparison with the archival *Chandra* observation, we also performed the same simulation with just 10 ks of exposure. Results are shown in Fig.22, left and central panel.

The *Chandra* observation used for our simulation represents a worst-case scenario, since M82 X-1 is in a high-luminosity state ($L_{\rm X} = 5.6 \times 10^{40}$ erg s$^{-1}$, in the 0.5–30 keV range), potentially leading to pile-up emission (as it is the case for *Chandra* data). For this reason, we do not extract M81 X-1 spectrum. However, our simulations prove that the spectrum of M82 X-2 is not contaminated or affected by the piled-up emission of M82 X-1, since its best-fit parameters are consistent with those from the literature. Pile-up can also be an issue for M82 X-2, which ranges in luminosity from $5 \times 10^{39}$ erg s$^{-1}$ to almost $2 \times 10^{40}$ erg s$^{-1}$ [87]. We cannot know beforehand in which state the source will be caught, but *AXIS* higher frame rate (5 Hz compared to *Chandra*'s 1/3.2 Hz) and its PSF and pixel sampling (1.5$''$ and 0.55$''$, respectively, compared to Chandra's 0.8$''$ PSF and 0.5$''$ pixels) should minimize the effects of pile-up, despite the higher effective area of the mission. There are two possible strategies to mitigate pile-up. One is to observe the source off-axis, but this would be a waste of *AXIS* potentiality to study the whole X-ray population of M82. Alternatively, M82 X-2 could be periodically monitored with *AXIS* or other X-ray satellites (as done for some ULXs with, e.g., *Swift*/XRT) and a longer observation could be triggered (as Target of Opportunity) when the flux level of the source assures a pile-up level, for instance, <5%. Compared to other X-ray missions, monitoring with *AXIS* will have the advantage of immediately resolving M82 X-1 and X-2 and safely choosing the best moment when both sources have the most convenient flux levels. In this case, monthly visits for a few ks should be sufficient.

**[Joint Observations and synergies with other observatories in the 2030s:]** While *AXIS* will be essential in disentangling M82 X-1 and X-2, other future X-ray missions can help gain a better understanding of the ULX population. In particular, *NewAthena* is expected to detect many more ULXs, enabling a deeper characterization of their properties and facilitating direct comparisons with the ULXs in M82.

**[Special Requirements:]** (pileup, e.g., Monitoring (Daily, Hourly, etc), TOO (<X hrs), TAMM) Possible monthly monitoring of M82 X-1 and X-2, with observations lasting a few ks each.

## d. Accreting White Dwarfs

### *15. Ultracompact white dwarf binaries in the AXIS era*

**First Author:** Liliana Rivera Sandoval (UTRGV, liliana.riverasandoval@utrgv.edu)
**Co-authors:** Nathan Steinle (U. of Manitoba), Samar Safi-Harb (U. of Manitoba), Kevin Burdge (MIT)
**Abstract:** Ultracompact white dwarf binaries, also known as AM CVns are H-deficient systems with orbital periods shorter than 70 min. The primary is a white dwarf that accretes mass from its low-mass companion, which can be another white dwarf or a He-rich star. In those binaries, the mass transfer rate decreases as the binary evolves towards longer orbital periods. Due to their short orbits, many AM CVns are expected to be detected by gravitational wave interferometers such as LISA as resolved sources. Systems with periods shorter than 20 minutes are persistent, and systems with periods longer than 20 minutes exhibit outbursts. We aim to use *AXIS* to identify and characterize both types of systems. The rapid response and high sensitivity of *AXIS* will permit the observation of the X-ray emission during outbursts. Furthermore, *AXIS* will allow the identification of magnetic systems of which very few are currently known. Existing and upcoming facilities, such as Roman, Rubin, eROSITA, etc will inform our searches.
**Science:**

*Magnetic AM CVns*

AM CVn are binaries with typical orbital periods less than an hour, in which a white dwarf (WD) accretes either from another white dwarf or from a helium star. These binaries are relatively rare and have been mostly studied in the optical. In AM CVns the X-rays can come from the WD accretor or the inner part of the accretion disk. For systems with orbital periods shorter than 10 min (HM Cnc and V407 Vul) there is no space to create a disk and the X-rays are likely coming from a hot spot created by the impact stream on the accretor. Currently, around 80 AM CVn systems have been spectroscopically confirmed, but of those only 2 systems have been suggested to have a magnetic nature [341]. This contrasts with the fraction of magnetic systems identified among the H-rich cataclysmic variables, which is around 30%. The origin of this discrepancy is unclear, but it might be related to the evolution of AM CVns for which different channels have been proposed: the evolved CV channel, the He-star donor channel, and the WD donor channel. Furthermore, to explain some of the observables during outbursts, such as the suppression of short and weak outbursts, it has been suggested that disk truncation (e.g., due to a magnetic accretor) might play an important role. Therefore, magnetic systems might indeed be more common than current observations indicate. Besides the apparent paucity of magnetic sources, another problem is the total number of AM CVns: current estimates of the number density $(5.5 \pm 3.7) \times 10^{-7}$ pc$^{-3}$ of AM CVns suggest that even within 1 kpc there should be more than 1000 AM CVns [501]. Telescopes like Roman and Rubin are expected to discover many AM CVns systems through their short periods and transient behavior, thus allowing to investigate the abundance of magnetic AM CVns further.

Contrary to neutron stars and black holes in ultra short orbits, the so-called ultracompact X-ray binaries (with $L_X < 1 \times 10^{36}$ erg s$^{-1}$ in quiescence), AM CVns have typically low X-ray luminosities, in the range of $L_X \sim 1 \times 10^{30-31}$ erg s$^{-1}$. Except for the systems with $P_{orb} \lesssim 10$ min, which can have $L_X$ of the order of $1 \times 10^{33}$ erg s$^{-1}$ [551]. This makes the AM CVns difficult to detect with current observatories because very long integrations are needed to obtain sufficient photons for spectroscopic analysis. In this context, *AXIS* is expected to revolutionize our understanding of AM CVns given its large effective area, which even with short integrations (of the order of $\sim$ 20 ks) will provide enough photons for adequate spectral analysis (see Figure 23).

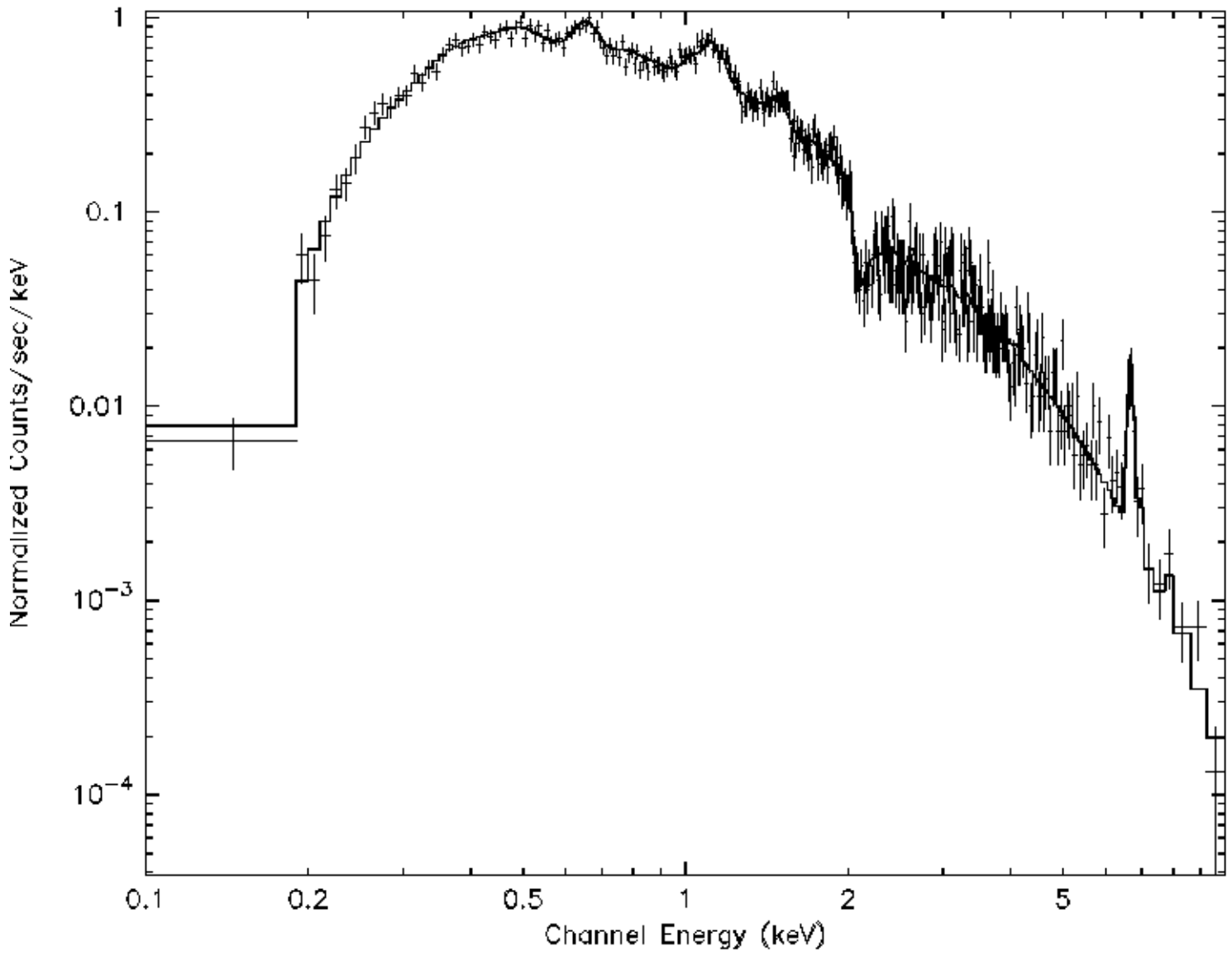

**Figure 23.** Spectrum for a faint AM CVn with *AXIS*. The simulation includes an absorbed thermal model (VMEKAL) as in [477,493] for faint optical state. This 20 ks exposure shows that short exposures lead to much higher quality spectrum than current observatories with *AXIS* errors of the order 0.15 keV compared to current errors of more than 1 keV in the plasma temperature.

*Persistent/Transient AM CVns*

A major open question regarding the understanding of the evolution and accretion of AM CVns is the value of the mass transfer and mass accretion rates, both in transient and persistent systems. In X-rays, systems with orbital periods shorter than 30 minutes appear to be over-luminous compared to expectations from models [56]. A possible explanation is that the mass accretion rate in these systems is large enough to make the boundary layer around their accreting white dwarfs optically thick, leading to the suppression of X-ray emission due to the thermalization of X-ray photons into UV radiation. This phenomenon has been directly observed in some AM CVns in outbursts [477,493]. However, the X-ray emission of AM CVns in high state and quiescence is still poorly understood due to instrumental limitations.

Another problem related to accretion in AM CVns is that outbursts of long-period systems exhibit an unexpected dichotomy: some systems display superoutbursts with durations of less than 10 days [443], while others last for years [e.g. 492]. To understand the mechanism behind these events, it is then fundamental to determine the mass transfer rates in both kinds of systems. To study both types of outbursts a sensitive telescope is needed, and particularly for the short duration superoutbursts, a fast response telescope is necessary. *AXIS* will address both problems. With a large effective area and a fast response to target of opportunity, similar to Swift's, it will enable the detailed study of AM CVs in outbursts for the first time. Together with facilities like Roman, JWST, Rubin and HST, *AXIS* will also allow the investigation of multi-wavelength correlations.

**[Exposure time (ks): ]** The exposure time depends on the distance and brightness of the target; we require at least 20 ks per target, see Fig. 23.

*16. Hunting and monitoring the orbital period change in double degenerate ultra-compact XRBs*

**First Author:** C. Maitra (MPE, cmaitra@mpe.mpg.de)
**Co-authors:** G. Vasilopoulos (NKUA), Ş. Balman (Istanbul Univ.)
**Abstract:** Double degenerate systems (DD) containing a pair of white dwarfs are the most compact binary systems known and can theoretically have orbital periods shorter than 5 minutes [400,583]. The evolution of these systems is driven by the loss of angular momentum due to the emission of gravitational radiation, and they are highly anticipated targets for space-based gravitational wave detection with LISA. They are likely the progenitors of at least some type Ia supernovae and may also represent a substantial fraction of supersoft X-ray sources. Recently, four such objects have been discovered in X-rays from orbital periods in the range of 5–23 min [351]. Monitoring is crucial to understand the evolutionary scenarios of white dwarfs in the most compact binary systems since orbital evolution in such systems results from an interplay between strong gravitational radiation and the mass accretion rate from the donor, and both the magnitude and the sign of the change in orbital period will provide constraints on this rare class of objects which are one of the prime targets of future space-based gravitational wave detectors like LISA. Thanks to the improved angular resolution and sensitivity of *AXIS*, apart from monitoring the period evolution in known systems, the proposal also aims to find new sources, which are expected to be numerous, especially in the Galactic plane.
**Science:** Double degenerate ultra-compact binary systems comprise two compact objects (white dwarfs, neutron stars, or black holes) at the post-common envelope phase of binary evolution. Double-degenerate interacting white dwarf systems (DDs from now on) are particularly interesting from a cosmological perspective as possible progenitors of SNe Ia. In addition, these systems, which are expected to be in tight binary orbits, can merge due to the loss of orbital angular momentum and are expected to be strong emitters of gravitational waves (GWs) detectable by future space-based gravitational wave detectors like, for example, the Laser Interferometer Space Antenna (LISA).

DDs have been identified as a subclass of heterogeneous objects classified as supersoft X-ray sources (SSSs). They are characterised by very soft X-ray spectra with kT $\sim$ 15–80 eV and a wide range of luminosities. The most luminous ($\sim$$10^{36}$–$10^{38}$ erg s$^{-1}$) ones can be explained by stable nuclear burning white dwarfs (WDs), which in most cases accrete H-rich matter from a companion star [588], or as central stars of planetary nebulae (PNe) [364], and magnetic cataclysmic variables (mCVs), including polars and soft intermediate polars [see e.g., 24,95,214,416]. Many of these systems were discovered in the direction of the Magellanic Clouds. A low Galactic foreground absorption in their direction makes them ideal laboratories for detecting and investigating SSSs [348].

Four sources, RX J0806.3+1527 [HM Cnc, 64,266], its twin RX J1914.4+2456 [V407 Vul, 214], 3XMM J051034.6−682640 [217], and eRASSU J060839.5−704014 [351] discovered as SSSs in ROSAT and XMM data, are considered classical examples of DDs with periodicities of 5.4 min, 9.5 min, 23.6 min, and 6.2 min, respectively. Some of these periodicities were also found in the light of their optical companions (HM Cnc: Ramsay et al. [475], Israel et al. [267], V407 Vul: Ramsay et al. [474], 3XMM J051034.6−682640: Ramsay et al. [476]). The X-ray flux drops to zero between pulses, and no other periods are seen [132,172,217]. Several models have been proposed to explain the X-ray emission from these systems. The two main accretion models in this regard include mass transfer from a Roche-lobe-filling WD to either a magnetic (polar-like) or a non-magnetic (Algol-like) accretor. In the polar-like model [132], the magnetic field of the accreting WD inhibits the formation of an accretion disc and matter reaches the magnetic polar cap. In the Algol-like type, also known as a "direct impact" accretion model [357,400], a light companion is assumed so that a disc would not form, resulting in the stream directly hitting the surface of the accreting WD. However, models invoking accretion predict an orbital widening for the two degenerate WDs, in contrast to what is observed in RX J0806.3+1527 and RX J1914.4+2456, although solutions to circumvent this issue have been proposed [see 274, and references therein]. The main alternative to the accretor model is the "unipolar inductor" model [e.g. 139]. This model involves a magnetic primary WD and a (non-magnetic) secondary that does not fill its Roche lobe. In this case, if the spin period of the primary and the orbital period are not synchronous, then the secondary crosses the primary's magnetic field as it moves along the orbit. The resulting electromotive force drives an electric current between the two WDs (assuming the presence of ionised material between them), whose dissipation heats the polar caps on the primary. This method has, however, been highlighted as not being efficient enough to explain the observed X-ray flux [310]. Nonetheless, the emission mechanism for the population of DDs remains an open question, and more observable systems are required to fine-tune the models.

The proposal first aims to search for new DDs, particularly in the crowded regions of the Galactic plane, owing to the uniform high-resolution PSF and large effective area of *AXIS*, which is uniquely suited for this capability. These will contribute as a major foreground component in the LISA detectors. The final goal is to monitor the orbits to constrain the orbital evolution. The orbital evolution in such systems results from an interplay between strong gravitational radiation, which removes orbital angular momentum, thus reducing the orbital separation (increasing the orbital frequency), and the mass accretion rate from the donor, which acts to increase the orbital separation when a lower mass component transfers mass to its heavier companion. Phase coherent timing solution through monitoring of both HM Cnc and V407 Vul revealed that their orbital period is decaying, consistent with expectations for loss of angular momentum due to gravitational radiation. However, the Lx of these systems is much lower compared to GR driven mass transfer rate. The decreasing orbital period is also difficult to reconcile with a stable mass transfer between the two WDs, especially in the scenario of direct impact accretion. These have led to alternate models such as unipolar inductor, or mass transfer from a purely degenerate helium donor or a non-degenerate outer envelope of the WD to the primary WD [310]. Unifying models that explain the interplay between GWs and mass transfer in DDs is an open problem, and monitoring of DD is crucial with *AXIS*.

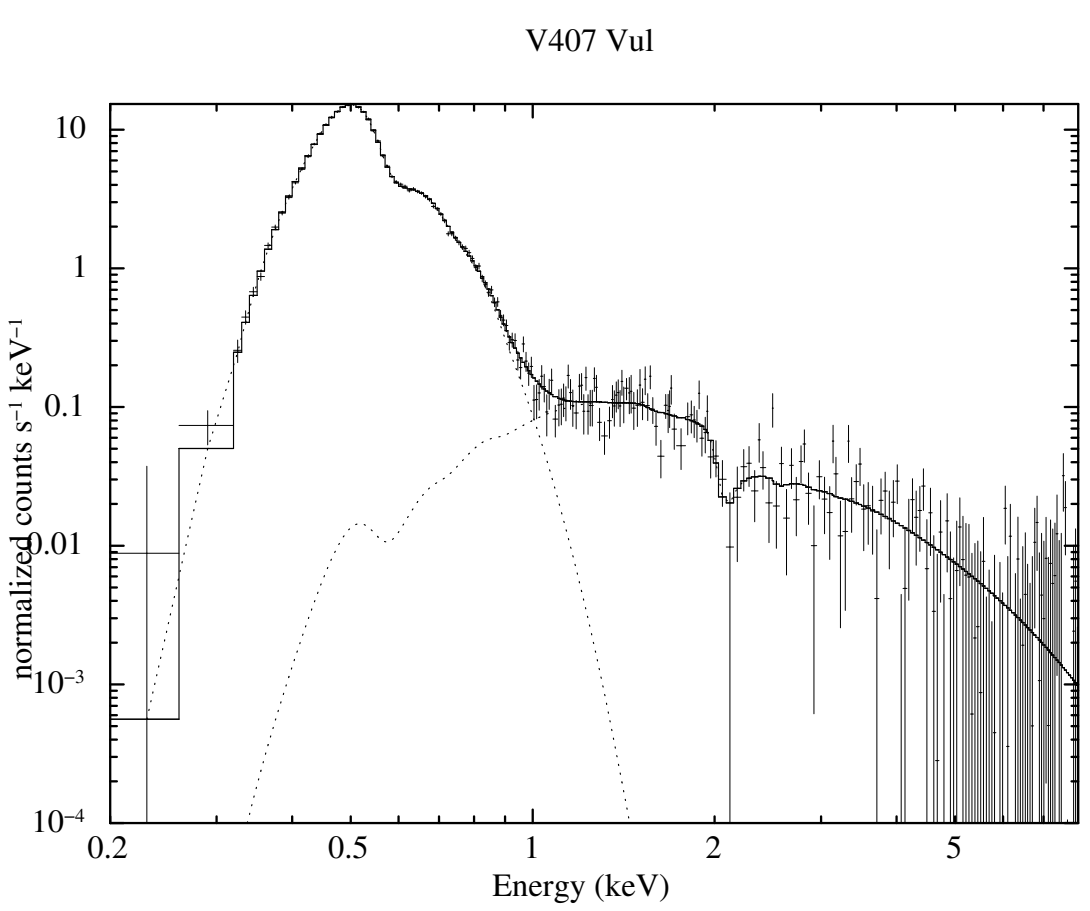

**Figure 24.** *on-axis spectrum for V407 Vul with AXIS. The model includes a blackbody with a bremsstrahlung tail.*

**[Exposure time (ks):]** 60 ks in total for each target, 4 known targets.

**Observing description:** We request 15 ks each visit with a cadence of 1, 2, and 3 weeks for each object. The long observing windows provided by L2 are suitable for this purpose. The exposure and cadence are designed to allow for phase connection of the photons. The total exposure time of 60 ks will also enable us to extract a high-quality spectrum and search for the elusive hard X-ray tail expected to originate from the magnetised bremsstrahlung component in these sources. The good time resolution and large effective area of *AXIS* are uniquely suited for this capability. The good and uniform spatial resolution will enable us to identify more of these objects, especially in the Galactic plane, which is prone to source confusion. The *AXIS* survey of the Galactic plane will be instrumental in that regard. See Table 4.

**[Joint Observations and synergies with other observatories in the 2030s:]** These systems will be a major foreground component in LISA, and *AXIS* observations can be used to systematically search for new ultra-compact double degenerate systems in regions of interest defined by LISA. In addition,

**Table 3.** Properties of DD ultra-compact binaries

| Source | Orbital Period [min] | kT [eV] | $L_x$ [erg/s] | $\log(f_x/f_{opt})$ | g′-r′ | $^f$*AXIS* c/s |
|---|---|---|---|---|---|---|
| Hm CnC | 5.4 | 65 | $^a4.5\times10^{34}$ | $^e$2.3 (2.4) | -0.44 | 3.9 |
| V407 Vul | 9.5 | 43 | $^b5\times10^{35}$- $4\times10^{36}$ | $^e$1.6 (3.5) | 0.99 | 2.5 |
| 3XMM J051034.6−682640 | 23.6 | 69 | $^a5\times10^{32}$ | $^e$1.1 (1.2) | -0.03 | 0.7 |
| eRASSU J060839.5−704014 | 6.2 | $^d$110 | $^c6\times10^{31}$-$1.5\times10^{33}$ | $^e$1.6 (1.9) | 0.25 | 1.3 |

**Table 4.** $^a$: for d=5 kpc; $^b$: for d=4–5 kpc, Gaia geometric distance $4.8^{+1.8}_{-1.6}$ kpc [**?** ]; $^c$: for d=1–5 kpc; $^d$: also a hard bremmstrahlung component detected; $^e$: computed with the information of the mean g′ magnitude and maximum observed $F_x$ (un-absorbed and reddening-corrected). The g′-r′ colours were obtained after reddening correction. The luminosities were corrected for absorption.

Rubin/Roman/ALMA/JVLA observations are expected to play a key role in further confirming the nature of the systems identified by *AXIS*.

*17. The AXIS-LISA synergy: a new era of population studies of ultra-compact binaries*

**First Author:** Nathan Steinle (University of Manitoba, nathan.steinle@umanitoba.ca),
**Co-authors:** Samar Safi-Harb (U. of Manitoba), Liliana Rivera Sandoval (UTRGV), Kevin Burdge (MIT), Tyrone Woods (U. of Manitoba)
**Abstract:** *AXIS* will unearth the population of ultra-compact binaries (UCBs) in the Milky Way galaxy and will allow for high-resolution follow-up to place unprecedented constraints on their properties. Among the many types of UCBs that *AXIS* will discover are AM CVns, ultra-compact X-ray binaries, hot sub-dwarfs with a white dwarf companion, etc., but these can be challenging to classify. This overlaps significantly with the science targets of the Laser Interferometer Space Antenna (LISA), a future space-based detector sensitive to the milli-Hz regime of gravitational waves that will launch in ∼2035. While the foreground composed of Galactic UCBs will pose a significant noise source for LISA, some of these UCBs, known as verification binaries (VBs), will be individually resolvable by LISA and will be observed prior with electromagnetic methods. As *AXIS* will observe the Galactic population of UCBs, it will naturally provide VBs as a guaranteed set of gravitational wave sources for LISA and will be important for testing the LISA instrument. This will crucially depend on the types of UCBs that *AXIS* uncovers and the constraints that *AXIS* is able to place on their parameters. Conversely, LISA will be able to constrain the parameters of the VBs as well, and possibly discover UCBs beyond the detectability of *AXIS*, providing an independent measurement. Consequently, to mutually maximize the multi-messenger potential of both *AXIS* and LISA, data analysis tools will be developed to confidently classify sources in the *AXIS* population, which will enable the procurement of a database of VBs for LISA science and modeling frameworks will be created to relate the binary parameters with *AXIS* observables. In turn, this will ensure joint *AXIS*-LISA observational campaigns of outstanding UCBs, illuminating a new view of the Galactic UCB population.
**Science:** Among the plethora of systems that *AXIS* will observe, ultra-compact binaries (UCBs) constitute a large diversity of sources [467], including AM CVn systems composed of a white dwarf star and a helium-rich donor star (e.g., white dwarf or semi-degenerate helium star); ultra-compact X-ray binaries (UCXBs) composed of a neutron star and a white dwarf star; hot sub-dwarfs with a white dwarf companion (sdBs); black hole–white dwarf binaries (BHWD) composed of a stellar-mass black hole accreting material from a white dwarf star; and double white dwarf (DWD) systems composed of two detached or mass-transferring white dwarf stars. Classifying the UCBs in *AXIS* data will be an exciting challenge. *AXIS* will classify different ultra-compact binary sources based on a combination of observational signatures, including their X-ray properties, orbital periods, spectral features, variability patterns, and multi-wavelength data [509]. With a UCB population from *AXIS*, this data challenge will be aided by other messengers, such as gravitational waves (GW), and will require novel data analysis methods.

LISA [13] is a planned space-based gravitational wave observatory adopted by ESA and in collaboration with NASA with a planned launch in the mid-2030s. LISA will detect milliHz GWs with impressive sensitivity to a wide variety of sources, such as stellar-mass compact binaries in the Milky Way, i.e. potentially tens of thousands of Galactic UCBs. Most of these UCBs will form the *galactic confusion noise*, i.e. the stochastic foreground signal composed mainly of white dwarf binary inspirals residing near the peak of LISA sensitivity. A few thousands will be sufficiently loud or persistent for LISA to resolve individually, a part of the LISA global fit problem, and many if not all will also be observed in X-rays with *AXIS*.

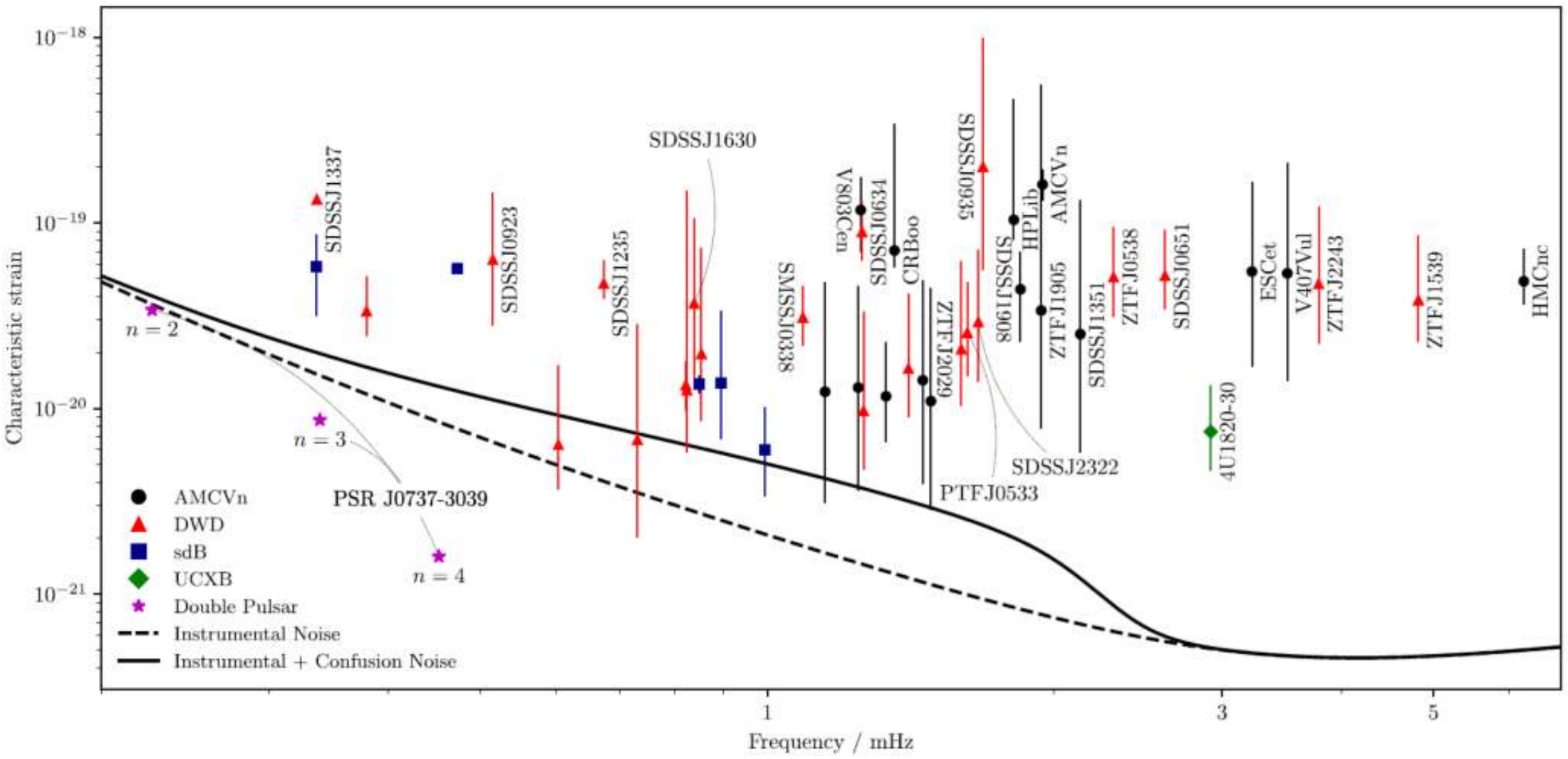

**Figure 25.** Taken from Figure 1 of [177]: the characteristic strain of 43 Verification Binaries (VBs) candidates compared to the LISA instrumental noise, averaged over LISA orbital modulations and polarisation angle. For those with reported uncertainties, error bars indicate the given measurement uncertainties in VB component masses, distances, inclinations, and frequencies (these sources of uncertainty being in decreasing order of significance). Also shown are the first few harmonics of the double pulsar PSR J0737-3039, which is on an eccentric orbit with period 2.45 hours; although it's close, this is not expected to be detectable with LISA.

The so-called 'verification binaries' (VBs) are compact binary systems that have already been observed electromagnetically and serve as guaranteed sources of gravitational waves for LISA. At leading order, the GW strain amplitude for a compact binary is [50,136],

$$h \propto \frac{(m_1 m_2)^{5/3} f_{\rm orb}^{2/3}}{d} \,, \tag{1}$$

where $m_1$ and $m_2$ are masses of the binary components, $d$ is the distance, and $f_{\rm orb}$ is the binary orbital frequency. Eq. (1) shows how VBs are likely to have higher frequencies as they will be the loudest sources. The sky location, orbital inclination, and GW polarization are also important parameters. In unison with *AXIS*, LISA will immediately detect $\gtrsim$10 VBs and many more over its lifespan as nearly constant-frequency sources because they generally have slow orbital decay where GW waveform harmonics accumulate in a narrow frequency band. VBs will allow us to verify LISA's sensitivity and calibrate its instruments shortly after it comes online. For example, as their parameters (e.g., masses, period, inclination, sky location) are already measured, implying their corresponding GW signals can be predicted and used to validate LISA's performance, data analysis assumptions, and pipelines. The types of UCBs detectable by *AXIS* are expected to provide VBs for LISA, e.g.'s, AM CVns, DWDs, BHWDs, sdBs, etc. Many systems are accreting binaries whose X-rays originate from accretion shocks or disk heating, but detached DWDs with measured orbital decay rates may provide many VBs.

*AXIS* will increase the number of known VBs by uncovering the Galactic UCB population to ensure multi-messenger coordination and will solve classic uncertainties in models of UCBs. Figure 25, taken from [177], demonstrates how many types of UCBs described above can emit narrow-band GWs from individually resolvable sources around the peak of LISA sensitivity. Characterizing the type of UCB observed by *AXIS* is crucial for accurately modeling the GW waveform for LISA, which will provide measurement uncertainties of the binary parameters.

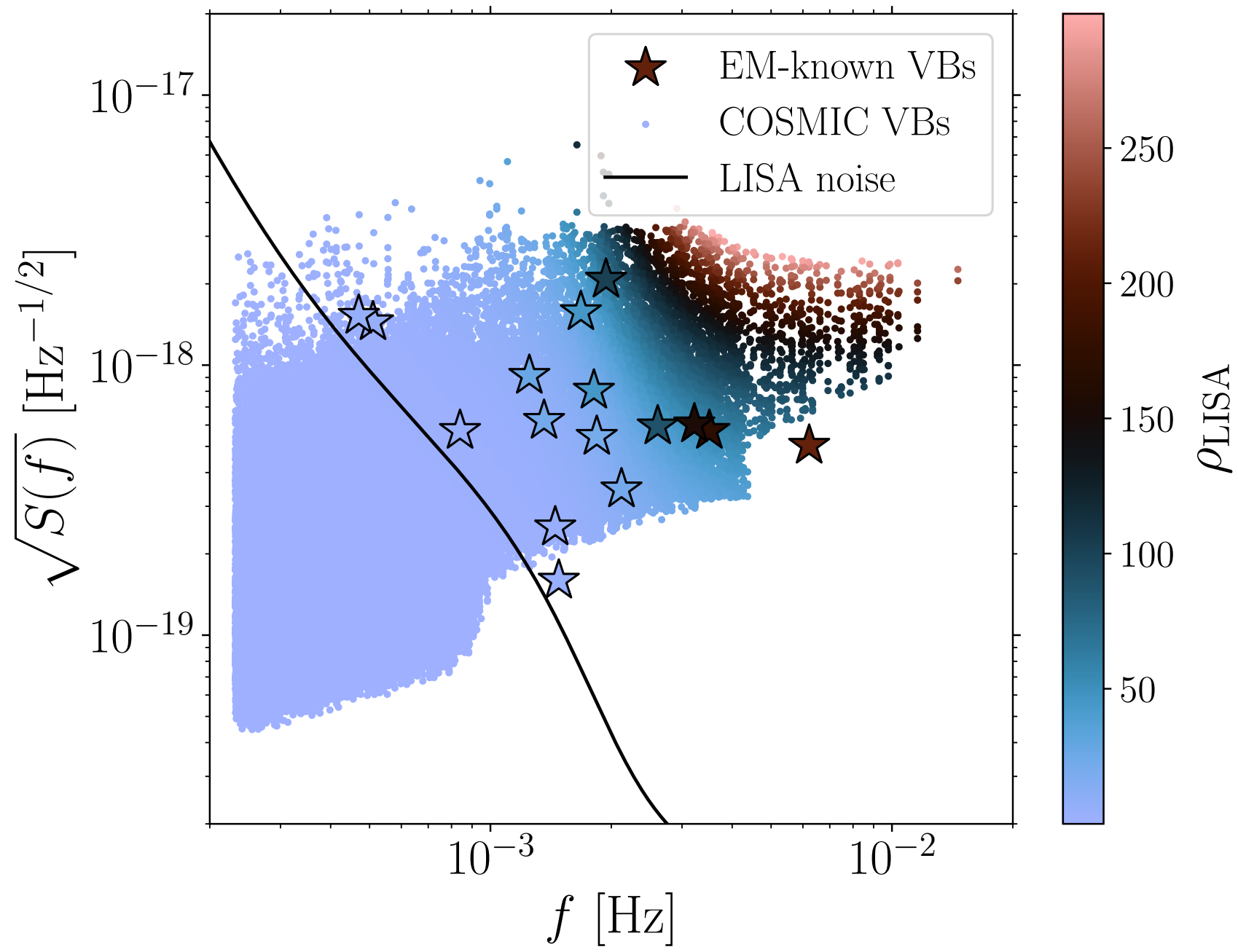

**Figure 26.** The amplitude spectral density as a function of the gravitational wave frequency of 16 known VBs (shown by the large stars) and of a theoretical population of binary white dwarfs (shown by the dots, i.e., COSMIC binaries) assuming a uniform distribution of distances, and the signal-to-noise ratio of the binaries with LISA shown by the colorbar. The strains and SNRs of the EM known VBs are computed the same way as for the theoretical COSMIC VBs, but with binary parameters taken from [304]. The black solid line is the default LISA sensitivity curve from LEGWORK. Throughout, we assume a 5-year LISA mission.

To highlight this area of significant complementarity between *AXIS* and LISA, we use the binary population synthesis code COSMIC [86] to obtain a population of Galactic DWDs *The* COSMIC *Way* by converging on the desired range of UCB parameters, which are then passed to LEGWORK [599] to compute the source signal-to-noise ratio (SNR) $\rho$ of each binary. We also use LEGWORK to compute the binary GW strain amplitude $h$, which is converted to the characteristic strain as $h_{\mathrm{c}} = h\sqrt{N}$ [379] where $N = f_{\mathrm{orb}} t_{\mathrm{LISA}}$ is the number of cycles the binary orbits during the LISA mission duration $t_{\mathrm{LISA}} = 5$ yr. The amplitude spectral density (i.e., the square root of the power spectral density $S(f)$) of each of the GW signals is thus given by $\sqrt{S(f)} = h_{\mathrm{c}}\,(2f_{\mathrm{orb}})^{-1/2}$ [379], and of the LISA detector noise from [498] including the galactic confusion foreground. These are shown as functions of GW frequency in Figure 26 by the data points and the black solid line, respectively, and the colorbar shows the SNR of each source. A few EM-known VBs are plotted (shown by stars) and colored by their LISA SNR as computed from [304], which utilized a fuller LISA simulation compared to the approximations employed in LEGWORK, resulting in different dependence of the SNR on the binary parameters as compared to the theoretical binaries (shown by dots) where we use LEGWORK to compute the SNR.

These results motivate the development of new models that map the binary properties to the *AXIS* X-ray observables to compare *AXIS* predictions of the VBs properties with the LISA predictions. The spatial distribution of Galactic UCBs is a complex interplay between the evolution of the binary system (such as mass transfer and natal kicks) and the evolution of the galaxy, resulting in preferences for certain types of UCBs in specific regions of the Galactic center and plane. Modeling realistic Galactic binary populations is generally a vast open problem, e.g., see [314,570], which multi-messenger strategies with future facilities will resolve. This necessitates models that translate the binary properties (i.e., intrinsic parameters such as masses and period, and extrinsic parameters such as sky location and inclination) directly into *AXIS* observables (i.e., flux or luminosity, sky location, etc.). While such a model is currently beyond reach, we motivate its development by utilizing existing codes to demonstrate the multi-messenger *AXIS*-LISA synergy for VBs.

**Observing description:** To uncover the population of Galactic UCBs, *AXIS* will provide: high-resolution imaging of dense environs such as globular clusters or near the Galactic center; detection of X-ray orbital modulation to confirm short-period binaries by precision time tracking; and X-ray spectroscopy will help us to determine the composition of the accreted material, the nature of the magnetic fields and high-energy processes, and the nature of the accretor (i.e., white dwarf vs. neutron star) which will be crucial for identifying the helium-dominated donors in AM CVn stars or the neon-rich spectra typical of UCXBs; monitoring of the orbital evolution informing predictions of when LISA will detect them; and multi-wavelength synergies that provide complementary information. Coupling *AXIS* with LISA, we can tightly constrain mass ratios, inclinations, and likely evolutionary histories of the VBs.

We can also use *AXIS* to maximize the scientific output of other detectors. For example, *AXIS* could detect a faint X-ray source in a globular cluster with a $\sim$10–minute orbital modulation; if follow-up with Roman/JWST shows a lack of hydrogen lines in the spectrum, the system is consistent with an AM CVn; then Gaia gives a distance of $\sim$5 kpc to the vicinity of the source; LISA analysts model its GW strain and confirm whether it will be detectable for LISA; *AXIS* monitors the system during its passage through the LISA detection window, providing time-resolved X-ray emission and GW constraints for precious insight into mass transfer.

We will also devise *AXIS*-LISA synergies that take advantage of *AXIS*'s design. For example, *AXIS* will perform a wide-field survey that will discover many VBs, providing a new sky map of VBs. LISA (assuming it is already online) can then conduct an initial search over that sky map to identify the loudest sources, providing zeroth-order multi-messenger constraints. Then *AXIS* follows up with higher resolution observations of the loudest LISA sources and the brightest *AXIS* sources and we iterate these steps to provide higher order multi-messenger constraints and narrow down specific astrophysical uncertainties to make ground-breaking/fundamental UCB multi-messenger population discoveries. The ability to classify *AXIS* objects can then be iteratively improved with LISA, e.g. LISA identifies tight binaries helping distinguish between double white dwarfs and AM CVns. Alternatively, if LISA is not online concurrently with *AXIS*, then the iteration above can be modified so that the telescope that is offline is limited in its multi-messenger scope to what is immediately observable by the online telescope. This takes advantage of the fact that both *AXIS* and LISA can provide sources for the other to follow up on (concurrently or at later times). Also, LISA can inform *AXIS* observations of faint X-ray sources that are loud in the LISA window.

**[Exposure time]:** We expect to detect known and new systems in the Galactic Plane Survey. Follow up observations for spectroscopy will require about 20 ks per target. Depending on the number of targets discovered, we expect to require around 100 ks per year.

**[Joint Observations and synergies with other observatories in the 2030s:]** Rubin/LSST, GOTO, Gaia, Theia, Square Kilometre Array (SKA), ELTs, NewAthena's X-IFU.

*18. AXIS prospects for state transitions and accretion physics in Cataclysmic Variables and related objects*

**First Author:** Şölen Balman (Istanbul Univ.; solen.balman@istanbul.edu.tr)
**Co-authors:** (with affiliations)
**Abstract:** Accreting white dwarf binaries (AWDs) comprise cataclysmic variables (CVs), symbiotics, AM CVns, and other related systems that host a primary white dwarf (WD) accreting from a main sequence or evolved companion star. AWDs are a product of close binary evolution; thus, they are important for understanding the evolution and population of X-ray binaries in the Milky Way and other galaxies. AWDs are essential for studying astrophysical plasmas under various conditions, including accretion physics and processes, transient events, matter ejection and outflows, compact binary evolution, and nuclear processes leading to explosions. AWDs are also closely related to objects in the late stages of stellar evolution, and other accreting objects in compact binaries; they even share common phenomena with young stellar objects, active galactic nuclei, quasars, and supernova remnants. *AXIS* with its superb sensitivity in the 0.3-10.0 keV range with moderate spectral resolution and good timing capability of $\sim$ 20 ms resolution, is a promising new mission for the field of AWDs. Given that most AWDs have low X-ray flux/luminosity (aside from Supersoft X-ray sources), *AXIS* is a great opportunity to study X-ray emission and transient properties and population characteristics of these systems with improved instrumentation and observational capabilities in regards to the present missions such as *XMM-Newton*, *Chandra*, and *Swift*. This is a proposal for an organized approach for CV observations (and some AWDs) with *AXIS*.
**Science:**

In most CVs, accretion is via Roche lobe overflow, and disk accretion occurs from a donor star that is either a late-type main sequence star or sometimes a slightly evolved star [607]. CVs have orbital periods of 1.4–15 hrs, with a few exceptions of up to 3 days. Another class of AWDs is the symbiotics, where the donors are red giants or AGB stars. Accretion in such systems is mainly driven by powerful winds, and most systems show disk formation. Symbiotics have orbital periods ranging from several hundred days to several years. AM CVns, another related system to CVs, are ultracompact systems with binary periods between 5 and 65 min that have passed the CV period minimum. AM CVn stars also display Roche lobe overflow with the possibility of stream impact accretion leading to hot spots on the WDs. For general characteristics and X-ray emissions of CVs and related systems, see the recent review by Balman et al. [46]. There are about 450 X-ray emitting CVs and candidates known at a flux limit of $> 1.0\times10^{-14}$ erg cm$^{-2}$ s$^{-2}$ [605].

In non-magnetic CVs the transferred material forms an accretion disk that reaches the WD. Standard accretion disk theory [529] predicts half of the accretion luminosity to emerge from the disk and the other half from the boundary layer (BL) very close to the WD [339]. During low-mass accretion states, the BL is optically thin emitting in the hard X-rays [397,461], and for high accretion rate states ($\dot{M}_{acc}\geq10^{-(9-9.5)}$M$_{\odot}$), it is optically thick emitting in the soft X-rays and EUV (kT$\sim10^{(5-5.6)}$ K; Hertfelder & Kley [241], Popham & Narayan [462]). The standard disk is often found inadequate to model high state CVs in the UV, as well as some eclipsing quiescent dwarf nova. It generates a spectrum that is bluer than the observed UV spectra, indicating that a hot optically thick inner flow of BL is non-existent [330,472]. As a result, recent standard disk models have employed a truncated inner disk (e.g., [202]), which adequately models the UV data. The accretion flows in high state CVs and dwarf novae are potentially well explained in the context of radiatively inefficient (advective) hot flows as opposed to standard optically thick accretion flows, that form in the inner disk, which seem to explain most of the complexities in the X-rays and other wavelengths [45,48,49]. Shock formation in ADAF flows (advection-dominated accretion flows) around accreting WDs has been calculated [141], and a shock may occur, e.g., $\sim$1.3$\times10^{9}$ cm (short distance) from the WD, which can explain the hard X-ray emission and the eclipse effects.

Magnetic cataclysmic variables (MCVs) have two sub-classes according to the degree of synchronization of the binary, comprising about 25-30% of the CV population. Polars have strong magnetic fields of 14-230 MG, which cause the accretion flow to directly channel onto the magnetic pole/s of the WD inhibiting the formation of an accretion disk. The magnetic and tidal torques cause the WD rotation to synchronize with the binary orbit. Intermediate Polars (IPs), which have a weaker field strength of 5-30 MG, are asynchronous systems to a large extent (mostly, Pspin/Porb$\sim$ 0.1) [145,388]. Polars show strong orbital variability at all wavelengths [525], whereas IPs may be disk-fed, diskless, or in a hybrid mode in the form of disk-overflow which may be diagnosed by spin, orbital and sideband periodicities at different wavelengths [145,238,406]. The accretion flow in MCVs, close to the WD poles, reach supersonic velocities producing a stand-off shock above the WD surface [8]. The post-shock region is hot (kT$\sim$ 10-50 keV) and cools via thermal Bremsstrahlung (hard X-rays) and/or cyclotron radiation emerging in the optical/nIR band. Hard X-rays are partially thermalized (and/or reflected) by the WD surface and re-emitted in the soft X-rays and/or EUV/UV domains creating the blackbody components of emission (kT$\sim$ 20-60 eV), see [145,179,523]. The complex geometry and emission properties of MCVs make these ideal laboratories to study in detail the accretion processes in moderate magnetic field environments, also help understanding the role of magnetic fields in close-binary evolution. Below is an itemized summary of what important things *AXIS* can do for AWDs:

1. Dwarf Novae (DN) outbursts occur when continuous or sporadic mass accretion is interrupted every few weeks to months by intense accretion of days to weeks with duty cycles months to years ($E_{out}$=$10^{39-40}$ erg, $\Delta$m=2-6 ). The three main types of DN are: 1) U Gem Type: Porb $>$ 3hrs; No Superoutburts seen, 2) Z Cam Type: with Standstills, 3) SU UMa Type: Porb $<$ 2hrs; Superoutbursts detected. DN outbursts are explained in the context of Disk Instability Model [DIM; 225,317,318]. Several inconsistencies exist regarding DIM and DN observations in quiescence and outburst [45]. For example, there is usually a delay in the rise of the X-ray light curves or the X-rays are completely suppressed during the optical peak. Moreover, in the high state (outburst), a BL is expected to be optically thick with a blackbody emission; however, this is detected in only a few systems (5 systems with 5-25 eV; see Balman et al. 46). A hard X-ray component is detected during the outburst in all incidences ($kT_{max}$=1-10 keV, $\leq$ several$\times 10^{31}$ erg/s) which is suggested to belong to the advective hot flows in the disk that pertains throughout the outburst and quiescence [49,164]. In general, there is a lack of systematic X-ray observations to resolve existing controversies, particularly in the soft X-rays, where *AXIS* can be influential due to its swift response for transients and high sensitivity in the 0.3-10.0 keV regime. Moreover, quiescent and outburst DNe show the existence of warm absorber effects (see Balman et al. 46, Balman et al. 49) that need high sensitivity of *AXIS* for detection at the expense of high spectral resolution. Figure 27, on the left is an *AXIS* spectral simulation of a DN in outburst (10 ksec) using a composite model of *tbabs*$\times$*(BBODY*+*CEVMKL*+*power*+*GAUSS)* from Balman et al. [49] for outbursting DN ($kT_{BB}$=27 eV, $kT_{max}$=1.2 keV). The simulated spectrum is fitted with *tbabs*$\times$*zxipcf*$\times$*(CEVMKL)*, showing a definite discrepancy between warm absorber (zxipcf) effects and a blackbody emission and a power law tail in the harder X-rays. The blackbody component in the left hand panel has flux $1.5\times10^{-11}$ erg cm$^{-2}$ s$^{-1}$ that has been increased by a factor of 10 times in the right panel, these correspond to detection of a standard BL emission in outburst at 4 kpc and 1 kpc distance, respectively. This demonstrates that DIM model can be tested by *AXIS*. The detection of the alleged power-law emission is at $3\sigma$–$4\sigma$. Note that the power law detection requires plasma emission at temperatures less than 3–4 keV (the harder the plasma emission, the less the detectability of the power law becomes unless close to equal normalizations).
2. multi-wavelength emission of high state CVs shows strong inconsistencies and discrepancies from standard disk flow i.e., X-ray observations show hard X-rays only (largely belonging to a non-equilibrium ionization plasma Balman et al. 48 with $\leq$ several$\times10^{32}$ erg/s; 5-15 keV) with the existence of power law tails ($\Gamma$=1.3-2.0) and radio observations detect some of the Nova-likes and DN (in outburst) [124,243]. It is important to understand the origin of the SED and the disk structure in nonmagnetic CVs [202,249,405]. Advective hot flows in the inner disks with a dominant hard X-ray emission from a non-equilibrium ionization plasma can be the solution [45,48,49]. However, the extent and limits, or constraints, on such flow characteristics require a large, dedicated data set that can be supplied by *AXIS*. Simulation on the left (10 ksec) in Figure 28 indicates that *AXIS* can resolve a collisional equilibrium plasma from a nonequilibrium ionization plasma, but the power law tails will not be properly distinguished from thermal plasma emission for plasma temperatures higher than 3-4 keV (the two components may be disentangled if norms of the components are comparable with about a factor of 4 or less). For this, synergies with missions using a harder X-ray band than 10 keV will need to be taken into account. Simulation on the right (Figure 28) shows the Fe XXVI P Cygni profile of a high state CV detected in Balman et al. [48] with *NuSTAR* as would have been detected by *AXIS* in an exposure of 15 ksec, half the exposure time of *NuSTAR*.

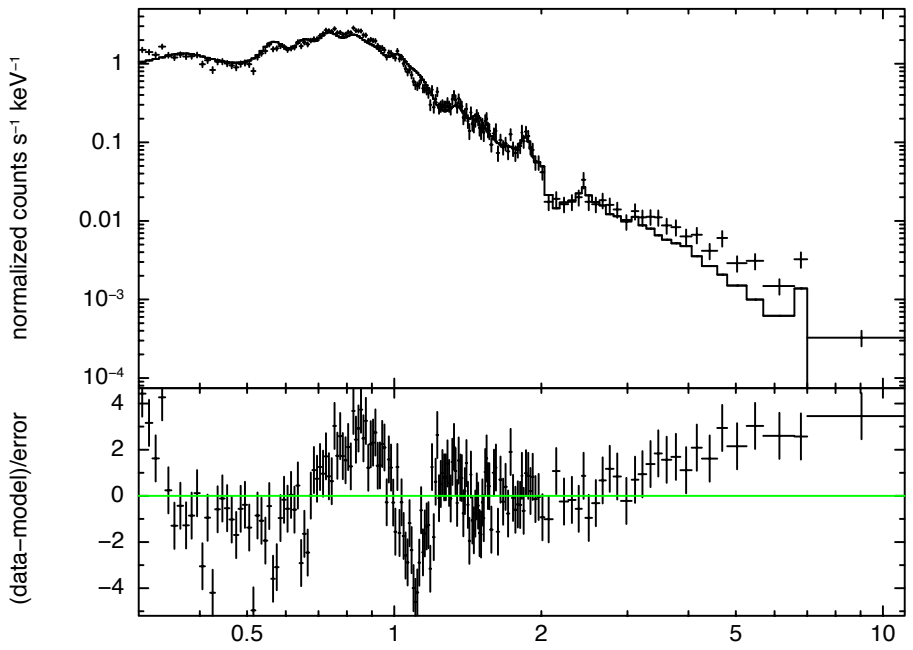

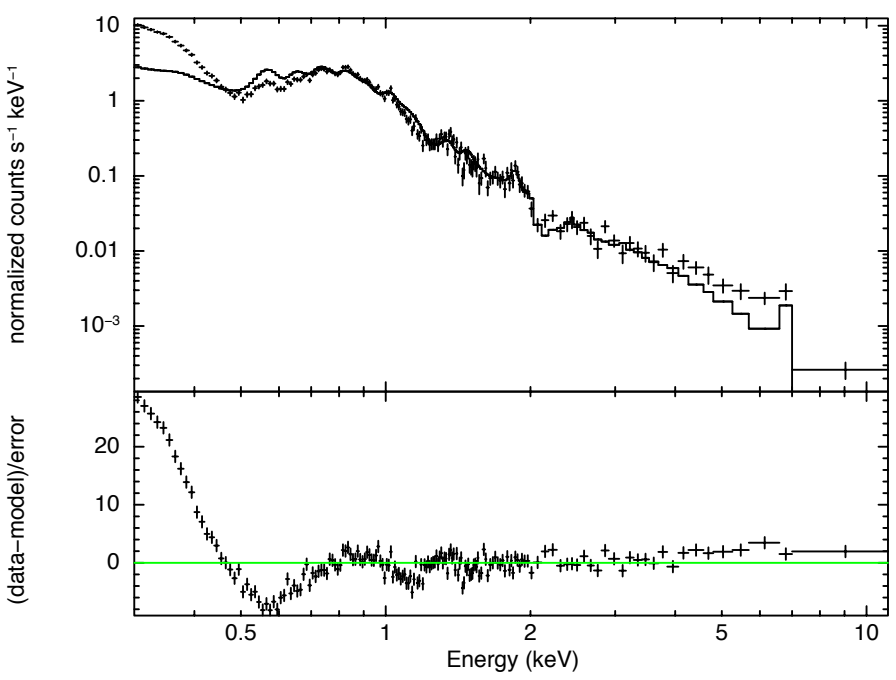

**Figure 27.** On the left and right are 15 ksec *AXIS* spectral simulation of a high state DN in outburst with model *tbabs*×*(BBODY+CEVMKL+Power)* (k$T_{BB}$=27 eV, k$T_{max}$=1.2 keV) fitted with only a CEVMKL (XSPEC). The difference is the blackbody flux of $1.5\times10^{-11}$ erg cm$^{-2}$ s$^{-2}$, on the left and $1.5\times10^{-10}$ erg cm$^{-2}$ s$^{-2}$ on the right. The harder band indicates significant detection of the power law.

3. Broadband noise and its variations in accretion flows have been a diagnostic tool for understanding the structure of accretion disks, together with accretion history and state changes. CVs demonstrate band-limited noise (mainly 1-6 mHz) in the optical, UV, and X-ray energy bands [47], which can be adequately explained within the framework of the model of propagating fluctuations yielding break frequencies. The broadband noise studies indicate that an (advective) hot flow structure resides inside nonmagnetic CV disks, mainly detected in the X-ray regime [45,48,164]. However, the statistical quality of the X-ray data has not been adequate to derive flux vs. rms correlations or study hard or soft lags in the X-rays, which could shed more light on the physics of the X-ray-emitting regions. *AXIS* has the timing resolution necessary and superb sensitivity in the 0.3-10.0 keV regime, which will enable organized spectral-timing studies of CVs and AWDs, in general.
4. High state, low state, and active phase statistics and observations of accreting nonmagnetic WDs are lacking, and systematic studies with observational monitoring are required to understand this population and its evolutionary aspects. For example, Symbiotic or AM CVn systems show poorly categorized active phases that are thought to involve disk outbursts, where only a few examples have been relatively well-observed in the X-rays [337,338,493,494]. AM CVns have $L_x$ = $10^{30-32}$ erg/s (3.5-8.7 keV). AM CVn active states are not well understood, but long active states (one very short) are detected, which are 5-10 times the predicted values by DIM. *AXIS* with its superb sensitivity and transient response should be able to reveal many such active states (and follow them), permitting the study of evolution through these states and resolving the origin of such states. WD Symbiotics exhibit active states of long duration, which share similarities with DNe; however, for example, the absence of soft X-ray-emitting BL emission is detected [46].

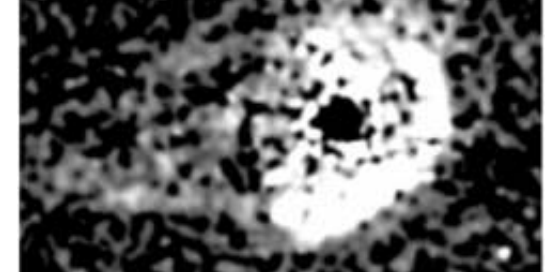
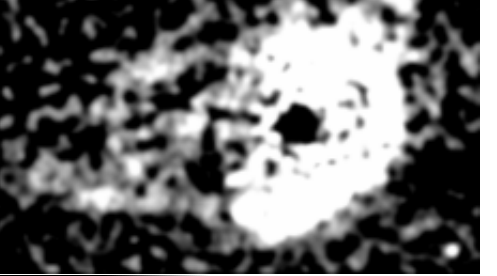

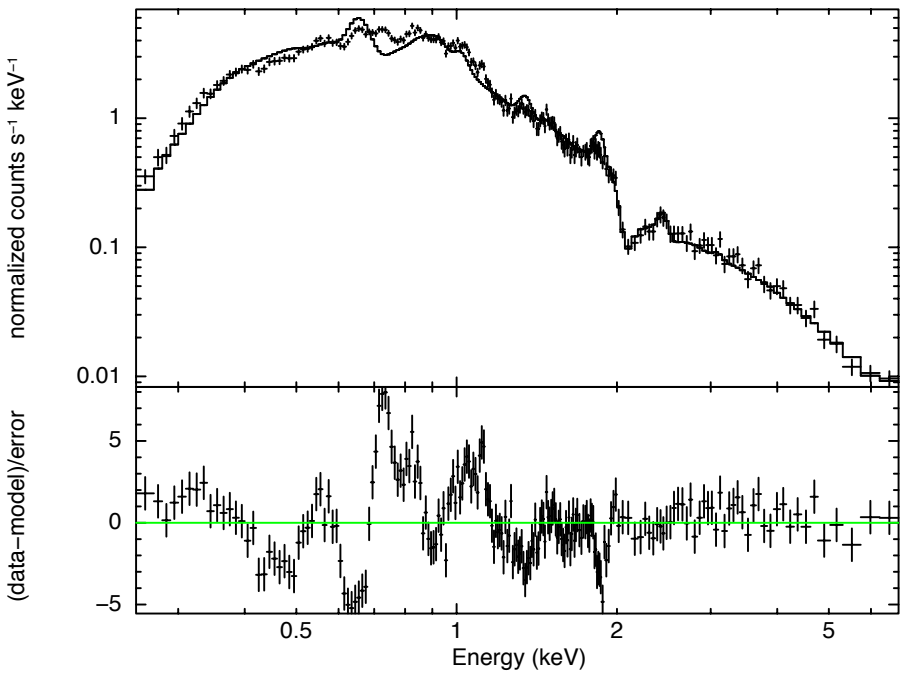

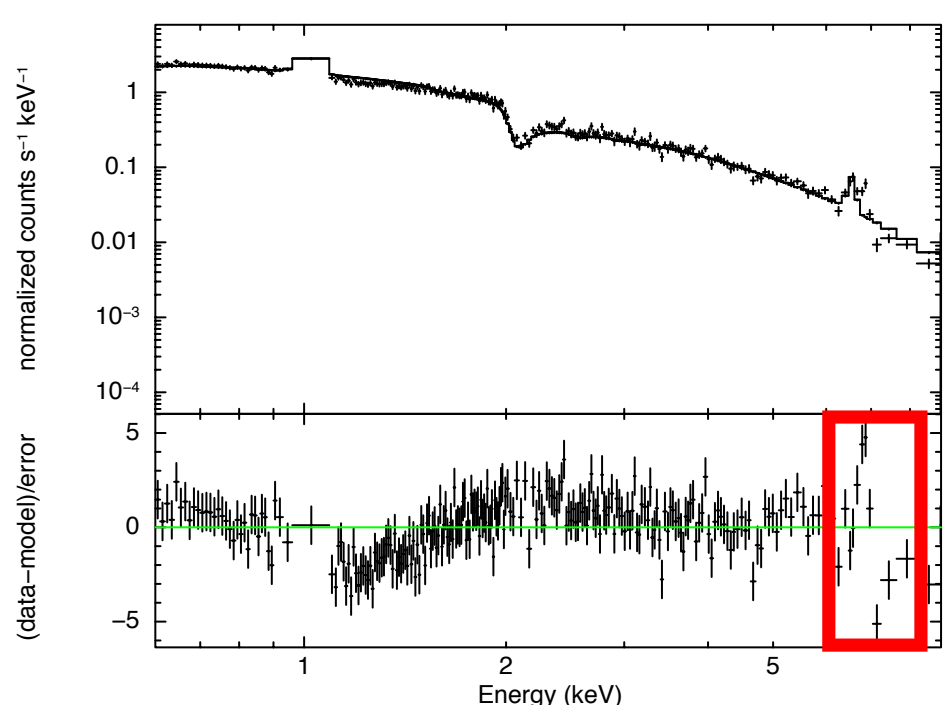

**Figure 28.** On the left is a 10 ksec *AXIS* spectral simulation of a high state CV with model *tbabs×(VNEI+Power)* fitted with CEVMKL (XSPEC) that is a collisional equilibrium plasma model. Residual variations are at a high sigma level and indicate that we will be able to detect plasma equilibrium differences. On the right is a similar composite model simulation for 15 ksec fit using *tbabs*gabs×(Bremss+4gaussians)*, which is fitted without the Fe XXVI absorption and emission lines (Fe XXV emission and reflection line is fitted). The P Cygni profile of the Fe XXVI line is significant.

5. Period bouncers are late stages of CV evolution that are expected to be (40–70)% of the CV population. There are only a handful of detections, with a few showing X-ray emission. Some of them seem nonmagnetic with one magnetic system. X-rays indicate, e.g., a plasma emission of 3-5 keV with $10^{28-30}$ erg/s (a few $\times 10^{-14}$ erg s$^{-1}$ cm$^{-1}$) and an accretion rate of (3-9)$\times 10^{-14}$ $M_{\odot}$yr$^{-1}$. See Figure 29 for a 20 ksec *AXIS* simulation. These extremely low luminosity and accretion rate objects can only be studied effectively with *AXIS* and particularly the sensitivity limit of the GPS should reveal many of these systems and resolve the controversy on the existence of such systems. If not, CV evolutionary stages will have to be reconsidered.

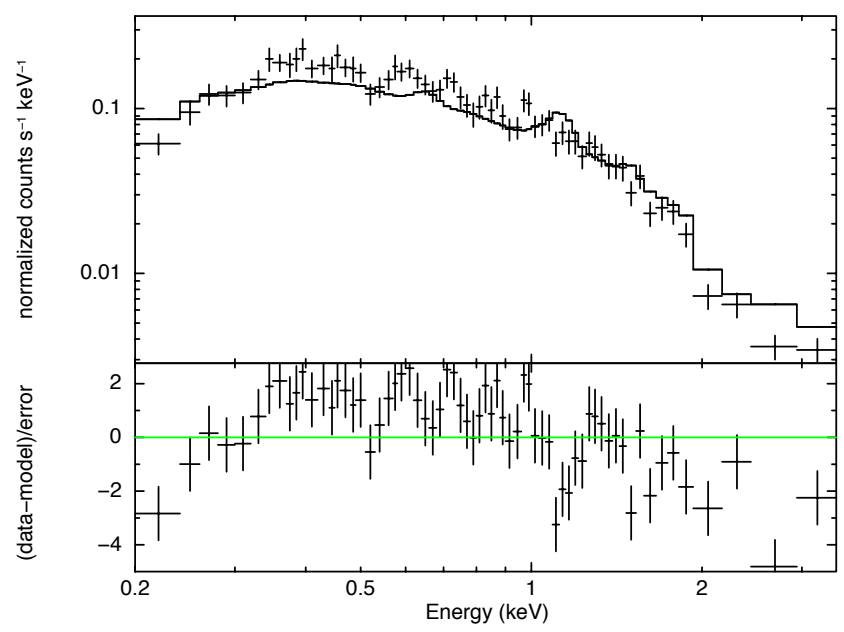

**Figure 29.** A 20 ksec *AXIS* spectral simulation of a period bouncer. The fits indicate a low metallicity of 0.04, and the simulated spectrum is fitted with a solar metallicity, and the residual variations indicate that we will be able to parse abundances with *AXIS*, which would be of interest.

6. In MCVs (Polars and Intermediate polars), a stand-off shock near the magnetic poles occurs emitting X-rays and cool via Bremstrahlung ($kT_{max}$=10-50 keV, $\leq$ several$\times 10^{34}$ erg/s;) and/or Cyclotron cooling in the optical-IR depending on the strength of the magnetic field of the WD along with reprocessing of hard X-rays into soft X-ray blackbody emission components. IPs are the hardest

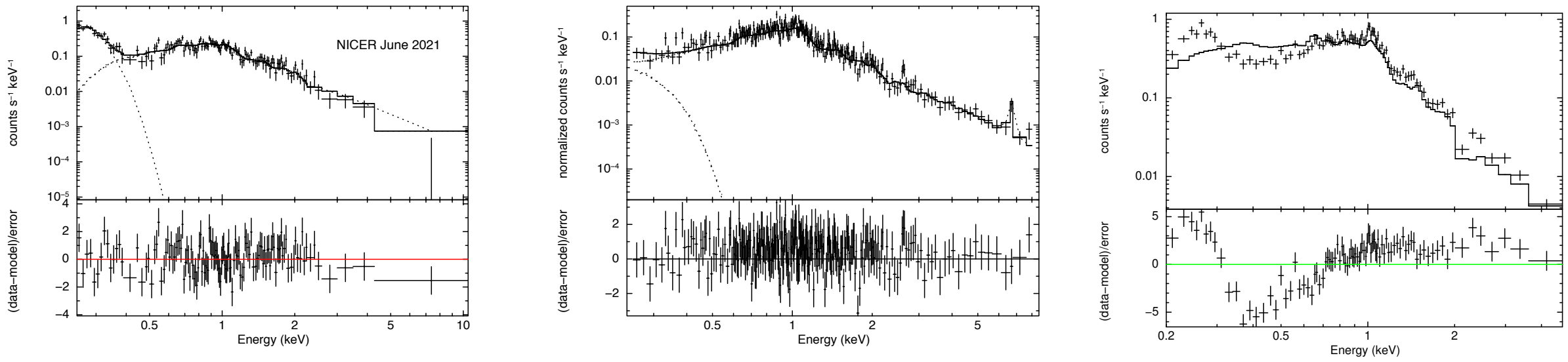

**Figure 30.** On the left is the fitted NICER spectrum of SGRAJ 213151 (25 ksec) using the model *phabs×(BBODY+CEVMKL)*. The middle panel is the 10 ksec *XMM-Newton* simulation of the same spectrum where the blackbody emission is not detected. The right panel shows a 10 ksec *AXIS* spectral simulation of a low-state MCV (SGRAJ 2123151) with the same model as in NICER spectrum fitted with CEVMKL (XSPEC) only. Residual variations show that we will be able to detect a 15 eV blackbody emission with *AXIS* in a much shorter exposure than NICER.

X-ray emitters that have been detected with the *Swift* BAT and *INTEGRAL* ISGRI. Both MCV species indicate complex absorption intrinsic to the system. Studying the complex absorption in MCVs has been conducted, but has not been fruitful, as such complex absorption (ISM+Cold+Ionized/Warm) models [47,170,264,388] need good spectral resolution and eminent soft X-ray sensitivity to study this aspect in phase average and/or in a phase-resolved manner. This will reveal the location of absorbing regions for the emission regions and the accretion geometry of MCVs to an unattained level. Along with adequate spectral resolution, *AXIS* has the appropriate sensitivity and will provide a detailed inspection of most of the MCVs, revealing unprecedented information laying the grounds for theoretical model improvements. Such attempts on timing and spectral analysis to construct WD mass distribution and physical models of accretion columns have been successful to some extent using a small sample of MCVs [see 531,553].

7. The high sensitivity of *AXIS* will allow for the first time detailed X-ray monitoring of high & low state changes of polars where low states are defined up to 100× difference in luminosities with kT $\leq$ 5 keV, and $L_x \leq 10^{30}$ erg/s [425,522–524]. Figure 30 shows the first time detection of a blackbody emission (15-18 eV) during a low state with a NICER observation, on the left. The middle panel shows that *XMM-Newton* EPIC pn would not have detected it. In contrast, the right-hand panel shows that *AXIS* will detect such soft components in less than half the exposure time of NICER, and the harder 3-6 keV component will be detected at a good S/N ratio. Few IPs have been detected to show changes to short-lived (compared to polars) low states by about a factor of 10 decrease in the luminosity, where studies indicate that the accretion geometry and the disk structure completely changes [285,331]. The systematic collection of data by *AXIS* to study the response near the WD and geometrical changes to accretion rate variations from the donor can be resolved in this way. State changes of MCVs, as well as new state changes, particularly of IPs, can be systematically studied using a large dataset and multiple sources.
8. IPs can be disk-fed, stream-fed, in a hybrid mode, and polars have funnels that show pole switching, accretion mode changes with differing accretion regions, ballistic impact accretion and all MCVs are expected to show QPOs (with no detections in polars, so far) [62,63,77,238,359,407]. These properties require power spectral studies of both periodic and aperiodic variations to understand the accretion process, physics, and modeling of these phenomena. For this purpose, *AXIS* can provide systematic, high-sensitivity long-term data with good spectral resolution and timing resolution, which would yield unprecedented results.

9. X-ray studies of pre-polars (kT $\leq$ 1 keV, $\leq 10^{29}$ erg/s, rate $\leq 10^{-13}$ $M_\odot$/yr), have not been conducted properly by any mission to date [46, and references therein]. The emission is suggested to result from the secondaries, but remains inconclusive since these systems are very faint (also at low temperatures) and have not been well analyzed or studied in detail. This population is of importance to MCV/CV evolutionary models.

**[Exposure time (ks):]** 10 ksec (minimum) to several 10 ksec. For accretion rates about or less than $10^{-13}$ $M_\odot$/yr the exposures need to be larger than 20 ksec.
**Observing description:** All CVs and related systems are detected in X-rays and candidates. This proposal also includes newly detected systems with *AXIS* GPS and *AXIS* survey observations or serendipitous detections.

ToO is a necessity for transient systems. High and low states or DN outbursts will need to be observed via ToO. The exposures will then need to be specified. It will vary between 10-30 ksec. Trigger will be made with optical monitoring (ground-based) and/or other space observatory survey missions.
**[Joint Observations and synergies with other observatories in the 2030s:]**

Synergies with ELTs, Rubin (LSST), and SKA will be crucial for multi-wavelength approaches to identification and study. NewAthena synergies will be important when high spectral resolution (i.e., line diagnosis) is necessary. Moreover, there will be a need for synergies with X-ray missions that have a higher X-ray energy band, such as NuSTAR or SVOM, if they last until the *AXIS* mission time. Other possibilities include missions like THESEUS, which observes out to 200–300 keV, selected by ESA for the next Phase, along with only two other M7 mission candidates. Space missions like UVEX are planned for similar times with *AXIS* and the soon to launch Roman space telescope (provided it lasts out to *AXIS* mission times) will be of importance, particularly the three survey modes with quick and high cadence monitoring of the Galactic plane and the high-latitude survey are important for *AXIS* transient phenomenon studies on AWDs.
**[Special Requirements:]** (pileup, e.g., Monitoring (Daily, Hourly, etc), TOO (<X hrs), TAMM)
–ToO for DN less than a day.
–monitoring depends on the class of the DN, daily or weekly.
–ToO for high states or low states within the week response. If any particular low or high state that requires a faster response will be less than a day. But flaring activity may require a response within several minutes. All constraints should be determined according to scientific necessity.

*19. Cataclysmic Variable populations in large surveys*

**First Author:** Craig Heinke (U. Alberta, heinke@ualberta.ca)
**Co-authors:** Şölen Balman (IU, solen.balman@istanbul.edu.tr), Nazma Islam (NASA GSFC and UMBC, nislam@umbc.edu)

**Abstract:** Cataclysmic variables (CVs) are our most common and accessible laboratories for accretion physics and binary evolution. *AXIS* will enable significant advances in our studies of CV populations and accretion physics. Much of this advance will come from the identification of hundreds of X-ray faint CVs in the planned *AXIS* Galactic Plane (and Bulge) survey [509], combined with optical and infrared imaging and spectroscopic surveys. These faint CVs will include dwarf novae (DNe), nova-likes (NLs), and nonmagnetic CVs in general, magnetic CVs in low states, "period bouncers" (CVs that have evolved to very low donor masses past the period minimum), low luminosity Intermediate Polars, low-state polars, pre-polars and AM CVn systems (accreting WDs with 5-65 min orbital periods).

With a large and well-defined sample, we will illuminate the binary evolution of CV systems, helping us understand the angular momentum loss mechanisms, space density, and formation rates of CVs of different types, and the endpoints of CV evolutionary pathways (donor detachment, destruction, or merger). With a large sample, we will also unravel the evolution of magnetic CVs, which has recently been postulated to include direct connections between long-period radio transients, white dwarf pulsars, and nonmagnetic CVs [e.g. 500,521].

**Science:**

Cataclysmic variables (CVs) generally evolve from longer to shorter periods as they transfer mass from a donor star to the white dwarf, driven by loss of angular momentum. Above 3 hours, the angular momentum loss is understood to be driven by magnetic braking from the secondary's stellar wind [e.g. 292]. At an orbital period of 3 hours, stars become completely convective; in the standard model, magnetic braking stops (as the magnetic dynamo at the convective/radiative interface disappears), and the secondary shortly fails to fill its Roche lobe. The secondary, which was slightly out of thermal equilibrium during mass transfer, relaxes to a denser state, and the orbit slowly decreases through gravitational radiation, until the donor fills its Roche lobe again at a period just over 2 hours (thus leaving a period gap of 2-3 hours with few CVs). Now, the CV evolves more slowly, as the angular momentum loss is lower; therefore, population syntheses predict a pile-up of CVs at short periods, as seen in the SDSS [193]. Assuming gravitational radiation is the only angular momentum loss mechanism does not match the data; so, a reduced magnetic braking must still be active here [292]. When CVs with hydrogen-rich donors reach about 80 minutes, the donors become degenerate, and further mass loss leads to an expanding orbit, making 80 minutes a "period minimum" for CVs [e.g. 193].

This evolutionary picture predicts that CVs which have evolved past the period minimum, "period bouncers", should make up 40-80% of all CVs [e.g. 59,293]. However, only a small fraction of the predicted CVs have been found, e.g., 7-14% in a volume-limited Gaia sample [424]. However, these period bouncers tend to be faint and are in a quiescent state with low accretion rates, making them hard to spot. At low accretion rates, the low-density accretion flow efficiently converts to hard bremsstrahlung X-rays, making X-ray surveys an efficient way to find faint CVs; e.g., the eROSITA survey [387].

The *AXIS* Galactic Plane survey (reaching $F_X$(0.5-2.5 keV)= $10^{-15}$ erg/cm$^2$/s with 6 ks exposures, over a 1$^o$ by 45$^o$ swath of the Plane; [482]) will enable a deep search for CVs, including period-bouncers (as well as thousands of brighter CVs). Typical period bouncers have bolometric $L_X \sim 10^{29-30.4}$ erg/s [387]. Thus, we anticipate being able to detect them at distances up to 0.9 kpc even for $L_X = 10^{29}$ erg/s. With a predicted CV space density of $10^{-5}$ pc$^{-3}$, we anticipate finding on the order of 30 CVs within 900 pc, and thousands more farther into the Galaxy. With *AXIS*'s arcsecond spatial resolution, it will be possible to identify the counterparts in deep optical (e.g. Rubin/LSST) and infrared imaging, verifying that they are CVs. This deep census of CVs should help us to understand whether the expected population of period bouncers exists, or whether additional physics is required to destroy period bouncers during their evolution.

In addition, the deep Galactic Plane survey will find thousands of brighter CVs at larger distances. This will enable us to study the X-ray luminosity functions of CVs as a whole, and of interesting subsets. For instance, intermediate polars (where the strong magnetic field of the white dwarf interrupts a disk, channeling accretion to the magnetic poles) appear to show two populations, around $L_X = 10^{33}$ and $10^{31}$ erg/s; this survey will enable us to constrain the population of the fainter group efficiently. The survey will also discover supersoft sources (at close distances in our Galaxy, or in surveys of other galaxies, see Chapter 20), ultracompact binaries and other unusual X-ray behavior in CVs (see Section II.d).

A key unresolved issue is determining the true space density of the CV population, especially at the Galactic Center, and assessing its contribution to the unresolved Galactic Ridge Emission [300,471]. There is an under-representation of Intermediate Polars (IPs) in the ROSAT Bright Survey compared to Polars [470]. That is because Polars are luminous, soft X-ray sources, whereas IPs have a higher intrinsic absorption and a harder X-ray spectrum. The higher effective area of *AXIS* and moderate timing capabilities would be crucial in detecting and confirming several IPs, as well as estimating the true space density and population of CVs.

**[Exposure time (ks):]** This science case explains an outcome of the Galactic Plane Survey, and needs no additional exposure time.

**Observing description:** This will make use of the Galactic Plane Survey, described elsewhere.

**[Joint Observations and synergies with other observatories in the 2030s:]** Classifying the identified systems will require the use of optical, infrared, and radio surveys, including Rubin/LSST, Roman, H$\alpha$ surveys (e.g. [104]), and the SKA.

**[Special Requirements:]** (e.g., Monitoring (Daily, Hourly, etc), TOO (<X hrs), TAMM) The science here does not specifically require TOOs, but, likely, unusual CV detections during the Galactic Plane survey (especially if carried out via fast-tiling mode, see section 21 will be used to trigger TOOs for various science cases (e.g. detections of dwarf nova outbursts, nova explosions, low states of intermediate polars).

*20. Studying the populations of novae and supersoft X-ray sources with AXIS*

**First Author:** Nazma Islam (NASA GSFC, UMBC, nislam@umbc.edu)
**Co-authors:** Craig Heinke (U. Alberta, heinke@ualberta.ca), Marina Orio (UW-Madison, orio@astro.wisc.edu), Şölen Balman (IU, solen.balman@istanbul.edu.tr), Chandreyee Maitra (MPE, cmaitra@mpe.mpg.de)

**Abstract:** .

Nova eruptions occur in interacting white dwarf (WD) binaries as a result of thermonuclear runaways due to shell burning in electron-degenerate material [621] in accreting white dwarfs (WDs). They are extremely luminous at all wavelengths ( see [116]). A "fireball" phase, a very luminous soft X-ray flash lasting for a few hours, is followed by optical brightening and by the emission of X-rays from shocked gas for several weeks. As the ejecta expand and become optically thin, the WD transitions to a very luminous supersoft X-ray state (SSS), characterized by an effective temperature of several hundred thousand K. The bolometric luminosity is close to the Eddington limit for 1 $M_\odot$ star ($10^{38}$ erg $s^{-1}$), for a time ranging from days to many months. *AXIS* will observe novae in X-rays all phases of the outburst *providing important physical observations not only on individual sources, but also on the Galactic and nearby galaxies' populations of novae and other classes of SSS*. During the "shock phase", many novae will be discovered as hard X-ray sources in the Galactic bulge - too heavily absorbed in the optical range for discovery, constraining the Galactic nova rate. *AXIS* will detect many novae in external galaxies in the SSS phase. A large number of luminous SSS in external galaxies, discovered with Chandra and XMM-Newton (e.g. [160,415,604]) have never been identified at other wavelengths. Some are thought to be massive WDs in which the mass accretion rate from the companion is very high, above $10^{-8}$ $M_\odot$ $yr^{-1}$, undergoing either non-ejecting thermonuclear flashes or "quiet" burning [244,628], but other SSS may also be black hole transients; it has also been suggested that the most luminous ones are intermediate mass black holes [333]. Monitoring surveys of nearby galaxies will constrain this population and identify the actual novae. The parameters of the nova SSS phase, such as duration and effective temperature, will be linked to the properties of the Galaxy (abundances, star formation rate, etc.) to constrain its evolution. *AXIS* will also throw light on the root cause of the short period modulations of the SSS, from $\simeq$35 s to half an hour [401]. Because these oscillations have variable amplitude, the Lomb-Scargle method underestimates the uncertainty in the derived periods. An apparent drift in the periodicity indicates QPOs, and appears to be due to stellar pulsations rather than to the spin period of the WD. Long ($\geq$ 10 ks), uninterrupted exposures with *AXIS* will reveal the nature of these modulations. Finally, *AXIS* is expected to enable the detection and spatial resolution of a few nova shells in X-rays, thereby guiding models of chemistry, hydrodynamics, and ejection mechanisms.

**Science:**

A nova is a transient astronomical event marked by the sudden emergence of a bright, seemingly "new" star that gradually dims over the course of weeks or months. All known novae occur in interacting binary systems containing a white dwarf (WD), accreting matter from a non-degenerate companion, which is either still close to the main sequence (with an orbital period of hours, i.e. a **cataclysmic variable**) or is already a red giant or AGB star (orbital period of years, namely **a symbiotic**). Novae may also occur in **massive binaries containing WDs**, with typical orbital periods of a few weeks. However, the optical amplitude in this case is small, making the outburst difficult to discover. As the accreted matter continues to accumulate on the WD surface, it becomes electron degenerate and ignites nuclear burning in a shell (usually CNO burning of hydrogen, rarely helium burning); this eventually becomes explosive in a thermonuclear runaway, causing the system to rapidly increase in luminosity (see [621] and references therein). A very fast wind follows, ejecting the bulk of accumulated matter and producing shocks in the ejecta or between the ejecta and the circumstellar environment. During the eruption, there is a sudden and dramatic increase in brightness. The return to a quiescent state occurs over timescales varying from a few weeks to over a year. Panel (a) of Figure 31 shows the schematic diagram of the various processes generating copious X-ray flux during the nova eruption. There is an initial "fireball" phase in which a rapid expansion of the outer layers occurs on the onset of the thermonuclear runaway and the start of H-burning. The expanding layers form the fireball; the star is extremely hot and emits far-UV and soft X-rays. This phase of a nova has been serendipitously observed so far in the Galactic nova YZ Reticuli, about 11 hours before the optical maximum [297], but the excellent soft X-ray response of *AXIS* may allow more discoveries of this early phase.

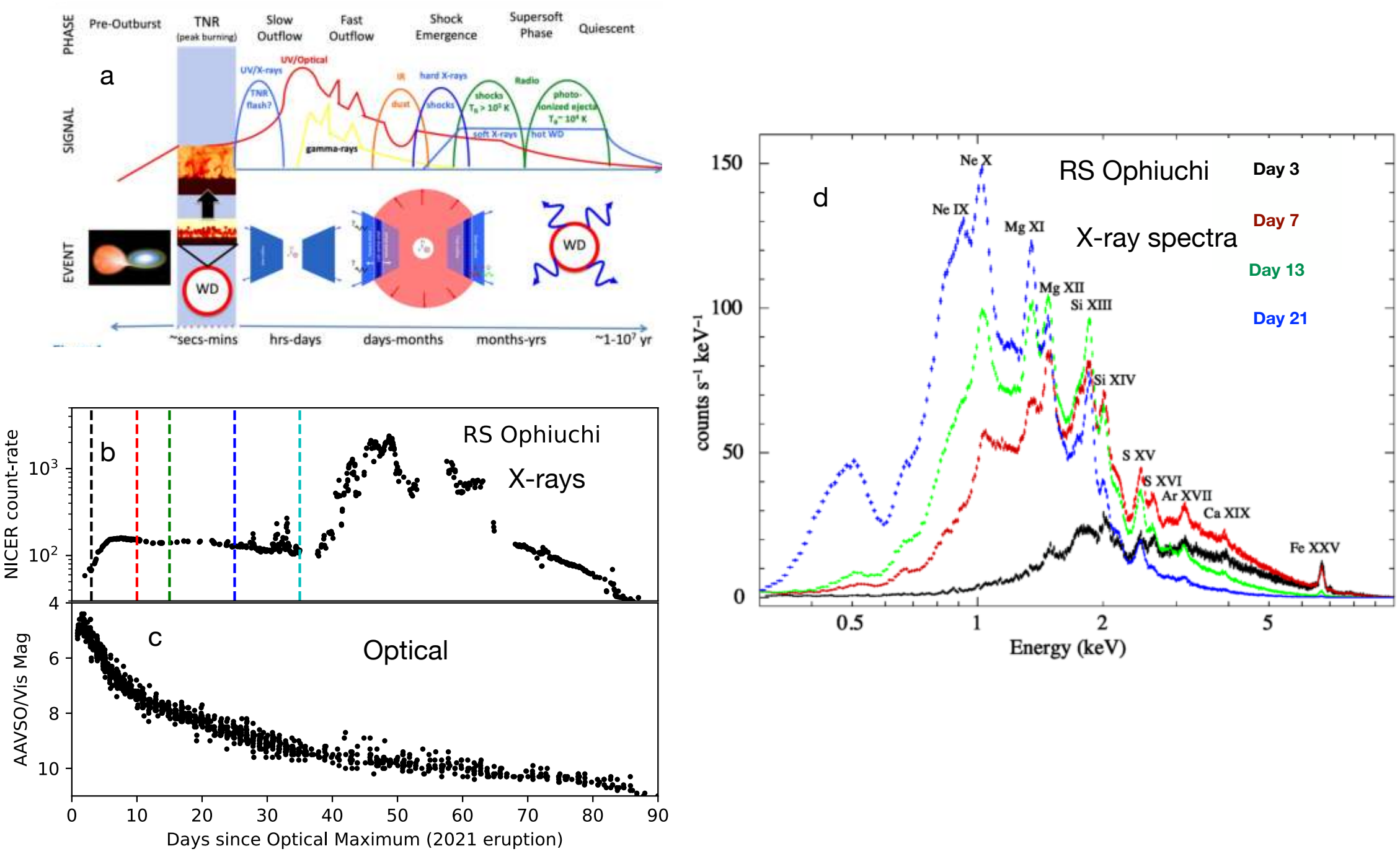

**Figure 31.** Evolution of a Nova Eruption. Panel (a) shows the schematic diagram of a nova, illustrating its various stages of emission (described in detail in the text). Taken from [116]. Panels (b) and (c) show the lightcurves of the recurrent nova RS Ophiuchi during its 2021 eruption in X-rays (monitored with NICER) and in optical (monitored by AAVSO in the Visual band), respectively. Panel (d) shows the evolution of the X-ray spectra of RS Ophiuchi, observed with NICER, over various stages of emission. It shows an evolution of the hot shocked gas and the emission lines from different medium Z elements.

The hard X-ray emission follows, due to hot, shocked gas. Panel (b) and (c) of Figure 31 show the light curves of the recurrent nova RS Ophiuchi during its 2021 eruption observed in X-rays with the Neutron Star Interior Composition Explorer (NICER) and the light curves provided by the American Association of the Variable Star Observers (AAVSO) in the visual band. The plots show that the optical light curves declined steadily after the optical maximum was reached within hours. However, the X-rays painted a different picture. The X-ray flux began to increase four days after the optical maximum and was dominated by hard emission from the hot, shocked gas. As shown in panel (d) of Figure 31, the X-ray spectra extracted from the NICER observations evolved as the eruption progressed. The X-ray spectra could be modeled with a non-equilibrium ionization collisional plasma or a multi-temperature collisional equilibrium plasma [265,418,423]. The evolution of the plasma temperature and the strength of the emission lines of medium Z elements such as Mg, Si, Ne, Fe [265,417,418] suggested an X-ray emitting shock driven by a blast wave moving into an asymmetric circumstellar environment [536].

Novae are truly multi-wavelength emitters [116]: the rise in the optical luminosity occurs as most of the accreted material is driven by an optically thick wind expanding and carrying the ejecta into the surrounding circumstellar environment, at velocities of the order of thousands of kilometers per second. When the ejecta expand and become optically thin to the soft X-rays, the extremely luminous SSS is predominant in most novae, due to still ongoing thermonuclear burning in a shell near the WD surface. Radio emission is attributed to both thermal and synchrotron emission; non-thermal emission is observed at earlier times in the eruption [118]. To date, 20 novae have also been observed as transient GeV gamma-ray sources with Fermi/LAT, and RS Oph was also detected at TeV energies [5]. To obtain a complete picture of the nova eruption, it is essential to identify where particle acceleration occurs and what energy processes drive emission across different wavelengths. Early detections of the X-rays from novae with *AXIS* will enable us to study the correlation between hard X-ray and gamma-ray emission and the simultaneous emission in radio, optical, and gamma-ray bands. The high spectral resolution and rapid response time of *AXIS* for alerts will be invaluable for studying evolving shock conditions and the properties of the X-ray-emitting plasma. We would like to emphasize in this respect that hard X-ray transients in the Galactic center region are thought to be, in large part, novae [389]. *X-ray monitoring of the bulge will also inform on the census and improve the current understanding of the nova population in the Galaxy.*

The SuperSoft State (SSS) of novae is due to atmospheric emission with an extremely hot, H-burning layer underneath [621] and lasts as long as the shell burning continues. The SSS state is often modeled by an atmospheric model, given by the publicly available grid of the Tübingen non-local thermodynamic equilibrium atmosphere, TMAP [479], and a velocity- and thermally broadened emission spectrum from the collisionally ionized diffuse gas `bvapec` model. However, due to the lower effective area of Swift-XRT, a blackbody model was used instead of an atmosphere model [423]. The larger effective area of *AXIS* at lower energies, compared to Swift-XRT, will be essential for accurately modeling the SSS state. However, SSS may also be non-novae WDs that are accreting and burning "quietly"). There are more than 450 SSS only in the Chandra-ACIS catalog of galaxies within 40 Mpc of [604]. The population of luminous SSS has, in fact, been observed not only in the Local Group, but also within $\simeq$10 Mpc, in directions of low column density [see 604]. In external galaxies, the transient SSS may be novae, while the probably rarer, persistent SSS may be instead non-ejecting, nuclear burning WDs [244,628], which are extremely interesting as possible SNe Ia progenitors. A careful monitoring cadence coupled with the large effective area of *AXIS* below 2 keV, compared to other X-ray telescopes, will be extremely useful to detect these SSS, with the moderate spectral resolution that should allow distinguishing black holes (expected to emit like a blackbody) from burning WDs, with an actual stellar atmosphere. More complete SSS population studies will definitely clarify which SSS in nearby galaxies observed with Chandra were, or may have been, novae in outburst, and how common non-ejecting burning is, possibly leading to thermonuclear supernovae from single-degenerate systems.

The SSS often shows dramatic X-ray variability, both aperiodic and periodic, after the nova outburst. The X-ray lightcurve of RS Ophiuchi, in panel b of Figure 31, shows several short X-ray flares in the SSS phase. These are likely due to different ionization states, or clumpiness and varying abundances in the red giant wind. *AXIS* monitoring will certainly provide more necessary information to understand this intriguing phenomenon fully. Moreover, periodic modulations of the order of minutes are also often measured in the SSS light curves [see 401]. The periods are as short as only $\sim$35 sec for RS Ophiuchi – [418] and KT Eridani – [439]). Periods of the order of several minutes seem to be usually due to the WD rotation period of magnetic WDs, even if it is not clear why they are measurable during nuclear burning. A puzzling issue has been the reason for an apparent, irregular drift of these periods (e.g., [401]), which seemed to point to stellar pulsations as the root cause of the modulations. However, the analysis done on V4743 Sgr and V1716 Sco [163] shows evidence that this can be an artifact of varying amplitude modulations measured in the short interrupted exposures of Swift and NICER, of the order of only 1000 s. Longer, uninterrupted exposures are necessary, and they will be possible with *AXIS*.

Finally, *AXIS* will discover several nova remnants in X-rays. Chandra has been useful to some extent, and a handful of remnants of old novae have been resolved and studied (see [46]). *AXIS*, both with pointed observations and with the GPS, is expected to perform much better in this regard, thanks to its superb effective area and sensitivity, and because its spatial resolution does not degrade at off-axis angles. An X-ray flux upper limit of $10^{-12}$ erg s$^{-1}$ cm$^{-2}$ [43] has been estimated for several nova shells; the exposure times necessary for detection are at least 50-100 ks with ACIS. Given the shorter exposure timescales for *AXIS* GPS, the *AXIS* limiting flux and spatial resolution allow for the recovery of several more remnants, revealing nova morphology and leading to improved models of hydrodynamics for nova explosions, evolution, and ejection mechanisms.

**Exposure times**: For Galactic novae, we request 10–20 ks exposures repeated daily or every few days from the onset of the eruption, aiming to cover all eruption phases, depending on the target, over a span of several weeks to a few months. Alerts will be communicated through an ATel or GCN, and we request prompt initiation of observations to capture the fleeting "fireball" phase. A monitoring survey of M31 and M33 should be conducted with 30 ks exposures repeated monthly for at least a year; a few selected nearby galaxies should be monitored with 60-80 ks long exposures. Exposure times of the nova remnants will be of the order of few tens of ks. To illustrate *AXIS*'s capabilities, we simulated a 1 ks exposure of RS Oph in its SSS state during its 2021 eruption. The spectra from Day 36, since the beginning of the eruption, are modeled using an atmospheric model plus two `bvapec` models with the parameters provided in [423]. The simulated spectra are shown in Figure 32. Observations of bright Galactic Novae are proposed in Continuous Clocking Mode or Windowed Mode, which would mitigate pile-up. For monitoring observations of Novae in nearby galaxies, it is expected that the count rates would be lower, and pile-up will not be a concern. Total Exposure time: $\sim$500 ks.

**Observing description**: The Galactic novae will only be TOOs; we propose repeated monitoring observations of M31, M33, and possibly NGC300, M81, M55, M101 - for all these galaxies, the cadence and exposure times should be coordinated also with the proposers of the other scientific targets in this white paper.

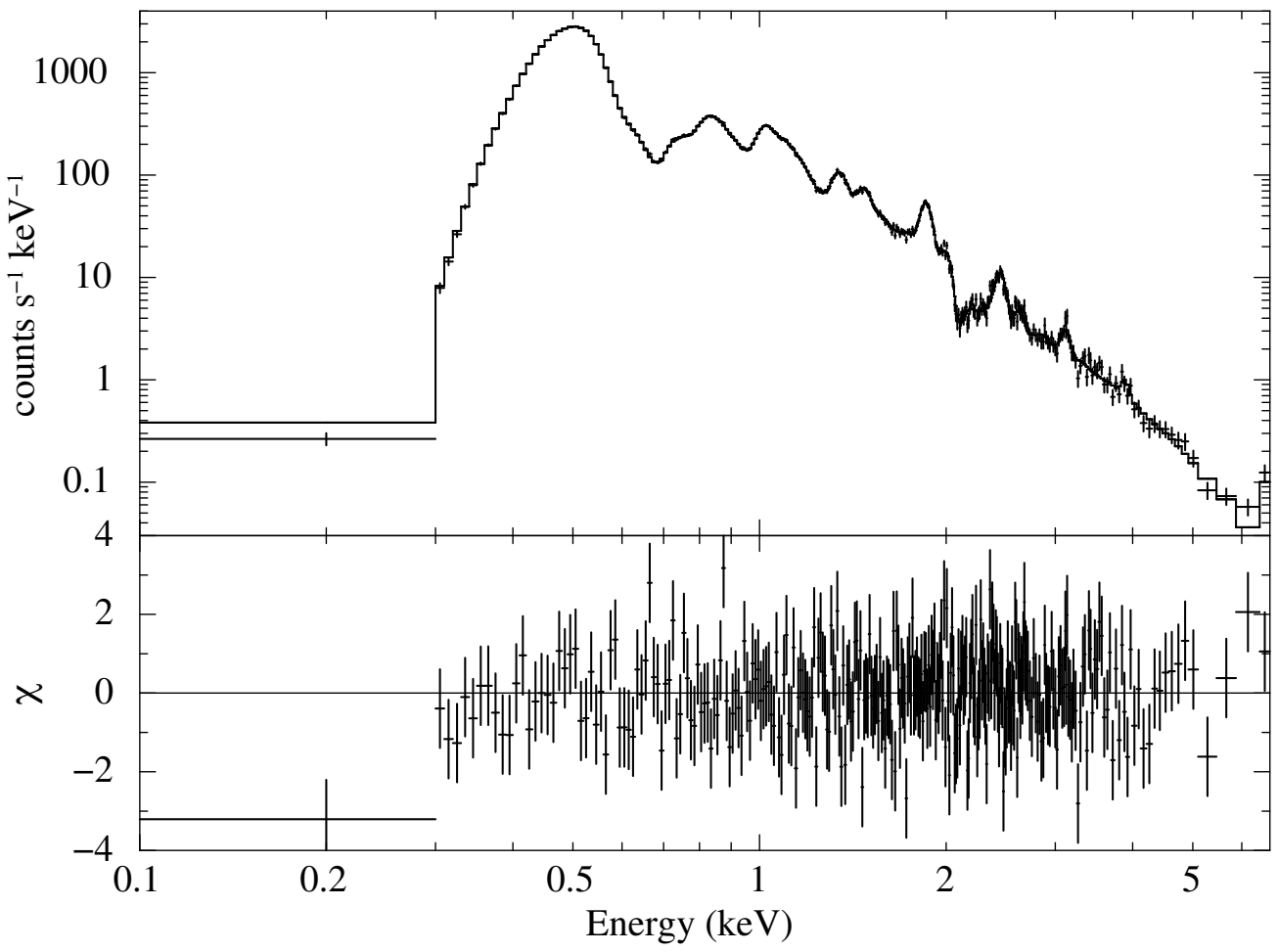

**Figure 32.** *AXIS* simulation of 1 ks exposure of RS Oph in the SSS. The model used is the atmospheric model [479] plus two BVAPEC models.

**Joint Observations and synergies with other observatories in the 2030s:** Novae are multi-wavelength sources, so there will be large synergy with other telescopes! We note, of course, that a vast number of new novae will be known thanks to Rubin/LSST.

# Part III
# Survey Science

## e. Galactic Plane Survey

*21. Fast tiling to find transient X-ray sources in AXIS Galactic (and nearby galaxy) surveys*

**First Author:** Craig Heinke (U. of Alberta, heinke@ualberta.ca)
**Co-authors:** (with affiliations)
**Abstract:**

*AXIS*'s design as an observatory with a wide field of view and fast reaction time enables it to deliver unprecedented efficiency in detecting X-ray transients in the Galactic Plane, Bulge, and Center. We suggest designing major *AXIS* surveys to maximize the detection of faint X-ray transients. *AXIS*'s sensitivity and accurate localization will enable significant leaps in our understanding of several transients.

The key idea is to spend the majority of the time on each field in short tiling observations, spread out over the 5-year planned lifetime. In so doing, we can dramatically increase our sensitivity to a wide range of X-ray transients fainter than the sensitivity of all-sky X-ray monitors. For our Galaxy, this would include magnetar outbursts, low-luminosity X-ray transients, cataclysmic variable outbursts, extreme stellar flares, protostar accretion episodes, and unusual objects such as long-period radio transients and transitional millisecond pulsars. For nearby galaxies, this would maximize our sensitivity to low-mass and high-mass X-ray binary transients.

**Science:**

Monitoring the X-ray sky for X-ray transient activity has produced a wide variety of exciting results. Our understanding of accretion onto (stellar-mass) black holes and neutron stars is largely driven by our identification of transient accretion outbursts through regular monitoring. We can use *AXIS* to effectively search for faint X-ray transients by tiling large areas with short exposures regularly. High-priority areas will accumulate long exposures (enabling the study of very faint sources) while allowing for the detection of brighter transients by monitoring on timescales comparable to their outburst length.

The key point is *AXIS*'s ability to rapidly slew and settle, within 30 seconds for adjacent fields. This enables pointed observations to go down to 200 s exposures (still reaching $F_X \sim 10^{-13}$ erg/s/cm$^2$ for moderate absorption, or $F_X \sim 6 \times 10^{-14}$ erg/s/cm$^2$ for low absorption studies of nearby galaxies) while maintaining $\sim$ 87% observing efficiency.

An example is the Galactic Plane survey, a core survey for *AXIS*. For 6 ks total exposure along the Galactic Plane, for example, divided into 230-second chunks as part of large tile mapping, this would enable up to 26 observations spread over 5 years, or an observation roughly once every other month of every Galactic Plane field for 5 years, with a sensitivity of $L_X \sim 10^{33}$ erg/s to sources at 8 kpc. Dividing the observations up like this could increase the yield of interesting transients by a factor up to 26, for objects with outbursts typically shorter than $\sim$ 2 months. (Alternatively, the total exposure time could be divided between one longer exposure, e.g., 3 ks, to detect periodicities in faint persistent sources, and short exposures with a less frequent cadence.) Detection of a transient would trigger a deeper (e.g., 2 ks) *AXIS* observation within a day, to better characterize its spectrum, variability, and position, and potentially can trigger other multi-wavelength facilities. For example, the Swift Galactic Plane survey found 151 sources with $F_X > 10^{-12}$ erg/s/cm$^2$, of which at least 11 varied by over an order of magnitude [410]. The Swift Galactic Plane survey was limited in its search for variability, as it was largely single pointings. Nevertheless, variable sources found by it included low-mass X-ray binaries, high-mass X-ray binaries, pulsars, and the young star cluster Westerlund 1[410].

The SMC is rich in high-mass Be X-ray binaries, which typically undergo ∼-month-long outbursts every orbit (often 100s of days). The first year of the S-CUBED Swift survey [284, , see Fig. 33] found many new transients above $5 \times 10^{36}$ erg/s (160 new, of 265 detected), largely Be XRBs, detecting pulse periods and in many cases orbital periods. Assuming 18 ks/tile (as suggested for the SMC survey, §aa above), we could perform 78 200-second exposures, >1 per month, for 5 years. Each exposure would reach $\sim 10^{36}$ erg/s, allowing near-completeness in studying the SMC's Be XRB population. A similar repeated tiling of the LMC would reveal a different mix of variables, less dominated by Be X-ray binaries but with a higher proportion of LMXBs.

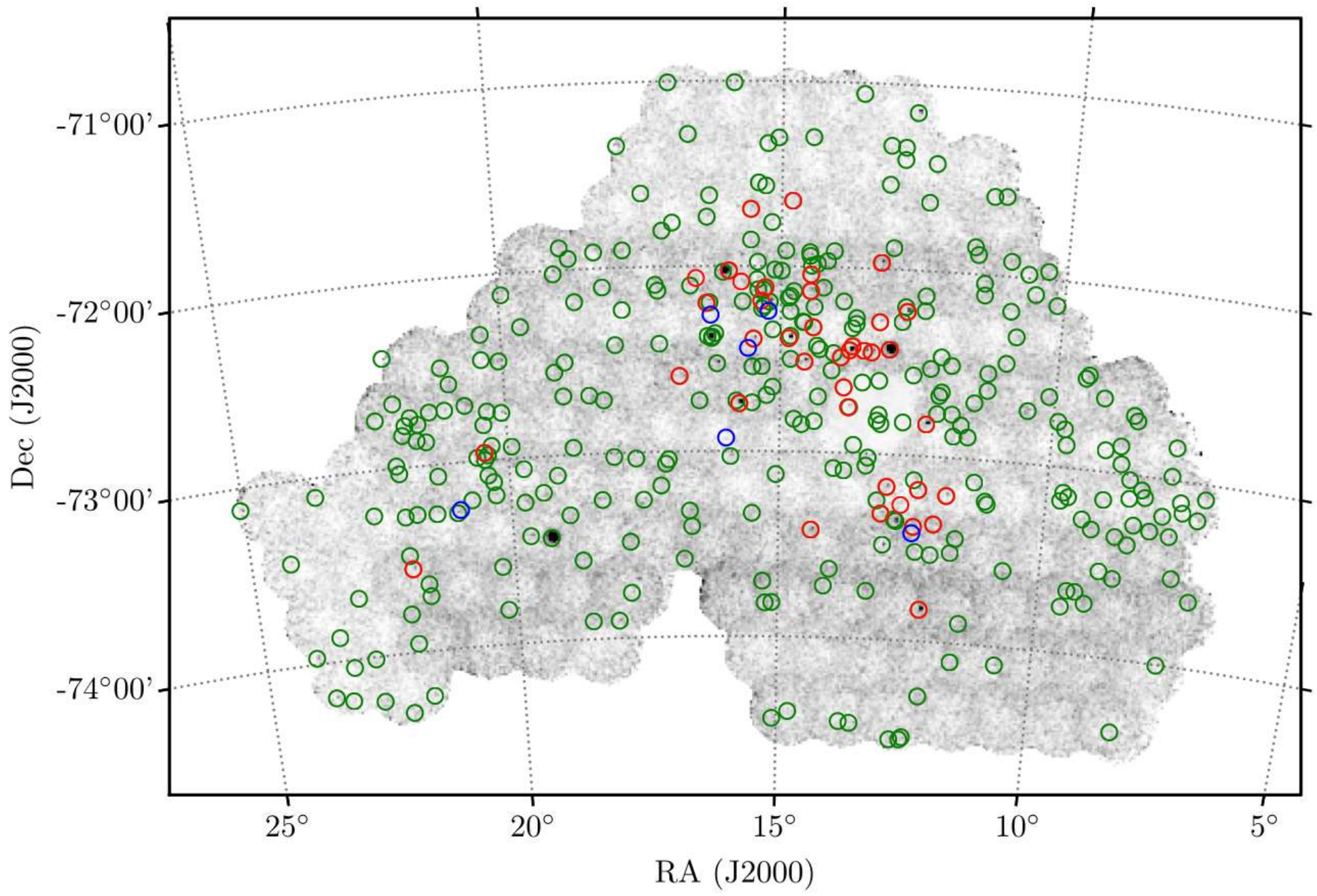

**Figure 33.** Combined S-CUBED map of the SMC by Kennea et al. [284]. Variations in the background indicate varying exposure times. Circles are detected X-ray sources, over half of which are transients.

M31 and M33, at slightly larger distances, provide the opportunity to observe large populations of transient X-ray binaries (see §m and y above). M31 has been a rich target for Chandra [616], XMM-Newton [**?** ], and NuSTAR [320,378] observations, as has M33 [617]. For instance, repeated X-ray observations of M31 have constrained the nature of X-ray binaries [320], supersoft X-ray source numbers [161], extended nuclear burning in novae [444], formation of transient X-ray binaries in globular clusters [436], and allowed identification and tracking of ULX transients in M31 [281,369].

For M31 and M33 (at 700 pc), 200-second observations can reach to $L_X = 3 \times 10^{36}$ erg/s, enabling us to catch X-ray binaries on the rise. Detections can then be followed up with deeper ToO observations to collect X-ray spectra, search for pulsations, etc.

The Galactic Bulge has proven to be an ideal place to study the characteristics of transient X-ray binaries. Using all-sky, or wide-field, X-ray monitors has allowed detailed studies of bright X-ray binaries. However, even the deepest surveys of the Bulge, including the RXTE PCA scans [557], the Integral bulge monitoring program [305], and the Swift bulge survey [37, see Fig. 34], have not been deep or frequent enough to determine the nature of lower-luminosity ($< 10^{36}$ erg/s) X-ray binary transients. We know (from frequent monitoring of, e.g., the Galactic Center) that some low-luminosity transient outbursts are produced by transients that also produce brighter outbursts [e.g. 150]. However, there also seems to be a population of faint transients that have never reached "normal" X-ray binary luminosities [e.g. 390].

Such intrinsically fainter objects could be much older LMXBs, with lower-mass donor stars, or suffer interference with their mass transfer, or some other phenomenon [232].

Monitoring of the Galactic Bulge region at 200 seconds/pointing will reach $L_X \sim 10^{33}$ erg/s (at 8 kpc). This ensures we could identify essentially any outbursts from faint X-ray binaries. If we were to survey a 5 by 5 square degree region of the Galactic Bulge monthly for 5 years, this would amount to roughly 3 Ms of total time.

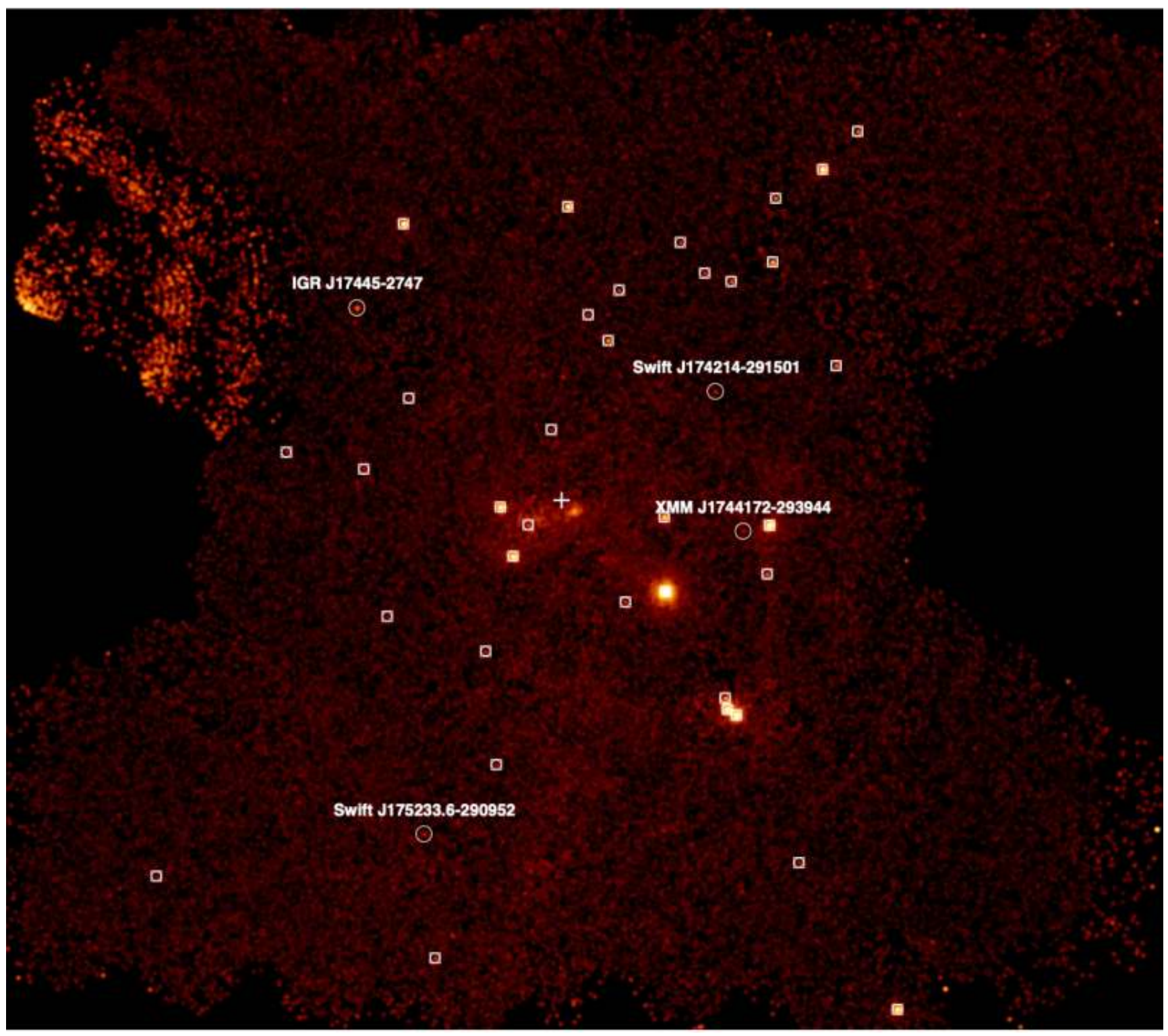

**Figure 34.** Combined image of the first year of the Swift Bulge Survey (4 by 4 degrees), with some detected sources indicated by squares, or circles with their names.

To make maximum use of this sensitivity to transients, we'll need to develop software to process new *AXIS* survey data through pipelines and identify new transients above certain thresholds. We can pass detections through machine learning programs to identify the likely nature of the source (e.g. stellar coronal activity–likely to be the majority; X-ray binaries; magnetars; cataclysmic variables), working with multi-wavelength datasets (e.g. relating to the most recent Rubin/LSST surveys of the region to confidently identify cataclysmic variable outbursts).

We will then design trigger programs to observe different source classes, accompanied by appropriate peer-reviewed follow-up programs. These may include further *AXIS* pointings, optical/IR observations (e.g., using Gemini's SCORPIO simultaneous optical/IR imaging in 8 bands, followed by spectroscopy, for X-ray binaries), and/or radio (e.g., SKA follow-up for brightening LMXBs, or magnetars).

**[Exposure time (ks):]** This science case modifies existing (Galactic Plane) and proposed (SMC, LMC, M31, M33) survey plans for *AXIS*.

We also recommend a deep survey of the Galactic Bulge, which is supported by other science cases (§p, s, af, ak). We suggest a total exposure time for the Galactic Bulge survey of 3 Ms.

**Observing description:** The suggested Galactic Bulge survey would cover 5 square degrees (roughly symmetrically, likely avoiding scattered light around the bright LMXB GX 1+1 to Galactic NE) around the Galactic Center. Monthly tilings of this region would provide 200 seconds of exposure time per pointing, totaling 60 ks of time each month (excluding two months per year when the Bulge is near the Sun in the sky).

In addition, we anticipate that the director's discretionary time will likely be needed to follow up discoveries, averaging perhaps 3 ks per month to follow 2 targets at 1-2 ks each, along with additional time on optical/IR and radio telescopes.

We assume the FOV of 24 arcminutes diameter, angular resolution of 2". The key point is that we aim at a slew plus settle time for adjacent *AXIS* pointings of $<$30 seconds, as obtained for Swift and suggested as possible by Northrop Grumman.

**[Joint Observations and synergies with other observatories in the 2030s:]**

X-ray transient detection with AXIS will require multi-wavelength observations to produce the highest-impact science. The nature of X-ray binaries and cataclysmic variables can be verified through optical and infrared observations. This may rely on already-existing surveys (e.g. Rubin/LSST, Roman) or on follow-up observations. Initial follow-up could use 8-meter class facilities (e.g. Gemini, Keck, VLT), and particularly exciting discoveries could use thirty-meter class telescopes for follow-up. Some objects (e.g. magnetars, jets, transitional millisecond pulsars) will require radio follow-up, with the SKA, Jansky VLA, Green Bank Telescope, etc. Some discoveries would benefit immensely from high-spectral-resolution NewAthena observations.

**[Special Requirements:]** Effectively using *AXIS* as a transient monitor will require maintaining the ability to rapidly slew and settle, with a target time of less than 30 seconds for adjacent fields, as currently indicated by Northrop Grumman. Performing efficient tiling will require preparing software to perform this tiling, as has already been done for Swift [284]. We note that this tiling capability will also be essential for tiling observations of distant extragalactic multi-messenger transients, e.g., neutrino or gravitational wave signals.

Making efficient use of the detected transients will require fast data processing and transient detection algorithms. Transient detection should then permit triggering of follow-up observations, with *AXIS* itself and of multi-wavelength facilities including Gemini (e.g., with SCORPIO, Robberto et al. 495), the Jansky VLA, the Square Kilometre Array, etc.

*22. Discovery of XRBs and their optical counterparts in the AXIS Galactic Plane Survey and beyond*

**First Author:** Avi Shporer (MIT, shporer@mit.edu)
**Co-authors:** Lee Townsend (Southern African Large Telescope and South African Astronomical Observatory, South Africa)

**Abstract:** X-ray binaries (XRBs) are vital for studying extreme gravity, accretion physics, and compact object evolution. XRBs also provide insights into stellar evolution, relativistic jets, and gravitational wave sources, making them crucial in high-energy astrophysics. XRBs are detected through X-ray observations, which reveal their distinct spectral features. Thanks to its wide field and angular resolution *AXIS* is expected to detect new XRBs, in the *AXIS* Galactic Plane Survey. The search for XRBs will be complemented by new deep optical and IR surveys that are expected to be carried out in the next few years by future observatories, including the Roman Galactic Plane Survey. The latter will be used to detect the XRB optical counterpart, or donor. Of special interest are eclipsing XRBs, where the optical counterpart eclipses the compact object and/or the accretion disk, as they provide the opportunity to gain more detailed information about the system. Identifying X-ray eclipses will be achieved by developing and applying periodic variability detection methods to the X-ray data, which will enable the detection of other X-ray variability processes on time scales of days and weeks. Combining *AXIS* data with new deep optical surveys can also be done for other possible *AXIS* surveys (e.g., M33, M31).

**Science:** X-ray binaries (XRBs) consist of a compact object, either a neutron star or a black hole, accreting material from a companion star — a donor, also referred to as an optical counterpart. where the accretion process generates intense X-ray emission. Detecting and studying their optical counterparts is essential for determining key system parameters such as the mass of the compact object, the nature of the donor star, and the orbital characteristics. Therefore, the search for X-ray binary optical counterparts is a crucial aspect of high-energy astrophysics.

Optical counterparts of XRBs are typically identified through deep imaging and spectroscopic surveys, in the optical and near-infrared (IR), often guided by precise X-ray localization. The optical/near-IR emission in these systems arises from a combination of factors: intrinsic luminosity of the companion star, reprocessed X-ray radiation heating the outer layers of the accretion disk or donor star, and Doppler effects from high-velocity interactions. Low-mass X-ray binaries (LMXBs; [181]) typically have faint, evolved donor stars and are best observed during quiescence, when the accretion disk is dim. High-mass X-ray binaries (HMXBs; [334]), on the other hand, have massive, luminous companions such as O- or B-type stars, making them more easily detectable in optical and near-IR wavelengths.

Eclipsing XRBs are systems in which the optical counterpart eclipses the compact object and/or the accretion disk. While rare, such systems are especially valuable, as the eclipses provide additional details about the system [445,447].

Identifying optical counterparts allows us to measure their radial velocities (RVs) and infer mass functions, ultimately leading to constraints on the nature of the compact object – especially in the search for stellar black hole candidates [e.g., 65,419]. Figure 35 presents an example of this approach.

The *AXIS* Galactic Plane Survey (GPS) will reach a sensitivity of about $10^{-15}$ erg cm$^{-2}$ s$^{-1}$ at 0.5–10 keV for a 1 deg $\times$ 50 deg field, and $10^{-16}$ erg cm$^{-2}$ s$^{-1}$ at 0.5–10 keV for a 4 deg $\times$ 6 deg field covering the Galactic Bulge. At a distance of about 3 kpc it will detect objects down to $L_X = 10^{30}$ erg/s. The increased sensitivity and wider field compared to previous X-ray observations will lead to the detection of many new X-ray sources, by an order of magnitude or more than previously known, including newly detected XRBs. *AXIS* data will also enable a time series study of the newly detected and currently known XRBs. We will use these data to search for X-ray variability, to detect eclipses and variability induced by other processes (e.g., accretion rate fluctuations, disk instability, X-ray pulsations, quasi-periodic oscillations).

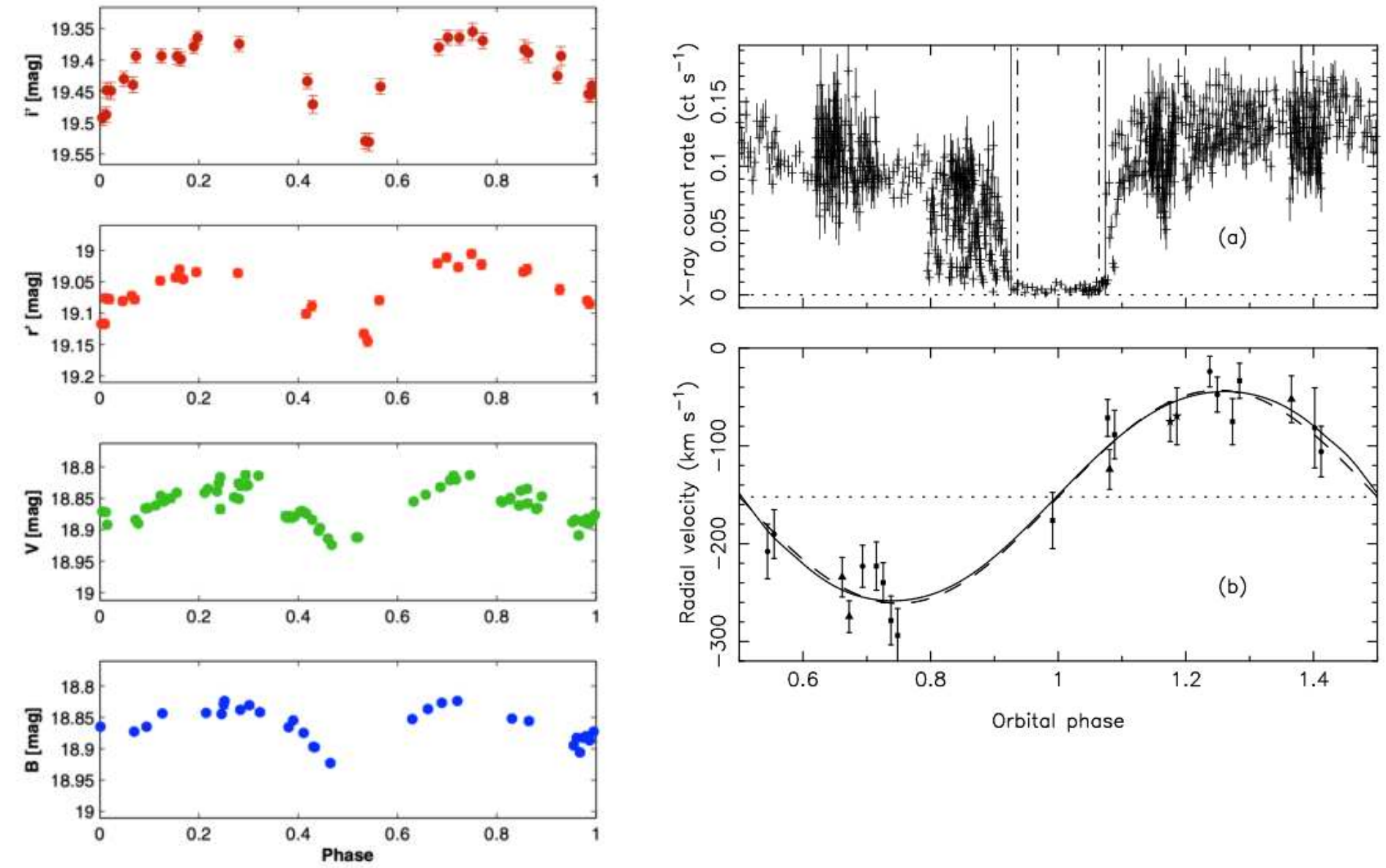

**Figure 35.** M33 X-7 optical, X-ray, and radial velocity (RV) time series. **Left:** Optical variability, in the (top to bottom) $i'$, $r'$, $V$ and $B$ bands. The multi-band light curve shows variability originating primarily from the donor's ellipsoidal distortion, at the same period and phase as the X-ray eclipse, thereby identifying this HMXB optical counterpart (Figure from [533]). **Right:** X-ray eclipse (top), and phase folded RVs of the donor (bottom), providing the mass ratio between the two objects in the system (Figure from [419]). The latter, combined with an estimate of the donor's mass based on its spectrum, lead to the mass measurement of the compact object of $15.65 \pm 1.45\ M_\odot$ [419], making it a stellar mass black hole.

We will match the positions of the XRBs observed by *AXIS* GPS with those of targets in optical/near-IR surveys, to identify XRB optical counterparts. The optical data will be from surveys that have already concluded (e.g., OGLE) or are now ongoing, and from new surveys that will be done over the next several years. One of the latter is the Roman Galactic Plane Survey, to be carried out by NASA's Nancy Grace Roman Space Telescope and is expected to cover about 1,000 deg$^2$ with depths of 23–25.5 mag at 1.06–2.13 microns [512]. Once an optical counterpart is identified with high probability, it will be studied using the optical survey data itself or by collecting new data on specific targets (commonly referred to as follow-up observations), including time-series photometric data and spectroscopy.

We will also apply variability search to the optical/near-IR data. One source of photometric variability is the tidal distortion of the optical counterpart in short-period systems, leading to detectable ellipsoidal variability, as shown in Figure 35 left panel. This variability is also expected in non-eclipsing systems.

This approach of searching for XRB optical counterparts and studying the variability of both X-ray data and photometric data can also be applied to other *AXIS* surveys, such as those of nearby galaxies.

**[Exposure time (ks):]** Galactic Plane Survey

**Observing description:** This science case does not require dedicated *AXIS* observations, but instead focuses on enhancing the *AXIS* Galactic Plane Survey scientific yield, and potentially that of other *AXIS* surveys.

**[Joint Observations and synergies with other observatories in the 2030s:** Roman Space Telescope, Rubin Observatory

## f. Extragalactic (nearby galaxies)

### *23. Intermediate mass black holes in extragalactic star clusters*

**First Author:** Kristen Dage, Curtin University (kristen.dage@curtin.edu.au)
**Co-authors:** Kwangmin Oh (MSU), Stephen Zepf (MSU)
**Abstract:** Intermediate mass black holes (IMBHs) are a crucial missing component in our understanding of how supermassive black holes form. NBODY simulations suggest an intimate relationship between IMBHs and star clusters; however, current astronomical facilities have not been successful in identifying a population of unambiguous IMBH candidates in star clusters. With the combined power of Rubin Observatory, AXIS, and the SKA or ngVLA, it will be possible to place comprehensive constraints on the elusive population of IMBHs.
**Science:** In the last year alone, new results from the James Webb Space Telescope have shown that supermassive black holes exist very early on in the Universe, even within just one-tenth of the entire age of the Universe. These observations show that massive black holes must form within about a few million years from the Big Bang, through the seed of an intermediate-mass black hole (IMBHs; $10^2 < M < 10^6$). A leading theory is that these IMBH seeds can be formed quickly (within a few million years) because of dynamical interaction in star clusters. Thought to be the building blocks of supermassive black holes, IMBHs are thus likely to provide crucial clues about the formation and evolution of their supermassive counterparts.

Extreme mass ratio inspirals from IMBHs will be detected by the Laser Interferometer Space Antenna, out to Redshift 2 [271]. However, we know very little about IMBHs from an observational standpoint. They are a major open question in dynamical modeling of star clusters: numerical simulations of dense star clusters predict two primary pathways through which IMBHs could form. The first is via runaway stellar mergers (e.g., [463]), on timescales of less than 10 Myr. The second scenario grows IMBHs slowly through repeated mergers of smaller black holes (e.g., [372]).

Recent, state-of-the-art simulations by [615] using the MOCCA Monte Carlo code provide explicit predictions for the formation of IMBHs in globular clusters (GCs). Their models show that IMBHs can form either through runaway stellar mergers within a few million years or through successive black hole mergers over gigayear timescales. When accreting from stellar companions, these IMBHs can produce ultraluminous X-ray sources (ULXs) with luminosities reaching nearly $10^{40}$erg/s. The simulations predict that non-tidally filling GCs host significantly larger ULX populations due to the prolonged retention of compact objects and enhanced dynamical interactions. Additionally, the presence of an IMBH influences cluster dynamics by altering the retention and ejection of stellar remnants, leading to the formation of field ULXs that originated in GCs.

Such bright X-ray sources, particularly ULXs exceeding $\sim 10^{39}$erg/s, have so far only been observed in extragalactic globular clusters, with detections in systems such as NGC 4472, NGC 1399, and NGC 4649. These findings suggest that GC-hosted IMBHs may be responsible for a fraction of the bright ULXs observed in the field. Simulations indicate that IMBH-powered ULXs form continuously over a cluster's lifetime, whereas stellar-mass black hole ULXs are more common in the early evolutionary stages of a cluster. The results also support the idea that a significant fraction of field ULXs originate from GC escapers, which retain their dynamical signatures despite being ejected. Future high-resolution X-ray missions, such as *AXIS*, will be crucial for identifying and characterizing these sources, thereby bridging the gap between theoretical predictions and observational constraints to refine our understanding of IMBH formation and ULX evolution in dense stellar environments.

Oh et al. [411] argued that an ultra-compact X-ray binary (UCXB) with a helium white dwarf (He WD) donor can explain a ULX in a GC of NGC 1399, characterized by its strong [NII] and [OIII] emission lines without detectable hydrogen. This system, referred to as GCU7, has exhibited a stable X-ray luminosity ($\sim 10^{39}$erg/s) for over two decades, suggesting a persistent accretion state [17,263]. The absence of hydrogen and the presence of high-ionization nebular lines indicate a hydrogen-poor accretion flow, consistent with a He WD donor. These features make GCU7 one of the most compelling extragalactic UCXB candidates, and a new angle to study these gravitational wave-emitting systems. However, confirming the nature of such systems requires precise spatial resolution and *deep X-ray sensitivity*, particularly in crowded globular cluster environments.

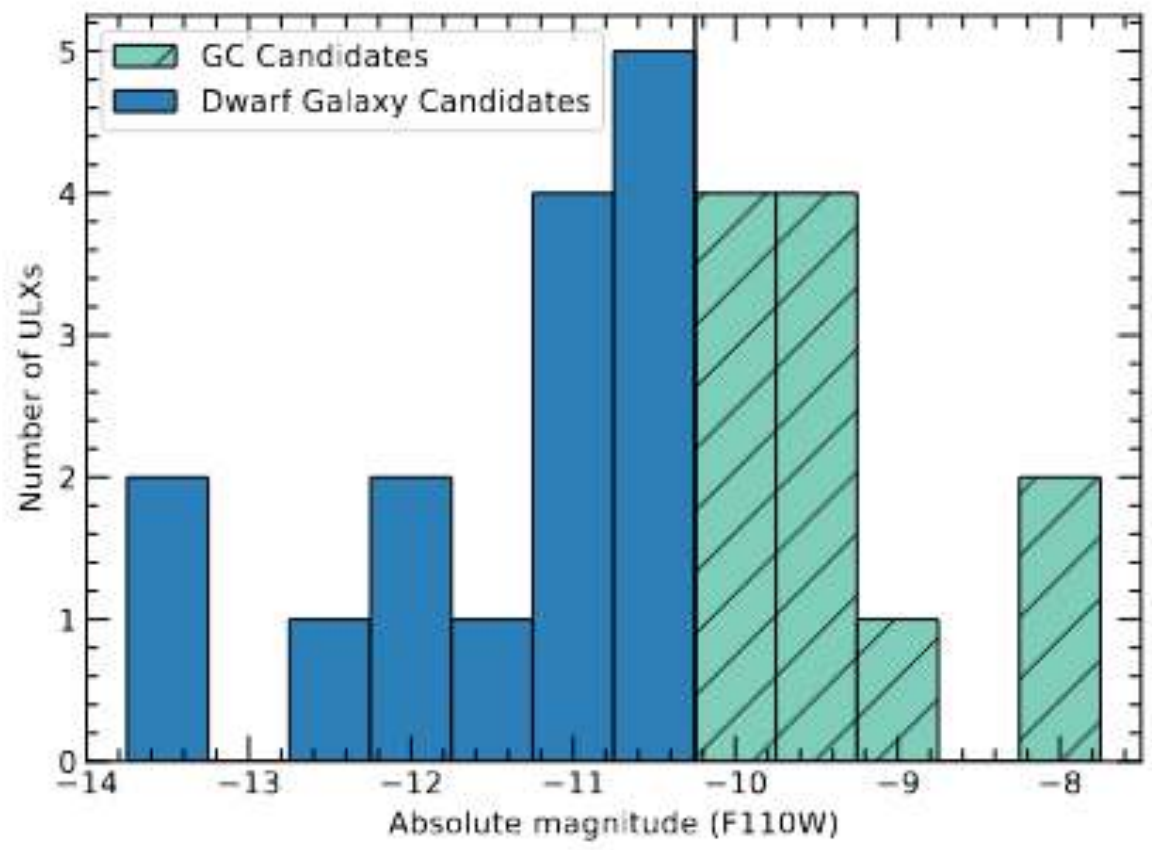

**Figure 36.** ULXs detected in distance (70 Mpc) dwarf galaxies and star clusters. Figure from [566].

The superb spatial resolution of *AXIS* is essential for discovering and characterizing these unique sources, as precise imaging is required to resolve their optical counterparts in dense stellar environments. In systems like GCU7, where distinguishing between an active binary and surrounding cluster members is challenging, *AXIS* will provide the necessary X-ray sensitivity to separate the ULX from contaminating field sources. Furthermore, by enabling high-resolution X-ray observations of extragalactic globular clusters, *AXIS* will help identify and classify dynamically formed compact binaries, shedding light on their formation mechanisms and long-term evolution. While detailed multi-wavelength follow-up will still be needed to confirm the donor type and accretion state, ***AXIS* will be the key mission enabling the discovery of new UCXBs in GCs**, bridging the gap between theoretical predictions and observed compact binaries in dense stellar environments.

The Vera C. Rubin Observatory's Legacy Survey of Space and Time (Rubin/LSST; first light in April 2025) is poised to uncover swathes of new extragalactic globular clusters, and is predicted to detect several million GCs over the lifetime of the ten-year survey, out to z=0.05 [586]. By selecting new GC populations discovered by Rubin for X-ray follow up with *AXIS*, with further characterization by SKA or ngVLA, we have an ideal population to hunt for IMBHs in star clusters, which are **not clearly detected, despite the clear theoretical predictions** [137].

Because GCs are older populations of stars, they are the ideal environment to probe low mass X-ray binary (LMXB) formation and evolution. Our proposed program would therefore also benefit population studies endeavouring to understand the origins of LMXBs. Given that simulations, e.g. by [615], suggest that GCs may significantly contribute to field populations of X-ray binaries, it is therefore essential to understand these key populations of LMXBs in GCs to understand their formation mechanisms. **[Exposure time (ks):]** 60 ks per field

**Observing description:** Based on previous studies of LMXBs in extragalactic star clusters, even out to distances of 70 Mpc, 60 ks per field will be able to detect the brightest X-ray sources—i.e., the most probable

IMBH candidates [566], see Figure 36. For closer systems at 10 or 20 Mpc, a 60-ks exposure will guarantee the detection and characterization of a large population of field and cluster XRBs.
**[Joint Observations and synergies with other observatories in the 2030s:** Rubin Observatory, LISA, ngVLA, SKA.
**[Special Requirements:]** None

*24. An AXIS survey of the Magellanic Clouds*

**First Author:** C. Maitra (MPE, cmaitra@mpe.mpg.de), G. Vasilopoulos (NKUA, gevas@phys.uoa.gr)
**Co-authors:** M. Sasaki (FAU), T. Woods (U. Manitoba), Yosi Gelfand (NYUAD), Jun Yang, Breanna Binder, G. Israel, Labani Mallick (U. of Manitoba/CITA); Samar Safi-Harb (U. Manitoba), Martin Mayer (FAU), F. Zangrandi (FAU), S. Points (NSF NOIRLab/CTIO), Lee Townsend (Southern African Large Telescope and South African Astronomical Observatory, South Africa)
**Abstract:** The Large and Small Magellanic Clouds (LMC & SMC) are the nearest gas-rich irregular dwarf galaxies, where recent star formation has been driven by tidal interactions between the two galaxies and the Milky Way. At distances of just 50–60 kpc, with minimal Galactic foreground extinction and absorption, they provide an ideal laboratory for studying X-ray populations that are often inaccessible in the Milky Way. Understanding the diverse emission components of these nearby galaxies is essential for interpreting the unresolved X-ray emission in more distant systems. Covering nearly 200 square degrees of the sky, surveys of the MCs have always been challenging. Optical and radio surveys of the ISM in the MCS have revealed interstellar structures of sizes ranging from a few parsecs to over 1000 parsecs. X-ray surveys by XMM-Newton and eROSITA have provided valuable insights into bright point sources (limiting $L_x \sim 10^{34}$erg/s) and supernova remnants (SNRs, limiting $L_x \sim 10^{35}$erg/s). *AXIS*, with its unique combination of high angular resolution, large effective area, and low background, would significantly enhance our understanding of the X-ray universe in these galaxies. With short 5 ks exposures *AXIS* could detect sources at $10^{34}$erg/s (or lower depending on background), while for regions of particular interest, larger observing times can be obtained with stacked observations. An *AXIS* survey of the MCs would enable the detection of faint X-ray binaries and their luminosity function, reveal supersoft sources (cataclysmic variables or Be/white dwarf binaries) that are often lost in diffuse emission (especially in dense star-forming regions like the Tarantula Nebula), and provide unprecedented spatial detail on SNRs and superbubbles, shedding light on stellar winds and supernova feedback. Additionally, *AXIS* could enable the long-sought identification of the compact remnant of SN 1987A in X-rays. The study of the X-ray population of the MCs could also build upon synergies with multi-wavelength surveys, such as the ongoing optical spectroscopic survey by SDSS-V LVM and the upcoming optical and UV surveys such as UVEX and the Vera Rubin Observatory. These discoveries would build upon the legacy of previous X-ray missions, offering new insights into the life cycles of stars and the evolution of galaxies.
**Science:** The Magellanic Clouds are our closest star-forming galaxies with global star formation rates of 0.02–0.05 $M_\odot$ $yr^{-1}$ and 0.07–0.16 $M_\odot$ $yr^{-1}$ for the Small Magellanic Cloud (SMC) and the Large Magellanic Cloud (LMC), respectively [180]. This makes them a unique laboratory to study the population of high-energy sources. At a distance of 50 kpc for the LMC and 60 kpc for the SMC, sources down to a few $10^{33}$ erg $s^{-1}$ can be observed with *XMM-Newton* in relatively short exposures. The SMC is known to host a large population of Be/X-ray binaries (BeXRBs) that is associated with high star formation activity 25-40 Myr ago [21,215]. The high-mass X-ray binary (HMXB) population in the LMC is associated with a star formation period at an earlier epoch and at a lower HMXB formation efficiency and shows a higher abundance of supergiant systems than the SMC. An important asset for studying the source population of the entire LMC was the launch of eROSITA, the X-ray instrument on board the Spektrum-Roentgen-Gamma (SRG) mission [469]. The seven telescopes ($0.2-8$ keV energy band) surveyed the complete sky four times over two years. A fifth survey was conducted partially, covering the northern part of the LMC. The first survey (eRASS1) started in Dec. 2019 while eRASS5 was halted in Feb. 2022 until further notice. *eROSITA* scanned the sky in great circles which cross at the ecliptic poles. Given the 4 hr scanning period, a source typically passes 6 times through the $1^o$ (diameter) field of view of the telescopes within one day, repeating this every half year. The individual snapshots last $\sim$40 s, yielding typically 240 s exposure per survey for most of the sky. This led to the detection and discovery of point-like sources up to a limiting $L_x \sim 10^{34}$erg/s, Kaltenbrunner at al. 025, in prep) and supernova remnants [limiting $L_x \sim 10^{35}$erg/s 640]. The SMC was scanned less frequently and accumulated a total exposure of 1-2 ks during the eROSITA surveys.

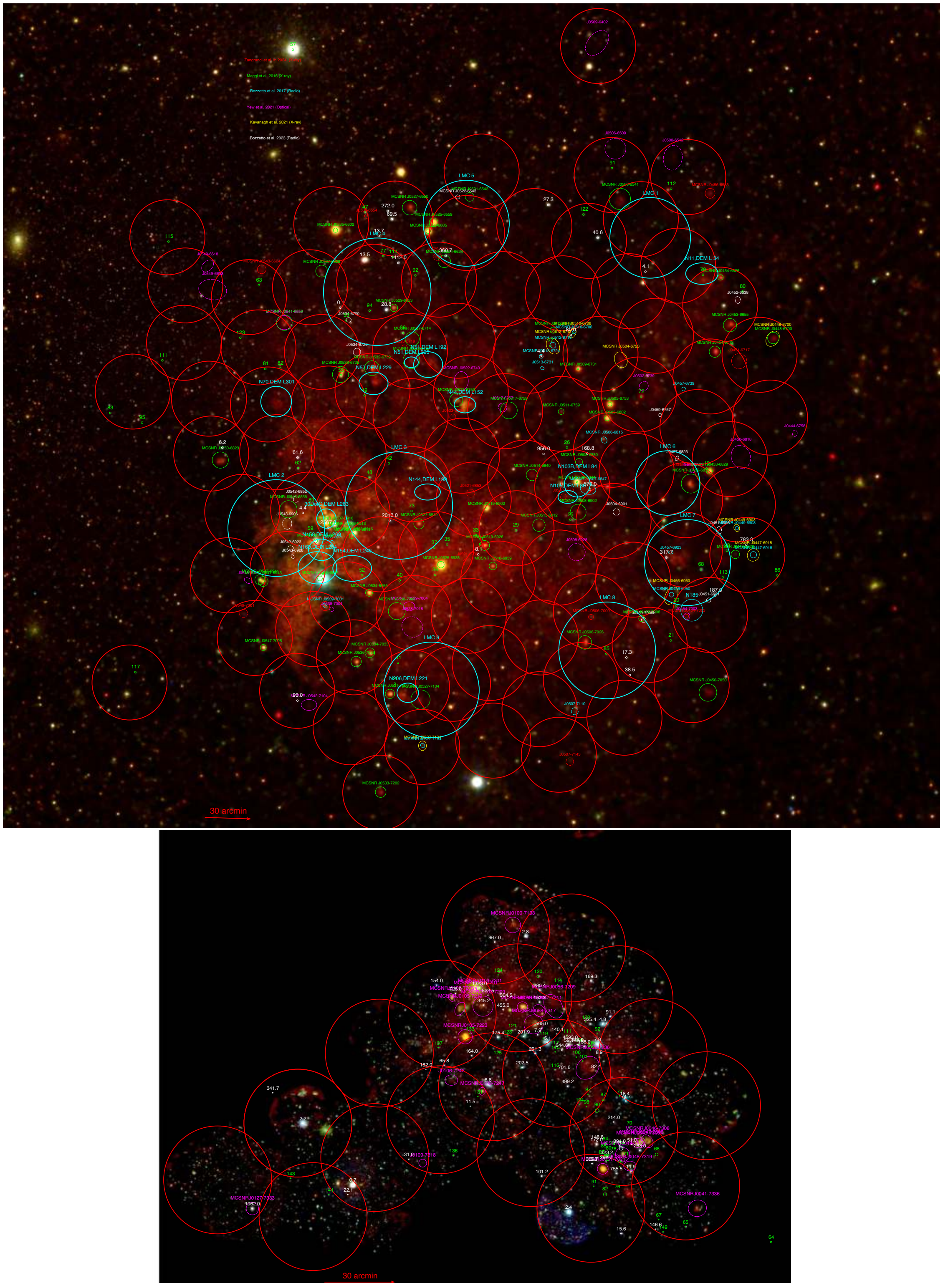

**Figure 37.** Top: eRASS1 RGB (R: 0.2–1.0 keV, G: 1.0–2.0 keV, B: 2.0–4.5 keV) image of the LMC showing the proposed *AXIS* pointings (in red), centered on the main body of the LMC and supergiant shells and superbubbles (cyan), and regions with known HMXBs (small white and green circles denoting pulsars and HMXB candidates respectively) and SNRs. The SNRs are also marked in colors denoted in the figure. Bottom: XMM Newton RGB (R: 0.2–1.0 keV, G: 1.0–2.0 keV, B: 2.0–4.5 keV) of the SMC showing the proposed *AXIS* pointings (in red), centred on regions with known HMXBs (small white and green circles denoting pulsars and HMXB candidates respectively) and SNRs (magenta).

The MC are an ideal laboratory to study a population of point as well as extended sources and have been associated with discoveries of many unique objects such as a He-burning white dwarf [207], a Be white dwarf nova [355] and HMXBs associated with their parent supernova remnants [349,350]. The main class of objects associated with the MCs are described below.

**The most numerous hard X-ray sources in the MCs are HMXBs.**. HMXBs are divided into supergiant X-ray binaries (SgXRBs) and Be/X-ray binaries (BeXRBs). In BeXRBs, the Be star ejects matter in the equatorial plane, leading to the formation of a decretion disc. Type-I outbursts occur usually on a periodic basis as the NS passes through the disc ($L_{\rm x} \gtrsim 10^{36}$erg/s; lasting typically 20% of the orbit). Alternatively, disc instabilities can provide extra material causing brighter type-II outbursts ($L_{\rm x} \gtrsim 10^{37}$ erg s$^{-1}$; lasting several orbits). To date ~130 BeXRBs and 1 SgXRB are known in the SMC. Including recent discoveries, for 70 of them the spin period of the NS is known [215], and for more than 50 systems orbital periodicities have been established [572]. Most systems are located in the Bar of the SMC. Some are found in the Eastern Wing and the Magellanic Bridge (MB) towards the LMC, i.e., regions with different star-formation (SF) history and lower metallicity. Within the LMC, 28 HMXB pulsars are known (mostly BeXRBs with only 2 SgXRB) and ~50 additional candidate HMXBs. **Supersoft X-ray Sources** The general scenario for an SSS is thermonuclear burning on the surface of an accreting white dwarf (WD), which can be stable for certain accretion rates. Classical examples are Cal 83 and Cal 87 in the LMC, which are close binaries with a WD accreting at high rates and stable hydrogen burning [588]. SSSs are also associated with cataclysmic variables (CVs), symbiotic stars, and post-outburst optical novae. Super-soft X-ray emission at lower luminosity can originate from some CVs, cooling neutron stars (NSs), PG 1159 stars (hot cooling isolated WDs), planetary nebulae, and as the elusive class of Be+WD X-ray binaries, opening up a new window into finding these systems. New candidates for Be+WD systems were found in the LMC [218] and SMC [123,355]. Due to the low foreground absorption in the direction of the LMC, very interesting SSS have also been discovered in the foreground of the MCs, e.g. double degenerate systems [351] and intermediate polars [348]. **Supernova Remnants and hot ISM** are a major source of matter feedback into the ISM. Supernova explosions that are correlated in space and time generate super-bubbles (SBs), typically hundreds of parsecs in extent. SNRs and SBs are among the prime drivers controlling the morphology and the evolution of the ISM. Observing their properties is crucial in understanding the galactic matter cycle. The published results from MC SNRs have demonstrated the capabilities of the EPIC instruments in morphological and spectral studies. This already includes a new SNR in the LMC discovered in our first LMC survey observation [343]. Survey observations and XMM-Newton follow-up programs to study SNRs in the LMC have revealed a large number of low surface brightness SNRs in the restricted area covered by XMM [282,344]. The eROSITA survey of the LMC has further contributed to the discovery of new candidate SNRs and the confirmation of the nature of some existing candidates [640]. In the SMC the sample of SNRs was studied by [345].

The *AXIS* survey of the MCs combined with state-of-the-art optical and radio surveys like SDSS-V, 4MOST, Vera Rubin, and MeerKat will address the following questions:

1. What is the population of HMXBs in the MCs, and how does it compare to the Milky Way and other galaxies with different metallicity environments?
2. How are HMXBs formed and switched on as accreting pulsars? What is the role of torque on HMXBs? What role does the magnetic field of the NS and the donor star play in this regard?
3. What constitutes the faintest population of HMXBS? Where are the elusive Be-WD and Be-BH systems? Are all faint HMXBs NSs in propeller state?

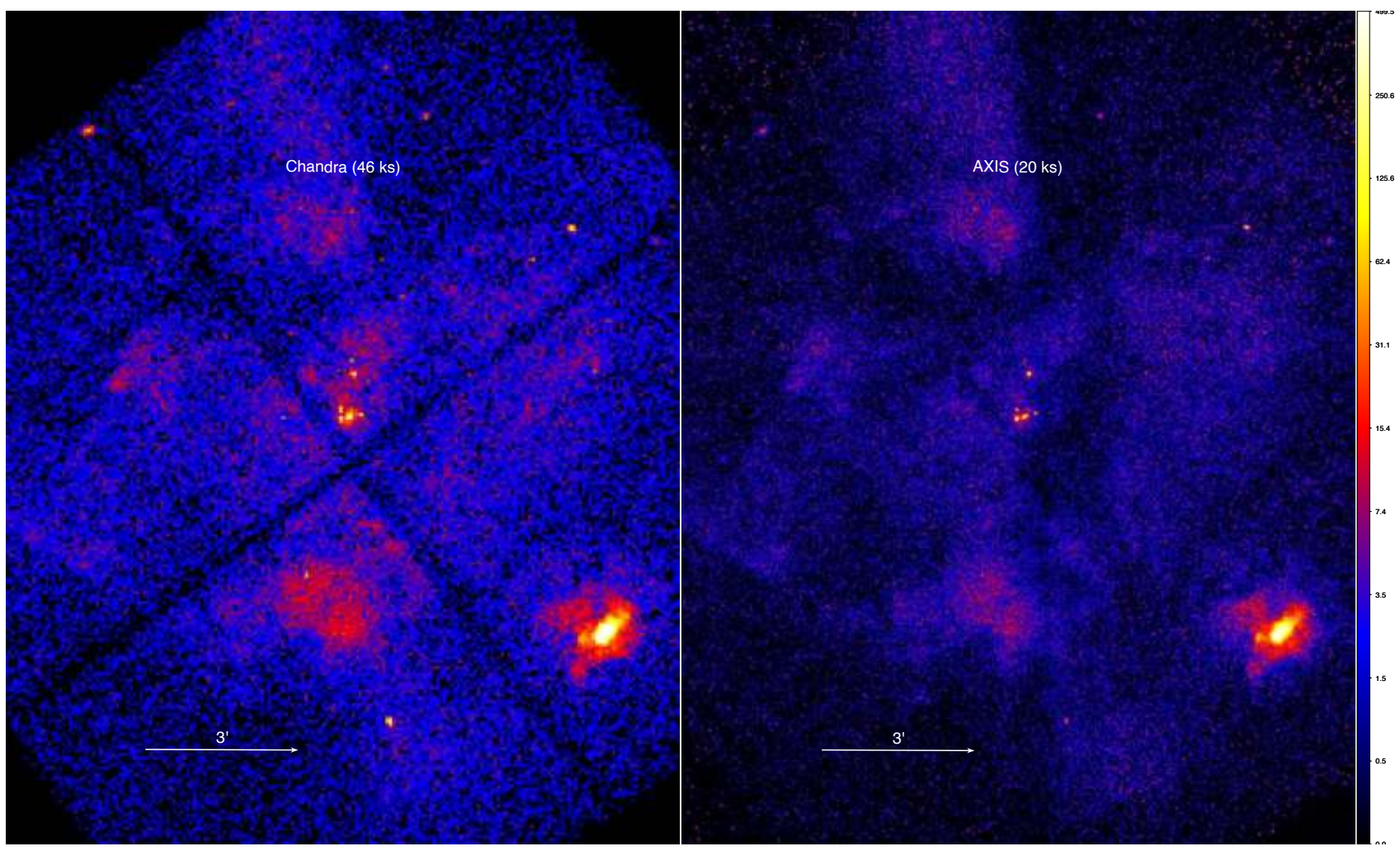

**Figure 38.** *Left: Chandra image of 30 Doradus with an exposure of 46 ks (Obsid 07263) compared to Right: SIXTE simulation of the same region with an exposure of 20 ks.*

4. What is the mass distribution of WDs, and what are the progenitors of SNe type Ia?
5. What is the population of different SNRs in the MCs, and how does it differ between the LMC and SMC? Where are the compact objects associated with the core-collapse SNRs?
6. Where is the elusive compact object in SN 1987A? The timescales of *AXIS* will be ideal in that regard since the reverse shock has started to move in the SNR, and the interior is expected to get brighter, and this could destroy the putative dust cloud obscuring the compact object.
7. What role do massive stars play in the ISM, and how was the 30 Doradus formed? What constitutes the diffuse component in 30 Doradus?
8. What is the nature of hot ISM in the MCs?

**[Exposure time (ks):]** LMC 97 pointings and a total of 1.56 Ms. SMC 19 pointings with a total of 342 ks.
**Observing description:** We request 3×5 ks visits on 94 pointings in the LMC centered on the main body and on supergiant shells and regions with known HMXBs and SNRs. We request deeper pointings around the 30 Doradus region (2×25 ks) to be able to resolve fainter sources in the crowded region contaminated with hot ISM. In the case of SMC, we request a 3×6 ks visits on all pointings. Multiple visits of the same region will also enable one to retain variability information of the sources. Assuming a SSS in the SMC with a luminosity of $10^{34}$ erg/s, typical foreground absorption and $kT = 50$ eV we expect a count rate of 0.1 c/s in 0.2–2 keV accumulating a count of 500 in 5 ks enough for a secure detection and enough photons in the stacked observation to look for periodic signals and detailed spectral signatures. In case of a HMXB of a luminosity of $10^{34}$ erg/s, with powerlaw index of 1 the expected count rate is 0.03 c/s (0.2–5.0 keV) accumulating a count of 150 in 5 ks enabling a detection, where further stacked observation can be used to characterize the source in detail.

A survey of the Magellanic Clouds with *AXIS* will allow us to study the sources with 1) consistently high angular resolution, and 2) high sensitivity, particularly at energies $>$ 2 keV. With its uniform high-resolution PSF and large effective area $>$ 2 keV, *AXIS* has the unique capabilities to perform such a survey and resolve faint X-ray sources as well as diffuse large-scale structures.
**[Joint Observations and synergies with other observatories in the 2030s:]** SDSS-V LVM spectroscopic survey, Rubin/LSST, UVEX, ELTs, NewAthena
**[Special Requirements:]** Mosaic observations of the LMC and SMC with 3 visits on each region with weekly/bi-weekly cadence. A total of 97 LMC pointings and 19 SMC pointings are requested.

*25. M33 survey*

**First Author:** Manami Sasaki (Friedrich-Alexander-University Erlangen-Nürnberg, manami.sasaki@fau.de)
**Co-authors:** Avi Shporer (MIT, shporer@mit.edu), Paul Draghis (MIT), Mark Reynolds (OSU), Chandreyee Maitra (MPE)

**Abstract:**
M33 is the third-largest galaxy in the Local Group after M31 and the Milky Way and is a late-type spiral galaxy with a high star-formation rate. The intermediate inclination angle of ∼56 degrees and modest extinction make it the ideal target for the study of the global structure of the entire galaxy. As M33 is located at a distance of ∼900 kpc, 2 arcsec subtends ∼9 pc in M 33. Therefore, a deep survey of M33 with *AXIS* will allow us to study the distribution of X-ray sources and the properties of the interstellar medium (ISM) on scales from parsecs to kiloparsec and will be a vital complement to deep surveys already carried out with, e.g., LOFAR or JWST.

**Science:**
Galaxies evolve because stars are born, live, and die. Stars evolve due to nuclear fusion in their interiors and change their environment through their radiation, stellar winds, and eventually, supernova (SN) explosions. The radiation from massive stars ionizes their environment, creating HII regions, while supernova explosions form supernova remnants (SNRs). The shock waves of stellar winds and SNRs combined will form interstellar structures called superbubbles. These shocks are primary sources of energy, momentum, and matter for the ISM and are sites of acceleration of galactic cosmic rays. At the end of the life of a star, a white dwarf, a neutron star, or a black hole is formed. As stars are typically formed in stellar clusters and often form multiple systems, especially binary systems, compact objects often have a companion. Compact objects in close binary systems are ideal for studying various binary synthesis and evolution paths, as well as complex astrophysical processes such as accretion, the formation and evolution of strong magnetic fields, and the acceleration of particles. Accreting black holes and neutron stars in binary systems, which are called X-ray binaries (XRBs), and SNRs are bright X-ray sources and are the best laboratories to study physical processes in extreme environments. Additionally, depending on the underlying population of stars and binary systems, the sources exhibit distinct signatures. They can thus be used to study the stellar populations and the star-formation history in a galaxy.

M33 is the second-closest spiral galaxy and has a high star-formation rate. Therefore, it has been observed in various surveys in, e.g., radio, optical, or X-rays. Surveys carried out with the X-ray Multi-Mirror Mission (XMM-Newton) [446,618] and the Chandra X-ray Observatory [208,456] have made a detailed study of the X-ray source population possible. These studies have shown that the population of XRBs in M33 is dominated by high-mass X-ray binaries (HMXBs), as expected for galaxies with a high star-formation rate. M33 is thus the ideal target to study the properties of HMXBs and their formation and evolution histories [194,309,321]. Another population of interest is novae, especially those hosting massive white dwarfs [158,528]. *AXIS* will be instrumental in the follow-up of the post-nova supersoft X-ray phase, which would provide important constraints on the mass of the white dwarf and its environment. X-ray observations have also led to the detection of a large number ($>$ 100) SNRs in M33 [195,336] and have shown that metallicity and star-formation history in galaxies as well as the local environments have significant effects on the detectability and X-ray luminosities of SNRs in different galaxies. Moreover, several interesting sources have been found such as the eclipsing black hole binaries M33 X-7 and [PMH2004] 47 [419,445,447], ultra-luminous X-ray source (ULX) M33 X-8 [301,609], or the non-thermal superbubble IC131 [582].

Even though there have been surveys of M33 both with Chandra and XMM-Newton, Chandra only covered the central part of the galaxy, while the lower spatial resolution of XMM-Newton did not allow for resolving and classifying the X-ray sources well. A new X-ray survey of M33 with *AXIS* will allow us to get a census of the properties of SNRs in the entire galaxy down to a limit of $L_X = 10^{34}$ erg/s and to study the star-formation history of M33 as written in the record of compact X-ray sources. It will extend on previous X-ray M33 observations, and will benefit from new observations in, e.g., radio, inrared, or optical. In particular, new optical observations expected to be carried out by future observatories will enable the identification of XRBs and accreting white dwarfs. We will study the overall 'ecology' of the ISM in a spiral galaxy, specifically the energy input in SNRs, HII regions, and superbubbles, and their relationship to the colder phases of the ISM.

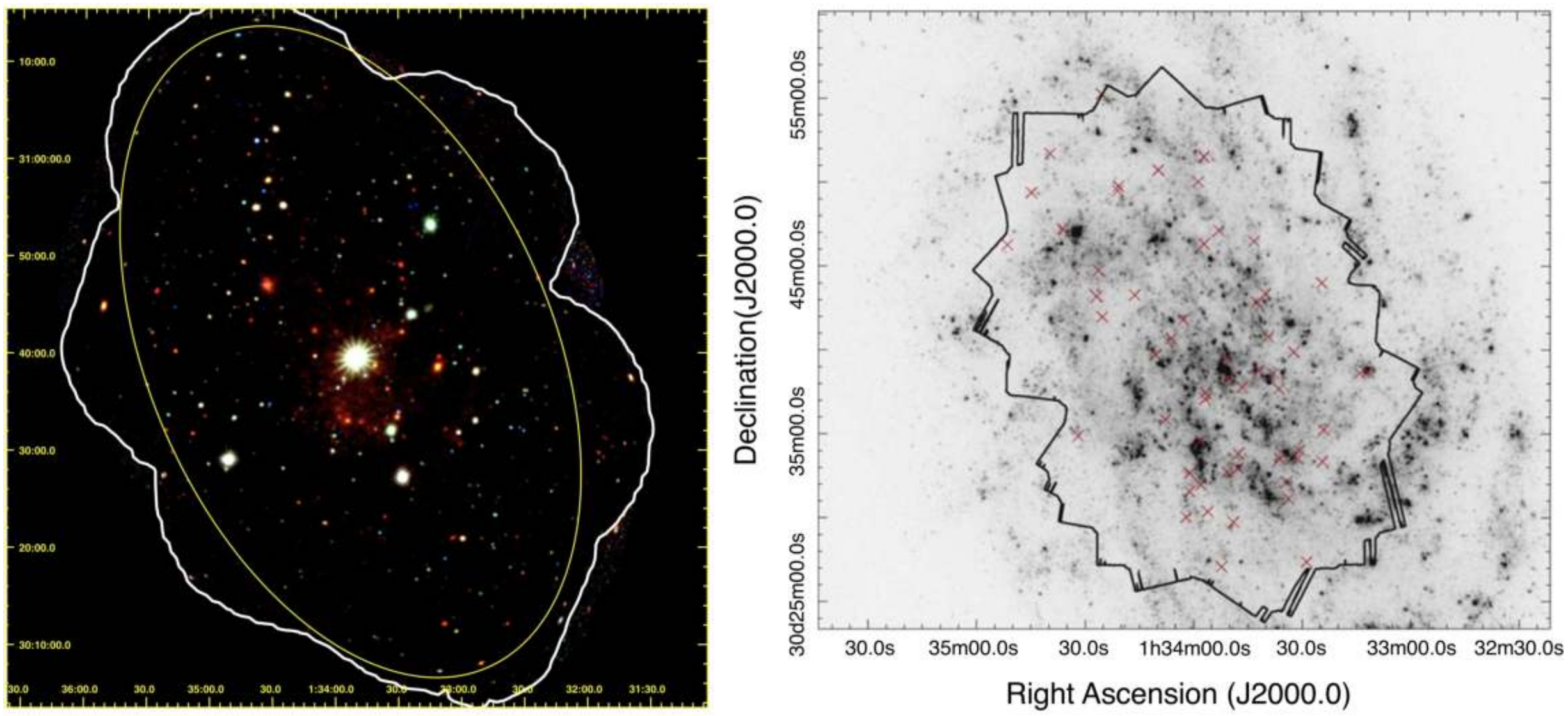

**Figure 39.** X-ray mosaic image of M33 from the XMM-Newton survey [left, taken from 618] and an ultraviolet (UV) image taken with GALEX [323] with the outline of the sky coverage of the Chandra observations [right, taken from 309]. In the XMM-Newton image, the 0.2 – 1.0 keV band is shown in red, 1 – 2 keV in green, and 2 – 4.5 keV in blue. The yellow ellipse indicates the $D_{25}$ region of M33, and the white line outlines the observed areas. In the UV image (right) the positions of the brightest X-ray sources are marked with red crosses. Note that only the inner $\sim$20′ × 30′ region has been covered with Chandra.

By performing a deep survey of the entire M33 combined with multi-wavelength analysis, we will address the following questions:

- How are HMXBs formed and how do they evolve?
- How do the magnetic fields in compact objects form and decay?
- How does the accretion mechanism evolve?
- What processes/mechanisms make HMXBs to gamma-ray sources?
- What is the SN rate in M33?
- What is the population of the different SNR types? How is the distribution related to the properties of the galaxy?
- How do massive stars affect the ISM, and what effect does it have again on star formation?
- How do particles gain relativistic energies and become galactic cosmic rays?
- What is the nature of the diffuse X-ray emission in galaxies?

**Exposure time:** 525 ks

**Observing description:**

With an apparent extent of 40′ × 70′, the galaxy M33 can be fully covered by a 3 × 5 pointing grid. Split into 3 observations of 5 ks each and one observation for 20 ks for each position, the survey will allow variability studies and transient detection as well as long-duration observations of extended sources and diffuse emission. The long exposures will be also useful for follow-ups of transient sources.

**Joint Observations and synergies with other observatories in the 2030s:**

With CTAO and SKA approaching in the gamma and radio bands, a better understanding of the population of high-energy sources, such as pulsars/PWNe, X-ray binaries, novae, SNRs, and superbubbles, will be crucial for understanding the thermal and non-thermal sources. As these sources are best observed in X-rays, high-resolution X-ray observations are necessary for the understanding of the MWL sources. Moreover, the identification of the sources will also be crucial once high-resolution spectroscopy with NewAthena/X-IFU becomes possible, allowing for the disentanglement of the components in the spectra.

## g. Other Galactic

### *26. AXIS Galactic Center Monitoring Program*

**First Author:** Shifra Mandel (Columbia University, ss5018@columbia.edu)
**Co-authors:** Kaya Mori (Columbia University), Tong Bao (INAF-Merate), Arash Bodaghee (Eureka Scientific), Daryl Haggard (McGill), Gabriele Ponti (INAF-Merate; MPE), Amruta Jaodand (Harvard CfA), Mason Ng (McGill), Gianluca Israel (INAF), Paul Draghis (MIT), Jeremy Hare (NASA GSFC), Mark Reynolds (The Ohio State University), Ceaser Stringfield (Columbia), Walid Majid (JPL)
**Abstract:** The Galactic center (GC) is home to the largest known concentration of exotic X-ray sources ever identified in our Galaxy, including an overabundance of compact objects, X-ray transients, high-energy X-ray filaments, molecular clouds, and the supermassive black hole, Sagittarius A*. With its uniform high-resolution PSF and large >2 keV effective area, *AXIS* promises to deliver unprecedented insights into the Galactic center X-ray emission and the processes that generate it. We propose a Galactic center monitoring program that will observe the central $12' \times 12'$ of the Galaxy daily, and cover a wider $\sim 1^\circ$ diameter region every week. Spending just 1 ks per day observing the Galactic center with *AXIS* will allow us to [1] detect and classify X-ray transients down to faint luminosities ($L_X \lesssim 10^{34}$ erg s$^{-1}$ cm$^{-2}$); [2] uncover the properties of X-ray flares from Sgr A* and investigate their emission mechanism; [3] map the distribution of compact object binaries within the dense nuclear star cluster; [4] constrain the masses of white dwarfs in magnetic cataclysmic variables, improving our understanding of their origin and evolution; and [5] open a window into the history of Sgr A* activity by studying the echoes of its past flares that are reflected from giant molecular clouds. Our proposed monitoring survey will be optimized to detect a variety of periodic signals and non-periodic flares; over time, these observations will yield an exceptionally rich dataset with cumulative exposures totaling 1 Ms and provide a groundbreaking legacy dataset with comprehensive spectral, spatial, and timing information for the GC.

**Science:**

The Galactic Center (GC) is a crowded hub teeming with a diverse array of X-ray sources: compact objects, X-ray binaries (XRBs), molecular clouds, magnetic filaments, and hot gas outflows. At the heart of this activity lies Sagittarius A* (Sgr A*), the supermassive black hole (SMBH) at the center of our Galaxy. Sgr A* profoundly influences the gas dynamics and the formation of binary systems within the central few parsecs, where stars and XRBs tend to congregate [383,391]. Despite its typically faint X-ray emission, Sgr A* sporadically unleashes X-ray flares. Evidence suggests a more active past for Sgr A*, potentially giving rise to the Fermi GeV bubbles [110] and X-ray chimneys [457]. As recently as a few hundred years ago, it likely illuminated the GC molecular clouds that reflected its X-ray emission [122,458].

Given that compact object binaries are copious X-ray emitters, X-ray observations are crucial for probing the nature and distribution of high-energy sources within the GC. However, considerable difficulties are inherent in studying this extremely crowded region, situated a distant 8 kpc away. The high density of sources necessitates high-angular-resolution X-ray telescopes to resolve them. Extensive surveys by Chandra, XMM-Newton, and Suzaku over the past two decades have mapped many of the X-ray sources within the GC, including thousands of point sources detected by Chandra in the central $2^\circ \times 0.8^\circ$ [392,603,652]. $\sim$ 20 X-ray transients were detected within $\sim 24'$ of Sgr A*, many through the continuous X-ray monitoring of the GC with Swift-XRT (Figure 40, right) [153].

While Chandra has previously imaged the central molecular zone (CMZ), *AXIS* is nearly 10x more sensitive than Chandra and will detect even fainter sources, both compact and diffuse. More significantly, the *AXIS* PSF is roughly uniform throughout its $27' \times 27'$ FoV – in contrast to Chandra's significantly degraded one at off-axis angles $> 4'$. Therefore, *AXIS* is capable of mapping the distributions of point sources in the central $\sim$ 0.5 deg with greater precision than ever before, opening a new window into the history and dynamics of the GC.

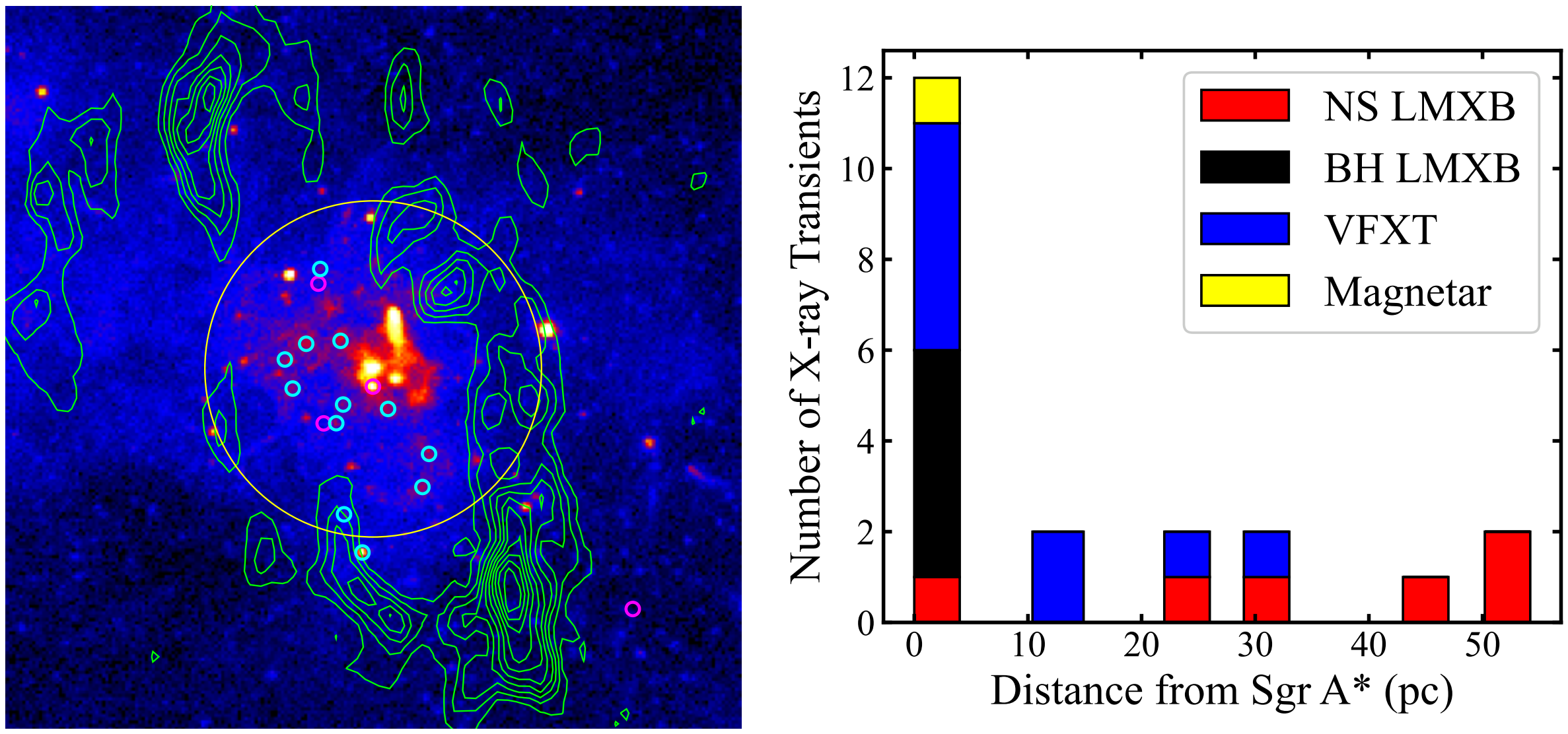

**Figure 40.** *Left:* Candidate BH-LMXBs in the GC. 12 quiescent sources identified by [222,382] are in cyan; transients in magenta. Green contours denote the circumnuclear disk. The yellow circle has a radius of 1 pc. *Right:* Distribution of X-ray transients detected within $r \sim 60$ pc of the GC, shown in projected distance from Sgr A* [353,383].

*X-ray transients:*

X-ray transients identified in the past two-and-a-half decades span a variety of sources, including candidate black hole (BH) low-mass X-ray binaries (LMXBs); neutron star (NS) LMXBs; very faint X-ray transients (VFXTs); and famously, a magnetar (Figure 40, right) [381]. As Figure 41 illustrates, over a dozen transients have been detected in the central $2' \times 2'$ alone. The high source density means that $\sim$arcsecond localization is often necessary to determine whether an X-ray outburst is recurrent or from a new source. To fully understand the evolution of transient outbursts, which can undergo changes on short timescales, frequent monitoring is required, starting from the initial rise from quiescence.

Very faint X-ray transients (VFXTs) are a particularly interesting subset of X-ray transients. With peak luminosities of $L_X \sim 10^{34}$–$10^{36}$ erg s$^{-1}$[153] and relatively short durations, their outbursts can easily be missed, hence they are not well understood. VFXTs are thought to be LMXBs with particularly low mass accretion rates due either to magnetically truncated accretion disks, or H-poor companion stars; in the latter scenario, VFXTs would be ultra-compact X-ray binaries (UCXBs) [587]. Our proposed study will enable us to detect more VFXTs, particularly near the fainter end of the luminosity distribution. We will be able to observe how their outbursts evolve, and search for periodic signals that might indicate whether they are UCXBs.

*AXIS* will be particularly useful in determining the outburst recurrence rates for different types of transients in the GC. Previous studies have indicated that LMXBs with NS vs BH primaries exhibit different outburst recurrence rates, with NS LMXBs often exhibiting much more frequent outbursts [380,383]. To confirm this correlation, we need frequent, sensitive observations of the GC that can a) detect faint outbursts, b) determine source localization down to sub-arcsecond scale (allowing us to see whether a previous transient occurred at the outburst location), and c) monitor for type I X-ray bursts or pulsations that indicate a NS primary. *AXIS* will be able to meet all those requirements. Constraining the outburst behavior of different types of sources will allow us to use those properties as identifiers in future surveys.

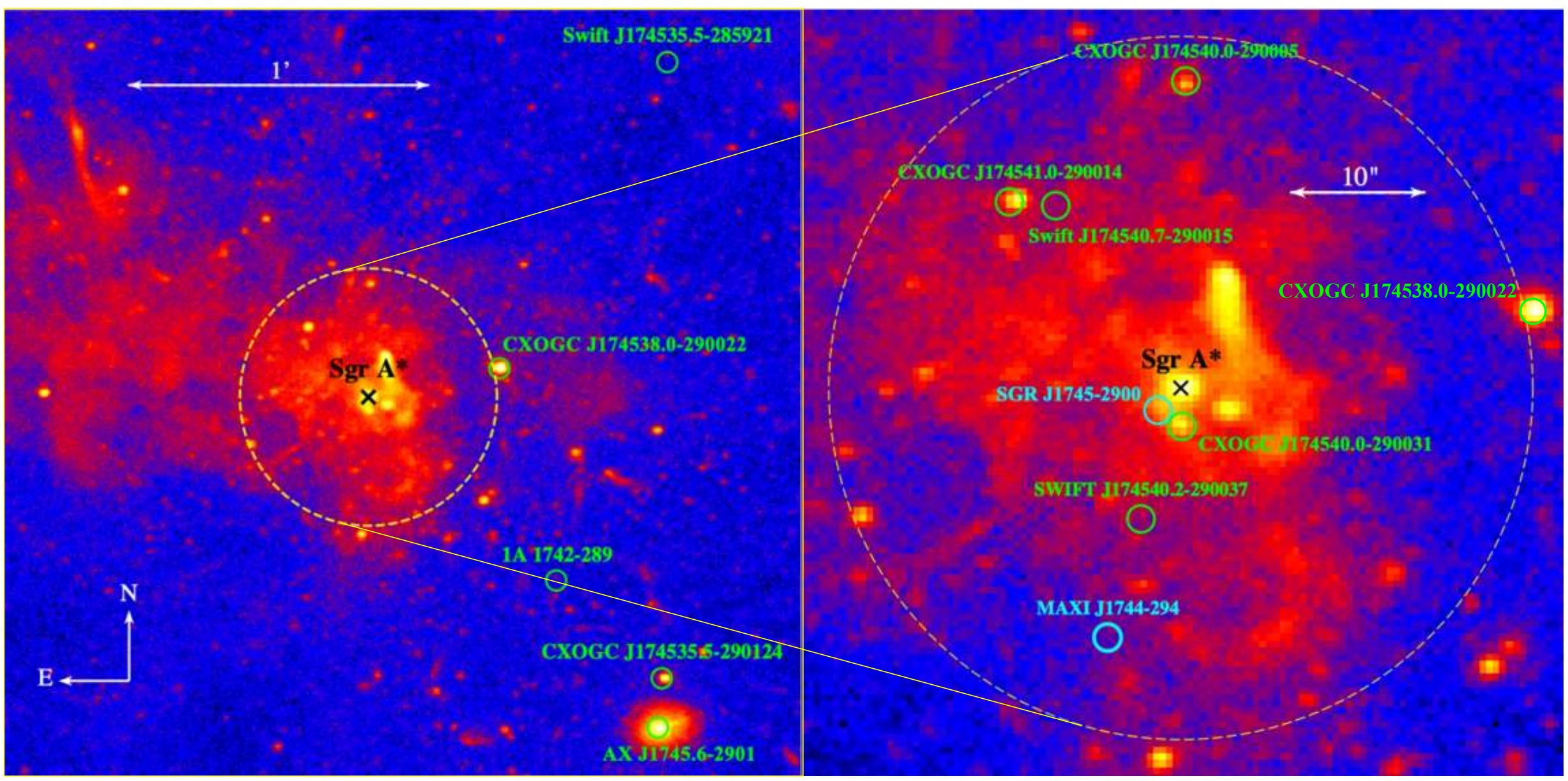

**Figure 41.** *Chandra* image showing the transients detected in the central $2' \times 2'$ [153,353,382]. Inset on the right shows the sources within a radius of $r < 1$ pc from Sgr A* (dashed circle). Sub-arcsecond angular resolution is often required to determine whether an outburst is from a new or recurring transient due to the proximity between sources.

*Quiescent sources:*

Thousands of X-ray point sources have been identified in the central degree of the Galaxy. Based on analytical studies and dynamical simulations, hundreds of BH LMXBs are believed to be among those sources [197,384], along with a similar number of NS LMXBs and an even more robust population of magnetic cataclysmic variables (CVs) [223]. Notably, 12 candidate BH-LMXBs have been identified based on observational data [222,383]. High crowding has previously hampered our understanding of the X-ray source distributions in the central degree. With its uniform high angular resolution, *AXIS* will allow us to not only resolve more X-ray point sources in the innermost region of our Galaxy, but also understand their intrinsic luminosity and spectral distributions. Over time, our survey data will allow us to constrain the periods of a substantial fraction of these sources; when combined with spectral data, this will help us differentiate between different types of X-ray point sources.

*Sgr A* flares:*

Sgr A* exhibits daily X-ray flares. While existing studies offer some clues about the cause of these flares, sensitive observations with a high cadence are required to help constrain the luminosity distribution of those flares and allow us to determine how they vary. With its increased sensitivity, *AXIS* will enable detailed studies of fainter flares, which are the most common.

Understanding the physical mechanisms behind Sgr A*'s flaring requires simultaneous observations across the electromagnetic spectrum. Currently, the exact mechanism driving this variability remains unknown, with several competing theories including accretion flow instabilities, magnetic reconnection, jetted ejections, and expanding plasma clouds.

Chandra observations have shown that flaring events follow a power-law distribution. *AXIS* will significantly advance the study of these flares by enabling detailed analysis of their X-ray spectra and structure, even for weaker flares. The rare bright flares offer the best opportunities for understanding the broadband emission mechanism; high-time-resolution spectroscopy with *AXIS* will complement the capabilities of other 2030s telescopes during these events.

*Molecular clouds:*

Past observations revealed that molecular clouds in the vicinity of Sgr A* display intense and fluctuating X-ray fluorescence. This emission is believed to be generated by the reflection of bright flares from Sgr A* that occurred at a time when the SMBH was more active. Tracking the changes in this emission over longer timescales will allow us to infer the timeline of past activity from Sgr A*. *AXIS* will provide exceptionally detailed spatial information about the origin of the Fe K-shell emission, allowing us to investigate the distribution and structure of dense molecular gas in the GC.

**[Exposure time (ks):]** 350 ks; 1 ks/day

**Observing description:**

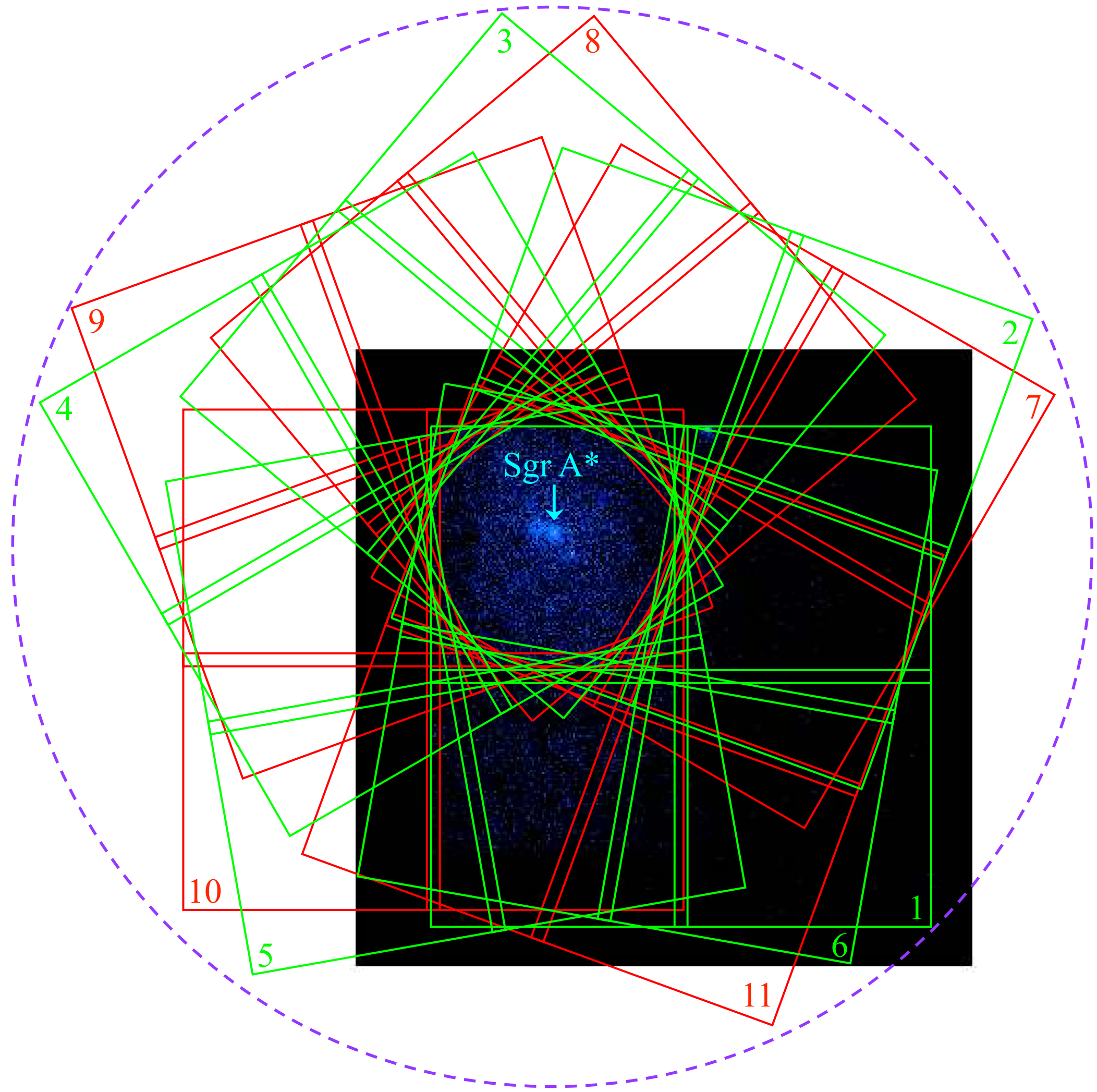

**Figure 42.** Observing plan for GC monitoring program. 1 ks exposures with one chip centered on Sgr A*, rotated by 70° daily to cumulatively cover a region of radius $r \sim 30'$ (purple dashed circle). The footprints of the first 11 observations are highlighted in green and red.

We propose a daily monitoring campaign of the GC with 1 ks exposures. The FoV will be rotated by 70° every day, while Sgr A* will remain centered on one of the detector chips at all times. This will allow us to cover a region of radius up to $r \sim 30'$ from Sgr A* with a ~weekly cadence, while the more active central $r \sim 12'$, including Sgr A*, will be covered daily. Figure 42 illustrates a sample of the proposed observing pattern (11-day subset).

**[Joint Observations and synergies with Roman:]**

The Nancy Grace Roman Space Telescope (Roman) will observe the GC at a twice-daily cadence as part of its Galactic Bulge Time Domain Survey (GBTDS). The Roman survey is projected to yield precise photometric measurements for $> 3$ million stars at an unprecedented high cadence and depth (up to F146 $\sim 23 - 24$ mag) [564]. This will enable us to monitor the GC for any NIR counterparts to X-ray flares or outbursts. Even for quiescent X-ray sources, our long-term monitoring will allow us to track their variability over time and detect periodic variations. We will then be able to match those X-ray sources to NIR counterparts with identical periods (e.g., ellipsoidal modulation due to tidal distortion of the donor star) that Roman observes. For NIR counterparts to X-ray sources that are successfully identified in Roman's photometric data, follow-up spectroscopy can be used to confirm the identification of those sources. For example, recombination emission lines from accretion disks around compact objects can be used to infer radial velocity, and by extension, a mass function for the binary [103,135]. This, in turn, can allow us to constrain the compact object mass, determining, e.g., whether an X-ray binary hosts a BH or NS.

*27. AXIS view of variability in globular cluster X-ray sources*

**First Author:** Jeremy Hare (NASA GSFC, CUA, CRESST II, jeremy.hare@nasa.gov)
**Co-authors:** Arash Bahramian (CIRA), Oleg Kargaltsev (GWU), Hui Yang (IRAP), Craig Heinke (U. Alberta)
**Abstract:** It has long been known that globular clusters host an overabundance of X-ray binaries per unit mass compared to the Galactic disk. This increase in binaries is attributed to the numerous dynamical interactions that occur in the cores of globular clusters. Given the large amount of crowding in the center of globular clusters, studying the fainter X-ray sources can only be achieved by X-ray observatories with high angular resolution. Chandra played a critical role in uncovering the low luminosity end of the X-ray luminosity function in globular clusters, where several classes of interesting compact object binaries reside. Here we propose a uniform survey of 100 ks observations of 10 globular clusters with *AXIS*, which, given *AXIS*'s large effective area, will allow us to accomplish three primary goals. The first is probing the X-ray luminosity function below a luminosity of $10^{30}$ erg s$^{-1}$ across these clusters, which span a range of ages, metallicities, and densities. Searching for sources at lower luminosities will also allow us to detect or place tight constraints on any IMBH candidates located in the centers of these clusters. The second goal is to probe the transient and variability properties of GC source populations, which we will accomplish by splitting the observations into chunks of 10-20 ks randomly spaced over a 2-year time interval. The third goal is to localize better and understand the X-ray source population in the outskirts of these globular clusters, where it was often difficult to locate potential optical/IR companions to X-ray sources due to Chandra's PSF degradation at large off-axis angles. While *AXIS* has a broader PSF than Chandra, its significant increase in effective area (and relatively uniform PSF over the entire field of view) will allow for more accurate source localizations.
**Science:** The number of X-ray binaries per unit mass in globular clusters is enhanced due to the dynamical formation channels at play in the cluster cores [121]. These dynamical interactions lead to the capture and exchange of compact objects by the more numerous main sequence (or giant) stars. Additionally, as the cluster collapses under its own gravity, binaries lend support against this collapse by exchanging kinetic energy with nearby stars, resulting in a shortening of their orbital periods. This leads to an increase in the interactions between compact objects and their companion stars, leading to an excess of observed X-ray binaries.

Sub-arcsecond angular resolution is necessary to detect and separate the fainter X-ray source population due to the large amount of crowding of X-ray sources in the clusters' centers. To this end, Chandra has played an essential role and has allowed us to understand the dominant source populations in globular clusters at luminosities below $10^{32}$ erg s$^{-1}$ (see e.g., [220,234,235]). The most abundant classes found in GCs are quiescent low-mass X-ray binaries (qLMXBs) hosting neutron stars, millisecond pulsars (both isolated and in binaries; MSPs), cataclysmic variables (CVs), and active binaries (ABs). Interestingly, several black hole binary candidates have been identified in globular clusters [36,117,374,545], one of which recently produced radio jets after undergoing an outburst [38,448].

Here we propose deep 100 ks *AXIS* observations of 10 GCs (see Table 5). The clusters were chosen based on a combination of their distance, size, density, and number of X-ray sources detected by Chandra. Several clusters were also chosen as they have known X-ray binaries with interesting properties, such as the BH X-ray binary candidate in Glimpse-C01 [38,448]. These observations will enable a deep and efficient survey of these Galactic GCs down to luminosities of about $10^{(28-29)}$ erg s$^{-1}$, depending on the GC distance. Discovering more of these compact object binaries allows for a better understanding of the dynamical production and destruction of them, as a function of the host GC's encounter rate, metallicity, cluster structure, and other properties (see e.g., [459,460]). Furthermore, pushing to lower luminosities is crucial for discovering new unique objects, such as BH LMXBs, several of which have quiescent X-ray luminosities $< 10^{30}$ erg s$^{-1}$ [230,545]. By finding more of these BH systems (with the help of deep radio observations), we can begin to understand their formation rate in comparison to binaries hosting NSs and how they dynamically move within the cluster (e.g., is there an overabundance towards the GC center)? This has major implications for the rate of BH-BH mergers being detected by LIGO (see e.g., [19,20,642]). Additionally, there are still over 100 GC MSPs that remain undetected down to luminosities of $10^{30}$ erg s$^{-1}$ [649], which our survey will detect or set even deeper limits on. We will also be able to place very deep limits on any X-ray emission from putative intermediate mass BHs in the cluster centers, such as the candidate discovered in Omega Cen [219].

*AXIS* maintains a relatively consistent PSF across its entire 450 arcmin$^2$ field of view [482]. Therefore, this survey will also allow us to detect and more precisely localize off-axis sources in the outskirts of the GC, away from the GC center. We anticipate that, between now and the time this survey is conducted, many of these GCs will have extensive JWST coverage in multiple filters, while HST data exists for all of the selected clusters. Additionally, new wide-field observatories such as Rubin/LSST and Roman will provide coverage of the outskirts, as well as the crowded cores, of these clusters. These data will allow us to search for potential optical/IR counterparts to these X-ray sources, and will be crucial for separating out background AGN and foreground stars (see e.g., [227,235,526]). They will allow us to place constraints on the companions in these binaries spanning a large range in companion masses. It will also allow for studies of the spatial distributions of these X-ray binaries and to answer questions such as: do BH LMXBs sink to the cluster center? Do NS and BH LMXB spatial distributions differ?, etc.

We also anticipate that between now and the time this survey is conducted, many additional MSPs will be identified by existing and new radio surveys with improved sensitivity and localization capabilities (see, e.g., [488,645]). There are also likely to be newly identified BH binary candidates in these GCs. *AXIS* will allow us to go 1-2 orders of magnitude fainter in luminosity than Chandra and will also provide enhanced sensitivity to faint X-ray sources in the outskirts of these GCs, which is a relatively unexplored population. Given *AXIS*'s low slewing overhead, we request that these observations be split into 5 to 10 10-20 ks observations randomly spaced over two years. This will not impact the detection of the persistent sources and will allow us to probe source variability and search for transient sources on a year-long timescale. Several fast X-ray transients were serendipitously discovered by Chandra (see e.g., [233,237]), but *AXIS* will allow for a systematic study of these sources and allow for constraints to be placed on their rates. This will also allow us to search for correlations between the rate of X-ray transients and other GC properties. Additionally, this survey will allow us to constrain the accretion and outburst duty cycle in a large sample of X-ray binaries with well-measured distances beyond Chandra. These measurements are important as they can be used to study the cooling of the NS crusts after an accretion episode (see e.g., [154]).

**Exposure time (ks):** 1 Ms

**Observing description:** We propose 100 ks *AXIS* observations of each cluster listed in Table 5. To calculate the limiting luminosity reached for each cluster in this survey, we assumed a 5 sigma detection limit of $3{\times}10^{-17}$ erg cm$^{-2}$ s$^{-1}$ in the 0.5-2 keV band, assuming an absorbed power-law spectrum with a photon index of 2. We will also be able to reach even lower luminosities for thermally emitting sources (e.g., qLMXBs). *AXIS*'s PSF remains relatively constant across its field-of-view and only degrades slightly. Therefore, even for off-axis sources in the outskirts of these GCs, we will reach limiting luminosities of about a factor of 2.5 larger than those listed in Table 5. Figure 43 shows a simulated 10 ks *AXIS* image of 47 Tuc, which demonstrates *AXIS* will be able to detect the vast majority of GC sources in each 10 ks exposure. We request that these observations occur in 10-20 ks chunks randomly spaced in time and spanning a two-year interval. This will enable us to investigate the variable and transient source population in GCs, which can comprise a significant fraction of sources.

**Joint Observations and synergies with other observatories in the 2030s:** The *AXIS* data will be combined with existing and future HST, Roman, JWST, and radio observations, which can be used to identify the nature of the X-ray source through detection of pulsations in radio, or the identification of the optical/IR counterpart to the X-ray source. In the outskirts of the cluster, Rubin/LSST data can be used to search for variable sources with potential X-ray counterparts.

| Cluster Name | Distance (kpc) | $5\sigma$ limiting X-ray luminosity 0.5–2 keV (erg/s) |
|---|---|---|
| 47 Tuc | 4.45 | $7\times10^{28}$ |
| Terzan 5 | 5.9 | $10^{29}$ |
| M22 | 3.2 | $4\times10^{28}$ |
| Glimpse-C01 | 3.4 | $4\times10^{28}$ |
| M62 | 6.8 | $2\times10^{29}$ |
| Omega Centauri | 4.84 | $8\times10^{28}$ |
| M4 | 1.85 | $10^{28}$ |
| NGC 6304 | 5.9 | $10^{29}$ |
| NGC 6397 | 2.4 | $2\times10^{28}$ |
| M28 | 5.5 | $10^{29}$ |

**Table 5.** List of Globular Clusters proposed to be observed by *AXIS* in this observing program.

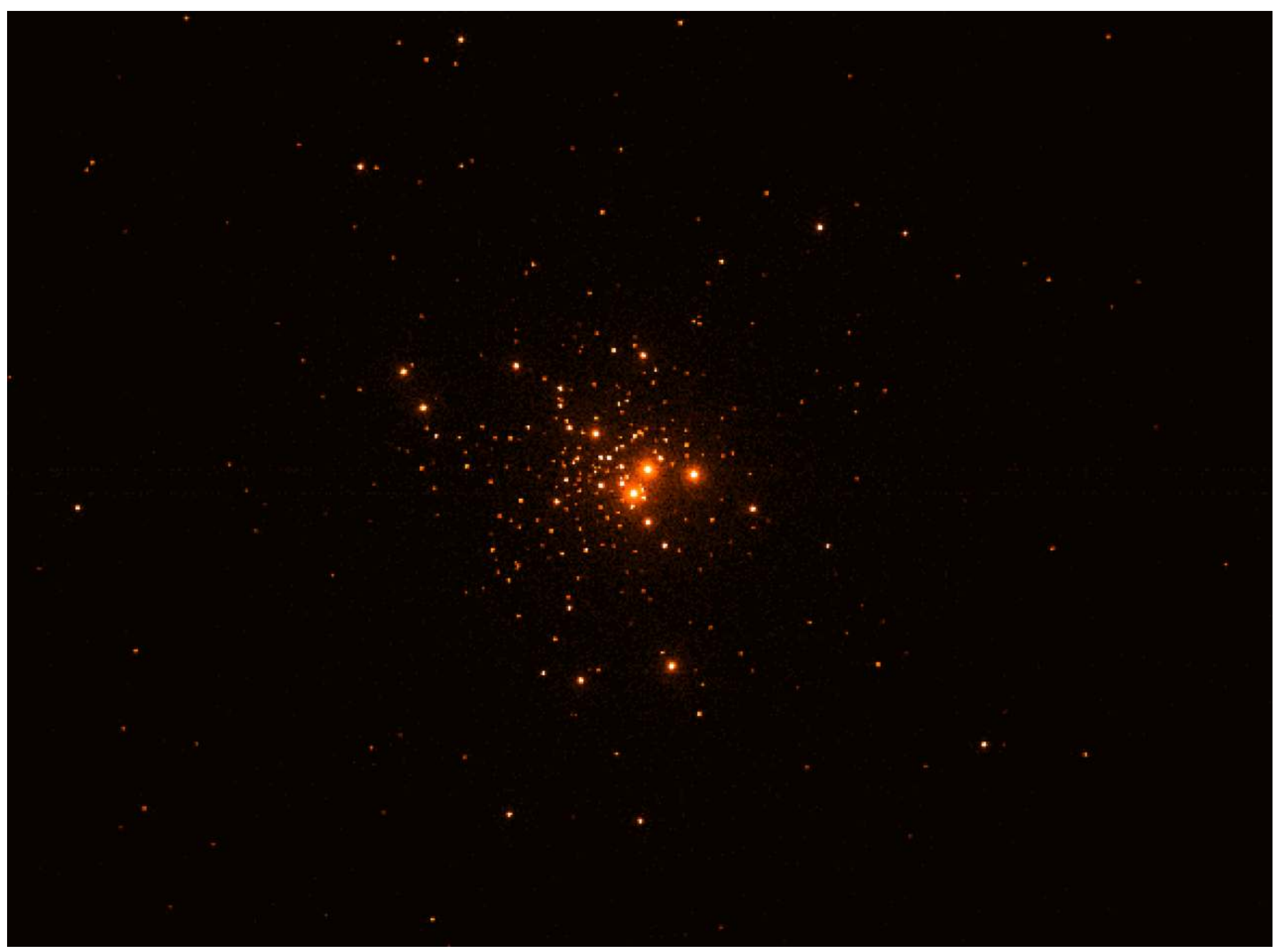

**Figure 43.** Simulated 10 ks image of the globular cluster 47 Tuc as observed by *AXIS*.

*28. A survey of compact binaries in the globular cluster Omega Cen*

**First Author:** Liliana Rivera Sandoval (UTRGV, liliana.riverasandoval@utrgv)
**Co-authors:** Kristen Dage (Curtin), Kwangmin Oh (MSU)
**Abstract:** Omega Centauri is the brightest globular cluster in our galaxy. However, its high mass, kinematics, and complex stellar populations make it stand out from the rest of the globular clusters in the Milky Way, suggesting it is the remnant nucleus of a dwarf galaxy. Omega Cen has also been a prime target for studying compact binaries due to its intriguing nature. Current observations have identified around 275 X-ray sources down to a limiting luminosity of $1 \times 10^{30}$ erg s$^{-1}$ within a FOV of $17' \times 17'$ towards the center of the cluster [239]. We propose to capitalize on the wider FOV of *AXIS*, its greater sensitivity, and timing capabilities to reach an X-ray limiting luminosity at least an order of magnitude fainter than current observations. This survey will allow us to identify and characterize for the first time various exotic binaries in the cluster harboring black holes, neutron stars and white dwarfs, including the expected abundant population of accreting white dwarf binaries, such as cataclysmic variables and AM CVns (H-deficient systems with periods shorter than 70 minutes), as well as binary star systems with significant magnetic activity also known as active binaries. With a robust sample of compact binaries and multi-wavelength information obtained through synergies with ground- and space-based telescopes, we will be able to compare them to state-of-the-art models to test the effects of dynamical interactions in crowded stellar environments, as well as mass transfer and mass accretion. Furthermore, reaching an X-ray luminosity of $1 \times 10^{29}$ erg s$^{-1}$ will allow us to test for the presence of an intermediate mass black hole in the core of the cluster.
**Science:**

*The compact binary population in Omega Cen*

Investigating the population of compact binaries (CBs), which are short-period systems containing stellar remnants, within globular clusters (GCs) is essential for understanding the broader evolution of globular clusters. In particular, binaries serve as a critical energy source that helps prevent core collapse in GCs. On the other hand, it is also important to understand the role of clusters as environments where compact binaries can form, potentially leading to events like Type Ia supernovae, kilonovae, black hole (BH) mergers, and sources of gravitational waves detectable by LISA. In this context, X-rays, in particular, are effective probes of the interaction processes in these compact binaries, often revealing systems undergoing mass accretion.

Omega Cen is one of the clusters that has been better studied in optical and X-rays with current facilities such as HST, MUSE, Chandra, and Swift. Around 275 X-ray sources have been identified towards the cluster down to a limiting luminosity of $L_X = 1 \times 10^{30}$ erg s$^{-1}$ at 5.2 kpc, among which $60 \pm 20$ have been determined to belong to the cluster [239]. More recently, 19 pulsars have been identified [108,138], which coincide with known X-ray sources [650]; likely, there is a much larger undiscovered pulsar population associated with currently unclassified X-ray sources. Furthermore, based on current statistics, CVs in Omega Cen seem to be underabundant with respect to the field [220]. This is somewhat surprising, considering the cluster's large mass. However, recent theoretical and observational studies suggest that dynamics plays an important role in destroying the progenitors of CVs [58,491]. These studies have also shown that as the half-mass relaxation time increases in a globular cluster, the fraction of CVs located outside the half-light radius increases, reaching a maximum limit of approximately 50%. Omega Cen has an extremely long relaxation time, of the order of $10^{9-10}$ years, meaning that likely a very large population of CVs beyond its half-light radius remains to be discovered. It is expected that most CVs in globular clusters, including Omega Cen, have luminosities below $L_X = 1 \times 10^{30}$ erg s$^{-1}$. This means that to further assess the importance of dynamics in the CV population, an X-ray survey below that threshold needs to be conducted.

Another population of accreting white dwarfs that is expected to be abundant and produce X-rays in globular clusters is the population of AM CVns (H-deficient, He-rich systems with orbits typically shorter than 1 hr). However, up to now, no single system has been fully confirmed. This contrasts with binaries harboring neutron stars and black holes, such as the LMXBs and UCXBs, which are overabundant in GCs, though in Omega Cen, only one LMXB has been identified. The lack of LMXBs in Omega Cen remains an open question, possibly due to its distinct dynamical history.

Together with state of the art simulations, such as Monte Carlo Cluster Simulations (MOCCA), which efficiently model and capture the long-term dynamical evolution and complex interactions between the stars we can precisely reconstruct the interactions from compact objects within dense stellar environments like Omega Cen's, systematically exploring the possible presence of an intermediate-mass black hole (IMBH), as well as formation channels and important dynamical parameters from various compact binaries. Comparing simulation outputs with observational data will allow us to refine the initial conditions and evolutionary history of Omega Cen.

Armed with its very large FOV, large effective area, and almost homogeneous PSF, *AXIS* will allow us to explore the compact binary population of Omega Cen beyond the central parts currently explored using Chandra. Previous studies with XMM-Newton have also been performed to a larger area but reaching even shallower luminosities $L_X = 1.3 \times 10^{31}$ ergs s$^{-1}$ in the $0.5 - 5$ keV range [196]. *AXIS* will permit the identification of faint CVs, AM CVns, pulsars, active binaries, the missing symbiotic stars in GCs, UCXBs, and quiescent LMXBs (which can have X-ray luminosities as low as a few times $1 \times 10^{30}$ erg s$^{-1}$). It will be possible to determine their short- and long-term variability, as well as their periods (e.g., the short orbital periods of AM CVns and ultracompact X-ray binaries). Also, by obtaining large amounts of X-ray photons it will be possible to model their X-ray spectra properly. Facilities like HST, MUSE, Roman, VLA, MeerKAT, and Rubin will allow us to identify and further characterize the counterparts to these X-ray sources.

In Figure 44, we show a comparison of simulated spectra for a bright CV in Omega Cen as expected from the proposed 120 ks on- and off-axis observations with *AXIS*. We used parameters as determined by [239] with Chandra ACIS-I. With *AXIS* we obtain $\sim$ 20300 and $\sim$ 17300 counts in the energy range $0.5 - 6$ keV for on- and off-axis observations, respectively. This contrasts with the $\sim$ 2250 counts obtained with Chandra in 222ks in the same energy band, and demonstrates that *AXIS* will allow us to carry out detailed studies of compact binaries in the cluster.

*Intermediate mass black holes in Omega Cen*

As the largest globular cluster in the Milky Way, Omega Cen has been scrutinized for almost two decades as a potential home to an intermediate-mass black hole [e.g. 219,408]. Kinematics have suggested an IMBH with masses ranging from 2000-8000 solar masses [438], but evidence for X-ray emission is sparse, with central X-ray emission limited to an unabsorbed flux of $< 1 \times 10^{30}$ erg s$^{-1}$ [221]. This suggests that any IMBH present is either less massive than suggested by kinematics, or accreting very inefficiently (Mahida et al in prep) $-$ which can be probed by deeper X-ray constraints. One of the major limitations of kinematic black hole mass measurements is that it is not possible to distinguish whether the measured mass is a single massive object or many, less massive objects. Only detection of electromagnetic emission can definitively verify a central IMBH in Omega Cen, and *AXIS*'s spatial resolution is essential to test this.

Numerical simulations of star clusters are also an important tool to interpret the signatures of candidate black holes [54]. However, the process of determining the initial conditions that can evolve into present-day clusters is non-trivial. It requires benchmarking through a detailed study of the compact binaries present in the cluster. A complete census of compact binaries in Omega Cen is therefore highly valuable, not only to understand the effects of dynamics on compact binaries but to understand the evolution of massive globular clusters as Omega Cen.

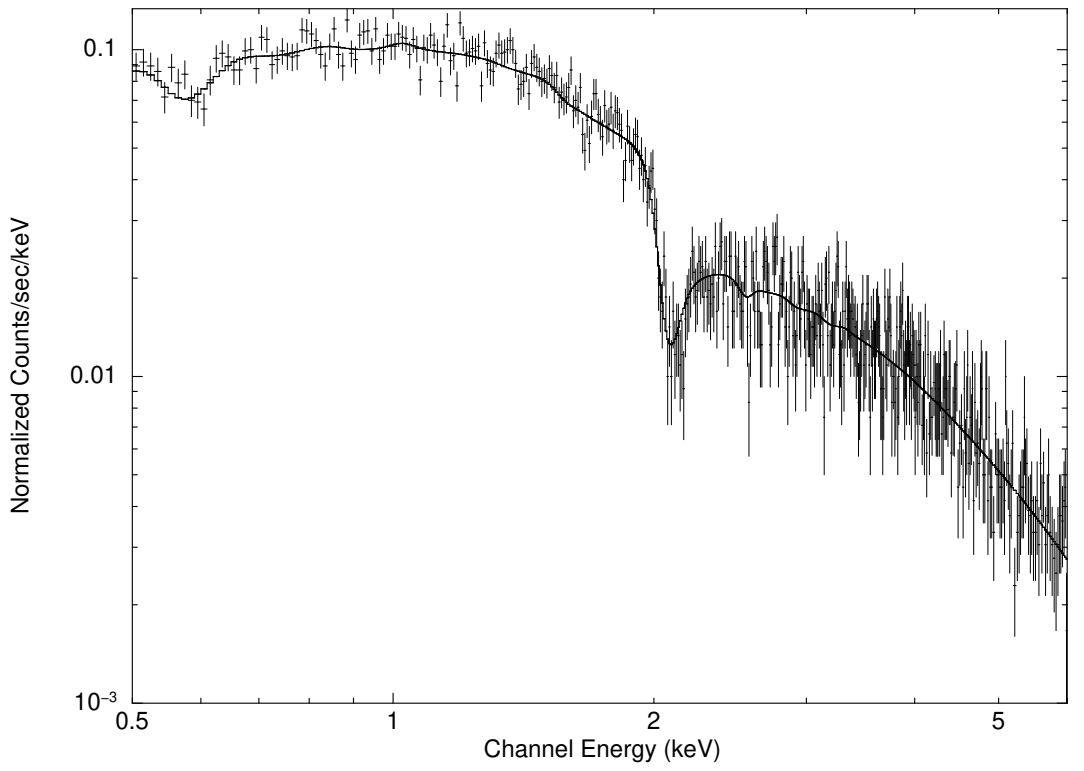

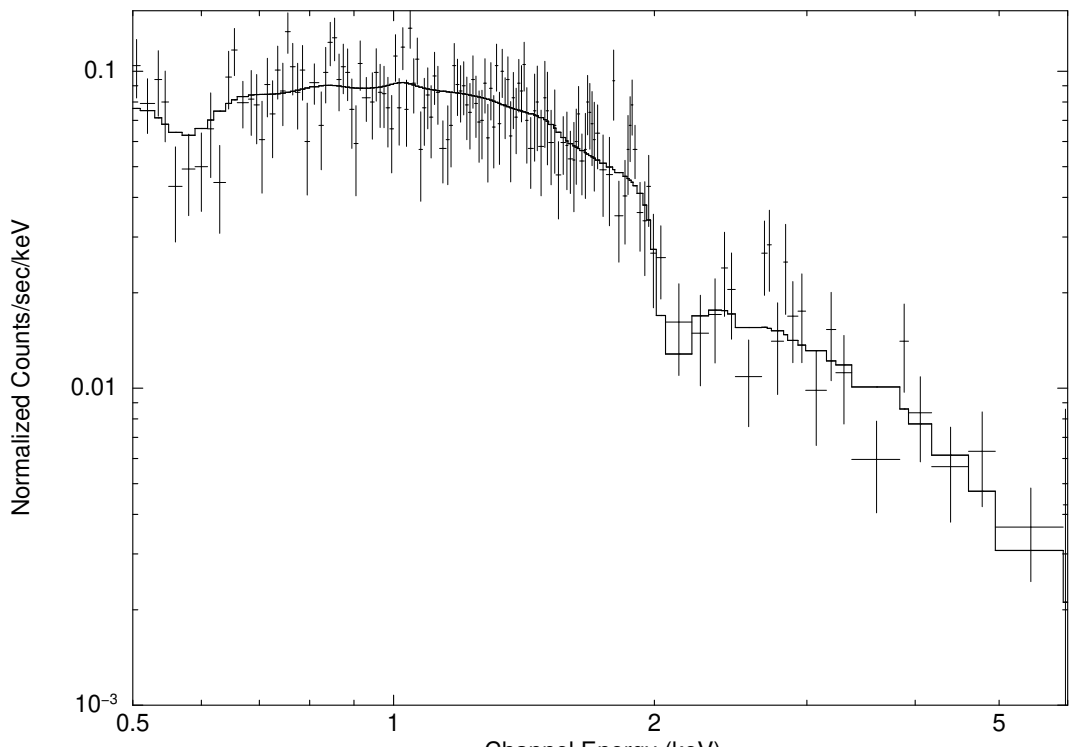

**Figure 44.** Webspec simulated spectra of a cataclysmic variable with *AXIS* using an exposure time of 120ks. The model includes an absorbed thermal plasma model (zphabs*vmekal) with a temperature of 34 keV, $n_H = 16 \times 10^{20}$ cm$^{-2}$, and abundances corresponding to Omega Cen in the energy range $0.5 - 6$ keV. Left: on-axis. Right: off-axis simulation. The parameters used are based on CV identifications by [239] using 222ks of Chandra ACIS-I observations.

*Omega Cen as a precursor to study low-density globular clusters*

Currently, most of the X-ray sources identified in globular clusters have been detected using Chandra, thanks to its sub-arcsecond on-axis resolution. However, identifying the optical counterparts of these X-ray sources remains a challenging task due to extreme stellar crowding, especially in the cores of the clusters. Even with facilities such as HST, multiple stars within the Chandra error circle often appear to be plausible counterparts.

While *AXIS* is expected to have a larger on-axis PSF than Chandra, its much flatter PSF across the field of view makes it a powerful tool for studying globular clusters, particularly those with a high number of known X-ray sources and those with relatively low stellar densities, such as Omega Cen.

This survey of Omega Cen will not only enhance our understanding of this particular cluster, but also serve as a precursor to investigate the compact binary population in other low-density globular clusters that have been poorly explored by Chandra or other X-ray telescopes so far. This will allow us to expand the parameter space for globular cluster studies beyond the traditionally studied clusters. Studying low stellar density globular clusters will also facilitate the identification of counterparts to *AXIS* X-ray sources at UV, optical, IR, and radio wavelengths compared to denser clusters.

**[Exposure time (ks):]** 120 ks

**Observing description:** To reach a luminosity of $1 \times 10^{29}$ erg s$^{-1}$ at the distance of the cluster (assumed to be 5.2 kpc), we need to reach a flux of at least $3 \times 10^{-17}$ erg cm$^{-2}$ s$^{-1}$. Based on the *AXIS* sensitivity curve that flux can be reached in the $0.5 - 2$keV band with an exposure time of around 120 ks when considering a power law model with $\Gamma = 2$ on-axis. Under the same conditions, a luminosity of around $2 \times 10^{29}$ erg $s^{-1}$ will be reached over the entire FOV. Fainter fluxes can be reached for a thermal model. We propose an initial deep exposure of 40ks, allowing us to reach luminosities of $\sim 2.6 \times 10^{29}$ erg $s^{-1}$. This will permit the detection of faint objects early on for multi-wavelength follow-up as well as period determination. To identify longer-term variability (e.g., due to outbursts), longer-term monitoring would be necessary. Considering that different populations of CBs have different recurrence times, a good compromise would be to monitor the cluster every 2-3 months with integrations of $\sim$ 20 ks such that at the end of a year we would reach a luminosity of $1 \times 10^{29}$ erg $s^{-1}$, an order of magnitude deeper than with current observations.

**[Joint Observations and synergies with other observatories in the 2030s:]** To further classify and characterize systems, we will need synergies with Roman, HST, LSST, JWST, VLA, and other optical/IR/radio observatories.

# Part IV
# Diffuse Emission

### h. Supernova Remnants

*29. Rapid X-ray time variability at supernova remnant shocks*

**First Author:** Federico Fraschetti (Harvard, CfA; federico.fraschetti@cfa.harvard.edu)
**Co-authors:** Martin Mayer (FAU Erlangen, Germany), Anne Decourchelle (CEA, Saclay), Samar Safi-Harb (U. of Manitoba), Gilles Ferrand (U. of Manitoba/RIKEN)
**Abstract:** Galactic Supernova Remnants (SNRs) have been used as a space laboratory to study the time evolution of high-speed, non-relativistic, magnetized plasma plunging into turbulent Interstellar Medium (ISM). SNRs are also regarded as major source of Galactic Cosmic Rays (CRs) via the shock waves driven by the super-Alfvenic plasma outflowing in the aftermath of the SN explosion. The rapid time variability (on a few-year timescale) of thin, non-thermal X-ray rims in SNR shocks, assumed to be produced by shock-accelerated relativistic electrons via synchrotron emission, was interpreted as a tracer of magnetic field evolution in the plasma behind the large-scale forward shock. Moreover, the inner regions of SNRs have shown comparable time-scale variability in the X-ray flux at a number of inward-moving structures, which was interpreted as the post-shock magnetic field rapidly amplifying due to the shock-crossing of inhomogeneities within the remnant. However, other mechanisms of amplification requiring an energetically non-negligible population of multi-TeV ions (CRs) are consistent with observations. The rapid fading of the X-ray rims could be due to a damping of the magnetic field itself in the downstream of the shock; $X-$ray observations from the past decade were not able to favor either scenario. Such rapid time variability has been sampled in only a limited number of instances for historical SNRs, making it challenging to identify the relative role of these amplification mechanisms. *AXIS* will overcome this limitation with a much shorter time exposure than needed for Chandra.
**Science:**

Galactic Supernova Remnants (SNRs) have been used as a space laboratory to study the time evolution of high-speed, non-relativistic, magnetized plasma plunging into turbulent Interstellar Medium (ISM). $X-$ray observations have been essential in interpreting the non-thermal processes occurring in these systems. SNRs are regarded as a major source of the CRs approximately up to the "knee" ($\sim 10^{15}$ eV), or $\sim 10^{18}$ eV, of the overall CRs spectrum. The general consensus is that charged particles are accelerated by the shock wave driven by the super-Alfvenic plasma outflowing in the aftermath of the SN explosion. The connection between CRs and magnetic field at SNRs has been subject of intense theoretical and observational investigation for decades: CRs feedback on the dynamics of the expanding shock modifies the upstream plasma dynamics (density compression and magnetic field at the shock); however, the rate of the small-scale dynamo triggered downstream of the shock by the ineherent (namely pre-existing to the SN explosion) fluctuations as the shock plunges into circumstellar and interstellar medium might exceed the CRs acceleration rate. Only observations combining $\sim$ arcsec spatial resolution with sufficiently large field of view can help significantly boost the interpretation of multi-wavelength observations and disentangle the two effects described above that are likely to cooperate synergycally.

Turbulence in the ISM is ubiquitously observed via fluctuations in the amplitude and phase of scattered radio waves and spreads over 10 orders of magnitude in length scale, from hundreds of km to hundreds of parsecs [29], consistently with *in-situ* measurements of Voyagers spacecraft in the pristine ISM [94]. The power spectrum of the fluctuations surprisingly follows a single power law throughout the range of scales with a slope $\sim 5/3$ as predicted by Kolmogorov for homogeneous turbulence with a unique outer injection scale [537]. The interpretation of SNR's non-thermal X-ray observations is likely to require accounting for the ISM turbulence.

The rapid time variability of thin, non-thermal X-ray rims in SNR shocks, assumed to be produced by shock-accelerated relativistic electrons via synchrotron emission, has been regarded as a tracer of magnetic field evolution in the post-shock plasma. In the SNR RX J1713.7 – 3946 [584], the year time scale variability of X-ray emission was explained via extremely short synchrotron cooling time of relativistic electrons in a compressed and amplified magnetic field at the forward shock. As a second instance, high spatial resolution Chandra multi-epoch observations of SNR Cassiopeia A used proper motion analysis to show evidence of a surprisingly large time variability of the X-ray flux (up to 50%) in several regions located toward the center of the remnant [519] over a time period of 15 years. These variations were interpreted as a turbulent dynamo amplification of the magnetic field as the shock traverses the inhomogeneous ISM [182].

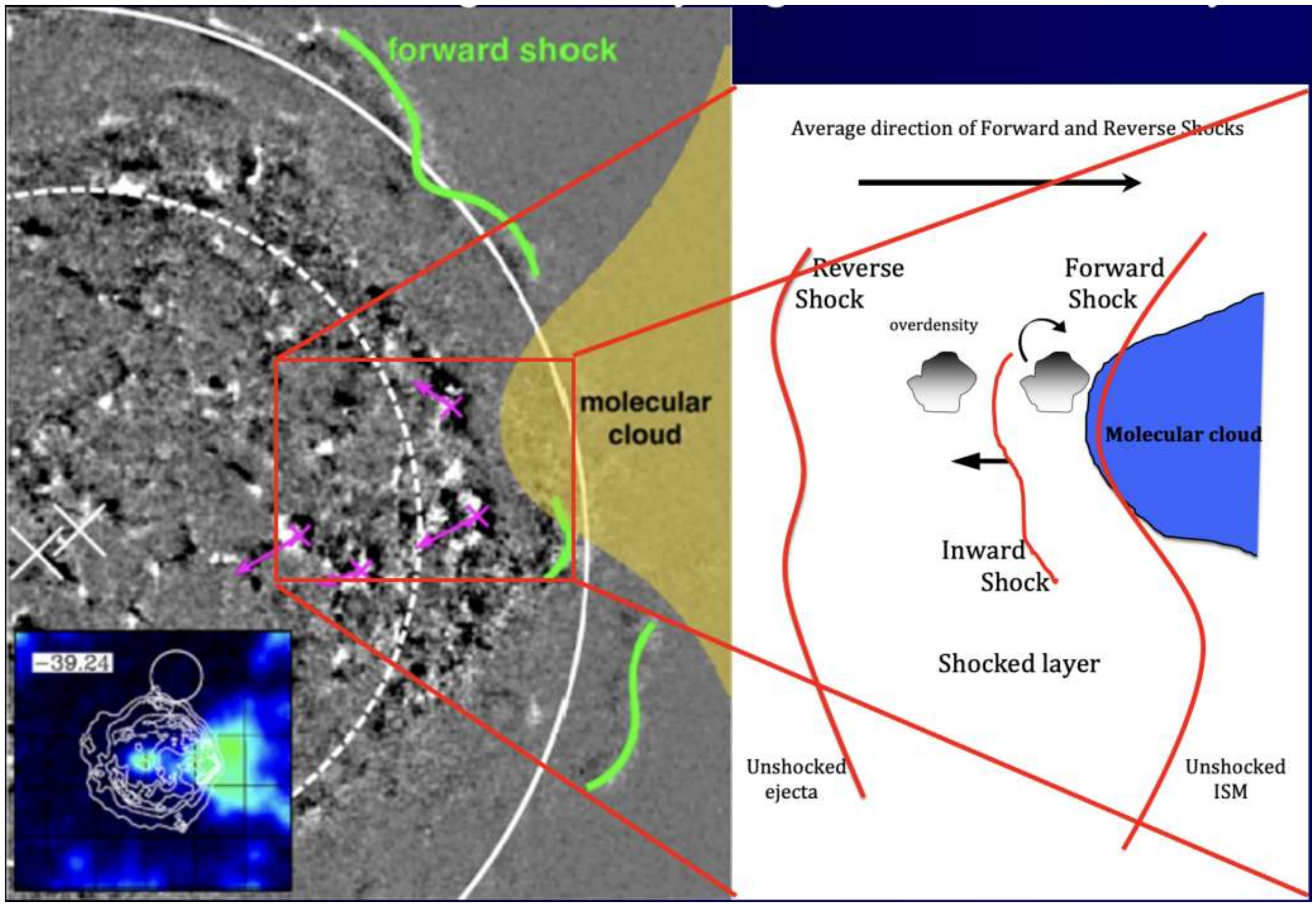

**Figure 45.** Left: Cartoon for the shock–cloud interaction inside the Cassiopeia-A SNR [519]. Magenta arrows show the proper-motion directions of the inward shocks. Thick green lines indicate some forward shocks. Dashed and solid white circles indicate the radii of the reverse shock and forward shock. The small inset on left bottom shows the CO-map with the velocity of -39.24 km $s^{-1}$. Right: Cartoon illustration of the scenario for inward shocks in Cassiopeia-A SNR [182]: the inward shock recedes toward the center of Supernova explosion and crosses outward overdensity clumps thereby generating vorticity and magnetic field enhancement in the downstream fluid. The arrows indicate the direction of the shocks motion in the observer frame.

[487] collected radio, infrared and X-ray observations of 6 young SNRs (G1.9+0.3, Cassiopeia A, Kepler, Tycho, SN 1006, and RCW 86, spanning an age from 100 to 1800 years) to infer, via a spatial average determination of the relevant shock parameters (radio luminosity, plasma density, and shock velocity $V_{sh}$), the energy ratios in non-thermal electrons and magnetic field to the ram pressure, $\epsilon_e$ and $\epsilon_B$, respectively.

The broad range of variability and lack of trends of $\epsilon_e$ and $\epsilon_B$ as a function of $V_{sh}$ is likely to conceal, as discussed by [487], dependence on additional factors, e.g., neutrals in the upstream or pre-existing turbulence. A multi-epoch (2001 and 2013) analysis of Chandra data for several thermal and non-thermal filaments in the South-Western limb of RCW 86 was conducted by [555] to determine proper motion and flux time-variations. The synchrotron emission from the thermal-dominated filaments is highly correlated with the ambient density and only weakly correlated with the shock speed (ranging from ∼200 to ∼2,000 km/sec) and shock magnetic obliquity (ranging from ∼0 to ∼ 60°). The interpretation by [555] is that, in the framework of a shock-cloud interaction scenario, local turbulence conditions determine the particle acceleration more than previously thought.

The mechanism of the magnetic field amplification, one or two orders of magnitude beyond the expectation from the magnetohydrodynamic jump conditions, is an open question: late-time (10-100 years after SN explosion) shock evolution generates turbulence ahead of the shock via the streaming of energized ions that excite modes therein at a variety of length-scales, both resonant and non-resonant [57], enhancing the turbulent magnetic field ahead of the shock. In several observational instances, the relative role of this amplification mechanism compared with the small-scale dynamo has not been robustly identified. In addition, the rapid fading of the X-ray rims could be due to a damping of the magnetic field itself in the downstream of the shock; $X-$ray observations from the past decade were not able to favor either scenario [481,571].

The rapid (∼ years) time scale of evolution of thin non-thermal structures was interpreted as magnetic field amplification. Due to magnetic enhancement at the shock and the quadratic dependence of the synchrotron power on the magnetic strength, X-rays observations could only probe the downstream region very close to the shock. The exquisite spatial resolution of *Chandra* has allowed, over the past 25 years, the identification of thin X-ray filaments, or rims, in a large number of SNRs. The small thickness of the filaments was interpreted via the rapid synchrotron cooling time of the electrons in the highly amplified magnetic field. Since the precise location of the rims was not determined, both mechanisms described above —small-scale dynamo and CRs-induced instabilities —can operate. Only a larger sample of X-ray rims with near-arcsecond spatial resolution, comparable to *Chandra*, can provide deeper insight into this scientific question. Thus, these observations need the capabilities of AXIS.

Limitations from the Chandra telescopes (effective area), if overcome by AXIS, could help address the origin of the non-thermal $X-$ray rims with a much shorter time exposure than needed for Chandra. With an effective area at 1 keV about 10 times larger than *Chandra*, *AXIS* is expected to reduce the 30 ks exposure in SNR RX J1713.7 – 3946 [584] down to about 10 ks

**[Exposure time (ks):]** 10 ks per exposure per source; with 3 exposures separated by 3 to 5 years and for 2 or 3 sources (to be identified), this amounts to a total exposure of ∼60–90 ks.

**[Joint Observations and synergies with other observatories in the 2030s:]** Synergies with multi-wavelength facilities will be instrumental: (N)IR observations (JWST or ALMA) can confirm detection of clumpy medium in front of the expanding shock. Gamma-ray observations (HESS, Fermi, VERITAS), if hadronic origin, can probe both the maximal energy of the accelerated ions as well as surrounding clumpy ISM via pion decay.

*30. Expansion and dynamics of supernova remnants*

**First Author:** Martin Mayer (FAU Erlangen, Germany; mgf.mayer@fau.de),
**Co-authors:** Manami Sasaki (FAU), Samar Safi-Harb (U. of Manitoba), Werner Becker (MPE, Germany), Naomi Tsuji (ICRR), Federico Fraschetti (Harvard, CfA), Gilles Ferrand (U. of Manitoba / RIKEN), Daniel Castro (CfA, USA), Anne Decourchelle (CEA, Saclay)
**Abstract:** Many Galactic supernova remnants (SNRs) are young enough for their expansion to be observable on timescales of years to decades. High-resolution X-ray observations spanning sufficiently long temporal baselines provide a means to measure the velocity of the forward shock of SNRs expanding into the surrounding interstellar medium (ISM), as has been demonstrated for several young SNRs ($\lesssim 2000$ yr) with Chandra. The *AXIS* observatory will enable the astronomical community to build upon this legacy, exploiting temporal baselines around 30 years in conjunction with archival Chandra data, to precisely measure forward shock speeds for numerous SNRs. On one hand, such observations will yield accurate constraints on the expansion rate, and thereby the likely age, for older (i.e., less rapidly expanding) and more distant SNRs, with a fraction of the exposure time required by Chandra. On the other hand, for young SNRs, this will enable the detection of temporal variations in shock velocity. This may be particularly relevant in regions of interaction with dense circumstellar material, where the shock wave experiences strong deceleration, resulting in significant deviations from analytical evolutionary models. Especially in such dense environments and in young SNRs, long temporal baselines are also expected to reveal intrinsic variability in the X-ray emission, caused by fluctuations in density or magnetic field.
**Science:** Supernovae (SNe) are among the most violent explosions in the universe, releasing energies around $10^{51}$ erg into the surrounding interstellar medium (ISM), mostly in the form of kinetic energy of ejecta. While their optical emission typically fades within a few months, the impact of the resulting supernova remnant (SNR) on the ISM lasts for around $10^5$ yr, as the shock wave of the explosion expands outward. This blast wave compresses and heats the surrounding ISM, which in turn decelerates the expansion over time. While no Galactic SN has been directly observed for over 400 years, around 300–400 Milky-Way SNRs [175,205] are known, many of which exhibit shock waves propagating at speeds of the order of 1000 km/s. Given these rapid expansion speeds, morphological changes are often detectable in imaging of SNRs, provided that observations are sufficiently spaced in time. In the optical band, such observations are usually the most precise, given the available spatial resolution, but require the presence of optically bright features. As these typically trace very dense clumps or filaments cooling radiatively, motion is almost uninhibited by interaction with ISM, allowing for precise age determinations of SNRs [176,178,565,573]. Alternatively, synchrotron emission from SNR shells is pronounced in the radio band, where the propagation of SNR shock fronts through the ISM accelerates particles [e.g., 67,206].

In contrast to the optical and radio bands, X-ray observations trace the thermal emission from very hot and tenuous shocked gas ($T \sim 10^7$ K) in an SNR. Given its low density compared to radiative optically visible clumps, the expansion of SNRs in the X-ray band appears more decelerated than in the optical, providing clues not only on their ages [e.g. 6,79,81,361,577], but also their evolutionary stages, and the density and structure of the surrounding ISM [e.g, 251,431,594]. In addition to thermal emission, for young and rapidly expanding SNRs ($v \gtrsim 2000$ km/s), X-ray synchrotron emission may be produced by electrons accelerated to TeV energies, tracing the propagation of the SNR blast wave.

In the X-ray band, high-resolution imaging was revolutionized by the launch of *Chandra* in 1999, with its unparalleled sub-arcsecond point spread function (PSF) enabling the measurement of proper motion in X-ray bright filaments in SNRs for the first time. The *AXIS* mission has the unique potential to extend the legacy of *Chandra* with a relatively small observational effort, by building on its more than 25 years of archival data. Being the first X-ray imaging instrument matching the imaging resolution of *Chandra*, *AXIS* observations in the 2030s, combined with those of its predecessor, will allow for establishing temporal baselines of $> 30$ yr for probing morphological changes in expanding SNRs. A key factor enabling this effort is not only the uniform PSF of *AXIS*, but also its ability to detect many serendipitous point sources, allowing for a precise astrometric alignment of X-ray images with external reference frames. On one hand, *AXIS* will improve the achievable sensitivity in the detection of small displacements of expanding SNRs. While baselines around 10 yr, typical for *Chandra*, allowed for establishing the expansion of rather young ($\lesssim 2000$ yr) SNRs [e.g. 79,81,577], combined observations will likely allow for measuring the slower expansion of well-evolved remnants ($\sim 10\,000$ yr). Naturally, such a measurement of the expansion rate of an SNR can provide good estimates of its age, in combination with evolutionary models [e.g., 576], as well as extremely reliable upper limits, assuming free expansion. Furthermore, precise shock velocity measurements are crucial for constraining the processes of collisionless shock heating [e.g., 198,315,626] and particle acceleration [e.g., 485,517,578,653], which can be probed in conjunction with measurements of thermal and nonthermal X-ray spectra, respectively. Finally, if available, comparisons with expansion measurements from the radio or optical [e.g., 201,251,431,594,619] can constrain inhomogeneities in the surrounding ISM, and the resulting asymmetries induced in dense clumps, hot gas, and energetic electrons.

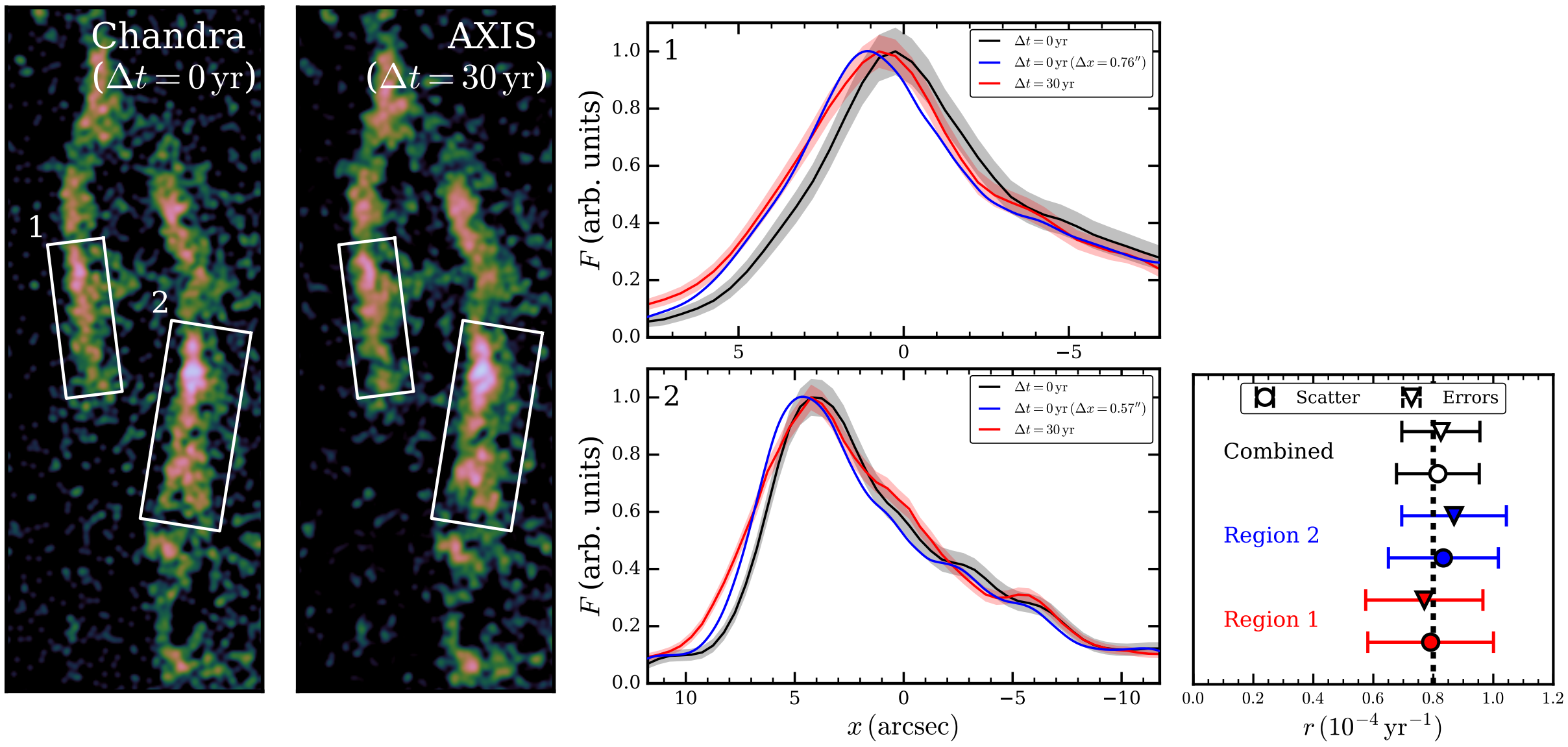

**Figure 46.** Simulated expansion measurement of Kes 79. The left panels show the two input images of *Chandra* and *AXIS* in the $0.5 - 7.0$ keV band. The boxes mark the two regions in which the motion of the filaments is traced. The center panels display the quantitative expansion measurement, comparing the integrated flux profiles. In each panel, the black (red) lines represent the smoothed early (late) count-rate profiles with $\Delta t = 30$ yr. The shaded region indicates the estimated error, and the blue line shows the early-time profile, shifted to ideally match the late-time data. The right panel shows the summarized constraints on the fractional expansion rate $r$, obtained from the two regions individually, and combined. Triangle markers indicate the weighted mean, and errors determined directly from fitting the two profiles to each other, whereas circles mark the mean and observed scatter across 100 simulations.

We illustrate the feasibility of such an expansion measurement for an evolved SNR, using Kes 79 as a representative example. Given its age of $\tau \approx 5000$ yr [651], and an evolution assumed to be following the Sedov-Taylor solution [e.g., 576], its fractional expansion rate $r$ is expected to be on the order of

$$r = \frac{\mathrm{d}R}{R\,\mathrm{d}t} \approx \frac{2}{5\tau} = 8 \times 10^{-5}\,\mathrm{yr}^{-1}, \tag{2}$$

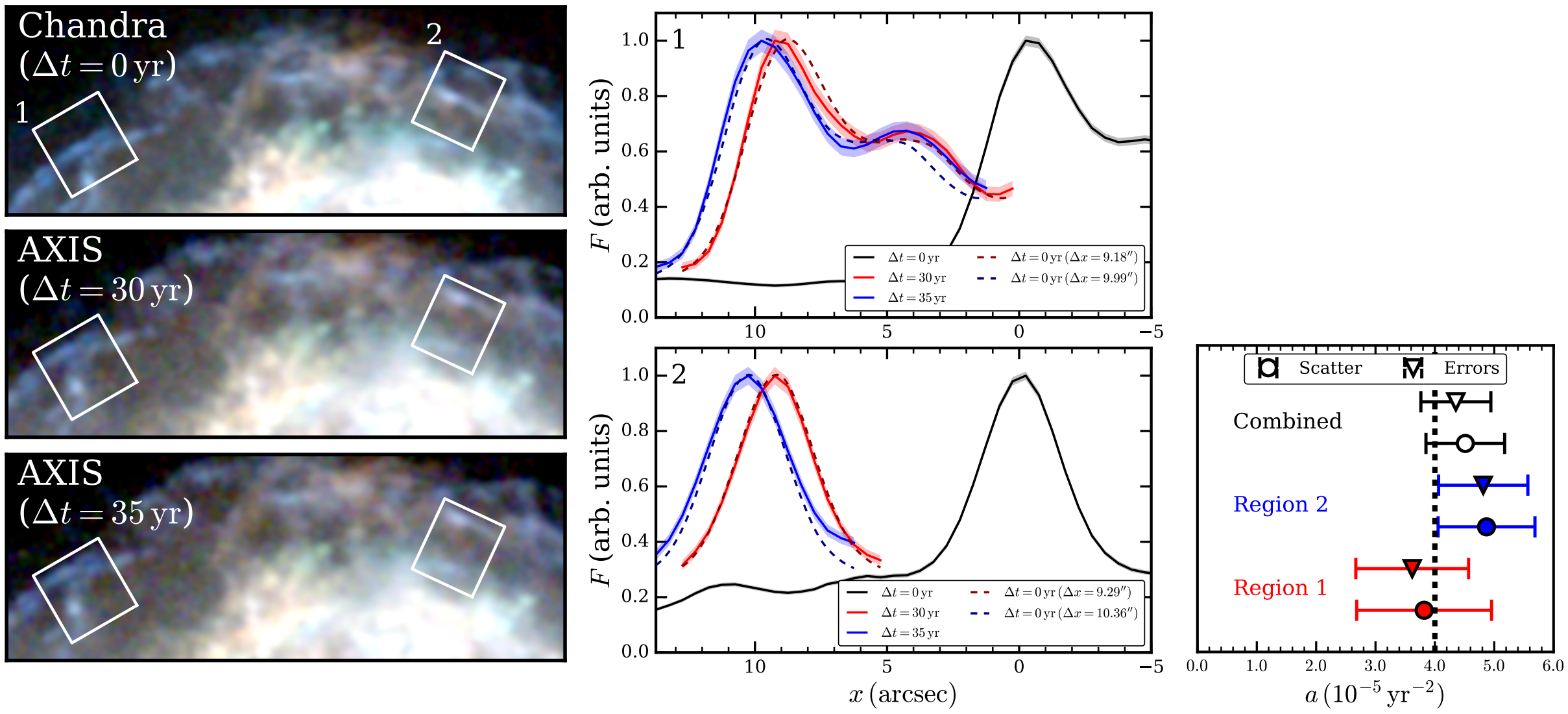

**Figure 47.** Simulated expansion deceleration measurement of Cas A. The left panels show the three input images of *Chandra* and *AXIS* in a false-color representation ($0.5 - 1.2, 1.2 - 2.0, 2.0 - 7.0$ keV). The boxes mark the two regions in which the motion of the filaments is traced in the hard band. The center panels display the quantitative expansion measurement, in which the emission profiles are compared. In each panel, the black line traces the smoothed count-rate profile from the early epoch. In contrast, the red and blue lines indicate the simulated late profiles with $\Delta t = 30$ yr and $\Delta t = 35$ yr, with shaded areas indicating estimated errors. The corresponding dashed lines show the early-time profile, shifted to match the two later epochs ideally. The right panel shows the summarized result of our effort, in the form of constraints on the normalized deceleration rate $a$, as in Fig. 46.

where $R(t)$ describes the evolution of the shock-wave radius over time. In combination with a 30 ks early *Chandra* observation [527,554], we demonstrate the potential of a typical temporal baseline by simulating a 15 ks *AXIS* observation $\Delta t = 30$ yr later, using the `SIXTE` simulator [142]. We artificially inject the expansion into our data by assuming the morphology observed with *Chandra*, and stretching the image by a factor $r\,\Delta t = 2.4 \times 10^{-3}$. We visualize the resulting images and the output of a simplistic expansion measurement [as in 361] in Fig. 46. We focused our efforts on two relatively straight filamentary features located east of Kes 79, where we compared the projected flux profiles perpendicular to the extent of the filament and determined the offset that minimized relative differences. We estimated the intrinsic error of our expansion rate constraint by performing 100 statistically independent simulations and using their standard deviation to constrain our uncertainty. This effort demonstrates that we would be able to recover the input fractional expansion rate ($r = 8 \times 10^{-5}\,\mathrm{yr}^{-1}$) accurately, rejecting the null hypothesis of no expansion at a high significance of $6\sigma$, and obtaining a five times smaller error on $r$ than previous efforts based on *Chandra* alone [361].

For younger rapidly expanding SNRs, the $\geq$ 30-year baselines enabled by *Chandra* and *AXIS* will enable the detection of deviations from expansion at constant speed. Deceleration of the forward shock is naturally expected even in one-dimensional evolutionary models of SNRs, but may be much larger if the shock wave encounters density peaks, for instance, at interstellar clouds. An interesting example, in which this process may be pronounced, is the archetypical 350-year-old SNR Cas A [176]. *Chandra*-based constraints on its normalized deceleration rate

$$a = -\frac{\mathrm{d}^2 R}{R\,\mathrm{d}t^2} \tag{3}$$

were recently published, reporting significant deceleration and acceleration ($|a| \sim 5 \times 10^{-5}\,\mathrm{yr}^{-2}$) in different regions [596]. Constraints on shock front deceleration are not only crucial for the determination of an SNR's evolution, explosion site, and age, but they may also provide constraints on the interstellar densities probed by the shock front. Such constraints can be compared to the observed distributions of cold gas, traced, for instance, by HI emission, or to the hot plasma densities inferred from X-ray spectra, thereby improving our understanding of shock propagation on small scales. In contrast, a potentially accelerating forward shock could be a smoking gun for an SNR breaking out of a circumstellar shell [596].

We illustrate the potential of performing such a measurement for Cas A in conjunction with an archival early *Chandra* dataset [252], by using `SIXTE` [142] to simulate two 10 ks *AXIS* observations, $\Delta t = 30\,\mathrm{yr}$ and 35 yr after the original observation, respectively. Similarly as above, we injected our "signal" into the simulation by stretching the input image following the forward shock expansion rate $r \approx 2.0 \times 10^{-3}\,\mathrm{yr}^{-1}$ [596], but reduced this amount corresponding to a constant deceleration at $a = 4 \times 10^{-5}\,\mathrm{yr}^{-2}$. Given the synchrotron-dominated nature of the sharp filaments tracing the forward shock [595], we limited our experiment to the hard band $2.0 - 7.0\,\mathrm{keV}$, where we measured the forward-shock offset between the original and the two respective simulated images, as visualized in Fig. 47. As displayed, the deceleration rate determined by the three measurements recovers the injected value within the errors, with a detection significance around $6\sigma$. This demonstrates that, given the presence of a single baseline *Chandra* observation, changes in shock velocity can be observed by *AXIS* within its nominal mission lifetime of five years.

A somewhat related science case enabled by such long baselines is the search for variability in the diffuse X-ray emission of an SNR. Rapid temporal changes of nonthermal synchrotron emission are expected as a product of magnetic field fluctuations, and are discussed in detail in Sect. 29. In contrast, variability in thermal emission is particularly important in young SNRs in strong ionization non-equilibrium (due to shock heating or adiabatic cooling), or blast waves encountering rapid density changes. Thus far, there have been few detections of such fluctuations in thermal emission [360,430,504], but the 30-year baseline and high spatial resolution available with *AXIS* and *Chandra* will greatly increase our ability to search for brightness changes in individual SNR features.

**Exposure time:** $\sim$35 ks; $10 - 20$ ks per observation

**Observing description:** In principle, any SNR younger than $\sim 10^4$ yr is a suitable target, if sufficiently deep *Chandra* observations are available to resolve compact clumps or filaments whose motion can be traced. Since the main strength of this effort lies in the combination of data over very long temporal baselines, observations matching the sensitivity of archival *Chandra* data are sufficient to achieve our goal of obtaining precise image-plane velocities of both the shocked ambient and ejecta material. Given the enormous improvement in effective area with respect to *Chandra*, $10 - 20$ ks exposures with *AXIS* should typically be sufficient to match or exceed the quality of archival images. In order to measure linear SNR expansion velocities, a single observation at any point during the *AXIS* mission lifetime would be sufficient. For the goal of measuring deviations from linear expansion in young SNRs, or standalone expansion measurements with *AXIS*, multiple observations would be necessary, ideally spaced by at least five years.

**Joint Observations and synergies with other observatories in the 2030s:** Synergies exist with multi-wavelength facilities which establish sufficiently long temporal baselines for expansion measurements, for instance in the optical or radio bands. Additionally, line-of-sight velocity measurements carried out via high-resolution X-ray spectroscopy, e.g., with *NewAthena* X-IFU [440], could be combined with image-plane constraints, to obtain precise 3D velocity fields of SNRs [as in 201].

### i. Other Remnants (Novae, Kilonovae)

*31. An AXIS lens on kilonovae through their remnants*

**First Author:** Nathan Steinle (U. of Manitoba, nathan.steinle@umanitoba.ca)
**Co-authors:** Austin MacMaster (U. of Manitoba), Isabel Sander (U. of Manitoba), Neil Doerksen (U. of Manitoba), Samar Safi-Harb (U. of Manitoba), Chris Fryer (LANL), Aya Bamba (U. Tokyo), Yukikatsu Terada (Saitama U.), Shin-ichiro Fujimoto (NITKC), Liliana Rivera Sandoval (UTRGV), Kevin Burdge (MIT)
**Abstract:** The coincident detection of gravitational and electromagnetic radiation from the coalescence of the binary neutron star (BNS) GW170817 ushered in a new era of multi-wavelength, multi-messenger astrophysics, where Chandra's continued monitoring uniquely tracked the evolution of the kilonova for nearly the last decade. Future gravitational wave (GW) facilities are expected to detect more BNSs, and follow-up high-resolution X-ray observations will be needed for their crucial role in understanding these extreme astrophysical events. As the kilonova ejecta and surrounding interstellar medium interact, the kilonova evolves into the kilonova remnant (KNR) phase, producing thermal X-ray emission arising from shock-heated material as well as the decay of heavy r-process elements. *AXIS*, thanks to its ToO capabilities, superior sensitivity, and high-angular resolution across its large FoV, will open a new discovery window into KNRs and their diversity of engines analogous to the case of supernova remnants. Utilizing new continuum and line emission models of KNRs, we propose that the *AXIS* mission will reveal sources in the Galactic plane and track the evolution of kilonovae, such as the nearby GW170817 event, into the KNR phase, thereby probing the emerging class of KNRs for the first time. Furthermore, we demonstrate how these *AXIS* sources will form a population of KNRs through binary population synthesis and hydrodynamical simulations of the KNRs' physical evolution. *AXIS* will directly probe the joint BNS-KNR population, such as providing crucial information for ToOs of gravitational wave detectors and estimating the dependence of KNR rates on BNS rates, which will ultimately revolutionize multi-messenger astronomy.
**Science:** The gravitational waves (GWs) from the first binary neutron star (BNS) merger GW170817 were accompanied with electromagnetic (EM) counterpart signals [2]. In the neutron star merger process, significant amounts of neutron star material, along with other components such as the post-merger accretion disk wind, were ejected at high velocities ($\sim$0.1—0.3c). These ejecta underwent rapid neutron capture (r-process) nucleosynthesis, and the radioactive decay of newly formed heavy elements (like gold, platinum, lanthanides) powered the kilonova emission [119,165]. For GW170817, the kilonova AT2017gfo was first seen about 10 hours after the GW detection and faded from optical view after $\sim$2—3 weeks. Subsequently, GW170817 has been detected in the X-ray and radio regimes and has been monitored since [10,409]. Radio and X-ray observations some 940 days post-merger suggest energy injection by a long-lived central engine and the onset of a kilonova afterglow arising from the interaction of the sub-relativistic merger ejecta with the surrounding medium [574]. The ejecta mass from GW170817 was constrained to be $\sim 0.05\ M_\odot$, see [404] and references therein. Continued combined Chandra X-ray and radio observations of GW170817 over the past several years have confirmed the presence of a structured jet, with orientation influencing observed emissions; detected persistent X-ray emission, consistent with off-axis gamma-ray burst afterglows; and provided constraints on late-time emissions, ruling out significant re-brightening in radio wavelengths and indicating a possible additional energy injection at late times [280,354]. Chandra, Hubble, and radio telescopes continue to observe the vicinity of GW170817 for emission from shocked material or signs of a compact object.

Just as supernovae evolve into supernova remnants, the ejected material in a kilonova ultimately evolves into a diffuse remnant – which we refer to as a kilonova remnant (KNR) – resulting from the interaction of the merger ejecta with the interstellar or circumstellar medium. However, compared to a supernova, the kilonova ejecta is more neutron-rich and less massive, causing a faster evolution into the remnant phase when compared to supernova remnant evolution [507]. This leads to key differences in the light curves, spectra, and chemical yields of the kilonova remnant evolution [7].

As the kilonova evolves into its remnant phase, the possibility of thermal X-rays arises from the interaction between the merger ejecta and the surrounding interstellar medium [507], similar to that observed in supernova remnants. The shorter evolution timescales for KNRs imply that we can trace their evolution over a ToO spanning a few hundred days to a few years, corresponding to multiple *AXIS* observational cycles. *AXIS*, thanks to its ToO capabilities and high-angular resolution across its large FoV [482,509], will provide a new lens for viewing the remnants of kilonovae, probing the engines that cause the kilonova and the resulting outflows. In this science case we demonstrate (i) how *AXIS* will reveal for the first time thermal X-ray emission from KNRs arising from the interaction of the merger ejecta with the surrounding medium and the emission from r-process nucleosynthesis products, and (ii) how the properties of the KNR shock evolution are related to the masses of a BNS population.

To model the line emission, we employ full nuclear-network numerical calculations based on [563]. Recent observations of thermal kilonova emission from ejecta produced in BNS mergers have established BNS mergers as a promising r-process site, possibly even more important than the standard case involving core-collapse supernovae. In the *AXIS* X-ray band, it is not possible to detect direct gamma rays from r-process nuclei [563], but we can expect X-rays via the internal conversion process. Figure 48 shows X-ray emission from r-process nuclei in nearby Galactic KNRs and the evolution of the X-ray emission as the KNR ages. Although shallow Galactic plane survey exposures of ∼6 ksec may not reach the sensitivity needed, *AXIS* has the potential to make the first detection of the r-process signal from KNRs with a deep exposure. In addition, the detection of lines around 4 keV becomes the first direct observational evidence of the neutron-rich environment in the r-process site. Therefore, this is a unique probe for studying the r-process from KNRs and establishing synergy of late-time X-ray observations with GW observations in the multi-messenger era.

*AXIS* will also be sensitive and crucial to probe GW170817-like events. Firstly, as demonstrated by Chandra, the angular resolution of *AXIS* will be essential to resolve the GW event from nearby sources or the host galaxy's engine. Furthermore, utilizing the existing Chandra observations of the peak X-ray emission from GW170817, we estimate an *AXIS* predicted count rate of $1.5{\times}10^{-2}$ counts per second (i.e., ∼10 times higher than the Chandra ACIS count rate), thus enabling us to not only better constrain the spectrum which revealed hints of thermal X-ray emission [507], but also be sensitive to GW sources that are 10 times fainter with a flux of $\sim 10^{-15}$ erg cm$^{-2}$ s$^{-1}$. Furthermore, with the planned Galactic Plane Survey, a 6 ksec snapshot will yield ∼100 counts from a GW170817-like source spectrum, sufficient for a detection and initial spectral characterization.

In coordination with future GW detectors, *AXIS* will reveal the population of BNSs and KNRs in a multi-messenger network that will require new theoretical modeling frameworks for GW detectors and for *AXIS* to constrain the system's total evolution jointly, from the stellar progenitors to the later-time remnants. This would require a model for translating the BNS properties (i.e., intrinsic parameters like masses and period, and extrinsic parameters like sky location and inclination) into *AXIS* observables (i.e., flux or luminosity). Although such a model is currently beyond reach, we motivate its development by utilizing existing codes to demonstrate the multi-messenger KNR view afforded with *AXIS*. We obtain a BNS population with the rapid binary population synthesis code COMPAS [562], and we use their BNS properties (chirp mass, mass ratio, and inclination) as inputs for the EM transient code MOSFiT [210] to obtain the corresponding kilonova ejecta masses ($m_{\rm ejecta}$) and velocities ($v_{\rm ejecta}$). Then we use these ejecta masses and velocities as inputs for the KNR evolution, which we model with a modified version of the hydrodynamical simulation of supernova remnant evolution [185] with the energetics rescaled for KNR physics, analogous to Fig. 5 of [507]. We employ multivariate linear least-squares regression on the inputs and outputs from the hydrodynamical simulation to compute a fit for the relation between two main initial KNR parameters and three properties of the KNR shock evolution (velocity, position, and density), i.e., a map $(m_{\rm ejecta}, v_{\rm ejecta}) \rightarrow (v_{\rm shock}, r_{\rm shock}, \rho_{\rm shock})$ assuming the same interstellar-medium density for each system. Combining these ingredients, we arrive at an end-to-end pipeline for correlations between the progenitor BNS and resultant KNR properties, as shown in Figure 49. This demonstrates how different combinations of the BNS masses produce non-trivial distributions of ($m_{\rm ejecta}$ and $v_{\rm ejecta}$), translating into unique predictions for the KNR shock properties. The evolution of the shock (shown by red, blue, and green corresponding to 100, 1000, and 10,000 days, respectively) highlights the classic sequence of free expansion, adiabatic, and radiation-dominated phases and its dependence on the BNS mass parameters. Ultimately, one would compute *AXIS* and GW detectability of these systems, where frameworks such as this will provide precious insights, e.g., complicated dependence of the observable rates of KNRs on BNS merger rates and astrophysical progenitor evolutionary channels.

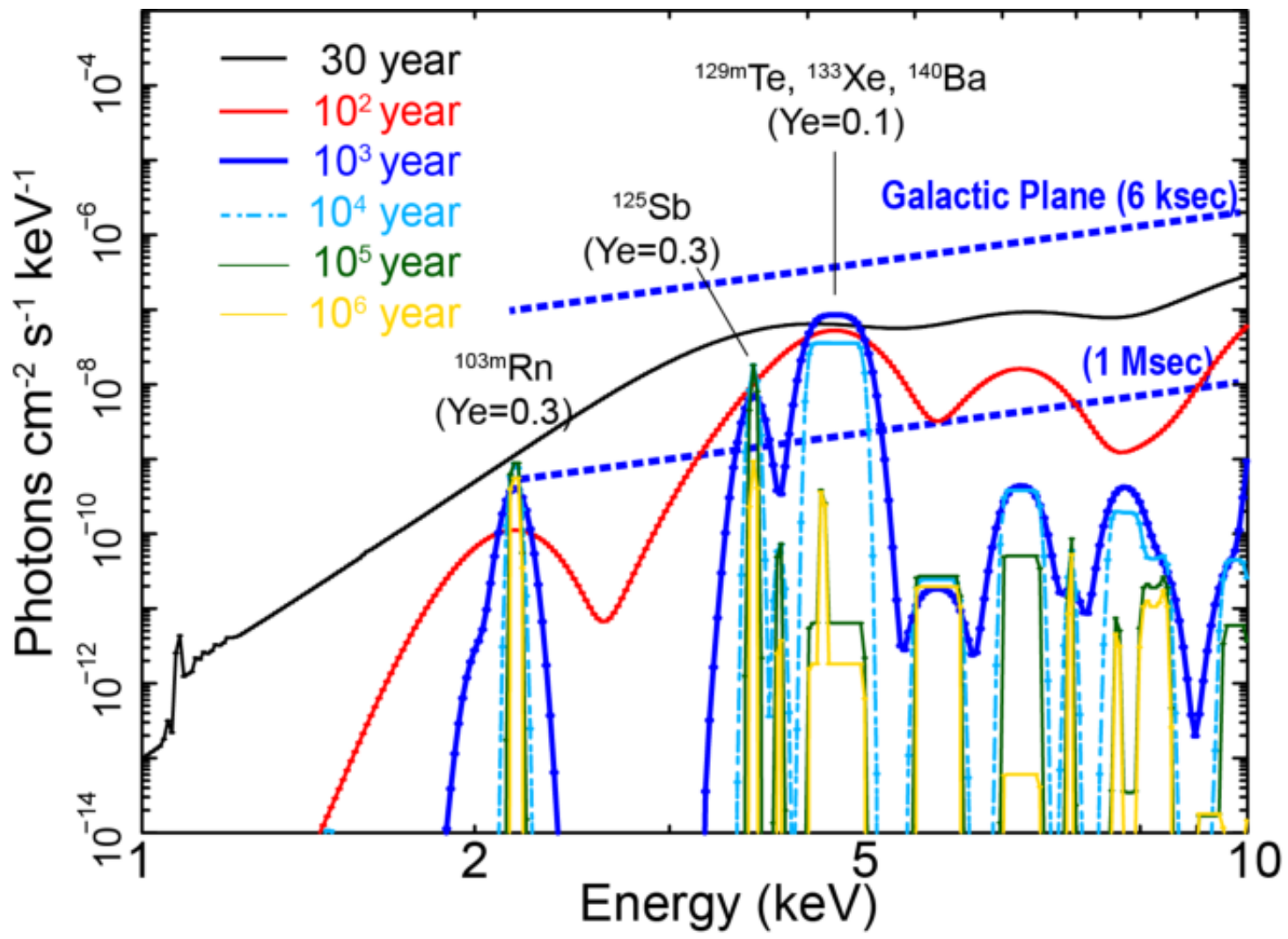

**Figure 48.** The evolution of the thermal X-ray emission (1–10 keV) arising from r-process nucleosytnesis in kilonova remnants at an assumed distance of 1 kpc and of ages ranging from 30 yr to $10^6$ yr. Based on [563]. *AXIS* sensitivity with 6 ksec (equivalent to an exposure with the Galactic Plane Survey) and 1 Msec exposures are also shown.

**Observing description:** For this scientific objective, we intend to employ the *AXIS* Galactic Center/Plane Survey to search for possible KNRs without additional request for observation. Utilizing a single snapshot of the Galactic plane survey with a 6 ksec exposure for each pointing, the sensitivity reaches only very nearby objects (at less than 1 kpc), as estimated in Figure 1. The accumulated data from the Galactic plane survey will reveal indications of the X-ray lines from the r-process nuclei and will establish constraints on the event rate of Galactic KNRs. In this context, we anticipate a total exposure time significantly exceeding 100 k (a mimimum of 200 ks and ideally 500 ks), depending on the survey plan. This survey will enable us to detect the $^{103m}$Rn, $^{125}$Sb, $^{129m}$Te, $^{133}$Xe, and $^{140}$Ba lines, as represented in Figure 1. Notably, the line peak fluxes exhibit low dependency on the age of KNRs, although the line widths become narrower by age, as demonstrated by [563]. Consequently, the line width also represents the typical event rate of neutron star mergers within our Galaxy, as the line fluxes themselves. Moreover, the line ratio will provide the diagnostics of the electron fraction ($Y_e$) environment; the $^{129m}$Te, $^{133}$Xe, and $^{140}$Ba are produced in a low $Y_e$ environment (i.e., a very neutron-rich environment), whereas the $^{125}$Sb is formed in a higher $Y_e$ environment. A low background capability is critical for this research as we aim to detect very faint

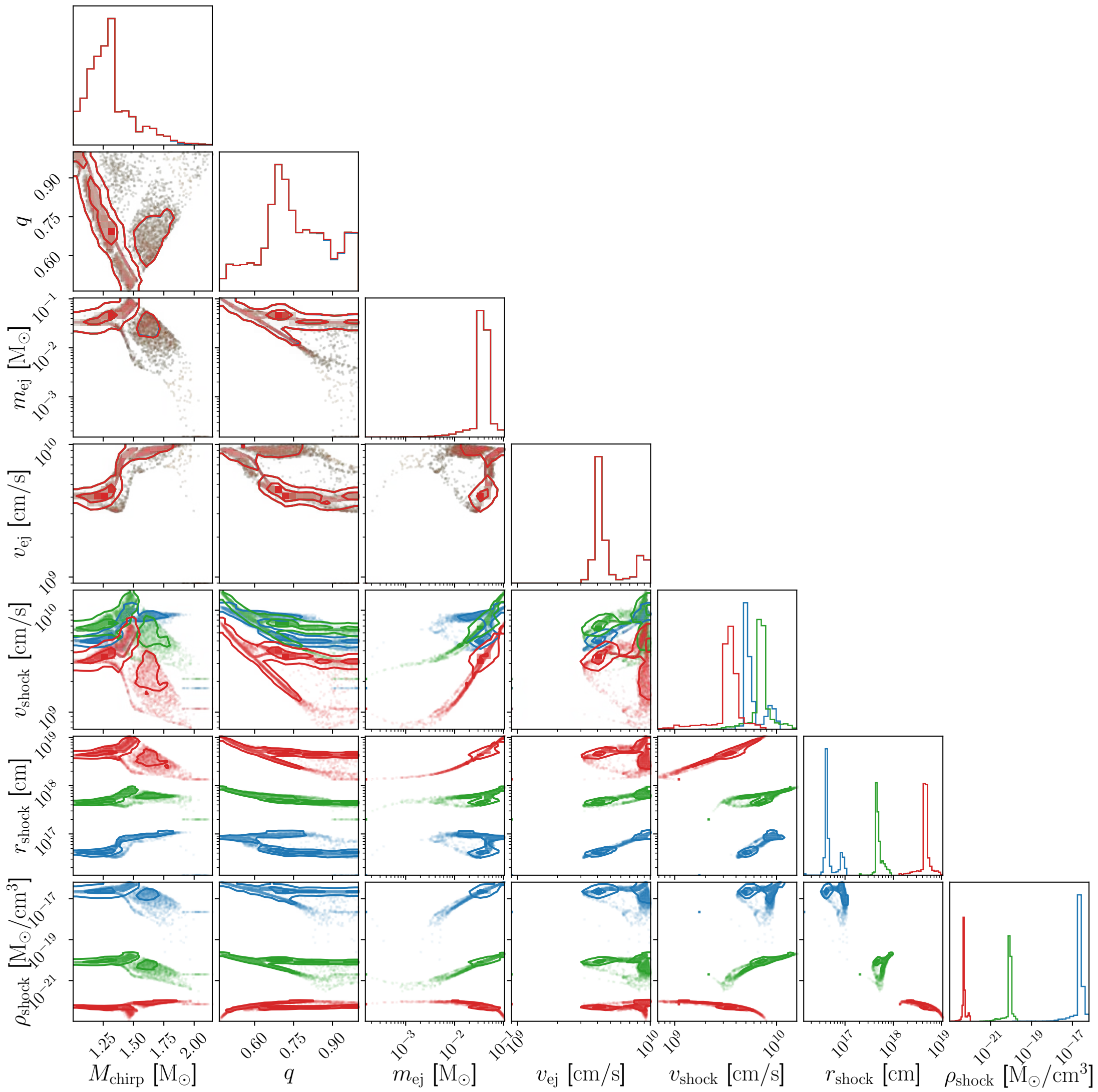

**Figure 49.** A population of BNSs from COMPAS population synthesis with kilonova ejecta mass and velocity computed from the MOSFiT transient fitter, which we then use as input parameters for regression on the hydrodynamical evolution of the KNR shock (shown by red, blue, and green corresponding to 100, 1000, and 10,000 days, respectively)

spectral lines; high accuracy of the estimation of the non-X-ray background and the galactic ridge emission are required.

**[Joint Observations and synergies with other observatories in the 2030s:]**

The kilonova science case is synergistic with the TDAMM science case on BNS mergers with future GW detectors. Late-time follow up (100s of days to years post-BNS detection) observations in both X-rays and radio wavelengths (e.g., SKA, ngVLA) would be valuable. The KNR science will require a low background and deep exposures (200–500 ks). The thermal emission science will be synergistic with XRISM and NewAthena's X-IFU.

*32. AXIS capabilities for extended emission in novae and planetary nebulae*

**First Author:** Şölen Balman (Istanbul Univ.; solen.balman@istanbul.edu.tr)
**Co-authors:** (with affiliations)
**Abstract:** Accreting WDs (AWDs) have been observed to show evidence for extended emission in the form of ejected shells, equatorial rings, collimated, sometimes time-dependent outflows and jets, and hot wind bubbles. These structures are relevant to phenomena observed in many supernova remnants, stellar winds, circumstellar media, and mass ejections in black hole or neutron star X-ray binaries. Imaging studies revealed that several different morphologies have been extracted using observational data. However, extended emissions in the X-ray regime have been photon-starved and recovered the least, thus not well-studied except 10–20 of them, for all species. *AXIS*' spatial resolution is worse than *Chandra*'s on-axis, but is much more efficient as it does not degrade at large off-axis angles. The more important aspect is the high sensitivity with regard to the telescope's effective area. Depending on the exposures from 20 ksec out to $\sim$ 100 ksec, *AXIS* potentially can retrieve extended emissions from AWDs (e.g., nova remnants, planetary nebulae) and permit studies of morphology and energetics that will shed light into their driving mechanisms, hydrodynamics of outbursts, ejections, outflows/winds, shock formation (e.g., internal shocks) and ISM structure as outflows/ejections interact with the circumstellar media around them.
**Science:**

Classical novae (CNe) outbursts occur in AWDs as a result of an explosive burning of accreted material on the WD surface (Thermonuclear Runaway—TNR) causing the ejection of $10^{-7}$ to $10^{-3}$ $M_\odot$ of material at velocities up to several thousand kilometers per second [74]. The classical nova systems have an initial low-level accretion phase ($\leq 10^{-10}$ $M_\odot$ yr$^{-1}$), where recurrent novae (RNe) that are found in outburst with 20-100 years of re-occurrence time generally show higher accretion rates at the level of $10^{-8}$–$10^{-7} M_\odot$ yr$^{-1}$ [23].

The circumstellar interaction models for classical and recurrent novae indicate the existence of forward and reverse shocked material traversing into the circumstellar medium and ejecta. By direct imaging method in the **optical**, the ejecta of about 40-year-old novae have been spatially resolved with evidence/detection of polar blobs and equatorial rings in many of them [74]. The asymmetry in the expansion velocities is fastest at the poles and slowest at the equatorial regions, together with the common envelope phase, results in a prolate asymmetry. Recent works that catalogue morphological studies of nova remnants (classical or recurrent) recover 66 novae younger than 150 years (out of 550 detected to date) and find diffuse emission in 45 sources in the optical with ground-based imaging and archival searches [516]. This study also finds that the remnants are mostly round in shape, with 56% being ellipticals, and 13% having bipolar structures and equatorial enhancements.

The physical mechanisms creating these optical remnants (i.e., shocks) indicate that nova remnants can emit X-rays, just like SNRs. The X-ray luminosity should be about $10^{26}$-$10^{33}$ erg/s on the onset of cooling for nova remnants. An archival search on 250 classical and recurrent nova candidates using *Chandra*, *XMM-Newton*, *ROSAT* and *ASCA* databases [43], no significant extended emission was detected which placed an upper limit of $F_x < 10^{-12}$ erg/s/cm$^2$ (unabsorbed) except for a few known detections (for 50–100 ksec of Chandra time). This makes nova remnants an obvious candidate for research using *AXIS* to understand the TNR, ejection mechanisms, ejecta hydrodynamics, geometry and energetics, and the particle acceleration sites.

Owing to the superb spatial resolution of *Chandra*, extended emission has been investigated with the expectation that the expelled fast ejecta would interact with the surrounding environment [see 46]. The full remnant ($\sim$1–1.5 arc min, $\sim$0.18 pc) of the nova shell of GK Per was observed and resolved at approximately 100 yrs post the explosion [42]. This study revealed nonthermal emission from the shell, with a photon index of 2.3 and a luminosity of $4.5 \times 10^{30}$ erg/s, along with two thermal components (kT =

0.1–0.3 keV and kT = 0.5–2.6 keV) with an absorbing region in between them. The nonthermal emission is due to diffusive nonlinear particle acceleration in the southwest region, where the remnant interacts with a molecular dust cloud and is decelerated. The remnant shows a wedge-like shape with a faster flow in the northwest to southeast direction at expansion velocities $\sim$2600 km $s^{-1}$. The top panels in Figure 51 show the morphology in different energy bands. A distinct emission line of neon, He-like Ne IX, is detected, revealing several emission clumps/blobs. The X-ray luminosity of the forward shock is 4.3 $\times$ $10^{32}$ erg/s. The shocked mass, the X-ray luminosity, and comparisons with other wavelengths suggest that the remnant started cooling and is most likely in a Sedov–Taylor phase. A follow-up paper on GK Per 10 years later, [559], finds that the flux decline is evident in fainter regions. The mean decline is 30–40% in the 0.5–1.2 keV energy band (without cooling) with a typical expansion as 0.″14 $yr^{-1}$, while the fading of X-rays is due largely to expansion. Figure 50 shows on the right the Chandra image of the remnant in the full energy band which is used to simulate an *AXIS* image with a 20 ksec exposure that yields 5-6 c/s using the spectrum of the nova shell and indicates that even at 1/5 the exposure time of Chandra, more morphology will be driven and studied. Balman [42] indicates that several issues remained unresolved (aside from cleaning the pile-up from the central source)–for example, what is the morphology and total emission above 2 keV like? The variations of spectra across the remnant were not well constrained, e.g., what is the change is photon index/spectral index from power law model or rather the *srcut* model in the south-west quadrant (what is the geometry of the particle acceleration region and how does it vary spatially and energetically, how does that comply with the magnetic field configuration/polarization if applicable). The overall abundance variations across the remnant and the spectral emission from localized regions, collection of blobs/knots (do they contain a particular element, are they neon knots? If so, have the ejection been in the form of knots/blobs rather than a shell ?). How does the remnant evolve in time, given that the *AXIS* observation would be about 35 years after the first epoch observation of GK Per.

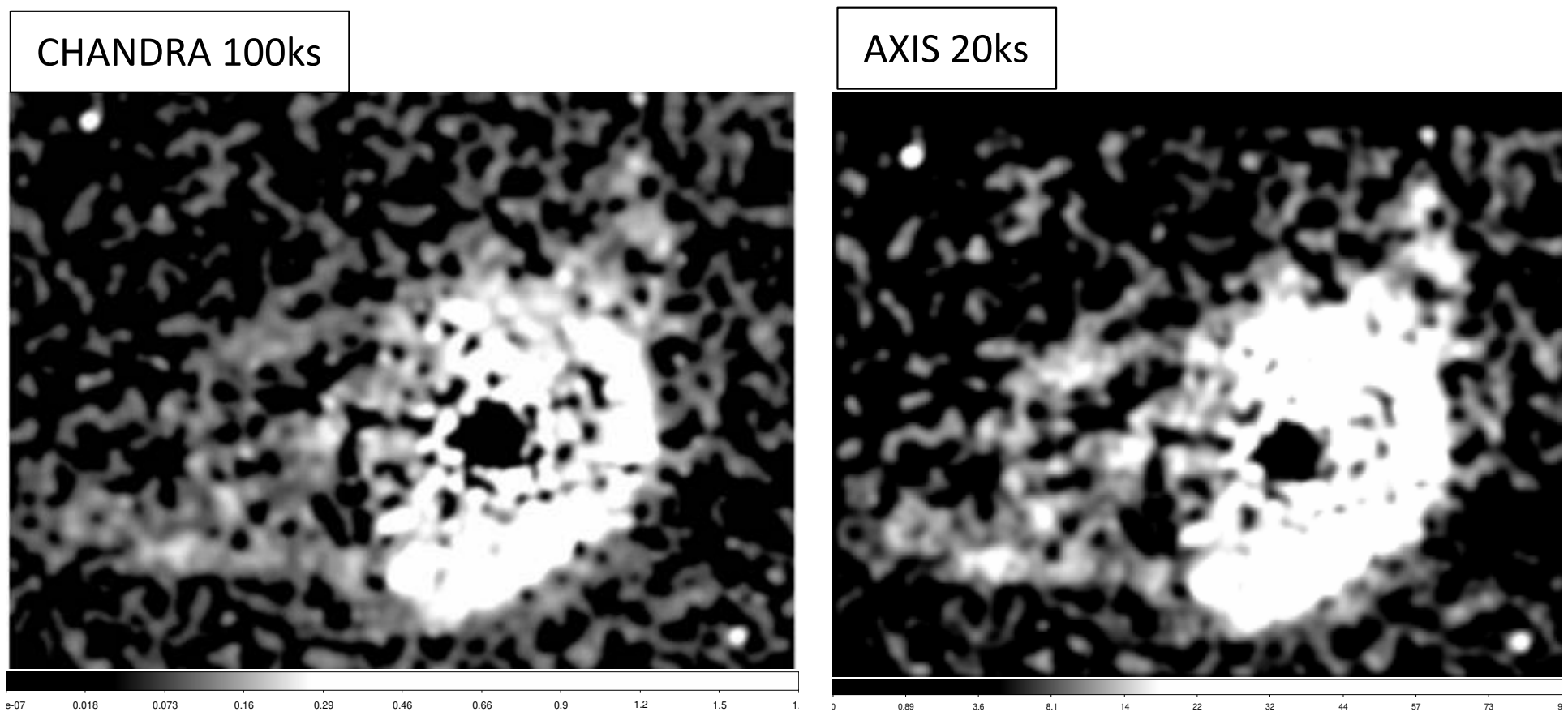

**Figure 50.** The left panel is the nova remnant of GK Persei (1901) as detected by Chandra in 100 ks with a 1 arcsec spatial resolution. North is up, and East is to the left. On the right is the *AXIS* simulation using SIXTE with 20 ks exposure (resolution of 2.5 arc sec).

A sub-pixel-level deconvolution revealed extended soft X-ray emission expanding from 1.3 to 2.0 arcsec (2009 to 2011) detected 5 years after the 2006 outburst of the recurrent nova RS Oph [377]. The emission was asymmetric, oriented in the East–West direction. *Chandra* observations were used to drive an expansion velocity of 4600 km $s^{-1}$ (D/2.4 kpc), consistent with the optical and radio observations. The X-ray-emitting plasma was not cooling, but expanding freely in the polar directions into a cavity left by likely the 1985 eruption. On the other hand, since RS Oph had 7-9 eruptions (time span around 100 yrs), thus the circulstellar medium of the source is complex and diffuse low level of X-ray emission from cooling shocked-ejecta in regard to other eruptions can be recovered using ∼100 ksec of *AXIS* time.

The recurrent nova T Pyx was observed with *Chandra* before the last outburst in 2010 (in quiescence), and a small extended emission region at a S/N of 7–10 has been recovered using 100 ksec observation utilizing deconvolution at the sub-pixel level (unabsorbed $F_x$=5$\times$10$^{-15}$–5$\times$10$^{-14}$ erg/s/cm$^2$). The extended emission revealed an elliptical/ring-like shape with an outer radius of ∼1 arcsec reminiscent of likely the 1966 outburst [44]. Though this remnant is too small for *AXIS*, yet again, because this is a recurrent nova that has a complex circumstallar medium (bright optical remnant is 15 arc sec) with several outbursts (6 in 120 yrs), diffuse low level of X-ray emission from cooling shocked-ejecta in regard to other eruptions can be recovered using around 100 ksec of *AXIS* time.

Finally, [568] finds extended emission associated with the nova remnant of DQ Her (eruption in 1934) in the form of a bipolar jet-like structure extending 32 arcsec NE to SW direction using Chandra ACIS-S (70 ksec, see Figure 51). The XMM-Newton observation (42 ksec) also revealed the additional presence of a diffuse X-ray emission from a hot bubble filling the nova shell (see Figure 51, right). The bipolar feature can be modeled by an optically thin plasma emission at low temperature (similar to first component of GK Per and T Pyx) and a power-law component with a photon index of 1.1 ± 0.9 and a luminosity of (2.1 ± 1.3) × 10$^{32}$ erg/s. Given the luminosity ranges derived for the bipolar jet structure of DQ Her (10$^{30-32}$ erg/s), a similar exposure time as in GK Persei will yield resolution of the morphology of the jets and the ejecta separately (now, it is all marginally discovered).

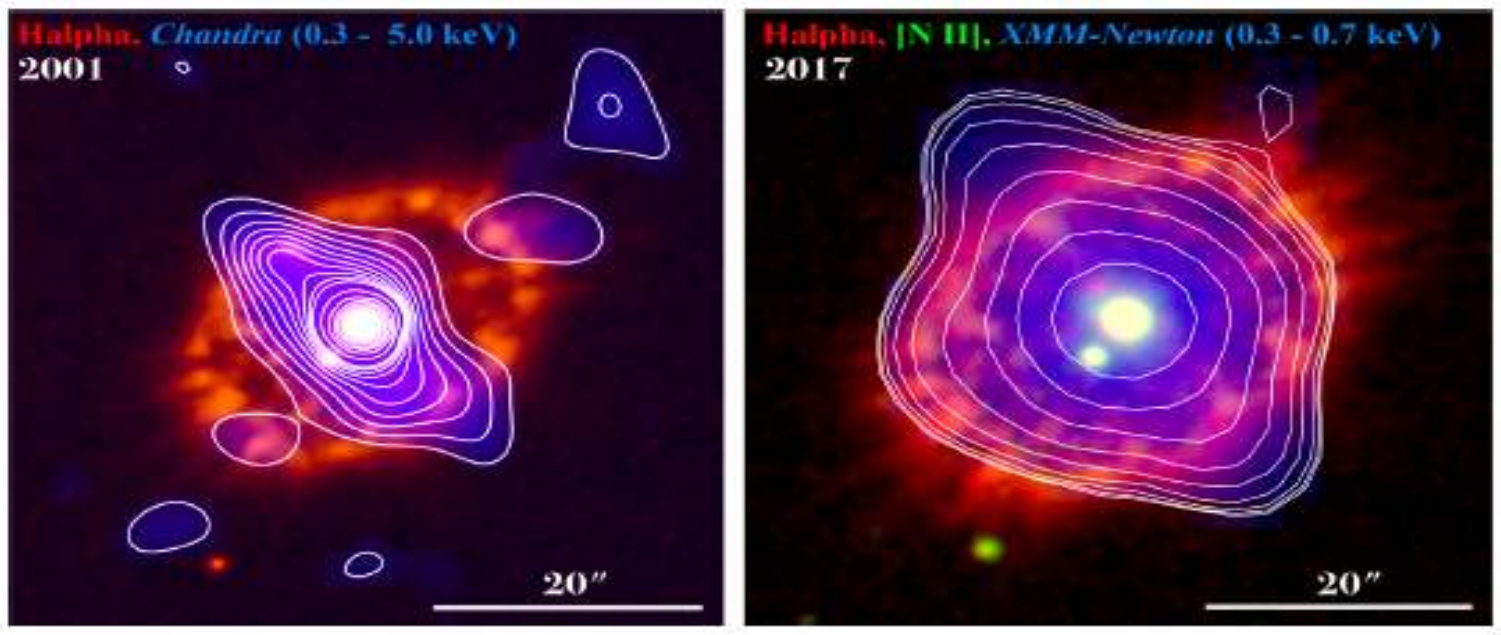

**Figure 51.** The left-hand panel shows an elongated jet-like structure associated with the shell of DQ Her. The contours show the Chandra ACIS-S flux overlaid on the optical image in H$\alpha$, obtained with the William Herschel Telescope. The right-hand panel shows the Nordic Telescope (NOT) H$\alpha$ image and the XMM-Newton EPIC pn contours of the extended emission. [568].

Planetary nebulae (PNe) are the late stages of low-mass stellar evolution [307]. PNe morphologies involve processes that reside in many other astrophysical systems, e.g., the wind-blown bubbles of massive stars and shocks in novae ejecta. Understanding of their X-ray properties and morphologies is important for multi-wavelength structure of PNe and the ejection mechanisms and the hydrodynamics of the interacting winds [40,306]. An observationally consistent model is of a hot bubble and swept-up shell and the extinction caused by the swept-up shell. The Chandra Planetary Nebula Survey, ChanPlaNS, (volume-limited, $R_{neb} \leq 0.4$ pc, young PNe that lie within ∼1.5 kpc), was designed to investigate X-ray emission from PNe, which increased the detected PNe to 59 objects by 2014 [183]. ChanPlaNS diffuse X-ray detection rate was about 27% and the point source detection rate was 36%. Diffuse X-ray emission was mainly associated with young (≤5000 yr), compact (≤ 0.15 pc radius) PNe that exhibited closed elliptical structures and high electron densities ($n_e$ >10$^3$ cm$^{-3}$). The X-ray fluxes of diffuse emission is in a range of (3.0–10.0)× 10$^{-14}$ erg/s/cm$^2$ with X-ray temperature ∼ 0.1 keV [183,209]. These limits make PNe excellent targets for diffuse emission studies with *AXIS*, given the sizes (a few tens of arc secs) and the low luminosities.

**[Exposure time (ks):]**

20 – 100 ksec. This will be a range of exposures given the diversity of morphology and science that need to be extracted from the data. In general, 10–30 ksec will suffice for basic detection and spectral analysis of the entire nebulosity.

**Observing description:**

Low background will be useful, but can be accounted for with adjusting the exposure times.

**[Joint Observations and synergies with other observatories in the 2030s:]**

Classical and recurrent nova remnants, PNe, and extended emission in symbiotics have always been performed over the entire electromagnetic spectrum to the best sensitivity or availability of telescopes, ground- or space-based. This is evident since a full understanding of the associated physical phenomena and morphological studies requires multi-wavelength analysis. Synergies with ELTs, LSST, SKA, and ALMA. NewAthena may be used when high spectral resolution is required.

**[Special Requirements:]** (pileup, e.g., Monitoring (Daily, Hourly, etc), TOO (<X hrs), TAMM). No particular requirements.

## j. Pulsar and Magnetar Wind Nebulae

*33. AXIS peering into pulsar wind nebulae–some of the most extreme particle accelerators in the Universe*

**First Author:**
Oleg Kargaltsev (GWU, kargaltsev@email.gwu.edu), Samar Safi-Harb (U. of Manitoba, samar.safi-harb@umanitoba.ca)
**Co-authors:** Isabel Sander, Austin MacMaster (U. Manitoba), Xiying Zhang (U. Barcelona), Seth Gagnon (George Washington U.), Noel Klingler (NASA's GSFC), Kaya Mori (Columbia U.), Joseph D Gelfand (NYUAD), Hongjun An (Chungbuk National University), Jeremy Hare (NASA's GSFC), Jooyun Woo (Columbia U.), Pol Bordas (U. Barcelona), Cole Treyturik (U. Manitoba), Bettina Posselt (U. of Oxford), Daniel Castro (CfA)
**Abstract:** Pulsar-wind nebulae (PWNe) harbor some of the most extreme particle accelerators known to exist in the Galaxy – young rotation-powered pulsars powering ultra-relativistic magnetized winds. Their study sheds light on the physics of jets and relativistic outflows, which are ubiquitous in high-energy astrophysics. A key characteristic that demands high-angular resolution and sensitivity is the termination shock where the pulsar's wind meets the surrounding medium — the hosting supernova remnant (SNR) or the interstellar medium (ISM). Thanks to its large effective area and excellent angular resolution across the FoV, *AXIS* will be able to both accurately measure the synchrotron spectra of particles injected at the termination shock and map the SED evolution as a function of distance from the pulsar on much larger spatial scales. The former will strongly constrain yet unknown particle acceleration mechanisms, while the latter will reveal the evolution of the magnetic field with distance and particle transport, including the transition from an advection to a diffusion regime, the roles of magnetic turbulence and magnetic field anisotropy, and the kinetic particle escape into the ISM. Theoretical modeling and understanding of these complex physical processes will strongly depend on the availability of high-quality measurements in the *AXIS* X-ray band, as radio observations capture the old and cooled particle population. In contrast, hard X-ray and gamma-ray observations lack the necessary imaging resolution. As strong positron sources, pulsar winds may also account for the observed positron excess near Earth; however, confirmation of this hypothesis relies on the reliable calibration of particle transport modes using multi-wavelength data. *AXIS* will expand this sample at least tenfold, enabling a transformative, multi-wavelength synergy with facilities for the next decade: the ngVLA, SKA, and its pathfinders (radio) and CTAO+LHAASO+SWGO (gamma-ray).
**Science:** PWNe are a dominant source of leptonic cosmic rays and may even power the highest energy hadronic cosmic rays. In fact, PWNe dominate the Galactic TeV sky and may even be responsible for the positron excess seen at Earth. Structured on arcsecond-scales, PWNe emit featureless power-law spectra that trace the particle acceleration, cooling, and transport (see [414,486] reviews). In general, PWN properties are influenced by the parent pulsar properties (including its velocity, age, spin-down power, and magnetosphere geometry) and by the surrounding medium (including host SNR, ISM, and ambient photon fields). Some of the fundamental questions involving QED processes in strong fields, relativistic shock and magnetic field re-connection physics, wave-particle interaction, magnetized relativistic turbulence, diffusion of relativistic particles in magnetized medium, and ISM properties are listed below.

- To explain X-ray emission from PWNe, particle energies must be boosted by a factor of $10^3 - 10^4$ between the pulsar's magnetosphere and the region outside of the pulsar wind termination shock. Neither the acceleration mechanism nor the actual acceleration site is currently known. *Measuring particle-injection spectra through the slope of the uncooled synchrotron component in the vicinity of the termination shock probes the acceleration mechanism.*

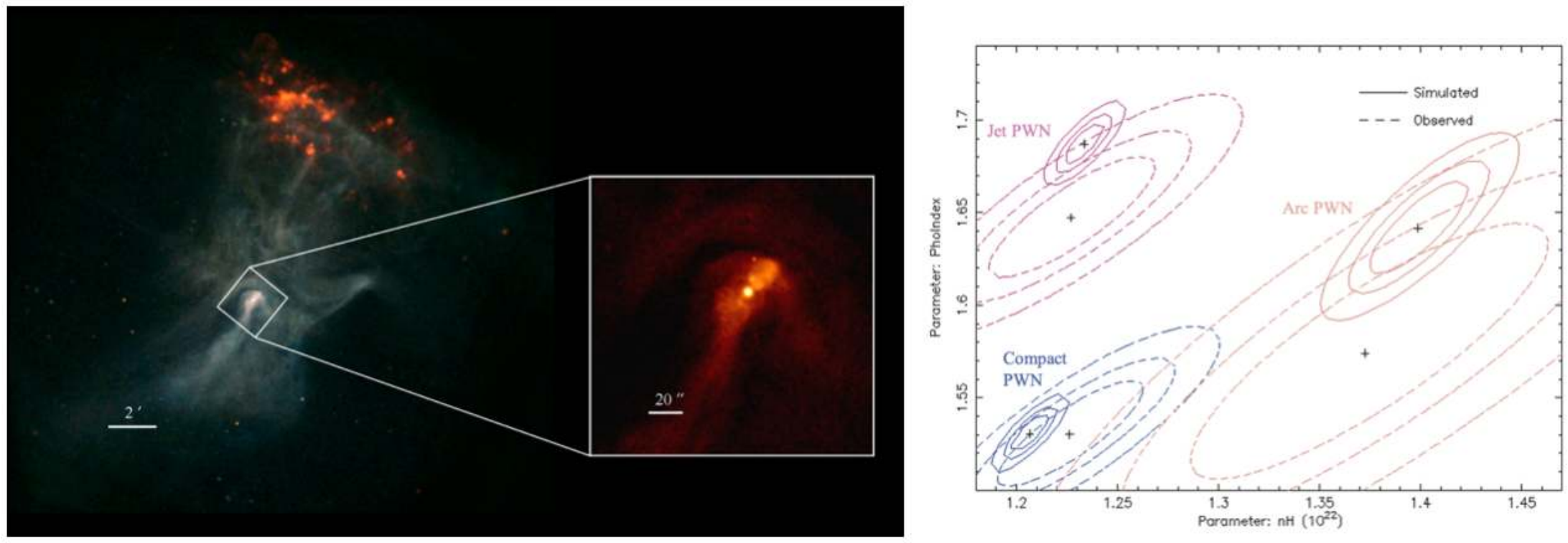

**Figure 52. Left:** AXIS 300 ks simulated image of 'Hand of God' – the SNR MSH 15–52 and its PWN associated with PSR B1509–58 – demonstrating the rich and complex large- and small-scale structures to be probed with AXIS. **Right:** Confidence contours for the column density and photon index of the high-resolution structures of the PWN (shown as an inset in the left panel), showing how *AXIS* will provide far tighter constraints on spectral properties compared to Chandra.

- One of the elusive but fundamental parameters in modeling pulsar magnetospheres is the pair cascade multiplicity, which is directly related to the $e^{+}/e^{-}$ pair cascade physics. Since most of the pairs produced in the magnetospheric cascade are advected into the PWN, *measuring the brightness of emission produced by freshly injected particles in the vicinity of the termination shock directly constrains pair multiplicity.*
- High-resolution Chandra images reveal that many PWNe exhibit highly anisotropic pulsar winds, dominated by equatorial tori and polar jets (Figure 2 in [277]). These structures reflect intrinsic pulsar properties—such as velocity and the angle between spin and magnetic axes—and reveal the spin axis orientation relative to the observer. Measuring the angle between spin axis and velocity helps probe supernova kick mechanisms and explosion physics [55]. *Imaging fainter PWNe can expand the sample of pulsars with independent constraints on magnetospheric geometry.*
- Particle escape from PWNe can play a crucial role in seeding our Galaxy with relativistic electrons and positrons reaching PeV energies. The naive hydrodynamics-based picture of a PWN as a wind bubble confined by the contact discontinuity that separates the shocked pulsar wind from the shocked ISM has been severely undermined by recent discoveries of *TeV halos* and "pulsar filaments" (or "misaligned outflows"). *Detecting faint but very extended structures associated with the kinetic transport of ultra-relativistic particles in ISM magnetic field sheds light on ISM field structure, particle-wave interactions, and turbulence scales can serve as an excellent test-bed for relativistic plasma astrophysics.*

We propose deep observations of several PWNe probing different evolutionary stages of supernova remnant evolution, from objects well within their hosting SNRs (typified by PSR B1509–58 in MSH 15–52, Fig. 57) to objects leaving their SNRs (PSR B1853+01 in W44, Fig. 53) to pulsars having already left the SNR and whose winds are interacting with the interstellar magnetic fields (PSR J1101-6101/SNR MSH 11–61A, aka lighthouse nebula, Fig. 54). Furthermore, many unidentified TeV sources are believed to be associated with relic or evolved PWNe; we select the TeV source HESS J1809—193 potentially associated with a relic nebula powered by PSR J1809-–1917 (Fig. 55).

Firstly, the SNR MSH 15–52 and associated PSR B1509–58 are shown in Fig. 57 (see e.g., [189]). This is a middle-aged SNR powered by an energetic pulsar, which is powering a PWN resolved into high-resolution structures (a compact nebula, an arc, and a jet) thanks to Chandra. The PWN is also interacting with the SNR ejecta in the north. At this evolutionary stage, the system is an ideal example of ongoing interaction between the pulsar wind and the SNR ejecta, and where *AXIS* will provide a unique opportunity to (a) measure the proper motion of the SNR thermal knots over a $\geq$30-year timescale, and (b) constrain the spectral indices of all these components accurately enough to constrain the particle acceleration and propagation models.

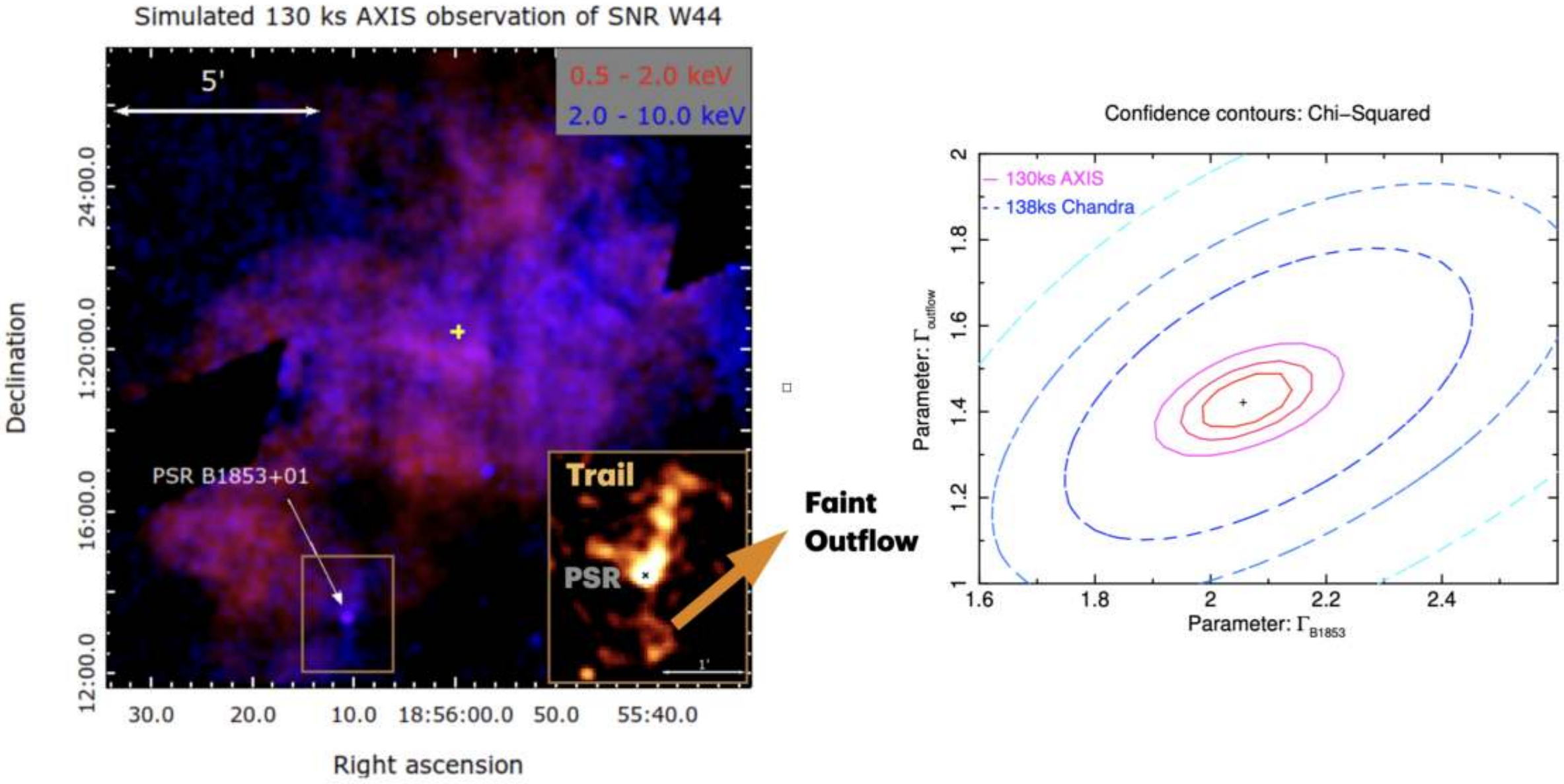

**Figure 53. Left:** *AXIS* 130 ks simulated image of the SNR W44 housing the compact PWN surrounding PSR B1853+01. The PWN stands out in the hard X-ray band, whereas the SNR emission dominates in the soft X-ray band. *AXIS* sensitivity and sharp PSF across its FoV will be ideal to resolve all these structures and perform a spatially resolved spectroscopic study. **Right:** Spectral indices Confidence levels for the pulsar versus faint outflow to show how *AXIS* will enable us to constrain the spectral parameters much better than Chandra with a modest exposure.

Secondly, the SNR W44 represents a case where a compact and faint PWN was discovered in hard X-rays (above 2 keV) close to the SNR edge (see Fig. 53). PSR B1853+01 powers the PWN ($\tau_c \sim 20$ kyr, $\sim 7'$ from the remnant's geometric center) and presents several high-resolution compact and faint structures discovered by Chandra: (a) an arcminute trail behind (north of) the pulsar, suggesting that the pulsar is moving supersonically south out of the SNR shell, (b) a faint "outflow", oriented almost parallel to the direction of the pulsar's proper motion, protruding from the bow shock region. This "outflow" feature could be interpreted as a signature of high-energy particles (up to about 100 TeV energy) escaping from the bow shock region of the nebula [648]. In a 130-ks observation, *AXIS*'s superior sensitivity and excellent PSF across the whole field will allow us to better characterize this system by performing spatially resolved spectroscopy of different regions, especially in the faintest "outflow" region (as shown in Fig. 53). *AXIS* will also be able to detect much fainter yet more extended structures, e.g., a possible "X-ray halo" surrounding this pulsar wind nebula. In summary, a more detailed characterization of this source will advance our understanding of particle acceleration and escape processes in PWNe, as well as the subsequent diffusion of high-energy particles into the ambient medium.

For the remarkable Lighthouse nebula (left panel in Figure 54), a 500-ks *AXIS* observation will enable us to collect $\sim$7 times more photons than in a comparable-length CXO ACIS observation. This will enable a much more detailed view of the internal structure of the escaping particle beam and informative spectral maps with a factor of $\sim 3$ times smaller uncertainties for the photon index compared to CXO ACIS for the same exposure. The right panel in Figure 54 shows a simulated *AXIS* spectral map. This unique PWN is a key for understanding the ISM magnetic field structure and ultrarelativistic particle transport in the ISM.
**[Observing description:]**

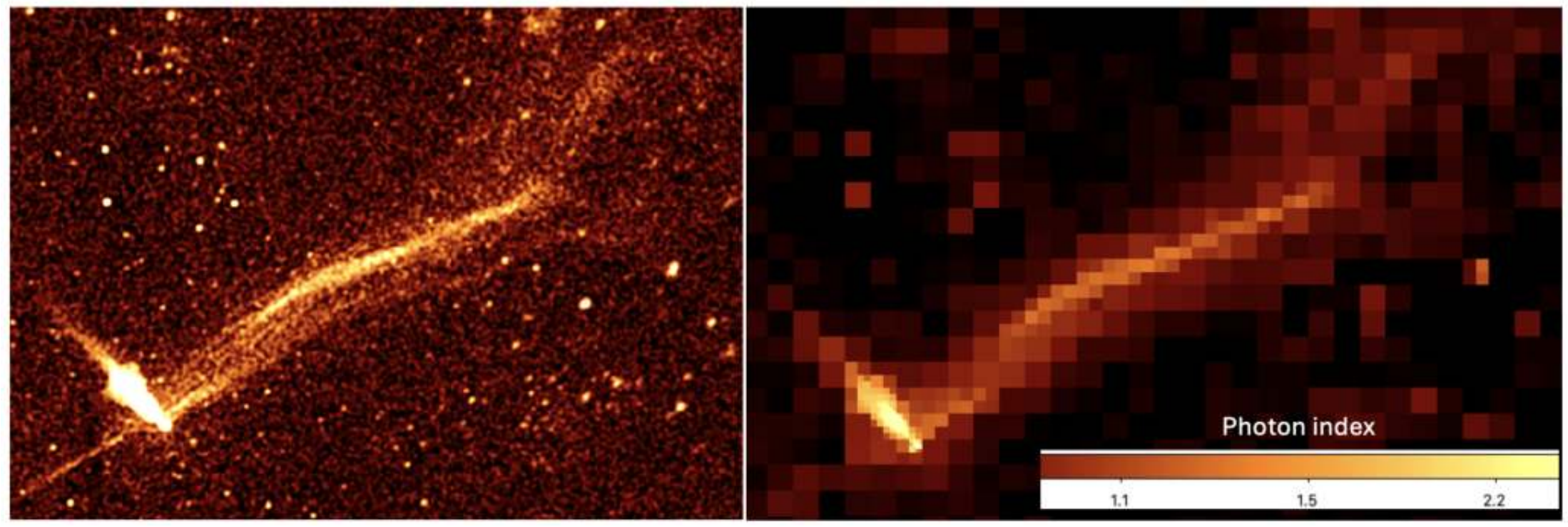

**Figure 54.** 380 ks CXO image (**left** [288], binned to 1.5″ resolution (to be representative of *AXIS*'s resolution) and simulated spectral map (**right**) for 500-ks *AXIS* observation and a hypothetical scenario where the spectrum is softer in brighter regions due to a stronger magnetic field causing stronger radiative cooling. Typical uncertainty of photon index in spectral map is $\delta\Gamma \approx 0.03$.

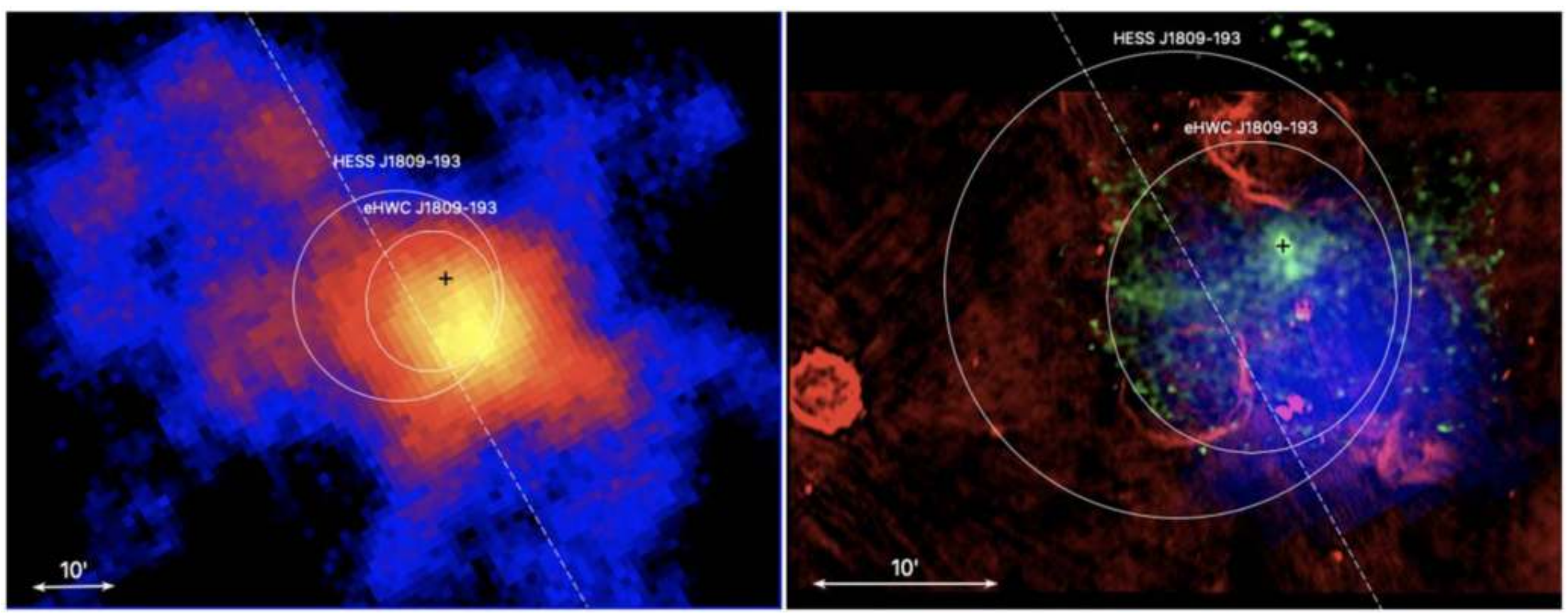

**Figure 55. Left:** TeV image of PSR J1809–1917 region form the HESS Galactic Plane Survey [212]. **Right:** Multi-wavelength image of the same area. Red: radio (JVLA, 1.4 GHz; Castelletti et al. [105]), green: X-ray (Chandra ACIS, 0.5-–8 keV), blue: TeV $\gamma$-rays (the same image as in the left panel). Adapted from Figure 6 of [289].

Deep, 500-ks, *AXIS* observation of the PWN of PSR J1809–1917 and HESS J1809–193 (see Figure 55) can probe the true extent of X-ray emission and provide spatially-resolved spectroscopy with $\approx 1'$ angular resolution, enabling meaningful mutli-zone modeling of multi-wavelength emission for older, more evolved, PWNe and shedding light on relative importance of diffusion versus advection, energy-dependence of diffusion coefficient and its spatial anisotropy, and constraining ambient magnetic field strength.

*AXIS* will deliver the arc-second resolution, large field of view, high effective area coupled with low detector background, and an ability to measure spatially-resolved spectra of the faint diffuse emission in the $\sim$0.5–10 keV band – the keys ingredients needed to advance our understanding of PWNe and address important astrophysical questions mentioned above.

**[Joint Observations and synergies with other observatories in the 2030s:]**

These observations are synergistic with those from radio (ngVLA, SKA) and high-energy gamma-ray telescopes (e.g., HAWC, CTAO, LHAASO, SWGO). SKA observations of the proposed SNRs are expected to match the arcsecond resolution of *AXIS* images. HESS J1809–193 is expected to be one of the important targets for CTAO because of its large angular extent, which should enable spatially-resolved spectroscopy in TeV. Together with *AXIS*'s spatially-resolved spectroscopy in X-rays, this data will be perfect for testing multi-zone models of particle transport and interaction with the SNR reverse shock.

**[Special Requirements:]** None.

*34. AXIS as a discovery tool for variability in pulsar and magnetar wind nebulae, the magnetar population, and their connection to Fast-Radio-Bursts*

**First Author:**
Oleg Kargaltsev (GWU, kargaltsev@email.gwu.edu), Samar Safi-Harb (U. of Manitoba, samar.safi-harb@umanitoba.ca)
**Co-authors:** Austin MacMaster (U. Manitoba), George Younes (NASA's GSFC and UMBC), Isabel Sander (U. Manitoba), Seth Gagnon (George Washington U.), Bettina Posselt (U. of Oxford), Alice Borghese (ESA/ESAC)
**Abstract:** Pulsar wind nebulae (PWNe) are among the most powerful particle accelerators in the Universe, shedding light on the interaction of ultra-relativistic winds with their surroundings. Chandra has significantly advanced our understanding of these fascinating objects, uncovering their variability and dynamic timescales, and pioneering the emerging field of magnetar wind nebulae (i.e., wind nebulae around ultra-magnetized neutron stars or magnetars). Characterizing recurring structural changes in bright PWNe may reveal evidence of free precession in neutron stars and help determine their characteristic timescale. Studying nebulae powered by magnetars can aid in identifying numerous dormant or transient magnetars within our Galaxy, particularly through the envisioned *AXIS* Galactic Plane Survey (GPS). Discovering these objects will enable us to probe the diversity of neutron stars and their winds, trace their evolutionary pathways, and explore the connection between Galactic magnetars and extragalactic Fast Radio Bursts.
**Science:** Ultra-relativistic pulsar winds transport particles created in the magnetosphere to large distances from the neutron star (NS). Studying variability of PWNe emission sheds light on the long-term changes of pair cascades and other NS phenomena, such as magnetospheric restructuring, crust cracking, or free precession. Chandra observations of bright PWNe have shown that some of their structures are remarkably dynamic. In particular, the jets of PWNe powered by Crab, Vela, and J1811–1925 pulsars exhibit corkscrew-like motion that may arise temporarily from plasma instabilities or, if persistent, could indicate free precession of the NS. Detecting such precession would offer a new method to estimate the strength of gravitational wave signals from isolated NSs – a crucial input for the new generation of gravitational wave observatories.

There is also growing evidence that highly magnetized rotation-powered pulsars can occasionally exhibit a magnetar-like behavior, releasing large amounts of plasma that generate hybrid pulsar/magnetar wind nebulae. Determining the nature of this emission using higher-quality spectra is crucial for distinguishing it from dust scattering halos, which often accompany bright magnetar flares. Furthermore, the discovery of a compact, faint, non-thermal nebula around a Galactic transient magnetar suggests that such nebulae may persist well beyond the brief active phase of the magnetar. However, the timescales over which these magnetar wind nebulae evolve remain poorly understood. Due to their faintness, such nebulae are often undetectable with Chandra unless a known magnetar is present. *AXIS*, with its superior sensitivity, high angular resolution, and rapid Target of Opportunity (ToO) capabilities, will be able to detect many of these elusive objects. *AXIS* will enable us to: (a) capture snapshots of variable and potentially variable PWNe, (b) follow up on nebulae around known magnetars or pulsars exhibiting magnetar-like activity, and (c) uncover a population of magnetar wind nebulae (MWNe) and previously unrecognized, possibly older, dormant magnetars that flare only rarely.

Firstly, we propose long-term monitoring of the Vela PWN jet, which has shown remarkable dynamics in a series of Chandra observations (see Fig. 56). Most intriguing is its rotating corkscrew structure, which could be either a temporary appearance due to the development of kink-like instability or the result of free precession of the neutron star. A more recent finding of a distinct crock-screw structure in a deep Chandra image [80] of a longer jet from a much younger and much more energetic pulsar J1811–1925 (in G11.2–0.3) supports the latter interpretation. Confirming the corkscrew pattern and measuring its rotation period, estimated from Chandra data to be ∼122 days, is crucial for probing possible free precession of the Vela pulsar. By optimizing the binning in Chandra images, we find that *AXIS* will have sufficient angular resolution to track the motion of jet material. 20 ks *AXIS* exposures will be a factor of 2 shorter than those of Chandra, while providing a factor of 2 improvement in S/N and better statistics for fitting the precessing ballistic jet model (see [169]). We propose a total of 20 observations separated by $12 \pm 2$ day intervals. This will provide monitoring over a 240-day interval, which is twice as long as the precession period of ≈ 122 days proposed by [169].

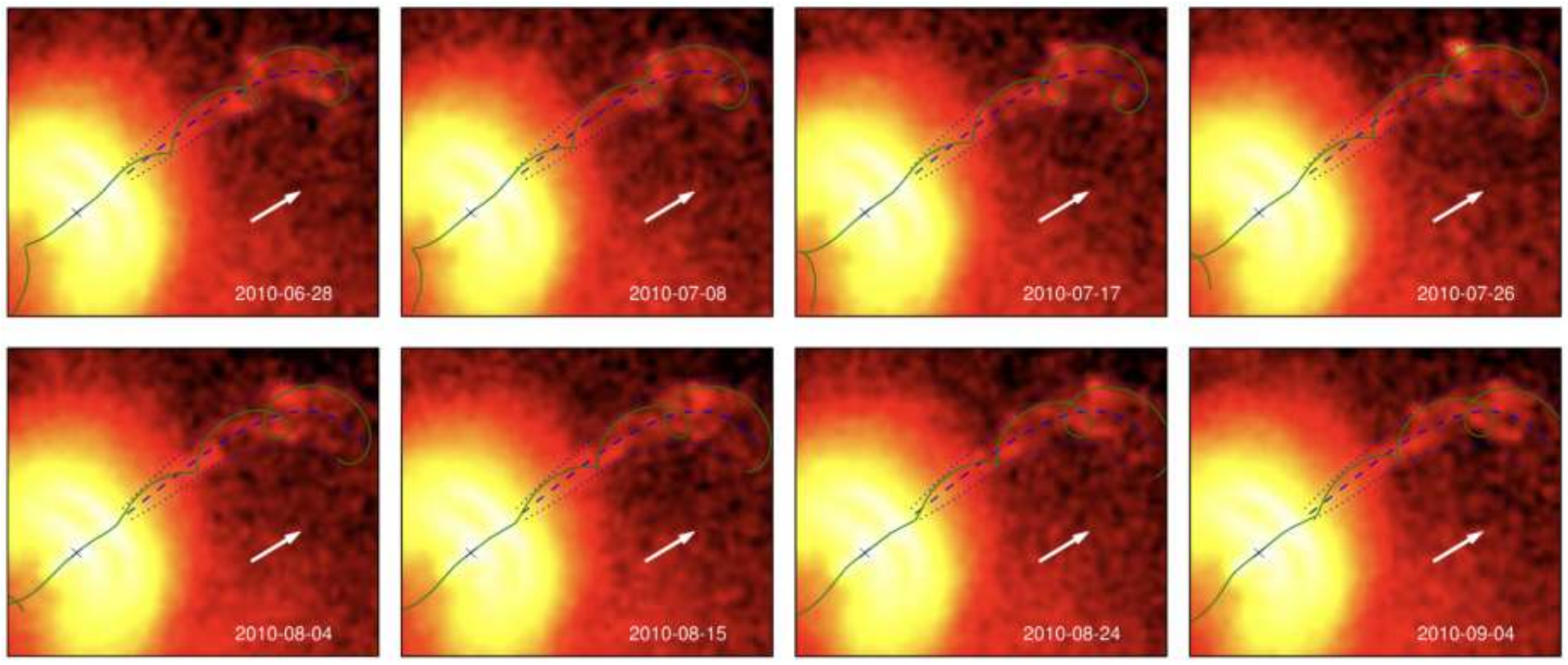

**Figure 56.** Dynamics of the Vela pulsar jet revealed by Chandra. The green curve shows the model trajectory for material ballistically launched along the spin axis precessing with ≈122 day period. Adapted from [169].

Secondly, for the magnetar wind nebula (MWN) science, we propose a 100-ks *AXIS* study of the faint X-ray nebula associated with the transient magnetar Swift J1834.9–0846 [630,631]. Although a few transient magnetars are known, this source is unique because Chandra discovered the surrounding emission before the magnetar became active [279]. Hence, we have a good estimate of the MWN extent and flux during the magnetar's quiescent state, which is its primary state. Magnetars likely become less transient as they age, but also their persistent flux level decreases – hence we have a non-negligible number of 'orphan' MWNe that harbor old, dormant magnetars which *AXIS* is prime to detect with its superior sensitivity, either directly or indirectly through the detection of their faint, non-thermal nebulae. Therefore, we will conduct a detailed characterization of the MWN associated with Swift J1834.9–0846, including a morphological and spectroscopic study, which will inform searches for similar objects in the planned *AXIS* Galactic Plane Survey and other observations conducted in the Galactic plane. A 100-ks *AXIS* observation will enable us to collect about 2,000 photons from the MWN and measure its spectrum in 4 concentric annuli from 15 spectral bins (S/N=4 per bin) for each annulus. These measurements will, for the first time, probe cooling in MWNe and constrain the particle transport and magnetic field strength in MWNe.

Additionally, Chandra has opened a new window into connecting the diversity of neutron stars through the discovery and monitoring of magnetar-like outbursts from a handful of highly magnetized neutron stars commonly believed to be rotation-powered, or from the discovery of wind nebulae around transient magnetars (see e.g., [69,70,72,509]). In particular, if it wasn't for Chandra, the X-ray counterpart and compact and faint nebula associated with the radio high-B pulsar J1119–6127 would have been missed. This object presents the canonical example of a commonly believed rotation-powered pulsar displaying as well magnetar-like behavior following the discovery of a magnetar-like outburst in 2019. Chandra observations and monitoring of this source led to the discovery of variability in both morphology and spectrum of the associated nebula, revealing for the first time the dual nature of this source and associated nebula. The question remains on what powers the X-ray emission from the nebula, and how many such sources are we missing? *AXIS* will answer these questions thanks to its ToO capabilities, superior sensitivity, and power to discern the pulsar spectrum from the surrounding nebula and contaminating point sources in the field. As shown in Fig. 58, *AXIS* is needed to not only resolve the PSR from its associated compact nebula, but also constrain the spectral properties which are needed to determine the powering mechanism in pulsar/magnetar wind nebulae: rotation, magnetism, both, or a yet unknown mechanism.

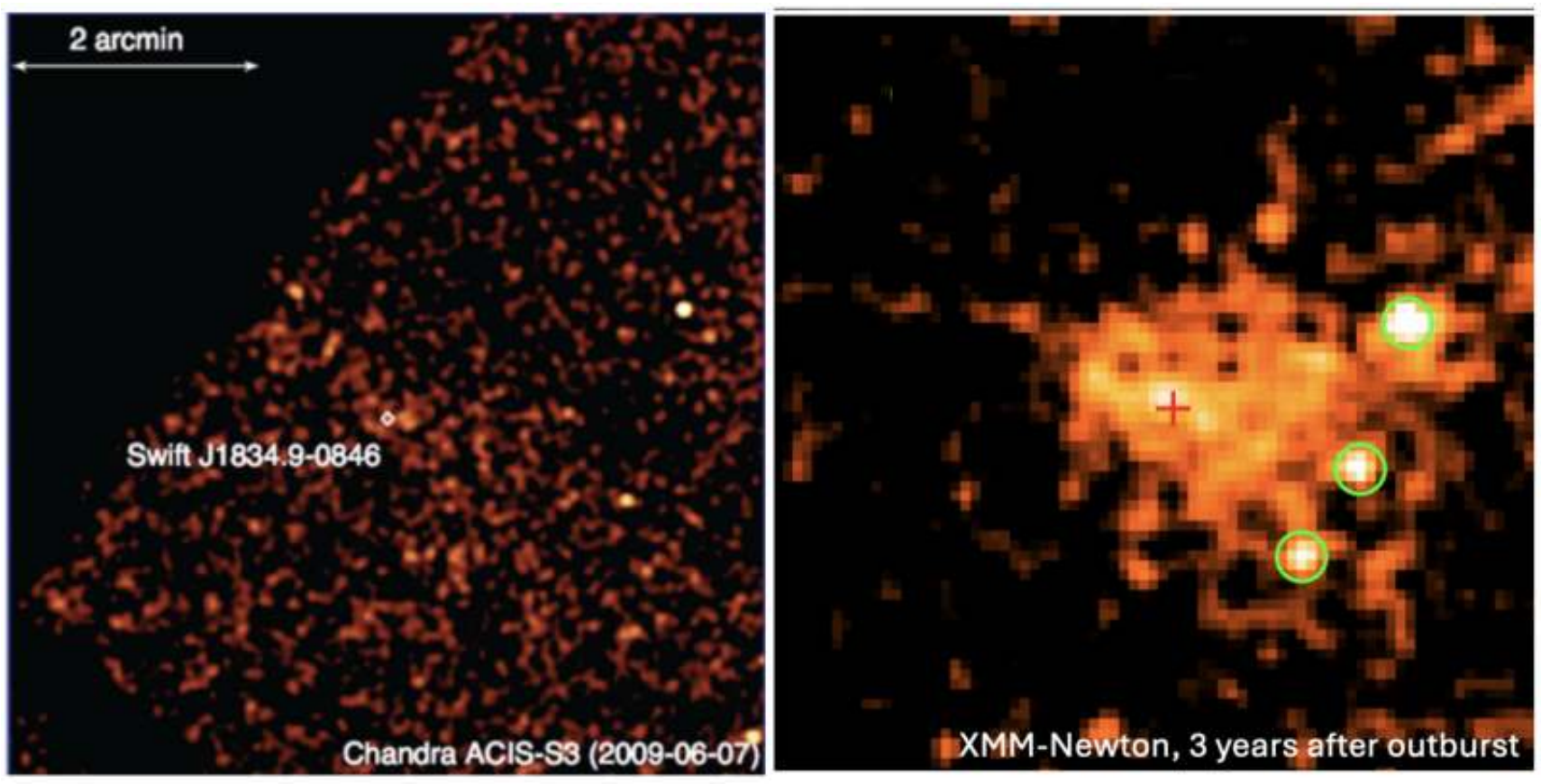

**Figure 57.** 46-ks Chandra ACIS (pre-outburst) and 180-ks XMM-Newton (post-outburst) observations of the region around the transient magnetar Swift J1834.9–0846 showing faint extended emission of $\sim 2'$ size. Adapted from [279,631]. In the right panel, the red cross shows the position of the magnetar, which had already faded below the level of the extended emission. The green circles show X-ray point sources detected in the field. *AXIS*'s sensitivity and excellent resolution maintained off-axis will enable us to resolve all point sources in the FoV and accurately determine the spectrum of the diffuse emission from the MWN.

The *AXIS* Galactic Plane Survey (GPS) will be able to detect sources like J1119–6127 in the bursting phase. A 6 ks *AXIS* GPS exposure will constrain model uncertainties at the same level as a 55 ks ACIS observation. Furthermore, *AXIS* will detect sources of J1119-like spectral properties down to a flux $4.24\times10^{-15}$ ergs cm$^{-2}$ s$^{-1}$.

**The magnetar population and their connection to fast Radio Bursts**

Recent advances in population synthesis modeling of the magnetar population predict the presence of over 70 such objects that are readily detectable by *AXIS* within its planned Galactic Plane Survey (see [520] and references therein). Such discoveries, which are unbiased towards the younger, more active sources (this is currently the only channel for magnetar discovery), would triple the number of known magnetars and allow for more robust inferences on their birth rate and evolutionary history. This has large implications for the physics of magnetar formation, their field decay through magneto-thermal evolution, and their place within the larger isolated neutron star zoo. Moreover, such discoveries would inform us about the rate of bursting magnetars in the Galaxy, which is connected to the more open question of the number of magnetars powering the mysterious class of extragalactic (repeating and non-repeating) Fast Radio Bursts (FRBs).

*AXIS*, through its large effective area and agile scheduling capabilities, will enable detailed studies of the internal and external dynamics governing the variety of transient activity of magnetars, spanning millisecond to year timescales, including their FRB-like emission. The magnetic energy density of magnetars can at times exceed the crustal yield strain, forcing the crust to respond plastically, depositing heat in the crust and leading to external twisted magnetic field bundles (e.g., [316,328]). This theory is at the core of magnetar outbursts and bursting behavior, and only recently strong evidence of such behavior has been detected through near-daily NICER observations of a magnetar in outburst [634,635]. Yet, NICER can only track magnetar outbursts in exquisite detail down to a sensitivity limit of $\gtrsim 1.0 \times 10^{-13}$ erg cm$^{-2}$ s$^{-1}$, below which only sparse XMM-Newton and Chandra observations exist. *AXIS* can provide a much more complete coverage compared to the latter down to sensitivities of around $1.0 \times 10^{-15}$ erg cm$^{-2}$ s$^{-1}$, or over 2 orders of magnitude deeper than NICER. This will help track the late-time evolution of failed crustal areas and associated twisted field bundles, and inform us, for the first time, on twist and plasma migration as well as crustal relaxation throughout the complete life cycle of the outburst.

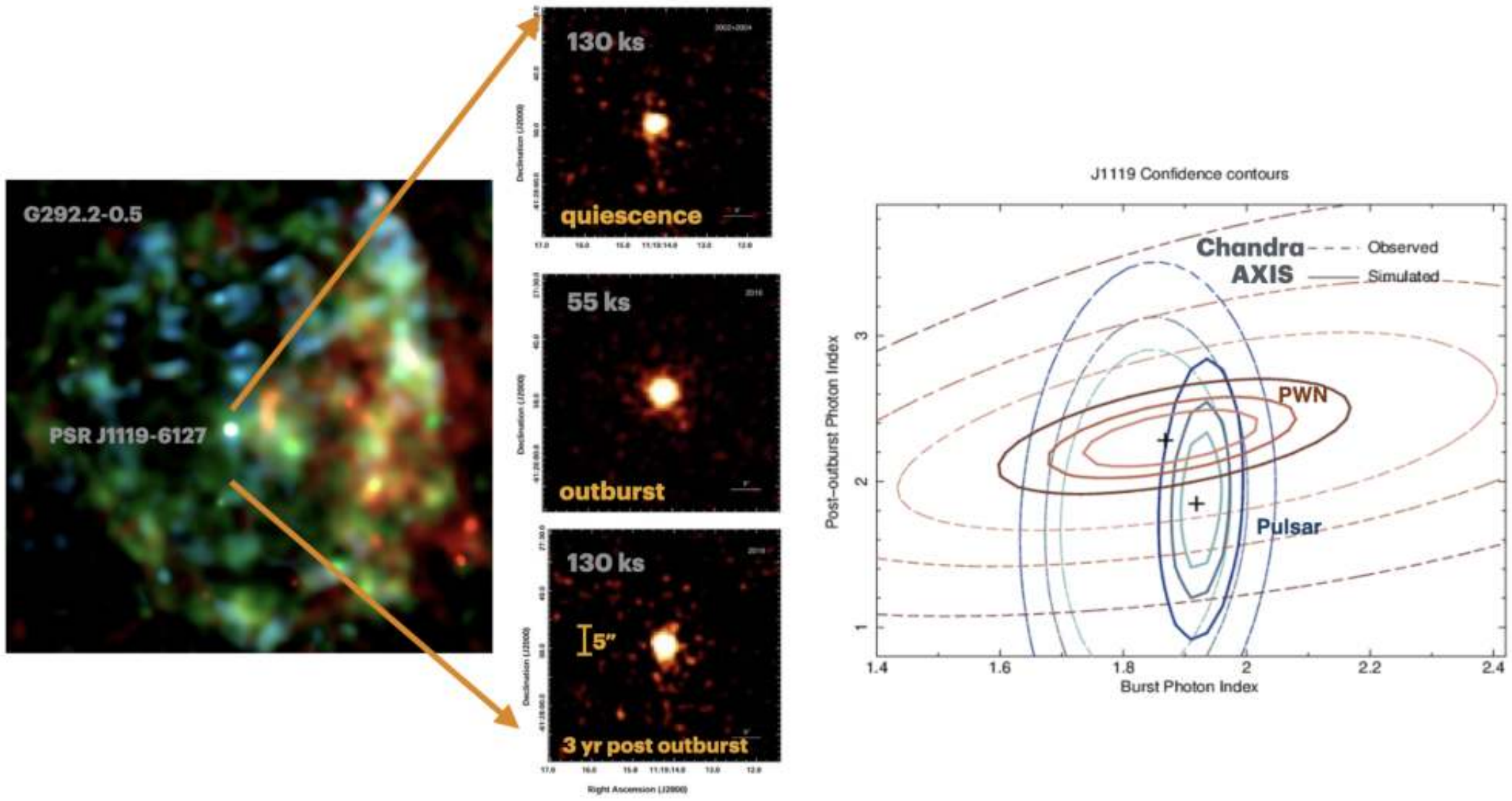

**Figure 58.** X-ray image of the SNR G292.2–0.5 hosting the highly magnetized neutron start J1119–6127 which exhibited a magnetar-like outburst in 2019. The middle panel shows the zoomed-in images of the compact arcseconds-scale PWN before, during, and 3 years post outburst where the PWN changed its morphology and spectral properties as monitored by Chandra (based on [70,71,203]). The right panel shows the confidence contour levels for the simulated-with-AXIS pulsar and PWN spectral indices during and post outburst (using the same exposures as Chandra shown in the middle panel), illustrating the need for AXIS to not only resolve the pulsar from its pulsar/magnetar wind nebula, but also to constrain the spectral index which sheds light on the powering emission mechanism in these bursting sources.

Fast radio burst-like emission has been detected from at least one magnetar in the Galaxy so far, SGR 1935+2154 [73,114], during a period of intense X-ray bursting activity (so-called burst storms, [632]). The brightest radio burst had a fluence comparable to the tail-end of extragalactic FRBs. This radio burst, and a few fainter ones, occurred simultaneously to millisecond X-ray bursts (e.g., [365]). It is unclear, on the other hand, why the majority of X-ray bursts occur in the absence of any radio bursting activity (e.g., [39,329,633]). It is critical to increase the sample size of simultaneous radio and X-ray burst detection spanning a range of radio and X-ray luminosities, which is possible only through rapid follow-up ($\sim$1 day) of recently active magnetars with a sensitive instrument like *AXIS*. This would allow a sample study of the population, especially the fraction of radio to X-ray burst energy over a range of brightness, which would inform on the emission mechanism (e.g., [346,598]), and will have large implications on the origin and plausible progenitors of extragalactic FRBs (e.g., [437]).

The direct detection of a magnetar burst storm could also inform us about the composition of the neutron star interior, an open question in neutron star physics (e.g., [106]). Recently, a dense monitoring campaign with NICER and NuSTAR around the time of strong bursting activity and a burst storm from the magnetar SGR 1935+2154 revealed the triggering of a spin-up glitch shortly before the most intense burst period [248]. This was accompanied by a jump in spin-down rate to $10^{-9}$ Hz s$^{-1}$, two orders of magnitude larger than any ever observed. This episode lasted for around 9 hours before another spin-up glitch brought the spin ephemeris to its pre-outburst level. This extreme spin evolution in such short time-scales implies that a large fraction of the magnetar interior (>10%) must be in a superfluid state, including a non-negligible fraction of the core. *AXIS* provides the necessary monitoring capabilities, effective area, and timing resolution for studying these effects in other magnetars, and in more detail. It is worth noting that an X-ray burst associated with an FRB-like burst occurred amid the enhanced spin-down period, which NuSTAR missed, albeit targeting SGR 1935+2154 due to Earth-occultation [248]. The proposed *AXIS* orbit ensures continuous source coverage during an observation, a critical criterion during periods of intense magnetar activity.

In summary, *AXIS* and the planned GPS will help uncover the class of dormant magnetars which not only connects the diversity of neutron stars and reveals the physics of MWNe, but also sheds light on the Galactic source population which is now believed to be connected to the mysterious class of extragalactic Fast Radio Bursts.

**Exposure time:** 1430 ks and the Galactic Plane Survey.

**[Observing description]**

The high angular resolution and low background are crucial for studying PWN emission and resolving it from the point source (pulsar or magnetar).

Need ToO observations of magnetar-like outbursts from neutron stars, within hours to days, and more long-term monitoring. Pile-up may be an issue for the brightest sources, although it is less of an issue than Chandra has been.

**[Joint Observations and synergies with other observatories in the 2030s:]**

This program is synergistic with radio observations, particularly for magnetar-like outbursts. For magnetar wind nebulae, SKA or ngVLA will either detect radio emission from them, or strongly constrain their SED. Jointly fitting radio and X-ray spectra will be much more informative than fitting either of them separately.

**k. PeVatrons, Extended Jets & Outflows**

*35. AXIS Galactic PeVatron Identifier (AGPI) program*

**First Author:** Kaya Mori (Columbia University, kaya@astro.columbia.edu)
**Co-authors:** Sabrina Casanova (Instytut Fizyki Jądrowej PAN), Hongjun An (Chungbuk National University), Giulia Brunelli (INAF Bologna), Stephen DiKerby (Michigan State University), Gilles Ferrand (University of Manitoba/RIKEN), Joseph Gelfand (NYU Abu Dhabi), Jeremy Hare (NASA GSFC), Oleg Kargaltsev (GWU), Naomi Tsuji (ICRR, University of Tokyo), Samar Safi-Harb (University of Manitoba), Isabel Sander (University of Manitoba), Ceaser Stringfield (Columbia University), Jooyun Woo (Columbia University) and Shuo Zhang (Michigan State University)
**Abstract:** Galactic PeVatrons are recognized as the most energetic and enigmatic phenomena in the Milky Way. The recent discovery of $\sim$50 ultra-high-energy (UHE; > 100 TeV) gamma-ray sources and TeV-PeV neutrino emission in the Galactic Plane has provided compelling evidence of the existence of Galactic PeVatrons capable of accelerating cosmic rays to $10^{15}$ eV and beyond. In the coming decade, the ground-based LHAASO and HAWC observatories, joined by SWGO covering the southern sky, are expected to uncover over 100 Galactic PeVatrons. However, most the Galactic PeVatrons remain unidentified due to the degeneracy of the hadronic-leptonic emission mechanisms in the UHE band. As demonstrated for numerous gamma-ray sources, multi-wavelength observations, particularly in the X-ray band, are crucial for determining the nature of Galactic PeVatrons, which are often classified as either leptonic or hadronic sources, also thanks to due to superior angular resolution of X-ray observations with respect to that of UHE telescopes. We propose an *AXIS* legacy survey of eight UHE sources. Due to the unknown and potentially diverse nature of Galactic PeVatrons, it is most critical to survey the whole TeV source region with a large FOV, high-angular resolution, and broad energy band. Our "*AXIS* Galactic PeVatron Identifier" (AGPI) program will identify X-ray counterpart candidates and determine their X-ray properties most effectively with the upcoming Cherenkov Telescope Array Observatory (CTAO) in the TeV band (which will have superior angular and energy resolution with respect to current gamma-ray observatories) and next-generation radio observatories (e.g., SKA and the ngVLA). High angular resolution across the FOV will allow for sub-arcsecond localization for potential X-ray counterparts of UHE particle accelerators, thus enabling accurate cross-matching to multi-wavelength counterparts from various surveys at lower frequencies and enabling machine-learning classification of the multi-wavelength information. *AXIS* can detect point-like sources, such as pulsars and TeV gamma-ray binaries, and resolve pulsar wind nebulae, supernova remnants interacting with molecular clouds, and other diffuse X-ray emissions. *AXIS* will also be able to determine whether diffuse X-ray sources are of thermal or non-thermal (synchrotron) origin spectroscopically. *AXIS* and CTAO will be a golden duo to determine gamma-ray emission and particle acceleration mechanisms of Galactic PeVatrons and their environments (e.g., ambient magnetic field and density distributions) through morphological and SED studies. Our AGPI program will provide the most sensitive X-ray probe of UHE sources and significantly advance our understanding of Galactic PeVatrons, ultimately resolving the origin of local cosmic rays detected up to the knee ($\sim$3 PeV).
**Science:** Cosmic rays (CRs) are highly energetic ($>$ 1 GeV) particles which impinge on the Earth from all directions in the sky with an energy density of about 1 eV/cm$^3$. That is comparable to the energy density of the Galactic magnetic fields, radiation fields, and turbulent motions of the interstellar gas. The CR population measured at Earth is mainly composed of protons and heavier nuclei, with electrons and positrons, representing roughly 1% of the flux at GeV energies. The CR spectrum, which extends over a vast range of energies from $10^9$ to $10^{21}$ eV, has a prominent feature, the so-called knee, at a few PeV (PeV =$10^{15}$ eV). Thanks to the increasing precision of CR measurements, several features in the spectra of all CR species are being discovered, among which the hardening of the proton and heavy nuclei spectra at about 200 GeV/n in rigidity, the different CR species showing discrepant spectral hardening indices. The knee remains a very prominent feature in the CR spectrum and is thought to mark the onset from Galactic into extra-Galactic CRs, but could also be a transport feature or due to the convolution of different source spectra. The nature of the astrophysical sources that accelerate particles up to and beyond several PeVs, the so-called *PeVatrons or super-PeVatrons*, is yet to be clarified.

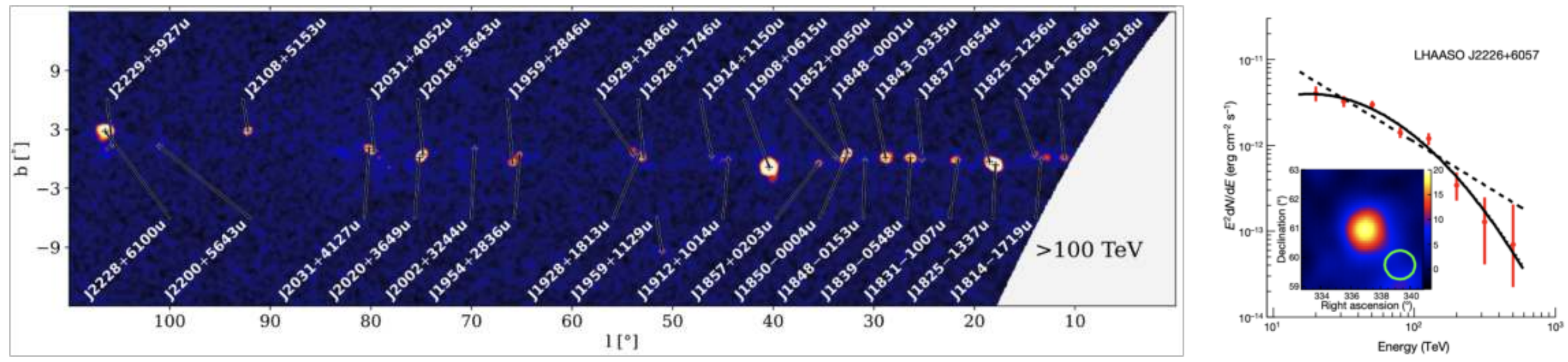

**Figure 59.** *Left:* LHAASO > 100 TeV map with UHE sources within the Galactic disk [101]. *Right:* TeV spectrum and image of LHAASO J2226+6057 [100]. The PSF size ($\theta = 0.49^\circ$) is shown as a green circle.

Since interstellar magnetic fields bend the CR's trajectories, we can only indirectly locate CR accelerators through associated gamma-ray or neutrino sources. In particular, a powerful tracer for CR populations distant from the Earth is gamma rays. These are emitted by two dominant emission processes, namely the decay of neutral pions, $\pi_0$, produced when CR hadrons collide with ambient gas in the ISM, which is commonly called the *hadronic* production mechanism, and inverse Compton (IC) scattering of CR electrons off radiation fields, which is called the *leptonic* production mechanism.

In the last few years, the hunt for the most extreme CR factories in the Galaxy, the PeVatrons, has become a major focus in high-energy astrophysics with the advent of extensive air shower arrays. The search for PeVatron sources has been given a new impulse by the survey of the Galactic Plane in the so-called UHE (ultra-high-energy; $E_\gamma > 100$ TeV) band carried out with the currently-operating HAWC, LHAASO and Tibet AS-$\gamma$ Observatories. In the future, similar searches will be undertaken by the Southern Wide-field Gamma-ray Observatory (SWGO) [250] and by the upcoming Cherenkov Telescope Array Observatory (CTAO) [18], which will benefit from higher angular resolution and help compare TeV gamma-ray sources and known particle accelerator candidates (e.g., SNRs, PWNe, compact object binaries) or targeting molecular clouds.

Recent discoveries of $\sim 50$ UHE sources in the Galactic Plane established the existence of *Galactic PeVatrons* that can accelerate particles to PeV energies (e.g., [100]; Figure 59). In addition, the IceCube collaboration unveiled Galactic diffuse neutrino emission, providing direct proof of hadronic PeVatrons in our galaxy [258]. The photon energies reported by the LHAASO and HAWC collaborations are so high that the attenuation of these high-energy gamma-rays, due to pair production of gamma-rays interacting with background photons from both the cosmic microwave background (CMB) and interstellar radiation fields (ISRF), discussed among others by [644], revealed itself thanks to the capability of extensive air shower arrays to observe at the highest energies, detect relatively distant sources, and measure the gamma-ray spectrum with relatively small statistical uncertainties. The solution to the question of the maximum acceleration energy achieved in the Galaxy and the major contributors to Galactic CR flux must necessarily come from high-significance spectra of multi-TeV to PeV photons produced by CRs close to the knee energy, colliding with the ambient gas. The spectrum of the highest energy gamma-ray radiation from different PeVatron candidates contains crucial information on the source acceleration mechanisms and the contribution of the different sources to the formation of the knee in the local CR spectrum.

However, identifying the UHE sources is challenging due to the poor angular resolution of extensive air shower arrays ($\Delta\theta \sim 0.2-0.7^\circ$ [325]), which hinders cross-matching to known Galactic sources. Imaging

Air Cherenkov Telescope (IACT) observations by H.E.S.S., VERITAS, and MAGIC at $E_\gamma < 50$ TeV can help resolve the UHE sources with $< 0.1^\circ$ angular resolution, but only a fraction of the UHE sources are associated with known IACT sources. In general, TeV and GeV observations alone cannot distinguish between the leptonic and hadronic cases. Currently, while some of the UHE sources have been successfully associated with known particle accelerators such as SNRs, PWNe, and microquasars, a majority of them remain unidentified. Below the gamma-ray band, X-ray telescopes have played a crucial and unique role by detecting synchrotron radiation originating from primary and secondary electrons in the leptonic and hadronic accelerators, respectively. While radio emission originates from GeV particles, non-thermal X-ray emission traces the most energetic (TeV–PeV) primary/secondary $e^{\pm}$, which emit synchrotron photons in the X-ray band ($E_\gamma^{\rm syn} = 4(E_e/100\rm TeV)^2(B/10\ \mu G)$ keV) for typical ISM and molecular cloud (MC) B-fields ($B \sim 1{-}10^3\ \mu$G). Hard X-rays and UHE gamma rays trace the same electron population. Combining these information helps constraining both the electron spectrum and the magnetic field in these sources.

X-ray observations of Galactic TeV sources by Chandra, XMM-Newton, Suzaku and NuSTAR played complementary and essential roles in identifying pulsars, PWNe, X-ray filaments/knots, and diffuse X-ray emission from MCs (e.g., [646]). In addition, high-resolution CO line observations can detect MCs as targets for hadronic interactions [579]. In general, finding X-ray and CO line counterparts is the most robust way to identify particle acceleration sites and target MCs, and a combination of morphology and spectral energy distribution (SED) data in the X-ray and TeV band allow us to determine particle acceleration, cooling and propagation mechanisms (e.g., [429,508]). Given its arcsecond angular resolution, large effective area, and FOV, *AXIS* is uniquely suited for identifying UHE sources in conjunction with the upcoming TeV telescopes such as CTAO.

**Exposure time (ks): 560 (total). 8 targets $\times$80 ks**

**Observing description:**

We propose to conduct an *AXIS* X-ray survey of eight unidentified UHE sources detected above $\sim 100$ TeV. They are the best candidates for Galactic PeVatrons, and we aim to detect and characterize diffuse (synchrotron) X-ray emission. High-quality *AXIS* X-ray morphological and spectral information, as demonstrated by one of the UHE sources below through our simulations, will be incorporated into multi-wavelength SED studies to identify source types (e.g., PWNe) and gamma-ray emission mechanisms (e.g., leptonic vs. hadronic processes).

To demonstrate the unique capabilities of *AXIS* for exploring Galactic PeVatrons, we present simulations of an 80-ksec *AXIS* observation of 1LHAASO J0343+5254u, which is one of the unidentified UHE sources detected by LHAASO. Within the LHAASO source, a recent XMM-Newton observation detected a diffuse X-ray source extending to $r \sim 1.8'$ with an X-ray flux of $1.3 \times 10^{-12}$ erg cm$^{-2}$ s$^{-1}$ [162]. The diffuse X-ray source was suggested to be a PWN due to its centrally-peaked morphology, non-thermal X-ray spectrum, and marginal spectral softening with increasing radius from the center. In addition, Nobeyama CO observations revealed five distinct MCs surrounding the LHAASO source, and they are potential hadronic interaction sites [579]. These X-ray and CO line detections still do not lead to identifying the UHE source as originating from either the leptonic or hadronic origins associated with the PWN-like diffuse X-ray source and the dense cloud, respectively.

Figure 60 shows simulated *AXIS* images of the PWN candidate (left) and dense molecular clouds (right), representing the leptonic and hadronic origin of the UHE emission, respectively. Within the PWN candidate, we will search for its neutron star and PWN's fine structures, such as a torus and/or jet, which can only be resolved by *AXIS* with arcsecond angular resolution, serving as a smoking gun to establish the PWN hypothesis. We will also perform spatially-resolved spectral analysis within the PWN, confirm the synchrotron burn-off effect, and provide multi-zone X-ray emission information for sophisticated multi-wavelength SED modeling [429]. On the other hand, detecting diffuse X-ray emission from the MCs (produced via synchrotron radiation from secondary electrons) will establish the hadronic origin. In either case, we will incorporate *AXIS* X-ray spectra to multi-wavelength SED data and determine the underlying particle energy spectra with high precision (e.g., $E_{\rm max}$ and power-law indices). Repeating these processes for other unidentified UHE sources will allow us to estimate their contributions to the local CR spectra around the knee. Other UHE sources may require multiple pointings to cover their more extensive TeV emission regions.

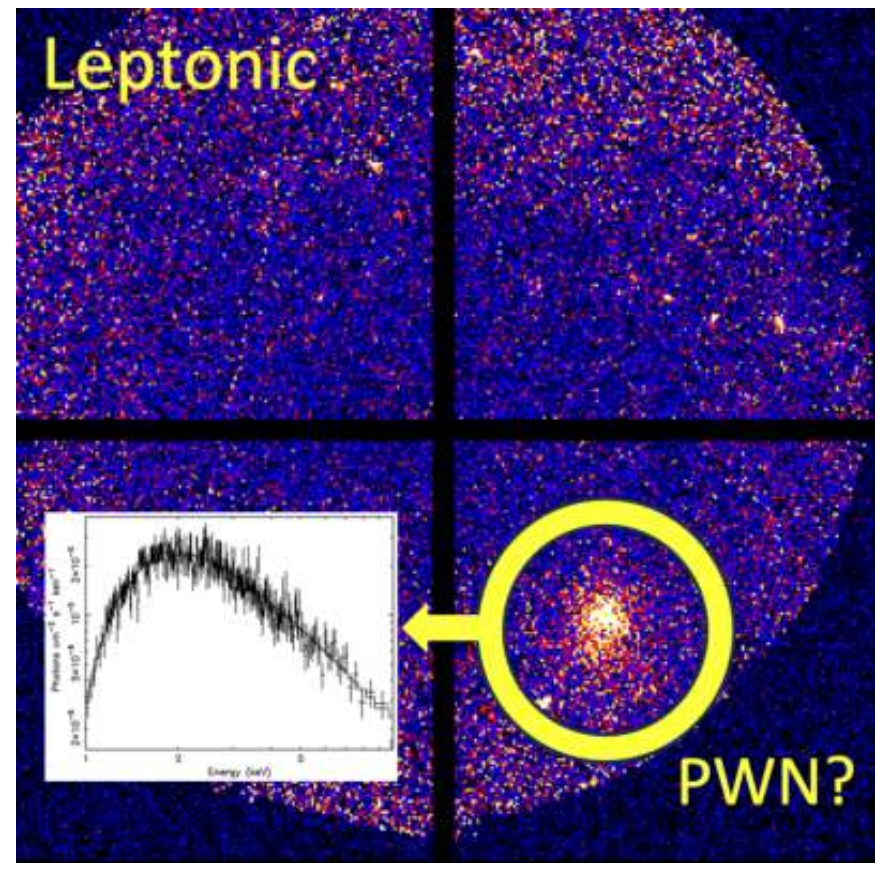

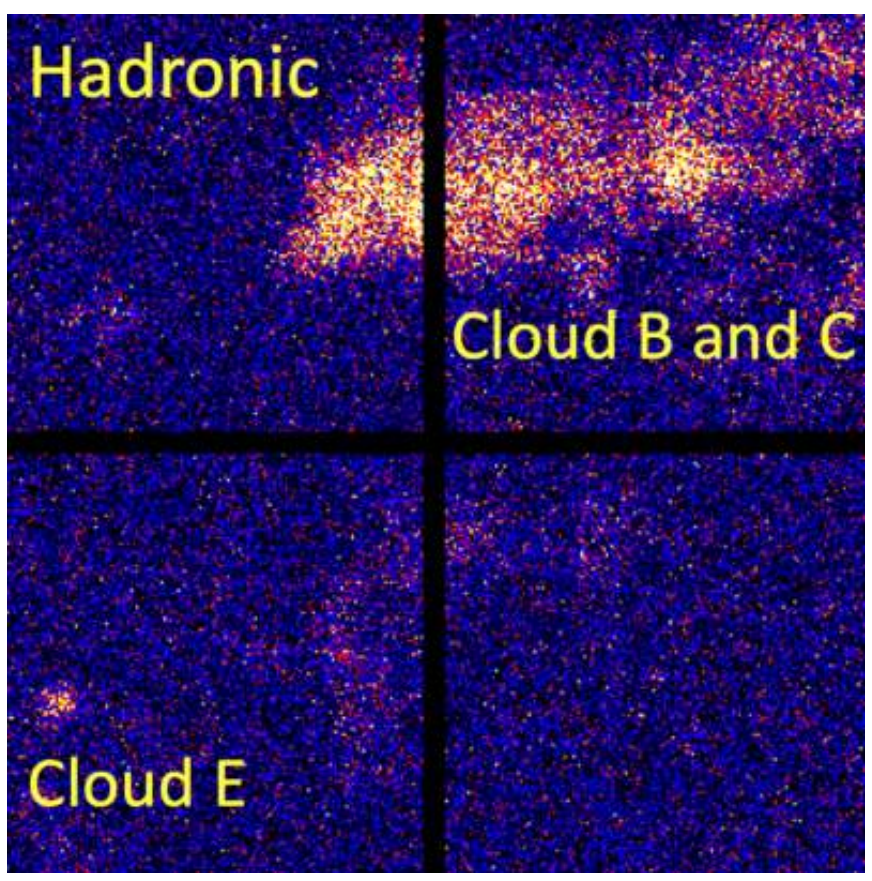

**Figure 60.** *Left:* Simulated AXIS image and spectrum of the PWN candidate potentially associated with 1LHAASO J0343+5254. The simulated image was generated from the XMM-Newton observation of the LHAASO source [162]. *Right:* Simulated 80-ks *AXIS* image of the MCs named in [579]. Note that the XMM-Newton observation has not covered these MCs. These two plots represent the leptonic and hadronic origins of UHE emission.

**Joint Observations and synergies with other observatories in the 2030s:** CTAO, LHAASO, HAWC, Tibet-AS$\gamma$, SWGO, IceCube, KM3NeT, P-ONE, HUNT, SKA, ngVLA, ALMA2 and LST.

*36. Particle acceleration in microquasars*

**First Author:** Naomi Tsuji (ICRR, ntsuji@icrr.u-tokyo.ac.jp)
**Co-authors:** Kaya Mori (Columbia University), Brydyn Mac Intyre, Samar Safi-Harb (University of Manitoba), Laura Olivera-Nieto (MPIK), Sabrina Casanova (Instytut Fizyki Jądrowej PAN), Federico Fraschetti (Harvard, CfA), and Martin Mayer (FAU Erlangen)

**Abstract:** The origin of Galactic cosmic rays (CRs) below a few PeV (the so-called knee) remains a long-standing question in astrophysics. While supernova remnants are widely accepted as the primary accelerators of Galactic CRs, they face increasing challenges in accounting for the maximum energy (knee) of Galactic CRs. Alternatively, microquasars — black hole or neutron star systems with jets — have gained significant attention as potential PeV CR accelerators (PeVatrons), particularly following recent detections of very-high-energy (VHE; $E > 0.1$ TeV) and ultra-high-energy (UHE; $E > 100$ TeV) gamma rays. Remarkably, the detection of gamma rays up to 0.8 PeV from the microquasar V4641 Sgr would have established a new picture that microquasars may contribute to the production of CRs up to or even beyond the knee. However, the limited capabilities of the current gamma-ray telescopes prevent a detailed investigation of the exact location of the acceleration site and the acceleration mechanism.

We propose *AXIS* observations of gamma-ray-emitting microquasars to understand the particle acceleration mechanism and assess their potential as PeVatrons. An extensive X-ray survey of the microquasar SS 433 and its associated nebula, W50, has demonstrated the important role of analyzing X-ray counterparts, which reveal non-thermal X-ray spectral properties in detail within the gamma-ray lobes. Furthermore, high-angular-resolution Chandra observations have provided invaluable insights into microquasar jets. Combined with *AXIS* observations in 2030s, this will establish a crucial baseline for future studies and enable long-term (>10-yr) investigations of spectral evolution and proper motion. Currently, only five microquasars (SS 433, V4641 Sgr, GRS 1915+105, MAXI J1820+070, and Cygnus X-1) are confirmed as TeV-PeV gamma-ray sources. However, future gamma-ray surveys with existing telescopes and next-generation instruments (e.g., SWGO, planned for the 2030s) may expand this sample. *AXIS* observations will be timely and essential for obtaining multi-wavelength data on these newly identified gamma-ray-emitting microquasars, further advancing our understanding of their role as cosmic-ray accelerators.

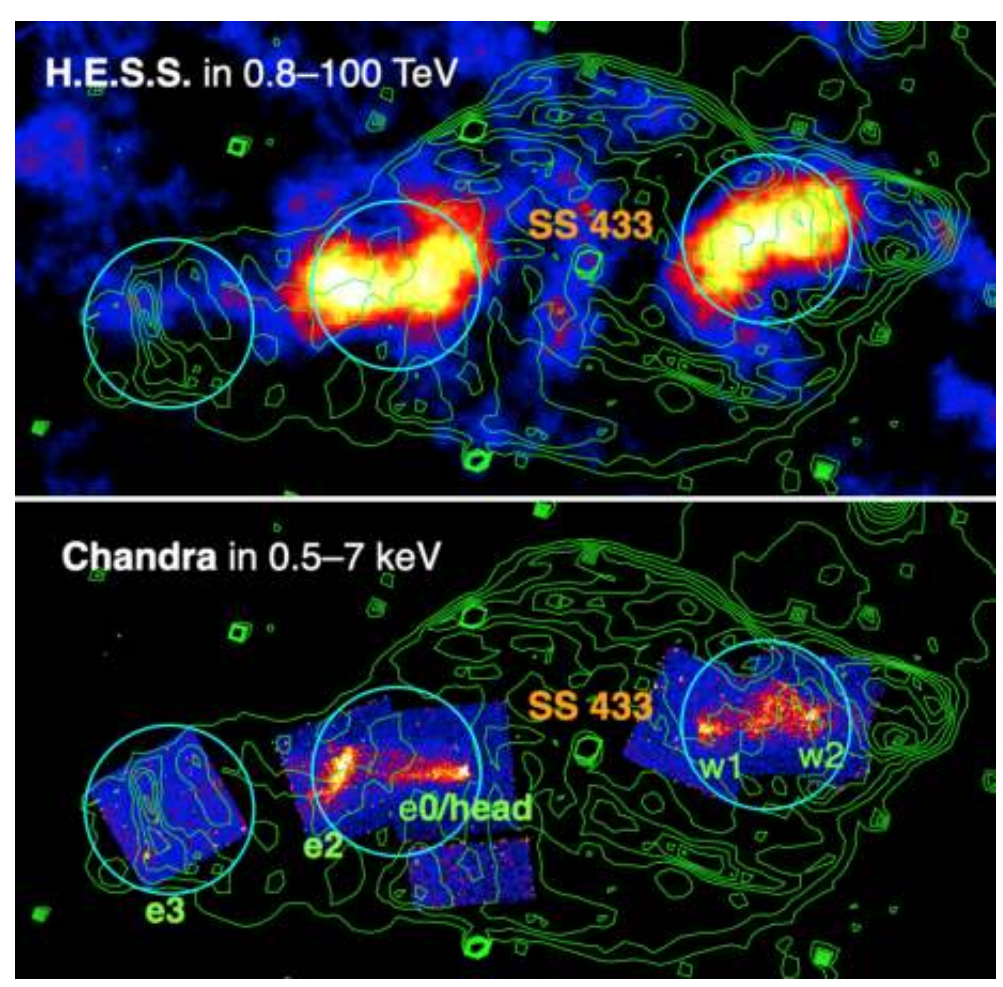

**Figure 61.** 0.8–100 TeV gamma-ray image by H.E.S.S. (top; [242]) and 0.5–7 keV X-ray image by Chandra (bottom; [580]). The contours indicate 1.4 GHz radio continuum emission. The circles show the FoVs of AXIS.

**Science:**

The SS 433/W50 system represents the first microquasar detected in the TeV gamma-ray band. Combined with the extensive X-ray data, it serves as an ideal laboratory for probing particle acceleration mechanisms in microquasars. The TeV gamma rays, spatially well correlated with nonthermal X-rays, are emitted from knot-like structures located within the eastern and western lobes, likely powered by jets launched from SS 433 (see Figure 61) [e.g., 4,283,508,580]. H.E.S.S. revealed that the higher energy gamma-ray radiation ($E > 10$ TeV) originates from the innermost knots (e0/head and w1) [242], which exhibit the hardest X-ray spectra with $\Gamma \approx 1.5$, indicating these regions as the particle acceleration sites.

Figure 62 illustrates the spatial profiles of column density, photon index, and flux along the western lobe in SS 433/W50 [283]. Such arcsecond-scale (sub-pc-scale at a distance of $d =$ 5.5 kpc) spectral profiles can only be resolved with Chandra and *AXIS*, enabling detailed modeling of particle propagation and cooling along the jets [e.g., 552]. Mapping the spectral features facilitates three key science objectives:

1. Identify acceleration sites: Regions with the hardest spectra, which are w1 and e0 in SS 433, are likely locations of active particle acceleration. This can also be confirmed with gamma-ray observations by imaging atmospheric Cherenkov telescopes (IACTs) with great angular resolution, such as H.E.S.S. and CTAO.
2. Constrain propagation properties: The gradual softening and flux decrease toward the outer lobes in SS 433 suggest that accelerated particles (electrons) are injected at the innermost knot and subsequently propagate and cool along the jets. By applying such a theoretical model [552] to the observed X-ray spectral profile, the flow speed and magnetic field were constrained to be 0.65–0.26$c$ and 5–10 $\mu$G, respectively [283].
3. Unveil the nature of knot formation: Regions with enhanced $N_{\rm H}$ that spatially coincide with molecular clouds, such as the area between w1 and w2 in SS 433 [627], would indicate jet-cloud interactions, which may affect the formation of the X-ray knots. Furthermore, to reconcile with the observed abrupt increases in $\Gamma$ and flux, the w2 knot can be interpreted as a site of enhanced magnetic field with $\sim$50 $\mu$G [283]. The amplification of the magnetic field can be tested by monitoring the flux variability, as energetic electrons rapidly cool in strong magnetic fields. Future observations with *AXIS*, e.g. in 2035, will enable to obtain $\sim$10-yr ($\sim$30-yr) time interval since the existing X-ray data in 2023-2024 (2003-2004). Detecting such variability can demonstrate the magnetic field amplified up to $\sim 100$ $\mu$G, providing crucial constrains on knot formation mechanisms.

**Figure 62.** The $N_H$, $\Gamma$, and flux profiles along the jets in SS 433, overlaid with the theoretical model with different values of the magnetic field [283,552]. The advection speed is assumed to be 0.065$c$.

Figure 63 shows the simulated 50-ks *AXIS* image of the eastern lobe and projected flux profile of the e0/head knot. Proper motion measurement using Chandra data over a $\sim$20-year baseline revealed a possible outward shift of 0.18 arcsec/yr (corresponding to $v \sim 0.016c$ at $d$ = 5.5 kpc), although this measurement is currently dominated by systematic uncertainties, conservatively resulting in an upper limit of 0.50 arcsec/yr ($v < 0.04c$) [580]. Based on these findings, the knot is expected to shift by 2 arcsec for $v \sim 0.016c$ between the Chandra observations in 2023-2024 and a future *AXIS* observation in 2035 as shown in Figure 63. We also show a case with the higher velocity of 0.065$c$, which is 1/4 of the jet speed launched from SS 433. The *AXIS*'s great angular resolution will enable to measure these motions along the lobes; from the innermost knots (e0 and w1) to the outer knots (e2–e3 and w2), which will provide new insights into the dynamics of the 100-pc-scale jets in this microquasar. Furthermore, the proper motion measurement at the innermost knots is crucial for understanding the acceleration mechanism, since the observed speed can be converted into the shock speed $v_{\rm sh}$ and the acceleration efficiency on the assumption of the DSA-type acceleration process. For example, a proper-motion speed of $v \sim 0.016c$ would imply $v_{\rm sh} \sim 10{,}000$ km/s and the most efficient (Bohm-limit) acceleration, assuming the downstream velocity of $> 0.025c$ [242].

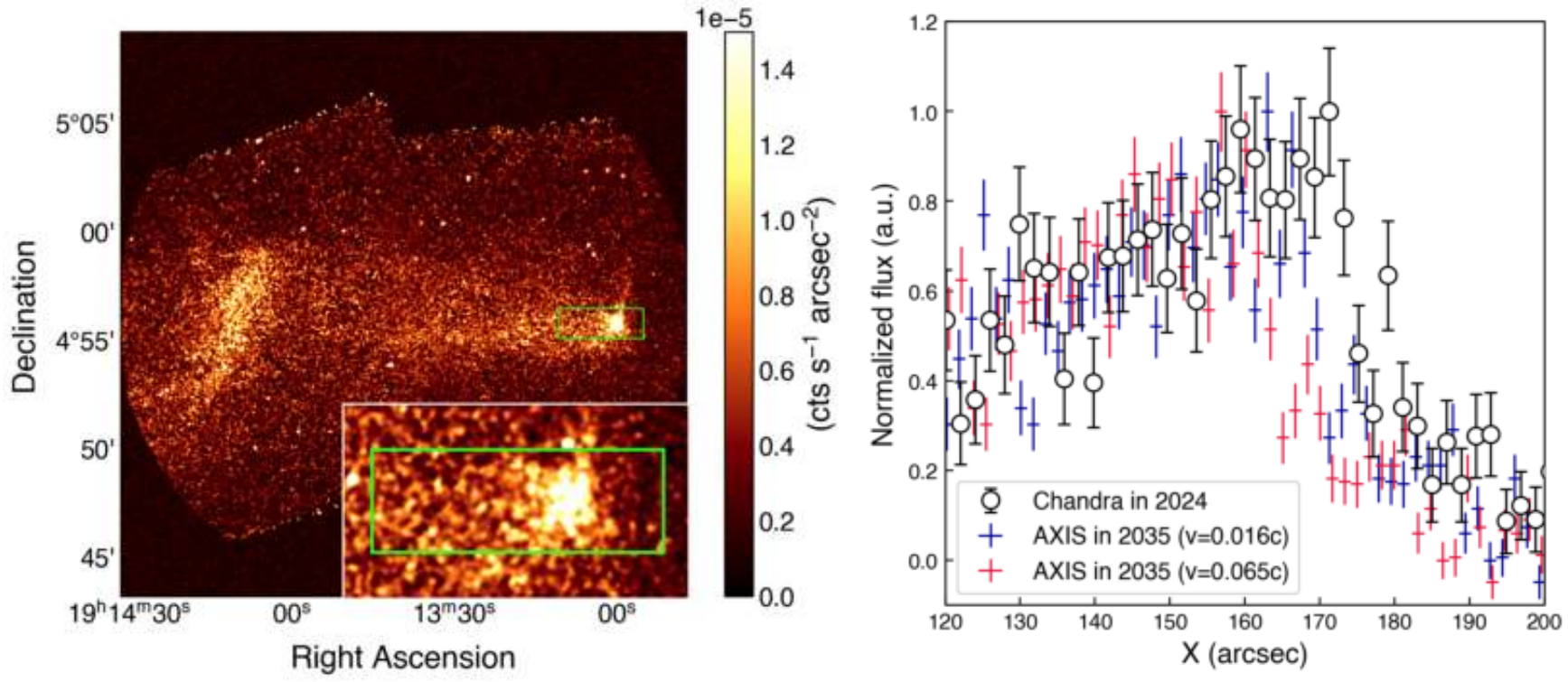

**Figure 63.** Left: simulated 50-ks AXIS image at the eastern lobe (inset: zoom-in image of the e0/head knot). Right: Projection profiles at the e0/head knot, assuming the velocities with 0.016*c* and 0.065*c*, with *AXIS* in 2035. The profile of the Chandra data in 2024 is also shown. The background, estimated from the non-knot region, is subtracted.

With *AXIS*, we aim to expand this study to other microquasars. The spatially resolved X-ray spectral analysis, which is hardly feasible in the gamma-ray band, will play a crucial role for probing acceleration sites, propagation mechanisms, and the physical origin of knot formation in microquasars. It should be noted that observations will be conducted during quiescent phases to minimize contamination from the central compact object. To date, five microquasars have been detected at the TeV gamma-ray range (SS 433, V4641 Sgr, GRS 1915+105, MAXI J1820+070, and Cygnus X-1). This number is expected to at least double in the 2030s, thanks to observations by the existing and next-generation observatories. We plan to perform follow-up *AXIS* observations on these newly identified sources, enabling a systematic, high-resolution X-ray study of particle acceleration in Galactic microquasars.

**[Exposure time (ks):]** 700 ks
**Observing description:**

Table 6 summarizes our target sources — five microquasars currently reported as TeV gamma-ray emitters. The diffuse X-ray emission around SS 433 has been extensively studied, making it an ideal reference case for this study. We propose three pointings to sufficiently cover the nonthermal X-ray knots (e2, e0/head, w1, and w2) in gamma-ray emitting lobes, as well as the termination region in e3, where the thermal emission is dominated (Figure 61). On the other hand, the gamma-ray emitting regions of the other microquasars were not fully explored in X-ray. Thus, we present preliminary X-ray surface brightness estimates derived from the observed X-ray flux in SS 433 and V4641 Sgr, along with the corresponding distances [283,326,556]. The ratio of X-ray to gamma-ray fluxes can be used to constrain the magnetic field strength in the leptonic gamma-ray scenario. To obtain the spatially resolved (arcsecond-scale) spectra and achieve the proposed scientific objectives, the required exposure and number of pointings are determined based on the column density, the gamma-ray extension, and the preliminary estimates of the X-ray surface brightness. In addition to the known gamma-ray microquasars, we allocate an additional 150 ks for follow-up observations of other gamma-ray-emitting microquasar candidates, which are expected to be discovered at the VHE-UHE gamma-ray bands with future observations.

**[Joint Observations and synergies with other observatories in the 2030s]:** We plan to coordinate joint observations with IACTs in the 2030s, including CTAO, ASTRI Mini-Array, and LACT, among others.

**Table 6.** Target microquasars and observation plan.

| **Microquasar** | **Coordinate** ($\ell$, $b$) | **Distance** (kpc) | $N_{\rm H}$ ($10^{22}$ cm$^{-2}$) | **Extension** (deg) | $F_\gamma^{(a)}$ | $F_X^{(b)}$ | ***AXIS* Observation** |
|---|---|---|---|---|---|---|---|
| SS 433 | (39.69, −2.244) | 4.6 | 0.5–3 | 0.70 | 0.50 | $\sim$2–30 | Three pointings × 50 ks |
| V4641 Sgr | (6.774, −4.789) | 6.2 | 0.2–1.8 | 0.5 | 3.9 | $\sim$1 | Three pointings × 70 ks |
| GRS 1915+105 | (45.37, −0.219) | 9.4 | 1.4 | 0.33 | 0.17 | $\sim$0.4? | One pointing × 80 ks |
| MAXI J1820+070 | (35.85, 10.16) | 2.96 | 0.1 | <0.28 | 0.13 | $\sim$4? | One pointing × 50 ks |
| Cygnus X-1 | (71.33, 3.067) | 2.2 | 0.7 | <0.22 | <0.01 | $\sim$8? | One pointing × 60 ks |
| Others if detected | — | — | — | — | — | — | 150 ks |

[a] $F_\gamma$ is a flux at 100 TeV in units of $10^{-12}$ erg cm$^{-2}$ s$^{-1}$.
[b] $F_X$ is surface brightness in 2-10 keV in units of $10^{-14}$ erg cm$^{-2}$ s$^{-1}$ arcmin$^{-2}$. For GRS 1915+105, MAXI J1820+070, and Cygnus X-1, the entire gamma-ray extent was not covered in X-ray, thus $F_X$ was preliminarily estimated assuming the faintest flux case in SS 433 and V4641 Sgr and the distance.

Note that simultaneous observations are not required. In addition, our program benefits from strong synergies with other gamma-ray observatories capable of continuous all-sky monitoring, including HAWC, LHAASO, Tibet AS-$\gamma$, SWGO, and ALPACA.

*37. Hypernebulae*

**First Author:** Navin Sridhar (Stanford University); nsridhar@stanford.edu
**Co-authors:** (with affiliations)

**Abstract:**

The evolution of massive star binaries is a crucial topic for a wide variety of areas in astrophysics, including local- and galaxy-scale feedback, heavy-element nucleosynthesis, and some of the most energetic transients in the universe, such as supernovae, gamma-ray bursts, and compact binary merger gravitational wave sources. In compact object binaries, when an evolved post-main-sequence donor star grows faster upon mass-loss than its Roche radius, the mass-transfer can become unstable, giving rise to a runaway increase in mass-loss and the likely engulfment of the accretor in the envelope of the donor star (common envelope). Before common envelope events, the compact objects are typically fed above their 'Eddington accretion rate', and therefore, lose a significant fraction of the transferred mass in the form of powerful disk winds, jets, and circumbinary disks. This event is expected to give birth to a new class of compact ($<$ a few pc) sources called *hypernebulae* [538]. Embedded within the hypernebula are relativistic electrons heated at the termination shock of the faster wind/jet from the inner accretion flow. These sources, which can last for decades to millennia, are not only radio-synchrotron-bright but also appear as supersoft X-ray sources and could act as signposts to imminent common envelope merger transients. *AXIS* could play a crucial role in the discovery and characterization of hypernebulae, which could be the keystone to understanding binary stellar evolution.

**Science:**

Let's consider a 'fiducial' system comprised of a stellar mass compact object, say a black hole with mass $M_\bullet = 10\,M_\odot$, and an evolved companion star of mass $M_\star = 30\,M_\odot$ that is transferring matter at an Eddinton-normalized rate $\dot{M}/\dot{M}_{\rm Edd} = 10^5$. A large-scale quasi-spherical bubble is inflated by the slow winds (with speeds $v_{\rm w}/c \sim 0.03$ and luminosity $L_{\rm w} \sim 0.5\dot{M}v_{\rm w}^2$) from the accretion disk, and a faster outflow/jet is launched with speeds $v_{\rm j}/c \sim 0.3$ and a luminosity $L_{\rm j} = \eta L_{\rm w}$. These outflows interact with the circumstellar medium (CSM) with a number density $n_{\rm csm} = \rho_{\rm csm}/\mu m_{\rm p} \sim 10\,{\rm cm}^{-3}$ via a forward shock (FS).

After the radiative transition, the total luminosity $L_{\rm fs}$ radiated behind the FS will follow its kinetic luminosity,

$$L_{\rm fs} \approx \frac{9\pi}{8}\rho_{\rm fs}R^2 v_{\rm fs}^3 \approx 6\pi R^2 v_{\rm fs} n k T_{\rm fs} \approx 7\times10^{41}\,{\rm erg\,s^{-1}}\,L_{\rm w,42}, \tag{4}$$

where $\rho_{\rm fs}[R] = \rho_{\rm csm}$, and we use the notation. This FS also heats the gas to a temperature set by the shock jump conditions,

$$kT_{\rm fs} = \frac{3}{16}\mu m_{\rm p} v_{\rm fs}^2 \approx 130\,{\rm keV}\left(\frac{L_{\rm w,42}}{n_1}\right)^{2/5}\left(\frac{t}{70\,{\rm yr}}\right)^{-4/5}, \tag{5}$$

where $v_{\rm fs} = {\rm d}R/{\rm d}t = 3R/5t$ is the FS velocity, corresponding to the deceleration phase.

The free-free emission from the shock will peak at photon energies $h\nu_{\rm X} \sim kT_{\rm fs}$, typically in the X-ray band. One can estimate the X-ray luminosity as the portion of the shock's total luminosity emitted via free-free emission:

$$L_{\rm X,fs} \approx \left(\frac{\Lambda_{\rm ff}}{\Lambda}\right) L_{\rm fs} \simeq 10^{-9}\left(\frac{T}{\rm Kelvin}\right)^{0.2} L_{\rm fs}, \tag{6}$$

where $\Lambda$ and $\Lambda_{\rm ff}$ are the total cooling function and free-free emission's contribution to it. Note that this is a conservative lower limit on $L_{\rm X,fs}$ because it assumes all line emission is radiated in the optical/UV instead of X-rays.

Fig. 64, adopted from [538], shows the FS X-ray luminosity for the same parameters and ranges of $\dot{M}$ and system age as in [538]. The highest X-ray luminosity $L_{\rm X} \sim 10^{40}\,{\rm erg\,s^{-1}}$ is attained for systems accreting

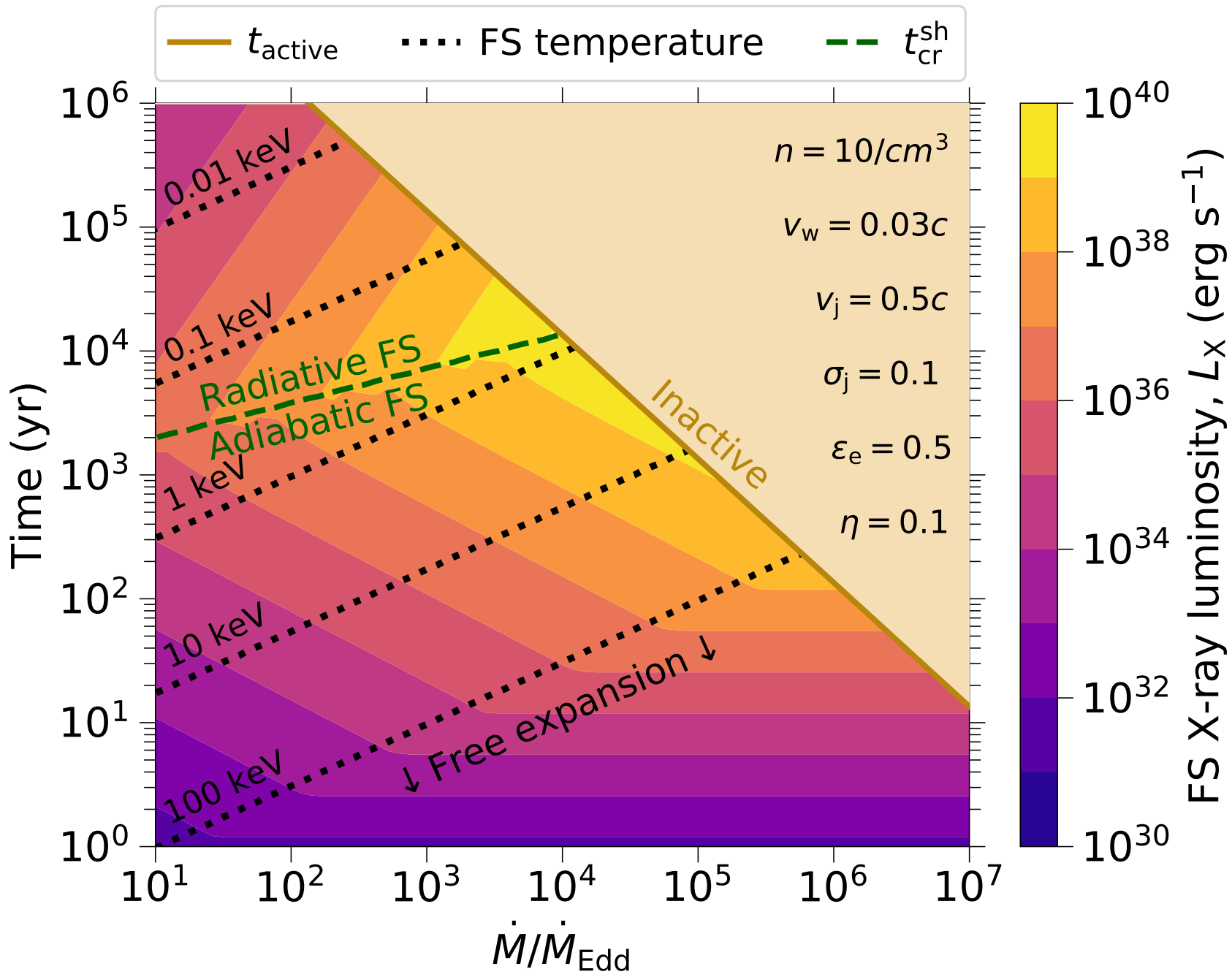

**Figure 64.** Colored contours show the thermal X-ray luminosity (Eq. 6) powered by the forward shock of hypernebulae as a function of the mass transfer rate $\dot{M}/\dot{M}_{\rm Edd}$ for the fiducial parameters of [539]. The forward shock initially expands freely, before beginning to decelerate at later times. Black dotted contours denote the temperature of the gas heated behind the forward shock (Eq. 5).

at $\dot{M}/\dot{M}_{\rm Edd} \sim 10^4$, after the FS becomes radiative at time $t \sim 10^4$ years. The peak emission temperature drops from $kT \sim$100 keV during the early shock-free expansion times ($t \sim t_{\rm free}$) to $kT \sim 0.1 - 1$ keV (through the AXIS band) conveniently by the time the X-ray luminosity peaks. Assuming a hypernebula at 100 Mpc, the flux of the free-free X-ray emission from the pc-scale bubble would be $\sim 10^{-15}$ erg s$^{-1}$ cm$^{-2}$, which AXIS can detect with an exposure time of 2 ks.

**Broader astrophysical implications:** The flaring jets of super-Eddington accreting stellar-mass compact objects, if pointed at our line of sight, could be sources of a sub-population of repeating fast radio bursts (FRB; [539]). Some FRB sources are accompanied by spatially- coincident, compact (<few pc), bright ($\nu L_\nu \gtrsim 10^{29}$ erg/s) persistent radio sources (PRS). If such FRBs are powered by accreting engines, the hypernebula surrounding them could generate the PRS and contribute large and time-variable rotation measure (RM) to the FRB pulses, consistent with those seen from FRBs 20121102 and 20190520B. Furthermore, the ions accelerated at the termination shock—where the collimated fast disk winds/jet collide with the slower, wide-angled wind-fed shell—can generate high-energy neutrinos via hadronic (pp) reactions, and photohadronic ($p\gamma$) interactions with the disk thermal and Comptonized nonthermal background photons. If the hypernebula birth rate follows that of common envelope events with a compact object, their volume-integrated neutrino emission could explain up to ∼25% of the high-energy diffuse neutrino flux observed by the IceCube Observatory [540]. The large optical depth through the hypernebula to Breit-Wheeler ($\gamma\gamma$) interaction attenuates the escape of GeV-PeV gamma-rays co-produced with the neutrinos, rendering these gamma-ray-faint neutrino sources consistent with the Fermi observations of the isotropic gamma-ray background. *AXIS* could play a crucial role in the discovery and characterization of hypernebulae, which could be the keystone to understanding binary stellar evolution.

- **Angular resolution**: The angular size of a hypernebula of size $s$ at a distance of $d$ is $d\theta = 0.2'' \left(\frac{s}{1\,\mathrm{pc}}\right)\left(\frac{d}{10\,\mathrm{Mpc}}\right)$. Assuming an angular resolution of $2''$ for *AXIS*, we note that all hypernebulae would appear as unresolved point sources offset from their host galactic nuclei.
- **Exposure time (ks):** $\sim$100 ks.
- **Joint Observations and synergies with other observatories in the 2030s:** Being radio-synchrotron-bright sources, it is estimated that tens of thousands of hypernebulae may already be present in current radio surveys like VLASS [538]. Furthermore, the line cooling contribution of the shock-ionized plasma of the hypernebula would predominantly be emitted in UV/optical bands, with a peak optical counterpart with a luminosity of $10^{39}\,\mathrm{erg\,s^{-1}}$, corresponding to a V-band apparent magnitude of $\sim$24 for sources located at $\sim$100 Mpc. *AXIS* may be well-suited to coordinate with current optical surveys, such as Pan-STARRS, and future surveys, like Rubin (LSST), to identify and characterize hypernebulae in their survey data.

1. Abbott, B. P., Abbott, R., Abbott, T. D., et al. 2017, ApJ, 848, L12
2. Abbott, B. P., et al. 2017, ApJ, 848, L12
3. Abbott, B. P., Abbott, R., Abbott, T. D., et al. 2018, PhRvL, 121, 161101
4. Abeysekara, A. U., Albert, A., Alfaro, R., et al. 2018, Nature, 562, 82
5. Acciari, V. A., Ansoldi, S., Antonelli, L. A., et al. 2022, Nature Astronomy, 6, 689
6. Acero, F., Katsuda, S., Ballet, J., & Petre, R. 2017, A&A, 597, A106
7. Acharya, S. K., Beniamini, P., & Hotokezaka, K. 2025, A&A, 693, A108
8. Aizu, K. 1973, Progress of Theoretical Physics, 49, 1184
9. Albert, J., Aliu, E., Anderhub, H., et al. 2009, ApJ, 693, 303
10. Alexander, K. D., Margutti, R., Blanchard, P. K., et al. 2018, ApJ, 863, L18
11. Alpar, M. A., Cheng, A. F., Ruderman, M. A., & Shaham, J. 1982, Nature, 300, 728
12. Altamirano, D., Linares, M., Patruno, A., et al. 2010, MNRAS, 401, 223
13. Amaro-Seoane, P., Audley, H., Babak, S., et al. 2017, arXiv e-prints, arXiv:1702.00786
14. Ambrosino, F., Miraval Zanon, A., Papitto, A., et al. 2021, Nature Astronomy, 5, 552
15. An, H., Romani, R. W., Johnson, T., Kerr, M., & Clark, C. J. 2017, ApJ, 850, 100
16. Andersson, N. 2007, Ap&SS, 308, 395
17. Angelini, L., Loewenstein, M., & Mushotzky, R. F. 2001, ApJ, 557, L35
18. Angüner, E. O., Cassol, F., Costantini, H., Trichard, C., & Verna, G. 2019, in International Cosmic Ray Conference, Vol. 36, 36th International Cosmic Ray Conference (ICRC2019), 618
19. Antonini, F., & Gieles, M. 2020, Ph. Rev. D, 102, 123016
20. Antonini, F., Gieles, M., Dosopoulou, F., & Chattopadhyay, D. 2023, MNRAS, 522, 466
21. Antoniou, V., & Zezas, A. 2016, MNRAS, 459, 528
22. Anumarlapudi, A., Kaplan, D. L., Rea, N., et al. 2025, MNRAS, 542, 1208
23. Anupama, G. C. 2013, in IAU Symposium, Vol. 281, Binary Paths to Type Ia Supernovae Explosions, ed. R. Di Stefano, M. Orio, & M. Moe, 154
24. Anzolin, G., de Martino, D., Bonnet-Bidaud, J. M., et al. 2008, A&A, 489, 1243
25. Archibald, A. M., Bogdanov, S., Patruno, A., et al. 2015, ApJ, 807, 62
26. Arf, K., & Schwenzer, K. 2024, Ph. Rev. D, 110, 123042
27. Armas Padilla, M., Degenaar, N., Patruno, A., et al. 2011, MNRAS, 417, 659
28. Armas Padilla, M., Wijnands, R., & Degenaar, N. 2013, MNRAS, 436, L89
29. Armstrong, J. W., Rickett, B. J., & Spangler, S. R. 1995, ApJ, 443, 209
30. Bañados, E., Venemans, B. P., Mazzucchelli, C., et al. 2018, Nature, 553, 473
31. Bachetti, M., Harrison, F. A., Walton, D. J., et al. 2014, Nature, 514, 202
32. Bachetti, M., Heida, M., Maccarone, T., et al. 2022, ApJ, 937, 125
33. Baglio, M. C., Coti Zelati, F., Campana, S., et al. 2023, A&A, 677, A30
34. Baglio, M. C., Coti Zelati, F., Di Marco, A., et al. 2025, ApJ, 987, L19
35. Bahramian, A., & Degenaar, N. 2023, in Handbook of X-ray and Gamma-ray Astrophysics, 120
36. Bahramian, A., Heinke, C. O., Tudor, V., et al. 2017, MNRAS, 467, 2199
37. Bahramian, A., Heinke, C. O., Kennea, J. A., et al. 2021, MNRAS, 501, 2790
38. Bahramian, A., Tremou, E., Tetarenko, A. J., et al. 2023, ApJ, 948, L7
39. Bailes, M., Bassa, C. G., Bernardi, G., et al. 2021, MNRAS, 503, 5367
40. Balick, B. 1987, AJ, 94, 671
41. Ballhausen, R., Lorenz, M., Fürst, F., et al. 2020, A&A, 641, A65
42. Balman, Ş. 2005, ApJ, 627, 933
43. —. 2006, Advances in Space Research, 38, 2840
44. —. 2014, A&A, 572, A114
45. —. 2020, Advances in Space Research, 66, 1097

46. Balman, Ş., Orio, M., & Luna, G. J. M. 2025, Universe, 11, 105
47. Balman, Ş., & Revnivtsev, M. 2012, A&A, 546, A112
48. Balman, Ş., Schlegel, E. M., & Godon, P. 2022, ApJ, 932, 33
49. Balman, Ş., Schlegel, E. M., Godon, P., & Drake, J. J. 2024, ApJ, 977, 136
50. Barack, L., & Cutler, C. 2004, Ph. Rev. D, 69, 082005
51. Barkov, M. V., & Bosch-Ramon, V. 2016, MNRAS, 456, L64
52. Barrett, P., Dieck, C., Beasley, A. J., Mason, P. A., & Singh, K. P. 2020, Advances in Space Research, 66, 1226
53. Basko, M. M., & Sunyaev, R. A. 1976, MNRAS, 175, 395
54. Baumgardt, H., He, C., Sweet, S. M., et al. 2019, MNRAS, 488, 5340
55. Bear, E., & Soker, N. 2018, ApJ, 855, 82
56. Begari, T., & Maccarone, T. J. 2023, JAAVSO, 51, 227
57. Bell, A. R. 2004, MNRAS, 353, 550
58. Belloni, D., Giersz, M., Rivera Sandoval, L. E., Askar, A., & Ciecielåg, P. 2019, MNRAS, 483, 315
59. Belloni, D., Schreiber, M. R., Pala, A. F., et al. 2020, MNRAS, 491, 5717
60. Beniamini, P., Wadiasingh, Z., Hare, J., et al. 2023, MNRAS, 520, 1872
61. Bernardini, F., & Cackett, E. M. 2014, MNRAS, 439, 2771
62. Bernardini, F., de Martino, D., Falanga, M., et al. 2012, A&A, 542, A22
63. Bernardini, F., de Martino, D., Mukai, K., Falanga, M., & Masetti, N. 2019, MNRAS, 489, 1044
64. Beuermann, K., Thomas, H. C., Reinsch, K., et al. 1999, A&A, 347, 47
65. Bhalerao, V. B., van Kerkwijk, M. H., & Harrison, F. A. 2012, ApJ, 757, 10
66. Bhattacharya, D., & Datta, B. 1996, MNRAS, 282, 1059
67. Bietenholz, M. F. 2006, ApJ, 645, 1180
68. Blandford, R. D., & Begelman, M. C. 1999, MNRAS, 303, L1
69. Blumer, H., & Safi-Harb, S. 2020, ApJ, 904, L19
70. Blumer, H., Safi-Harb, S., Borghese, A., et al. 2021, ApJ, 917, 56
71. Blumer, H., Safi-Harb, S., & McLaughlin, M. A. 2017, ApJ, 850, L18
72. Blumer, H., Safi-Harb, S., McLaughlin, M. A., & Fiore, W. 2021, ApJ, 911, L6
73. Bochenek, C. D., Ravi, V., Belov, K. V., et al. 2020, Nature, 587, 59
74. Bode, M. F., & Evans, A. 2008, Classical Novae, Vol. 43
75. Bogdanov, S., & Halpern, J. P. 2015, ApJ, 803, L27
76. Bogdanov, S., Deller, A. T., Miller-Jones, J. C. A., et al. 2018, ApJ, 856, 54
77. Bonnet-Bidaud, J. M., Mouchet, M., Falize, E., et al. 2020, A&A, 633, A145
78. Borghese, A., & Esposito, P. 2023, in Handbook of X-ray and Gamma-ray Astrophysics, 146
79. Borkowski, K. J., Miltich, W., & Reynolds, S. P. 2020, ApJ, 905, L19
80. Borkowski, K. J., Reynolds, S. P., & Roberts, M. S. E. 2016, ApJ, 819, 160
81. Borkowski, K. J., Reynolds, S. P., Williams, B. J., & Petre, R. 2018, ApJ, 868, L21
82. Bozzo, E., Romano, P., Ferrigno, C., & Oskinova, L. 2022, MNRAS, 513, 42
83. Brandes, L., & Weise, W. 2025, Ph. Rev. D, 111, 034005
84. Braun, R., Bonaldi, A., Bourke, T., Keane, E., & Wagg, J. 2019, arXiv e-prints, arXiv:1912.12699
85. —. 2019, arXiv e-prints, arXiv:1912.12699
86. Breivik, K., Coughlin, S., Zevin, M., et al. 2020, ApJ, 898, 71
87. Brightman, M., Walton, D. J., Xu, Y., et al. 2020, ApJ, 889, 71
88. Brightman, M., Harrison, F. A., Barret, D., et al. 2016, ApJ, 829, 28
89. Brown, E. F., Bildsten, L., & Rutledge, R. E. 1998, ApJ, 504, L95
90. Brown, E. F., & Cumming, A. 2009, ApJ, 698, 1020
91. Brown, G. E., & Bethe, H. A. 1994, ApJ, 423, 659
92. Buccheri, R., Bennett, K., Bignami, G. F., et al. 1983, A&A, 128, 245
93. Bult, P., Altamirano, D., Arzoumanian, Z., et al. 2022, ApJ, 935, L32
94. Burlaga, L. F., Florinski, V., & Ness, N. F. 2018, ApJ, 854, 20

95. Burwitz, V., Reinsch, K., Beuermann, K., & Thomas, H. C. 1996, A&A, 310, L25
96. Caleb, M., Heywood, I., Rajwade, K., et al. 2022, Nature Astronomy, 6, 828
97. Caleb, M., Lenc, E., Kaplan, D. L., et al. 2024, Nature Astronomy, 8, 1159
98. Campana, S. 2009, ApJ, 699, 1144
99. Campana, S., Parmar, A. N., & Stella, L. 2001, A&A, 372, 241
100. Cao, Z., Aharonian, F., An, Q., et al. 2024, ApJS, 271, 25
101. Cao, Z., Aharonian, F. A., An, Q., et al. 2021, Nature, 594, 33
102. Carpano, S., Haberl, F., Maitra, C., & Vasilopoulos, G. 2018, MNRAS, 476, L45
103. Casares, J. 2015, ApJ, 808, 80
104. Casares, J., & Torres, M. A. P. 2018, MNRAS, 481, 4372
105. Castelletti, G., Giacani, E., & Petriella, A. 2016, A&A, 587, A71
106. Chamel, N., & Haensel, P. 2008, Living Reviews in Relativity, 11, 10
107. Chen, K., & Ruderman, M. 1993, ApJ, 402, 264
108. Chen, W., Freire, P. C. C., Ridolfi, A., et al. 2023, MNRAS, 520, 3847
109. Chené, A.-N., Vasilopoulos, G., Oskinova, L. M., & Martínez-Vázquez, C. 2025, ApJ, 983, 53
110. Cheng, K. S., Chernyshov, D. O., Dogiel, V. A., Ko, C. M., & Ip, W. H. 2011, ApJ, 731, L17
111. Cherenkov Telescope Array Consortium, Acharya, B. S., Agudo, I., et al. 2019, Science with the Cherenkov Telescope Array
112. Chernyakova, M., & Malyshev, D. 2020, in Multifrequency Behaviour of High Energy Cosmic Sources - XIII. 3-8 June 2019. Palermo, 45
113. Chernyakova, M., Abdo, A. A., Neronov, A., et al. 2014, MNRAS, 439, 432
114. CHIME/FRB Collaboration, Andersen, B. C., Bandura, K. M., et al. 2020, Nature, 587, 54
115. Cho, P. B., Halpern, J. P., & Bogdanov, S. 2018, ApJ, 866, 71
116. Chomiuk, L., Metzger, B. D., & Shen, K. J. 2021, ARA&A, 59, 391
117. Chomiuk, L., Strader, J., Maccarone, T. J., et al. 2013, ApJ, 777, 69
118. Chomiuk, L., Linford, J. D., Aydi, E., et al. 2021, ApJS, 257, 49
119. Chornock, R., Berger, E., Kasen, D., et al. 2017, ApJ, 848, L19
120. Clark, C. J., Breton, R. P., Barr, E. D., et al. 2023, MNRAS, 519, 5590
121. Clark, G. W. 1975, ApJ, 199, L143
122. Clavel, M., Terrier, R., Goldwurm, A., et al. 2013, A&A, 558, A32
123. Coe, M. J., Kennea, J. A., Evans, P. A., & Udalski, A. 2020, MNRAS, 497, L50
124. Coppejans, D., & Knigge, C. 2020, arXiv e-prints, arXiv:2003.05953
125. Cornelisse, R., Verbunt, F., in't Zand, J. J. M., Kuulkers, E., & Heise, J. 2002, A&A, 392, 931
126. Corral-Santana, J. M., Casares, J., Muñoz-Darias, T., et al. 2016, A&A, 587, A61
127. —. 2013, Science, 339, 1048
128. Coti Zelati, F., & Borghese, A. 2024, arXiv e-prints, arXiv:2412.12763
129. Coti Zelati, F., Rea, N., Pons, J. A., Campana, S., & Esposito, P. 2018, MNRAS, 474, 961
130. Coti Zelati, F., Papitto, A., de Martino, D., et al. 2019, A&A, 622, A211
131. Coti Zelati, F., Hugo, B., Torres, D. F., et al. 2021, A&A, 655, A52
132. Cropper, M., Harrop-Allin, M. K., Mason, K. O., et al. 1998, MNRAS, 293, L57
133. Cruise, M., Guainazzi, M., Aird, J., et al. 2025, Nature Astronomy, 9, 36
134. Cumming, A., Brown, E. F., Fattoyev, F. J., et al. 2017, Physical Review C, 95, 025806
135. Cúneo, V. A., Casares, J., Armas Padilla, M., et al. 2023, A&A, 679, L11
136. Cutler, C., & Flanagan, É. E. 1994, Ph. Rev. D, 49, 2658
137. Dage, K. C., Tremou, E., Otahola, B. C., et al. 2025, ApJ, 979, 82
138. Dai, S., Johnston, S., Kerr, M., et al. 2023, MNRAS, 521, 2616
139. Dall'Osso, S., Israel, G. L., & Stella, L. 2006, A&A, 447, 785
140. Dall'Osso, S., Perna, R., & Stella, L. 2015, MNRAS, 449, 2144
141. Datta, S. R., Dhang, P., & Mishra, B. 2021, ApJ, 918, 87

142. Dauser, T., Falkner, S., Lorenz, M., et al. 2019, A&A, 630, A66
143. De Luca, A., Caraveo, P. A., Mereghetti, S., Tiengo, A., & Bignami, G. F. 2006, Science, 313, 814
144. De Luca, A., Mereghetti, S., Caraveo, P. A., et al. 2004, A&A, 418, 625
145. de Martino, D., Bernardini, F., Mukai, K., Falanga, M., & Masetti, N. 2020, Advances in Space Research, 66, 1209
146. de Naurois, M. 2007, Ap&SS, 309, 277
147. de Ruiter, I., Rajwade, K. M., Bassa, C. G., et al. 2025, Nature Astronomy
148. Degenaar, N., Page, D., van den Eijnden, J., et al. 2021, MNRAS, 508, 882
149. Degenaar, N., & Wijnands, R. 2009, A&A, 495, 547
150. —. 2010, A&A, 524, A69
151. Degenaar, N., Wijnands, R., Cackett, E. M., et al. 2012, A&A, 545, A49
152. Degenaar, N., Wijnands, R., & Miller, J. M. 2014, ApJ, 787, 67
153. Degenaar, N., Wijnands, R., Miller, J. M., et al. 2015, Journal of High Energy Astrophysics, 7, 137
154. Degenaar, N., Wijnands, R., Bahramian, A., et al. 2015, MNRAS, 451, 2071
155. Degenaar, N., Ootes, L. S., Page, D., et al. 2019, MNRAS, 488, 4477
156. Degenaar, N. D. 2010, PhD thesis, University of Amsterdam, Netherlands
157. Del Santo, M., Sidoli, L., Mereghetti, S., et al. 2007, A&A, 468, L17
158. Della Valle, M., & Izzo, L. 2020, A&A Rev., 28, 3
159. Deller, A. T., Archibald, A. M., Brisken, W. F., et al. 2012, ApJ, 756, L25
160. Di Stefano, R., Kong, A., & Primini, F. A. 2010, New A Rev., 54, 72
161. Di Stefano, R., Kong, A. K. H., Greiner, J., et al. 2004, ApJ, 610, 247
162. DiKerby, S., Zhang, S., Ergin, T., et al. 2025, ApJ, 983, 21
163. Dobrotka, A., & Ness, J. U. 2017, MNRAS, 467, 4865
164. Dobrotka, A., Ness, J. U., Nucita, A. A., & Melicherčík, M. 2023, A&A, 674, A188
165. Drout, M. R., Piro, A. L., Shappee, B. J., et al. 2017, Science, 358, 1570
166. Dubus, G. 2013, A&A Rev., 21, 64
167. Dupree, A. K., Strassmeier, K. G., Calderwood, T., et al. 2022, ApJ, 936, 18
168. Durant, M., Kargaltsev, O., Pavlov, G. G., Chang, C., & Garmire, G. P. 2011, ApJ, 735, 58
169. Durant, M., Kargaltsev, O., Pavlov, G. G., Kropotina, J., & Levenfish, K. 2013, ApJ, 763, 72
170. Dutta, A., & Rana, V. 2022, ApJ, 940, 100
171. Eldridge, J. J., Stanway, E. R., Xiao, L., et al. 2017, PASA, 34, e058
172. Esposito, P., Israel, G. L., Dall'Osso, S., & Covino, S. 2014, A&A, 561, A117
173. Fabbiano, G. 1989, ARA&A, 27, 87
174. Fender, R. P., Gallo, E., & Jonker, P. G. 2003, MNRAS, 343, L99
175. Ferrand, G., & Safi-Harb, S. 2012, Advances in Space Research, 49, 1313
176. Fesen, R. A., Hammell, M. C., Morse, J., et al. 2006, ApJ, 645, 283
177. Finch, E., Bartolucci, G., Chucherko, D., et al. 2023, MNRAS, 522, 5358
178. Finkelstein, S. L., Morse, J. A., Green, J. C., et al. 2006, ApJ, 641, 919
179. Fischer, A., & Beuermann, K. 2001, A&A, 373, 211
180. For, B. Q., Staveley-Smith, L., Hurley-Walker, N., et al. 2018, MNRAS, 480, 2743
181. Fortin, F., Kalsi, A., García, F., Simaz-Bunzel, A., & Chaty, S. 2024, A&A, 684, A124
182. Fraschetti, F., Katsuda, S., Sato, T., Jokipii, J. R., & Giacalone, J. 2018, Physical Review Letters, 120, 251101
183. Freeman, M., Montez, Jr., R., Kastner, J. H., et al. 2014, ApJ, 794, 99
184. Fruchter, A. S., Stinebring, D. R., & Taylor, J. H. 1988, Nature, 333, 237
185. Fryer, C. L., Benz, W., & Herant, M. 1996, ApJ, 460, 801
186. Fürst, F., Walton, D. J., Stern, D., et al. 2017, ApJ, 834, 77
187. Fürst, F., Walton, D. J., Harrison, F. A., et al. 2016, ApJ, 831, L14
188. Fürst, F., Walton, D. J., Israel, G. L., et al. 2023, A&A, 672, A140
189. Gaensler, B. M., Arons, J., Kaspi, V. M., et al. 2002, ApJ, 569, 878
190. Gallo, E., Fender, R. P., Miller-Jones, J. C. A., et al. 2006, MNRAS, 370, 1351

191. Galloway, D. K., Morgan, E. H., & Chakrabarty, D. 2008, in American Institute of Physics Conference Series, Vol. 1068, A Decade of Accreting MilliSecond X-ray Pulsars, ed. R. Wijnands, D. Altamirano, P. Soleri, N. Degenaar, N. Rea, P. Casella, A. Patruno, & M. Linares (AIP), 55
192. Galloway, D. K., Morgan, E. H., Krauss, M. I., Kaaret, P., & Chakrabarty, D. 2007, ApJ, 654, L73
193. Gänsicke, B. T., Dillon, M., Southworth, J., et al. 2009, MNRAS, 397, 2170
194. Garofali, K., Williams, B. F., Hillis, T., et al. 2018, MNRAS, 479, 3526
195. Garofali, K., Williams, B. F., Plucinsky, P. P., et al. 2017, MNRAS, 472, 308
196. Gendre, B., Barret, D., & Webb, N. A. 2003, A&A, 400, 521
197. Generozov, A., Stone, N. C., Metzger, B. D., & Ostriker, J. P. 2018, MNRAS, 478, 4030
198. Ghavamian, P., Schwartz, S. J., Mitchell, J., Masters, A., & Laming, J. M. 2013, Space Science Reviews, 178, 633
199. Gilfanov, M., Fabbiano, G., Lehmer, B., & Zezas, A. 2022, *Handbook of X-ray and Gamma-ray Astrophysics*, 105
200. Gladstone, J. C., Roberts, T. P., & Done, C. 2009, MNRAS, 397, 1836
201. Godinaud, L., Acero, F., Decourchelle, A., & Ballet, J. 2023, A&A, 680, A80
202. Godon, P., Sion, E. M., Balman, Ş., & Blair, W. P. 2017, ApJ, 846, 52
203. Gonzalez, M., & Safi-Harb, S. 2003, ApJ, 591, L143
204. Grebenev, S. A. 2017, Astronomy Letters, 43, 464
205. Green, D. A. 2025, Journal of Astrophysics and Astronomy, 46, 14
206. Green, D. A., Reynolds, S. P., Borkowski, K. J., et al. 2008, MNRAS, 387, L54
207. Greiner, J., Maitra, C., Haberl, F., et al. 2023, Nature, 615, 605
208. Grimm, H. J., McDowell, J., Zezas, A., Kim, D. W., & Fabbiano, G. 2007, ApJS, 173, 70
209. Guerrero, M. A., Ruiz, N., Hamann, W. R., et al. 2012, ApJ, 755, 129
210. Guillochon, J., Nicholl, M., Villar, V. A., et al. 2018, ApJS, 236, 6
211. Gusinskaia, N. V., Jaodand, A. D., Hessels, J. W. T., et al. 2025, MNRAS, 536, 99
212. H. E. S. S. Collaboration, Abdalla, H., Abramowski, A., et al. 2018, A&A, 612, A1
213. H. E. S. S. Collaboration, Abdalla, H., Adam, R., et al. 2020, A&A, 633, A102
214. Haberl, F., & Motch, C. 1995, A&A, 297, L37
215. Haberl, F., & Sturm, R. 2016, A&A, 586, A81
216. Haberl, F., Zavlin, V. E., Trümper, J., & Burwitz, V. 2004, A&A, 419, 1077
217. Haberl, F., Israel, G. L., Rodriguez Castillo, G. A., et al. 2017, A&A, 598, A69
218. Haberl, F., Maitra, C., Greiner, J., et al. 2020, The Astronomer's Telegram, 13789, 1
219. Häberle, M., Neumayer, N., Seth, A., et al. 2024, Nature, 631, 285
220. Haggard, D., Cool, A. M., & Davies, M. B. 2009, ApJ, 697, 224
221. Haggard, D., Cool, A. M., Heinke, C. O., et al. 2013, ApJ, 773, L31
222. Hailey, C. J., Mori, K., Bauer, F. E., et al. 2018, Nature, 556, 70
223. Hailey, C. J., Mori, K., Perez, K., et al. 2016, ApJ, 826, 160
224. Hallinan, G., Ravi, V., Weinreb, S., et al. 2019, in Bulletin of the American Astronomical Society, Vol. 51, 255
225. Hameury, J. M., Knigge, C., Lasota, J. P., Hambsch, F. J., & James, R. 2020, A&A, 636, A1
226. Hare, J., Kargaltsev, O., Pavlov, G., & Beniamini, P. 2019, ApJ, 882, 74
227. Hare, J., Kargaltsev, O., & Rangelov, B. 2018, ApJ, 865, 33
228. Hare, J., Pavlov, G. G., Garmire, G. P., & Kargaltsev, O. 2023, ApJ, 958, 5
229. Hare, J., Pavlov, G. G., Kargaltsev, O., & Garmire, G. P. 2023, Research Notes of the American Astronomical Society, 7, 52
230. Hare, J., Yang, H., Kargaltsev, O., et al. 2021, The Astronomer's Telegram, 14499, 1
231. Hartman, J. M., Patruno, A., Chakrabarty, D., et al. 2009, ApJ, 702, 1673
232. Heinke, C. O., Bahramian, A., Degenaar, N., & Wijnands, R. 2015, MNRAS, 447, 3034
233. Heinke, C. O., Cohn, H. N., & Lugger, P. M. 2009, ApJ, 692, 584
234. Heinke, C. O., Edmonds, P. D., Grindlay, J. E., et al. 2003, ApJ, 590, 809
235. Heinke, C. O., Grindlay, J. E., Edmonds, P. D., et al. 2005, ApJ, 625, 796
236. Heinke, C. O., & Ho, W. C. G. 2010, ApJ, 719, L167

237. Heinke, C. O., Altamirano, D., Cohn, H. N., et al. 2010, ApJ, 714, 894
238. Hellier, C. 1995, in Astronomical Society of the Pacific Conference Series, Vol. 85, Magnetic Cataclysmic Variables, ed. D. A. H. Buckley & B. Warner, 185
239. Henleywillis, S., Cool, A. M., Haggard, D., et al. 2018, MNRAS, 479, 2834
240. Herold, H. 1979, Ph. Rev. D, 19, 2868
241. Hertfelder, M., & Kley, W. 2015, A&A, 579, A54
242. H.E.S.S. Collaboration. 2024, Science, 383, 402
243. Hewitt, D. M., Pretorius, M. L., Woudt, P. A., et al. 2020, MNRAS, 496, 2542
244. Hillman, Y., Orio, M., Prialnik, D., et al. 2019, ApJ, 879, L5
245. Ho, W. C. G., Elshamouty, K. G., Heinke, C. O., & Potekhin, A. Y. 2015, Physical Review C, 91, 015806
246. Ho, W. C. G., Zhao, Y., Heinke, C. O., et al. 2021, MNRAS, 506, 5015
247. Hu, C.-P., Ueda, Y., & Enoto, T. 2021, ApJ, 909, 5
248. Hu, C.-P., Narita, T., Enoto, T., et al. 2024, Nature, 626, 500
249. Hubeny, I., & Long, K. S. 2021, MNRAS, 503, 5534
250. Huentemeyer, P., BenZvi, S., Dingus, B., et al. 2019, in Bulletin of the American Astronomical Society, Vol. 51, 109
251. Hughes, J. P. 2000, ApJ, 545, L53
252. Hughes, J. P., Rakowski, C. E., Burrows, D. N., & Slane, P. O. 2000, ApJ, 528, L109
253. Hui, C. Y., & Li, K. L. 2019, Galaxies, 7, 93
254. Hurley-Walker, N., Zhang, X., Bahramian, A., et al. 2022, Nature, 601, 526
255. Hurley-Walker, N., Rea, N., McSweeney, S. J., et al. 2023, Nature, 619, 487
256. Hurley-Walker, N., McSweeney, S. J., Bahramian, A., et al. 2024, ApJ, 976, L21
257. Hynes, R. I., Charles, P. A., Garcia, M. R., et al. 2004, ApJ, 611, L125
258. Icecube Collaboration, & Abbasi, R. 2023, Science, 380, 1338
259. Illiano, G., Papitto, A., Ambrosino, F., et al. 2023, A&A, 669, A26
260. Inoue, Y., Tanaka, Y. T., & Isobe, N. 2016, MNRAS, 461, 4329
261. in't Zand, J. J. M., Cornelisse, R., & Méndez, M. 2005, A&A, 440, 287
262. in't Zand, J. J. M., Jonker, P. G., & Markwardt, C. B. 2007, A&A, 465, 953
263. Irwin, J. A., Brink, T. G., Bregman, J. N., & Roberts, T. P. 2010, ApJ, 712, L1
264. Islam, N., & Mukai, K. 2021, ApJ, 919, 90
265. Islam, N., Mukai, K., & Sokoloski, J. L. 2024, ApJ, 960, 125
266. Israel, G. L., Panzera, M. R., Campana, S., et al. 1999, A&A, 349, L1
267. Israel, G. L., Hummel, W., Covino, S., et al. 2002, A&A, 386, L13
268. Israel, G. L., Belfiore, A., Stella, L., et al. 2017, Science, 355, 817
269. Israel, G. L., Papitto, A., Esposito, P., et al. 2017, MNRAS, 466, L48
270. Iwasawa, K., Norman, C., Gilli, R., Gandhi, P., & Peréz-Torres, M. A. 2023, A&A, 674, A77
271. Jani, K., Shoemaker, D., & Cutler, C. 2020, Nature Astronomy, 4, 260
272. Johnston, S., Manchester, R. N., Lyne, A. G., et al. 1992, ApJ, 387, L37
273. Kaaret, P., Feng, H., & Roberts, T. P. 2017, ARA&A, 55, 303
274. Kaplan, D. L., Bildsten, L., & Steinfadt, J. D. R. 2012, ApJ, 758, 64
275. Kargaltsev, O., Hare, J., Volkov, I., & Lange, A. 2023, ApJ, 958, 79
276. Kargaltsev, O., Klingler, N. J., Hare, J., & Volkov, I. 2022, ApJ, 925, 20
277. Kargaltsev, O., & Pavlov, G. G. 2008, in American Institute of Physics Conference Series, Vol. 983, 40 Years of Pulsars: Millisecond Pulsars, Magnetars and More, ed. C. Bassa, Z. Wang, A. Cumming, & V. M. Kaspi (AIP), 171
278. Kargaltsev, O., Pavlov, G. G., Durant, M., Volkov, I., & Hare, J. 2014, ApJ, 784, 124
279. Kargaltsev, O., Kouveliotou, C., Pavlov, G. G., et al. 2012, ApJ, 748, 26
280. Katira, A., Mooley, K. P., & Hotokezaka, K. 2025, MNRAS, 539, 2654
281. Kaur, A., Henze, M., Haberl, F., et al. 2012, A&A, 538, A49
282. Kavanagh, P. J., Sasaki, M., Bozzetto, L. M., et al. 2015, A&A, 583, A121
283. Kayama, K., Tanaka, T., Uchida, H., et al. 2022, PASJ, 74, 1143

284. Kennea, J. A., Coe, M. J., Evans, P. A., Waters, J., & Jasko, R. E. 2018, ApJ, 868, 47
285. Kennedy, M. R., Garnavich, P. M., Littlefield, C., et al. 2017, MNRAS, 469, 956
286. King, A., & Lasota, J.-P. 2019, MNRAS, 485, 3588
287. King, A. R., & Wijnands, R. 2006, MNRAS, 366, L31
288. Klingler, N., Hare, J., Kargaltsev, O., Pavlov, G. G., & Tomsick, J. 2023, ApJ, 950, 177
289. Klingler, N., Yang, H., Hare, J., et al. 2020, ApJ, 901, 157
290. Kluźniak, W., & Lasota, J.-P. 2015, MNRAS, 448, L43
291. Knevitt, G., Wynn, G. A., Vaughan, S., & Watson, M. G. 2014, MNRAS, 437, 3087
292. Knigge, C., Baraffe, I., & Patterson, J. 2011, ApJS, 194, 28
293. Kolb, U. 1993, A&A, 271, 149
294. Koliopanos, F., Vasilopoulos, G., Godet, O., et al. 2017, A&A, 608, A47
295. Koljonen, K. I. I., & Linares, M. 2023, MNRAS, 525, 3963
296. Kong, A. K. H., Takata, J., Hui, C. Y., et al. 2018, MNRAS, 478, 3987
297. König, O., Wilms, J., Arcodia, R., et al. 2022, Nature, 605, 248
298. Kosec, P., Pinto, C., Walton, D. J., et al. 2018, MNRAS, 479, 3978
299. Kraft, R. P., Burrows, D. N., & Nousek, J. A. 1991, ApJ, 374, 344
300. Krivonos, R., Revnivtsev, M., Churazov, E., et al. 2007, A&A, 463, 957
301. Krivonos, R., Sazonov, S., Tsygankov, S. S., & Poutanen, J. 2018, MNRAS, 480, 2357
302. Krumholz, M. R., Crocker, R. M., Bahramian, A., & Bordas, P. 2024, Nature Astronomy, 8, 1284
303. Kulkarni, S. R., Harrison, F. A., Grefenstette, B. W., et al. 2021, arXiv e-prints, arXiv:2111.15608
304. Kupfer, T., Korol, V., Shah, S., et al. 2018, MNRAS, 480, 302
305. Kuulkers, E., Shaw, S. E., Paizis, A., et al. 2007, A&A, 466, 595
306. Kwok, S. 2024, Galaxies, 12, 39
307. Kwok, S., Purton, C. R., & Fitzgerald, P. M. 1978, ApJ, 219, L125
308. Kyer, R., Roy, S., Strader, J., et al. 2025, ApJ, 983, 112
309. Kyer, R., Albrecht, S., Williams, B. F., et al. 2024, ApJ, 961, 168
310. Lai, D. 2012, ApJ, 757, L3
311. Laktionov, R. 2024, Study of X-ray emission from nearby normal galaxies using eROSITA All-Sky Survey data, https://www.sternwarte.uni-erlangen.de/docs/theses/2024-02_Laktionov.pdf
312. Lam, C. Y., Lu, J. R., Udalski, A., et al. 2022, ApJ, 933, L23
313. Lam, C. Y., Abrams, N., Andrews, J., et al. 2023, arXiv e-prints, arXiv:2306.12514
314. Lamberts, A., Blunt, S., Littenberg, T. B., et al. 2019, MNRAS, 490, 5888
315. Laming, J. M., Raymond, J. C., McLaughlin, B. M., & Blair, W. P. 1996, ApJ, 472, 267
316. Lander, S. K., Andersson, N., Antonopoulou, D., & Watts, A. L. 2015, MNRAS, 449, 2047
317. Lasota, J.-P. 2001, New A Rev., 45, 449
318. —. 2008, New A Rev., 51, 752
319. Lattimer, J. M. 2012, Annual Review of Nuclear and Particle Science, 62, 485
320. Lazzarini, M., Hornschemeier, A. E., Williams, B. F., et al. 2018, ApJ, 862, 28
321. Lazzarini, M., Hinton, K., Shariat, C., et al. 2023, ApJ, 952, 114
322. Leahy, D. A., Elsner, R. F., & Weisskopf, M. C. 1983, ApJ, 272, 256
323. Lee, J. C., Gil de Paz, A., Kennicutt, Jr., R. C., et al. 2011, ApJS, 192, 6
324. Lehmer, B. D., Eufrasio, R. T., Tzanavaris, P., et al. 2019, ApJS, 243, 3
325. LHAASO collaboration. 2021, arXiv e-prints, arXiv:2101.03508
326. LHAASO Collaboration. 2024, Ultrahigh-Energy Gamma-ray Emission Associated with Black Hole-Jet Systems
327. Li, K.-L., Strader, J., Miller-Jones, J. C. A., Heinke, C. O., & Chomiuk, L. 2020, ApJ, 895, 89
328. Li, X., Levin, Y., & Beloborodov, A. M. 2016, ApJ, 833, 189
329. Lin, L., Zhang, C. F., Wang, P., et al. 2020, Nature, 587, 63
330. Linnell, A. P., Godon, P., Hubeny, I., Sion, E. M., & Szkody, P. 2010, ApJ, 719, 271
331. Littlefield, C., Garnavich, P., Kennedy, M. R., et al. 2020, ApJ, 896, 116

332. Liu, J. 2011, ApJ Supplements, 192, 10
333. Liu, J., & Di Stefano, R. 2008, ApJ, 674, L73
334. Liu, Q. Z., van Paradijs, J., & van den Heuvel, E. P. J. 2006, A&A, 455, 1165
335. Long, K. S., Dodorico, S., Charles, P. A., & Dopita, M. A. 1981, ApJ, 246, L61
336. Long, K. S., Blair, W. P., Winkler, P. F., et al. 2010, ApJS, 187, 495
337. Luna, G. J. M., Nelson, T., Mukai, K., & Sokoloski, J. L. 2019, ApJ, 880, 94
338. Luna, G. J. M., Mukai, K., Sokoloski, J. L., et al. 2018, A&A, 619, A61
339. Lynden-Bell, D., & Pringle, J. E. 1974, MNRAS, 168, 603
340. Lyne, A. G., Stappers, B. W., Keith, M. J., et al. 2015, MNRAS, 451, 581
341. Maccarone, T. J., Kupfer, T., Najera Casarrubias, E., et al. 2024, MNRAS, 529, L28
342. Maccarone, T. J., & Patruno, A. 2013, MNRAS, 428, 1335
343. Maggi, P., Haberl, F., Bozzetto, L. M., et al. 2012, A&A, 546, A109
344. Maggi, P., Haberl, F., Kavanagh, P. J., et al. 2016, A&A, 585, A162
345. Maggi, P., Filipović, M. D., Vukotić, B., et al. 2019, A&A, 631, A127
346. Mahlmann, J. F., Philippov, A. A., Levinson, A., Spitkovsky, A., & Hakobyan, H. 2022, ApJ, 932, L20
347. Maiolino, R., Scholtz, J., Witstok, J., et al. 2024, Nature, 627, 59
348. Maitra, C., & Haberl, F. 2022, A&A, 657, A26
349. Maitra, C., Haberl, F., Maggi, P., et al. 2021, MNRAS, 504, 326
350. Maitra, C., Haberl, F., Filipović, M. D., et al. 2019, MNRAS, 490, 5494
351. Maitra, C., Haberl, F., Vasilopoulos, G., et al. 2024, A&A, 683, A21
352. Malyshev, D., Chernyakova, M., Santangelo, A., & Pühlhofer, G. 2019, Astronomische Nachrichten, 340, 465
353. Mandel, S., Gerber, J., Mori, K., et al. 2025, arXiv e-prints, arXiv:2503.21139
354. Margutti, R., & Chornock, R. 2021, ARA&A, 59, 155
355. Marino, A., Yang, H. N., Coti Zelati, F., et al. 2025, ApJ, 980, L36
356. Markoff, S., Falcke, H., Yuan, F., & Biermann, P. L. 2001, A&A, 379, L13
357. Marsh, T. R., & Steeghs, D. 2002, MNRAS, 331, L7
358. Marsh, T. R., Gänsicke, B. T., Hümmerich, S., et al. 2016, Nature, 537, 374
359. Mason, P. A., Morales, J. F., Littlefield, C., et al. 2020, Advances in Space Research, 66, 1123
360. Matsuda, M., Uchida, H., Tanaka, T., Yamaguchi, H., & Tsuru, T. G. 2022, ApJ, 940, 105
361. Mayer, M. G. F., & Becker, W. 2021, A&A, 651, A40
362. McClintock, J. E., Garcia, M. R., Caldwell, N., et al. 2001, ApJ, 551, L147
363. McGowan, K. E., Zane, S., Cropper, M., Vestrand, W. T., & Ho, C. 2006, ApJ, 639, 377
364. Mereghetti, S., Krachmalnicoff, N., La Palombara, N., et al. 2010, A&A, 519, A42
365. Mereghetti, S., Savchenko, V., Ferrigno, C., et al. 2020, ApJ, 898, L29
366. Mereminskiy, I. A., Grebenev, S. A., & Sunyaev, R. A. 2017, Astronomy Letters, 43, 656
367. Merloni, A., Lamer, G., Liu, T., et al. 2024, A&A, 682, A34
368. Metzler, Z., & Wadiasingh, Z. 2025, arXiv e-prints, arXiv:2503.10511
369. Middleton, M. J., Sutton, A. D., Roberts, T. P., Jackson, F. E., & Done, C. 2012, MNRAS, 420, 2969
370. Middleton, M. J., Fragile, P. C., Bachetti, M., et al. 2018, MNRAS, 475, 154
371. Miller, J. M., Swihart, S. J., Strader, J., et al. 2020, ApJ, 904, 49
372. Miller, M. C., & Hamilton, D. P. 2002, MNRAS, 330, 232
373. Miller, M. C., Lamb, F. K., Dittmann, A. J., et al. 2019, ApJ, 887, L24
374. Miller-Jones, J. C. A., Strader, J., Heinke, C. O., et al. 2015, MNRAS, 453, 3918
375. Mineo, S., Rappaport, S., Levine, A., et al. 2014, ApJ, 797, 91
376. Mondal, S., Belczyński, K., Wiktorowicz, G., Lasota, J.-P., & King, A. R. 2020, MNRAS, 491, 2747
377. Montez, R., Luna, G. J. M., Mukai, K., Sokoloski, J. L., & Kastner, J. H. 2022, ApJ, 926, 100
378. Moon, H., Wik, D. R., Antoniou, V., et al. 2024, ApJ, 970, 167
379. Moore, C. J., Cole, R. H., & Berry, C. P. L. 2015, Classical and Quantum Gravity, 32, 015014
380. Mori, K., Mandel, S., Hailey, C. J., et al. 2022, arXiv e-prints, arXiv:2204.09812

381. Mori, K., Gotthelf, E. V., Zhang, S., et al. 2013, ApJ, 770, L23
382. Mori, K., Hailey, C. J., Mandel, S., et al. 2019, ApJ, 885, 142
383. Mori, K., Hailey, C. J., Schutt, T. Y. E., et al. 2021, ApJ, 921, 148
384. Morris, M. 1993, ApJ, 408, 496
385. Motch, C., Haberl, F., Dennerl, K., Pakull, M., & Janot-Pacheco, E. 1997, A&A, 323, 853
386. Motch, C., Pakull, M. W., Soria, R., Grisé, F., & Pietrzyński, G. 2014, Nature, 514, 198
387. Muñoz-Giraldo, D., Stelzer, B., & Schwope, A. 2024, A&A, 687, A305
388. Mukai, K. 2017, PASP, 129, 062001
389. Mukai, K., Orio, M., & Della Valle, M. 2008, ApJ, 677, 1248
390. Muno, M. P., Pfahl, E., Baganoff, F. K., et al. 2005, ApJ, 622, L113
391. Muno, M. P., Baganoff, F. K., Bautz, M. W., et al. 2003, ApJ, 589, 225
392. Muno, M. P., Bauer, F. E., Baganoff, F. K., et al. 2009, ApJS, 181, 110
393. Munoz-Sanchez, G., de Wit, S., Bonanos, A. Z., et al. 2024, A&A, 690, A99
394. Munoz-Sanchez, G., Kalitsounaki, M., de Wit, S., et al. 2024, arXiv e-prints, arXiv:2411.19329
395. Murphy, E. J., Bolatto, A., Chatterjee, S., et al. 2018, in Astronomical Society of the Pacific Conference Series, Vol. 517, Science with a Next Generation Very Large Array, ed. E. Murphy, 3
396. Mushtukov, A. A., Suleimanov, V. F., Tsygankov, S. S., & Poutanen, J. 2015, MNRAS, 454, 2539
397. Narayan, R., & Popham, R. 1993, Nature, 362, 820
398. Narayan, R., & Yi, I. 1994, ApJ, 428, L13
399. —. 1995, ApJ, 452, 710
400. Nelemans, G., Portegies Zwart, S. F., Verbunt, F., & Yungelson, L. R. 2001, A&A, 368, 939
401. Ness, J. U., Beardmore, A. P., Osborne, J. P., et al. 2015, A&A, 578, A39
402. Ng, C. Y., Ho, W. C. G., Gotthelf, E. V., et al. 2019, ApJ, 880, 147
403. Ng, M., Ray, P. S., Sanna, A., et al. 2024, ApJ, 968, L7
404. Nicholl, M., Margalit, B., Schmidt, P., et al. 2021, MNRAS, 505, 3016
405. Nixon, C. J., & Pringle, J. E. 2019, A&A, 628, A121
406. Norton, A. J., Hellier, C., Beardmore, A. P., et al. 1997, MNRAS, 289, 362
407. Norton, A. J., & Mukai, K. 2007, A&A, 472, 225
408. Noyola, E., Gebhardt, K., & Bergmann, M. 2008, ApJ, 676, 1008
409. Nynka, M., Ruan, J. J., Haggard, D., & Evans, P. A. 2018, ApJ, 862, L19
410. O'Connor, B., Kouveliotou, C., Evans, P. A., et al. 2023, ApJS, 269, 49
411. Oh, K., Dage, K. C., Bobrick, A., et al. 2025, MNRAS, 537, 3884
412. Ohnaka, K., Hofmann, K. H., Weigelt, G., et al. 2024, A&A, 691, L15
413. Olejak, A., Belczynski, K., Bulik, T., & Sobolewska, M. 2020, A&A, 638, A94
414. Olmi, B. 2023, Universe, 9, 402
415. Orio, M., Nelson, T., Bianchini, A., Di Mille, F., & Harbeck, D. 2010, ApJ, 717, 739
416. Orio, M., Gendreau, K., Giese, M., et al. 2022, ApJ, 932, 45
417. Orio, M., Behar, E., Luna, G. J. M., et al. 2022, ApJ, 938, 34
418. Orio, M., Gendreau, K., Giese, M., et al. 2023, ApJ, 955, 37
419. Orosz, J. A., McClintock, J. E., Narayan, R., et al. 2007, Nature, 449, 872
420. Özel, F., & Freire, P. 2016, ARA&A, 54, 401
421. Page, D., Prakash, M., Lattimer, J. M., & Steiner, A. W. 2011, PhRvL, 106, 081101
422. Page, D., & Reddy, S. 2013, PhRvL, 111, 241102
423. Page, K. L., Beardmore, A. P., Osborne, J. P., et al. 2022, MNRAS, 514, 1557
424. Pala, A. F., Gänsicke, B. T., Breedt, E., et al. 2020, MNRAS, 494, 3799
425. Pandel, D., & Córdova, F. A. 2005, ApJ, 620, 416
426. Papitto, A., & de Martino, D. 2022, in Astrophysics and Space Science Library, Vol. 465, Astrophysics and Space Science Library, ed. S. Bhattacharyya, A. Papitto, & D. Bhattacharya, 157
427. Papitto, A., Hessels, J. W. T., Burgay, M., et al. 2013, The Astronomer's Telegram, 5069, 1

428. Papitto, A., Ambrosino, F., Stella, L., et al. 2019, ApJ, 882, 104
429. Park, J., Kim, C., Woo, J., et al. 2023, ApJ, 945, 66
430. Patnaude, D. J., & Fesen, R. A. 2007, AJ, 133, 147
431. —. 2009, ApJ, 697, 535
432. Patruno, A. 2010, in Proceedings of High Time Resolution Astrophysics - The Era of Extremely Large Telescopes (HTRA-IV). May 5 - 7, 28
433. Patruno, A., Altamirano, D., Hessels, J. W. T., et al. 2009, ApJ, 690, 1856
434. Pavlov, G. G., Chang, C., & Kargaltsev, O. 2011, ApJ, 730, 2
435. Pavlov, G. G., Hare, J., Kargaltsev, O., Rangelov, B., & Durant, M. 2015, ApJ, 806, 192
436. Peacock, M. B., Maccarone, T. J., Kundu, A., & Zepf, S. E. 2010, MNRAS, 407, 2611
437. Pearlman, A. B., Scholz, P., Bethapudi, S., et al. 2025, Nature Astronomy, 9, 111
438. Pechetti, R., Kamann, S., Krajnović, D., et al. 2024, MNRAS, 528, 4941
439. Pei, S., Orio, M., Ness, J.-U., & Ospina, N. 2021, MNRAS, 507, 2073
440. Peille, P., Barret, D., Cucchetti, E., et al. 2025, Experimental Astronomy, 59, 18
441. Pelisoli, I., Marsh, T. R., Buckley, D. A. H., et al. 2023, Nature Astronomy, 7, 931
442. Perna, R., Raymond, J., & Narayan, R. 2000, ApJ, 541, 898
443. Pichardo Marcano, M., Rivera Sandoval, L. E., Maccarone, T. J., & Scaringi, S. 2021, MNRAS, 508, 3275
444. Pietsch, W., Fliri, J., Freyberg, M. J., et al. 2005, A&A, 442, 879
445. Pietsch, W., Haberl, F., Sasaki, M., et al. 2006, ApJ, 646, 420
446. Pietsch, W., Misanovic, Z., Haberl, F., et al. 2004, A&A, 426, 11
447. Pietsch, W., Haberl, F., Gaetz, T. J., et al. 2009, ApJ, 694, 449
448. Pike, S. N., Negoro, H., Tomsick, J. A., et al. 2022, ApJ, 927, 190
449. Pinto, C., Middleton, M. J., & Fabian, A. C. 2016, Nature, 533, 64
450. Pinto, C., & Walton, D. J. 2023, arXiv e-prints, arXiv:2302.00006
451. Pintore, F., Zampieri, L., Stella, L., et al. 2017, ApJ, 836, 113
452. Pintore, F., Belfiore, A., Novara, G., et al. 2018, MNRAS, 477, L90
453. Pintore, F., Motta, S., Pinto, C., et al. 2021, MNRAS, 504, 551
454. Pintore, F., Pinto, C., Rodriguez-Castillo, G., et al. 2025, A&A, 695, A238
455. Plotkin, R. M., Gallo, E., & Jonker, P. G. 2013, ApJ, 773, 59
456. Plucinsky, P. P., Sasaki, M., Gaetz, T. J., et al. 2006, in AAS Meeting Abstracts, Vol. 209, 87.01
457. Ponti, G., Morris, M. R., Churazov, E., Heywood, I., & Fender, R. P. 2021, A&A, 646, A66
458. Ponti, G., Terrier, R., Goldwurm, A., Belanger, G., & Trap, G. 2010, ApJ, 714, 732
459. Pooley, D., & Hut, P. 2006, ApJ, 646, L143
460. Pooley, D., Lewin, W. H. G., Anderson, S. F., et al. 2003, ApJ, 591, L131
461. Popham, R. 1999, MNRAS, 308, 979
462. Popham, R., & Narayan, R. 1995, ApJ, 442, 337
463. Portegies Zwart, S. F., Baumgardt, H., Hut, P., Makino, J., & McMillan, S. L. W. 2004, Nature, 428, 724
464. Posselt, B., & Pavlov, G. G. 2018, ApJ, 864, 135
465. —. 2022, ApJ, 932, 83
466. Posselt, B., Pavlov, G. G., Suleimanov, V., & Kargaltsev, O. 2013, ApJ, 779, 186
467. Postnov, K. A., & Yungelson, L. R. 2014, Living Reviews in Relativity, 17, 3
468. Potekhin, A. Y., Zyuzin, D. A., Yakovlev, D. G., Beznogov, M. V., & Shibanov, Y. A. 2020, MNRAS, 496, 5052
469. Predehl, P., Andritschke, R., Arefiev, V., et al. 2021, A&A, 647, A1
470. Pretorius, M. L., Knigge, C., & Schwope, A. D. 2013, MNRAS, 432, 570
471. Pretorius, M. L., & Mukai, K. 2014, MNRAS, 442, 2580
472. Puebla, R. E., Diaz, M. P., & Hubeny, I. 2007, AJ, 134, 1923
473. Qu, Y., & Zhang, B. 2025, ApJ, 981, 34
474. Ramsay, G., Cropper, M., Wu, K., Mason, K. O., & Hakala, P. 2000, MNRAS, 311, 75
475. Ramsay, G., Hakala, P., & Cropper, M. 2002, MNRAS, 332, L7

476. Ramsay, G., Marsh, T. R., Kupfer, T., et al. 2018, A&A, 617, A88
477. Ramsay, G., Wheatley, P. J., Rosen, S., Barclay, T., & Steeghs, D. 2012, MNRAS, 425, 1486
478. Rana, V., Loh, A., Corbel, S., et al. 2016, ApJ, 821, 103
479. Rauch, T., Orio, M., Gonzales-Riestra, R., et al. 2010, ApJ, 717, 363
480. Rea, N., Hurley-Walker, N., Pardo-Araujo, C., et al. 2024, ApJ, 961, 214
481. Ressler, S. M., Katsuda, S., Reynolds, S. P., et al. 2014, ApJ, 790, 85
482. Reynolds, C. S., Kara, E. A., Mushotzky, R. F., et al. 2023, in Society of Photo-Optical Instrumentation Engineers (SPIE) Conference Series, Vol. 12678, UV, X-Ray, and Gamma-Ray Space Instrumentation for Astronomy XXIII, ed. O. H. Siegmund & K. Hoadley, 126781E
483. Reynolds, C. S., Miller, E. D., Hodges-Kluck, E., et al. 2024, in Society of Photo-Optical Instrumentation Engineers (SPIE) Conference Series, Vol. 13093, Space Telescopes and Instrumentation 2024: Ultraviolet to Gamma Ray, ed. J.-W. A. den Herder, S. Nikzad, & K. Nakazawa, 1309328
484. Reynolds, M. T., Reis, R. C., Miller, J. M., Cackett, E. M., & Degenaar, N. 2014, MNRAS, 441, 3656
485. Reynolds, S. P. 1998, ApJ, 493, 375
486. Reynolds, S. P., Pavlov, G. G., Kargaltsev, O., et al. 2017, Space Science Reviews, 207, 175
487. Reynolds, S. P., Williams, B. J., Borkowski, K. J., & Long, K. S. 2021, ApJ, 917, 55
488. Ridolfi, A., Gautam, T., Freire, P. C. C., et al. 2021, MNRAS, 504, 1407
489. Riggio, A., Burderi, L., Di Salvo, T., et al. 2013, The Astronomer's Telegram, 5086, 1
490. Riley, T. E., Watts, A. L., Bogdanov, S., et al. 2019, ApJ, 887, L21
491. Rivera Sandoval, L., Belloni, D., & Ramos Arevalo, M. 2024, ApJ, 964, L20
492. Rivera Sandoval, L. E., Heinke, C. O., Hameury, J. M., et al. 2022, ApJ, 926, 10
493. Rivera Sandoval, L. E., & Maccarone, T. J. 2019, MNRAS, 483, L6
494. Rivera Sandoval, L. E., Maccarone, T. J., Cavecchi, Y., Britt, C., & Zurek, D. 2021, MNRAS, 505, 215
495. Robberto, M., Roming, P. W., van der Horst, A. J., et al. 2018, in Society of Photo-Optical Instrumentation Engineers (SPIE) Conference Series, Vol. 10702, Ground-based and Airborne Instrumentation for Astronomy VII, ed. C. J. Evans, L. Simard, & H. Takami, 107020I
496. Roberts, M. S. E. 2013, in IAU Symposium, Vol. 291, Neutron Stars and Pulsars: Challenges and Opportunities after 80 years, ed. J. van Leeuwen, 127
497. Roberts, T. P. 2007, Ap&SS, 311, 203
498. Robson, T., Cornish, N. J., & Liu, C. 2019, Classical and Quantum Gravity, 36, 105011
499. Rodriguez, A. C. 2025, A&A, 695, L8
500. Rodriguez, A. C., El-Badry, K., Hakala, P., et al. 2025, arXiv e-prints, arXiv:2501.01490
501. Rodriguez, A. C., El-Badry, K., Suleimanov, V., et al. 2025, PASP, 137, 014201
502. Rodríguez Castillo, G. A., Israel, G. L., Belfiore, A., et al. 2020, ApJ, 895, 60
503. Romani, R. W., & Sanchez, N. 2016, ApJ, 828, 7
504. Rutherford, J., Dewey, D., Figueroa-Feliciano, E., et al. 2013, ApJ, 769, 64
505. Rutledge, R. E., Bildsten, L., Brown, E. F., Pavlov, G. G., & Zavlin, V. E. 1999, ApJ, 514, 945
506. Rutledge, R. E., Bildsten, L., Brown, E. F., et al. 2002, ApJ, 580, 413
507. Safi-Harb, S., Doerksen, N., Rogers, A., & Fryer, C. L. 2019, Journal of the RAS of Canada, 113, 7
508. Safi-Harb, S., Mac Intyre, B., Zhang, S., et al. 2022, ApJ, 935, 163
509. Safi-Harb, S., Burdge, K. B., Bodaghee, A., et al. 2023, arXiv e-prints, arXiv:2311.07673
510. Salvaggio, C., Wolter, A., Belfiore, A., & Colpi, M. 2023, MNRAS, 522, 1377
511. Salvaggio, C., Wolter, A., Pintore, F., et al. 2022, MNRAS, 512, 1814
512. Sanderson, R. E., Hickox, R., Hirata, C. M., et al. 2024, arXiv e-prints, arXiv:2404.14342
513. Sanna, A., Burderi, L., Riggio, A., et al. 2016, MNRAS, 459, 1340
514. Sanna, A., Ferrigno, C., Ray, P. S., et al. 2018, A&A, 617, L8
515. Sanna, A., Bult, P., Ng, M., et al. 2022, MNRAS, 516, L76
516. Santamaría, E., Guerrero, M. A., Ramos-Larios, G., Toalá, J. A., & Sabin, L. 2025, MNRAS, 539, 246
517. Sapienza, V., Miceli, M., Bamba, A., et al. 2022, ApJ, 935, 152

518. Sathyaprakash, R., Roberts, T. P., Walton, D. J., et al. 2019, MNRAS, 488, L35
519. Sato, T., Katsuda, S., Morii, M., et al. 2018, ApJ, 853, 46
520. Sautron, M., McEwen, A. E., Younes, G., et al. 2025, arXiv e-prints, arXiv:2503.11875
521. Schreiber, M. R., Belloni, D., & van Roestel, J. 2023, A&A, 679, L8
522. Schwarz, R., Schwope, A. D., Vogel, J., et al. 2009, A&A, 496, 833
523. Schwope, A. D., Brunner, H., Buckley, D., et al. 2002, A&A, 396, 895
524. Schwope, A. D., Knauff, K., Kurpas, J., et al. 2024, A&A, 690, A243
525. Schwope, A. D., Beuermann, K., Buckley, D. A. H., et al. 1998, in Astronomical Society of the Pacific Conference Series, Vol. 137, Wild Stars in the Old West, ed. S. Howell, E. Kuulkers, & C. Woodward, 44
526. Servillat, M., Dieball, A., Webb, N. A., et al. 2008, A&A, 490, 641
527. Seward, F. D., Slane, P. O., Smith, R. K., & Sun, M. 2003, ApJ, 584, 414
528. Shafter, A. W., Darnley, M. J., Bode, M. F., & Ciardullo, R. 2012, ApJ, 752, 156
529. Shakura, N. I., & Sunyaev, R. A. 1973, A&A, 500, 33
530. Shaw, A. W., Degenaar, N., Maccarone, T. J., et al. 2024, MNRAS, 527, 7603
531. Shaw, A. W., Heinke, C. O., Mukai, K., et al. 2020, MNRAS, 498, 3457
532. Shaw, A. W., Heinke, C. O., Maccarone, T. J., et al. 2020, MNRAS, 492, 4344
533. Shporer, A., Hartman, J., Mazeh, T., & Pietsch, W. 2007, A&A, 462, 1091
534. Shternin, P. S., Ofengeim, D. D., Heinke, C. O., & Ho, W. C. G. 2023, MNRAS, 518, 2775
535. Shternin, P. S., Yakovlev, D. G., Heinke, C. O., Ho, W. C. G., & Patnaude, D. J. 2011, MNRAS, 412, L108
536. Sokoloski, J. L., Luna, G. J. M., Mukai, K., & Kenyon, S. J. 2006, Nature, 442, 276
537. Spangler, S. R. 2001, Space Science Reviews, 99, 261
538. Sridhar, N., & Metzger, B. D. 2022, ApJ, 937, 5
539. Sridhar, N., Metzger, B. D., Beniamini, P., et al. 2021, ApJ, 917, 13
540. Sridhar, N., Metzger, B. D., & Fang, K. 2024, ApJ, 960, 74
541. Stanway, E. R., & Eldridge, J. J. 2018, MNRAS, 479, 75
542. Stappers, B. W., Archibald, A. M., Hessels, J. W. T., et al. 2014, ApJ, 790, 39
543. Stobbart, A. M., Roberts, T. P., & Warwick, R. S. 2004, MNRAS, 351, 1063
544. Stoop, M., van den Eijnden, J., Degenaar, N., et al. 2021, MNRAS, 507, 330
545. Strader, J., Chomiuk, L., Maccarone, T. J., Miller-Jones, J. C. A., & Seth, A. C. 2012, Nature, 490, 71
546. Strader, J., Li, K.-L., Chomiuk, L., et al. 2016, ApJ, 831, 89
547. Strader, J., Chomiuk, L., Cheung, C. C., et al. 2015, ApJ, 804, L12
548. Strader, J., Swihart, S., Chomiuk, L., et al. 2019, ApJ, 872, 42
549. Strader, J., Swihart, S. J., Urquhart, R., et al. 2021, ApJ, 917, 69
550. Strader, J., Ray, P. S., Urquhart, R., et al. 2025, ApJ, 980, 124
551. Strohmayer, T. E. 2021, ApJ, 912, L8
552. Sudoh, T., Inoue, Y., & Khangulyan, D. 2020, The Astrophysical Journal, 889, 146
553. Suleimanov, V. F., Doroshenko, V., & Werner, K. 2019, MNRAS, 482, 3622
554. Sun, M., Seward, F. D., Smith, R. K., & Slane, P. O. 2004, ApJ, 605, 742
555. Suzuki, H., Katsuda, S., Tanaka, T., et al. 2022, ApJ, 938, 59
556. Suzuki, H., Tsuji, N., Kanemaru, Y., et al. 2025, The Astrophysical Journal, 978, L20
557. Swank, J., & Markwardt, C. 2001, in Astronomical Society of the Pacific Conference Series, Vol. 251, New Century of X-ray Astronomy, ed. H. Inoue & H. Kunieda, 94
558. Swihart, S. J., Strader, J., Chomiuk, L., et al. 2022, ApJ, 941, 199
559. Takei, D., Drake, J. J., Yamaguchi, H., et al. 2015, ApJ, 801, 92
560. Tauris, T. M., & Savonije, G. J. 1999, A&A, 350, 928
561. Tavani, M., & Arons, J. 1997, ApJ, 477, 439
562. Team COMPAS: Riley, J., Agrawal, P., Barrett, J. W., et al. 2022, ApJS, 258, 34
563. Terada, Y., Miwa, Y., Ohsumi, H., et al. 2022, ApJ, 933, 111
564. Terry, S. K., Hosek, Jr., M. W., Lu, J. R., et al. 2023, arXiv e-prints, arXiv:2306.12485

565. Thorstensen, J. R., Fesen, R. A., & van den Bergh, S. 2001, AJ, 122, 297
566. Thygesen, E., Sun, Y., Huang, J., et al. 2023, MNRAS, 518, 3386
567. Tiengo, A., Esposito, P., Mereghetti, S., et al. 2013, Nature, 500, 312
568. Toalá, J. A., Guerrero, M. A., Santamaría, E., Ramos-Larios, G., & Sabin, L. 2020, MNRAS, 495, 4372
569. Tomsick, J., Boggs, S., Zoglauer, A., et al. 2024, in 38th International Cosmic Ray Conference, 745
570. Toubiana, A., Karnesis, N., Lamberts, A., & Miller, M. C. 2024, A&A, 692, A165
571. Tran, A., Williams, B. J., Petre, R., Ressler, S. M., & Reynolds, S. P. 2015, ApJ, 812, 101
572. Treiber, H., Vasilopoulos, G., Bailyn, C. D., Haberl, F., & Udalski, A. 2025, A&A, 694, A43
573. Trimble, V. 1968, AJ, 73, 535
574. Troja, E., van Eerten, H., Zhang, B., et al. 2020, MNRAS, 498, 5643
575. Trudolyubov, S. P. 2008, MNRAS, 387, L36
576. Truelove, J. K., & McKee, C. F. 1999, ApJS, 120, 299
577. Tsuji, N., & Uchiyama, Y. 2016, PASJ, 68, 108
578. Tsuji, N., Uchiyama, Y., Khangulyan, D., & Aharonian, F. 2021, ApJ, 907, 117
579. Tsuji, N., Takekawa, S., Mori, K., et al. 2025, ApJ, 983, 22
580. Tsuji, N., Inoue, Y., Khangulyan, D., et al. 2025, arXiv e-prints, arXiv:2510.06431
581. Tsygankov, S. S., Mushtukov, A. A., Suleimanov, V. F., & Poutanen, J. 2016, MNRAS, 457, 1101
582. Tüllmann, R., Long, K. S., Pannuti, T. G., et al. 2009, ApJ, 707, 1361
583. Tutukov, A., & Yungelson, L. 1996, MNRAS, 280, 1035
584. Uchiyama, Y., Aharonian, F. A., Tanaka, T., Takahashi, T., & Maeda, Y. 2007, Nature, 449, 576
585. Urias, L., Maccarone, T. J., Antoniou, V., Mandel, I., & Vinciguerra, S. 2021, Research Notes of the AAS, 5, 209
586. Usher, C., Dage, K. C., Girardi, L., et al. 2023, PASP, 135, 074201
587. van den Eijnden, J., Degenaar, N., Pinto, C., et al. 2018, MNRAS, 475, 2027
588. van den Heuvel, E. P. J., Bhattacharya, D., Nomoto, K., & Rappaport, S. A. 1992, A&A, 262, 97
589. Vasilopoulos, G., Haberl, F., Carpano, S., & Maitra, C. 2018, A&A, 620, L12
590. Vasilopoulos, G., Koliopanos, F., Haberl, F., et al. 2021, ApJ, 909, 50
591. Vasilopoulos, G., Lander, S. K., Koliopanos, F., & Bailyn, C. D. 2020, MNRAS, 491, 4949
592. Vasilopoulos, G., Petropoulou, M., Koliopanos, F., et al. 2019, MNRAS, 488, 5225
593. Veledina, A., Nättilä, J., & Beloborodov, A. M. 2019, ApJ, 884, 144
594. Vink, J. 2008, ApJ, 689, 231
595. Vink, J., & Laming, J. M. 2003, ApJ, 584, 758
596. Vink, J., Patnaude, D. J., & Castro, D. 2022, ApJ, 929, 57
597. Volkov, I., Kargaltsev, O., Younes, G., Hare, J., & Pavlov, G. 2021, ApJ, 915, 61
598. Wadiasingh, Z., & Timokhin, A. 2019, ApJ, 879, 4
599. Wagg, T., Breivik, K., & de Mink, S. 2022, The Journal of Open Source Software, 7, 3998
600. Walton, D. J., Mackenzie, A. D. A., Gully, H., et al. 2022, MNRAS, 509, 1587
601. Walton, D. J., Fürst, F., Bachetti, M., et al. 2016, ApJ, 827, L13
602. Walton, D. J., Fürst, F., Heida, M., et al. 2018, ApJ, 856, 128
603. Wang, Q. D., Gotthelf, E. V., & Lang, C. C. 2002, Nature, 415, 148
604. Wang, S., Qiu, Y., Liu, J., & Bregman, J. N. 2016, ApJ, 829, 20
605. Wang, X., & Takata, J. 2025, arXiv e-prints, arXiv:2504.10794
606. Wang, Z., Rea, N., Bao, T., et al. 2024, arXiv e-prints, arXiv:2411.16606
607. Warner, B. 1995, Cambridge Astrophysics Series, 28
608. Weng, S.-S., Qian, L., Wang, B.-J., et al. 2022, Nature Astronomy, 6, 698
609. West, L. A., Lehmer, B. D., Wik, D., et al. 2018, ApJ, 869, 111
610. Wijnands, R. 2008, in American Institute of Physics Conference Series, Vol. 1010, A Population Explosion: The Nature & Evolution of X-ray Binaries in Diverse Environments, ed. R. M. Bandyopadhyay, S. Wachter, D. Gelino, & C. R. Gelino (AIP), 382
611. Wijnands, R. 2014, in 40th COSPAR Scientific Assembly, Vol. 40, E1.18

612. Wijnands, R., Degenaar, N., & Page, D. 2017, Journal of Astrophysics and Astronomy, 38, 49
613. Wijnands, R., in't Zand, J. J. M., Rupen, M., et al. 2006, A&A, 449, 1117
614. Wijngaarden, M. J. P., Ho, W. C. G., Chang, P., et al. 2019, MNRAS, 484, 974
615. Wiktorowicz, G., Giersz, M., Askar, A., Hypki, A., & Helstrom, L. 2025, A&A, 696, A90
616. Williams, B. F., Garcia, M. R., Kong, A. K. H., et al. 2004, ApJ, 609, 735
617. Williams, B. F., Gaetz, T. J., Haberl, F., et al. 2008, ApJ, 680, 1120
618. Williams, B. F., Wold, B., Haberl, F., et al. 2015, ApJS, 218, 9
619. Williams, B. J., Chomiuk, L., Hewitt, J. W., et al. 2016, ApJ, 823, L32
620. Wilson-Hodge, C. A., Malacaria, C., Jenke, P. A., et al. 2018, ApJ, 863, 9
621. Wolf, W. M., Bildsten, L., Brooks, J., & Paxton, B. 2013, ApJ, 777, 136
622. Wu, Y. X., Yu, W., Li, T. P., Maccarone, T. J., & Li, X. D. 2010, ApJ, 718, 620
623. Xing, Y., & Wang, Z. 2015, ApJ, 808, 17
624. Xu, Y.-D. 2011, ApJ, 729, 10
625. Yakovlev, D. G., & Pethick, C. J. 2004, ARA&A, 42, 169
626. Yamaguchi, H., Eriksen, K. A., Badenes, C., et al. 2014, ApJ, 780, 136
627. Yamamoto, H., Okamoto, R., Murata, Y., et al. 2022, PASJ, 74, 493
628. Yaron, O., Prialnik, D., Shara, M. M., & Kovetz, A. 2005, ApJ, 623, 398
629. Yoneda, H., Makishima, K., Enoto, T., et al. 2020, PhRvL, 125, 111103
630. Younes, G., Kouveliotou, C., Kargaltsev, O., et al. 2012, ApJ, 757, 39
631. —. 2016, ApJ, 824, 138
632. Younes, G., Güver, T., Kouveliotou, C., et al. 2020, ApJ, 904, L21
633. Younes, G., Baring, M. G., Kouveliotou, C., et al. 2021, Nature Astronomy, 5, 408
634. Younes, G., Lander, S. K., Baring, M. G., et al. 2022, ApJ, 924, L27
635. —. 2025, arXiv e-prints, arXiv:2502.20079
636. Yuan, F., & Narayan, R. 2014, ARA&A, 52, 529
637. Yukita, M., Swartz, D. A., Tennant, A. F., Soria, R., & Irwin, J. A. 2012, ApJ, 758, 105
638. Yungelson, L., & Lasota, J. P. 2008, New A Rev., 51, 860
639. Yungelson, L. R., Lasota, J. P., Nelemans, G., et al. 2006, A&A, 454, 559
640. Zangrandi, F., Jurk, K., Sasaki, M., et al. 2024, A&A, 692, A237
641. Zdziarski, A. A., Neronov, A., & Chernyakova, M. 2010, MNRAS, 403, 1873
642. Zevin, M., Samsing, J., Rodriguez, C., Haster, C.-J., & Ramirez-Ruiz, E. 2019, ApJ, 871, 91
643. Zezas, A., Fabbiano, G., Baldi, A., et al. 2007, ApJ, 661, 135
644. Zhang, J. L., Bi, X. J., & Hu, H. B. 2006, A&A, 449, 641
645. Zhang, L., Freire, P. C. C., Ridolfi, A., et al. 2023, ApJS, 269, 56
646. Zhang, S., Hailey, C. J., Mori, K., et al. 2015, ApJ, 815, 132
647. Zhang, S. N., Feroci, M., Santangelo, A., et al. 2016, in Society of Photo-Optical Instrumentation Engineers (SPIE) Conference Series, Vol. 9905, Space Telescopes and Instrumentation 2016: Ultraviolet to Gamma Ray, ed. J.-W. A. den Herder, T. Takahashi, & M. Bautz, 99051Q
648. Zhang, X., Bordas, P., Safi-Harb, S., & Iwasawa, K. 2025, arXiv e-prints, arXiv:2507.13278
649. Zhao, J., & Heinke, C. O. 2022, MNRAS, 511, 5964
650. —. 2023, MNRAS, 526, 2736
651. Zhou, P., Chen, Y., Safi-Harb, S., et al. 2016, ApJ, 831, 192
652. Zhu, Z., Li, Z., & Morris, M. R. 2018, ApJS, 235, 26
653. Zirakashvili, V. N., & Aharonian, F. 2007, A&A, 465, 695

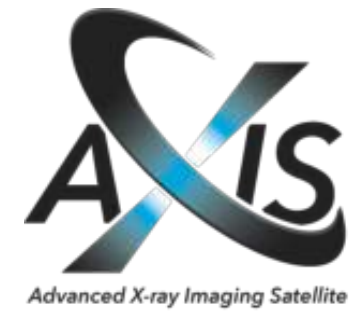

# AXIS Stars and Exoplanets GO Science

**T. Ayres[1], B. A. Binder[2], J. A. Caballero[3], R. Cilley[4], M. F. Corcoran[5], L. Corrales[4], S. Dillmann[5], J. J. Drake[7,8], G. M. Duvvuri[9], C. Espaillat[10], A. Feinstein[11], J. Fernández Fernández[12], H. M. Günther[13], S. J. Gunderson[13], K. Hamaguchi[14], G. W. King[4], M. Kounkel[15], C. M. Lisse[16], Y. Nazé[17], L. M. Oskinova[18], R. Osten[19], D. A. Principe[13], K. G. Stassun[9], P. J. Wheatley[12], S. Wolk[7]**

[1] CASA/CU Boulder [2] Cal Poly Pomona [3] Centro de Astrobiología (CSIC-INTA) [4] University of Michigan [5] Catholic University of America [6] Stanford University [7] Harvard & Smithsonian Center for Astrophysics [8] Lockheed Martin, USA [9] Vanderbilt University [10] Boston University [11] Michigan State University [12] University of Warwick [13] Massachusetts Institute of Technology [14] NASA Goddard Space Flight Center [15] University of North Florida [16] Johns Hopkins University [17] FNRS/University of Liége [18] Potsdam University [19] Space Telescope Science Institute

---

**a. Star formation and accretion**

*1. Magnetic Activity at Star Birth: Searching for X-rays from Class 0 Protostars with AXIS*

**Science Area:** Stars & Exoplanets
**First Author:** David A. Principe, Massachusetts Institute of Technology, daveprincipe1@gmail.com
**Co-authors:** Moritz Günther, Massachusetts Institute of Technology
**Abstract:** Stars are born deeply embedded in molecular clouds where they rapidly accrete material to form protostellar disks and eventually planets. The earliest stage of a star's life is the so-called 'Class 0' phase, which occurs during the first ∼100,000 years of formation for low mass stars (< 3 $M_{\odot}$). At this stage, most of the stellar system's mass resides in an envelope surrounding a newly formed protostar and its protostellar disk. Two outstanding questions that have important ramifications for star and planet formation are whether or not material is accreted onto the star in a spherically symmetric fashion and whether X-ray emission is present to drive chemistry in the earliest years of a star's life. The presence of a strong magnetic field, as indicated by the detection of X-rays at the location of the central protostar, would indicate that accretion is not spherically symmetric but instead funneled along magnetic field lines onto discrete stellar hotspots. Such accretion is observed for stars at later stages of pre-main-sequence (pre-MS) evolution, but not for the earliest stages of star formation, due to the embedded nature of their formation. Due to the limited sensitivity of Chandra and XMM, only a handful of definitive Class 0 protostars have been detected in X-rays. The high effective area of AXIS, combined with its ∼1 arcsecond spatial resolution stable over a large field of view, makes it the ideal instrument to search for X-ray emission from Class 0 protostars, which form in stellar clusters. Detection of Class 0 protostars with X-rays will indicate whether aspherical accretion is present at the earliest stages of star formation. Such confirmation will affect stellar evolution models and our understanding of how early planets may form in such an environment.
**Science:** Over the course of approximately 10 million years, young, pre-main sequence (pre-MS) stars undergo complex and drastic changes starting from the initial collapse of a molecular cloud clump and ending with the formation of a pre-MS star and planetary system. For low-mass stars (below about 3 solar masses) this process can be characterized in roughly four 'stages' of pre-MS evolution ranging from so-called Class 0 objects at birth to Class III objects in the final stage of young stellar evolution.

Class 0 protostars represent the first ∼100,000 years of a star's life. They are generally defined as the period where > 50% of a star's eventual mass still resides in the molecular envelope surrounding the central protostar [6]. After ∼100,000 years, enough of the envelope mass is accreted into the disk and eventually onto the star to enter the Class I phase of pre-MS evolution. Observations of young stars in the Class 0 and Class I phase show the presence of outflows and jets [43], likely powered by magnetic fields in the disk, which remove angular momentum and facilitate continued mass-buildup of the central star via accretion from the disk.

While the Class 0 stage of pre-MS evolution represents only ∼0.1% of a star's evolution towards the main sequence, important questions about this stage of evolution remain that have implications for stellar evolution models and our understanding of the critical physics of how stars gain mass. While we know the central protostar must be accumulating mass, it is not yet clear whether accretion is spherically symmetric onto the protostar or whether it is, for example, funneled along magnetic field lines leading to accretion 'hot spots' onto the surface. A clear signature of the latter 'hot spot' accretion would be the presence of a strong magnetic field extending from the central protostar (Fig. 1 Left). Unfortunately, the Class 0 stage is notoriously difficult to observe due to the embedded nature of the protostellar system in its parent molecular cloud. High spatial and spectral resolution instruments like ALMA and the VLA excel at studying the large-scale (1000s of au) disk and envelope structure surrounding these stars because the material is generally cold and thus emits primarily in the submm/radio. Yet little is observationally known about the central protostar which is rapidly gaining mass and providing an important source of light to heat and ionize the surrounding material as it collapses to build a star-disk system.

Determining Whether Class 0 Protostars Have Strong Magnetic Fields with AXIS:

Young, low-mass ($\lesssim$3 $M_\odot$) pre-MS stars are ubiquitous X-ray sources with X-ray detections of Class I to Class III stars ranging in the 10s of thousands [67,71]. X-rays from young stars are primarily generated in stellar coronae, which are heated to 1-10 million degrees K as a result of stellar magnetic dynamos [70]. Such dynamos are believed to be powered by a combination of the protostars large convective zone and their rapid ($\sim$days) rotation periods during pre-MS evolution. A direct result of a stellar magnetic dynamo is a large-scale magnetic field that, in the case of pre-MS stars, can extend out into the disk region and facilitates the accretion of disk material funneled along magnetic field lines onto the central star. Observations show a clear correlation between magnetic field strength and X-ray luminosity for the Sun, older active stars and pre-MS stars [132]. Moreover, a stellar activity and rotation relationship exists which clearly demonstrates a relationship between a star's fractional X-ray luminosity ($L_X/L_{bol}$) and its magnetic 'Rossby' number, which is a star's convective turnover time divided by its rotation period (Fig. 1 Right). X-ray detected Class I-III pre-MS stars occupy the 'saturated' region of this plot < Rossby number $\sim$1.4 where, regardless of the Rossby number, the star's fractional X-ray luminosity ($L_X/L_{bol}$) will be $\sim$1E-3. One theory as to why these stars 'saturate' is because the entire stellar surface is covered in coronal loops and there is no more room for new loops to allow fractional X-ray luminosities greater than $\sim$1E-3 [70].

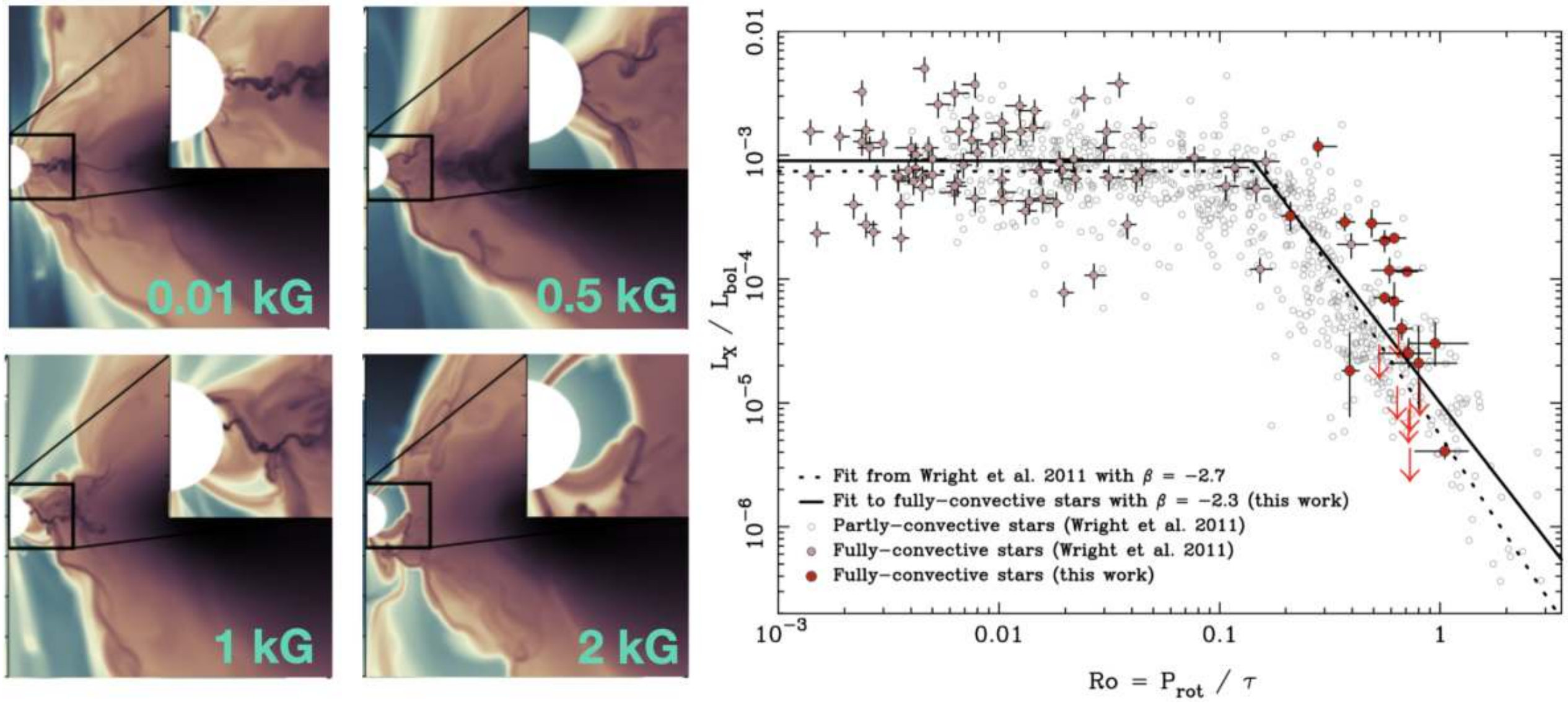

**Figure 1.** *Left:* Magnetohydrodynamic simulations of protostellar accretion showing density distribution for different protostellar magnetic field strengths (lower right of panels). Figure has been modified from the original in [62]. *Right:* The coronal activity-rotation relationship probing magnetic activity in solar and late-type stars from [Figure 3 in 174]. A star's fractional X-ray luminosity remains 'saturated' at $\sim 1 \times 10^{-3}$ for Rossby values up to 1.14 before decreasing with increased rotation periods ($P_{rot}$). Tau is the time it takes for a convective cell to rise and fall in the star.

Coronal X-rays from young stars offer a powerful tool to understand the magnetic environment of the protostar itself. While soft X-rays ($\lesssim$1-2 keV) are readily absorbed by intervening material (e.g., ISM, molecular clouds, or circumstellar disks) the central star's hard X-rays ($\gtrsim$2 keV) pierce through cloud material even for highly absorbed sources like those residing deep in molecular clouds. Likely as a result, only a handful of Class 0 protostars have X-ray detections even though dozens have fallen in the field of view of archival X-ray observations. Theoretical models show that magnetic dynamos may exist at this young age [17].

Building a sample of Class 0 protostars from these ALMA/VLA campaigns [161], we aim to pursue the first deep, targeted X-ray observations of Class 0 protostars to date. Constraining the X-ray emission from these sources will achieve two goals:

1. detecting X-ray emission at the earliest stage of star formation will solidify the idea that strong magnetic fields can exist at ages < 100,000 years and material accreted onto the star is likely funneled into hot spots. Stars accrete most of their mass at this phase and how it falls onto the star has important implications for stellar evolution models [11,62,66].
2. measuring $L_X/L_{bol}$ for Class 0 sources will identify whether they follow the well-known activity-rotation relationship which all low-mass stars follow. Since this relationship is a result of a star's underlying magnetic dynamo, $L_X/L_{bol}$ values much lower than 1E-3 could indicate a different type of dynamo, such as a primordial dynamo, is producing X-rays at this stage of evolution whereas $L_X/L_{bol}$ values of 1E-3 would indicate these sources are magnetically saturated like they are at later stages of evolution. The presence and strength of X-rays at this very young age impacts circumstellar chemistry and planet formation [116].

**Exposure time (ks): 250 ks**

**Observing description:** We propose a single AXIS pointing centered at RA=05:35:27.7 DEC=-05:07:04.3 to observe protostars in the Orion Molecular Cloud (OMC 2-3) for 250 ks. This field of view includes up to 25 Class 0 protostars, 13 Class I stars, and hundreds of Class IIs and IIIs and represents the highest number of known Class 0 protostars in any 24x24 arcminute region (the FOV of a single AXIS pointing; Fig. 2). To date, there have been only a handful of positive X-ray detections of Class 0 protostars. OMC 2-3 is one of the largest and nearest sites of active star formation and includes HOPS-393 and HOPS-91, two X-ray-detected Class 0 protostars. The other 23 Class 0 stars are not detected. HOPS-393 was previously identified as having undergone a multi-year outburst as evidenced by a significant brightening in the mid-IR between 2004 and 2008 [148]. Deep 90 ks Chandra X-ray observations of the source in 2017 only detected 28 counts from the source during a brief $\sim$12 ks flare [69]. The Class 0 protostar HOPS-91 was detected with 17 counts in 2001 during a 90 ks Chandra observation [source 10 in 164]. Both of these Class 0 protostars displayed a very hard (3-4 keV), highly absorbed (1.4 - 7.0 x $10^{23}$ $cm^{-2}$) thermal plasma.

Using the AXIS response and effective area, we simulated spectra to determine an exposure time that would detect a protostar with $L_{bol}$ = 1 $L_\odot$, that would eventually form a 0.2 $M_\odot$ star, to a sensitivity of $L_X/L_{bol}$ = $10^{-5}$ (Fig 1 right). A 250 ks observation would detect such a source with 8 counts assuming a conservative plasma temp of 2 keV and NH of 3x$10^{23}$ $cm^{-2}$. The majority of Class 0 protostars in the field will be brighter than this, given their median $L_{bol}$ of 6.8 and potential to be even brighter than $L_X/L_{bol}$ of $10^{-5}$. The high sensitivity, resolution, and stable PSF over the entire field of view make AXIS the only X-ray observatory capable of carrying out this science. Stars form in clusters and a 1-2 arcsecond PSF is required to differentiate multiple nearby sources.

**Specify critical AXIS Capabilities:** Young, pre-MS stars form in groups that often require high angular resolution ($\lesssim$ 2 arcsec) to spatially resolve multiple members. Star-forming regions extend over large portions of the sky so efficiently detecting X-rays from multiple Class 0 protostars require a stable PSF over the large AXIS field of view and low background. Class 0 protostars are embedded in molecular cloud material and thus high sensitivity $\gtrsim$2 keV is required.)

**Joint Observations and synergies with other observatories in the 2030s:** AXIS observations of Class 0 protostars have strong synergies with instruments into the 2030s. Near- and mid-IR spectroscopy with JWST and submm observations with ALMA can constrain the chemical makeup of circumstellar material, some of which is expected to be affected by X-ray irradiation from the central protostar. To learn about the

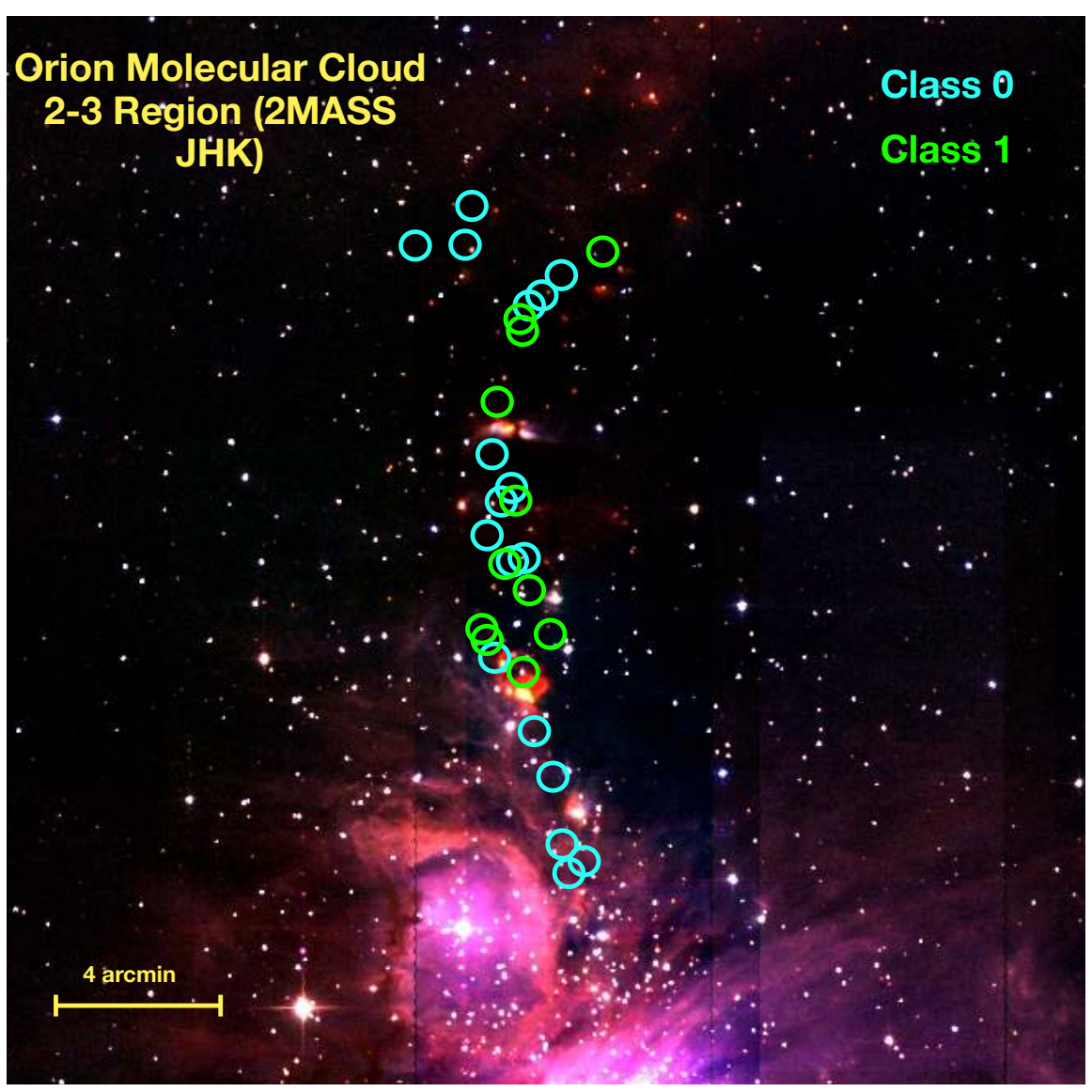

**Figure 2.** A 24x24 arcmin 2MASS JHK color image of the Orion Molecular Cloud 2-3 region. Hundreds of young, pre-MS stars are visible in this image directly north of the famous Orion Nebula Cluster. The youngest protostars from [161] (Class 0 and I) reside in a molecular cloud filament not visible in the 2MASS image.

underlying magnetic dynamo at the earliest stages of stellar evolution, cm observations with the Square Kilometer Array (SKA) can be used to constrain coronal magnetic field strength for comparison with that derived from X-rays.
**Special Requirements:** None

*2. The Fastest Components in Jets from Young Stars*

**Science Area:** Stars and Exoplanets
**First Author:** Hans Moritz Günther (MIT, hgunther@mit.edu)
**Abstract:** Young stars are surrounded by circumstellar material, they drive wide-angle outflows from the disk, they accrete material onto planets and stars, and some of them also drive highly collimated jets. Depending on the mass and evolutionary state of the host star, these jets can be close in or extend several pc. Typically, most of the mass is in relatively slow outflow components with tens of km/s speeds, but many jets show a fast component that carries only 0.1-1% of the outflow mass, but reaches $> 300$ km/s - sufficient speed to generate X-rays in collimation shocks or clumps launched at different times run into each other. Since the fastest components are launched deepest in the gravitational well, they probe matter launched from the star itself or the innermost regions of the accretion disk. How those jets are accelerated and collimated is an open question. Magnetic fields clearly play a role, but beyond that, we need to measure shock temperatures, abundances, and speeds to distinguish stationary collimation shocks, moving working surfaces, and to finally understand how those jets form. On larger scales, the bow shocks at the tip of jets drive into the gas and dust remaining in the star-forming region, inject angular momentum, and provide a diagnostic to measure densities, temperatures, and abundances of the primordial material. For either application, the large collecting area of AXIS can provide sufficient signal in 50-100 ks per target, and in some cases, that large field of view might be sufficient to detect new, as-yet-unknown X-ray jets from neighboring sources in the same star-forming region.

**Science:** Young stars often shed angular momentum through outflows: from slow-moving, wide-angle disk winds to fast and highly collimated jets launched close to the star. In some cases, those jets can be traced to several pc from the star. X-ray emission can be seen from the termination shock where the jet runs into the interstellar medium [e.g. 9,68,137], from inner working surfaces in the shock [23], and also from the inner region very close to the launching region [24,74]. Inner jets (within an arcsec of the source) in particular have been shown to be X-ray emitters in a number of young stars. For younger stars that are still deeply embedded in molecular clouds, a jet that pierces an opening into the infalling envelope may be observable even while the central proto-star is not, due to obscuration by its natal cloud. In those cases [e.g. HH154 52], all the observed X-rays come from jet-induced shocks; the observed emission is spatially offset from the central source and is so soft that it cannot come from the embedded star. Stellar jets provide feedback on the molecular cloud, pierce holes that make the inner star visible to us, and carry away angular momentum that allows accretion to proceed. Understanding jets is absolutely crucial to make progress on several science questions identified in the decadal survey for "Interstellar Medium and Star and Planet Formation". In particular, F-Q2a asks:"What processes are responsible for the observed velocity fields in molecular clouds?". To answer this question, we need to quantify the mass and momentum that stellar jets inject into the clouds. For question F-Q3b: "How do protostars accrete from envelopes and disks, and what does this imply for protoplanetary disk transport and structure?", we need to quantify the angular momentum removed from the system by winds and outflows and determine at what radius that occurs. Since X-rays probe the fastest outflow component of the jet, which are launched deep inside the gravitational well of the star, possibly even from the inner disk interface or the stellar surface, they are crucial to get a complete picture of the stellar jets and their impact on star formation and molecular clouds.

Class II sources are typically evolved star-disk systems that have cleared out their surrounding molecular environment. Observing Class II sources with jets provides an opportunity to investigate jet-launching mechanisms by revealing the structure closer to the launching region. The best-studied object of this type is the CTTS DG Tau, which has 450 ks of cumulative observations with Chandra [72,73,75], and reveals two spatially distinct components that are X-ray emitters in the outflow. Proper motion has been observed in some X-ray jets [e.g., 150], but for DG Tau, this question remains unsettled. Figure 3 shows a 350 ks Chandra image of DG Tau, which was obtained in a single year. This image resolves this system into three components: the young stellar corona, an inner jet region, and an outer jet region.

The inner jet component is barely 0.2" offset from DG Tau itself (30-50 au), but its spectrum is much softer than the star because the star is deeply embedded, but the jet is not. That allowed Schneider & Schmitt [151] to compare the position of soft and hard photons and measure the offset, even though it is much less than the size of the Chandra PSF. Similar techniques can be employed with AXIS, but thanks to the much higher collecting area, measurements like this will no longer be limited by the low count rate. AXIS will thus be able to repeat the Chandra measurement several times during the mission timeline and probe the evolution of the jet. Different hypotheses have been presented to predict the time revolution of DG Tau's inner jet. Bonito et al. [25] presents a model of a diamond shock, and Günther et al. [76] suggest a model where hydrodynamic pressure of the outer wind component re-collimates the inner, faster, less dense region and focuses a standing recollimation shock. In those scenarios, the shock position should remain mostly steady over time, but the shock intensity should change with changes in the accretion (and thus mass outflow) rate. Only AXIS has the collecting area and angular resolution to perform these measurements repeatedly and for multiple stars. While DG Tau is so far the only known example where the jet shock is spatially resolved in imaging, the same spectral shape consisting of a hard, highly absorbed stellar corona and a much softer, much less absorbed shock component is also seen in e.g. GV Tau and DP Tau [74]. It seems likely that these objects also have X-ray-emitting jets. Unfortunately, with the major loss of the Chandra/ACIS soft energy effective area, those observations are no longer feasible with Chandra, even for DG Tau, which is the brightest of the known X-ray sources that exhibit evidence for a corona-jet system. AXIS is the only X-ray telescope that can definitively prove or disprove this idea through the most direct method: imaging.

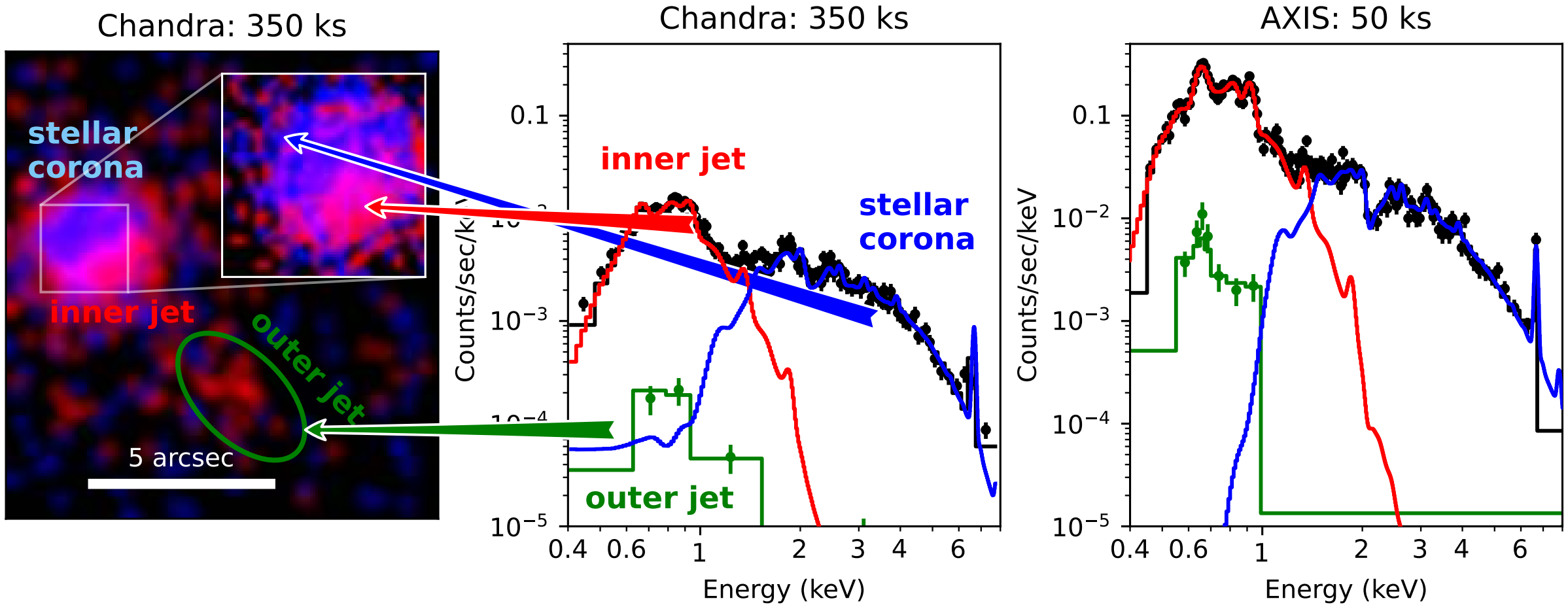

**Figure 3.** The CTTS DG Tau in X-rays. The left panel shows a 350 ks Chandra image combining multiple observations taken in 2010. The image is color-coded with photons below 1.2 keV shown in red and photons above 1.2 keV in blue. Three spatially distinct components are visible. The DG Tau corona is absorbed, and only photons $> 1.2$ keV are visible (blue) at this location. There is a softer inner jet component (red), that is offset about 0.3 arcsec from the stellar position, but the two components overlap due to Chandra's PSF. The two components can be spectrally resolved by fitting with two different temperatures and absorption components (black data with red and blue model components in the middle panel). There is also an outer jet component that is several arcsec (650 AU) away from the central star. We expect substructure in this region, but given the low count rate in Chandra, we only know that it is spatially extended. This feature will be resolved with AXIS, and a spectrum can be extracted separately (green). The right panel shows a simulation for a 50 ks AXIS observation. The jet (green) spectrum is binned to 10 counts/bin, and the stellar/inner jet spectrum to 30 counts/bin. The superior collecting area of AXIS will allow us to fit abundances and multi-temperature models, search for sub-structure, and look for proper motion of the inner and outer jet component.

In addition to the inner jet shock, DG Tau also shows an even fainter, outer region with X-ray emission. This region is clearly resolved by Chandra and will be clearly resolved by AXIS. Unfortunately, 350 ks of Chandra observation yield only 50 photons from this region. Based on observations in the UV and optical, where the signal is better and HST can resolve finer details, we know that there is considerable substructure, including a bow shock with hot and cold material. However, the X-rays must come from a different component in the shock, since neither temperatures nor velocities of the components seen in the IR, optical, and UV are sufficient to explain the fitted X-ray temperature of 0.2 keV. Assuming that a bow shock or termination shock is responsible, the mass loss rate seen in the X-rays is about $10^{-10}$ $M_{\odot}$ $\mathrm{yr}^{-1}$ [77], a factor of 1000 below the mass flux seen in the optical [79]. At the same time, the velocity must be at least 300 km/s to explain the observed X-ray spectrum. With AXIS, we will be able to obtain an exposure deep enough to pinpoint the location of the X-ray emission and associate it with the jet structure. We can also repeat the observations over the lifetime of the AXIS observatory to follow the proper motion. There are two possible explanations for the X-ray emission in the outer jet. Either the gas has been heated as it went through a standing shock close to DG Tau and is now adiabatically and radiatively cooling, or it is going through a shock wave at its current location. In the latter case, the shock wave could be caused by an obstacle, such as a previous, slower outflow blob, or it could be a shock wave traveling along the jet. AXIS observations of DG Tau can distinguish all those possibilities. If the gas has previously been heated and is cooling adiabatically, we expect the jet to have a cone opening angle. We know that the outer shock is resolved with Chandra, but we have insufficient signal to constrain its shape. AXIS will deliver an order of magnitude more counts and an arcsec-scale PSF - enough to detect an opening angle of 5 degrees at 5″ from the central source. Without expansion, the density of the emitting material is so high that it cools much faster and no X-ray emitting material could reach beyond 1″. Thus, if AXIS finds the outer X-ray jet to be highly collimated, the jet material must be heated in place by a shock. AXIS imaging can pinpoint the shock location to determine if it is associated with a known, cooler structure visible in other wavelengths, which the X-ray jet collides with. AXIS spectroscopy (see Fig. 3) determines the temperature of the plasma, and thus the shock speed (Chandra indicates about 300 km/s, however, that fit is very uncertain since it is based on about 40 photons only). AXIS observations spaced by a year or more would reveal proper motion. At the distance of DG Tau, 300 km/s corresponds to $0.5''$ $\mathrm{yr}^{-1}$, and with just 50 ks of AXIS observing, the centroid of the jet emission can be determined through sub-pixel resolution algorithms. A simple comparison between proper motion and shock speed shows if the shock plows through the moving jet material or if the material passes a stationary shock wave (e.g., caused by an obstacle or the magnetic field). In other words, AXIS will be able to finally reveal what physical mechanism powers X-rays from young stellar jets.

Beyond known sources, AXIS will be a discovery engine for new stellar jets. Its PSF remains sharp over a much larger field-of-view than Chandra's PSF. This enables us to search for resolved extended emission over much larger regions, including those with dozens to hundreds of other young stars with potential jets, in a single observation. At the same time, the increased effective area collects enough photons to securely identify jets in star-forming regions and measure the jet speed via shock temperatures. Currently, there are hundreds of jets known in the optical and IR, and only about a dozen X-ray-detected jets. Given the low soft X-ray sensitivity of Chandra, we cannot test whether this means that special conditions are needed to form the fast, X-ray emitting outflow components, or if our sensitivity is simply too low to detect them in most cases. With AXIS, we can answer the question: Does every stellar jet have an inner fast component? If so, that opens the door to a unified theory of how angular momentum is removed in the innermost regions of the disk, allowing for ongoing accretion and disk clearing.

**Exposure time (ks):** 750 ks

**Observing description:** DG Tau is the best-studied object of the class of small stellar jets. Fig. 3 (right) shows a simulation of the 50 ks AXIS observation, which would spatially resolve the arcsec-scale jet and might even resolve structure within that jet (if present). With about 100 photons, we can fit the shock temperature, and thus the shock speed. At the same time, a 50 ks exposure will provide more than enough counts to fit the inner (unresolved) jet collimation shock with sufficient signal to determine relative abundances of Ne/O or Ne/Fe.

Since jets are dynamic objects, the cooler components have to be re-observed within a year or so of the AXIS observations.

*Based on today's X-ray data, the target list for this proposal is:* DG Tau, HH 168, Cepheus A, HH 154, HH 2, RY Tau, and HD 163296. However, jets evolve over time scales of years, and some of the sources observed with Chandra over the last two decades may have cooled below X-ray-emitting temperatures. On the other hand, AXIS, with its wide field of view and excellent PSF over a large range of that field, will be a discovery engine for X-rays from stellar jets.

There are three crucial diagnostics that we can extract from X-ray data about the innermost and fastest components of stellar jets:

- Spatial distribution (correlate with data from other wavelengths)
- Spectrum (to fit the shock speed)
- Time evolution (distinguish stationary collimation shocks from moving working surfaces)

The observing program is thus structured to have a single observation of the targets in the targets list above with 50 ks per target plus a few targets that will be discovered with AXIS in surveys of star forming regions: 10 targets of 50 ks = 500 ks.

Half of those targets will be selected for re-observation 2-3 years later, based on the flux of their resolved X-ray emission for an additional $5 * 50$ ks = 250 ks of observing time.

**Specify critical AXIS Capabilities:** effective area in the 0.5-1 keV band: These objects are faint and to determine meaningful physical properties, we need to get enough counts to measure locations, shapes, and spectra (to fit the shock temperature and abundances). Specific simulations are presented in the text above, but it cannot be overstated that the one property this science requires is increasing the number of collected photons by at least a factor of 10, ideally by a factor of 50, compared to Chandra.

**Joint Observations and synergies with other observatories in the 2030s:** JWST and ground-based IFUs, e.g. on TMT, GMT, Gemini: Optical and IR IFUs can reveal forbidden emission lines and measure their origin, flux, proper motion and line-of-sight motion. All stellar jets are relatively faint in X-rays and much brighter in the optical/IR, so these observations can be conducted with existing instruments on 8-m class telescopes. However, JWST and ground-based 30-m telescopes, expected in the 2030s, will reveal more spatial and kinematic detail.

**Special Requirements:** None

### *3. Measuring Line-of-sight Oxygen Abundance in Circumstellar Disks with AXIS*

**Science Area:** Stars & Exoplanets

**First Author:** David A. Principe (Massachusetts Institute of Technology), principe@mit.edu

**Co-authors:** Moritz Günther (Massachusetts Institute of Technology), Lia Corrales (University of Michigan)

**Abstract:** Planets form out of circumstellar disk material in the first ∼10 million years of a young star's life. During this stage of pre-main-sequence stellar evolution, the central protostar irradiates the surrounding environment, providing the necessary energy for many of the diverse chemicals observed in planet-forming disks. Young, low-mass stars ($\lesssim$3 $M_{\odot}$) are rapid rotators with deep convective zones which manifest in a magnetic dynamo capable of heating plasma to millions of degrees Kelvin. As such, young stars are ubiquitous X-ray sources with tens of thousands of detections from Chandra, XMM-Newton, eROSITA, and ROSAT. Many of these young stars have had their circumstellar disks spatially resolved with mm/submm instruments, such as ALMA and the VLA, leading to stringent constraints on disk inclinations. This information, combined with X-ray observations from the central protostar, provides a powerful tool for understanding the line-of-sight oxygen abundance in disks. For star/disk systems that are inclined between $\sim$30-90 degrees, X-rays from the central protostar must pass through the disk and are attenuated primarily by metals such as C, O, and N. In particular, oxygen is a crucial component of many molecules in the evolution of circumstellar disks and the formation of planets. The oxygen photo-electric absorption edge at 540 eV provides a tool to directly constrain the line-of-sight oxygen content, but only high effective area instruments like AXIS are capable of observing this for typical fluxes of nearby young stars. Since stars form in clusters where high spatial resolution and a stable PSF over the entire field of view are important, AXIS makes it feasible to measure line-of-sight oxygen content in disks for dozens of these sources in nearby star-forming regions like the Orion Nebula Cluster (ONC) proposed here. Such observations will constrain protostellar disk chemistry at the epoch of planet formation.

**Science:** Over approximately 10 million years, young, pre-main sequence (pre-MS) stars undergo complex and drastic changes starting from the initial collapse of a molecular cloud clump and ending with the formation of a pre-MS star and planetary system. This process can be characterized in roughly four 'stages' of pre-MS evolution, ranging from Class 0 objects at birth to Class III objects in the final stage of young stellar evolution.

Class 0 protostars represent the first $\sim$100,000 years of a star's life, a period of time where > 50% of a star's eventual mass still resides in the molecular envelope surrounding the central protostar [6]. After $\sim$100,000 years, enough of the envelope mass is accreted into the disk and eventually onto the star to enter the Class I phase of pre-MS evolution. Observations of young stars in the Class 0 and Class I phases reveal the presence of outflows and jets [43], likely powered by magnetic fields in the disk, which transport angular momentum and facilitate the continued mass buildup of the central star through accretion from the disk. Over the next $\sim$300,000 years, outflows from the star and continued accretion from the envelope onto the disk remove remaining envelope material and the star-disk system enters the Class II phase of pre-MS evolution. At this stage, the central star has accreted the majority of its eventual mass. It begins to irradiate the circumstellar disk, eventually culminating in its removal and the cessation of any ongoing gaseous planet formation. The removal of the gas disk ends the accretion phase as the star now has its final mass and it enters the Class III stage of pre-MS evolution. Depending on the star's mass, it will reside in the Class III phase for $\sim$50 million years before eventually beginning hydrogen fusion in its core and entering the main sequence.

At some point during the Class 0-II evolutionary phases, planet formation occurs in the circumstellar disks composed of gas and dust surrounding these pre-main sequence stars. The central protostar and its surrounding circumstellar environment play a crucial role in the formation mechanisms that give rise to planets. The molecular cloud material out of which both stars and planets form have some initial chemistry and as the star progresses in pre-MS evolution, stellar irradiation provides heat to the surrounding environment, providing pathways for chemical evolution. X-rays in particular can affect abundances of molecules [e.g., H2O, CO2, OH, CH4, HCN, NH3; 116]. High spatial resolution and sensitive instruments in the mm/submm regime like the Atacama Large Millimeter Array (ALMA) have revolutionized our understanding of circumstellar disk morphology and chemistry over the past decade. ALMA has imaged and spatially resolved hundreds of nearby circumstellar disks due to its sensitivity to the thermal emission of cm-sized dust grains. ALMA has also imaged disks in molecular hydrogen tracers such as isotopologues of carbon monoxide (CO) and a host of other molecular emission lines, demonstrating that disks are chemically and morphologically complex [7].

Many of the chemicals identified using mm/submm observations in circumstellar disks contain oxygen, a critical component to the formation of planets [116,118]. For example, ALMA studies of 1-10

Myr disks show that some are one to two orders of magnitude underabundant in gas-phase CO compared to that of the ISM [180]. One theory is that the CO gas is frozen out onto dust grains at large distances from the star or deep in the disk, where the environment is cold.

Young pre-MS stars are ubiquitous X-ray sources due to stellar dynamos generated by interior convection and fast rotation periods during youth [70]. Such dynamos heat coronae to millions of degrees K, where they generate X-rays. Stellar coronae can host several plasma components of different temperatures, with a typical kT $\sim$ 1 keV, which emits from $\sim$0.3 - 8.0 keV. While soft ($\lesssim$ 2 keV) photons are readily absorbed by intervening material (e.g., circumstellar disks, molecular clouds, ISM), hard X-rays (> 2 keV) pierce through dense material allowing for their detection even for very embedded objects. The primary sources of soft X-ray absorption are metals C, N, and O, which have photoionization cross sections < 1 keV [172]. In particular, oxygen has a strong K-edge absorption feature at 0.54 keV. Depending on the viewing geometry of the circumstellar disk to our line of sight, disk-absorption of X-rays from the central star can be used to constrain the line-of-sight abundance of oxygen (Figure 4), the 3rd most abundant atom in the galaxy and a vital element in water, rocks, and simple organics like CO, H2CO, CH3OH—and thus critical for the chemical evolution of disks, planets, and life as we know it [10,51,63].

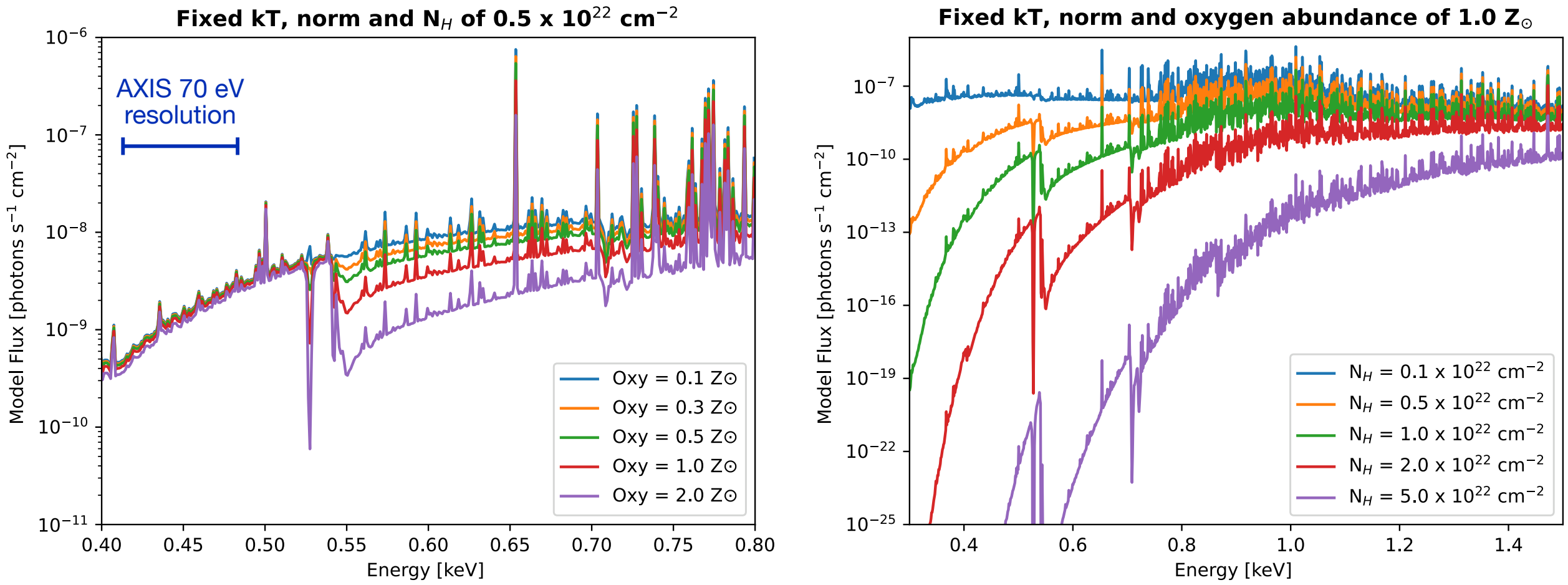

**Figure 4.** One temperature thermal plasma models (kT = 1keV and norm = 1E-4) with intervening absorption (tbabs*apec) for a variety of conditions. Left: Models with fixed NH but varying levels of oxygen abundance showing strong absorption at $\sim$0.54 keV for O abundances 0.3 Solar or greater. Right: Models with fixed oxygen abundance but varying levels of NH demonstrating that the 0.54 keV feature is visible with NH$\gtrsim$ 0.5 x $10^{22}$ cm$^{-2}$ but overall soft flux drops off rapidly with NH$\gtrsim$1x$10^{22}$ cm$^{-2}$. Therefore, the best targets to constrain O would have NH values between $\sim$0.5-1 x $10^{22}$ cm$^{-2}$.

We propose a 150 ks observation of the Orion Nebula Cluster, one of the closest large sites of star formation ( 400 pc). Over the past 25 years, the ONC has been observed for $\sim$1 Ms with Chandra ACIS-I imaging [67] and $\sim$2 Ms with ACIS-S and the HETG [152]. It is a rich field of $\sim$1600 pre-MS X-ray-emitting young stars. The ONC is a young stellar cluster and thus the majority of its stars still host circumstellar disks randomly oriented to our line of sight. Several high spatial resolution radio observations with ALMA programs have spatially resolved many circumstellar disks to obtain disk inclinations (orientations), mass, and radii, producing a spectacular dataset for follow-up observations with the high sensitivity and PSF of AXIS.

**Exposure time (ks): 250**

**Observing description:** We propose to utilize AXIS's excellent spatial resolution and stable PSF to obtain spectra for $\sim$1600 pre-MS stars in the Orion Nebula Cluster (ONC; Fig. 5) in a single pointing of 150 ks.

To determine the feasibility of the program, we simulated a 150 ks absorbed three-temperature plasma spectrum using AXIS responses ($N_H$ = 0.5×10$^{22}$ cm$^{-2}$, O = 2.0 Solar, kT1 = 0.5 keV, kT2 = 1.0 keV, kT3 = 1.6 keV, plasma abundance = 0.6 Solar, norm1 = 3.5×10$^{-5}$, norm2 = 8×10$^{-5}$, and norm3 = 3.5×10$^{-5}$). The unconvolved model is shown in the bottom panel of Figure 6. Since we will not know the true multi-temperature distribution of a young star when we observe it, we fit the simulated spectrum that was generated using a three-temperature model with a model that contains only two temperature components. The goal was to ensure we could still robustly constrain oxygen abundance even if we were fitting the model with fewer temperatures.

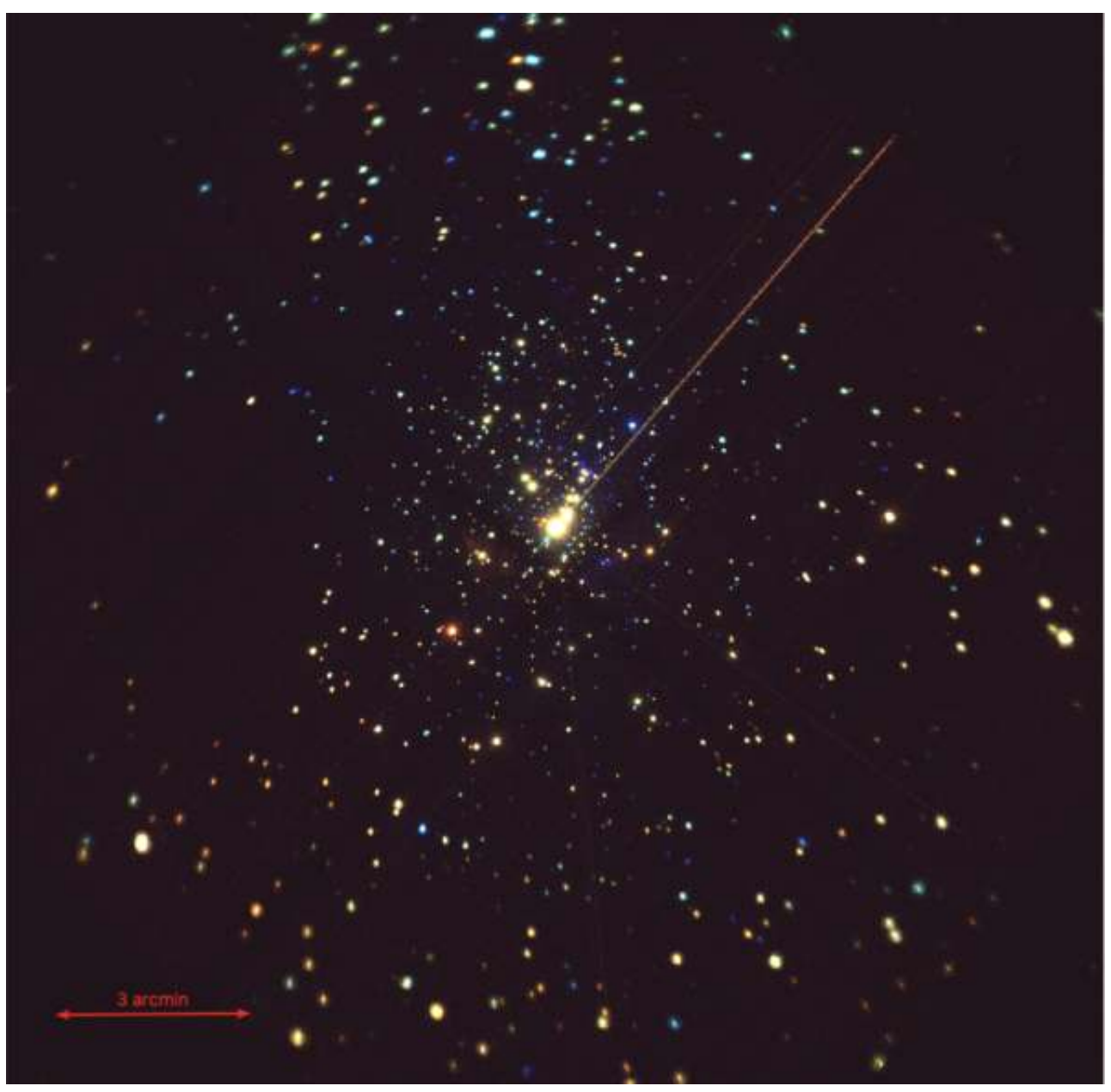

**Figure 5.** A deep three-color∼1 Ms merged ACIS-I Chandra observation of the Orion Nebula Cluster (ONC) using archival data. The ONC is an ideal stellar cluster for AXIS due to its target-rich environment with over 1600 young pre-MS stars in a ∼16x16 arcmin field of view [67]. Many circumstellar disks in the cluster have been spatially resolved with high resolution instruments like ALMA and the average $N_H$ for ONC sources is ∼0.5 x 10$^{22}$ cm$^{-2}$ [Fig 8 in 55] which is sufficient to constrain line-of-sight oxygen abundances (Figs 4 and 6).

The simulated data and the two temperature model fits can be seen in the top panel of Figure 6. The reduction in flux between 0.54-0.7 keV is a result of the oxygen absorption edge region due to intervening gas and dust (Fig. 4). The 'true' oxygen abundance of the simulated data was 2.0 Solar, and the fit obtained an abundance of 1.88 [-0.21 +0.21] (90% confidence bounds). The other parameter values were also within the errors of their true values. The observed (absorbed) flux of the source is 6.5 $\times$ 10$^{-14}$ erg s$^{-1}$ cm$^{-2}$ from 0.3-8.0 keV which translates to an intrinsic X-ray luminosity at the distance of the ONC of 5 $\times$ 10$^{30}$ erg s$^{-1}$. The luminosity distribution of young stars in the ONC [Fig 5 in 55] demonstrate that there are more than $\sim$ 80 sources observable with AXIS that would produce at least similar counts as the simulated spectrum.

While AXIS is not expected to spectrally resolve the oxygen edge, the flux decrease due to changes in the oxygen abundance of the absorber remains significant when using an X-ray instrument with high soft sensitivity, such as AXIS. This feature is not observable with Chandra or XMM due to their effective areas at this energy.

**Specify critical AXIS Capabilities:** Young, pre-MS stars form in groups that often require high angular resolution ($\lesssim$ 2 arcsec) to spatially resolve multiple members. Star-forming regions extend over large portions of the sky so efficiently detecting X-rays from multiple pre-MS stars require a stable PSF over the large AXIS field of view. The oxygen edge feature at 0.54 keV requires the high effective area of AXIS at soft energies ($\gtrsim$ 0.4 keV).

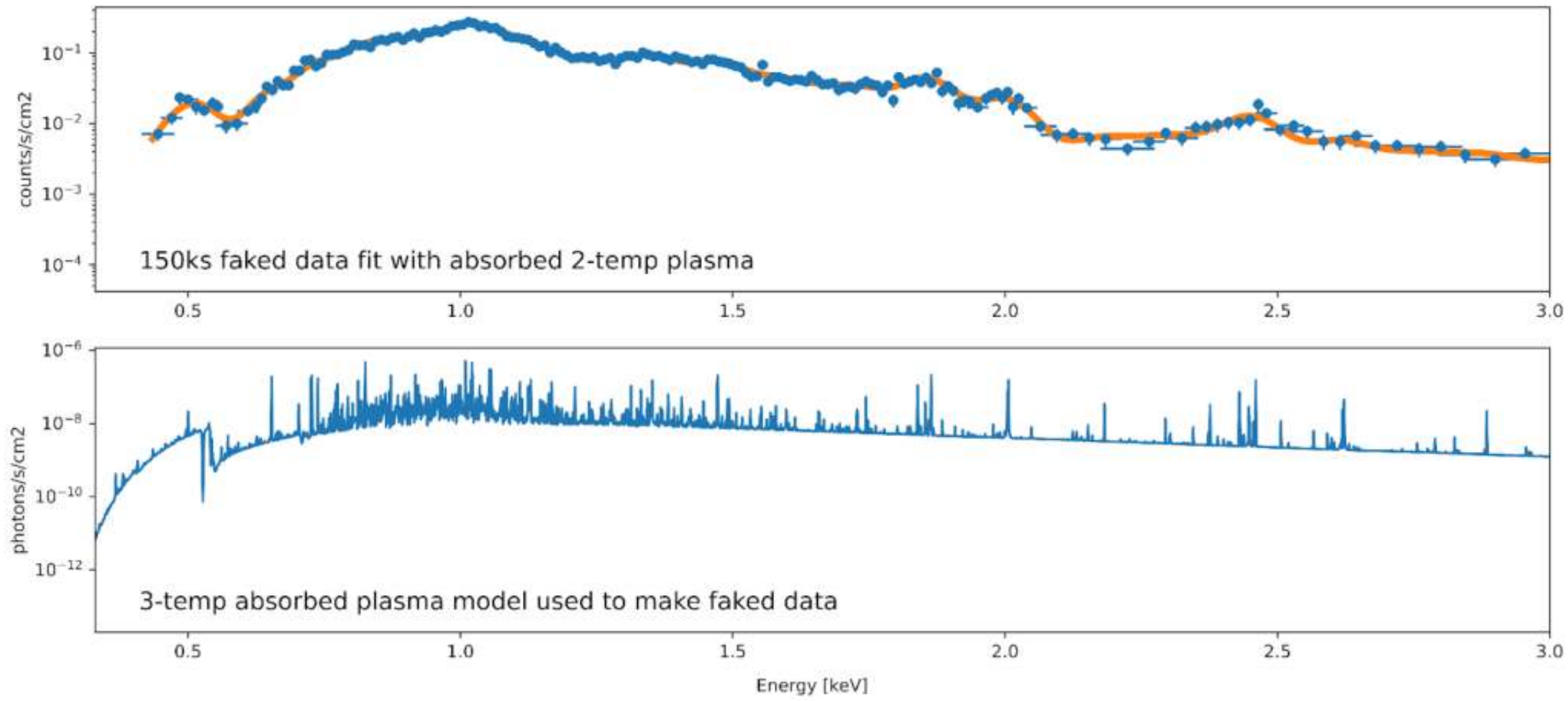

**Figure 6.** A 150 ks simulated spectrum (top) of a typical pre-MS stellar coronae with an absorption component that includes oxygen abundance. The 'dip' associated with the oxygen edge is clearly seen in the simulated spectrum (top) and the unconvolved model at 0.54 keV (bottom).

**Joint Observations and synergies with other observatories in the 2030s:** Strong constraints on circumstellar disk inclination (orientation) are necessary to interpret the line-of-sight oxygen abundance feature detectable with AXIS. While modern instruments such as ALMA, VLA, and VLT-SPHERE have produced large catalogs of spatially resolved disks with such inclination constraints, telescopes planned to operate in the 2030s, such as ALMA, WFIRST, ngVLA, and the ELT, will increase the number of potential AXIS targets by hundreds. Telescopes with spectroscopic capabilities, such as ALMA, ngVLA, JWST, and the ELT, which are capable of detecting circumstellar disk chemistry, will provide opportunities with AXIS to create chemical profiles of individual circumstellar disks and search for temporal evolution.

**Special Requirements:** None

**b. Massive stars and binaries**

*4. An AXIS Look at the Massive Star Population of our Closest Neighbour*

**Science Area:** Stars & Exoplanets

**First Author:** Yaël Nazé, University of Liège, Belgium, ynaze@uliege.be

**Co-authors:** Lidia M. Oskinova (Potsdam University, Germany), Michael F. Corcoran (Catholic University of America, USA), Jeremy J. Drake (SAO, Lockheed Martin, USA)

**Abstract:** A dedicated survey of the main clusters in the LMC with AXIS will, for the first time, precisely characterize the X-ray properties of massive stars at intermediate metallicity. These observations will allow a detailed study of feedback and interactions, shedding new light on massive star evolution.

**Science:** Massive stars are true cosmic engines. In only a few million years, these stars emerge from molecular clouds and end their lives in supernova explosions. In between, these hot stars eject an intense flow of ionizing radiation and dense stellar winds. As most massive stars reside in multiple systems, binary interactions frequently occur, resulting in additional ejections and a range of mass-transfer phenomena. Additionally, the remnants themselves are numerous: massive stars are thus the progenitors of X-ray binaries and double compact systems (hence gravitational wave events). Despite their brief lives, massive stars therefore impact their galactic environment in a tremendous way and their study is thus in line with several main AXIS science themes (progenitors of core-collapse supernovae and black holes, influence on their surroundings - including low-mass stars with planets, galactic feedback).

Over the past few decades, a large observational effort led to the characterization of the X-ray emission from massive stars in the Milky Way. Several sources of high-energy emission have been identified.

First, there are embedded wind shocks. The pressure of intense UV radiation fields in these stars acting on metal line opacity "lifts" their outer layers and accelerates them, generating a dense and fast (1% of c) outflow. This process is unstable, leading to shocks between wind clumps which in turn generate X-rays. This high-energy emission has been demonstrated to follow a tight, so-called "canonical", relationship : $L_X/L_{BOL} \sim 10^{-7}$ [108,122,140]. Small-amplitude modulations around this value have been recorded due to large-scale structures in the winds of some massive stars (e.g. $\zeta$ Pup - Nichols et al. 114, EZ CMa - Huenemoerder et al. 80). Such an emission provides invaluable information on a realistic understanding of wind-driven mass loss, shedding light on an important feedback ingredient.

X-ray emission can also arise in magnetically confined stellar winds. Large–scale magnetic fields have been detected in about 10% of massive stars. These fields can channel the winds towards the magnetic equator, leading to shocks which can produce additional X-ray emissions. Such X-ray bright confined winds are tightly linked to the magnetospheric properties [110]: they allow us to probe the processes modifying the feedback in massive stars (e.g., magnetospheric ejection or fall-back) and stellar rotation.

Finally, most, if not all, massive stars lie in multiple systems. The presence of two strong stellar winds leads to wind-wind collisions, which also generate X-ray emission (for a review, see Rauw 138). This additional emission varies in a phase-locked manner, as the separation changes between stars (in eccentric systems) and/or as the absorption along the line-of-sight varies (as the collision is alternatively seen through one wind or the other). Such variations thus directly probe the wind strengths.

Additional X-rays are also seen in other massive binaries, e.g., nova-like emissions in Be+WD systems (e.g. Gaudin et al. 64). The mere existence of such systems can only be explained if mass transfer has occurred in the past; therefore, they constitute important probes of the binary interaction processes that are now thought to be widespread.

The various X-ray emissions sensitively depend on the radiatively driven winds and the properties of the winds (mass loss rates, wind velocities) depend sensitively on metallicity. Currently, nearly all the detailed X-ray studies of high-mass stars were performed at near-solar metallicity. An X-ray survey of massive stars in the LMC provides a critical test of wind properties at low metallicity, and provides a critical test of wind feedback models. This study will extend existing massive star X-ray surveys to more numerous, lower mass OB stars and provide important constraints on massive star wind-driven feedback in the young, low-metallicity Universe.

In the last few decades, several optical and UV investigations of massive stars in the LMC have been undertaken. The massive star population of e.g. the Tarantula and N11 regions is now well known: spectral types, luminosities, temperatures, masses, and multiplicity are precisely constrained. However, the key to understanding the feedback and evolution of these systems is to evaluate the level of high-energy emission. X-ray surveys provide an efficient way to study massive star winds in the LMC, since X-rays serve as a proxy for wind properties in single, binary, and magnetized massive star systems.

There were several preliminary X-ray investigations of massive stars in the LMC, notably in N11 and Tarantula Nebula [40,111]. The limited sensitivity only allowed general conclusions to be reached. In particular, the canonical X-ray–luminosity relationship seems to be followed for LMC stars, although the uncertainty on the relationship exponent is no better than 0.65 dex when combining all sources. Clearly, the increased sensitivity of AXIS, and its excellent wide-field spatial resolution, will allow for the first time to precisely determine individual X-ray properties for large numbers of massive stars, thereby improving global measures by at least an order of magnitude. Targeting clusters allows to gather information at once for a whole set of massive stars, so that all their various X-ray emission processes will be observed.
**Exposure time (ks): 900**
**Observing description:** In 300ks, AXIS reaches a flux of $3 \times 10^{-17}$erg cm$^{-2}$s$^{-1}$, which corresponds to Lx of $10^{31}$erg s$^{-1}$ at the LMC distance. For the canonical relationship, this X-ray limit corresponds to being able to detect all massive stars down to early B V stars. Of course, many massive stars are brighter in the X-ray range, due to e.g. colliding winds or magnetic phenomena which often increase the X-ray flux by one order of magnitude. This ensures an in-depth study of any LMC population of massive stars.

We propose to observe three main massive star clusters in the LMC: the Tarentula Nebula, N11, and N44, resulting in a total exposure time of 900 ks. The observation should be split into several (up to 10) exposures, scattered over a year, to allow for the variability study of the brightest objects.
**Specify critical AXIS Capabilities:**

1. The angular resolution is of paramount importance to resolve stars, diffuse gas, and supernova remnants in star-forming regions of the MCs.
2. Low background is very important to achieve detections of faint soft sources, such as OB stars in the SMC.
3. The soft response of AXIS is perfectly suited for X-ray detections of massive stars and their associated diffuse emission.
4. The large AXIS's FOV is necessary to achieve the detections of hundreds of stars in a single snapshot, as required by our science objectives. Furthermore, the FOV enables the discovery and study of diffuse emission, supernova remnants, and X-ray binaries - i.e., the entire ecosystems of young massive star-forming regions.
5. The spectral resolution is sufficient to establish the temperature of the hot plasma typically observed in massive stars.

**Joint Observations and synergies with other observatories in the 2030s:** The LMC clusters have already been extensively studied, but dedicated, complementary observations of specific systems might be achieved with ELT and NewAthena.
**Special Requirements:** None

*5. Determining X-ray Properties Of Massive Stars In The Small Magellanic Cloud Galaxy*

**Science Area: Stars and Exoplanets**
**First Author:** Lidia M. Oskinova, Potsdam University, Germany, *lida@astro.physik.uni-potsdam.de*
**Co-authors:** Yaël Nazé (Liége University, Belgium), Mike Corcoran (Catholic University of America, USA), Jeremy J. Drake (Lockheed Martin, USA)
**Abstract:** Massive stars of OB and WR spectral types are dominant sources of stellar feedback in low-metallicity star-forming galaxies such as those detected by the JWST at high redshifts. Establishing properties of massive stars and their powerful stellar winds in low-metallicity galaxies is one of the most pressing questions in modern astrophysics. While there is indirect evidence for the presence of X-rays in winds OB and WR stars at metallicities lower than 20% solar, so far X-ray detections are limited to a handful of peculiar systems. The closest galaxy with such low metallicity is the Small Magellanic Cloud (SMC). We propose AXIS observations of the two youngest and most massive star-forming regions in the SMC: the giant H II region N66 and the supergiant shell SGS 1. These regions contain many hundreds of OB stars, the majority of which are congregated in two dense star clusters, NGC 346 and NGC 602. The field of view, angular resolution, and sensitivity of AXIS for soft X-rays make it the only telescope suitable for the proposed pioneering study of X-rays from low-metallicity massive stars. With 2 Ms of exposure time, we will detect X-rays from hundreds of OB stars spanning a broad range of masses and evolutionary stages, as well as hot bubbles around star clusters and supernova remnants, unveiling the high-energy properties of entire ecosystems in low-metallicity starbursts. These observations will open a new page in understanding massive stars, their stellar winds, and their feedback in low-metallicity galaxies, and in young Universe in general.

**Science:** Stars that are born with masses exceeding $10\,M_\odot$ start their lives on the main sequence with OB spectral types. The majority of massive stars end their lives by becoming cool red supergiants (RSG), exploding as supernova type SN II, and producing neutron stars. However, very massive stars and those in close binaries evolve into very hot, hydrogen-depleted, Wolf-Rayet (WR) type stars which explode as supernovae type SN Ibc or gamma-ray bursts, and produce black holes.

Massive stars are rapidly evolving on a timescale of millions of years. During most of their evolution, they are hot, UV bright, and possess strong stellar winds. Stellar winds expand with velocities $\sim 1\%$ of the speed of light, remove stellar envelopes, and thus strongly influence stellar evolution and masses of resulting black holes. Stellar winds are radiatively driven by scattering of UV photons in resonance lines of metal ions [32,130]. Ionization of winds by X-rays plays an important role in establishing wind temperature and ionization structure and strongly affects the metal ions responsible for wind driving [16]. Hence, both metallicity and X-rays plays a key role in governing the strengths of stellar winds.

Among the most significant open questions in modern astrophysics is establishing the properties of massive stars in low-metallicity galaxies. This is necessary to understand the first galaxies being observed by JWST, the mass spectrum of black holes provided by gravitational wave observatories, and the rich phenomenology of high-redshift supernovae and gamma-ray bursts. To address these fundamental questions, new generations of X-ray telescopes, such as AXIS, are needed.

Extensive X-ray observations of massive stars and young massive star clusters in low-metallicity galaxies are addressing all key science questions driving the AXIS mission. Besides, X-ray observations play an outstanding role in uncovering the physics of massive stars and their stellar winds. The *Chandra* and *XMM-Newton* observatories made dramatic progress in unveiling the properties of X-ray emission of massive stars. It is established that OB and WR stars emit X-rays, which are thought to be generated by the same mechanisms responsible for stellar wind driving. Typical X-ray luminosities (in 0.2– 12 keV energy band) of massive stars are in the range $10^{30} - 10^{32}$ erg s$^{-1}$. X-ray spectra reveal that stellar winds are permeated by hot thermal plasma with temperatures of a few MK. X-ray light curves show mild variability associated with stellar rotation. These findings strongly question current stellar wind models and established theories and motivate their further development [112,123].

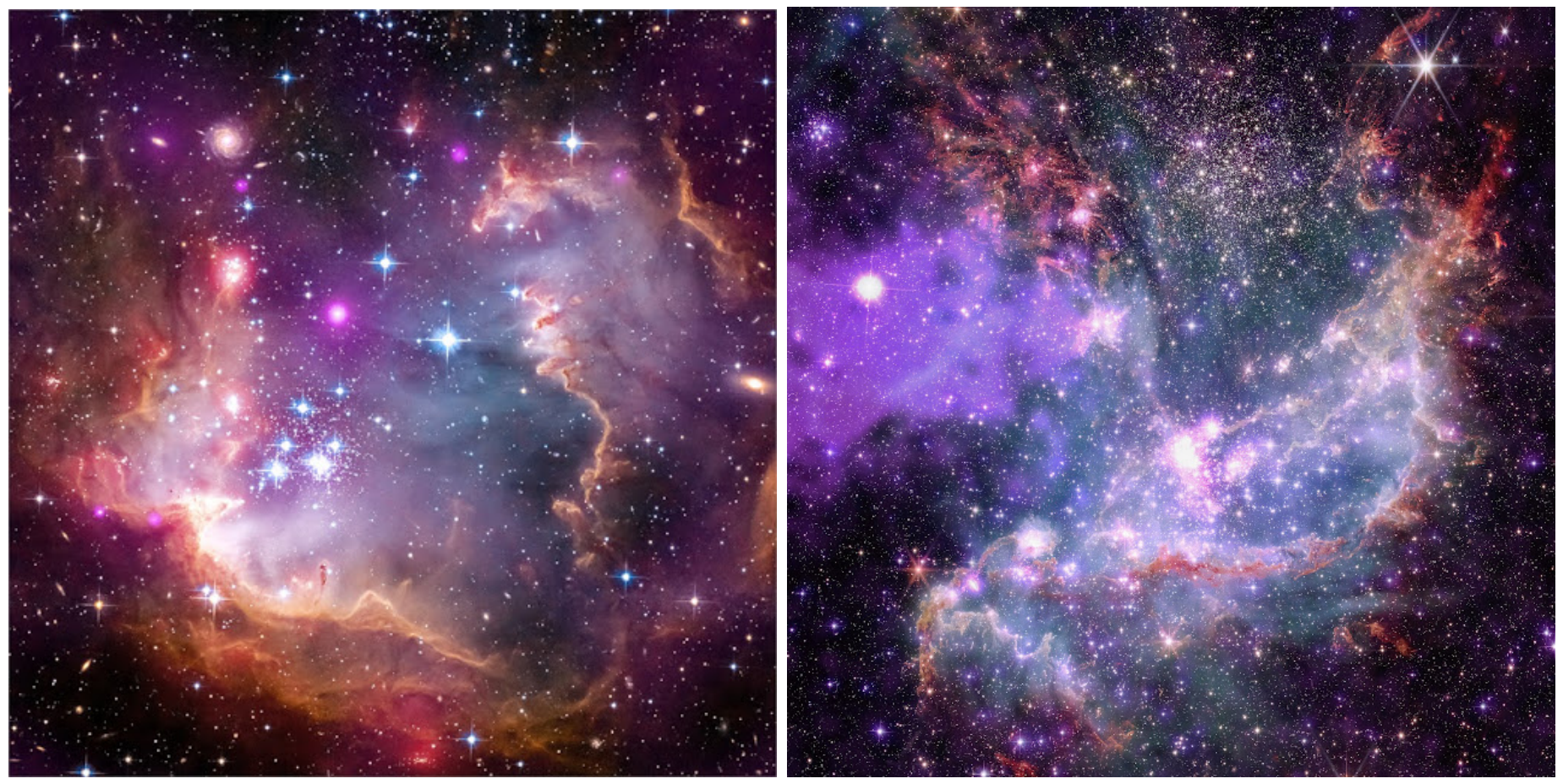

**Figure 7.** Color composite images of proposed targets: *Chandra*'s ACIS-I X-ray images are in purple; blue and green show the optical *HST* images; the *Spitzer* and *JWST* images in near infrared are in red. On the left is the cluster NGC 602, while the cluster NGC 346 is shown on the right. Image sizes are $3.5' \times 3.5'$ ($62 \times 62$ pc at $d = 61$ kpc). Point-like X-ray sources are background AGNs and a colliding wind binary in the vicinity of NGC 346. Diffuse X-ray emission is associated with unresolved stars and supernova remnants. No individual massive OB and WR stars (clearly seen in the optical) have been detected in X-rays so far (Credit: ESA/NASA/STScI).

Current X-ray observations primarily target massive stars in our own Galaxy. This is because even modern X-ray telescopes are not sufficiently powerful to measure X-rays from individual stars in other galaxies. The deepest observation of a low-metallicity star cluster so far is a 1 Ms Chandra's study of the 30 Dor region in the Large Magellanic Cloud (LMC) galaxy. Chandra showed that X-ray properties of OB stars in the LMC are similar to those in our own Milky Way [40]. However, 30 Dor region is polluted by supernovae and its metallicity is only slightly lower than solar. Some massive binaries, where the winds from both components collide, producing bright X-rays, have been detected in the LMC; however, the emerging picture of the high-energy properties of low-metallicity massive stars is still highly incomplete.

The nearest galaxy with the metallicity of only 20% of solar (i.e. comparable to the average metallicity of the Universe at cosmic noon, $z \sim 2$ is the SMC. Previous Chandra and XMM-Newton observations and surveys of the SMC detected only a couple of brightest stars in this galaxy (Fig. 1) [109,124].

Given the exceptional importance of understanding low-metallicity massive stars, the Hubble Space Telescope's largest-ever science program, ULLYSES, was devoted to collecting UV spectra of OB stars and WR stars in the LMC and SMC. These UV spectra display lines of high ions, such as N V and O VI, which could be produced only by X-ray photoionization. Thus, while there is convincing indirect evidence for the presence of X-rays in the winds of low-metallicity massive stars, they have never been detected, and the properties of their X-ray emission are not known.

The combination of high angular resolution, a broad field of view, and sensitivity in the soft part of the X-ray spectrum makes AXIS pivotal for the first-ever detection and study of low-metallicity massive stars. The key science goals of these observations are (1) to establish the dependence of X-ray luminosities of OB and WR stars on stellar type; (2) to determine the temperature of hot plasma filling stellar winds by means of X-ray spectroscopy; (3) to use X-ray emission to empirically measure mass-loss rates of these stars. To achieve these science goals X-ray detections of hundreds of OB and WRs is necessary.
**Exposure time (ks): 2000**

**Observing description:** To address these important science goals and revolutionize our understanding of metal-poor massive stars, we propose pioneering AXIS observations of two fields in the SMC galaxy containing the young massive star clusters NGC 346 and NGC 602 (in the order of priority, Fig. 1).

NGC 346: More than half of all OB-type stars in the SMC galaxy are congregated in the N66 star-forming region and the NGC 346 cluster (2-4 Myr old), which lies at its heart. XMM-Newton detected X-ray emission from this region but was unable to resolve the cluster, whereas Chandra is not sufficiently sensitive. The low metallicity, large stellar content, and supplementary observations by HST and JWST make this star-forming region a primary target for AXIS.

NGC 602: The cluster is at the rim of the only supergiant shell in the SMC galaxy, SGS 1, located in the Wing of the SMC. This area has even lower metallicity than N66 region. SGS 1 contains dozens of 1-2 Myr old star clusters, numerous OB stars, and a WR star. The previous deep Chandra observations (300 ks) succeeded in detecting only unresolved subclusters of young stars in this region. The large field–of-view of AXIS is perfectly suited to study massive stars in the Wing of the SMC.

Two objectives drive the exposure time estimates: to measure the X-ray flux and to establish the spectral shape. We adopt the limiting source flux for a detection at $10^{-17}\,\mathrm{erg\,s^{-1}\,cm^{-2}}$, corresponding to the $L_X = 10^{31}\,\mathrm{erg\,s^{-1}}$ and allowing detection at $3\sigma$ level of all O-type stars, and the earliest type B stars. The line-of-sight absorption is very low and is known from the UV. Thus, we request 1 Ms exposure time for each field, i.e. 2 Ms time.

**Specify critical AXIS Capabilities**

1. The large AXIS's FOV is necessary to achieve the detections of hundreds of stars, as required by our science objectives. Furthermore, the FOV enables the discoveries and studies of diffuse emission, supernova remnants, and X-ray binaries - i.e., the whole ecosystems of young massive star-forming regions.
2. The angular resolution is of paramount importance to resolve stars, diffuse gas, and supernova remnants in star forming regions.
3. Low background is very important to achieve detections of faint soft sources, such as OB stars in the SMC. (4) The soft response of AXIS is perfectly suited for X-ray detections of massive stars and diffuse emission.
4. The spectral resolution is sufficient to establish the temperature of hot plasma.

**Joint Observations and synergies with other observatories in the 2030s:** To obtain the first picture of high-energy properties of low-metallicity massive stars, the synergies with ELT (spectroscopy of stars), LSST (transients), and SKA (diffuse emission and star formation) are extremely helpful. newAthena will not resolve dense star clusters. The synergies with newAthena will allow us to strongly improve our understanding of massive star feedback on galactic scales. Determining realistic X-ray properties of low-metallicty massive stars is an important synergy with the JWST studies of high-redshift metal-poor galaxies.

**Special Requirements:** None

*6. Superflares on Giant stars*

**Science Area:** Stars and Exoplanets

**First Author:** Hans Moritz Günther (MIT, hgunther@mit.edu)

**Co-authors:** Girish M. Duvvuri (Vanderbilt University)

**Abstract:** Late-type stars have convective envelopes and the convective motion in these envelopes can generate a large-scale magnetic field. This is the dynamo mechanism active in our Sun, as well as in low-mass young stars, cool main sequence stars, and also cool giants. Differential rotation can cause the field lines to twist until they reconnect and release the stored magnetic energy, heating the plasma to extremely high temperatures. We observe the cooling of this plasma as soft X-rays. Since the vast majority of stars in the universe are cool, understanding how flares work, how the plasma cools, and how it falls back onto the star or how it interacts with any other matter in the system (binary star components, exoplanets, primordial disks, dust etc.) is needed to count, model, or simulate the evolution of most of the stars in the universe. In main-sequence stars, flares typically last only hours, and their brightness is limited. Giant stars with their larger size and much deeper convection zone can support flares that last for several weeks; thus, we can study the cooling processes in much more detail there. We suggest a science program that is centered on TOO observations of giant flares from giant stars. With additional preparatory work, those observations could be triggered by optical white-light flares from giants, as discovered, for example, with Rubin, or by X-ray flares detected with all-sky monitors. AXIS's large collecting area makes it possible to obtain a series of well-resolved spectra to fit individual temperature components, their abundances, and their evolution with time. A fast TOO response is only needed for the initial observations at the flare peak; later observations can follow the cooling with observations every few days. So far, only a single superflare has been observed with sufficient signal to differentiate the evolution of different temperature components. With AXIS, we can gain a statistical sample of the largest stellar flares in the universe to test theories of element fractionation in flares, cooling flow models, and ideas for the shape of stellar magnetic fields.

**Science:**

Stars of different spectral types produce X-ray emission through different mechanisms. Cool stars have convective envelopes, and the turnover of plasma in these envelopes generates magnetic fields that rise to the surface. Differential rotation can twist field lines and add more magnetic energy. Those magnetic fields thread the stellar corona, an optically thin layer of hot (several MK) gas that can reach out to several stellar radii for very active stars. In general, stars with faster rotation and faster turnover time will generate stronger magnetic fields. The processes that heat the corona are not fully understood, even in our Sun, where spacecraft such as the Parker Solar probe can take in situ measurements. Magnetic waves might play a role, indicated by the fact the there is an abundance difference between corona and photosphere: On star with low activity levels, elements with high first ionization potential (FIP) such as Ne are depleted in the corona compared to the photosphere; in stars of higher activity level the opposite is true (the "Inverse FIP" or IFIP effect). The corona is not only steadily heated, but can also erupt in powerful flare-outbursts. The frequency of flares generally follows a power-law with an exponent close to -2, where more powerful flares are exceedingly rare [see reviews by 28,71] . Parker [129] first described the basic elements of a flare: The flare's energy ultimately comes from the magnetic field. When two of the twisted field lines cross, they can reconnect and release the magnetic energy, accelerating electrons to non-thermal speeds. Those electrons emit in the radio and impact the stellar surface, where they cause hard X-ray Bremsstrahlung, heating plasma in the photosphere that streams upward, funneled by the magnetic field, and fills the magnetic loop. A shock wave heats that plasma, which begins to cool through soft X-ray radiation. The soft X-rays are thus delayed with respect to the radio and hard X-rays [113], but they also last much longer as we can observe their cooling time. From the peak temperature of the flare and the cooling timescale, we can estimate the physical size of the flare based on models [e.g. 85,141]. From the relationship between temperature and volume emission measure during the decay period, we can determine whether the flare is cooling radiatively or if it is continuously reheated [see review by 142]. Obviously, the brighter the flare, the more detailed models we can fit for the emission of the cooling plasma.

Observations of the brightest flares offer us an opportunity to study the dynamic evolution of the plasma in the corona in terms of temperature and abundances (grating data would also reveal densities, but that is not critical for the interpretation). Flares can have an immediate impact on other bodies in the stellar system: They can ionize circumstellar gas and dust, cumulatively strip away planetary atmospheres, and sterilize planetary surfaces. In advanced civilizations such as ours, large flares and solar storms can disrupt satellites, wireless communication, and collapse electric and communication networks. The famous Carrington event of 1859 is the strongest solar storm recorded with scientific instruments and rendered telegraph lines inoperable, despite being several orders of magnitude weaker than the superflares observed on the most active stars [37,78,85]. Flares on main-sequence stars typically last up to a few hours, but on more evolved giant stars with their much larger radii and thus more extended convection zones, the flares have been observed to cool down over several weeks [see list compiled from literature by 85].

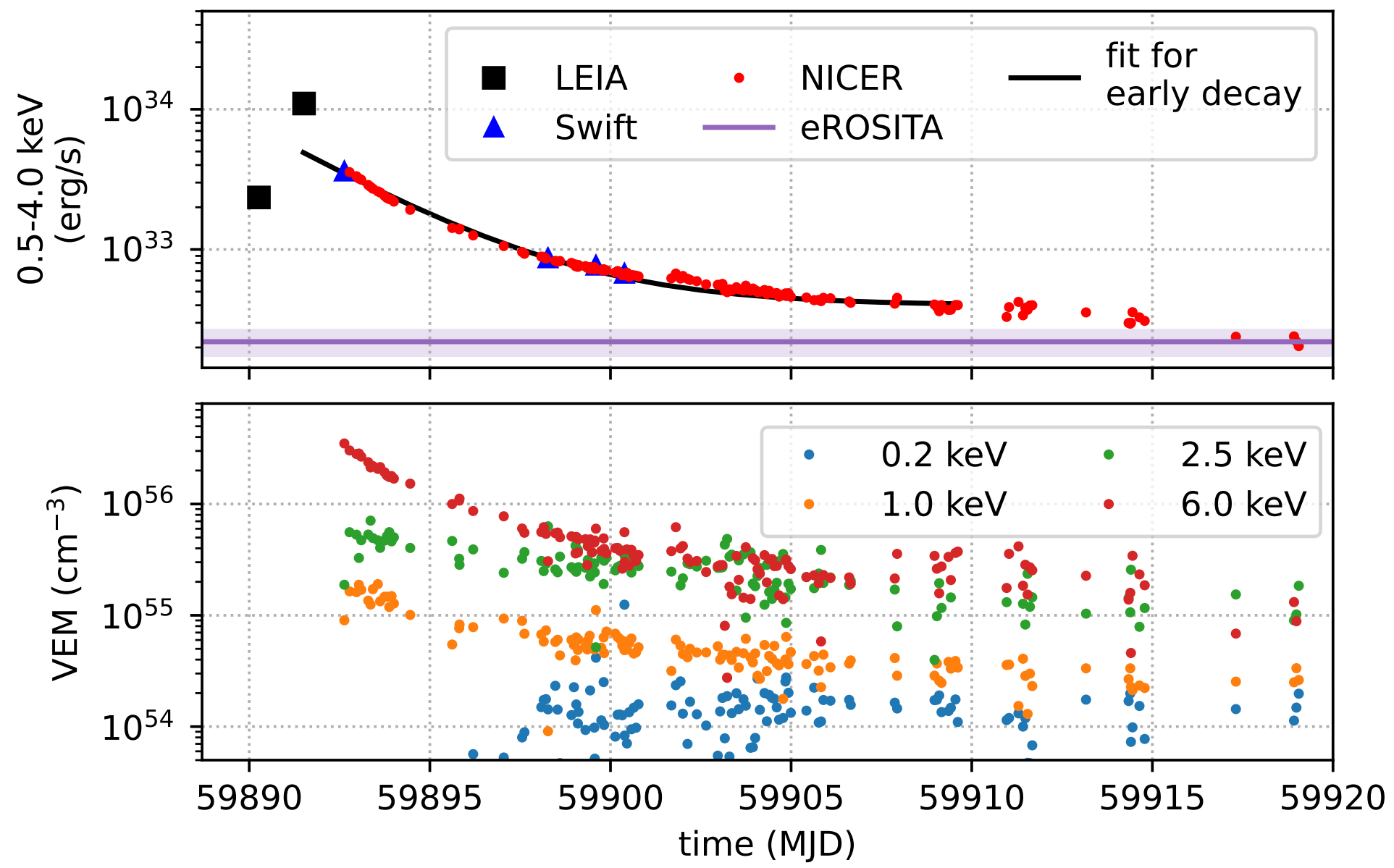

**Figure 8.** Lightcurve for HD 251108. Where error bars are invisible, the statistical uncertainties are smaller than the plot symbols. Top: absorbed X-ray flux in the 0.5–4.0 keV range. The average flux in eRASS is marked by a purple line, and the observed variability (shaded) is also consistent with the XMM-Newton and ROSAT fluxes. The black line shows the best-fit exponential decay fitted to the early decay phase. Bottom: Volume emission measure for a fit with four components at fixed temperature and abundance. The hotter components decay faster, while the 0.2 keV component is mostly stable. Error bars are omitted for clarity. Data replotted from Günther et al. [78]

Many giant stars are in binaries (RS CVn stars), but flares have also been observed in apparently single stars [121]. Either way, flares on cool subgiants or giants offer us an opportunity to study the physics of flare evolution in unprecedented spectral and temporal resolution. Fig. 8 shows the best X-ray data from a giant flare that was ever observed. This figure is based on over 100 NICER observations with tens of thousands of counts in most of the spectra, providing the data to study the evolution of several plasma temperature components independently.

Typically, the flare emission decays exponentially. While secondary flares and changes in heating can occur, the science does not require a time sampling as dense as shown in Fig. 8. On the other hand, it is important to obtain data near the peak of the flare. AXIS' fast TOO response time enables observations that can catch the flare peak, and then sample flexibly every few days in the decay phase. With tens of thousands of counts per observation, we can reliably fit multi-temperature, multi-abundance models to study the structure of large flares in unprecedented detail to answer the following question:

- How does flare plasma cool? The theoretical models of Reale [141] provide diagnostics in terms of a single temperature component, as more detailed observations were not feasible at the time. With bright super-flares and AXIS large collecting area, we can perform much more detailed tests of the cooling behavior and test if existing simulations actually describe flare cooling satisfactorily.

- Do the abundances change in flares compared to the quiescent corona? Favata & Schmitt [54] claim to see changes in a super-flare on Algol, while Günther et al. [78] do not detect changes in HD 251108. Because abundances are often degenerate with the temperature structure of a plasma, well-exposed spectra are required to answer this question. The mechanism by which elements are enhanced or depleted in the corona compared to the photosphere is still under discussion, but is likely related to some type of magnetic wave [see, e.g. 106, for recent work]. Knowing if flares impact abundances can constrain those scenarios.
- Is plasma re-heated during the cooling phase? Observations indicate that this is sometimes the case, but to understand how this process is controlled, we need to correlate X-ray observations with tracers of the chromosphere and photosphere, such as the location of stellar spots. The extreme brightness of giants makes optical observations easy to obtain even with small telescopes.

As far as we know, large and small flares all share the same physical mechanisms for energy release, shock heating, cooling, and interaction with the transition region and the photosphere. Everything we learn about how flares work, should thus also apply to the smaller and much more frequent flares on pre-main sequence and main-sequence cool stars, where the shorter flare duration and luminosity limits the signal and thus the detail that can be extracted from spectral and temporal modeling with models such as the simulations by Reale [141] or the RADYN/F-CHROMA models by Carlsson et al. [31].
**Exposure time (ks):** 300 ks (5 targets with 60 ks each, split into an initial 30 ks TOO, followed by $10 \times 3$ ks monitoring)
**Observing description:** Flares on giants can be detected in the optical (e.g. with Rubin) or with all-sky X-ray monitors. Once a flare on a giant above certain threshold flux is detected, AXIS will be triggered to slew onto the target to catch the rise phase of the flare, which we expect to be of order 1 ks, based on scaling the rise phases of main-sequence stars to the overall longer flare duration, however no resolved rise phase from an X-ray superflare on a giant is available in the literature. AXIS will then continue to observe the early cooling phase of the flare for about 30 ks. After that, *monitoring* with repeat observations every few days is sufficient.

We based the exposure time estimate on the superflare on HD 251108 observed with NICER by Günther et al. [78]. AXIS will have about 2-3 times the effective area of NICER. In the early phases of a super-flare, even 2-3 ks of observation will deliver tens of thousands of counts to allow fitting of a multi-temperature plasma model. In later phases, the count rate decreases, and thus the exposure time needs to increase. Thus, the total exposure time per target is set mostly by the cadence of observations, given AXIS overheads. For a flare with a flux similar to HD 251108, AXIS shall observe the rise (if triggered and slewed fast enough) and the early decay phase for about 30 ks. After that, repeat observations of 3 ks every few days are sufficient, giving a *total observing time per target of order 60 ks. A science program that triggers on five flares to obtain a statistical sample would thus need 300 ks.*

X-rays allow us to diagnose the hot plasma confined in the magnetic flare loop. However, flares also interact with features on the photosphere, such as spots and plages, in the transition region between the photosphere and the corona. Obtaining a comprehensive picture of plasma evolution also requires monitoring with optical/IR spectroscopy. Given the high flux of giant stars, small optical telescopes are sufficient for this purpose.

- TOO observations: Flares will be discovered by all-sky surveys, preferably in X-rays (but optical would also work). AXIS needs to slew onto target within 4 hours to track the rise phase, or within 12 hours to catch the peak of the flare.
- Flexible scheduling: Follow-up observations during the flare decay phase need to be scheduled every few days for a month after the flare trigger.

- Effective area: We know how light curves from flares evolve, but in order to fit separate temperature components and different abundances at the same time, we need of order 10000 counts per spectrum. For typical brightnesses and distances of super-giant flares, this requires an effective area at least an order of magnitude better than XMM-Newton or Chandra.

**Joint Observations and synergies with other observatories in the 2030s:** Rubin or X-ray all-sky surveys like MAXI or eROSITA will be used to trigger TOO
**Special Requirements:** This program critically requires fast TOO capabilities (<4 hrs).

*7. Probing High-Energy Phenomena in Colliding Wind Binaries within Massive Star Clusters with AXIS*

**Science Area:** Stars and Exoplanets
**First Author:** Pragati Pradhan, Embry-Riddle Aeronautical University, pradhanp@erau.edu
**Co-authors:** Sean Gunderson, MIT Kavli Institute; Kenji Hamaguchi, UMBC
**Abstract:** This proposal aims to utilize *AXIS*'s high spatial, spectral, and timing resolution to investigate the high-energy phenomena in colliding wind binaries within star clusters, such as Westerlund 1, the Tarantula Nebula, and Cyg OB2. By modeling their X-ray spectra with multi-temperature plasma models (APEC, VAPEC, BVAPEC), we will resolve soft X-ray excesses to explore radiative processes in dense stellar winds. AXIS will also enable detailed studies of short-term variability, shedding light on the interplay between local absorption and temperature fluctuations. Additionally, we will examine orbital phase-dependent emission patterns to refine our understanding of wind collision zones and their evolution, advancing our knowledge of these complex high-energy systems in star-forming regions.
**Science:** The life cycle of a massive star begins when a gigantic nebula collapses under its own gravity and nuclear fusion begins at its core. Such stars will burn at incredibly high temperatures and have masses that are more than ten times that of the Sun. They have short lifespans and end their lives in only a few million years through supernova explosions. Low-mass stars, like our Sun, have been around for billions of years. Most massive stars are created in binary systems, and the interactions between the two stars guide their stellar evolution. Owing to powerful winds colliding between the two massive stars, they are bright emitters of X-rays. X-ray emission from wind collisions in massive binaries, especially those with non-zero eccentricity, provides unique physical constraints on mass-loss parameters (velocities and densities) at specific distances from the star, and on how these parameters change with stellar separation and orbital phase. Strong X-ray emission has even been used to reveal "single" stars as binary systems, advancing our understanding of the role of companions in massive star evolution and the impact of duplicity on the final evolutionary end-product. See [139] for review.

X-ray emission in these systems is produced by a strong bow shock where the two fast ($v_\infty \gtrsim 1000$ km s$^{-1}$) radiatively driven winds from the stars collide in the space between them, producing hot gas at temperatures upward of $10^7$ K. In addition, some wind-wind collisions also produce dust at temperatures of $\sim$ 1000 K when the Wolf-Rayet (WR) star—a rare, evolved massive star that has burned all of the hydrogen in its core—is carbon-rich. These systems thus provide an astrophysical laboratory for studying cooling processes in shocked astrophysical plasmas and the formation of molecules. This means that understanding the X-ray emission from a variety of sources with known mass-loss rates and at varying separations is an important basis for interpreting X-ray emission in general.

With the knowledge of merely the two visual-binary WR colliding wind stars, WR 140 [159] and $\gamma^2$ Velorum [90], our detailed knowledge of high-energy X-ray emission from these sources is limited. The focus of this proposal is to measure X-ray emission from other colliding wind binary systems at various phases in their orbits. These X-rays provide crucial information about the density near the shock apex—where dust forms through electromagnetic interactions—and also reveal the relative wind velocities through temperature measurements. Therefore, the variation of X-ray emission with orbital phase provides a unique opportunity to constrain the physical environment of the stars, which can be done with *AXIS*.

The *AXIS* mission provides an unprecedented opportunity to probe the high-energy environments of massive stars within star clusters, such as Westerlund I, drawing inspiration from recent advancements in the field, such as the work of Anastasopoulou et al. [5]. Westerlund I, home to numerous massive stars—including peculiar objects like the colliding wind binary "A"—represents a compelling case study for investigating the physical processes governing these stellar systems (see Fig. 9). *AXIS* could target other rich clusters and associations, such as Cygnus OB2, the Tarantula Nebula, as well as eROSITA-identified fields, including iconic systems like $\eta$ Carinae, to establish a broader understanding of massive star X-ray populations.

This proposal leverages *AXIS*'s advanced spatial and spectral resolution to also resolve the soft X-ray excess observed in systems like $\gamma^2$ Velorum, providing deeper insights into the radiative processes occurring in dense stellar winds. Additionally, *AXIS*'s exceptional timing capabilities will enable precise studies of variability on short timescales, revealing how local absorption and temperature fluctuations affect both soft and hard X-ray emissions. Earlier studies with *Chandra* were limited due to restricted soft excess coverage below $\sim 2$ keV. We also propose investigating the orbital phase-dependent behavior of these systems to assess whether their emission patterns remain consistent over time and to identify the factors driving any observed deviations. Such analyses will enhance our understanding of wind collision zones and their evolution, shedding light on the mechanisms driving high-energy phenomena in these systems.

**Exposure time (ks): 250**

**Observing description:** Based on Chandra observations from the EWOCS project, which total approximately 1.1 Ms of exposure time, we estimate that AXIS would require significantly less exposure time to achieve a comparable signal-to-noise ratio due to its enhanced sensitivity. AXIS is expected to be 5–10 times more sensitive than Chandra in the soft X-ray band, meaning it would require only about 1/5th the exposure time to reach similar depth. This implies that AXIS could achieve equivalent results with an exposure time of approximately 220 ks. Depending on the scientific goals, such as resolving fainter sources or improving resolution, slightly longer exposures of up to 250 ks may be warranted to exploit AXIS's advanced capabilities fully.

In the 2030s, synergy between *NuSTAR*, or its successor, and *AXIS* will revolutionize our understanding of high-energy processes in massive stars. *AXIS*, with its advanced sensitivity and spatial resolution in the soft X-ray band, will provide a powerful platform for probing non-thermal emission in colliding wind binaries, revealing particle acceleration mechanisms in shocks and their role in driving high-energy stellar wind processes. These capabilities will complement *NuSTAR*'s coverage of the hard X-ray band, enabling a more complete characterization of the high-energy spectrum and uncovering signatures of inverse Compton scattering or hard X-ray tails from shocked plasmas.

Furthermore, *AXIS*'s ability to identify and characterize magnetic stars, such as $\theta^1$ Ori C, and other peculiar stars, such as $\gamma$ Cas analogs, will shed light on the X-ray properties linked to magnetically confined winds and accretion processes. By combining *AXIS* and *NuSTAR* observations with multiwavelength data, we can explore the intricate interplay between massive star winds, magnetic fields, and high-energy emissions, advancing our understanding of stellar physics and the broader role of massive stars in shaping galactic environments.

To maximize the scientific return, a well-structured observing plan would target a diverse sample of massive star systems, including colliding wind binaries, magnetic stars, and $\gamma$ Cas analogs. *AXIS* observations should focus on capturing high-resolution spectra and images in the soft X-ray band (0.2–12 keV) with exposure times of approximately 220–250 ks per target to ensure sufficient depth for detailed spectral and photometric analysis. Coordinated *NuSTAR* observations in the hard X-ray band (3–79 keV) should be scheduled concurrently or within a short temporal window to capture complementary high-energy emission signatures. Key systems could be observed over multiple epochs to investigate variability and transient phenomena, providing critical insights into dynamic processes such as shock evolution and magnetic reconnection. This multi-instrument strategy, supplemented by data from optical, infrared, and radio telescopes, will enable comprehensive studies of the high-energy environments of massive stars and their influence on surrounding interstellar media.
**Specify critical AXIS Capabilities:**

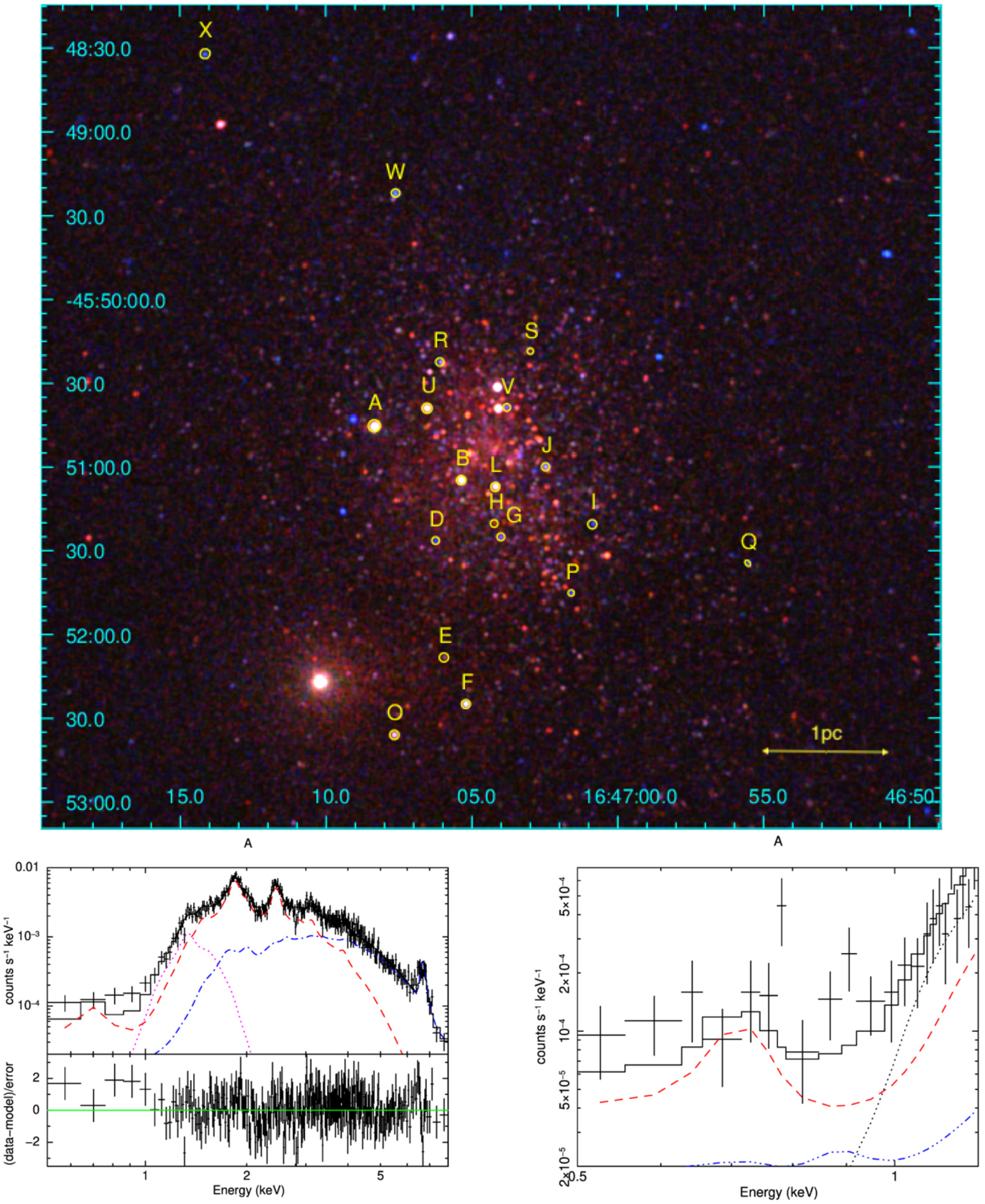

**Figure 9.** Top: A smoothed, color-composite *Chandra* image of the central $4 \times 4$ arcminutes of Westerlund 1, showing X-ray emission in different energy bands: soft (0.5–2.0 keV) in red, medium (2.0–4.0 keV) in green, and hard (4–8.0 keV) in blue. Yellow circles mark 19 of the 20 detected WR stars.
Bottom Left: Spectral modeling of star A. The combined X-ray spectrum of star A, fitted with the best-fit model comprising three `vapec` components (assuming solar composition). The residuals of the fit are displayed below the spectrum in terms of sigma, with $1\sigma$ error bars.
Bottom Right: The same spectrum as in the left panel, focusing on the low-energy range ($< 1.5$ keV) and grouped with 5 counts per bin for clarity. Figure from [5].

- **Field of View (FOV):** The wide FOV allows AXIS to observe entire massive star clusters (e.g., Westerlund 1), capturing multiple CWBs simultaneously for population studies.
- **Angular Resolution:** Sub-arcsecond angular resolution enables AXIS to resolve individual CWBs in crowded, massive star-forming regions and disentangle nearby point sources from diffuse background emission.
- **Energy Band Sensitivity:** AXIS's soft X-ray response (0.2–12 keV) is ideal for probing the thermal emission from shock-heated plasma in CWBs, particularly in the 1–4 keV range where many such systems emit most strongly.
- **L2 Orbit and Continuous Viewing:** Placing AXIS at the Earth-Sun L2 point avoids Earth occultations and allows near-continuous observation of targets, eliminating data gaps crucial for time-resolved monitoring of CWBs.

**Joint Observations and synergies with other observatories in the 2030s:**

- **WFIRST** will provide wide-field near-infrared imaging and spectroscopy, aiding in the identification of obscured massive binaries and monitoring of variability across large spatial scales.
- **JWST** will deliver high-resolution infrared spectroscopy, enabling detailed studies of wind composition, shock chemistry, and—in dust-producing systems such as some WR+O binaries—the characterization of episodic dust formation.
- **ATHENA** will provide high-throughput, high-resolution X-ray spectroscopy for detailed plasma diagnostics of the hot shocked winds. In contrast, *AXIS* will offer superior spatial resolution and faster response times, ideal for resolving compact colliding wind systems and tracking phase-dependent X-ray variability.

**Special Requirements:** None

**c. Stellar activity**

*8. Dedicated Monitoring of Stellar Flares: Evolution of Activity, Coronal Mass Ejections, and Angular Momentum*

**Science Area:** Stars & Exoplanets
**First Author:** Keivan G. Stassun (Vanderbilt University)
**Co-authors:** Breanna Binder (Cal Poly Pomona), Jose Caballero, Lia Corrales (University of Michigan), Jeremy Drake (SAO, Lockheed Martin), Girish M. Duvvuri (Vanderbilt University), Catherine Espaillat (Boston University), Adina Feinstein (Michigan State University), Hans Moritz Günther (MIT), Marina Kounkel (University of North Florida), Rachel Osten (STScI), David Principe (MIT), Peter Wheatley (Warwick), Scott Wolk (SAO)
**Abstract:** We describe a science case for the dedicated monitoring, more or less continuously over a timespan of $\sim$weeks, of stellar flares and associated coronal mass ejections in Carina, a canonical massive star-forming region. Following the example of the Chandra Orion Ultradeep Project (COUP), we anticipate transformative gains in our understanding of the evolution of stellar activity (and its relation to stellar rotation), flare energy distributions, coronal mass ejection mass distributions, and the evolution of stellar angular momentum, as well as implications for the evolution of the habitability of stellar environs.
**Science:** Solar-type pre-main-sequence (PMS) stars typically evince magnetic activity in a variety of forms, including strong and time-variable coronal emission at X-ray wavelengths. The strong magnetic fields associated with this ubiquitous activity are thought to be central to a number of key physical processes in PMS stars, including specifically the transfer of mass and of angular momentum from and to the circumstellar environment. X-ray activity observed in PMS stars bears a striking resemblance to solar X-ray activity, albeit scaled up by several orders of magnitude in both energy and frequency of occurrence. Young, solar-type stars appear, then, to have flaring magnetic field configurations that behave like, and can be interpreted as, scaled-up solar-type flaring fields.

Much of the first few Myr of a PMS star's evolution is influenced by its flaring corona, including its interaction with its protoplanetary disk, loss of mass and angular momentum via coronal mass ejections (CMEs), and the high-energy environment of its forming planets. And it is during the first few Myr, when magnetic fields and associated flares and CMEs are at their most powerful, that PMS stars can serve as laboratories to understand the underlying physics.

Chandra's Orion Ultradeep Project [COUP; 55] was arguably one of the most singular achievements of the Chandra Observatory. Its nearly continuous 1 Ms observation of the Orion star-forming region yielded X-ray light curves for 1300 PMS stars, with which two major advances were made:

1. Flare and CME physics for the most powerful events. Favata et al. [53] were able to fit in detail 32 ultra-powerful flares with solar cooling-loop models that allowed the measurement of the temperatures, densities, and lengths of the flaring magnetic loops.
2. Rotational modulation to map the structure of magnetic fields. Flaccomio et al. [57] were able to measure rotationally modulated coronal holes and other structures for 23 PMS stars, extending what is possible from visible-light modulation studies that probe only the footpoints (starspots) of the fields.

COUP provided compelling evidence that young, solar-type stars exhibit magnetic activity akin to the Sun, albeit at much higher energies and flare frequencies. More recently, Aarnio et al. [1] developed an empirical model linking PMS X-ray flare energies to CME mass-loss rates. Their study demonstrated that PMS stars may experience CME-driven mass loss at rates up to $10^{-9}$ $M_{\odot}$ $yr^{-1}$, suggesting a significant role in stellar angular momentum evolution. This work confirmed that flaring and CMEs in PMS stars follow solar-like scaling laws but extend over orders of magnitude in energy.

However, **Orion is not the optimal massive star-forming region if Carina is accessible**. Chandra was limited to observing Orion's central ∼**5′ region**, covering its densest stellar cluster. In contrast, **Carina is much more massive and far richer in young stars**, but Chandra's limited field of view (FOV) and sensitivity prevented a similar study there. With its **25× larger FOV**, AXIS will unlock Carina's potential, providing an enormous increase in sample size over COUP. The **central 25′ region of Carina** alone will yield a dramatic boost in the number of PMS stars studied (see Figure 10). Given COUP's 1 Ms exposure resulted in only ∼**20 rotational modulation signals and** ∼**30 powerful flares/CMEs**, AXIS will deliver an exponentially larger dataset, enabling statistical robustness far beyond Orion's reach.

The AXIS X-ray observatory presents an extraordinary opportunity to build upon COUP's legacy by targeting the **Carina Nebula**, an even richer and younger star-forming region (∼2 Myr old) with a vastly larger sample of PMS stars. Compared to Chandra, AXIS provides:

- **Much greater sensitivity:** AXIS can detect fainter X-ray sources and resolve weaker flare events, enabling the study of a broader range of stellar activity levels.
- **A much larger field of view:** The Carina Nebula hosts over 10,000 young stars, providing a vastly improved statistical sample over Orion.

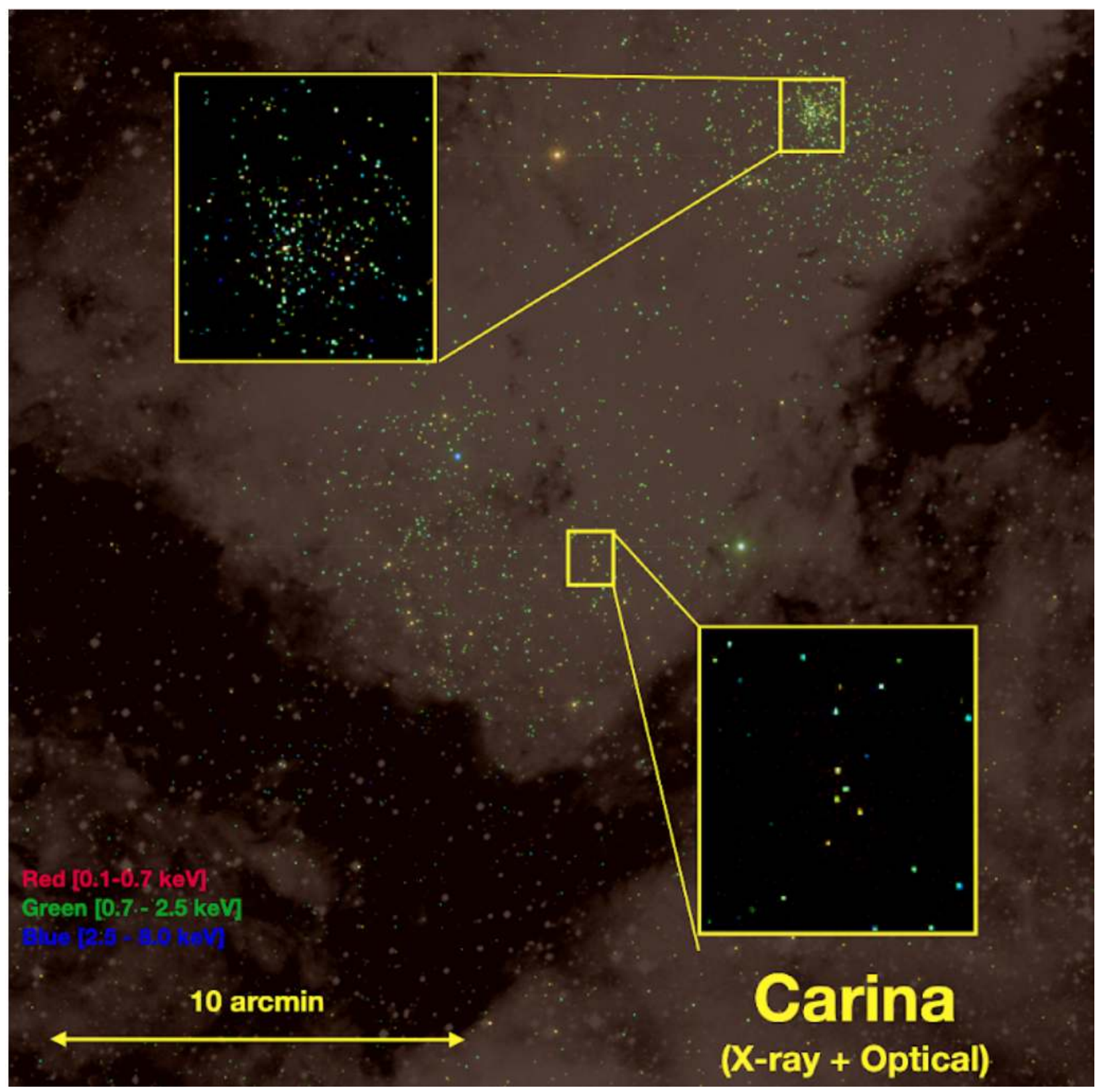

**Figure 10.** Simulated AXIS observation of the Carina star-forming complex. Simulated 50 ks three-color AXIS observation of point sources in a portion of the Carina star-forming complex. The X-ray image is overlaid with a DSS optical image (white), demonstrating the utility of X-rays in piercing the molecular cloud material from which stars are born. With stars having ages 1–6 Myr, the entire Carina complex has one of the highest concentrations of massive stars in the Galaxy, in addition to more than 100,000 young, low-mass stars. A Chandra legacy survey of the Carina complex utilized 1.2 Ms over 22 pointings, yielding $\sim$14,000 X-ray detections, 90% of which had fewer than 50 counts. A single AXIS pointing in this portion of the star-forming complex can detect the 6000 stars previously detected by Chandra and thousands of new, young, low-mass stars that were previously undetected, easily reaching M3-type dwarfs around the peak of the stellar IMF.

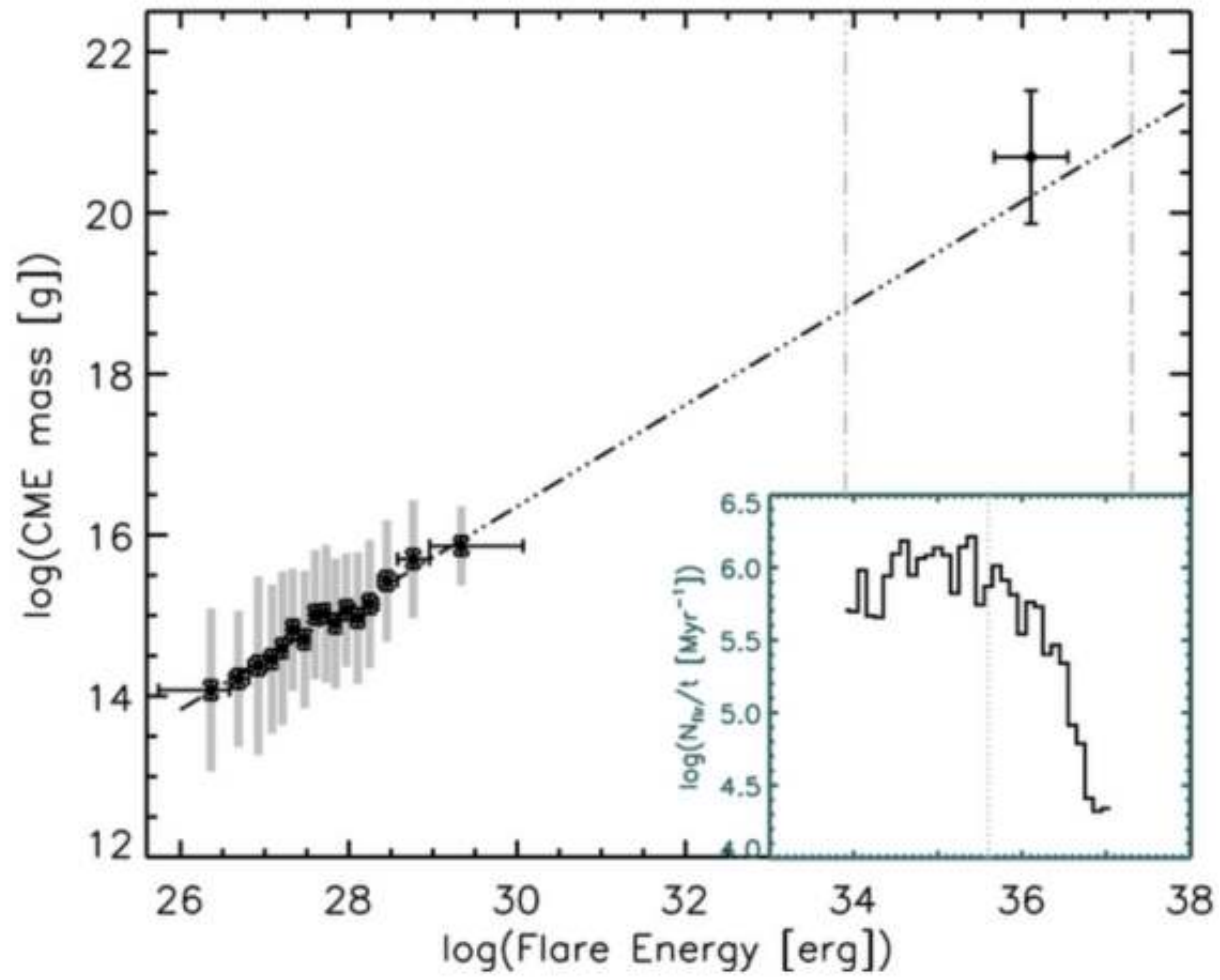

**Figure 11.** Solar relationship between flare energy and CME mass, transformed from the flux/mass relationship of Aarnio et al. [2]. The line is a fit to the solar data over the energy range of observed solar flares, $10^{26} - 10^{30}$ erg, and then extrapolated to the regime of T Tauri star flare energies (represented by vertical lines). The single point at upper right shows the mean CME mass and flare energy for the 32 T Tauri star mega-flares observed by COUP [53] and the error bars represent the standard deviations of those mean values. Inset: the energy distribution of flare rates for T Tauri stars observed by COUP Albacete Colombo et al. [4]. The vertical dotted line at $10^{35.6}$ erg represents the energy above which the observed flare sample is complete. [Figure reproduced from Aarnio et al. [1] ]

- **High time resolution:** AXIS's temporal resolution allows for fine sampling of flare rise and decay phases, permitting the modeling of coronal loop dynamics.

These advantages will enable major advances in our understanding of:

- **Flare Physics and Magnetic Field Scaling:** With a larger sample of extreme PMS flares, AXIS can refine the empirical scaling laws that link young stellar activity to solar analogs. The study by Aarnio et al. [1] demonstrated that PMS CMEs span up to $10^7$ **times** the mass of solar CMEs, but there remains a **major gap in intermediate-mass CMEs (see Figure 11)**. AXIS will fill this gap by detecting flares and associated CMEs at energies ranging from solar-like events to the superflares identified by Favata et al. [53]. Taking young Sun-like stars as representative, combined with their cumulative TESS flare rate of $\sim$0.1 per day as a proxy for very bright X-ray flares, and assuming that young Sun-like stars are $\sim$10% of the sample, we expect to see $\sim$100 flares per day.
- **Coronal Mass Ejections (CMEs):** By tracking variability in the X-ray emission, AXIS can identify signatures of CMEs, revealing their role in mass and angular momentum loss in young stars. Aarnio et al. [1] estimated that PMS stars with **CME mass-loss rates exceeding** $10^{-10}$ $\mathbf{M_\odot}$ $\mathbf{yr^{-1}}$ may experience substantial spin-down over a few Myr. With its improved sensitivity, AXIS will systematically test this prediction across a wide stellar population by comparing current spin rates with the current mass loss rates.
- **Protoplanetary Disk Interaction:** Intense X-ray irradiation from PMS flares influences disk chemistry and ionization, impacting planet formation. AXIS will assess how flare-driven ionization affects inner disk evolution in conjunction with statistical and/or contemporaneous observations with JWST/ALMA.

**Exposure time (ks):** 1000 (i.e., 1 Ms, comparable to the COUP survey of Orion)

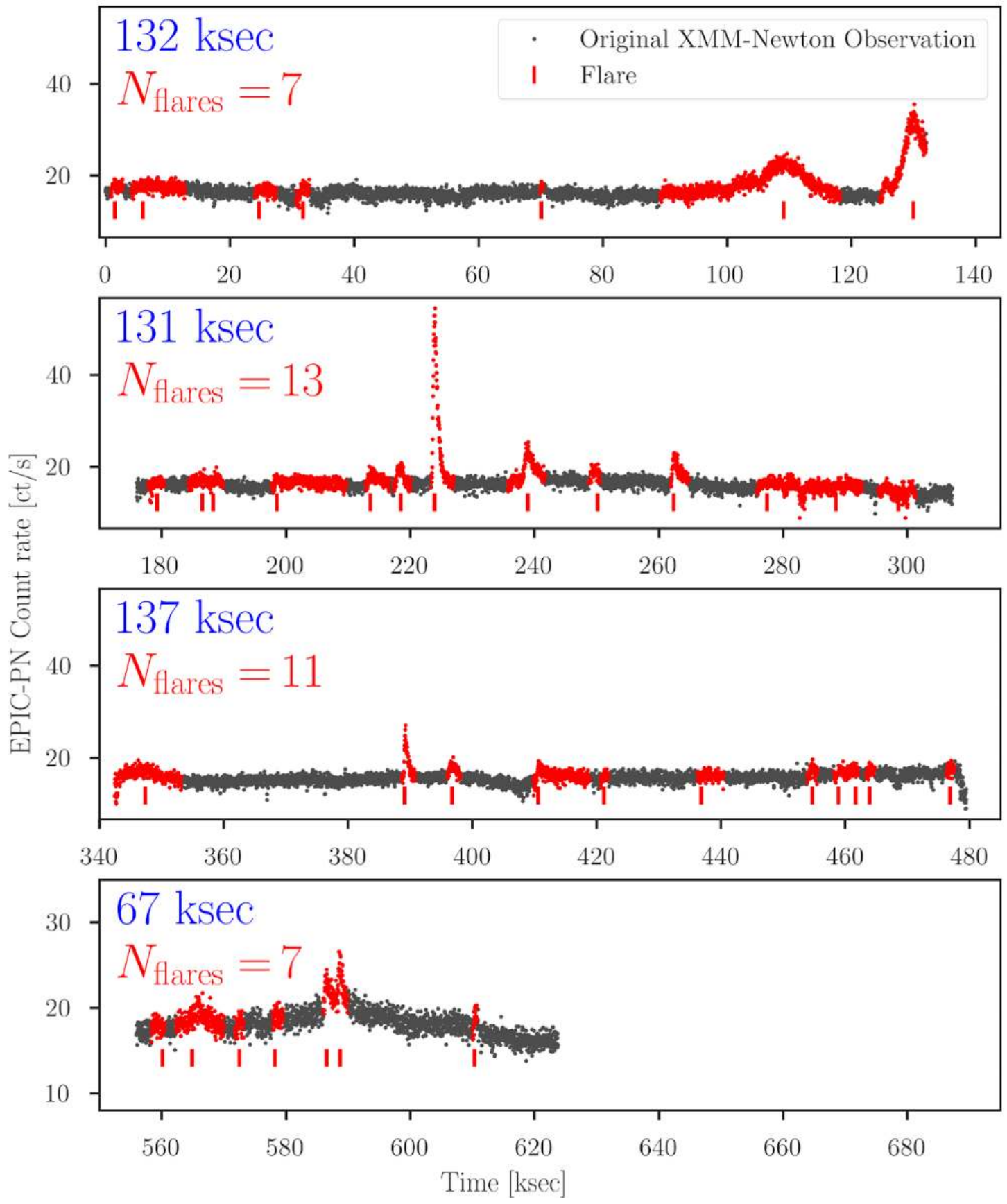

**Figure 12.** An AXIS version of this XMM-Newton ∼500 ksec flare monitoring campaign for AU Mic [162], a single well-known flare star, targeting the Carina Nebula field shown in Figure 13 would be sensitive to the largest of these flares for the brightest stars among ∼10,000 sources. AXIS's increased effective area and PSF stability would make it possible to determine robust flare statistics across a broad range of spectral types in each stellar association with a sufficiently deep field observation, something present X-ray observatories cannot manage without chaining multiple prohibitively long exposures.

**Observing description:** To maximize scientific impact, AXIS will execute a **continuous 1 Ms stare at the Carina Nebula**, mirroring COUP's successful strategy. A simulated light curve based on XMM-Newton observations of the flaring M-dwarf star AU Mic is shown in Figure 12. This will ensure:

- **Uninterrupted monitoring of flare activity**, capturing both short-duration impulsive events and longer-duration super-flares.
- **High signal-to-noise X-ray light curves for thousands of PMS stars**, allowing robust statistical analysis of magnetic activity across different stellar masses and evolutionary stages.
- **Cross-matching with multi-wavelength surveys**, leveraging data from JWST (infrared), ALMA (radio), and Gaia (optical) to correlate X-ray flares with disk properties and stellar rotation rates.

By applying the COUP approach to a richer and larger star-forming region, AXIS will provide an unparalleled dataset, bridging the gap between solar-like X-ray activity and the extreme flares observed in young stars, revolutionizing our understanding of stellar magnetic activity and its broader astrophysical consequences.

**Joint Observations and synergies with other observatories in the 2030s:** TESS, UVEX, ULTRASAT, PLATO + X-ray flares can lead to changes in disk chemistry that JWST and ALMA can trace.

**Special Requirements:** Close to continuous scheduling (distribute 1 Ms of observing time over no more than 1.1 Ms of real time).

*9. Intrinsic Variability of the Stellar Age-Activity Relation*

**Science Area:** Stars & Exoplanets
**First Author:** Keivan G. Stassun (Vanderbilt University)
**Co-authors:** Lia Corrales (University of Michigan), Girish M. Duvvuri (Vanderbilt University), George W. King (University of Michigan), Marina Kounkel (University of North Florida), David Principe (MIT)
**Abstract:** The empirical relationship between stellar rotation and X-ray activity serves as one of the primary probes of the fundamental physics underlying coronal behavior, as well as positioning X-ray luminosities as a useful secondary diagnostic of stellar ages (by extension of the rotation-age—or gyrochronology—relationship). In recent years, the rotation-age relationship has become increasingly well characterized across stellar types and age. However, the rotation-activity relation, while reasonably well characterized in the average, remains uncertain due to the significant scatter in existing rotation-activity studies. The degree to which the observed scatter is intrinsic to the physics underlying the relationship, versus the result of unresolved time-variability in stellar X-ray emission, has yet to be established. By conducting a large-scale study of the relationship between stellar rotation and activity among star clusters/groups across a wide range of ages—and importantly at multiple snapshots in time—will permit a determination of the true nature of the scatter in the rotation-activity relation, and in turn this will permit a determination of the true scatter in activity-age relationships.
**Science:** The X-ray emission from stars has important ramifications for the structure and evolution of planetary atmospheres, and therefore habitability. Steady X-ray emission is produced by rotationally driven dynamos that heat stellar coronae to MK temperatures. Importantly, X-rays can also be produced in extremely energetic, impulsive events (flares), and coronal mass ejections associated with powerful flares can also have impacts on planetary atmospheres. At the youngest ages, active accretion onto stars from their protoplanetary disks can also produce enhanced and impulsive high-energy emission. Understanding the evolution of stellar X-ray emission (and associated flares, coronal mass ejections, and accretion) as a function of stellar age and mass is important for developing a complete picture of what planetary atmospheres experience from their host stars over the stellar lifetimes.

The relevant AXIS parameters that make this program feasible are:

- PSF (1 keV): Ability to cleanly resolve stars in clusters, with typical separations of a few arcsec.
- Effective area: Ability to detect M3 stars (FX=8.4e-17erg/s/cm2 ; 0.2-2.4keV) at a distance of 2.5 kpc

**Exposure time (ks):** $\lesssim$ 3 Ms
**Observing description:** We wish to use a set of open clusters with large numbers of stellar members observable in a single AXIS pointing, and having known ages spanning the epoch that is most important for evolution of exoplanet atmospheres, from the youngest clusters ($\sim$1 Myr) to a few hundred Myr, sampled uniformly in log age. We used the AXIS exposure simulator together with a nominal relation for X-ray luminosity versus stellar age and mass from Poppenhaeger & Wolk [135] to predict the number of counts observed for stars of a given mass, age, distance, and extinction. The table below summarizes the relevant properties of the selected clusters, and the final column gives the estimated exposure time needed to achieve a minimum of 100 counts for the faintest sources we need to observe (spectral type $\sim$M3, representing roughly the low-mass peak of the stellar IMF); brighter stars will of course yield higher counts and even more robust statistics.

The goal of at least 100 counts for each source is motivated both by (1) a desire to achieve at least SNR > 10 on the source X-ray flux measurements and (2) the need to construct X-ray light curves with which to time-resolve flares and extract the physical properties of flaring plasmas and associated coronal mass ejections. The light curve of a source with at least 100 counts can be binned into a minimum of 10 time bins with at least 10 counts each, allowing statistically meaningful measurements of variability amplitudes and the opportunity to resolve and characterize flares. For example, the Chandra Orion Ultradeep Project [55] found that at young ages, strong X-ray flares have durations of $\sim$10 ks; divided into 10 time bins of 1 ks would permit such flares to be well resolved and characterized.

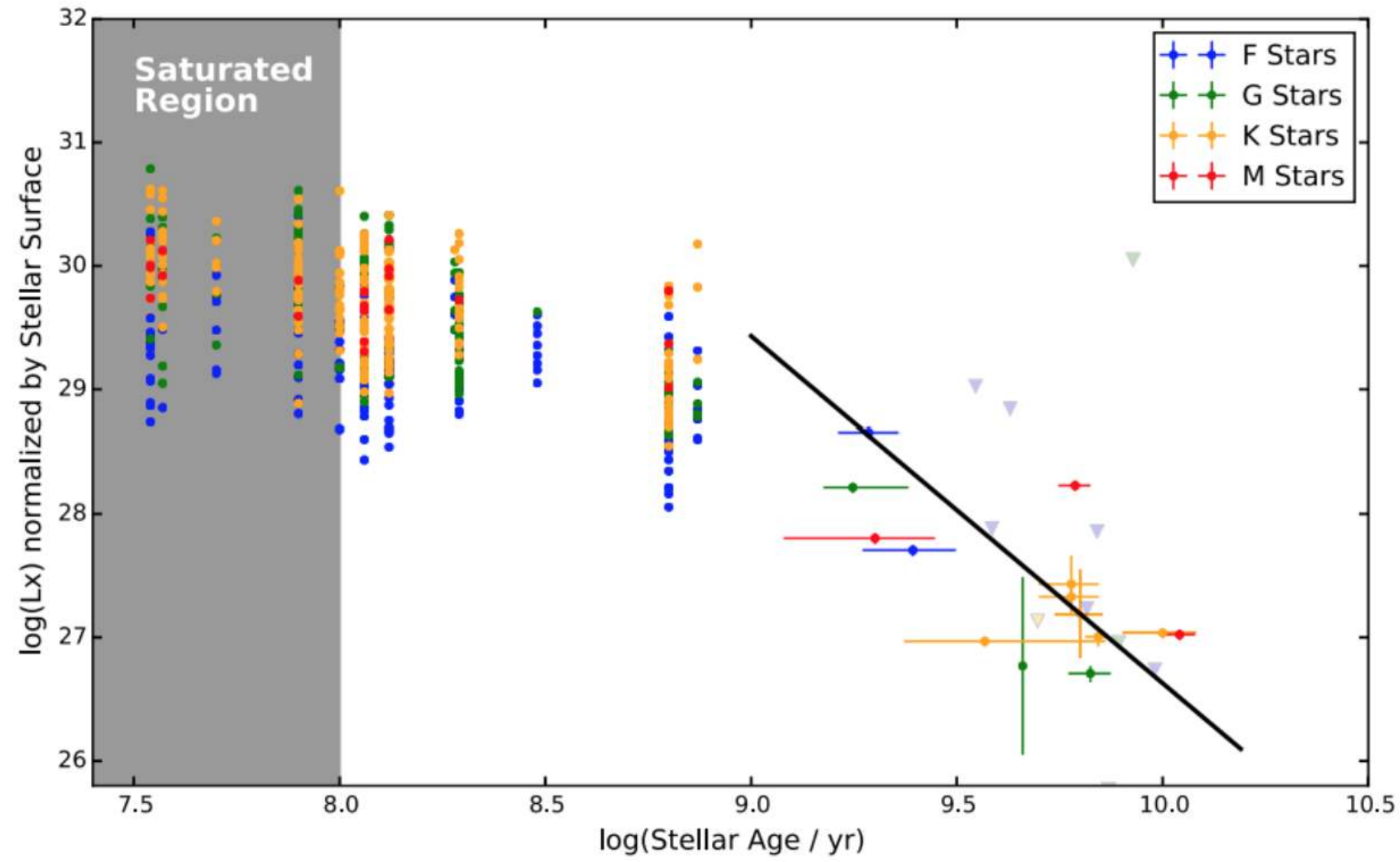

**Figure 13.** Figure 3 from Booth et al. (2017) demonstrates that there is significant (factor of $\sim$10) dispersion in the measured luminosities of saturated stars. It is unknown whether this represents the dispersion in the true quiescent luminosity of these stars or is instead the result of flares and variability. With the resolving power of AXIS, we will be able to answer this question by observing open clusters across a range of ages over the course of the mission lifetime. The collecting area of AXIS will also enable statistically significant analyses of smaller M dwarf stars, which are the most abundant type of star in the Universe and the vast majority of terrestrial planet-hosting stars known today.

As shown below, the total exposure time required for the entire set of clusters is $\sim$1 Ms. This would be tripled to permit three visits to each cluster, spaced by several months, to measure variability on long timescales. Such an observing plan would require $\sim$3 Ms.

**Table 1.** Open star clusters suitable for studying the evolution of stellar activity with age.

| Cluster | RA | Dec | Radius (deg.) | $N_*$ | $d$ (pc) | log Age | $\log L_X^a$ | $\log F_X^a$ | $\log N_H$ | $A_V$ | $ks^b$ |
|---|---|---|---|---|---|---|---|---|---|---|---|
| Collinder 69 | 83.792 | 9.813 | 0.989 | 620 | 416 | 7.05 | 29.3 | -14.0 | 0.45 | 0.25 | 45 |
| NGC 6124 | 246.332 | -40.661 | 0.401 | 1273 | 654 | 8.147 | 28.5 | -15.2 | 3.47 | 1.93 | 125 |
| NGC 6494 | 269.237 | -18.987 | 0.292 | 724 | 742 | 8.477 | 28.1 | -15.7 | 1.64 | 0.91 | 220 |
| NGC 7654 | 351.195 | 61.590 | 0.161 | 1003 | 1586 | 7.764 | 28.9 | -15.6 | 3.33 | 1.85 | 315 |
| NGC 4755 | 193.415 | -60.371 | 0.086 | 595 | 1992 | 7.216 | 29.2 | -15.5 | 2.20 | 1.22 | 160 |
| Carina | 161.265 | -59.684 | – | $10^5$ | 2300 | 6.7 | 29.4 | -15.4 | 2.25 | 1.25 | 165 |
| ONC | 83.750 | -5.483 | – | 1600 | 415 | 6.3 | 29.5 | -13.8 | 0.90 | 0.5 | 4 |

[a] Estimates for the unabsorbed luminosity and flux of an M3 star in the cluster, based on age.
[b] Exposure to achieve 100 counts of an M3 star in the cluster.

**Joint Observations and synergies with other observatories in the 2030s:** The Transiting Exoplanet Survey Satellite (TESS) provides high precision stellar light curves in the optical, providing rotation, starspot, and flare information. The ultraviolet capabilities of ULTRASAT and UVEX can provide complementary information about stellar activity from UV flares, which don't necessarily correlate with X-ray flares.
**Special Requirements:** 10 ks (minimum contiguous exposure time to fully capture typical flares of duration $\sim$few hours)

*10. Stellar Magnetic Activity Cycles in the X-ray*

**Science Area:** Stars and Exoplanets
**First Author:** Thomas Ayres (CASA/CU Boulder, thomas.ayres@colorado.edu)
**Co-authors:** Girish M. Duvvuri (Vanderbilt University, girish.duvvuri@vanderbilt.edu)
**Abstract:** The 11-yr ebb and flow of sunspots at the solar surface is the superficial expression of a poorly understood deep-interior cycling magnetic "dynamo." Besides the photospheric dark spots, other familiar indicators of the decadal solar cycle are chromospheric Ca II H & K emissions and coronal X-rays. Analogous activity cycles have been identified in dozens of late-type (F–K) stars by long-term surveys of Ca II emission, extended more recently to M stars using precision optical photometry. While the Sun's cycle range is a mere < 10% in Ca II, and nearly imperceptible in visible photometry, the amplitude blooms to a factor of 4 in coronal soft X-rays, and spikes to 100, or more, in hard X-rays. The Sun's dramatic high-energy cycle is an enabler of Space Weather (SW), occasionally battering Planet Earth with harsh radiation and particle streams. Further afield, host-star SW analogs can severely impact the habitability of their exoplanets. Only about 20 cool stars currently have long-baseline X-ray observations sufficient to identify dynamo cycles. As a result, the high-energy phenomena, especially the variability modes, are tempting targets for an advanced facility like AXIS. The soft X-ray sensitivity, spectral capability, and high spatial resolution of AXIS are ideal to probe stellar coronae both broadly in surveys of general properties, as well as narrowly for individual objects in the time domain. AXIS can fill out regions of stellar cycles parameter space that will provoke magnetic dynamo theories (and the so-far unsolved coronal heating problem), possibly in new directions. Further, stacking of repeated observations across years will build an assortment of quasi-deep fields that could yield serendipitous sources that previous all-sky X-ray surveys lacked the depth to discover and spectroscopically dissect.
**Science:** The 11-year sunspot cycle is one of many "activity" phenomena thought to be instigated by an oscillating magnetic "dynamo" deep inside the Sun [128]. Other important activity signatures are the Ca II 3968 Å ("H") and 3933 Å ("K") chromospheric emission lines and coronal (0.1 – 10 keV) X-rays. These exotic emissions arise from extended "active regions" surrounding the much smaller photospheric dark spots. The classical approach to identify activity levels and cycles of other stars is to monitor the Ca II emission index of suitable targets, over time scales of decades for the long-term variability part. In this way, HK cycles have been captured in about 50 F–K stars. An alternative cycle proxy involves precision optical photometry to implicitly follow changes in starspot areas over time. This approach has extended cycle detections down to the M dwarfs, which often display "saturated" chromospheric and coronal fluxes (Wright et al. 2011). The Sun's activity cycle is rather modest in Ca II HK, displaying only a $\lesssim$ 10% range from cycle minimum to maximum in recent decades [47]. At the same time, the Sun's cycle is striking in X-rays, from factors of 4 to >100, depending on the energy bandpass (the higher the energy, the larger the intensity swing: 156).

It also is important to recognize that generalized dynamo phenomena are ubiquitous in the cosmos – even AGN accretion disks have a magnetic corona, and a coronal-heating problem [34] – so investigating the processes underpinning stellar magnetism has relevance beyond the Sun and stars.

The solar X-ray flux is closely tied to "Space Weather" (SW), a catch-all for energetic coronal phenomena such as flares, coronal mass ejections, bursts of charged particles, and high-speed wind streams. Contemporary SW impacts our local heliosphere, especially technologically vulnerable Planet

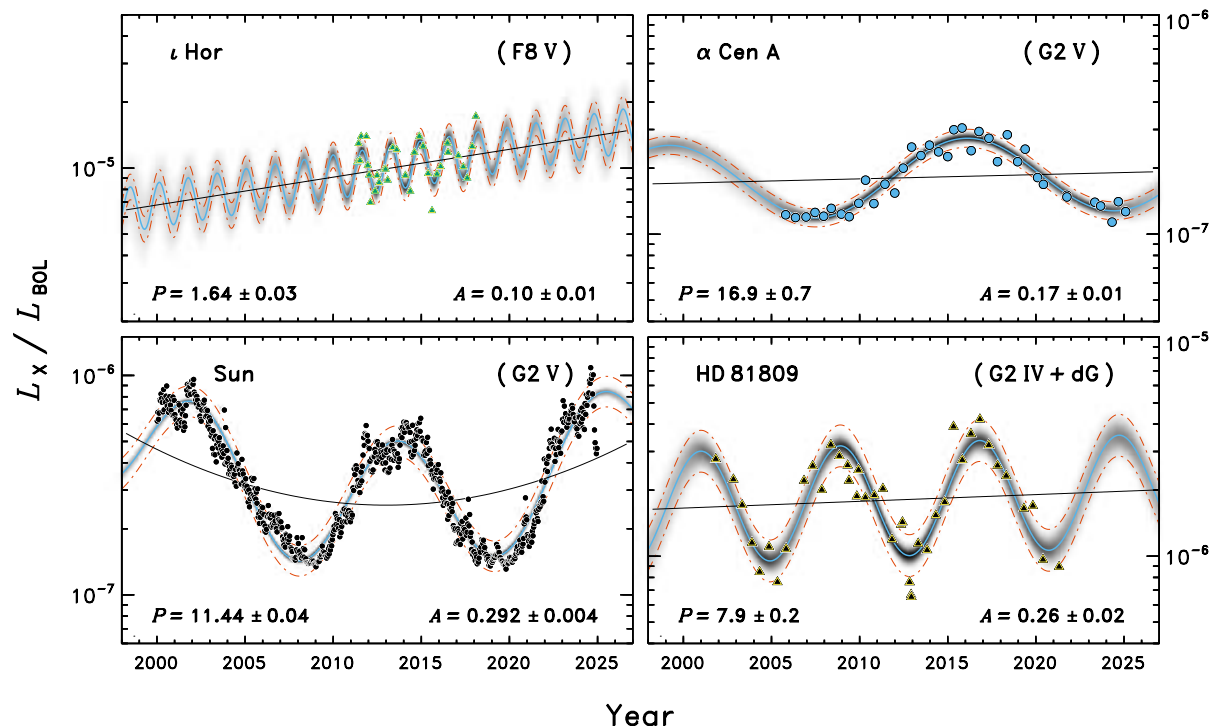

**Figure 14.** Adapted from Figure C1 of Ayres [8] showing the X-ray activity cycles of the Sun compared to three other stars. Activity cycles are easier to detect and characterize in the X-ray compared to most other magnetic activity metrics, the sample is only limited by the lack of dedication of resources to repeat observations and long-term monitoring.

Earth; and likewise host stars for their exoplanets. In fact, the high-energy radiation of active hosts can sterilize surfaces of rocky planets with thin atmospheres, while associated coronal winds can persistently strip away volatiles until little or no atmosphere remains [3]. Mars is the poster child of SW-induced planetary trauma in our own Solar System. Consequently, assessing magnetic-related SW phenomena, especially atmospheric escape, over the course of planetary system lifetimes is critical for understanding exoplanet evolution, demographics, and potential habitability. But, the degree of SW-related impacts in the early Solar System, and young stellar analogs, owing to the heightened magnetic activity levels in stellar youth, is still unsettled.

On the theory side, an important historical recognition was that apparent parallel trends of $L_{\mathrm{CaII}}/L_{\mathrm{bol}}$ vs. $P_{\mathrm{rot}}$ seen among the different spectral types (different surface temperatures) could be collapsed together by replacing the stellar rotation period with the so-called Rossby number [117]. The latter – a key parameter in stellar and terrestrial dynamo theories – is a ratio of the stellar rotation period PROT to a temperature-sensitive "convective turnover time" $\tau_C$ , and reflects the importance of Coriolis forces for twisting zonal plasma flows, ultimately creating a regenerative dynamo with an oscillating magnetic field. A curious aspect of the Rossby number is that both the rotation period and turnover time (around 1 month each in the Sun) are hundreds of times shorter than the cycle duration (a decade). This enormous disparity of time scales creates headaches for dynamo theorists.

Despite the large amplitude of the Sun's X-ray cycle, and the value of characterizing stellar high-energy counterparts, fewer than 20 stars at present have the baseline of repeated observations from the long-lived X-ray "Big Three" (Chandra, XMM-Newton, Swift) needed to find and characterize activity cycles. Just this small sample has already displayed a diversity of behaviors: high-contrast, decadal-class X-ray modulations are found exclusively at low-to-medium $L_X/L_{\mathrm{bol}}$; higher X-ray intensity stars tend to display lower amplitude, faster variations, if cycling at all; whereas the highest activity cases show stable ("saturated") long-term X-ray trends, but punctuated by persistent flaring in the short term. Some X-ray cycles are brief, 2-3 yr; others are longer than the Sun's 11 yr; and there are "flat-activity" cases as well (See Figure 14 for some examples).

The prevailing wisdom a few years ago was that stellar cycles seemed to organize themselves into two distinct, diverging branches in a "Böhm-Vitense" diagram comparing cycle duration, $P_{\mathrm{cyc}}$, against rotation period, $P_{\mathrm{rot}}$ [21]. A steeply rising "Active" (A) branch had relatively fast rotators, but long cycle periods; whereas a shallow "Inactive" (I) branch had slower rotators, but shorter cycles. Curiously, a few stars displayed double cycles: one on the A branch paired with another on the I branch. Oddly, the Sun, and solar twin Alpha Cen A, sat mid-way between the branches, in a sparsely populated region all to their own. It was suggested later [104] that the Sun was in a transitional stage of its dynamo evolution, having run up the shallow Inactive branch from left (short $P_{\rm rot}$) to right (long $P_{\rm rot}$) for much of its multi-Gyr lifetime, but now was rapidly segueing upward between the I and A branches toward ever longer $P_{\rm cyc}$, ultimately to a flat-activity state. (Foreshadowed, perhaps, by the Sun's "Maunder Minimum" in the 1600s when sunspots were rarely seen for almost seven decades 46). Complicating matters, the addition of new stellar cycle examples in recent years (e.g., 26) has made the original A and I branches appear less cohesive, giving rise to alternative classification schemes [22], coupled with dueling views of theoretical dynamo behavior [84,157].

If the Sun truly is in a transitional dynamo state, that would be bad news for theorists, because many of the observed properties of the solar cycle – latitudinal differential rotation, meridional circulation, Hale polarity reversal, sunspot emergence butterfly pattern, Joy's law of sunspot polarity tilts, and so forth – are known either exclusively from the Sun [33], or if stellar proxies are available, they typically are less informative. Theorists factor the exquisitely measured solar traits into their models, but could be led astray if the solar dynamo properties refer to an abnormal, transitory state, rather than the normal operating condition [143]. Additional observational perspectives are needed to bring clarity to an otherwise murky picture based on relatively small existing samples. The difference with stellar cycles, compared to typical Astronomy endeavors, is that a new perspective cannot be achieved simply by a few nights at the telescope, but rather requires long-term dedication.

X-rays can play an outsized role in assessing dynamo behavior because of the responsiveness of the high-energy emissions to coronal variability. Here is where AXIS could play a central role. The facility could obtain short snapshots of dozens or hundreds of late-type coronal stars on a regular basis (every few months or half year) all over the sky without significantly impacting other programs, potentially in-between slews to other scheduled targets. A sample of stars with a range of Rossby numbers and well-characterized activity cycles would be a powerful tool to inform both dynamo theory and the coronal magnetic heating problem. The high spatial resolution of AXIS would be valuable because there are many close visual binary systems with separations between 2″–10″ that would be excellent targets (generally a warmer plus cooler star, with well-known parameters, including especially the stellar masses). If AXIS operated for a decade or more, new cycles could be captured for most of the interesting cases, $P_{\rm cyc}$= 2–10 yr, and revious X-ray examples could be followed-up to find longer cycles and characterize amplitude changes and/or phase shifts; further fodder for theory.

**Exposure time (ks):** 1 ks per pointing for 15 targets, with a goal of 20 pointings per year; 5 ks per pointing for 10 targets, with a goal of at least 5 pointings per year per target: 550 ksec per year of AXIS operations.

**Observing description:** This is a notional target list divided into two categories. For both, we estimate count rates from ROSAT PSPC, Chandra HRC-I, and XMM MOS, where available, and assume that AXIS is approximately 20 times more sensitive in the energy ranges relevant to stellar coronae:

- 15 individual targets (some are binaries that can be done in one pointing); 19 separate stars altogether. Candidates were taken from a recent survey by Boro Saikia et al. [26] and Ayres [8]. A >10 sigma detection usually can be obtained in 1 ks, often much less. Exposures longer than 5 ks are useful for flare filtering, but several shorter exposures per target per year can serve the same purpose, and enable better variability sampling for most of the objects that do not flare frequently (the exception is the dM flare star Proxima Cen).
- 5 stars each from the young Pleiades (100 Myr) and Hyades (600 Myr) open clusters. These stars are too faint for a realistic multi-cycle program with Chandra and/or XMM, or ground-based HK, but can be done in 1 ksec pointings with AXIS to achieve S/N 10. Based on cycle periods measured in fast-rotating F-star Iota Hor and K-star Eri, the cluster members should have relatively short $P_{cyc} < 5$ yr. The targets were chosen to fill in regions of $P_{cyc}$ vs. $P_{rot}$ space that are currently poorly populated. The expected increased amplitude of X-ray cycles (compared to, say, HK), combined with the sensitivity of AXIS, should enable easy $P_{cyc}$ detections, informing dynamo theory beyond the currently available sample.

For the first category, suppose that each target is observed for 1 ks several times per year (in effectively a "snapshot" mode). That would require around 300 ks of exposure per year (ignoring overheads such as slews). The program could be run in a "filler mode" much like the HST SnapSHOT observations, to optimize the observation schedule.

**Table 2.** Category 1: Long stellar activity cycle

| **Name** | **Type** | $d$ [pc] | $P_{rot}$ [d] | $P_{cyc}$ [yr] | **Survey** | **Visual Sep** | **Rate** [ct s$^{-1}$] |
|---|---|---|---|---|---|---|---|
| $\kappa^1$ Cet[a] | G5V | 9.28 | 9.2 | 5.7 | MW | >10″ | 20 |
| 9 Cet[b] | G3V | 21.3 | 7.8 | 9.1 | MW | >10″ | 10 |
| HR 222[c] | K2V | 7.44 | 38.5 | 8.5 | MW | — | 1 |
| GJ 688[c, h] | K3V | 10.1 | 36.4 | 7.2 | MW | <2″ | 1 |
| $\sigma$ Drac | K0V | 5.76 | 29.0 | 7.1 | MW | — | 5 |
| HR 3750[c, i] | G2IV+ | 31.0 | 40.2 | 7.9 | XM | <2″ | 0.5 |
| $\mathcal{E}$ Eri[c] | K2V | 3.22 | 11.5 | 2.4 | XM | — | 30 |
| 18 Sco[d] | G1V | 14.1 | 22.7 | 11.4 | MW | >10″ | 0.1 |
| 61 Cyg AB | K5V + K7V | 3.5 | 35.7 + 37.8 | 6.8 , 11.5 | XM | >10″ | 1 , 1 |
| $\xi$ Boo AB | G7V + K5V | 6.75 | 6.2 + 12 | ? , ? | XC | 2″ – 10″ | 30 , 5 |
| 70 Oph AB | K0V + K4V | 5.11 | 19.7 + (34) | 5.1 , ? | MW | 2″ – 10″ | 15 , 3 |
| $\alpha$ Cen AB | G2V + K1V | 1.34 | 26 + 40 | 16.9 , 7.6 | XC | 2″ – 10″ | 15 , 30 |
| Procyon[e] | F5IV | 3.51 | (19) | – | XC | 2″ – 10″ | 30 |
| Proxima Cen[f] | M5V | 1.3 | 84 | – | XMS | >10″ | 10 |
| $\tau$ Cet[g] | G8V | 3.65 | 34 | – | MW | — | 1 |

**Table notes:**

a Secondary cycle of 22.0 yr
b Secondary cycle of 22.3 yr
c Well-defined cycle
d Near solar twin, probable cycle
e X-ray dark white dwarf companion
f Variability dominated by flares, no clear long-term cycle detected so far
g Old Sunlike G star
h SB1 with period of 84 days
i Close (sep=0.4″), fainter dG companion; disagreement over which component is cycling

For the second category – X-ray fainter with expected short $P_{cyc}$ – each target should have at least five pointings of 5 ksec per year across the lifetime of the mission for adequate temporal sampling. The compact sizes of each cluster will enable AXIS to slew to each target efficiently. The total time required for this program each observing year would be about 550 ksec (again ignoring overheads).

**Relevant AXIS capabilities:**

- High spatial resolution to resolve close visual binaries (two targets for price of one pointing).
- Soft X-ray sensitivity: stellar coronae (particularly for older, less active main sequence stars) emit a large fraction of their X-ray flux at energies < 0.5 keV.
- Rapid slews to minimize overhead.

**Table 3.** Category 2: Short stellar activity cycle (Pleaides)

| Name | Type | $d$ [pc] | $P_{\rm rot}$ [d] | Rate [ct s$^{-1}$] |
|---|---|---|---|---|
| BD+22 574 | F8V | 144 | 2.2 | 0.29 |
| HD 283067 | F8V | 137 | 2.2 | 0.28 |
| TYC 1800-2144-1 | F9V | 135 | 2.9 | 0.20 |
| BD+22 548 | F9V | 121 | 2.9 | 0.30 |
| BD+23 527 | G0V | 133 | 3.6 | 0.14 |

**Table 4.** Category 2: Short stellar activity cycle (Hyades)

| Name | Type | $d$ [pc] | $P_{\rm rot}$ [d] | Rate [ct s$^{-1}$] |
|---|---|---|---|---|
| HD 285773 | K0V | 45.1 | 10.7 | 0.36 |
| HD 28878 | K2V | 48.3 | 9.8 | 0.36 |
| HD 28977 | K2V | 52.0 | 9.2 | 0.29 |
| HD 284552 | K5V | 44.0 | 11.8 | 0.14 |
| HD 285804 | K7V | 42.3 | 13.2 | 0.08 |

- Uniform spatial resolution allows the possibility to capture several targets in the same field, especially for the compact Pleiades cluster.

**Joint Observations and Synergies with other Observatories in 2030s:** The UVEX survey will provide at least 3 separate measurements of FUV/NUV fluxes for most of these stars to compare to AXIS X-ray pointings

**Special requirements:** Frequent scheduling across the lifetime of the mission. An operational mode analogous to the Hubble SnapSHOT program is a possible way to enable this.

**d. Planetary Science and Exoplanets**

*11. X-ray Characterization of Exoplanet Hosts and Habitable Worlds Observatory Targets*

**Science Area:** Stars and Exoplanets
**First Author:** Girish M. Duvvuri (Vanderbilt University, girish.duvvuri@vanderbilt.edu)
**Co-authors:** Breanna A. Binder (Cal Poly Pomona), George King (University of Michigan), Lia Corrales (University of Michigan)
**Abstract:** The X-ray and ultraviolet radiation from exoplanet host stars erodes the atmospheres of exoplanets by photoevaporation and, in some cases, is capable of driving a hydrodynamic outflow [98,107,125,149,175]. The sample of exoplanet hosts with well-characterized X-ray spectra has mostly been populated by campaigns that focused on the planet and/or narrow regions of stellar parameter space (e.g., HAZMAT 153, MUSCLES 60, MEATS 12, JWST transmission spectroscopy targets 177), and there is a need for a more systematic survey that spans both the stellar parameter space of both age and activity (and possibly metallicity) to determine whether there is a "cosmic shoreline" dictating the fate of planet atmospheric retention [178]. Complementarily, Habitable Worlds Observatory (HWO) has a tiered target list for its search for a directly imaged Earth-sized exoplanet in the habitable zone of its star [102], but 6 targets in the top tier lack X-ray data entirely, and the best data available for 32 top-tier targets comes from ROSAT (HWO Precursor Science Report). Even among those HWO target stars with XMM-Newton or Chandra data, about half were too faint for reliable spectroscopy with the existing archival observations [19]. AXIS' soft X-ray sensitivity and effective area make it possible to accurately measure the X-ray spectra of older, less active FGKM dwarfs, addressing both these needs in exoplanet science.

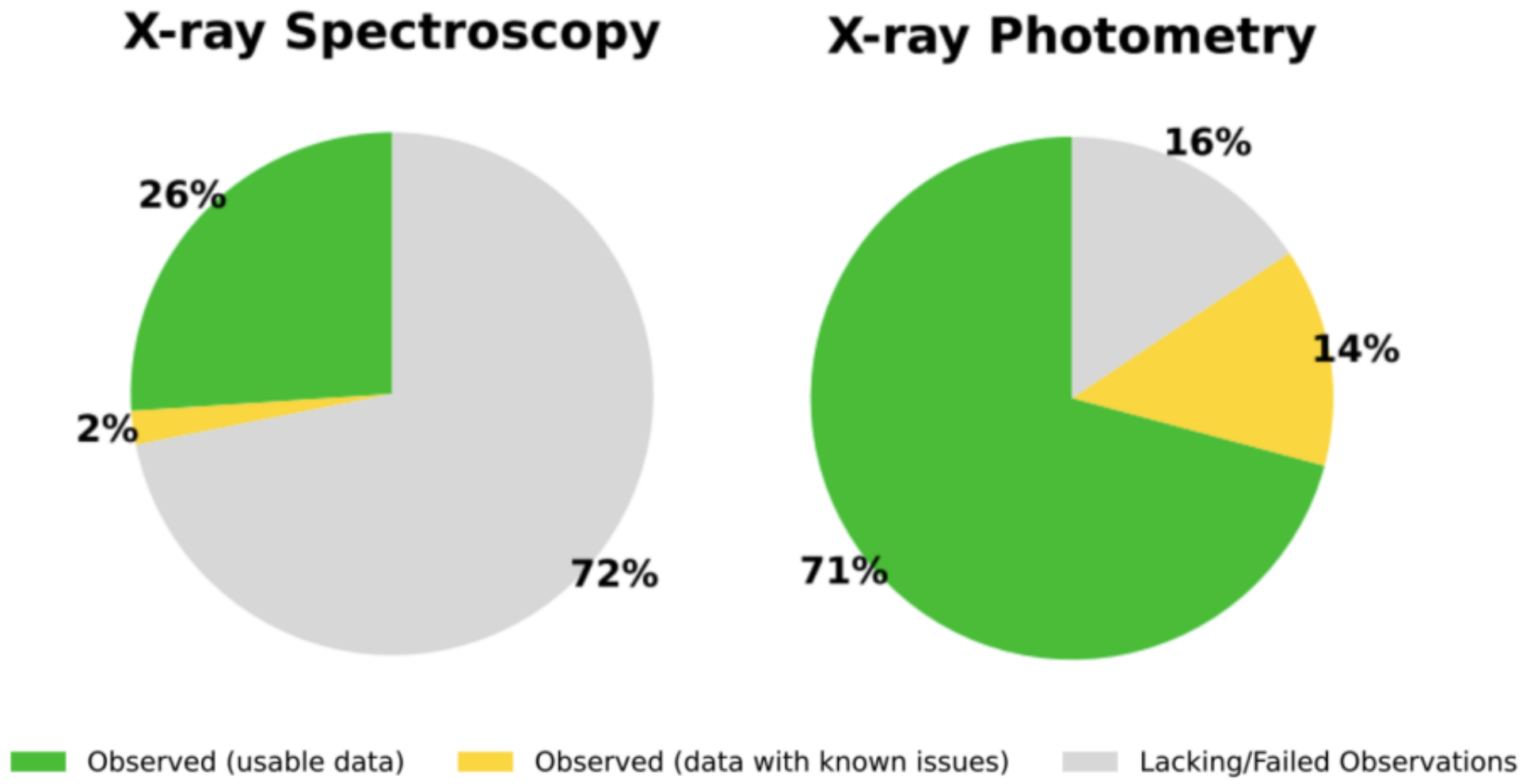

**Figure 15.** Adapted from Figure 5 of the Stellar High Energy Emission section of the *HWO Target Stars and Systems: A Survey of Archival UV and X-ray Data* [131]. The sectors represent the percentage of 98 high-priority *HWO* targets with good quality data (green), poor quality data (orange), and no data at all (grey) for both X-ray spectroscopy and photometry. AXIS has the sensitivity to complete the sample of spectroscopic data for these targets and others important for exoplanet science.

**Science:** The early days of exoplanet detection characterization largely treated stars as simple, well-understood objects, while the planets' atmospheres and properties were the only mystery. Now, the planets are not the primary problem; their host stars' complexity gets in the way. This is a comically exaggerated description, but the core idea that many questions in planet formation, evolution, and demographics are plagued by uncertainties in the stars' influence on their planets holds true. In preparation for Habitable Worlds Observatory (HWO), a mission charged with directly imaging an Earth-sized planet in the liquid water "habitable zone" of its star, Mamajek & Stapelfeldt [102] prepared a table (that is still evolving) of potential targets that might provide that exo-Earth detection. This table is divided into priority tiers based on the ease of detection, and its top tier, Priority A, sample is full of stars visible to the naked eye, and many of the most well-studied stars in history. But even for these exemplars of stellar astrophysics, not all of them have well-measured X-ray spectra (see Figure 15). Binder et al. [19] went through archival X-ray data for potential HWO and Extremely Large Telescope (ELT) targets and found a number of them were too faint to analyze their time variability or their spectrum.

The X-ray and ultraviolet (XUV, 1 – 912 Angstroms) irradiation of exoplanets drives photoevaporative escape of H, He, C, and O [58,98,107,125,175,179]; influencing both the primordial H/He dominated atmosphere and the secondary atmosphere that might form on a terrestrial exoplanet by outgassing. The extreme ultraviolet (EUV), ranging from 100 – 912 Angstroms, has no operational observational resources available and a funded mission in development is a CubeSat capable of limited EUV photometry (MANTIS, 81), not spectroscopy. In the absence of data, the EUV spectrum must be inferred and empirical approaches anchor their predictions using X-ray spectra. While scaling relations using integrated X-ray fluxes exist to predict EUV emission [87,149], these predictions are for the EUV flux integrated over large 100 – 800 Angstrom-wide bandpasses. Such fluxes are useful inputs to models that only require energies, but more detailed modeling that cares about the shape of the EUV spectrum and its impact on individual species requires at least low-resolution EUV spectra to resolve particular features [45,149]. Such spectra are possible to make from semi-empirical models of the stellar upper atmosphere, but their precision relies on having soft 0.1 – 0.5 keV X-ray spectra where emission lines formed at different temperatures can be resolved. AXIS can provide those soft X-ray spectra.

The sample of exoplanet hosts with measured X–ray spectra has largely been populated by campaigns that are motivated by the planet; the relative (perceived) ease of atmospheric characterization small planets around small stars drove a wave of observational surveys focused on M dwarfs (HAZMAT 153, MUSCLES 60) and another set of observations were motivated by planetary systems scheduled to be observed by JWST (MEATS 12, JWST transmission spectroscopy targets 177). Filling out the HWO Priority A stars without good X-ray data is more focused on the star, but more importantly the selection function for these stars is well-defined. A survey project to understand atmospheric escape at the population level could start from the HWO table and relax some of the constraints regarding direct imaging detectability, then identify the set of exoplanet hosts that satisfy the newly defined selection function. Some care would be required to account for how the likelihood of exoplanet detection affects the sample, but nonetheless this star-first sample of X-ray spectra would be an incredibly valuable tool for understanding whether there is in fact a "cosmic shoreline". This idea of a boundary separating worlds with air from those that don't was developed in the context of the Solar System, comparing the lifetime irradiation of a world to its surface gravity, then modifying the irradiation to be specific to the XUV [178]. A more holistic version of this requires accounting for the stellar mass-dependence of activity evolution and then seeing how the JWST detections of planets consistent with bare rock are distributed relative to those planets where clear atmospheric absorption has been detected. To construct this diagram with the current set of exoplanet host X-ray spectra would yield a statistically uninterpretable mess, but AXIS' contributions to the unoccupied regions of planetary system parameter space would make this experiment viable.

**Exposure time (ks):** 1 – 50 ksec per HWO target, $\sim$ 50 ksec per exoplanet host target.

**Observing description:** Table 5 describes five HWO Priority A stars where existing X-ray data only provide a flux or flux limit but do not permit spectral analysis [19].

**Table 5.** Five HWO Priority A stars of interest

| Star | Distance | Spectral Type | $\log_{10} L_X$ (upper limit) | $\log_{10} F_X$ (upper limit / 100) |
|---|---|---|---|---|
| Iota Per | 10.51 | F9.5V | 27.23 | -14.89 |
| Gamma Pav* | 9.27 | F9V | 26.46 | -15.55 |
| 40 Eri A* | 5.04 | K0.5V | 26.36 | -15.12 |
| 47 UMa | 13.8 | G0V | 27.31 | -15.05 |
| 18 Sco | 14.13 | G2V | 26.15 | -16.23 |

Using the fluxes and limits provided in [19], and assuming the true fluxes are 100 times fainter than the limit, these targets have $\log_{10}$ X-ray fluxes ranging from -14.5 – -16.5. Assuming the 0.1 keV blackbody spectrum model for these fluxes, the exposure time chart suggests that AXIS would need between 1 – 50 ksec to measure their spectra. The factor of 100 is conservative, but other cool star exoplanet hosts as part of the atmospheric escape survey will likely be further than these stars, and the 50 ksec exposure time will likely be appropriate for the bulk of those targets. Many of the Priority A and B HWO targets are in binaries, and the spatial resolution of AXIS will be required to resolve the sources cleanly enough to measure their spectra independently.

**[Joint Observations and synergies with other observatories in the 2030s ]** AXIS will complement Ultrasat and UVEX, which would provide FUV/NUV fluxes for the targets currently missing this information.

**[Special Requirements:]** No special requirements.

*12. Studying Exoplanet Atmospheres via X-ray Transits with AXIS*

**Science Area:** Stars and Exoplanets

**First Author:** Lía Corrales (University of Michigan), liac@umich.edu

**Co-authors:** Raven Cilley (University of Michigan), George King (University of Michigan), Girish Duvvuri (Vanderbilt), Scott Wolk (Harvard & Smithsonian Center for Astrophysics), Jorge Fernández Fernández (University of Warwick)

**Abstract:** Measuring the effective X-ray radius of planets during transit provides important constraints on atmospheric escape models and the structure of host stars' coronae. To date, the hot Jupiter HD 189733 b remains the only exoplanet with a detected X-ray transit. AXIS will be more sensitive to the soft X-ray emission of stars than both XMM-Newton and Chandra, providing enough signal to capture X-ray transits from additional targets. Recent work by Cilley et al. [36] evaluated the probability of detecting X-ray transits with the collecting area of next-generation X-ray telescopes, including AXIS. We build on this suite of probabilistic simulations to develop a feasible AXIS observing plan to yield high-impact results in the topic of exoplanet atmospheric evolution.

**Science:** One of the most remarkable results in exoplanet science today has been the discovery of direct absorption from gas escaping short-period Jupiters and sub-Neptunes. X-ray and extreme ultraviolet (together, XUV) radiation from the host star is believed to be a major driver of this hydrodynamic escape [88,91,125,160,168,176]. In turn, atmospheric escape plays a significant role in shaping the observed properties of short-period planets: the "radius valley," an observed dearth of planets with radii of 1.5 to 2 $R_E$ [61,98,127], and the "Neptune desert," the observed lack of short-period Neptunes [103,126]. Directly observing the extent and physical conditions of planetary atmosphere outflows is needed to study the outflow launching mechanisms and develop a full picture of planet formation and evolution.

Extended and escaping layers of gas have been observationally confirmed in many exoplanets through transit spectroscopy, including detections of hydrogen via Ly-$\alpha$ absorption [48,49,93], various metal absorption lines [15,59,94,168,169], and metastable helium absorption [e.g., 92,120,154]. Unfortunately, the utility of the Ly-$\alpha$ feature (and metal lines in the FUV) is limited by the strong absorption of FUV light by the interstellar medium. Meanwhile, the conditions for producing metastable He absorption in the infrared are highly sensitive to the stellar NUV and EUV luminosity and variability, as well as the density of the extended atmosphere [119,170]. Thus, many planets expected to undergo photoevaporation do not necessarily produce observable metastable He absorption [42]. Heavy metals like Fe and Mg are also not going to escape as easily as the lighter H and He atoms, and may not be an ideal tracer for all but the hottest gas giant systems [41,99–101].

X-rays are absorbed by the inner-shell electrons of all abundant elements such as H, He, C and O. At low spectral resolution, the X-ray photoelectric absorption effect is roughly the same for low-ionization states (e.g. OI, OII and OIII). This makes X-ray transits a more agnostic tracer of the physical extent of planetary atmospheres compared to individual absorption lines. X-ray transit shapes can also give us information about the structure of the stellar corona [86], revealing vital information about the host stars themselves. The first and only planet transit observed in the X-ray is that of hot Jupiter HD 189733 b. Using the XMM-Newton and Chandra X-ray observatories, Poppenhaeger et al. [134] detected an X-ray transit that was 3-4 times deeper than the optical. This provides evidence for a diffuse extended H/He atmosphere indicating atmospheric escape.

AXIS will be more sensitive to the soft X-ray emission of stars than both XMM-Newton and Chandra, providing enough collecting area to capture X-ray transits from more planets. The 1.5" imaging resolution of AXIS is also important for distinguishing the target exoplanet host from background point sources and companion stars [e.g., 134,136]. Recent work by Cilley et al. [36] evaluated the probability of detecting X-ray transits with the collecting area of next-generation X-ray telescopes, including AXIS. We build on this suite of probabilistic simulations to develop a feasible AXIS observing plan to yield high-impact results on the topic of exoplanet evaporation.

Cilley et al. [36] identified a "Top 15" list of known transiting planets that are the most likely to be detectable in the X-ray under the ideal conditions of stacking ten uninterrupted observations. This ranking assumes that the apparent X-ray radius of the planet is equal in size to the optical radius, which is true for a planet with a bare-rock surface and no appreciable atmosphere. However, planets with appreciable atmospheres are more likely to appear larger in the X-ray than in the optical. because X-rays have a higher absorption cross-section and are fully absorbed at higher altitudes, such as the exosphere or outer layers of an escaping atmosphere. This makes X-ray transits a more sensitive probe of a tenuous atmosphere than optical and infrared observations. Additionally, the majority of known transiting planets have short orbital periods and are thus subject to high-energy irradiation, which can trigger photoevaporative escape. X-rays transits measure the extent of these outflows, providing strong constraints on evaporation models. We focus on two of the top 15 targets in the study by Cilley et al. [36] that provide excellent laboratories for testing models of photoevaporative escape.

*Potential Targets*

- **HIP 65 A b** is the second most likely planet to be detected via X-ray transit, after HD 189733 b. This planet has a mass of about 3.2 $M_J$ and crosses the optical stellar disk at a grazing incidence angle, where the coronal X-ray emission is brightest, leading to a high-signal X-ray transit [86]. HIP 65 A b is also on an ultra-short period of 1 day, subjecting it to incredibly high irradiation, but also subjecting it to tidal dissipation forces that will cause it to spiral into its host star within 80-1000 Myr [115].
- **HIP 67522 b** is a Jupiter-sized but very low density planet ($\lesssim 0.1$ g cm$^{-3}$) that has been observed in transmission by JWST, which revealed strong molecular absorption features that were 30-50% deeper than the baseline transit depth, implying that it has a thin and extended atmosphere [158]. HIP 67522 b orbits a particularly young 17 Myr-old star, and has the highest estimated mass loss rate of all planets considered.
- As a benchmark for comparison, we also simulate AXIS observations of the **HD 189733 b** transit, which has already been confirmed with Chandra and XMM [134, Wheatley et al., in prep]

*A note about AU Mic b:* This is one of the smaller planets potentially observable in X-ray transit, but only if it has a large effective radius due to atmospheric escape. Studies to date have shown only marginal evidence for escape via Ly-$\alpha$ absorption, and such observations are hindered by high levels of stellar activity [56,146,147]. The exceptionally high X-ray brightness of AU Mic makes the scenario of witnessing strong atmospheric escape via X-ray transit accessible to modern-day observatories.

**Exposure time (ks):** 600 ks is sufficient to monitor 20 exoplanet transits for 30 ks each. Typical short-period exoplanet transits occur every 3-4 days, and the transit takes 3-4 hours (10-15 ks) to complete. Similar exposure coverage of the exoplanet host star's light curve during the out-of-transit period is required to obtain a reasonable baseline of comparison. Stacking 20 transits of the same target is sufficient to study extended outflows at the 3-sigma level for the targets considered below. Even non-detections will enable valuable constraints on atmospheric escape.

**Observing description:** It is extremely difficult to predict the apparent size of an escaping atmosphere, as evidenced by the unexpectedly extreme 50% Lyman-alpha transit observed from Neptune-sized GJ 436 b [mass loss rate of $10^8 - 10^9$g s$^{-1}$, 49]. Nonetheless, we can take some reasonable guesses based on the size of each planet's Roche lobe. We simulated AXIS observations under the assumption of no appreciable atmosphere ($R_X = R_{\rm opt}$, most difficult to observe), an atmosphere filling the Roche lobe (easiest to observe), and a point in between.

Following the methodology of Cilley et al. [36], we stacked 20 simulated observing datasets for each potential transit target, including HD 189733 b. To account for stellar activity, we randomly removed sections of the light curve of each target to simulate data gaps induced by masking stellar flares. We used the flare distributions of AU Mic [162] for simulations of HD 67522 b, and the flare distributions of HD 189733 b [133] for HIP 65 A b. We then measured the statistical significance (sigma-level) to which we could recover the transit model from the stacked light curve. We repeated this task 100 times and recorded the median sigma-level of significance from these hypothetical AXIS observing campaigns, reported in Table 6.

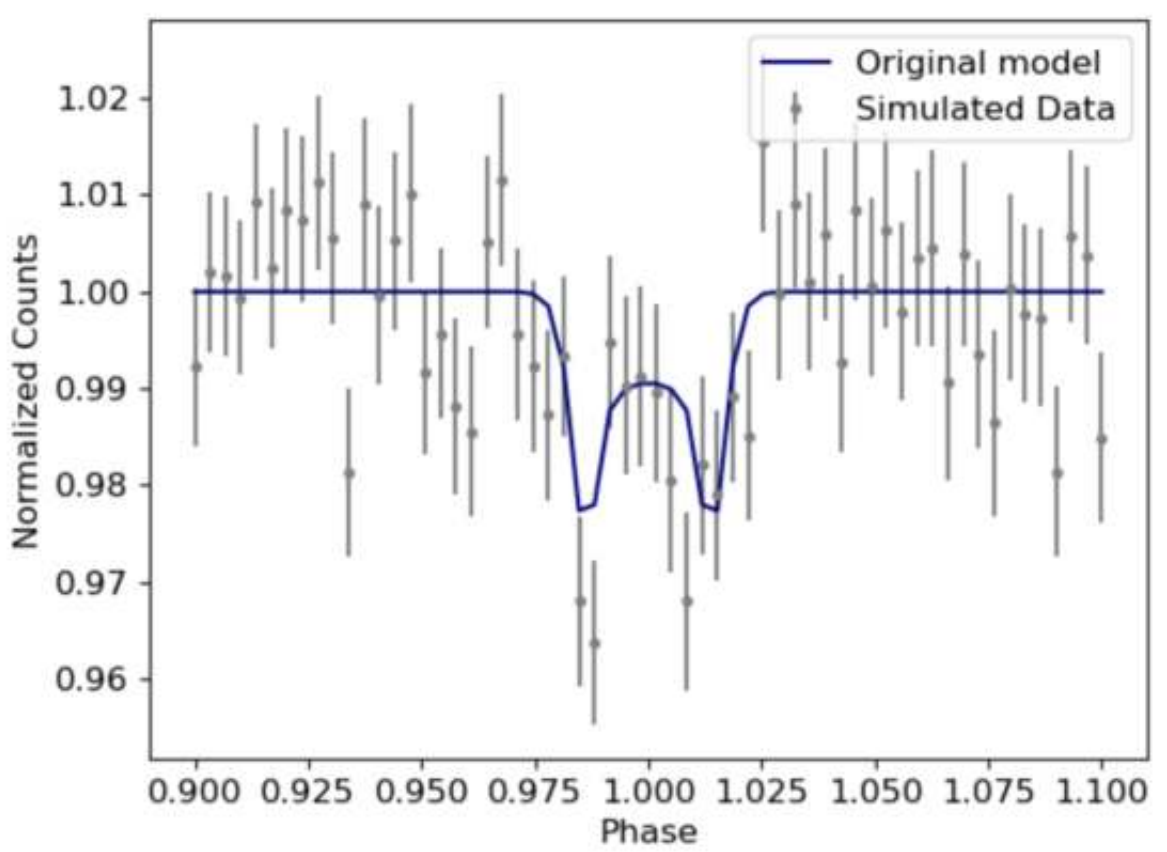

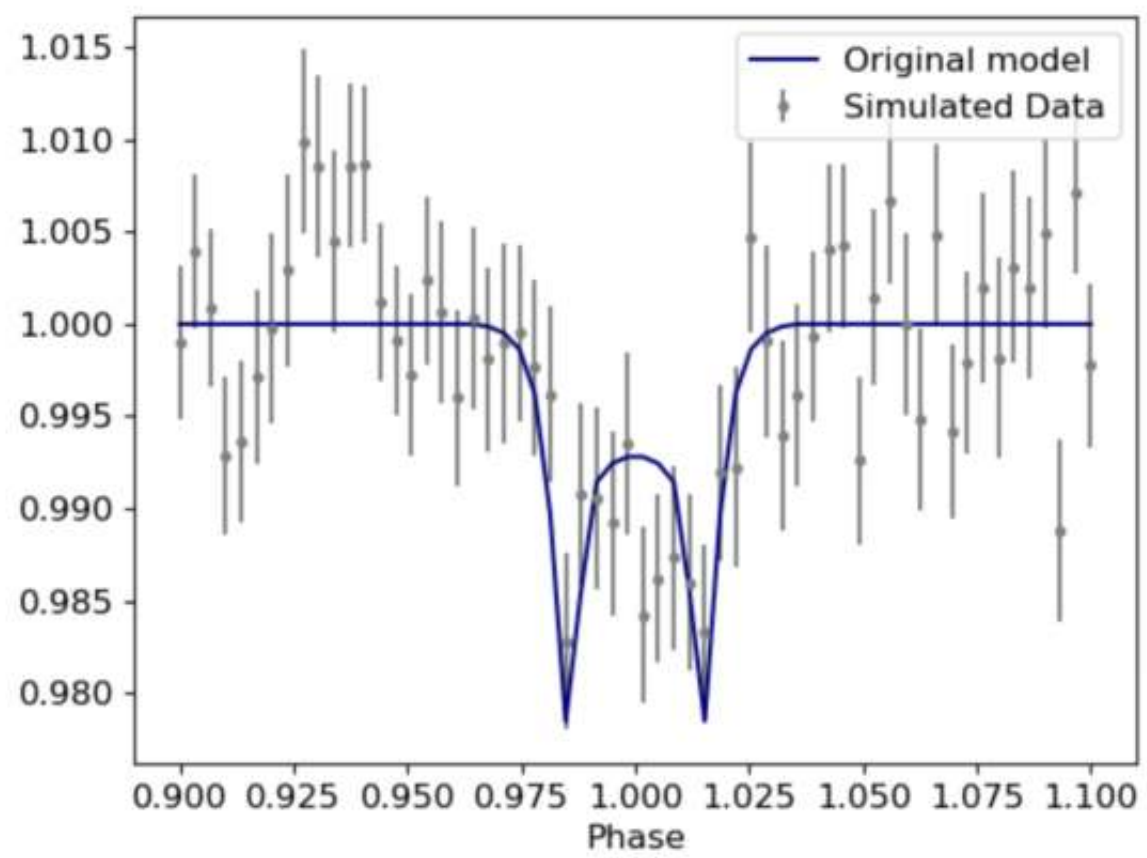

**Figure 16.** Simulated stacked AXIS light curve for HD 189733 b (left) with $R_X = R_{opt}$ and HIP 67522 b (right) with $R_X = 2.75 \times R_{opt}$.

**Table 6.** Detection significance after stacking 20 AXIS transit light curves

| **Planet** | $R_X$ | **Significance ($\sigma$)** | **AXIS ct/s** | **Estimated Mass-Loss (g s$^{-1}$)** |
|---|---|---|---|---|
| HD 189733 b | $R_{opt}$<br>$2 \times R_{opt}$<br>$3 \times R_{opt} = R_L$ | $> 4.4$<br>$> 4.4$<br>$> 4.4$ | 1.97 | $3 \times 10^{10}$ [89] |
| HIP 65 A b | $R_{opt}$<br>$1.4 \times R_{opt} = R_L$ | 3.9<br>$> 4.4$ | 0.16 | $7.0 \times 10^{11}$ |
| HIP 67522 b | $R_{opt}$<br>$1.5 \times R_{opt}$<br>$2.75 \times R_{opt} = R_L$ | 1.5<br>3.0<br>$> 4.4$ | 6.79 | $2 - 6 \times 10^{12}$ [158] |

The strong photoelectric absorption edges of C and O at 0.3 and 0.5 keV, respectively [167,172], provide an opportunity to study the photoevaporative efficiency of these elements by dividing transit observations into the relevant energy bands. If count rates are sufficient, we may also be able to extract a two-point X-ray transmission spectrum to study the abundance and vertical extent of C and O in an extended atmosphere. These will provide information on the thermal structure and composition of the upper atmospheres of these exoplanets.

**Specify critical AXIS Capabilities:**

- Soft X-ray sensitivity ($E < 2$ keV), where stars are brightest
- Angular resolution to resolve stellar binaries and field sources
- Ability to obtain an uninterrupted $\sim 30$ ks exposure
- Fast read out and PSF sub-sampling to avoid pileup

**Joint Observations and synergies with other observatories in the 2030s:** This study is synergistic with JWST (see HIP 67522 b) and ground-based ELTs studying exoplanet atmospheres in transmission or with phase curves. Complementary UV transit data – where metal absorption lines are expected – might also be obtained by UVEX (launching 2030).

**Special Requirements:** Uninterrupted $\sim$ 30 ks exposure at specific phases of the planetary orbit

*13. The AXIS search for planets transiting white dwarfs*

**Science Area:** Stars and Exoplanets
**First Author:** Tansu Daylan, Washington University, tansu@wustl.edu
**Co-authors:** Lia Corrales, University of Michigan Steven Dillmann, Stanford University Keivan Stassun, Vanderbilt University
**Abstract:** With its high effective area, the Advanced X-ray Imaging Satellite (AXIS) will offer unprecedented sensitivity to characterizing white dwarfs in soft X-rays. White dwarfs may harbor transiting planets if close-in planets around Sun-like stars survive post-main-sequence evolution, new planets form around the resulting white dwarfs, or outer planets migrate inward, potentially resulting in brief and deep transits in their light curves. Planetary transits will be individually detectable by AXIS for two exceptionally bright white dwarfs that are relatively hot and nearby. Moreover, periodic transits will be detectable via phase folding over a larger set of more than 20 white dwarfs, given sufficiently long temporal baselines in serendipitous observations. In addition, the Galactic Plane Survey (GPS) will discover hot white dwarfs in the galactic disk that could be adequately bright in soft X-rays for transit searches. While the resulting number of searchable white dwarfs is not sufficient to ensure a discovery considering the small geometric transit probabilities and potentially low occurrence rate of intact planets around white dwarfs, AXIS will uniquely complement optical and UV surveys to search for planets transiting white dwarfs, advancing their demographic studies and probing mechanisms of planet retention, inward migration, and secondary planet formation beyond main sequence.

**Science:** The Advanced X-ray Imaging Satellite [AXIS; 144] will provide unprecedented capabilities for studying planetary systems. While the ability of AXIS to characterize planetary hosts is covered in other sections, its unique capabilities leave a vast parameter space to explore. Here, we present exoplanet science cases that can constitute part of the AXIS Guest Observer program.

More than 97% of all low-to-intermediate mass stars in the Universe, including our Sun, will evolve toward a white dwarf at the end of their lifetimes. Unable to generate heat via nuclear fusion after the Asymptotic Giant Branch (AGB) phase, young white dwarfs cool from initial temperatures of more than 100,000 K to temperatures as low as ∼3000 K. The planetary architecture of a star is significantly affected by its evolution beyond the main sequence. The decreasing mass of the system causes planetary orbits to relax to higher semi-major axes, potentially leading to resonance crossings that destabilize the system's architecture. Significant increases in the stellar wind strengths also erode or completely evaporate close-in planets. As a result, planets may be engulfed, contaminating the atmosphere of the remnant star, or they may be ejected. Overall, post-main-sequence evolution may significantly reduce the abundance of planets around Sun-like stars. Nevertheless, the dense cores of planets that survive the stellar wind, or those that migrate inwards later, would continue to orbit the white dwarf close-in with a relatively high probability of transit. Moreover, post-main-sequence evolution can also lead to secondary planet formation. Therefore, truly contextualizing the Earth around the Sun also requires probing the ability of Sun-like stars to maintain their planetary systems or produce new planets, thereby placing the Earth within the cosmological context of stellar evolution. The discovery and characterization of planets around white dwarfs can help close an outstanding knowledge gap in exoplanet research, providing key insights into planet formation that would otherwise be impossible to obtain.

Our Sun, like other stars with convective envelopes, heats and sustains an extensive corona that emits optically thin bremsstrahlung extending into soft X-rays. Lower-mass stars can typically exhibit even stronger soft X-ray emission, despite having lower bolometric luminosities. A useful parameter that characterizes the high-energy emission of a star is $f_X$, the fraction of its soft X-ray luminosity, $L_X$, to its bolometric luminosity, $L_{Bol}$. For our Sun, $L_X$ is estimated as $\sim 2 \times 10^{27}$ erg/s, which leads to an $f_X$ of $6 \times 10^{-7}$ since the bolometric luminosity of the Sun is $3.8 \times 10^{33}$ erg/s. Therefore, at 10 parsecs, the soft X-ray flux of a Sun-like star would be $\sim 10^{-13}$ erg/s/cm$^2$. White dwarfs are ∼100 times smaller than the Sun and have $f_X$ even smaller than the Sun due to the absence of a convective surface and corona. However, the blackbody emission of white dwarfs peaks across optical to UV (i.e., ∼10 eV) with a Wien tail that extends to soft X-rays (0.2-0.5 keV) for sufficiently nearby, hot, and low-density white dwarfs that can give them soft X-ray brightnesses of $\sim 10^{-16} - 10^{-14}$ erg/s/cm$^2$.

AXIS will be sensitive to a limiting flux of $\sim 10^{-16}$ erg/s/cm$^2$ in a 20 ksec (5.6 hour) exposure. Therefore, it will only be sensitive to the brightest white dwarfs in the tail of the brightness distribution, making them rare targets amenable to transit searches. Several factors can cause a white dwarf to remain hot or become hotter. First, a young white dwarf with an age less than ∼500 million years would be expected to still be hot. Second, the white dwarf can be accreting material from a companion or a disrupted planet. Moreover, the white dwarf could also be the product of a recent merger of two white dwarfs, such as ZTF J190132.9+145808.7 [30]. However, if the white dwarf is accreting material from a stellar companion or is itself the remnant of a recent binary merger, it is unlikely to have retained a planet in a stable orbit, making these targets less interesting for transit searches. Another reason a white dwarf can be hot is if it is in the Q-branch, where the star's cooling becomes significantly delayed [**?** ].

A simulated image of a white dwarf being transited by a Jupiter-sized planet is illustrated in the right panel of Figure 17. For Earth-sized exoplanets with a size comparable to the white dwarf or even larger planets transiting a white dwarf, the transit is typically deep (i.e., ∼100% of baseline flux) and the light curve is V-shaped due to a typically large impact factor. The duty cycle of transits, i.e., duration of the primary transit divided by the orbital period, is typically small (∼0.2%) for white dwarfs due to their high densities. Therefore, transits of a white dwarf by a planet are brief, ∼200-second events for a fiducial orbital period of 1 day. Transits shorter than ∼200 seconds require planets to orbit much closer to the white dwarf, in which case they would not remain intact. A close-in planet can only remain intact outside the Roche limit,

$$d_R = AR_S \left( \frac{\rho_S}{\rho_p} \right)^{1/3} \tag{1}$$

within which the planet gets disrupted into a ring, where $R_S$ is the stellar radius, $\rho_S$ is the stellar density, $\rho_p$ is the planetary density, and $A$ is 1.3 and 2.4 for rocky and gaseous planets, respectively. The Roche limit corresponds to an orbital period of $\sim$ 10 hours for white dwarfs.

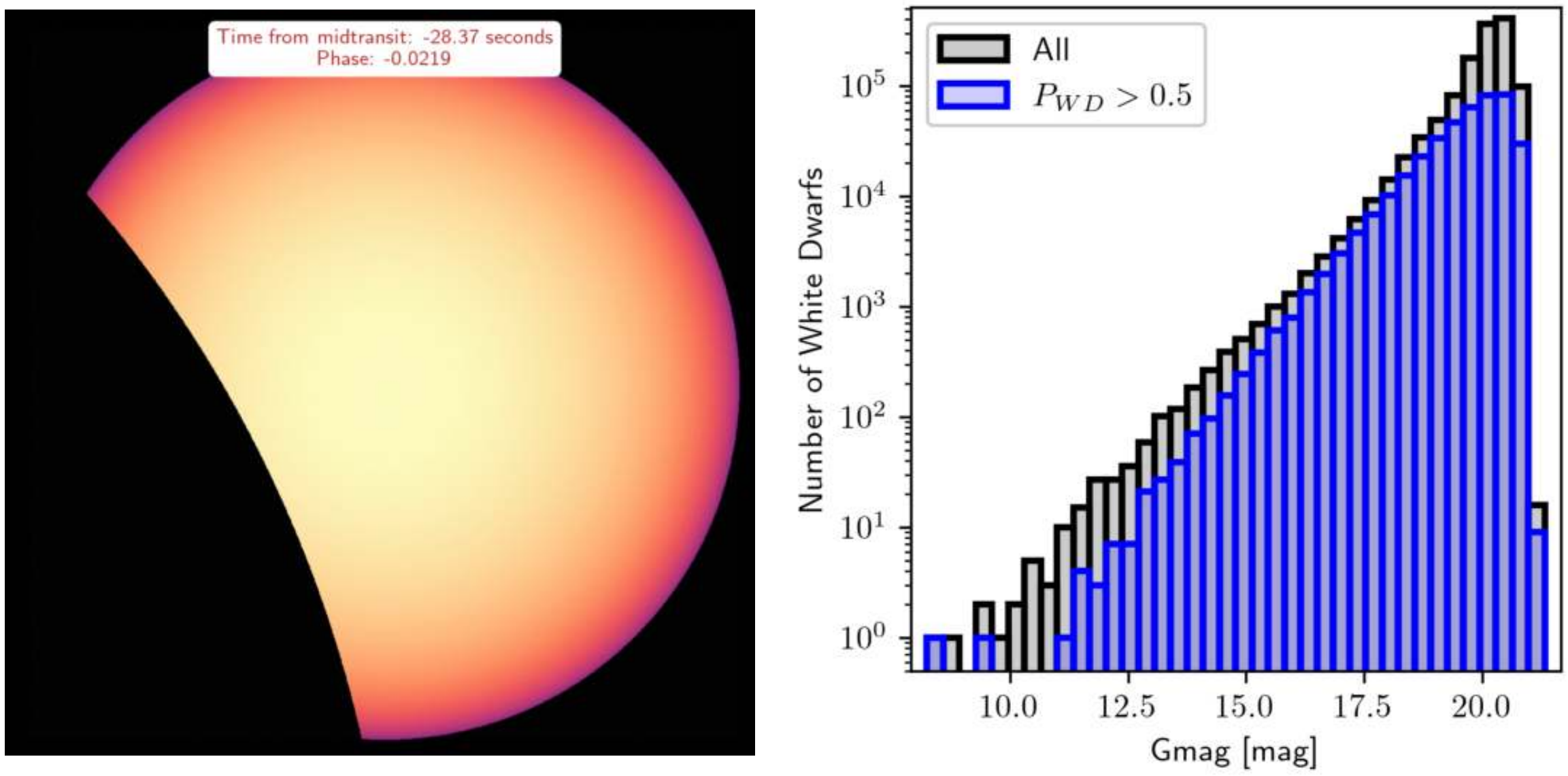

**Figure 17.** Left: A simulated image of a white dwarf during the ingress of a Jupiter-sized transiter planet. Right: G magnitude distribution of white dwarf candidates from [65]. Black shows the full catalog, whereas the blue histogram indicates those high-confidence candidates with a probability of less than 50% of being false identifications as white dwarfs.

Historically, planetary signatures around evolved stars have been observed as metal lines in the atmospheres of white dwarfs [181] and as disrupted planetary material orbiting white dwarfs [165]. However, the recent discovery of an intact transiting planet, WD 1856 b, using the Transiting Exoplanet Survey Satellite [TESS; 145] suggests that close-in intact planets must exist around white dwarfs [166]. Nevertheless, the abundance of these transiters around white dwarfs remains mostly unexplored due to the challenges with TESS photometry in crowded fields. Additional discoveries can be transformative for the exoplanet field, enabling the demographic study of exoplanets around post-main-sequence stars and revealing the rate of planet retention during stellar evolution, the conditions that determine inward migration after white dwarf formation, and potential signatures of secondary planet formation. In particular, the abundance of intact planets immediately outside the Roche limit may be indicative of the rate of planetary engulfment.

**Observing description:** To be sensitive to an individual transit of 200-second duration, one would require 9 counts (i.e., a signal-to-noise ratio of 3) in 200 seconds, which would lead to targets of interest to be brighter than $\sim 4 \times 10^{-15}$ erg/s/cm$^2$. This brightness limit and the inherent faintness of white dwarfs impose a constraint on the plausible target list for a transit search around white dwarfs in X-rays. The left panel of Figure 17 shows the catalog of 1.3 million white dwarf candidates based on Gaia EDR3 [65]. The distribution indicates that only two white dwarfs, WDJ041521.80-073929.20 and WDJ064509.30-164300.72, can support a targeted search for individual transits. However, it will be possible to detect repeating transits on fainter white dwarfs by phase-folding the light curve, as long as the temporal baseline of the observations is sufficiently long to cover multiple transits. For example, a 1 Msec observation can push the brightness limit to $\sim 4 \times 10^{-16}$ erg/s/cm$^2$, yielding $\sim$20 targets suitable for a transit search. While a 20 Msec survey would be challenging to justify for conducting transit searches alone, a multifaceted GO program that precisely characterizes the soft X-ray variability of these observationally favorable white dwarfs would constitute a compelling, compound science case.

In addition, some of the observations that can be used to search white dwarf transits may be enabled by serendipitous observations of the sufficiently bright white dwarfs collected within the field of view during target of opportunity (TOO) observations and other GO surveys that will constitute 10% and 70% of the available mission time, respectively. The minimum exposure time needed would be $\sim$500 seconds to be able to measure the transit baseline, which would be satisfied by most observations. Furthermore, the Galactic Plane Survey (GPS) will cover the galactic plane with a biannual cadence as part of the directed science, leading to the discovery of 1 million new objects, including previously unconfirmed white dwarfs, which are relatively bright and suitable for transit searches.

Because the transit duration scales proportionally to the orbital period, transits with longer orbital periods can produce longer transit durations at the expense of limiting the observation to at most a single transit. While the signal-to-noise ratio boost from single, long-duration transits could also extend detectability to fainter white dwarfs, the small geometric transit probability of these longer-period transiters would render them relatively unlikely. Furthermore, single transits across white dwarfs, which are typically faint, would require spectroscopic follow-up with 10-meter-class telescopes to confirm that they are due to a periodic transiter. The limitations of this demanding follow-up make the search for repeating transits a more feasible approach for AXIS to make a transit discovery.

**Specify critical AXIS Capabilities:** AXIS will be particularly useful for discovering and characterizing exoplanets transiting white dwarfs due to its high ($\sim$3600 cm$^2$) effective area for soft (0.2-2 keV) X-rays with a reasonably good (1.5 arcsecond FWHM) point spread function (PSF). Furthermore, the L2 orbit provides an observational baseline that enables long light curves to be collected.

**Joint Observations and synergies with other observatories in the 2030s:** AXIS observations of white dwarfs will be synergistic with several upcoming surveys that have sensitivity to white dwarfs, including the Legacy Survey of Space and Time (LSST) at the Rubin Observatory [82], the Roman Space Telescope [155], and ULTRASAT [14].

LSST will soon start collecting deep photometry over the entire southern sky. Because the LSST observation strategy is not optimized for transits, it will sample phase curves of transiting systems at random phases, resulting in thousands of photometric outliers whose periodicity will be challenging to establish. While Rubin will be a much deeper survey than AXIS, it will also be crowded, making white dwarf transits particularly challenging to confirm solely from LSST data. AXIS will have a PSF Full Width at Half Maximum FWHM) comparable to that of the seeing of the Simonyi Telescope. The detection of even a single transit by AXIS will enable periodic photometric outliers from LSST to be confirmed as transiting white dwarfs.

Furthermore, ULTRASAT, which is currently being developed for an expected launch in 2028, will have a passband in the near-ultraviolet (NUV) range, between 230-290 nm, making it a particularly useful survey for white dwarfs. Being a wide-angle survey, however, its $5''$ pixels will suffer from source blending and confusion. As a result, even marginal detections will enable AXIS to confirm planet candidates transiting hot white dwarfs.

As an infrared telescope, the Roman Space Telescope will yield lower contrast for white dwarfs compared to ULRASAT and even the optimal bands of LSST. Nevertheless, its high-cadence galactic bulge survey will offer a unique opportunity to search for transits. AXIS will be able to confirm marginal or potentially blended transit detections around hot white dwarfs.

**Special Requirements:** This science case does not require Target of Opportunity (TOO) observations. Conducting the observations of a given target continuously would be optimal for minimizing aliasing in periodicity searches. However, such periodicity degeneracies can also be lifted using light-curve data from other surveys, such as LSST, Roman, and ULTRASAT.

*14. Solar System Science with AXIS*

**Science Area:** Stars & Exoplanets

**Author:** Scott J. Wolk (Center for Astrophysics | Harvard & Smithsonian), swolk@cfa.harvard.edu

**Co-authors:** Carey M. Lisse (Johns Hopkins University, Carey.Lisse@jhuapl.edu)

**Abstract:** AXIS's high spatial resolution, combined with its large effective area and broad energy response, will be very impactful for solar system science. Below, we give four short examples highlighting the wide range of science and physics that can be explored. Regarding Mars, Chandra detected both scattering from the planet and Charge Exchange from the escaping atmosphere. Due to the increased effective area at low energy, AXIS is increased 10-fold the S/N of the escaping photons. AXIS will allow an independent measurement of the Martian atmospheric loss rate to add to the MAVEN mission and determine if there is a seasonal asymmetry in the orientation of the escaping atoms. For comets and other sources of charge exchange distributed throughout the heliosphere, AXIS's enhanced low-energy resolution will allow us to discern the ionization state and speed of the solar wind's ions as they travel to the heliopause, at a precision not possible with Chandra. Regarding Jupiter, spatial resolution and effective area are both required to understand the $>5$ keV coronal emission detected by XMM-Newton, but not understood. Finally, come Neptune and the Kuiper belt objects (KBOs). 175 ksec of Chandra observations were used to detect Pluto's weak atmosphere and cometary tail of Pluto based on eight oxygen photons in a 5-$\sigma$ detection, and for Neptune, scattering from its atmospheric disk with 20 photons in 100 ksec for a 12-$\sigma$ detection. Similar observations with AXIS could achieve S/N $> 50$ in 75ksec of observing time, and allow for discernment of these outer system's extended atmospheres and mass loss rates.

**Science:** In this presentation, we will detail two observations. One of a random comet and one of Jupiter. As of now, six planets, the moon, and Pluto have been observed by Chandra and XMM, along with over a dozen comets.

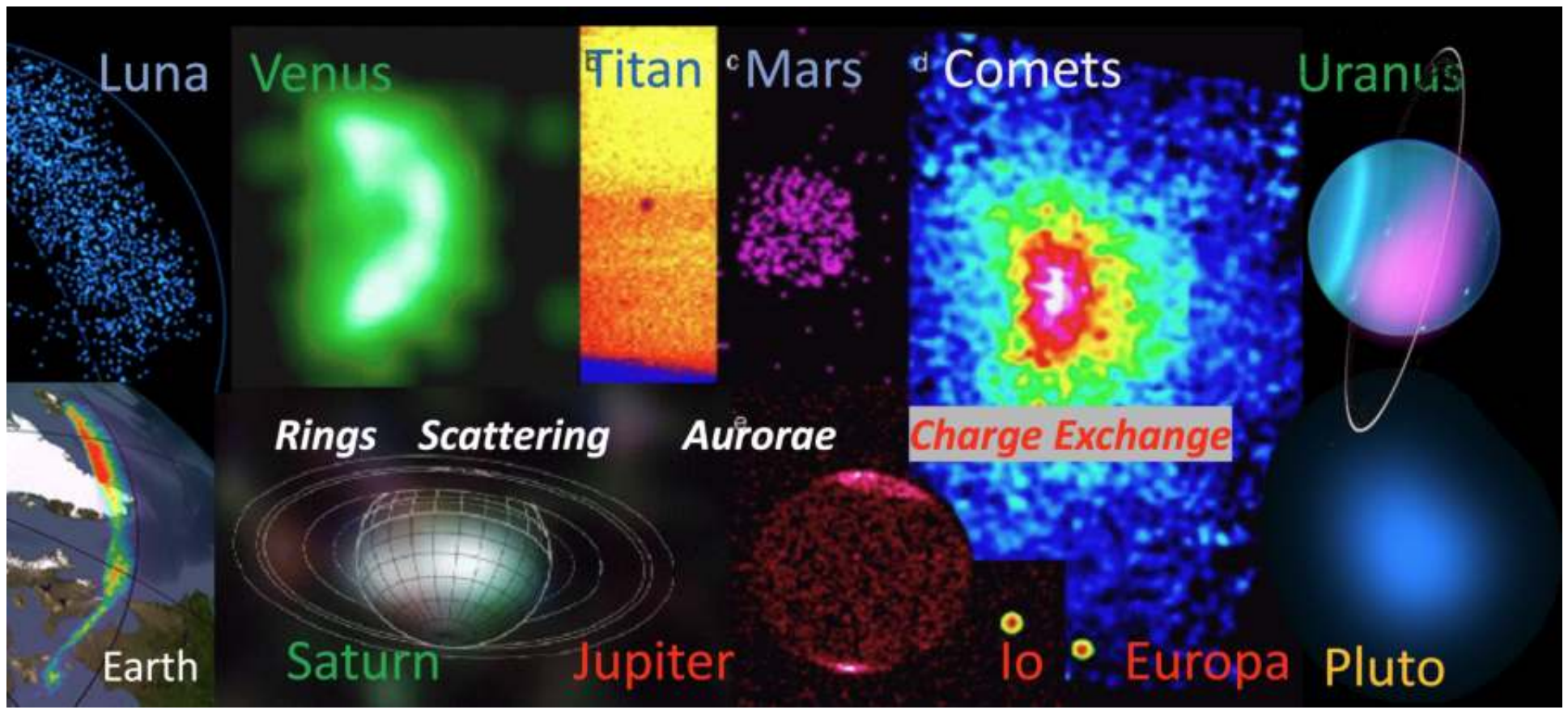

**Figure 18.** Some Chandra observations of Solar system objects indicate a range of physical processes. Most of the images (Jupiter system excepted) are photon-limited in their resolution, meaning that AXIS will be able to deliver sharper images in addition to higher signal-to-noise.

*Example 1 - Comets:* The first comet was detected by in X-rays by ROSAT. These were photometric observations of comet C/Hyakutake 1996 B2 in 1996 as it sped through the ecliptic plane close to Earth. The emission was found in a crescent shape offset sunward of the nucleus, which seemed to expand and change morphology as the comet retreated southward of the ecliptic away from the Sun and Earth [96]. Later, with the very first comet observed by Chandra in 2000, comet C/1999 LS4, the mystery of how icy objects with internal temperatures 30 K and surface temperatures on the order of 300K could emit appreciable amounts of X- ray radiation was solved [95]. Using ACIS-S spectroscopy, line emission, especially the OVII triplet at 560-580 eV, due to Charge Exchange Emission (CXE) between highly stripped hydrogenic and heliogenic solar wind ions and gas was found in the comet's extended neutral coma atmosphere. This CXE exhibited an unusually high OVII (f+i)/r ratio [13], and weaker lines due to CV, NVI, and OVIII, due to the preponderance of stripped CNO ions in the solar wind.

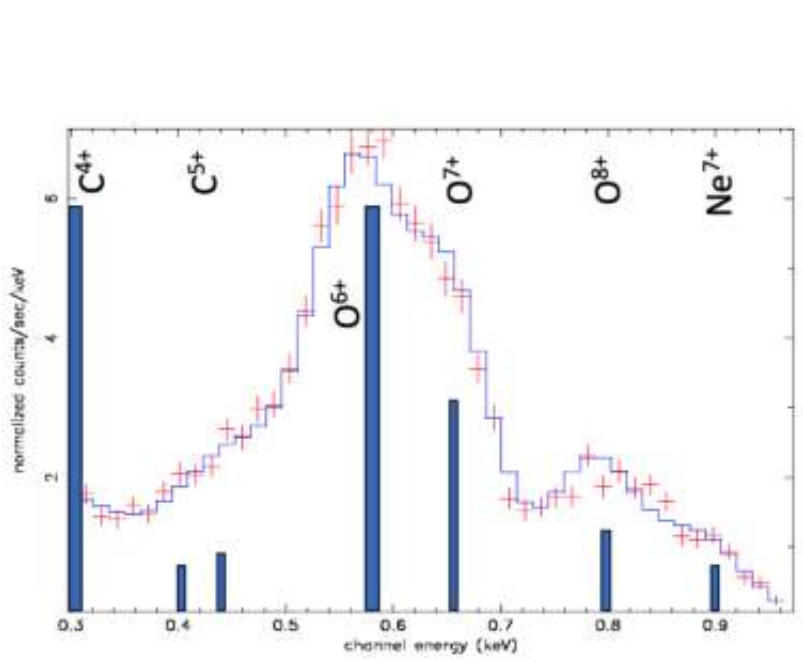

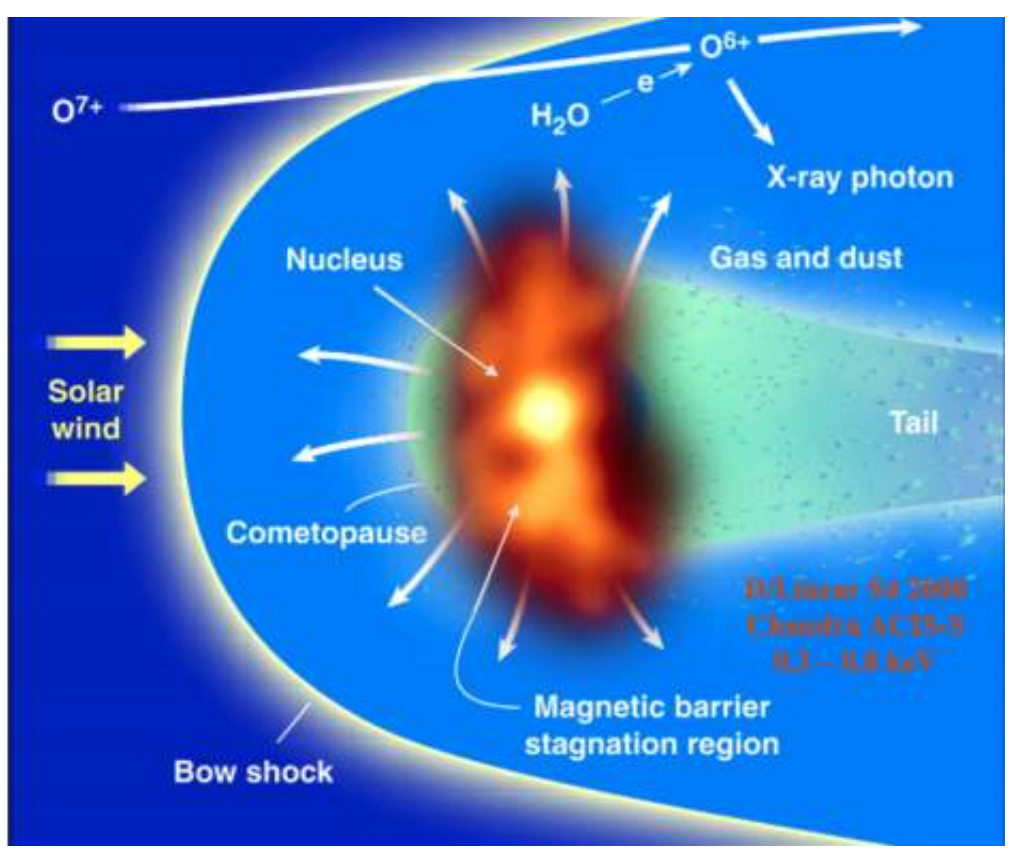

**Figure 19.** Right: A graphical description of the CXE process. The Chandra image of D/Linear S4 is overlain. Left: The inferred lines (blue bars) superimposed on the CCD Spectrum (Raw data are red "+" and the model fit is in blue) [adapted from 38].

With subsequent comets observed by Chandra, this model for CXE x-rays from comets near the ecliptic has been verified versus in situ ACE solar wind ion measurements [20,35,97,173]. After explaining the cause of the emission, one of the primary goals of observing cometary CXE has been to utilize X-ray measurements to understand CXE in other environments where hot, highly ionized gas encounters cold neutral gas, such as the heliosphere/ISM interface, at the boundaries of HII regions, or in massive star-forming areas. To do this well requires observations of diverse comet-neutral gas sources encountering as wide a range of solar wind environments as possible.

Our understanding of cometary X-ray emission and the structure of the solar wind argues that while any specific measurements should be typical (e.g. a highly active comet that is collisionally thick to the hot equatorial wind or, low gas comet in the cool polar wind), the conditions are highly variable and often unusual, as the comet encounters varying admixtures of the cold polar wind. We can use early Chandra observations to baseline a future comet's behavior, allowing us to compare it directly to the other low-latitude comets we have observed and to SWFO-L1/ACE in situ measures of the solar wind. This allows us to directly test the theory; we may expect a given comet to encounter some of the low-density and low-ionization-temperature polar solar wind, which should be entirely lacking in OVIII ions, deficient in OVII ions, and rich in OVI ions. This solar wind composition should lead to a spectrum with no OVIII lines and very weak OVII lines, It should also lead to unusually strong OVI emission. We will also be able to search for weaker line emission due to highly stripped C, N, Fe, Mg, Ne, and Si solar wind ions that are normally dominated by OVII and OVIII at 500 – 800 eV in these transition winds.

All existing comet observations by Chandra are "photon-starved." The meaning here is that while we may have detected 1000 photons, from a given comet, those photons were spread over $\sim$ 10,000 pixels (a box 50" on a side). In this case, to detect four photons (S/N$\sim$ 3) requires 40 pixels which is a box 3.5" on a side. So, the best possible resolution in the example case is 3.5". For the same comet and same exposure time with AXIS, we will achieve 10,000 counts across the same number of pixels and achieve instrumental resolution.

*Example 2 - Jupiter:* Jupiter's X-ray emissions were first discovered in 1979 by the Einstein Observatory [105], with later observations conducted by ROSAT [163], the Chandra X-ray Observatory [171], and XMM-Newton [83]. These observations identified two distinct sources of X-ray emission on the planet: the equatorial disk and the polar auroral regions. The X-ray emission from the disk is primarily generated by fluorescent scattering of solar X-rays in Jupiter's upper atmosphere, making it highly sensitive to variations in solar activity [18,27,39,44].

Observations from Chandra's ACIS-S instrument revealed that Jupiter's auroral X-ray spectrum consists primarily of line emissions from highly charged oxygen ions, including a significant portion of fully stripped ions. Additional line emissions at lower energies (0.31–0.35 keV and 0.35–0.37 keV) may originate from sulfur and/or carbon [50]. If carbon is the source, it would indicate a solar wind origin. The presence of highly charged ions suggests that significant acceleration must have occurred, regardless of whether the ions originated in the solar wind or Jupiter's magnetosphere. Rather than showing regular periodic oscillations, the auroral X-ray emissions exhibited chaotic variability, with power concentrated in the 20–70 minute range. Similar variability patterns were observed in Ulysses radio data collected at 2.8 AU from Jupiter during the same period [50]. One proposed explanation for this shift from organized to chaotic variability is pulsed magnetic reconnection at the dayside magnetopause, where magnetospheric and magnetosheath field lines interact, a mechanism suggested by Bunce et al. [29]. This process, which occurs at Jupiter's magnetospheric cusps, could accelerate ions regardless of whether they originate from the solar wind or the planet's magnetosphere. However, the ACIS observations were constrained by optical light contamination, leading to all subsequent high-spatial resolution follow-up studies being conducted with Chandra's HRC instrument, which lacks the energy resolution needed to distinguish line emission from continuum emission.

Current state-of-the-art issues include:

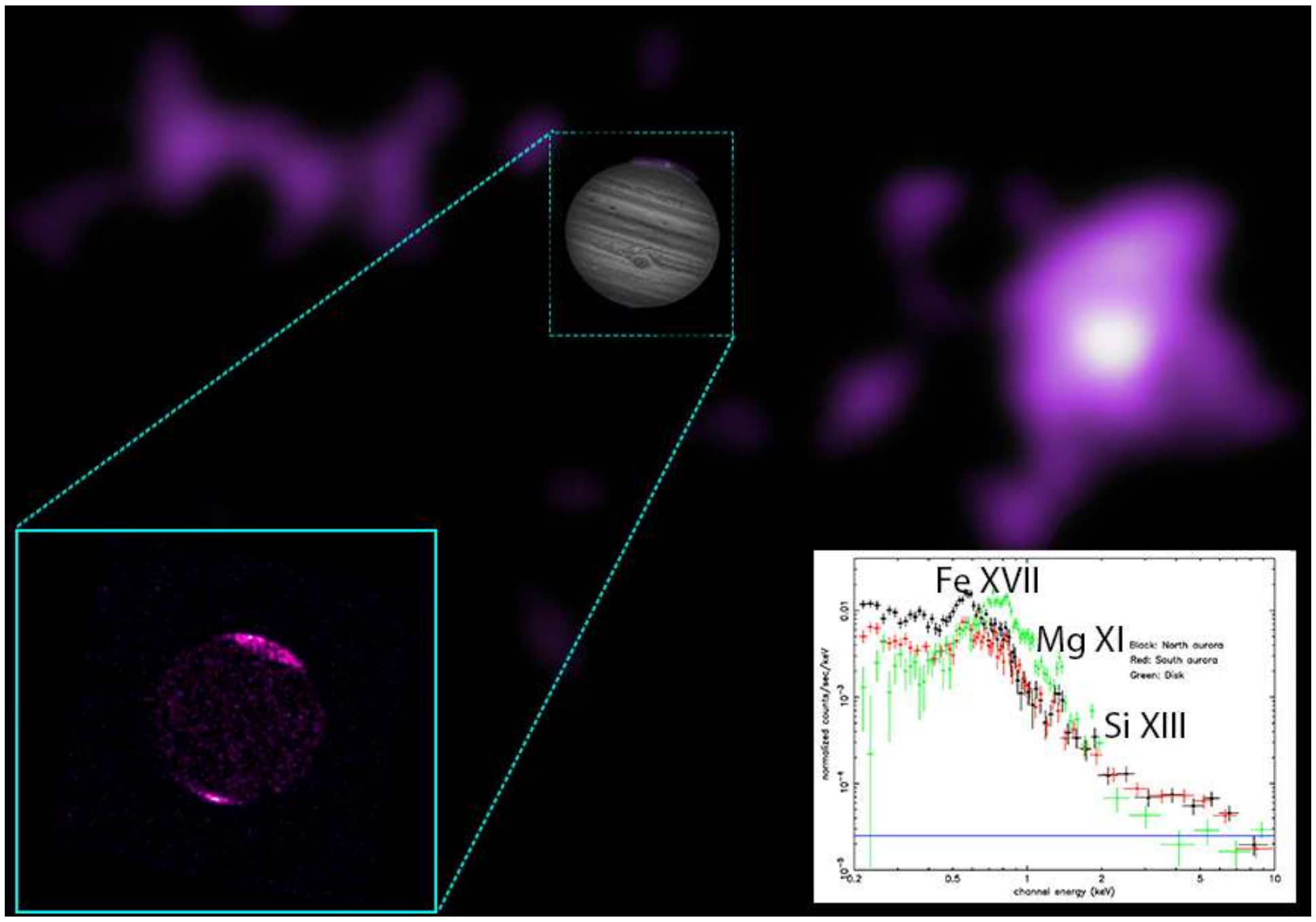

**Figure 20.** A multiscale smooth image of Jupiter reveals the Io plasma Torus and auroral emission from Jupiter. The left inset also shows disk emission, and the right inset shows spectral differences between the disk and the poles.

1. Various statistically significant quasi-periodic pulsations with a 25/40-minute period, which are likely linked to the arrival of a solar-wind compression.
2. Correlating disk X-ray emissions with surface magnetic field strength. This allows AXIS to monitor surface fields in the absence of *in situ* instruments and to augment and calibrate in situ observations.
3. X-rays map close to the magnetopause from noon to dusk, with the center of the averaged hot spot emissions mapping to noon, suggesting that the X-ray driver(s) may be linked with ultra-low-frequency wave activity along the magnetopause.
4. Emission along the Io plasma Torus.
5. Emission from the moons, especially Io and Europa.

**Exposure time (ks): 600**
**Observing description:**

- For a bright comet 10-20 ks will be able to achieve 1.5″ resolution. We expect about 2 of these in the 5-year baseline mission.
- Jupiter, the atomic unit for a Jupiter observation is one Jovian day, about 36 ks. One can imagine multiple coordinated observations with the Europa Clipper. The day length on Saturn is similar.
- Venus and Mars are relatively bright in X-rays, but about 10 ks will still be required to obtain enough photons to resolve the disks fully. Specifically, the 2033, 2035, and 2037 apparitions of Mars at opposition will be of considerable interest, as will the related quadrature observations.
- Uranus, Neptune, and Pluto each have required about 150 ks to obtain a handful of photons. In the case of Pluto, we strongly believe the emission is extended, so these objects will need to be resolved. I would plan on 100ks each.

The above leads to a rough estimate of 600 ks of observation of Solar System objects for a 5-year primary mission.

All solar system body positions in the sky need to be calculated from the perspective of AXIS. They all move to various degrees during observation. For bright comets, 10 ks should be sufficient to achieve 1.5″ resolution. We expect about 2 of these in the 5-year baseline mission. These can be rapidly moving about 1′/ks. So, over the 10 ks, we should not need to repoint, but software needs to be available to reproject the photons into the fixed planetary frame of reference.

**Specify critical AXIS capabilities:**

- Sun avoidance $<$ 46 degrees to observe Venus.
- An optical blocking without a "red leak."
- No Moon or Earth avoidance zone if you wish to study Earth Aurorae or lunar emission, which might include lunar geology and CXE in the "Earth-glow."
- The guide system should have a gyro hold (without a guide camera) for the Earth and Moon, and/or a tracking mode for moving planets and comets. This requires an ability to reconstruct pointing post facto based on gyro data.
- An ability to observe along the ecliptic, including opposition and quadrature. Both *Chandra* and *XMM* have difficulties at various points in this.
- A guide camera can handle bright sources in the field of view. Even if we are pointing at the moon or Earth on gyros, Venus, Jupiter, and Mars can swamp the *Chandra* guide camera.
- Low energy sensitivity, especially 300-500 eV.

**Joint Observations and synergies with other observatories in the 2030s:** NASA's Europa Clipper is en route and will orbit Jupiter starting in 2030. NASA's Dragonfly quadcopter drone mission to Titan is planned to arrive at Saturn in 2034. ESA's ExoMars Rover is expected to be in operation starting in 2030. NASA's DAVINCI, VERITAS, and ESA's ENVISION are all scheduled to launch towards Venus in 2031. There is also the stated objective of the current administration to land humans on Mars in the 2030s.

**Special Requirements:** We expect each reachable planet to be observed multiple times as planetary characteristics change according to their own seasons, literally. There is also the impact of changing view angle, solar cycles, and possible coordination with a planetary orbiter.

1. Aarnio, A. N., Matt, S. P., & Stassun, K. G. 2012, ApJ, 760, 9
2. Aarnio, A. N., Stassun, K. G., Hughes, W. J., & McGregor, S. L. 2011, , 268, 195
3. Airapetian, V. S., Barnes, R., Cohen, O., et al. 2020, International Journal of Astrobiology, 19, 136
4. Albacete Colombo, J. F., Caramazza, M., Flaccomio, E., Micela, G., & Sciortino, S. 2007, A&A, 474, 495
5. Anastasopoulou, K., Guarcello, M. G., Flaccomio, E., et al. 2024, A&A, 690, A25
6. Andre, P., Ward-Thompson, D., & Barsony, M. 2000, in Protostars and Planets IV, ed. V. Mannings, A. P. Boss, & S. S. Russell, 59
7. Andrews, S. M., Huang, J., Pérez, L. M., et al. 2018, ApJ, 869, L41
8. Ayres, T. 2025, AJ, 169, 281
9. Bally, J., Feigelson, E., & Reipurth, B. 2003, ApJ, 584, 843
10. Banzatti, A., Pascucci, I., Bosman, A. D., et al. 2020, ApJ, 903, 124
11. Baraffe, I., Vorobyov, E., & Chabrier, G. 2012, ApJ, 756, 118
12. Behr, P. R., France, K., Brown, A., et al. 2023, AJ, 166, 35
13. Beiersdorfer, P., Boyce, K. R., Brown, G. V., et al. 2003, Science, 300, 1558
14. Ben-Ami, S., Shvartzvald, Y., Waxman, E., et al. 2022
15. Ben-Jaffel, L., & Ballester, G. E. 2013, A&A, 553, A52
16. Bernini-Peron, M., Sander, A. A. C., Najarro, F., et al. 2025, A&A, 697, A41
17. Bhandare, A., Kuiper, R., Henning, T., et al. 2020, A&A, 638, A86
18. Bhardwaj, A., Branduardi-Raymont, G., Elsner, R. F., et al. 2005, , 32, L03S08
19. Binder, B. A., Peacock, S., Schwieterman, E. W., et al. 2024, ApJS, 275, 1
20. Bodewits, D., Christian, D. J., Torney, M., et al. 2007, A&A, 469, 1183
21. Böhm-Vitense, E. 2007, ApJ, 657, 486
22. Bonanno, A., & Corsaro, E. 2022, ApJ, 939, L26
23. Bonito, R., Orlando, S., Miceli, M., et al. 2010, A&A, 517, A68
24. —. 2011, ApJ, 737, 54
25. Bonito, R., Orlando, S., Peres, G., et al. 2010, A&A, 511, A42
26. Boro Saikia, S., Marvin, C. J., Jeffers, S. V., et al. 2018, A&A, 616, A108
27. Branduardi-Raymont, G., Bhardwaj, A., Elsner, R. F., & Rodriguez, P. 2010, A&A, 510, A73
28. Brun, A. S., & Browning, M. K. 2017, Living Reviews in Solar Physics, 14, 4
29. Bunce, E. J., Cowley, S. W. H., & Yeoman, T. K. 2004, Journal of Geophysical Research (Space Physics), 109, A09S13
30. Caiazzo, I., Burdge, K. B., Fuller, J., et al. 2021, Nature, 595
31. Carlsson, M., Fletcher, L., Allred, J., et al. 2023, A&A, 673, A150
32. Castor, J. I., Abbott, D. C., & Klein, R. I. 1975, ApJ, 195, 157
33. Charbonneau, P. 2020, Living Reviews in Solar Physics, 17, 4
34. Cheng, H., Liu, B. F., Liu, J., et al. 2020, MNRAS, 495, 1158
35. Christian, D. J., Bodewits, D., Lisse, C. M., et al. 2010, ApJS, 187, 447
36. Cilley, R., King, G. W., & Corrales, L. 2024, AJ, 168, 177
37. Cliver, E. W., & Dietrich, W. F. 2013, Journal of Space Weather and Space Climate, 3, A31
38. Cravens, T. E. 2002, Science, 296, 1042
39. Cravens, T. E., Clark, J., Bhardwaj, A., et al. 2006, Journal of Geophysical Research (Space Physics), 111, A07308
40. Crowther, P. A., Broos, P. S., Townsley, L. K., et al. 2022, MNRAS, 515, 4130
41. dos Santos, L. A., Ehrenreich, D., Bourrier, V., et al. 2019, A&A, 629, A47
42. Dos Santos, L. A., García Muñoz, A., Sing, D. K., et al. 2023, AJ, 166, 89
43. Dunham, M. M., Stutz, A. M., Allen, L. E., et al. 2014, in Protostars and Planets VI, ed. H. Beuther, R. S. Klessen, C. P. Dullemond, & T. Henning, 195
44. Dunn, W. R., Branduardi-Raymont, G., Carter-Cortez, V., et al. 2020, Journal of Geophysical Research (Space Physics), 125, e27219

45. Duvvuri, G. M., Pineda, J. S., Berta-Thompson, Z. K., et al. 2021, ApJ, 913, 40
46. Eddy, J. A. 1976, Science, 192, 1189
47. Egeland, R., Soon, W., Baliunas, S., et al. 2017, ApJ, 835, 25
48. Ehrenreich, D., Lecavelier Des Etangs, A., Hébrard, G., et al. 2008, A&A, 483, 933
49. Ehrenreich, D., Bourrier, V., Wheatley, P. J., et al. 2015, Nature, 522, 459
50. Elsner, R. F., Ramsey, B. D., Waite, J. H., et al. 2005, Icarus, 178, 417
51. Ercolano, B., Weber, M. L., & Owen, J. E. 2018, MNRAS, 473, L64
52. Favata, F., Bonito, R., Micela, G., et al. 2006, A&A, 450, L17
53. Favata, F., Flaccomio, E., Reale, F., et al. 2005, ApJS, 160, 469
54. Favata, F., & Schmitt, J. H. M. M. 1999, A&A, 350, 900
55. Feigelson, E. D., Getman, K., Townsley, L., et al. 2005, ApJS, 160, 379
56. Feinstein, A. D., France, K., Youngblood, A., et al. 2022, AJ, 164, 110
57. Flaccomio, E., Micela, G., Sciortino, S., et al. 2005, ApJS, 160, 450
58. Foley, B. J., & Smye, A. J. 2018, Astrobiology, 18, 873
59. Fossati, L., Haswell, C. A., Froning, C. S., et al. 2010, ApJ, 714, L222
60. France, K., Loyd, R. O. P., Youngblood, A., et al. 2016, ApJ, 820, 89
61. Fulton, B. J., Petigura, E. A., Howard, A. W., et al. 2017, AJ, 154, 109
62. Gaches, B. A. L., Tan, J. C., Rosen, A. L., & Kuiper, R. 2024, A&A, 692, A219
63. Gaidos, E. 2015, ApJ, 804, 40
64. Gaudin, T. M., Coe, M. J., Kennea, J. A., et al. 2024, MNRAS, 534, 1937
65. Gentile Fusillo, N. P., Tremblay, P. E., Cukanovaite, E., et al. 2021, Monthly Notices of the Royal Astronomical Society, 508
66. Geroux, C., Baraffe, I., Viallet, M., et al. 2016, A&A, 588, A85
67. Getman, K. V., Flaccomio, E., Broos, P. S., et al. 2005, ApJS, 160, 319
68. Grosso, N., Feigelson, E. D., Getman, K. V., et al. 2006, A&A, 448, L29
69. Grosso, N., Hamaguchi, K., Principe, D. A., & Kastner, J. H. 2020, A&A, 638, L4
70. Güdel, M. 2004, A&A Rev., 12, 71
71. Güdel, M., & Nazé, Y. 2009, A&A Rev., 17, 309
72. Güdel, M., Skinner, S. L., Audard, M., Briggs, K. R., & Cabrit, S. 2008, A&A, 478, 797
73. Güdel, M., Skinner, S. L., Briggs, K. R., et al. 2005, ApJ, 626, L53
74. Güdel, M., Telleschi, A., Audard, M., et al. 2007, A&A, 468, 515
75. Güdel, M., Audard, M., Bacciotti, F., et al. 2011, in Astronomical Society of the Pacific Conference Series, Vol. 448, 16th Cambridge Workshop on Cool Stars, Stellar Systems, and the Sun, ed. C. Johns-Krull, M. K. Browning, & A. A. West, 617
76. Günther, H. M., Li, Z.-Y., & Schneider, P. C. 2014, ApJ, 795, 51
77. Günther, H. M., Matt, S. P., & Li, Z. Y. 2009, A&A, 493, 579
78. Günther, H. M., Pasham, D., Binks, A., et al. 2024, ApJ, 977, 6
79. Hartigan, P., Edwards, S., & Ghandour, L. 1995, ApJ, 452, 736
80. Huenemoerder, D. P., Gayley, K. G., Hamann, W. R., et al. 2015, ApJ, 815, 29
81. Indahl, B., & Wilson, D. 2022, MANTIS: Monitoring Activity from Nearby sTars with uv Imaging and Spectroscopy, NASA Prop. ID 22-APRA22-121
82. Ivezić, , Kahn, S. M., Tyson, J. A., et al. 2019, The Astrophysical Journal, 873
83. Jansen, F., Lumb, D., Altieri, B., et al. 2001, A&A, 365, L1
84. Käpylä, P. J. 2022, ApJ, 931, L17
85. Karmakar, S., Naik, S., Pandey, J. C., & Savanov, I. S. 2023, MNRAS, 518, 900
86. King, G. W., Corrales, L. R., Wheatley, P. J., Cilley, R. C., & Hollands, M. 2024, AJ, 168, 262
87. King, G. W., Wheatley, P. J., Salz, M., et al. 2018, MNRAS, 478, 1193
88. Koskinen, T. T., Lavvas, P., Harris, M. J., & Yelle, R. V. 2014, Philosophical Transactions of the Royal Society of London Series A, 372, 20130089

89. Kubyshkina, D., Fossati, L., Erkaev, N. V., et al. 2018, A&A, 619, A151
90. Lamberts, A., Millour, F., Liermann, A., et al. 2017, MNRAS, 468, 2655
91. Lammer, H., Selsis, F., Ribas, I., et al. 2003, ApJ, 598, L121
92. Lampón, M., López-Puertas, M., Lara, L. M., et al. 2020, A&A, 636, A13
93. Lecavelier Des Etangs, A., Ehrenreich, D., Vidal-Madjar, A., et al. 2010, A&A, 514, A72
94. Linsky, J. L., Yang, H., France, K., et al. 2010, ApJ, 717, 1291
95. Lisse, C. M., Christian, D. J., Dennerl, K., et al. 2001, Science, 292, 1343
96. Lisse, C. M., Dennerl, K., Englhauser, J., et al. 1996, Science, 274, 205
97. Lisse, C. M., Christian, D. J., Wolk, S. J., et al. 2013, Icarus, 222, 752
98. Lopez, E. D., Fortney, J. J., & Miller, N. 2012, ApJ, 761, 59
99. Lothringer, J. D., Fu, G., Sing, D. K., & Barman, T. S. 2020, ApJ, 898, L14
100. Malsky, I., Rogers, L., Kempton, E. M. R., & Marounina, N. 2023, Nature Astronomy, 7, 57
101. Malsky, I., & Rogers, L. A. 2020, ApJ, 896, 48
102. Mamajek, E. E., & Stapelfeldt, K. R. 2024, in LPI Contributions, Vol. 3068, LPI Contributions, 2089
103. Mazeh, T., Holczer, T., & Faigler, S. 2016, A&A, 589, A75
104. Metcalfe, T. S., Finley, A. J., Kochukhov, O., et al. 2022, ApJ, 933, L17
105. Metzger, A. E., Luthey, J. L., Gilman, D. A., et al. 1983, J. Geophys. Res., 88, 7731
106. Murabito, M., Stangalini, M., Laming, J. M., et al. 2024, PhRvL, 132, 215201
107. Murray-Clay, R. A., Chiang, E. I., & Murray, N. 2009, ApJ, 693, 23
108. Nazé, Y. 2009, A&A, 506, 1055
109. Nazé, Y., Hartwell, J. M., Stevens, I. R., et al. 2002, ApJ, 580, 225
110. Nazé, Y., Petit, V., Rinbrand, M., et al. 2014, ApJS, 215, 10
111. Nazé, Y., Wang, Q. D., Chu, Y.-H., Gruendl, R., & Oskinova, L. 2014, ApJS, 213, 23
112. Nebot Gómez-Morán, A., & Oskinova, L. M. 2018, A&A, 620, A89
113. Neupert, W. M. 1968, ApJ, 153, L59
114. Nichols, J. S., Nazé, Y., Huenemoerder, D. P., et al. 2021, ApJ, 906, 89
115. Nielsen, L. D., Brahm, R., Bouchy, F., et al. 2020, A&A, 639, A76
116. Notsu, S., van Dishoeck, E. F., Walsh, C., Bosman, A. D., & Nomura, H. 2021, A&A, 650, A180
117. Noyes, R. W., Hartmann, L. W., Baliunas, S. L., Duncan, D. K., & Vaughan, A. H. 1984, ApJ, 279, 763
118. Öberg, K. I., Murray-Clay, R., & Bergin, E. A. 2011, ApJ, 743, L16
119. Oklopčić, A. 2019, ApJ, 881, 133
120. Oklopčić, A., & Hirata, C. M. 2018, ApJ, 855, L11
121. Oláh, K., Seli, B., Kővári, Z., Kriskovics, L., & Vida, K. 2022, A&A, 668, A101
122. Oskinova, L. M. 2005, MNRAS, 361, 679
123. —. 2016, Advances in Space Research, 58, 739
124. Oskinova, L. M., Sun, W., Evans, C. J., et al. 2013, ApJ, 765, 73
125. Owen, J. E., & Jackson, A. P. 2012, MNRAS, 425, 2931
126. Owen, J. E., & Lai, D. 2018, MNRAS, 479, 5012
127. Owen, J. E., & Wu, Y. 2017, ApJ, 847, 29
128. Parker, E. N. 1975, in Role of Magnetic Fields in Physics and Astrophysics, ed. V. Canuto, Vol. 257, 141
129. Parker, E. N. 1988, ApJ, 330, 474
130. Pauli, D., Oskinova, L. M., Hamann, W. R., et al. 2025, A&A, 697, A114
131. Peacock, S., Wilson, D. J., Richey-Yowell, T., et al. 2025, arXiv e-prints, arXiv:2509.08999
132. Pevtsov, A. A., Fisher, G. H., Acton, L. W., et al. 2003, ApJ, 598, 1387
133. Pillitteri, I., Argiroffi, C., Maggio, A., et al. 2022, A&A, 666, A198
134. Poppenhaeger, K., Schmitt, J. H. M. M., & Wolk, S. J. 2013, ApJ, 773, 62
135. Poppenhaeger, K., & Wolk, S. 2015, in IAU General Assembly, Vol. 29, 2245990
136. Poppenhaeger, K., & Wolk, S. J. 2014, A&A, 565, L1
137. Pravdo, S. H., & Tsuboi, Y. 2005, ApJ, 626, 272

138. Rauw, G. 2022, in Handbook of X-ray and Gamma-ray Astrophysics, ed. C. Bambi & A. Sangangelo, 108
139. Rauw, G., & Nazé, Y. 2016, Advances in Space Research, 58, 761, x-ray Emission from Hot Stars and their Winds
140. Rauw, G., Nazé, Y., Wright, N. J., et al. 2015, ApJS, 221, 1
141. Reale, F. 2007, A&A, 471, 271
142. —. 2014, Living Reviews in Solar Physics, 11, 4
143. Rempel, M. 2008, in Journal of Physics Conference Series, Vol. 118, Journal of Physics Conference Series (IOP), 012032
144. Reynolds, C., Kara, E., Mushotzky, R. F., & Ptak, A. 2023,
145. Ricker, G. R., Winn, J. N., Vanderspek, R., et al. 2015, Journal of Astronomical Telescopes, Instruments, and Systems, 1, 014003
146. Rockcliffe, K. E., Newton, E. R., Youngblood, A., et al. 2023, AJ, 166, 77
147. —. 2025, AJ, 169, 321
148. Safron, E. J., Fischer, W. J., Megeath, S. T., et al. 2015, ApJ, 800, L5
149. Sanz-Forcada, J., Micela, G., Ribas, I., et al. 2011, A&A, 532, A6
150. Schneider, P. C., Günther, H. M., & Schmitt, J. H. M. M. 2012, A&A, 542, A123
151. Schneider, P. C., & Schmitt, J. H. M. M. 2008, A&A, 488, L13
152. Schulz, N. S., Huenemoerder, D. P., Principe, D. A., et al. 2024, ApJ, 970, 190
153. Shkolnik, E. L., & Barman, T. S. 2014, AJ, 148, 64
154. Spake, J. J., Sing, D. K., Evans, T. M., et al. 2018, Nature, 557, 68
155. Spergel, D., Gehrels, N., Baltay, C., et al. 2015, arXiv e-prints, arXiv:1503.03757
156. Stern, R. A. 1998, in Astronomical Society of the Pacific Conference Series, Vol. 154, Cool Stars, Stellar Systems, and the Sun, ed. R. A. Donahue & J. A. Bookbinder, 223
157. Strugarek, A., Beaudoin, P., Charbonneau, P., & Brun, A. S. 2018, ApJ, 863, 35
158. Thao, P. C., Mann, A. W., Feinstein, A. D., et al. 2024, AJ, 168, 297
159. Thomas, J. D., Richardson, N. D., Eldridge, J. J., et al. 2021, Monthly Notices of the Royal Astronomical Society, 504, 5221
160. Tian, F., Toon, O. B., Pavlov, A. A., & De Sterck, H. 2005, ApJ, 621, 1049
161. Tobin, J. J., Sheehan, P. D., Megeath, S. T., et al. 2020, ApJ, 890, 130
162. Tristan, I. I., Notsu, Y., Kowalski, A. F., et al. 2023, ApJ, 951, 33
163. Trumpler, J. 1993, in Frontiers of Astronomy in 1990s. Proc. Workshop held in Beijing, 16
164. Tsuboi, Y., Koyama, K., Hamaguchi, K., et al. 2001, ApJ, 554, 734
165. Vanderburg, A., Johnson, J. A., Rappaport, S., et al. 2015, Nature, 526
166. Vanderburg, A., Rappaport, S. A., Xu, S., et al. 2020, Nature, 585, 363
167. Verner, D. A., & Yakovlev, D. G. 1995, A&AS, 109, 125
168. Vidal-Madjar, A., Lecavelier des Etangs, A., Désert, J. M., et al. 2003, Nature, 422, 143
169. Vidal-Madjar, A., Désert, J. M., Lecavelier des Etangs, A., et al. 2004, ApJ, 604, L69
170. Wang, L., & Dai, F. 2021, ApJ, 914, 98
171. Weisskopf, M. C., Tananbaum, H. D., Van Speybroeck, L. P., & O'Dell, S. L. 2000, in Society of Photo-Optical Instrumentation Engineers (SPIE) Conference Series, Vol. 4012, X-Ray Optics, Instruments, and Missions III, ed. J. E. Truemper & B. Aschenbach, 2
172. Wilms, J., Allen, A., & McCray, R. 2000, ApJ, 542, 914
173. Wolk, S. J., Lisse, C. M., Bodewits, D., Christian, D. J., & Dennerl, K. 2009, ApJ, 694, 1293
174. Wright, N. J., Newton, E. R., Williams, P. K. G., Drake, J. J., & Yadav, R. K. 2018, MNRAS, 479, 2351
175. Yelle, R., Lammer, H., & Ip, W.-H. 2008, Space Science Reviews, 139, 437
176. Yelle, R. V. 2004, Icarus, 170, 167
177. Youngblood, A., Bean, J., Behr, P., et al. 2024, in American Astronomical Society Meeting Abstracts, Vol. 243, American Astronomical Society Meeting Abstracts, 202.16
178. Zahnle, K. J., & Catling, D. C. 2017, ApJ, 843, 122
179. Zahnle, K. J., Catling, D. C., & Claire, M. W. 2013, Chemical Geology, 362, 26

180. Zhang, K., Schwarz, K. R., & Bergin, E. A. 2020, ApJ, 891, L17
181. Zuckerman, B., Koester, D., Reid, I. N., & Hunsch, M. 2003, The Astrophysical Journal, 596